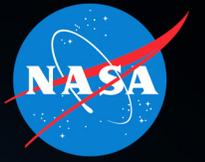

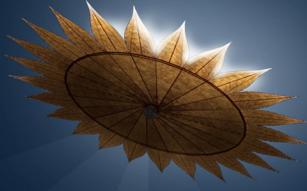

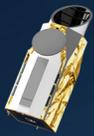

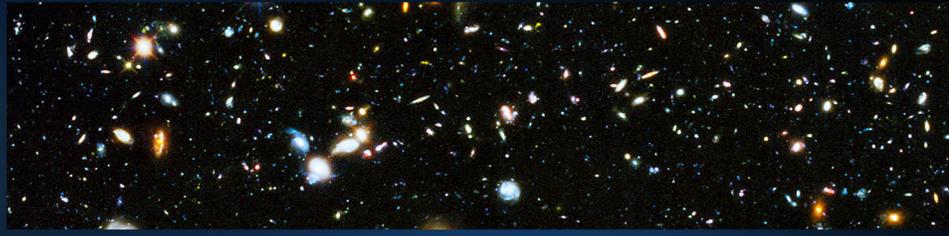

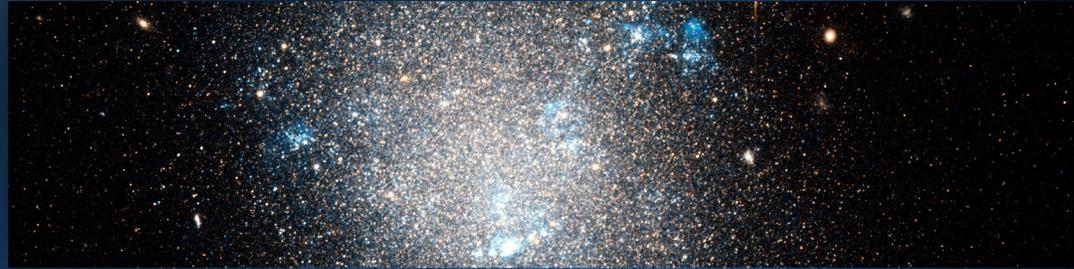

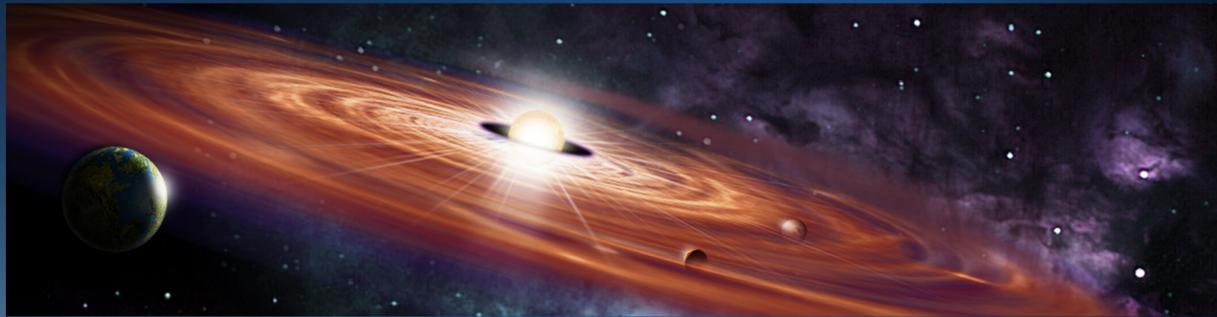

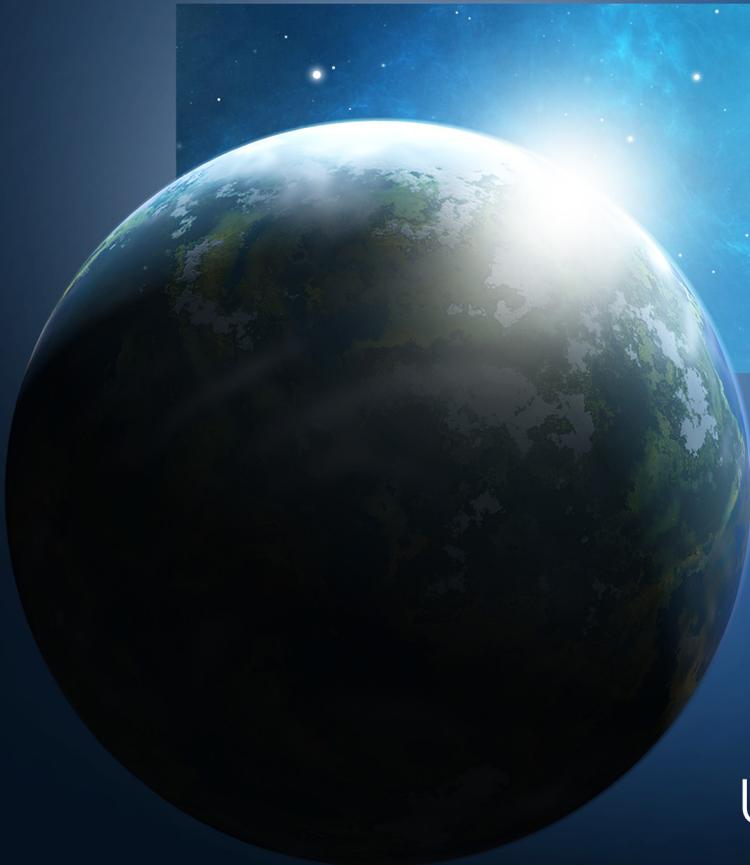

# HabEx

**Habitable Exoplanet Observatory**

Exploring New Worlds,
Understanding Our Universe

# HabEx

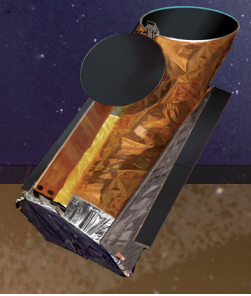

HABITABLE
EXOPLANET
OBSERVATORY

## EXPLORING NEW WORLDS – UNDERSTANDING OUR UNIVERSE

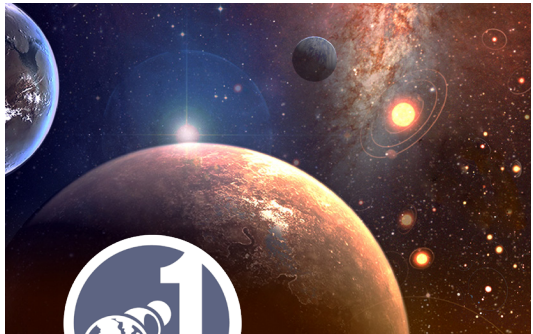
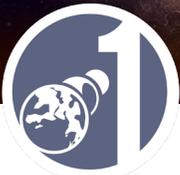

### GOAL 1

**To seek out nearby worlds and explore their habitability,**

*HabEx* will search for Habitable Zone Earth-like planets around sunlike stars using direct imaging and will spectrally characterize promising candidates for signs of habitability and life.

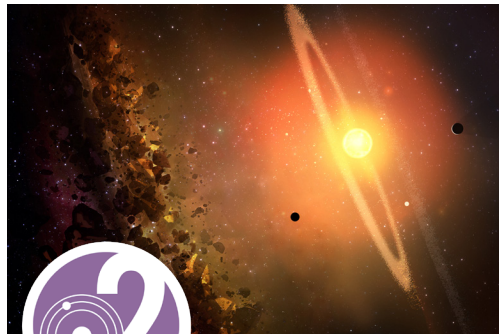
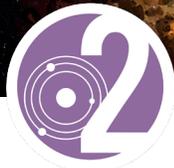

### GOAL 2

**To map out nearby planetary systems and understand the diversity of the worlds they contain,**

*HabEx* will take the first "family portraits" of nearby planetary systems, detecting and characterizing both inner and outer planets, as well as searching for dust and debris disks.

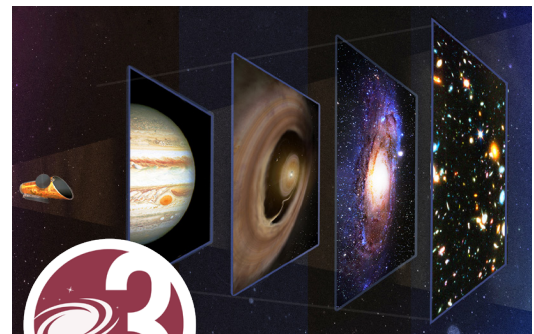
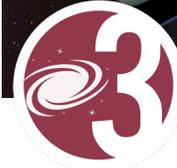

### GOAL 3

**To enable new explorations of astrophysical systems from our solar system to galaxies and the universe by extending our reach in the UV through near-IR,**

*HabEx* will have a community-driven Guest Observer program to undertake revolutionary science with a large-aperature, ultra-stable UV through near-IR space telescope.

---

*HabEx* is a great observatory for the 2030's, balancing exoplanet surveys with a community-driven, guest observer astrophysics and solar system science program.

*HabEx* is also highly efficient, with exoplanet observations occurring in parallel with deep field observations using the astrophysics instruments.

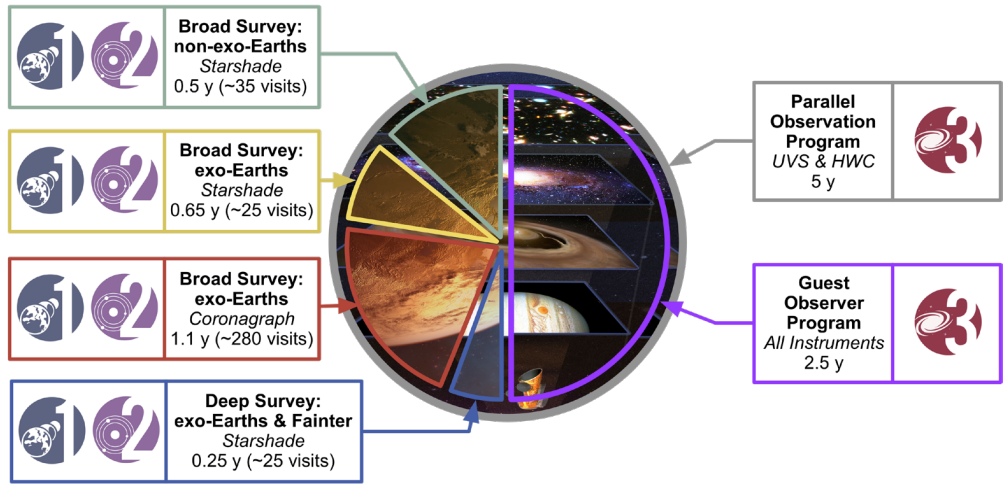

**Broad Survey: non-exo-Earths**
*Starshade*
0.5 y (~35 visits)

**Broad Survey: exo-Earths**
*Starshade*
0.65 y (~25 visits)

**Broad Survey: exo-Earths**
*Coronagraph*
1.1 y (~280 visits)

**Deep Survey: exo-Earths & Fainter**
*Starshade*
0.25 y (~25 visits)

**Parallel Observation Program**
*UVS & HWC*
5 y

**Guest Observer Program**
*All Instruments*
2.5 y

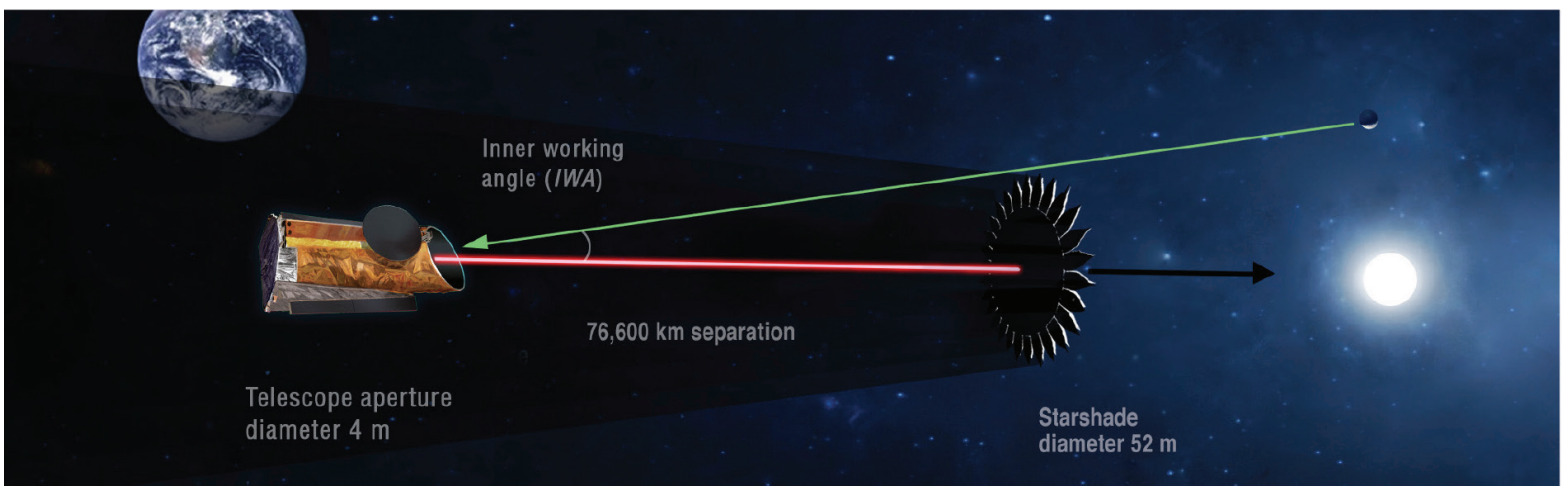

Inner working angle (*IWA*)

76,600 km separation

Telescope aperture diameter 4 m

Starshade diameter 52 m

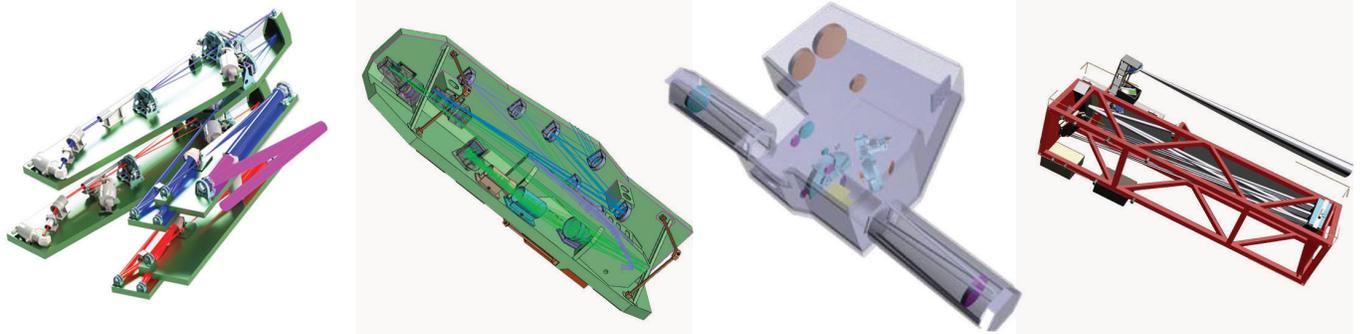

| | Coronagraph (HCG) | Starshade (SSI) | Workhorse Camera (HWC) | UV Spectograph (UVS) |
|---|---|---|---|---|
| **Purpose** | Exoplanet imaging and characterization | Exoplanet imaging and characterization | Multipurpose, wide-field imaging camera and spectrograph for observatory science | High-resolution, UV imaging and spectroscopy for observatory science |
| **Instrument Type** | Vector Vortex charge 6 coronagraph with:<br>- Raw contrast: 2.5 x $10^{-10}$ at the IWA<br>- $\Delta$ mag limit = 26.5<br>- 20% instantaneous bandwidth<br>- Imager and spectograph | 52 m diameter starshade occulter with:<br>- 76,600 km separation (Visible)<br>- Raw contrast: 1 x $10^{-10}$ at the IWA<br>- $\Delta$ mag limit = 26.5<br>- 107% instantaneous bandwidth<br>- Imager and spectograph | Imager and spectrograph | High-resolution imager and spectrograph |
| **Channels** | Visible: 0.45–0.975 μm<br>- Imager + IFS with $R$ = 140<br>Near-IR: 0.975–1.8 μm<br>- Imager + IFS with $R$ = 40 | UV: 0.2–0.45 μm<br>- Imager + grism with $R$ = 7<br>Visible: 0.45–0.975 μm<br>- Imager + IFS with $R$ = 140<br>Near-IR: 0.975–1.8 μm<br>- Imager + IFS with $R$ = 40 | Visible: 0.37–0.975 μm<br>- Imager + grism with $R$ = 1,000<br>Near-IR: 0.95–1.8 μm<br>- Imager + grism with $R$ = 1,000 | UV: 115–320 nm (with 115–370 nm available at $R \leq$ 1,000)<br>$R$ = 60,000; 25,000; 12,000; 6,000; 3,000; 1,000; 500; imaging |
| **Field of View** | IWA: 2.4 $\lambda/D$ = 62 mas at 0.5 μm<br>OWA: 32 $\lambda/D$ = 830 mas at 0.5 μm | IWA: 58 mas at 0.3–1.0 μm<br>OWA: 6 arcsec (Vis. broadband imaging)<br>OWA: 1 arcsec (Visible IFS) | 3 x 3 arcmin² | 3 x 3 arcmin² |
| **Features** | 64 x 64 deformable mirrors (2)<br>Low-order wavefront sensing and control | Formation flying, sensing, and control | Microshutter array for multi-object spectroscopy<br>- 2 x 2 array, 171 x 365 apertures | Microshutter array for multi-object spectroscopy<br>- 2 x 2 array, 171 x 365 apertures |

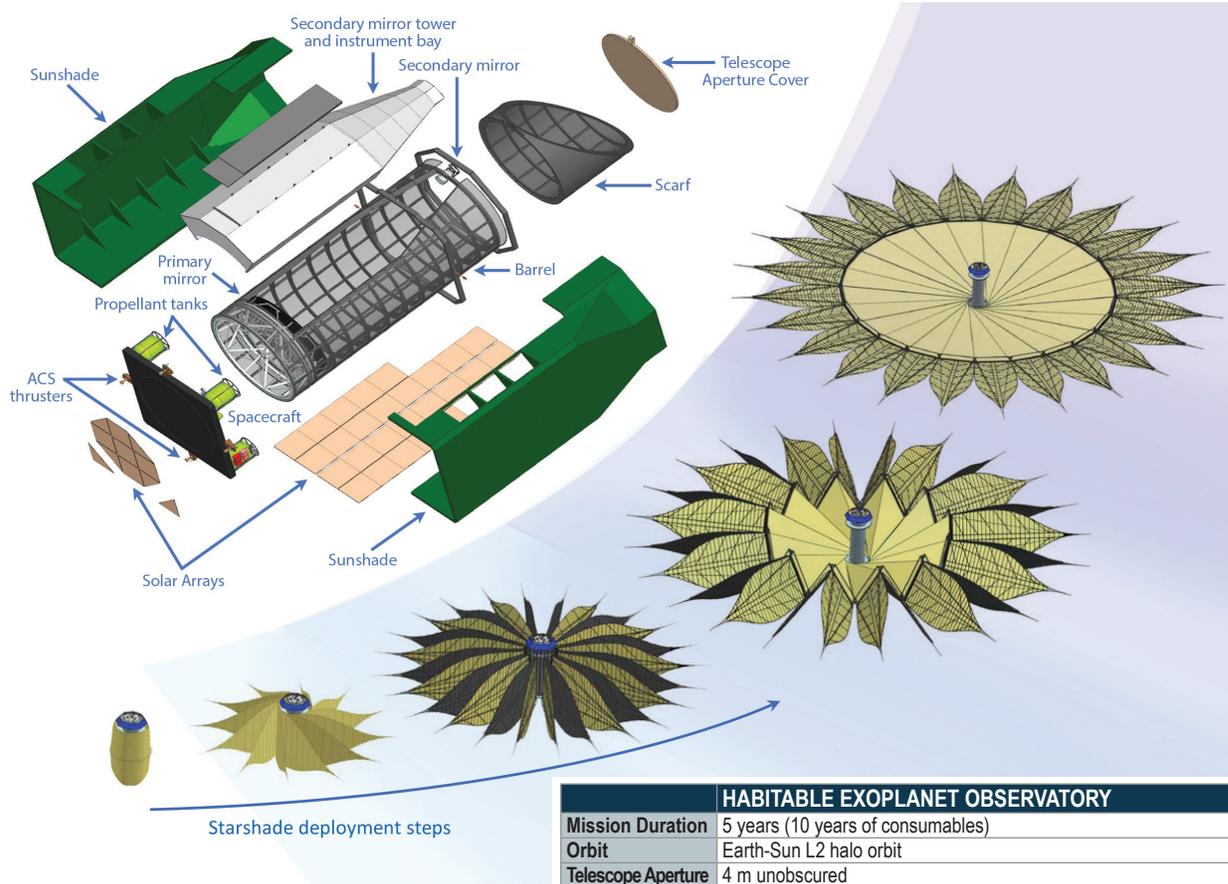

Starshade deployment steps

| HABITABLE EXOPLANET OBSERVATORY | |
|---|---|
| **Mission Duration** | 5 years (10 years of consumables) |
| **Orbit** | Earth-Sun L2 halo orbit |
| **Telescope Aperture** | 4 m unobscured |
| **Telescope Type** | Off-axis three-mirror anastigmat |
| **Primary Mirror** | Monolithic; glass-ceramic substrate; Al + $MgF_2$ coating |
| **Instruments (4)** | Exoplanet science: Coronagraph (HCG), Starshade (SSI)<br>Observatory science: UV spectrograph (UVS), Workhorse Camera (HWC) |
| **Attitude Control** | Slewing; hydrazine thrusters; Pointing: microthrusters |

# How to Read the HabEx Final Report

The HabEx Report captures the results of the HabEx Science and Technology Definition Team to assess the science discoveries, potential mission architectures, risks, and technology pathways to achieving HabEx's science goals. As a result, it is over 500 pages in length with the expectation of a wide audience. The below offers a high-level overview of the report's themes.

**A Summary of HabEx and its evaluated architectures** is included in the *Executive Summary*, *Chapter 1*, and *Chapter 13*. *Chapter 10* focuses on the tradespace underlying the nine mission architectures.

**HabEx's science** is detailed in *Chapters 3*, *4*, and *12*. *Appendix C* describes the estimation of one of HabEx's key science metric, exoplanet yield. *Appendix D* is the target list for HabEx's exoplanet survey.

**HabEx's science requirements** and the derivation of its implementation requirements appear in *Chapter 5*.

**The implementation of HabEx's baseline, preferred architecture** is detailed in *Chapters 6*, *7*, and *8*, which focus on the telescope flight system, starshade flight system, and general mission architecture, respectively. Alternative architectures are described in *Appendices A* and *B*.

**HabEx's technology** is detailed in *Chapter 11* and *Appendix E*. Industry technology white papers are included in *Appendix F*.

**HabEx's management plan, cost, risk, and schedule** is detailed in *Chapter 9*. *Appendix G* reports the result of an Independent Cost Estimate and Schedule Assessment study.

Additional sections include reader aids, like the Fact Sheet, Frequently Asked Questions, Glossary (*Appendix H*), Acronyms (*Appendix I*), and References (*Appendix J*).

# THE HABITABLE EXOPLANET OBSERVATORY: FREQUENTLY ASKED QUESTIONS

**1.    Is HabEx an exoplanet-only mission?**

No. HabEx is designed to be the Great Observatory of the 2030s, capable of a broad range of science serving the entire astrophysics community. As a successor to Hubble Space Telescope (HST), HabEx will have two observatory science instruments that will provide unique capabilities from the ultraviolet (UV) through the near-infrared (near-IR), enabling groundbreaking solar system, Galactic, and extragalactic science from the vantage of space. Observatory science will represent more than 50% of HabEx's prime 5-year mission and will replace and improve upon the capabilities that will be lost at the end of HST's life.

Importantly, HabEx is capable of parallel observations with its observatory science instruments during exoplanet observations, generating archival observations, such as deep fields, concurrently with long exoplanet observations to inspire and support generations of scientists to come. HabEx's use of thrusters, instead of reaction wheels, and its ultra-stable structure enable fast slews, 180° in less than five minutes, to support multi-messenger and target of opportunity science.

**2.    How does HabEx protect the budgets of smaller astrophysics missions, like Explorer- and Probe-class missions?**

HabEx is designed to fit into a NASA astrophysics funding profile that allows for investments in less costly missions and in individual investigator grants. Initiating the funding for HabEx development once the Wide Field Infrared Survey Telescope (WFIRST) development has ramped down allows for two Probe-class (~$1B) missions over the 10-year development time of HabEx, maintaining the pace of two Explorer missions per decade, protecting individual investigator grants, and other existing astrophysics commitments.

Additionally, the NASA astrophysics budget is sized to include a Flagship mission. Without a Flagship on the program of record, the overall NASA astrophysics budget may shrink.

**3.    Is high-contrast coronagraphy compatible with UV-capable telescopes?**

Yes. Maintaining the capability of Guest Observer UV science beyond HST was a priority for HabEx. Serendipitously, detailed optical polarization studies have demonstrated that the aluminum-coated mirrors, required for UV observations, actually provide better coronagraphic performance than silver-coated mirrors.

**4.    How is HabEx's observational efficiency 90%?**

There are several unique aspects to HabEx that enable it to observe at very high efficiency. First, HabEx's use of thrusters and ultra-stable structure permits fast slew and settle times. Second, by using phased array antennas HabEx can downlink science data while performing observations. Third, by locating HabEx in an Earth-Sun L2 orbit, there are small keepout zones and no eclipses to restrict field of regard and operations. Finally, long observing times for exoplanetary science—within its allocated 50% of the prime mission—boost HabEx's total observational efficiency.

**5.   Is HabEx's 4 meter aperture large enough to directly image and characterize Earth analogs?**

Yes. It is now known from NASA's Kepler that small planets are common and the frequency of rocky planets in the habitable zone is roughly 25%. Also, results from the recent Large Binocular Telescope Interferometer (LBTI) exozodiacal light survey indicate relatively small amounts of habitable zone dust around nearby sunlike stars.

Folding in these uncertainties, using 50% of the prime 5-year mission, the baseline HabEx Observatory has a 98.6% chance of detecting and characterizing at least one rocky planet in the habitable zone of a sunlike star.

**6.   How will HabEx identify potentially habitable planets if their spectra look different from our own planet?**

HabEx will empirically define the "habitable zone," rather than relying on a preconceived notion of it. HabEx can discriminate between Modern, Proterozoic, and Archean Earth-like atmospheres. Furthermore, because HabEx is able to get spectra from 0.2 to 1.8 microns at a reasonable signal-to-noise ratio and resolution, HabEx is capable of detecting any molecular species that would be abundant in a wide variety of environments, including both those that are habitable or uninhabitable.

**7.   Does the HabEx Observatory have two spacecraft?**

Yes, the baseline HabEx observatory consists of telescope and starshade spacecraft. The baseline telescope spacecraft carries a 4-meter telescope and four astrophysics instruments, including an internal coronagraph, as its payload. The telescope flies in formation with the starshade spacecraft, which is used as an external occulter in the line-of-sight with a starshade instrument on the telescope. The internal coronagraph and external starshade occulter act as HabEx's two starlight suppression technologies.

**8.   Why does the baseline HabEx Observatory use two starlight suppression technologies, a starshade and a coronagraph?**

As described in the report, starshades and coronagraphs are exceptionally complementary. The coronagraph is nimble and ideal for exo-Earth blind searches and exoplanet orbit determination, while the starshade is ideal for wide-field mapping of planetary systems and exoplanet spectral characterization.

**9.   What has been accomplished to show that starshades can be used for observing exoplanets?**

Due to focused investments by NASA over the last 5 years, starshade technology has matured rapidly. In particular, the Starshade to TRL 5 (S5) Technology Development Plan led by NASA's Exoplanet Exploration Program Office (ExEP) has matured all starshade-related technologies to TRL 4 or higher today. HabEx is able to leverage the work being done by S5, which is expected to continue maturing the relevant technologies to TRL 5 three years in advance of the start of the baseline HabEx starshade flight system development.

**10.  How does HabEx align the telescope and starshade?**

Several missions, including NASA's Gravity Recovery and Climate Experiment Follow-On (GRACE-FO), have already successfully demonstrated precision formation flying (albeit at smaller separations). An S-band radio system on each HabEx spacecraft and a set of laser beacons allow the two to share position knowledge. This is used in a control algorithm that takes into account information from both vehicles to make attitude control decisions. Through NASA's S5 efforts, this formation flying technology has been advanced and is currently at Technology Readiness Level 5 (TRL 5).

## 11. How does HabEx's coronagraph achieve ~$10^{-10}$ instrument contrast to observe exo-Earths?

HabEx made four major design decisions to achieve the contrast and wavefront stability required to observe exo-Earths.

1. HabEx is designed to change pointing using thrusters instead of reaction wheels, which are a major source of jitter in spacecraft. The HabEx microthruster approach has heritage in ESA's Gaia and Laser Interferometer Space Antenna (LISA) Pathfinder missions, where the specific technology being baselined by HabEx was flown in the latter.
2. The HabEx observatory baselined a massive, monolithic primary mirror, which provides extreme thermal stability.
3. All residual telescope mirror motion is monitored and adjusted for using HabEx's laser metrology system.
4. HabEx's internal coronagraph is a vector vortex charge 6 coronagraph, which is much less sensitive than other coronagraph masks to common low-order telescope aberrations arising from environmental disturbances.

As a result of these strategies, HabEx meets its wavefront stability requirement with 106% margin.

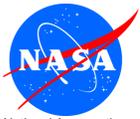

National Aeronautics and
Space Administration

**Jet Propulsion Laboratory**
California Institute of Technology
Pasadena, California

# HabEx

## Habitable Exoplanet Observatory

## Final Report

Jet Propulsion Laboratory
for
Astrophysics Division
Science Mission Directorate
NASA

August 23, 2019

## The Habitable Exoplanet Observatory Study Team

| STDT Community Chairs | |
|---|---|
| Scott Gaudi, Ohio State University | |
| Sara Seager, Massachusetts Institute of Technology | |

| Study Scientist | |
|---|---|
| Bertrand Mennesson, NASA Jet Propulsion Laboratory | |

| Deputy Study Scientist | |
|---|---|
| Alina Kiessling, NASA Jet Propulsion Laboratory | |

| Study Manager | |
|---|---|
| Keith Warfield, NASA Jet Propulsion Laboratory | |

| Science and Technology Definition Team Members | |
|---|---|
| Kerri Cahoy, Massachusetts Institute of Technology | Tyler Robinson, Northern Arizona University |
| John T. Clarke, Boston University | Leslie Rogers, University of Chicago |
| Shawn Domagal-Goldman, NASA Goddard Space Flight Center | Paul Scowen, Arizona State University |
| Lee Feinberg, NASA Goddard Space Flight Center | Rachel Somerville, Rutgers University |
| Olivier Guyon, University of Arizona | Karl Stapelfeldt, NASA Jet Propulsion Laboratory |
| Jeremy Kasdin, Princeton University | Christopher Stark, Space Telescope Science Institute |
| Dimitri Mawet, California Institute of Technology | Daniel Stern, NASA Jet Propulsion Laboratory |
| Peter Plavchan, George Mason University | Margaret Turnbull, SETI Institute |

| Design Team Members | |
|---|---|
| Rashied Amini, Report Manager, NASA Jet Propulsion Laboratory | Mary Li, NASA Goddard Space Flight Center |
| Gary Kuan, Design Team Lead, NASA Jet Propulsion Laboratory | Doug Lisman, NASA Jet Propulsion Laboratory |
| Stefan Martin, Payload Lead, NASA Jet Propulsion Laboratory | Milan Mandic, NASA Jet Propulsion Laboratory |
| Rhonda Morgan, Lead Technologist, NASA Jet Propulsion Laboratory | John Mann, NASA Jet Propulsion Laboratory |
| David Redding, HabEx Starshade-Only Architectures Lead, NASA Jet Propulsion Laboratory | Luis Marchen, NASA Jet Propulsion Laboratory |
| H. Philip Stahl, Telescope Lead, NASA Marshall Space Flight Center | Colleen Marrese-Reading, NASA Jet Propulsion Laboratory |
| Ryan Webb, Flight Systems Lead, NASA Jet Propulsion Laboratory | Jonathan McCready, North Carolina State University |
| Oscar Alvarez-Salazar, NASA Jet Propulsion Laboratory | Jim McGown, NASA Jet Propulsion Laboratory |
| William L. Arnold, a.i. solutions | Jessica Missun, Ball Aerospace |
| Manan Arya, NASA Jet Propulsion Laboratory | Andrew Miyaguchi, Northrop Grumman Corporation |
| Bala Balasubramanian, NASA Jet Propulsion Laboratory | Bradley Moore, NASA Jet Propulsion Laboratory |
| Mike Baysinger, Jacobs | Bijan Nemati, University of Alabama |
| Ray Bell, Lockheed Martin | Shouleh Nikzad, NASA Jet Propulsion Laboratory |
| Chris Below, L3 Harris | Joel Nissen, NASA Jet Propulsion Laboratory |
| Jonathan Benson, Ball Aerospace | Megan Novicki, Northrop Grumman Corporation |
| Lindsey Blais, NASA Jet Propulsion Laboratory | Todd Perrine, NASA Jet Propulsion Laboratory |
| Jeff Booth, NASA Jet Propulsion Laboratory | Claudia Pineda, NASA Jet Propulsion Laboratory |
| Robert Bourgeois, Collins | Otto Polanco, NASA Jet Propulsion Laboratory |
| Case Bradford, NASA Jet Propulsion Laboratory | Dustin Putnam, Ball Aerospace |
| Alden Brewer, L3 Harris | Atif Qureshi, Maxar |

| Design Team Members, continued | |
|---|---|
| Thomas Brooks, NASA Marshall Space Flight Center | Michael Richards, Northrop Grumman Corporation |
| Eric Cady, NASA Jet Propulsion Laboratory | A.J. Eldorado Riggs, NASA Jet Propulsion Laboratory |
| Mary Caldwell, Jacobs | Michael Rodgers, Synopsys |
| Rob Calvet, NASA Jet Propulsion Laboratory | Mike Rud, NASA Jet Propulsion Laboratory |
| Steven Carr, Colins | Navtej Saini, NASA Jet Propulsion Laboratory |
| Derek Chan, Ball Aerospace | Dan Scalisi, L3 Harris |
| Velibor Cormarkovic, NASA Jet Propulsion Laboratory | Dan Scharf, NASA Jet Propulsion Laboratory |
| Keith Coste, NASA Jet Propulsion Laboratory | Kevin Schulz, NASA Jet Propulsion Laboratory |
| Charlie Cox, United Technologies Aerospace Systems | Gene Serabyn, NASA Jet Propulsion Laboratory |
| Rolf Danner, NASA Jet Propulsion Laboratory | Norbert Sigrist, NASA Jet Propulsion Laboratory |
| Jacqueline Davis, NASA Marshall Space Flight Center | Glory Sikkia, Maxar |
| Larry Dewell, Lockheed Martin | Andrew Singleton, Jacobs |
| Lisa Dorsett, ARCS | Stuart Shaklan, NASA Jet Propulsion Laboratory |
| Daniel Dunn, Collins | Scott Smith, NASA Marshall Space Flight Center |
| Matthew East, L3-Harris | Bart Southerd, Collins |
| Michael Effinger, NASA Marshall Space Flight Center | Mark Stahl, NASA Marshall Space Flight Center |
| Ron Eng, NASA Marshall Space Flight Center | John Steeves, NASA Jet Propulsion Laboratory |
| Greg Freebury, Tendeg, LLC | Brian Sturges, ARCS |
| Jay Garcia, Jacobs | Chris Sullivan, L3 Harris |
| Jonathan Gaskin, University of North Carolina | Hao Tang, University of Michigan |
| Suzan Greene, Ball Aerospace | Neil Taras, L3 Harris |
| John Hennessy, NASA Jet Propulsion Laboratory | Jonathan Tesch, NASA Jet Propulsion Laboratory |
| Evan Hilgemann, NASA Jet Propulsion Laboratory | Melissa Therrell, Jacobs |
| Brad Hood, Ball Aerospace | Howard Tseng, NASA Jet Propulsion Laboratory |
| Wolfgang Holota, Holota Optics | Marty Valente, Arizona Optical Systems |
| Scott Howe, NASA Jet Propulsion Laboratory | David Van Buren, NASA Jet Propulsion Laboratory |
| Pei Huang, Ball Aerospace | Juan Villalvazo, NASA Jet Propulsion Laboratory |
| Tony Hull, University of New Mexico | Steve Warwick, Northrop Grumman Corporation |
| Ron Hunt, Jacobs | David Webb, NASA Jet Propulsion Laboratory |
| Kevin Hurd, NASA Jet Propulsion Laboratory | Thomas Westerhoff, SCHOTT |
| Sandra Johnson, Ball Aerospace | Rush Wofford, Northrop Grumman Corporation |
| Andrew Kissil, NASA Jet Propulsion Laboratory | Gordon Wu, Ball Aerospace |
| Brent Knight, NASA Marshall Space Flight Center | Jahning Woo, NASA Jet Propulsion Laboratory |
| Daniel Kolenz, NASA Jet Propulsion Laboratory | Milana Wood, NASA Jet Propulsion Laboratory |
| Oliver Kraus, MPIA | John Ziemer, NASA Jet Propulsion Laboratory |
| John Krist, NASA Jet Propulsion Laboratory | |
| Additional Contributing Scientists | |
| Giada Arney, NASA Goddard Space Flight Center | Lucas Macri, Texas A&M University |
| Jay Anderson, Space Telescope Science Institute | Mark Marley, NASA Ames Center |
| Jesús Maíz-Apellániz, Centro de Astrobiología, Spain | William Matzko, George Mason University |
| James Bartlett, NASA Jet Propulsion Laboratory | Johan Mazoyer, NASA Jet Propulsion Laboratory |
| Ruslan Belikov, NASA Ames Center | Stephan McCandliss, Johns Hopkins University |
| Eduardo Bendek, NASA Jet Propulsion Laboratory | Tiffany Meshkat, California Institute of Technology/IPAC |
| Brad Cenko, NASA Goddard Space Flight Center | Christoph Mordasini, University of Bern, Switzerland |

| Additional Contributing Scientists, continued | |
|---|---|
| Ewan Douglas, Massachusetts Institute of Technology | Patrick Morris, Caltech/IPAC |
| Shannon Dulz, University of Notre Dame | Eric Nielsen, Stanford University |
| Chris Evans, UKATC, Royal Observatory of Edinburgh, UK | Patrick Newman, George Mason University |
| Virginie Faramaz, NASA Jet Propulsion Laboratory | Erik Petigura, Caltech |
| Y. Katherina Feng, University of California, Santa Cruz | Marc Postman, Space Telescope Science Institute |
| Harry Ferguson, Space Telescope Science Institute | Amy Reines, University of Montana |
| Kate Follette, Amherst College | Aki Roberge, NASA Goddard Space Flight Center |
| Saavik Ford, CUNY/American Museum of Natural History | Ian Roederer, University of Michigan |
| Miriam García, Centro de Astrobiología, CSIC-INTA, Spain | Garreth Ruane, NASA Jet Propulsion Laboratory |
| Marla Geha, Yale University | Edouard Schwieterman, University of California, Riverside |
| Dawn Gelino, Caltech/NExScI | Dan Sirbu, Bay Area Environmental Research Institute |
| Ylva Götberg, Carnegie Observatories | Christopher Spalding, Yale University |
| Sergi Hildebrandt, NASA Jet Propulsion Laboratory | Harry Teplitz, Caltech/IPAC |
| Renyu Hu, NASA Jet Propulsion Laboratory | Jason Tumlinson, STScI |
| Knud Jahnke, Max Planck Institute for Astrophysics, Germany | Neal Turner, NASA Jet Propulsion Laboratory |
| Grant Kennedy, University of Warwick, UK | Jessica Werk, University of Washington |
| Laura Kreidberg, Harvard & Smithsonian Center for Astrophysics | Aida Wofford, Space Telescope Science Institute |
| Andrea Isella, Rice University | Mark Wyatt, University of Cambridge, UK |
| Eric Lopez, NASA Goddard Space Flight Center | Amber Young, Northern Arizona University |
| Franck Marchis, SETI Institute | Rob Zellem, NASA Jet Propulsion Laboratory |
| Report Production Team | |
| Samantha Ozyildirim, Lead Documentarian, NASA Jet Propulsion Laboratory | Joseph Harris, NASA Jet Propulsion Laboratory |
| Laura Generosa, Raytheon | David Levine, NASA Jet Propulsion Laboratory |
| Grace Ok, NASA Jet Propulsion Laboratory | Randall Oliver, Raytheon |
| Ex-Officio Non-Voting Members | |
| Martin Still, NASA Headquarters | |
| Douglas Hudgins, NASA Headquarters | |
| International Ex-Officio Non-Voting Members | |
| Christian Marois, NRC Canada (Canadian Space Agency, CSA, Observer) | |
| David Mouillet, IPAG Grenoble (Centre National d'Etudes Spatiales, CNES, Observer) | |
| Timo Prusti, ESA (European Space Agency, ESA, Observer) | |
| Andreas Quirrenbach, Heidelberg University (Deutschen Zentrums für Luft- und Raumfahrt, DLR, Observer) | |
| Motohide Tamura, University of Tokyo (Japanese Aerospace Exploration Agency, JAXA, Observer) | |
| Pieter de Visser, Netherlands Institute for Space Research (SRON) | |

## Acknowledgments

The HabEx Study Team would like to thank the following for their efforts in supporting the development of the HabEx Report:

- Technical reviewers, for their dedication and feedback: Randall Bartman, David Bearden, Charles Beichman, Ruslan Belikov, Andrew Coffey, Alyssa Deardroff, Alan Didion, Benji Donitz, Aigneis Frey, Anthony Freeman, Todd Gaier, Gregory Garner, Paul Graf, Ravi Kumar Kopparapu, Amanda la Venture, Roger Lee, Elizabeth Luthman, Charley Noecker, Jane Rigby, Hari Subedi, Mark Swain, Neil Zimmerman, and Steven Zusack.

- Members of the Independent Cost Estimate and Independent Schedule Assessment Team, for their independent assessment: Fred Doumani, Erica Beam, and Jerry Shen

- Everyone involved in unlimited release review, for ability to meet our schedule needs: Alex Abramovici, Amanda Beckett, Douglas Isbell, Peter Kahn, Young Lee, Kathleen Lynn, Sunjay Moorthay, and Danette Zuniga.

- Aki Roberge (NASA Goddard Space Flight Center) for her unflagging efforts to promote cooperation and collaboration between the LUVOIR and HabEx study teams.

## Disclaimer

Pre-Decisional Information – For Planning and Discussion Purposes Only

This research was carried out at the Jet Propulsion Laboratory, California Institute of Technology, under a contract with the National Aeronautics and Space Administration.

The cost information contained in this document is of a budgetary and planning nature and is intended for informational purposes only. It does not constitute a commitment on the part of JPL and Caltech.

# Table of Contents

**NOTE:** For reader convenience, links within this proposal are provided for certain content (i.e., TOC entries and cross references). After following a link, you can return to the page you were previously viewing by pressing **Command + Left Arrow (Mac)** or **Alt + Left Arrow (PC)**, or by using the Acrobat Page Navigation toolbar.

## Appendices

# EXECUTIVE SUMMARY

For the first time in human history, technologies have matured sufficiently to enable an affordable space-based telescope mission capable of discovering and characterizing habitable planets like the Earth orbiting nearby bright sunlike stars. Such an observatory can be equipped with instruments that provide a wide range of capabilities, enabling unique science not possible from ground-based facilities. This science is broad and exciting, ranging from new investigations of our own solar system to understanding the life cycle of baryons and its impact on the formation and evolution of galaxies, to addressing fundamental puzzles in cosmology.

The Habitable Exoplanet Observatory, or HabEx, has been designed to be the Great Observatory of the 2030s, a successor to the Hubble Space Telescope (HST) with enhanced capabilities and community involvement through a competed and funded Guest Observer (GO) program. This GO program—which shall represent 50% of HabEx's prime 5-year mission—will include competed novel observations, parallel and serendipitous observations, and archival research. After HabEx's 5-year prime mission, HabEx is capable of undertaking an extended mission of at least five additional years without servicing, during which the GO program would represent 100% of observing time.

HabEx is a space-based 4 m diameter telescope with ultraviolet (UV), optical, and near-infrared (near-IR) imaging and spectroscopic capabilities, replacing and enhancing those lost at the end of HST's lifetime. During its 5-year prime mission, HabEx has three driving science goals described in **Table ES-1** and **Figure ES-1**.

**Table ES-1.** HabEx Science Goals encompass direct detection and characterization of exoplanets and their systems along with wider-reaching questions about the nature of our universe.

| HabEx Science Goals | |
|---|---|
| 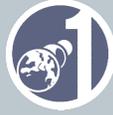 | *To seek out nearby worlds and explore their habitability* |
| 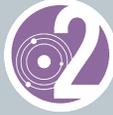 | *To map out nearby planetary systems and understand the diversity of the worlds they contain* |
| 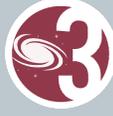 | *To enable new explorations of astrophysical systems from the solar system to galaxies and the universe by extending our reach in the UV through near-IR* |

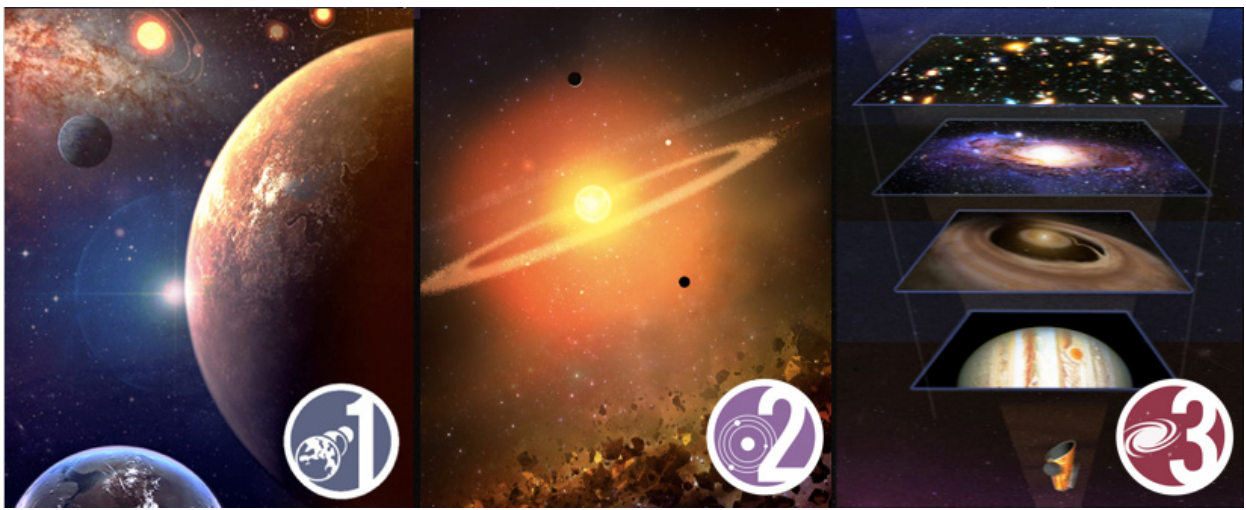

**Figure ES-1.** During its 5-year prime mission, the HabEx Observatory will divide equally its observing time between exoplanetary science and competed, guest observer-directed observations. Because all HabEx instruments can observe simultaneously, a Parallel Observing Program will generate archival observations, such as deep fields, to inspire and support generations of scientists to come. With its large aperture, stiff structure, and rapid slews, HabEx will observe at ~90% efficiency.

## HabEx Science

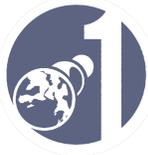

**HabEx will seek out nearby worlds and explore their habitability.** A pervasive and fundamental human question is: Are we alone? Astronomy has recast this elemental inquiry into a series of questions: Are there other Earths? Are they common? Do any have signs of life? Space-based direct imaging above the blurring effects of our atmosphere is the only way to discover and study exo-Earths candidates (EECs) in reflected light, e.g., the only way to detect and take spectra of Earth-sized planets in Earth-like orbits in reflected light (at near-UV, optical, and near-IR wavelengths) about sunlike (F, G, and K-type) stars.

With unparalleled high-contrast direct imaging capabilities, HabEx will spectrally characterize dozens of rocky worlds, including validating (obtaining orbits and 0.3–1 μm spectra) about 8 EECs. It will also detect and characterize over a hundred larger planets around mature stars (**Figure ES-2**). With around six visits to each system, using the coronagraph, HabEx will measure the orbits of EECs to a few percent in both the semi-major axis and inclination, and will determine the eccentricity to

~0.02 (**Figure 3.1-2**), allowing a robust determination of whether or not the EEC resides in the theoretical habitable zone (HZ). Of particular interest for investigations of EECs, HabEx will be sensitive to Rayleigh scattering, water vapor ($H_2O$), molecular oxygen ($O_2$), and ozone ($O_3$). It will detect all three gases down to column densities as low as 1% of modern Earth levels. In addition, HabEx will detect other atmospheric gases for context, such as methane and carbon dioxide, determining if they have concentrations higher than modern Earth. For our nearest neighbors, HabEx will also search for evidence of surface liquid water oceans on exo-Earth candidates by searching for specular reflection, or glint (*Section 3.1.4*).

In order to set the requirements needed to meet HabEx's first goal of seeking out nearby worlds and exploring the habitability, four specific objectives are defined in *Chapter 3*.

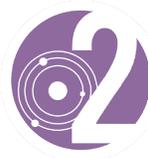

**HabEx will map out nearby planetary systems and understand the diversity of the worlds they contain.** With high-contrast $12 \times 12$ arcsec$^2$ (equivalent to $36 \times 36$ AU$^2$ at a distance of 3 pc) observations using the starshade, HabEx will be the first observatory capable of providing nearly complete "family portraits" of our nearest neighbors. HabEx will characterize full individual planetary systems, including exoplanet analogs to Earth, Saturn and Jupiter, and analogs to the zodiacal and Kuiper dust belts. For many of these planets, HabEx will not only obtain multi-epoch broadband spectra from 0.3–1.0 μm (and in some cases from 0.2–1.8 μm), but for those with periods of <10 years, HabEx will also determine the orbital parameters with a typical precision of <5% on the inclination, 25% on the semi-major axis, and measure the eccentricity to an uncertainty of 0.1. HabEx is also expected to find and spectrally characterize a diversity of planetary systems that bear little resemblance to our system, including those with worlds that have no analogs in our solar system, but are known to be common in other planetary systems, including super-Earths and sub-Neptunes (**Figure ES-3**). In general, given that HabEx's requirements are set by the characterization of EECs, the spectra of all planets that are brighter than an EEC will have much higher signal-to-noise ratio (**Figure ES-3**).

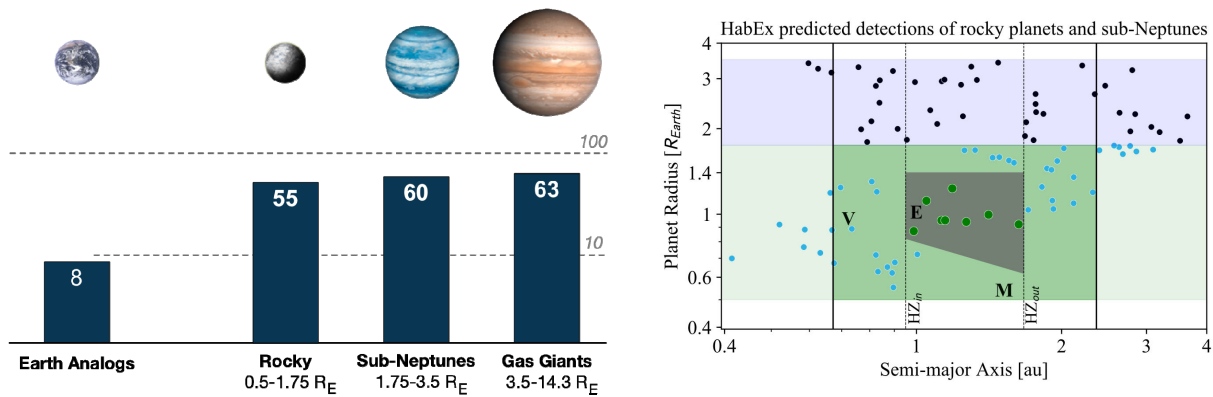

**Figure ES-2.** *Left panel (note log scale):* Under the design reference mission and survey strategy, assuming nominal occurrence rates (see *Section 8.2* for details on the DRM), HabEx will detect over 150 exoplanets with a diversity of sizes and temperatures (*Section 3.2.2*). It will obtain broad spectra (from at least from 0.3–1 μm) of the majority of these planets, including ~37 rocky planets proximate to the HZ (*Right panel*, dark green region), and will get orbits and spectra of ~8 planets with radii and separations consistent with the adopted conservative definition of exo-Earth candidates (EECs; *Right panel:* grey shaded region). HabEx will empirically constrain the HZ region concept, both in terms of planet separations and radii. *Right panel*: HabEx small planet characterization space close to the HZ. Note: in the right panel, the semi-major axis boundaries are for a solar twin. For other host stars, they have been scaled to maintain a constant bolometric insolation.

Discoveries of nearby planetary systems will provide detailed architectures, addressing open topics ranging from planetary system formation, planetary migration, and to the role of gas giants in the delivery of water to inner system rocky worlds. HabEx will test theories on planetary diversity, investigate planet-disk interactions, and place our solar system into detailed context for the first time.

As with HabEx's first goal, four specific objectives are also defined in *Chapter 3*.

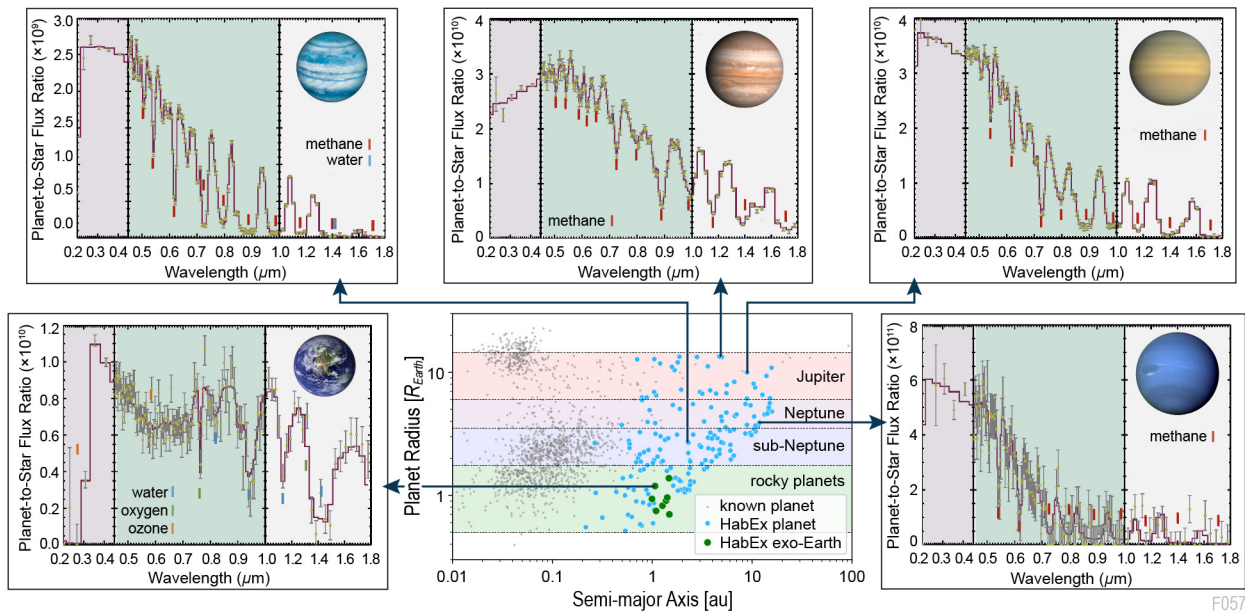

**Figure ES-3.** HabEx will discover and characterize over 150 new exoplanets (cyan points), from small exo-Earths candidates (green points) to gas giants, populating previously unexplored regions of parameter space, including planets that have no analogues in our solar system. The majority of these worlds will be well-characterized, with relatively high signal-to-noise ratio (SNR ≥ 10), resolution $R \sim 140$ spectra from 0.45–1.0 μm. By moving the starshade further away from telescope, HabEx will obtain crude spectra with resolution $R \sim 7$ in the UV channel from 0.20–0.45 μm, and by moving the starshade closer from the telescope, HabEx will obtain spectra with resolution $R \sim 40$ in the near-IR channel from 0.95–1.8 microns. Thus, HabEx can obtain complete spectral coverage of most of the planets discovered from 0.2–1.8 μm. Using multiple visits with the coronagraph, HabEx will also obtain reasonably precise orbits (inclinations measured to <5%, semi-major axes to 25%, eccentricities with an uncertainty of 0.1) for those with periods of less than roughly 10 years and eccentricities less than roughly 0.3 within the nominal 5-year mission lifetime. These orbital uncertainties improve significantly for shorter period orbits, particularly those with periods less than the mission lifetime. Thus, HabEx can obtain "family portraits" of a diversity of worlds in nearby planetary systems.

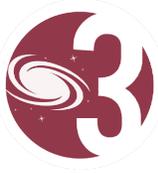 **HabEx will enable new explorations of astrophysical systems from our own solar system to galaxies and the universe by extending our reach in the UV through near-IR.** HabEx will be NASA's Great Observatory in the 2030s. Observing with a large aperture from above the Earth's atmosphere in an era when neither the Hubble Space Telescope (HST) nor the James Webb Space Telescope (JWST) are operational, HabEx will provide the highest-resolution images yet obtained at UV and optical wavelengths. HabEx will also provide an ultra-stable platform and access to wavelengths inaccessible from the ground.

These capabilities allow for a broad suite of unique, compelling science that cuts across the entire NASA astrophysics portfolio and includes topics as diverse as the life cycle of baryons (**Figure ES-4**), the sources of the metagalactic ionizing background, the origins of the elements from the first generations of stars and supernovae, the local expansion rate of the universe, dark matter models, the formation of Galactic globular clusters, the atmospheres of transiting exoplanets via transit spectroscopy, interactions between the Sun and the giant planets in our solar system (**Figure ES-5**), and the structure of protoplanetary transition disks.

Of course, knowing which of the scientific questions that motivate HabEx's GO program will still be relevant in the 2030s, is not possible.

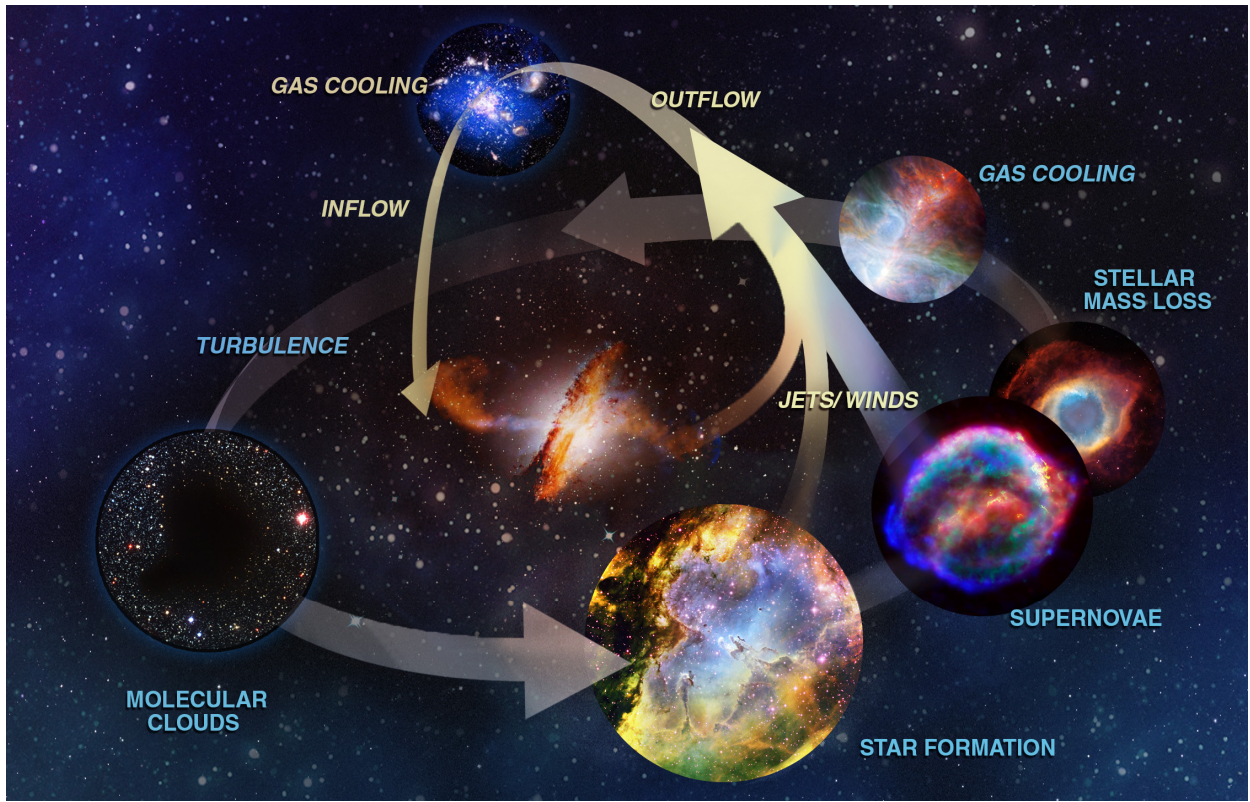

**Figure ES-4.** With its multiplexing capability via microshutter arrays, high spectral (up to $R = 60,000$) and spatial resolution (0.025″) in the UV (115–320 nm), and its order-of-magnitude larger effective collecting area in the 150–300 nm range relative to HST, HabEx will revolutionize our understanding of the life cycle of baryons. HabEx will probe the properties and structure of the intergalactic and circumgalactic medium, including both galactic outflows and inflows. It will determine the sources of the metagalactic ionizing background, probe the origin of the elements created in the first stars and supernovae, and probe the structure of protoplanetary transitions disks. Thus, HabEx will study the complete cycle of the inflow of pristine gas into galaxies, the incorporation of that gas into stars, the explosions of these stars as SNe, which pollute the intergalactic medium with many of the elements of life, as well as drive galactic outflows. Finally, it will complete the cycle by studying how those outflows subsequently cool and are re-incorporated into stars and their protoplanetary disks, ultimately forming planets, and perhaps life.

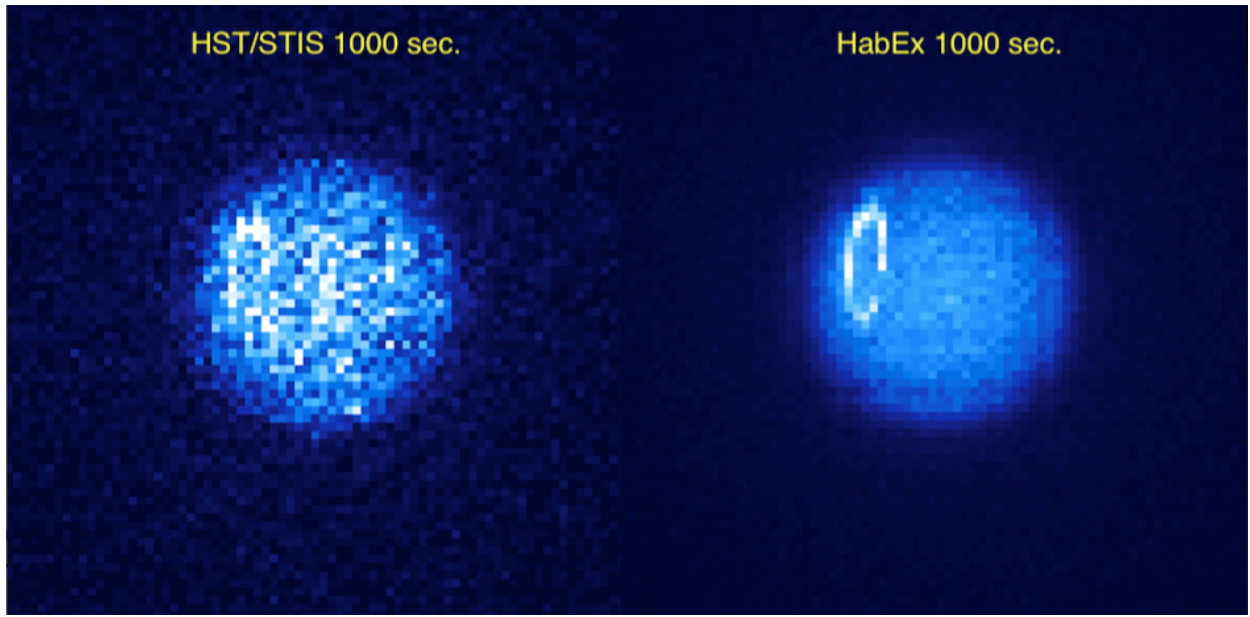

**Figure ES-5.** With its large collecting area, high resolution, and large effective area in the UV (Figure ES-9), HabEx will be able to examine the detailed morphology and time-variable nature of aurorae of the giant planets in our solar system, probing the physics of star-planet interactions. *Left:* A simulated image of Uranus with STIS on HST, assuming an exposure time of 1000 s. *Right:* Same as the left panel, except assuming a 1000 s exposure with UVS on HabEx. The larger aperture and effective area of HabEx not only enable more detailed studies of the aurorae of Jupiter and Saturn, including changes on timescales of considerably less than an hour, but also enable studies of the morphology and changes of aurorae of Uranus and Neptune on timescales of hours, which is difficult or impossible with HST.

However, by designing HabEx to have capabilities that significantly extend and enhance those of any current or planned mission, the community's imagination and future priorities can be relied on to maximize the science return of the mission. Nevertheless, in order to set requirements on the HabEx mission that will enable such a broad portfolio of exciting science, nine objectives are defined in *Chapter 4*.

### Baseline HabEx Implementation

The HabEx Observatory design utilizes an off-axis, monolithic 4 m diameter telescope, diffraction-limited at 0.4 μm, launched on an SLS Block 1B launch vehicle to an Earth-Sun L2 orbit, and a 52 m starshade spacecraft, separately launched on a Falcon Heavy launch vehicle, also to an Earth-Sun L2 orbit (**Figure ES-6**). The nominal distance between the spacecraft and the starshade is ~76,600 km (**Figure ES-7**), but the starshade can be moved closer or further away from the telescope to increase the range of wavelengths covered by the telescope and starshade (*Section 3.3.1.1*) during high-contrast observations.

HabEx has two starlight suppression systems: a coronagraph and a starshade, each with their own dedicated instruments for direct imaging and spectroscopy of exoplanets. HabEx also has two general purpose instruments: a UV imaging spectrograph and a visible through near-IR imaging spectrograph.

The overall HabEx design has been optimized for high-contrast direct imaging at small angular separations and broad spectroscopy of Earth-sized and larger exoplanets (**Figure ES-8**). The off-axis monolithic primary mirror avoids the significant challenges faced by obscured and/or segmented mirrors in achieving both high contrast in direct imaging *and* high planet light throughput with a coronagraph. The Earth-Sun L2 orbit provides a stable thermal and gravitational environment, ideal for high-contrast imaging and formation flying. The dual starlight suppression capabilities provide a flexible approach for optimized exoplanet searches and detailed characterization of individual exoplanets and their planetary systems, and is more resilient to uncertainties.

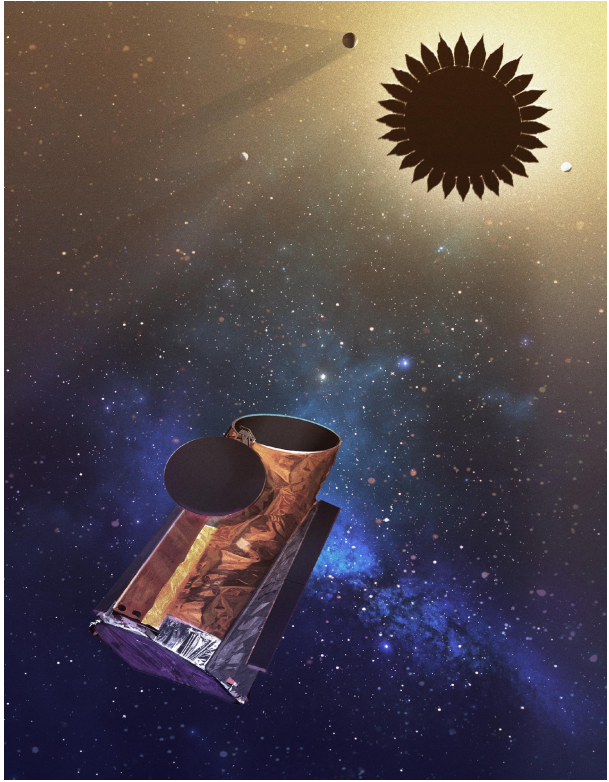

**Figure ES-6.** HabEx is the Great Observatory of the 2030s, consisting of a telescope and starshade flying in formation. The telescope includes four instruments: two for direct imaging and spectroscopy of exoplanets, the starshade and the coronagraph, and two facility instruments, a UV spectrograph and a UV through near-IR camera and spectrograph.

The coronagraph is nimble, residing inside the telescope, allowing for efficient multi-epoch surveying of multiple target stars to identify new exoplanet and EECs and also measure their orbits. However, the coronagraph has a narrow annular high-contrast field of view (FOV) with a spectroscopy bandpass limited to 20%, implying that obtaining broadband spectra of detected exoplanets is generally mission-time expensive, as observations must be taken in serial.

Compared to the coronagraph, the starshade provides a wider FOV and broader instantaneous wavelength coverage. However, it is fuel limited due to the distance and thrust required to transit the starshade occulter between target stars. For the nominal mission concept, the starshade can take up to two weeks to travel from one target to another, and has enough fuel for roughly 100 distinct pointings.

Importantly, this hybrid (coronagraph plus starshade) approach to direct exoplanet detection and characterization is a powerful combination, taking advantage of the complementary strengths of each instrument and significantly increasing the resultant yields of *well-characterized* planets that have both high-quality broadband spectra and orbits.

### The Four HabEx Instruments

**HabEx Coronagraph (HCG) Instrument.** The coronagraph mask suppresses starlight within the telescope to reveal light from proximate exoplanets. HabEx uses a vector vortex coronagraph because of its high resilience to common low-order wavefront aberrations, which translates into significantly less stringent requirements on telescope thermal and mechanical stability than other coronagraph designs. The coronagraph has a 62 mas inner working angle (IWA$_{0.5}$, a reasonable proxy for the minimum detectable exoplanet separation from its star) and includes a blue channel with a camera and integral field spectrograph (IFS) covering 0.45–0.67 μm, a red channel with a camera and IFS covering 0.67–1.0 μm, and an IR imaging spectrograph that covers 0.975–1.8 μm.

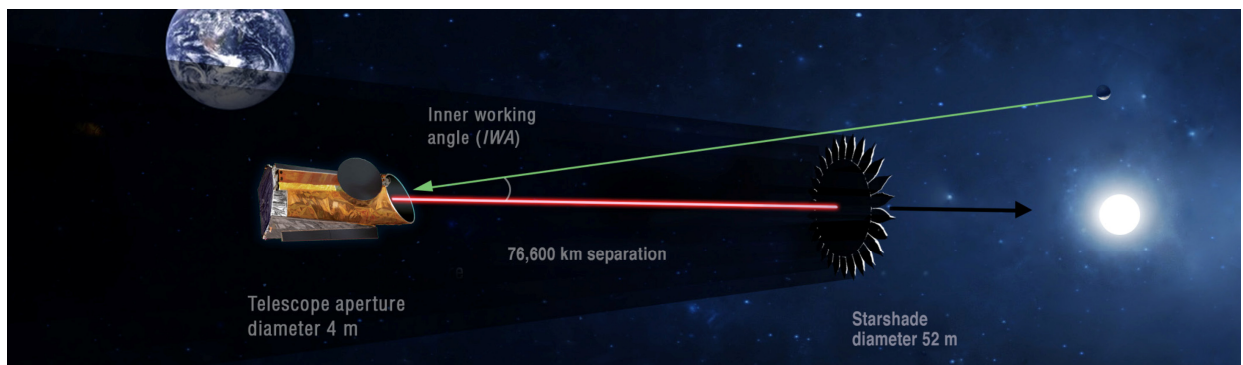

**Figure ES-7.** The HabEx telescope flying in formation with the starshade. The telescope can detect and characterize exoplanets viewable near or beyond the angular radius of the starshade as seen from the telescope, which defines the inner working angle (IWA).

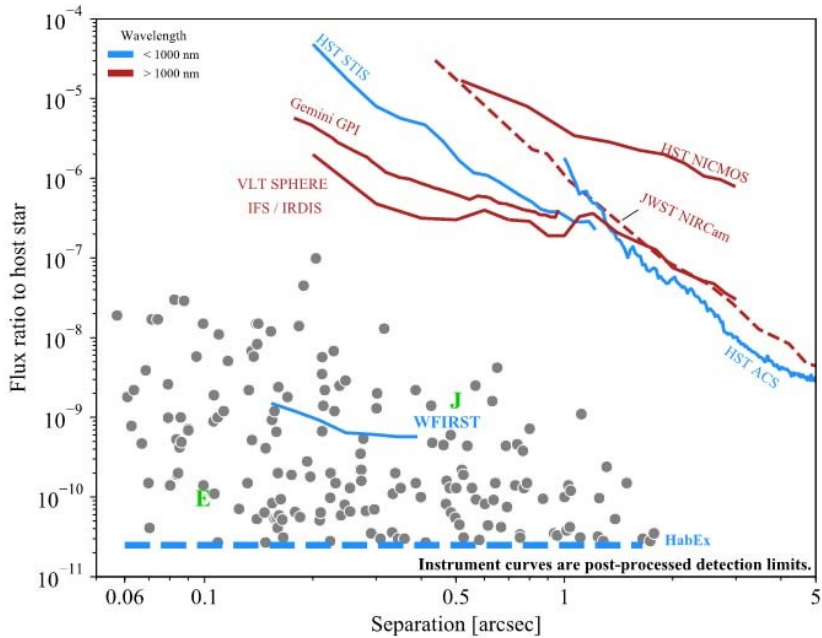

**Figure ES-8**. HabEx will detect and characterize newly discovered exoplanets at low planet-to-star flux ratios (grey points), enabling the first detailed studies of Earth-like planets in the habitable zone. Earth (E) and Jupiter (J) are shown for a solar system analog at 10 pc.

**HabEx Starshade Instrument (SSI).** The starshade external occulter blocks starlight before it enters the telescope, allowing light from an off-axis exoplanet to be observed. The HabEx 52 m diameter starshade will fly in formation with the telescope at a nominal separation of 76,600 km (**Figure ES-7**). The starshade advantages include a high throughput, small IWA$_{0.5}$, with an outer working angle (OWA) limited only by the instrument FOV and an ultra-broad bandwidth available for high contrast spectroscopy. The HabEx starshade has a 58 milliarcsecond (mas) IWA$_{0.5}$ at 1 μm and, a 6 arcsec OWA for broadband imaging, and offers deep enough starlight suppression for spectroscopy over an instantaneous bandwidth of 0.3–1.0 μm. The starshade may also operate at two additional separations from the telescope. At a larger separation of 114,910 km, it covers bluer wavelengths at 0.2–0.67 μm with a constant IWA$_{0.5}$ of 39 mas, providing unique access to the deep ozone features expected in Earth-like atmospheres. At a smaller separation of 42,580 km, it covers redder wavelengths at 0.54–1.8 μm with an IWA$_{0.5}$ of 104 mas at 1.8 μm, enabling sensitivity to multiple water vapor features.

The SSI has three channels: a near-UV/blue channel covering 0.2–0.45 μm with a grism, a visible channel covering 0.45–0.975 μm with an IFS and camera, and a near-IR channel covering 0.975–1.8 μm with an IFS and camera.

**UV Spectrograph/Camera (UVS).** The UVS covers 115–320 nm with a FOV of 3 × 3 arcmin$^2$ and multiple spectroscopic settings up to resolutions of 60,000. Additionally, a grating set contains a mirror to provide imaging capability. The UVS has more than 10 times the effective area of HST's Cosmic Origins Spectrograph (COS; **Figure ES-9**, right panel) from roughly 150–300 nm. Not only does the UVS provide improved angular resolution and throughput relative to HST, it also includes a microshutter array, allowing multiplexed UV slit spectroscopy for the first time in space. Together with the multiplexing capability, the UVS will be significantly more capable than COS over most of range of the UVS (115–320 nm).

**HabEx Workhorse Camera (HWC) and Spectrograph.** The HWC is an imaging multi-object slit spectrograph with two channels covering wavelengths from the visible through near-IR and a spectral resolution of 1,000. The visible channel covers 0.37–0.95 μm and the near-IR channel covers 0.95–1.8 μm. The HWC, with its larger 3×3 arcmin² FOV and higher resolution (**Figure ES-9**, left panel), will provide capabilities similar to, but significantly more sensitive than, HST's Wide-Field Camera 3 (WFC3) or Advanced Camera for Surveys (ACS).

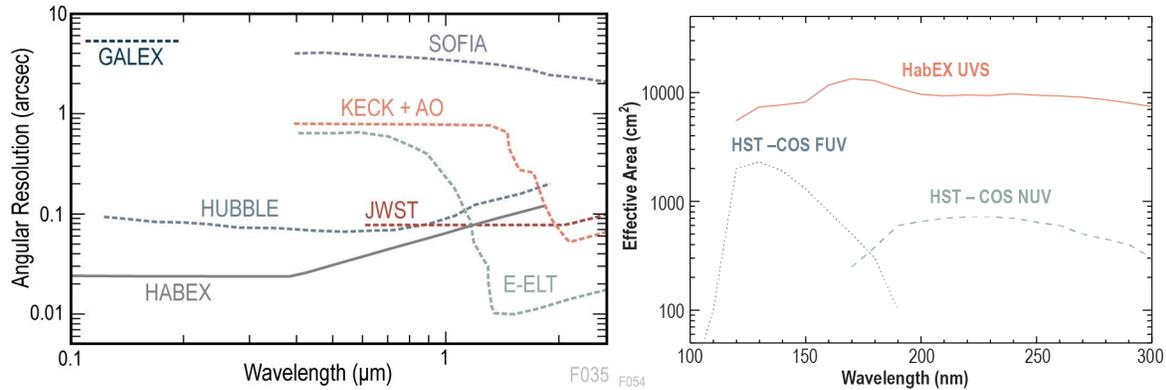

**Figure ES-9.** *Left:* HabEx will provide the highest-resolution UV/optical images of any current or planned facility, enabling a broad suite of observatory science. Opportunities range from studies of solar system objects, the Milky Way Galaxy, nearby resolved stellar populations, high-redshift galaxies, and large-scale structure. Note: the assessment assumes that the extremely large telescopes will only achieve their theoretical diffraction limit around ~1 μm. *Right:* With more than ten times the effective area of HST-COS at wavelengths from 150–300 nm, combined with a microshutter array, the HabEx UVS provides several orders of magnitude improved efficiency for UV spectroscopic studies both by an order of magnitude increase in effective area, and multiplexing capabilities.

Both the UVS and HWC can be used in parallel with the HCG and SSI instruments, as well as in parallel with each other.

### The HabEx Observational Strategy

A notional distribution of observing time for all HabEx programs during the 5-year prime mission (ten years of consumables), is shown in **Figure ES-10**.

Approximately 50% of the prime mission time is dedicated to exoplanet observations via two ambitious exoplanet surveys designed to take full advantage of the dual starlight suppression instruments.

A *broad* survey of roughly 42 nearby, mature stars will be optimized for discovery of small HZ exoplanets, even though a broad range of planets will also be detected and characterized. This survey utilizes the coronagraph's pointing agility to revisit the target stars over multiple epochs for discovery, confirmation of physical association with the host star, measurement of orbital periods and semi-major axes to within 25%, and eccentricities to within 0.1 for nearly all detected planets with periods shorter than 10 years. For HZ planets detected, orbits are short enough that inclination, semi-major axis and eccentricity are determined to within a few percent. Broadband (0.3–1.0 μm) spectra of all of these planetary systems are then obtained by the starshade. Systems with EECs detected receive at least 3 starshade visits to cover an even broader spectroscopic wavelength range (up to 0.2–1.8 μm) and / or measure spectra at different orbital phases.

A *deep* survey utilizing the starshade for multi-epoch broad spectroscopic observations of roughly eight of the nearest sunlike stars will provide even more detailed information about our nearest neighbors, with access to even smaller planets and star-planet separations than in the broad survey.

The primary difference between the surveys is that the deep survey will systematically search for and spectrally characterize fainter planets around the very nearest stars, integrating down to a planet-to-star flux ratio consistent with a Mars-sized planet around a sunlike star. In comparison, the individual exposure times for the broad survey are set to maximize the overall the number of EECs detected and validated through orbit determination and spectroscopy. The broad survey planet-to-star flux ratio detection limit will generally be higher than in the deep survey.

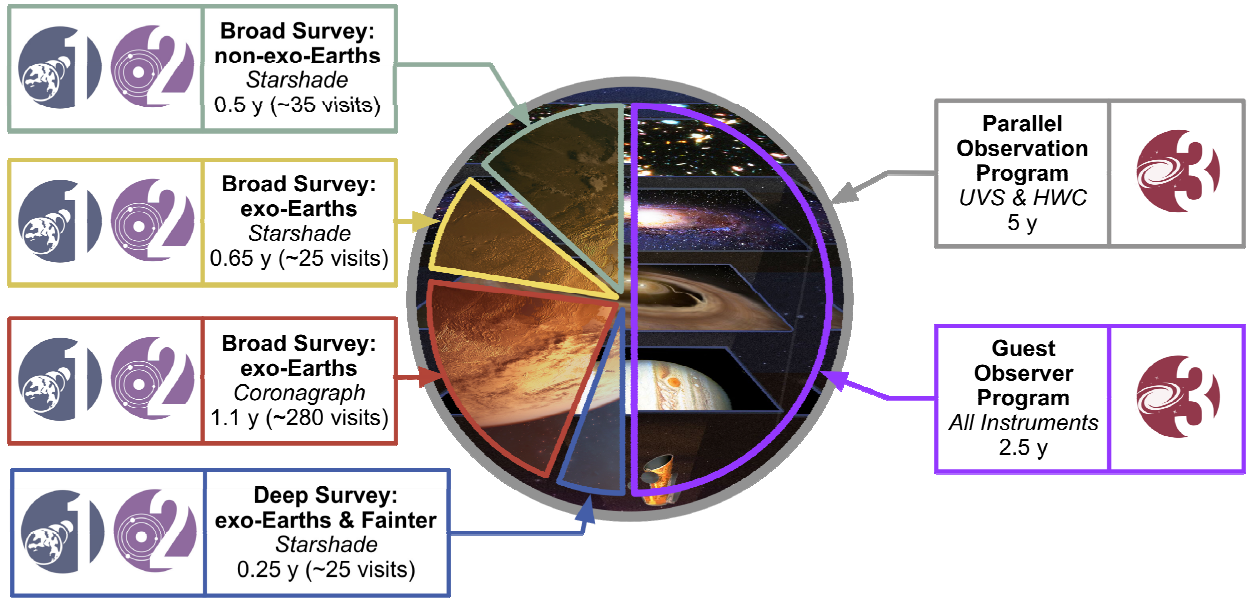

**Figure ES-10.** HabEx is an observatory for the greater astrophysics community, balancing exoplanet and general astrophysics observing programs. All of the instruments can be operated in parallel, and thus HabEx is also capable of parallel deep field observations with one or both observatory science instruments during the entire mission duration.

HabEx does not rely on any prior knowledge or contemporaneous independent observations provided by other ground- or space-based facilities. New observatories, however, are expected to be operational by the time HabEx launches and may provide additional data on the target systems, enabling more robust HabEx target prioritization and scheduling (see *Chapter 12* for details).

Through joint scheduling of exoplanet and general astrophysics observations and engineering design, HabEx is capable of about 90% observational efficiency. The Guest Observer program will be community driven and competitively selected and will likely include solar system, exoplanet, Galactic, and extragalactic studies.

Two kinds of opportunities will exist to guest observe with HabEx. Standalone observations across the sky are scheduled during starshade retargeting transits, and concurrent observations utilizing the ability to observe in parallel with the UVS and/or the HWC while observing with any one of the other instruments. Thus, parallel UVS and HWC observations will enable images and/or spectra in two separate $3 \times 3$ arcmin$^2$ fields of view during observations by the SSI or HCG, providing, e.g., two HST-like ultra-deep fields in the vicinity of the exoplanet target stars. Similarly, the HWC can obtain parallel observations during observations by the UVS, and vice versa. Thus, the four instruments on HabEx are highly multiplexed, thereby greatly improving the mission's scientific productivity.

### The HabEx Philosophy

HabEx was designed to be a Great Observatory that can be realized in the 2030s. To achieve this, the guiding philosophy of this study was the recognition that any recommendation by the Astro2020 Decadal Survey must balance scientific ambition with programmatic and fiscal realities, while simultaneously considering the impact of its development on the greater astronomy community's need for a broad portfolio of science investigations. The HabEx study therefore aimed to develop a mission capable of the most compelling science possible, while still adhering to likely cost, schedule, and risk constraints.

The preferred HabEx architecture was thus chosen to be technically achievable within this time frame, leveraging the investment in several enabling technologies made over the last decade or more. HabEx adopted a conservative design with substantial margins, utilizing moderate to high technological maturity resulting in manageable development risk.

Preventing the portfolio-disrupting type of cost and schedule growth experienced on the James Webb Space Telescope was also a major consideration for HabEx. Fortunately, the kind of optimism of the "faster, better, cheaper" paradigm that led to JWST's initial estimate no longer exists. In the new reality of Technical, Risk, and Cost Evaluation (TRACE) reviews, HabEx has worked to keep cost and schedule estimates pinned to actual missions wherever possible. Reserves and margins are added to actual costs and schedule durations. In addition, the costs for the baseline concept were independently estimated by the JPL Cost and Pricing Office.

HabEx is ambitious, offering humankind the first opportunity to glimpse into worlds like our own and uncover signs of life outside of our solar system. It provides far better than Hubble-like imaging and spectroscopic capabilities to support community science observations in many different fields. But it also balances scientific ambition with the programmatic and fiscal realities that constrain space exploration.

### HabEx Alternate Designs and Architecture Trades

Finally, a tradespace analysis was performed to give the Decadal Survey the maximum flexibility on sizing a HabEx mission to fit into whatever funding situation emerges (see *Chapter 10*). In addition to the baseline concept, HabEx has evaluated the comparative science yield, number of new technologies and cost for eight other architectures (see *Chapter 10* for details on the architectures). Together with the baseline option, these nine options span three aperture sizes (4 m, 3.2 m, and 2.4 m) and three starlight suppression methods (coronagraph-only, starshade-only, and a hybrid that uses both), offering some understanding of the performance, cost, and risk sensitivities of the tradespace.

The results of such a sensitivity analysis are shown in **Table ES-2** where the different architecture capabilities are compared against HabEx scientific Goals and Objectives (see *Chapters 3, 4, and 5*). The colors of the cells indicate whether the architecture is expected to achieve the science objective baseline requirements (green), threshold requirements (yellow), or neither (orange) during the HabEx prime 5-year mission. Scientific advances are still possible in the orange-colored cells, although these advances fall below the threshold requirements established by the HabEx study team.

Of the eight other architectures examined as part of this study, two were developed in detail; not as descope or back-up options—the Science and Technology Definition Team (STDT) only prioritizes the preferred, baseline option—but as points of reference for the TRACE team to be able to calibrate the cost and risk for all nine HabEx options to their standards. The two designs are captured in *Appendix A* and *Appendix B*. In so doing, the entire tradespace will be usable by the Decadal Survey in their recommendation.

### Why Now? Scientific Readiness

HabEx stands available as an achievable mission to directly image exo-Earths as a result of decades of scientific and technological achievement. There has been tremendous progress in the discovery of exoplanets over the last 20 years. In particular, astronomers have discovered that small rocky planets around main sequence stars are common. One implication of these discoveries is the prospect for atmospheric characterization of rocky planets orbiting M dwarf stars in the near-term. These systems have been and will be identified by surveys such as MEarth, Search for habitable Planets

Table ES-2. The HabEx study investigated a total of nine different architectures. All of the architectures include both the UV Spectrograph (UVS) and the HabEx Workhorse Camera (HWC). Three telescope apertures were considered—4 m, 3.2 m, or 2.4 m—with three instrument options for each aperture: both starlight suppression technologies (H), coronagraph-only (C), or starshade-only (S). The study team's preferred architecture is the 4H, although all of the architectures are capable of meeting a subset of the science requirements flowing from the HabEx science goals and objectives in Table 5.1-1. The colors indicate whether that architecture meets the baseline requirements of the objective (green), the threshold requirements (yellow), or neither (orange). Besides the 4H, architectures 4C and 3.2S were developed in some depth. See *Chapter 10* for more detail.

| HabEx Science Goals & Objectives | | 4H | 4C | 4S | 3.2H | 3.2C | 3.2S | 2.4H | 2.4C | 2.4S |
|---|---|---|---|---|---|---|---|---|---|---|
| **Habitable Exoplanets** 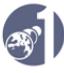 | O1 Exo-Earth candidates around nearby sunlike stars? | green | green | green | green | green | yellow | green | orange | orange |
| | O2 Water vapor in rocky exoplanet atmospheres? | green | green | green | green | yellow | yellow | yellow | orange | orange |
| | O3 Biosignatures in rocky exoplanet atmosphere? | green | yellow | yellow | green | yellow | orange | yellow | orange | orange |
| | O4 Surface liquid water on rocky exoplanets? | green | green | green | green | orange | orange | green | orange | orange |
| **Exoplanetary Systems** 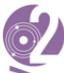 | O5 Architectures of nearby planetary systems? | green | green | orange | green | yellow | yellow | green | yellow | green |
| | O6 Exoplanet atmospheric variations in nearby systems? | green | yellow | yellow | yellow | orange | orange | yellow | yellow | yellow |
| | O7 Water transport mechanisms in nearby planetary systems? | green | yellow | yellow | green | yellow | yellow | green | orange | orange |
| | O8 Debris disk architectures in nearby planetary systems? | green | yellow | yellow | green | yellow | yellow | green | yellow | green |
| **Observatory Science** 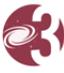 | O9 Lifecycle of baryons? | green | green | green | yellow | yellow | yellow | yellow | yellow | yellow |
| | O10 Sources of reionization? | green | green | green | yellow | yellow | yellow | yellow | yellow | yellow |
| | O11 Origins of the elements? | green | green | green | yellow | yellow | yellow | yellow | yellow | yellow |
| | O12 Discrepancies in measurements of the cosmic expansion rate? | green | green | green | yellow | yellow | yellow | yellow | yellow | yellow |
| | O13 The nature of dark matter? | green | green | yellow | green | yellow | yellow | orange | orange | orange |
| | O14 Formation and evolution of globular clusters? | green | green | green | green | green | green | yellow | yellow | yellow |
| | O15 Habitable conditions on rocky planets around M-dwarfs? | green | green | green | green | green | green | green | yellow | yellow |
| | O16 Mechanisms responsible for transition disk architectures? | green | green | yellow | green | green | green | yellow | yellow | green |
| | O17 Physics driving star-planet interactions, *e.g.* auroral activity? | green | green | green | green | green | green | green | yellow | green |

EClipsing ULtra-cOOl Stars (SPECULOOS), and NASA's Transiting Exoplanet Survey Satellite (TESS). The atmospheres of these planets will be characterized by follow-up observations from ground-based Extremely Large Telescopes (ELTs) and space-based missions such as JWST and eventually even HabEx. In parallel, steady progress in high-contrast direct imaging technology has been very impressive, with the first direct detection of bright self-luminous exoplanets announced in 2008, and the characterization of closer-in self-luminous planets since then.

This progress points to the next logical step: the discovery and detailed characterization of Earth-like worlds and complete planetary systems around nearby sunlike stars. We now know, thanks primarily to NASA's Kepler mission, that small planets orbiting sunlike stars are not rare. Indeed, although difficult to measure and requiring some extrapolation, we know that the frequency of rocky planets in the habitable zones of sunlike stars is likely not very small, nor very large. The best estimates indicate that ~25% of all sunlike stars host a rocky planet at the right distance to its host star to have liquid water, assuming appropriate atmospheric conditions.

HabEx will start this journey of exploration, providing the first detailed images and spectra of the full range of exoplanets orbiting nearby mature stars, and searching for signs of habitability and life on all of the small rocky worlds detected.

### Why Now? Technological Readiness

All technologies necessary for a HabEx mission are currently at Technology Readiness Level (TRL) 4 or higher. This is due in large part to the highest priority medium-scale investment recommendation for space in the 2010 Decadal Survey. A "New Worlds Technology Development Program" in "preparation for a planet-imaging mission beyond 2020" has matured many critical technologies to a point where a HabEx mission start in 2025 is finally feasible. Dramatic progress has occurred in four key technology areas that make the current design possible:

- High-contrast imaging at small angular separations using broadband coronagraphs
- Starshade modeling advances and technology demonstrations (**Figure ES-11**)
- Manufacturing of large aperture monolithic mirrors
- Microthrusters for fine pointing

Recent coronagraph demonstrations including the vector vortex design are pushing within a few multiples of required contrast levels at the appropriate IWA for habitable zone direct imaging. The starshade technology "S5" task, led by NASA's Exoplanet Program Office, has reached TRL 5 for the technology gap in formation flying, and will advance all other starshade-related gaps to TRL 5 by 2023. Starshade test articles have reached the $10^{-10}$ broad-band contrast level at small separations required by HabEx. Ongoing work on thermally stable materials, structurally suitable designs and new coating facilities make a 4-meter-class mirror, suitable for coronagraphy, within reach for HabEx. Finally, a major breakthrough for HabEx, and space-based observatories in general, is the advancement of microthrusters to flight applications made by the European Space Agency (ESA) through their Gaia and LISA-

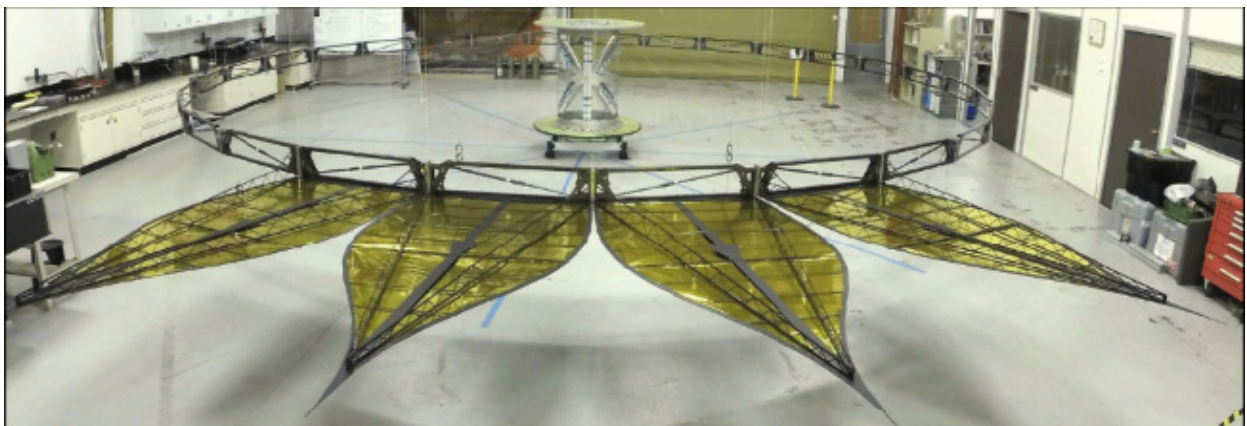

**Figure ES-11.** Prototype starshade truss with petals demonstrated by the Starshade to TRL 5 (S5) effort.

Pathfinder missions. This technology allows HabEx to counteract the effects of solar pressure without the use of reaction wheels, eliminating the primary vibrational disturbance on the telescope platform.

Advancing the capabilities of tomorrow's space telescopes over those that are operating today requires new technology, and new technology brings risk. Where possible, HabEx has looked to use the low risk solution (e.g., monolithic mirrors, HST coatings). Where new technology is truly enabling for the mission, HabEx has looked to characterize the technology as completely and openly as possible. Rather than rolling up coronagraph technologies into a "high contrast coronagraph" gap, the individual gaps are recognized at the component level (i.e., deformable mirrors [DMs] and Zernike wavefront sensor mask) and at the system level (i.e., coronagraph architecture). It is important to recognize both. Identifying technologies at the system-level only hides the number of technological advances truly needed. Identifying at the component-level only presumes system-level maturation based on the component TRL which is not necessarily assured. Accordingly, HabEx currently recognizes thirteen technologies at TRL 4 and three technologies at TRL 5 (see **Table ES-3**). These numbers of new technologies reflect both the completeness and maturity of this study's understanding of the technology developments needed to reach the concept's science goals. All new missions are required to advance all technologies to TRL 5 before the mission start and to TRL 6 before the end of the formulation phase of the project. HabEx will advance all but two technologies to at least TRL 5 by the end of 2023, with two technologies expected to reach TRL 6 by that same date. The two technologies that will require until 2024 to finish TRL 5 qualification are the 4 m mirror, and the micro-channel plate (MCP) detectors for the UVS instrument. The cost of a 4 m prototype has prevented work ahead of the Decadal Survey recommendation, and the time required to cast, grind and polish the mirror limits the TRL completion date. Similarly, the MCP also assumes a technology development start in 2022, leading to TRL 5 completion in 2024. Earlier funding could advance this date by a year. Even by the release of the 2020 Decadal Survey report, the current number of TRL 4 technologies will be reduced to ten with the current funded development efforts.

To reduce technology risk further, HabEx will be qualifying two key technologies with full-scale prototypes. In addition to the 4 m primary mirror (required for TRL 5), a 52 m starshade prototype will be built to advance the starshade from TRL 5 to TRL 6. Details on the HabEx technologies and their maturation plans can be found in *Chapter 11* and *Appendix E*. Furthermore, the redundancy in the concept's direct imaging capability allows some mitigation in technology development risk as well. Should problems surface on starshade or coronagraph technology development, their inclusion in the mission and the scope of the HabEx exoplanet science could be reevaluated.

**Table ES-3.** HabEx enabling technologies, rolled up by category. Detailed ("unrolled") breakdowns of the individual technologies, description of the current state of the art and capabilities needed can be found in *Chapter 11,* and the technology roadmap in *Appendix E.* As of 2019, the HabEx baseline design requires thirteen TRL 4 technologies, and three TRL 5 technologies. Following the technology roadmaps, by 2023, HabEx will carry two TRL 4 technologies and twelve TRL 5 technologies. Two technologies will be at TRL 6 and will have completed technology development one year before the start of Phase A.

| # of Enabling Tech. | 2019 | | 2020 (estimated) | | 2023 (estimated) | | |
|---|---|---|---|---|---|---|---|
| Category | TRL 4 | TRL 5 | TRL 4 | TRL 5 | TRL 4 | TRL 5 | TRL 6 |
| Starshade | 3 | 2 | 2 | 3 | 0 | 4 | 1 |
| Large Mirror | 2 | 0 | 2 | 0 | 1 | 1 | 0 |
| Metrology | 0 | 1 | 0 | 1 | 0 | 0 | 1 |
| Coronagraph | 3 | 0 | 2 | 1 | 0 | 3 | 0 |
| Detectors | 4 | 0 | 4 | 0 | 1 | 3 | 0 |
| Microthrusters | 1 | 0 | 0 | 1 | 0 | 1 | 0 |
| Total | 13 | 3 | 10 | 6 | 2 | 12 | 2 |

While HabEx recognizes a large number of new technologies, these technologies are advancing rapidly, will be qualified at both the component and system levels, will be qualified at full scale where applicable, and all the technology work will be completed before most of the NASA investment in the mission is made. HabEx's necessary technologies are close to ready and the risks are manageable.

### *Beginning a New Era for Astrophysics with HabEx*

HabEx is a cost-effective, modest risk, high-impact science mission. HabEx will leverage recent advancements in starlight suppression technologies to utilize both a coronagraph and starshade to explore new worlds, assess their habitability, and map our nearest neighbor planetary systems to understand the diversity of the worlds they contain.

While the HabEx mission architecture is capable of direct imaging and spectral characterization of a broad range of exoplanets, including Earth analogs orbiting nearby sunlike stars, HabEx also provides unique capabilities for UV through near-IR astrophysics and solar system science from the vantage of space, moving UV capabilities to the next level after HST retires. HabEx is a worthy UV/optical successor to HST in the 2030s with significantly improved sensitivity and spatial resolution stemming from HabEx's significantly larger 4 m diameter aperture, improved detector technology, exquisite wavefront control, and a more thermally stable orbit.

# 1 THE CASE FOR A GREAT OBSERVATORY FOR ASTROPHYSICS AND EXOPLANETARY SCIENCE

This report presents a detailed mission concept for Habitable Exoplanet Observatory, or HabEx, a large space-based strategic mission that would serve as a NASA Great Observatory for the 2030s and beyond. This telescope will operate in the tradition of the Hubble Space Telescope (HST) and the James Webb Space Telescope (JWST), by providing community access to unique observational capabilities that are not currently available or planned. This mission concept has been funded and developed by NASA as input for the Astro2020 Decadal Survey.

This chapter motivates the case for HabEx as an excellent candidate for the next Great Observatory to follow on after the Wide Field Infrared Survey Telescope (WFIRST). HabEx will operate in ultraviolet (UV) to near-infrared (NIR) wavelengths, providing imaging and spectroscopic capabilities that are substantial improvements over what is currently available with HST.

HabEx will be an observatory for the entire community, with a competed and funded community-driven Guest Observer (GO) program that will be defined and driven by the most compelling science questions of the time, most of which cannot even be imagined yet. As one example, HabEx will automatically obtain parallel ultraviolet, optical, and near-infrared observations of unprecedented depth and resolution, rivaling and even surpassing that of the Hubble Ultra Deep field, but over a much wider area. These parallel observations, as well as serendipitous and archival observations, will open up new areas of discovery space, and provide abundant targets for focused ground-based observing programs with the next generation of giant segmented telescopes.

These capabilities will be exploited through a community driven, competed, and funded GO program, serendipitous or parallel observations, and archival research.

HabEx will replace and enhance many capabilities that will be lost when HST reaches the end of its operational lifetime. HabEx is also responsive to the Astro2010 New Worlds New Horizons (NWNH) Decadal Survey, which anticipated this gap in capability, noting "Key advances could be made with a telescope with a 4-meter-diameter aperture with large field-of-view and fitted with high-efficiency UV and optical cameras/spectrographs operating at shorter wavelengths than HST" (NRC 2010, p. 220). HabEx observes at wavelengths as short as HST and will have provide over a factor of 10 improvement in effective sensitivity over HST. Enhancing options would allow HabEx to extend its wavelength range to shorter wavelengths with additional investments to develop coating and detector technologies.

As described below, technological and scientific advances over the past decade make it feasible and fiscally advantageous to define a mission that directly images and characterizes Earth-like planets orbiting nearby sunlike stars. It will be able to determine whether these planets are likely to be habitable, and may even find evidence that these planets are inhabitable. It will therefore also address one of the most profound questions of humankind: **Is there life outside the solar system?**

Combining two primary science motivators, HabEx mission will have three primary science goals, identified in **Table 1-1**.

This chapter motivates these three primary science goals, which were developed to be responsive to the extraordinary revolutions in the fields of Galactic and Extragalactic

**Table 1-1.** HabEx science goals.

| HabEx Science Goals | |
|---|---|
| 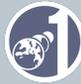 | *To seek out nearby worlds and explore their habitability* |
| 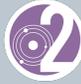 | *To map out nearby planetary systems and understand the diversity of the worlds they contain* |
| 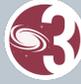 | *To carry out observations that open up new windows on the universe from the UV through near-IR.* |

Astrophysics, Exoplanets, and Planetary Science over the past few decades, revolutions we which briefly review below. These revolutions have raised a large number of scientific questions and have led to a number of scientific hypotheses, from which we have defined the scientific objectives of the mission. The telescope, instrument, and mission functional requirements, outlined in *Chapter 5*, ultimately flow directly from the science goals and objectives.

While the idea of a large strategic mission in the tradition of HST, one capable of transformative general observatory science in the UV/optical/NIR that can also direct detect and characterize Earth-like planets orbiting nearby sunlike stars in reflected light, is not new (e.g., Kasdin 2009), we argue that it is only now realizable. A confluence of several technological developments and scientific results over the past one or two decades has made now the optimal time to start focused technology development to enable such a mission to have a new start in the mid-2020s.

As such, this chapter also addresses two essential questions that define the timeline, scope, and range of architectures considered in this study: Why should we start the development of such a Great Observatory now? What motivates the family of mission designs that we present and consider, and what motivates the specific preferred architecture that is presented in detail in later chapters. In other words, what motivates different architecture choices and their subsequent fiscal, technological, and mission schedule requirements?

This chapter begins by briefly reviewing the revolutions in astrophysics, exoplanets, and planetary science over the past few decades. These revolutions have set the stage for the primary science goals and resulting science objectives of HabEx by defining a foundation of scientific knowledge in these fields and raising many unexpected and compelling scientific questions. Addressing these questions then necessitates the development of an observatory with the proposed capabilities of HabEx, and motivates the specific HabEx observatory architecture presented here.

Because we cannot know which current, compelling questions will remain unanswered by the 2030s, we have designed HabEx to have capabilities well beyond those of any current or planned mission. Thus, we are confident that HabEx will be the Great Observatory that can address these current questions and the myriad of new questions that cannot even be anticipated today.

## 1.1 The Scientific Motivation for HabEx: Building upon Revolutions

There has been enormous and truly extraordinary progress over the past 25 years in many areas of astrophysics. We have detected thousands of planetary systems orbiting other stars and now appreciate the diversity that these systems encompass. As a result of this observed diversity, we now have a completely different perspective on our own solar system. We are also now in the age of precision cosmology, where we have precise and accurate determinations of the contents and growth history of the universe. These revolutions have largely laid the foundation of knowledge in these fields but have nevertheless left us with a wide variety of questions that are needed to fully appreciate and understand the complexity of the phenomena that we observe.

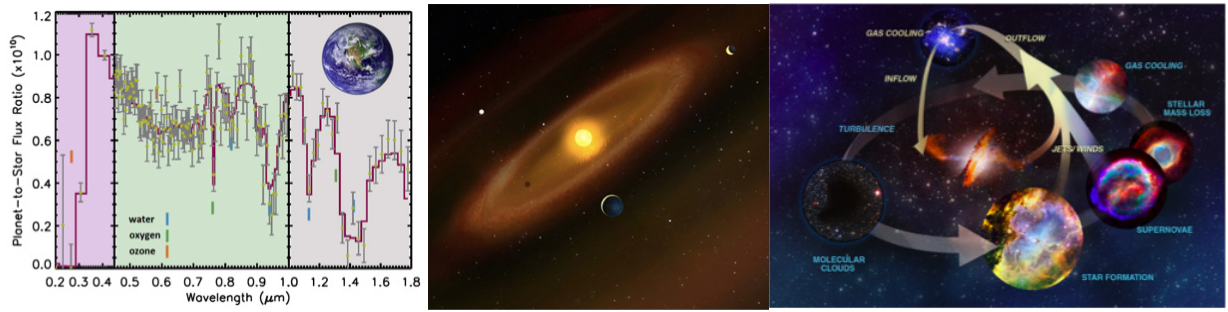

**Figure 1-1.** The HabEx mission will have three primary science goals. *Left:* **Seek out nearby worlds and explore their habitability.** HabEx would be the first mission capable of detecting blue skies, oxygen, and water vapor on habitable planets around nearby stars, as shown in this simulated HabEx spectrum of an Earth-like planet in the habitable zone (see *Chapter 2* for details). *Center:* **Map out nearby planetary systems and understand the diversity of the worlds they contain.** HabEx will obtain complete 'family portraits' of nearby planetary systems, including orbits and broadband spectra for many of the planets in these systems, enabling us to planet our solar system into context. *Right:* **Carry out observations that open up new windows on the universe from the UV through near-IR.** With a UV sensitivity over 14 times greater than HST and multiplexing capabilities, HabEx will be significantly more efficient for investigations of the life cycle of baryons.

These questions then drive our three primary science goals, which HabEx will address via a large range of secondary science objectives:

- Do rocky planets continuously orbiting within the habitable zone (HZ) of nearby sunlike stars exist? If so, do these planets harbor potentially habitable conditions, such as an atmosphere containing water vapor? Do these planets contain signs of life (*e.g.*, biosignature gases)? Do they contain liquid water oceans?

- What are complete architectures of planetary systems? How do the compositions of the atmospheres of planets orbiting nearby sunlike stars vary as a function of planet size and separation from their star? Is the presence of outer giant planets related to the existence of water vapor in the atmospheres of rocky planets orbiting in the habitable zones of nearby sunlike stars? What is the range of architectures of dust belts in exoplanetary systems, and what is the physical mechanisms by which these dust belts interact with planets in these systems?

- What is the complete life cycle of baryons and where are the missing baryons in the local universe? What are the sources responsible for the metagalactic ionizing background over cosmic time? What are the properties and end states of the first generations of stars, and what role do they play in the origins of elements? What are the mechanisms driving the formation and evolution of galactic globular clusters? Is there a need for new physics to explain the disparity of local and cosmic microwave background measurements of the expansion rate of the universe? Can we understand the small-scale behavior of dark matter by studying the gas and resolved stellar populations of nearby dwarf galaxies? Do rocky planets orbiting in the habitable zones of nearby mid-type M stars have potentially habitable conditions? What is the range and physical cause of the architectures of transition disks? What is the physics governing star-planet interactions and how do they drive the phenomenological behavior of aurorae in our giant planets?

HabEx will uniquely revolutionize astronomy, by providing crucial capabilities needed to address the questions listed above, some of which are the most compelling questions about the universe that humanity is asking today. Although other, new facilities will begin to explore these questions, it is likely that only HabEx will have the capabilities required to fully answer many of them. Indeed, by identifying the gaps in capabilities that will *not* be filled by existing or planned facilities, the design of HabEx has been optimized to be a uniquely capable, powerful, versatile, and yet technologically feasible observatory, which enable the search for habitable worlds and signatures of life, but will also address questions that cannot even be anticipated today.

### 1.1.1 Direct Imaging of Nearby Exoplanetary Systems

HabEx will be the first large strategic mission capable of directly imaging and characterizing mature planets as small as that of the Earth in reflected light orbiting the most nearby stars (distances of less than roughly 15 pc). In particular, it will be able to detect, obtain orbits for, and spectrally characterize over a broad band, Earth-sized planets orbiting in the habitable zones of nearby sunlike stars. This requires being able to detect and characterize a planet that is roughly 10 billion times fainter than a star that is a close as 0.06 arcseconds away. HabEx will accomplish this using technology that has only recently become available or will be available in the near future. Furthermore, it will utilize two starlight suppression technologies, which are highly complementary and enable a much broader and richer range of science capabilities than each alone. These capabilities will enable HabEx, for the first time, to address the first two of its three primary science goals.

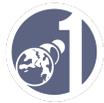 **Seek out nearby worlds and explore their habitability**

Do rocky planets continuously orbiting within the habitable zone of nearby sunlike stars exist? If so, do these planets harbor potentially habitable conditions, such as an atmosphere containing water vapor? Do these planets contain signs of life (e.g., biosignature gases)? Do they contain liquid water oceans?

By answering these questions, *HabEx will determine, for the first time, whether or not potentially habitable Earth-like worlds orbit nearby sunlike stars, and whether or not they have biosignatures in their atmospheres.* This is quite a remarkable statement, given that just over three decades ago, it was not known whether other stars hosted planetary systems at all, let alone solar-like systems, Earth-like planets, or life-bearing worlds. It was not until technological developments in the 1980s achieved the capability to detect planets around other stars, that the quest to answer these questions in a scientific manner became possible.

The primary difficulty of directly imaging an Earth analog in reflected light is that at optical wavelengths, the luminosity of the Earth is only about one 10 billionth that of the Sun. This, compared to the fact that the parent stars of Earth analogs are separated by angular distances as small as 0.06 arcseconds (for the HabEx target sample), make direct detection of Earth analogs exceptionally challenging.

The first detections of exoplanets came in the late 1980s using radial velocity (RV) and pulsar timing techniques (Campbell et al. 1988; Latham et al. 1989; Wolszczan and Frail 1992). However, the first broad demographic survey of exoplanets to provide statistics of a large number of planetary systems over a broad region of parameter space was NASA's Kepler mission. Kepler used ultra-precise photometry to find small planets transiting their host stars. The primary goals of Kepler were to find Earth-sized and presumably rocky planets in the habitable zones of sunlike stars, and to quantify their frequency.

For Kepler's goal of quantifying $\eta_{Earth}$, typically defined as the frequency of rocky planets orbiting sunlike stars in their habitable zones, there exists a broad range of results from a large number of the different studies by different authors that have attempted to answer this question using the Kepler data. This broad range is due to several facts. First, the stellar variability of sunlike stars turned out to be larger than was originally assumed, and partially as a result, and partially due to the fact that its primary mission was cut short due to the loss of a reaction wheel, Kepler only detected a few candidate Earth-like planets in the habitable zones of sunlike stars. Second, transit surveys are subject to severe selection biases, such that careful modeling is required to uncover the underlying true frequency. The habitable zones of sunlike stars happen to lie just at the edge of Kepler's survey sensitivity, and so the frequency of such planets is quite sensitive to these biases. Finally, it is only recently that a thorough end-to-end quantification of the survey sensitivity has been performed (Burke et al. 2015). Thus, any estimate of $\eta_{Earth}$ derived from Kepler necessarily comes with the caveats that it is based on a small number of detected candidates, large corrections for incompleteness, and modest extrapolation.

Nevertheless, the net conclusion is that $\eta_{Earth}$ is not unity (i.e., not every sunlike star hosts a potentially habitable planet), nor is it very small. Our best estimates are that $\eta_{Earth}$ is in the range of 8–70% (1σ confidence range; Belikov 2017).

This result has profound implications for missions whose primary purpose is to directly image and characterize potentially Earth-like planets and search for life. The yield of these

**Why Direct Imaging of Sunlike Stars from Space?**

In general, there are two primary methods of characterizing the atmospheres of exoplanets: direct imaging with starlight suppression, and transit spectroscopy of favorably aligned systems where the planetary orbit crosses our line of sight to the host star. For direct imaging in reflected light, one measures a spectrum of the host star that has been filtered through the planetary atmosphere twice, whereas for direct imaging in thermal light, one measures the thermal emission from the planet that has been filtered through the atmosphere once. In both wavelength ranges, the constituents of the planetary atmosphere are imprinted on the spectrum. Studying the brightness of the planet as a function of phase also reveals aspects of the planetary atmosphere, and potentially can detect the presence of a surface ocean (*Section 2.1.4*). Transit spectroscopy reveals information about the thermal emission of the planet filtered through the planetary atmosphere via eclipse spectroscopy, the atmospheric constituents of the planet via transmission spectroscopy, and the brightness of the planet as a function of longitude via phase curves.

Direct imaging from space is the only way to systematically spectroscopically observe atmospheres of resolved rocky planets in reflected light orbiting in the habitable zones of nearby sunlike stars. Such planetary atmospheres are inaccessible to ground-based direct imaging in reflected light, where current and likely future instrumentation is unlikely to be able to reach the necessary contrast ratios at the required close separations (e.g., **Figure ES-9**) and transmission spectroscopy is not viable because of the low likelihood of transit and the small ratio of the area of the atmospheric annulus to the area of the star, resulting in an exceptionally small signal.

Transit spectroscopy of potentially habitable terrestrial planets orbiting low-mass stars, such as M-class dwarfs, is an attractive approach for several reasons. First, the habitable zones of low-mass stars are closer to their parent star than for sunlike stars, and therefore the transit probability is higher. Second, the larger planet-to-star radius ratio increases the amplitude of the transit signal. This is often referred to as the "small star opportunity," and is the motivation behind ground-based surveys such as $M_\oplus$ (Charbonneau et al. 2009) and Search for habitable Planets EClipsing ULtra-cOOl Stars (SPECULOOS; Gillon et al. 2017), and NASA's space-based Transiting Exoplanet Survey Satellite (TESS; Ricker et al. 2015). These surveys have and will discover rocky planets orbiting in the habitable zones of their low-mass host. These planets can then be followed up with, e.g., JWST or even HabEx to search for signatures of habitability and biosignatures. However, these same arguments imply that transit spectroscopy is essentially impossible with current technology for small planets orbiting larger, sunlike stars, at least at optical and near-infrared wavelengths.

missions is primarily contingent on three things: 1) the ability to resolve the planet from the host star (also parameterized by the inner working angle, IWA); 2) the ability to collect enough photons to be able to obtain a spectrum that can robustly identify biosignatures; and 3) the frequency of potentially habitable planets (*i.e.*, $\eta_\oplus$), which statistically sets how far one must look to find a potentially habitable planet. The first two are essentially proportional to the aperture of the telescope. Thus, the larger the value $\eta_\oplus$, the smaller the aperture required to have access to a given number of targets. Because $\eta_\oplus$ is in the regime of tens of percent, it is possible to achieve the goal of reliably detecting and characterizing roughly ten potentially habitable planets with relatively small telescope apertures.

The estimate of $\eta_\oplus$, one of the primary results of Kepler, has only recently become available. As we will describe, this estimate, along with the several other recent scientific and technological advances, has enabled the design of a relatively low-risk, low-cost mission, which can answer the science questions presented above and address the first of the three primary goals: to seek out nearby worlds and explore their habitability.

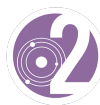 *Map out nearby planetary systems and understand the diversity of the worlds they contain*

What are complete architectures of planetary systems? How do the compositions of the atmospheres of planets orbiting nearby sunlike stars vary as a function of planet size and separation from their star? Is the presence of outer giant planets related to the existence of water vapor in the atmospheres of rocky planets orbiting in the HZs of nearby sunlike stars? What is the range of architectures of dust belts in exoplanetary systems, and what is the physical mechanism by which these dust belts interact with planets in these systems?

By answering these questions, *HabEx will determine the nearly complete architectures of planetary systems, including the interplay of these architectures with* the atmospheric properties of the planet they contain. This will provide the ground truth for models of planet formation and evolution. *HabEx will also place results of the first primary goals in context, by addressing how the habitability of Earth-like planets is connected to the larger planetary architecture.*

Although its primary goal was to detect and determine the abundance of Earth analogs, Kepler far exceeded expectations. It is now known from Kepler that small, short-period planets, planets with radii less than that of Neptune and periods less than roughly 100 days, are exceptionally common. Most of these planets are "super-Earths" or "sub-Neptunes," planets with radii between that of the Earth and Neptune. Intriguingly, despite the fact that Kepler found these planets to be very common, they have no analogue in our solar system. Indeed, Kepler revealed that, on average, most stars host at least one such planet.

While Kepler has revolutionized the field of exoplanets, and has provided our first estimates of $\eta_{Earth}$, it was not capable of answering the first question above: What are the complete architectures of planetary systems? This is because Kepler is basically insensitive to planets beyond 1 astronomical unit (AU), the mean distance between the Earth and the Sun. Yet, Kepler has found nearly 4,500 planetary candidates with orbital semi-major axes less than ~1 AU and radii from roughly that of the Earth to roughly that of Jupiter, indicating that most planetary systems *do not* look like our own. Other methods, such as radial velocity, are sensitive to Jupiter analogs, but in order to complete the statistical census of planetary systems, a method that is sensitive to more distant, lower-mass planets is required. NASA's WFIRST, the next flagship mission after JWST, will perform a microlensing survey that will be sensitive to planets with mass greater than that of Earth and orbital separations from 1 AU to infinity, i.e., including free-floating planets. However, WFIRST microlensing will rarely detect multiple planets in a given system and so will not provide information about the complete architecture of individual systems. The results from Kepler and

WFIRST will be combined to provide a more complete statistical census of planetary systems containing planets with mass greater than the Earth and separations from zero to free floating planets, including analogs to all those in our solar system except Mercury.

However, it is important to emphasize that this compendium will be statistical in nature. While WFIRST will almost certainly discover thousands of planets with masses from that of roughly the moon to the mass of Jupiter, and determine their orbital separations distribution, it will not do so for planets with separations significantly smaller than 1 AU. Therefore, it will not be able determine, for instance, if terrestrial worlds in the habitable zones of sunlike stars typically harbor giant planets on longer period orbits, as in our own solar system. As such, it will not be clear whether or not specific architectures like our own, with rocky planets within 2.5 AU concurrent with giant planets beyond 5 AU, are common or rare.

There are a variety of reasons why this subtlety is important. For example, there is considerable controversy over the origin of the water on the Earth. It is somewhat of a cosmic paradox that the material in the regions of protoplanetary disks that correspond to the habitable zone in mature systems is well inside the 'snow line,' the distance from the star where water ice is stable in the near-vacuum of the protoplanetary disk. Thus, planets that form in the habitable zones of their parent stars are almost certainly largely devoid of water. The liquid water that is now on Earth, which we believe is requisite for all life, was thus likely delivered from beyond the 'snow line.' While the details of the physics of water delivery, and even the determination of the primary reservoir of Earth's water, are still an area of active research, most researchers agree that the existence of giant planets beyond the snow will have a significant effect on water delivery to the Earth.

Barring dramatic improvements in radial velocity or astrometric techniques of detecting exoplanets, spaced-based direct imaging is likely the only method that can provide nearly complete architectures of planetary systems, including both potentially habitable inner planets and giant planets beyond the snow line. Furthermore, it is essentially the only technique that will also enable the characterization of the spectra of the majority of these planets in reflected light.

By carefully considering of the trade space of mission architectures, we have determined that a mission capable of searching for potentially habitable worlds and biosignatures can also address the science questions addressed above. It was this discovery that drove the design choice of the baseline HabEx 4H, "hybrid" architecture, which combines a more traditional coronagraphic starlight suppression system with a formation-flying starshade occulter. The Starshade Instrument along with the starshade occulter is able acquire instantaneous broadband spectra over a wide field of view, enabling such spectra for essentially all the detectable planets in a system, as well as the determination of the orbital elements of those planets with orbits of less than roughly 10 AU. However, the starshade flight system, which carries the starshade occulter as its payload, is propellant-limited and can only support ~100 targetings before requiring refueling.

Thus, with the addition of a starshade, the preferred HabEx architecture can naturally address the second of its three primary goals: To map out nearby planetary systems and understand the diversity of the worlds they contain.

### 1.1.2 Broad and Unique Observational Capabilities in the UV through Near-IR

In addition to its ability to detect and characterize exoplanets using advanced starlight suppression technologies, as a great observatory with unique and unprecedented capabilities and instrumentation (e.g., **Figure 1.1-1**), HabEx will open up entire new vistas of scientific inquiry in galactic and extragalactic astrophysics, as well as planetary science.

Highlighted here are a few compelling science themes that HabEx would be uniquely capable of addressing. These are the subset of the themes that were used to define the functional requirements of the HabEx observatory and science instrumentation via the science traceability matrix (**Table 5.1-1**). However, HabEx will achieve vastly more science than is introduced here. In particular, this aspect of the HabEx Observatory would be community-led, through a competitive, funded GO program that would encompass 50% of the primary 5-year mission, and all of any extended mission. In additional, because all four instruments can be operated simultaneously, HabEx will be able to use the two general observatory instruments to obtain deep parallel observations during the direct imaging exoplanet surveys. Such a GO program would leverage advances in our understanding and the community's imagination to produce the most scientifically productive mission.

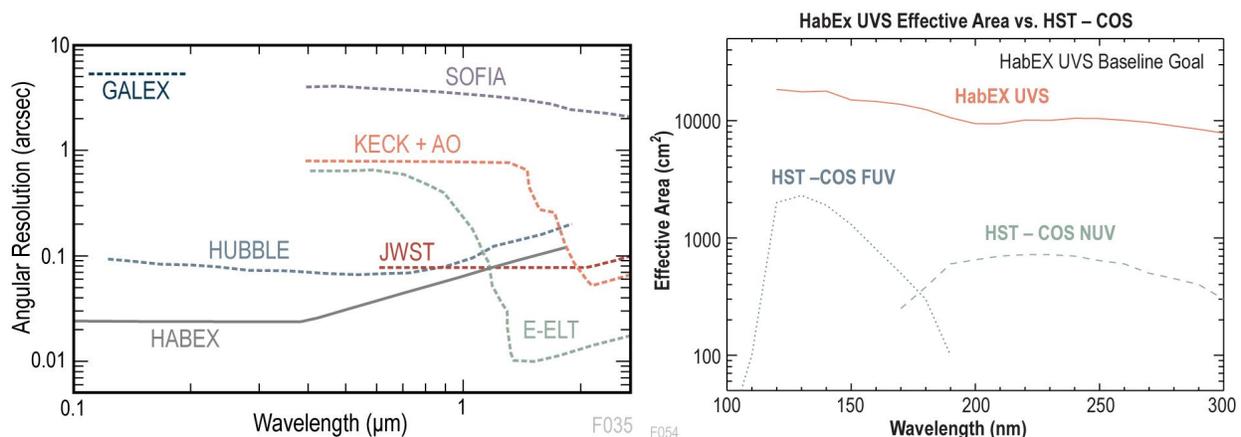

**Figure 1.1-1.** *Left:* HabEx's 4 m, unobstructed, ultra-stable aperture provides unprecedented spatial resolution, and (*right*) effective area at UV and optical wavelengths. These capabilities are not replicated by any currently planned ground- or space-based observatory. This will enable not only the direct detection and characterization of a broad range of exoplanets, including Earth-like planets around sunlike stars, but also unique and exciting astrophysical research, including solar system, galactic, and large-scale structure science. See *Chapters 3* and *4* for details.

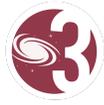

*Carry out observations that open up new windows on the universe from the UV through near-IR*

What is the complete life cycle of baryons and where are the missing baryons in the local universe? What are the sources responsible for the metagalactic ionizing background over cosmic time? What are the properties and end states of the first generations of stars, and what role do they play in the origins of elements? What are the mechanisms driving the formation and evolution of galactic globular clusters? Is there a need for new physics to explain the disparity of local and cosmic microwave background measurements of the expansion rate of the universe? Can we understand the small-scale behavior of dark matter by studying the gas and resolved stellar populations of nearby dwarf galaxies? Do rocky planets orbiting in the HZs of nearby mid-type M stars have potentially habitable conditions?

### Cosmology, Galactic and Extragalactic Astrophysics, and Transiting Exoplanets

HabEx will transform our understanding of the universe by building upon the dramatic progress over the past three decades of research in observational cosmology and galactic and extragalactic astronomy. Although the existence of dark matter has been known for decades (e.g., Zwicky 1933; Rubin and Ford Jr 1970), the detailed accounting of the contents and overall geometry of the universe—e.g., whether it is positively curved, negatively curved, or flat—was not known until the mid-1990s. Even one of the most fundamental properties of the universe, its age, was poorly constrained until the mid-1990s.

The situation is radically different today, with fairly precise measurements of many of the key cosmological parameters. This is due to many breakthroughs, made possible by highly successful observational campaigns, space-based missions, and the enormous efforts of many astronomers. These include the conclusion that the universe is currently accelerating (Riess et al.

1998; Perlmutter et al. 1999), and thus the majority (~70%) of the energy density of the universe today is composed of a mysterious component with negative pressure, now known as dark energy. They also include the exquisite measurement by space-based missions of the cosmic microwave background anisotropies by space-based missions such as the Wilkinson Microwave Anisotropy Probe (WMAP; Spergel et al. 2003) and Planck (Ade et al. 2014). Meanwhile, wide-field ground-based surveys, such as the Sloan Digital Sky Survey (SDSS), provided complementary constraints, such as the measurement of the scale baryonic acoustic oscillations (Eisenstein 2005). Similarly, weak lensing measurements provided estimates of the mass density of the universe and growth of structure.

Because of these and other observations, the overall geometry of the universe is now known to unprecedented precision. Exceptionally accurate measurements of the fraction of the universe's total energy density associated with its key components (i.e., relativistic particles, baryons, dark matter, and dark energy) have now been achieved. Indeed, astronomy is now in the era of precision cosmology.

This era has provided the foundation of our understanding of the large-scale structure and contents of the universe. However, while providing the cosmic "skeleton," it has also invited ever more fine-grained questions about the detailed physics that determines both the large and small-scale structure and evolution of galaxies and stellar systems.

In particular, with regards to baryonic physics, there are many gaps in our understanding. Fundamentally, we do not understand how baryons settle into dark matter haloes, become galaxies, form stars that eventually birth planetary systems, which themselves ultimately (perhaps) give rise to life. Further, we do not understand how these stars ultimately die and expel baryons, including newly created heavy elements, which subsequently become ejected from galaxies in massive outflows, or are trapped in their dark matter halos and are eventually resequestered into new stars, planets, and ultimately become the building blocks of life. In short, we do not understand the life cycle of baryons, which is inextricably tied to cycle of the life. HabEx will provide the capabilities to better understand the evolution of baryons, from the birth of galaxies to the birth of life, and how this process repeats itself over cosmic time.

Refinements and new observations have also led to questions about the full range of validity of the standard cosmological model itself. For example, now that a precise measurement of the mass density of baryons in the universe exists, it has become clear that a detailed census of baryons in the local universe falls short of this total by some ~30%. Where are the local missing baryons? The most recent measurements of the local Hubble constant (the current expansion rate of the universe) appear to be inconsistent with those derived from the Cosmic Microwave Background (CMB) at the ~4.4$\sigma$ level (Riess et al. 2019). Is this an exciting sign of new physics, or simply evidence for systematics in the CMB and/or local Hubble constant measurements? Is the fact that many nearby low-mass, dark-matter dominated galaxies appear to have flattened cores, rather than having the cuspy cores predicted by the most vanilla flavors of cold dark matter, providing clues as to the nature of dark matter, or is it simply due to missing physics in galaxy formation models?

Perhaps more importantly, HabEx will provide unique and unprecedented capabilities never before realized, including the highest resolution images at wavelengths from roughly 0.3 to 1 µm, the largest UV collecting area of any previous satellite, and multi-object spectroscopy from the UV to the near-IR over a relatively large field of view. As a result of these capabilities, HabEx will achieve vastly more science than we can possibly imagined or anticipated based on our current understanding of the universe, its contents, and its evolution.

## Solar System Science

While one of the primary goals of HabEx is the study and characterization of exoplanets, particularly potentially habitable planets, the planets that can be studied in the most detail are those in our own solar system. The bodies in our solar system exhibit diverse and complex behaviors, which are difficult to interpret from first principles. Importantly, improved understanding of these phenomena can then be used to complement and enhance the knowledge gleaned from HabEx's exoplanet exploration and characterization surveys. The representative questions posed above are exemplars of this complementarity between the exoplanet and solar system science enabled by HabEx, but they are certainly only a subset of the solar system applications of HabEx.

To place this into context, it is important to recognize that, concurrent with the rapid progress in the fields of exoplanets and cosmology over the past few decades, our understanding of the contents of the solar system, as well as our models for its formation and evolution, have also undergone dramatic revision. This is due to a combination of ground- and space-based observations and surveys, as well as an impressive fleet of missions that have performed in situ explorations of many of the planets in our solar systems, their satellites, and other small bodies such as Ceres and Vesta.

In many ways, the discovery of the trans-Neptunian belt (Jewitt et al. 1992) heralded the beginning of a transformative era in our understanding of the contents of the solar system. This transformation was further fueled by the discovery of exoplanetary systems, and the recognition that an improved understanding of our solar system will inform our understanding of exoplanetary systems. In turn, the sheer number and diversity of exoplanetary systems places our solar system in context and informs our understanding of its formation and evolution.

Although we now have a much better accounting of the contents of our solar system, from near-Earth objects with sizes of roughly ~100 m, to the giant planets, we have much to learn about the detailed physical processes at work in these bodies. Many of these can be addressed by the suite of imaging and spectroscopic capabilities afforded by HabEx, such as planet-star magnetic interactions, atmospheric escape, cryovolcanism, and the processes by which volatiles are delivered to the inner bodies in the solar system (including the origin of Earth's water).

## 1.2 The Technological Basis for HabEx: Dramatic Progress in the Past Few Decades

### 1.2.1 High-Contrast Coronagraphy

**History of Coronagraphy.** The idea of using an optical device to suppress the glare of a central bright object to study fainter surrounding structures dates back to French astronomer Bernard Lyot, who invented the coronagraph in 1930 to observe the solar corona. With the advent of adaptive optics (AO), wavefront sensing, and control to correct the turbulent effects of the Earth atmosphere, increasingly powerful visible/near-IR coronagraphs came in operation in the 1990s. Bright, self-luminous exoplanets were directly imaged and characterized shortly afterwards for the first time using coronagraphy (e.g., Marois et al. 2008; Lagrange et al. 2008). Ground-based instruments can now detect exoplanets $10^6$ times fainter than their host star at separations of ~0.5 arcsec in the near-IR (Macintosh et al. 2015).

**Recent Technological Advances.** The required improvement over the current state of the art to optically detect planets orbiting solar-type stars that are $10^3$ to $10^4$ times fainter than can currently be detected has been the topic of a vigorous research and development effort throughout the science community since the early 2000s. In particular, JPL testbeds have demonstrated adequate levels of narrow-band starlight suppression (~$6 \times 10^{-10}$, with ten times better stability; Trauger and Traub 2007) at relevant angular separations using unobscured apertures such as the one considered for the baseline HabEx design. The latest laboratory demonstrations, funded under NASA's Strategic Astrophysics Technology / Technology Development for Exoplanet Missions (SAT/TDEM) program have since concentrated on suppressing starlight over broader wavelength ranges and wider regions of the science image, in particular using multiple deformable mirrors for simultaneous correction of amplitude and phase corrugations, and improved coronagraph designs to reach IWAs closer to the diffraction limit (Trauger et al. 2015; Guyon et al. 2014; Serabyn and Trauger 2014; Serabyn et al. 2019).

Finally, work completed by the WFIRST Coronagraph Instrument (CGI) also helped move the HabEx coronagraph technologies forward (**Figure 1.2-1**). Large deformable mirrors with 48×48 elements have now been shown to work in the expected thermal environment and a more demanding vibrational environment. Notably, the CGI wavefront sensing and correction system has also already been shown to provide remarkable performance in the laboratory, with pointing and low-order wavefront drift residual errors adequate for reaching contrasts of ~$10^{-9}$ at $3\,\lambda/D$ with WFIRST. Interestingly, because the baseline HabEx 4 m telescope is optimized for coronagraphy with its off-axis primary mirror and slower beam, it is more tolerant to aberrations than the 2.4 m WFIRST. As a result, the level of low-order wavefront control demonstrated by the WFIRST CGI may already be adequate for reaching contrasts of ~$10^{-10}$ at $2.4\,\lambda/D$ with HabEx.

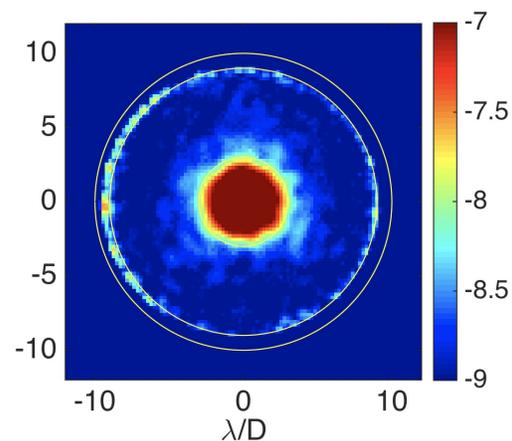

**Figure 1.2-1.** The WFIRST coronagraph testbed has achieved a 360° dark hole, with a contrast ratio of $10^{-9}$ from $3\,\lambda/D$ to $8\,\lambda/D$ at wavelength, $\lambda$, where $D$ is the telescope diameter. Results shown are unpolarized light with a 10% bandwidth centered at 0.55 µm. (Jun-Byoung Seo et al., private communication, Dec. 2016).

## 1.2.2    High Contrast with a Starshade

**History of Starshades.** The idea of using a starshade to image planets was first proposed in 1962 by Lyman Spitzer at Princeton (Spitzer 1962). In this landmark paper, he proposed that an external occulting disk could be used to block most of the starlight from reaching the telescope, thus enabling the direct imaging of planets around nearby stars. He realized that diffraction from a circular disk would be problematic for imaging an Earth-like planet due to an insufficient level of light suppression across the telescope's pupil. He posited that a different edge shape could be used instead, foreshadowing today's approach.

It was the seminal paper by Cash (2006) that showed that an occulter consisting of an opaque solid inner disk surrounded by petals forming an offset hypergaussian function, tip-to-tip about 60 inch diameter, created a broadband, deep shadow. With a small IWA and reasonable manufacturing tolerances, this design finally allowed for the possibility of an affordable solution.

Designs based on a solid inner disk and shaped petals form the basis of several variations in the apodization function. Vanderbei et al. (2007) developed a nonparametric, numerically generated approach to petal shape design. The resulting numerical designs allow for optimization considering engineering constraints, such as petal tip and valley width, petal length, and overall diameter, while preserving desired science performance.

In 2008, two teams were selected under the Astrophysics Strategic Mission Concept Study (ASMCS) to study starshades. Cash et al. (2009) developed the New Worlds Observer mission concept, while Kasdin (2009) developed the Telescope for Habitable Exoplanets and Intergalactic/Galactic Astronomy (THEIA) concept. Both missions used 4 m aperture telescopes coupled with a starshade to achieve the sensitivity required to characterize Earth-like planets in the habitable zones of their parent stars. More recently, the Exo-S report (Seager et al. 2015) presents two probe-class exoplanet direct imaging mission concepts, a rendezvous mission designed to work with the WFIRST 2.4 m telescope, and a dedicated mission with the co-launch of a 1.1 m telescope and a starshade.

**Recent Technological Advances.** A number of key starshade technologies have already been demonstrated to a high level through the TDEM component of NASA's SAT program since 2009, including manufacturing starshade petals (Kasdin et al. 2012) and verifying deployment mechanisms (Kasdin et al. 2014) at the required precision, the development of stray light mitigation techniques through modeling and sharp-edge materials development (Casement et al. 2016), and starlight suppression demonstration and model validation through field experiments (Glassman et al. 2016).

Further starshade technology work has been advanced through the "Starshade to Technology 5" (S5) project. Under that project, formation flying sensing has reached Technology Readiness Level (TRL) 5. Contrast performance modeling and validation and starshade scattered sunlight for petal edges (Martin et al. 2013) will reach TRL 5 before the National Academies issue their Decadal Survey report. Subscale deployment and shape stability testing will bring the overall starshade to TRL 5 by 2022, before the start of the HabEx project as currently planned

## 1.2.3    Large Mirror Technology Advances

The baselined HabEx architecture utilizes a 4 m, off-axis, monolithic primary design. Off-axis monolithic primary designs highly benefit coronagraphs; providing higher throughput, smaller IWA, and larger contrasts. This is provided that the primary can be made sufficiently stiff, and the mirror coatings sufficiently smooth. Directly imaging exo-Earths with the preferred HabEx design only requires modest improvements relative to existing mirror coating technology. SCHOTT now routinely makes 4 m blanks for the microlithography industry and cast a 4.2 m secondary mirror for the European Extremely Large Telescope (E-ELT) earlier this year. The primary challenge for the Zerodur® mirror had been its open back design, making it less stiff than closed-back mirrors. This first issue was addressed by an experienced space telescope manufacturer, UTC Aerospace Systems/Collins Aerospace (UTAS), with their method for measuring surface figure during manufacturing while in Earth's gravity. UTAS has been able to produce precision surfaces using this approach. The second issue is addressed in the HabEx design by eliminating the primary source of vibrational disturbance in traditional bus design, specifically the reaction wheels.

### 1.2.4 Spacecraft Pointing and Vibration Control Advances

HabEx's stability requirement is set by the coronagraph error budget, but technologies to meet these requirements have been developed and demonstrated in space for other ultra-stable observatories. Like the European Space Agency's (ESA's) highly successful Gaia astrometry mission and the ESA/NASA Laser Interferometer Space Antenna (LISA) Pathfinder mission, HabEx has replaced reaction wheels with microthrusters for tight pointing control. The microthrusters only offset the effects of solar pressure; Earth-Sun L2 station keeping and slewing will be handled by a conventional monopropellant hydrazine propulsion system. A microthruster system has been launched on four missions to date, and microthrusters are also baselined for ESA's upcoming Euclid and LISA missions. Gaia has already reached four years of successful L2 operations and will have completed its 5-year baseline mission by the time the HabEx final report is submitted. Like Gaia, HabEx is using a phased array antenna, further reducing environmental vibration and enabling continuous science downlink, even during observations.

In summary, the preferred HabEx design was developed to both minimize risk while maintaining capabilities. This was accomplished by minimizing the number of low TRL technologies, as well as relying as much as possible on heritage from previous missions and mission designs.

## 1.3 The Motivation behind the HabEx Design: A Timely, Compelling Mission in the Context of Realistic Cost, Technology, Risk, and Schedule Constraints.

### 1.3.1 The HabEx Mission Study Guiding Philosophy

Early on in the development of the HabEx mission, it was recognized that any recommendation by the Astro2020 Decadal Survey will have to balance scientific ambition with the programmatic and fiscal realities that currently constrain space exploration. Furthermore, the impact of the prolonged development of any large strategic mission like HabEx on the greater astronomy community must also be considered. Missions that are too large in scope or take too long to develop may limit the ability of NASA to fund a broad portfolio of science investigations. Therefore, from the beginning, the HabEx study was guided by the philosophy of developing a mission capable of the most compelling science possible while still adhering to likely cost, technology, risk, and schedule constraints. Simply put, HabEx was designed to be a Great Observatory that can be realized in the 2030s.

Indeed, the preferred HabEx architecture was chosen to be technically achievable within this time frame by leveraging the extant technologies that were matured through investments recommended by the 2010 Decadal Survey, as well as through the development of related technologies for the WFIRST. The technologies that HabEx requires have thus benefited from years of investments by NASA and others. They also have planned demonstrations on precursor missions, assuring that TRL 5 will be met by 2025. Furthermore, HabEx is conservatively designed with substantial margins. HabEx therefore benefits from lower levels of risk and higher levels of technical maturity than many previous large strategic mission concepts. As an example, no new institutions, fabrication, or test facilities are required for HabEx, dramatically reducing the risk of cost and schedule growth.

### 1.3.2 Why Now?

Initiating the development of a mission like HabEx was largely impossible even a decade ago. Yet the science community has gained the scientific knowledge in many areas that now motivate the three primary science goals outlined in the following chapters and the science traceability matrix (**Table 5.1-1**). This is no truer that in the area of exoplanet direct imaging, where we now know, from focused investments from NASA and other agencies, that are rocky planets in the habitable zones sufficiently common that they can be discovered and characterized around the nearest stars with a modest aperture telescope. We also know that these stars are not too dusty as to prevent the detection and characterization of Earth analogs.

The past three decades have also witnessed enormous progress in the technologies needed to answer both the general astrophysics questions that are now waiting to be answered, but also the technologies needed to detect and characterize nearby Earth analogs. Specifically, dramatic technological progress in four key areas, accomplished over the last three decades, make HabEx possible today: high-contrast imaging with coronagraphs, starshade-specific technology developments, manufacturing of large monolithic mirrors, and vibration control using microthrusters for fine spacecraft pointing.

### 1.3.3 The Motivation Behind the HabEx Preferred Architecture

The preferred HabEx architecture is a 4 m, monolithic, off-axis telescope that employs two starlight suppression technologies: a coronagraph and a starshade. By carrying two such technologies, the HabEx mission hedges against the risk that one will not perform to the required specifications. The two starlight suppression technologies also make the hybrid version of HabEx much more powerful than a version containing one or the other. HabEx also has two additional instruments, each capable of both imaging and spectroscopy. Together they are capable of observing from ultraviolet to near-IR. Neither instrument rely upon on any nascent

**Why a Coronagraph _and_ a Starshade?**

During the course of this study, it became clear that the combination of both the coronagraphic and starshade suppression technologies was far more powerful than either one alone. The basic reason for this is clear: each have their own strengths and weaknesses, which are almost completely complementary. Because it is located in the telescope itself, a coronagraph is very nimble, and is therefore able to observe many systems, or individual systems many times. As a result, it can survey many stars to search for, e.g., Earth-like planets, and obtain orbits to demonstrate that candidate Earth analogs indeed have orbits in the HZ. However, coronagraphs generally have poorer raw contrast, less throughput, larger IWAs, smaller fields of view, and, most crucially, are only able to take spectra in relatively narrow bandwidths of ~20%. Therefore, obtaining broadband spectra of a planet with a coronagraph is very time consuming. On the other hand, starshades generally have excellent raw contrast, relatively small IWAs, OWAs limited by focal plane size, and relatively large fields of view. Most importantly, starshades are largely achromatic, thereby allowing for very large instantaneous broadband spectra of all of the planets within the integral field spectrograph's field of view. It is therefore able to get complete 'family portraits' of planetary systems, including orbits and broadband spectra of most. However, because the starshade is located nearly 100,000 km away from the telescope, it takes two weeks on average to reposition (or transit) from one target to another. As a result, starshades typically have a limited number of transits, making it very difficult to survey a large number of stars for Earth analogs, and difficult to obtain orbits for candidate Earth analogs. Notably, when used in combination, each method essentially negates the major weakness of the other method. For example, for the primary HabEx goal of finding and characterizing potentially Earthlike planets orbiting nearby sunlike stars, the coronagraph will be used to survey a relatively large number of stars, and then determine orbits for promising candidates. Once these are confirmed to be in the HZ of their parent stars, the starshade can be repositioned to the system to obtain a complete, broadband spectrum of the candidate Earth analog, and search for signatures of habitability as well as biosignatures. This hybrid strategy will also allow one to obtain broadband spectra and orbits of the majority of the other planets in the fields of view of the coronagraph and starshade instruments. Notably, although up until recently, starshade technology development was lagging behind that of coronagraph technology develop, due to focus investments by NASA, this is no longer true, and two technologies are now roughly on par in terms of their development.

(TRL 3 or less) technologies. Nevertheless, these instruments on HabEx will provide the community with imaging and spectroscopic capabilities orders of magnitude better than the Hubble Space Telescope, on a stable and quiet platform that will be able to observe more than 85% of the time. Discoveries made in the 2020s and 2030s will require follow-up observations with a mission such as HabEx, which provides a strong complement to upcoming observatories by uniquely enabling UV observations, as well as higher angular resolution than any planned space- or ground-based facility between roughly 0.3–1 μm.

### 1.3.4 A Buffet of Alternate Architectures: The Complex HabEx Tradespace

Acknowledging that the constraints that must be considered by the Astro2020 Decadal Survey may be difficult to anticipate, or may even change over time, this study also considers eight other architectures. By doing so, this report is responsive to one of the recommendations in the National Academy of Sciences report "Powering Science: NASA's Large Strategic Missions," which states that "NASA should ensure that robust mission studies that allow for trade-offs (including science, risk, cost, performance, and schedule) on potential large strategic missions are conducted prior to the start of a decadal survey. These trade-offs should inform, but not limit, what the decadal surveys can address" (p. 57).

The baseline HabEx mission design is the result of numerous trade studies to best meet science-driven observing requirements with available design options. In particular, we considered monolithic versus segmented primaries, fast versus slow primaries, on-axis versus off-axis secondaries, reaction wheels versus microthrusters, and coronagraphic and starshade starlight suppression systems, as well as hybrid systems that employ both. Each of these trades came with benefits and drawbacks, as well as different risk postures and cost and schedule constraints. A discussion of these architecture trades is presented in *Chapter 10*.

Although, for the reasons briefly mentioned above and described in detail later in the report, we ultimately chose to focus on one preferred architecture, with a relatively slow *f*/2.5 4 m monolithic primary mirror, off-axis secondary, and hybrid starlight suppression system. We emphasize that this last choice (to employ both a coronagraph and a starshade) is crucial for achieving all of our science goals, and makes our preferred architecture much more powerful than an architecture with only one of these starlight suppression systems alone.

Nevertheless, all nine architectures still enable groundbreaking science, including the direct imaging and characterization of exoplanets. The other eight architectures are presented in *Chapter 10* and offer flexibility in budgeting and phasing, such that HabEx may still be compatible with a balanced portfolio, even for the most pessimistic fiscal projections. Thus, by evaluating all nine design architectures for science return and cost, this study provides the Astro2020 Decadal Survey additional flexibility in its decision making. As a specific example, six HabEx architectures utilize a starshade, making it possible to phase the starshade launch to occur after a telescope launch. This provides the Decadal and NASA options with additional schedule flexibility, which may therefore be compatible with a broader range of budgetary realities.

### 1.4 HabEx: A Great Observatory Enabling the Golden Era of Astronomy and Solar System Science

The time for HabEx is now. Due to rapid technological advances resulting from focused investments, as well as strategic investments in ever more capable ground- and space-based observatories, HabEx is well positioned to be the great observatory of the 2030s, following in the footsteps of such transformative missions as the Hubble Space Telescope and the soon-to-be-launched JWST. It will also be the first facility capable of directly imaging and characterizing Earth-like planets orbiting in the habitable zones of sunlike stars, and thus will be the first facility capable of discovering potentially habitable words, and searching for evidence of life, e.g., biosignatures, in these worlds.

Enormous progress over the past three decades has produced revolutions in our understanding of the inventory of own solar system, the diversity other planetary systems, and indeed the contents and history of the entire universe. However, as described above, these revolutions have raised as many questions as answers, some of which are stated explicitly above. HabEx has been designed to answer a diverse set of science objectives organized around three primary science goals.

Between now and the expected launch of HabEx, many missions and facilities will come online that will begin to answer these questions. By building upon newfound understanding in these fields, leveraging recent technological advances, and by identifying gaps in these areas of science inquiry that will *not* be filled by existing or planned facilities, HabEx will play a unique and crucial role in addressing many key scientific questions that cut across the full range of the NASA astrophysics and solar system portfolios. In particular, by optimizing the design of HabEx to provide unique, powerful, and unprecedented, yet capabilities attainable with relatively low risk technology development, the HabEx Observatory will play a critical role in the next era of discovery, characterization, and understanding in this vast array of topics in astronomy, from cosmology to the study of solar systems, ours and others.

# 2  ASTRONOMY IN THE 2030S

HabEx is envisioned as a visionary observatory for the 2030s, providing unprecedented capabilities that will enable bold scientific investigations that are not possible with the suite of instruments expected in that decade. Between now and the expected launch of HabEx in the mid-2030s, many new observatories will come online, both space-based and ground-based, small and large. Some of these facilities are currently under construction or being planned, and some have yet to be envisioned. Notable examples include the James Webb Space Telescope (JWST), the Wide Field Infrared Survey Telescope (WFIRST), and the next generation of giant ground-based telescopes, with apertures larger than 20 m. These facilities span a range of capabilities and driving science goals. This section reviews the expectations for how the astrophysics science landscape is likely to develop over the next 15 years, leading up to and concurrent with the launch of the HabEx Observatory, to demonstrate how the mission will complement results from those facilities and enable science well beyond what can already be expected from current and upcoming facilities.

## 2.1  Exoplanet Science before the Launch of HabEx

Planetary systems consist of an enormous diversity of planets: gas giant planets, ice giants, sub-Neptunes, super-Earths, rocky terrestrial planets, and belts of small bodies that generate debris particles. Ongoing research, upcoming developments in ground-based facilities, and the launch of new space missions will continue to advance our knowledge of the variety and nature of these exoplanetary system components over the next decade and a half. Even so, a flagship exoplanet direct imaging and spectroscopy mission like HabEx will provide unique capabilities. The following subsections set the likely context for exoplanet science at the time of the launch of HabEx.

### 2.1.1  Exoplanets from Stellar Reflex Motion

Radial velocity (RV) surveys have detected 762 exoplanets as of August 2019,[1] with a median orbital period of roughly 1 year. The median RV semi-amplitude of these detections is 38 m/s. To date, only 16 exoplanets have been reported with RV semi-amplitudes below 1 m/s. The planets with the lowest claimed RV amplitudes to date are tau Ceti e and f (Tuomi et al. 2013), both at roughly 0.4 m/s. The current best measurement precision is expected to improve toward 10 cm/s through the development of a new generation of RV spectrographs on large telescopes, observations at higher cadence, and improved calibration methods (Fischer et al. 2016). Stellar RV jitter arising from spots and activity sets a natural noise floor near 2 m/s (Bastien 2014). Through careful averaging, filtering, and detrending of the data, the noise from stellar activity may be mitigated, allowing for RV detections of planets with semi-amplitudes below 1 m/s. Control of systematics at levels considerably better than 10 cm/s level will be required to both identify Earth analogs in the habitable zones (HZs) of sunlike HabEx targets and measure their masses. It is unclear at this writing whether future RV performance improvements will extend to these levels by 2035 for a significant fraction of the stars in the HabEx target sample (see *Chapter 12* for an extended discussion of the prospects for improving the precision and accuracy of the RV method).

Most of the known Jupiter-mass planets within a few AU of stars with types F5 and later come from RV surveys, but these surveys generally lack sensitivity to Neptune-mass planets outside a few tenths of an AU (Fulton et al. 2016). By 2035, a dedicated observing program with instruments and capabilities available today could achieve sensitivity to Saturn-mass and greater planets with orbital periods up to 20 years, and to super-Earths ($\sim$8 $M_\oplus$) with periods of several years. Complementary measurements of stellar astrometric wobbles by the European Space

---

[1] http://exoplanetarchive.ipac.caltech.edu

Agency (ESA) Gaia all-sky survey will be available by 2022. Gaia should detect and measure the full orbits for planets of Jupiter mass or larger with periods <5 years across most of the HabEx sample. Altogether, a complete census along with an accurate measurement of the orbital elements of inner giant planets of nearby stars should be well in-hand by 2035.

## 2.1.2 A Nearly Complete Statistical Census of Exoplanets: Kepler and WFIRST

Transit surveys have detected 3100 exoplanets as of August 2019,[2] largely thanks to NASA's Kepler mission. By 2035, this number is expected to increase by at least a factor of 10, considering expected results from the transit survey missions Transiting Exoplanet Survey Satellite (TESS) (Ricker et al. 2015) and PLAnetary Transits and Oscillations of stars (PLATO). TESS is now operating and has detected nearly a 1,000 transiting exoplanet candidates.[3] PLATO has a planned launch date in 2026. Nearly all transit detections are for short orbital periods, <1 year; the median orbital period of current transit detections is 9 days. The TESS and PLATO surveys will enable mass, radius, and density constraints on the detected exoplanets, which will also be suitable for follow-up transit spectroscopy. In particular, TESS should complete the survey of bright (V < 9) field stars across the sky. With the RV follow-up that will be possible for such targets, the frequency of planets as a function of their mass, and the planetary mass-radius relationship, should be well established for short orbital periods by the time HabEx launches in 2035. Note that both of these missions will have difficulty detecting rocky planets in the habitable zones of sunlike stars due to their long orbital periods and small transit signals.

The WFIRST observatory, planned for launch in 2025, will include a microlensing survey for exoplanets. With its dramatically higher data quality over ground-based microlensing surveys, WFIRST microlensing will discover thousands of exoplanets orbiting beyond the snow line with masses as small as Mars. The results will robustly define the mass and separation distribution of planets orbiting beyond a few AU in a representative sample of the Galactic stellar population. It will also enable the determination of the compact object mass function (including free-floating planets) over nearly eight orders of magnitude from objects with the mass of Mars to ~30 solar mass black holes. In combination with the results from Kepler, WFIRST will enable the completion of the statistical census of exoplanets with mass greater than the Earth and separations from zero to infinity (**Figure 2.1-1**). This will be the ultimate empirical dataset with which to validate planet formation models that span the full scale of planetary systems, and set expectations for the range of planet sizes and orbital locations that HabEx will study.

## 2.1.3 Characterizing Exoplanet Atmospheres: Transiting Planets

There are three primary methods of studying the atmospheres of transiting exoplanets. First, one can measure a thermal emission or reflection spectrum by measuring the drop in the combined planet plus star flux as the exoplanet passes behind the star. Second, when the planet passes in front of the star the absorption spectrum of the planet's atmosphere can be measured. The starlight is filtered through the intervening planetary atmosphere and absorbed at wavelengths corresponding to atomic and molecular transitions in that atmosphere. Finally, one can measure the variation in the combined stellar and planetary brightness as the planet orbits its star. To date, spectral observations of giant exoplanets in transit have confirmed identifications of Na I, $H_2O$, and Ly $\alpha$, but is also clear that high-altitude hazes are likely extant in a significant fraction of irradiated planets (Deming et al. 2009; Sing et al. 2016). These hazes obscure molecular features, and appear be particularly significant for sub-Neptune planets (Kreidberg et al. 2014; Lavvas et al. 2019). The physical mechanisms that dictate whether or not a planet will exhibit hazes remains unclear.

---

[2] http://exoplanetarchive.ipac. caltech.edu

[3] https://exoplanetarchive.ipac.caltech.edu/docs/counts_detail.html

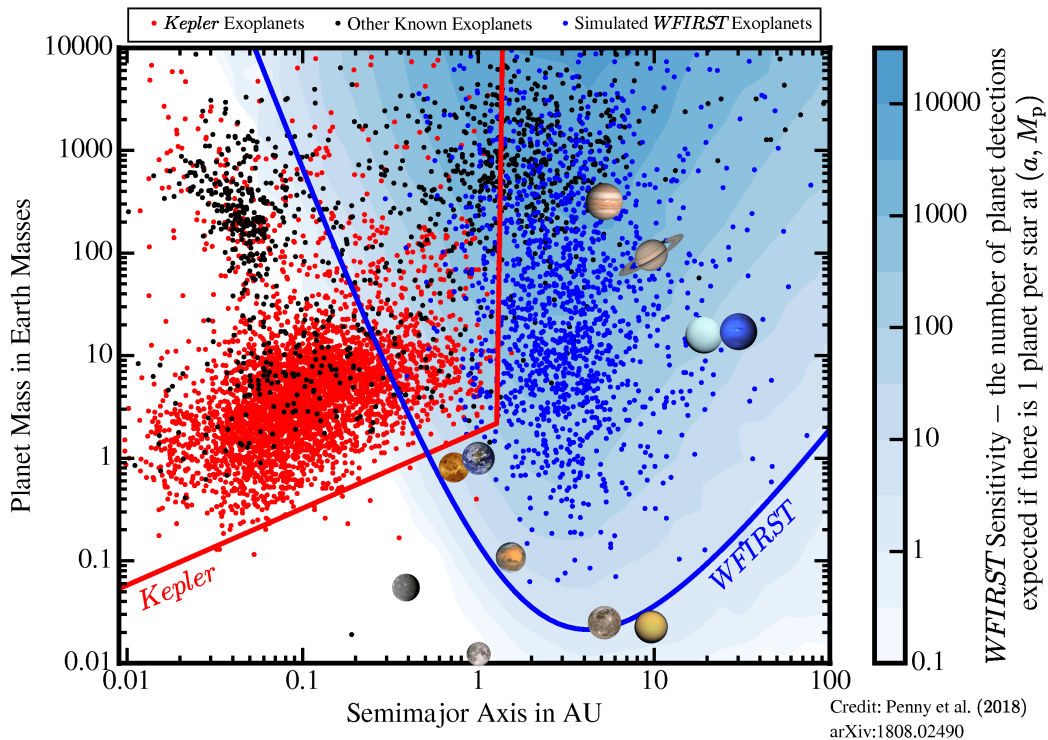

**Figure 2.1-1. Together, Kepler and WFIRST will complete a statistical census of exoplanets – but critical discovery space remains.** The red points show a subset of the detections from Kepler, whereas the black points show planets detected by other techniques. The blue points show a prediction of the planets that will be detected by WFIRST. The red and blue curves show the sensitivity limits of Kepler and WFIRST, where three planets would be detected if every star hosted a planet at that mass and semi-major axis. The small images are planets in our solar system, as well as several giant moons. HabEx will uniquely populate the empty discovery space at the bottom of the figure. Figure based on Penny et al. (2019).

By focusing on red dwarf stars, the K2 (Howell et al. 2014) and TESS (Ricker et al. 2015) missions, along with specially designed ground-based surveys such as MEarth (Charbonneau et al. 2009) and Search for habitable Planets EClipsing ULtra-cOOl Stars (SPECULOOS),[4] can detect transiting rocky planets with transit depths of 0.1–1.0%, JWST will obtain spectra of a small sample of super-Earth atmospheres and a larger sample of mini-Neptune atmospheres. Mid-IR wavelengths should penetrate haze layers that have hampered the detection of absorption features in near-infrared (IR) transit spectra to date. These observations should be able to establish definitive trends in atmospheric composition and cloud properties as a function of planet or host star properties. Overall, the JWST mission should provide spectra for dozens of warm to hot (mildly to highly irradiated) exoplanets, measuring their temperatures, albedos, and composition with greater sensitivity than ever before (Cowan et al. 2015). ESA's newly selected Atmospheric Remote-sensing Infrared Exoplanet Large-survey (ARIEL) mission, scheduled for launch in 2028, is expected to obtain transit spectra of up to a thousand short-period giant planets but will not have the sensitivity to study rocky planets.

Potentially more intriguing is that JWST could—with an optimal target, a large amount of observing time, and some luck—detect habitable conditions on a rocky transiting exoplanet in the HZ of a nearby mid-to-late-type red dwarf star. Detections of biosignatures (such as $O_2/O_3$ in disequilibrium with $CH_4$) will be difficult, and may only be possible for bright red dwarfs hosting large (but not too massive) exoplanets in their HZ. This will also require excellent control of systematics to definitively measure the exceptionally small (tens of parts per million) signal, as well as a concurrent major investment in observing time to reach the statistical precision needed to detect such signals. JWST spectroscopy of these "small black shadows" found by the aforementioned surveys, as well as TESS, around red dwarf stars will provide humanity's first opportunity to search for life outside the solar system. One of the greatest strengths of searching for habitable planets and biosignatures of transiting temperate terrestrial planets orbiting low-mass stars is that the targets will already be known. Indeed, at least two systems have already been identified that host transiting rocky planets in the habitable zones of their parent stars, specifically LHS 1440 (Dittmann et al. 2017) and TRAPPIST-1 (Gillon et al. 2017). HabEx will explore a second, complementary path of direct-imaging spectroscopy of "pale blue dots" around sunlike stars.

---

[4] http://www.amaurytriaud.net/Main/SPECULOOS/index.html

### 2.1.4 Characterizing Exoplanet Atmospheres: Direct Imaging

Forty-four potential planetary-mass companions have been imaged around young stars in the near-IR, although less than a dozen have masses securely below 13 Jupiter masses. For a few of these, detections of $CH_4$, $CO$, and $H_2O$ have been achieved. The Gemini Planet Imager (GPI) and the Very Large Telescope (VLT) Spectro-Polarimetric High-contrast Exoplanet REsearch (SPHERE) ground-based coronagraphs have been operating since 2014, detecting fewer self-luminous exoplanets than anticipated. Their best planet/star contrast achieved to date on a typical science target is $\sim10^{-6}$ at a separation of 0.4 arcsec (Macintosh et al. 2015).

Construction has commenced on two Extremely Large Telescopes (ELTs) in Chile: the European ELT (E-ELT) set for completion in 2024, and the Giant Magellan Telescope (GMT) scheduled for first-light in 2024. Presumably, these telescopes will eventually be equipped with the extreme adaptive optics (AO) systems needed to enable high-contrast coronagraphic imaging, and such systems will likely be built and operating by 2035. For broadband direct imaging, they will be limited by the Earth's atmosphere to contrasts no better than $10^{-8}$ in the near-IR, which should be sufficient to spectrally characterize a modest sample of warm giant planets detected by RV surveys.

The use of high-dispersion spectroscopy to isolate exoplanet signals from starlight has recently shown significant advances (Snellen et al. 2014; Birkby et al. 2017). This method cross-correlates an observed spectrum with a model template spectrum of the exoplanet. It relies on there being a large number of molecular absorption features in the exoplanet atmosphere spectrum to provide a measurable correlation signal. Performance models suggest Earth-like planets located in the habitable zones of a small sample of nearby red dwarf stars can be found when this technique is combined with coronagraphy on the ELTs (Snellen et al. 2015; Wang et al. 2017). The newly discovered low-mass planets in the habitable zones of Proxima Centauri (Anglada-Escudé et al. 2016) and Ross 128 (Bonfils et al. 2017) will be prime targets. Many years of ELT work with spectral template cross-correlation will have taken place by the time HabEx launches. The contrast and IWA capabilities for these detections ($\sim3\times10^{-8}$, $\sim30$ mas) will limit the results to near-IR ($>1$ μm) studies of the HZs of perhaps a dozen of the nearest red dwarf stars.

Finally, it may also be possible to detect some planets in thermal emission at wavelengths of $\sim10$ μm. Although the sky background due to the atmosphere is exceptional high at these wavelengths, the very high diffraction limit implies that, if these ELTs can operate with AO at these wavelengths, they can suppress the background to the point that some planets will be detectable in thermal emission, including known RV giant planets, and perhaps even a handful of small temperature super-Earth planets (Quanz et al. 2015).

In conclusion, while there are many promising avenues to explore potentially habitable planets around low-mass, red dwarf stars in the next few decades, spectral characterization of the reflected light of rocky planets in the HZs of sunlike stars will require a space mission, as exemplified by HabEx.

**Figure 2.1-2** clearly illustrates that point by placing HabEx in the context of existing and future facilities. Only a mission like HabEx may provide the huge performance improvement required to extend current characterizations of bright self-luminous giant exoplanets to mature rocky Earth-like planets orbiting sunlike stars.

### 2.1.5   Circumstellar Disks and Dust

Almost 300 resolved disks around nearby stars are known today. Their internal structures are of great interest, as they can be driven by perturbations from unseen planets. Over the last several years, there has been a surge in the number of resolved disks in continuum and line emission, due primarily to the Atacama Large Millimeter Array (ALMA). Rings, gaps, and spirals have been observed in the disks of HL Tauri (Ricci et al. 2015), TW Hydrae (Andrews et al. 2016), Elias 27 (Pérez et al. 2016) and in a large survey of 20 targets (Andrews et al. 2018), perhaps suggesting the presence of forming planets. Protoplanetary disks in the nearest star-forming regions (distances of ~150 pc) are ideal ALMA targets, as their optical depths give them high surface brightness in the submillimeter continuum. For these targets, ALMA's ultimate spatial resolution of 0.01 arcsec will be achievable. An exciting prospect for the 2030s is the follow-up of ALMA disk images with AO coronagraphic imaging on ground-based ELTs: imaging the protoplanets within the disk gaps and directly observing the planet/disk interaction. The ~20 mas inner working angles (IWAs) provided by the ELTs will be enabling for such studies. By the time HabEx launches, ALMA will have thoroughly explored the nearby populations of protoplanetary disks and defined the key targets for follow-up imaging in reflected light with HabEx.

Debris disks are distinct from protoplanetary disks as they are found around older main-sequence stars that have almost certainly ceased giant planet formation. As in our own solar system, they are likely signposts of extant planets and their ongoing dynamical sculpting of the reservoirs of small bodies by collisions, creating belts of material such as the main asteroid belt and the Kuiper belt. These collisions supply the small dust particles that reflect the starlight, ultimately revealing the existence of these belts and perhaps planets. Many are located relatively nearby with distances of only ~25 pc. They are optically thin with a much lower dust content and much fainter submillimeter continuum emission than protoplanetary disks. It is therefore challenging even for ALMA to resolve their detailed structure. ALMA will map a limited number of the brightest debris disks (i.e., those with total fractional luminosity greater than $10^{-4}$ of their host star) at 0.1 arcsec resolution (e.g., **Figure 2.1-3** and Macintosh et al. 2015). Scattered light observations with large diffraction-limited telescopes provide comparable resolution, but not comparable sensitivity, and show a strong detection bias towards debris disks inclined close to edge-on. Nevertheless, interesting systems have been discovered coming out of the GPI and SPHERE missions (e.g., Currie et al. 2015; Bonnefoy et al. 2016). There are hundreds of nearby (unresolved) Kuiper debris disks with a fractional luminosity of less than $10^{-4}$ observed by Herschel, Spitzer, and Wide-field Infrared Survey Explorer (WISE) studies that neither ALMA nor AO coronagraphy are able to directly detect. JWST will image some of these in thermal emission around a small sample of nearby luminous stars, but only with 0.3" resolution at $\lambda = 20$ μm. In 2035, most of the nearby debris disks detected by Spitzer, Herschel, and WISE will still be too faint for the available detection methods, and thus will be ripe for a space observatory like HabEx—with the sensitivity, contrast floor, and resolution—to make the first high-resolution images.

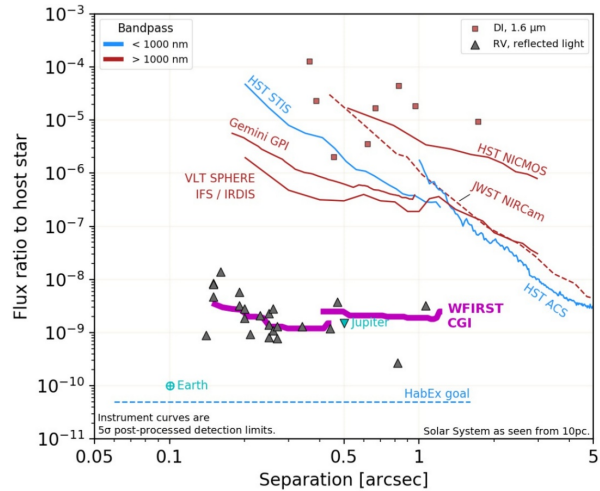

**Figure 2.1-2.** The HabEx planet-to-star flux ratio performance goal in the context of known planets and existing and planned high-contrast direct imaging instruments. Shown is the flux ratio between a planet and its star (points for individual planets) or between the dimmest source detectable (*solid and dashed lines*, assuming a 5σ detection after post-processing) and its star (for instrument performance curves) versus the projected separation in arcsec. The *black triangular points* are estimated reflected light flux ratios for known gas giant RV-detected planets at quadrature, with assumed geometric albedo of 0.5. Red squares are 1.6 μm flux ratios of known self-luminous directly imaged (DI) planets. *Cyan points* represent the Earth and Jupiter at 10 pc. Figure courtesy of K. Stapelfeldt, T. Meshkat & V. Bailey.

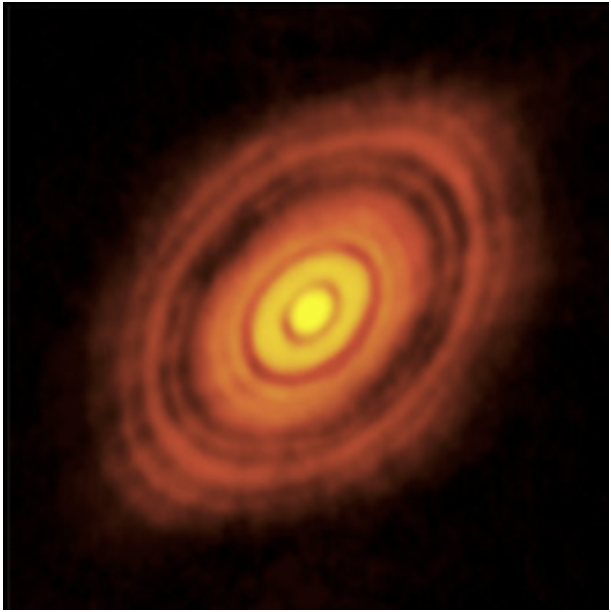

**Figure 2.1-3.** ALMA image of the young star HL Tau and its protoplanetary disk. This best image ever of planet formation reveals multiple rings and gaps that herald the presence of emerging planets as they sweep their orbits clear of dust and gas. Credit: ALMA (NRAO/ESO/NAOJ); C. Brogan, B. Saxton (NRAO/AUI/NSF).

For habitable zone dust in nearby stars, the Large Binocular Telescope Interferometer (LBTI) has recently completed a survey of 38 targets. The median exozodi level is no greater than 26 zodis (at 95% confidence; Ertel et al. 2018), and could be substantially less. The design of the WFIRST technology demonstration coronagraph should achieve comparable sensitivity to dust in the HZ of nearby stars. If a WFIRST exozodi science program takes place, it could complete the survey of southern hemisphere HabEx targets inaccessible to LBTI, and provide constraints on the dust albedo for the exozodiacal clouds LBTI did detect. Together these two datasets would allow HabEx to prioritize its targets and improve estimates of the needed integration times (Howell et al. 2014).

## 2.2 The Broader Astrophysics Landscape in the 2030s

Between now and the 2030s, astronomers will commission a wide array of impressive facilities and instruments. These include both surveys that will image large swaths of the sky with unprecedented sensitivity at optical (e.g., Large Synoptic Survey Telescope (Feng et al. 2017 [LSST]), near-IR (e.g., Euclid, WFIRST), and X-ray energies (e.g., eROSITA), as well as facilities with more limited fields of view, optimized for detailed follow-up studies (e.g., JWST, ELTs). Below is a brief discussion of some of the key questions expected to be left partially or wholly unanswered in the 2030s, and how the unique discovery space afforded by a space-based 4 m class UV-to-near-IR telescope will address these questions.

### 2.2.1 Key Science Questions for the 2030s

Several billions of dollars are currently being spent on ground- and space-based facilities with a primary goal of mapping large swaths of the universe in order to study the history of cosmic expansion and address fundamental questions of cosmology. As a byproduct of these studies, many classes of rare, exciting astronomical sources are expected to be found, from dwarf galaxies in the nearby universe, to a hundred-fold increase in the census of strong gravitational lenses, to quasars at redshift z ~ 10 and beyond. These discoveries will demand a range of follow-up studies, some of which will be amenable to ground-based facilities available in that era, but many of which will require space-based follow-up. Besides the exoplanet characterization questions addressed in other portions of this report, a multitude of key science questions are expected to remain unanswered into the 2030s, including, but not limited to, the missing baryon problem, the nature of dark matter, the history of cosmic acceleration, the history of cosmic reionization, the nature of the seeds of supermassive black holes, the sources and physics of gravitational wave events, a detailed understanding of the formation and evolution of galaxies, as well as a range of investigations of bodies within the solar system. It is beyond the scope of this document to detail this extraordinarily broad range of science, covering all non-exoplanet astrophysics, including fundamental physics, cosmology, and planetary science. To highlight a few here would do an injustice to the breadth of scientific terrain demanding study. Instead, *Chapter 4* discusses several of these questions in greater detail, and what instrumentation on a 4 m class near-ultraviolet to near-infrared (UVOIR) space-based telescope would be required to make significant progress in addressing these questions. In short, HabEx would be an extremely powerful successor to the Hubble Space Telescope (HST), providing a unique and important platform for studies that span the breadth of astronomical research.

---

**HabEx Discovery Space**

• Highest angular resolution UV/optical images

• Access to wavelengths inaccessible from the ground

• Ultra-stable platform

---

### 2.2.2 Discovery Space for the 2030s

With no more servicing missions planned, HST is expected to degrade into disservice sometime in the 2020s, thereby shutting off access to the UV portions of the electromagnetic spectrum (e.g., 115–320 nm), as these wavelengths are absorbed by the Earth's atmosphere and no future, sensitive UV observatories are currently approved. Many key diagnostic features are in this energy range, particularly from highly ionized species in hot plasmas. This energy range is essential for studying the hot phase of the interstellar medium (ISM), intergalactic absorption, as well as for understanding the physics of a range of objects, from planets to galaxies to gravitational wave sources. Access to the UV will be essential to the astronomical community for studying the wide array of sources to be found between now and the 2030s. Furthermore, with marked improvements in technology, the grasp of a UV instrument built in the 2030s will greatly exceed a simple scaling with aperture size. With its next-generation UV instrument and large-aperture in space, HabEx would fill this gap in sensitive UV capability, thereby realizing tremendous discovery potential.

There is a similarly large discovery potential for a next-generation visible/near-IR satellite. First light is expected to occur for several 30 m class, ground-based ELTs by the 2030s—specifically the GMT, the Thirty Meter Telescope (TMT), and the E-ELT. Since it is widely recognized that AO will remain infeasible at optical wavelengths for the foreseeable future (i.e., well past the 2030s), the greatest gains for these facilities will occur at longer wavelengths, where diffraction-limited AO-assisted observations of point sources provide gains that scale as aperture diameter, $D$, to the fourth power (i.e., $D^4$) rather than the simple seeing-limited $D^2$ gains provided by the larger aperture. Accordingly, significant effort is going into designing the AO systems for these telescopes, which will allow the full gains from these large apertures to be realized. Indeed, all the first-light instruments for the E-ELT are diffraction-limited, AO-fed infrared instruments, while GMT and TMT include first-light plans for both diffraction-limited, AO-fed infrared instruments and seeing-limited optical instruments. Therefore, the sharpest imaging at optical wavelengths will remain a domain best achieved from space for the foreseeable future.

Finally, space-based observations provide a platform significantly more stable than ground-based observatories, which is essential for a range of science applications, from sensitive weak lensing studies, which require an exceptionally stable, well-characterized point spread function (PSF), to astrometric studies that require a stable, well-characterized focal plane, to studies that require extremely accurate and stable photometry or spectrophotometry.

Much of the extraordinary progress in astrophysics over the past 20 years has been enabled by combining HST's exquisite resolution and stability, with the light-gathering power of larger-aperture 10 m-class telescopes, such as Keck and the VLTs. Often these resources were employed in tandem, with HST providing high-resolution imaging and the ground-based facilities providing spectroscopy (e.g., the Hubble Deep Field). We expect the 2030s to witness similar, but considerably more powerful synergies between HabEx and the ELTs.

# 3 DIRECT IMAGING AND CHARACTERIZATION OF EXOPLANETARY SYSTEMS

Humanity has reached an era where the long-standing scientific desire to seek and investigate new worlds and diverse planetary systems is now matched by the technological ability to fulfill that desire. There exists a wide variety of techniques to detect and characterize exoplanets, each with their own strengths, limitations and biases (e.g., Seager 2010; Perryman 2014; Wright and Gaudi 2013).

Exoplanet detection techniques can be subdivided into direct detection methods, sensitive to the exoplanet's radiation, and indirect detection methods, detecting the influence of the planet on its host star. The primary detection techniques are radial velocity (RV), transits, astrometry, microlensing, timing, and direct imaging. Spectra can generally be obtained for planets that transit or are detected by direct imaging. The uniqueness of space-based direct imaging of planets in reflected light, even in the mid-2030s (*Chapter 2*), resides in the unmatched capability to obtain near-ultraviolet (UV) to near-infrared (IR) spectra of temperate rocky planets around sunlike stars (FGK dwarfs), and search for atmospheric biosignatures and the presence of surface liquid water (*Section 3.1*). At the same time, such observations will bring detailed "family portraits" of images and spectra for planets with a wide range of sizes and semi-major axes, as well as extended dust structures in nearby exoplanetary systems, thereby putting our own solar system in context for the first time (*Section 3.2*). The HabEx exoplanet science objectives, survey strategy, and starlight suppression instruments are optimized to take full advantage of the diversity possible with such observations from space.

After identifying HabEx's main exoplanet science objectives and deriving the top-level functional requirements associated with them (*Sections 3.1* and *3.2*), detailed science yield simulation results are presented for the baseline HabEx 4 m hybrid architecture (*Section 3.3*). Over a nominal prime mission of five years and assuming a representative observing strategy with a 50/50 time split between exoplanet direct imaging and other observatory science themes (*Chapter 4*), the HabEx Observatory will discover and spectrally characterize ~50 planetary systems within 15 pc of the Sun. Assuming no prior knowledge of planets in these systems—a worst-case scenario—HabEx direct imaging exoplanet surveys are estimated to detect and spectrally

---

| Main Exoplanet Science Goals and Objectives | |
|---|---|
| 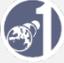 **Goal 1: To seek out nearby worlds and explore their habitability.** | 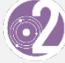 **Goal 2: To map out nearby planetary systems and understand the diversity of the worlds they contain.** |
| **O1:** To determine if rocky planets continuously orbiting within the habitable zone (HZ) exist around sunlike stars, surveying enough stars to detect and measure the orbits of at least 30 exo-Earths if each observed star hosted one. | **O5:** To determine the architectures of individual planetary systems around sunlike stars. |
| **O2:** To determine if planets identified in O1 have potentially habitable conditions (an atmosphere containing water vapor). | **O6:** To determine the variation of planetary atmospheric compositions as function of planet size and semi-major axis in planetary systems around nearby sunlike stars. |
| **O3:** To determine if planets identified in O1 contain biosignature gases (signs of life) and to identify gases associated with, or incompatible with, known false positive mechanisms. | **O7:** To determine if the presence of giant planets is related to the presence or absence of water vapor in the atmospheres of rocky planets detected in O1. |
| **O4:** To determine if any planets identified in O1 also contain water oceans. | **O8:** To constrain the range of possible dust-belt architectures and determine the interplay between planets, small bodies, and dust around nearby sunlike stars. |

characterize over 150 planets over a wide range of surface temperatures and planetary radii, nearly evenly split between rocky planets, sub-Neptunes and gas giants (*Section 3.3*). Remarkably, for the nominal observing strategy and planet occurrence rates assumed (*Appendix C*), HabEx will spectrally characterize and measure the orbits of ~15 habitable zone (HZ) rocky planets around sunlike stars, including ~8 exo-Earth candidates. The wavelength coverage from UV (0.2 μm) to near-IR (1.8 μm), and the spectral resolution (R = 140 in the 0.45–1 μm range) allow HabEx to capture the absorption bands of key molecular species, which can be used to distinguish between different types of exoplanets. These features include, but are not limited to, water vapor bands, oxygen and ozone features, carbon dioxide, and methane bands. All of these features are critical to assessing the habitability of and searching for life on these worlds.

## 3.1    GOAL 1: To seek out nearby worlds and explore their habitability

HabEx shall first search for exo-Earth candidates (EECs), i.e., point sources with separations and fluxes consistent with Earth-size planets orbiting in the HZ of their host stars, and then follow a path of increasingly deeper characterization (**Figure 3.1-1**).

HabEx observations shall successively confirm physical association, determine the planet orbit (*Section 3.1.1*), search for atmospheric water vapor (*Section 3.1.2*), look for atmospheric biosignatures, assess the likelihood of false positives (*Section 3.1.3*), and then conduct even finer characterization in favorable cases, searching for e.g., a "vegetation edge," and the presence of surface water oceans (*Section 3.1.4*).

### 3.1.1    Objective 1: Are there Earth-sized planets orbiting in the habitable zones of nearby stars?

A key requirement for HabEx is that it be able to detect, constrain the orbit and eventually measure the spectrum of at least one HZ exo-Earth with a high degree of confidence despite current uncertainties on the occurrence rate of EECs and distribution of exozodiacal dust (exozodi) brightness levels around nearby main sequence stars. The baseline requirement is that the HabEx design and observing strategy deliver a >95% chance of detecting and spectrally characterizing at least one EEC. The main observational parameter that controls this probability of success is the "cumulative completeness" of the search over the mission life. The search completeness of an individual target star corresponds to the conditional probability of detecting an HZ exo-Earth if there is one around that target. The "cumulative completeness" (allowed to be >1) is defined as the sum of such detection probabilities over the full sample of stars observed. In addition to merely detecting EECs, i.e., point sources with fluxes and separations consistent with HZ exo-Earths, HabEx shall be able to confirm physical association of any such EECs with the host star, determine their orbital parameters and place constraints on their radii. The concept of cumulative completeness can be extended to include orbit determination, which requires a minimum of 4 detections at different epochs rather than a single detection (**Figure 3.1-2**). In order to detect and measure the orbit of one EEC with a >95% probability, detailed yield calculations (*Section 3.3*) show that a cumulative completeness of >20 EECs is required. If the occurrence rate of EECs in the HabEx sample was *e.g.*, $\eta_{Earth} = 0.24$ (Belikov 2017), this would translate into a minimum of ~5 exo-Earths nominally detected with orbits determined and available for spectral characterization (*Sections 3.1.2, 3.1.3,* and *3.1.4*).

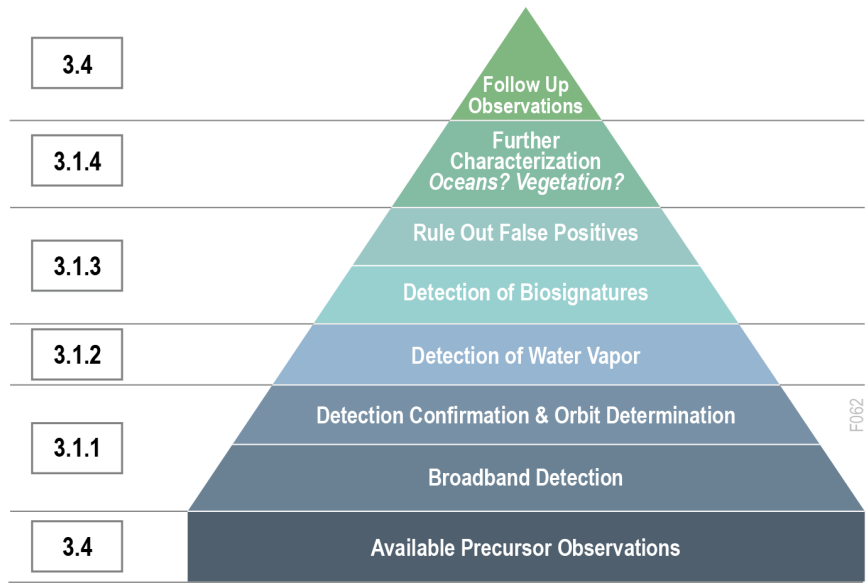

**Figure 3.1-1.** Exo-Earth candidates' (EECs) successive characterization steps (from pyramid base to top) and corresponding sections of this chapter. Precursor and follow-up observations are conducted by other facilities than HabEx (*Section 3.4*).

The "cumulative completeness" figure-of-merit offers the advantage of being a sole characteristic of the instrument and mission observational strategy, rather than an uncontrollable property of the universe, such as the exact distribution and occurrence rate of HZ exo-Earths around nearby stars. A cumulative completeness threshold can then be used effectively to drive instrument design. It is a function of many mission parameters, with telescope diameter, inner working angle (IWA; the approximate inner bound of the high contrast search area for starlight suppression instruments), point source detection limits, and overall survey time driving the calculation. The parameters are degenerate, but yield simulations (*Section 3.3* and

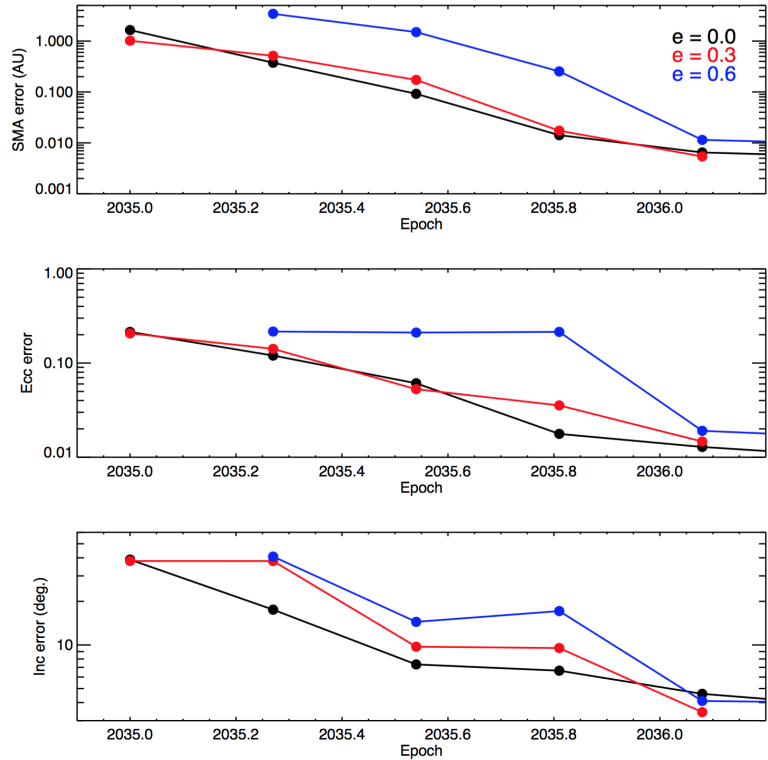

**Figure 3.1-2.** Orbital parameter retrieval simulation. For circular orbits, three well-spaced detections can generally constrain the semi-major axis, eccentricity, and inclination angle of an exo-Earth in a 1 AU orbit to better than 10% (1σ). In the general case, up to four well-spaced detections are required. Credit: Eric Nielsen.

*Appendix D*) demonstrate that a cumulative completeness of ≥20 EECs is achievable using a high-contrast imaging instrument designed to provide a visible planet-to-star flux ratio detection limit of ≤$10^{-10}$ at an apparent separation of ≤80 mas and devoting a total of 2 years of time to the exo-Earths survey (including detection, orbit determination and spectral characterization times).

### 3.1.1.1 Confirming Physical Association

Following the detection of a point source, revisits are required to determine whether the point source is an orbiting planet or a distant background object. For the closest stars (<5 pc) the stellar parallax and proper motion will generally be sufficient to confirm common proper motion in two epochs. Two epochs will also suffice for stars between 5–15 pc, except in rare cases where the candidate orbital motion is along the background track, where a third epoch will be required to break the degeneracy.

#### 3.1.1.2 Orbit Determination

The orbital semi-major axis, combined with the host star's luminosity, determines the stellar irradiance incident on the planet. The stellar irradiance on the planet, in turn:

- Provides an estimate as to whether or not the planet is inside the HZ (e.g., Kopparapu et al. 2013);

- Is needed to infer the reflectivity of the planet from its apparent brightness relative to the star; and

- Is a main input in atmospheric modeling and spectral retrieval.

For spectral retrieval studies, the incident flux on a planet should be measured to better than 10%. Orbital eccentricity affects both the instantaneous and orbit-averaged stellar irradiance incident on a planet. Eccentricity may also provide constraints on the formation and dynamical evolution of the planet's orbit. Measuring the orbit of potentially habitable planets is also important to place the planet in the context of the overall architecture of the planetary system (including inner and outer planets and dust/debris disks). In some multi-planet systems, dynamical stability considerations may allow one to refine the range of possible orbital solutions and, in some cases, allow for constraints on the planet masses (e.g., HR8799 planetary system; Fabrycky and Murray-Clay 2010).

Orbit determination via direct imaging requires more revisits to the target than simply confirming a planet's physical association with its host star. Based on simulations of orbit fitting, up to four well-spaced detections, with position uncertainty ≤5 mas rms (**Figure 3.1-2**), are required to achieve 10% precision measurements on the three, key planetary orbital parameters: semi-major axis, eccentricity, and inclination relative to the sky-plane.

Details of the simulations are as follows. The fiducial cases were chosen to span a variety of planets of interest, with semi-major axis of 0.5, 1, 2, and 5 AU, inclination angle of 30°, 50°, and 80°, planet radius of 0.5, 1, 2, 4, and 11 Earth radii, and eccentricity of 0.0, 0.3, and 0.6 (for a total of 180 cases). Planets were assigned the same time of periastron, position angle of nodes, and argument of periastron, and are all taken to orbit a solar mass star at 10 pc. Planet position measurements were generated with observations every 3.2 months, with a 5 mas rms Gaussian measurement uncertainty added. At each epoch, the simulation calculates whether the planet is detected using a high contrast imaging system using an IWA of 80 mas and a minimum detectable planet-to-star flux ratio of $10^{-10}$ at V band. Finally, utilizing the rejection sampling algorithm Orbits for the Impatient (OFTI; Blunt et al. 2017), each orbit is fit, progressively adding more epochs. **Figure 3.1-2** shows results for a representative set of orbits at 1 AU with an inclination angle of 30°. For the circular orbit case, a precision of 10% is achieved on the key parameters of semi-major axis, eccentricity, and inclination after three detections. In the general case, 4 detections are required. As orbital periods get longer, e.g., as in the case of giant outer planets (*Section 3.2.3*) detections must be spread over a longer time span.

It is worth noting that for planets discovered during precursor observations using the RV technique, fewer direct imaging epochs are needed to recover the orbital parameters. If the RV orbital parameters are well constrained at the time of the HabEx observation (in particular, assuming that the argument of periastron and time of periastron are well-constrained via, e.g., RV measurements taken during the HabEx mission), a single well-timed observation can determine the last missing parameter, the inclination angle, to within 10°. Even in unfavorable cases where the RV orbital parameters are poorly constrained, or the HabEx measurements are not optimally placed, two to three direct detection epochs will suffice to recover the orbital phase and constrain the inclination to within 10°.

#### Planet Radius Constraints

Visible broadband photometry alone cannot determine planet size due to the degeneracy between planetary radius and geometric albedo (hereafter simply referred to as albedo), e.g., a given planet could be either small with high reflectivity or large with low reflectivity.

The fundamental measurements of HabEx are planet position and the wavelength-dependent planet-to-star flux ratio, $F_p/F_s$ $(\lambda)$ at $\geq 4$ epochs, such that:

$$\frac{F_p}{F_s} = A_g(\lambda)\Phi(\lambda,\alpha)\left(\frac{R_p}{d}\right)^2,$$

where $A_g$ is the albedo, $\lambda$ is wavelength, $\Phi$ is the scattering phase function, $a$ is the phase angle (i.e., the star-planet-observer angle), $R_p$ is the planetary radius, and $d$ is the planet's distance from the host star.

To demonstrate that the planet size can be constrained, consider first that the planet illumination phase and separation will be well known from the orbit fitting. The primary remaining degeneracy is the $A_g$ $R_p^2$ product. Based on solar system analogs and transiting exoplanets, $A_g$ can reasonably be assumed to be between 0.06 and 0.96. Earth-size HZ planets with a lower albedo, if they exist, would actually be impossible to detect in the first place. With this generous albedo range, the corresponding values derived for e.g., a planet the size of the Earth would span 0.5 to 2 times Earth's radius.

The scattering phase function, $\Phi$, plays only a minor role in the planet-to-star flux ratio because it is dominated by the geometry of the illumination phase (for illumination phase angles <100°). This statement is supported by both measured and modeled scattering phase functions, which show relatively little spread at a given phase (Sudarsky et al. 2005) and are typically slowly varying functions of phase angle. Thus, phase uncertainties in the planet-to-star flux ratio equation are unlikely to be a dominant noise source when constraining planetary size. The weak influence of the phase function is further highlighted by the planet-to-star flux ratio expression, which shows that $R_p \propto \Phi^{-1/2}$. The main point here is that given a minimum of 4 broadband photometric measurements at different orbital phases, the planetary radius can be determined with a factor of 2 of its true value. Any planet with estimated semi-major axis and radius consistent with an Earth in the HZ, given measurement uncertainties in each parameter, will be a prime target for spectroscopic follow-up.

It is worth noting that in the case of a high signal-to-noise ratio (SNR) observation of a planet with a thin atmosphere, the optical spectrum provides even better radius constraints (Feng et al. 2018) than multi-epoch photometry alone, because the shape and strength of key reflected light atmospheric features encode information about the planet radius. Optically thick Rayleigh scattering atmospheres have geometric albedos (and planetary phase functions) that are independent of the surface optical properties and that are extremely well-described by radiative transfer theory (Madhusudhan and Burrows 2012). Here, then, the most uncertain term in the planet-to-star flux ratio expression is the planetary radius (assuming the orbit has been determined via astrometry). Thus, detections of Rayleigh scattering slopes between 0.45–0.70 µm have the ability to break the so-called "radius-albedo" degeneracy, and can provide constraints on the planetary size (**Figure 3.1-3**). Recently, Bayesian techniques, which originated from the Earth sciences (Rodgers 2000), have been adapted for use in interpreting simulated direct imaging reflected light observations of exoplanets. Detailed atmospheric retrieval studies on simulated visible-wavelength spectra of rocky exoplanets (e.g., Lupu et al. 2016) quantify the ability to estimate planet radius based on reflected-light spectral observations. In the case of an Earth-twin, detailed atmospheric retrieval studies (Feng et al. 2018) of a visible spectrum from 0.4–1.0 μm indicate that at $R \geq 140$ and SNR $\geq 10$ the planetary radius can be retrieved with a 1σ uncertainty <60% (**Figure 3.1-4**), and the spectral information that constrains the planet radius is found between 0.45–0.70 μm.

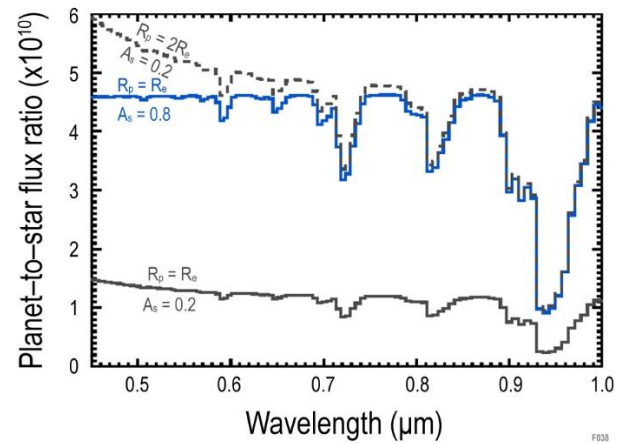

**Figure 3.1-3.** Spectroscopy constrains surface albedo and radius of exoplanets with the same optical brightness due to wavelength-dependent scattering in the atmosphere. Credit: T. Robinson.

### Objective 1 Requirements

| Parameter | Baseline | Threshold |
|---|---|---|
| **Probability of detecting at least one EEC** | >95% | >90% |
| **Number of EECs detected if each target had exactly one exo-Earth ("EEC Cumulative Completeness")** | ≥20 | ≥12 |
| **Inner working angle (IWA$_{0.5}$) (0.5 μm)** | ≤80 mas | ≤105 mas |
| **Minimum planet-to-star flux ratio detectable at IWA$_{0.5}$** | ≤10$^{-10}$ | ≤10$^{-10}$ |
| **Number of detections along orbit** | ≥4 | ≥4 |
| **Star-planet separation measurement accuracy (1σ)** | ≤5 mas rms | ≤5 mas rms |
| **Minimum wavelength range** | 0.45–0.55 μm | 0.45–0.55 μm |
| **Detection signal-to-noise ratio (SNR)** | ≥7 | ≥7 |

### 3.1.2 Objective 2: Are there Earth-like planets with atmospheres containing water vapor?

"Follow the water." Both in and beyond the solar system, this mantra is key to humanity's search for habitable environments beyond Earth. Water ($H_2O$) is central to life on Earth. Water's ability to act as a polarized solvent that undergoes hydrogen bonding gives it a unique role for all Earth-based life. As a result, water is one of the few requirements shared by all life on Earth, and life is ubiquitous on Earth where appropriate amounts of water can be found. When applied to exoplanets, this search for water has been formalized with the concept of the "habitable zone"—the region around a star for which models predict that liquid water oceans are stable at the planet's surface. Beyond this zone, oceans are predicted to either freeze over or lead to steam-dominated atmospheres that trigger a runaway greenhouse effect. HabEx—with its ability to search for water vapor and surface ocean features—has the ability to turn the habitable zone from a theoretical construct into a hypothesis that is testable through observations. Detecting atmospheric water vapor will be easier, but less diagnostic, than detecting surface water oceans. Objective 2 solely focuses on the detection of atmospheric water vapor. Atmospheric water vapor has five broad spectral absorption features from 0.7–1.5 μm (**Figure 3.1-5**), which can all be detected with a resolution of $R \geq 40$ (DesMarais et al. 2002). Detection of just one water vapor spectral feature is sufficient to securely identify water vapor in the planet's atmosphere. Moreover, detailed measurements of the shapes of multiple water vapor features provide constraints on water vapor atmospheric abundances. For example, Feng et al. (2018) were able to achieve constraints on the atmospheric water vapor abundance for an exo-Earth given simulated HabEx-like visible-wavelength observations at SNRs of 10 and larger. In order to search for two water vapor absorption features or more in the atmospheres of *all* EECs detected and with orbits determined in Objective 1, which required an IWA ≤ 80 mas at 0.5 μm, HabEx is required to provide an IWA comparable or better at a wavelength of 1 μm. In order to empirically define the HZ boundaries, it is also required that HabEx be able to search for water in an equal number of rocky planets detected slightly outside the nominal HZ (within a factor of 2 in each direction) and/or with estimated radii outside the $\sim 0.7\ R_\oplus$ to $1.4\ R_\oplus$ range adopted for EECs.

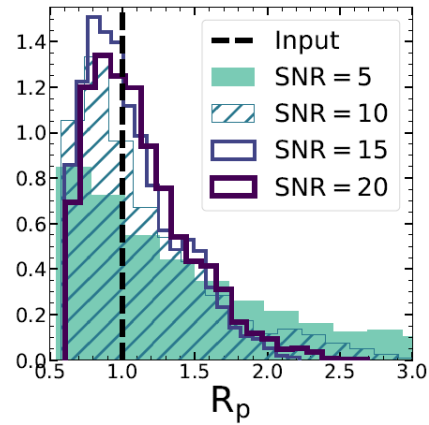

**Figure 3.1-4.** Moderate signal-to-noise (SNR = 10–15) spectroscopy constraints exoplanet radius to a 1σ uncertainty <60% in this simulation of the posterior probability distribution function of the radius of an exo-Earth. Credit: K. Feng.

It is further required that HabEx be able to detect and measure the abundance of water vapor if its column density exceeds the level of modern Earth if it were placed *anywhere* in the HZ. The minimum value is reached at the HZ outer edge and is $\sim 0.4\ \mathrm{g/cm^2}$, significantly lower than modern Earth water column value of $2.9\ \mathrm{g/cm^2}$, which is believed to have remained fairly constant throughout its geological history (excepting, potentially, snowball Earth episodes). Additionally, planetary systems without giant planets may be much richer in water in their inner regions than our solar system (e.g., Morbidelli and Raymond 2016). Despite the fact that roughly 70% of the surface of Earth is covered by water, it is important to recognize that the Earth is relatively dry. Ganymede, Europa, Titan, and even Triton, although 30–40 times less massive than Earth, are all believed to have more water by volume than Earth.

Small (temperate) planets do not hold onto atmospheric water vapor without a liquid ocean reservoir because stellar UV radiation dissociates water vapor in the atmosphere and the hydrogen then escapes into space. This, combined with the strong runaway climate feedbacks that occur outside the HZ, links the presence of water vapor to the presence of liquid water oceans on rocky worlds. However, moist water-dominated atmospheric states can exist on rocky planets without habitable conditions at the surface (Goldblatt 2015), and slightly larger planets such as sub-Neptunes—which may lie within the range of radius measurement uncertainties—are massive enough to hold onto hydrogen, and water vapor naturally occurs due to atmospheric equilibrium chemistry. Therefore, the ocean glint observations associated with Objective 4 (*Section 3.1.4*) are required to increase the confidence that liquid surface water is present on a planet for which atmospheric water vapor has been detected.

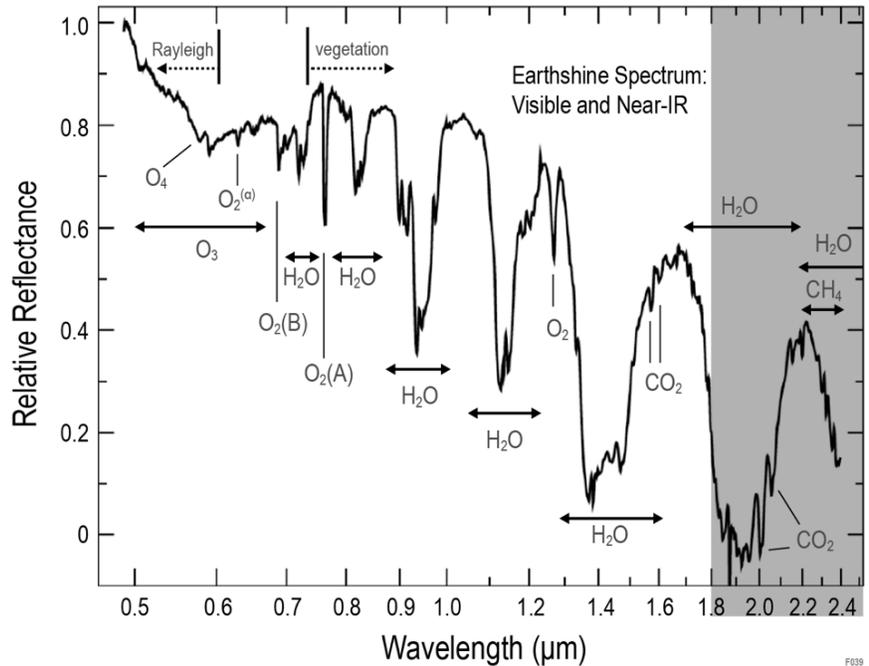

**Figure 3.1-5.** The visible and near-IR spectral range provides a wealth of key molecular features for Earth-like planets, as shown in this disk integrated, high-resolution spectrum of Earthshine. Credit: Turnbull et al. (2006).

### *Objective 2 Requirements*

| Parameter | Baseline | Threshold |
|---|---|---|
| IWA$_{0.5}$ | ≤80 mas (1.0 μm) | ≤105 mas (0.75 μm) |
| Minimum Planet-to-star flux ratio detectable at IWA | ≤10$^{-10}$ | ≤10$^{-10}$ |
| Wavelength range | ≤0.7 μm to ≥1.0 μm | ≤0.7 μm to ≥1.0 μm |
| Spectral resolution, *R* | ≥ 35 (SNR ≥ 10) | ≥ 35 (SNR ≥ 5) |
| Minimum H$_2$O column density detectable, | 0.4 g/cm$^2$ (modern Earth at HZ outer edge) | 2.9 g/cm$^2$ (modern Earth) |

### 3.1.3 Objective 3: Are there Earth-like planets with signs of life?

How common is life in the universe? The lack of measurements that directly respond to this question allows for a wide range of answers, all of which are driven by speculation and gross extrapolation. For example, some look at the history of life on Earth, which extends almost as far back as our ability to find biosignatures, as a sign that life arises easily and therefore must be common, but such extrapolations have never been tested by observations of other potentially Earth-like worlds. HabEx will conduct such observations and begin to constrain this problem.

The biosphere of Earth, both its modern-day state and its diversity over Earth history, is our baseline for understanding biology. Thus, Earth's biosphere is also the main driver of our expectations for the biosignatures that HabEx shall be capable of detecting. While we expect a variety of biosignatures to eventually be found on worlds beyond Earth, a lack of such known biospheres makes it difficult to derive performance requirements for any future telescope. Instead, HabEx shall be designed with three life-searching priorities in mind: 1) capability to detect the gaseous byproducts from a biosphere that, like modern Earth, is driven by oxygen-producing photosynthesizers; 2) the capability to detect the gaseous byproducts from a biosphere that, like Archean Earth, is driven by methane-producing chemosynthesizers; 3) the ability to discriminate between such biospheres and global nonbiological processes that can produce the same gaseous byproducts ("false positive" mechanisms). In order to draw further conclusions, it will be necessary to examine the relationship of these detections and non-detections to markers of our current theories on habitability, such as the presence of water vapor in the atmosphere (*Section 3.1.2*), signals from ocean surfaces (*Section 3.1.4*), and any other ancillary information available, from precursor or follow-up observations (*Section 3.4*).

#### 3.1.3.1 Biosignature Searches

The most significant and most detectable signals of life on modern Earth atmosphere are the byproducts of oxygenic photosynthesis, specifically the presence of large quantities of atmospheric molecular oxygen ($O_2$; **Figure 3.1-5**). $O_2$ has accumulated to the level of 20% by volume of atmosphere on modern Earth, resulting in a strong spectral feature at 0.76 μm. $O_2$ also leads to the accumulation of ozone ($O_3$) in the atmosphere, which has a strong cutoff feature at 0.33 μm and a broad, shallow feature at 0.55 μm (**Figure 3.1-5**). All $O_2$ and $O_3$ features are evident in the globally averaged spectrum of modern Earth, and would also appear in Earth's spectrum over the last ~500 million years that $O_2$ concentrations have constituted 10–20% of Earth's atmosphere. $O_2$ and its photochemical byproduct $O_3$ have actually been present at significant quantities ever since "the great rise of oxygen" ~2.5 billion years ago, when $O_2$ rose to about 1% of modern-day levels (**Figure 3.1-6**). **HabEx is required to be able to detect both $O_2$ and $O_3$ around all EECs detected, down to concentration levels covering the *full* 2.5 billion year history of Earth's oxygenated atmosphere (Table 3.1-1).** This means reaching the lowest estimates for the Proterozoic Earth concentration: $2\ g/cm^2$ for $O_2$ and $8 \times 10^{-5}\ g/cm^2$ for $O_3$ (Reinhard et al. 2017; Schwieterman et al. 2018). In order to detect both $O_2$ and $O_3$, the minimum spectral range for HabEx biosignature searches is then 0.3–0.8 μm, with a spectral resolution $R \geq 5$ for ozone at < 0.35 μm, and $R \geq 70$ for molecular oxygen at 0.76 μm.

HabEx shall also be capable of detecting biospheres that are not based on oxygen-producing synthesizers, at least on a subset of detected EECs. For example, evidence of life on Earth extends back at least 3.5 billion years, while prior to 2.5 billion years ago (in the Archean Era), geochemical evidence places constraints on surface $O_2$ concentrations to have been many orders of magnitude lower than those on modern-day Earth (Farquhar et al. 2000). During the times when $O_2$ was lower than today, it is believed that $CH_4$ concentrations were significantly higher. The lower amount of $O_2$ in the atmosphere would have drastically decreased the chemical sinks for atmospheric $CH_4$. Thus, biological fluxes of $CH_4$ on the scale seen on Earth today may have led to higher, more detectable concentrations of $CH_4$ in the past (e.g., Rugheimer et al. 2015). Simulations of the atmosphere and spectra for an Archean-like biosphere suggest that it would have had significant levels of $CH_4$, as well as a haze that is the photochemical byproduct of $CH_4$. Accordingly, HabEx is required to detect the lower end of estimated Archean $CH_4$ concentrations (0.1 g/cm²) via spectroscopic measurements in the 0.8–1.7 μm range with a spectral resolution $R \geq 30$ (DesMarais et al. 2002). Additionally, high $CH_4$ concentrations may lead to the photochemical production of an organic haze, which would impact the slope of the planet's spectrum shortward of ~0.6 μm. In order to detect this effect, HabEx is required to conduct either near UV-visible photometry or low-resolution ($R$>~10) spectroscopy from 0.45–0.6 μm (these requirements are not driving and are superseded by other requirements).

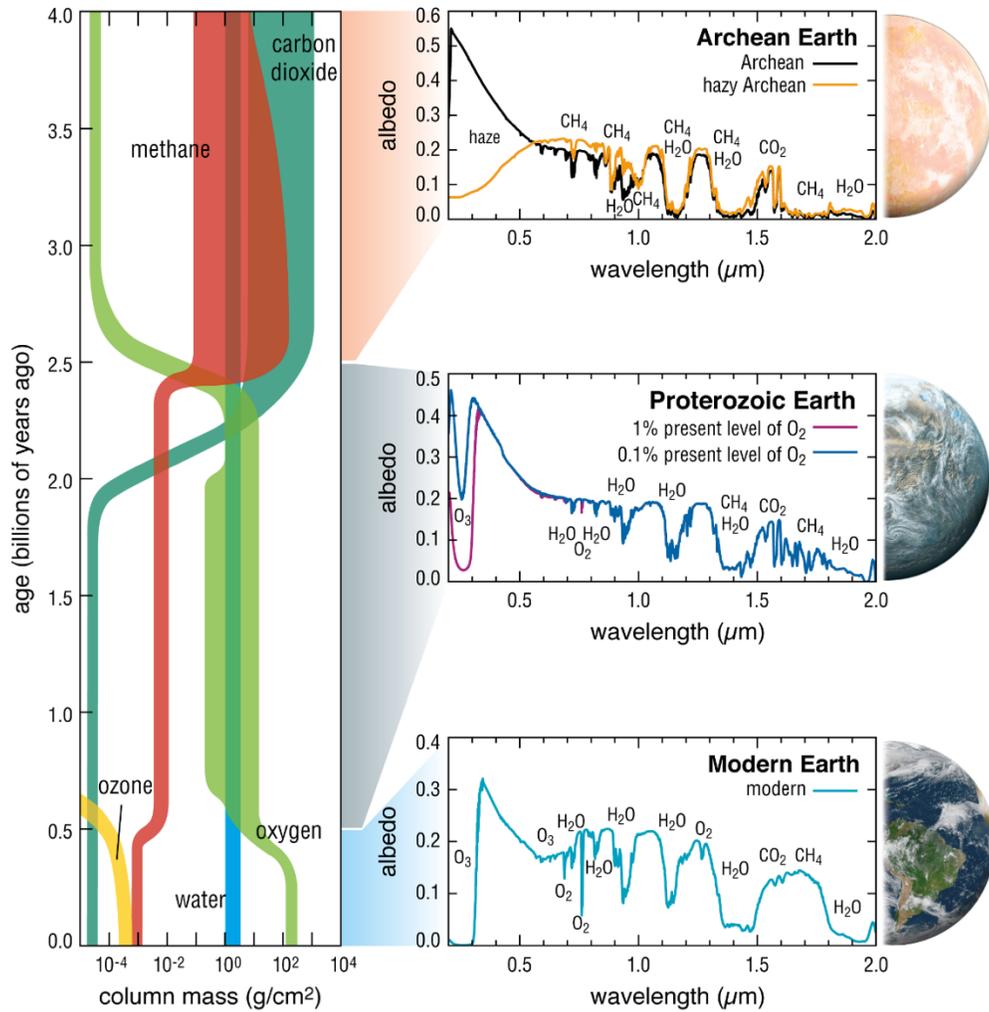

**Figure 3.1-6.** $O_2$, $O_3$, $H_2O$, $CH_4$ and $CO_2$ concentrations over Earth's history, during Archean, Proterozoic, and modern Earth eras. HabEx is required to be able to detect the gaseous byproducts from oxygen-producing synthesizers (all EECs) or methane-producing synthesizers (some EECs), if present at concentration levels similar to Earth over the last 3.5 Gy of its history. This covers part of the Archean Era as well as the full Proterozoic and Modern Eras during which life has been present on Earth. Credit: Britt Griswold, Giada Arney, and Shawn Domagal-Goldman.

### 3.1.3.2 Checking for False Positive Mechanisms

Detecting the gaseous byproducts of a biosphere is much easier than conclusively demonstrating that those gases arise from biological activity. In order to meet this higher threshold, observations must also rule out "false positives" where non-biological processes create some of the features found on biospheres. There

**Table 3.1-1.** Key atmospheric species, spectral features, and their column mass evolution over Earth's history. HabEx is required to spectrally characterize the full 0.3–1 μm wavelength range at once for all EECs detected, with extensions down to 0.2 μm and up to 1.8 μm on favorable targets.

| Species | Wavelength (μm) | Column Mass (g cm$^{-2}$) | Notes |
|---|---|---|---|
| Ozone (O$_3$)[1] | 0.32, 0.58 | Modern: $7.2\times10^{-4}$<br>Proterozoic: $8\times10^{-5}$–$6\times10^{-4}$ | Biosignature gas; greenhouse gas |
| Oxygen (O$_2$)[1] | 0.76, 1.27 | Modern: $2.4\times10^{2}$<br>Proterozoic: 2–20 | Biosignature gas |
| Methane (CH$_4$) | 0.89, 1.0, 1.15, 1.4, 1.69 | Modern: $9.1\times10^{-4}$<br>Proterozoic: $10^{-3}$–$10^{-2}$<br>Archean: $10^{-1}$–$10^{2}$ | Biosignature gas; greenhouse gas |
| Water vapor (H$_2$O)[2] | 0.72, 0.82, 0.94, 1.13, 1.41 | Modern: 2.9 | Key habitability indicator |
| Carbon dioxide (CO$_2$) | 1.21, 1.44, 1.59 | Modern: $5.2\times10^{-1}$<br>Proterozoic: $5\times10^{-1}$–$10^{-2}$<br>Archean: 10–$10^{3}$ | Greenhouse gas; potential biosignature; key species for habitability maintenance via carbonate-silicate cycle |

[1]: The Archean Earth atmosphere was not oxygenated, so column abundances for O$_2$ and O$_3$ for this period would have been negligible.
[2]: As Earth has remained habitable throughout its geological history (excepting, potentially, snowball Earth episodes), the water vapor column mass would be similar for the Archean through Modern Eras.

are two major ways in which a false positive for life could arise. First, there could be spectral confusion, where a molecule other than the biosignature gas absorbs at the same wavelength. For a putative detection of molecular oxygen at 0.76 μm, this type of false positive can be eliminated by looking for additional O$_2$ features at other wavelengths (e.g., 0.69 μm and 1.27 μm) and by searching for absorption from O$_3$, which is a photochemical byproduct of O$_2$ (**Figure 3.1-6**). The second kind of false positive is where a gas (such as O$_2$) exists in the exoplanet's atmosphere, but originates from non-biological processes. These non-biological processes are dependent on the stellar context, because UV and high-energy stellar emissions drive many of the processes that can create detectable O$_2$ concentrations without oxygenic photosynthesis being active at the planet's surface.

Fortunately, most of the known abiotic O$_2$ and O$_3$ generation processes are thought to be inefficient for planets in orbit around sunlike (FGK) stars (Harman et al. 2018), and HabEx's observational designs drive it towards a target list of sunlike stars (**Figure 3.3-11**). The only known abiotic source of O$_2$ on planets around FGK stars is based on theoretical studies of planets with low background atmospheric pressures (Wordsworth and Pierrehumbert 2014). These low pressures can prevent water from condensing into a cloud deck, allowing water vapor to reach the upper atmosphere where water can be photolyzed, liberating H atoms which can then escape to space. The loss of H leaves behind O atoms that are then free to recombine to form O$_2$ and O$_3$. Eventually, this can even cause oxidation of the mantle, which would allow detectable O$_2$ and O$_3$ concentrations to be sustained over geological timescales (Wordsworth et al. 2018).

Fortunately, there are multiple ways to identify this non-biological O$_2$ buildup process on an exoplanet. Planets for which the water vapor escape process is complete would be deficient in H atoms, and lack signs of water vapor in their atmospheres (and therefore would not be considered habitable planet candidates). This type of planet would also lack water clouds, darkening the spectral continuum albedo and excluding cloud-induced spectral variability. Finally, any planets currently undergoing water loss (and exhibiting water spectral features), would have either low atmospheric pressures or high partial pressures of O$_2$. The low atmospheric pressures could be identified from spectral data throughout the UV-Vis-NIR (Feng et al. 2018). A high O$_2$ partial pressure could be identified from the O$_2$-O$_2$ dimer features that would appear at visible (0.57 and 0.63 μm) and near-infrared wavelengths (1.06 and 1.27 μm); this is what dictates the minimum O$_2$-O$_2$ detection requirement of 0.4 bar partial pressure in O$_2$-O$_2$, corresponding to twice the modern Earth value.

There is a much larger list of nonbiological mechanisms for generating $O_2$ on planets in orbit around M-type stars (Meadows et al. 2018). Although unlikely or impossible on the majority of HabEx's target stars, HabEx will be able to identify or rule out non-biological $O_2$-generating mechanisms on any accessible planets orbiting M-type stars. A list of these mechanisms includes: photochemical processes on planets with $CO_2$ concentrations orders of magnitude greater than modern-day Earth (Domagal-Goldman et al. 2014) or on desiccated planets (Gao et al. 2015); massive H loss driven by excessive XUV and FUV radiation from the host star (Luger and Barnes 2015; Ramirez and Kaltenegger 2014); or short-lived $O_2$ or $O_3$ resulting from the photochemistry in the immediate aftermath of a stellar flare (Segura et al. 2003). HabEx will be able to constrain these mechanisms by its ability to detect $CO_2$ concentrations greater than $5 \times 10^3$ g/cm$^2$ via its absorption features at 1.21, 1.44, and 1.59 μm, $H_2O$ column masses greater than 0.4 g/cm$^2$ via multiple water vapor features between 0.7 to 1.4 μm (Objective 2), and the UV characterization of the host star with the HabEx UV instrument.

In the case of an Archean-like biosphere driven by methane-producing chemosynthesizers, the atmospheric $CH_4$ must be distinguished from non-biological $CH_4$, such as seen on Titan. To that end, HabEx shall measure or provide useful upper limits on $CO_2$ concentrations. That is because high Archean-like $CH_4$ concentrations are not sustainable in $CO_2$-rich atmospheres unless there are massive amounts of $CH_4$ only consistent with biological production (Arney et al. 2016; Krissansen-Totton et al. 2018). HabEx is hence required to be able to detect Archean-like $CO_2$ concentrations ($5 \times 10^3$ g/cm$^2$, close to the inferred range high end) on a subset of detected EECs, covering a wavelength range that includes $CO_2$ features at 1.21, 1.44, and 1.59 μm with a resolution greater than 11 (DesMarais et al. 2002).

HabEx will be able to further support biosignature gas identification with more detailed analyses of the planets for which it identifies biosignatures and eliminates false positives. If HabEx identifies planets with $O_2$ or $O_3$, the "life" hypothesis would be specifically associated with oxygenic photosynthesis. HabEx could test that hypothesis by searching for photosynthetic pigments, or red-edge effects. It could also test the broader hypothesis of there being a global biosphere by conducting a detailed search for the global liquid water oceans that are thought to be required to support such widespread life on the planet. The approach HabEx will take to confirm the presence of oceans is the subject of the next section.

### Objective 3 Requirements

| Parameter | Baseline | Threshold |
|---|---|---|
| **IWA$_{0.5}$ (0.8 μm)** | ≤80 mas | ≤105 mas |
| **Minimum planet-to-star flux ratio detectable at IWA** | ≤10$^{-10}$ | ≤10$^{-10}$ |
| **Wavelength range** | ≤0.3 μm to ≥1.7 μm | ≤0.3 μm to ≥1.7 μm |
| **Spectral resolution, *R*** | **O$_3$:** ≥ 5 0.30–0.35 μm <br><br>**O$_2$:** ≥ 70 (0.75–0.78 μm) <br><br>**CH$_4$:** ≥ 10 (1.52–1.66 μm) <br><br>**CO$_2$:** ≥ 11 (1.62–1.78 μm) <br><br>**O$_4$:** ≥ 40 (at 0.63μm) ≥ 22 (at 1.06 μm and 1.27 μm) | **O$_3$:** ≥ 5 (0.30–0.35 μm or 0.53–0.66 μm) <br><br>**O$_2$:** ≥ 70 (0.75–0.78 μm) <br><br>**CH$_4$:** ≥ 10 (1.52–1.66 μm) <br><br>**CO$_2$:** ≥ 11 (1.62–1.78 μm) <br><br>**O$_4$:** ≥ 50 (at 0.57 μm) ≥ 40 (at 0.63 μm) |
| **Minimum column density detectable** | **O$_3$:** 8×10$^{-5}$ g/cm$^2$ (Proterozoic Earth low end) <br><br>**O$_2$:** 2 g/cm$^2$ (Proterozoic Earth low end) <br><br>**CH$_4$:** 10$^{-1}$ g/cm$^2$ (Archean Earth low end) <br><br>**CO$_2$:** 5×10$^3$ g/cm$^2$ (5× Proterozoic Earth high end) | **O$_3$:** 7.2×10$^{-4}$ g/cm$^2$ (Modern Earth) <br><br>**O$_2$:** 2.4 10$^2$ g/cm$^2$ (Modern Earth) <br><br>**CH$_4$:** 10$^2$ g/cm$^2$ (Archean Earth high end) <br><br>**CO$_2$:** 10$^4$ g/cm$^2$ (10× Proterozoic Earth high end) |
| **Minimum partial pressure detectable** | **O$_4$:** 0.4 bar (2× Modern Earth) | **O$_4$:** 0.8 bar (4× Modern Earth) |

### 3.1.4 Objective 4: Are there Earth-like planets with water oceans?

In order to confirm the presence of liquid water, and to test the idea that water and life are inextricably linked, HabEx will search the surfaces of EECs with atmospheric water vapor detected (Objective 2), regardless of whether or not biosignatures were identified. Earth's surface water is actually just a thin veneer, although there are debates about how much water is stored in the Earth's interior (Fei et al. 2017). This suggests a fine-tuning issue: if the Earth were significantly drier, life might not have been able to thrive; if it were significantly wetter, then it might have been a water world, loosely defined as planets of 50% or more water by mass, planets with deep oceans and no continents. Water worlds may be poor places for life to originate since they suppress the carbon cycle, thought to be essential to maintaining the Earth's temperature and therefore habitability, although this conclusion is controversial (Kite and Ford 2018). In any case, correlating the detection of biosignatures with the presence and extent of surface water oceans will help constrain the amount of water required to produce habitable environments.

As stated above, water vapor is strongly suggestive evidence for, yet does not conclusively demonstrate the presence of, liquid water oceans on the surface of the planet. There are two ways that have been proposed to directly detect surface liquid water.

Specular reflection (glint), a disproportionate increase in the brightness of a planet in a crescent phase, indicates a liquid surface (Williams and Gaidos 2008; Oakley and Cash 2009). The main challenge with detecting glint is that its signal occurs when the planet is at a high illumination phase, resulting in a smaller apparent separation from the host star. For illumination phases >140°, Earth ocean glint results in a substantial (≥50%) increase in apparent albedo compared to the case where only the effect of clouds is accounted for (**Figure 3.1-7**, e.g., Robinson et al. 2014). Ocean glint also causes an apparent reddening of the planet (**Figure 3.1-8**). To detect glint on an Earth-like exoplanet requires at least two broadband photometric measurements at different epochs. One observation is required at an illumination phase between 120° and 160°, for which the continuum planet-to-star flux ratio is constant in the ∼7 to ∼9 × 10$^{-11}$ range for an Earth-twin (**Figure 3.1-7**). Statistically, 87% of the systems observed will be seen at an inclination > 30° from pole-on. This means that the majority of exo-Earth candidates characterized through objectives 1 to 3 are expected to be observable at illumination phases >120°, offering the possibility to look for glint from water oceans. Glint detection is a powerful observational technique to confirm liquid water on a planet surface. Observations combining evidence for the presence of bodies of liquid at the surface with the detection of water vapor in the atmosphere could reliably point to oceans on other worlds.

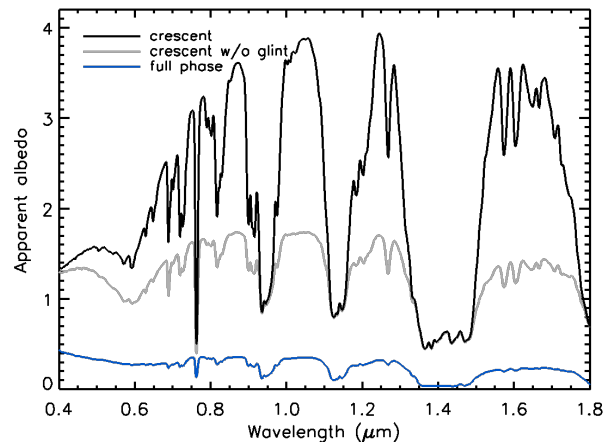

**Figure 3.1-8.** Apparent albedo of Earth at full phase (*blue*), and at crescent phase (150°) both with ocean glint (*black*) and without (*grey*). Ocean glint causes Earth-like planets to appear brighter and redder at crescent phase than predicted without oceans (adapted from Robinson (2018) based on the validated model described in Robinson et al. (2011)). Regions in the spectrum where glint increases the planet brightness most prominently are found in the continuum in between water absorption features. They appear around 0.87, 1.1, 1.25, and 1.6 μm. Apparent albedo is defined as the albedo a Lambert sphere (with radius equal to the planetary radius) would need to reproduce the observed brightness of the planet, and values larger than unity imply forward scattering.

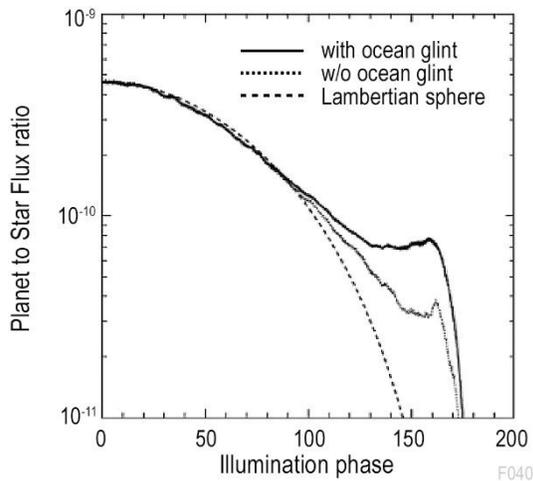

**Figure 3.1-7.** Broadband 1.0–1.1 μm Earth-to-Sun flux ratio observed as a function of illumination phase and averaged over a spin rotation period. The brightness enhancement due to ocean glint becomes prominent for illumination phases > 120°. Similar signal enhancement occurs around 0.87, 1.25, and 1.6 μm, i.e., in the continuum in between water absorption features. Credit: Robinson et al. (2014).

The second related method for detecting oceans is through the polarization that liquid surfaces imprint on reflected light. Depending on the exact nature of the planet observed, ranging from no atmosphere and covered by a calm ocean, to a thick Rayleigh scattering atmosphere with or without clouds, simulations show that the polarization fraction peak varies widely, and will be reached at different illumination phases (Zugger et al. 2010) and wavelengths (Sterzik et al. 2019). Consequently, polarization adds another dimension to the search for surface liquid water oceans, which can be used in combination with unpolarized light curves and planet-to-star flux ratio measurements at different illumination phases to more robustly confirm extrasolar oceans than glint detection alone. Given that the observed peak linear polarization fraction of the Earthshine is only 10–20% in the visible to near-IR range (Sterzik et al. 2019), polarimetric studies of water oceans remain challenging in the case of a true Earth analog and may only be possible in favorable cases (e.g., nearby exo-Earths with significant polarization fraction or super Earths). The only additional requirement derived here is that the baseline HabEx architecture enable polarimetric studies of exoplanets. Polarimetry is required to a greater extent for the characterization of circumstellar dust disks (Objective 8) and is further described in *Section 3.2.4*.

**Objective 4 Requirements**

| Parameter | Baseline | Threshold |
|---|---|---|
| IWA$_{0.5}$ (0.87 µm) | ≤ 64 mas | ≤ 129 mas |
| Planet-to-star flux ratio detection limit | ≤7 × 10$^{-11}$ with SNR > 7 | ≤7 × 10$^{-11}$ with SNR > 7 |
| Phase coverage | For EECs within 10 pc with apparent orbital inclination ≥ 50°, broadband photometry at 0.87 µm and at ≥140° phase | For EECs within 5 pc with apparent orbital inclination ≥ 50 deg, broadband photometry at 0.87 µm and at ≥140° phase |
| Polarimetric capability | Yes | No |

## 3.2　GOAL 2: To map out nearby planetary systems and understand the diversity of the worlds they contain

The HabEx mission shall have unprecedented capabilities to characterize exoplanets spanning sub-Earths to giants. Critically, HabEx shall obtain spectroscopic observations of entire planetary systems (above our planet-to-star flux ratio detection limit) while performing a 'deep dive' on a handful of nearby sunlike stars. Here, and for more distant systems where we detect an exo-Earth candidate, longer duration integrations to obtain moderate SNR spectra of rocky worlds also yield moderate to extremely high SNR spectra of all brighter targets in the system, including sub-Neptune-sized, Neptune-sized, and Jupiter-sized planets. Such whole-system observations enable HabEx to answer key questions about the architectures of exoplanetary systems: how stellar irradiation, planetary size, and metallicity influence atmospheric composition and properties; how the presence of a Jovian planet and overall planetary architecture influences the potential habitability of rocky worlds in a given system; and how the interactions between planets, planetesimals, and dust leads to emergent structures in debris disks.

### 3.2.1　Objective 5: What are the architectures of nearby planetary systems?

The solar system has a striking architecture compared to other planetary systems that have been glimpsed so far. The Sun hosts no planets inward of 0.3 AU, four terrestrial planets within 1.5 AU, followed by the asteroid belt at 2–3 AU. Beyond the 'snow line' at roughly 2.7 AU, where water ice was stable in the Sun's protoplanetary disk in a near vacuum, is located Jupiter at 5.2 AU, the most massive and dynamically dominant planet in the solar system. Beyond Jupiter, the remaining three outer planets become less massive and more metal rich. The size-ordering of solar system planets, with small planets close-in and gas-giants farther out is quite notable. Additionally, the solar system architecture is "dynamically cold," with all eight planets in our solar system having nearly circular, nearly co-planar orbits well aligned with the rotation of the Sun. Beyond the orbit of Neptune lies another belt of relatively small objects, the Kuiper belt, which is estimated to be substantially more massive than the asteroid belt (e.g., Delsanti and Jewitt 2006).

Models that successfully explained the gross features of our solar system architecture were developed in a number of papers published before the current era of exoplanet demographic science (e.g., Kokubo and Ida 1998; Pollack et al. 1996). These models invoked in situ formation of the planets within the Sun's protoplanetary disk to explain the features of our solar system, with little-to-no migration of the planets from their birth sites. While these works were generally viewed as successful explanations of the structure of our solar system, other work had identified a number of problems with these ideas (e.g., Weidenschilling 1995).

### 3.2.1.1 Observational Surprises

The discovery of 51 Peg b (Mayor and Queloz 1995), a giant exoplanet with a mass similar to Jupiter's but with an orbital period of only ~4 days (the first example of what is now known as a 'hot Jupiter'), led to the questioning of the entire paradigm of solar system formation models. Models generically predict that giant planets (primarily made of hydrogen and helium) must form beyond the snow line (e.g., Lin et al. 1996). Thus, the discovery of the planet 51 Peg b implied that large-scale planet migration must be considered, at least for some planetary systems.

More recently, NASA's Kepler mission (Borucki et al. 2010) made the startling discovery that super-Earth and sub-Neptune exoplanets on relatively close-in orbits are extremely common in our Galaxy, with nearly every sunlike star possessing one such planet, on average (Petigura et al. 2018). Given the lack of such planets orbiting the Sun, Kepler has thus demonstrated that the solar system is not a typical outcome of planet formation in at least one key respect.

### 3.2.1.2 Constraining Planet Formation Models by Getting Full Family Portraits

Current planet formation models offer different predictions regarding how the presence of a distant Jovian could impact the formation of a super-Earth or sub-Neptune on a closer-in orbit, and how Jovian planets may help sculpt tightly packed inner planetary systems (Hands and Alexander 2016). Strict in situ models (e.g., Chiang and Laughlin 2013; Schlaufman 2014) predict no correlation between distant giant planets and close-in sub-Neptunes, while models involving significant migration predict a strong anti-correlation (e.g., Batygin and Laughlin 2015; Izidoro et al. 2015). HabEx shall address theories for why nature so efficiently produces super-Earths and sub-Neptune exoplanets. Understanding the outer planetary system architecture for its effect on the impact history of inner potentially habitable planets is also of high importance. Not only does the outer architecture have consequences for the ability to pass comets inward, deliver water (*Section 3.2.3*), and possibly create the conditions necessary for the development of life (e.g., Patel et al. 2015), but small bodies may also bring volatiles inward that could replenish a primary atmosphere or strip the planet from its atmosphere. Exploring the link between distant giant planets and inner small planets in the same systems (Zhu and Wu 2018) is then critical to planet formation models and understanding the potential uniqueness of our own solar system.

Similar fundamental questions exist in the inner parts of individual systems. For instance, while it is known from Kepler that close-in super Earths and sub-Neptunes exist around roughly half of stars, it is not clear whether systems of solar system–like terrestrial planets exist outside of these close-in systems. If the two

architectures—centrally packed with no HZ terrestrial planets outside vs. centrally hollowed with HZ terrestrial planets outside—are mutually exclusive (e.g., Volk and Gladman 2015; Spalding 2018) it would imply that our own solar system was fortunate to obtain the planetary orbital arrangement of the terrestrial planets. In this way, understanding how close-in super-Earths and sub-Neptunes come to be informs us of the contingencies driving the formation of our own habitable planetary system architecture.

Also of particular interest to planet formation studies (e.g., Li and Winn 2016; Spalding and Batygin 2015) is the ability to directly measure spin-orbit misalignment between orbiting planets and the central star. Numerous techniques exist to estimate the projected inclination of a given star (e.g., asteroseismology, long-baseline optical interferometry (Le Bouquin et al. 2009), and spectroscopy), but this stellar inclination cannot be converted to a spin-orbit misalignment unless the planetary orbital plane is also known. Spin-orbit misalignments studies are currently limited to hot Jupiters (Winn et al. 2005) and a few directly imaged young planets in wide orbits (Wright et al. 2011). Space-based high contrast direct imaging of a broad range of planet sizes and separations would bring considerable new insight into that third dimension of planetary systems architectures.

### 3.2.1.3 Observational Requirements

The HabEx mission shall address these representative questions both in a statistical sense and through in-depth observations of individual systems.

For the former, HabEx shall image a large number of planetary systems around a variety of stellar hosts during its survey for exo-Earth candidates. Here, detecting at least 30 planets in each of the terrestrial (<1.75 R⊕), sub-Neptune (1.75–3.5 R⊕), and giant planet (Neptune-like and Jupiter-like) size regimes would provide a strong statistical sampling of planetary systems. Critically, the sample derived by HabEx could be inter-compared to, or combined with, complementary samples from transit, radial velocity, and microlensing surveys.

In addition to its statistical characterization of planetary systems, HabEx shall provide complete observations of a smaller number of "deep dive" very nearby systems, with high sensitivity from rocky to giant planets in ~0.3–30 AU orbits (**Figure 3.2-1**). The broad range of accessible semi-major axis is key for these deep-dive systems.

Characterizing the architectures of full planetary systems requires a deep integration over a wide range of angular separations. The nearest systems form the most favorable targets for these observations, since nearby targets require shorter integration times. However, detecting outer planets in these systems drives outer working angle (OWA; the outer bound of the high-contrast search area) requirements for high-contrast imaging and spectroscopy. In the fiducial case of a nearby solar system analog at 5 pc, detecting the equivalent of Neptune at 30 AU requires HabEx to have an OWA ≥ 6 arcsec. Similarly, capturing a range of planet sizes in the system requires access to planets smaller and significantly fainter than Earth. For a 0.6 $R_\oplus$ planet at 1 AU seen at quadrature in the same system, HabEx requires a detection limit for the planet-to-star flux ratio of ≤4 × 10⁻¹¹ at V-band, which is equivalent to a Δmag of 26.0. Multi-epoch (≥4) broadband visible imaging at SNRs greater than seven is required to provide accurate centroid measurements of planetary PSFs, constrain planetary orbital parameters and radii (*Section 3.1.1*).

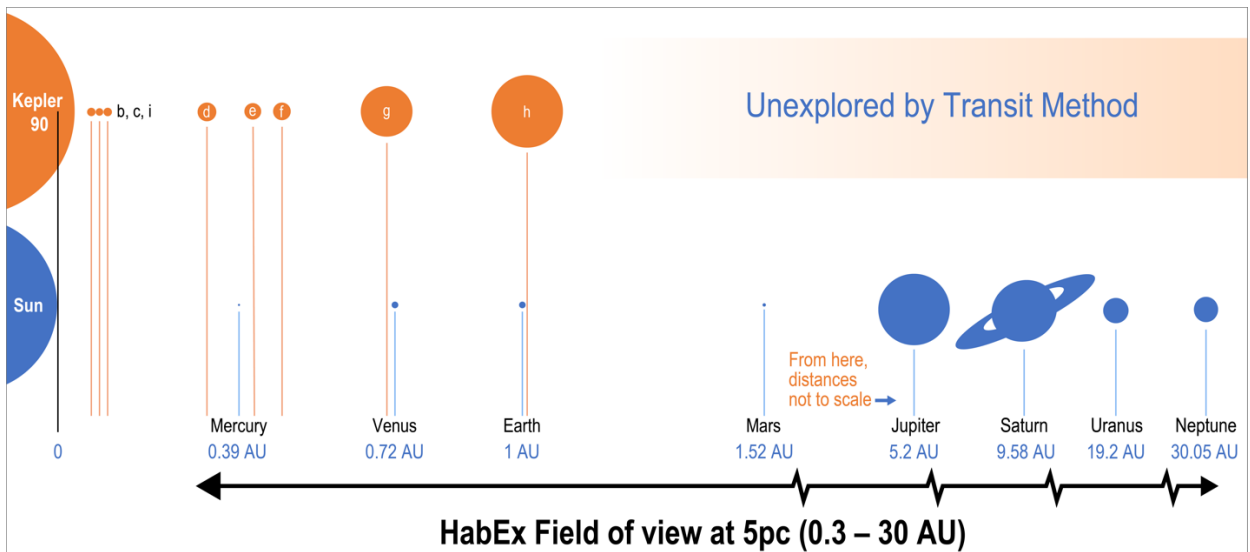

**Figure 3.2-1.** Comparison of the solar system distribution (size and separation) of planets to that of Kepler 90, a sunlike star with a very dissimilar, tightly packed inner planetary system. If such systems exist around nearby stars, HabEx shall provide a more complete view by including the population of outer planets.

A relaxed threshold requirement on the OWA would be to at least enable detections of planets in Jupiter-like orbits around targets at a distance of 10 parsecs (yielding an OWA of ≥0.5 arcsec) with the same multi-epoch imaging requirements for determining planetary radii.

Taken altogether, through a combination of planetary system observations, planetary orbit and size estimations, HabEx shall either confirm or disprove models of planet formation and orbital evolution. It shall detail the structures (including at larger orbital separations) of nearby planetary systems, thereby shedding light on whether or not our own solar system architecture is rare, while also exploring the unique planetary arrangements of our nearest stellar neighbors.

*Objective 5 Requirements*

| Parameter | Baseline | Threshold |
|---|---|---|
| **Number of rocky planets (<1.75 $R_\oplus$) detected** | >30 | >15 |
| **Number of sub-Neptunes (1.75–3.5 $R_\oplus$) detected** | >30 | >15 |
| **Number of giant planets (<3.5 $R_\oplus$) detected** | >30 | >15 |
| **Planet-to-star flux ratio detection limit** | ≤4 × 10⁻¹¹ with SNR > 7 at <5 pc | ≤4 × 10⁻¹¹ with SNR > 7 at <5 pc |
| **OWA (0.5 µm)** | ≥6 arcsec | ≥0.5 arcsec |
| **Wavelength range** | Visible | Visible |

### 3.2.2 Objective 6: How diverse are planetary atmospheres?

#### 3.2.2.1 Planet Diversity

A wide variety of atmospheric properties are represented across the terrestrial, sub-Neptune and giant planet categories.

Giant exoplanets have 'primary' atmospheres, formed by accretion during the formation of the planetary system, and are predominantly hydrogen and helium with some degree of metal-enrichment. In contrast, the atmospheres of super-Earths may be largely 'secondary,' formed by planetesimal impacts (de Niem et al. 2012), active geological and/or biological processes as is the case for Earth, or may be so metal-enriched that the primary species is water (Fortney et al. 2013).

Despite the wide predicted range for exoplanet atmospheric diversity, key underlying physical processes should govern these atmospheres in manners that HabEx shall investigate. First, larger, more massive planets should have atmospheres that are less metal enriched than small, less massive planets in the same system (Mansfield et al. 2018) and **Figure 3.2-2**). This is a prediction of the core accretion model of planet formation, where Jovian planets accrete more gaseous material from the protoplanetary disk and, thus, have atmospheric compositions with metallicities more like their host star (Fortney et al. 2013; Lopez and Fortney 2014). Second, planetary orbital distance should have a strong impact on planetary atmospheric composition, both through the evaporation of increasingly volatile species with decreasing orbital distance (Öberg et al. 2011; Kopparapu et al. 2018) and through the enhancement of photochemical processes at small orbital distances. HabEx shall confirm (and further detail) these critical concepts by observing near UV to near-IR exoplanet spectra and retrieving atmospheric composition as a function of planet size and orbital distance.

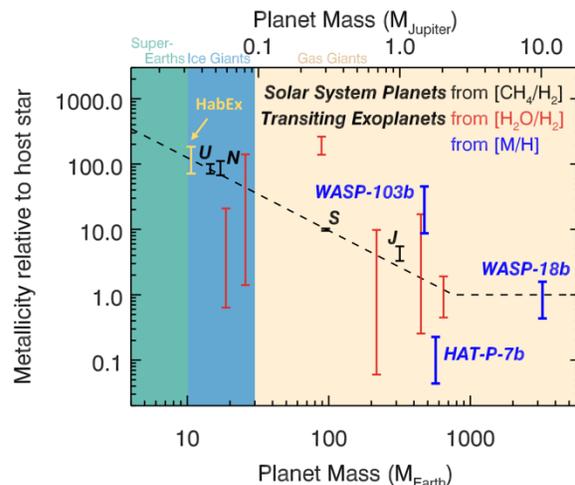

**Figure 3.2-2.** Metallicity versus planet mass for the solar system giant and ice giant planets as well as for some exoplanets. The black dotted line is a fit to the values for the solar system planets, forced to plateau at 1 once the planet metallicity equals the stellar metallicity. The HabEx estimated error bar is also shown. HabEx will investigate whether exoplanets across a range of sizes/types also follow the trend of increasing metallicity for progressively smaller (less massive) worlds. Figure taken from Mansfield et al. (2018).

### 3.2.2.2 Near UV to Near-IR Spectral Information

Examples illustrating the anticipated spectral diversity of a range of sub-giant exoplanets spectra observed in the near UV-visible range (0.2–0.7 µm) are shown in **Figure 3.2-3**. A number of salient features are seen, such as bright clouds on Venus, the effect of ozone in the UV, Rayleigh scattering for Earth-like planets, and distinct $CH_4$ and $H_2O$ vapor features in cool giant and sub-giant planets. The variety of planet types will provide different input to understanding evolution than spectra of hot Jupiters. Cool Neptune-size exoplanets are expected to show strong signatures of $CH_4$ and water vapor—key chemical species that are the dominant forms of carbon and oxygen, respectively. Complex photochemistry in the chemically reduced atmospheres of Neptunes and sub-Neptunes may lead to haze formation, reddening the spectra of these planets. Some terrestrial planets, including Earth and super-Earth sized exoplanets, may possess atmospheres substantially thicker than that of our Earth, which would be indicated by strong Rayleigh scattering features. Alternatively, rocky worlds like Mars, which have experienced atmospheric loss or erosion may appear as barren rock, presenting few spectral signatures beyond their red color.

While the precise mix of which atmospheric species to expect is unknown, the absorption bands of key species are well known and drive the instrument design. Strong water vapor bands, $O_2$ and $O_3$ features, $CO_2$ bands, and more are found in the 0.2–1.7 µm wavelength range, as shown in **Figures 3.2-4** and **3.2-5**. This wavelength coverage is broad enough to detect and distinguish between deep $CO_2$ atmospheres, $H_2O$-rich steam atmospheres, the $CH_4$-rich atmospheres of Jovians through sub-Neptunes, and, critically, the atmospheres of $O_2$-containing Earth-like planets.

In particular, spectral retrievals of $H_2O$ vapor can help constrain the value of surface gravity, which is not well constrained by $CH_4$ alone (Lupu et al. 2016; MacDonald et al. 2018). $H_2O$ abundance of a cool gas giant planetary atmosphere is a powerful indicator of formation conditions. Inferring $H_2O$ in the solar system giant planets is challenging, due to condensation depleting the upper atmosphere of water vapor. Substantially warmer hot Jupiter exoplanets readily allow detections of $H_2O$ via transmission spectroscopy, but such signatures are often diminished by the presence of clouds of other species. In contrast, highly scattering $H_2O$ clouds can brighten planets significantly in reflected light in the 0.45 to 1.0 μm range making $H_2O$ manifestly detectable in reflected light spectra of cool giant planets only marginally warmer than Jupiter (MacDonald et al. 2018 and **Figure 3.2-6**).

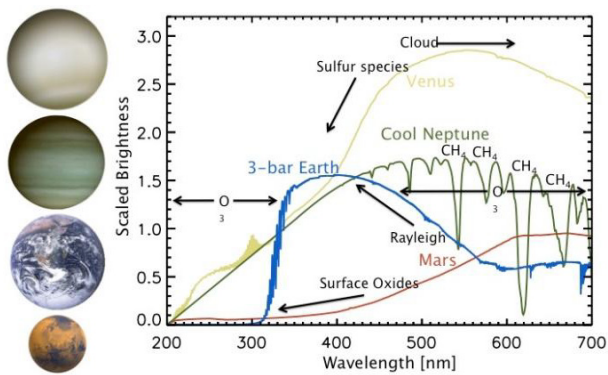

**Figure 3.2-3.** HabEx will begin to map out the true diversity of exoplanets as terrestrial through Neptune-like planets are expected to show a wide diversity in their atmospheric spectral features.

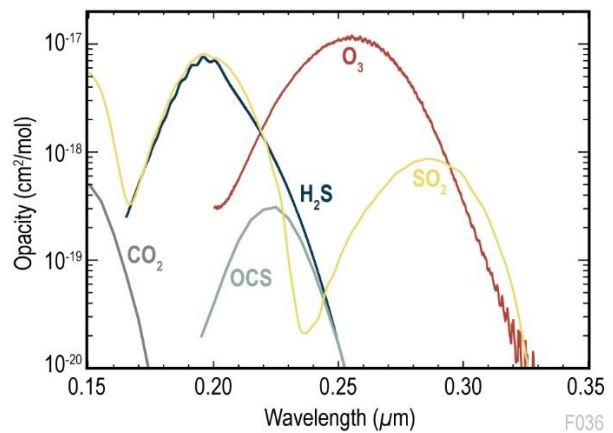

**Figure 3.2-4.** Key molecular absorption features appear in the UV/optical (0.15–0.35 µm), including $O_3$. UV opacities of key species shown are assembled from the Virtual Planetary Laboratory Molecular Spectroscopic Database.

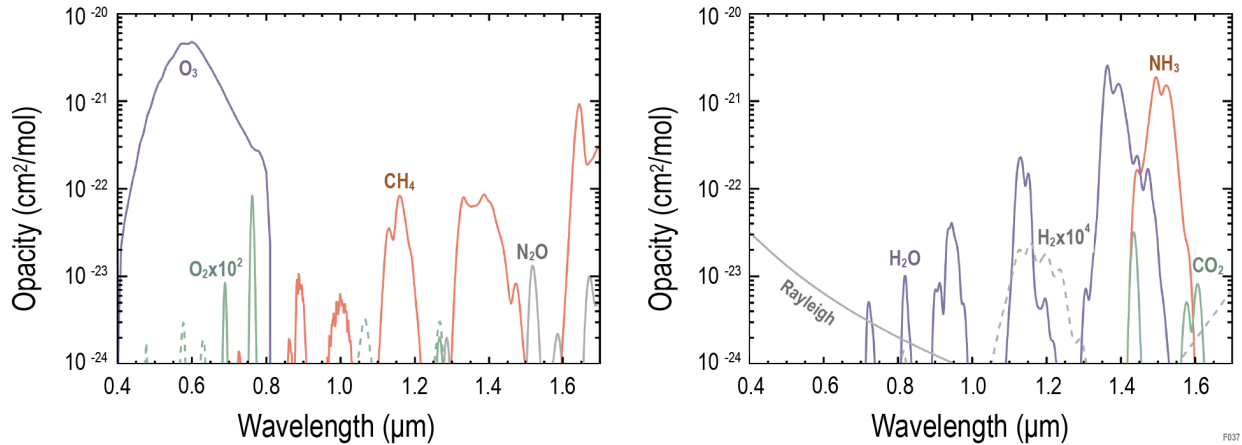

**Figure 3.2-5.** Optical to near-IR coverage 0.4–1.7 μm ensures sensitivity to key molecular absorption features including $O_2$, $H_2O$, and $CH_4$. Opacities of key molecular species are assembled from the HITRAN 2012 database, following Meadows and Crisp (1996)Meadows and Crisp (1996).

*Water Vapor and Methane Abundances*

Initial results from the Feng et al. (2018) spectral retrieval framework show that HabEx-like SNR of 20 spectroscopy spanning the visible and near-infrared on a warm mini-Neptune can yield constraints on the atmospheric water vapor and methane abundances to within a factor of two of the true value, thereby yielding strong constraints on the atmospheric metallicity (**Figure 3.2-7**). Such a measurement would be able to distinguish the metal enrichments of a Jupiter-like planet from a Neptune-like planet, whose metallicities differ by roughly an order of magnitude. Such spectral measurements and retrieval will help constrain elemental ratios such as C/O and O/H, which are critical for understanding planet formation mechanisms, because they govern the distribution and formation of chemical species in the protoplanetary disc. The gaseous C/O ratio in planets can deviate from the stellar value depending on different parameters (temperature, pressure, oxidation state, etc.) and processes during planet formation, including the initial location of formation of the planetary embryos, the migration path of the planet, and the

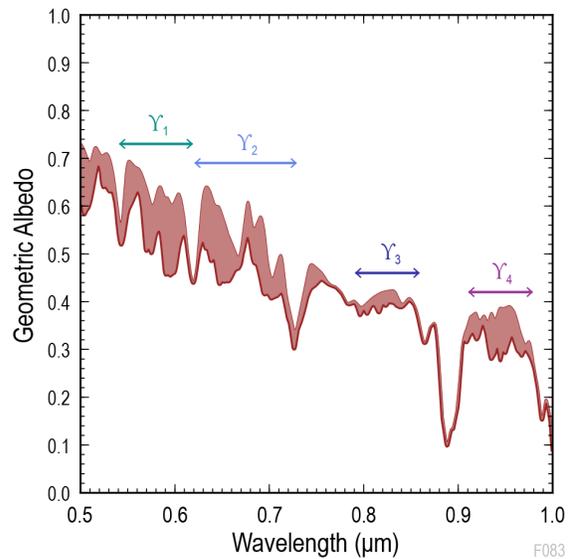

**Figure 3.2-6.** $H_2O$ vapor can strongly affect spectral features in reflection spectra of giant planets. The bold, bottom line of the red shaded area shows the geometric albedo spectrum of a giant planet model with an effective temperature of 180 K. The light red curve shows the geometric albedo of an identical model with $H_2O$ vapor opacity disabled, such that the enclosed red shaded region is caused by $H_2O$ absorption. Four spectral regions with prominent $H_2O$ features are indicated as $\gamma_1, \gamma_2, \gamma_3$ and $\gamma_4$ (adapted from MacDonald et al. (2018).

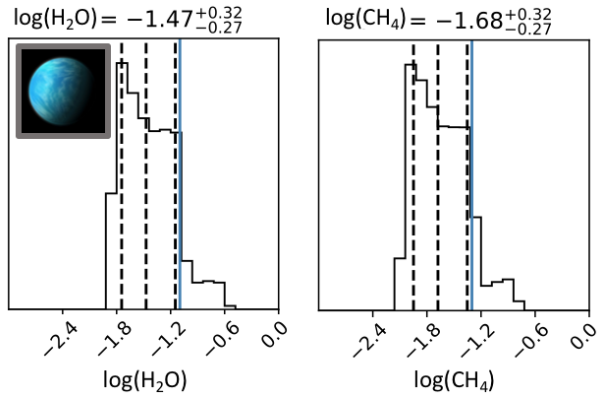

$\log(H_2O) = -1.47^{+0.32}_{-0.27}$   $\log(CH_4) = -1.68^{+0.32}_{-0.27}$

$\log(H_2O)$   $\log(CH_4)$

**Figure 3.2-7.** Inferred posterior distributions of water vapor (*left*) and methane (*right*) abundances for a retrieval analysis of a sub-Neptune observed with SNR of 20 spectroscopy, following Feng et al. (2018). Vertical blue lines indicate the known 'truth' value from the input atmospheric model. Dashed vertical lines indicate the mean value retrieved as well as ±1σ boundaries. Critically, retrieved values are within a factor of two of the true value. Inset image credit: Marc Ward.

evolution of the gas phase of a protoplanetary disc (Öberg et al. 2011; Ali-Dib et al. 2014; Madhusudhan et al. 2014; Thiabaud et al. 2015). Spectral retrieval and analysis also can help constrain parameters such as the effective temperature, surface gravity, atmospheric composition, or radius of the detected planets (e.g., Marley et al. 2012; Bonnefoy et al. 2014; Galicher et al. 2014).

*Molecular Hydrogen Abundance*

Finally, molecular hydrogen has a strong collision-induced absorption band at ~1.15 μm, and a weaker one at 0.8 μm (**Figure 3.2-5**). The depths of these features grow with the square of absorber number density, and can become quite significant for $H_2$ partial pressure above roughly 0.5 bar for an atmosphere with mean molecular weight equivalent to $N_2$ or heavier. The strength of a collision-induced absorption feature depends on the bulk atmospheric mean molecular weight, making such features easier to detect in higher mean molecular weight atmospheres (for a fixed absorbing gas partial pressure). This provides an opportunity for HabEx to both constrain the $H_2$ abundance in certain planetary atmospheres as well as to constrain background pressures in these atmospheres. Knowing whether planets have a primordial $H_2$ atmosphere is key for their classification, as illustrated by the ongoing discussion on the transition from solid planets to planets with (some) H/He (Fulton et al. 2017).

### 3.2.2.3    Requirements

While the types of planets and their atmospheres are likely to surpass our models and imaginations, a broad near UV to near-IR spectral range as well as access to a broad range of planet semi-major axis are key to planet diversity studies. HabEx main baseline requirements for this objective are to provide spectroscopic capabilities from 0.2 to 1.7 μm and an OWA ≥ 6 arcsec. In the near UV, a spectral resolution of $R \geq 5$ is required to differentiate between the different opacity sources (**Figure 3.2-4**). At a spectral resolution of $R \geq 70$ in the visible, the sharpest features (i.e., the $O_2$ bands) are resolved, and a near-IR resolution of ≥ 10 would reach numerous broad infrared bands for $CO_2$, $CH_4$, and $H_2O$ in particular (**Figure 3.2-5**). Previous work demonstrates that spectroscopic observations at this minimum resolution and characteristic SNRs of 10 (or larger) is sufficient for species detection for directly imaged terrestrial and gas giant exoplanets in reflected light (Nayak et al. 2017; Feng et al. 2018).

A threshold science case emphasizes the strongest visible-wavelength spectral features for a limited collection of gases ($CO_2$, $CH_4$, and $H_2O$), where atmospheric characterization remains feasible for characteristic spectral SNRs above 10 (Nayak et al. 2017), as well as identification of Rayleigh scattering in the 0.45–0.7 μm region.

### Objective 6 Requirements

| Parameter | Baseline | Threshold |
|---|---|---|
| **Wavelength Range** | ≤0.3 μm to ≥1.7 μm | ≤0.45 μm to ≥1.0 μm |
| **Spectral Resolution, R** | $O_3$: ≥5 (0.30–0.35 μm) $O_2$: ≥70 (at 0.76 μm) $H_2O$, $CO_2$, $CH_4$, $H_2$: ≥12 (1.0–1.7 μm) Rayleigh Scat.: ≥5 (0.45–0.7 μm) | $CO_2$: ≥100 (at 0.87 μm) $H_2O$: ≥17 (at 0.94 μm) $CH_4$: ≥32 (at 0.89 μm) $H_2$: ≥8 (at 0.80 μm) Rayleigh Scat.: ≥5 (0.45–0.7 μm) |
| **IWA_0.5** | ≤130 mas at 1.7 μm | ≤130 mas at 1.0 μm |
| **OWA** | ≥6 arcsec | ≥0.5 arcsec |

### 3.2.3 Objective 7: Do outer giant planets impact the atmospheric water content of small planets inside the snow line?

The prevailing theory in planetary science is that the architecture of our solar system had a substantial influence on the habitability of Earth. In particular, Jupiter may have regulated the dynamical delivery of water to Earth by perturbing the orbits of small bodies from beyond the snow line (Raymond et al. 2004). The HabEx mission shall provide key opportunities to test the generality of this profound theory. Raymond et al. (2004) argue that if Jupiter had a significantly higher eccentricity, little water would have been delivered to the inner solar system, leaving Earth too dry for habitability. Conversely, Faramaz et al. (2017) argue that mean-motion resonances with Jovian mass planets located exterior to a Kuiper-belt analog and on moderately eccentric ($e \gtrsim 0.1$) orbits could scatter planetesimals onto cometary orbits with delays of the order of several 100 Myr, resulting in continuous delivery of water to Earth over Gy time-scales. Depending on whether they reside inside or outside of cold reservoirs of planetesimals, outer Jovian planets in eccentric orbits may then drive the concentrations of water in the atmosphere of rocky worlds. Additionally, if there were no giant planets at all beyond the orbit of the Earth, Earth would have been bombarded by migrating, water-rich planetesimals from beyond the snow line, leaving a water world (Morbidelli and Raymond 2016).

High-contrast observations with HabEx shall be able to investigate these dynamical predictions by correlating the abundance of atmospheric water vapor—and other volatiles, e.g., de Niem et al. 2012—found on detected rocky planets (*Section 3.1.2*) with the presence/absence of a Jovian planet with specific orbital characteristics (semi-major axis, eccentricity and inclination). In the inner/ outer planet scenario envisioned here, an EEC was detected at least four times by HabEx to constrain its orbit (*Section 3.1.1*), requiring an average number of 8 separate visits. The orbit of an outer jovian planet at 5 AU can be adequately constrained by these observations, assuming it is also detected ≥4 times among the

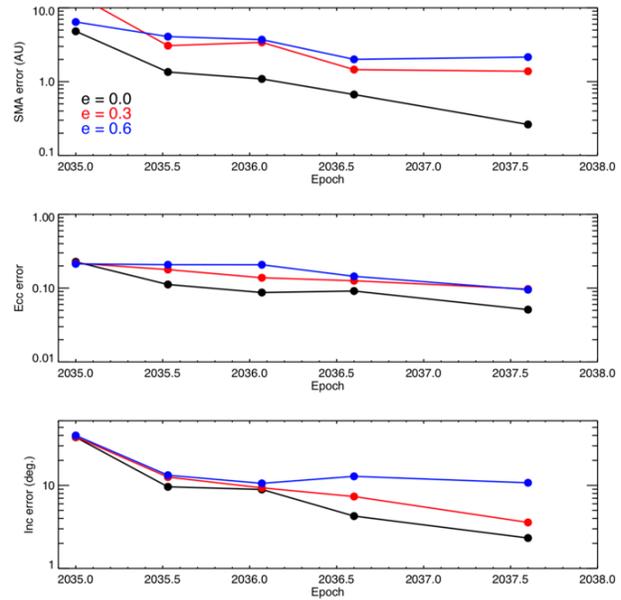

**Figure 3.2-8.** Orbital parameter retrieval simulation (same as in *Section 3.1.1*) for a Jupiter-size planet with a 5 AU semi-major axis, a 30 deg inclination and 3 different eccentricities. Four coronagraph-based detections separated by 6.4 months are simulated, plus a starshade follow-up detection a year later. The 3 main orbital parameters are derived to better than 10% (1σ) in the zero-eccentricity case, and generally well enough to distinguish a circular orbit from an eccentric one or reveal any significant orbital misalignment with other inner planets detected. Credit: Eric Nielsen.

8 visits to the EEC host star, and still detected a year later during the EEC spectroscopic observation (**Figure 3.2-8**). This requires an IWA small enough to detect water in inner rocky planets (which is 80 mas, from Objective 2) and an OWA large enough to detect and constrain the orbit of outer Jovian planets at 5–10 AU distances between the snow-line and an outer dust belt.

For a nearby stellar target at 3 pc, this 5–10 AU distance corresponds to an OWA greater than 3 arcsec. The minimum required wavelength range of 0.45–1.0 μm, taken from Objectives 2 and 5, enables planet-size constraints as well as identification of potentially habitable worlds through water vapor detection.

### Objective 7 Requirements

| Parameter | Baseline | Threshold |
|---|---|---|
| **IWA$_{0.5}$ at 1.0 μm** | ≤80 mas | ≤105 mas |
| **OWA at 0.45 μm** | ≥3 arcsec | ≥0.5 arcsec |
| **Wavelength range** | 0.45–1.0 μm | 0.45–1.0 μm |

### 3.2.4 Objective 8: How do planets, small bodies, and dust interact?

In addition to planets, circumstellar dust is a key component of an exoplanetary system that can be directly imaged. This section concentrates on the faint debris disk structures—zodiacal dust and Kuiper belt analogs—that will be imaged and characterized as part of HabEx observations focused on addressing scientific Goals 1 and 2. Additional dedicated observations of optically thick protoplanetary disks and bright extended debris disks (essentially around early-type stars not part of HabEx EEC surveys) are covered in the Observatory Science program (*Chapter 4*). There are three main science questions that drive HabEx capabilities in terms of faint debris disks observations.

**Is the solar system's two-belt architecture common?** In the standard solar system formation paradigm, the regions of low planetesimal density are just inside the snow line (which is attributed to the presence of Jupiter which forms just beyond it) and at the outer edge of the solar system. Alternatively, stochastic processes could leave planetesimal belts in stable regions between any pair of planets, yielding belts at apparent unimportant locations.

A comprehensive understanding of planetary system architectures requires measuring the location, density, and spatial distribution of dust and planetesimal belts around nearby mature stars. To detect and characterize a broad range of dusty debris disks in nearby systems, whether inner exozodiacal dust, exo-Kuiper belt analogs, or other dust structures located in between, HabEx requires the capability to detect disks at solar dust density level in the HZ, and at ~10 times the density of the Kuiper belt in the outer regions. That requirement shall be met per spatial resolution element, meaning that even lower dust density levels will be accessible at lower spatial resolution than provided by the telescope PSF. For sunlike stars located between 5 and 10 pc, this translates into a surface brightness detection limit per spatial resolution element of ~22 Vmag/ arcsec$^2$ at 0.1" (e.g., Roberge et al. 2012) and 26 Vmag/ arcsec$^2$ at 3"

(e.g., Kuchner and Stark 2010), with a high contrast imaging region extending from ~0.1" (1 AU at 10 pc) to ~6" (30 AU at 5 pc). In comparison, the best sensitivity limit reached so far for such studies, using HST/STIS for deep visible light imaging of debris systems around solar analogs (Schneider et al. 2014; Schneider et al. 2016) is about 15 $V_{mag}$/arcsec$^2$ at an IWA of 0.3"—corresponding to 5 AU for the closest disks detected- and 22 $V_{mag}$/arcsec$^2$ at 5". Assuming dust spatial distribution and physical characteristics similar to that of the solar system zodiacal cloud, the Large Binocular Telescope Interferometer (LBTI) mid-infrared survey sensitivity of 100 zodis around individual sunlike stars would correspond to a surface brightness of 17 $V_{mag}$/arcsec$^2$ at 0.1–0.3" separations, still far from the HabEx requirement, and with little to no spatial information on the actual brightness distribution on sub-AU scales.

**How is dust produced and transported in debris disks?** Observing much fainter debris disks than currently possible crosses an important threshold in disk physics. Bright disks—all those currently imaged—are collision dominated; the dust grains observed are mainly destroyed by collisions with other grains before their orbits are influenced by radiation forces. Disks with optical depths less than ~$v_{Keplerian}/c$ (with $v_{Keplerian}$ the local Keplerian speed and $c$ the speed of light) are predicted to be transport dominated (Kuchner and Stark 2010), meaning that grain-grain collisions are rare enough that grains can flow radially inwards under the influence of radiation drag forces until they are sublimated in the star's corona or ejected from the system by an encounter with a planet. Understanding the origin of warm exozodi dust present in or near the HZ of main sequence stars is of particular interest as they are essentially three main scenarios invoked for its generation (**Figure 3.2-9**):

1. In-situ collision(s) between parent bodies that reside where the dust is observed. These collisions may be between many bodies in an asteroid-like belt, or single giant impacts between larger bodies (Kennedy and Wyatt 2013);

2. Inward transport of dust from an outer asteroid or Kuiper-belt-like region via Poynting-Robertson (P-R) drag (Wyatt 2005; Kennedy and Piette 2015); and

3. Inward transport of comets scattered by planets from an outer asteroid or Kuiper-belt-like source region (Bonsor et al. 2012).

These scenarios can result in significantly different total brightness levels and optical depth radial profiles in the inner exozodiacal region. Recent in-situ dust production is identifiable by bright, localized, and possibly asymmetric structure, but over time becomes similar in appearance to the second P-R drag scenario (i.e., the disk becomes transport dominated, but the dust does not extend very far beyond the HZ). The P-R drag scenario yields a smoothly varying and theoretically predicted dust profile interior to the source belt, that may be depleted interior to massive planets (Bonsor et al. 2018). Delivery by comets can yield dust levels significantly above the P-R drag scenario (Wyatt et al. 2007), and should yield a radially broad dust distribution interior to the radius where sublimation becomes significant. In order to distinguish between these different scenarios, HabEx shall conduct spatially resolved observations of radial and azimuthal exozodi structures, as well as outer disk belts and planets that may be at stake in shaping them.

Dust transport also critically depends on the grain size, which defines the surface area to mass ratio and thus the strength of radiative forces on the grain. Grain sizes can be constrained effectively by measuring scattering phase functions over a wide range of wavelengths, both in total intensity and in polarized light (e.g., Perrin et al. 2015). Measuring the scattering phase function in polarized light is especially informative in terms of grain properties, as it is only sensitive to the polarization strength, while total intensity measurements may also reflect spatial variations in dust density. As such, HabEx requires the capability to obtain spatially resolved polarimetry and broad-band disk colors over the visible spectral range (0.45–1 μm). While spatially resolved visible spectroscopy ($R \geq 20$) of the disk inner 1" region is not a formal requirement, especially in the common case of dust grains large enough to be featureless in the visible, it may still be beneficial in helping disentangle faint inner planets and their characteristic absorption features from any bright resonant dust structures in the disk.

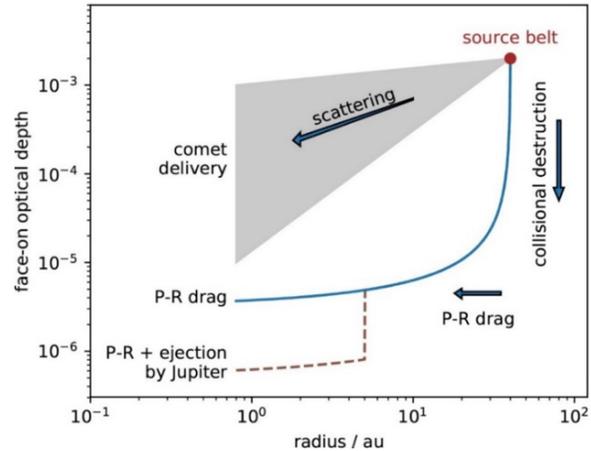

**Figure 3.2-9.** Cartoon example of exozodi delivery processes from an exterior source debris disk and the distinct optical depth profiles expected. The dust level is depleted by destructive collisions as it spirals in under P-R drag (*blue line*) and can be additionally depleted as it passes planets (*dotted line*). Conversely, cometary scattering can result in a very wide range of dust levels in the habitable zone, including higher optical depth regions (*gray swath*) incompatible with P-R drag alone. Image courtesy of G. Kennedy, Univ. of Warwick.

**Can debris disks really be used as a planet detection and characterization technique?** Debris disks inner cavities, warps, offsets, non-axisymmetric features, spirals, clumps and separated rings revealed in scattered light images (Schneider et al. 2014) of micron-sized to millimeter-sized dust particles are considered signposts of existing planets (e.g., Moro-Martín et al. 2008). A connection between debris disk features and the presence of a massive perturbing planet was successfully demonstrated with the very extended bright asymmetric debris disk observed around β Pictoris (Burrows et al. 1995; Mouillet et al. 1997) and the self-luminous giant planet subsequently imaged around it (Lagrange et al. 2008). While this extraordinary case remains one of a kind, the connection between debris disks and planets has also been seen in several of the currently known directly imaged planetary systems besides β Pictoris, such as HR 8799, HD 95086, HD 106906, Fomalhaut, and 51 Eridani (see, e.g., Bowler (2016) for a recent review). Using the largest sample of debris disks systems directly surveyed for long-period giant planets to date, Meshkat et al. (2017) recently found that the occurrence rate of long-period giant planets in dusty systems is about ten times higher than in dust-free systems (at current detection limits), providing tentative empirical evidence for a planet–disk connection.

But is it really the case? and if comparable structures exist in the inner exozodi region, do they indeed sometimes correlate with the presence or absence of specific types of planets (Raymond et al. 2011; Bonsor et al. 2012; Bonsor et al. 2018), and if so, in which case? To help establish the "ground truth" of disk-planet interactions, HabEx must obtain spatially resolved images of debris disks of variable dust levels and of the exoplanets embedded in them, over a broad range of planetary masses and separations. Accessing sub-jovian mass planets is paramount, as Saturns can already impede dust migration by PR drag (Bonsor et al. 2018), and super-Earths and Neptunes can play a significant role in comet scattering (e.g., Wyatt et al. 2017; Marino et al. 2017). To spatially resolve exozodi spatial structures around mature stars and *at the same time* directly reveal perturbing planets in them, HabEx observations shall reach detection limits per resolution element of $<\sim 10^{-10}$ in the visible at separations $<\sim 0.1"$. These contrast requirements are less stringent than those driven by the exo-Earth characterization objectives listed in *Section 3.1*.

It may also be possible to empirically determine which observed exozodi features (e.g., clump orbital motion, grain size distribution) correlate with the presence of planets. By measuring such correlations, HabEx will make it possible to truly use dust disk observations as an indirect planet detection technique and reveal the presence of planets too small and/or faint to image directly in reflected light (Shannon et al. 2015). In this regard, HabEx debris disks observations will have the potential to further characterize the diversity of planets as well, e.g., to set constraints on the outward migration of extrasolar "Neptunes" (Moro-Martin et al. 2008) through the signature they may imprint on the outer disk structures.

## Objective 8 Requirements

| Parameter | Baseline | Threshold |
|---|---|---|
| Wavelength range | Visible | Visible |
| Surface Brightness Sensitivity (0.5 µm) | 22 mag/arcsec$^2$ (0.1 arcsec) 26 mag/arcsec$^2$ (3 arcsec) | 22 mag /arcsec$^2$ (0.1 arcsec) 23.5 mag /arcsec$^2$ (1 arcsec) |
| IWA$_{0.5}$ (0.5 µm) | ≤100 mas | ≤125 mas |
| OWA | ≥ 6 arcsec (0.5 µm) | ≥ 1 arcsec (0.8 µm) |
| Polarimetric capability | Broadband images recorded in ≥3 different polarization states | None |

## 3.3 Exoplanet Science Yield Estimate

Assuming 50% of the 5-year primary mission is dedicated to direct imaging and characterization, HabEx will revolutionize exoplanet science by searching ~50 nearby stars for potentially Earth-like exoplanets during the deep and broad surveys, discovering nearly two hundred diverse exoplanets in the process. Precisely estimating the expected exoplanet science yield necessitates modeling the execution of such a mission, which in turn requires constraints on several key astrophysical parameters, such as exoplanet occurrence rates, as well as a high-fidelity simulator of exoplanet imaging missions. A decade ago, such modeling was not possible. Now, the Kepler Mission has constrained the frequency of Earth-sized potentially habitable planets around sunlike stars (e.g., Burke et al. 2015), the Keck Interferometer Nuller (Mennesson et al. 2014) and the Large Binocular Telescope Interferometer (Ertel et al. 2018) have placed constraints on the presence of warm dust around nearby stars. At the same time, exoplanet mission yield optimization algorithms and design reference mission simulators have advanced dramatically (e.g., Stark et al. 2014; Stark et al. 2015; Savransky et al. 2016).

This section assumes the HabEx baseline architecture. It makes use of a 4 m, off-axis primary mirror and of a hybrid coronagraph and starshade starlight suppression techniques enabling direct imaging and spectroscopy of Earth-sized and larger exoplanets. The HabEx

**Table 3.3-1**. Summary of HabEx direct imaging performance (see detailed specifications in *Sections 6.3* and *6.4*).

| | Coronagraph | Starshade |
|---|---|---|
| **Instrument Type** | Vector vortex charge 6 coronagraph with:<br>• Raw instrument contrast [1]: $2.5 \times 10^{-10}$ at $IWA_{0.5}$<br>• $\Delta$mag limit[3] = 26.5 (equiv. to $2.5 \times 10^{-11}$ flux ratio)<br>• 20% instantaneous bandwidth<br>• Imager and spectrograph | 52 m diameter starshade with:<br>• Raw instrument contrast [2]: $1 \times 10^{-10}$ at $IWA_{0.5}$<br>• $\Delta$mag limit = 26.5<br>• 107% instantaneous bandwidth<br>• Imager and spectrograph |
| **Channels** | • Vis (Blue, Channels A & B): 0.45–0.67 μm<br>  Imager + IFS with $R$ = 140<br>• Vis (Red, Channels A & B): 0.67–1.0 μm<br>  Imager + IFS with $R$ = 140<br>• NIR (Channel B): 0.975–1.8 μm<br>  Imager + slit spectrograph with $R$ = 40 | • UV: 0.2–0.45 μm<br>  Imager + grism with $R$ = 7<br>• Vis: 0.45–1.0 μm<br>  Imager + IFS with $R$ = 140<br>• NIR: 0.95–1.8 μm<br>  Imager + IFS with $R$ = 40 |
| **Field of View** | $IWA_{0.5}$: 2.4 $\lambda/D$ = 62 mas (0.5 μm)<br>OWA: 0.83 arcsec (0.5 μm) | $IWA_{0.5}$: 58 mas (0.3–1.0 μm)<br>OWA: 6 arcsec (Vis Broad-band Imaging)<br>OWA: 1 arcsec (Vis IFS) |

Notes:
[1] Achieved in the visible on a 1 mas diameter star (see full coronagraph error budget in Figure 5.2-1).
[2] Excluding solar glint (see full starshade error budget in Figure 5.2-2)
[3] Magnitude difference between central star and faintest detectable planet. Image processing techniques make it possible to detect planets with planet-to-star flux ratios up to 10× below the raw instrument contrast

coronagraph and starshade instrument main specifications (**Table 3.3-1**) trace directly from the science requirements established in the two previous sections (*Sections 3.1* and *3.2*). Full details on the coronagraph and starshade instruments can be found in *Sections 6.3* and *6.4*. The starshade occulter is described in *Chapter 7*.

One key characteristic of HabEx dual starlight suppression system is that **the starshade is specifically designed to provide the same IWA over the whole 0.3–1.0 μm region as the coronagraph for broadband detection at 0.5 μm. This means that all of the planets imaged by the coronagraph can be characterized spectroscopically from 0.3–1.0 μm at once by the starshade** (within the 100 starshade transits available and assuming proper timing), with spectroscopic capabilities designed to search for atmospheric biosignature gases and the presence of surface liquid water on HZ rocky planets. The wavelength coverage from UV (0.2 μm) to near-IR (1.8 μm) and the spectral resolution ($R$ = 140 in the 0.45–1.0 μm range, $R$ = 40 above 1.0 μm) captures the absorption bands of key molecular species, which can be used to distinguish between different types of exoplanets. Strong water vapor bands, oxygen and ozone features, carbon dioxide and methane bands, and more are part of the HabEx wavelength range. Both the coronagraph and starshade instruments have visible through near-IR capabilities, including imaging and integral field spectroscopy (IFS). The starshade instrument also includes a UV grism for low-resolution spectroscopy ($R$ = 7) down to 0.2 μm. The high-contrast imaging field of view extends from an IWA of 58 mas and a (maximum) OWA of 6 arcsec. This enables detection of planets over a broad range of orbital semi-major axes and temperatures.

The quantity and quality of exoplanet science that the HabEx mission concept can produce can be estimated using established exoplanet yield calculation and target prioritization methods (Stark et al. 2014). The science yield expected from the baseline 4 m HabEx concept and exoplanet surveys is summarized in this section. As an illustrative case, a total of 2.5 years (including overheads) was assumed for exoplanets direct imaging and spectral characterization, i.e., devoting 50% of the prime mission duration. A complete description of the exoplanet science yield estimates, the techniques used, assumptions made, and justification for the adopted exoplanet observing strategy can be found in *Appendix C*. Only the science yield of the high-contrast imaging instruments is presented here. In addition, the HabEx Observatory has two broad-purpose instruments dedicated to a Guest Observer (GO) program (*Chapter 4*): the HabEx Workhorse Camera (HWC), and the Ultraviolet Spectrograph (UVS).

### 3.3.1 Exoplanet Observing Programs and Operations Concepts

HabEx is designed to obtain three primary exoplanet data products: (1) multiband photometry to detect planets and dust disks; (2) precise astrometry to measure exoplanet orbits and determine if a planet resides in the HZ; and (3), spectra to assess chemical compositions. HabEx will obtain these measurements on all planetary systems observed, via two primary observation programs: the *broad* exoplanet survey and the *deep* exoplanet survey.

#### 3.3.1.1 HabEx Broad Exoplanet Survey

Under the notional time allocation assumed, HabEx will devote 1.75 years of wall clock time to conduct a broad survey of 42 nearby stars (*Appendices C* and *D*) optimized for the detection (1.1 year) and spectral characterization (0.65 year) of EECs. Another 0.5 year will be spent to spectrally characterize planetary systems in which no EECs were detected. This broad survey will hence have three components:

1. First, HabEx will obtain multi-epoch coronagraph images of all ~42 target stars in the broad survey, with the goal of detecting EECs, measuring their colors, and constraining their orbits. We expect each star will be observed ~6 times on average, with low priority stars observed a few times and high priority targets observed up to ~10 times. The yield simulations predict that if all target stars within 12 pc had an exo-Earth in their HZ, ~70% of them would have such planets detected and their orbits determined using this scenario. In other words, this initial phase is characterized by a ~70% average "HZ search completeness" for observed stars within 12 pc. These initial multi-epoch coronagraph observations (**Figure 3.3-1**) will allow HabEx to prioritize targets and optimize scheduling the phases of planets for the next step in this program, especially for smaller inner planets in fast orbits.

2. The next step in this survey uses the starshade to further characterize the most interesting systems identified in the previous phase.

Detailed simulations of starshade slews and consumables, discussed in the design reference mission section in *Chapter 8*, indicate that all planetary systems can be imaged and spectrally characterized. Starshade observations will begin with a 12"×12" broadband visible image, possibly revealing additional outer planets as well as outer dust belts inaccessible to the coronagraph smaller field of view. **Figure 3.3-2** shows a simulated starshade visible image of a putative five-planet system around β CVn, a representative sunlike target star at 8.6 . The whole exoplanetary system can be imaged, revealing its exozodi and exo-Kuiper belts, an Earth analog at 1 AU, a sub-Neptune at 2 AU, Jupiter, Saturn and Neptune analogs at 5, 10, and 15 AU respectively.

For all systems with exo-Earth candidates detected, the starshade visible IFS will obtain planetary spectra from 0.3–1.0 μm *in a single observation*, with $R = 7$ from 0.3–0.45 μm and $R = 140$ from 0.45–1.0 μm, with an SNR of 10 or higher per spectral channel. This 0.3–1.0 μm portion of the spectrum is obtained by placing the starshade at its nominal distance of 76,600 km.

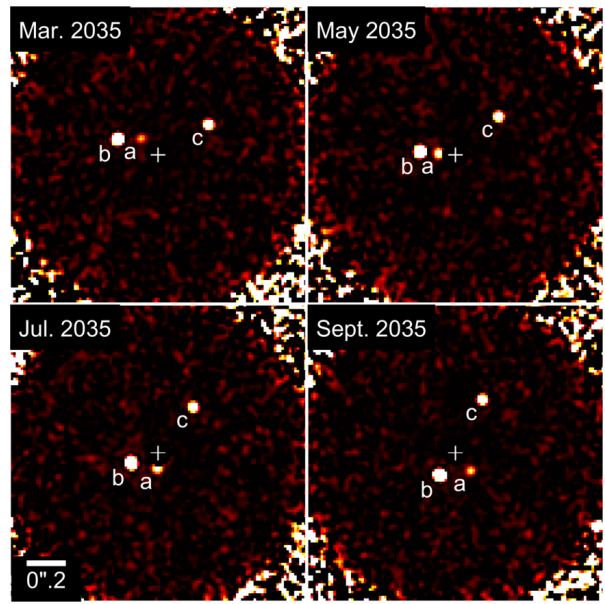

**Figure 3.3-1.** Simulated HabEx coronagraph instrument detection of an exo-Earth (a), a sub-Neptune planet (b) and a Jupiter analog (c) around β CVn - a sunlike star at 8.6pc- at four observing epochs; orbital inclination 60°; exo-Earth semi-major axis: 1 AU; wavelength range: 0.45–0.55 µm. The inner centro-symmetric exozodi dust component has been fitted and subtracted. Credit: G. Ruane.

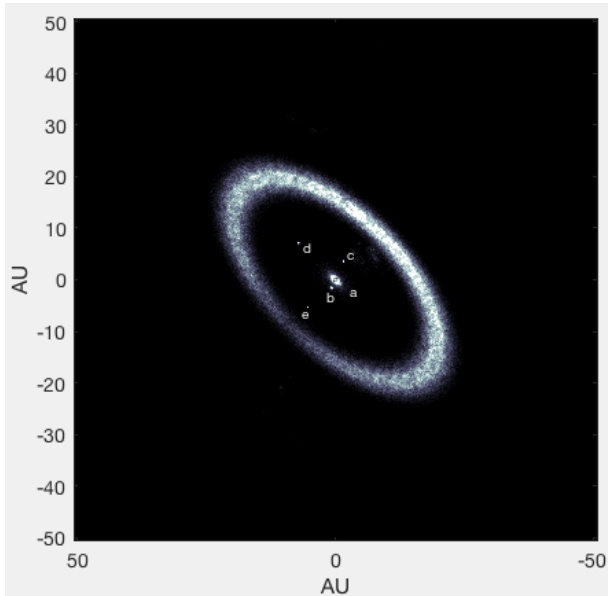

**Figure 3.3-2.** Simulated HabEx starshade instrument detection of an exo-Earth (a), a sub-Neptune (b), Jupiter (c) and Saturn (d) analogs, and a close-in Neptune (e) around β CVn. Inner dust belt (zodiacal dust analog within 1 AU) and outer dust belt (Kuiper belt analog around 30 AU), both with five times the density of solar system level, are clearly visible together with some background galaxies. Same system as in Figure 3.3-1 but now with a field of view 12″×12″ (~ 100 × 100 AU) revealing the outer planets and outer dust belt. Credit: S. Hildebrandt.

For a few select high-priority exo-Earths, multi-epoch visible spectra will be obtained, and further UV and near-IR characterization will be performed by maneuvering the starshade occulter to two different distances: 114,900 km to cover 0.2–0.67 μm (with improved 39 mas IWA) and 42,600 km to cover 0.54–1.8 μm (with 104 mas IWA). Simulated near-UV to near-IR spectra of the individual β CVn planets (same hypothetical system as in **Figure 3.3-2**) obtained by the HabEx starshade after 370 hours of total exposure time are shown in **Figure 3.3-3**. Spectra of the 5 planets are measured simultaneously by the IFS.

3. Planetary systems with no exo-Earth candidate found during the coronagraph

broad survey will *all* be spectrally characterized by the starshade as well, and be prioritized according to the types and number of planets detected. Planets detected slightly outside of the nominal EEC zone will for instance be high priority targets for water searches and empirical tests of the HZ concept (objective O2). Whether they contain EECs or not, systems with a large number of planets detected will provide the best targets for studying the diversity of planetary architectures (Objective O5) and atmospheres (Objective O6).

As shown by Stark et al. (2016), coronagraphs excel at orbit determination, but take longer to provide a spectrum with broad wavelength coverage. Starshades on the other hand, excel at quickly providing spectra, but can only constrain the orbits for a handful of targets due to the cost of slewing the starshade. The HabEx broad exoplanet survey is designed to fully capitalize on the complementary strengths of both instruments, combining them to provide higher yield and better characterization than either one alone (*Appendix D*).

### 3.3.1.2  HabEx Deep Exoplanet Survey

For the remaining 3 months of available exoplanet observations, HabEx will perform a "**deep survey**" of ~8 nearby (3–6 pc) high-priority sunlike stars with low exozodi levels (**Table 3.3-2**). These stars will be selected based on the very high search completeness that can be achieved through observations at even a single epoch with relatively short exposure times, for a broad range of planet types and physical separations (*Appendix D*). For this program, HabEx will use the starshade only to observe each star an average of three times. During each deep survey observation, HabEx will:

1. Obtain a deep broadband image limited by the systematic noise floor of $\Delta$mag = 26.5 to search for faint objects. This corresponds to a planet-to-star flux ratio of $2.5 \times 10^{-11}$, similar to a Mars-size planet seen at a gibbous phase in the HZ of a sunlike star. These deep broadband searches can be made quickly given the relative closeness of the target stars.

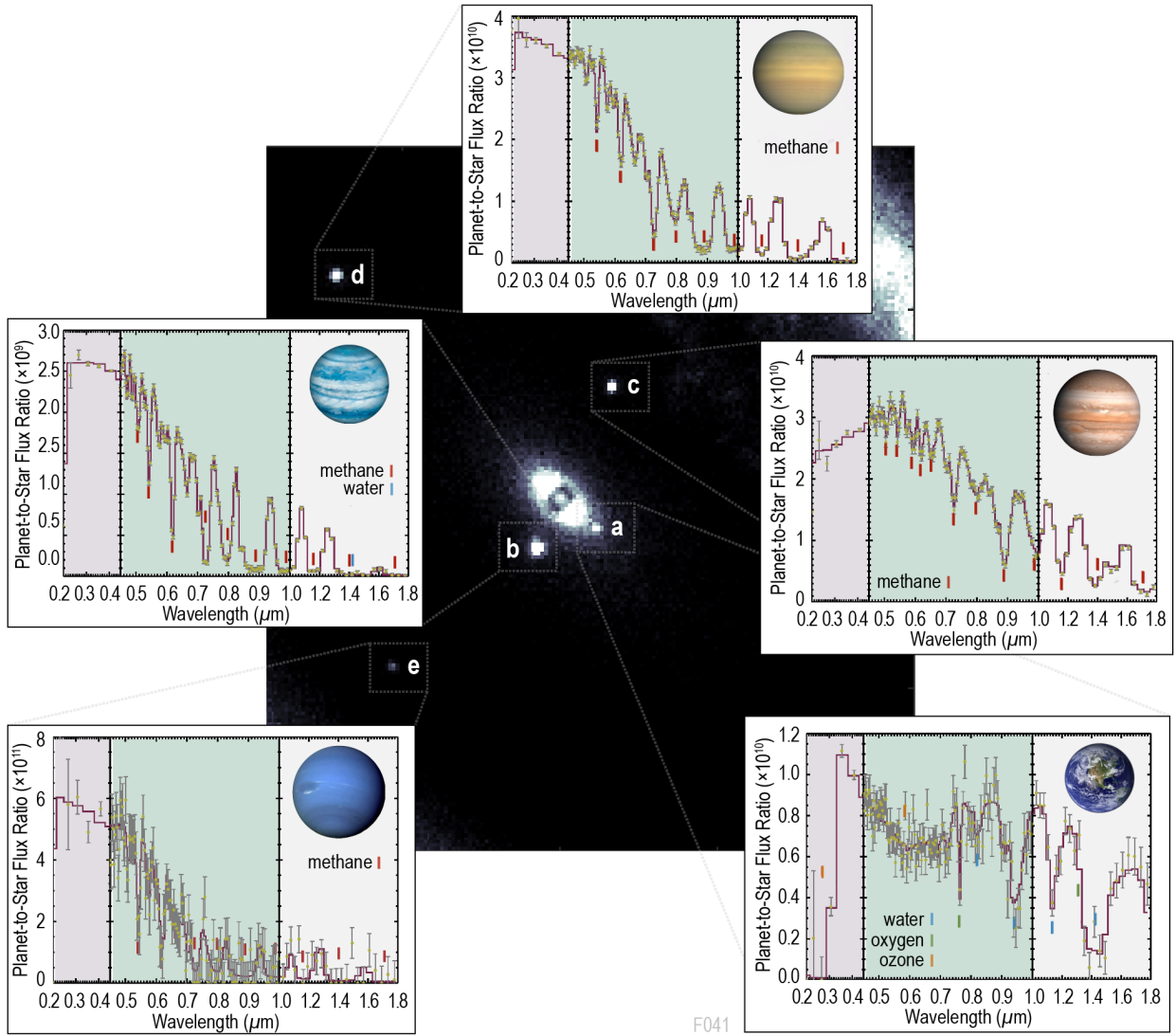

**Figure 3.3-3.** Simulated starshade instrument measurements of near-UV to near-IR spectra for all planets found in the 2"×2" inner area of Figure 3.3-2. Any bright planet located farther out can be spectrally characterized separately by offsetting the starshade IFS field center.

Obtain an R = 7 (grism) spectrum from 0.3–0.45 µm using the starshade UV channel and an R = 140 spectrum from 0.45–1.0 µm using the starshade visible channel IFS. The exposure times will be determined to enable detection of an Earth-twin at quadrature with an SNR = 10 per spectral channel, regardless of whether an exo-Earth candidate exists in the planetary system. Once again, these spectra can be obtained relatively quickly given the targets' distance and low exozodi levels (*Appendices C* and *D*).

These multi-epoch deep exposures and spectra will provide an unprecedented reconnaissance of ~8 of our closest neighbors, probing the atmospheric compositions of individual planets and revealing the overall architecture of their planetary systems including interplanetary dust structures in exquisite detail. The list of deep survey stars remains illustrative: the exact number and identity of these high-priority targets will be chosen based on additional knowledge about specific systems available by the time of the HabEx observations and gained during execution of the mission.

**Table 3.3-2.** Many of the eight deep survey targets, nearby sunlike stars, have captured the public's imagination for centuries.

| Star | Type | Dist. (pc) | V-mag | Age (Gyr) | Notes |
|------|------|-----------|-------|-----------|-------|
| τ Ceti | G8V | 3.7 | 3.5 | 5.8 | **Astronomy:** closest solitary G-star, 4 confirmed planets (2 in HZ) plus debris disk<br>**Popular culture:** homeport of *Kobayashi Maru* in *Star Trek* and location of *Barbarella* (1968) |
| 82 Eridani | G8V | 6.0 | 4.3 | 6.1–12.7 | **Astronomy:** 3 confirmed planets (all super-Earths) plus dusk disk |
| 40 Eridani | K1V | 5.0 | 4.4 | | **Astronomy:** triple-system, with white dwarf and M-dwarf<br>**Common name:** Keid<br>**Popular culture:** in *Star Trek*, host star to Vulcan |
| GJ 570 | K4V | 5.8 | 5.6 | | **Astronomy:** quadruple-system, with 2 red dwarfs and brown dwarf |
| σ Draconis | K0V | 5.8 | 4.7 | 3.0 ± 0.6 | **Astronomy:** 1 unconfirmed planet (Uranus-mass)<br>**Common name:** Alfasi<br>**Popular culture:** visited in *Star Trek* episode "*Spock's Brain*" (1966) |
| 61 Cygni A | K5V | 3.5 | 5.2 | 6.1 | **Astronomy:** wide-separation binary<br>**Common name:** Bessel's star |
| 61 Cygni B | K7V | 3.5 | 6.1 | | **Popular culture:** home system of humans in Asimov's *Foundation* series |
| ε Indi | K5V | 3.6 | 4.8 | 1.3 | **Astronomy:** triple-system, with 2 brown dwarfs<br>1 unconfirmed planet (Jupiter-mass) |

### 3.3.1.3   Notional Time Allocation during Prime Mission

HabEx exoplanet yield estimates are based on a 5-year primary mission, assuming a 50/50 split mission between exoplanet direct imaging surveys and "observatory science" investigations in general astrophysics and solar system science (*Chapter 4*). The notional time split assumed between the broad and deep exoplanet surveys, and between the coronagraph and starshade instruments are illustrated in **Figure 3.3-4**. Out of the 2.5 years of total time allocated to the exoplanet surveys, 0.5 year is dedicated to a broad survey of systems with no exo-Earth detected by the coronagraph, leaving 2 years of "total exo-Earth survey time" for the EEC search and spectral characterization. This time allocation is meant to serve as a proof of concept, but it is worth noting that the number of EECs detected and characterized is a weak function of total exo-Earth survey time as long as it is greater than about 2 years (**Figure 3.3-12**), so that a large fraction of the mission time will be devoted to non-exoplanet science. Time fractions shown include all wavefront control and pointing overheads, but not starshade slew times, since coronagraphic and general astrophysics observations are conducted while the starshade is slewing from target to target. The pie chart also includes the anticipated parallel observations (e.g., deep field imaging and spectroscopy) that can be conducted with the HWC and UVS instruments which can be operated—simultaneously—during long exoplanet direct imaging exposures. It is expected that Guest Observer (GO) investigations would cover 100% of an extended mission.

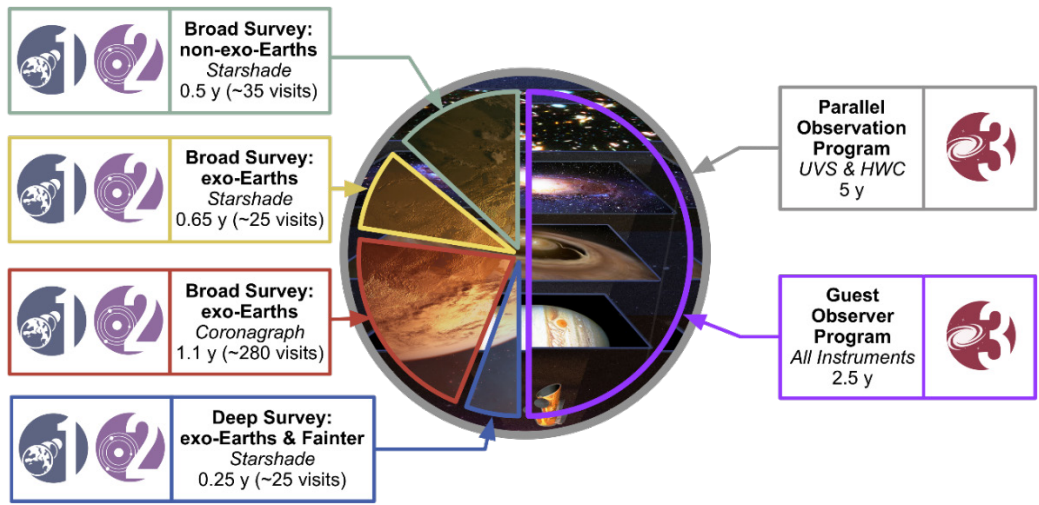

**Figure 3.3-4.** HabEx notional time allocation for a 5-year primary mission. Time is evenly split between exoplanet imaging surveys and "observatory science". The broad-survey uses both the coronagraph (for multi-epoch imaging) and the starshade (for spectroscopy). The deep survey only uses the starshade for imaging and spectroscopy. The total exo-Earth survey time is 2 years.

### 3.3.2 Exoplanet Yields for the Baseline 4-Meter Concept

The most demanding requirements coming from exoplanet science objectives O1–O8 are summarized in **Table 3.3-3**. All requirements are met with significant margin by the baseline HabEx 4H architecture (*Chapters 6, 7, and 8*) using the operations concept described in *Section 3.3.1*. Projected capabilities are based on detailed end-to-end models of instrument performance and error budgets for both starlight suppression systems (*Chapter 5*), and on nominal occurrence rates for all planet types (*Appendix C*).

#### 3.3.2.1 Overall Numbers of Planet Detections

While searching for and characterizing exo-Earth candidates, HabEx will detect nearly two hundred other planets, from hot rocky worlds to cold gaseous planets. **Figure 3.3-5** shows the nominal number and types of exoplanets expected to be detected during the HabEx broad coronagraphic survey (1.1 years), using the default occurrence rates derived from Kepler data (Belikov 2017; Dulz et al. 2019), the nominal exozodi distribution that best fits LBTI measurements (Ertel et al. 2018), and the planet

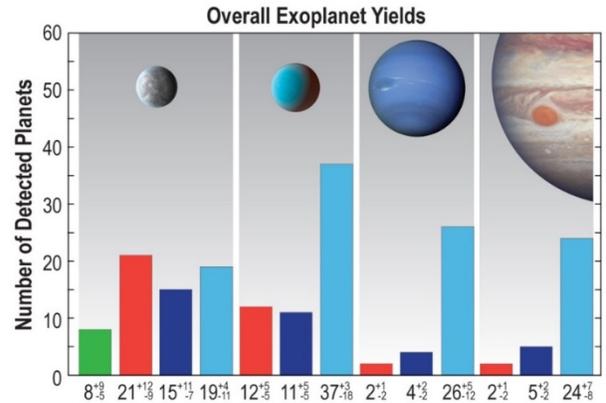

**Figure 3.3-5.** HabEx is expected to detect over 150 exoplanets over a wide range of surface temperatures and planetary radii: 55 rocky planets, among which ~8 could be potential Earth analogs (*green bar*), 60 sub-Neptunes, and 63 gas giants. Yield mean values and uncertainties for each planet type are indicated at the bottom of the plot.

size and temperature classification scheme recently proposed by Kopparapu et al. (2018). The boundaries of this planet classification scheme are computed using the known chemical behavior of $ZnS$, $H_2O$, $CO_2$ and $CH_4$ gases and condensates at different pressures and temperatures in a planetary atmosphere. Red, blue, and cyan bars (**Figure 3.3-5**) indicate hot, warm, and cold planets, respectively. The green bar shows the predicted yield of exo-Earth candidates, which is a subset of the warm rocky planets. Using these assumptions and instrument performance models consistent with its detailed telescope and coronagraph design specifications (*Appendix C and Chapter 5*), it is estimated that HabEx will detect and characterize the orbits of 55 rocky planets

**Table 3.3-3.** Comparison of HabEx driving baseline requirements compared to projected capabilities.

| Parameter | Baseline Requirement | Projected Capability (HabEx 4H) |
|---|---|---|
| Probability of detecting, determining the orbit and measuring 0.3–1.0 μm spectrum of at least one EEC | >95% | 98.6% |
| Number of EECs detected if each target had exactly one ("EEC Cumulative Completeness") | ≥20 | 32 |
| Number of rocky planets detected | ≥30 | 55 |
| Number of sub-Neptunes detected | ≥30 | 60 |
| Number of giant planets detected | ≥30 | 63 |
| $IWA_{0.5}$ (at 1 μm) for EEC spectroscopy | ≤80 mas | 58 mas (starshade) |
| $IWA_{0.5}$ (at 0.87 μm) for EEC ocean glint | ≤64 mas | 58 mas (starshade) |
| OWA (0.5 μm) | ≥ 6" | 6" (starshade broadband imaging) |
| Minimum planet-to-star flux ratio detectable at IWA | ≤4 × 10⁻¹¹ | 2.5 × 10⁻¹¹ (Δmag =26.5) |
| Minimum wavelength range | 0.3 μm to 1.7 μm | 0.2 μm to 1.8 μm |
| Spectral Resolution, $R$ | ≥5 from 0.3–0.45 μm<br>≥70 from 0.45–1.0 μm<br>≥22 from 1.0–1.7 μm | 7 from 0.3–0.45 μm<br>140 from 0.45–1.0 μm<br>40 from 1.0–1.8 μm |

(radii between 0.5–1.75 R⊕), 15 of them located in the HZ and 8 small enough (<1.4 R⊕) to be possible HZ Earth analogs (EECs), 60 sub-Neptunes (1.75–3.5 R⊕) and 63 gas giants (3.5–14.3 R⊕). All of the planets discovered by HabEx occupy a region currently unexplored of the radius vs. separation parameter space (**Figure 3.3-6**). The yields are based on optimizing the observation plan for the detection and characterization of exo-Earth candidates; a search optimized for gas giants could yield significantly more of these planets.

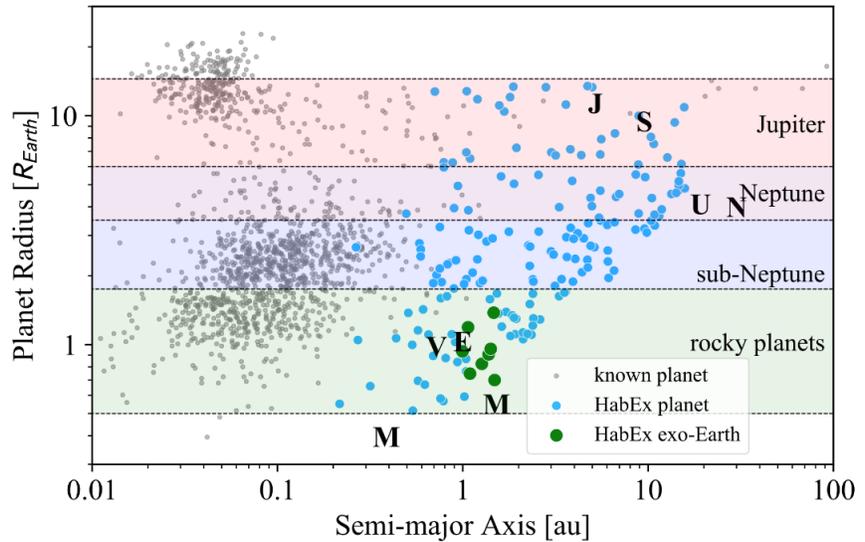

**Figure 3.3-6.** Expected distribution of HabEx discovered planets (*in blue or green*) as a function of physical separation (at quadrature) and radius. Random draw consistent with overall yield numbers per planet type (**Figure 3.3-5**), and adopted probability distributions in radius and physical separation. Credit: T. Meshkat.

### 3.3.2.2 Planetary Characterization: Orbits and Spectra

Both the broad and deep surveys rely on multiple visits to individual stars, and HabEx will hence obtain multiband, multi-epoch photometry for all planets discovered. For the vast majority of planets detected on orbits shorter than ~15 years, HabEx will determine the main orbital parameters (semi-major axis, inclination and eccentricity) and measure phase-dependent color variations.

HabEx will use the starshade to perform a total of ~85 slews for spectral characterization of *all* ~50 deep and broad survey target stars from the exo-Earth surveys. The deep survey program will use ~25 of these starshade observations on ~8 targets. The remaining ~60 slews will be used to spectrally characterize *all ~42 planetary systems* from the coronagraph broad survey. These broad survey targets will be classified in 3 tiers:

1. Systems with EECs detected and orbits determined. The timing of starshade slews and observations will be optimized for multiple (~3 in average) spectral characterizations of Earth analogs found in these highest priority systems;

2. Systems with no strict EECs found but with small (<4 R⊕) planets detected in the vicinity

of the HZ; such targets will be used to further explore the HZ concept and are the next highest priority for timing spectroscopic observations with the starshade; and

3. Systems with only larger and/or outer planets detected. All of these systems will also be spectrally characterized with the starshade, but they will not drive the timing of such observations. However, it is worth noting that such large and/or outer planets may remain detectable over longer durations after initial detection with the coronagraph, so that timing is less critical.

### 3.3.2.3 Planetary Spectra: EECs

For the *combined deep and broad surveys*, the expected yield of detected and spectrally characterized (0.3–1.0 μm) exo-Earth candidates for the baseline HabEx mission is $8^{+10}_{-5}$, where the range expresses the non-Gaussian, 1σ spread set by uncertainties in all major known astrophysical sources: the frequency of exo-Earths ($\eta_{\text{Earth}}$), uncertainties in the exozodi distribution, and finite sampling uncertainties associated with the planetary systems and exozodi levels of individual stars (see *Appendix C*).

For each exo-Earth candidate characterized, the spectra will reveal the presence of Rayleigh scattering in the planet's atmospheres, as well as water vapor molecular oxygen and ozone, if present with the same column density as modern Earth (fairly constant over most of Earth history) or even significantly lower in the case of $O_2$ and $O_3$. Indeed, for EECs orbiting sunlike stars within 15 pc, both $O_2$ and $O_3$ will be detectable down to Proterozoic levels, with their column densities measured in less than 1,000 hours through HabEx 0.3–1 µm spectra (**Figure 3.3-7**). Characterization all the way to 1.8 µm—e.g., looking for methane and carbon dioxide—will only be possible for targets within ~10 pc, due to HabEx IWA limitations at longer infrared wavelengths. For EECs at that distance or closer, methane could be detected if present at Archean or high Proterozoic levels. HabEx could also constrain the biological nature of this methane by detecting any high levels of carbon dioxide in the atmosphere.

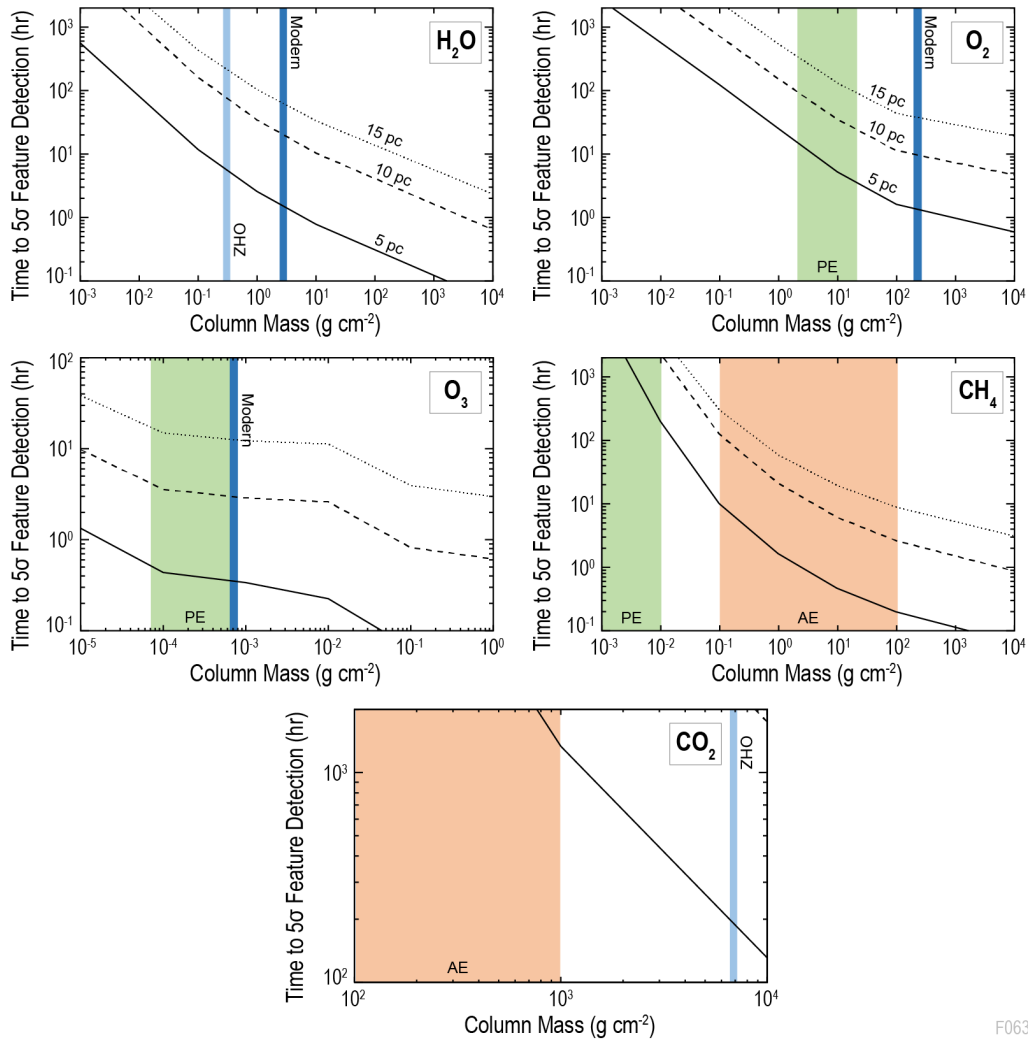

**Figure 3.3-7.** Integration time required with HabEx 4 m baseline concept for gas spectral feature detection. Integration time is shown as a function of gas column mass (i.e., the average gas mass overlying a unit area of the planet, or, alternatively, the gas mass density integrated over altitude) for an Earth-sized planet seen at quadrature around a sunlike star at either 5 pc (*solid line*), 10 pc (*dashed line*) or 15 pc (*dotted line*). Included species are water vapor, molecular oxygen, ozone, methane, and carbon dioxide. Color-coded vertical bars or areas indicate known column masses for modern Earth, or Proterozoic Earth (PE) or Archean Earth (AE). Finally, 'OHZ' indicates the carbon dioxide column mass required to maintain habitability for a world at the outer edge of the HZ and the associated water vapor column mass above this near-frozen surface. Detection time at different column masses comes from integrating signal and noise information over the 0.3 to 1.0 µm range for water vapor, oxygen and ozone. For methane and carbon dioxide detections, the simulated spectral range extends to 1.8 µm and IWA restrictions will further limit the detectability of planets within 0.1 arcsec of the star. Credit: T. Robinson.

**Figure 3.3-8** shows a simulated HabEx starshade spectrum of a modern Earth-twin at quadrature around a sunlike star located at 10 pc. The starshade provides a broad instantaneous spectral range from 0.3–1.0 μm with a single exposure and instrument setting; spectral coverage reaching bluer (down to 0.2 μm) or redder (up to 1.8 μm) wavelengths require additional observations with the starshade, more distant or closer to the telescope, respectively.

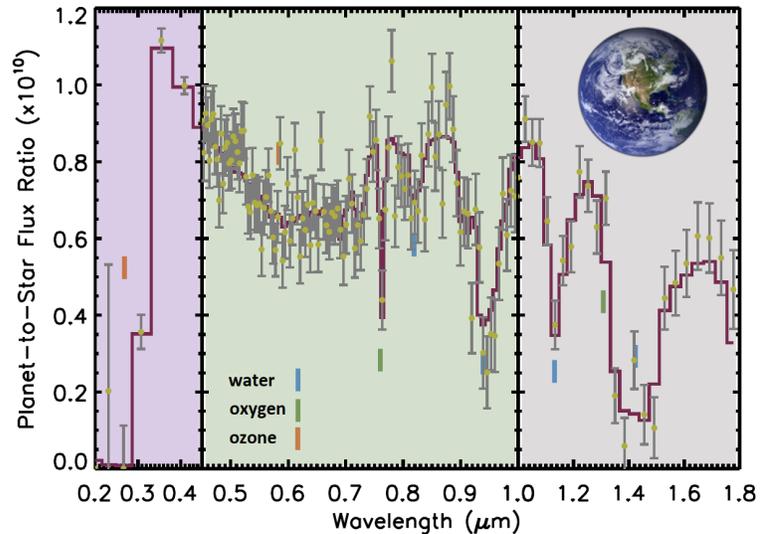

**Figure 3.3-8.** Shown is a simulated 370 h starshade observation of an Earth twin seen at quadrature around β CVn, a sunlike star located at 8.6 pc with a simulated exozodi level of 5 "zodis." HabEx will clearly detect the ozone cutoff below 0.33 μm, atmospheric Rayleigh scattering at blue wavelengths, and multiple signatures of molecular oxygen and water vapor.

### 3.3.2.4    Planetary Spectra: Small Planets

Zooming in on the region of small planets (0.4–4 R⊕) detected in the HZ vicinity (**Figure 3.3-9**) clearly illustrates HabEx's ability to not only detect and characterize EECs lying in some—necessarily pre-conceived—region of parameter space (grey shaded region), but also explore nearby regions of the radius vs. separation diagram. Through the detection, orbital determination, and broad spectral characterization of such planets, HabEx will measure the planet sizes and stellar irradiation levels where physical transitions in the atmospheres occur, and thus will empirically test and define the concept of a habitable zone. In particular, it will probe the region believed to represent the transition between large rocky planets (super-Earths) and sub-Neptunes, at larger physical separations than previously accessed by Kepler (Fulton et al. 2017). HabEx will provide high SNR spectra of these slightly larger planets allowing for instance clear distinction between rocky types and sub-Neptunes (**Figure 3.3-10**, top panel).

### 3.3.2.5    Planetary Spectra: Giant Planets

Excitingly, dozens of giant planet targets for HabEx will be spectrally characterized at extremely high SNR (≥20), through the pairing of the large starshade FOV with deep integrations required to characterize EECs. For giant planets in distant orbits, both the starshade and coronagraph spectrographs will be able to get high quality visible to near-IR spectra (lower two panels of **Figure 3.3-10**).

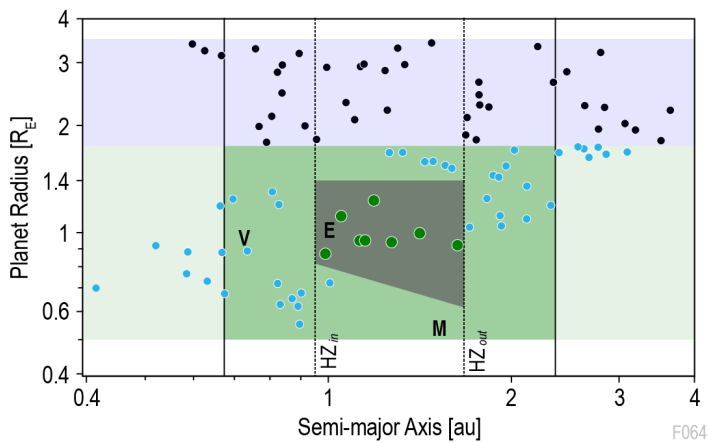

**Figure 3.3-9.** HabEx is expected to detect and obtain spectra of ~100 small (<3.75 R⊕) planets orbiting close to the nominal HZ. This includes dozens of rocky worlds (defined as <1.75 R⊕, blue dots) with insolation levels within a factor of 2 what is received at the inner and outer HZ edges ("proximate HZ region" shaded in dark green). The subset of 8 HZ rocky worlds located in the smaller EEC region (<1.4 R⊕) are indicated with green dots. Credit: T. Meshkat.

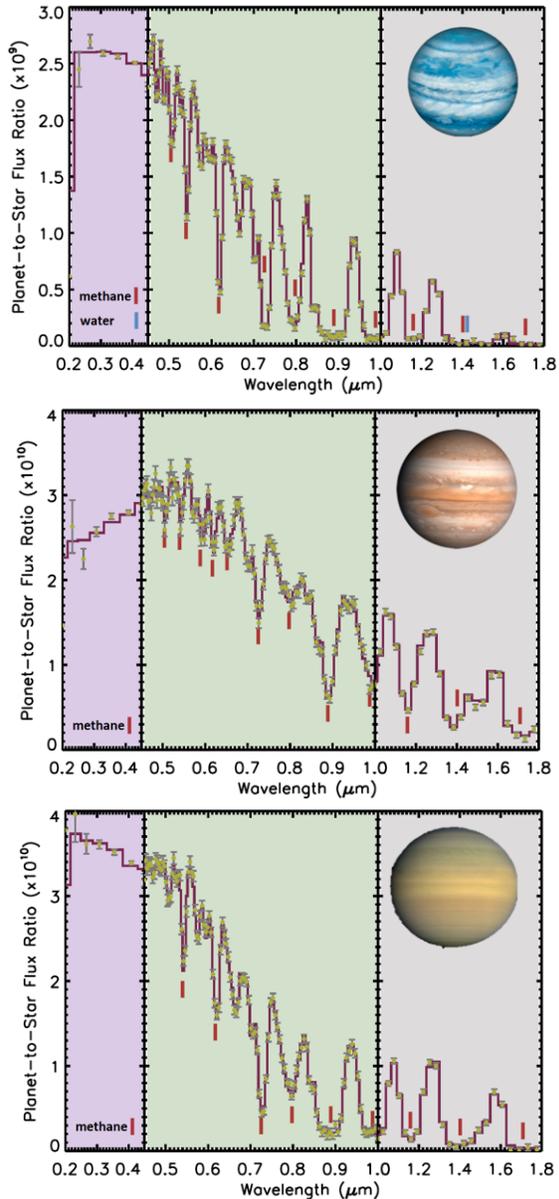

**Figure 3.3-10.** Starshade-based spectra of a 2 AU sub-Neptune (*top panel*), as well Saturn and Jupiter analogs (*middle and bottom panels*). Planets orbiting β CVn, a sunlike star at 8.6 pc (same simulation assumptions as Figure 3.3-3). Water and methane absorption features are all detected at high SNR.

### 3.3.2.6   Target List Characteristics

A larger "master list" of 150 HabEx target stars is given in *Appendix D* (**Table D-1**), assuming that the full 5-year duration of the HabEx prime mission is devoted to exo-Earth detection, orbit determination and spectral characterization.

For the 2-year-long HabEx exo-Earth surveys considered in this section, a shorter list of

50 highest priority stars is obtained by randomly assigning exozodi levels to each target in the master list (consistent with the distribution inferred by the LBTI exozodi survey), and by using the "AYO" algorithm (Stark et al. 2015; *Appendix C*) to optimize the total number of EECs detected by the coronagraph (at 0.50 μm) and spectrally characterized by the starshade (covering at least 0.3-1.0 μm). In this case, the exact target list and optimum sequence of observations varies depending on the exozodi level assigned to each individual stars, which can be measured during or prior to the mission. A representative baseline list of 50 HabEx target stars for the assumed combined deep and broad surveys is given in *Appendix D* (**Table D-2**). Over 95% of the stars in this high priority list are brighter than 7th magnitude and all are within 15 pc. A summary of target characteristics is given in **Figure 3.3-11**, showing that HabEx deep and broad surveys will cover a wide variety of spectral types, with an overwhelming majority of FGK stars. This list is prioritized for high exo-Earth search completeness. As a result, nearly all A stars are discarded (because of prohibitive flux-ratio requirements), and only a few nearby *early-type* M-stars are selected (due to prohibitive IWA requirements). For the same reasons, the highest completeness for HZ planets is obtained for G and K stars, which provide the best trade-off between contrast and angular separation requirements.

HabEx baseline architecture achieves an average EEC search completeness of ~65% for target stars within 15 pc (**Table D-2**), and reaches a cumulative completeness of 32 EECs over the 50-star sample. This is comfortably above the baseline requirement of detecting and characterizing >20 EECs if all targets had one (Objective O1). The required cumulative EEC search completeness of 20 is actually reached when observing all targets closer than β CVn (HIP 61317), the fiducial solar-type target star at 8.6 pc used in HabEx simulated images and spectra (**Figures 3.3-1**, **3.3-2**, and **3.3-3**). Looking at a large ensemble of draws with different exozodi levels per individual target, the completeness requirement is reached at a distance of 8 to 9 pc.

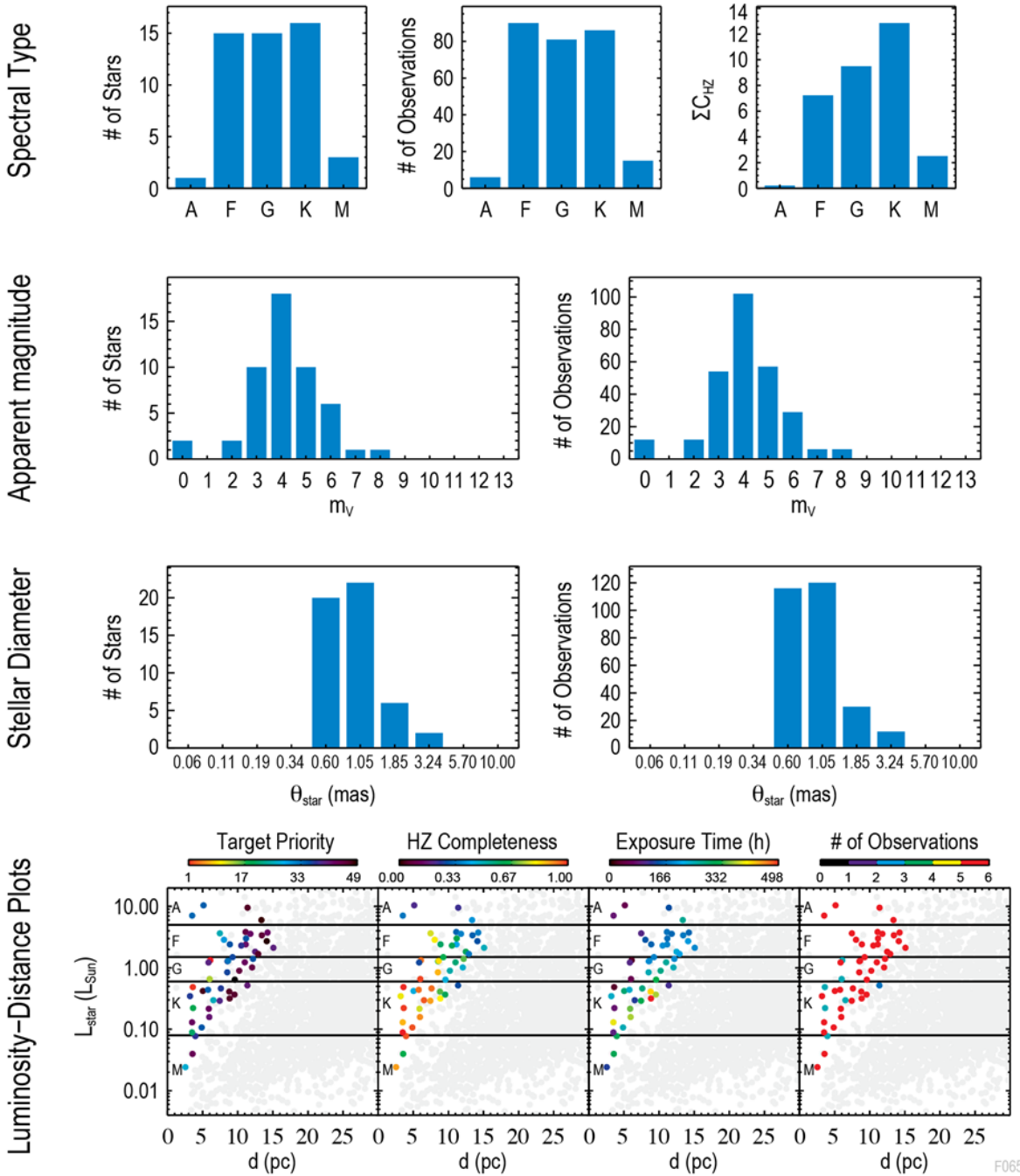

**Figure 3.3-11. Example list of target stars** surveyed by HabEx during the prime mission, assuming a random exozodi draw from the nominal distribution of exozodi levels derived from LBTI. The exact number of stars surveyed depends on the exozodi levels drawn around each potential target. A total of ~50 stars is expected to be surveyed: 8 during the deep survey (starshade only) and 42 during the broad survey (multi-epoch coronagraphic searches and planet orbit determination, followed by planet spectral characterization with the starshade). Based on the HabEx survey strategy, the upper-right panel shows the number of HZ Earth-like planets that would be characterized around stars of different types, assuming each star had one such planet.

### 3.3.3 Exoplanet Science Yield Dependencies to Astrophysical and Instrument Parameters

For mission design purposes, it is important to investigate how the nominal planet yield estimates depend on the adopted instrument and astrophysical parameters, and to determine the probability of success in the face of uncertainties in these parameters.

#### 3.3.3.1 Sensitivity of EEC Yield

**Figure 3.3-12** shows the sensitivity of the "exoplanet yield", narrowly defined here as the number of EECs detected and spectrally characterized between 0.3 and 1.0 µm, as a function of several key mission/instrument and astrophysical parameters, using nominal HabEx assumptions for all other parameters (including an exo-Earth occurrence rate of 0.24) and the same yield optimization algorithm used for the 4 m nominal HabEx architecture.

*Telescope Diameter*

As the telescope diameter $D$ increases, the coronagraph IWA (scaling as $1/D$) improves. In order to follow the hybrid architecture's basic principle (coronagraph visible detection IWA should be equal to the starshade spectral characterization IWA at 1 µm), the starshade IWA must be improved as well, while operating at the same Fresnel number to retain high contrast capabilities. As a result, the starshade size must scale as $D$ and its distance as $D^2$. While the coronagraph detects more EECs at telescope diameters larger than the 4 m baseline thanks to its reduced IWA, there is some diminishing return as the starshade becomes larger, heavier and more distant (unless starshade refueling is possible).

*Total Exoplanet Survey Time*

Similarly, the HabEx surveys operate near a "knee" in the total exposure time curve, beyond which there are diminishing returns on investment. Because of IWA limits, extending the EEC survey time, e.g., from 2 to 4 years, would only increase the EEC yield by 25%. This saturation effect would not occur as fast for outer planets, and significantly more would be detected and characterized as the mission gets extended past the prime mission.

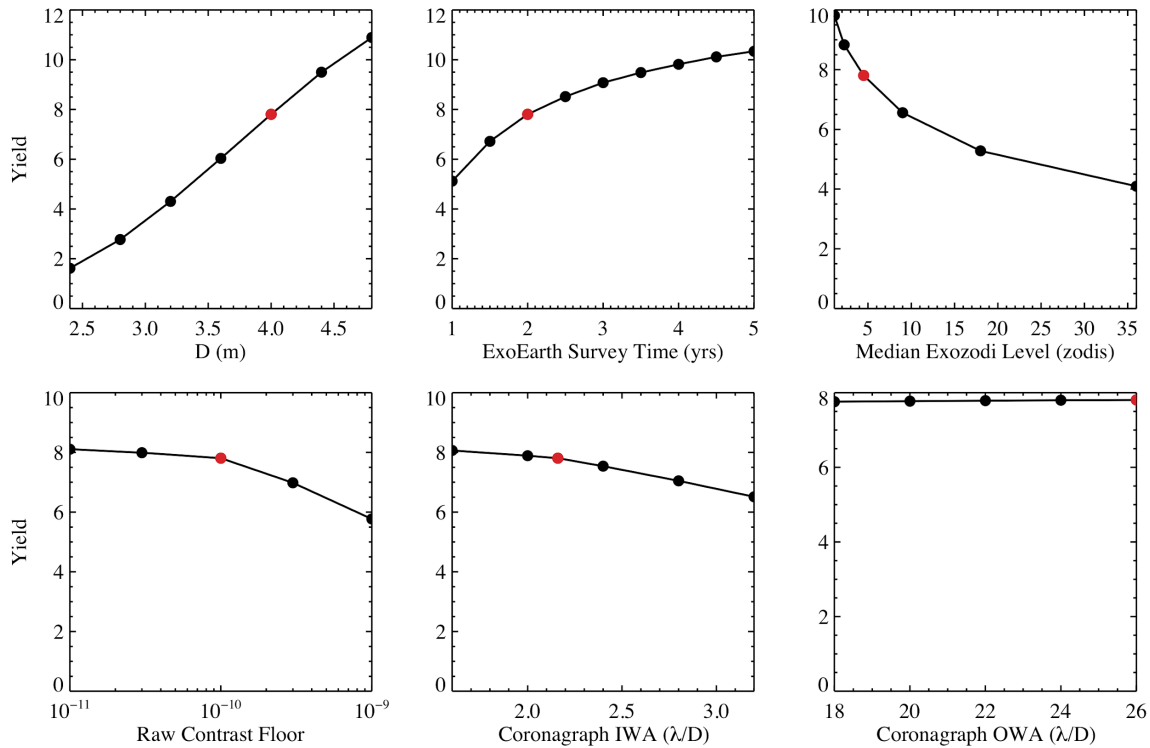

**Figure 3.3-12.** Number of exo-Earths detected and spectroscopically characterized over the full 0.3–1.0 µm range as a function of key mission/instrument and astrophysical parameters. In all cases, a hybrid coronagraph / starshade architecture is assumed. Nominal values corresponding to the HabEx 4 m baseline design and are shown in red. The nominal exo-Earth survey time (including spectral characterization of all EECs detected) is 2 years. The total exoplanet survey time is 6 months longer, allowing for the spectral characterization of targets with no EECs detected. See text for details.

### Exozodi Levels

HabEx exhibits the expected (Stark et al. 2014) sensitivity of yield to high exozodi levels ($\propto$ exozodi$^{-0.23}$) in the case where the exozodi's impact is represented by a pure shot noise penalty. In reality, resonant structures in bright exozodi clouds (~20 zodis or more; Defrère et al. 2012) may further impact the detectability of EECs.

### Raw Contrast Floor

Planet yields are estimated using the raw contrast performance predicted by end-to-end structural thermal optical performance (STOP) models (*Section 6.9*) for both the coronagraph and starshade instruments (*Sections 6.3* and *6.4*). However, to account for possible modeling residual uncertainties, the raw contrast adopted at a given separation is always defined conservatively as the worst of two values: the model predicted performance at that location and some constant "raw contrast floor," defined as the best instrument contrast reachable at any separation from the star. HabEx EEC yield is found to be fairly insensitive to the raw contrast floor as long as its value remains of the order of $10^{-10}$.

It is worth noting that the raw contrast is an instrumental performance parameter. It is different from the minimum planet-to-star flux ratio detectable, which can be significantly lower than the instrument raw contrast, as illustrated by the ground-based detections of exoplanets that are significantly fainter than residual starlight speckles using advanced post-processing techniques (e.g., Lafrenière et al. 2007; Soummer et al. 2012) .

### Inner and Outer Working Angles

Decreasing the coronagraph IWA does not improve the yield of spectrally characterized planets significantly. That is because the nominal starshade IWA (matching the coronagraph IWA at 0.5 μm) is already down to 1.2 $\lambda/D$ at 1 μm and cannot be further reduced while still resolving the planetary system at that wavelength.

Over the range of OWA considered, which are all >18 $\lambda/D$, the OWA is found to have no impact on EEC yield, as expected for the characterization of inner HZ planets.

#### 3.3.3.2 Exoplanet Yield Uncertainties and Probability of Success

The exoplanet yield results presented in the previous section were derived under the nominal survey duration, astrophysical and engineering parameter assumptions listed in *Appendix C*. Hence they represent mean values. In reality, however, yields may vary from the expected values, as shown by the uncertainties in **Figure 3.3-5** (x-axis values), due to astrophysical uncertainties and the actual distribution of planets around nearby stars. Yield uncertainties are estimated by simulating a 3×3 grid of possibilities, sampling the nominal and ±1σ distributions for all planet type occurrence rates and exozodi levels. Given that the shape of the 2D probability distribution is unknown, a normal distribution was assumed and the likelihood of each scenario was weighted appropriately. To sample the Poisson noise associated with the exozodi levels of individual stars, each simulation is performed 20 times with different random exozodi level draws. To sample the Poisson noise associated with the planetary systems around individual stars, each simulation is performed an additional 50 times with different random planet draws based on the adopted occurrence rates. Yield uncertainties shown in **Figure 3.3-5** are 1σ spread and include all known major sources of astrophysical uncertainty and astrophysical Poisson noise.

While HabEx mission success is broadly defined as meeting all exoplanet, observatory science, and solar system science objectives established in this chapter and the next, the focus here on its most challenging and technically driving science goal ("Goal 1"), i.e., HabEx's ability to detect, determine the orbit and spectrally characterize at least one EEC *during its prime mission exoplanet surveys*. **Figure 3.3-13** shows that probability of "success" as a function of telescope diameter, folding in uncertainties in the exo-Earth occurrence rate value, uncertainties in the distribution of exozodi levels—consistent with LBTI survey results (Ertel et al. 2018; Ertel et al. 2019) and finite sampling (Poisson) fluctuations. At the 0.24 nominal exo-Earth occurrence rate assumed and for the baseline 4 m hybrid HabEx mission, the number of exo-Earth spectra obtained is ~8, and the probability of obtaining none is 1.4%. While the telescope diameter selection was also driven by independent technical considerations (*Chapter 6*), it was determined that a ~4 m telescope provides a reasonable yield of exo-Earths and a high probability of success.

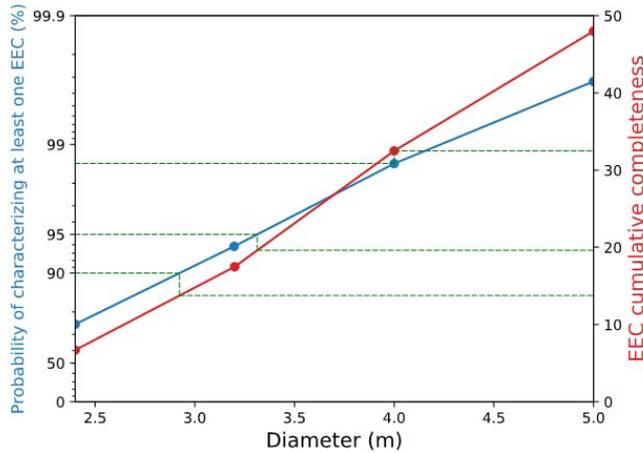

**Figure 3.3-13.** *Blue curve (from left y-axis)*: probability of characterizing at least one EEC as a function of telescope diameter, folding in all sources of astrophysical uncertainties (exo-Earth occurrence rate, distribution of exozodi brightness levels), as well as finite sampling (Poisson) uncertainties. *Red curve (from right y-axis)*: EEC cumulative completeness vs telescope diameter. For all sizes considered, the same hybrid starlight suppression system and observing strategy are assumed with 2 years of total mission time split between EEC search, EEC orbit determination and EEC spectroscopy over full 0.3–1 μm range. A 95% (90%) probability of success corresponds to a search completeness of ~20 (12) EECs. For the baseline 4m architecture and reference mission, the probability of measuring at least one EEC orbit and spectrum is 98.6% and the EEC cumulative completeness is 32.

In order to compute the probability of success, a full yield optimization was conducted for each telescope diameter assumed. In each case, targets, visit durations, and timings are optimized to maximize EEC total completeness and spectral characterization over 2 years, just as for the nominal 4 m hybrid architecture. For each telescope diameter, two figures of merit are computed:

1. The number of EECs that would have been characterized if all targets had one i.e., the cumulative EEC completeness (**Figure 3.3-13**, *red curve, right y-axis*), a best-case scenario; and,

2. The probability of characterizing at least one EEC, which now folds in all astrophysical uncertainties (**Figure 3.3-13**, *blue curve*).

From these results, it appears that to guarantee a 95% (90% threshold) probability of success, a cumulative EEC completeness >20 (12) is required. Because the cumulative EEC completeness is solely a characteristic of the instrument, it can in turn be used to drive high-level instrument performance. While planet yield is a degenerate function of many parameters, two have the largest impact: the minimum separation at which a planet may be detected (approximated by the instrument IWA), and the minimum planet-to-star flux ratio detectable at that separation. The cumulative EEC completeness obtained by the baseline HabEx 4H architecture as a function of these two top-level parameters is shown in **Figure 3.3-14**. In order to meet the cumulative completeness baseline requirement of 20 EECs, HabEx must be able to detect all planets outside of 80 mas, and with flux ratios larger than $10^{-10}$ ("B" point in **Figure 3.3-14**). Other solutions exist, with smaller planet-to-star flux ratios accessible at larger separations or vice versa. The baseline HabEx 4 m operating point is significantly better, with an IWA of 62 mas for the coronagraph, and a planet-to-star flux ratio detection floor of $2.5 \times 10^{-11}$ ($\Delta$mag limit = 26.5).

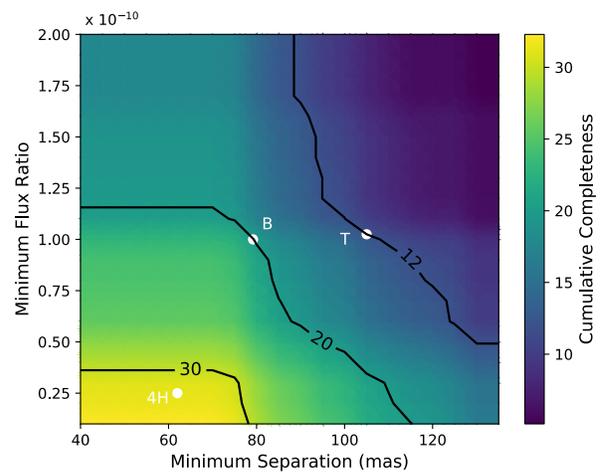

**Figure 3.3-14.** Cumulative completeness of EECs characterized by HabEx baseline (4 m hybrid) architecture as a function of minimum planet-to-star flux ratio detectable and minimum angular separation accessible. Completeness iso-contours at 12, 20, and 30 are indicated, together with HabEx baseline requirements (*B*), threshold requirements (*T*) and actual design point (*4H*). The EEC cumulative completeness is defined as the total number of EECs that would be characterized if all targets had one EEC.

## 3.4    Further Exoplanet Science Yield Enhancements

The exoplanet science yield outlined in the previous section is actually conservative in at least two different ways. First it assumes that no prior knowledge of planets or exozodi levels in individual systems is available by the time HabEx launches. Second, it assumes that stars in binary systems are, for the most part, unobservable by HabEx.

In reality, substantial gains in survey efficiency and overall exoplanet characterization potential may be provided through ancillary observations of nearby planetary systems prior to HabEx launch or during its prime mission. In addition, the successful development of new multi-star wavefront control techniques, particularly well-adapted to the HabEx hybrid starlight suppression system, would significantly improve the HabEx target sample size and increase its projected yield. All these potential gains, and what it would take to realize them, are discussed in details in *Chapter 12*. A summary of findings and top-level recommendations is given hereafter.

### 3.4.1    Ancillary Observations of Nearby Planetary Systems

The HabEx exoplanet survey target selection, prioritization and observation timings are all made assuming no information from precursor or contemporaneous observations of planets or exozodi dust belts in these systems is provided by other facilities, whether ground- or space-based. However, new observatories are expected to be operational by the time HabEx launches, providing additional data on the target systems, enabling more robust HabEx target prioritization and scheduling.

#### 3.4.1.1    Precision Radial Velocity and Astrometry

Contemporaneous radial velocity or astrometric observations, at precisions not achievable today, may confirm a small planet's location in the HZ. This would reduce the required number of HabEx direct imaging visits. Similarly, simultaneous precision astrometry observations of the host star and HabEx direct imaging observations are expected to improve the planet mass and orbit determination precision (Guyon et al. 2013). Over the next 5 years, Gaia astrometric measurements are also expected to provide additional information about the presence / absence of outer massive planets in some of the nearby systems targeted by HabEx. A high-precision mass estimation would aid in interpreting HabEx spectra and improve the characterization of the observed exoplanet atmosphere.

To maximally enhance the exoplanet science achieved with HabEx, the study strongly endorses the top-level recommendation of the NAS ESS report that "NASA and NSF should establish a strategic initiative in extremely precise radial velocities (EPRVs) to develop methods and facilities for measuring the masses of temperate terrestrial planets orbiting Sun-like stars." We note that, because the efforts needed to determine whether or not it is possible to reach a systematic precision of ~1 cm/s from the ground will almost certainly take many years, and that the stars that will be targeted by HabEx will need to be monitored for many years if and when this precision is demonstrated, it is critical that NASA and NSF begin planning for this initiative as soon as possible.

#### 3.4.1.2    Exozodi Surveys at Visible to Near Infrared Wavelengths

Precursor knowledge of individual exozodi levels down to 10× smaller uncertainties than currently available (Ertel et al. 2018), i.e., 10 times the solar zodi level would also make the exoplanet survey more efficient. Yield calculations show that given the exozodi distribution derived from the LBTI survey data, HabEx would need to skip ~25 nearby high-exozodi stars that are otherwise good targets. Because only one observation is necessary to measure the exozodi level, and HabEx will perform ~300 observations to depths consistent with the detection of Earth-analogs, the 30 shallow observations necessary to measure the exozodi levels of these skipped stars will require less than a few percent of the survey time.

However, the LBTI survey was conducted in the mid-infrared and a solar dust density profile was assumed to derive exozodi level estimates. In order to accurately estimate the exozodi background to be faced by HabEx around *individual* targets, i.e., to go beyond statistical knowledge and current model-dependent wavelength extrapolations (*Section 12.6*), high contrast resolved exozodi observations are required in the actual HabEx visible to near infrared wavelength range, with sensitivity down to ≤10 times the solar zodi level. This requires spatially resolved visible and infrared observations at higher contrast, angular and temporal resolutions than currently available from space and from the ground (e.g., Mennesson et al. 2019a).

As precursor exozodi observations are concerned, the main recommendation is therefore to foster instrumentation developments for (i) high-contrast space-based imaging systems in the visible, and (ii) ground-based high-contrast near-IR interferometric systems, using separate telescopes and aperture masking on ELTs. For instance, visible observations with ≳1 m space-based telescopes at contrast levels below ~$10^{-7,-8}$ per spatial resolution element—as specified for the WFIRST coronagraph instrument (Mennesson et al. 2018)—would be able to detect exozodi at <10× the solar level for the first time, and map their spatial structure with <50 mas resolution (Mennesson et al. 2018; Mennesson et al. 2019b) in the visible. Such observations will be important to optimize the target selection and observing efficiency of HabEx (or LUVOIR) direct imaging surveys.

### 3.4.2 High Contrast Observations of Binary Stars

Given that roughly half of all solar type stars are in binary systems, multi-star wavefront control and starlight rejection technologies have the potential to significantly increase HabEx viable target sample and thus the number of planets that can detected and characterized within a given IWA limit or total observing time. Of particular importance is the fact that some of these technologies have been shown to be applicable to the alpha Centauri system, which, if not for the fact that it is a binary, would easily be the best target for direct imaging searches for planets. However, many of these technologies are at relatively low TRL 3 levels. There are plans to bring several of these technologies to TRL 4 and we encourage continued investment in these technologies.

# 4 OBSERVATORY SCIENCE

Following in the tradition of NASA astrophysics flagships, such as the Hubble Space Telescope (HST), the James Webb Space Telescope (JWST), and the Wide Field Infrared Survey Telescope (WFIRST), HabEx is designed to be the next Great Observatory, with at least 50% of the primary mission and all of an extended mission dedicated to a competed, funded, annual Guest Observer (GO) program. The foundational principle behind **HabEx Goal 3**, *to enable new explorations of astrophysical systems from the solar system to galaxies and the universe by extending our reach in the ultraviolet through near-infrared*, is the requirement that HabEx deliver unique science, not possible from ground- or space-based facilities in the 2030s, when the mission would launch. The driving science for HabEx Goal 3 is broad and exciting, addressing the full range of primary NASA strategic priorities, from the solar system, to Cosmic Origins (COR), to Physics of the Cosmos (PCOS). Notably, Goal 3 includes significant Exoplanet Exploration (ExEP) science beyond HabEx Goals 1 and 2 (discussed in *Chapter 3*), including transit spectroscopy, direct imaging of protoplanetary disks, and studying reflected light from exoplanets around non-sunlike stars. The HabEx Observatory Science program includes all science related to Goal 3. Specifically, this encompasses community-driven, competed, funded programs for Guest Observations, parallel observations, and archival research. It is expected that HabEx will serve a very similar role to that played by HST in the astronomical community and the world at large for decades: a flexible and powerful tool producing an extremely broad range of exciting astrophysics, and fueling the public's interest in science, the cosmos, and NASA.

HabEx Observatory Science relies on three unique capabilities that define its discovery space. First, a large-aperture space telescope is required to provide the highest resolution imaging at ultraviolet (UV) and visible wavelengths. While adaptive optics (AO) on the

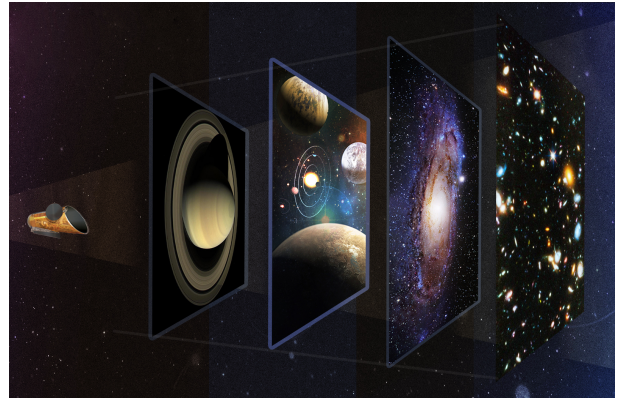

| Science Goals and Objectives | |
|---|---|
| **Goal 3: To enable new explorations of astrophysical systems from the solar system to galaxies and the universe by extending our reach in the UV through near-IR.** | |
| O9: | To probe the lifecycle of baryons by determining the processes governing the circulation of baryons between the gaseous phase of the intergalactic medium (IGM), circumgalactic medium (CGM), and galaxies. |
| O10: | To determine the sources responsible for initiating and sustaining the metagalactic ionizing background (MIB) across cosmic time. |
| O11: | To probe the origin of the elements by determining the properties and end states of the first generations of stars and supernovae. |
| O12: | To address whether there is a need for new physics to explain the disparity between local measurements of the cosmic expansion rate and values implied by the cosmic microwave background (CMB) using the standard Λ cold dark matter (ΛCDM) cosmological model. |
| O13: | To constrain dark matter models through detailed studies of resolved stellar populations in the centers of dwarf galaxies. |
| O14: | To constrain the mechanisms driving the formation and evolution of Galactic globular clusters. |
| O15: | To constrain the likelihood that rocky planets in the habitable zone around mid-to-late-type M-dwarf stars have potentially habitable conditions (defined as water vapor and biosignature gases in the atmosphere). |
| O16: | To constrain the range of possible structures within transition disks and to probe the physical mechanisms responsible for clearing the inner regions of transition disks. |
| O17: | To probe the physics governing star-planet interactions by investigating auroral activity on gas and ice giant planets within the solar system. |

Extremely Large Telescopes (ELTs) is expected to provide angular resolutions of 10 milliarcseconds (mas) in the near-infrared (near-IR), AO shortward of 1 μm is significantly more challenging and is not expected to be standard operations on the timeframe of HabEx. Impressive results are starting to come from the ground, such as 22–28 mas angular resolution at visible wavelengths using the Very Large Telescope (VLT; Schmid et al. 2018). However, such observations require short exposures with very bright (V < 9) guide stars. Well-corrected, wide-field, high-resolution UV/visible imaging of faint targets is not expected to be feasible from the ground before the 2040s. Second, a space-based telescope can observe at wavelengths that are inaccessible from the ground, including the UV and in visible-to-near-IR atmospheric absorption bands. Finally, an orbit far above the Earth's atmosphere that is free from the large thermal swings that are inherent to HST's low-Earth orbit enables an ultra-stable platform that is capable of undertaking science observations ranging from precision astrometry to the most sensitive weak lensing maps ever obtained.

In order to address HabEx Goal 3, two capable instruments are included in the HabEx design: the UV Imager and Spectrograph (UVS; *Section 6.5*) and the HabEx Workhorse Camera and Spectrograph (HWC; *Section 6.6*). The designs and requirements of these instruments flow down from specific science objectives, listed in the right column of the previous page, that were selected because they set the driving requirements for the instruments, though with this aspect of the mission competed, some of these science objectives may not ever be executed. Above and beyond the HabEx Goal 3 science objectives, possible additional science applications are vast, with many not yet anticipated; a small selection are presented in *Section 4.10*. The HabEx Objectives presented in *Sections 4.1–4.9* were selected as examples of compelling science to define specific instrument requirements, simultaneously motivating capable instruments with a broad usage potential.

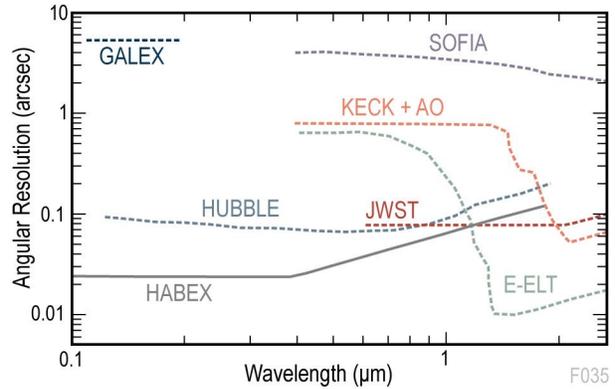

**Figure 4-1.** The HabEx baseline design will provide the highest resolution UV/visible images of any current or planned facility over fields of several arcminutes, enabling a broad suite of unprecedented Observatory Science.

The HabEx UVS and HWC instruments both rely on low-risk, flight-proven technology and the HabEx baseline concept (*Chapter 6*) is designed to provide the highest resolution UV/visible images ever obtained (**Figure 4-1**). The UVS instrument is an evolved version of HST's Cosmic Origins Spectrograph (COS), taking advantage of several decades of improvement in detector and optics technology since HST. The HWC instrument is an evolved version of the dual-beam Wide-Field Camera 3 (WFC3) on HST. The HWC will provide imaging and multi-slit spectroscopy in two channels: a visible channel and a near-IR channel.

The remainder of this section is structured as follows. *Sections 4.1–4.9* present the science objectives for HabEx Goal 3, providing scientific rationales and derivations of instrument requirements. *Section 4.10* presents additional scientific opportunities provided by HabEx, including those based on a parallel observation program using the UVS and HWC instruments and a rapid response program for targets of opportunity.

## 4.1 Objective 9: What is the Life Cycle of Baryons?

**Objective 9:** To probe the lifecycle of baryons by determining the processes governing the circulation of baryons between the gaseous phase of the intergalactic medium (IGM), circumgalactic medium (CGM), and galaxies.

### 4.1.1 Rationale

Despite decades of efforts, approximately one-third of the baryons in the local universe remain unaccounted (**Figure 4.1-1**). Notably, stars only account for <10% of the baryons in a typical galaxy. The "missing baryons" are thought to be predominantly in the form of diffuse, hot gas around and between galaxies, but many fundamental questions remain open about this gas, even within the very local universe. This material, the intergalactic medium (IGM; i.e., the gas between galaxies) and the circumgalactic medium (CGM; i.e., the gas external to, but near galaxies), is the fuel from which stars ultimately form, and, later in their lives, the material that galaxies redistribute and make more metal rich through supernovae and violent mergers. Inflows and outflows of the CGM are inextricably linked to key issues, such as star formation, galactic structure, and galactic morphological transformation.

Therefore, studying and understanding this gas is essential for understanding the life cycle of baryons in the cosmos, and for developing a more complete picture of galaxy formation and evolution (**Figure 4.1-2**). However, this presents observational challenges since the bulk (60%) of the CGM is predicted to be extremely hot, with the key diagnostic transitions at UV and X-ray energies and thus inaccessible to the ground. Furthermore, the CGM is roughly a million times less dense than the IGM ($10^3-10^5$ cm$^{-3}$, compared to ~$10^{10}$ cm$^{-3}$), and thus empirical studies of the CGM have relied primarily on its absorption signatures in the rest-frame UV spectra of bright background quasars.

Observations to date, largely based on statistical studies built out of samples of single sightlines per vast galaxy halo, show that the CGM is significantly metal-enriched, and that it is dynamic and short-lived. These results, largely based on observations by the COS on HST, show that metal-enriched, under-pressurized 'clouds' at galactocentric distances greater than 75 kpc appear to have no means of long-term survival, yet are commonly found in statistical studies of quasar absorption lines. In addition, vast reservoirs of neutral hydrogen surround both star-forming and passive galaxies alike, hinting that the lack of a gas supply cannot entirely explain the low levels of star formation in passive galaxies.

To make significant progress in constraining and understanding the cosmic baryon cycle over the past 10 Gyr, it is necessary to:

- Complete the census of baryons in the local universe;

- Measure the amount of gas and heavy elements around $z < 1$ galaxies;

- Determine the dynamical state and origin of the various components of the IGM, i.e., determine what fraction of the IGM is primordial, and what fraction is due to outflowing material, recycled accretion, or other physical causes; and

- Measure UV gas morphology at ≤1 kpc scales to determine the centers and sizes of massive star forming regions.

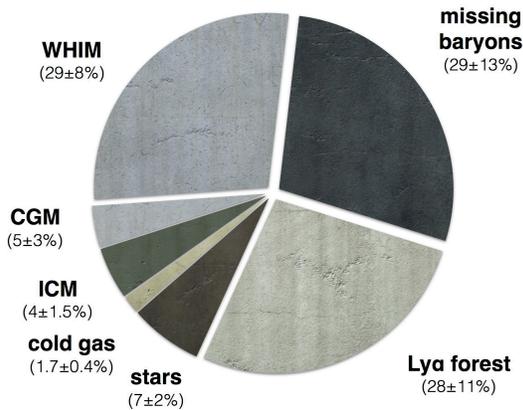

**Figure 4.1-1.** Approximately one-third of the baryons in the local universe are unaccounted for, likely tied up in a hot gas phase (Shull et al. 2012). WHIM: warm/hot interstellar medium, CGM: circumgalactic medium, ICM: intracluster medium.

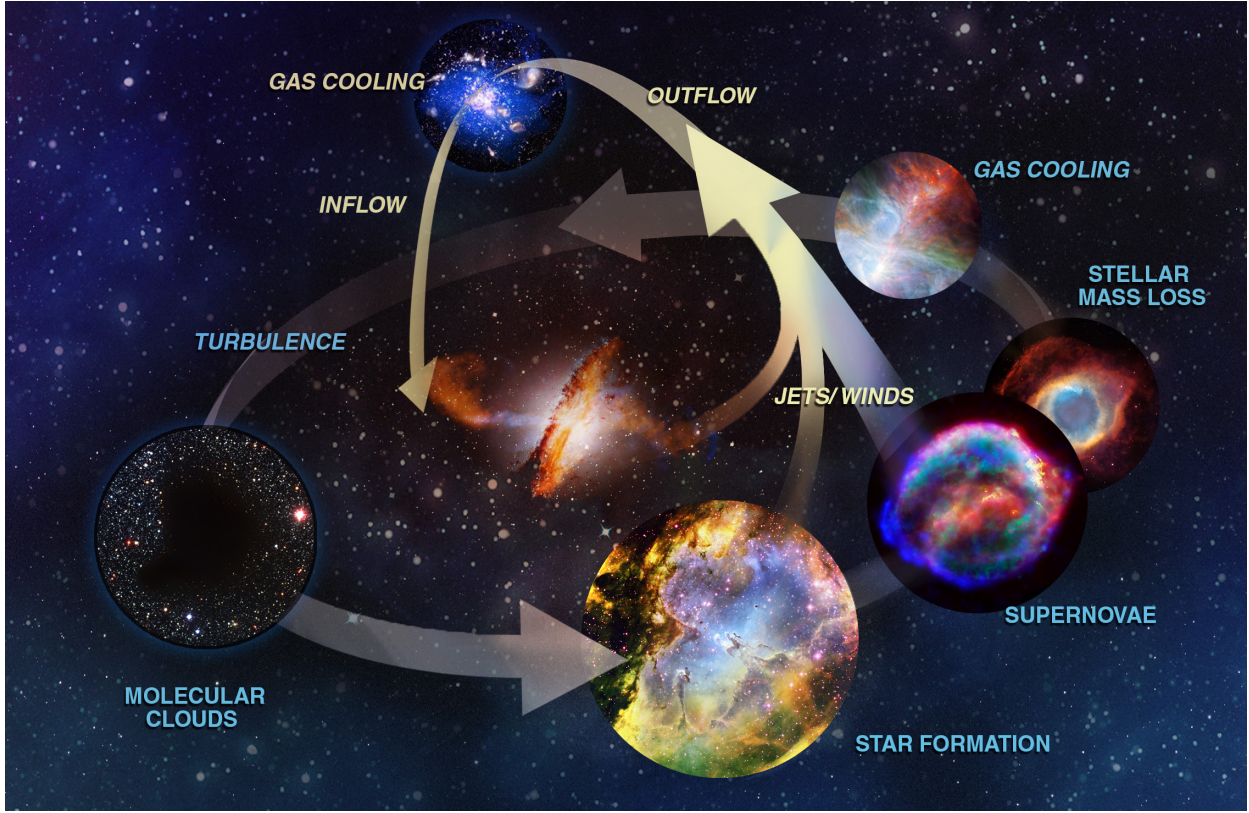

**Figure 4.1-2.** With its improved UV sensitivity and multiplexing capabilities, HabEx will be more than an order of magnitude more efficient for investigations of the lifecycle of baryons.

These measurements, which include inferring the temperatures, densities, metallicities, and structure of the CGM and IGM in a range of environments and over cosmic time, will inform the processes governing the circulation of baryons and address questions about the structure and origin of the absorbing material in halos of galaxies, the same material that will eventually fuel the formation of new stars and planets.

### 4.1.2 Requirements

Sensitive studies of the hot IGM present specific observational challenges. Foremost, at least in the local universe, the observations must be obtained at wavelengths inaccessible from the ground since the Earth's atmosphere absorbs and scatters photons blue-wards of 320 nm. UV astronomers distinguish between near-UV observations (120–360 nm) and far-UV observations (90–120 nm). As shown in **Figure 4.1-3**, the density of diagnostic spectral features is highest at the blue end of the far-UV, blue-ward of 100 nm. While an instrument sensitive to these energies in the observed frame would provide strong benefits, cosmological redshifting enables access to these rest-frame wavelengths without needing extreme-far-UV sensitivity. Setting a blue cutoff of 115 nm, identical to HST/COS, provides access to ion absorption lines, which enables gas density and temperature determinations through radiative transfer modeling.

Multiplexing capabilities, with ≥10 sightline probes around a single galaxy, will provide significant gains relative to current studies, allowing efficient two-dimensional studies of

### Objective 9 Requirements

| Parameter | Requirement |
|---|---|
| Observing modes | Multi-object spectroscopy and imaging |
| Spectral range | ≤115 nm to ≥320 nm |
| Field of view | ≥2.5 x 2.5 arcmin² |
| Spectral resolution | $R \geq 60,000$<br>SNR ≥ 5 per resolution element on targets of AB ≥ 20 mag (GALEX FUV) in exposure times of ≤12 h |
| Angular resolution | ≤0.3 arcsec |

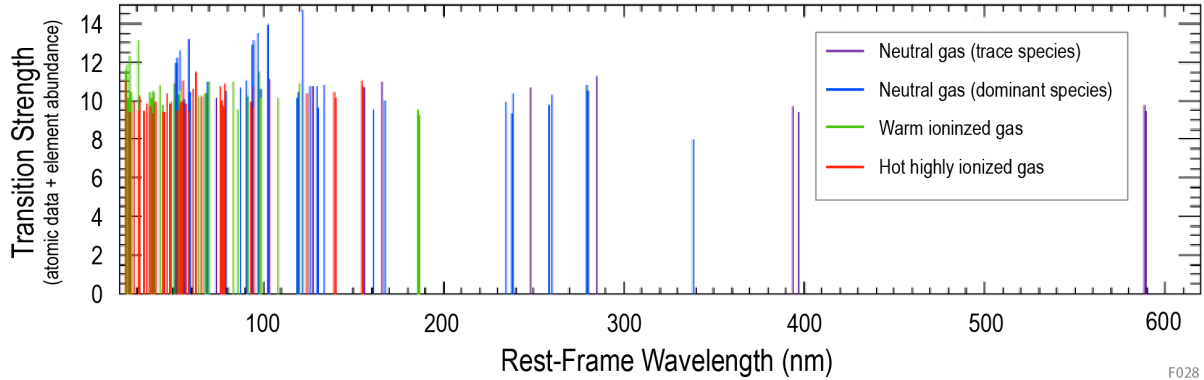

**Figure 4.1-3.** While visible and near-UV observations offer access to neutral gas, observations of near-UV and far-UV features are required to probe the more dominant warm and hot components of the IGM. This graphical representation shows the wealth of diagnostic lines the far-UV and near-UV offer to astrophysical investigation, comparing transition strength to rest-frame transition wavelength (Tripp 2019).

multiple sightlines in each galaxy observed. These sightlines can be a combination of background quasars, background galaxies, and "down-the-barrel" sources within the targeted galaxy (e.g., Barger et al. 2016). Achieving this significant scientific gain will require both the new capability of a multi-object UV spectrograph, as well as greater UV sensitivity than provided by HST/COS in order to have the requisite surface density of sufficiently bright background quasars for absorption line studies. The latter will require either a larger aperture mirror than HST, or a UV-optimized design—or, ideally, both. Based on experience with HST/COS, a signal-to-noise requirement (SNR) ≥ 5 is required to detect absorption lines from the stronger ionic species (e.g., C IV 154.9 nm, C III] 190.9 nm, Si III 189.2 nm) given their typical column densities (see Werk et al. 2013; Bordoloi et al. 2014). Motivating the wide bandpass, absorption measurements from multiple ionic species is necessary to model the temperature and density of the gas with radiative transfer codes such as CLOUDY (Ferland et al. 2017).

Typical nearby galaxies have sizes ≤2 arcmin, requiring a field of view (FOV) of at least 2.5 arcminutes on a side for imaging and at least 1 arcminute on a side for multi-object slit spectroscopy. Sub-arcsecond (≤0.3 arcsec, full-width half-maximum; FWHM) angular resolution is required to resolve both star formation in the host galaxy and filamentary structure in the IGM and CGM. To enable associating gas with galaxies requires measuring the gas kinematic velocities to a precision of ≤5 km/s, corresponding to a spectral resolving power of R ≥ 60,000.

## 4.2 Objective 10: What Caused the Reionization of the Early Universe?

**Objective 10:** To determine the sources responsible for initiating and sustaining the metagalactic ionizing background (MIB) across cosmic time.

### 4.2.1 Rationale

Most of the hydrogen in the universe became ionized over a relatively short period of time around 13 billion years ago, during the Epoch of Reionization (EOR). During this period, primordial gas clouds devoid of heavy elements began to collapse into proto-galaxies, forming the first stars and black holes. Identifying whether the first stars or black holes were primarily responsible for initiating the EOR is still a major open question in cosmology. This question can be generalized to understanding the evolving population of sources responsible for the metagalactic ionizing background (MIB) as a function of cosmic time.

The timing and duration of the EOR is crucial to the subsequent emergence and evolution of structure in the universe (see e.g., Madau et al. 1999; Ricotti et al. 2002; Robertson et al. 2015). The relative roles played by star-forming galaxies, low-luminosity active galactic nuclei (AGN), and quasars (i.e., high-luminosity AGN) in contributing ionizing photons to the MIB depends on the number density of each source type as measured by its luminosity function, convolved with the average number of ionizing photons that escape from each source type (e.g., Madau and Haardt 2015). Ionizing radiation escaping from quasars and AGN is relatively easily to detect. The situation is altogether different for fainter star-forming galaxies, for which there are only a handful of detections of the ionizing photon escape fraction, with escape fraction values of ~3–20% reported at redshifts $z < 3$ (e.g., Shapley et al. 2006; Siana et al. 2010; Heckman et al. 2011; Leitherer et al. 2016). Quasar counts suggest high-luminosity AGN did not play a dominant role during the EOR (e.g., Hopkins et al. 2007), but the potentially crucial contribution from star-formation in low-luminosity galaxies is still highly uncertain due to a poor understanding of the processes that allow ionizing radiation to escape from galaxies into the IGM (e.g., Finkelstein et al. 2019). In particular, how the ionizing photon escape fraction depends on parameters such as galaxy metallicity, gas fraction, dust content, star formation history, mass, luminosity, over-density, and quasar proximity is key to understanding the EOR.

While JWST will determine the faint end of the galaxy luminosity function during the EOR, understanding the EOR and the evolution of the MIB also requires understanding the escape fraction of these galaxies over cosmic time. Estimates for the average escape fraction required to sustain an ionized IGM by z = 6 are in the range of approximately 10–30% (see review by Finkelstein 2016, and references therein), depending on assumptions regarding the variation in the escape fraction across the galaxy luminosity function.

It is generally assumed that low metallicity star-forming galaxies at the faint end of the luminosity function ($M_{UV} > -18$) have the highest escape fraction. Measuring escape fractions requires rest-frame UV observations, i.e., photons emitted below the rest frame ionization edge of neutral hydrogen at 91.2 nm, also known as Lyman continuum (LyC) photons. It is standard in the field to refer to the Lyman continuum escape fraction as $f_{900}$ (where the subscript refers to the wavelength in Angstroms rather than nm). Both the LyC escape fraction and the related Lyman-alpha (Ly$\alpha$) escape fraction, corresponding to the fraction of unabsorbed ionizing photons blue-wards of rest frame 121.6 nm, are challenging to observe at high redshift due to absorption by foreground clouds in the IGM (e.g., Inoue and Iwata 2008). Instead, escape fractions are best measured in the local universe.

### 4.2.2 Requirements

Understanding the cosmic evolution of the sources responsible for the MIB requires the direct measurement of LyC escape fractions down to values of ≤1% for numerous galaxies and AGN, primarily at lower redshifts (z ≤ 0.5), but also sampling more distant galaxies. **Figure 4.2-1** shows the surface density of galaxies and estimated flux detection requirements for escaping LyC photons from star-forming galaxies as a function of redshift (using the galaxy luminosity functions from Arnouts et al. 2005). Determining the LyC luminosity function evolution with redshift requires surveying ≥25 galaxies with UV magnitudes of AB ≥28 mag. To make such observations feasible in a reasonable amount of observatory time requires multi-object (≥25 galaxies simultaneously) UV spectroscopy, reaching these depths in exposure times of ≤ 10 h per field.

*Objective 10 Requirements*

| Parameter | Requirement |
|---|---|
| Observing mode | Multi-object spectroscopy |
| Spectral range | ≤119 nm to ≥ 240 nm (low resolution) ≤154 nm to ≥ 304 nm (high resolution) |
| Spectral resolution (Low) | $R \geq 80$ SNR ≥ 3 per 3 nm interval on galaxies of AB ≥ 28 mag in exposure times of ≤10 h per field |
| Spectral resolution (High) | $R \geq 3,000$ SNR ≥ 3 per 0.1 nm interval on galaxies of AB ≥ 24 mag in exposure times of ≤15 h per field |

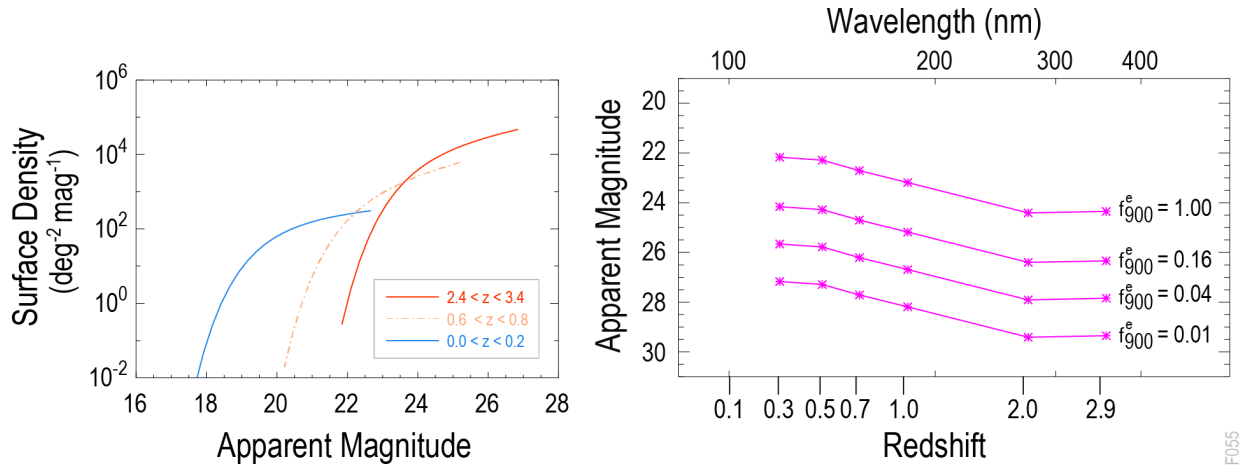

**Figure 4.2-1.** *Left panel*: Surface densities as a function of observer's frame apparent magnitude for galaxy populations with different redshift ranges, estimated following Arnouts et al. (2005). There are 100s-10,000s of galaxies per square degree. *Right panel*: The purple asterisks show the characteristic apparent magnitude (AB) of galaxies shortly blue-wards of LyC for different escape fractions as a function of redshift (bottom axis; top axis indicates the wavelength considered). Credit: S. McCandliss.

Spectroscopy and photometry provide complementary approaches to this problem. Spectroscopy offers the opportunity to examine in detail the variation of the flux escaping at wavelengths below the Lyman edge, along with the compositional properties of the objects. Photometry can go deeper and offer higher spatial resolution at the expense of spectral information; however, spectroscopy provides an avenue for training photometry. The investigation here concentrates on spectroscopy. Measuring the Lyman continuum escape fraction, $f_{900}$, only requires course spectral resolution, $R \geq 80$. Detailed understanding of the Ly$\alpha$ escape fraction, corresponding to rest frame 121.6 nm, requires higher spectral resolution, $R \geq 3,000$, in order to measure the profile of the spectral absorption. The requirements on source brightness (AB $\geq 24$ mag) are correspondingly relaxed for this secondary question.

**Figure 4.2-1** demonstrates that the HabEx baseline design using the UVS (*Section 6.5*) will be able to place $\leq 1\%$ escape fraction limits on dozens of galaxies in a single 3×3 arcmin² FOV over a redshift range from $0.4 < z < 1.0$, probing far below the knee of the galaxy luminosity function (i.e., studying galaxies at the faint end of the luminosity function).

## 4.3 Objective 11: What Are the Origins of the Elements?

**Objective 11:** To probe the origin of the elements by determining the properties and end states of the first generations of stars and supernovae.

### 4.3.1 Rationale

Understanding the origin of the elements is one of the major challenges of modern astrophysics. Most elements heavier than the iron group (i.e., atomic number $Z > 30$) are formed by neutron-capture reactions, when an atomic nucleus and one or more neutrons collide and merge to form a heavier nucleus. High neutron densities ($N_n > 10^{20}$ cm$^{-3}$), as occur in neutron star-neutron star mergers and supernovae, cause atomic nuclei to undertake rapid process, or r-process, neutron capture. The visible brightness of the binary neutron star merger event GW170817 (Abbott et al. 2017b) indicated that such mergers are the key sources of r-process element nucleosynthesis (**Figure 4.3-1**; e.g., Drout et al. 2017; Tanvir et al. 2017). For example, most of the platinum and gold in the universe is believed to have been formed in neutron star-neutron star mergers (Horowitz et al. 2018).

A number of open questions still remain about the frequency, detailed physics, and yields of the primary astrophysical events associated with r-process nucleosynthesis: binary neutron star mergers and supernovae. In principle, the most effective way to characterize these would be through detailed element-by-element compositional studies of their ejecta since, per event, neutron star mergers are expected to produce heavier r-process elements more abundantly than supernovae. However, since the freshly produced r-process material is ejected at a few tenths of the speed of light, spectral transition lines blur together, obscuring the true composition.

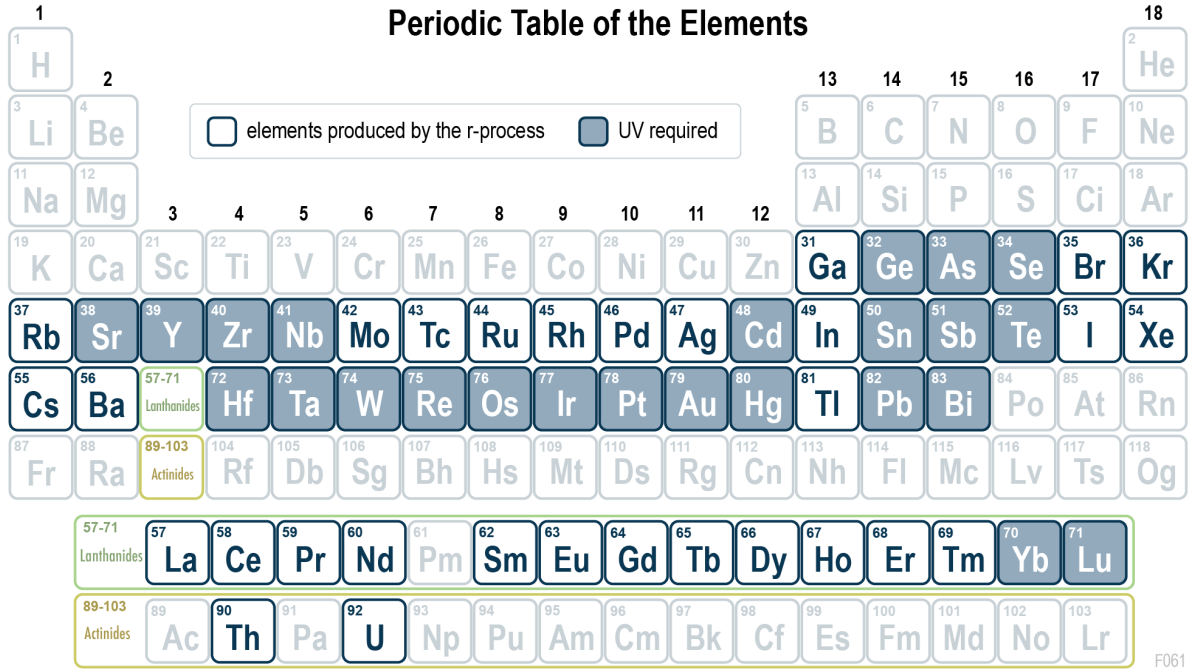

**Figure 4.3-1.** Most stable elements heavier than iron are formed by rapid, or r-process, neutron capture, which primarily occurs in neutron star-neutron star mergers and supernovae. High-resolution UV spectroscopy is essential for measuring the abundances of many such r-process elements, which enables the determination of the relative importance of these energetic events in the origins of the elements.

The most metal-poor r-process-enhanced stars provide a more effective avenue for investigation as each one reflects the yield of an individual r-process event that occurred in the early universe. Such stars preferentially reside in the Galactic halo and comprise ~3–5% of all metal-poor halo stars in the Milky Way (Barklem et al. 2005; Hansen et al. 2018; Roederer et al. 2018). The UV and visible spectra of these stars present a rich collection of absorption lines of r-process elements (e.g., Sneden et al. 2003; Mello et al. 2013). Visible spectra can reveal that a star is highly enhanced in r-process elements, and some characterization can be achieved based only on abundances derived from visible spectra (e.g., Hayek et al. 2009; Mashonkina et al. 2010; Sakari et al. 2018). However, elements that are key to discriminating between r-process nucleosynthesis models have no detectable absorption lines in the visible domain. By contrast, observations in the UV can detect >20 r-process elements (**Figure 4.3-1**; e.g., Roederer and Lawler 2012; Roederer et al. 2012) that provide the most sensitive constraints on r-process nucleosynthesis models and the conditions of the r-process events (e.g., Schmid et al. 2018; Lorusso et al. 2015; Shibagaki et al. 2016).

### 4.3.2    Requirements

High-resolution UV spectroscopy provides access to key diagnostic elements and features in metal-poor, r-process enhanced stars (e.g., Sneden et al. 1998; Roederer and Lawler 2012), including tin (key line: Sn II 175.8 nm), iodine (I I 178.2 nm), bismuth (Bi II 179.1 nm), antimony (Sb I 181.1 nm), mercury (Hg I 194.2 nm), arsenic

### Objective 11 Requirements

| Parameter | Requirement |
|---|---|
| Observing mode | Spectroscopy |
| Spectral range | ≤ 170 nm to ≥ 310 nm |
| Spectral resolution | $R \geq 24,000$<br>SNR ≥ 100 in the continuum per resolution element on stars of AB ≥14 mag in times of ≤ 10 h |

(As I 197.3 nm), selenium (Se I 207.4 nm), cadmium (Cd I 228.8 nm), tellurium (Te I 238.5 nm), platinum (Pt I 265.9 nm), gold (Au I 267.5 nm), and germanium (Ge I 303.9 nm). The key diagnostic lines drive a minimum required wavelength range of 170–310 nm. **Figure 4.3-2** illustrates an example of using UV spectroscopy to measure the strength of arsenic in a metal-poor sub-giant star. In order to separate close lines, a minimum spectral resolving power of $R \geq 24,000$ is required, with a minimum SNR of ≥100 per resolution element in order to detect weak lines in low-metallicity sources.

As members of the Galactic halo population, all but a few r-process-enhanced stars are too distant for practical UV spectroscopy with the HST/Space Telescope Imaging Spectrograph (STIS), which can only collect high-quality UV spectra of the handful of brightest stars in the solar neighborhood. The practical magnitude limit for collecting high-resolution UV spectra with STIS is AB ~10 mag. Accumulating a sample of hundreds of r-process-enhanced stars requires the ability to observe stars with a magnitude limit of AB ≥ 14 mag in exposure times of ≤10 hours. Such a sample will enable measurements of the variation in r-process element creation and Galactic enrichment both spatially and across cosmic time, thereby determining the properties and end states of the first generations of stars and supernovae.

## 4.4 Objective 12: What Is the Local Value of the Hubble Constant?

**Objective 12:** To address whether there is a need for new physics to explain the disparity between local measurements of the cosmic expansion rate and values implied by the cosmic microwave background (CMB) using the standard Λ cold dark matter (ΛCDM) cosmological model.

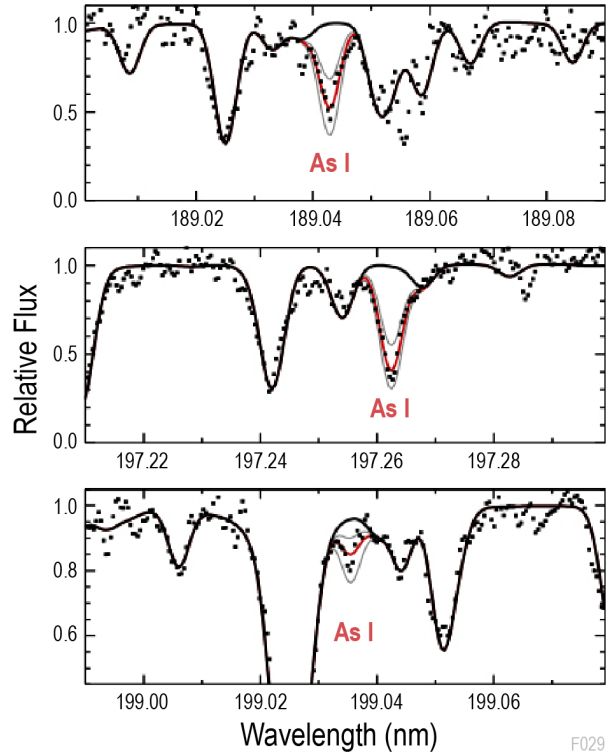

**Figure 4.3-2.** The HabEx UV Spectrograph's (*Section 6.5*) sensitivity makes observations of spectral lines from elements synthesized by r-process nucleosynthesis in neutron star-neutron mergers or supernovae ~50 times more efficient than with HST/STIS. The red curve in each panel fits detections of arsenic (*As*) in a metal-poor sub-giant (*black points*), while the thin grey curves show abundance values by a factor of 2, and the bold black curve shows a model spectrum with no arsenic. Credit: Roederer and Lawler (2012).

### 4.4.1 Rationale

Recent measurements of the local value of the Hubble constant, $H_0$ (i.e., the local expansion rate of the universe), have been controversial, and hint at possible new physics. While the Planck satellite reports a Hubble constant of $H_0 = 67.36 \pm 0.54 \text{ km s}^{-1} \text{ Mpc}^{-1}$ based on measurements of the cosmic microwave background (Akrami et al. 2018) recent measurements of the local value of the Hubble constant by HST are consistently higher. Riess et al. (2016) report on an extensive HST/Wide-Field Camera 3 (WFC3) visible and near-IR imaging program to obtain Cepheid distances of 11 nearby galaxies that hosted recent type Ia supernovae (SNe Ia), more than doubling the sample of reliable SNe Ia having a Cepheid-calibrated distance. That work finds a value of $H_0 = 73.24 \pm 1.74$ km s$^{-1}$ Mpc$^{-1}$, which is 3.4$\sigma$ higher than the Planck value. More recently, Riess et al. (2018b) reported on spatial scanning HST/WFC3 observations to obtain geometrical parallax distances to eight long-period Milky Way Cepheid variables, raising the number of long-period Cepheids with significant parallax measurements to 10. That work finds a value of $H_0 = 73.48 \pm 1.66$ km s$^{-1}$ Mpc$^{-1}$, increasing the tension with the Planck value to 3.7$\sigma$. Additionally, including HST photometry and Gaia parallaxes for 50 Galactic long-period Cepheids, Riess et al. (2018a) further refine the local value of the Hubble constant and raise the tension to 3.8$\sigma$ (**Figure 4.4-1**), while the most recent results, including HST observations of additional Cepheids in the Large Magellanic Cloud, have raised this discrepancy to 4.4$\sigma$ (Riess et al. 2019). Compared to a decade ago, the uncertainties in these measurements are impressively low: the era of *precision* cosmology has certainly arrived, though, at first glance, perhaps not yet entirely the era of *accurate* cosmology. Importantly, however, the HST programs measure the local value of the Hubble constant, while Planck observes the surface of last scattering of the cosmic microwave background at high redshift (z ~ 1,100) and infers the local value of the Hubble constant based upon an assumed cosmology. Potentially, the discrepancy arises from the assumption of a "vanilla" ΛCDM cosmology, i.e., a cosmology with the simplest dark energy equation of state, with a temporally invariant cosmological constant, Λ. More complicated cosmologies can naturally explain this apparent discrepancy between local measurements of the Hubble constant and the value inferred from the cosmic microwave background, with one plausible explanation being an additional source of dark radiation in the early universe.

Expanding upon this work will be achieved by improving calibrations for the lowest rungs on the cosmic distance ladder. While the current tensions are suggestive and intriguing, it is important to test whether the Hubble constant discrepancies exist at greater than the 5$\sigma$ level by achieving greater than 1% accuracy in measurements of the local value. This is the level required to necessitate serious consideration of new physics and to identify testable hypotheses to distinguish between the possible models. The required precision photometry is not achievable from the ground. From space, WFIRST, with the same aperture as HST, will only be able to improve upon HST to the extent that more SNe Ia occur within the small volume of the local universe in which Cepheid variables are accessible to a 2.4 m telescope. JWST will be able to achieve some of this science, but fewer accessible SN Ia will have been identified when JWST launches in 2021, and JWST is highly inefficient for cadenced observations given its slow slew and settle times.[1]

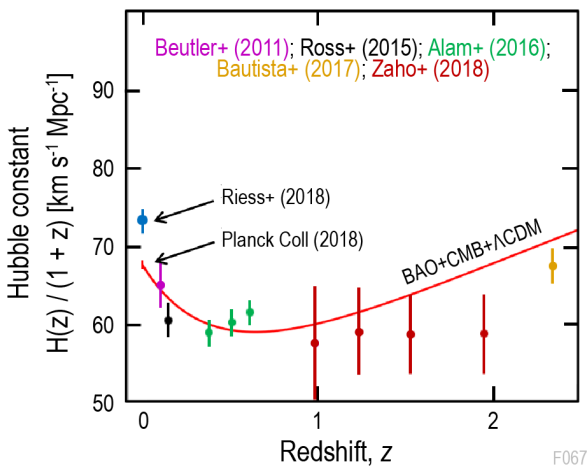

**Figure 4.4-1.** Recent measurements of the local value of the Hubble constant (blue point; Riess et al. 2018a) are discrepant with values inferred in the distant universe. The red line shows the best fit ΛCDM cosmology to the *Planck* cosmic microwave background measurements combined with baryon acoustic oscillations (BAO), assuming a ΛCDM. Current tensions are suggestive of new physics. HabEx will vastly improve local measurements of the Hubble constant and determine if this tension exists at greater than the 5$\sigma$ level, as would be required to rule out the simplest cosmological constant (Λ) dark energy models.

### 4.4.2 Requirements

Achieving the requisite precision in measurements of the local value of the Hubble constant, reaching sub-percent accuracies, will require a large increase in the number of SN Ia

---

[1] https://jwst-docs.stsci.edu/

*Objective 12 Requirements*

| Parameter | Requirement |
|---|---|
| Observing mode | Broadband imaging |
| Wavelength range | Visible: broad-band filters (*V*, *I*)<br>Near-IR: broad-band filter(s) (*J*, *H*) |
| Field-of-view | ≥2 × 2 arcmin$^2$ |
| Signal-to-noise | SNR ≥ 10 for point sources of H ≥<br>28 mag in exposure times of ≤2 h |

host galaxies observed with accurate Cepheid distances. Currently, only 19 such SN Ia host galaxies have precision Cepheid distances. The most distant host galaxy with a distance modulus uncertainty less than 0.05 mag is NGC 3370, at 30 Mpc (Riess et al. 2016). By pushing this horizon out to ≥ 50 Mpc, a concerted program for a mission launched in the mid-2030s will nearly quintuple the volume accessible to precision Cepheid distances. Typical Cepheids at this distance will have near-IR magnitudes of $H < 28$ mag (Vega; Riess et al. 2018b) and are well within the reach of a 4 m space telescope with integration times of <2 h. Cepheids are best identified at visible wavelengths, where their variability is greatest. Typical surveys require a minimum of 10 visible imaging epochs spread over days to months to reliably identify Cepheid variable stars and measure their periods and phases. This implies a visible imager with filtered imaging (e.g., *V*-band, *I*-band, and/or a broad visible band) with a field of view comparable to nearby galaxies, at least 2 arcmin on a side. While rapid slews are not required, these observations will be significantly more efficient if obtained by an agile telescope capable of efficient slews with short settle times. Many Cepheids in galaxies at these distances will also be found from ground-based survey telescopes such as the Large Synoptic Survey Telescope (LSST), albeit with lower photometric precision and more source confusion. For precision cosmology measurements, extremely accurate near-IR photometry of Cepheids at known phases in their light curves is essential. This will require a near-IR imager with *J*- and/or *H*-band filtered imaging with a field-of-view comparable to nearby galaxies, at least 2 arcmin on a side.

Riess et al. (2016) used HST to identify and measure Cepheids in 20 nearby galaxies with a mean exposure time of ~4 h per galaxy. Assuming a comparable exposure time, but reaching to much greater volumes given a significant improvement in sensitivity relative to HST (i.e., particularly for unresolved sources), a survey of 100 galaxies could be accomplished in a few weeks of observations. This would increase the number of well-calibrated Cepheid distances to galaxies known to host type-Ia supernovae by a factor of several, thereby decreasing the uncertainties in the local value of the Hubble constant. Such data would also be valuable for a range of nearby galaxy science, such as resolved studies of their stellar populations.

## 4.5 Objective 13: What is the Nature of Dark Matter?

**Objective 13:** To constrain dark matter models through detailed studies of resolved stellar populations in the centers of dwarf galaxies.

### 4.5.1 Rationale

Dark matter comprises most (~85%) of the matter density and about a third of the total energy density in the universe (Akrami et al. 2018), but beyond that, little is known about its nature. Is dark matter a single particle, or is there a whole dark periodic table of particles? Standard dark matter only interacts with itself and with normal matter (i.e., baryons) through gravity (and perhaps through the weak nuclear force). However, particle physics allows for many other possibilities. For example, it is possible that dark matter could be self-interacting, rather than being collisionless as in standard models (e.g., Spergel and Steinhardt 2000). Dwarf galaxies in the Local Group (e.g., **Figure 4.5-1**) provide promising laboratories for probing the nature of dark matter because, unlike larger galaxies which are mostly comprised of baryonic matter (e.g., stars and gas) near their centers, most dwarf galaxies are overwhelmingly dominated by dark matter all the way to their centers (e.g., Simon and Geha 2007; see van Dokkum et al. 2019 for counterexamples). If galaxies formed from pure "vanilla" cold dark matter (i.e., with no stars or gas, as well as no additional dark matter self-interactions; e.g., Tegmark et al. 2004), theory robustly predicts that their density profiles should monotonically increase from the outer regions all the way to their centers—i.e., that their density profiles should have "cusps" at their center (e.g., Navarro et al. 1997). While measuring the dark matter density profile is not possible directly, this has generally been done through the proxy of measuring the density profiles of resolved stellar populations in the galaxies. However, there has been much debate and consternation over the fact that such observations have instead found that the limited number of dwarf galaxies accessible to current telescopes instead have "cores"—i.e., their density profiles plateau to a constant value at the center (e.g., Burkert 1995; Oh et al. 2011). **Figure 4.5-2** illustrates how the density profiles of dwarf galaxies depends on the nature of dark matter.

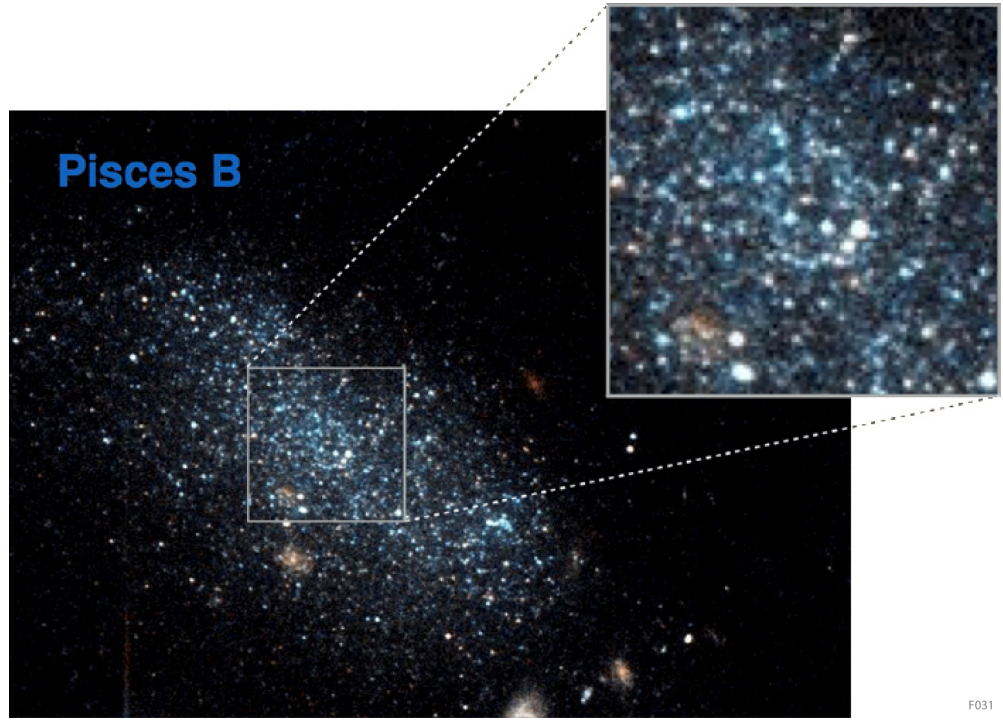

**Figure 4.5-1.** With high-resolution images of dwarf galaxies in the local universe, HabEx will study the nature of dark matter, as well as measure the distances and star formation histories of local analogs of the first galaxies. Shown here is a Hubble image of the nearby dwarf galaxy Pisces B at a distance of 8.9 Mpc (Tollerud et al. 2016). Compared to HST, HabEx will resolve fainter stars in galaxies like Pisces B, and obtain images like the one shown here for galaxies over a ~10× larger volume.

There are two main proposed solutions to explain these observations. Either (1) dark matter is not "vanilla," or (2) the large amounts of energy created by massive stars as they explode in supernovae removes the dark matter from the cusps, thus flattening them out into cores. There is a very large parameter space of "non-vanilla" dark matter models that are considered equally plausible, or natural, to particle physics theorists, and astrophysical observations are likely the most efficient way to narrow down this large parameter space (e.g., Weinberg et al. 2015). Theory groups largely agree on one clean prediction: if the flat density profile galaxy cores are created by non-vanilla dark matter, they should be seen universally in all galaxies. On the other hand, if the flat core profiles are created by stars and supernovae, then pristine "cusps" should be seen surviving in galaxies with truncated star formation histories (Read et al. 2016), because they did not have vigorous enough star formation to produce the requisite energy to remove the dark matter from the central regions and therefore destroy the cusps. It should also be possible to see correlations of the central galactic density profiles with galaxy properties, such as the ratio of the mass of stars to the mass of dark matter. Progress toward constraining dark matter models will require a larger sample of measured resolved stellar density profiles in the innermost regions of dwarf galaxies in order to determine whether the flattened dark matter density profiles at the centers are caused by baryon-dark matter interactions.

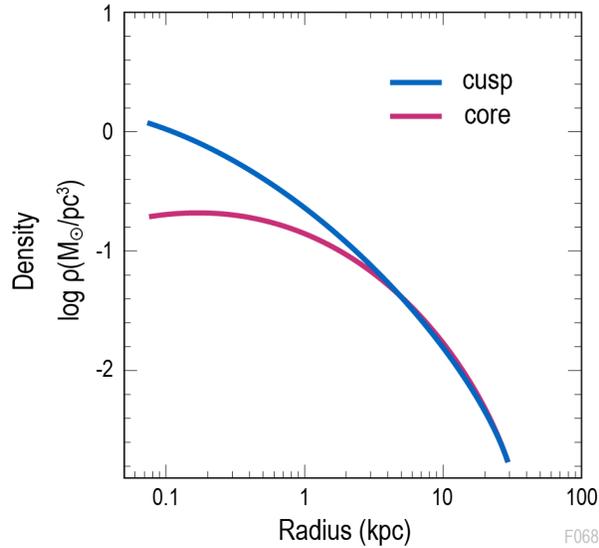

**Figure 4.5-2.** By spatially resolving the inner 500 pc of a sample of dwarf galaxies with a range of star formation histories, HabEx will determine if the flattened "core" profiles seen in many galaxies are indicative of self-interacting dark matter, or simply due to supernovae feedback clearing out the inner dark matter in galaxies with standard cold dark matter, which theory predicts should have "cuspy" density profiles.

### 4.5.2 Requirements

Advancement in this field requires improved understanding of the distribution and kinematics of stars in the crowded central regions of dwarf galaxies. The typical size of a nearby dwarf galaxy sets a FOV requirement of $2 \times 2$ arcmin$^2$. For the inner (<500 pc) regions of dwarf galaxies with stellar masses of $\sim 10^6$ M$_\odot$, HabEx must be able to distinguish a core stellar density of $\sim 0.5$ M$_\odot$/pc$^3$, predicted, for instance, by self-interacting dark matter models, from the cuspy stellar density of $\sim 1.5$ M$_\odot$/pc$^3$, predicted from standard cold dark matter models (Robles et al. 2017). At a distance of 100 kpc, this inner region corresponds to an angular size of $\sim 1$ arcsec. Requiring 20 resolution elements across this region implies a minimum angular resolution of 50 mas. Requiring the robust detection of main

### Objective 13 Requirements

| Parameter | Requirement |
|---|---|
| Observing mode | Broadband imaging |
| Wavelength range | Visible broadband filter (V) |
| Angular resolution | ≤0.05 arcsec |
| Field-of-view | ≥2 × 2 arcmin$^2$ |
| Sample size | ≥10 dwarf galaxies |
| Signal-to-noise | SNR ≥ 5 for point sources of V ≥ 30 mag in exposure times of ≤2 h per dwarf galaxy |

sequence stars down to stellar class K at this distance implies very deep, high-resolution photometry, e.g., achieving a limiting apparent magnitude of V ≥ 30 mag. In order to observe a statistical sample of dwarf galaxies in a feasible timeframe sets a requirement of SNR ≥ 5 for V = 30 mag point sources in ≤2 h. Well-corrected imaging reaching comparable depths with the ELTs is not feasible; photometry of crowded fields at these depths is only possible from space. Because of stochastic effects and differences between dwarf galaxies (e.g., star formation history, and dynamical interaction history with the Milky Way), the requirement is that ≥10 dwarf galaxies be imaged for this study.

Furthermore, there is a well-known degeneracy between the stellar density profile and the velocity anisotropy of stars. While the radial velocity of individual stars in the dwarf galaxy central regions are best done at near-IR wavelengths from the ELTs, visible imaging with an ultra-stable space-based platform could provide unprecedented measurements of the proper motions of faint stars in these regions. Using HST spatial scanning, Riess et al. (2018a) report parallax measurement precisions as low as 29 µas (SNR = 14) for Milky Way Cepheids. In crowded fields, the HST Proper Motion (HSTPROMO) survey report proper motions <10 µas/yr in Galactic globular clusters (Libralato et al. 2018). For the nearest dwarf galaxies, the expected proper motions are of order 1 µas/yr. Though no requirements are levied for this potential additional avenue of research, a 4 m-class ultra-stable telescope orbiting at L2 should achieve such astrometric precisions, particularly were the mission lifetime to extend beyond 10 years, as is typical for non-cryogenic NASA missions.

## 4.6 Objective 14: How Do Globular Clusters Form and Evolve?

**Objective 14:** To constrain the mechanisms driving the formation and evolution of Galactic globular clusters.

### 4.6.1 Rationale

Astrophysics has outgrown the old, traditional picture in which globular clusters are simple stellar systems, built from a uniform population of stars that formed in a short burst, and thus with similar ages and metal abundances. Building on extensive observational programs, largely space-based photometric surveys (e.g., Piotto et al. 2015) and ground-based spectroscopic surveys (e.g., Carretta et al. 2009), it is now clear that most globular clusters have multiple stellar populations with significant abundance spreads. While not as common in young (<2 Gyr old) globular clusters, multiple stellar populations appear ubiquitous in old (>5 Gyr old) globular clusters. These old systems typically show significant star-to-star variations in the lowest-mass elements far beyond what is expected from stellar evolutionary processes alone. Specifically, the basic pattern is enriched abundances of helium, nitrogen, and neon, coupled with depleted abundances of oxygen and carbon (e.g., MacLean et al. 2014). This pattern is unusual outside of globular clusters, rarely seen in either open clusters or in the field. These observations are suggestive of a second generation of stars in globular clusters, though a more circumspect approach is to refer to stars with heightened metal abundances as "enriched stars" in order to avoid the implication of a genetic link with the more pristine populations. Indeed, the simple conceptual picture of a second generation of stars formed out of material polluted by the ejecta of the first generation fails to match an increasing number of observational constraints. The existence of multiple stellar populations in globular clusters, recently reviewed in Bastian and Lardo (2018), constitutes one of the major unsolved problems in globular cluster and stellar populations research.

Improved understanding of the mechanisms driving the formation and evolution of globular clusters relies on a combination of observations. Globular cluster stellar populations are investigated using a combination of imaging and low-resolution spectroscopy. For example, **Figure 4.6-1** presents synthetic spectra of two red giant stars with differing abundance patterns that is typical of globular clusters, demonstrating the clear differences between the two spectra and illustrating the power of spectroscopy for distinguishing between multiple stellar populations. The kinematics of the stellar populations are studied with a combination of multi-epoch, high-resolution imaging (for stellar motions transverse to the line-of-sight; e.g., Bellini et al. 2014; Bellini et al. 2018) and high-resolution spectroscopy (for radial motions). For example, in their studies of NGC 362, Libralato et al. (2018) find kinematic differences in the velocity dispersions of multiple stellar populations, providing key observational data for testing whether the stellar populations correspond to first- or second-generation stars.

Current systematic effects due to source crowding limit HST imaging programs at

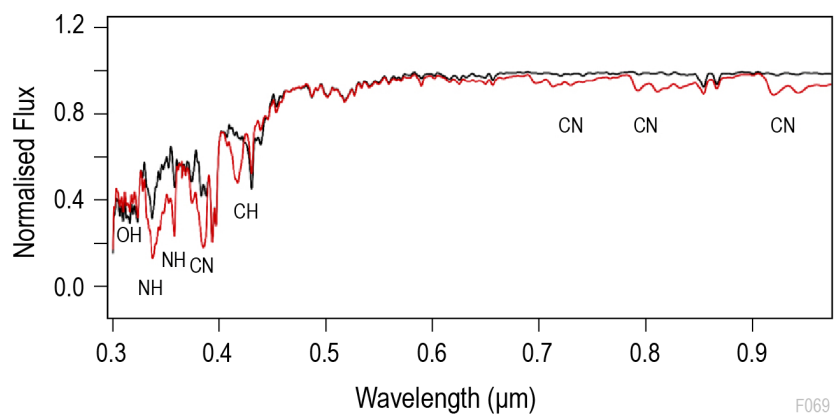

**Figure 4.6-1.** UV-to-visible spectra are critical for distinguishing stellar populations seen in globular clusters. Shown here are high-SNR synthetic spectra of a low-metallicity (*black*) and high-metallicity (*red*) red giant branch star. Strong molecular absorption bands are labelled. After Sbordone et al. (2011).

*Objective 14 Requirements*

| Parameter | Requirement |
|---|---|
| Observing mode | Multi-object spectroscopy |
| Spectral range | UV: ≤150 nm to ≥320 nm<br>Visible: ≤0.37 µm to ≥1.0 µm |
| Multi-object spectroscopy | $R ≥ 1,000$<br>SNR ≥ 3 per 0.5 nm effective resolution element for V ≥ 25 mag in exposure times of ≤10 h per instrument |
| Sample size | ≥400 stars |

relatively bright photometric limits (e.g., ~1% photometry at UV magnitudes of ~19; Nardiello et al. 2018). In terms of precision UV-to-visible spectroscopy of the crowded nuclear regions, HST has single slit spectroscopy and slitless spectroscopic capabilities, neither of which are amenable to spectroscopic campaigns in dense regions. While precision radial velocities (<1 km/s) are likely best done in the near-IR with the next generation of ELTs and AO, spectroscopic studies of globular cluster stellar populations are best done in the UV-to-visible from space with low-resolution, multi-object slit spectroscopy.

### 4.6.2    Requirements

A next-generation UV-to-near-IR telescope with improved capabilities for photometric, astrometric, and spectroscopic observations in dense, crowded fields is required to make transformative gains in studying the formation and evolution of globular clusters. The imaging (i.e., photometric and astrometric) requirements for this science are defined by science objectives discussed earlier in this chapter. This science, however, does drive new spectroscopic capabilities. Low-resolution ($R ≥ 1,000$) spectroscopy across the 0.15 µm to 1.0 µm wavelength range is required to distinguish stellar populations through the detection of key atmospheric features for a range of metallicities (e.g., **Figure 4.6-1**). Such studies will require studying a statistically significant number of stars (≥ 400) within a single globular cluster, with stars separated by ≤0.2 arcsec in the dense cluster nuclear regions. In order to achieve this science in a feasible amount of observatory time requires UV-to-near-IR spectroscopy, reaching

SNR ≥ 3 per 0.5 nm effective resolution element at V ≥ 25 mag in less than 10 h of total exposure time. This implies the capability of multi-object slit spectroscopy.

## 4.7    Objective 15: Are there Potentially Habitable Planets around M-dwarf Stars?

**Objective 15:** To constrain the likelihood that rocky planets in the habitable zone around mid-to-late-type M-dwarf stars have potentially habitable conditions (defined as water vapor and biosignature gases in the atmosphere).

### 4.7.1    Rationale

Given the thousands of planets discovered by NASA's Kepler and the Transiting Exoplanet Survey Satellite (TESS), exoplanet transit spectroscopy is currently the primary technique to study exoplanet atmospheres. Even in the era of direct imaging, this will remain an essential technique, particularly for characterizing rocky planets in the habitable zone (HZ; see *Section 3.1.2* for a description of the HZ) of low-mass stars, since such systems provide the opportunity to measure their large transit depths to high statistical significance. When an exoplanet passes in front of its host star, it appears slightly larger at wavelengths where the atmosphere is more strongly attenuated (e.g., within molecular absorption bands). Thus, the wavelength-dependent transit depth provides a spectrum of the planet's atmosphere, detecting diagnostic features of its physical properties (e.g., the existence of clouds and/or hazes) and atomic and molecular constituents. Similarly, when the planet passes behind the star, the difference in flux compared to the combined system provides a low-resolution emission spectrum of the planet which is sensitive to the planet's vertical thermal profile. When studying rocky planets orbiting in the HZ and investigating their potential habitability (i.e., through the detection of water vapor and biosignature gases in their atmosphere), it is essential that transit spectroscopy be done from space to avoid the challenges of disentangling telluric molecular absorption from that of the exoplanet.

While transit spectroscopy is challenging for small, rocky planets due to their small sizes and faint fluxes, it becomes easier when they are orbiting less massive host stars, e.g., mid-to-late-type M-dwarf stars. Such worlds are known to be common (Dressing and Charbonneau 2015) with key targets already discovered transiting nearby ultra-cool dwarf stars (e.g., Gillon et al. 2017; Bonfils et al. 2017) and many such systems predicted to be discovered with TESS (Sullivan et al. 2015; Barclay et al. 2018). Compared to rocky planets orbiting in the HZ of sunlike stars, terrestrial planets orbiting M-dwarf stars will have a larger radius relative to their host star, making them more amenable to spectroscopic investigation. In addition, having a HZ that is closer to the host star increases the likelihood of a transit and decreases the amount of time between subsequent transits (due to shorter orbital periods), facilitating the observation of more transits over a period of time compared to HZ planets orbiting a sunlike star. Conversely, the faintness of low-mass (i.e., M-dwarf) host stars and their close-in HZs make direct imaging techniques particularly challenging for studying rocky planets.

Great strides in exoplanet transit spectroscopy are expected from JWST (e.g., Deming et al. 2009; Cowan et al. 2015; Greene et al. 2016), particularly for giant planets. However, as an infrared satellite, JWST will not have access to several key molecular bands (e.g., the strong ozone bands). The Atmospheric Remote-sensing Infrared Exoplanet Large-survey (ARIEL) visible–IR mission, recently selected as ESA's next medium-class (M4) mission, will provide a first-of-its-kind dedicated spectroscopic survey of thousands of transiting exoplanets. However, it is a relatively small (0.64 m$^2$) space-based telescope, and thus will focus on planets ranging from warm sub-Neptunes to super-Jupiters, but it will not have the SNR to probe terrestrial planets, even orbiting low-mass stars. Progress on probing the potential habitability of exoplanets, particularly rocky exoplanets around mid-to-late-type M-dwarfs, will require a more sensitive and stable visible–near-IR facility in space.

### 4.7.2    Requirements

Rocky exoplanets transiting mid-to-late-type M-dwarfs are expected to present transit features throughout the visible and near-IR, with key features requiring a spectral range from 0.5 μm to 1.7 μm. These features can cause modulations in the transit depth of ~10–100 ppm (Meadows 2017). Currently, the best HST performance is ~12 ppm rms (Line et al. 2016; Tsiaras et al. 2016) and current transit observations with HST/WFC3 and Spitzer/Infrared Array Camera (IRAC) have yet to reach the noise floor, suggesting that next-generation missions will achieve precisions of ~ 10 ppm rms (= 100,000 SNR) (Zellem et al. 2019). Depending on abundances and atmospheric scale height, the modulation of these features can exceed 100 ppm for late-type M-dwarfs, although the overall lower luminosity of these stars will require longer integration times (or stacked transits; Barstow and Irwin 2016).

The key requirement for such an investigation is the ability to obtain low-resolution spectroscopy through a very stable system. Since the primary features under investigation are molecular, a resolving power of a few dozen to $R \sim 100$ suffices for most science cases, depending upon the species under investigation. Binning data obtained with significantly higher resolution, such as $R \sim 1{,}000$, provides efficiency benefits for avoiding detector non-linearity and saturation for extremely bright stellar targets. Detecting the 1.38 μm water feature requires a resolution of $R \geq 10$ and the broad ozone feature requires $R \geq 10$ at ~0.6 μm.

### Objective 15 Requirements

| Parameter | Requirement |
|---|---|
| **Observing mode** | Wide slit or slitless spectroscopy |
| **Wavelength range** | ≤0.5 to ≥1.7 μm |
| **Resolution and SNR** | Assuming average transit durations of 1 h for a 1 R$_\oplus$ HZ planet with an Earth-like atmosphere orbiting a late M-dwarf |
| | $H_2O$: $R \geq 10$ at 1.4 μm with SNR/$\sqrt{h}$ ≥ 32,000 per spectral bin |
| | $O_3$: $R \geq 10$ at 0.6 μm with SNR/$\sqrt{h}$ ≥9,500 per spectral bin |
| **Sample size** | ≥5 systems |

One of the most promising Earth analogs currently known is TRAPPIST-1e (Gillon et al. 2017). Discovered by the transit photometry method, TRAPPIST-1e is one of three rocky planets in the HZ of the ultracool (faint) M-dwarf star TRAPPIST-1 (spectral class M8V). In order to detect the molecular features listed above for a rocky planet in such a system with an Earth-like atmosphere that is orbiting its host star with 1 h transits require $SNR/\sqrt{hour\ (h)} \geq 32,000$ per spectral bin for $H_2O$ and $SNR/\sqrt{h} \geq 9,500$ per spectral bin for $O_3$. These SNRs requirements conservatively assume that data reduction methods (post-processing) can achieve 30% of the photon noise limit, which is motivated by the current performance of HST/WFC3 and Spitzer/IRAC (Zellem et al. 2019). Given the relative faintness of Trappist-1 at visible wavelengths (V = 18.8) required for $O_3$ detection, for a true Earth analog, only water vapor would be detectable by HabEx Workhorse Camera (HWC; *Section 6.6*) in the atmosphere of Trappist-1e, and it would require the observations of 50-100 transits.

However, TESS is predicted to discover 10 potentially habitable worlds (here defined as terrestrial, temperate planets) orbiting nearby M-dwarf stars (Barclay et al. 2018). These stars have a mean V-mag of 15.0 and are thus 34 times brighter than TRAPPIST-1 at visible wavelengths. Assuming TESS discovers such a planet, then HabEx could observe this system in just 3% of the time it takes to observe TRAPPIST-1e at similar precisions and higher precision would be accessible in reasonable exposure times. Assuming the baseline 4 m architecture and HWC throughput performance, HabEx would robustly detect several key molecules in the atmosphere of such a planet in the visible-near-IR wavelength range. Detected molecules include ozone and water vapor, which would be detected by stacking 50 transits, corresponding to 50 h of in-transit integration (simulations of **Figure 4.7-1**). A 10 ppm noise floor is included, motivated by state-of-the-art transit spectroscopy with Hubble/WFC3 (Zellem et al. 2019) and assuming no noise floor reduction

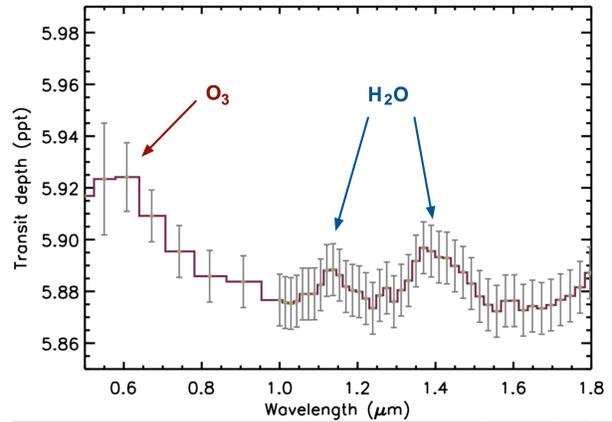

**Figure 4.7-1.** HabEx will identify molecules in the atmospheres of Earth-like planets from transit spectroscopy of eclipsing planets. Shown above is a simulated transit spectrum based on 50 transits of an Earth-like planet around a late M-dwarf (with a V-mag of 15, as predicted to be discovered by TESS), assuming an Earth-like atmosphere (ppt = parts- per-thousand). Simulated spectral resolution per bin is $R = 10$ shortward of 1 μm, revealing the presence of the $O_3$ Chappuis band (0.53–0.66 μm), and $R = 70$ longward of 1 μm, revealing water bands at 1.13 μm and 1.41 μm. Absorption bands are seen as an increase in transit depth. An irreducible 10 ppm rms noise floor is assumed in addition to shot noise.

after binning the data to even lower spectral resolution.

Finally, one significant challenge for transit spectroscopy is slit losses: any variation in flux reaching the spectrograph due to changes in telescope pointing or PSF shape will manifest as increased scatter in the transit light curves (e.g., Zellem et al. 2014). Exoplanet transit spectroscopy therefore requires either slitless spectroscopy or slit spectroscopy with a wide slit (width ≥5 times the PSF full-width at half-maximum; e.g., Pearson et al. (2018) to minimize slit losses.

## 4.8    Objective 16: Constraining Planet Formation Mechanisms

**Objective 16:** To constrain the range of possible structures within transition disks and to probe the physical mechanisms responsible for clearing the inner regions of transition disks.

### 4.8.1    Rationale

Protoplanetary disks are the first stage in the evolution of planetary systems. Thousands have been identified at infrared through millimeter wavelengths (e.g., Williams and Cieza 2011). Their young, pre-main sequence, stellar hosts are generally fainter (7 < V < 15) and more distant ∼140 pc) than the mature sunlike stars surveyed as part of HabEx's science Goals 1 and 2 (*Chapter 3*). Of particular interest among young circumstellar disks are "transition disks," which still have a substantial gas component, separating them from older dust-dominated debris disks. Transition disks show cleared inner regions of a few tens of AU in size, separating them from younger uncleared protoplanetary disks which are optically thick and therefore less amenable to high contrast imaging at visible wavelengths. Ubiquitous asymmetries observed in transitional disks suggest that they are almost certainly sites of on-going planet formation, thus they represent a key period in the coevolution of disks and planets. Since the transition disk phase is short, only about 30 have been identified in nearby star forming regions at magnitudes brighter than R = 13 mag (Kate Follette, private communication). Consequently, this entire well-defined, magnitude-limited sample can be observed in a reasonable time with a targeted Guest Observer campaign.

The absence of dust in the innermost, hottest, regions of a transition disk results in a deficit in the broadband spectral energy distribution (SED) at near-IR wavelengths, which enables them to be distinguished from regular protoplanetary disks via spectrophotometry. Transition disks have been directly resolved at a ∼few AU scales in millimeter-wavelength thermal continuum, millimeter-wavelength gas emission lines, and visible/near-IR scattered light. Sub/millimeter observations (e.g., Andrews et al. 2018) are limited to probing the cold, large-grain dust distributions, while high-contrast, high-spatial resolution visible and near-IR imaging is necessary to probe the small-grain dust distribution. High-contrast imaging in the visible also allows for detection of shocked emission from accretion onto the star and forming protoplanets.

At visible and near-IR wavelengths, HST and ground-based AO have had some success imaging transition disks. However, the majority of the inner transition disk cavities inferred by SED measurements have not directly resolved due to limitations in achievable point source-to-star contrast and angular resolution. HST's sensitivity and stability has made it the preferred instrument for low surface brightness disk features. However, with an inner working angle of 0.25 arcsec (using the BAR5 occulter on STIS), narrow outer disk gaps and nearly all transition disk cavities are inaccessible. Ground-based observations using AO have a point source-to-star contrast performance that remains limited to $10^{-3}$ to $10^{-4}$ in the ∼0.05–0.2 arcsec regime where most transitional disk cavities lie, primarily due to the relative faintness of transition disk targets (most at R > 10 mag). Furthermore, PSF subtraction techniques do not preserve disk structures to high-fidelity, resulting in controversy about the nature of protoplanet detections (e.g., Sallum et al. 2015b; Currie et al. 2019). A first improvement in distinguishing the disk from planet emission is to use differential imaging in and out of accretion lines, which enables investigations of the relative importance of shock and internal luminosity during planet formation, as illustrated by the recent claimed detections of accreting protoplanets in Hα (e.g., Sallum et al. 2015a; Wagner et al. 2018; Haffert et al. 2019). However, in order to clearly separate protoplanets from disk structures, and to detect less massive planets accreting at lower rates, higher-contrast, higher-spatial-resolution imaging is necessary. Indeed, planet formation and evolution models (e.g., Mordasini et al. 2017; **Figure 4.8-1**) indicate that accretion line (Hα) luminosity detection limits of ∼$10^{-8}$ L⊙ are required to detect forming protoplanets over a broad range of masses (down to 0.05 $M_{Jup}$ or 15 $M_{\oplus}$) and gas accretion rates. The corresponding point source-to-star flux ratio that is required to reach that sensitivity is ∼$10^{-6}$ at Hα, which is 100 to 1,000 times better than currently achieved from the ground.

Such high-contrast high-resolution visible observations, in the continuum and in accretion lines, will enable empirical exploration of the correlation between transition disk structures, the abundance of dust and gas, and the presence of when, where, and which planets emerge in protoplanetary disks will put more direct and key observational constrains on formation models. It will enable direct comparisons with theoretical predictions of planetary gas and solid accretion rates, and the timescale of planetary orbital migration through the protoplanetary disks. Similarly, since predictions for the luminosity of a planet at a given mass and age differ by several orders of magnitude at very early ages depending on the models used, measuring the brightness of very young giant exoplanets—in and out of accretion lines—provides crucial information to distinguish between different models of gas accretion onto planets (e.g., Marleau et al. 2017; Berardo et al. 2017).

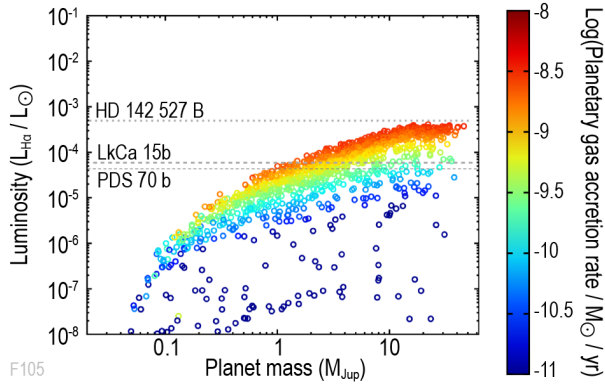

**Figure 4.8-1.** Hα luminosity predicted for giant planets during the formation phase, assuming a cold accretion model, for a variety of planet masses and accretion rates (from Mordasini et al. 2017, figure adapted by K. Follette). Also indicated are the Hα luminosities measured by ground-based facilities for candidate protoplanets LkCa 15 b and PDS 70 b and the low mass stellar companion HD 142 527 B. Reaching point source-to-star contrasts of ≤10⁻⁶ would provide access to Hα luminosities of ~10⁻⁸ L_☉, i.e., to sub-Jupiter-mass planets over a broad range of accretion rates.

### 4.8.2 Requirements

Most nearby transition disks are between 100–150 pc away and are embedded in gas and dust structures. In order to correlate observed disk structures with the presence of protoplanets and to derive meaningful constraints on the relative weights of core accretion and disk instability formation scenarios, a statistical number (≥20) of transition disks must be observed. This sets a requirement for high-planets (e.g., **Figure 4.8-2**). Repeating such observations at different disk ages and over a wide range of accretion rates will also inform formation mechanisms. For instance, core accretion and disk instability models predict different formation efficiencies and timescales at given star-planet separations. Directly observing

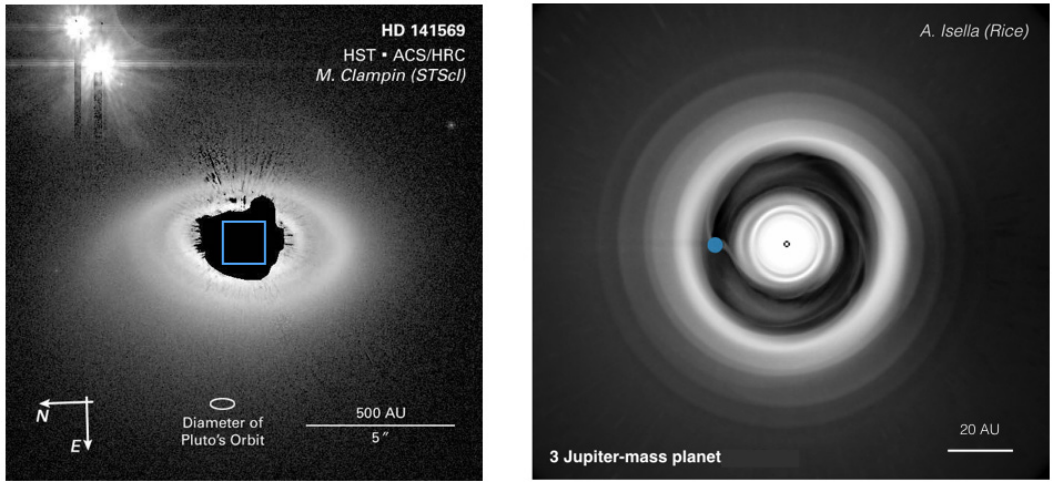

**Figure 4.8-2.** *Left panel*: Visible image of the HD 141569 A protoplanetary disk observed with HST / Advanced Camera for Surveys (ACS; Clampin et al. 2003) showing extended structures scattering light at large separations, but saturation in the inner 50 AU. This hybrid disk with a large debris component was later determined to be a transition disk with an ~11 AU inner clearing using sub-millimeter observations (Goto et al. 2006). *Right panel*: A theoretical model (A. Isella, private communication) protoplanetary disk with a planet three times more massive than Jupiter (blue dot). This model illustrates the rich structure expected to be found within 1.5 arcsec of the central star (blue box in the left panel, which is inaccessible to HST). HabEx is designed to explore this inner region at sufficiently high contrast and spatial resolution to reveal such structures at ≥0.1 arcsec from the central star, as well as any ≥0.05 M_Jup planets actively accreting in this region.

**Objective 16 Requirements**

| Parameter | Requirement |
|---|---|
| Observing mode | 2D broadband & narrowband imaging |
| R mag | ≥13 |
| Point source-to-star flux ratio at the IWA | ≤$10^{-6}$<br>≥2 orbital positions with ≤5 mas uncertainty |
| IWA at Hα (0.656 μm) | ≤100 mas to detect a protoplanet in a 15 AU orbit around a star located at 150 pc with SNR ≥ 7 in ≤ 50 h using narrow-band photometry |
| Surface brightness at the IWA | Broadband: R ≥ 20.5 mag/arcsec²<br>Narrowband Hα: R ≥ 18.0 mag/arcsec² |
| Sample size | ≥20 systems |

contrast observations of R ≥ 13 mag, as transition disk host stars are typically 1,000 times fainter than the mature sunlike stars that are the focus of HabEx's Goals 1 and 2 (*Chapter 3*). For a star located at 150 pc, detection of a protoplanet orbiting at 15 AU requires an inner working angle (IWA) of ≤100 mas. The point source-to-flux ratio requirement is ≤$10^{-6}$ when observing in Hα, corresponding to a planet mass detection limit of ~0.05 $M_{Jup}$ at gas accretion rates as low as $10^{-11}$ $M_{\odot}/yr$ (**Figure 4.8-1**), probing a whole new regime of low-mass forming protoplanets.

A robust protoplanet detection within a feasible integration time requires SNR ≥ 7 in times of ≤ 50 h using a narrow-band (~10 nm) filter centered around Hα. In practice this requirement is no more challenging than the one set for broad-band detection of significantly fainter exo-Earths around nearby main sequence stars (*Section 3.1.1*). To ensure the detection is not a background object requires measurement at a nearby continuum wavelength, which can also be used to constrain the mass of the object (e.g., Wagner et al. 2018). Finally, detecting additional emission lines (e.g., Hβ at 0.486 μm) through the UV to visible domain could provide further accretion diagnostics and allow for better understanding of accretion processes onto protoplanets (e.g., Robinson and Espaillat 2019).

Detection of structures within the transition disk at the same sensitivity as potential protoplanets, requires the same IWA and contrast ($10^{-6}$ per resolution element) as above. For a sunlike star at 150 pc, assuming a 4 m-class telescope, this contrast level corresponds to a surface brightness detection limit of R ≥ 18 mag/arcsec², which shall be met for narrow-band observations using a ~0.01 μm filter centered around Hα. For broadband observations at a nearby continuum wavelength, deeper contrast and a fainter surface brightness detection limit of R ≥ 20.5 mag/arcsec² will place additional constraints on the nature of the companion. It will possibly even enable direct detection from the planet's photosphere, which would break the degeneracy between planet mass and accretion rate for Hα.

These requirements are more relaxed than the V ≥ 22 mag/arcsec² detection limit set for broadband visible observations of debris disks (*Section 3.2.4*).

## 4.9 Objective 17: How Do Stars and Planets Interact?

**Objective 17:** To probe the physics governing star-planet interactions by investigating auroral activity on gas and ice giant planets within the solar system.

### 4.9.1 Rationale

The discovery of thousands of exoplanets orbiting nearby stars is a historic advance, with broad implications ranging from fundamental questions about the development of life, to detailed astrophysics questions surrounding this new scientific terrain. In terms of the latter, there is a strong desire to characterize and understand these exoplanets, how they formed, and how they interact with their host stars. To inform such studies, the planets within our own solar system are the ones that can be studied most closely, thereby providing important laboratories to understand the basic physical principles of how planets form and evolve. Planetary aurorae are key examples of star-planet interactions and some fundamental, outstanding questions include how strong planetary magnetic fields drive auroral activity in response to changes in stellar winds, and the extent to which planetary auroral activity is driven by stellar winds as opposed to plasma processes in the planetary magnetosphere.

Planetary aurorae are the results of charged particles in the stellar wind and in the planet's magnetosphere being strongly perturbed by the local magnetic and plasma environments, and thereby precipitating into the planet's upper atmosphere. In addition to being seen on Earth, this phenomenon has also been seen on all of the gas giant planets in our solar system (e.g., **Figure 4.9-1**). Solar system aurorae cover a wide range of physical scales and conditions, thereby providing an important testing ground for probing star-planet interactions in exoplanetary systems. For example, the physical processes that control aurorae on different scales of time and planet size, different levels of stellar winds, different planetary rotation rates and different magnetic field strengths are still unknown. On Earth, the solar wind flow time to pass the planet is a few minutes, and auroral storms develop in a complex interaction with the southward-pointing interplanetary magnetic field. On Jupiter and Saturn, the flow time is hours to days. Jupiter sometimes responds to changes in the solar wind, other times not at all, while Saturn's auroral activity responds to every solar wind pressure front. The question remains whether auroral activity on Saturn is controlled solely by solar wind pressure, or if the interplanetary magnetic field direction important. Another open question is whether Saturn's aurora is similar to the Earth's, or whether it has a different interaction with the solar wind.

Including flyby missions, observations to date have investigated aurorae on Earth, Jupiter, Saturn, but only weakly on Uranus. Extending high-resolution investigations to Uranus and Neptune (e.g., **Figure 4.9-2**) will enable access to different configurations of internal magnetic fields that are highly tilted and offset from the planets' rotation axes. This will provide solar system analogs to help understand the large number of super-Earths and sub-Neptunes recently discovered by Kepler.[2]

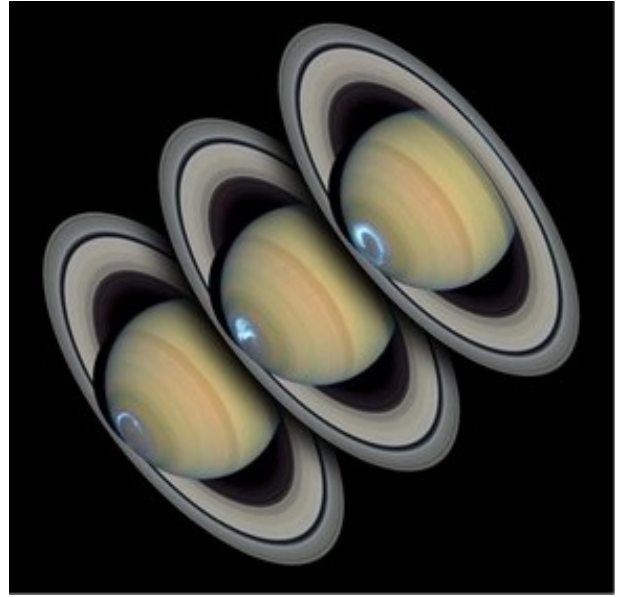

**Figure 4.9-1.** Shown here are HST UV images of Saturn's aurorae and changes during an auroral storm. To date, high-resolution studies of aurorae with HST have been restricted to Earth, Jupiter, and Saturn. A space-based UV telescope with a higher sensitivity than HST is required to undertake these fundamental investigations on Uranus and Neptune.

### 4.9.2   Requirements

Planetary aurorae are best studied at UV wavelengths, where the bulk of the emission is produced and the level of reflected sunlight is low, i.e., the highest contrast that can be obtained when observing the sunlit face of a planet. Observing at wavelengths of 115–162 nm provides to access to $H_2$ Lyman (Ly) and Werner band emissions, in addition to H Lyα.

Observations of planetary aurorae on the ice giants, Uranus and Neptune, with ≤300 km transverse and vertical resolution will resolve their auroral ovals and provide the diagnostic potential to undertake the studies outlined above. Assuming observations close to perigee,

___
[2] http://exoplanetarchive.ipac.caltech.edu

*Objective 17 Requirements*

| Parameter | Requirement |
| --- | --- |
| **Observing mode** | Time-resolved imaging |
| **Wavelength range** | ≤115 nm to ≥162 nm<br>SNR ≥ 3 for an auroral surface brightness ≤100 Rayleigh in an exposure time of ≤10 m |
| **Field-of-view** | ≥1 × 1 arcmin² |
| **Angular resolution** | ≤0.05 arcsec |
| **Tracking** | Non-sidereal |

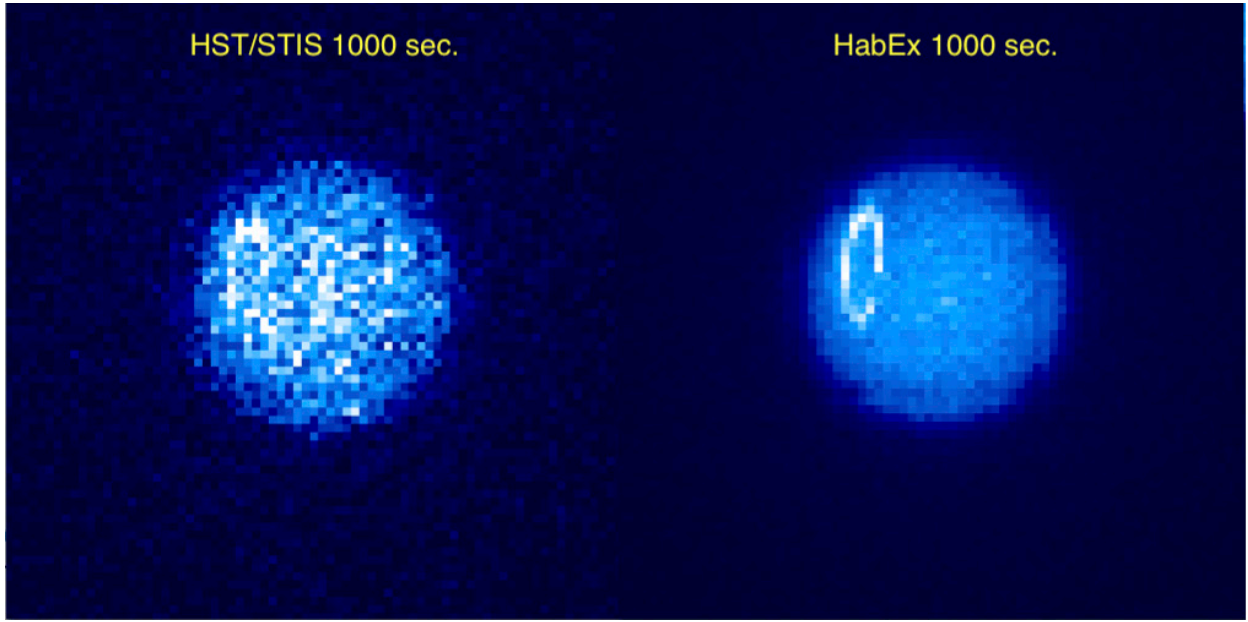

**Figure 4.9-2.** Observations with the HabEx baseline architecture and using the UVS (*Section 6.5*) HabEx will enable significant improvements in studies of aurorae over HST/STIS. In this simulation of Uranus airglow and aurora, the disc emission is resonantly scattered solar Lyα, with a maximum brightness of 1.6 kilo-Rayleighs. The reflected emission is based on a radiative transfer model. The auroral oval is approximately the size and location of the expected northern oval based on Voyager measurements of the internal magnetic field, with a maximum brightness of 2 kilo-Rayleighs but variable along the oval.

this implies a minimum angular resolution of ≤0.05 arcsec is required for UV auroral studies. In order to undertake observations in at time that minimizes blurring due to planetary rotation requires UV imaging reaching SNR ≥ 3 for an auroral surface brightness of ≤100 Rayleigh in an exposure time of ≤10 minutes.

Jupiter at perigee is 45 arcsec in diameter, which motivates a minimum required imaging FOV of ≥1 arcmin on a side. In addition, solar system planetary studies require a capability to undertake observations with non-sidereal tracking.

## 4.10    Additional Science

History has demonstrated that the breadth of scientific possibilities with a highly-capable observatory is well beyond what was envisioned at the time the observatory was designed. As one key example, HST has far exceeded the science envisioned when it was initially launched in 1990. Every year since its launch, the number of HST papers being published over an enormous range of scientific investigations has been growing and was almost 1,000 in 2018, with over

20,000 authors listed in HST's lifetime (STScI 2019). As such, it is infeasible to discuss the full range of science possible with an observatory that has the capabilities detailed in *Chapters 3* and *4* of this report: the science enabled by HabEx will be driven the imagination and ingenuity of the community. The following text briefly presents a few illustrative examples of such science, with the explicit understanding that these represent just the tip of the iceberg for what is possible with a highly-capable, next-generation great observatory like HabEx.

### 4.10.1    Intermediate-Mass Black Holes in Globular Clusters

#### Introduction

A key open question in globular cluster studies is whether they host intermediate-mass black holes (IMBHs), with masses ~100–10,000 $M_\odot$. While both stellar mass (<100 $M_\odot$) and supermassive (>$10^5$ $M_\odot$) black holes are well represented in observational studies, IMBHs, recently reviewed by Mezcua (2017), reside in a mass gap with few, if any, uncontentious examples. Finding a large population of IMBHs would be transformative, providing an evolutionary link in the growth of supermassive black holes, and shedding light on the formation channels of seed black holes in the early universe. Specifically, theorists are hard-pressed to explain how black holes can grow to masses of a few times $10^9$ M$_\odot$ in the few hundred Myr available for the most distant quasars known (e.g., Bañados et al. 2018).

Extrapolating observed galaxy trends, such as the empirical correlation between the stellar velocity dispersion of a galaxy bulge and the mass of the supermassive black hole at its center (e.g., Magorrian et al. 1998), suggests that globular clusters might be the hosts of IMBHs and fruitful hunting grounds for finding such objects. However, globular clusters have several properties that are quite distinct from galaxies, most notably that globular clusters are baryon dominated, while dwarf galaxies of comparable mass are generally dominated by dark matter. This suggests that extrapolating galaxy trends to globular clusters may not be appropriate.

With high masses and old ages, globular clusters should have produced a large number of stellar mass black holes. Given the lack of ongoing star formation, these black holes rapidly become the most massive objects in the cluster, more than four times more massive than the stars in the cluster. Mass segregation leads to the black holes forming a dense nucleus, decoupled from the dynamics of the rest of the cluster (Spitzer 1969). While in principle this dense nucleus could merge to form an IMBH, dynamical interactions within the nuclear cluster are expected to eject the majority of black holes in less than a Hubble time (Kulkarni et al. 1993; Sigurdsson and Hernquist 1993), suggesting globular clusters might only host stellar mass black holes at their core, without those black holes merging into an IMBH.

To date, seemingly contradictory results span from reported discoveries of IMBHs in globular clusters (e.g., Pooley and Rappaport 2006; Kızıltan et al. 2017) to reported discoveries of stellar mass black holes in globular clusters (e.g., Zepf et al. 2008; Giesers et al. 2018). The latter

## Section 4.10.1 Program at a Glance

| |
|---|
| **Science Goal:** Determine if globular clusters host intermediate-mass black holes (IMBHs). |
| **Program Details:** Multi-epoch imaging of globular cluster cores to measure proper motions of stars in central regions and determine if a massive central source is present. |
| **Instrument(s) + Configuration(s):** HWC + broadband UV/blue imaging (e.g., F336W, F475W). |
| **Key Observational Requirements:** High spatial resolution imaging with extremely high stability. Observe the same system over multiple years. |

are difficult to reconcile with the existence of an IMBH in those clusters since dynamical interactions should eject stellar mass black holes within a Gyr for any cluster hosting an IMBH. A further complication comes from the observational challenges in distinguishing globular clusters from the stripped nuclei of accreted dwarf galaxies. In summary, the current state of affairs is uncertain. Significant observational efforts are underway to find IMBHs in globular clusters, while theoretical and empirical arguments suggest globular clusters might or might not host black holes in that mass range.

### The Role of HabEx

An IMBH at the center of globular cluster would leave a distinct signature in the dynamics of stars in the central region, detectable in the motions of the stars. Motion along the line-of-sight will require spectroscopy at very high resolution at visible-to-near-IR wavelengths, ideally at high spatial resolution in order to isolate individual stars. Such measurements are well suited to the next generation of ground-based ELTs. Motions perpendicular to the line of sight, i.e., proper motions, are essential for maximizing the value of the radial motions and will require very high spatial resolution imaging from an ultra-stable platform with well-measured geometrical distortions. HabEx is ideally suited to undertake these proper motion measurements, which are not feasible with ground-based AO given their time-variable geometric distortions and smaller corrected FOVs.

### Science Program

To achieve this science will require multi-epoch imaging of nearby globular cluster cores over a period of several years. Sensitivity is key, implying broadband imaging. However, due to crowding and the dominating brightness of red giant stars, shorter wavelength observations at UV or blue wavelengths are preferred. HST has begun such efforts, e.g., Mann et al. (2019) find no evidence of an IMBH in 47 Tuc. A HabEx program would extend such studies to significantly more systems.

#### 4.10.2 Constraints on Massive Stellar Binary Evolution Models

##### Introduction

Massive star binary systems are the progenitors of gravitational wave events and supernovae. Around 70% of all massive stars form in close binaries, where strong interactions between the two stars occur across their entire lifetimes (Sana et al. 2012). Mass loss, accretion, and coalescence of the stars are the outcomes of such interactions and define evolutionary paths that can completely change the future evolution of the stars. Massive star binary evolution is necessary to produce explosive end-of-life events such as gravitational wave events or stripped-envelope supernovae—and yet the first and most common stellar interaction phases are currently not well probed and modelers are forced to adopt crude approximations for the evolution that the stars follow after their interactions. These approximations impact population studies of stars, leading to uncertain predictions for rates of gravitational waves or supernovae.

One of the most common types of interaction is envelope-stripping, when the more massive star in a binary system swells to fill its Roche-lobe (i.e., the region around a star in a binary system where surrounding material will be bound to the star by gravity) and initiates mass transfer to its binary companion, eventually losing its entire hydrogen-rich outer envelope. The

### Section 4.10.2 Program at a Glance

| |
|---|
| **Science Goal:** Study massive star binaries to understand their interactions (e.g., mass loss, accretion), which eventually give rise to coalescence and gravitational wave events. |
| **Program Details:** UV spectroscopy of stripped stars in massive star binary systems in order to their study surface composition as an indicator of stellar interior. |
| **Instrument(s) + Configuration(s):** Intermediate-resolution UVS spectroscopy ($R \geq 1,000$). |
| **Key Observational Requirements:** Sensitive UV spectroscopic capabilities. |

result is a small, hot, and long-lived stripped star that shows enhanced abundances of helium and processed material from the carbon-nitrogen-oxygen (CNO) cycle. This product of massive star binary interaction is likely the most distinct after the first interaction phase since it differs entirely from what is expected in the evolution of single stars.

Stripped stars are expected to be long lived and therefore common, but only a handful of them have been observed to date (e.g., Gies et al. 1998; Groh et al. 2008; Peters et al. 2015; Peters et al. 2013; Wang et al. 2017). The lack of detections is likely due to their more massive, bright companions outshining them in visible wavelengths (e.g., Götberg et al. 2017; Götberg et al. 2018). However, due to their high temperatures, stripped stars are expected to be much brighter in the UV than their more massive stellar companions (**Figure 4.10-1**), motivating future studies at UV wavelengths.

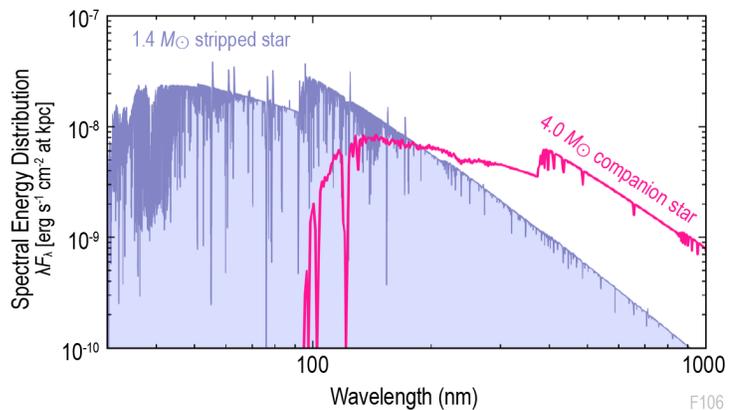

**Figure 4.10-1.** Despite being lower mass, the extreme heat of a stripped secondary star (*blue*) in a post-interaction binary system can outshine the more massive primary (*pink*) at UV wavelengths. Figure credit: Y. Götberg.

### The Role of HabEx

The HabEx UVS (*Section 6.5*) provides a unique opportunity to directly measure the properties of massive star binary systems, which will enable insight into their evolution over time, including constraining predictions of future evolution. UV spectroscopy probes the stellar surface composition, providing a better understanding of stellar interiors. In addition, UV lines are remarkably powerful tracers of stellar winds (e.g., see *Section 4.10.3*), the mechanisms and processes for which are poorly understood for hot, helium-rich stars.

UV spectra of a large sample of stripped stars will provide, for the first time, a statistically meaningful probe of stellar wind mass loss rates for helium stars, which will enable a better physical understanding of stellar wind mass loss. The mass loss rate of stripped stars affects how much they swell up as they evolve, and therefore determines whether they will enter the next interaction phase that ultimately leads to a gravitational wave event. Similarly, insight into mass loss rates will constrain how much of the hydrogen-rich envelope is left at the end of life, and therefore which type of supernova the star will ignite.

### Science Program

To achieve this science will require moderate-resolution UV spectroscopy ($R \geq 1,000$) of at least a dozen stripped stars from massive star binaries. The sensitivity improvements of HabEx as compared to HST/COS will extend such studies to significantly more systems, thereby making important progress in understanding the evolution of massive star binary systems and enabling higher precision forecasts for gravitational wave and supernova events.

### 4.10.3  Winds from Massive Stars

#### Introduction

Massive stars are the chemical and dynamical agents of the universe. Through powerful stellar winds and supernova explosions, they inject kinetic energy into the ISM and their extreme temperatures ionize hydrogen, producing spectacular HII structures such as the Tarantula nebula in the Large Magellanic Cloud (LMC). Massive stars evolve and die on fast timescales, which translates into rapid chemical enrichment of the host galaxy. The most massive specimens are progenitors of extremely energetic events, such as super-luminous supernovae and long gamma-ray bursts, that can be used to probe the high-redshift regime of the universe. The final stages of massive star binary systems correspond to the only gravitational wave events so far detected (e.g., Abbott et al. 2016; Abbott et al. 2017a). In addition, massive stars are the most plausible candidates to have commenced the re-ionization of the universe.

Many astrophysical phenomena thus require a good description of the evolutionary stages of massive stars in terms of duration, effective temperature ($T_{eff}$), luminosity, outflows, and expected end-products. Sophisticated models that account for the complicated physics of massive stars exist. However, the crucial parameter of the mass loss rate to radiation-driven winds remains poorly constrained and uncertain.

Massive stars shed mass throughout their lives (**Figure 4.10-2**, left) but mass lost during the earlier hydrogen- and helium-burning phases has a more profound impact on evolution because of the long duration of these stages. These stages correspond to stars with temperatures of $T_{eff}$ = 20,000–50,000 K, and the intense UV radiation field transfers energy and momentum into the metallic ions of the atmosphere, creating radiation-driven winds. Mass removal by the wind alters the conditions at the stellar core, and thus the rate of nuclear reactions, the duration of the main-sequence,

#### Section 4.10.3 Program at a Glance

| |
|---|
| **Science Goal:** Study winds from massive stars as key necessary input for massive star models. |
| **Program Details:** Measure UV spectra of massive stars to determine stellar wind velocity, abundance, and mass-loss rate. Low-metallicity stars are preferred. |
| **Instrument(s) + Configuration(s):** UVS spectroscopy ($R \geq 12,000$). |
| **Key Observational Requirements:** Sensitive UV spectroscopic capabilities. Multi-object capabilities would improve efficiency. |

and the nature of later stages of the massive star evolution. Theory predicts that radiation-driven winds depend on the luminosity and chemical composition of the star (e.g., Leitherer et al. 1992). This has been confirmed observationally (e.g., Mokiem et al. 2007; Garcia et al. 2014; **Figure 4.10-2**, right top and bottom).

### The Role of HabEx

The HabEx UVS (*Section 6.5*) will enable significant progress in understanding radiation-driven winds from young, massive stars. UV is the preferred wavelength regime for such studies, particularly for low-metallicity (≤1/5 $Z_\odot$), massive star winds, since these are expected to be weaker than the winds from high-metallicity stars. The UV contains diagnostics sensitive to 100 times smaller mass loss rates than visible lines. Specifically, this science requires a spectral range of 115–200 nm, with spectral resolution

$R \geq 12,000$ to measure the chemical abundance and determine its relation to mass loss rates and wind terminal velocity. The 120–180 nm interval contains numerous lines of iron-group elements, which are the main drivers of mass loss. Abundances of these elements are inaccessible to any other spectral range except X-rays (e.g., Garcia et al. 2014). The UV also contains spectral lines of carbon, nitrogen, oxygen, silicon, and sulfur, whose abundance may be complicated or impossible to constrain from analyses based solely on data at visible wavelengths (depending on the $T_{eff}$). Abundances for the former three, in particular, represent a crucial test of mixing mechanisms implemented in the models of stellar evolution to bring fresh products of nuclear reactions from the stellar core to the surface.

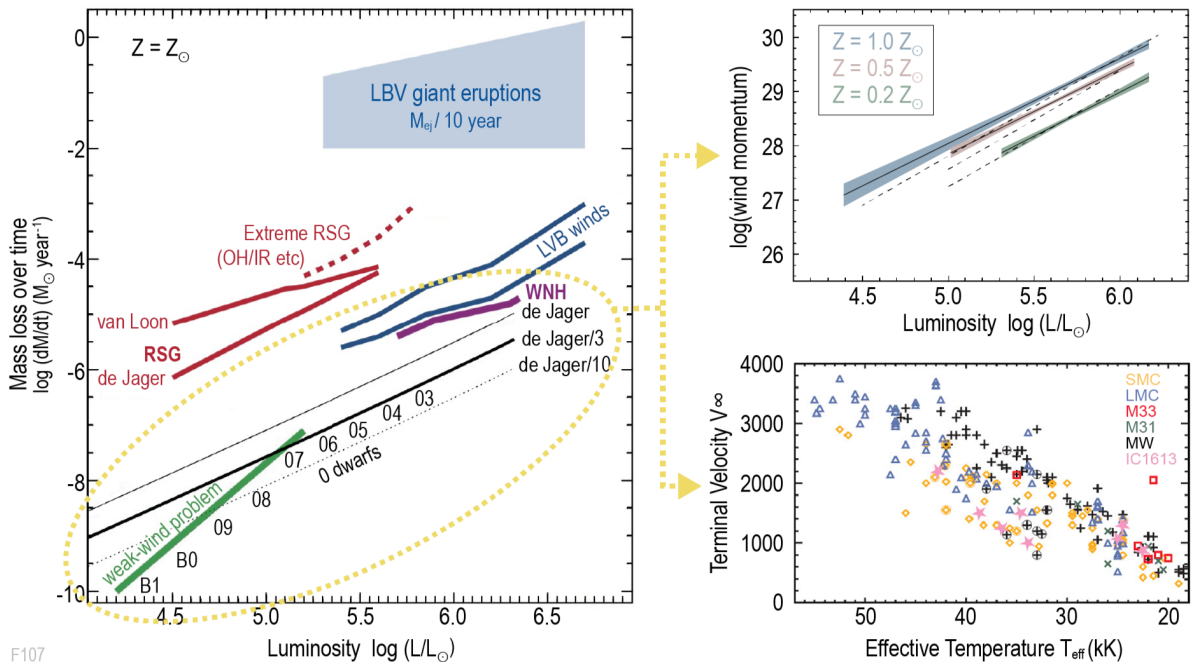

**Figure 4.10-2.** Understanding winds from massive stars is essential for testing models of massive star evolution, which impacts a range of astrophysics, from galactic chemical enrichment to the sources of gravitational wave events. *Left panel:* Summary of mass-loss regimes during the different life stages of massive stars, adapted from Smith (2014). Even though the advanced stages of massive stars (e.g., Red Super Giant, RSG, and Luminous Blue Variable, LBV) experience very high mass loss rates, most stellar mass is shed during the hydrogen-burning stages, e.g., O-dwarfs, O-super-giants, and nitrogen-rich Wolf-Rayet (WNH) stars. *Right-top:* Predicted relation between the wind momentum, which acts as a proxy for the mass loss rate, and stellar luminosity as a function of metallicity (Mokiem et al. 2007). *Right-bottom:* Terminal velocity, V∞, measurements of OB-type stars in different metallicity environments, from Garcia et al. (2014), showing empirical evidence that higher metallicity galaxies, such as the Milky Way (MW), produce stronger winds than lower metallicity galaxies, such as the Small and Large Magellanic Clouds (SMC, LMC).

**Science Program**

This science requires high-resolution ($R \geq 12{,}000$) UV spectroscopy of massive stars in low-metallicity galaxies in the Local Group (and potentially beyond). The Magellanic Clouds provide the closest targets, though several other low-metallicity dwarf galaxies are known that are located at distances beyond the reach of HST/COS, but would be accessible to an instrument with improved UV spectroscopic sensitivity. Multi-object spectroscopy would increase the efficiency of this science.

### 4.10.4   UV Extinction towards OB-Star Populations

**Introduction**

Interstellar extinction is the absorption and scattering of photons as they travel through the ISM to an observer, which leads to a change in the shape of the observed spectrum and a decrease in the observed brightness of the star. The ISM is responsible for causing extinction and the composition of the ISM (e.g., abundance of different molecular species and grain sizes), which varies between galaxies, determines what extinction occurs. Extinction is known to be variable along different lines-of-sight and to be dependent on wavelength (e.g., Cardelli et al. 1989). However, studies of wavelength-dependent extinction laws have not yet reached consensus. For example, the works of Cardelli et al. (1989) and Fitzpatrick and Massa (2007) on UV-to-IR extinction are in contradiction. The former finds a tight correlation between UV and visible/IR extinction laws, leading to the "one-parameter" family of extinction laws. The latter found that, with the exception of a few curves, the UV and IR portions of extinction curves are not correlated with each other.

**Section 4.10.4 Program at a Glance**

| |
|---|
| **Science Goal:** Measure the UV extinction of OB stars over a range of environments, enabling measurements of reddening laws from UV to near-IR wavelengths. |
| **Program Details:** UV spectroscopy of OB-star clusters and associations. |
| **Instrument(s) + Configuration(s):** UVS multi-object spectroscopy ($R \geq 1000$). |
| **Key Observational Requirements:** Sensitive UV multi-object spectroscopic capabilities. |

More recently, higher-precision visible and near-IR extinction laws for both 30 Doradus and the Milky Way have been measured (Maíz Apellániz et al. 2014). However, large uncertainties continue to prevail in the UV. This is largely due to archival International Ultraviolet Explorer (IUE) and HST data not having sufficient quality (or quantity) to accurately characterize the UV extinction.

Sight lines to O-type stars experience a more diverse extinction than those towards late-type stars, making O-type stars ideal targets for a detailed investigation into UV extinction laws. While extinction along sight lines to late-type stars is predominantly caused by the average, diffuse ISM, O-type stars are affected by a range of dust grain sizes, from the small grains of molecular clouds to the large grains of HII regions. **Figure 4.10-3** shows the ISM structure in two different HII regions around Milky Way O-type stars for which extinctions have been measured.

**The Role of HabEx**

UV observations (with complementary visible-IR observations) for several dozen O-type stars are required to adequately determine the UV extinction laws. However, there are currently only a few tens of stars that have been observed in the UV (and the counterpart visible-IR observations are incomplete). Such data will test the one-parameter family of extinction laws and determine how tight the correlation is between UV and visible-IR extinction laws over a much larger range of environments than is currently possible.

HabEx is ideally suited to provide the necessary data to accurately characterize extinction towards OB-star clusters. Investigating associations across a variety of environments and large range of extinctions requires agile multi-object UV spectroscopic capabilities, as provided by UVS (*Section 6.5*), and multi-object visible-near-IR spectroscopic capabilities either from the ground or with the HWC (*Section 6.6*). The capability for multi-object spectroscopy is particularly essential for sites with highly variable extinction (e.g., Carina, 30 Doradus).

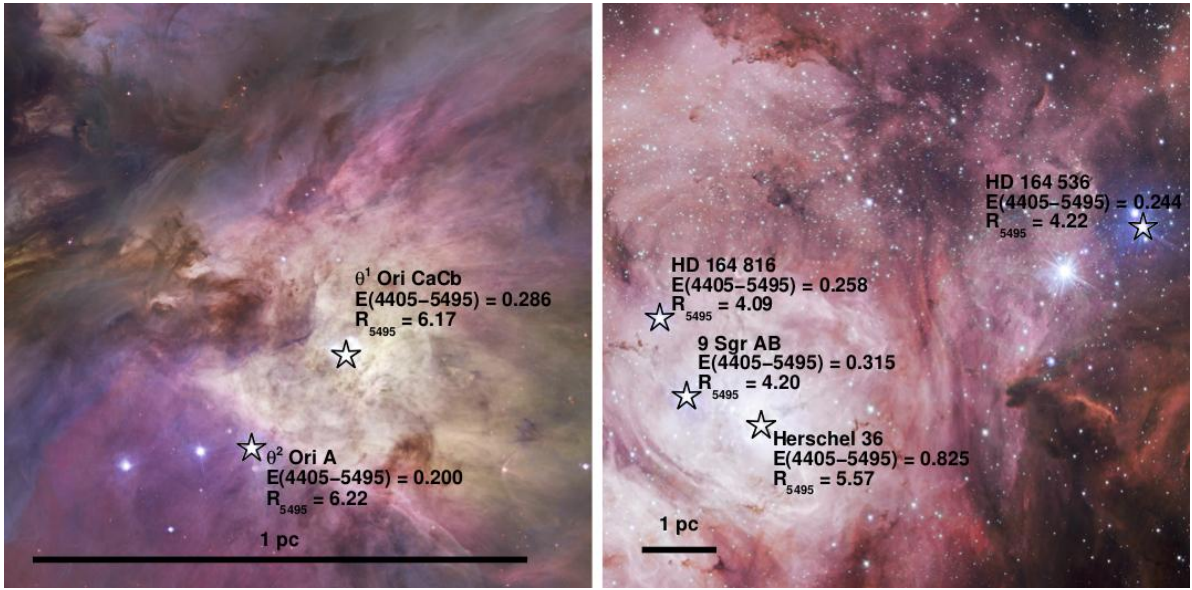

**Figure 4.10-3.** O-type stars experience varied extinction based on the size and composition of the ISM along the line-of-sight. Extinctions have been measured for the O-type stars in these images, which are surrounded by ISM that is rich in HII. The sample size of observed stars (currently a few tens) needs to be at least doubled in order to adequately measure the UV extinction laws. *Left panel:* The Orion Nebula. *Right panel:* M8. Figure credit: Maíz Apellániz and Barbá (2018).

### Science Program

To measure the UV extinction of OB stars over a range of environments will require multi-object UV spectroscopy of multiple OB-star clusters and associations with a range of environments, extinctions, and metallicities, reaching to fainter stars than are accessible with HST. Correlating UV extinction with visible and near-IR extinction provides added value, with the latter provided either by the HWC (*Section 6.6*) or from the ground.

### 4.10.5 Chemical and Kinematic Properties of the ISM towards OB-Star Populations

### Introduction

Current samples of ISM spectra along the line-of-sight towards OB-star clusters and associations in nearby galaxies are incomplete. Obtaining a representative and statistically significant sample of spectra across far more sight lines is essential for understanding how massive star feedback affects galaxy evolution.

Rest-frame UV spectroscopy in the wavelength range from 115–300 nm plays a unique role in chemical abundance and kinematic studies of the ISM for several reasons. First, the nebular abundance of carbon, which is the second most abundant metal by mass in the universe and an element that is essential to life as we know it, is best derived from UV collisionally excited emission lines (CELs). This is because the recombination lines of carbon are weak at visible wavelengths. HST has measured interstellar carbon and oxygen abundances through low-resolution COS and STIS spectra for only ~30 galaxies to date (**Figure 4.10-4**; e.g., Berg et al. 2016; Peña-Guerrero et al. 2017). These current data are insufficient since the UV spectra are often too noisy around the required CIII] 190.7 nm + CIII] 190.9 nm features.

### Section 4.10.5 Program at a Glance

| |
|---|
| **Science Goal:** Measure the chemical and kinematic properties of the ISM towards OB star associations. |
| **Program Details:** UV spectroscopy of OB-star clusters and associations. |
| **Instrument(s) + Configuration(s):** UVS multi-object spectroscopy at moderate to high spectral resolution. |
| **Key Observational Requirements:** Sensitive UV multi-object spectroscopic capabilities, $R \geq 16{,}000$. Multi-object spectroscopy and spatial resolution that is improved relative to HST would both provide added scientific value. |

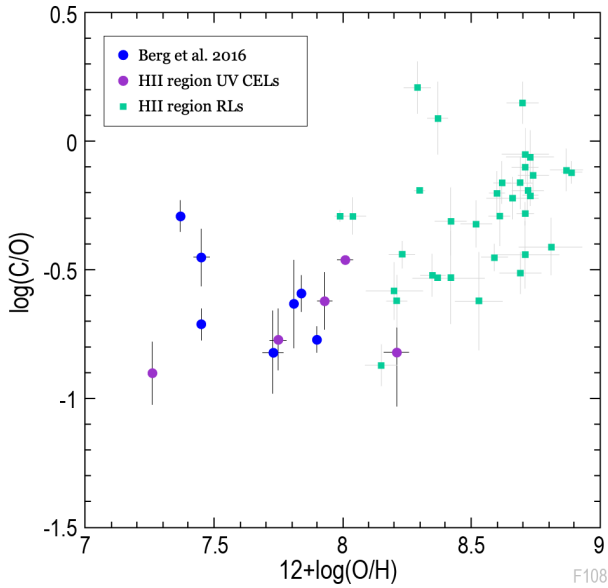

**Figure 4.10-4.** Carbon-to-oxygen ratio vs. oxygen abundance for star-forming galaxies. The purple points represent abundances derived from UV collisionally excited emission lines (CELs) with strengths of 3σ or greater. The green points represent abundances derived from recombination lines (RLs) in star-forming galaxies. Improved sensitivity and spatial resolution over HST/STIS are required to obtain sufficiently high SNR spectra and imaging to resolve carbon enrichment regions with enough precision to constrain the role of massive star feedback in galaxy evolution. Figure adapted from Berg et al. (2016).

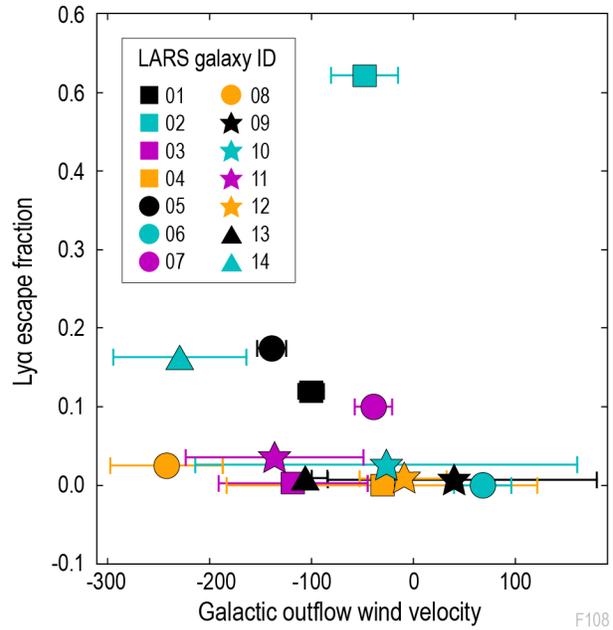

**Figure 4.10-5.** Galactic wind outflow velocity vs. escape fraction of Lyα photons for the 14 galaxies in the Lyα reference sample (LARS). A larger sample (≥ 100) of galaxies is required to more precisely constrain models for how ISM kinematics affect the escape of Lyα photons from galaxies. Figure credit: Rivera-Thorsen et al. (2015).

Another reason why UV spectroscopy is important for this study is that neutral-gas abundances of silicon, magnesium, iron, and aluminum can be obtained via unsaturated UV absorption line spectroscopy. Such abundances are important for assessing the (in)homogeneities of the multiphase ISM in galaxies where the bulk of metals can be hidden in the neutral phase (e.g., James et al. 2014). Observations of multiple sightlines through larger samples of galaxies are required to complete the census of the chemical composition of the multiphase ISM.

Finally, UV provides access to the important Lyα transition of hydrogen. The small sample of galaxies from the Lyα reference sample (LARS; Hayes et al. 2013) shown in **Figure 4.10-5** indicate that galactic wind outflows appear to be necessary for Lyα escape from galaxies. However, a galactic wind outflow does not seem to guarantee Lyα escape (Rivera-Thorsen et al. 2015). A larger, statistical sample of galaxies will enable more detailed studies into the processes that may be competing with the galactic wind outflows to prevent the Lyα escape.

For kinematic studies of the ISM towards massive OB-stars, as done by HST/STIS for the Carina nebula (Walborn 2012), very high resolution, $R > 100,000$, UV spectroscopy is preferred. However, the bulk ISM kinematics towards OB-star populations can be investigated at a more moderate resolution of $R \geq 16,000$. The latter data are useful for studying the properties of large-scale ISM/galactic outflows in star-forming galaxies (e.g., Chisholm et al. 2016b; Chisholm et al. 2016a; Chisholm et al. 2017; Chisholm et al. 2018).

### The Role of HabEx

The HabEx UVS (*Section 6.5*) will enable observations of rest-frame UV spectra of the ISM toward OB-star clusters for a representative, statistically significant sample of nearby galaxies. At key UV wavelengths between 115–300 nm, HabEx's improved spectral resolution and spatial sensitivity over HST will enable the separation of light from multiple OB-star clusters, thus enabling a spatially resolved view of carbon enrichment in CEL studies. In addition, the multiplexing capabilities of the UVS are essential to undertaking a thorough investigation of neutral gas abundances to better understand the chemical composition of the multiphase ISM. Multiplexing will also enable studies of a larger sample of galaxies to provide insight into outflows in star-forming galaxies and how ISM kinematics affect the escape of Lyα photons from galaxies (Wofford et al. 2013; Rivera-Thorsen et al. 2015).

### Science Program

This science requires high-resolution ($R \geq 16,000$) UV spectroscopy of OB-star populations for studies of chemical abundances and bulk ISM kinematics. Multi-object spectroscopy enables sufficient sightlines to examine a sufficiently large ($\geq 100$) sample of galaxies to constrain models for how massive star feedback affects galaxy evolution.

### 4.10.6  Solar System Planetary Exospheres

#### Introduction

The outermost, gravitationally bound, low-density gas around a planet or satellite (moon) is referred to as the *exosphere*. Studying exospheres in our own solar system provides insight into the wider range of exoplanetary exospheres, which

## Section 4.10.6 Program at a Glance

| |
|---|
| **Science Goal:** Study solar system planetary exospheres, as our nearest proxy for understanding exoplanet exospheres and the physical principles of planetary atmosphere loss. |
| **Program Details:** UV imaging of solar system planets at a range of seasons. |
| **Instrument(s) + Configuration(s):** UVS imaging. |
| **Key Observational Requirements:** Non-sidereal tracking. |

form the interaction region of a planetary atmosphere with the space environment. Exospheres are best observed in the UV, where the strongest transitions occur and reflected solar continuum is weak.

Within the solar system, the physical principles and processes that govern the loss of an atmosphere into space vary strongly from planet to planet. For the Earth, atmospheric loss is predominantly due to the high-energy tail of the velocity distribution of particles in the atmosphere exceeding the escape speed, and thereby being lost into space. This process is referred to as Jeans escape. At Mars and Venus, hot hydrogen gas populations are likely to dominate the exosphere loss. Also, for Mars, large annual variations exist, implying a strong seasonal control of the escape flux (e.g., Bhattacharyya et al. 2015; **Figure 4.10-6**). At Mercury, solar radiation pressure and solar wind proton charge exchange may dominate. At Uranus, a high-temperature hydrogen gas corona affects ring particle lifetimes. At Pluto, there is the potential for hydrodynamic flow of escaping hydrogen, which could entrain heavier species. Many of these phenomena are still poorly understood since they are too faint to be observed with HST and progress requires observations with a facility that has higher sensitivity in the UV.

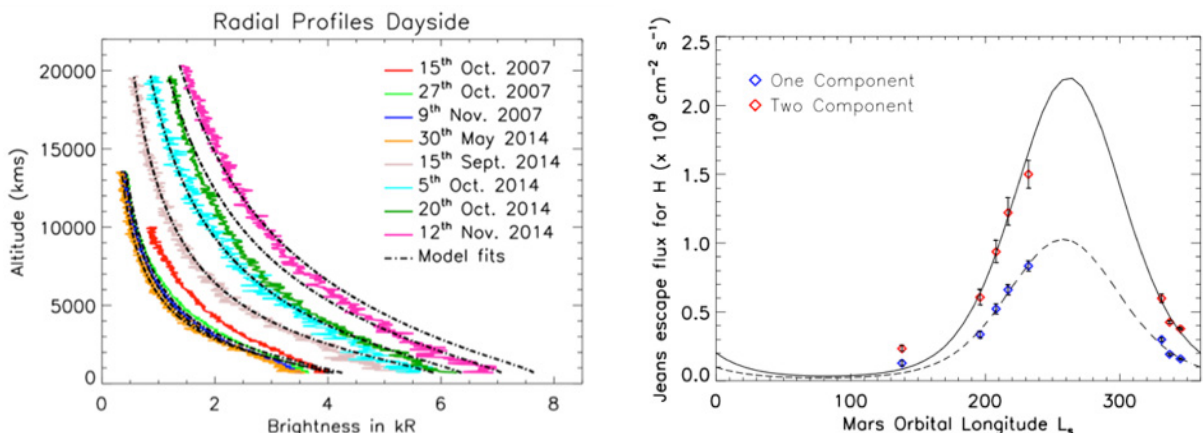

**Figure 4.10-6.** With access to hydrogen Lyα emission and high-sensitivity imaging in the UV, HabEx will expand the study of solar system exospheres by observing much fainter phenomena than has been possible to date. *Left panel:* HST altitude profiles of H Lyα emission from the Mars exosphere showing large changes over time. *Right panel:* Hydrogen escape flux vs solar longitude derived from observations using a radiative transfer model. Solar longitude corresponds to Martian season; the broad increase around 270 degrees roughly corresponds to perihelion and southern summer. Figure credit: Bhattacharyya et al. (2015).

### The Role of HabEx

Observing the exospheres of solar system planets and resolving effects like seasonal variation requires a space-based telescope with UV imaging and non-sidereal tracking capabilities and a modest spatial resolution requirement of ≤0.05 arcsec. An order of magnitude improvement in sensitivity over HST at UV wavelengths and a highly stable PSF is required to observe many faint signatures in the exospheres of planets in the solar system.

The HabEx UVS (*Section 6.5*) meets these requirements and will enable significant progress in understanding the processes governing planetary exospheres and their interaction with the space environment.

### Science Program

This scientific program requires multiple high-sensitivity UV imaging observations over a range of seasons for solar system planets including Mercury, Venus, Mars, and Uranus. These observations will provide a deeper understanding of the physical processes occurring in interactions of the exosphere with the space environment and also constrain models of atmospheric loss into space.

### 4.10.7 Solar System Cryovolcanism

#### Introduction

Cryovolcanism is an analog of the volcanism commonly observed on Earth, except that rather than molten rock (magma) being spewed by a volcano, the eruptions consist of volatiles such as water, ammonia, or methane. From HST and the Voyager flybys, several examples of cryovolcanism and cryoventing have been observed in the solar system, including on moons of gas giant planets, such as Jupiter's Europa (e.g., Roth et al. 2014b; Roth et al. 2014a; **Figure 4.10-7**), Saturn's Enceladus (e.g.,

### Section 4.10.7 Program at a Glance

**Science Goal:** Observe cryovolcanism in the solar system to understand what processes trigger eruptions and the range of amplitudes and environments in which it happens.

**Program Details:** UV monitoring of small bodies in the solar system.

**Instrument(s) + Configuration(s):** UVS imaging.

**Key Observational Requirements:** Non-sidereal tracking.

Dougherty et al. 2006; Hansen et al. 2006; Porco et al. 2006), and Neptune's Triton (e.g., Smith et al. 1989). Establishing statistics on the conditions in which cryovolcanism occurs, and what sets off the eruptions, is key to understanding the principles of volcanism in general. Current observations of eruptive plumes on Europa by HST are critical to the design and planning of the Europa flyby mission, and address the important question of extant life on Europa. HST only sees evidence for plumes ~10% of the time, and always observes at UV wavelengths with high sensitivity to small columns of gas.

### The Role of HabEx

With ~10 times greater UV sensitivity compared to HST, the HabEx UVS (*Section 6.5*) will vastly improve understanding of the processes governing cryovolcanism on small bodies such as Europa, Enceladus, and Triton by probing the duty cycle and distribution of cryovolcanic activity. HabEx will also have the capability of detecting previously unknown cryovolcanism on many other small bodies in the solar system—even Pluto is within the range of HabEx UV imaging capabilities.

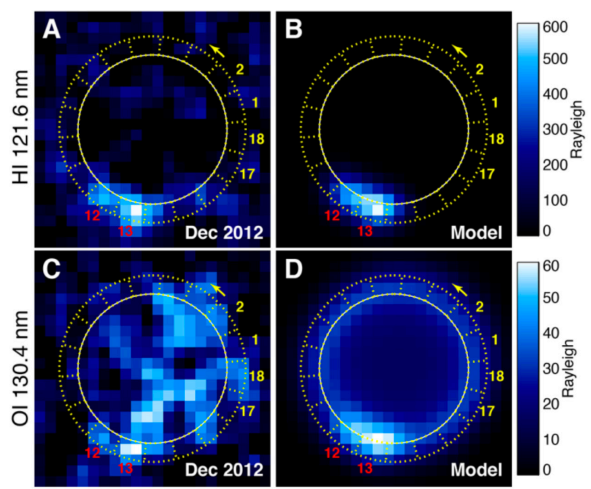

**Figure 4.10-7.** The HabEx UVS would investigate cryoplumes on bodies throughout the solar system. Shown here are HST far-UV images of oxygen airglow emission from cryoplumes on Europa. Figure credit: Roth et al. (2014b).

### Science Program

This scientific program would entail UV imaging of small bodies in the solar system where cryovolcanism has previously been seen or is suspected to occur (e.g., Europa, Enceladus, and Triton). To understand this transient process, multiple observations over time are required.

### 4.10.8 Exoplanet Atmospheric Escape with Transit Spectroscopy

#### Introduction

Atmospheric escape is a fundamental physical process that leads atmospheric constituents to become unbound from a planet and alters the composition of the remaining atmosphere. Understanding the relative roles of escape, outgassing, and accretion in a variety of exoplanet atmospheres is critical to understanding the origin and evolution of planetary atmospheres. UV observations of transiting exoplanets are required to both identify the escaping species and to constrain the

### Section 4.10.8 Program at a Glance

| |
|---|
| **Science Goal:** Constrain the origin and evolution of planet atmospheres. |
| **Program Details:** UV-to-near-IR transit spectroscopy of escaping exoplanetary atmospheres. |
| **Instrument(s) + Configuration(s):** HWC and UVS spectroscopy. |
| **Key Observational Requirements:** Sensitive UV-to-near-IR spectroscopic capabilities. |

physics of escape.

The first observations of atmospheric escape were obtained by Vidal-Madjar et al. (2003), who obtained HST/STIS UV transmission spectra of the close-in giant planet HD 209458b revealing that the planet possesses a highly extended hydrogen atmosphere due to heating by stellar X-ray and extreme-UV photons. Subsequent HST observations also detected various metals in the exospheres of giant planets including carbon, oxygen, and magnesium (Vidal-Madjar et al. 2004; Linsky et al. 2010), as well as escaping hydrogen from the warm Neptune-sized planet GJ 436b, which revealed a large tail of escaped planetary material (**Figure 4.10-8**; Ehrenreich et al. 2015).

Transit survey missions like NASA's TESS and ESA's upcoming PLATO will uncover thousands of transiting planets orbiting the brightest stars including hundreds of hot Jupiters and hot Neptunes around sunlike stars and dozens of hot transiting planets that are ideal for characterization with UV transit spectroscopy using the Lyα line at 121.5 nm. However, since stellar UV emission from active stars can be highly variable from one transit to another, it is important to obtain high SNR detections within an individual transit (Bourrier et al. 2017), which in turn drives the need for a larger effective collecting area compared to what is currently possible with HST STIS or COS.

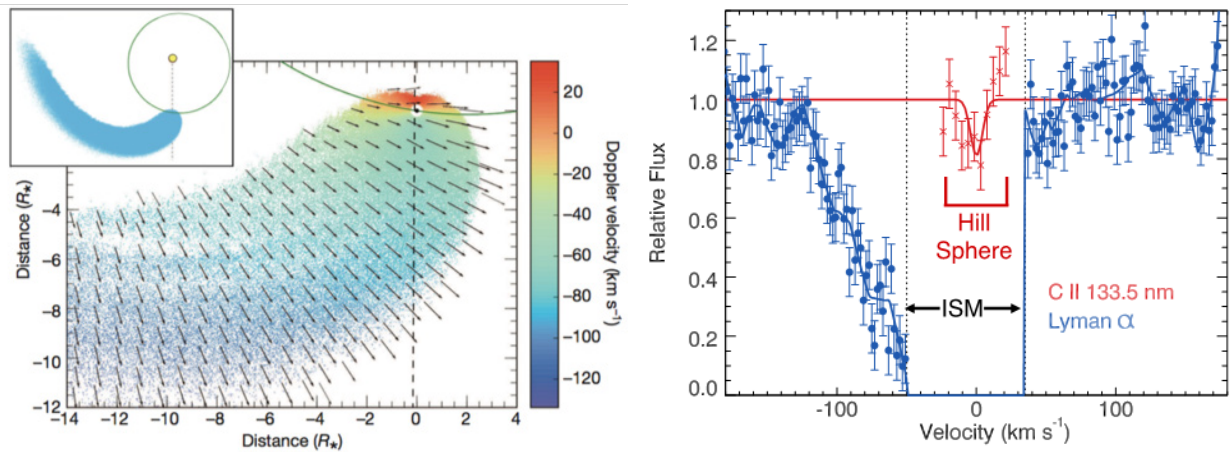

**Figure 4.10-8.** *Left panel:* Reconstruction of escaping hydrogen observed with HST in Lyα for the hot-Neptune GJ 436b (Ehrenreich et al. 2015). *Right panel:* Simulation of what the baseline design of HabEx using the UVS (*Section 6.5*) can achieve for this planet in just single transit. In addition to measuring atmospheric escape in Lyα for dozens of planets, HabEx will also measure UV metal lines like CII at 133 nm. Observing these lines at high-SNR provides key information about atmospheric composition and the structure of the escaping upper atmosphere, particularly for the low velocity material close to the planet where Lyα is completely hidden by the ISM. Figure Credit: E. Lopez.

In addition to Lyα, the UV region contains an array of strong lines for key metals like O I at 130.4 nm, C II at 133.5 nm, Si IV at 140.0 nm. Observing these lines has two key benefits; first, it opens up an important window into measuring the abundance of key metal species like carbon and oxygen in a part of the atmosphere that cannot be hidden by clouds; second, these metal lines solve a fundamental limitation of the Lyα line, which is that even for the nearest stars the core of the stellar Lyα line is completely made extinct by the ISM. As a result, any material that is moving at speeds below ~50 km/s relative to the exoplanet (i.e., the entire bound portion of the upper atmosphere) is completely hidden by the ISM. The UV metal lines listed above will solve this problem since these lines face much lower ISM extinction and revolutionary insights into the physics of atmospheric escape can be achieved by mapping these lines to obtain the detailed velocity, density, and temperature structures of the exoplanet upper atmospheres.

Atmospheric escape is also a key factor shaping the evolution and distribution of low-mass close-in planets (e.g., Owen and Wu 2013) and their habitability (e.g., Cockell et al. 2016). Indeed, many highly irradiated rocky planets (e.g., CoRoT-7b, Kepler-10b) might be the remnant cores of evaporated Neptune-mass planets (e.g., Lopez et al. 2012). As a consequence, atmospheric escape also has a major impact on our understanding of planet formation (e.g., Van Eylen et al. 2018).

### The Role of HabEx

HabEx will be an outstanding platform for studies of transiting planets of all types and will provide exceptional new science opportunities from hot Jupiters down to temperate, terrestrial mass planets. Transiting planet science is highly complementary to studies of directly imaged exoplanets since the transit technique primarily uncovers planets on orbits close to their host stars that are difficult to image directly, but comparatively likely to transit at high-SNR. Moreover, both the masses and radii of transiting exoplanets can be measured, enabling investigations of how atmospheric properties link to bulk composition and formation. In UV-to-visible wavelengths, HabEx will reach ~4× the SNR per transit compared to HST/STIS, while in the near-IR it will reach the same SNR per transit compared to JWST/NIRSpec.

JWST will thoroughly characterize the atmospheres of transiting Jupiter and Neptune-mass planets in the IR, resulting in a revolution in our understanding of hot planets orbiting close to their parent star. However, JWST lacks the coverage in the UV-to-visible that HST and HabEx possess. These wavelengths offer unique insights into planetary atmospheres that enable understanding of cloud properties and the physics of atmospheric photochemistry. In this respect, the HabEx baseline design (*Chapters 6, 7, and 8*) is a true successor to HST, with a larger aperture, higher throughput, and greater stability in its orbit at L2.

HabEx's high sensitivity at visible wavelengths will also enable detections of atomic species such as the alkali metals for a wide range of both terrestrial and non-terrestrial planets, many of which will otherwise be inaccessible due to the low throughput of HST/STIS and the wavelength limits of JWST. Furthermore, HabEx will be able to measure Rayleigh scattering slopes for planets of all sizes with high precision.

### Science Program

The scientific program would entail UV-to-near-IR spectroscopic observations of transiting exoplanets using the HWC (*Section 6.6*) and UVS (*Section 6.5*) instruments. **Figure 4.10-9** illustrates the SNR that HabEx would achieve with just a single transit for ten benchmark hot Jupiters

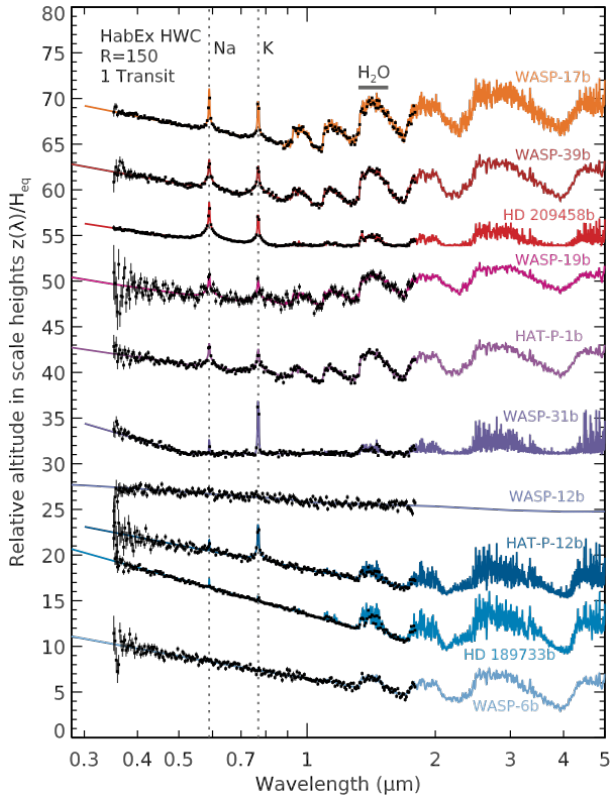

**Figure 4.10-9.** HabEx will achieve excellent SNR for hot transiting exoplanets in just a single transit, enabling measurements of molecular abundances and a more thorough understanding of clouds. This figure shows simulated transit spectra from the UV-to-near-IR for 10 benchmark hot Jupiters with model spectra from Sing et al. (2016) and simulated data for a single transit per channel with the HabEx HWC (*Section 6.6*).

previously observed with HST and Spitzer over many transits for each system.

The HabEx UVS would undertake thorough characterization of the escaping upper atmospheres of hot transiting gaseous exoplanets, providing key insights into the physics of atmospheric escape.

### 4.10.9 Parallel Observations

HabEx Goals 1 and 2 (*Chapter 3*) are focused on exoplanet science that is undertaken using starlight suppression instruments for direct imaging and spectroscopy. Since many of the observations for Goal's 1 and 2 would involve long, multiday exposures of nearby stars, this provides opportunities for ultra-deep parallel observations with other instruments on HabEx (*Sections 6.5* and *6.6*), which would increase the observatory's scientific efficiency. Notably, since the HabEx exoplanet direct imaging surveys would observe nearby stars, these targets are distributed roughly isotropically across the sky, and thus are amenable to both Galactic and extragalactic science goals.

Parallel observations with the exoplanet direct imaging program would enable multiple ultra-deep imaging fields, similar to the extremely successful Hubble Deep Fields and Hubble Ultra-Deep Survey. In addition, ultra-deep spectroscopic surveys done in parallel with the exoplanet direct imaging observations would enable a range of science, from ultra-deep probes of the IGM in the UV, to ultra-deep and/or highly complete spectroscopic surveys in the visible to near-IR.

The concept of operations for parallel observations is described in further detail in *Section 8.1.6*. In this scenario, both observatory science instruments (i.e., the UVS and HWC) can be operational at all times, i.e., in parallel with each other and with the high contrast exoplanet direct imaging instruments. The fields of view for each of the UV and visible–near-IR instruments are sufficiently offset from the direct imaging instrument fields of view to ensure that scattered light is not an issue. The inclusion of a fine steering mirror in the visible–near-IR instrument design also enables dithering during the deep exoplanet stares (*Section 6.6*).

### 4.10.10 Multi-Messenger Astrophysics / Targets of Opportunity

HabEx is a capable facility for Target of Opportunity (ToO) observations, able to rapidly respond to triggers for re-pointing, such as from multi-messenger astrophysics. As discussed in *Section 8.1.6*, the HabEx exoplanet direct imaging stability requirements motivate a pointing system based on thrusters, rather than reaction wheels. The telescope flight system's thrusters are able to slew and settle the telescope extremely rapidly. The nominal slew rate is approximately 0.15° per second, assuming a 90° slew at a 5% thruster duty cycle. However, the telescope flight system retains the capability of slewing 180° in less than 5 minutes should the need arise at 100% duty cycle. However, rapid slews will use significantly more propellant than standard slews, using up this limited resource. A future science team will need to develop criteria, priorities, and protocols for deciding when observations of different types will be interrupted for ToO observations. In addition, they will need to develop requirements on the uplink latencies for communicating ToO interrupts, and the speed with which ToO data will need to be returned to the ground.

The HabEx instrumentation includes several unique capabilities for ToO science. In particular, UV spectroscopic capabilities provide access to unique science not available from the ground or other facilities expected in the HabEx timeframe. Specifically, the current UV capabilities of HST and the Neil Gehrels Swift observatory (Swift) are not expected to be operational in the 2030s, and no new, large-aperture, space-based UV facilities are currently planned. As one example of unique HabEx ToO UV multi-messenger science for the 2030s, multiple theories predict bright early-time emission from binary neutron star mergers at UV wavelengths (Arcavi 2018). This was confirmed for the gravitational wave event GW170817, detected by Swift at UV wavelengths beginning 0.6 days after the gravitational wave trigger (Evans et al. 2017). Detailed study of this UV emission can explain its origin, for example, distinguishing blue kilonova models (Metzger et al. 2010) from cocoon shock models (Piro and Kollmeier 2018). The UV emission also provides key information for understanding the composition, abundance, and energetics of the ejecta, as well as clues as to how fast the neutron stars collapse into a black hole, which, in turn, can be used to constrain the neutron star equation of state (Piro et al. 2017).

# 5 SCIENCE TRACEABILITY MATRIX, ERROR BUDGETS, AND REQUIREMENTS

Mapping the science goals of a concept to key and driving requirements is necessary to connect the promised level of science return to a specific concept design and its associated cost and risk. This chapter breaks the mapping into three areas:

- A science traceability matrix (STM; *Section 5.1*), which relates science goals and objectives to high-level design requirements.
- Key error budgets (*Section 5.2*), which take key requirements from the STM and map them into lower-level implementation-specific requirements.
- Implementation requirements defined in a mission traceability matrix (MTM; *Section 5.3*).

The chapter concludes by summarizing the HabEx key and driving requirements and illustrating how they connect back to the STM and error budgets.

While a flight project will eventually identify and address hundreds of requirements, the focus of this study has been to identify the major requirements that shape the HabEx baseline design.

## 5.1 The Science Traceability Matrix

The science traceability matrix (STM) is an established framework to capture science goals and objectives and to flow these to high-level functional requirements. HabEx has developed an STM (**Table 5.1-2**) that encompasses the exoplanet direct imaging and observatory science goals detailed in *Chapters 3* and *4*. It is important to note that the STM as developed for HabEx is not specific to a single implementation; at most the STM payload requirements begin to shape the number and type of instruments needed to achieve the identified objectives. The intent of the STM is not to connect a pre-determined design to the HabEx science goals but rather, to connect the science to meaningful observational and payload requirements without prejudice.

In each row of the STM, the HabEx science goals and objectives connect to physical parameters and observables, and then to the necessary payload functional requirements. This ensures that as long as the HabEx architecture meets these requirements, the mission will be capable of delivering the data required to address the science goals and objectives.

The HabEx STM also attempts to capture baseline and threshold objectives, differentiating the level of science advancement for each within the science objectives. This bifurcation is continued into the requirements and provides understanding of the sensitivity of the HabEx science goals to major architecture decisions, and is discussed in more detail in *Chapter 10*.

### 5.1.1 HabEx Science Goals and Objectives

#### 5.1.1.1 Science Goals

The HabEx science goals are framed around the desire to identify and investigate nearby Earth-like exoplanets around sunlike stars, to undertake detailed investigations of our nearest neighbor planetary systems, and to enable a highly-capable community-led Guest Observer (GO) program that takes advantage of an ultra-stable, large-aperture, ultraviolet (UV) through near-infrared (near-IR) telescope in space. The HabEx science goals are identified in **Table 5.1-1**.

**Table 5.1-1.** HabEx science goals form the basis of the Science Traceability Matrix, Table 5.1-2.

| HabEx Science Goals |
|---|
| 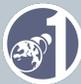 *To seek out nearby worlds and explore their habitability* |
| 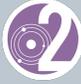 *To map out nearby planetary systems and understand the diversity of the worlds they contain* |
| 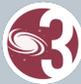 *To enable new explorations of astrophysical systems from the solar system to galaxies and the universe by extending our reach in the UV through near-IR* |

**Table 5.1-2.** Science Traceability Matrix. Baseline science objectives and requirements appear in black typeface, while threshold objectives and requirements appear in grey, italic typeface. Driving requirements appear in the payload functional requirements in blue, bold typeface.

| Goal | Science Objectives | Scientific Measurement Requirements | | Payload Functional Requirements | Baseline Projected Performance |
|---|---|---|---|---|---|
| | | Physical Parameters | Observables | | |
| To seek out nearby worlds and explore their habitability. | **O1:** To determine if rocky planets (0.5–1.75 $R_\oplus$) continuously orbiting within the habitable zone (HZ) exist around nearby sunlike stars, surveying enough stars to detect and measure the orbits of at least 20 exo-Earth candidates (EECs) if each observed star hosted one EEC. To detect and measure the orbit of at least one EEC with ≥95% confidence. *Threshold: To determine if rocky planets (0.5–1.75 $R_\oplus$) continuously orbiting within the habitable zone (HZ) exist around nearby sunlike stars, surveying enough stars to detect and measure the orbits of at least 12 exo-Earth candidates (EECs) if each observed star hosted one EEC. To detect and measure the orbit of at least one EEC with ≥90% confidence.* | Planet position with respect to the central star over time to determine the orbit semi-major axis, eccentricity, and inclination to ≤10% accuracy. Planet radius determined within a factor of 2 of the true value. | Star-to-planet separation measured at ≥4 different orbital positions. Broadband planetary flux centered at 0.5 μm measured at ≥4 different orbital positions. | **F1.1 Broadband high contrast visible imaging with an IWA$_{0.5}$ ≤ 80 mas at 0.5 μm.** **F1.2 Angular positional accuracy ≤5 milliarcseconds (mas) root mean square (RMS).** **F1.3 Ability to visit target star ≥4 times.** **F1.4 Signal-to-noise ratio SNR ≥ 7 on a point source that is ≥ $10^{10}$ times fainter than a solar twin star located at 9 pc (V = 4.6 mag) and at ≤80 mas from it using broadband photometry centered at 0.5 μm in an exposure time of ≤20 hours (h).** See **Figure 5.2-1** for the baseline coronagraph instrument error budget. *Threshold: IWA$_{0.5}$ ≤105 mas at 0.5 μm.* | Coronagraph broadband high contrast visible imaging with an IWA$_{0.5}$ = 62 mas at 0.5 μm. Angular positional accuracy: 0.7 mas at 100 Hz Ability to visit target star: 6 times, as necessary SNR = 7 on a point source that is $10^{10}$ times fainter than a solar twin star located at 9 pc (V = 4.6 mag) and at 80 mas from it using broadband photometry centered at 0.5 μm in an exposure time of 9.6 h. |
| | **O2:** To determine if planets identified in Objective 1 have potentially habitable conditions (an atmosphere containing water vapor). Also, to determine if rocky planets outside the "2D EEC zone" have potentially habitable conditions, surveying an equivalent number of rocky planets outside the 2D EEC zone to those within it. *Threshold: To determine if the planets identified in the threshold requirement of Objective 1 have potentially habitable conditions (an atmosphere containing water vapor). Also, to determine if rocky planets that do not fit into Objective 1 have potentially habitable conditions, surveying ≥1 rocky planet interior and ≥1 rocky planet exterior to the HZ.* | The abundance of atmospheric $H_2O$ if the column density is ≥0.4 g/cm² (modern Earth at the outer edge of the HZ). *Threshold: The abundance of atmospheric $H_2O$ if the column density is ≥2.9 g/cm² (Modern Earth).* | Planetary spectrum, including ≥2 $H_2O$ absorption features in the visible–near-IR. *Threshold: Planetary spectrum, including ≥1 $H_2O$ absorption feature in the visible.* | **F2.1** Visible–near-IR spectroscopy with an **IWA$_{0.5}$ ≤ 80 mas at 1.0 μm.** **F2.2** Spectral range ≤0.7 μm to ≥1.0 μm. **F2.3** Spectral resolution ($R$): $R$ ≥ 35 at 0.82 μm with SNR ≥ 10 and $R$ ≥ 17 at 0.94 μm with SNR ≥ 10. Or $R$ ≥ 17 at 0.94 μm with SNR ≥ 10 and $R$ ≥ 19 at 1.13 μm with SNR ≥ 10. *Threshold: Visible spectroscopy with an IWA$_{0.5}$ ≤ 105 mas at 0.75 μm. Spectral range ≤ 0.7 μm to ≥ 1.0 μm. $R$ ≥ 35 at 0.72 μm with SNR ≥ 5.* | Starshade UV–near-IR spectroscopy with an IWA$_{0.5}$ = 58 mas at 1.0 μm. Spectral range: 0.2–1.8 μm. $R$ = 7 with SNR = 10 from 0.2–0.45 μm. $R$ = 140 with SNR = 10 from 0.45–0.975 μm. $R$ = 40 with SNR = 10 from 0.975–1.8 μm. |
| | **O3:** To determine if planets identified in Objective 1, regardless of whether they meet the conditions in Objective 2, contain biosignature gases (signs of life) and, for a subset of them, to identify gases associated with, or incompatible with, known false positive mechanisms **(Figure 3.4-4)** **O3a:** To determine if planets identified in both Objective 1 and Objective 2 contain biosignature gases (signs of life) and to identify gases associated with, or incompatible with, known false positive mechanisms. **O3b:** To determine if planets identified in Objective 1, but not Objective 2, contain biosignature gases (signs of life) and to identify gases associated with, or incompatible with, known false positive mechanisms. | The abundance of atmospheric molecular species if the column density is: • $O_3$ ≥ 8 × 10⁵ g/cm² (low end of Proterozoic Earth levels). • $O_2$ ≥ 2 g/cm² (low end of Proterozoic Earth levels). • $CH_4$ ≥ 10⁻¹ g/cm² (low end of Archean Earth levels). *Threshold: The abundance of atmospheric molecular species if the column density is: • $O_3$ ≥ 7.2 × 10⁻⁴ g/cm² (modern Earth level). • $O_2$ ≥ 2.4 × 10² g/cm² (modern Earth level). • $CH_4$ ≥ 100 g/cm² (high end of Archean Earth levels).* | Planetary spectrum from the UV–near-IR, including the $O_3$ cutoff, $O_2$ absorption features, and $CH_4$ absorption features. *Threshold: Planetary spectrum in the UV–visible (or visible-only), including the $O_3$ cutoff (or the $O_3$ broad absorption feature in the visible), $O_2$ absorption features (in the visible), and $CH_4$ absorption features.* | **F3.1 UV–near-IR spectroscopy** IWA$_{0.5}$ ≤ 80 mas at 0.8 μm. **F3.2 Spectral range ≤0.3 μm to ≥1.7 μm.** **F3.3 SNR ≥ 10 per $R$ ≥ 70 spectral bin on a point source ≥10$^{10}$ times fainter than a solar twin star located at 9 pc (V = 4.6 mag) and at ≤80 mas from it using visible spectroscopy anywhere between 0.45–0.975 μm in an exposure time of ≤ 43 days.** **F3.4** • $O_3$: $R$ ≥ 5 from 0.3–0.35 μm with SNR ≥ 10 per spectral bin. • $O_2$: $R$ ≥ 70 from 0.75–0.78 μm with SNR ≥ 10 per spectral bin. • $CH_4$: $R$ ≥ 10 at 1.69 μm with SNR ≥ 10 per spectral bin. See feature detection in **Figure 3.3-7** and the starshade error budget in **Figure 5.2-2.** *Threshold: UV–visible (or visible-only) spectroscopy with an IWA$_{0.5}$ ≤ 105 mas at 0.8 μm. Spectral range ≤0.3 μm to ≥1.0 μm or ≤0.45 μm to ≥1.0 μm.* | Starshade UV–near-IR spectroscopy with an IWA$_{0.5}$ = 58 mas at 0.8 μm. Spectral range: 0.2–1.8 μm. SNR = 10 per $R$ = 70 spectral bin on a point source 10$^{10}$ times fainter than a solar twin star located at 9 pc (V = 4.6 mag) and at 80 mas from it using visible spectroscopy anywhere between 0.45–0.975 μm in an exposure time of 22.5 days or less. $R$ = 7 with SNR = 10 from 0.2–0.45 μm. $R$ = 140 with SNR = 10 from 0.45–0.975 μm. $R$ = 40 with SNR = 10 from 0.975–1.8 μm. |

**Table 5.1-2.** Science Traceability Matrix. Baseline science objectives and requirements appear in black typeface, while threshold objectives and requirements appear in grey, italic typeface. Driving requirements appear in the payload functional requirements in blue, bold typeface.

| Goal | Science Objectives | Scientific Measurement Requirements | | Payload Functional Requirements | Baseline Projected Performance |
|---|---|---|---|---|---|
| | | Physical Parameters | Observables | | |
| | | The abundance of atmospheric molecular species if the column density is:<br>• $CO_2 \geq 5 \times 10^3$ g/cm$^2$ (5× the high end of Archean Earth levels).<br>• $CH_4 \geq 10^{-1}$ g/cm$^2$ (low end of Archean Earth levels).<br>• $O_4 \geq 0.4$ bar (2× increase in $O_2$ from modern-day Earth).<br><br>*Threshold: The abundance of atmospheric molecular species if the column density is:*<br>*• $CO_2 \geq 10^4$ g/cm$^2$ (10× the high end of Archean Earth levels).*<br>*• No $CH_4$ detection threshold.*<br>*• $O_4 \geq 0.8$ bar (4× increase in $O_2$ from modern-day Earth).* | Planetary spectrum absorption features in the visible–near-IR, including $CO_2$, $CH_4$, and $O_4$.<br><br>*Threshold:*<br>*Planetary spectrum absorption features in the visible–near-IR, including $CO_2$ and $O_4$.* | • $O_3$: $R \geq 5$ from 0.3–0.35 μm with SNR ≥ 10 per spectral bin or $R \geq 5$ from 0.53–0.66 μm with SNR ≥ 10 per spectral bin.<br>• $O_2$: $R \geq 70$ from 0.75–0.78 μm with SNR ≥ 10 per spectral bin.<br>• $CH_4$: $R \geq 32$ from 0.88–0.91 μm with SNR ≥ 10 per spectral bin.<br>**F3.5** Spectral range ≤0.6 μm to ≥1.7 μm.<br>**F3.6**<br>$CO_2$: $R \geq 11$ at 1.59 μm with SNR ≥ 10 per spectral bin.<br>$CH_4$: $R \geq 10$ at 1.69 μm with SNR ≥ 10 per spectral bin.<br>$O_4$: $R \geq 40$ at 0.63 μm with SNR ≥ 10 per spectral bin and $R \geq 22$ at 1.06 μm and 1.27 μm with SNR ≥ 5 per spectral bin.<br><br>*Threshold:*<br>*Spectral range ≤0.5 μm to ≥1.6 μm.*<br><br>*• $CO_2$: $R \geq 11$ at 1.59 μm with SNR ≥ 10 per spectral bin.*<br>*• $O_4$: $R \geq 50$ at 0.57μm with SNR ≥ 5 per spectral bin and $R \geq 40$ at 0.63 μm with SNR ≥ 5 per spectral bin.* | |
| | **O4:** To determine if a subset of planets identified in Objective 2, with orbital inclinations >50 deg, contain water oceans. | Glint from surface oceans. | Planetary broadband photometry measured at ≥ 2 illumination phases, with ≥1 measurement at illumination phase ≤90 deg and ≥1 at ≥140 deg. | **F4.1** Broadband visible imaging with an **IWA$_{0.5}$ ≤ 64 mas at 0.87 μm.**<br>**F4.2** Photometric range extends to ≥0.9 μm.<br>**F4.3**<br>SNR ≥ 7 on a point source (exo-Earth twin at 140 deg illumination phase in a system inclined 50 deg from pole-on) that is ≥1.4 × 10$^{10}$ times fainter than a solar twin star located at 10 pc (V = 4.8 mag) and at ≤64 mas from it using visible broadband photometry centered at 0.87 μm in an exposure time of ≤100 hours.<br><br>*Threshold: IWA$_{0.5}$ ≤ 129 mas at 0.87 μm.*<br><br>*SNR ≥ 7 on a point source (exo-Earth twin at 140 deg illumination phase in a system inclined 50 deg from pole-on) that is ≥1.4 × 10$^{10}$ times fainter than a solar twin star located at 5 pc (V = 3.3 mag) and at ≤129 mas from it using visible broadband photometry centered at 0.87 μm in an exposure time of ≤100 hours.* | Starshade IWA$_{0.5}$ = 58 mas at 0.87 μm.<br><br>Photometric range: 0.2–1.8 μm.<br><br>SNR = 7 on a point source (exo-Earth twin at 140 deg illumination phase in a system inclined 50 deg from pole-on) that is 1.4 × 10$^{10}$ times fainter than a solar twin star located at 10 pc (V = 4.8 mag) and at 64 mas from it using visible broadband photometry centered at 0.87 μm in an exposure time of 62 hours. |
| To map out nearby planetary systems and understand the diversity of the worlds they contain. | **O5a:** Determine the detailed architecture of individual planetary systems from rocky to giant planets in the inner HZ to giant planets in Neptune-like orbits for ≥5 sunlike stars.<br><br>*Threshold: Determine the detailed architecture of individual planetary systems from rocky to giant planets on inner HZ to Jupiter-like orbits for ≥1 sunlike star.* | Planetary positions with respect to the central star.<br><br>Radii of the planets, within a factor of 2 of the mean estimate, for orbital periods ≤15 years. | Angular star-to-planet separation measured at ≥4 different orbital positions.<br><br>Broadband planetary flux centered at 0.5 μm measured at ≥4 different orbital positions. | **F5.1** Broadband visible imaging with an IWA$_{0.5}$ ≤ 80 mas at 0.5 μm and an **OWA ≥ 6 arcsec** (Neptune with a 30 AU orbit at 5 pc).<br>**F5.2** Angular positional accuracy ≤5 mas RMS.<br>**F5.3** SNR ≥ 7 on a point source that is ≥2.5 × 10$^{10}$ times fainter than a solar twin star located at ≥5 pc (V = 3.3 mag) and at ≤80 mas from it using visible broadband photometry centered at 0.5 μm in an exposure time of ≤ 60 h.<br><br>*Threshold: OWA ≥ 0.5 arcsec (Jupiter orbit at 10 pc).*<br>*Exposure time of ≤100 h.* | Starshade broadband visible imaging with an IWA$_{0.5}$ = 58 mas at 0.5 μm and an OWA = 6 arcsec.<br><br>Angular positional accuracy: 0.7 mas at 100 Hz<br><br>SNR = 7 on a point source that is 2.5 × 10$^{10}$ times fainter than a solar twin star located at 5 pc (V = 3.3 mag) and at 80 mas from it using visible broadband photometry centered at 0.5 μm in an exposure time of 36 h. |
| | **O5b:** To determine or refine the architectures of planetary systems over orbital distances that include the inner HZ to Saturn-like orbits, observing enough sunlike stars to detect over 30 planets of each type (≥30 rocky, ≥30 sub-Neptunes, and ≥30 giants), assuming nominal occurrence rates for each. | Planetary positions with respect to the central star.<br><br>Radii of the planets, within a factor of 2 of the mean estimate, for orbital periods ≤15 years. | Angular star-to-planet separation measured at ≥4 different orbital positions.<br><br>Broadband planetary flux centered at 0.5 μm measured at ≥4 different orbital positions. | **F5.4** Broadband visible imaging with an IWA$_{0.5}$ ≤ 80 mas at 0.5 μm and an OWA ≥ 2 arcsec (Saturn orbit at 5 pc).<br>**F5.5** Angular positional accuracy ≤5 mas RMS.<br><br>SNR ≥ 7 on a point source that is ≥2.5 × 10$^{10}$ times fainter than a solar twin star located at ≥ 5 pc (V = 3.3 mag) and at ≤80 mas from it using visible broadband photometry centered at 0.5 μm in an exposure time of ≤60 h. | Starshade broadband visible imaging with an IWA$_{0.5}$ = 58 mas at 0.5 μm and an OWA = 6 arcsec.<br><br>Angular positional accuracy: 0.7 mas at 100 Hz |

**Table 5.1-2.** Science Traceability Matrix. Baseline science objectives and requirements appear in black typeface, while threshold objectives and requirements appear in grey, italic typeface. Driving requirements appear in the payload functional requirements in blue, bold typeface.

| Goal | Science Objectives | Scientific Measurement Requirements | | Payload Functional Requirements | Baseline Projected Performance |
|---|---|---|---|---|---|
| | | Physical Parameters | Observables | | |
| | *Threshold: To determine or refine the architectures of planetary systems over orbital distances that include the inner HZ to Jupiter-like orbits, observing enough sunlike stars to detect over 15 planets of each type (≥15 rocky, ≥15 sub-Neptunes, and ≥15 giants), assuming nominal occurrence rates for each.* | | | *Threshold: OWA ≥ 0.5 arcsec (Jupiter orbit at 10 pc). Exposure time of ≤ 100 h.* | SNR = 7 on a point source that is 2.5 × 10^10 times fainter than a solar twin star located at 5 pc (V = 3.3 mag) and at 80 mas from it using visible broadband photometry centered at 0.5 μm in an exposure time of 36 h. |
| | **O6:** For planetary systems around sunlike stars, determine how planetary atmospheric compositions vary as a function of planet radius and star-planet separation, detecting and spectrally characterizing ≥30 sub-Neptune-sized (≤3.5 R_⊕) or smaller planets and ≥30 larger planets (>3.5 R_⊕). *Threshold: For planetary systems around sunlike stars, determine how planetary atmospheric compositions vary as a function of planet radius and star-planet separation, detecting and spectrally characterizing ≥15 sub-Neptune-sized (≤3.5 R_⊕) or smaller planets and ≥15 larger planets (>3.5 R_⊕).* | The abundance of atmospheric molecular species if the column density is: • O_3 ≥ 8 × 10^-5 g/cm^2. • O_2 ≥ 2 g/cm^2. • CO_2 ≥ 5 × 10^3 g/cm^2. • CH_4 ≥ 3 × 10^2 g/cm^2. • H_2O ≥ 0.4 g/cm^2. • H_2 partial pressure ≥1 bar in atmospheres with a mean molecular weight equivalent to N_2-dominated or heavier. *Threshold: The abundance of atmospheric molecular species if the column density is:* • *O_3 ≥ 7.2 × 10^-4 g/cm^2.* • *O_2 ≥ 2.4 × 10^2 g/cm^2.* • *CO_2 ≥ 10^4 g/cm^2.* • *CH_4 ≥ 1 g/cm^2.* • *H_2O ≥ 3 g/cm^2.* | Planetary spectra molecular absorption features in the UV–near-IR. *Threshold: Planetary spectra molecular absorption features in the visible.* | **F6.1** UV–near-IR multi-object spectroscopy with an IWA_0.5 ≤ 130 mas at 1.7 μm. **F6.2** Spectral range ≤0.3 μm to ≥1.7 μm. **F6.3** SNR ≥ 10 per spectral bin for: • O_3: R ≥ 5 at from 0.3–0.35 μm. • O_2: R ≥ 70 from 0.75–0.78 μm. • CO_2: R ≥ 11 at 1.6 μm. • CH_4: R ≥ 10 at 1.7 μm. • H_2O: R ≥ 10 at 1.4 μm. • H_2: R ≥ 8 at 0.8 μm and R ≥ 12 at 1.15 μm. *Threshold: Visible spectroscopy with an IWA_0.5: 130 mas at 1.0 μm. Spectral range ≥ 0.8 μm to ≥ 1.0 μm. SNR ≥ 10 per spectral bin for:* • *O_3: R ≥ 5 from 0.3–0.35 μm or R ≥ 5 from 0.53–0.66 μm.* • *O_2: R ≥ 70 from 0.75–0.78 μm.* • *CO_2: R ≥ 100 at 0.87 μm.* • *CH_4: R ≥ 32 at 0.89 μm.* • *H_2O: R ≥ 17 at 0.94 μm.* • *H_2: R ≥ 8 at 0.8 μm.* | Starshade UV–near-IR multi-object spectroscopy using an IFS with an IWA_0.5 = 104 mas at 1.7 μm. Spectral range: 0.2–1.8 μm. R = 7 with SNR = 10 from 0.2–0.45 μm. R = 140 with SNR = 10 from 0.45–0.975 μm. R = 40 with SNR = 10 from 0.975–1.8 μm. |
| | | Planetary positions with respect to the central star. | Angular star-to-planet separation measured at ≥4 different orbital positions. | **F6.4** Broadband visible imaging with an IWA_0.5 ≤ 80 mas at 0.5 μm and an **OWA ≥ 6 arcsec** (Neptune with a 30 AU orbit at 5 pc). **F6.5** Angular positional accuracy ≤ 5 mas RMS. | Starshade broadband visible imaging with an IWA_0.5 = 58 mas at 0.5 μm and an OWA = 6 arcsec. Angular positional accuracy: 0.7 mas at 100 Hz |
| | | Radii of the planets, within a factor of 2 of the mean estimate, for orbital periods ≤15 years. | Broadband planetary flux centered at 0.5 μm measured at ≥4 different orbital positions. | **F6.6** SNR ≥ 7 on a point source that is ≥2.5 × 10^10 times fainter than a solar twin star located at ≥5 pc (V = 3.3 mag) and at ≤80 mas from it using visible broadband photometry centered at 0.5 μm in an exposure time of ≤60h. *Threshold: OWA ≥ 0.5 arcsec (Jupiter orbit at 10 pc).* | SNR = 7 on a point source that is 2.5 × 10^10 times fainter than a solar twin star located at 5 pc (V = 3.3 mag) and at 80 mas from it using visible broadband photometry centered at 0.5 μm in an exposure time of 18 h. |
| | **O7:** To determine if the presence and orbital characteristics of cold giant planets (≥3.5 R_⊕) are related to the presence (or absence) of water vapor in the atmospheres of rocky planets detected in Objective 1. | Same as Objective 2 and Objective 5. | Same as Objective 2 and Objective 5. | **F7.1** Broadband visible imaging and visible–near-IR spectroscopy with an OWA ≥ 3 arcsec, which enables detection of Jupiters in ≥ 10 AU orbits around the closest (≤ 3.6 pc) single sunlike stars. **F7.2** Spectral range ≤0.7 μm to ≥1.5 μm. *Threshold: Broadband visible imaging and visible–near-IR spectroscopy with an OWA ≥ 0.5 arcsec, which enables detection of giant planets in Jupiter-like orbits at 10 pc.* | Starshade broadband visible imaging and visible–near-IR spectroscopy with an OWA = 6 arcsec. Spectral range: 0.2–1.8 μm. |
| | **O8:** To constrain the range of possible dust-belt architectures in planetary systems around sunlike stars, to explore the physical properties | Disk morphology. | Disk broadband images in ≥2 bands in the visible. | **F8.1** Broadband visible imaging with an IWA_0.5 ≤ 100 mas, which enables detection of HZ dust at ≥ 10 pc. **F8.2** Spectral coverage in ≥2 bands in the range ≥0.45 μm to ≥1.0 μm. | Starshade broadband visible imaging with an IWA_0.5 = 58 mas from 0.3–1.0 μm. OWA = 6 arcsec. |

**Table 5.1-2.** Science Traceability Matrix. Baseline science objectives and requirements appear in black typeface, while threshold objectives and requirements appear in grey, italic typeface. Driving requirements appear in the payload functional requirements in blue, bold typeface.

| Goal | Science Objectives | Scientific Measurement Requirements | | Payload Functional Requirements | Baseline Projected Performance |
|---|---|---|---|---|---|
| | | Physical Parameters | Observables | | |
| | of the dust grains (size, composition) in the debris disks, and to determine if debris disk spatial structures can be used as signposts of existing planets (identified in Objective 5), surveying ≥30 exoplanetary systems. | | | **F8.3** OWA ≥ 6 arcsec, which enables detection of planets in Neptune-like orbits around stars at ≤ 5 pc.<br><br>**F8.4** Surface brightness detection limit V ≥ 22 mag/arcsec² at the IWA$_{0.5}$ (equivalent to the solar zodiacal surface brightness at 1 AU).<br><br>**F8.5** Surface brightness detection limit V ≥ 26 mag/arcsec² at 3 arcsec (equivalent to 10× the Kuiper-belt surface brightness at 30 AU).<br><br>*Threshold:*<br>*OWA ≥ 1 arcsec at 0.8 μm.*<br>*IWA$_{0.5}$ ≤ 125 mas at 0.5 μm, which enables detection of HZ dust at ≥8 pc.*<br>*Surface brightness detection limit R ≥ 23.5 mag/arcsec² at 1 arcsec.* | Spectral coverage: 0.2–1.8 μm.<br><br>Surface brightness detection limit V = 22 mag/arcsec² at the IWA$_{0.5}$.<br><br>Surface brightness detection limit V = 26 mag/arcsec² at 3 arcsec. |
| | | Dust grain size and composition. | Disk broadband polarized images in ≥2 bands in the visible. | **F8.6** Single linear polarization measurements, with ≥3 different polarization states. | Polarizers included in starshade and coronagraph instruments. |
| **To enable new explorations of astrophysical systems from the solar system to galaxies and the universe by extending our reach in the UV through near-IR.** | **O9:** To probe the lifecycle of baryons by determining the processes governing the circulation of baryons between the gaseous phase of the intergalactic medium (IGM), circumgalactic medium (CGM), and galaxies. | Gas temperature between ≤10⁴ to ≤10⁶ K.<br><br>Gas density ≤ 0.1 cm⁻³.<br><br>Gas kinematics at ≤5 km/s. | Equivalent widths of ion absorption lines to provide gas density and temperature measurements through radiative transfer modeling. The requirement is the same species at a range of ionization states (e.g., CII, CIII, CIV; OI, OII, OIII, OIV, OV, OVI; SiII, SiIII, SiIV), measured with ≤0.005 nm uncertainties.<br><br>Central wavelengths of ion absorption lines to provide gas kinematic measurements for ≥2 transitions to probe kinematics as a function of ionization state.<br><br>*Threshold: Central wavelengths of ion absorption lines to provide gas kinematic measurements for 1 transition to probe kinematics as a function of ionization state.* | **F9.1** Multi-object spectroscopy with a field of view ≥2.5 × 2.5 arcmin².<br><br>**F9.2** Spectral range ≤115 nm to ≥320 nm.<br><br>**F9.3** R ≥ 60,000.<br><br>**F9.4** SNR ≥ 5 per resolution element on targets (e.g., QSOs) of AB ≥ 20 mag (GALEX FUV filter) in exposure times of ≤12 h.<br><br>*Threshold:*<br>*R ≥ 25,000; field of view ≥ 2 × 2 arcmin²; Exposure times of ≤ 20 h.* | UVS multi-object spectroscopy using a microshutter array with a field of view = 3 × 3 arcmin².<br>Spectral range: 115–320 nm.<br>R = 60,000.<br><br>SNR = 5 per resolution element on targets (e.g., QSOs) of AB = 20 mag (GALEX FUV filter) in exposure times of 10.9 h. |
| | | Galaxy gas morphology at ≤1 kpc scales. | UV imaging of the distribution of gas within the galaxy at z ≤ 0.2 to spatial resolutions of ≤1 kpc. | **F9.5** Filtered imaging in the wavelength range ≤115 nm to ≥320 nm; angular resolution ≤ 0.3 arcsec. | UVS Filtered imaging in the wavelength range 115–320 nm; angular resolution = 0.25 arcsec. |
| | **O10:** To determine the sources responsible for initiating and sustaining the metagalactic ionizing background (MIB) across cosmic time. | LyC emission escape fraction of ionizing photons down to values of ≤1% for a range of galaxies. | Continuum strength in the region around 91.2 × (1+z) nm per galaxy, for galaxies between z ≤ 0.4 and z ≥ 1.0.<br><br>Survey ≥25 galaxies per unit magnitude interval per unit redshift interval to determine the LyC luminosity function evolution with redshift. | **F10.1** Multi-object low-resolution spectroscopy.<br><br>**F10.2** Spectral range: ≤119 nm to ≥240 nm.<br><br>**F10.3** R ≥ 80.<br><br>**F10.4** SNR ≥ 3 per 3 nm interval on galaxies of AB ≥ 28 mag with exposure times of ≤10 h per field.<br><br>*Threshold: Same requirements, with exposure times of ≤30 h per field.* | UVS multi-object low-resolution spectroscopy using a microshutter array.<br>Spectral range: 115–320 nm.<br>R = 500.<br><br>SNR = 3 per 3 nm interval on galaxies of 28 mag with exposure times of 9.8 h per field. |
| | | Lyα escape fraction of ionizing photons to an accuracy | Continuum strength between rest frame ≤110 nm and ≥130 nm for | **F10.5** Multi-object spectroscopy.<br><br>**F10.6** Spectral range: ≤154 nm to ≥304 nm. | UVS multi-object spectroscopy using a microshutter array.<br>Spectral range: 115–320 nm. |

**Table 5.1-2.** Science Traceability Matrix. Baseline science objectives and requirements appear in black typeface, while threshold objectives and requirements appear in grey, italic typeface. Driving requirements appear in the payload functional requirements in blue, bold typeface.

| Goal | Science Objectives | Scientific Measurement Requirements | | Payload Functional Requirements | Baseline Projected Performance |
|---|---|---|---|---|---|
| | | Physical Parameters | Observables | | |
| | | commensurate with that for LyC above for each galaxy. | galaxies at redshifts between z ≤ 0.4 and z ≥ 1.4.<br><br>Shape of the Lyα emission/absorption profile for a subset of the LyC emitters observed above to provide information on the neutral hydrogen velocity and column density. | **F10.7 R ≥ 3,000.**<br>SNR ≥ 3 per 0.1 nm interval on galaxies of AB ≥ 24 mag in exposure times of ≤15 h per field.<br><br>*Threshold: Same requirements, with exposure times of ≤40 h per field.* | R = 3,000.<br><br>SNR = 3 per 0.1 nm interval on galaxies of 24 mag in exposure times of 8 h per field. |
| | **O11:** To probe the origin of the elements by determining the properties and end states of the first generations of stars and supernovae. | Abundances of metals including C, Si, P, S, Cr, Fe, Ni, and Zn in low-metallicity (Z ≤ -4 Z⊙) sunlike stars. | Strengths of r-process absorption transitions in the UV down to line depths of ≤1% of the continuum at 99.7% confidence. | **F11.1** Spectral range ≤170 nm to ≥310 nm.<br><br>**F11.2 R ≥ 24,000.**<br><br>**F11.3** SNR ≥ 100 in the continuum per resolution element on stars of AB ≥ 14 mag in exposure times of ≤10 h.<br><br>*Threshold:*<br>*SNR ≥ 100 in the continuum per resolution element with R ≥ 24,000 on stars of AB ≥ 13 mag in exposure times of ≤10 h.* | UVS Spectral range: 115–320 nm.<br><br>R = 25,000.<br><br>SNR = 100 in the continuum per resolution element on stars of AB = 14 mag in exposure times of 7.5 h. |
| | **O12:** To address whether there is a need for new physics to explain the disparity between local measurements of the cosmic expansion rate and values implied by the cosmic microwave background (CMB) using the standard Λ cold dark matter (ΛCDM) cosmological model. | Local value of the Hubble-Lemaître constant with 1% precision. | Cepheid-based distances to local (out to ≥50 Mpc) galaxies that have hosted recent (since 1995) SNIa with ≤10% precision at 99.7% confidence. | **F12.1 Broadband visible–near-IR imaging** (e.g., V-, I-, J-, and H-band), with F12.2.<br><br>**F12.2 Field of view ≥2 × 2 arcmin²,** which enables detection of multiple Cepheid stars in a single pointing.<br><br>**F12.3** SNR ≥ 10 for point sources of H ≥ 28 mag in exposure times of ≤2 h.<br><br>*Threshold: SNR ≥ 10 for point sources of H ≥ 27 in exposure times of ≤2 h.* | HWC broadband visible–near-IR imaging with a field of view of 3 × 3 arcmin².<br><br>SNR = 10 for point sources of H = 28 mag in exposure times of 2 h. |
| | **O13:** To constrain dark matter models through detailed studies of resolved stellar populations in the centers of dwarf galaxies. | Stellar density profiles of stars in the inner regions of dwarf galaxies (i.e., galaxies with stellar mass in the range 10^5.5 M⊙ – 10^6.5 M⊙). | Visible imaging of resolved stars in the central regions of dwarf galaxies (radius of ≤500 pc) with a precision of ≤0.5 M⊙/pc³ (3σ). | **F13.1** Broadband visible imaging (e.g., V-band) over a field of view comparable to nearby dwarf galaxy sizes (≥2 × 2 arcmin²), with (F13.2).<br><br>**F13.2 Angular resolution ≤ 0.05 arcsec.**<br><br>**F13.3** SNR ≥ 5 for point sources of V ≥ 30 mag in exposure times of ≤2 h per dwarf galaxy, for ≥10 dwarf galaxies.<br><br>*Threshold: Angular resolution ≤75 mas;*<br>*SNR ≥ 5 for point sources of ≥30 mag in exposure times of ≤6 h.* | HWC broadband visible imaging with a field of view of 3 × 3 arcmin² and an angular resolution of 0.03 arcsec.<br><br>SNR = 5 for point sources of V = 30 mag in exposure times of 1.5 h per dwarf galaxy. |
| | **O14:** To constrain the mechanisms driving the formation and evolution of Galactic globular clusters. | Key atmospheric line strengths for individual stars in crowded central regions of Galactic globular clusters in order to probe globular cluster stellar populations (e.g., ages and abundances as a function of cluster-centric radius). | UV and optical spectra of ≥400 stars within a single Galactic globular cluster, for stars separated by ≤0.2 arcsec. | **F14.1** Multi-object UV and **multi-object visible spectroscopy.**<br><br>**F14.2** UV spectral range ≤150 nm to ≥320 nm.<br><br>**F14.3 Visible spectral range ≤0.37 μm to ≥1.0 μm.**<br><br>**F14.4 R ≥ 1000.**<br><br>**F14.5** SNR ≥ 3 in the continuum per 0.5 nm effective resolution element on ≥400 stars of V ≥ 25 mag in a total exposure time of ≤10 h per instrument.<br><br>*Threshold: Same requirements in total exposure time of ≤30 h.* | UVS multi-object spectroscopy in the UV using a microshutter array. HWC multi-object spectroscopy in the visible using a microshutter array.<br>UVS UV spectral range: 115–320 nm.<br>HWC visible spectral range: 0.37–1.8 μm.<br>R = 1,000.<br><br>SNR = 3 in the continuum per 0.5 nm effective resolution element on 400 stars of V = 25 mag in a total exposure time of 6.5 h per instrument. |
| | **O15:** To constrain the likelihood that rocky planets in the HZ around mid-to-late-type M-dwarf stars have potentially habitable conditions (defined as water vapor and biosignature gases | Abundance of atmospheric H₂O if the column density is ≥2.9 g/cm² (modern Earth) | Near-IR planetary spectrum over with a wavelength range covering ≥ 2 H₂O absorption features. | **F15.1** Slit or slitless spectroscopy for a V ≥ 18.8 mag star.<br><br>**F15.2 Spectral range: ≤1.1 μm to ≥1.7 μm.**<br><br>**F15.3** H₂O: R ≥ 10 at 1.4 μm with **SNR/√hour (h) ≥ 32,000 per spectral bin.** | HWC spectroscopy for a V = 18.8 mag star.<br>Spectral range: 0.37–1.8 μm.<br><br>H₂O: R = 10 at 1.4 μm with SNR/√h = 41,000 per spectral bin. |

Table 5.1-2. Science Traceability Matrix. Baseline science objectives and requirements appear in black typeface, while threshold objectives and requirements appear in grey, italic typeface. Driving requirements appear in the payload functional requirements in blue, bold typeface.

| Goal | Science Objectives | Scientific Measurement Requirements | | Payload Functional Requirements | Baseline Projected Performance |
|---|---|---|---|---|---|
| | | Physical Parameters | Observables | | |
| | in the atmosphere) by surveying ≥5 systems, assuming an average transit duration of 1 h. *Threshold: To constrain the fraction of M-dwarf stars that may be habitable by surveying ≥2 systems, assuming an average transit duration of 1 h.* | | *Threshold: Near-IR planetary spectrum with a wavelength range covering ≥1 $H_2O$ absorption features.* | **F15.4 Precision ≤ 31.3 parts per million (ppm) / $\sqrt{h}$.** **F15.5 Noise floor ≤ 10 ppm.** *Threshold: SNR/$\sqrt{h}$ ≥ 25,500 per spectral bin at 1.4 μm. Precision ≤39.2 ppm/$\sqrt{h}$.* | Precision = 24.4 ppm/$\sqrt{h}$. Noise Floor = 10 ppm. |
| | | Abundance of atmospheric molecular species if the column density is: • $O_3$ ≥ 7.2 × $10^{-4}$ g/cm$^2$ (modern Earth) | Planetary spectrum: $O_3$ broad absorption feature in the visible. | F15.6 Slit or slitless spectroscopy for a V ≥ 15.0 mag star. Spectral range ≤0.5 μm to ≥0.8 μm. F15.7 $O_3$: R ≥ 10 at 0.6 μm with SNR/$\sqrt{h}$ ≥ 9,500, per spectral bin. Precision ≤105.3 ppm/$\sqrt{h}$. **F15.8 Noise floor ≤10 ppm.** *Threshold: $O_3$: R ≥ 10 at 0.6 μm with SNR/$\sqrt{h}$ ≥ 7,500, per spectral bin. Precision ≤133.3 ppm/$\sqrt{h}$.* | HWC spectroscopy for a V = 15.0 mag star. Spectral range: 0.37–1.8 μm. $O_3$: R = 10 at 0.6 μm with SNR/$\sqrt{h}$ = 12,000, per spectral bin. Precision = 83.3 ppm/$\sqrt{h}$. Noise Floor = 10 ppm. |
| | **O16:** To constrain the range of possible structures within transition disks and to probe the physical mechanisms responsible for clearing the inner regions of transition disks, surveying ≥20 transition disks. *Threshold: ≥5 transition disks.* | 2D surface brightness of the transition disk. Protoplanetary positions with respect to the central star. Planetary flux. | 2D broadband high contrast images of the transition disk. 2D narrowband high contrast images centered at Hα. Angular star-to-planet separation with ≤1 AU uncertainty. | **F16.1 2D broadband and narrowband high contrast imaging on star with apparent R ≥ 13 mag.** F16.2 IWA$_{0.5}$ ≤ 100 mas at Hα, which is small enough to detect a protoplanet at 15 AU from a star located at 150 pc. F16.3 Broadband high contrast visible image of a transition disk with a surface brightness detection limit of R ≥ 20.5 mag/arcsec$^2$ at IWA$_{0.5}$ with SNR ≥ 7 in ≤ 50 h. F16.4 Narrowband (≤ 10 nm around Hα at 0.656 μm) high contrast visible image of a transition disk with a surface brightness detection limit of R ≥ 18.0 mag/arcsec$^2$ at IWA$_{0.5}$ with SNR ≥ 7 in ≤ 50 h. F16.5 Point source-to-star flux ratio detection limit of ≤$10^{-6}$ with an SNR ≥ 7 in a narrowband filter centered at 0.656 μm with a total exposure time of ≤50h for sunlike star at 150 pc. *Threshold: IWA$_{0.5}$ ≤ 140 mas. Broadband surface brightness detection limit of ≥20.5 mag/arcsec$^2$ at IWA$_{0.5}$. Narrowband surface brightness detection limit of ≥18 mag/arcsec$^2$ at IWA$_{0.5}$. Point source-to-star flux ratio detection limit of ≤4 × $10^{-6}$.* | Coronagraph 2D broadband and narrowband high contrast imaging on star apparent R = 13 mag. IWA$_{0.5}$ = 82 mas at 0.656 μm with the coronagraph Broadband high contrast visible image of a transition disk with a surface brightness detection limit of 20.5 mag/arcsec$^2$ at IWA$_{0.5}$ with SNR = 7 in 23 h. Narrowband (10 nm around Hα at 0.656 μm) high contrast visible image of a transition disk with a surface brightness detection limit of 19.0 mag/arcsec$^2$ at the IWA$_{0.5}$ with SNR = 7 in 23 h. Point source-to-star flux ratio detection limit of $10^{-6}$ with an SNR = 7 in a narrowband filter centered at 0.656 μm with a total exposure time of 23 h. |
| | **O17:** To probe the physics governing star-planet interactions by investigating auroral activity on gas and ice giant planets within the solar system. | UV auroral emission in the upper atmosphere magnetic polar regions of the gas and ice giant solar system planets. | Time-resolved UV imaging of $H_2$ Lyman band, $H_2$ Werner band, and H Lyα. | F17.1 Wavelength range: ≤115 nm to ≥162 nm. F17.2 Field of view ≥ 1 × 1 arcmin$^2$ (to provide ≥300 km resolution on Uranus and Neptune, **transverse and vertical angular resolution ≤ 0.05 arcsec**). F17.3 SNR ≥ 3 for an auroral surface brightness ≤100 Rayleigh in an exposure time of ≤ 10 min. **F17.4 Non-sidereal tracking of ≥1 arcsec/min.** *Threshold: Same requirements in an exposure time of ≤30 min.* | UVS wavelength range: 115–320 nm. Field of view = 3 × 3 arcmin$^2$. Transverse and vertical angular resolution = 0.025 arcsec. SNR = 7.5 for an auroral surface brightness of 100 Rayleigh over 1 × 1 arcsec$^2$ on Uranus in an exposure time of 10 min. Non-sidereal tracking with a maximum slew rate of 42.6 arcmin/s. |

## 5.1.1.2 Science Objectives

The HabEx science objectives listed in **Table 5.1-2** are scientific unknowns that are derived from the HabEx goals. The scientific community often has theories, or hypotheses, for the solutions to these unknowns and these can be supported or refuted directly with measurements by HabEx. Using Goal 3, Objective 13 (*Section 4.5*) as an example, one scientific hypothesis is: "Flattened dark matter halo density profiles at the centers of nearby dwarf galaxies are caused by baryon-dark matter interactions." The measurements detailed in this row of the STM enable scientists to support or refute the hypothesis, which could lead to new hypotheses including alternate dark matter models if the original hypothesis is found to be false.

The 17 objectives used in this study represent only a portion of the science enabled by HabEx's capability. As earlier great observatories like the Hubble Space Telescope (HST) have proven, the scientific possibilities with a highly capable observatory are tremendous and the innovative nature of the scientific community will ensure that the range of science eventually undertaken with a next-generation great observatory like HabEx will far exceed the goals and objectives in the STM and what has even been considered by the community today.

### 5.1.2 Science Measurement Requirements

Each of the science objectives is associated with one or more physical parameters, the observation or measurement of which will help expand on the current understanding of the objective. Each parameter is in turn associated with an observable feature, which includes science-driven observational constraints. In the previous example of Objective 13, the physical parameter is the stellar density profile and the observable requirement is visible band imaging of resolved stars in the central regions of dwarf galaxies (radius of $\leq 500$ pc) with a precision of $\leq 0.5$ $M_\odot/pc^3$ ($3\sigma$). The science measurement requirements connect unresolved scientific questions linked to the science objectives, with the observations needed to address these questions.

### 5.1.3 Functional Requirements

In the fifth column of the STM, the observational requirements are translated into functional requirements. This step takes what needs to be observed to meet the science objectives, and identifies the high-level payload and mission requirements that need to be satisfied to make those observations. Since the STM does not presume a design implementation, the STM functional requirements address overarching requirements such as the number of observations, wavelength range, signal-to-noise ratio (SNR), and spectral resolution ($R$). These high-level requirements are used to define the concept's instrument complement and the most fundamental key performance requirements for the HabEx payload and mission designs.

### 5.1.4 Baseline and Threshold

**Table 5.1-2** lists both "baseline" and "threshold" objectives and requirements. The baseline defines a highly capable mission that meets all of the science objectives completely. The "threshold" objectives and requirements represent the maximum descope for each science objective, below which the science objective could not be achieved.

In the remainder of this chapter, the "baseline" requirements outlined in the STM are used as the basis that defines the payload for the baseline HabEx mission: a 4 m aperture telescope with the UV spectrograph (UVS), workhorse camera (HWC), in addition to both the coronagraph (HCG) and starshade (SSI) instruments in a "hybrid" configuration. In *Chapter 6*, these payloads are described—along with the baseline HabEx telescope flight system. In *Chapter 7*, the starshade occulter and flight system are described. In *Chapter 8*, the mission concept and shared elements, like the ground system, are described. *Chapter 10* examines the requirements of the STM in an evaluation of architecture trades, their ability to meet threshold requirements, their cost, and other details.

## 5.2    Error Budgets

Of all the functional requirements identified in the HabEx STM (**Table 5.1-2**), the most influential on the overall concept design is the raw instrument contrast required to detect exoplanets very close to their host stars and at very faint planet-to-star flux ratios (typically $10^{-10}$ for exo-Earth Candidate, EECs) within the maximum exposure times allowed for broadband detection (Objective 1) and spectroscopy (Objective 3). While the "detectable planet-to-star flux ratio" and the "raw instrument contrast" may take similar values, the two terms should not be confused. The former is a property of the starlight suppression instrument, while the latter is a property of the astrophysical source, independent of the instrument used to observe it (see *Appendix H* for definition of commonly used high contrast imaging terms). Exoplanets with planet-to-star flux ratios fainter than the local raw instrument contrast level set by residual starlight can for instance be detected using advanced point spread function (PSF)-subtraction techniques.

The raw instrument contrast (and contrast stability) design constraint is easily the most demanding requirement that HabEx will need to meet and it touches many areas of the overall flight system design. Since so many design factors within HabEx affect this performance, decomposing this requirement into a number of lower-level requirements is essential for developing the design. This decomposition is handled through two contrast-based error budgets for each of the two driving objectives: one addressing the coronagraph raw contrast performance and broad-band detection of EECs within a time allocation (**Figure 5.2-1**); and the other addressing the starshade raw contrast performance (**Figure 5.2-2**) and ability to measure an EEC spectrum within a single starshade continuous observability window of 43 days (for worst-case targets at low ecliptic latitude). The performance estimates of both error budgets are based on the scattered light level in the image plane as a function of telescope, instrument or system perturbations, on the modeled end-to-end optical throughput, and on a representative fiducial astrophysical scene. For both error budgets, the fiducial astrophysical scene consists of an exo-Earth seen at quadrature around a sunlike star located at 9 pc, in line with the minimum distance to be accessed (*Section 3.3.2.6* and **Table D-2**) in order to meet the HabEx EEC completeness baseline requirement and overall Goal 1 objectives.

It is important to note that unlike the STM, where there is a direct linkage between objectives and requirements, the error budget is an allocation of lower-level requirements where the only absolute objective is that the total error budget meets the STM's requirement to detect or spectrally characterize a fiducial planet at a given separation and planet-to-star flux ratio in some maximum allowed exposure time. As such, the error budget can be allocated in an almost infinite number of ways. For instance, an optical system with a low end-to-end throughput requires better raw instrument contrast to meet the time-to-SNR requirement than a higher throughput optical system does. Best practice favors using measured performance over modeled performance in the error budget, but both can be included.

### 5.2.1    Coronagraph Instrument Error Budget

The HabEx Coronagraph (HCG) instrument error budget centers around high contrast observations as captured in the STM Objectives 1 and 3 (**Table 5.1-2**). Raw contrast levels must be balanced against integration durations to find a workable error budget. The top-level error budget for the HCG is shown in **Figure 5.2-1.** The top-level planet-to-star flux ratio to be detected is set at the $10^{-10}$ level consistent with Earth-sized HZ planets around sunlike stars, and the raw instrument contrast requirement is set to $3.0 \times 10^{-10}$ at the fiducial planet separation. The instrument raw contrast requirement is disintegrated into allocations for photometric and systematic noise sources, which are then further disintegrated into verifiable performance requirements on key systems and hardware elements. The expected performances for each of these elements are also captured and rolled up through this same relational budget structure. This allows the calculation of the expected performance for both photometric and systematic noise, and ultimately the exposure time required to detect the fiducial planet at the specified SNR based on the current best estimates (CBEs) of the performance of the key hardware elements captured in the budget. This top-down process of defining requirements and bottom-up process of assessing likely performance is iterated to balance the error budget and create performance margins on all key systems and hardware. Discussion of the performance estimates identified as CBE in the error budget is covered in *Section 6.9*.

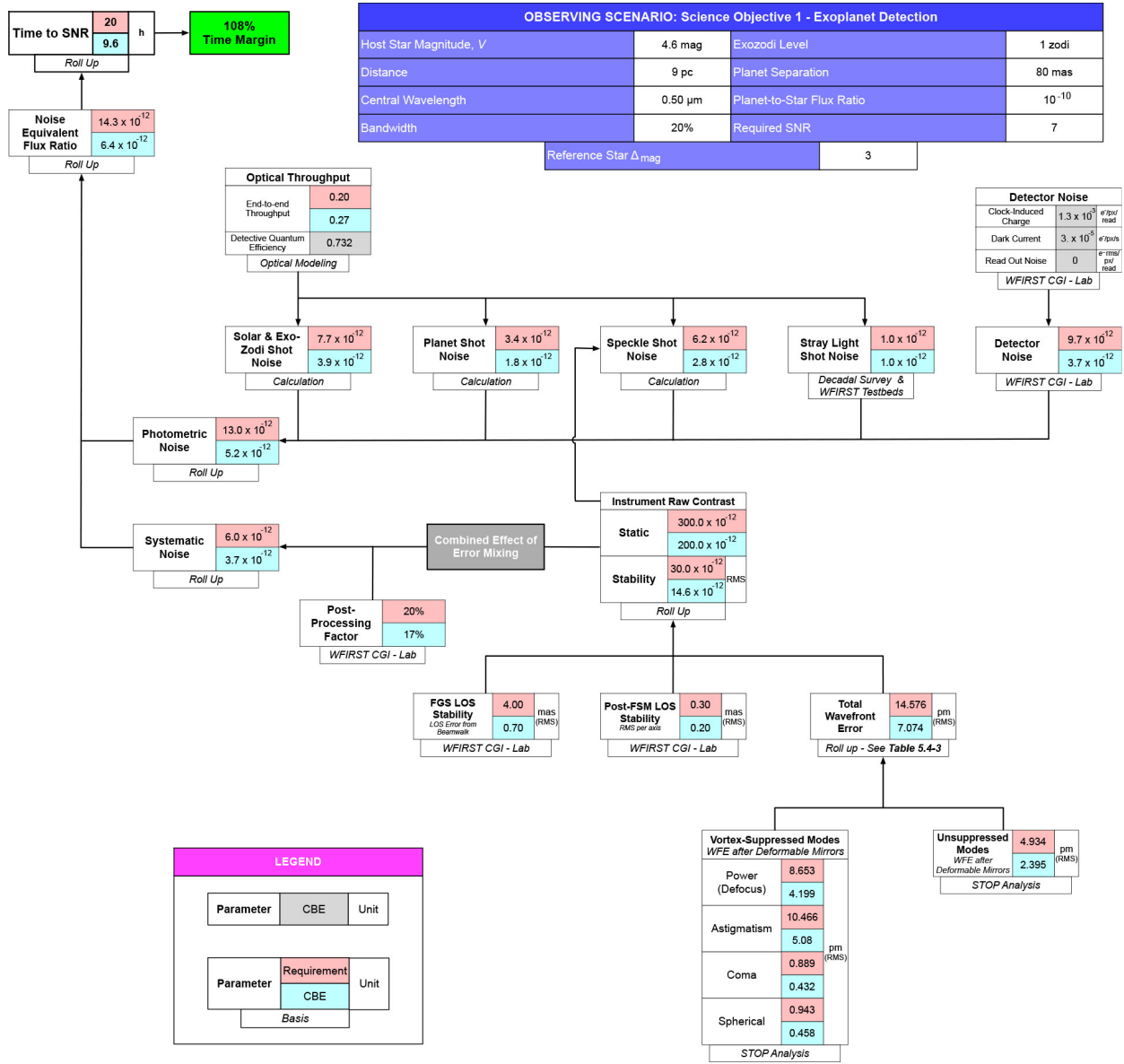

**Figure 5.2-1.** Coronagraph error budget for imaging an Earth-like planet around a sunlike star at 9 pc with a required delta magnitude of 25 (i.e., a planet-to-star flux ratio of 10⁻¹⁰). A coronagraph bandwidth of 20% around a central wavelength of 0.5 μm is assumed.

The two main sources of noise that affect the error budget are photometric noise and systematic effects. The sources of photometric error are shot noise from: the planet, residual starlight (termed "speckle"), and astronomical background such as zodiacal dust. Along with the shot noise sources, detector noise also contributes to the overall photometric noise. The systematic noise can be thought of as wavefront changes that manifest as time-based variations in the image's background speckles. Variation in the intensity of the local speckles cannot be taken out entirely by post-processing, producing a false signal that does not reduce over time and shows up as systematic noise floor and Δmag detection limit.

Optical throughput and detector requirements connect to the photometric noise portion of the error budget, while all other design requirements stem from the systematic noise portion of the budget. Since the systematic noise comes from the time-varying speckles, minimizing this variation introduces a number of stability requirements on the telescope and HCG systems. The physical events producing dynamic disturbance include rigid body tilting motion of telescope, internal motion of the telescope mirrors and instrument optics, and thermal distortions of the optics; all resulting from mechanical jitter, thermal drift, or line-of-sight (LOS) sensing errors. Additionally, assuming the selection of a vector vortex coronagraph mask (*Section 6.3*), wavefront error (WFE) requirements can be defined for vortex-suppressed Zernike modes and unsuppressed Zernike modes. This requirement is detailed in *Section 5.4.2.1*.

### 5.2.2    Starshade Error Budget

Observations with the starshade system must also address instrument contrast-driven performance requirements. Similar to the HCG, a detailed error budget (**Figure 5.2-2**) is used to allocate performance requirements to all key observation-related systems. As in the HCG error budget, the starshade system time-to-SNR spectroscopy error budget breaks down the high-level instrument raw contrast static and stability requirements into photometric and systematic error allocations. With the starshade system, the systematic error comes from perturbations in petal shape, petal size, petal position, and formation flying. Perturbations can be both correlated and uncorrelated. Thermal and mechanical distortions are the primary sources of starshade systematic error.

Similar to HCG's broadband photometric detection error budget, the starshade system spectroscopy error budget includes detector noise and zodiacal light. It also includes several unique contributing sources, namely micrometeroid holes in the external occulter, reflectance of astrophysical sources from the telescope-facing surface of the external occulter, and solar glint from the edges of the external occulter. The most significant noise contributor is expected to be solar glint. Optical edge scatterometry has been performed on both specular and diffuse edges (Martin et al. 2013; Casement et al. 2016) resulting in a baseline design employing a sharp, smooth edge (Steeves et al. 2018). The leakage and reflection are expected to be fainter than the exozodiacal light surrounding the target star, while solar edge scatter will have local components that are brighter than exozodiacal light, although this does not exceed the allowable allocation in the starshade error budget. The STM requirement for Objective 3 is to reach an SNR of 10 per $R = 70$ spectral bin anywhere between 0.45 μm and 0.975 μm in less than 43 days, the shortest window for continuous starshade observations. The error budget presented in **Figure 5.2-2** is computed at the red end of that range (precisely at the central wavelength of a broad $H_2O$ absorption band), which represents the most challenging case over the required spectral range, due to increased exozodi signal and lower detector QE.

The starshade system spreadsheet-based error budget used in **Figure 5.2-2** has been in development for nearly a decade (Shaklan et al. 2010; Shaklan et al. 2017). Error performance terms are determined by modeling the electric fields at the focal plane of the telescope using diffraction algorithms that have been independently developed (Shaklan et al. 2010) and tested against laboratory data. The analysis is performed as a function of both wavelength and working angle. Like HCG's error budget, starshade's key system requirements are allocated from a top-down disintegration of systematic and photometric errors, while an overall CBE performance estimate is generated with a bottom-up summation of all the significant modeled or measured errors. Along with details on the starshade system design, CBE performances are discussed in *Chapter 7*.

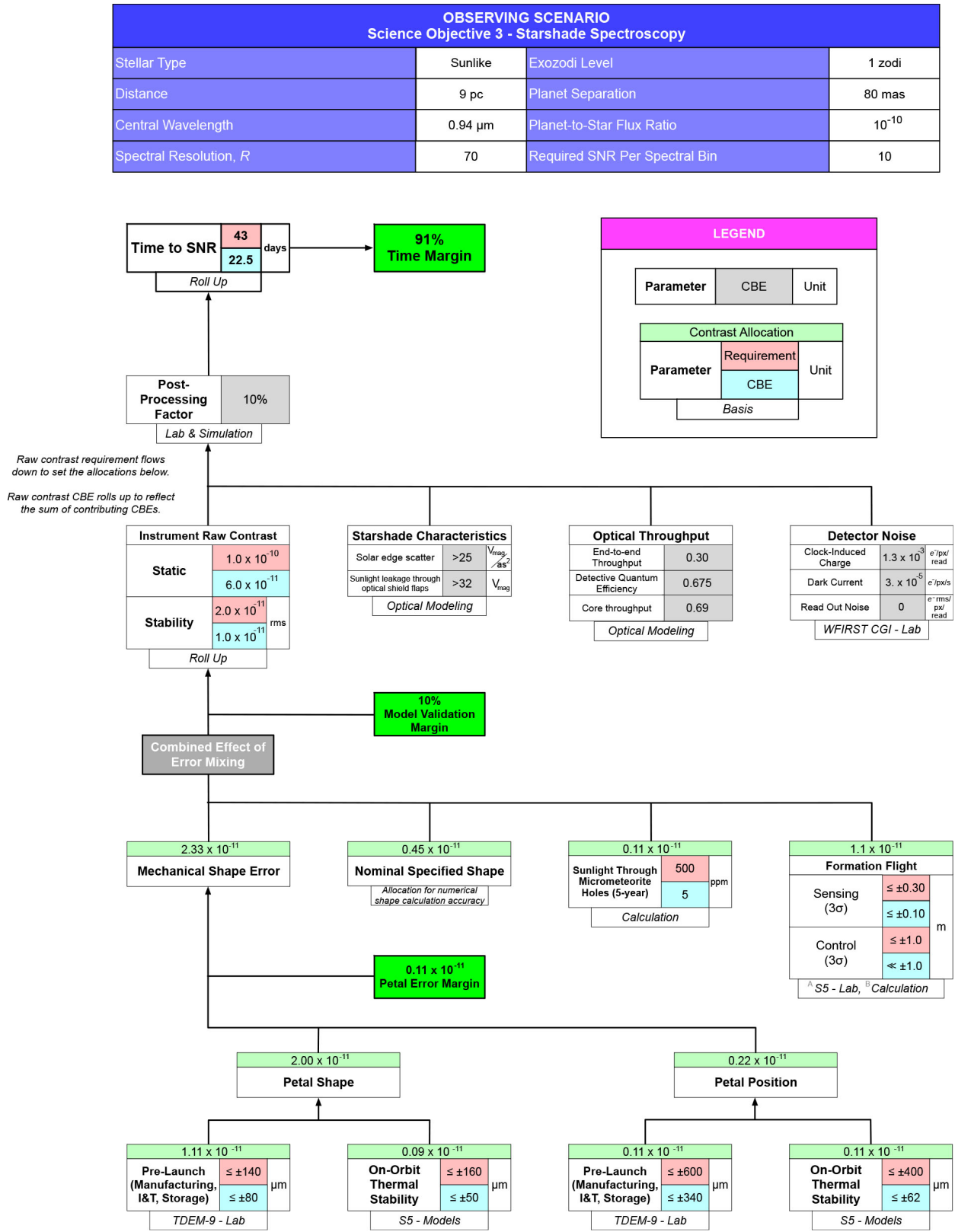

**Figure 5.2-2.** Starshade contrast error budget for performing spectroscopy of an Earth-like planet around a sunlike star at 9 pc at 0.94 μm with a resolution of R = 70 at SNR=10 per spectral bin.

## 5.3    Mission Traceability Matrix

Given the broad science mission of HabEx, the payload functional requirements presented in the STM (**Table 5.1-2**) levy overlapping constraints on mission and systems design. While the payload functional requirements in the STM explicitly define the HabEx systems requirements, additional requirements are also derived from the science measurement requirements in the error budgets in *Section 5.2* and through analysis of architectural assumptions, as described in *Chapter 10*, e.g. aperture size.

The Mission Traceability Matrix (MTM; **Table 5.3-1**) identifies the key systems design drivers for HabEx. To simplify the representation of these requirements, the MTM groups related requirements in order to define the high-level functional requirements for the HabEx mission and its systems. While baseline projected performance is defined as the final column in the STM and as defined in the HCG, SSI, and starshade system error budgets, the MTM does not describe the projected performance of mission systems. Note that some MTM requirements were derived based on payload requirements and architectural assumptions. For instance, the power requirement for the telescope flight system is driven by the telescope's temperature requirement and size, both of which are set by instrument requirements. For another example, the requirement for the starshade external occulter power is driven by trade study results assuming a certain number of Hall thrusters determined by simulations described in *Section 8.2*.

The two left-most columns of the MTM group payload functional requirements from the STM by the instrument defined in the baseline projected performance column of the STM, and results of the error budgets. The Telescope Requirements detail the requirements which flow from the columns to the right. For instance, the UVS observations will require a 260 K primary mirror. The following Mission Design Requirements column details derived requirements on the mission design. For instance, relatively disturbance-free environments are required for coronagraph and starshade observations. Here, the derived starshade system requirement for operating at Earth-Sun L2 ultimately drives the design as described in following chapters. The following three columns in the MTM describe the mission system and operational requirements that are derived from the columns to their left.

## 5.4    Key and Driving Requirements

The requirements on the HabEx mission ultimately determine the baseline concept designs described in *Chapters 6, 7,* and *8*, which in turn determine the necessary new technologies described in *Chapter 11*. These requirements are typically categorized as "key" and "driving" requirements. Key requirements flow from the science goals and objectives, and are tightly tied to the overall scientific capabilities of the observatory. Driving requirements "drive" the scope of the overall mission including its size, cost, duration. Some requirements can be both key and driving. For example, the diameter of the primary mirror of the telescope limits instrument spatial resolution, scales integration times, and factors into the inner working angle of the HCG, and it also drives the size and cost of the overall telescope flight system. This section summarizes the major requirements that determine the HabEx baseline concept presented in subsequent chapters.

### 5.4.1    Mission Requirements

The study's mission concept is largely constrained by three programmatic requirements (see **Table 5.4-1**). Programmatic constraints were established by early study guidance from NASA. The first programmatic requirement is that the mission must be serviceable. Serviceability of all future large space-based astrophysical observatories is established by law and given the significant investment required for a mission like HabEx, having the ability to extend and expand the facility's science return is practical. The second programmatic requirement is that the mission must be launched by an American-built launch vehicle that is likely to be available at the time of the HabEx mission. This requirement subsequently traces to requirements on the overall launch mass and volume for both the HabEx telescope and starshade flight systems. The third programmatic requirement–a minimum primary mission duration of 5 years–sets the concept reliability level and sizes on-board consumables. A primary mission duration of five years is consistent with many past, current, and developing large multi-purpose space observatories. In addition, this requirement also defines the minimum servicing cadence needed to maintain the facility. These programmatic requirements are driving requirements, but not key requirements, since they drive the cost, volume, and mass of the mission but do not come from the STM.

**Table 5.3-1.** The HabEx Mission Traceability Matrix (MTM) summarizes requirements derived from the STM and error budgets, described above.

| Payload Functional Requirements | Mission Functional Requirements | Telescope Requirements | Mission Design Requirements | Telescope Flight System | Starshade Flight System | Operations |
|---|---|---|---|---|---|---|
| F1.1, F1.2, F1.3, F1.4, F4.1, F5.2, F8.2, F8.6, F16.1 | Coronagraph (HCG) Observations | • ƒ/number: ≥2.25 • Unobscured • Monolithic Primary Mirror • Primary Mirror Diameter: ≥3.7 m • Bandpass: 0.45–1.70 μm | • Telescope must launch on SLS Block 1B with 8.4 m fairing • Earth-Sun L2 or drift away orbit • Ability to revisit target stars 6 times within 5 years | • Mass: ≤35,000 kg • Power: ≥4,500 W EOL • Must fit within 8.4 m SLS fairing • Launch 1st Mode (Lateral): >8 Hz • Launch 1st Mode (Axial): >15 Hz • Wavefront Error Stability: See Table 5.4-3 • LOS Error: ≤2.0 mas rms • LOS Stability (from beamwalk): ≤4.0 mas rms • PM Thermal Stability: ±1.1 mK | | "Digging the Dark Hole" and Reference Differential Imaging (see *Section 6.10*) |
| F2.1, F3.1, F3.2, F3.3, F5.1/6.4, F5.2/5.5/6.5, F5.3, F6.1, F8.6 | Starshade (SSI) Observations | • Bandpass: 0.30–1.70 μm | • Starshade must launch on a launch vehicle with a 5 m fairing and able to deliver >12,500 kg to L2 • Must operate at L2 orbit • Field of regard from 40°–85° off the sun-telescope line | • LOS Stability: ≤2.5 mas • Field of regard from 40°–85° off the sun-telescope line | • Mass ≤ 12,500kg • Power ≥ 35,000 W • Must fit in a Falcon H fairing • Pointing control ≤ 1° • Pre-Launch ▪ Petal Shape: ≤±140 μm ▪ Petal Position: ≤±600 μm • In-Flight ▪ Petal Shape: ≤±160 μm ▪ Petal Position: ≤±400 μm | • Telescope-Starshade Separation Distance: 42,580–114,900 km • Spacecraft Separation Distance Accuracy Along LOS: < ±250 km • Distance Sensing: <±25 km • Starshade Lateral Displacement from LOS: ≤±1.0 m • Starshade Lateral Formation Sensing: ≤±0.3 m |
| F9.1, F9.2, F9.3, F9.5, F10.1, F10.3, F10.7, F11.2, F17.2, F17.4 | UV Spectrograph (UVS) Observations | • Bandpass: 0.115–0.320 μm | | • Power: ≥4,500 W EOL • LOS Error: ≤2.5 mas • Primary Mirror Operating Temperature: ≥260 K • Non-Sidereal Tracking: 1 as/min | | • Multi-object spectroscopy |
| F12.1, F12.2, F13.2, F14.1, F14.3, F14.4, F15.2, F15.3, F15.4, F15.5/15.9 | HabEx Workhorse Camera (HWC) Observations | • Bandpass: 0.37–1.70 μm | | • LOS Error: ≤2.5 mas | | • Multi-object spectroscopy |

**Table 5.4-1.** HabEx Mission Requirements. K represents a key requirement and D represents a driving requirement.

| Parameter | Requirement | K | D | Source |
|---|---|---|---|---|
| Design Life | ≥5 years | | ✓ | Programmatic |
| Servicing | Must be serviceable | | ✓ | Programmatic |
| Possible Launch Vehicle | US-Only | | ✓ | MTM |
| Orbit | Earth-Sun L2 or Drift Away | | ✓ | MTM |
| Revisit Ability | 6 times in 5 years, as necessary | | ✓ | STM |

**Table 5.4-2.** Telescope Flight System Requirements. K represents a key requirement and D represents a driving requirement.

| Parameter | Requirement | K | D | Source |
|---|---|---|---|---|
| Instrument Complement | Exoplanet direct imaging and spectroscopy. Imaging and spectroscopy in the UV, visible, and near-IR. High-resolution spectroscopy in the UV. | ✓ | ✓ | MTM |
| Mass | ≤35,000 kg | ✓ | ✓ | MTM |
| Power | ≥4,500 W | ✓ | ✓ | MTM |
| Configuration | Must fit within 8.4 m SLS fairing | ✓ | ✓ | MTM |
| 1st Launch Mode (Lateral) | >8 Hz | ✓ | | MTM |
| 1st Launch Mode (Axial) | >15 Hz | ✓ | | MTM |
| Field of Regard | ≥40° | ✓ | | MTM |
| Slew Rate | ≥1 arcsec/min | ✓ | | STM |
| Raw Contrast | 3.00 10$^{-10}$ | ✓ | ✓ | Error Budget |
| Raw Contrast Stability | 3.00 10$^{-11}$ | ✓ | ✓ | Error Budget |
| LOS Error | ≤2 mas | ✓ | | MTM |
| LOS Error from Beamwalk | ≤4 mas | ✓ | | Error Budget |
| WFE Stability | See Table 5.4-3 | ✓ | | Error Budget |
| PM Thermal Stability | ±1.1 mK | ✓ | | MTM |

### 5.4.2   Telescope Flight System Requirements

Top-level requirements on the telescope flight system are listed in **Table 5.4-2**. The instrument complement requirement for the telescope comes directly from the STM (**Table 5.1-2**). At least one exoplanet direct imaging instrument is necessary to meet Science Goals 1 and 2. In the case of the HabEx baseline architecture, both the internally occulting coronagraph (HCG; *Section 6.3*) and a camera/spectrograph (SSI; *Section 6.4*) capable of working with the externally occulting starshade (*Section 7.1*) are included in the complement for reasons discussed in *Section 3.3* and *Appendix C*. In addition, the STM functional requirements call for high resolution spectroscopy in the UV, and a general imaging/spectroscopy capability in the UV–near-IR. These requirements are addressed by two separate instruments: the UV spectrograph and imager (UVS; *Section 6.5*) and the general purpose "workhorse" camera and spectrograph (HWC; *Section 6.6*).

Both the telescope and starshade system build their error budgets from the common planet-to-star flux ratio requirement taken from the STM. Both error budgets break out contrast and contrast stability as system-level requirements. These two requirements characterize the ultimate sensitivity of the two starlight suppression techniques and are the result of many design factors. As such, the error budgets disintegrate these requirements to lower-level requirements to enable the design and development of the major components of the light suppression systems. In the case of the HCG error budget, contrast and contrast stability are broken down into line-of-sight (LOS) stabilities and wavefront error requirements, which drive the telescope, HCG, and telescope flight system designs.

The telescope LOS stability is specific to the telescope itself; the HCG has a more demanding LOS requirement internal to the instrument. The telescope LOS requirement largely constrains the ridged body motion of the telescope's primary, secondary, and tertiary mirrors. This requirement influences many aspects of the telescope and flight system design including: the fine guidance system, telescope thermal control, primary mirror material and its fabrication and design, the use of microthrusters and laser metrology, and operational constraints. At less than 2 mas, the HabEx telescope will need to meet HST's very best pointing control on a routine basis.

### 5.4.2.1  Wavefront Error Stability Requirements

In addition to meeting the LOS requirements, controlling wavefront error (WFE) is another major consideration for the telescope, HCG, and telescope flight system designs. Wavefront error terms are conventionally stated using Zernike polynomials, which correspond to standard optical aberration terms. While the Zernike polynomial sequence is infinite, most of the WFE is captured in the lowest-order terms. The overall WFE error requirement comes from the HCG error budget. This requirement is further decomposed into a WFE budget with requirements levied on each of the first 19 Zernike terms (see **Table 5.4-3**). Budget allocations were made in conjunction with the telescope and HCG design to ensure that the design has adequate margin in all Zernike modes.

Wavefront error primarily comes from three sources: a coupling of LOS error into the wavefront; mechanical bending of the primary mirror; and thermal distortion of the primary mirror. Line-of-sight WFE instability occurs when LOS drift or jitter causes beamwalk on the secondary and tertiary mirrors. Since the mirrors are conics, beamwalk manifests itself as low-order astigmatism and coma. Inertial WFE instability occurs when the primary mirror is accelerated by mechanical disturbances causing it to react (i.e., bend) against its mounts. Changes in the primary mirror's bulk temperature or temperature gradient cause thermal WFE instability. If the mirror's coefficient of thermal expansion (CTE) is completely homogeneous and constant, then a bulk temperature change should only result in a defocus error, but any inhomogeneity in the mirror's CTE will result in a temperature dependent WFE. Additionally, since CTE is itself temperature dependent, any change in the mirror's thermal gradient will also result in a WFE.

**Table 5.4-3.** Allocation of wavefront error (WFE) stability requirements for different Zernike modes. The HabEx performance in meeting the WFE stability requirement is detailed in *Chapter 6.*

| Zernike | | WFE Budget [pm RMS] | Contrast Allocation [$10^{-12}$] |
|---|---|---|---|
| **Index** | | | |
| **n** | **m** | **Aberration** | | |
| n | m | **TOTAL RMS** | **14.576** | **30.00** |
| 1 | ±1 | Tilt | 1.427 | 0.001 |
| 2 | 0 | Power (Defocus) | 8.653 | 0.005 |
| 2 | ±2 | Astigmatism | 10.466 | 0.005 |
| 3 | ±1 | Coma | 0.889 | 0 |
| 4 | 0 | Spherical | 0.943 | 0.001 |
| 3 | ±3 | Trefoil | 4.824 | 13.666 |
| 4 | ±2 | 2nd Astigmatism | 0.429 | 2.238 |
| 5 | ±1 | 2nd Coma | 0.223 | 1.285 |
| 6 | 0 | 2nd Spherical | 0.058 | 0.441 |
| 4 | ±4 | Tetrafoil | 0.564 | 1.661 |
| 5 | ±3 | 2nd Trefoil | 0.533 | 3.358 |
| 6 | ±2 | 3nd Astigmatism | 0.043 | 0.441 |
| 7 | ±1 | 3nd Coma | 0.068 | 0.832 |
| 5 | ±5 | Pentafoil | 0.472 | 3.994 |
| 6 | ±4 | 2nd Tetrafoil | 0.06 | 0.493 |
| 7 | ±3 | 3nd Trefoil | 0.031 | 0.345 |
| 6 | ±6 | Hexafoil | 0.054 | 0.635 |
| 7 | ±5 | 2nd Pentafoil | 0.031 | 0.379 |
| 7 | ±7 | Septafoil | 0.021 | 0.218 |

An important feature of the vector vortex coronagraph (VVC) is its general insensitivity to WFE in low-order Zernike terms. Since most of the WFE is concentrated in these lower Zernike modes this VVC characteristic greatly simplifies the overall telescope and flight system designs. The coronagraphic mask acts as a filter for the Zernike modes of the wavefront. The higher the vortex charge, the more low-order error the coronagraph can tolerate, but the larger its IWA and the lower its throughput.

**Table 5.4-4.** Telescope Requirements. Note that WFE and LOS stability are included in the telescope flight system requirements, as they are responsible for maintaining a stable environment for the telescope itself. K represents a key requirement and D represents a driving requirement.

| Parameter | Requirement | K | D | Source |
|---|---|---|---|---|
| f/number | ≥2.25 | ✓ | ✓ | MTM |
| Obscured or Unobscured | Unobscured | ✓ | ✓ | MTM |
| Monolithic or Segmented PM | Monolithic | ✓ | ✓ | MTM |
| Angular Resolution (0.4 μm) | 50 mas | ✓ | ✓ | STM |
| PM Diameter | ≥3.7m | ✓ | ✓ | MTM |
| Bandpass | ≤0.115 μm to ≥1.7 μm | ✓ | | STM |

### 5.4.3  Telescope Requirements

The HabEx telescope requirements (**Table 5.4-4**) are connected to all of the HabEx science objectives (**Table 5.1-2**), but the most optically demanding science, and accordingly the science that has the largest impact on defining the HabEx telescope design, comes from high contrast imaging with the coronagraph in Goal 1. Most of the telescope's requirements were defined by the coronagraph error budget, or early architecture trades and simulations of telescope/coronagraph performance.

The telescope primary mirror diameter requirement was set at greater than 3.7 m based on an early aperture verses yield trade study indicating that an aperture of that size is needed to reduce the probability of *not* characterizing an exo-Earth to below 0.5%. Smaller aperture options were also investigated to explore the performance space, which is detailed in *Chapter 10* as architecture trades.

Another important telescope design parameter is the *f*/number. A slower telescope (i.e., larger *f*/number) is longer and consequently heavier, more costly, and less stable. However, a faster telescope can degrade coronagraph performance by introducing polarization crosstalk, which impacts coronagraph contrast. A study was conducted to evaluate the minimum acceptable telescope *f*/number for HabEx (**Figure 5.4-1**). The trade indicated that *f*/2.25 or greater meets the contrast requirements for the vector vortex coronagraph (VVC) charge 6.

Coronagraphy prefers both unobscured apertures and monolithic primary mirrors (as opposed to segmented primary mirrors). In combination, these two telescope architecture choices eliminate all possible diffraction sources in the optics ahead of the coronagraph.

The spectral range of the telescope optics is set by the combined spectral requirements of all 17 science objectives. The need to carry out UV science set the minimum operating temperature for the telescope optics at 260 K. Telescopes at colder temperatures face significant contamination issues (e.g., Bolcar et al. 2016). Power considerations will discourage a telescope operating temperature much above this design requirement. For infrared astronomy the telescope is cool enough to allow operation out to 1.8 µm, with an extension to 2.5 µm being possible for some targets, before background from the warm optics becomes a limiting factor.

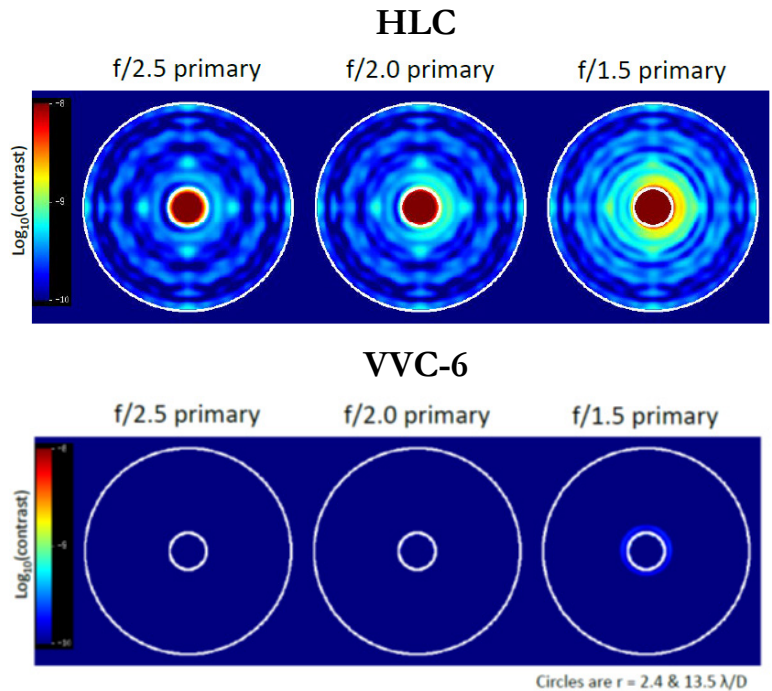

**Figure 5.4-1.** Simulated contrast vs. *f*/number for the vector vortex (VVC) and hybrid Lyot (HLC) coronagraphs over the 0.4–0.49 μm band. All simulations assume an HST-like aluminum coating with Mg-Fl overcoat on the primary and secondary mirrors.

Telescope throughput is set by shot noise requirements in the coronagraph error budget. In addition, the systematic noise branch of the error budget sets telescope mirror alignment requirements.

### 5.4.4    Coronagraph Requirements

The HabEx Coronagraph (HCG) requirements (**Table 5.4-5**) come from the coronagraph error budget (**Figure 5.2-1**) or directly from the STM. A spectral range covering at least 0.45–1.7 µm and a spectral resolution of at least $R = 100$ are necessary to ensure spectral detection and characterization of the most important spectral features. An IWA of at least 80 mas is needed to enable access to a sufficient number of habitable zones to meet the minimum exo-Earth yield. The HCG outer working angle (OWA) requirement is levied by the threshold requirements from Science Objectives 5, 6, and 7 in order to ensure that HabEx can characterize entire planetary systems as well as deliver science on exozodiacal dust.

Table 5.4-5. HabEx Coronagraph (HCG) requirements. *The HCG is designed to meet baseline requirements for Objectives 1, 2, & 16 and threshold requirements for Objectives 3–8 (Table 5.1-2). K represents a key requirement and D represents a driving requirement.

| Parameter | Requirement | K | D | Source |
|---|---|---|---|---|
| Spectral Range | ≤0.45 µm to ≥1.7 µm | ✓ | | STM |
| Spectral Resolution, $R$ | ≥5 (0.53–0.66 µm)* ≥50 (0.57 µm)* ≥40 (0.63 µm)* ≥35 (0.72 µm)* ≥70 (0.75–0.78 µm) ≥35 (0.82 µm)* ≥8 (0.8 µm)* ≥100 (0.87 µm)* ≥32 (0.88–0.91 µm)* ≥17 (0.94 µm) ≥11 (1.59 µm)* | ✓ | | STM O2, 3, 6 Threshold* |
| IWA (0.5 µm) | ≤80 mas | ✓ | ✓ | STM |
| OWA | ≥0.5 arcsec (0.5 µm)* ≥1.0 arcsec (0.8 µm)* | ✓ | | STM O5, 6, 7 Threshold* |
| Post-FSM LOS Stability | ≤0.3 mas RMS / axis | ✓ | | Error Budget |
| End-to-End Throughput | ≥20% | ✓ | | Error Budget |

The coronagraph error budget translates the contrast and contrast stability performance requirements on the end-to-end telescope flight system into a number of lower-level hardware requirements. The primary contrast-driven requirement solely addressed by the coronagraph is for a post-fine steering mirror (post-FSM) LOS stability to better than 0.3 mas per axis. This stability is necessary to keep the target star image sufficiently centered on the coronagraph's optical mask to meet required contrast levels. To reach the tight coronagraph LOS requirement, the coronagraph carries Zernike wavefront sensors (ZWFS) that can sense and correct LOS error on the order of 2.5 mas to less than 0.3 mas per axis using a FSM within the instrument.

Coronagraph throughput can have an impact on photometric noise and hence, can influence the overall contrast performance of the HCG. The error budget requires a minimum of 20% instrument throughput to meet minimum contrast performance.

### 5.4.5    Starshade Instrument Requirements

The payload functional requirements in the STM (**Table 5.1-2**) do not assume an instrument implementation. The HabEx baseline design has addressed these requirements by leveraging the strengths and complementarity of an internal coronagraph and an external starshade. The HCG is the ideal instrument to undertake the multiple visits required for an exo-Earth blind search and exoplanet orbit determinations. The SSI offers a wider field of view and bandpass for mapping and deep characterization of exoplanetary systems.

The SSI baseline requirements derived from the STM are shown in **Table 5.4-6**. The spectral range and resolution are specified in the STM's first eight science objectives. In principle, an unlimited OWA is possible for starshade observations, but in practice this is limited by engineering considerations. Only one SSI requirement is derived from the starshade error budget, which is end-to-end throughput, as all other requirements from the error budget are levied on the external starshade occulter and formation flight, described below.

**Table 5.4-6.** Starshade Instrument (SSI) Requirements. K represents a key requirement and D represents a driving requirement.

| Parameter | Requirement | K | D | Source |
|---|---|---|---|---|
| Spectral Range | ≤0.30 μm to ≥1.70 μm | ✓ | | STM |
| Spectral Resolution, *R* | ≥5 (0.3-0.35 μm)<br>≥40 (0.63 μm)<br>≥70 (0.75-0.78 μm)<br>≥8 (0.80 μm)<br>≥35 (0.82 μm)<br>≥100 (0.87 μm)<br>≥32 (0.89 μm)<br>≥17 (0.94 μm)<br>≥20 (1.06 μm)<br>≥19 (1.13 μm)<br>≥12 (1.15 μm)<br>≥10 (1.40 μm)<br>≥12 (1.59-1.60 μm)<br>≥10 (1.69-1.70 μm) | ✓ | | STM |
| OWA (0.5 μm) | ≥6 arcsec | ✓ | | STM |
| End-to-End Throughput | 22% | | ✓ | Error Budget |

### 5.4.6    UV Spectrograph Requirements

The UV Spectrograph (UVS) is designed for general purpose high-resolution UV imaging and spectroscopic observations. The UVS requirements (**Table 5.4-7**) come from the STM science Objectives 9 through 11 and Objective 17. The most demanding spectral resolution is set by the baryon science in Objective 9 (*Section 4.1*). This science involves mapping of the intergalactic medium and circumgalactic medium, which requires multi-object spectroscopy (MOS) over a modest-sized field of 2.5 × 2.5 arcmin².

**Table 5.4-7.** UV Spectrograph (UVS) Requirements. K represents a key requirement and D represents a driving requirement.

| Parameter | Requirement | K | D | Source |
|---|---|---|---|---|
| Spectral Range | ≤ 115 nm to ≥320 nm | ✓ | | STM |
| Spectral Resolution, *R* | Up to ≥60,000 depending on the measurement | ✓ | | STM |
| Angular Resolution | ≤300 mas | ✓ | | STM |
| FOV | ≥ 2.5 × 2.5 arcmin² | ✓ | | STM |
| Multi-object Spectroscopy | Yes | ✓ | | STM |

### 5.4.7    Workhorse Camera Requirements

The HabEx Workhorse Camera and spectrograph (HWC) requirements stem from Objectives 12 through 15 in the STM. The HWC requires a minimum 2.0 × 2.0 arcmin² field of view and a microshutter array to conduct MOS. The most demanding spectral resolution is set by the globular cluster science in Objective 14 (*Section 4.6*). The Hubble constant science in Objective 12 (*Section 4.4*) drives the photometric precision of the instrument. **Table 5.4-8** identifies the key HWC requirements.

### 5.4.8    Starshade Occulter & Flight System Requirements

The requirements on the starshade flight system (**Table 5.4-9**) flow from either the starshade contrast error budget (**Figure 5.2-2**) or the STM (**Table 5.1-2**). Specific spectral features addressing the exoplanet direct imaging science objectives in Objectives 2 through 8 in the STM set the SSI's spectral range. As with the HCG, an IWA of at least 80 mas at 1.0 μm is needed to enable access to a sufficient number of habitable zones. Both the IWA and the spectral band are tied to the size of the starshade and as such, are driving requirements.

As noted earlier, both the starshade system and the HCG error budgets start with the planet-to-star flux ratio requirement taken from the STM, then break it down to contrast and contrast stability requirements, which are further decomposed into lower-order requirements needed for the starlight suppression system component design. The error budget presented in **Figure 5.2-2** levies these additional requirements based on the overall need to meet a planet-to-star flux ratio of ≤10⁻¹⁰. Primary factors affecting the contrast performance are the external starshade occulter's shape accuracy and its position with respect to the target star. Allocations for the external occulter's edge scatter, surface reflectance, and micrometeroid damage are also identified in the budget. Meeting contrast levels constrains the external occulter's shape, and accordingly establishes a number of starshade external occulter petal and disk structural requirements. These precise manufacturing and thermal stability requirements largely scale with the diameter of the starshade occulter. The need for tight positioning of the starshade occulter during observations creates a set of translational requirements on the position of the occulter while formation flying (see **Table 5.4-10**). The starshade occulter tip-tilt with respect to the telescope's LOS is driven by engineering considerations, not contrast performance, so are not included in this section. Performance simulations have shown that the starshade occulter can be tilted by as much as 20° with respect to the telescope LOS without significant degradation in contrast performance. Light scattered from thruster exhaust from the starshade occulter spacecraft will significantly degrade contrast, so the starshade thrusters will not operate while collecting imaging or spectral data.

**Table 5.4-8.** HabEx Workhorse Camera (HWC) requirements. K represents a key requirement and D represents a driving requirement.

| Parameter | Requirement | K | D | Source |
|---|---|---|---|---|
| Spectral Range | ≤0.37 μm to ≥1.7 μm | ✓ | | STM |
| Spectral Resolution, *R* | Up to ≥1,000 depending on the measurement | ✓ | | STM |
| Angular Resolution | ≤25 mas | ✓ | | STM |
| FOV | ≥2 × 2 arcmin² | ✓ | | STM |
| Multi-Object Spectroscopy | Yes | ✓ | | STM |
| Noise Floor | ≤10 ppm | ✓ | | STM |

**Table 5.4-9.** Starshade Flight System Requirements. K represents a key requirement and D represents a driving requirement.

| Parameter | Requirement | K | D | Source |
|---|---|---|---|---|
| Observational Band | 0.30–1.7 μm | ✓ | ✓ | STM |
| IWA | ≤64 mas (0.87 μm) ≤80 mas (1.0 μm) | ✓ | ✓ | STM |
| Raw Contrast | ≤1 × 10⁻¹⁰ | ✓ | ✓ | STM |
| Raw Contrast Stability | ≤2 × 10⁻¹¹ | ✓ | | Error Budget |
| Formation Flying | See Table 5.4-10 | ✓ | | Error Budget |
| Pointing Control | ≤1° | ✓ | | MTM |
| Edge Scatter | V >25 mag/arcsec² | ✓ | | Error Budget |
| Sunlight Leakage | >32 V$_{mag}$ | ✓ | | Error Budget |
| Micrometeoroid holes | ≤500 ppm | ✓ | | Error Budget |
| Petal position (manufacture) | ≤±600 μm | ✓ | | Error Budget |
| Petal shape (manufacture) | ≤±140 μm | ✓ | | Error Budget |
| Petal position (thermal) | ≤±400 μm | ✓ | | Error Budget |
| Petal shape (thermal) | ≤±160 μm | ✓ | | Error Budget |

### 5.4.8.1 Starshade Formation Flight Requirements

The starshade must maintain its position within a designated radius of the telescope's LOS to the star by complying with the requirements outlined in **Table 5.4-10**. To maintain the necessary contrast levels for direct imaging Earth-sized planets in the habitable zone of nearby stars requires that the starshade stay within 250 km of the nominal separation distance from the telescope (for visible wavelengths), and within 1 m lateral displacement from the telescope-to-star sight-line. These requirements are derived from the starshade error budget.

**Table 5.4-10.** Formation Flight Requirements. K represents a key requirement and D represents a driving requirement.

| Parameter | Requirement | K | D | Source |
|---|---|---|---|---|
| LOS Separation Distance Accuracy | ≤± 250 km | ✓ | | MTM |
| Distance Sensing | ≤± 25 km | ✓ | | MTM |
| Lateral Displacement from LOS | ≤± 1 m | ✓ | | Error Budget |
| Lateral Displacement Sensing Accuracy | ≤± 0.3 m | ✓ | | Error Budget |

# 6 BASELINE TELESCOPE PAYLOAD AND BUS

The baseline HabEx 4 m observatory concept will be the largest, most stable, space telescope covering ultraviolet (UV), visible, and near-infrared (NIR) wavelengths ever built. With an unobscured 4 m diameter aperture, it is capable of collecting three times as many photons as the 2.4 m Hubble Space Telescope (HST). Its diffraction resolution limit is 25 milliarcseconds (mas) at 0.4 μm, compared to HST's 34.4 mas. HabEx is designed to be the most stable astronomical observing platform ever; capable of a routine telescope pointing stability of less than 1 mas compared with HST's best pointing stability of 2 mas (Nelan et al. 1998). Importantly, the HabEx telescope design is based on manufacturing capabilities and state-of-the-art telescope technologies presently available.

As summarized below, the baseline mission defines two flight elements: the HabEx telescope and starshade. This chapter details the HabEx telescope, its payload, and the supporting bus. *Chapter 7* details the starshade design and construction and its spacecraft bus. *Chapter 8* details the mission's concept of operations and design reference mission (DRM).

## 6.1 Baseline HabEx Mission Summary

The baseline HabEx mission is composed of two separate spacecraft flying in formation in an Earth-Sun L2 orbit. One spacecraft carries a 4 m off-axis telescope and four science instruments. The HabEx Coronagraph (HCG) and Starshade Instrument (SSI) perform exoplanet direct imaging and spectroscopy. The wide-field HabEx Workhorse Camera (HWC) and wide-field, high-resolution Ultraviolet imaging Spectrograph (UVS) are for general observatory science. The other spacecraft carries a 52 m starshade. With the starshade flying in formation on axis with the telescope, an external occulter observatory for exoplanet imaging and spectral characterization is formed. The starshade occulter suppresses the light from the target star while the telescope's SSI observes the planetary system surrounding the target star. To form this observatory, the starshade is positioned into the line of sight (LOS) between the telescope and the target star at a 76,600 km separation from the telescope, and maintains alignment using a positional control loop carried over a spacecraft-to-spacecraft radio link. Longer and shorter starshade ranges will also be used depending on the observing band. Starshade lateral and longitudinal position sensing are carried out by instrumentation on the telescope spacecraft and position control is handled by the propulsion system on the starshade spacecraft.

The telescope spacecraft is launched on a Space Launch System (SLS) Block 1B launch vehicle into a 780,000 km diameter orbit about Earth-Sun L2. The starshade will launch separately on a Falcon Heavy. The primary mission will run for five years, after which the telescope flight system will have enough propellant to continue operations for at least an additional five years. The starshade has propellant for at least five years of operations, after which it will remain at L2 until serviced. Serviceability is a formal requirement for all large astrophysics observatories. Both the starshade and telescope flight systems are able to be refueled and upgraded, however, the deployed starshade occulter cannot be replaced during servicing.

Exoplanet science observations are accomplished with two approaches. First, using an internal occulter, HCG. The second approach uses an external occulter, the starshade occulter, together with SSI, which is located on the telescope. These two instruments are complementary in nature. While SSI is capable of very high contrast imaging and spectroscopy over a large field of view (FOV), it is limited in the number of observations due to the large slew times of the starshade. The HabEx Coronagraph, on the other hand, is capable of faster slews, making many more observations possible, but has a narrower, high-contrast FOV with reduced spectrographic capability. Working together, the HCG performs the repeated planet detections required to determine planetary orbits, while the SSI achieves the high-resolution spectral profiles needed to characterize exoplanet atmospheric gases. In addition, the inclusion of imaging and spectroscopy capabilities within both instruments adds resiliency against both technical and programmatic risks. Loss of one exoplanet instrument does not eliminate the exoplanet science return of the mission since both HCG and SSI carry imaging and spectroscopy channels with similar capabilities. While SSI is better suited to spectroscopy and HCG to exoplanet searches, either can serve the purpose in the event of a failure in the other instrument. In addition, flexibility in the starshade development schedule is provided with the planned separate launches. Though not the baseline plan, a delayed starshade launch is possible and may still meet mission objectives. In that scenario, the HCG would carry out a planet detection survey in advance of the starshade's arrival at L2.

High contrast imaging with a coronagraph levies stringent requirements on many aspects of the telescope design. A coronagraph is most efficient working through an unobscured aperture. Without obscurations, the system throughput is maximized and diffracting edges within the pupil are avoided, improving the baseline contrast performance and reducing the control requirement placed on the deformable mirrors (DMs) as the incoming wavefront is corrected. Unobscured aperture coronagraph science yields are comparable to significantly larger diameter on-axis and segmented aperture yields. Accordingly, HabEx adopted an off-axis, unobscured monolithic telescope architecture to maximize coronagraph science yield while minimizing aperture diameter and telescope size and cost.

Maintenance of the coronagraph contrast performance also requires that the Optical Telescope Assembly (OTA) be ultra-stable in both internal LOS errors and optical surface figure errors (SFEs). These requirements derive from the coronagraph error budget and performance modeling discussed in *Section 5.2*. Laser metrology and control (MET) maintains the alignment of the secondary mirror (SM) and tertiary mirror (TM) assembly with respect to the primary mirror (PM), thus minimizing internally derived LOS error and resultant wavefront variation.

To ensure excellent LOS pointing and wavefront stability, the telescope includes precision thermal control, mirror positional control using laser metrology, and a fine guidance sensor system (FGS). Microthrusters are used for pointing control during observations. The use of microthrusters for fine attitude control avoids the introduction of high-frequency jitter typical of reaction wheel systems, which would be outside the control bandwidth of MET. Cold gas microthrusters are currently in use on the Gaia mission. Cold gas and colloidal microthrusters were previously used on the Laser Interferometer Space Antenna (LISA) Pathfinder mission. The alternative approach, reaction wheels and passive or active isolation at the levels of stability needed for high-contrast coronagraphy, has not yet been demonstrated in space nor on the ground. Microthrusters produce less than $0.3\,\mu N/\sqrt{Hz}$ noise power spectral density (PSD), which is as much as four orders of magnitude less than a reaction wheel. With such low jitter levels there is no need for vibration isolation between the payload and spacecraft. As shown in **Figure 6.1-1**, for a 100 Hz structure, picometer displacement levels are possible by limiting acceleration from noise sources to $10^{-6}\,g$ levels; while a 1 Hz structure will have ~10 $\mu$m of displacement.

The starshade spacecraft, detailed in *Chapter 7*, includes the 52 m deployable starshade occulter payload, solar electric propulsion (SEP)

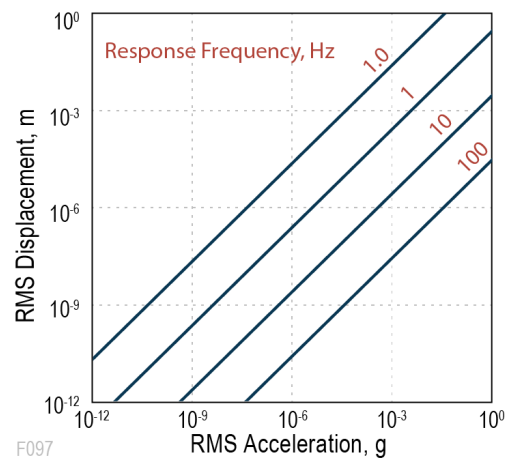

**Figure 6.1-1.** HabEx's use of microthrusters results in an ultrastable platform for coronagraph use. The figure shows the displacement of a structure versus acceleration for different structural frequencies. Adapted from Lake et al. (2002).

to move the starshade from target to target, a bipropellant propulsion system to hold alignment when observing, and a formation flying beacon and communication link. There are no science instruments on the starshade spacecraft. The starshade is spin-stabilized, rotating about the line of sight axis at a rate of 0.33 revolutions per minute (rpm).

## 6.2    Telescope Payload

The payload on the HabEx telescope flight system is comprised of the OTA and the four science instruments. The baseline telescope, shown in **Figure 6.2-1**, consists of the PM assembly, SM assembly, secondary mirror tower with integrated science instrument module, and stray-light tube with forward scarf. The scarf angle of 40° determines the closest angle of observation to the Sun. The tower and baffle tube are the optical bench, which maintains alignment between the PM, SM, and TM assemblies. The OTA is physically separate from the spacecraft, which includes the solar array sunshield.

The OTA and spacecraft connect only at the interface ring. This ring is also the interface between the payload and the SLS. The advantage to this approach is that the spacecraft does not carry the payload mass while on the ground or during launch. The telescope aperture cover is closed for launch to prevent contamination and to provide launch stiffness. In addition, launch locks connect the spacecraft solar panels to the tube to provide them with additional stiffness for launch.

Instruments are laterally mounted (i.e., mounted along the telescope's barrel between the PM and SM), rather than mounted behind the PM as is typical for on-axis telescopes. Lateral mounting facilitates instrument servicing and reduces the flight system's stack height and solar torque.

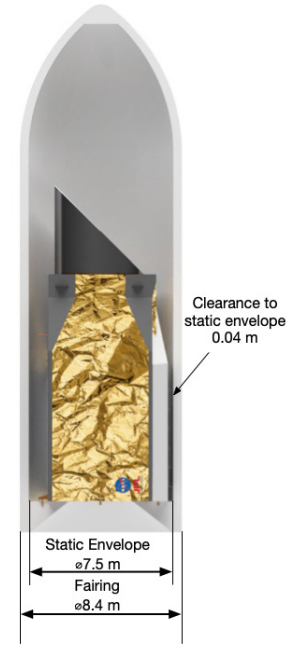

**Figure 6.2-2**. The HabEx telescope flight system fits in SLS 8.4 m fairing with a clearance of 0.04 m to the static envelope.

The baseline observatory is designed for the SLS Block 1B mass and volume capacities, and launch environment (Stahl et al. 2016; NASA 2018). As demonstrated in **Figure 6.2-2**, the flight system and payload fit inside the 8.4 m fairing of the SLS Block 1B Cargo. The telescope and spacecraft structure are designed to have a first

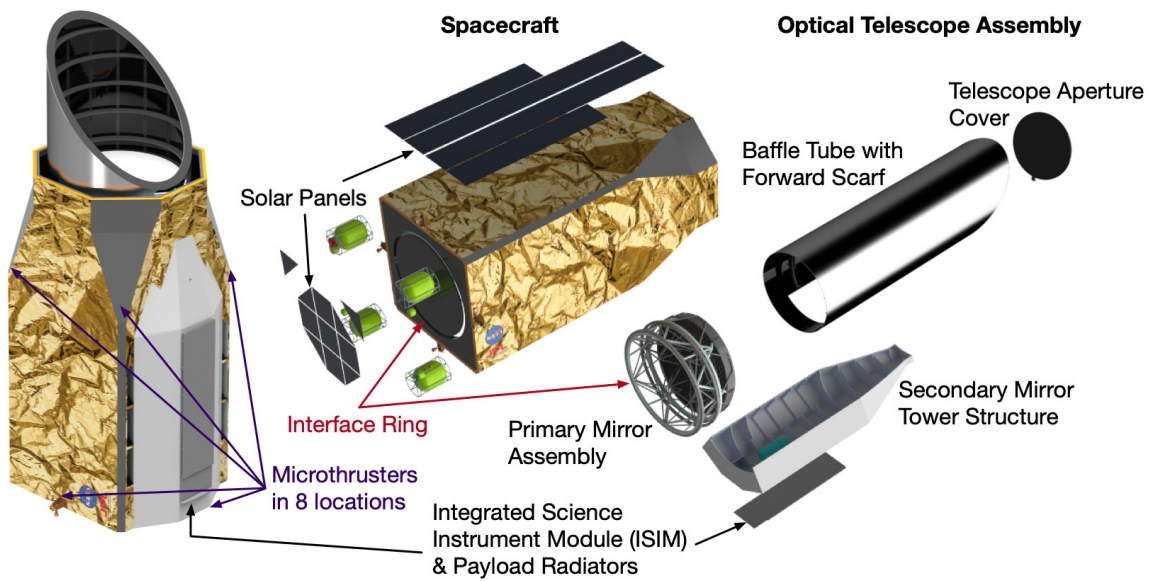

**Figure 6.2-1**. An exploded view of the baseline HabEx telescope flight system and its payload.

**Table 6.2-1.** Key HabEx telescope system baseline requirements compared to expected performance, based on Table 5.4-4. *Note: Please see *Appendix H* for the definition of margin used in this report.

| Parameter | Requirement | Expected Performance | Margin* | Source |
|---|---|---|---|---|
| *f*/number | ≥2.25 | 2.5 | Met by design | MTM |
| Obscured or Unobscured | Unobscured | Unobscured | Met by design | MTM |
| Monolithic or Segmented PM | Monolithic | Monolithic | Met by design | MTM |
| Angular Resolution (0.4 µm) | 50 mas | 25 mas | 100% | STM |
| PM Diameter | ≥3.4 m | 4.0 m | 8.1% | MTM |
| Bandpass | ≤0.115 µm to ≥1.7 µm | 0.115–2.5 µm | Met by design | MTM |

vibrational mode higher than 10 Hz to survive a 3.5 *g* axial and 1.5 *g* lateral launch load.

### 6.2.1 Optical Telescope Assembly

The SLS Block 1B Cargo lift capability is more than enough to allow for a multi-ton, monolithic primary mirror. A monolithic primary mirror is considerably less complex than a segmented telescope system, requiring fewer parts while reducing risk and potentially cost.

The HabEx telescope is a three-mirror anastigmat (TMA) design with a 4 m diameter primary mirror, and a secondary mirror that is positioned 2.5 m off-axis. Robb's method (Robb 1978) was used to obtain the initial parameters for the OTA design and the result was optimized in Zemax® to produce a collimated 50 mm beam at the output. The TMA design enables minimization of the three principal optical aberrations (spherical, coma, astigmatism). In addition, field curvature can be brought to zero by choosing appropriate mirror figures. Such a design provides a well-corrected wavefront over a wide FOV, allowing the science instruments to operate simultaneously in different areas of the sky.

**Table 6.2-1** shows the OTA design requirements and performance, **Table 6.2-2** identifies OTA design parameters, and **Figure 6.2-3** shows the telescope's optical layout. A 4 m PM directs light to the SM then onto the TM. The primary mirror has a relatively long *f*/number (which is the ratio of the OTA focal length to the PM diameter) of 2.5, in order to control polarization mixing for coronagraphy, as described in *Section 6.8.*

**Table 6.2-2.** The HabEx telescope is a three-mirror anastigmat (TMA) design and this table defines the dimensions of the primary (PM), secondary (SM), and tertiary (TM) mirrors.

| Optic | PM | SM | TM |
|---|---|---|---|
| Diameter | 4,000 mm | 450 mm | 680 mm |
| Radius of curvature | 19,800 mm | -1,953 mm | -2,168 mm |
| Thickness | 320 mm | 100 mm | 100 mm |
| Spacing to next optic | 9,030 mm | 9,080 mm | N/A |
| Conic constant | -1.00 | -1.55 | -0.99 |
| Coating | Protected Al | Protected Al | Multiple coatings |

The science and fine guiding instruments are arranged near the TM.

In the annular field TM design, the Cassegrain focus, formed by the PM and SM, is spatially larger than the exit pupil, formed by all three mirrors, and this results in widely separated ray bundles for each field on the TM. In addition, the design permits baffling of stray light from the telescope tube and from space by placing an aperture plate at the Cassegrain focus, reducing exposure to stray light for all the downstream instruments. In a three-mirror design, the output beam is directed towards the SM, so fold mirrors are normally used to bring the beam back behind the PM or TM. By placing a mirror between the TM and the exit pupil, the ray bundle from an individual field can be extracted and passed to the appropriate instrument. This design also allows different optical coatings on different areas of the TM, or separate tertiary mirrors with instrument-specific coatings, to aid transmission efficiency for some instruments. Since the telescope supports UV observations, a protected aluminum coating is required on at least the PM and SM. Sensitivity to mirror contamination in the UV drives the operating temperature of the PM. A temperature that is too low will result in the condensation of contaminants onto the PM surface, which will significantly reduce the UV throughput.

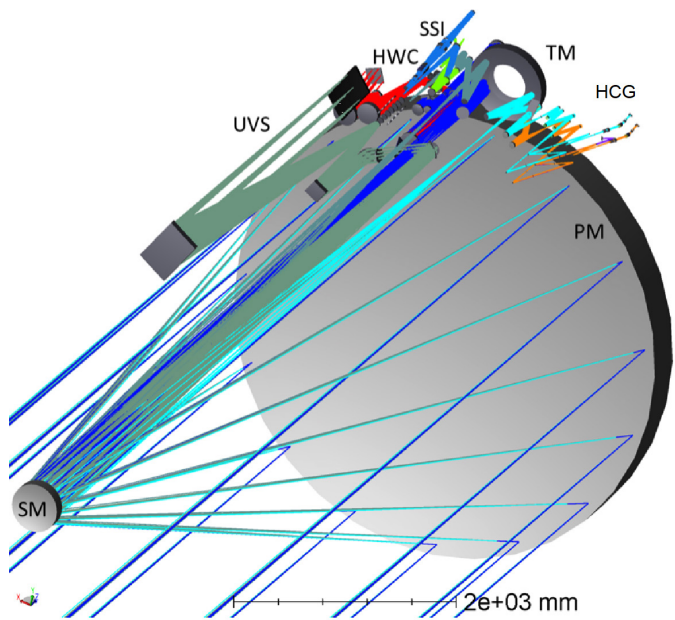

**Figure 6.2-3.** The HabEx telescope optical layout showing the primary mirror (PM), secondary mirror (SM), and tertiary mirror (TM). The four science instruments are mounted laterally along the barrel of the telescope between the PM and SM. The science instruments are the UV Spectrograph (UVS), Workhorse Camera (HWC), Starshade Instrument (SSI), and Coronagraph (HCG), respectively.

The instruments are laterally mounted, arranged on the side of the telescope near the TM as shown in **Figure 6.2-3**. Compared to mounting instruments behind the PM, the side mounting allows easier extraction of the instrument modules for servicing and takes advantage of existing volume created by the presence of the TM beside the PM. Furthermore, mounting the instruments laterally provides easy access to externally mounted radiators, which are needed to cool the detectors. Mounting the instruments behind the PM would make use of space available between the supporting structure and the bus.

However, this would increase the overall telescope height, which would make cooling paths longer and the extraction of the instrument modules, for the purpose of servicing, more complex.

**Figure 6.2-4** shows the HabEx instruments arranged near the TM with the rays from the SM traveling from left to right in the image. The instrument FOVs become separated at the TM. **Figure 6.2-5** shows the on-sky fields of view of the instruments. As can be seen in the figure, the UVS occupies the center of the field and the HCG views a smaller field to the left. The SSI views a region to the right and slightly upwards, while the HWC views a wide area to the right and slightly downwards. Arranged around the rest of the TM are the four FGS sensor areas.

After striking the TM, the light beams are collimated and converge towards a common on-axis pupil plane. However, prior to reaching that plane the beams are extracted by fold mirrors and directed to steering mirrors at the pupil plane that direct the light into the instruments. An exception is the UVS, which has its own tertiary mirror and the beam is extracted prior to the Cassegrain focus using a fold mirror.

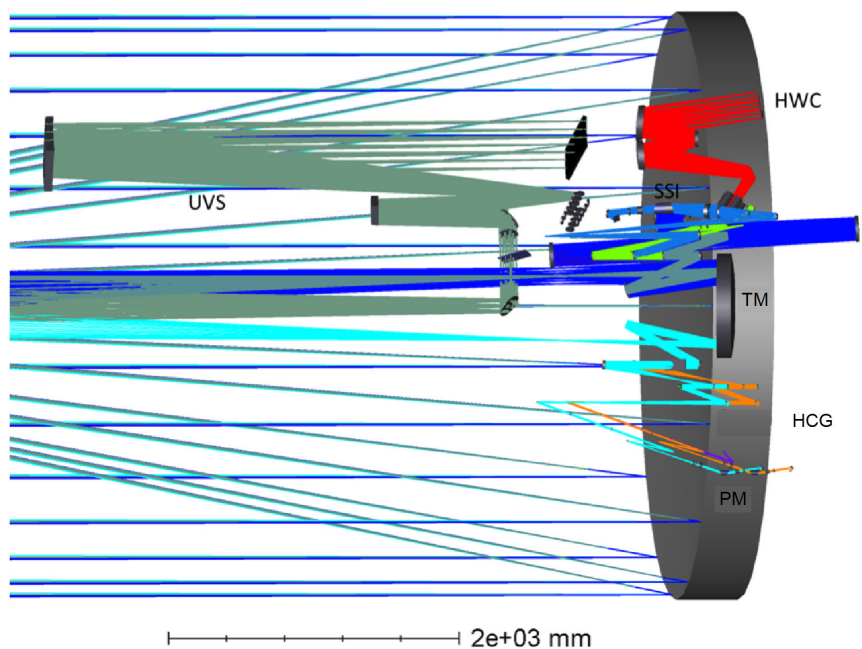

**Figure 6.2-4.** Optical paths of the HabEx instruments: the UVS, HWC, SSI, and HCG. Not shown: the FGS.

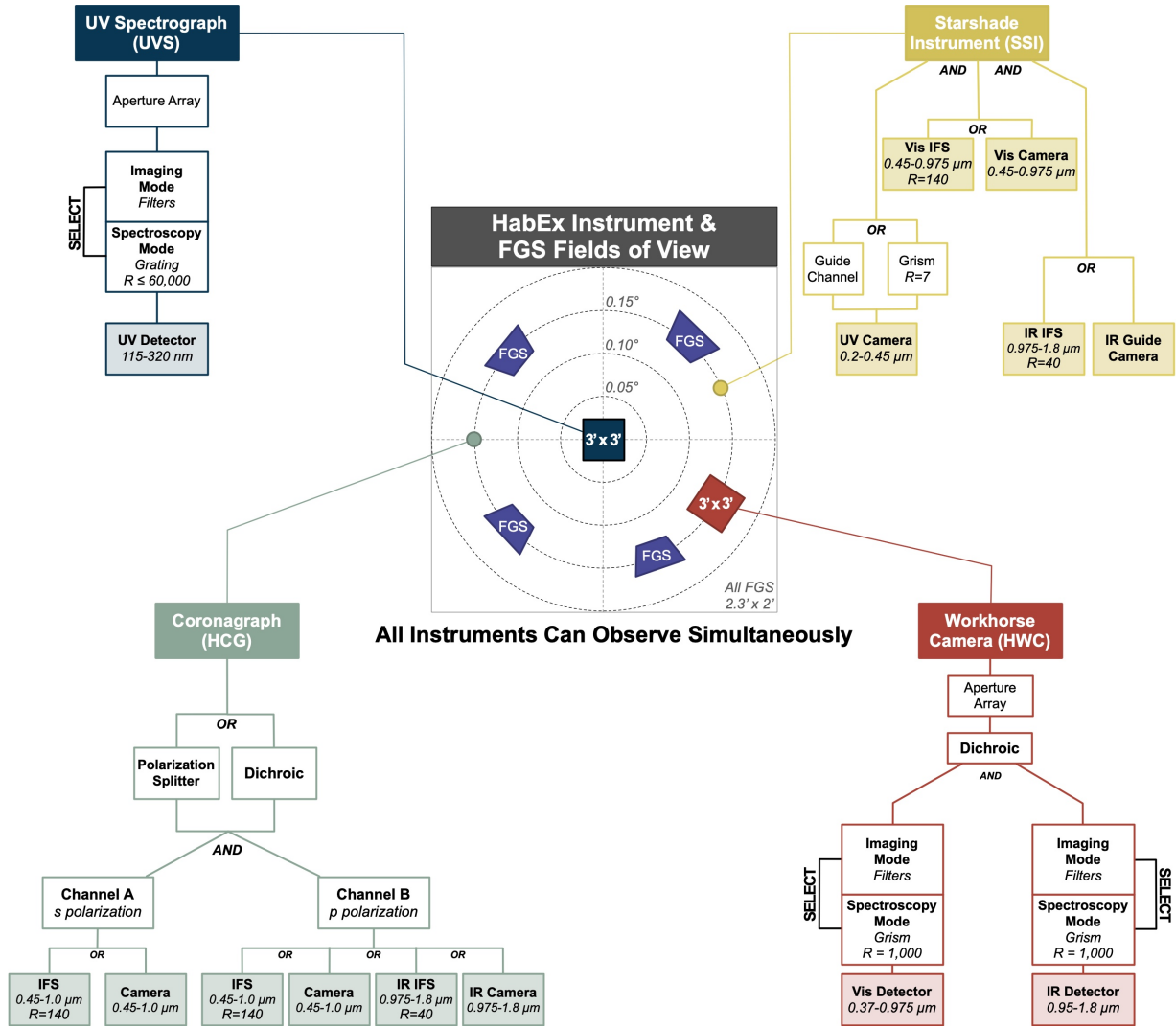

**Figure 6.2-5.** HabEx's four science instruments have multiple focal planes that cover wavelengths from the UV to the near-IR for both imaging and spectroscopy. This configuration enables parallel observations using the UVS and HWC while the exoplanet direct imaging instruments are observing, resulting in high observatory utilization. The lightly shaded boxes identify the focal planes. The FOVs for the four fine guidance sensors in the FGS) are denoted in violet in the central image.

### 6.2.2    On-Orbit Optical Alignment

Space telescopes are not launched in their final optical alignment. The mirrors in the OTA are retracted and held with launch restraints, which are released on-orbit so that the mirrors may be deployed and aligned. For example, once on orbit, the James Webb Space Telescope (JWST) primary mirror segment assemblies will be deployed 12.5 mm from their launch restraints to reach their operational positions (Chonis et al. 2018). Space telescopes are also not launched in their operational thermal state, but rather are launched "warm," nominally at 300 K, and cooled to their operational temperature once on-orbit.

Because of structural material coefficient of thermal expansion (CTE) this thermal change alters the optical position of the mirrors on the structure. The integrated thermal model predicts that structural changes will result in PM and SM rigid body displacements of 250–500 µm and 25–50 µm, respectively. In addition, the PM and SM are expected to experience rotations on the order of 20–50 µrad with changes in temperature.

To correct these displacements, an on-orbit sensing and positioning system establishes and maintains alignment between the primary, secondary and tertiary mirrors. Hexapod actuator systems on the primary and secondary mirrors, with a 20 mm range and sub-nanometer resolution, will deploy the optics to their initial positions. Changes to these positions to achieve final alignment is determined by the internal MET system. The MET system also senses and corrects mirror positions in response to slow thermal drifts. Finally, optical alignment can be confirmed by performing multi-field phase retrieval using UVS.

Low wavefront error (WFE) and high LOS stability requires an ultra-stable optomechanical telescope, which is described in detail in *Section 6.7*. The baseline telescope architecture achieves the required performance (**Table 6.2-1**) with the aid of the mass and volume capacities of the planned SLS Block 1B Cargo. The SLS mass capacity enables the design of an extremely stiff opto-mechanical structure that can align the PM, SM, and TM to each other and maintain that alignment. Furthermore, the SLS Block 1B volume capacity enables the use of a monolithic aperture off-axis primary mirror with no deployments.

The telescope structure is the foundation for the entire observatory, being the optical bench to which the optical components and science instruments are attached. It has the critical function of placing the primary, secondary, and tertiary mirrors at the locations specified by the optical design and keeping them at those locations with sufficient stability to meet the required performance specifications. This is accomplished both by making the structure as stiff as possible and by minimizing the disturbances to which the structure is exposed. As shown in **Figure 6.2-1**, the telescope is a standalone structure separate from the spacecraft. The spacecraft surrounds the telescope without physically touching it except at the interface ring, which also connects to the launch vehicle payload adapter fixture (PAF). This configuration minimizes the propagation of mechanical disturbances from the spacecraft into the telescope and provides thermal shielding of the telescope while minimizing heat leaks. Removing portions of the spacecraft's anti-sun structure that flank the detector radiators provides sufficient radiative cooling of the baffle tube. While the primary and secondary mirrors have active thermal control, the structure does not because the telescope includes a laser truss system which maintains alignment between the primary, secondary and tertiary mirrors. The laser truss senses and corrects slow thermal drifts and its noise uncertainty is sufficient to meet the LOS Jitter and LOS WFE stability specifications.

### 6.2.3 Instrument Payload

The telescope scientific payload consists of the four science instruments (**Figure 6.2-5**) plus ancillary equipment. The science instruments are the Coronagraph (HCG; *Section 6.3*), the Starshade Instrument (SSI; *Section 6.4*), the UV Spectrograph (UVS; *Section 6.5*), and the Workhorse Camera (HWC; *Section 6.6*).

The ancillary optical payload equipment consists of the metrology system (MET; *Section 6.8.5*) and a fine guidance sensor system (FGS; *Section 6.8.6*) to maintain the extreme pointing stability desirable for exoplanet observations. In addition, a passive cooling system is used to maintain specified temperatures at the focal planes (*Section 6.7*).

## 6.3 HabEx Coronagraph (HCG)

Coronagraphy was invented in the 1930s by French astronomer Bernard Lyot (Lyot 1939) to study the solar corona outside of natural eclipses. The Lyot coronagraph consists of a sequence of stops aimed at blocking the light diffracted from a central bright object (the Sun or a star), and allowing off-axis objects (the solar corona or exoplanets) to pass through. In its most basic form, the Lyot coronagraph consists of a hard edge circular occulter in the focal plane aligned with the central bright source, and the Lyot stop, which is a diaphragm in the relayed pupil plane downstream from the occulter. The occulter will physically block some of the sunlight or starlight and diffract almost all of the rest so that it propagates into a cone with a dark region at the core. The light is stripped off by a physical block at the Lyot stop (**Figure 6.3-1**). Off-axis light bypasses the center

**Vector Vortex Coronagraph Architecture**

For HabEx, four main coronagraph families have been considered: the shaped pupil (SP) and apodized pupil Lyot coronagraph (APLC), the phase-induced amplitude apodization complex mask coronagraph (PIAACMC), the hybrid Lyot coronagraph Trauger et al. 2016, and the vector vortex coronagraph (VVC; Mawet et al. 2005; Foo et al. 2005). The VVC family was found to present the most favorable trade-off between inner working angle (IWA), optical efficiency, and immunity to low-order aberrations. For this reason, the VVC was selected as the baseline coronagraph for HabEx.

The VVC is a special form of a more general vortex phase mask. The vortex coronagraph is a phase-based coronagraph that imprints a phase screw dislocation of the form $e^{in\theta}$ on the Airy diffraction pattern at the instrument focus, where $\theta$ is the azimuthal coordinate in the focal plane and $n$ is an integer number called the topological charge. The topological charge quantifies the number of times the phase ramp goes through a full $2\pi$ radian cycle. In the case of the vector vortex design, the phase screw is achieved by forming a waveplate with an azimuthally rotating principal axis. When the star is centered on the VVC mask, the screw dislocation forced upon the electric field generates a singularity or optical vortex. As the phase rotates rapidly around the singularity, the electric field traveling in the forward direction interferes across the vortex and sums to zero, creating a dark hole. Further off-axis, the electric field interferes constructively across the vortex creating a cone of light spreading outwards. As the light propagates to the downstream Lyot stop, the dark hole grows to fill the entire pupil and the surrounding cone of starlight is intercepted by the stop. Light from an off-axis source, such as a planet, falls on one side of the singularity and receives (to first order) only a uniform phase shift and therefore propagates as normal, passing through the Lyot stop.

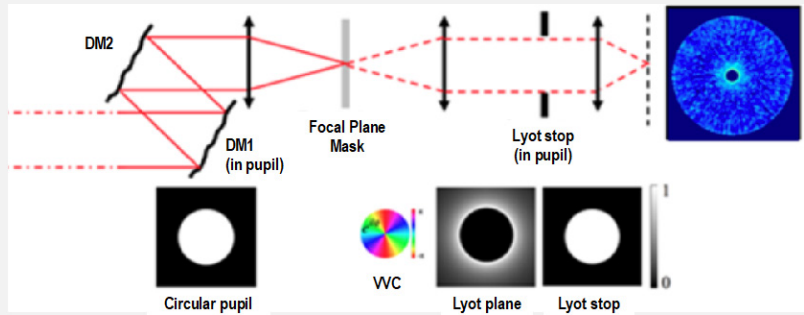

**Figure 6.3-1.** A simplified coronagraph layout assuming a vector vortex coronagraph (VVC) focal plane mask.

Achromatic vector vortex coronagraph masks have been made (Serabyn et al. 2019) and are relatively well-developed. These masks are used with a circularly polarized input beam; the input polarization is transmitted with the opposite circular polarization and mask errors result in a component of the output with the original circular polarization. The output can be cleaned up with a downstream polarizer, improving the contrast. For this reason, current VVCs employ a single input polarization. In future, it may be possible to make high-quality scalar VVC masks that would enable dual polarization operations, but HabEx has not assumed this. Mawet et al. (2005) demonstrated that perfect starlight rejection within the downstream geometric area can be achieved with an unobscured circular entrance aperture and VVCs of even numbered topological charges. Moreover, the choice of topological charge affects the propagation of low-order optical aberrations such as tilt and defocus (Mawet et al. 2010b; Ruane et al. 2017). Indeed, the higher the charge, the lower the sensitivity to low-order aberrations, but the larger the IWA.

The VVC has been used on several ground-based adaptively corrected telescopes. Its small IWA, layout simplicity, potential for achromaticity (as shown in liquid crystal vortices) and high throughput makes it an attractive solution for high contrast imagers. VVCs are currently in operation at Palomar (Serabyn et al. 2010; Mawet et al. 2010a; Mawet et al. 2011; Bottom et al. 2015; Bottom et al. 2016), the Very Large Telescope (VLT; Mawet et al. 2013; Absil et al. 2013), Subaru (Kühn et al. 2017), Keck (Absil et al. 2016; Serabyn et al. 2017; Mawet et al. 2017; Reggiani and TEAM 2017; Ruane et al. 2017; Guidi et al. 2018; Xuan et al. 2018; Mawet et al. 2019; Ruane et al. 2019), and the Large Binocular Telescope (LBT; Defrere et al. 2014).

of the mask and is propagated though the Lyot stop.

The principal challenges with coronagraphy and the associated mask design are the same regardless of type: optical efficiency, contrast achievable, and achromaticity. In association with a telescope, other factors may come into play such as performance characteristics under the influence of small input wavefront disturbances.

This section describes the HCG design, its performance, and its operation.

### 6.3.1 Design

A vector vortex coronagraph (VVC) design is shown schematically in **Figure 6.3-1** and the complete HCG schematic is shown in **Figure 6.3-2**. Collimated light from the telescope's TM enters the coronagraph instrument, with the first optic being a fine-steering mirror (FSM) placed at a pupil image plane. The FSM is used for pointing control and, if required, jitter suppression within the instrument. The collimated beam is then passed to a pair of DMs for wavefront correction. Following the DMs, the beam passes through polarizers and is focused using a parabolic mirror. A coronagraphic mask is placed at the focal point. The mask design will have a coating that will reflect out-of-band light to the Zernike wavefront sensor (ZWFS). The coating has a very small central structure, illustrated in **Figure 6.3-3**, that causes phase across the wavefront to be detected as an amplitude change. The ZWFS is used in conjunction with the DMs and FSM to sense and correct tip, tilt, and focus drift. Light transmitted through the mask is recollimated with a second parabolic mirror and reaches the Lyot stop located at the exit pupil. The Lyot stop blocks light at the perimeter of the beam, i.e., the rejected starlight. Finally, the light enters either a camera or an integral field spectrograph (IFS) where imaging and spectral measurements are completed.

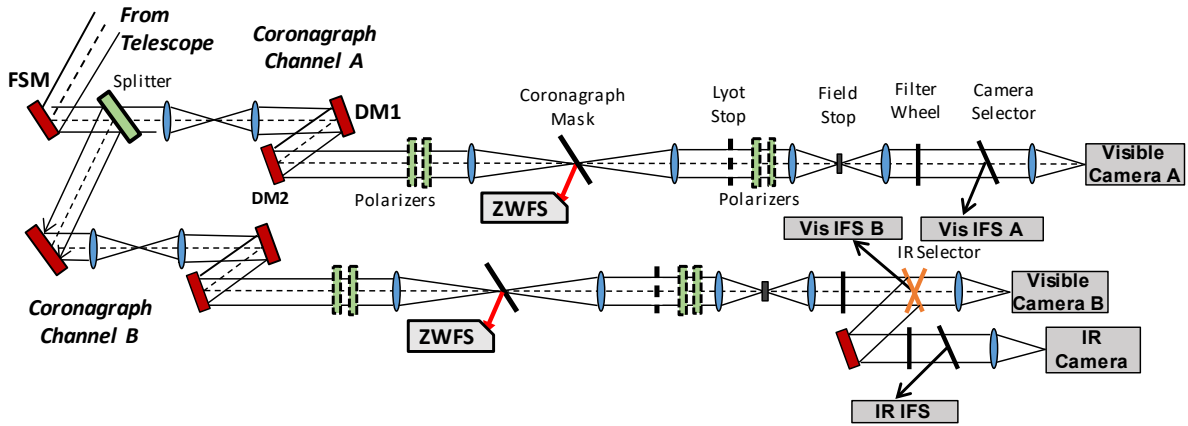

**Figure 6.3-2.** This HabEx coronagraph instrument (HCG) schematic shows the two-channel design with nearly identical channels. Channel B carries an IR integral field spectrograph (IFS) and an IR camera in addition to reproducing the equipment in Channel A. This identical two-channel design at visible wavelengths permits simultaneous dual polarization measurements.

**Table 6.3-1** shows the coronagraph design requirements and the HCG expected performance and margin, and **Table 6.3-2** shows the HCG functional design parameters. To provide the required wavelength range, the HCG has both visible and IR detectors. To provide the required imaging and spectral functions, the HCG includes cameras and IFSs.

Two near-identical coronagraph channels are specified within the coronagraph instrument to enable observing efficiency and these are identified as Channel A and Channel B. Both channels view the exact same field, shown in **Figure 6.3-2**. Depending on the observation strategy, light entering the instrument is split into two bands in different ways. In the first observational strategy, a polarization beam splitter transmits vertically polarized light to Channel A and reflects horizontally polarized light to Channel B. Within the two channels, dichroic filters set the optical bandwidth to 20% utilizing the same shortest blue filters as shown in **Table 6.3-2**. This strategy is preferred for planet detection at the smallest inner working angles. In

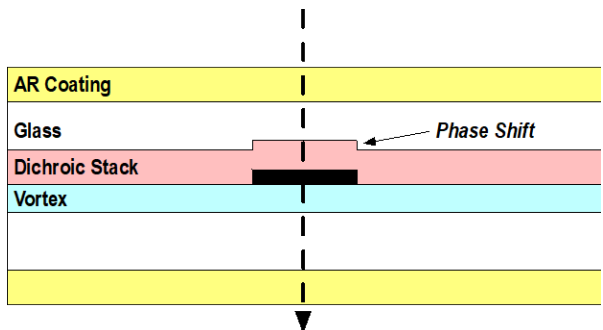

**Figure 6.3-3.** The HabEx vortex mask structure. Light, *dashed line*, enters from the top. Out-of-band light reflects off the dichroic coating to the Zernike wavefront sensor (ZWFS) detector. A ¼ wave phase step provides the reference signal. The black dot, which has a diameter of λ times the *f/#* (which is 2.5 for HabEx), masks the central defect of the vortex.

 

**Table 6.3-1.** Key HCG design requirements, based on Table 5.4-5. *Note: HCG is designed to meet baseline requirements for Objectives 1, 2, & 16 and threshold requirements for Objectives 3–8 (Table 5.1-2).

| Parameter | Requirement | Expected Performance | Margin | Source |
|---|---|---|---|---|
| Spectral Range | ≤0.45 μm to ≥1.7 μm | 0.45–1.8 μm | Met by design | STM |
| Spectral Resolution, *R* | ≥5 (0.53–0.66 μm)*<br>≥50 (0.57 μm)*<br>≥40 (0.63 μm)*<br>≥35 (0.72 μm)*<br>≥70 (0.75–0.78 μm)*<br>≥35 (0.82 μm)<br>≥8 (0.8 μm)*<br>≥100 (0.87 μm)*<br>≥32 (0.88–0.91 μm)*<br>≥17 (0.94 μm)<br>≥11 (1.59 μm)* | 140 (0.45–1.0 μm)<br>40 (0.975–1.8 μm) | Met by design | STM<br>O2, 3, 6<br>Threshold* |
| IWA (0.5 μm) | ≤80 mas | 62 mas | 29% | STM |
| OWA | ≥0.5 arcsec (0.5 μm)*<br>≥1.0 arcsec (0.8 μm)* | 0.83 arcsec (0.5 μm)<br>1.32 arcsec (0.8 μm) | 66% (0.5 μm)<br>32% (0.8 μm) | STM<br>O5, 6, 7, 8<br>Threshold* |
| Post-FSM LOS Stability | ≤0.3 mas RMS / axis | 0.2 mas RMS / axis | 200% | Error Budget |
| Instrument Throughput | ≥20% | 27% | 35% | Error Budget |

the second observational strategy, a polarization beam splitter transmits vertically polarized light to Channel A and reflects horizontally polarized light to Channel B. Within the two channels, dichroic filters set the optical bandwidth to 20% utilizing one short blue filter and one redder filter set as shown in **Table 6.3-2**. This approach allows for a faster retrieval of spectra.

Each coronagraph channel carries a camera and an IFS, selected by inserting a mirror. An IR channel in Channel B is selected similarly. It carries an IR fiber spectrograph with $R = 40$ and covers the range 0.975–1.8 μm in three 20% wavelength bands. An IR linear mode avalanche photodiode detector (LMAPD) is used on this channel.

Two DMs enable the correction of both the wavefront phase and amplitude. The first DM is placed at a pupil plane (correcting phase) and the second is normally placed ¼ Talbot length away (correcting amplitude). The second mirror distance has been reduced to ⅛ Talbot length because, in the presence of DM surface errors discussed in *Section 6.3.3.1*; this improves the wavefront correction across the spectral band at a small expense in correction efficiency. Following the coronagraph mask, small wavefront imperfections arising from upstream optical elements and mask defects interfere to create a wavelength dependent speckled light field at the focal plane. While a perfect correction of the wavefront error across the whole focal plane and across all wavelengths is not possible, the DMs are used to create a "dark hole" over a restricted annular area of the focal plane. The angular size of the dark hole depends on the number of DM actuators, as does the degree of correction and hence the achievable contrast.

**Table 6.3-2.** The HCG Channel A and B specifications. Note that Channel B has identical specifications to Channel A over the 0.45–1.0 μm band.

| Camera Channels | | |
|---|---|---|
| | Visible<br>(Channels A & B) | IR<br>(Channel B only) |
| FOV full-width | 1.5–2.7 arcsec | 3.1 arcsec |
| Wavelength Bands (μm) | 0.45–0.55<br>0.55–0.67<br>0.67–0.82<br>0.82–1.00 | 0.975–1.8 |
| Pixel Size | 11.6 mas | 29.9 mas |
| Telescope Resolution | 23 mas (at 0.45 μm)<br>42 mas (at 0.82 μm) | 49 mas (at 0.95 μm) |
| IWA (2.4 λ/D) | 56 mas (at 0.45 μm)<br>102 mas (at 0.82 μm) | 118 mas (at 0.95 μm) |
| OWA | 0.74–1.35 arcsec | 1.57 arcsec |
| Detector | 1×1 CCD201 | 1×1 LMAPD |
| Array width | 1024 × 1024 | 256 × 320 |
| Spectrograph Channels | | |
| | Visible<br>(Channels A & B) | IR<br>(Channel B only) |
| FOV | 1.5–2.7 arcsec | 3.1 arcsec |
| Spectral Resolution, *R* | 140 | 40 |
| Spectrometer Type | IFS | IFS |
| Detector Configuration | 1/4 CCD282 (EMCCD) | 2×2 LMAPD |
| Array Width | 2048 px | 2048 px |
| Other | | |
| Deformable Mirror | 2 mirrors<br>64×64 elements<br>0.4 mm pitch | |
| Polarization | Vertical (A channel)<br>Horizontal (B channel) | Horizontal (primarily)<br>Vertical (possible) |

The outer working angle (OWA) is a function of the number, $N$, of actuators across the DM and is given by $N\lambda/2D$, where $D$ is the diameter of the telescope and $N$ in this case is equal to 64. For the HCG, the OWA is 740 mas at 0.45 μm. With this OWA, the coronagraph can only meet the threshold performance on some of the HabEx science objectives, so the starshade is required to meet those baseline requirements. It should be noted that the OWA of a coronagraph instrument increases linearly with the number of actuators and as larger DMs become available in the future, access to larger OWAs will be facilitated. Note that HabEx has 1.8 times the number of actuators planned for the Coronagraph Instrument (CGI) on NASA's Wide Field Infrared Survey Telescope (WFIRST).

The coronagraph channels, depicted schematically in **Figure 6.3-2**, follow a similar layout to the WFIRST coronagraph design, while attempting to minimize the number of mirrors needed so as to maintain optical throughput. A more detailed engineering layout is shown in **Figure 6.3-4**. Following the common fine-steering mirror, the HCG A and B channels are separated in collimated space by a selectable dichroic beam splitter. An initial relay of two off-axis parabolic mirrors sets the magnification to place the pupil on the DM. Telecentricity is not preserved at the entrance but is restored after the relay so that DM1 is positioned at a pupil plane. Following a fold, the beam strikes DM2 and is then focused onto the coronagraphic mask. Light reflected by the mask is directed to a Zernike wavefront sensor. Following recollimation, the beam aperture is reduced at the Lyot stop. After the Lyot stop the light can be directed via selector mirrors to the IFS, to the camera, or in the case of the red channel, to the IR camera/spectrograph. A filter wheel allows selection of the appropriate wavelength bands as shown in **Tables 6.3-3, 6.3-4, 6.3-5,** and **6.3-6**. The filters shown in **Tables 6.3-4** and **6.3-5** are specifically chosen to reveal certain types of exoplanets in shorter exposure times than can be achieved with the IFS, by the method outlined in Krissansen-Totton et al. (2016b). **Table 6.3-6** shows the filters specific to the IR mode in Channel B. Polarizers are included in the beam train to allow operation with the vector vortex masks and also to allow selection of polarized light from the science targets, for example during disk imaging.

The optical throughput of the HCG is shown in **Figure 6.3-6**. Maximizing throughput from the large number of reflections inside the instrument is managed by specifying all protected silver surfaces for the mirrors (except for the telescope PM and SM, which are protected aluminum). At 0.7 μm, the transmittance for the camera mode is ~20% and the IFS mode is ~12%.

**Table 6.3-3.** The HCG visible channel filter set.

| Band # | Wavelength Start (μm) | Wavelength End (μm) | Bandwidth (%) |
|---|---|---|---|
| 1 | 0.45 | 0.55 | 20 |
| 2 | 0.495 | 0.605 | 20 |
| 3 | 0.585 | 0.715 | 20 |
| 4 | 0.7 | 0.86 | 20 |
| 5 | 0.82 | 1.0 | 20 |

**Table 6.3-4.** The HCG includes a filter set in the optimal photometric bands for identifying Earth-like exoplanets (Krissansen-Totton et al. 2016a).

| Band # | Wavelength Start (μm) | Wavelength End (μm) | Bandwidth (%) |
|---|---|---|---|
| 6 | 0.431 | 0.531 | 20 |
| 7 | 0.569 | 0.693 | 20 |
| 8 | 0.77 | 0.894 | 20 |

**Table 6.3-5.** The HCG narrowband filter set, which will be used for giant planet color characterization (Krissansen-Totton et al. 2016b).

| Band # | Wavelength Start (μm) | Wavelength End (μm) | Bandwidth (%) | Comments |
|---|---|---|---|---|
| 9 | 0.45 | 0.5 | 10 | Rayleigh + weak $CH_4$ |
| 10 | 0.51 | 0.57 | 10 | Weak $CH_4$ |
| 11 | 0.6 | 0.66 | 10 | Weak/medium $CH_4$ & $NH_3$ |
| 12 | 0.695 | 0.765 | 10 | Intermediate $CH_4$ & $H_2O$ |
| 13 | 0.85 | 0.94 | 10 | Strong $CH_4$ & $H_2O$ |

**Table 6.3-6.** The HCG Channel B IR filter set.

| Band # | Wavelength Start (μm) | Wavelength End (μm) | Bandwidth (%) |
|---|---|---|---|
| 14 | 1.0 | 1.2 | 20 |
| 15 | 1.19 | 1.46 | 20 |
| 16 | 1.45 | 1.8 | 20 |

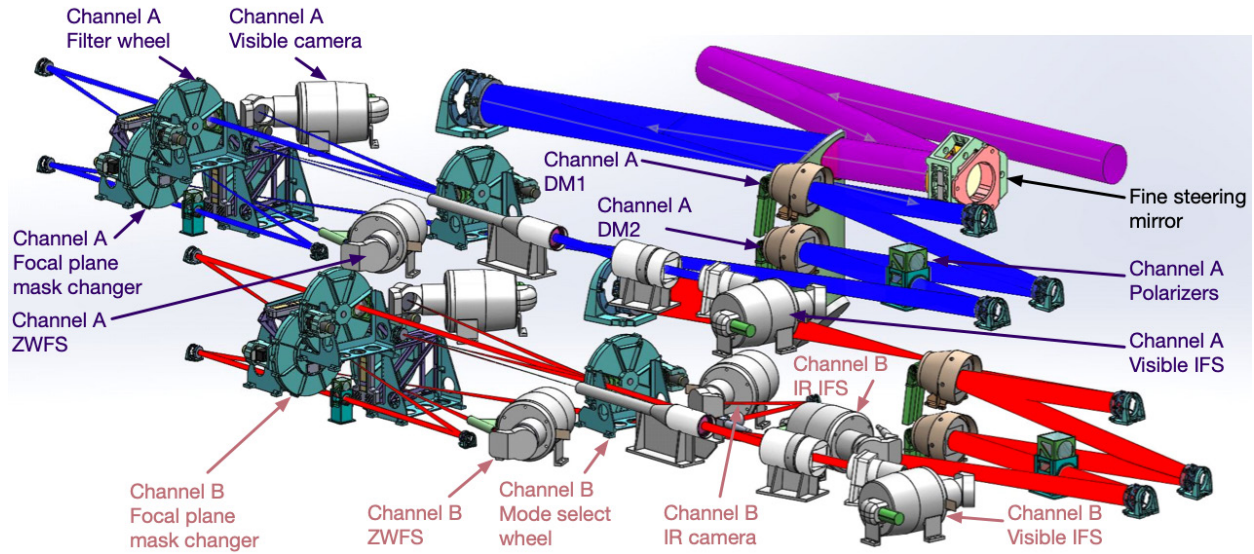

**Figure 6.3-4.** The two HCG channels shown with detector, filter wheels, and optical mount models: the incoming beam is in purple, which is split into Channel A (*blue beams*) and Channel B (*red beams*).

### 6.3.2 Coronagraph Optical Components

#### 6.3.2.1 Fine-Steering Mirror (FSM)

The FSM is used to stabilize the optical system LOS by keeping the target star image centered on the coronagraph mask as the spacecraft attitude wanders within the limits of its control capability. The FSM is located at the pupil image formed by the telescope's tertiary mirror. Placing the FSM at the pupil minimizes the beam walk downstream as the FSM steers the beam. Beam walk displaces the beam onto different parts of the downstream optics, and slight imperfections on surfaces subtly changes the wavefront error in the beam and therefore adversely affects the contrast achievable by a coronagraph.

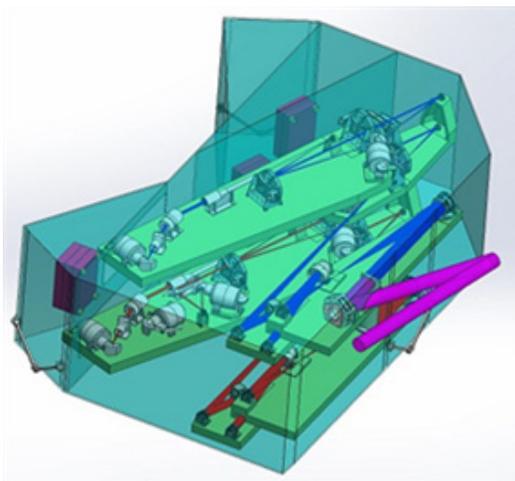

**Figure 6.3-5.** Coronagraph set up on optical benches inside its enclosure.

The maximum tip and tilt of the FSM is sufficient to handle small pointing biases from the spacecraft. It also has an angular resolution that is small compared to the pointing error corrections that are required. The FSM function is implemented as a two-axis tip/tilt stage carrying a plane fold mirror.

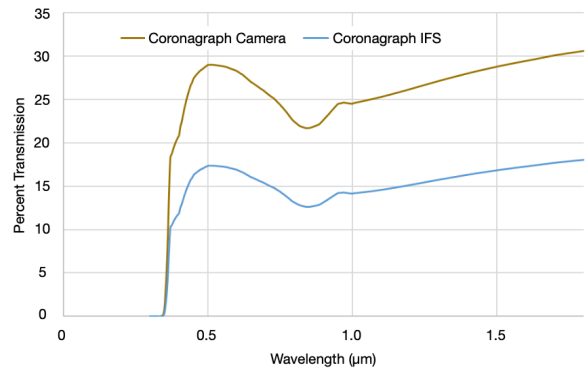

**Figure 6.3-6.** Optical transmission of the HCG camera and IFS modes across the visible and IR bands. Note that detector effects, such as quantum efficiency (QE), are excluded.

#### 6.3.2.2 Deformable Mirrors

Deformable mirrors are a key technology for a coronagraph instrument, enabling cancellation of optical aberrations to a very high level. Using two DMs enables phase and amplitude control over an annular field of view, providing efficient observing.

There are two potentially suitable DM technologies available. The first utilizes lead-magnesium-niobate (PMN) electrostrictive ceramic actuators on a 1 mm pitch to drive a continuous fused-silica mirror face sheet. This technology is currently baselined on the WFIRST CGI. The second is based on microelectro-mechanical systems (MEMS) DM technology. Each actuator in the MEMS DM can be individually deflected by electrostatic actuation to achieve the desired pattern of deformation without hysteresis. The MEMS technology is a backup to the PMN on WFIRST.

The number of DM actuators, in conjunction with the telescope aperture and wavelength of operation, determines the coronagraph OWA, while the pitch and number of actuators contribute to the overall instrument size. HabEx has baselined a 64×64 MEMS DM with an 0.4 mm actuator pitch; the actuator count allows the coronagraph to reach a 32 $\lambda/D$ OWA, while the small actuator pitch helps minimize the overall instrument size. Simulations show that the HabEx DM configuration is sufficient to provide wavefront control in both amplitude and phase, correcting minute wavefront errors due to fabrication and alignment inaccuracies in the system and enabling the required deep starlight suppression for the HCG.

#### 6.3.2.3 Coronagraphic Masks

The collimated beam reflecting off DM2 is brought to a focus by an off-axis parabolic mirror with a focal ratio of $f/30$. The focused star image has a point spread function (PSF) Airy disc diameter of 40 µm, at a 0.55 µm wavelength. The coronagraphic mask element is placed at this focal plane. To cover the entire HabEx bandwidths within both the visible and infrared channels, multiple masks are needed to provide the best starlight suppression over the full wavelength range. These masks are carried by a wheel mechanism, with the appropriate mask is rotated into position depending on the waveband selected for observation.

The VVC is a phase-mask coronagraph, requiring only a focal-plane mask and a standard circular Lyot stop. The vortex phase mask comes in various "topological charges," which parametrize the height of the phase ramp. The topological charge allows IWA to be traded against insensitivity to stellar size and low-order aberrations, with higher vortex mask charges bringing larger IWAs and less sensitivity to aberrations. The state-of-the-art vortex coronagraph masks have all been based on the vector vortex coronagraph concept, using one of the following technologies: liquid crystal polymers (LCP), photonics crystals, or subwavelength gratings. The best lab results for vortex masks have been obtained on the High Contrast Imaging Testbed (HCIT) with the LCP approach (Serabyn and Trauger 2014).

In the HabEx implementation, similar to WFIRST CGI, the VVC mask is slightly tilted with respect to the beam path and each type of mask has a reflective spot at the mask center, sending incident starlight into the FGS/ZWFS, the elements of which are discussed in *Section 6.8.6* and *Section 6.3.2.5*, respectively.

The optimal coronagraph mask for the HCG is a vortex mask of topological charge 6. This charge represents the optimal trade-off between IWA (2.4 $\lambda/D$ at 50% total off-axis throughput: IWA$_{0.5}$), and immunity to low-order aberrations (tip-tilt, defocus, astigmatism, coma, spherical; see **Table 5.4-4**). No other coronagraph mask considered for this report matched the charge 6 vortex's performance in this regard. The HLC mask has high technology readiness (TRL 5) and traceability to the WFIRST CGI, but is considered only a potential backup due to its relatively poor immunity to low-order wavefront errors.

#### 6.3.2.4 Lyot Stops

The Lyot stop design for the vortex coronagraph does not depend on wavelength, and so in principle, only a single Lyot stop is needed.

### 6.3.2.5 Zernike Wavefront Sensor

Initially, the WFE introduced by the optical system is corrected on two DMs included in the coronagraph beam train, and the DMs are adjusted to produce a very dark annular hole with some residual speckles. During observations, which may take many hours, the wavefront slowly evolves under small thermal changes producing changes in the speckles on the coronagraph focal plane. Since these speckle changes limit the ability to detect exoplanets they must be controlled. Changes to tip/tilt and focus will be detected at the ZWFS. This sensor enables detection of the wavefront error at sub-milliarcsecond levels, so that tip/tilt can be corrected by the FSM at the entrance to the coronagraph.

The ZWFS uses the rejected starlight from the VVC to sense the evolving low-order wavefront error. After the spacecraft slews to a target star and stabilizes, an acquisition process results in the star being centered on the coronagraph's occulting mask, and the starlight reflecting off the mask. This light is re-imaged by the optical elements onto the ZWFS detector, creating a pseudo-interferogram. Motion of the telescope creates a change in the interferogram and the resulting error signal is fed back to the FSM to correct.

The ZWFS sensor is similar to the WFIRST CGI's low order wavefront sensor (LOWFS; Shi et al. 2017), now called LOCAM. The ZWFS is based on the Zernike phase contrasting principle where a small ($\sim 1\,\lambda/D$ diameter) phase step, which produces a phase difference of $\sim \pi/4$, is placed at center of the starlight PSF. The combined light reflecting from the phase step ('phase dimple') and the surrounding part of the mask is then collimated to form a small image of the telescope pupil at the ZWFS camera. Interference between the light that reflects from inside and outside the phase dimple converts the wavefront phase error into intensity variations in the pupil image on the camera. The spatial sampling of the pupil image on the ZWFS camera depends on the spatial frequency of the WFE modes required to be sensed. There is a design trade between number of sensed modes, photons per pixel, and the ZWFS camera frame rate. On WFIRST CGI, LOCAM is running at a high frequency ($\sim 1$ kHz frame rate) in order to sense fast LOS jitter from vibration sources, such as the telescope's reaction wheels, and to correct it using the FSM at the coronagraph entrance. For HabEx, this high frame rate is not required since microthrusters do not produce high frequency disturbances. However, the ZWFS on HabEx will spatially sample at a higher rate than LOCAM to obtain improved wavefront sampling. This is because HabEx is using Boston Micromachines Corporation (BMC) DMs, which have some surface features, and proper sampling of the resulting wavefront error requires $\sim 1,024$ pixels across the pupil.

Like WFIRST CGI, a control loop using the ZWFS as the sensor operates on the FSM to correct the telescope's LOS to a level higher than that achieved by the FGS (*Section 6.8.6*). The ZWFS-sensed wavefront error modes beyond tip-tilt will be corrected using one of the DMs.

The WFIRST CGI LOWFS testbed has shown the ability to sense coma and astigmatism-level aberrations and has been used to demonstrate partial feed-forward compensation of these aberrations using the DMs.

The HabEx approach uses essentially the same optical layout as the WFIRST LOWFS, taking light from the coronagraph mask and directing it to a camera. Rather than using only the rejected starlight from the core of the beam, HabEx will use a dichroic filter layer on the surface of the mask to collect light outside the science band for wavefront sensing. For coronagraph observations in visible bands, for instance, UV or near-IR light will be directed to the ZWFS. Using light from the target star, this approach could potentially measure the full telescope plus coronagraph WFE, out to the spatial frequencies affected by DM control. **Table 6.3-8** shows the time needed to sense up to 21 Zernike modes at the ZWFS, assuming a conservative optical throughput of 14%. Even on relatively faint stars, the time to measure these modes is not prohibitive, given the very long observations times identified in the structural, optical, and thermal performance (STOP) modeling in *Section 6.9*. Note that the STOP modeling shows that the HabEx baseline design does not require wavefront control of these higher order modes. These modal measurements, however, are useful to observe for diagnostic purposes.

**Table 6.3-8.** The HCG ZWFS can rapidly sense up to 21 Zernike modes for all potential targets cataloged in *Appendix D*.

| Stellar Magnitude | Duration to Sensing Accuracy (s) | |
|---|---|---|
| | 10 pm | 1 pm |
| 0 | 0.03 | 3 |
| 3 | 0.48 | 48 |
| 6 | 7.7 | 780 |
| 9 | 120 | 11,880 |

### 6.3.2.6    Focal Planes

The detector arrays for the visible channels are electron-multiplying charge-coupled devices (EMCCDs), selected because of their exceptionally low effective read noise. The imaging focal planes (blue and red channels) consist of a single EMCCD per channel operated at 153 K. The chosen type is a modified CCD201 with delta doping and a thickened substrate together with a broadband "astro" coating giving response out to 1.0 μm. The pixel scale is shown in **Table 6.3-2**. The corresponding IFS focal planes consist of a modified, cut down version of CCD282, which is a 4k × 4k device with a 2k × 4k frame store at each end. The device format allows it to be cut in half along both axes forming a 2k × 2k sensor area with a 2k × 2k frame store. The IFS is also operated at 153 K to minimize dark current.

For the IR channel, an HgCdTe linear mode avalanche photodiode array (Saphira LMAPD) is baselined. The detector is cooled to 77 K to minimize dark current. It has a low read noise, with avalanche gains of 50 or more available.

Noise requirements and expected performance for all three detectors are summarized in **Table 6.3-9**.

### 6.3.3    End-to-End Modeling

Each optical surface in a coronagraph beam path contributes errors to the optical wavefront, ultimately limiting the achievable raw contrast.

**Table 6.3-9.** Focal plane detector noise requirements and expected performance. Note that these detectors are used on HCG and SSI.

| Detector Type | Parameter | Expected Performance |
|---|---|---|
| CCD-201 | Read Noise | 0.008 e-/s |
| | Dark Current | 3 10⁻⁵ e-/s |
| | CIC | 2 10⁻³ e-/f |
| CCD-282 | Read Noise | 0.008 e-/s |
| | Dark Current | 3 10⁻⁵ e-/s |
| | CIC | 2 10⁻³ e-/f |
| Saphira | Read Noise | 0.32 e-/s |
| | Dark Current | 0.005 e-s |

The raw contrast is the planet-to-star flux ratio at which the planet's peak intensity in the final image is the same as the intensity of the stellar speckles at the same location. Raw contrast is therefore a function of position in the image. To determine the expected raw contrast, an end-to-end optical model was developed using the PROPER Optical Propagation Library (Krist 2007) for a charge 6 vortex coronagraph using the HCG's bluest band (450–550 nm). PROPER is a field-based optical propagation modeling system. The assumptions and results of this simulation are outlined below.

### 6.3.3.1    Propagation Modeling

The optical model of the coronagraph was converted to an unfolded layout suitable for PROPER. An achromatic vortex charge 6 mask with phase having the azimuthal form $2\pi e^{i2\pi\,60}$ was modeled with a $0.6\,\lambda/D$ central spot to cover the defect at the central vortex singularity. The polarization input was provided from the Zemax™ model.

The primary and secondary mirror specifications are based on the surface quality measured for the WFIRST telescope. The primary mirror's wavefront error as a function of spatial frequency is given in **Table 6.3-10**. Otherwise, the surface error of each optic follows a similar power spectral density (PSD) to the optics manufactured for the Gemini Planet Imager (GPI) instrument (Macintosh et al. 2008), where the errors follow an $f^{2.5}$ mid-spatial frequency power law. A summary of the input wavefront errors, expressed in Zernike modes, is shown in **Table 6.3-11**.

**Table 6.3-10.** The telescope primary mirror wavefront error specification as a function of spatial frequency.

| Spatial Frequency (cycles/beam diam.) | Wavefront Error (nm RMS) |
|---|---|
| 0–4 | 12.1 |
| 4–10 | 4.2 |
| 10–30 | 2.6 |
| 30–80 | 1.7 |
| >80 | 0.2 |

The curved surfaces of the telescope mirrors introduces slightly different low order aberrations into each polarization state as well as cross-talk between input polarizations (Breckinridge and Chipman 2016). The optical layout of the telescope is specifically designed to reduce the cross-talk between polarizations to levels consistent with $\sim10^{-10}$ contrast imaging.

The baseline BMC 64×64 actuator MEMS DMs have relatively large, periodic surface errors (**Figure 6.3-7**), which generate bright speckles mainly outside the coronagraph field but also produce light near the source. Since these errors occur at the actuator level, they cannot be directly compensated by the DM. However, a BMC presentation from 2016 showed a factor of three improved surface finish at 3.3 nm RMS. BMC is actively developing this improved surface for use on their commercially-available MEMS DMs. For this analysis, the wavefront error introduced by each DM surface is assumed to have this reduced 3.3 nm RMS. As noted earlier, the DM mirrors were separated by ⅛ Talbot length. This adjustment was found to reduce the effect of amplitude errors generated by the DMs' surface errors.

It should be noted that an alternative DM made by Xinetics® has a smoother, flat, mirror-like surface. However, that mirror has a characteristic actuator drift that would need to be actively monitored and corrected. In addition, its actuator separation results in a Talbot length 2.5 times longer than with the BMC DM, which makes packaging more difficult and the overall instrument significantly larger.

Polarization aberrations were computed using both Polaris-M® and Zemax® for the HabEx optical layout including coatings (Al on the PM and SM, Ag on other mirrors). Broadband images were simulated using 9 wavelengths. The computed X-Y polarization WFE, which is the dominant polarization cross-term, is in agreement between the two modeling tools.

**Table 6.3-11.** Wavefront error produced by telescope and coronagraph optical surfaces.

| | Wavefront error |
|---|---|
| **Primary only (including gravity release)** | Z1–Z49: 12.4 nm RMS |
| | >Z49: 3.0 nm RMS |
| **At focal plane mask exit pupil** | Z1–Z49: 14.0 nm RMS |
| | >Z49: 5.2 nm RMS |
| **At focal plane mask exit pupil, after WFE flattening:** | Z1–Z49: 0.1 nm RMS |
| | >Z49: 2.5 nm RMS |

Using the PROPER model, design sensitivity analyses were carried out to assess the impact of certain requirements on the overall system performance. Ideally, the target star will be sufficiently far away that it appears as a point source. However, when the star is at least partially resolved by the telescope, the contrast is degraded. In addition, telescope jitter appears as a tilt error on the wavefront and may also degrade the contrast. The calculated static mean contrast as a function of stellar diameter is shown in the left panel of **Figure 6.3-8**. With a stellar diameter of 1 mas, the azimuthal average of the raw contrast at 2.4 $\lambda/D$ is better than the required $3 \times 10^{-10}$. The right panel of **Figure 6.3-8** shows the effects of jitter on contrast. With tip/tilt jitter of 0.3 mas RMS, the azimuthal average of the raw contrast at 2.4 $\lambda/D$ is also better than $3 \times 10^{-10}$. **Table 6.3-12** shows a summary of the effect of varying the primary mirror, secondary mirror, and DM surface quality, indicating that improving DM surface errors has the largest effect on contrast near the IWA and that efforts to improve the primary and secondary mirrors start to yield diminishing returns.

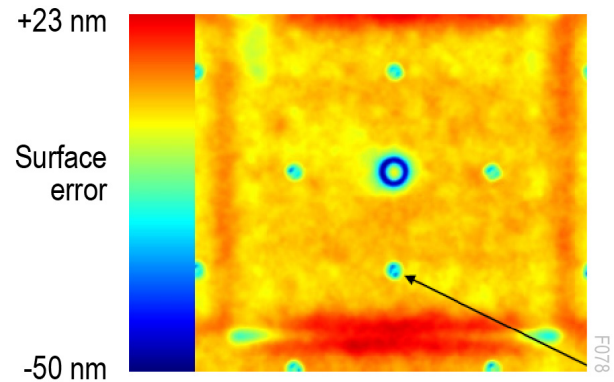

**Figure 6.3-7.** Surface deviation from flat for a 34×34 DM. The image shows one cell of the mirror and the arrow points to one of the small holes in the surface. The major source of surface figure error is the banded structures at the top and bottom of the image.

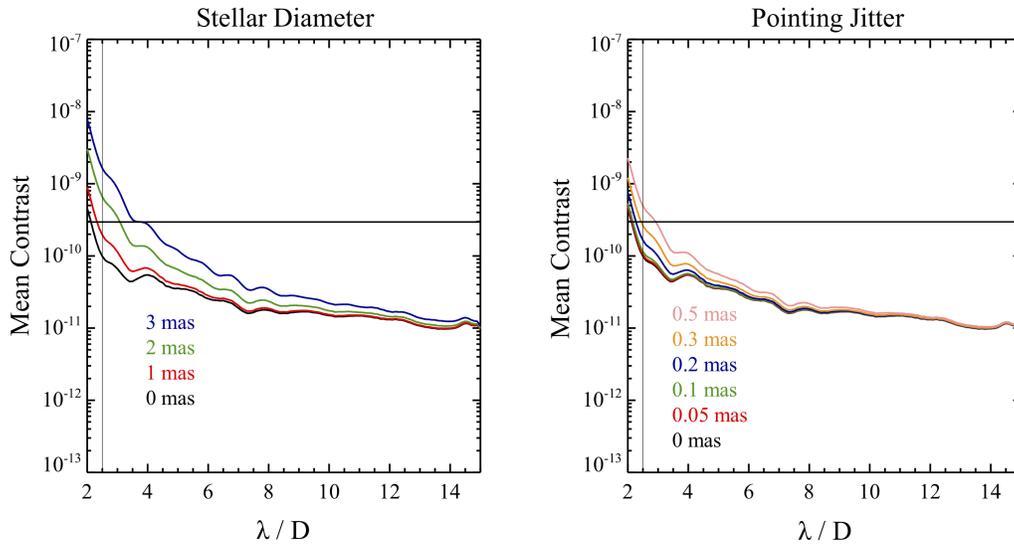

**Figure 6.3-8.** Post-coronagraph mask results for VVC charge 6: 0.45–0.55 μm, with surface and polarization errors and improved DM errors. *Left:* stellar diameters from 0–3 mas. *Right:* Pointing jitter from 0–0.5 mas.

**Table 6.3-12.** The effect on contrast at the IWA of varying surface quality of the principal optics. The values in the first three columns are WFE RMS in nm.

| Primary | Secondary | DM | Contrast at 2.4 λ/D |
|---------|-----------|-----|---------------------|
| 25.5 | 11.4 | 3.3 | 2.0 × 10⁻¹⁰ |
| 17.0 | 11.4 | 3.3 | 1.3 × 10⁻¹⁰ |
| 12.8 | 5.7 | 3.3 | 1.0 × 10⁻¹⁰ |
| 25.5 | 11.4 | 0.0 | 0.9 × 10⁻¹⁰ |

### 6.3.3.2 Exoplanet Observation Simulation

The end-to-end performance of the HCG was simulated as part of the PROPER modeling. The left panel of **Figure 6.3-9** shows one such simulation using a representative case with two simulated planets injected into the HCG's dark hole. A plot of the simulation's associated raw contrast as a function of separation angle is also included in the right panel of **Figure 6.3-9**. The simulated stellar intensity maps and planet PSFs used in the HCG performance simulations were also used as inputs for the yield calculations described in *Chapter 3*. Modeling using PROPER shows the performance of the HCG design. To understand the overall functional performance requires system-level modeling to capture the thermal and mechanical disturbances in the telescope system. A structural, thermal and optical performance (STOP) analysis was also carried out as part of this study and is described in *Section 6.9*. Overall contrast performance using a simulated observation sequence is discussed in that section.

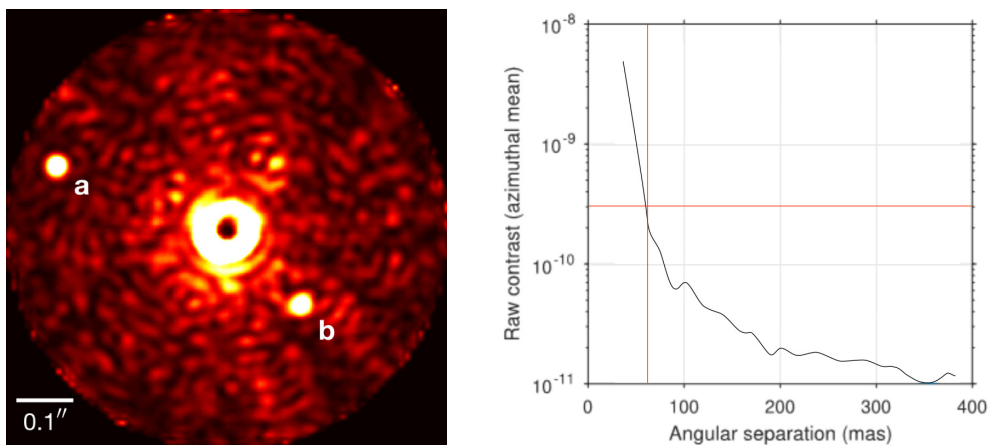

**Figure 6.3-9.** Representative result from the HabEx coronagraph end-to-end simulator using a charge 6 vortex coronagraph in band #1, 0.45–0.55 μm. *Left:* Simulated raw intensity assuming RMS tip/tilt jitter of 0.2 mas and a stellar diameter of 1 mas. Two planets, *a* and *b*, with planet-to-star flux ratios of 2 × 10⁻¹⁰ were injected at arbitrary positions. The intensity is shown out to an angular separation of 0.39 arcsec beyond which the raw contrast remains relatively constant. This image represents 1/3 of the full field of view since the coronagraph OWA is 0.74 arcsec in band #1. *Right:* The azimuthally averaged raw contrast corresponding to the simulated stellar intensity in the left panel.

## 6.4    HabEx Starshade Instrument (SSI)

The starshade system consists of the starshade itself and the Starshade Instrument (SSI). This section focuses on the SSI and its operation. In *Chapter 7*, the starshade flight system is discussed in detail together with the overall contrast performance of the telescope. The science-based requirements of the STM that are specific to SSI are given in **Table 6.4-1**. The SSI design accommodates the required broad spectral range with three optical channels covering the UV, visible, and IR. The starshade, represented by its block diagram in **Figure 6.4-1**, operates in the science bands illustrated in **Figure 6.4-2**. When operating in the central green band, the starshade has a starlight suppression band from 0.3–1.0 μm. To obtain suppression down to 0.20 μm in the UV, the starshade can be moved further away from the telescope, potentially achieving an IWA$_{0.5}$ of 47 mas. For IR science, the starshade moves closer to the telescope and the IWA will increase in proportion to the wavelength. **Table 6.4-2** shows the science bands, starshade/telescope separation, and IWAs.

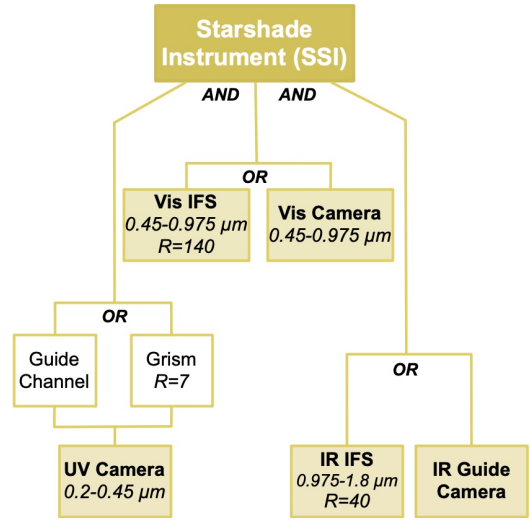

**Figure 6.4-1.** HabEx SSI paths. The UV, visible, and IR paths operate simultaneously. Within each channel, one of two functions can be selected, for example in the visible channel, either spectroscopy or imaging. Depending on the science wavelength selected, a guide channel is chosen from either the UV or IR band.

A high suppression (dark) conical shadow region exists behind the starshade and the telescope is placed as far back as possible within this shadow (114,900 km) while maintaining high starlight suppression. The telescope can move laterally ±1 m within the shadow and the tips of the starshade form an angle of 70 mas (IWA$_{tip}$) to the line of sight when operating in the 0.3–1.0 µm spectral band.

**Table 6.4-1.** Key SSI design requirements, based on Table 5.4-6. Note that IWA is a requirement levied upon the starshade flight system and occulter. The IWA requirement, performance, and margin are shown in Table 7.3-1.

| Parameter | Requirement | Expected Performance | Margin | Source |
|---|---|---|---|---|
| Spectral Range | ≤0.30 μm to ≥1.70 μm | 0.2–1.80 μm | Met by design | STM |
| Spectral Resolution, *R* | ≥5 (0.3–0.35 μm)<br>≥40 (0.63 μm)<br>≥70 (0.75–0.78 μm)<br>≥8 (0.80 μm)<br>≥35 (0.82 μm)<br>≥100 (0.87 μm)<br>≥32 (0.89 μm)<br>≥17 (0.94 μm)<br>≥20 (1.06 μm)<br>≥19 (1.13 μm)<br>≥12 (1.15 μm)<br>≥10 (1.40 μm)<br>≥11 (1.59–1.60 μm)<br>≥10 (1.69–1.70 μm) | 7 (0.20–0.45 μm)<br>140 (0.45–0.975 μm)<br>40 (0.975–1.80 μm) | Met by design | STM |
| OWA (0.5 μm) | ≥6 arcsec | 6 arcsec | Met by design | STM |
| End-to-End Throughput (450 μm) | 22% | 30% | 36% | Error Budget |

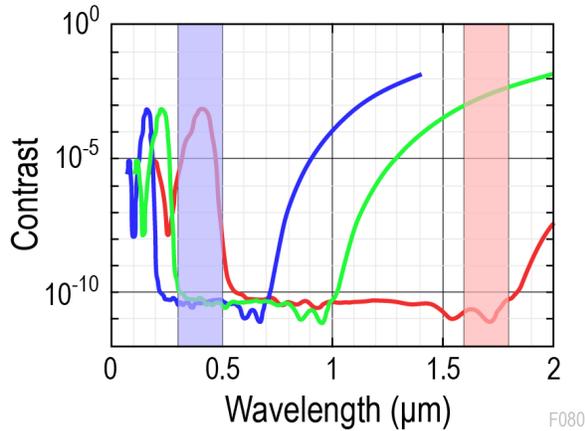

**Figure 6.4-2.** Starshade occulter contrast functions. *Green line*: the visible science band, with the red rectangle showing its guide band. *Red line*: the IR science band, with guiding in the blue rectangle. *Blue line*: the UV science band with guiding in the red rectangle.

**Table 6.4-3.** SSI focal plane design specifications of its three channels. Not that SSI IR Guide Channel pixel resolution identifies the resolution at the starshade occulter at 76,600 km.

| Cameras | UV Channel | Visible Channel | IR Guide Channel |
|---|---|---|---|
| FOV | 10" | 12" | - |
| Bandpass (µm) | 0.20–0.45 | 0.45–0.975 | 0.975–1.80 |
| Pixel Resolution | 14.2 mas | 14.2 mas | 12 cm |
| Angular Resolution | 21 mas | 21 mas | - |
| Detector | 1×1 CCD201 | 1×1 CCD201 | 1×1 LMAPD |
| Array width (pixels) | 1024 | 1024 | 256 |

| Spectrometers | UV Channel | Visible Channel | IR Channel |
|---|---|---|---|
| FOV | 10" | 2" | 4" |
| Bandpass (µm) | 0.20–0.45 | 0.45–0.975 | 0.975–1.80 |
| Spectral Resolution, $R$ | 7 | 140 | 40 |
| Spectrometer Type | Slit/grism | IFS | IFS |
| Detector | 1×1 CCD201 | 1×1 CCD282 | 2×2 LMAPD |
| Array width (pixels) | 1024 | 4,096 | 2,048 |

**Table 6.4-3** shows the instrument specifications flowed down from the STM. To provide full coverage of the wavelength range, SSI has UV, visible, and IR detectors. Each of the channels has imaging and spectral capabilities. **Table 6.4-4** shows the principal specifications of the starshade itself. **Figure 6.4-1** outlines the instrument's functionality.

In addition to the imaging and spectroscopic capabilities, SSI also detects the starshade occulter's position with respect to the line of sight to the target star. This is done using starlight outside the science band. The starshade shape is of the numerically optimized type rather than hyper-Gaussian, and produces a designed high-suppression wavelength band. Starshade science operation requires that both the science and formation flying cameras view the starshade simultaneously. The corresponding starshade guide bands (used for starshade station-keeping) are shown in **Figure 6.4-2**.

Light of both shorter and longer wavelengths is attenuated but leaks into the starshade occulter's shadow region and is used for starshade positioning. **Figure 6.4-2** shows the starshade transmission functions for the three planned science bands, 0.20–0.67 nm, 0.30–1.00 µm, and 0.540 µm to 1.80 µm. When performing science at longer wavelengths, shorter wavelength, out-of-band light is used for guiding and vice versa. Details on the guiding algorithms are included in the section on formation flying, *Section 8.1.7*.

**Figure 6.4-3** shows the optical throughput of the starshade as a function of radial distance from the center. Interestingly enough the starshade exhibits a small amplification near the tips. However, for the purpose of contrast specification, this gain is ignored and the throughput unity in the wider field is used as the

**Table 6.4-2.** SSI science bands and related starshade occulter separation distances with corresponding starshade guiding bands. Note that the guide channel band differs from the observation band in all cases.

| Science Band | UV | Visible | IR |
|---|---|---|---|
| Wavelength band (µm) | 0.20–0.67 | 0.30–1.0 | 0.54–1.8 |
| Starshade-telescope separation (km) | 114,910 | 76,600 | 42,580 |
| $IWA_{tip}$ (mas) | 47 | 70 | 126 |
| $IWA_{0.5}$ (mas) | 39 | 58 | 104 |
| Guide band (µm) | 1.6–1.8 | 1.6–1.8 | 0.30–0.45 |

**Table 6.4-4.** HabEx starshade occulter specifications. See *Chapter 7* for more details on the starshade occulter.

| Parameter | Specification |
|---|---|
| Overall diameter | 52 m |
| Petal length | 16 m |
| Disc diameter | 20 m |
| Number of petals | 24 |
| Design type | Numerically optimized |

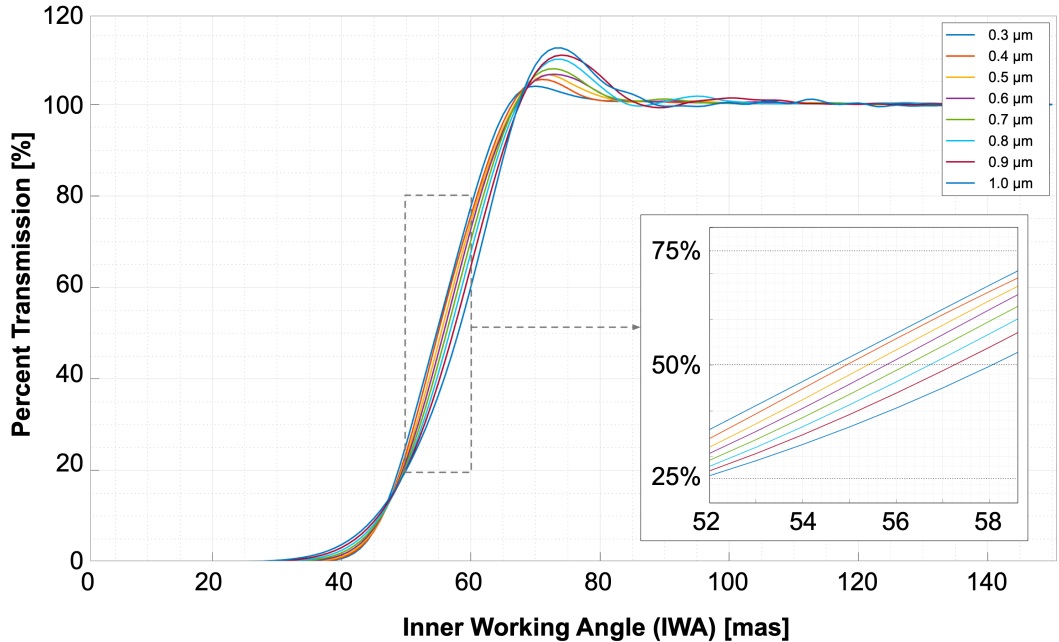

**Figure 6.4-3.** Starshade occulter throughput function with radial angle at a separation distance of 76,600 km. The *inset* identifies the small variation of IWA$_{0.5}$ as a function of wavelength.

maximum. Some planet light can evidently be seen between the petals of the starshade and this results in a gradual roll-off of throughput towards the center, closely approximating the geometric obscuration of the starshade as a function of radius. To specify the IWA$_{0.5}$ for the shade, the IWA$_{0.5}$ for the longest wavelength of the science band is used; as can be seen in the inset to **Figure 6.4-3**, this is at 58 mas at 0.975 μm wavelength.

### 6.4.1 Design

The starshade instrument contains six beam paths, where **Figure 6.4-4** shows an optical view and **Figure 6.4-5** shows a mechanical view. These beam paths accommodate the UV, visible, and infrared optical channels. Light entering the starshade camera is split by dichroic optics into UV, then visible, and IR beam paths, so all of these channels can be operated simultaneously as seen in **Figure 6.4-1**. Camera and spectrograph properties are shown in **Table 6.4-3**. The UV channel carries a simple slit spectrograph employing a grism with $R = 7$. The visible channel carries a broadband IFS capable of covering the wavelength range from 0.45–0.975 µm, plus an imaging camera for more rapid and wide-field system imaging. The infrared channel carries an IFS with $R = 40$ to enable disc and object spectroscopy. The operation of these channels is detailed below.

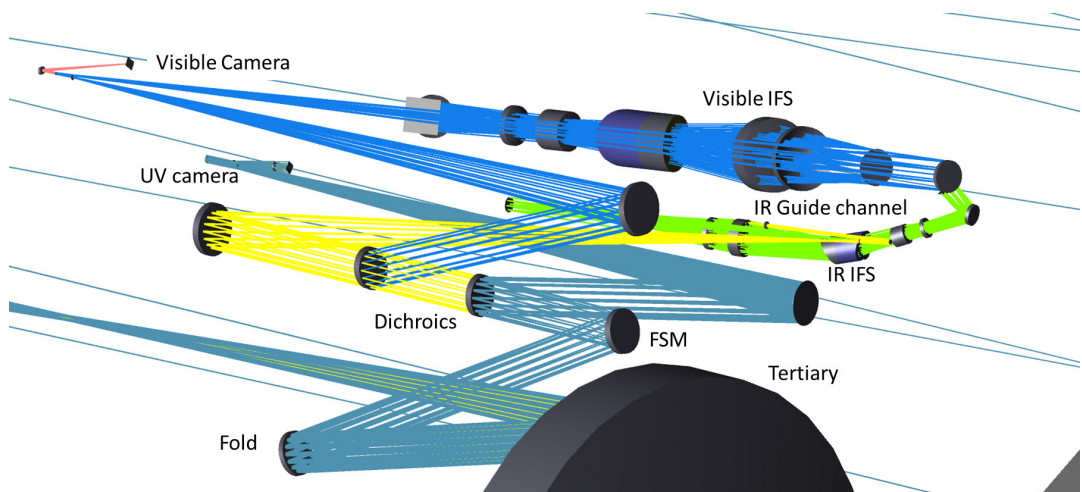

**Figure 6.4-4.** SSI light paths and components. The incoming beam is shown in *teal*. The incoming beam is split into three channels using two dichroics. Note that the incoming *teal beam* reflects off the first dichroic to enter the UV channel. *Blue* and *pink beams* go to the visible IFS and camera, respectively. *Green* and *yellow beams* go to the IR IFS and Guide Channel, respectively.

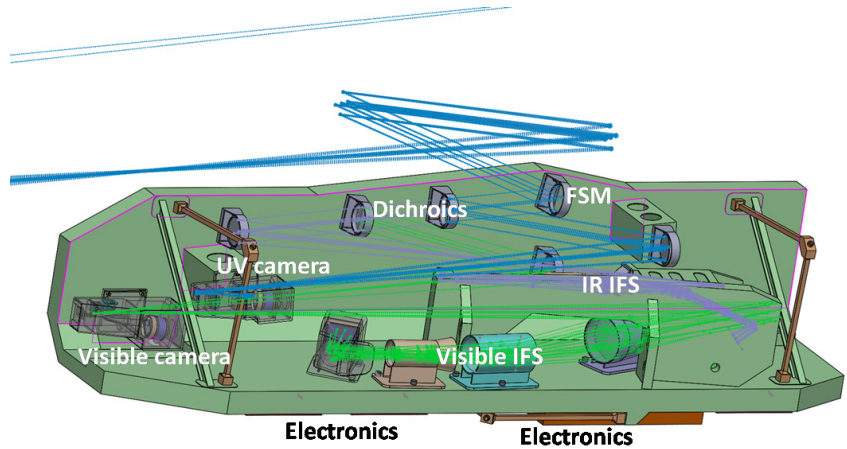

**Figure 6.4-5.** SSI mechanical assembly lies adjacent to the primary mirror. Note that the SSI enclosure is not depicted.

### 6.4.1.1 Visible Channel

The visible channel is the principal science channel and carries a camera and an IFS. The layout is shown schematically in **Figure 6.4-6**. Light from the telescope's tertiary mirror strikes a fold mirror and then a fast steering mirror at the entrance to SSI. It then passes through a dichroic optic, which reflects UV light. The remaining visible and infrared light passes to a second dichroic where the visible light is reflected to an off-axis parabolic mirror and then to a focus where field stops are inserted to limit the field of view, one for the imaging mode (11.9″ diameter) and a second for spectroscopy (1.9″ diameter). This focus is reimaged by an ellipsoidal mirror to the focal plane. A filter wheel is inserted after the ellipse with filters to select wavebands appropriate for different starshade-to-telescope distances. For example, with the starshade at the nominal distance for visible work, the filter would pass 0.45–1.00 µm light. With the starshade more distant as set up for UV science, the spectral range would be 0.45–0.67 µm (see **Table 6.4-2**).

Further science filters and polarizing optics for polarization studies could also be inserted here.

The imaging focal plane consists of a single EMCCD operated at 153 K. The chosen type is a modified CCD201 with delta doping and a thickened substrate together with a broadband "astro" coating giving response out to 1.00 µm (Nikzad et al. 2017). The pixel scale is as shown in **Table 6.4-3**. During a thruster firing, the sensor is read out at 1 kHz to keep the accumulated photon count appreciably below full well.

For spectroscopy, an additional ellipsoidal mirror is inserted into the beam following the first ellipse, producing a large increase in the $f/$number from 47 to 1,330. Via a fold mirror, this beam is focused onto a microlens array (MLA), which forms the entrance to the IFS. The IFS consists of the MLA, a matching multiple-aperture mask to restrict stray light, a set of lenses to collimate the beam, prisms to disperse the wavelengths and a second set of lenses to focus onto the focal plane. This type of IFS (**Figure 6.4-7**) is described by (McElwain et al. 2016). IFS operation can be visualized thus: for each MLA element, one spectrum is produced on the focal plane. Using an MLA, an array of spectra is produced and the optical system is designed to ensure that these spectra remain separate and do not overlap.

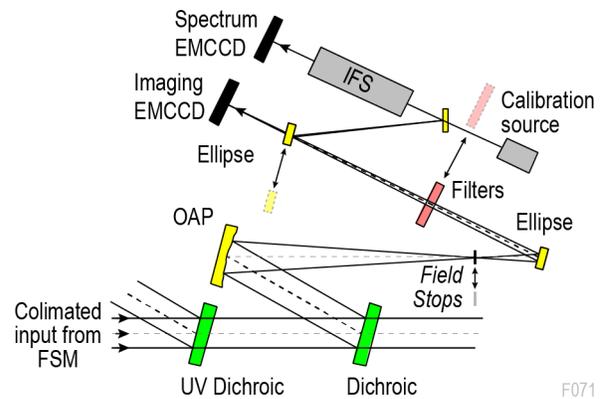

**Figure 6.4-6.** SSI visible channel schematic, representing the blue light path in Figure 6.4-4. Note that filters do not enter line of sight of the calibration source.

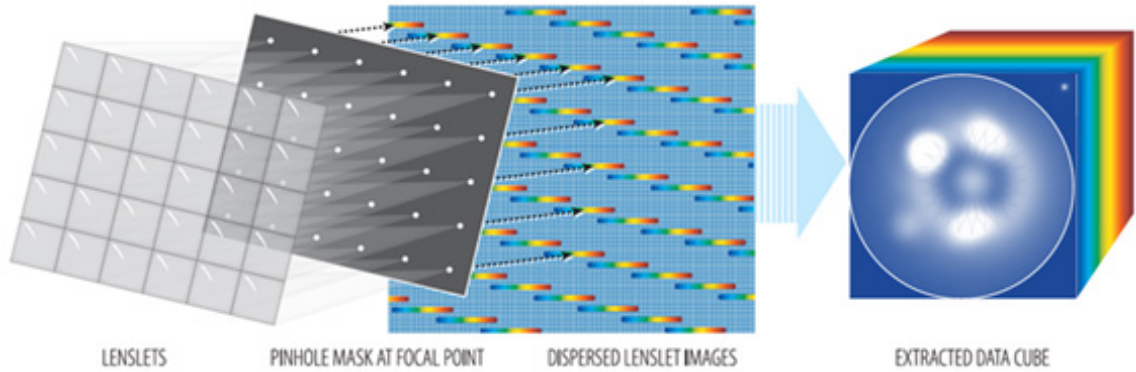

**Figure 6.4-7.** Schematic form of the PISCES IFS adapted from McElwain et al. (2016) showing the lenslet array, pinhole mask array, arrayed spectra on the focal plane and the resulting "data cube" consisting of a stacked series of spectral images. The image is based on a simulation of the starshade field and contains the suppressed starlight near center, the two solar glint lobes from the starshade edge, exozodiacal light and the planets Venus and Earth as they would be seen from 10 pc.

An image of the scene at one wavelength can be formed by using all the pixels on the focal plane that correspond to the same wavelength. A series of images known as "slices" can be assembled into a "data cube" with sides corresponding to the directions of the field of view and height corresponding to wavelength. Thus, the scene is reproduced in a stack of images representing narrow wavelength bands. To calibrate the images, it is necessary to provide a calibration source with at least one known wavelength in the band, and that illuminates the entire MLA. As shown in **Figure 6.4-6**, the calibration source light is injected when required through the fold mirror, which has a small leakage of about 2%.

The IFS focal plane consists of a large single electron multiplying CCD (Teledyne/e2v CCD282) operated at 163 K. The chosen type is a modified version of the off-the-shelf detector with delta doping and a thickened substrate together with a broadband "astro" coating giving sensitivity out to 1.0 μm. The format is an 8k × 4k array with frame store areas at both sides of the 8k length, and a 4k × 4k central imaging area. The spectral images are produced on the 4k square central area and moved into the frame stores before readout at high EMCCD gain.

### 6.4.1.2    UV Channel

Shown schematically in **Figure 6.4-8**, the UV channel carries a low-resolution spectrometer and is also used as the guide channel for IR science. Following the FSM at the entrance to the starshade instrument, light reaches a dichroic optic that reflects the UV component to an off-axis paraboloidal mirror and thence through a field stop to an ellipsoidal mirror. Following the ellipse is a filter wheel to allow filter selection. Two field stops are provided, one to allow a field of view up to 10.2" diameter, and another with 0.02" diameter to select individual objects. The field stops are mounted on a piezoelectric stage to allow selection and positioning. The beam is then refocused to the focal plane, passing through a filter placed at the intermediate exit pupil which removes light of wavelengths longer than 0.45 µm. At this exit pupil, a grism or zero-deviation prism can be introduced for low-resolution spectroscopy. With the grism removed, the camera forms an undispersed image. With the introduction of a mirror further downstream, the exit pupil is relayed to the focal plane, forming a pupil image suitable for starshade guiding. The pupil scale need not be large; a 32×32 pixel image is formed with each pixel covering a 12 cm square section of the entrance pupil.

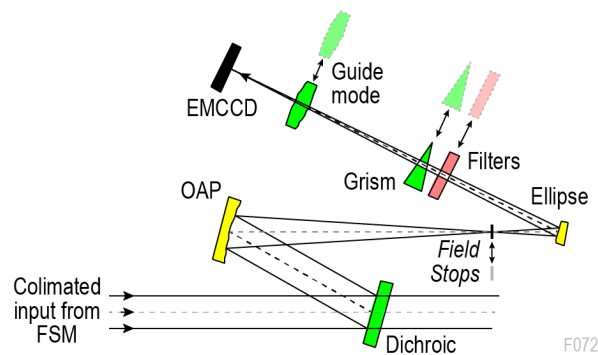

**Figure 6.4-8.** SSI UV channel schematic, representing the teal light path after reflecting off of the first dichroic in Figure 6.4-4.

The focal plane consists of a single EMCCD (CCD201) operated at 153 K. The chosen type is a modified version of the off-the-shelf item optimized for high UV sensitivity by deep depletion and delta-doping processes (Nikzad et al. 2012), together with a broadband coating to improve response down to 0.20 µm. Pixel scale is as shown in **Table 6.4-3**. The format is a 1k × 1k array with adjacent frame store.

### 6.4.1.3    Infrared Channel

The starshade instrument's infrared channel is the primary guide channel used for both visible and UV science, and also carries an IR IFS. When SSI is being used for imaging or taking spectra on the visible and UV channels, guiding is handled on the IR guide channel. IR light entering the instrument passes through both dichroics and is reflected off a paraboloidal mirror (**Figure 6.4-9**). Between the second dichroic and the paraboloid, a filter wheel operates to allow band selection. The subsequent layout follows a similar scheme to the UV channel with a focus, ellipsoid, and conditioning optics to reach the desired $f/$numbers. At the focus, a fixed field stop limits the field of view to 4″, slightly larger than the IFS MLA FOV.

The guide channel consists of a lens to relay the exit pupil following the ellipsoid to the focal plane with the magnification providing 32 pixels across the telescope aperture. The focal plane consists of a single linear mode avalanche photodiode (LMAPD) array detector based on an HgCdTe sensor (Saphira array by Selex). The avalanche gain-mode allows the effective read noise to be reduced (but not yet to the extent possible in EMCCDs). The detector is cooled to 77 K to minimize dark current. Note that an IR imaging mode could be easily provided in this layout, though such a mode is not called for in the STM.

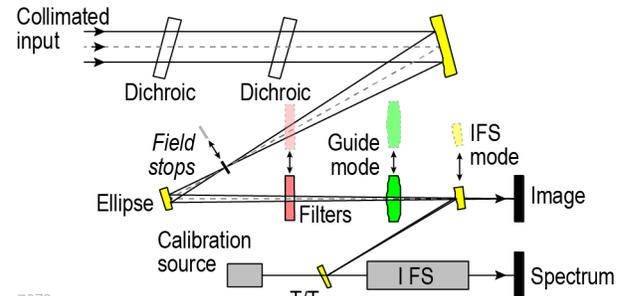

**Figure 6.4-9.** SSI IR channel schematic, representing the yellow light in Figure 6.4-4. *T/T* represents a tip/tilt mirror.

The science channel consists of a powered relay mirror inserted near the guide channel relay lens. This provides the necessary larger focal length to the MLA. Before reaching the MLA, the beam is folded at a plane mirror. A calibration source is provided behind this mirror, injecting through it, so that the position of the spectrum can be identified on the IFS focal plane. The IFS utilizes a planned variant of the Saphira detector with a 1k × 1k format and smaller pixels (12 µm) operated at 77 K. Four of these detectors are arrayed in a 2×2 format to provide a full FOV of 3.8″ × 3.8″ at $R = 40$. The IFS optics follow the same general design as for the visible IFS, with appropriate optical prescription changes. The field of view is Nyquist sampled by the lenslets and likewise, the spectrum is Nyquist sampled at the detector.

### *6.4.2    Performance and Analysis*

**Figure 6.4-10** shows the optical transmission of SSI channels compared to the HCG channels, not including quantum efficiency and related effects. Since SSI requires far fewer mirrors, efficiency is about twice as high at 44% at 0.70 µm so that spectra and images can be acquired in about half the time. Also, the UV channel has excellent performance down to 200 nm (and below).

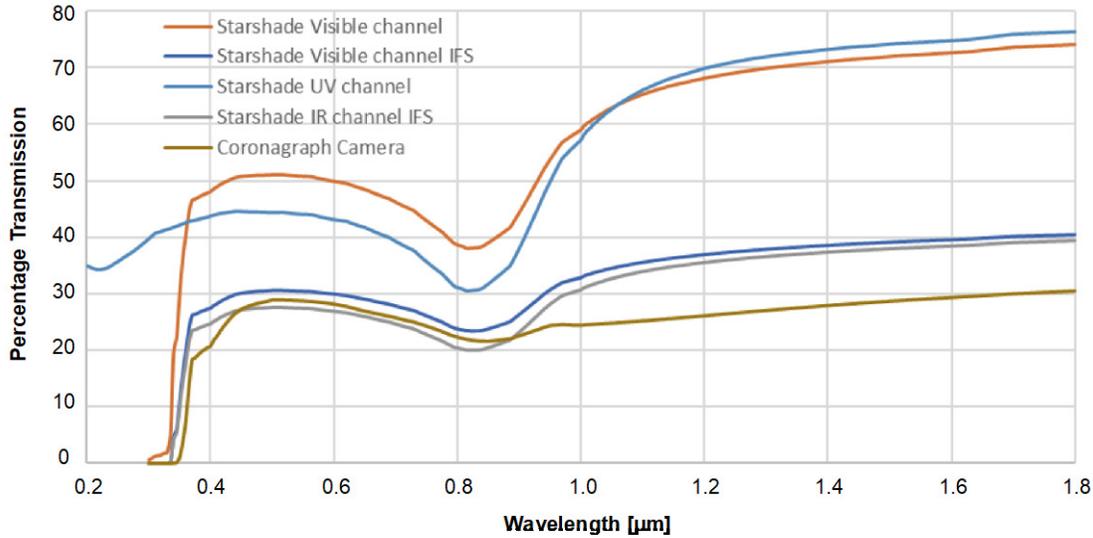

**Figure 6.4-10.** Transmission of HCG and the different SSI channels.

Facilitated by the narrow FOV (**Table 6.4-3**), SSI is diffraction-limited at the camera focal planes as well as at the input planes of the IFSs.

### 6.4.3   Operations

As noted earlier, out-of-band light that leaks around the starshade occulter and into SSI is used for guiding the starshade. This light manifests as a faint structured pattern within the starshade occulter's shadow. **Figure 6.4-11** shows an image of the starshade shadow structure in infrared light when the starshade is set up for visible science. The central bright spot appears directly on the line of sight between the center of the starshade and the star and thus forms the target for the guide system.

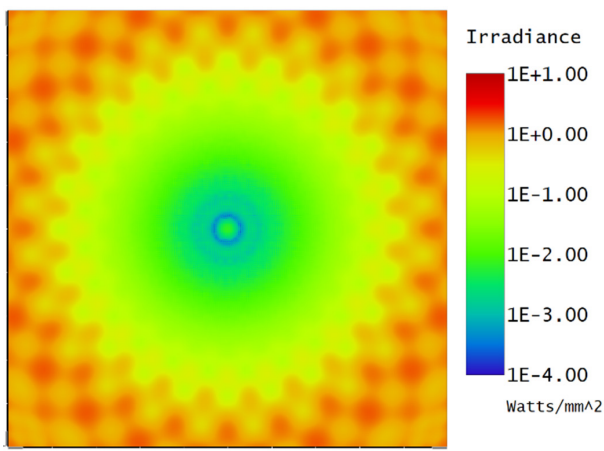

**Figure 6.4-11.** Starshade shadow in the infrared. Note the bright Arago, central spot in the center, whose shape is used to lock formation with the starshade occulter.

Outside this core, there are two faint rings and then the flux increases with a monotonic slope towards the edge. Outside this region, a pattern that is reflective of the starshade geometry appears with, in this case, 24 peaks around a circumference, corresponding to the starshade occulter's 24 petals. The central peak has a diameter approximately equal to $d\lambda/D_{SS}$ where d is the separation distance between the starshade and the telescope, $\lambda$ is the wavelength, and $D_{SS}$ is the starshade diameter. In the case modeled, the central peak is about 3 m in diameter. The diameter of the smooth, sloping region is approximately 25 m and a pronounced pattern exists outside this with a diameter of about 50 m.

#### 6.4.3.1   Formation Flight

Formation flight and starshade navigation, covered in detail in *Section 8.1*, utilizes this pattern to bring the starshade into line with the star in the acquisition and science modes. The UV and IR channels have guide camera modes, which project an image of the telescope pupil onto the respective focal planes. With a selected channel in science mode, an optic is introduced into the corresponding guide channel to place an image of the pupil on the guide CCD.

The starshade occulter's lateral position is sensed from an image of the light distribution at the telescope entrance pupil (see **Figure 6.4-10**) so that the pixel resolution is given in centimeters in **Table 6.4-3**. At the entrance pupil, the starshade shadow has some structure, typically with a much diminished Arago spot at the center. The lateral position of the telescope is sensed by comparing an image of this structure with a library of expected images. Starshade occulter and formation flight technologies are discussed in *Section 11.2*.

Within the patterned region, the starshade follows the gradient down to the center. Once centered, the system maintains the central spot in the telescope pupil by firing thrusters on the starshade every ~600 s. This alignment is precise because the target stars are bright and the attenuation by the starshade in the guide bands is poor.

The HabEx SSI concept of operations mitigates the impact of thruster firings through telescope-starshade coordination and SSI operation. When a thruster firing is commanded following telescope-starshade coordination, SSI observation enters a suspended data recording state for several seconds. This is because the thruster plumes are illuminated by the Sun and would contaminate the data. However, the plume exits the SSI FOV within 0.8 s. As discussed in *Section 7.3.5*, vibrations in the starshade occulter dampen out to acceptable levels within seconds of thruster firings. Within 10 s, the detectors are electronically cleared and resume data recording.

During the SSI suspended data recording state, the sensors are still read out at very high frame rates to keep the accumulated charge down and thus avoid contamination of the science data caused by charge persistence after data recording resumes.

## 6.5    HabEx Ultraviolet Spectrograph (UVS)

The UVS instrument is designed to enable high-resolution imaging and spectroscopy down to 115 nm in the UV. The driving science cases for the instrument are discussed in *Chapter 3* with their associated performance requirements summarized in the STM, *Chapter 5* and **Table 6.5-1**. The UVS will access a large number of diagnostic emission and absorption lines available at wavelengths between 115 nm and 320 nm, with an $R = 1,000$ spectroscopy mode extending the band to 370 nm. The science cases set a need for a wide field of view and the ability to perform multi-object spectroscopy (MOS) within that field. The science also calls for a range of spectral resolutions to enable measurement of both line shape and separation of specific lines in both emission and absorption. Spectral range must reach down to at least 115 nm.

### 6.5.1    Design

The UVS, whose optical layout is shown in **Figure 6.5-1**, utilizes a microshutter array (MSA) situated at the two-mirror Cassegrain focus to enable selection of objects of interest from a 3'×3' FOV. **Table 6.5-2** shows key design parameters for the UVS. With a maximum resolution of $R = 60,000$, the UVS needs a large set of gratings to cover the wavelength band (**Table 6.5-3**). The detector area is large, requiring about 30,000×17,000 pixels (or "pores" in the case of microchannel plate detectors) to cover the FOV. This area will be covered using a 3×5 array of approximately 100×100 mm microchannel plate (MCP) detectors, or alternatively, a larger array of delta-doped, UV-optimized CCDs. Both types of detector have similar performance and technology readiness levels.

Table 6.5-1. HabEx UVS requirements, expected performance, and margins, based on Table 5.4-7. *Note that through UVS can observe up to 370 nm in a low resolution spectroscopy mode.

| Parameter | Requirement | Expected Performance | Margin | Source |
|---|---|---|---|---|
| Spectral Range | ≤ 115 nm to ≥320 nm | 115–320 nm (with 115–370 nm available at $R = 1,000$)* | Met by design | STM |
| Spectral Resolution, *R* | Up to R ≥ 60,000 depending on the measurement | 60,000; 25,000; 12,000; 6,000; 3,000; 1,000; 500; imaging | Met by design | STM |
| Angular Resolution | ≤50 mas | 21 mas | 138% | STM |
| FOV | ≥2.5 × 2.5 arcmin² | 3 × 3 arcmin² | 20ß% | STM |
| Multi-object Spectroscopy | Yes | 342 × 730 apertures | Met by design | MTM |

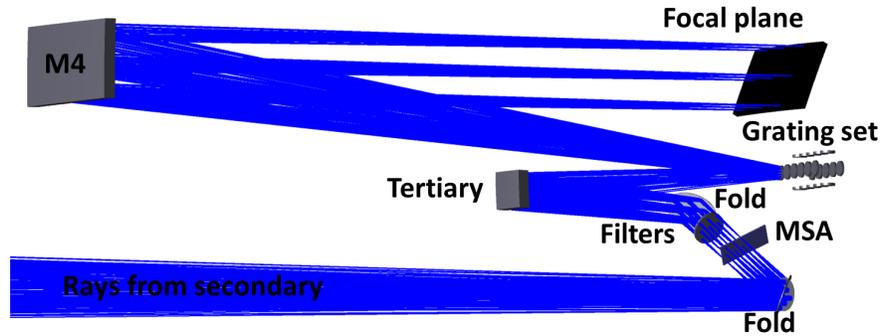

**Figure 6.5-1.** UVS light path and its components. Note that UVS picks off light from the SM, not the TM like all other HabEx instruments.

Because of the relatively low reflectance of available mirror coatings in the UV (the reflectivity of aluminum/MgF$_2$ is ~61% at 115 nm), it is important to minimize the number of reflections from the primary mirror to the focal plane. The currently baselined UVS design has seven reflections in its path compared with three and six in the narrow field of view COS instrument on Hubble. This design was chosen because of various integration conflicts with the other instruments. It contains two additional folds over the minimum required. In addition, five- and six-reflection designs were developed subsequent to the baseline definition and these would improve throughput considerably.

The wavelength band is covered by a set of 20 gratings plus one plane mirror. A set of seven optical filters (see **Table 6.5-4**) is also available. The set of gratings will be mounted on a cylinder forming four rows of optics. By rotating the cylinder about its axis and translating it, any of the gratings (and the one plane mirror used for imaging) can be selected.

The optical design is constrained by the telescope's first two mirrors and so a wide design space was explored to reach a solution. The best-corrected focus available is on-axis at the Cassegrain focus formed by the primary (M1) and secondary mirrors (M2), so this is where the MSA is located. The beam is folded out of the main telescope path and passes through the MSA and any filter placed in its path. The UVS tertiary mirror (M3) is an off-axis aspheric surface, producing an exit pupil at the grating. The planar grating carries an evenly spaced groove pattern with approximately 2 μm spacing for the shortest wavelength and highest dispersion. This surface can be fabricated using conventional optical polishing plus electron beam lithographic techniques. Following the grating, the beam travels ~3.7 m to M4 where it is reimaged onto the detector positioned behind the grating set. The resultant design yields the 3'×3' field, corrected at the telescope diffraction limit of 400 nm. Gratings are individually optimized for each waveband and the design includes one optic without grating lines, so that an undispersed UV image is formed.

**Table 6.5-2.** UVS design specification. *Note that through UVS can observe up to 370 nm in a low-resolution spectroscopy mode.

| Parameter | Specification |
|---|---|
| FOV | 3×3 arcmin$^2$ |
| Bandpass | 20 bands covering 115-320 nm. |
| Spectral Resolution, $R$ | 60,000; 25,000; 12,000; 6,000; 3,000; 1,000; 500; 1 |
| Telescope Resolution | Diffraction limited at 400 nm |
| Detector | 3×5 MCP array, 100 mm sq each |
| Array width | 17,000 × 30,000 pixels (pores) |
| Microshutter aperture array | 2×2 array of 171×365 apertures, 200×100 μm |

**Table 6.5-3.** Spectral bands for UVS instrument. Note that for $R$ = 500 and 1,000, UVS observes up to 370 nm.

| Resolution $R$ $\lambda/\Delta\lambda$ | $\lambda$ min nm | $\lambda$ max nm | $\Delta\lambda$ pm | Resolution $R$ $\lambda/\Delta\lambda$ | $\lambda$ min nm | $\lambda$ max nm | $\Delta\lambda$ Pm |
|---|---|---|---|---|---|---|---|
| 60,000 | 115 | 127 | 2.02 | 25,000 | 115 | 149 | 5.46 |
| 60,000 | 127 | 141 | 2.24 | 25,000 | 149 | 192 | 7.05 |
| 60,000 | 141 | 156 | 2.48 | 25,000 | 192 | 248 | 9.11 |
| 60,000 | 156 | 173 | 2.74 | 25,000 | 248 | 320 | 11.76 |
| 60,000 | 173 | 192 | 3.04 | 12,000 | 115 | 192 | 12.51 |
| 60,000 | 192 | 213 | 3.37 | 12,000 | 192 | 320 | 20.87 |
| 60,000 | 213 | 235 | 3.73 | 6,000 | 115 | 320 | 33.38 |
| 60,000 | 235 | 261 | 4.13 | 3,000 | 115 | 320 | 66.75 |
| 60,000 | 261 | 289 | 4.58 | 1,000 | 115 | 370 | 200.1 |
| 60,000 | 289 | 320 | 5.07 | 500 | 115 | 370 | 399.9 |

**Table 6.5-4.** UVS bandpass filters.

| Wavelength (nm) | Name |
|---|---|
| 248-314 | F300X |
| 206-239 | F218W |
| 213-259 | F225W |
| 251-290 | F275W |
| 240-243 | FQ232N |
| 245-249 | FQ243N |
| 281-285 | F280N |

With a Nyquist sampling criterion for the field of view at 0.4 µm, the pixel width is equal to $\lambda/2D$. In the spectral domain, the criterion for spectral elements to be resolved is the same so that a spectral resolution element $\Delta\lambda$ covers two pixels. For example, with $R = 60,000$ at 120 nm, $\Delta\lambda = 2$ pm (**Table 6.5-3**) and the number of spectral elements needed to cover the first band 12 nm wide is 6,000. Thus, a single spectrum on the detector will cover 12,000 pixels, resulting in the rectangular shape of the focal plane.

The detector is a photon-counting device consisting of an MCP array utilizing large-format plates (~100 mm width). The MCPs are glass capillary arrays (GCAs) consisting of thin-walled hexagonal tube assemblies, and are fabricated in very low-Pb glass for a low X-ray cross-section. The tubes (micropores) are arranged with a small angle typically ~15° off normal. Each plate consists of a micropore array of two layers with opposing pore angles for efficiency. Furthermore, the materials used are very pure and contain few radioactive isotopes, leading to a very low dark count. Atomic layer deposition is used to create the resistive and emissive layers (GaN and multi-alkali) of the cathode, producing improved performance specifications over conventional MCPs. Above the detectors, $100 \times 100$ mm$^2$ MgF$_2$ windows (possible with the large crystal boules now being made) are specified over vacuum-sealed detector units. Beneath the plate a lattice of wires forms the anode. Incoming UV photons produce a cascade of electrons with gain of 106 or more, and the charge cloud emerges at the base of the MCP assembly to impact the anode wires. High-speed analog-to-digital converters (ADCs), digitizing 8 bits at 10 MHz, collect charge from the wires. An application-specific integrated circuit (ASIC) postprocessor outputs a stream of data consisting of charge cloud centroid position (x and y), peak height, coincidence flag, and time stamp. In the case of two or more photons arriving at the same time, the postprocessor rejects the event based on peak height. Behind the anode is a plastic scintillator viewed by a miniature avalanche photodiode or photomultiplier tube that acts as a cosmic ray detector. In the case of a detection here coincident with an event on the wire grid, the coincidence flag is set and the event rejected. Thus, a clean signal can be generated in the presence of a cosmic ray background. The spatial digitization is at the micropore spacing. While the charge cloud spreads upon emission from the base of the plates, the large electron count allows localization of the event at the micropore level. The ADC rate (10 MHz), a determinant of photon flux, can handle up to approximately $10^7$ photons/sec with a small efficiency loss due to the coincidence of a portion of events. Most targets will be weak, so that typically hundreds of objects may be observed simultaneously.

Forming a 3×5 array of plates produces a 3'×3' field of view with the long axis accommodating the spectral dispersion. Since the plates are surrounded by the frame of the vacuum assembly there will be small gaps of coverage as is often the case with detector arrays.

The optomechanical design is shown in **Figure 6.5-2**. A cage structure supports the system. Most of the components are at one end, the exception being M4. At the base, the MCP array is positioned with its readout electronics. Light is collected by the first fold mirror and passes through the MSA assembly. This assembly can be removed for imaging as it slides on rails, producing an unobstructed field of view. Following this, a filter set is provided, together with an "open" setting. Following the second fold mirror is the tertiary mirror. The gratings are mounted on a cylinder which allows the large number of gratings to be accommodated in a small package. Electronics for the system are mounted on the exterior of the enclosure giving good access to radiators on the telescope exterior.

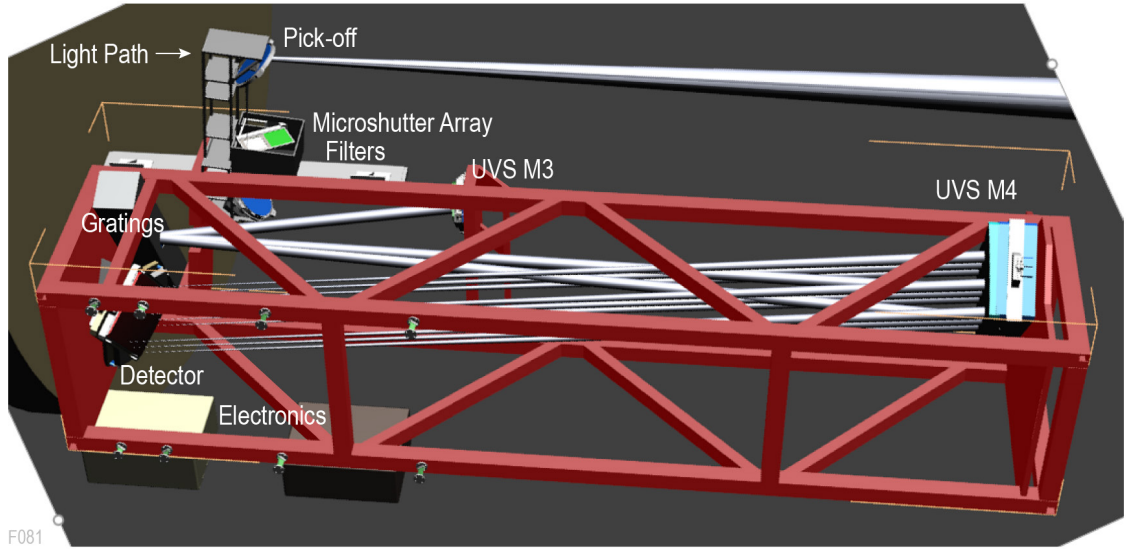

**Figure 6.5-2.** Optomechanical layout of the UVS. External radiators and enclosure have been removed.

### 6.5.2 Performance and Analysis

**Figure 6.5-3** shows the optical throughput of the UVS compared to two Hubble channels, considering the relative area of the two telescopes and ignoring any differences in detector performance. Compared with Hubble's COS instrument which has 3 reflections (FUV channel)

and 6 reflections (NUV channel), the UVS instrument has 7 reflections. COS benefits (at least in the FUV channel) by needing only a small FOV, which can be achieved with just three mirrors, while the optical design of the UVS needs a minimal 5 reflections to provide the 3'×3' FOV. However, for packaging with the other instruments, two additional fold mirrors were added. It's worth noting that in a redesign of the overall instrument layout, a 5-reflection design could be accommodated, yielding better throughput than COS NUV and greater effective area than COS FUV at all wavelengths longer than 117 nm. Finally, note that the 5-reflection UVS delivers 30% greater effective area than the 7 reflection UVS with throughput exceeding 50% at the longer wavelengths.

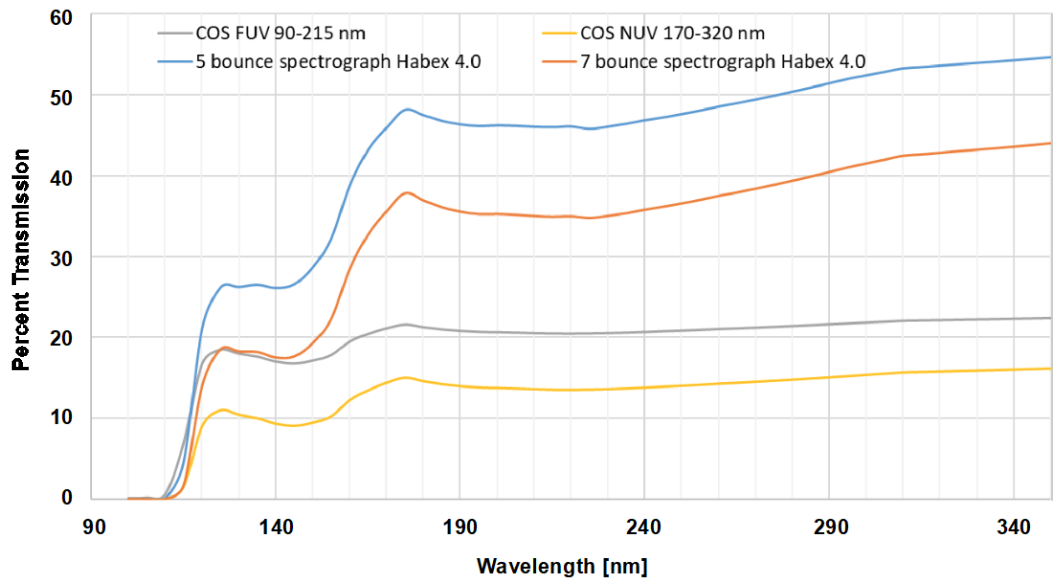

**Figure 6.5-3.** Effective transmission of UV spectrographs compared to the two Hubble COS channels. Transmission is relative to the HabEx collecting area (100% is the HabEx aperture area of 126,000 cm²). Detector quantum efficiency, grating efficiency, etc. are excluded but any transmissive optics are included. Also included here is the variant HabEx 3.2S UVS, which has five reflections but a smaller primary mirror.

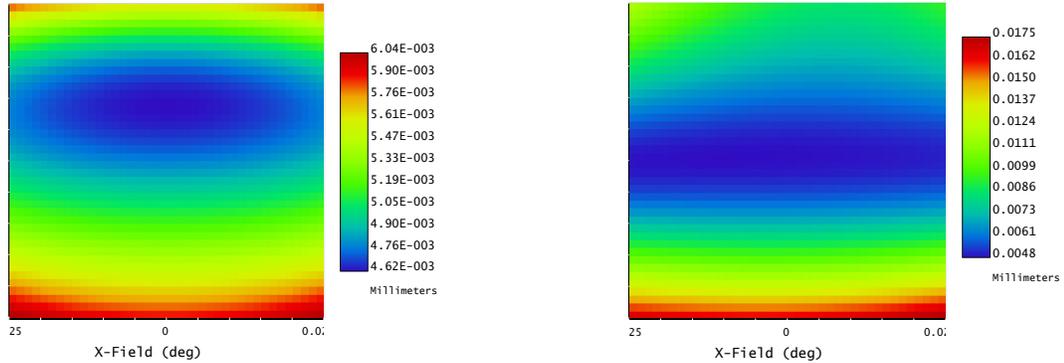

**Figure 6.5-4.** RMS spot size of the UVS design across the field, 100% of the field is diffraction limited. *Left:* Imaging mode, wavelengths 115–320 nm. *Right:* Spectroscopy mode, $R = 60,000$, 115–127 nm, 78% of the field is diffraction limited.

The left panel of **Figure 6.5-4** shows the RMS spot radius as a function of field for the imaging mode. The Airy radius is 9.7 µm, so the design is diffraction limited across the entire field. The right panel of **Figure 6.5-4** shows the RMS spot radius as a function of field for the design for the $R = 60,000$ spectroscopy mode at 127 nm. The Airy radius is 10.7 µm and the design is diffraction limited across 78% of the field.

## 6.6    HabEx Workhorse Camera (HWC)

The HabEx Workhorse Camera (HWC) is a general-purpose instrument providing visible through near-IR imaging and spectroscopy (**Figure 6.6-1**), with objectives ranging from solar system science to detailed studies of galaxies and quasars at the epoch of reionization to cosmology. The HWC would enable detailed follow-up of interesting targets, such as those identified from the wide-field surveys of the 2020s, such as Euclid, Large Synoptic Survey Telescope (LSST), and WFIRST. The science objectives addressed by the HWC are discussed in detail in *Chapter 4*, while the design requirements associated with these objectives are summarized in the STM, *Chapter 5*, and **Table 6.6-1**. Specifically, the instrument is designed to provide unique scientific capabilities compared to the facilities expected in the 2030s. For example, nearly all first-generation instruments on the new 30 m-class telescopes (e.g., TMT, GMT, and ELT) are near-IR instruments because ground-based adaptive optics (AO) are not expected to be effective for wavelengths much shorter than about 1 µm. The HWC would provide high-spatial resolution imaging, a stable platform for both photometry and morphology, and access to spectral regions inaccessible on the ground due to telluric absorption.

### 6.6.1    Design

Like Wide-Field Camera 3 (WFC3) on HST, the HWC design has two channels that can simultaneously observe the same field of view: a visible channel using delta-doped CCD detectors providing sensitivity from 0.37–0.975 µm, and a near-IR channel using H4RG10 HgCdTe arrays providing sensitivity from 0.95–1.80 μm. Beyond 1.80 μm thermal backgrounds dominate over most celestial targets.

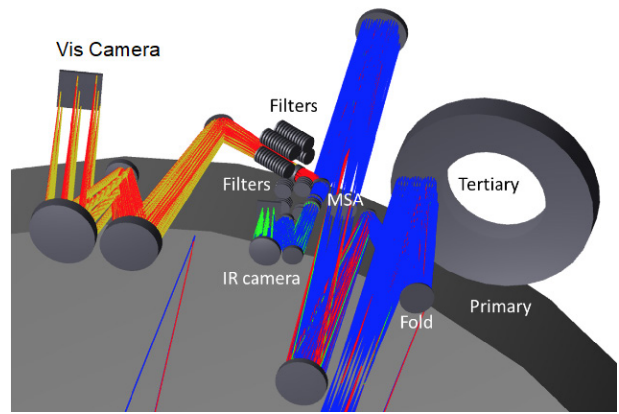

**Figure 6.6-1.** HWC light path and its components. The blue incoming beam is split using a dichroic into the IR channel (*green beam*) and the Vis channel (*red/yellow beam*).

**Table 6.6-1.** HWC requirements and expected performance, based on Table 5.4-8. *Note that UVS is designed to complete the spectral range up to 0.37 μm at R = 1,000 as HWC spectral range does not cover down to the 0.30 μm required for globular cluster measurements discussed in *Chapter 4*.

| Parameter | Requirement | Expected Performance | Margin | Source |
|-----------|-------------|---------------------|--------|--------|
| Spectral Range | ≤ 0.37 μm to ≥1.70 μm | 0.37–1.80 μm | Met by design | STM |
| Spectral Resolution, *R* | Up to ≥ 1,000 depending on the measurement | ≤1,000 | Met by design | STM |
| Angular Resolution | 50 mas | 25 mas | 138% | STM |
| FOV | ≥2 × 2 arcmin$^2$ | 3 × 3 arcmin$^2$ | 50% | STM |
| Multi-object Spectroscopy | Yes | 342 × 730 apertures | Met by design | STM |
| Noise Floor | ≤10 ppm | 10 ppm | Met by design | STM |

Both channels will have imaging and spectroscopic modes, and an MSA assembly provides for simultaneous slit spectroscopy of multiple sources, significantly reducing the background and source confusion compared to the slit-less spectroscopic modes available on HST. The two modes of operation share the same optical path and cameras. In the spectroscopic mode, the MSA and grism sets are introduced into the beam paths. The MSA is attached to a mechanism and thereby removable for imaging. **Table 6.6-2** shows the design parameters for the HWC's two channels. For imaging, the pixel magnification is chosen to Nyquist sample the PSF. To obtain sufficient field of view, the visible channel has a 3×3 array of 4k square CCD detectors, and the IR channel utilizes a 2×2 array of H4RG10 HgCdTe detectors.

**Figure 6.6-2** shows the schematic layout of the HWC instrument. Imaging and spectrograph planes are common, spectroscopy being achieved by inserting a grism. **Figure 6.6-3** shows the optical layout. After reflecting off the TM and the fold mirror, the input beam strikes a fine-steering mirror used for image dithering and small pointing adjustments and is normally fixed during an observation. The beam then passes through a relay formed by a pair of biconic paraboloidal mirrors. In spectroscopy mode, an MSA is inserted into the focal plane of the relay, enabling selection of particular targets. This MSA array is identical to the set installed in JWST's Near-Infrared Spectrograph (NIRSPEC). Following the relay, the beam passes to a dichroic where the visible light is separated from the IR light.

**Table 6.6-2.** HWC design specifications.

| | VIS Channel | IR Channel |
|---|---|---|
| FOV | 3'×3' | 3'×3' |
| Bandpass (μm) | 0.37–0.975 | 0.95–1.80 |
| Pixel Resolution | 15.5 mas | 24.5 mas |
| Angular Resolution | 30.9 mas | 49 mas |
| Design Wavelength | 0.6 μm | 0.95 μm |
| Detector | 3×3 CCD203 | 2×2 H4RG10 |
| Detector Array Width | 12,288 pixels | 8,192 pixels |
| Spectral Resolution, *R* | 1,000 | 1,000 |
| Microshutter Array | 2×2 arrays; 180×80 μm aperture size; 171×365 apertures | |

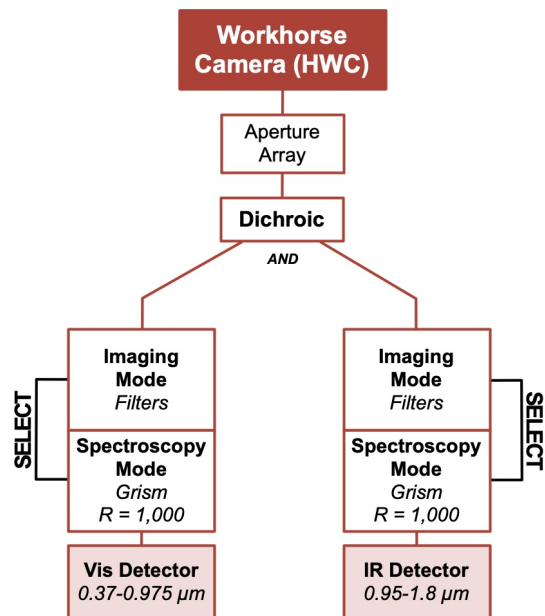

**Figure 6.6-2.** HWC uses a dichroic to split incoming light into Vis and IR channels. Each channel is capable of imaging and spectroscopy modes through filter or grating selection.

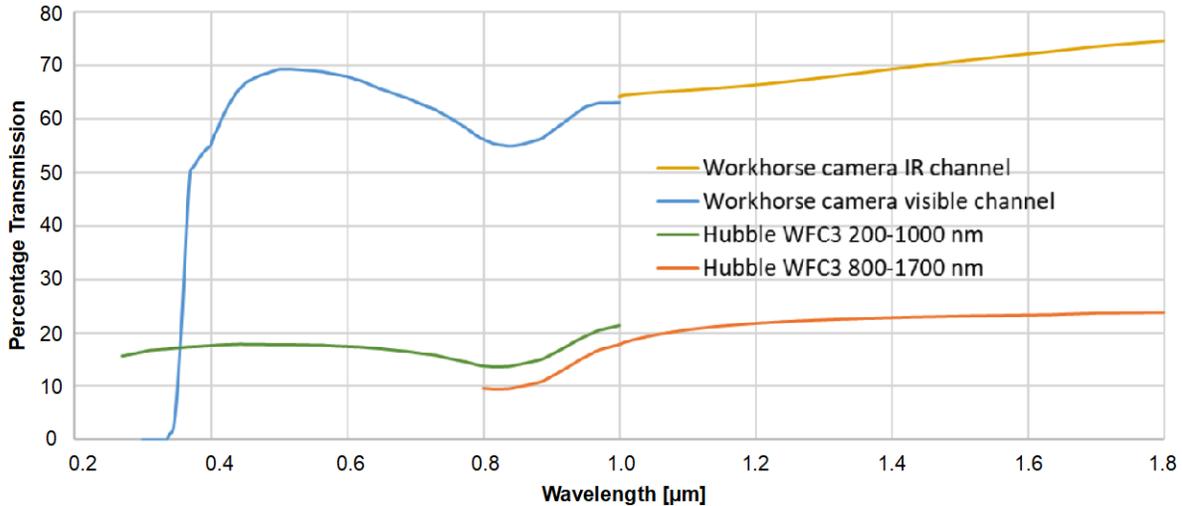

**Figure 6.6-3.** Relative throughput of HWC UV/Vis and IR channels compared to Hubble WFC3. Transmission is relative to the HabEx collecting area (100% is the HabEx aperture area of 126,000 cm²). Detector QE, grating efficiency, etc., are excluded but any transmissive optics are included.

#### 6.6.1.1    Visible Channel

At the dichroic, visible light is reflected and passes through a filter wheel to a camera. The filter wheel is mounted at a pupil plane and enables selection of different wavelengths of interest in the image. A grism is also placed in the wheel to allow spectroscopy in conjunction with the MSA at $R = 1,000$. The camera consists of a three-mirror relay and the focal plane itself. The performance is diffraction limited at 0.60 μm. The focal plane is designed for Nyquist sampling of the 3'×3' field at the same wavelength. The selected array is CCD203, a conventional low-noise CCD with 12 μm pixel size and a 4k × 4k format. A set of nine of these CCDs, cooled to 153 K, forms the focal plane. For UV performance these arrays would be deep depletion, delta-doped devices.

#### 6.6.1.2    Infrared Channel

At the dichroic, IR light from 0.95–1.80 μm is transmitted and passes through a filter wheel to a camera. As in the visible channel, the filter wheel is mounted at a pupil plane and enables selection of different wavelengths of interest in the image. Again, a grism is placed in the wheel to allow spectroscopy in conjunction with the MSA at $R = 1,000$. The camera consists of a three-mirror relay leading to the focal plane. Performance is diffraction limited at 0.95 μm. The focal plane is designed for Nyquist sampling of the 3'×3' field at the same wavelength. The selected array is the Teledyne H4RG10, a low-noise hybrid HgCdTe/CMOS bump-bonded array with 10 μm pixel size and a 4k × 4k format. These FPAs are currently being developed for WFIRST. A set of four FPAs cooled to 77 K forms the focal plane.

### 6.6.2    Performance and Analysis

**Figure 6.6-3** shows the optical throughput performance of the visible and infrared channels compared to WFC3, based on aperture area with detector differences excluded. Compared with WFC3, HWC has more reflecting surfaces but still posts better throughput by excluding UV operation. In HWC, only two mirror surfaces are Al and the rest protected silver, which has higher reflectivity above λ = 0.40 μm.

**Figures 6.6-4** and **6.6-5** show the diffraction spot sizes across the field. In the visible, the camera is near diffraction limited. In the IR, it is diffraction limited across the whole field of view. Compared with Hubble, HabEx has a larger aperture providing improved angular resolution (1.67× better). The aperture is also unobscured, so the characteristic diffraction from the secondary mirror's spiders seen in Hubble images will not appear.

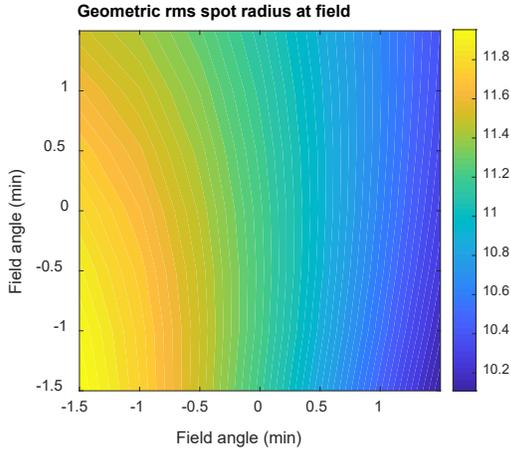

**Figure 6.6-4.** HWC diffraction spot size on the visible channel: the diffraction limited spot size is 19 µm so the entire focal plane is near-diffraction limited.

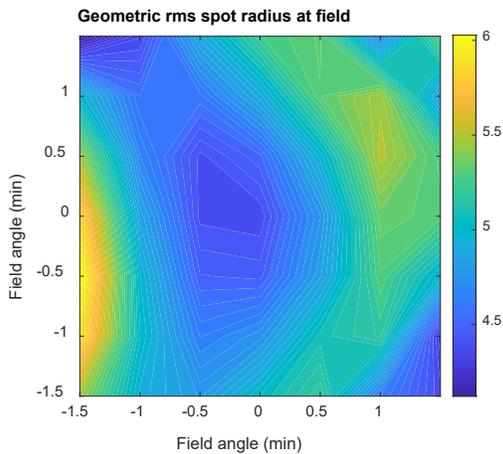

**Figure 6.6-5.** HWC geometric spot radius on the infrared channel (units of micrometer): the diffraction limited spot size is 23 µm: the entire focal plane is diffraction limited.

## 6.7 Instrument Thermal System

With the exception of the UVS, all other HabEx detector focal planes, and any sidecar electronics, require a cold environment than the primary mirror. HabEx passively cools these focal planes to achieve the required temperatures. The payload thermal design is based on a worst-case condition, where the telescope aperture is pointing anti-Sun with a solar angle is 180°. The temperature for each of the major detector components, including the FGS, and their worst-case heat lift requirements, current best estimate (CBE) and with margin, are shown in **Table 6.7-1**. There are heat lift requirements for the two temperature intercepts. At 77 K, the total lift requirement with margin is 9.3 W, where the contribution from the detectors is 5.1 W and structural losses is 4.2 W. At 153 K, the total lift requirement is 32.1 W, where the contribution from the detectors is 19.2 W and structural losses is 12.9 W. These structural losses also carry 50% margin from CBE. Separately, the UVS worst case heat lift requirement is 16.5 W at 270 K, which is negligible relative to spacecraft's thermal load and is assumed to be dissipated to the structure.

**Table 6.7-1.** HabEx's CBE detector heat lift requirements with 50% margin set the design point for the two-stage radiator. While this table identifies the heat lift margin, HabEx design can accommodate larger radiators as a form of configuration margin.

| Instrument | Temperature (K) | Worst-Case Lift, CBE (mW) | Worst-Case Lift, MEV (mW) |
|---|---|---|---|
| **HabEx Coronagraph (HCG) – Channel A** | | | |
| Vis IFS | 77 | 200 | 400 |
| Camera | 77 | 50 | 100 |
| **HabEx Coronagraph (HCG) – Channel B** | | | |
| Vis IFS | 153 | 200 | 400 |
| Vis Camera | 153 | 50 | 100 |
| IR IFS | 77 | 1,500 | 3,000 |
| IR Camera | 77 | 30 | 60 |
| **Starshade Instrument (SSI)** | | | |
| Vis IFS | 153 | 800 | 1,600 |
| Vis Camera | 153 | 50 | 100 |
| IR IFS | 77 | 30 | 60 |
| IR Guide Camera | 77 | 1,500 | 3,000 |
| UV Camera | 153 | 50 | 100 |
| **UV Spectrograph (UVS)** | | | |
| UV Detector | 270 | 11,000 | 22,000 |
| **HabEx Workhorse Camera (HWC)** | | | |
| UV/Vis Camera | 153 | 11,000 | 22,000 |
| IR Camera | 77 | 90 | 135 |
| **Fine Guidance System** | | | |
| FGS A | 153 | 50 | 100 |
| FGS B | 153 | 50 | 100 |
| FGS C | 153 | 50 | 100 |
| FGS D | 153 | 50 | 100 |

HabEx's payload thermal architecture is summarized in **Figure 6.7-1**. The coolest, 77 K components are individually isolated and heat strapped together to a 77 K thermal bench, while 153 K components are similarly isolated and strapped to a 153 K bench. Each of the benches are coupled to a heat pipes connecting them to their respective radiator. In order to minimize parasitic loads, the heat pipes will run inside a cooling tunnel (configuration shown in **Figure 6.7-2**) that nests the lowest temperature heat pipes in the center of the tunnel. In the same nesting concept, the manifold will also be nested inside a cooling box with the coldest manifold in the center. This cooling manifold will allow the combined capacity of separate passive cooler panels into effectively one large panel.

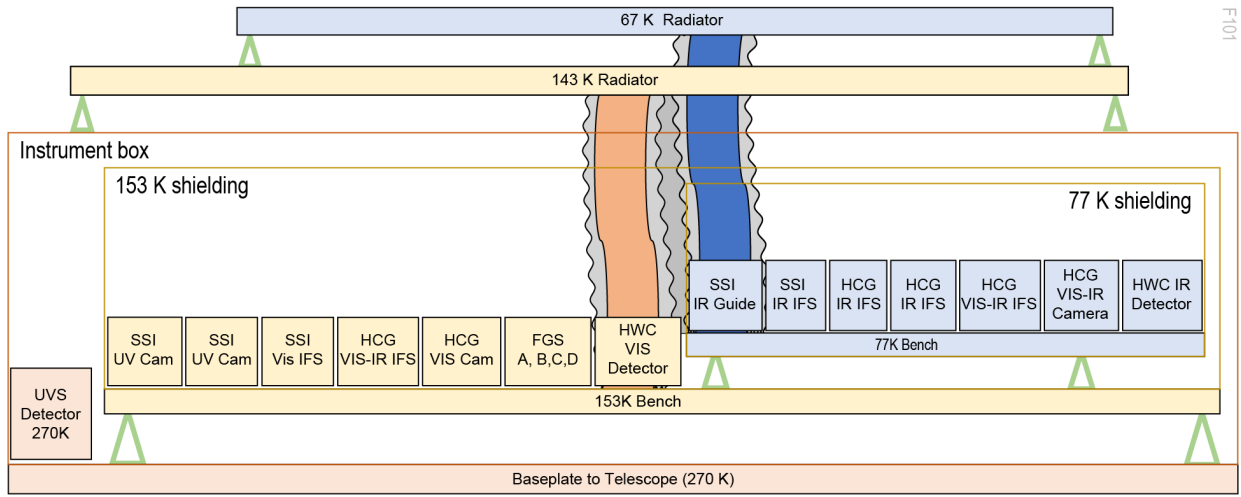

**Figure 6.7-1.** HabEx's instrument thermal system is a conservative architecture that utilizes the large available area for radiators to passively cool focal planes and sidecar electronics to meet their thermal requirements. Note that conduction path from benches to radiators are thermally isolated but mechanically configured within the same cooling tunnels. See Figure 6.7-2 for detailed view of the cooling tunnel.

In order to provide lift at 153 K, the first stage radiator rejects heat at 143 K while the second stage, mounted directly above the first, rejects to 67 K. The first stage radiator is 9.5 $m^2$, with 1.4 $m^2$ exposed to free space. The 67 K radiator sitting above the 143 K radiator is

8.1 $m^2$. The total footprint of the radiators, 9.5 $m^2$, is significantly smaller than the total available area on the anti-Sun side of the telescope, offering clearance to the telescope cover and configuration margin should the required radiator size grow.

## 6.8    Optical Telescope Assembly

As noted previously, the baseline telescope consists of the primary mirror assembly, secondary mirror assembly, tertiary mirror assembly, secondary mirror tower with integrated science instrument module, and stray-light baffle tube with forward scarf. The tower and baffle tube form an optical bench maintaining alignment between the PM, SM, and TM assemblies. The OTA is physically separate from the spacecraft so that the OTA and spacecraft connect only at the interface ring. The optical assemblies forming the OTA are discussed in the following subsections.

### 6.8.1    Primary Mirror Assembly

The PM assembly is an integrated optomechanical system consisting of the primary mirror, its mount, support structure and launch lock system, and an active thermal control system. This section discusses critical trades leading to the baseline PM assembly design, and that design itself.

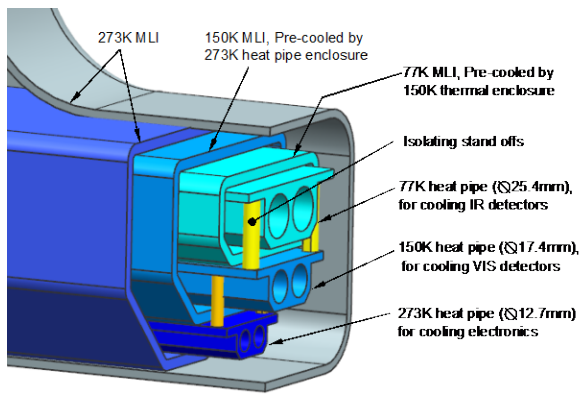

**Figure 6.7-2.** Cross-sectional view of the cooling tunnel that contains heat pipes connecting. Illustrated concept includes heat pipe for 273 K should UVS not be able to conduct heat to a sink attached to the detector.

### 6.8.1.1 Trade: Monolithic vs. Segmented Primary Mirror

The single most important telescope design decision is whether to make the primary mirror monolithic or segmented. It is much easier to achieve the ultra-stable wavefront required for coronagraphy with a monolithic mirror than with a segmented mirror. Habitable zone coronagraphy requires extreme wavefront stability in the mid-spatial frequency regime, i.e., 2–10 cycles per aperture. If the mirror were segmented, the segments would have to be aligned and phased to nanometer accuracy and their positions would have to be maintained with picometer stability (RMS). Also, segment edges causes diffraction in the telescope, degrading coronagraph contrast performance. A monolithic mirror avoids both of these issues.

A segmented design also requires a significant number of actuators for rigid body positioning and segment surface curvature adjustments, and a stiff backplane to react against the actuators. This adds considerable complexity to the primary mirror assembly.

To their advantage, segment mirrors may be lighter weight than an equivalent sized monolithic mirror, and the segments can be made stiffer than the monolith which eases segment fabrication requirements. Given that the telescope flight system is launching on the SLS Block 1B and has over 10,000 kg of launch margin, mass was not a major design consideration for HabEx.

In light of all these issues, HabEx adopted the monolithic design for its system simplicity, and superior contrast performance.

### 6.8.1.2 Trade: Primary Mirror Diameter

The STM sets the minimum mirror diameter at greater than 3.7 m due to exoplanet yield considerations, but the larger baseline design was selected because discussions with industry and space telescope experts indicated that nearly all the infrastructure necessary to build a 4m monolithic mirror for space applications was already in place. SCHOTT has existing infrastructure to melt, cast and machine lightweight Zerodur® mirror substrates up to 4.2 m (**Figure 6.8-1**; Westerhoff and Hull 2018). Similarly, Corning has infrastructure to assemble 4 m 'class' ULE® mirror substrates via either frit bond or low-temperature-fusion (LTF). Additionally, several organizations have the infrastructure to grind and polish 4 m class substrates into space mirrors, including Collins Aerospace, L3/Brashears, Harris Corporation, Arizona Optical Systems and University of Arizona, and RESOC (see *Appendix F* for more details). An example of this capability is shown in **Figure 6.8-2**: the 4.2 m SOAR primary mirror undergoing computer-controlled polishing at Collins Aerospace (Cox 2018). Additionally, several organizations are considering, planning or implementing the ability to coat 4 m to 6 m class mirrors by scaling up proven processes demonstrated on 2.5 m class mirrors, including: Collins Aerospace (Cox 2018), ZeCoat Corp (Sheikh 2018), and Harris Corp. With all the necessary infrastructure likely to be in place by the start of a HabEx mission, fabrication of a 4 m primary is a modest technology risk. See *Chapter 11* of details on the mirror and mirror coating technologies.

### 6.8.1.3 Trade: Primary Mirror Material

The coronagraph requires a very stable wavefront to reach the maximum contrast. The wavefront changes as thermal conditions change and so thermal properties of the mirror material properties have a great deal of influence. The mirror material choice determines the mirror CTE and CTE homogeneity properties ($\Delta$CTE), and material choice also affects mirror stiffness. Stiffness is largely a function of the mirror design, but the material chosen constrains what mirror design features are possible. Most large space telescope mirrors are fabricated from either Zerodur® or ULE® materials; both having highly stable thermal properties. HabEx evaluated both as part of the primary mirror material decision.

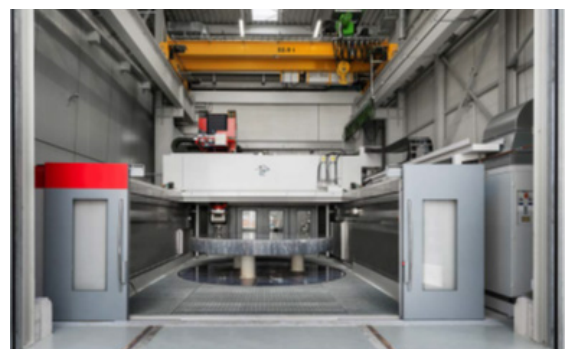

**Figure 6.8-1.** SCHOTT glass 5 m 5-axis CNC machine center loaded with a 4.5 m glassy Zerodur® blank (Nemati et al. 2017).

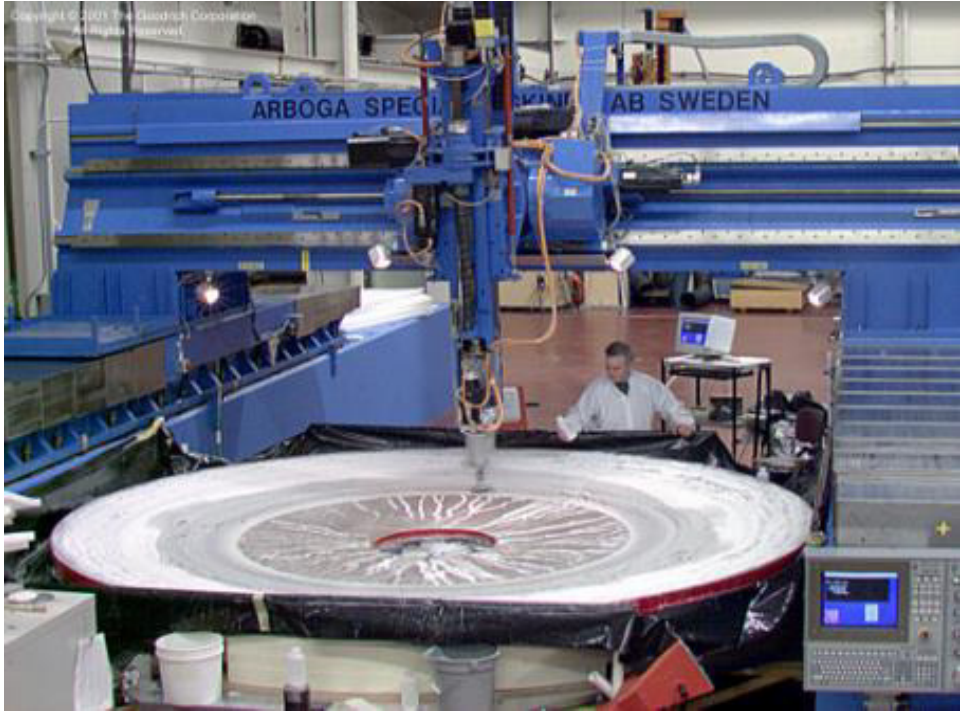

**Figure 6.8-2.** Collins Aerospace computer-controlled manufacture of 4.2 m SOAR primary mirror (Cox 2018).

CTE and CTE homogeneity determine how the mirror's shape deforms as a function of changes in bulk temperature or thermal gradient changes, introducing WFE. Zerodur® and ULE® mirrors are at TRL 9 with multiple examples currently flying in space. Both materials can be tailored for a certain zero CTE temperature and they have similar CTE homogeneity at ±5 ppb.

However, differences arise in material-dependent manufacturing methods and in mirror architecture—whether the mirror is open backed or closed back. Because Zerodur® is a ceramic, it must be machined from a single boule, resulting in an open-back architecture. By comparison, ULE® is a glass and can be assembled from multiple boules via frit bonding or LTF processes into a closed-back architecture, which results in significantly higher stiffness for a given mass. On the other hand, since Zerodur® mirrors are machined from a single boule, their CTE distribution can be smoother and more homogeneous than a ULE mirror.

The higher a mirror's stiffness, the easier it is to produce the smooth surface needed to achieve the required wavefront quality and the easier it is to handle (i.e., mount to machinery or turn over), which reduces fabrication risk. Mirror stiffness also impacts the WFE in two ways: first by coupling vibrations into the wavefront and second, via gravity sag. Inertial WFE is the deformation that a mirror experiences when it is accelerated against its mount. Telescope vibrations thus create a time-varying deformation of the mirror surface, in turn causing a varying wavefront. The stiffer the mirror, the less deformation it will undergo for a given acceleration. The alternative is to minimize the mirror's exposure to acceleration by minimize thruster noise or by isolating the mirror from such noise.

Gravity sag introduces a static WFE contribution. During fabrication, mirrors deflect under gravity when attached to their mount, whereas in space there is no such deflection. The change in the mirror's shape from 1 $g$ to 0 $g$ is called $g$-release. The primary mirror must be fabricated to its required on-orbit figure by characterizing and removing gravity sag from metrology data. Any difference between the estimated and actual on-orbit figure is termed *g*-release error and will form a component of the wavefront error budget. Thus, the higher the mirror's stiffness, the smaller will be its inertial wavefront error (produced by vibrations for example) and the less likely it is that the mirror will have significant *g*-release error.

While methods exist to characterize self-weight deflection and produce 0 *g* space mirrors, it is also possible to mitigate *g*-release error risk via an active mirror (primary or secondary or deformable mirror). As seen in **Figure 6.8-3**, analysis of an alternative ULE® HabEx primary mirror indicates that 15 actuators can reduce *g*-release error by a factor of 20, 25 can reduce *g* release error by a factor of 40, and 50 actuators can reduce *g*-release error a factor of 100 (Kissil 2018).

To first order, to minimize gravity sag for a 4 m mirror, the mirror should be as thick and as light as possible. For ULE® the state-of-art thickness is ~30 cm. Specialized abrasive waterjet machines can cut core elements as thick as 28 cm (Egerman et al. 2015). These elements can be frit bonded or low-temperature fused to form a mirror substrate. To make thicker mirrors, the Advanced Mirror Technology Development (AMTD) project successfully demonstrated a stack and seal process by manufacturing a 40 cm thick test mirror (Stahl et al. 2014; Stahl et al. 2013; Egerman et al. 2015). For Zerodur®, SCHOTT has demonstrated 42 cm thick substrates and they are working to produce 45 cm thick mirrors (Yoder and Vukobratovich 2015). Other design elements that impact stiffness include: 1) face-sheet thickness; 2) open, closed or partially closed back 3) thickness of the back-sheet if any; 3) geometry of the core structure (i.e., iso-grid, rectilinear-grid or hex-grid), pocket size, core wall thickness, etc. For example, a hex-grid is more mass efficient but less stiff than an iso-grid. Thus, a hex-grid is typically used for closed back mirrors (because the back sheet adds stiffness) which an iso-grid is typically used for open back mirrors. Also of importance is the mirror mount geometry, i.e., 6- point versus 3-point kinematic mounting.

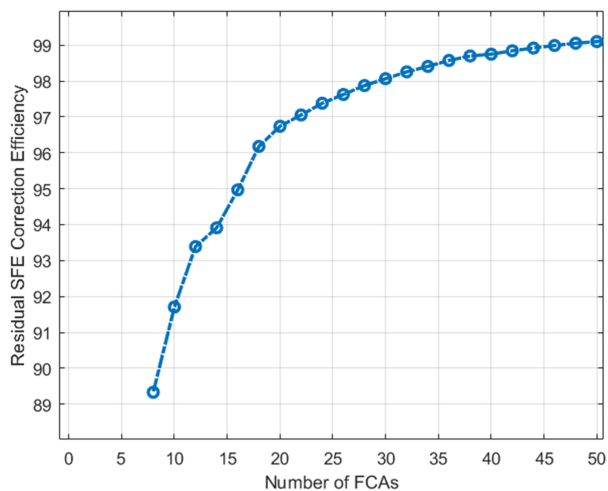

**Figure 6.8-3.** Correction efficiency (%) of alternative ULE® 4 m primary mirror design as a function of actuators.

As was done for HST, the gravity sag is dealt with passively through manufacturing measurement and polishing and includes actuators that can mitigate the risk of excessive *g*-release error (Yoder and Vukobratovich 2015).

Mirror stiffness is therefore an important design consideration in the development of the HabEx primary mirror. With reaction wheel-based ACS systems, the primary goal is to obtain the highest possible first mode frequency to minimize inertial WFE instability caused by reaction wheels. However, with low noise microthrusters baselined, the emphasis was on optimizing thermal stability, inertial WFE stability and demonstrated manufacturability. While both substrate materials are likely capable of meeting the HabEx design needs, Zerodur® was selected as the primary mirror material because the single boule nature of the Zerodur® mirror is expected to provide better thermal stability. Tests conducted at MSFC of a 1.2 m Zerodur® ELZM mirror have demonstrated better thermal stability than a 1.5 m ULE® mirror manufactured as part of the AMTD project (Brooks et al. 2018). Additionally, SCHOTT has demonstrated a routine ability to fabricate 4.2 m diameter Zerodur® substrates and turn them into lightweight structures via their extreme-lightweight Zerodur® Mirror (ELZM) machining process.

The baseline Zerodur® mirror assembly provides an excellent balance between mass and stiffness. The substrate has a flat back, open back geometry with a 42 cm edge thickness and mass of approximately 1,400 kg (**Figure 6.8-4**). The mirror's free-free first mode frequency is 88 Hz and its mounted first mode frequency is 70 Hz. Mass is important because it provides thermal capacity yielding a more stable mirror. Additionally, mass allows for local stiffening of the substrate to minimize gravity sag (Arnold and Stahl 2018). The mirror substrate geometry and hexapod mount designs were optimized to produce a uniform XYZ gravity sag deformation. The mirror is attached at three edge locations to a hexapod mount system. This geometry was selected to allow defocus and minimize spherical gravity sag based on vector vortex coronagraph aberration sensitivity. **Figure 6.8-5** shows the baseline mirror's predicted 1 $g$ peak-valley surface gravity sag in global telescope XYZ coordinate system.

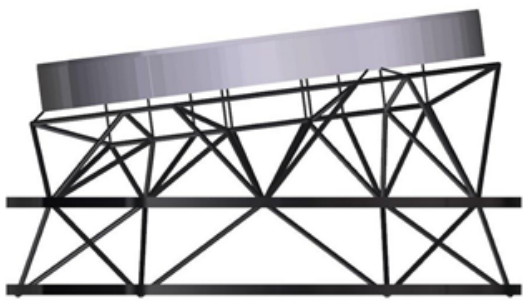

**Figure 6.8-4.** HabEx primary mirror assembly structure.

Finally, additional mass enables the mirror to have thicker structural elements which makes it easier to manufacture with lower risk and lower cost.

As a matter of completeness, other substrate designs considered include: an MSFC 45 cm thick closed-back ULE® mirror with total mass of 1,388 kg and first mode frequency of 180 Hz (Davis et al. 2017); a Harris Corp. 40 cm thick closed-back ULE® mirror with total mass of 440 kg and first mode frequency of 137 Hz developed under the AMTD project (Matthews et al. 2013); a SCHOTT AG Zerodur® ELZM on-axis 34 cm thick design with 718 kg and approximately 80 Hz first mode (Hull et al. 2013); and a Collins Corp. Zerodur® shaped-back mirror

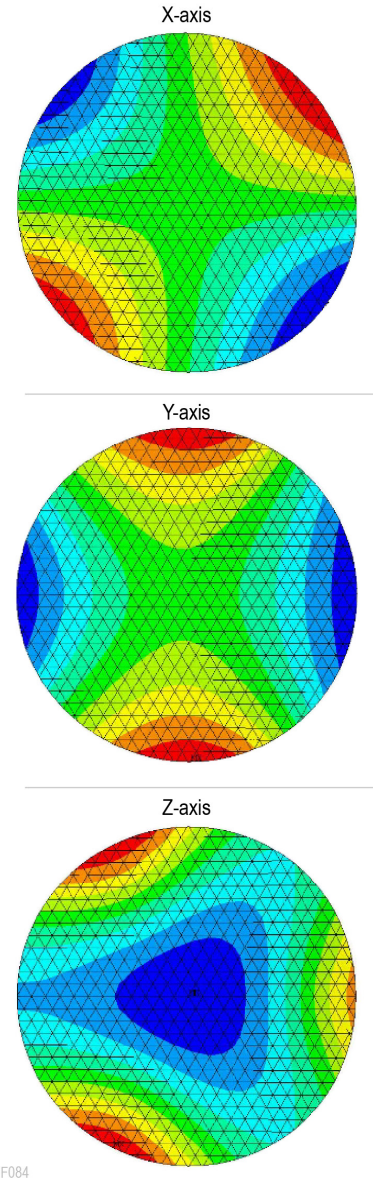

F084

**Figure 6.8-5.** Modeled results of PM gravity sag, with scale representing peak-to-valley. X-,y-, and z-axis gravity sag are 18.6, 18.4 and 12.6 µm rms, respectively.

with mass of 1,200 kg and first mode frequency of 120 Hz (Cox 2018).

### 6.8.1.4 Primary Mirror Support Structure

The primary mirror support structure is a simple truss. It is specified to be manufactured from TRL 9 M46J with a total mass of approximately 1,200 kg (**Figure 6.8-6**). To minimize WFE instability, the HabEx primary mirror hexapod supports and truss structure are designed for its rigid body and bending modes to be above 40 Hz. **Figures 6.8-6** and **6.8-7** show a 43.5 Hz rocking mode and a 50 Hz bouncing mode.

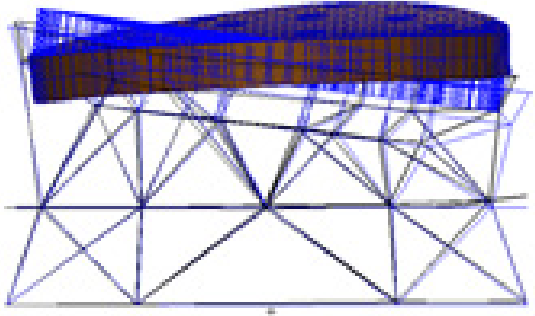

**Figure 6.8-6.** 43.5 Hz rocking mode.

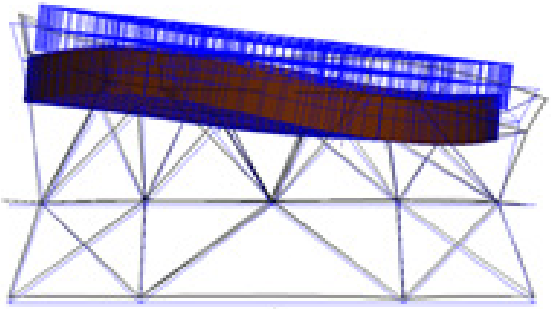

**Figure 6.8-7.** 50 Hz bouncing mode.

Finally, the PM truss structure is designed to accommodate a launch constraint system consisting of 18 axial and 12 radial launch locks (**Figure 6.8-8**). While Zerodur can withstand stresses up to 17.4 kpsi for short durations (Hartmann 2019), standard engineering practice is to limit the maximum launch load to 600 psi. The HabEx launch constraint system is predicted to expose no point on the mirror to greater than 300 psi (**Table 6.8-1**). Without the constraint system, launch stress of as much as 1,000 psi would be concentrated at the 3 hexapod attachment locations (**Figure 6.8-9**). Note that the launch constraint support structure is also used as a reaction structure for mirror actuators in the active low order mirror figure control system carried as a safeguard against errors in gravity release compensation.

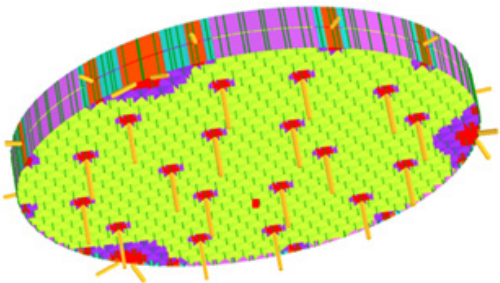

**Figure 6.8-8.** Primary mirror launch constraint systems has 18-axial and 12-radial launch locks.

**Table 6.8-1.** Baseline HabEx 4 m primary mirror launch stress.

| Acceleration Loads [g] | | | No-Lock Stress [psi] | Locked Stress [psi] |
|---|---|---|---|---|
| X | Y | Z | | |
| 0.5 | 0.0 | 6.0 | 995 | 197 |
| 0.0 | 0.5 | 6.0 | 959 | 160 |
| 2.0 | 0.0 | 3.5 | 702 | 297 |
| 0.0 | 2.0 | 3.5 | 657 | 233 |

### 6.8.1.5    Primary Mirror Actuators

The primary mirror is attached to 6 hexapod actuators and 30 launch lock mechanisms (18 axial and 12 radial). The secondary mirror is also attached to 6 hexapod actuators and launch lock mechanisms. These actuators and mechanisms are considered to be TRL 9 components given that similar actuators and mechanisms are currently flying on Hubble and Kepler and will fly on JWST and WFIRST. For example, JWST uses a two-stage stepper-motor actuator for both launch restraint and alignment. Its coarse stage has a range of 20 mm (12.5 mm of which is used to deploy from the launch restraints). The fine stage uses a mechanical gear stage to drive an eccentric cam shaft with a step size of 7.7 nm (Chonis et al. 2018). Given that the fine stage is mechanical, smaller step sizes required for HabEx can be achieved with a different gear ratio. Additionally, the AMTD project designed, built and characterized a fine-stage actuator with a range of 15 µm, step size of 0.8 nm, mass of 0.313 kg, and axial stiffness of 41 N/µm (**Figure 6.8-10**; Stahl et al. 2014).

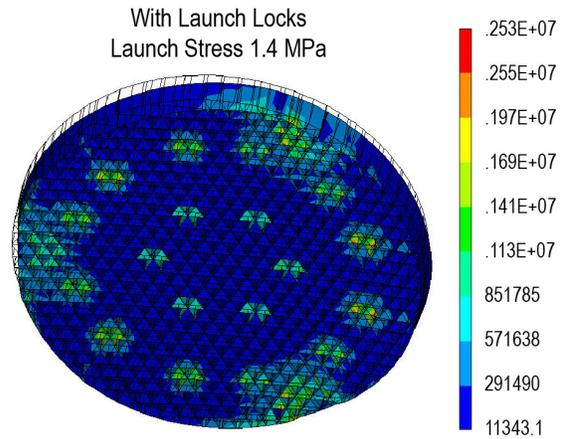

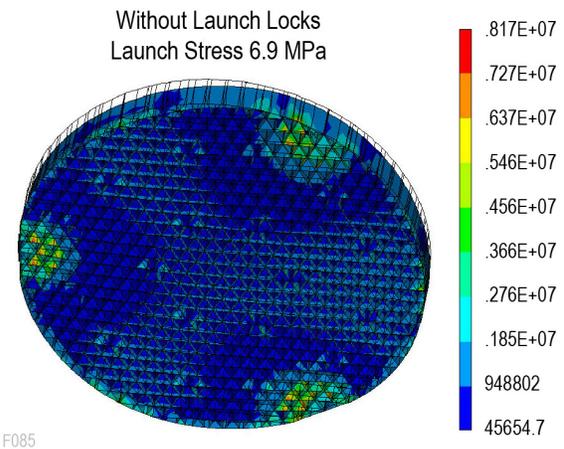

**Figure 6.8-9.** Launch locks redistribute launch stress from the three hexapod attachment locations to the entire mirror.

### 6.8.1.6    Primary Mirror Reflective Coating

The baseline reflectance coating for the primary and secondary mirrors is a Hubble-like aluminum coating with magnesium-fluoride protective overcoat. The tertiary mirrors will be coated as required for the instruments served. The coatings and deposition processes are TRL 9, having been used on flight programs since the 1970s.

### 6.8.1.7    Primary Mirror Thermal Control System

The primary mirror thermal control system is critical to the telescope's ability to achieve the required diffraction limited performance and wavefront stability. The function of the thermal control system is to uniformly set the primary mirror's front surface to the desired operating temperature and keep it at that temperature regardless of where the telescope points on the sky relative to the Sun. The precision to which the system can maintain the mirror's temperature determines the wavefront stability. A gradient in the mirror's temperature will introduce a static wavefront error and a temporal variation will introduce a varying wavefront error.

Like Hubble, HabEx will heat the primary and secondary mirrors to the desired operating temperature within a cooler environment (known as cold-biasing). The operating temperature is constrained by two competing requirements: near-IR science requires cool mirrors to minimize in-field thermal radiation from the mirror surfaces while UV science requires that the mirrors be free of any contamination, such as water ice or other out-gassed molecules, to maximize spectral throughput. HabEx has an operating temperature of 270 K for its mirrors, above the sublimation temperature for water ice. The amount of cold bias is also constrained by competing requirements. The greater the bias, the easier it is to control the mirror temperature but the more electrical power required to achieve that control. The ideal operating temperature is where the mirrors are minimally cold-biased for all potential sun orientation angles. The desired cold bias was achieved by providing some area open to space on the anti-sun side of the spacecraft.

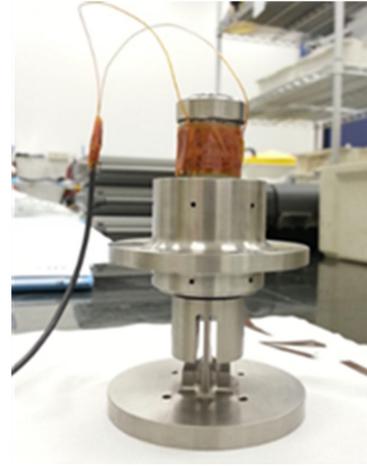

**Figure 6.8-10.** AMTD fine stage (0.8 nm step) actuator (Stahl et al. 2014).

HabEx utilizes a thermal control system with radial and azimuthal heater zones behind and around the perimeter of the primary mirror to compensate for PM thermal gradients. Radiative transfer to the primary mirror within the straylight tube can cause the middle of the mirror to be colder than the edge. Also, the thermal load into the sun-side of the telescope will change as the telescope boresight angle changes relative to the sun, thus changing the lateral temperature gradient experienced by the primary mirror. The thermal control system's radial heater zones add heat to the back of the mirror to compensate for the radial gradient, creating a uniform front surface temperature. The azimuthal heater zones add heat where necessary around the edge of the mirror to compensate for changes in the lateral thermal gradient.

The baseline HabEx active radial thermal control concept is an engineering scale-up of 0.7 m, 1.1 m and 1.5 m telescope mirror systems built by the Harris Corp. Zonal active thermal control of primary mirrors is currently TRL 9 at the 1.1 m size with systems currently flying on the commercial Spaceview™ telescopes. Additionally, under the Astrophysics Division funded Predictive Thermal Control Study (PTCS), Harris Corp. has built and delivered to NASA, a 1.5 m system with 37 thermal control zones (**Figure 6.8-11**). This system has 6 azimuthal heater zones in each of 5 radial and circumferential zones.

The primary mirror's RMS surface figure error stability is proportional to its CTE and the temperature change, and inversely proportional to its mass and thermal capacity (Brooks et al. 2015). Thus, larger mirror mass and the smaller CTE are preferred, leading to the choice of zero CTE materials such as Zerodur® and ULE® glass. Achieving an ultra-stable thermal wavefront requires the thermal control system to sense and correct fluctuations to the mirror's thermal environment faster than the mirror's response time to those changes. For example, using the generic mirror design illustrated in **Figure 6.8-12**, if the thermal sensors are uncertain to 50 mK, the control period needs to be about 50 s but if the sensors are uncertain to 5 mK, the control period can be 500 s. With the same 5 mK uncertainty, a factor of 10 improvement in wavefront control can be accomplished by controlling at 50 s.

For a set thermal sensing system control period, sensing noise determines wavefront stability performance. The current TRL 9 telescope thermal sensing noise capability is defined by the Spaceview™ telescope thermal control system sensors which have a noise of ~50 mK and control their 1.1 m telescope mirrors to a temperature of 100–200 mK (Havey, Keith, private communication, March 13, 2019). It is important to note that these control systems are Earth-viewing systems in low Earth orbit. These same systems would be much more thermally stable performing astrophysics science, stationed in a SE-L2 orbit. STOP analysis presented in *Section 6.9* shows that the more massive HabEx baseline primary mirror can be controlled to a temperature of ~1 mK with a system having the same 50 mK sensor noise as Spaceview, and a 30 s control period. The thermal enclosure system modeled uses commercial platinum resistance thermometers with ±5 mK reproducibility and ±10 mK long term stability (Cryotronics 2019). This shows that there is significant performance margin that could be obtained (although not required) by using sensors with lower noise. For example, the Harris-designed thermal control system for WFIRST uses 4-wire bridge-circuit thermistors with less than 4 mK noise (Havey, Keith, private communication, March 13, 2019).

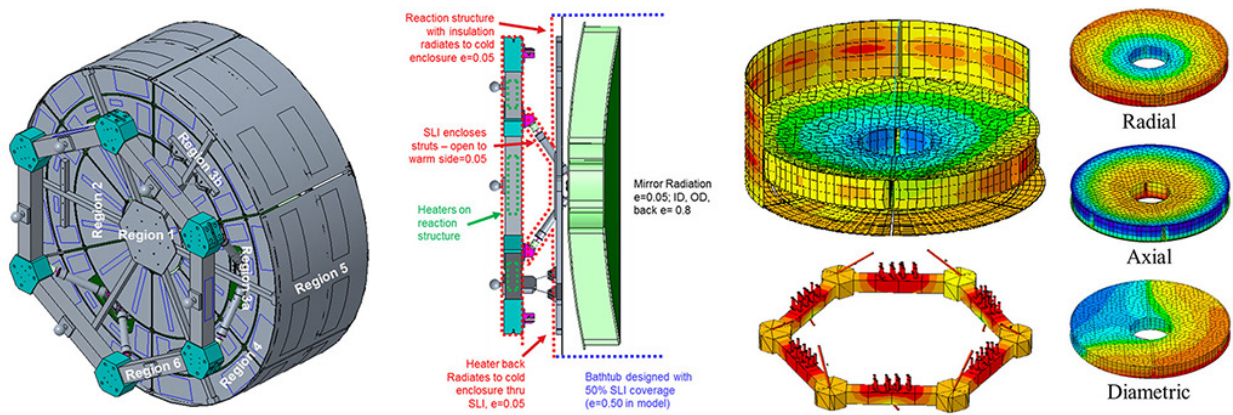

**Figure 6.8-11.** Predictive thermal control study zonal thermal control system technology demonstrator.

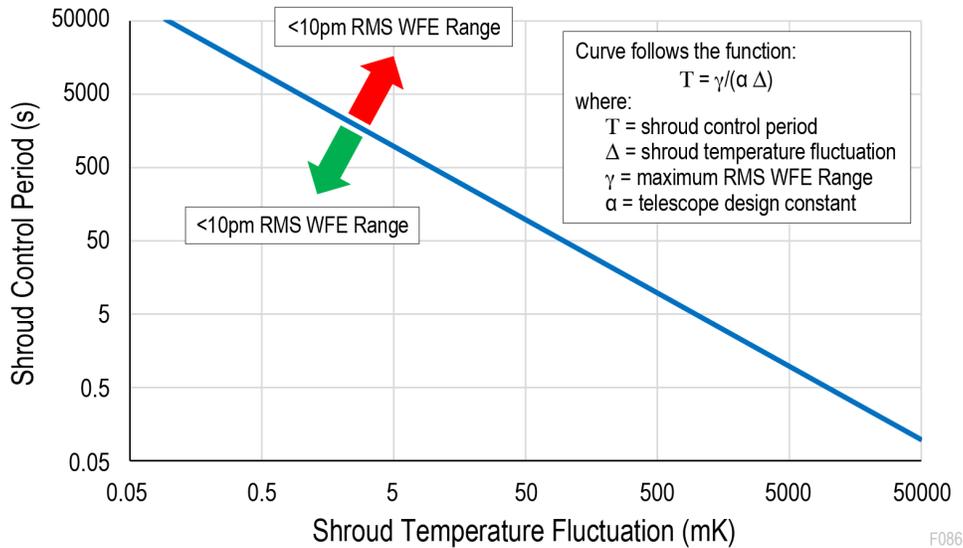

**Figure 6.8-12.** Thermal wavefront stability is achieved by balancing thermal sensing noise and control period (Hartmann 2019).

### 6.8.2 Secondary Mirror Assembly

Like the PM assembly, the SM assembly is an integrated opto-mechanical system consisting of a mirror substrate, its mount, support structure, launch lock system, and thermal control system. The mirror substrate is a 0.45 m diameter off-axis Zerodur® mirror. Zerodur® was selected based on its expected CTE homogeneity. The SM assembly mount, support structure, launch locks, and thermal control system are similar to those of the primary mirror assembly. Because the tertiary mirror is fixed, the SM assembly is actuated via a hexapod to maintain its optical alignment with the PM and SM. The SM assembly and supporting laser-truss system (not shown) are mounted to the top of the tower structure (**Figure 6.8-13**).

For a 0.45 m size mirror, a surface figure error of less than 7 nm RMS is readily achievable. The AMTD study demonstrated 5.4 nm RMS on a 0.43 m mirror with a 10 mm facesheet (Stahl et al.

2013). Furthermore, because the SLS has significant mass margin, the SM can have a thicker facesheet to minimize mid-spatial frequency error (important for coronagraphy). It is very likely that the SM total surface error could be better than 5 nm RMS. The design, manufacture, and verification of the SM assembly and its constituent components is considered to be well within the state of practice for space telescopes.

### 6.8.3 Tertiary Mirror Assembly

In the HabEx TMA design, the TM's location is fixed and PM, SM, and science instruments are aligned to the TM. This is particularly important for the science instruments because each uses a different portion of the main toroidal TM (the UVS uses a separate on-axis TM).

The volume surrounding the tertiary mirror (**Figure 6.8-14**) is crowded with incoming light beams, pick-off mirrors, and the four fine guidance sensors underneath. As such, it makes sense to include all of these optical elements in a single unit—the TM/FGS assembly.

The reflected light after the tertiary mirror is collimated, making alignment with the instruments simpler. The reflected light is diverted to the HCG, SSI, HWC, and the four FGS focal planes by flat, pick-off mirrors mounted above the TM.

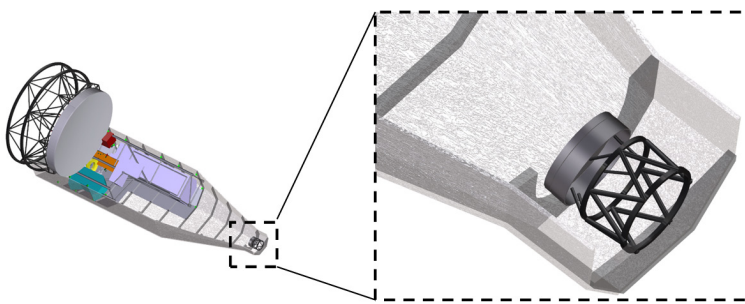

**Figure 6.8-13.** Secondary mirror assembly mounted to top of tower structure.

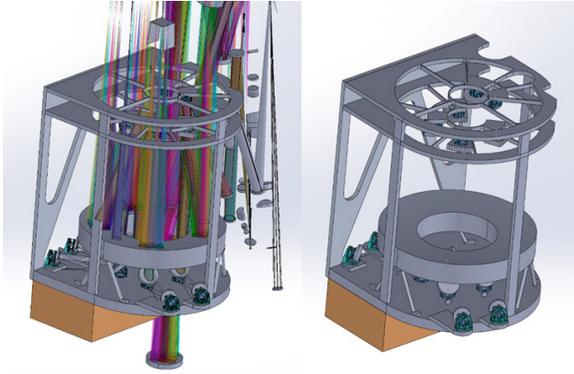

**Figure 6.8-14.** Tertiary mirror assembly, with and without light beams.

Laser metrology monitors the TM relative to the SM, which is, in turn, monitored relative to the PM. Additional MET beams between the TM and the PM improve the ability to sense, and correct, motion of the primary mirror with respect to the tertiary mirror.

The four fine guidance sensors are mounted underneath the TM. Each FGS has a steering mirror to enable a 4.7 arcmin² field of regard and the ability to optimally place a guide star on the detector. The FGS control and detector electronics are mounted in an enclosure directly underneath the bottom plate.

The tertiary mirror is the third powered optic of the three-mirror anastigmat afocal telescope. It is 720 mm in diameter with a 370 mm diameter hole in the center. The TM is constructed from ULE© using low temperature frit bonding techniques. A thicker front face sheet is required to ensure a high-quality, very low surface error footprint. The TM is mounted with three bipods to support it from the TM assembly structure.

### 6.8.4   Integrated Science Instrument Module

The integrated science instrument module (ISIM) is an optomechanical structure whose function is to maintain optical alignment of the science instruments relative to the tertiary mirror. The ISIM is also a key structural component of the overall tower structure. The ISIM is designed to be removed from the observatory on precision HST-style optical rails for servicing as a whole.

Once removed, individual science instruments can be replaced—again using precision HST-style optical rails.

### 6.8.5   Laser Metrology System (MET)

The laser metrology system provides sensing and control of the rigid body alignment of the telescope. In a closed loop with actuators, MET actively maintains alignment of the telescope front-end optics, thereby eliminating the dominant source of wavefront drift. With an internal laser source, which is bright compared to the starlight, MET is not photon-starved and can operate at high bandwidth. Furthermore, laser metrology maintains wavefront control even during attitude maneuvers such as slews between target stars.

Laser metrology for large coronagraph-equipped space-born observatories was first proposed for the Terrestrial Planet Finder Coronagraph (Shaklan et al. 2004). The early versions consisted of small optical benches populated with discreet optical beam splitters, retroreflectors and lenses. Recently, the optical bench has been miniaturized using planar lightwave circuit (PLC technology developed for the optical communications industry) resulting in a compact lightweight beam launcher (**Figure 6.8-15**).

The principle of operation is as follows: a laser beam is split into two and frequency shifted by different amounts (**Figure 6.8-15**) to obtain a 40 kHz difference frequency. The frequency shift is achieved using two acousto-optic modulators (AOMs). The beams travel via fiber optics to a beam launcher built using planar lightwave circuit (PLC) technology. The beam launcher transmits a collimated beam through free space to a corner cube at the secondary mirror (target). The reflected beam couples back into the beam launcher where it mixes with the other incoming beam (local beam). Similarly, the two source and reference beams are mixed in the PLC circuit. Both mixed (heterodyne) signals are sent to a phasemeter via fiber optics. The phasemeter detects the signals, producing two 40 kHz sine waves where the measured phase difference is proportional to the change in the distance to the corner cube (modulo $2\pi$). Each beam launcher needs only a few microwatts of laser light, so one laser will be sufficient to support all 18 beam launchers. Each beam launcher weighs less than 70 g with a single-digit [nm/°C] temperature (error) coefficient. The current generation MET system has 0.1 nm measurement error at 1 kHz sampling.

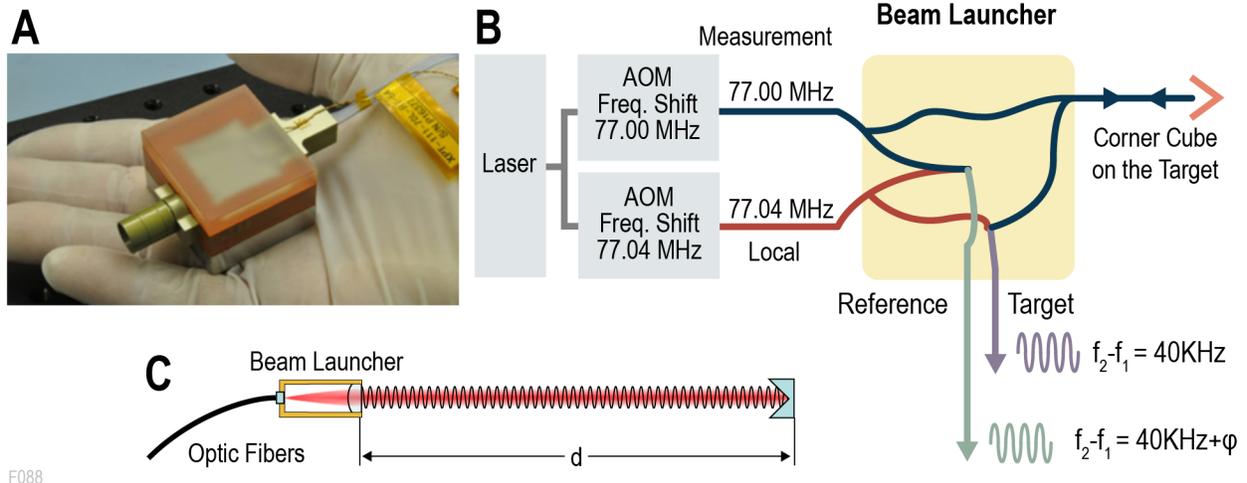

**Figure 6.8-15.** *A)* Planar lightwave circuit compact beam launcher. *B)* The heterodyne technique eliminates common mode phase shifts between the target phase and the reference phase measured at the phasemeter. *C)* Collimated laser beam launched to distant retroreflector.

The phasemeter output is sensed at 1 kHz, while the full control loop operates at 10 Hz. This control bandwidth is sufficient to counteract any thermally related disturbances in the structure. A greater control bandwidth is unnecessary given that the use of microthrusters during observations produces almost negligible high frequency mechanical disturbances.

### 6.8.5.1 HabEx Laser Metrology Truss

The HabEx laser metrology truss measures the distances between the telescope PM, SM, and TM. A laser metrology "truss" involves multiple single distance measurements arranged much like bipods. As shown in **Figure 6.8-16**, nine distances are measured between three points on the circumference of the primary mirror and three points on the secondary mirror. Similarly, another nine distances are measured between three points on the tertiary mirror assembly and three on the secondary mirror.

Each leg of this truss consists of a beam launcher and a retroreflector. From changes in the distances, relative rigid body motions of two of the mirrors with respect to the third may be derived from the geometric truss equation. A closed loop control system sends commands to rigid body actuators on the secondary and tertiary mirrors to counteract these motions and maintain the truss in its original state.

With an uncorrelated gauge error of 0.1 nm per gauge, and a laser truss based on the 4 m

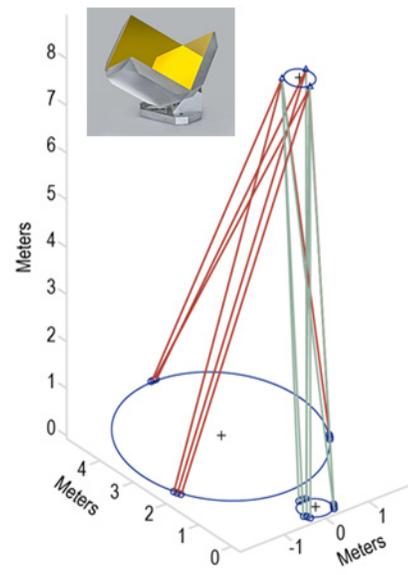

**Figure 6.8-16.** Model of the HabEx metrology truss. Positions of primary, secondary and tertiary mirrors are illustrated by circles. Inset, example of a hollow corner cube retroreflector.

HabEx off-axis telescope configuration, MET is capable of maintaining the position of M2 to less than 1 nm and 1 nrad, and M3 to less than 3 nm and 1 nrad (with the exception of M3 clocking at which is less than 5 nrad; see **Table 6.8-2**).

Laser metrology systems have flown on LISA Pathfinder and GRACE-Follow On. See *Chapter 11* and *Appendix E* for further details on the program to mature the technology required by MET.

### 6.8.6    *Telescope Fine Guiding Sensors*

The telescope pointing requirement is 2 mas RMS per axis at the FGS—about 1/10th of the 21 mas full width at half maximum (FWHM) of its diffraction-limited PSF at 0.4 μm wavelength. This amount of error reduces the Strehl ratio from the nominal 80% (diffraction limited) to 77.5%. Thus, the peak of the PSF for a chosen target will be reduced by only 3%: a small effect on observing efficiency. For the starshade instrument, the workhorse camera, and the UV spectrograph, this level of pointing is sufficient. For the coronagraph instrument, additional internal pointing refinement is required.

The HWC itself would be an adequate sensor for pointing the telescope 98% of the time. However, for the coronagraph it is an inadequate sensor for roll because there are too few well-separated stars and the instrument has insufficient angular resolution. Assuming the telescope rolls slightly around its optical axis, there is an induced motion of the off-axis instruments' fields of view on the sky, which appears as a tilt of the wavefront of any observed object. Even a small tilt is important for the coronagraph. This "tilt" will be detected by the coronagraph's ZWFS and corrected by the FSM. However, there remains a wavefront error caused principally by the beam footprint moving a small amount across the tertiary mirror. Different parts of the mirror have different residual surface irregularities, and the effect is to introduce an uncorrected wavefront error at the coronagraphic mask. The coronagraph's pointing requirements arise from the coronagraph error budget, shown in **Figure 5.2-1** and summarized in **Table 6.8-3**. In turn, these requirements derive fundamentally from the contrast degradation created as the input beam "walks" across the optics. This error appears as a variation in the speckle pattern in the coronagraph dark field and drives the requirements on the telescope LOS error and hence its roll.

**Table 6.8-2.** MET rigid body motion residuals for a 0.1 nm uncorrelated gauge uncertainty per gauge for the HabEx MET truss.

| DOF | Secondary Mirror | Tertiary Mirror |
|---|---|---|
| Θx (nrad) | 0.18 | 0.27 |
| Θy (nrad) | 0.20 | 0.40 |
| Θz (nrad) | 0.82 | 4.6 |
| Δx (nm) | 0.22 | 2.9 |
| Δy (nm) | 0.22 | 2.3 |
| Δz (nm) | 0.04 | 0.14 |

For these reasons, a dedicated fine guidance system (FGS) is included in the HabEx design, utilizing some of the unused annular field of the TM. Roll stability requirements depend particularly on: the surface quality on the mirrors that experience most beam walk (in this case the TM and the following fold mirror), the accuracy of the estimate of the PSF center, and the availability of bright guide stars. Current technology allows the production of extremely well-figured optics for UV lithography with 1 nm RMS surface figure error, so special optics can be made for the sensitive locations in the beam train (TM and the fold mirror). The challenge then is to accurately measure the positions of a sufficient number of guide stars by which the roll of the telescope can be measured. **Table 6.8-3** shows a roll sensitivity comparison between two sensors, one with a small field of view (about 3'×3', corresponding to the HWC), and one with a larger FOV across the well-corrected annular field of the telescope (0.3° across). The large FOV FGS has ~9× better roll resolution.

**Table 6.8-3.** Fine guidance sensor roll sensitivity comparison. HabEx baselines four small FGS to cover a larger area and reduce pointing uncertainty.

| | Small FGS | Large FGS | |
|---|---|---|---|
| **Angular roll resolution** | 5.8 | 0.65 | arcsec |
| **Angle between FGS and coronagraph** | 0.29 | 0.15 | deg |
| **On-sky tilt angle** | 29.6 | 1.70 | mas |

With the set of four FGS sensors, roll estimation accuracy will be below ~1 mas for stars 17th magnitude or brighter. To estimate the sky coverage of the FGS, a numerical model was used (GSFC 2018) that generates an average across the whole sky of the number of stars in a given field of view. This model (because it assumes a uniform distribution) shows that the FGS will see sufficiently bright stars ~100% of the time. Naturally, there are some parts of the sky where the density of stars is low, but a gradual degradation in performance would be expected since the FGS can guide on just two stars and obtain both pointing and roll.

## 6.9 Telescope Flight System Performance and Error Analysis

There are many system-level design aspects that must be considered to find a space telescope design that successfully meets requirements. The HabEx study carried out both mechanical and thermal modeling at the flight system level to verify these performances. In addition, the study also carried out integrated structural, thermal, and optical performance (STOP) modeling to assess one of the most important design requirements of all—coronagraph contrast performance evaluated as part of the flight system and under likely operational conditions. It should be noted that these performance evaluations were not only conducted as a simple verification of the baseline design, but were an integral part in developing the design. This section describes the modeling and simulation of key aspects of the telescope flight system conducted during study.

### 6.9.1 Key System-Level Requirements

Modeling and simulation efforts were aimed at verifying a number of key requirements at the system level. These requirements are both related to the flight system's observational capabilities as well as basic launch performance requirements. All are summarized in **Table 6.9-1**.

As noted earlier, the PM operating temperature requirement is driven by the need to carry out observations in the ultraviolet and the telescope's throughput sensitivity to mirror contamination, particularly in the ultraviolet. Testing during JWST integration and test (I&T) indicated significant mirror contamination at temperatures below 258 K (Bolcar et al. 2016) so a requirement of mirror temperatures greater than 260 K was set on HabEx. Heating mirrors to temperatures considerable above the requirement uses considerable power and would eventually limit performance at the longer observational wavelengths so a nominal telescope mirror operating temperature of 270 K was adopted for the baseline. This temperature provides sufficient margin for the mirror temperature without unduly increasing power generation requirements.

**Table 6.9-1**. Key telescope system requirements compared to expected performance, based on Table 5.4-2.

| Parameter | Requirement | Expected Performance | Margin | Source |
|---|---|---|---|---|
| Instrument Complement | Exoplanet direct imaging and spectroscopy Imaging and spectroscopy in the UV, visible and NIR. high-resolution spectroscopy in the UV. | HCG, SSI, UVS, HWC | Met by design | MTM |
| Mass | ≤35,000 kg | 18,420 kg | 90% | MTM |
| Power | ≥4,500 W | 6,980 W | 55% | MTM |
| Configuration | Must fit within 8.4 m SLS fairing | Fits in SLS Block 1B | Met by design | MTM |
| 1st Launch Mode (Lateral) | >8 Hz | 11 Hz | 38% | MTM |
| 1st Launch Mode (Axial) | >15 Hz | >85 Hz | 466% | MTM |
| Field of Regard | ≥40° | ≥40° | Met by design | MTM |
| Slew Rate | ≥1 as/min | Up to 42.6 arcmin/sec | 153400000% | MTM |
| Raw Contrast | $3.00 \times 10^{-10}$ | $2.00 \times 10^{-10}$ | 50% | Error Budget |
| Raw Contrast Stability | $3.00 \times 10^{-11}$ | $1.45 \times 10^{-11}$ | 107% | Error Budget |
| LOS Stability | ≤2 mas | 0.7 mas | 186% | MTM |
| LOS Stability (from Beamwalk, HCG-only) | ≤4 mas | 1 mas | 300% | Error Budget |
| WFE Stability | 4.3811 nm RMS | 3.09 nm RMS | 29% | Error Budget |
| PM Thermal Stability | ±2 mK | 0.15 mK | 633% | MTM |

Telescope thermal stability requirements arise from the coronagraph thermal stability WFE allocation shown in **Table 5.4-3**. Telescope mirror thermal-mechanical distortion contributes to contrast degradation within the coronagraph, so managing the thermal stability of the mirrors, particularly the PM, is a significant feature in the overall flight system design (see *Section 6.8.1.7* for details on the telescope's thermal design). The mirror thermal stability is a factor in calculating the overall WFE budget (*Section 6.9.3*).

Similarly, **Table 5.4-3** identifies telescope mirror jitter requirements shown as rigid body motion (RBM) WFE error allocations that must be met for the coronagraph to meet the STM-specified contrast levels. Again, of the telescope's mirrors, the PM will be the most susceptible due to its relatively low stiffness when compared to the other two mirrors.

Telescope LOS stability comes from the STM requirement on telescope diffraction limit combined with the telescope aperture. Both set the diameter of the imaging PSF that sets the telescope's LOS requirement. An LOS stability of less than 2 mas is needed to avoid significant image blur and an effective loss of throughput at the science image planes.

Probably the single most important and challenging exoplanet-related requirement is meeting the $\sim 10^{-10}$ contrast level needed for detection and characterization of earth-sized planets in the habitable zone. This requirement comes from the STM and, with respect to the coronagraph, applies over the entire "dark hole" imaging area. This is a system-level requirement because the entire flight system design must be taken into consideration to meet this contrast level.

Lastly, there are several other key requirements related to the launch vehicle. Flight system mass, launch vibrational modes and meeting the fairing volume constraint are all telescope system-level requirements derived from the Mission Traceability Matrix (MTM).

All of the system-level requirements examined as part of this study strongly bear on the overall flight system performance or flight system sizing and as such are primary design considerations requiring early estimation to verify the viability of the overall system design.

### 6.9.2  Modeling Approach

While the launch mass requirement could be verified with a simple tabulation, other requirements need some form of system-level modeling and simulation. Mirror temperature and thermal stability require thermal modeling, whereas the launch requirements, telescope LOS and mirror vibrational displacement require structural modeling. The most complex modeling effort was for the coronagraph contrast simulation which required the development of a STOP model to evaluate the effects of system-level thermal-mechanical distortions on optical performance.

Structural and thermal modeling were carried out with commercially-available finite element tools. Recognizing that there would be an eventual need to use the tools in concert for the STOP modeling work, the thermal and structural models were built in parallel.

**Figure 6.9-1** shows the necessary elements of a full system model. There are two disturbance sources, the change in the sun angle and the operation of the thrusters and microthrusters. The mechanical forces generated by the microthrusters operate primarily on the primary mirror, its support structure and the secondary mirror tower. Similarly, thermal inputs operate on the telescope itself as well as the mirrors. Thermal effects on the structure result primarily in changes in telescope focus and, through bending, in astigmatism (paths A and B in **Figure 6.9-1**). Since these length changes to the structure are measured by MET and corrected via the telescope alignment control system they make only a small contribution to the total wavefront error.

Thermal effects on the mirrors are represented in paths C, D, E, and F of the figure. Bulk temperature change and radial temperature gradient contribute focus to the telescope, in the former case by changing the distances between mirror surfaces and in the latter by bowing the mirror. CTE inhomogeneity produces numerous effects explained below, but the focus components of C, D, and E can be detected via the ZWFS and compensated in the initial DM setup. Lateral $\Delta T$ (F) produces astigmatism, passed through to the vortex mask.

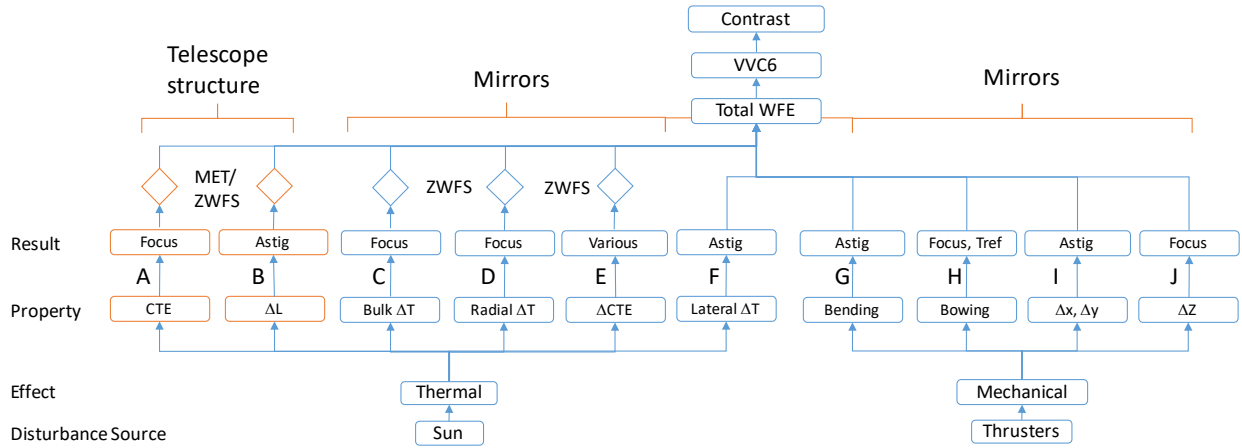

**Figure 6.9-1.** Principal elements of the thermo-mechanical system model. Thermal effects produce slow changes in the system, some of which may be compensated by MET or the ZWFS operating in concert with the DMs. Mechanical effects are too fast to compensate for but the larger forces are transient, while the continuous microthruster firings produce negligible forces.

Mechanical effects are shown on the right of **Figure 6.9-1** and they arise from the microthrusters and from the RCS thrusters when slewing from one star to another. The latter effect is temporary and only occasional, while the former is a small continuous effect as the telescope counters solar radiation pressure-induced torque. Lateral and vertical accelerations result in bending (G) and bowing (H) of the primary mirror, and also relative displacements of the mirrors (I and J) resulting in a small astigmatic response together with focus.

### 6.9.3 Wavefront Error Stability Budget

Mechanical jitter, thermal distortion, mirror CTE inhomogeneity and—through beam walk—MET error, can all contribute to the telescope system's WFE and its stability, which in turn can degrade the coronagraph's contrast performance. The WFE stability result of these thermal and mechanical effects can be understood through the optical model of the telescope which converts positional perturbations of the mirrors into a wavefront sensitivity matrix via the optical model

(top row of **Figure 6.9-2**; Nemati and Stahl 2019). Separately, a model of the coronagraph is used to produce a table of allowable amplitudes of these Zernike terms, converting WFE stability effects via a model of the vortex mask into coronagraph contrast (bottom row of **Figure 6.9-2**). The combination of the two models is used to produce an allowance for positional errors of the mirrors. Using this method, **Table 6.9-2** shows the generated requirements for the three TMA mirrors that meet a 2 mas LOS accuracy. The same modeling method can also be used to convert thermal and mechanical perturbations into WFE estimates.

**Table 6.9-3** shows the wavefront error budget in terms of the Zernike components with the corresponding contrast terms. It shows that baseline HabEx telescope's structural, thermal, optomechanical design meets the contrast requirements (**Figure 5.2-1**) with over 100% margin.

The overall contrast stability allocation from the coronagraph error budget is $3.0 \times 10^{-11}$. The error budget is constructed using the STOP model. The three contributors to the error budget are shown in the leftmost columns: rigid body motion (RBM), Inertial and Thermal. RBM instability is the WFE produced by rigid body motion of the primary and secondary mirror relative to the tertiary mirror excited by micro-thruster acceleration noise. As expected, RBM WFE instability is simply lower order optical alignment terms, which are well attenuated by the vector vortex. Inertial instability is the WFE produced when microthruster acceleration noise causes the mirrors to bend and deform their surface figures as they react against their mounts.

**Table 6.9-2.** Displacement and tip/tilt & rotation allowances for the telescope mirrors that achieve LOS stability <2 mas on sky per axis.

| Mirror | Displacement [nm] | | | Rotation [mas] | | |
|---|---|---|---|---|---|---|
| | ΔX | ΔY | ΔZ | Δθx | Δθy | Δθz |
| PM | 10.0 | 10.0 | 60.0 | 0.5 | 0.5 | 2.0 |
| SM | 10.0 | 10.0 | 60.0 | 5.0 | 5.0 | 20.0 |
| TM | 12.0 | 12.0 | 120.0 | 4.0 | 4.0 | 40.0 |

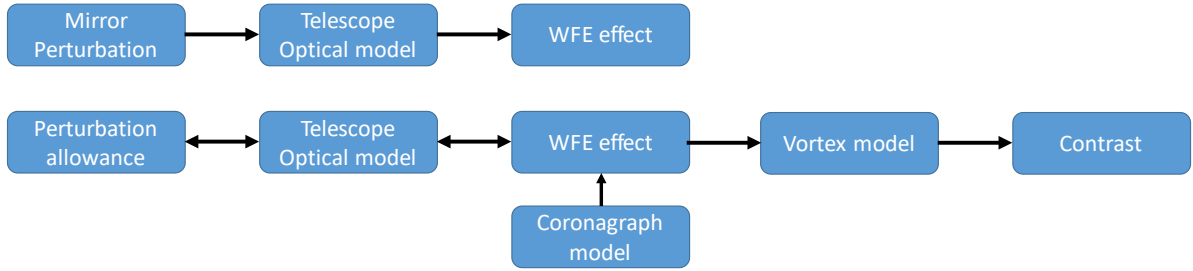

**Figure 6.9-2.** Calculating the effects of perturbations on the telescope optics and the generation of a perturbation error budget.

RBM and inertial WFE instability are assumed to be undetectable by MET and is thus uncontrolled. Thermal WFE instability arises from the deformation of the optics under thermal changes. And as shown in the earlier sections, there are three components to thermal WFE instability: CTE inhomogeneity, thermal gradients and residual RBM (where residual RBM is the amount of RBM produced by the structure that the MET system does not correct). The resulting total predicted wavefront error for each Zernike mode is calculated by root-sum-squaring the three components across the rows.

**Table 6.9-3.** WFE stability budget calculation. Estimated WFE from RBM, inertial, and thermal contributions are combined across each Zernike term, converted into raw contrast stability estimates for each Zernike, then totaled over all Zernike terms. The total contrast stability allocation (30.0 $10^{-12}$) taken from the coronagraph error budget is then allocated to Zernike terms with the same proportionalities as the estimated raw contrast stability performance per Zernike to the total raw contrast stability. This way of determining the WFE budget results in ~110% margin in each Zernike term.

| Zernike | | Predicted Performance Amplitude [pm RMS] | | | Total WFE [pm RMS] | WFE Budget [pm RMS] | Raw Contrast [10$^{-12}$] | Contrast Allocation [10$^{-12}$] | WFE Margin |
|---|---|---|---|---|---|---|---|---|---|
| Index | Aberration | RBM | Inertial | Thermal | | | | | |
| n | m | **TOTAL RMS** | 1.767 | 3.994 | 5.565 | 7.074 | 14.576 | 14.56 | 30.00 | |
| 1 | ±1 | Tilt | 0.681 | 0.123 | 0.026 | 0.693 | 1.427 | 0 | 0.001 | 106% |
| 2 | 0 | Power (Defocus) | 1.208 | 1.43 | 3.759 | 4.199 | 8.653 | 0.002 | 0.005 | 106% |
| 2 | ±2 | Astigmatism | 1.069 | 3.559 | 3.463 | 5.08 | 10.466 | 0.002 | 0.005 | 106% |
| 3 | ±1 | Coma | 0.24 | 0.099 | 0.345 | 0.432 | 0.889 | 0 | 0 | 106% |
| 4 | 0 | Spherical | 0.004 | 0.213 | 0.405 | 0.458 | 0.943 | 0 | 0.001 | 106% |
| 3 | ±3 | Trefoil | 0.012 | 1.039 | 2.098 | 2.341 | 4.824 | 6.633 | 13.666 | 106% |
| 4 | ±2 | 2nd Astigmatism | 0.004 | 0.178 | 0.108 | 0.208 | 0.429 | 1.086 | 2.238 | 106% |
| 5 | ±1 | 2nd Coma | 0.001 | 0.026 | 0.105 | 0.108 | 0.223 | 0.624 | 1.285 | 106% |
| 6 | 0 | 2nd Spherical | 0 | 0.028 | 0 | 0.028 | 0.058 | 0.214 | 0.441 | 107% |
| 4 | ±4 | Tetrafoil | 0 | 0.198 | 0.189 | 0.274 | 0.564 | 0.806 | 1.661 | 106% |
| 5 | ±3 | 2nd Trefoil | 0 | 0.112 | 0.233 | 0.259 | 0.533 | 1.63 | 3.358 | 106% |
| 6 | ±2 | 3nd Astigmatism | 0 | 0.021 | 0 | 0.021 | 0.043 | 0.214 | 0.441 | 105% |
| 7 | ±1 | 3nd Coma | 0 | 0.033 | 0 | 0.033 | 0.068 | 0.404 | 0.832 | 106% |
| 5 | ±5 | Pentafoil | 0 | 0.074 | 0.217 | 0.229 | 0.472 | 1.939 | 3.994 | 106% |
| 6 | ±4 | 2nd Tetrafoil | 0 | 0.029 | 0 | 0.029 | 0.06 | 0.239 | 0.493 | 107% |
| 7 | ±3 | 3nd Trefoil | 0 | 0.015 | 0 | 0.015 | 0.031 | 0.168 | 0.345 | 107% |
| 6 | ±6 | Hexafoil | 0 | 0.026 | 0 | 0.026 | 0.054 | 0.308 | 0.635 | 108% |
| 7 | ±5 | 2nd Pentafoil | 0 | 0.015 | 0 | 0.015 | 0.031 | 0.184 | 0.379 | 107% |
| 7 | ±7 | Septafoil | 0 | 0.01 | 0 | 0.01 | 0.021 | 0.106 | 0.218 | 110% |

The raw contrast produced by these WFE instabilities is then calculated for each term using vector vortex transmission factors appropriate for each Zernike mode. The resulting total predicted contrast is shown at the top of the column. A contrast error budget is created by allocating $30.0 \times 10^{-12}$ of the total contrast calculated in **Figure 5.2-1** to each Zernike term. Allocations are calculated by the simple process of allowing an equal margin of contrast, relative to predicted contrast, to each component with a target total of $30.0 \times 10^{-12}$. Finally, the contrast error budget can be converted back into an equivalent WFE stability error budget via the vector vortex transmission factors that relate contrast sensitivities to WFE.

Finally, note that the effect of rigid body motion of the entire telescope, controlled by FGS and ZWFS, was not included in this table as it has negligible effect on contrast ($0.17 \times 10^{-12}$).

### 6.9.4    Structural Model

The structural model is illustrated in **Figure 6.9-3**. It is made up of two primary hardware blocks: the telescope payload and the spacecraft bus. Each block transfers their respective launch loads through a common bulkhead; an annular ring structure constructed of aluminum/composite honeycomb panels with additional titanium ribs for added strength and stiffness. This arrangement connects the heavy telescope payload directly to the bulkhead and then to the launch vehicle interface. The spacecraft structure does not carry the telescope, allowing for a lighter spacecraft bus structure. As a result, the spacecraft and telescope structures are independent as seen in **Figure 6.9-4**. For this analysis, all flight system masses were current best estimates taken from the HabEx master equipment list (MEL) and assumed to include an additional 30% contingency.

The payload portion of the model consists of telescope Zerodur primary and secondary mirrors and a ULE tertiary mirror, a composite tube truss structure, a composite barrel structure, a barrel scarf, a composite secondary mirror support tower and instruments. All payload loads are transferred through the composite tube truss structure into the bulkhead, with the primary mirror, secondary support tower and telescope barrel interfacing directly with the truss structure. For the purposes of this analysis, instruments were modeled as point masses.

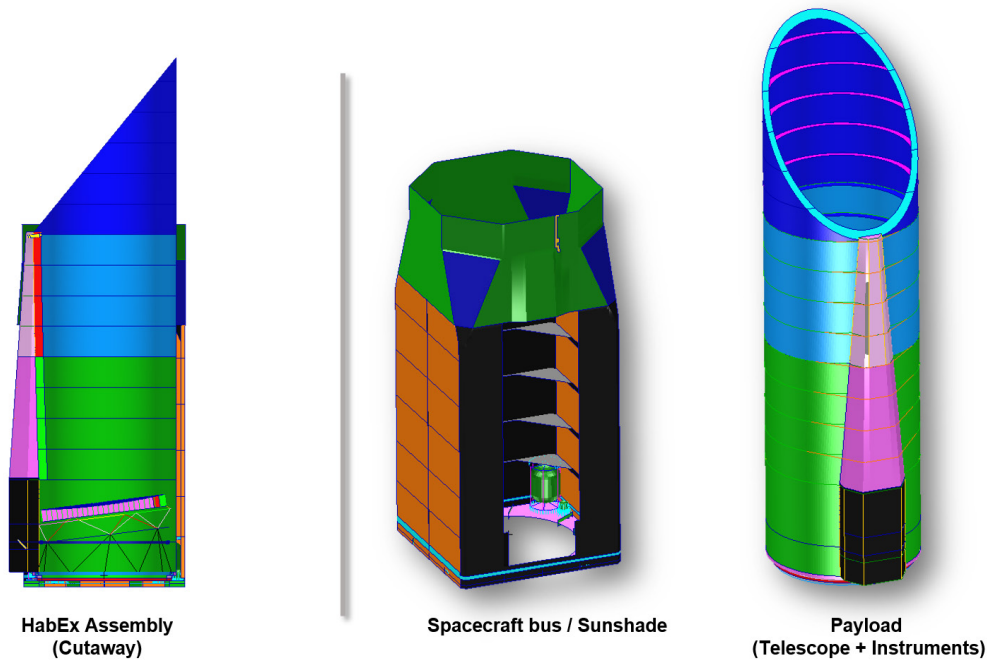

**HabEx Assembly
(Cutaway)**

**Spacecraft bus / Sunshade**

**Payload
(Telescope + Instruments)**

**Figure 6.9-3.** Structural model of the HabEx telescope. *Left* identifies the assembly cross-section of the combined bus/structure, *center,* and payload, *right,* assemblies.

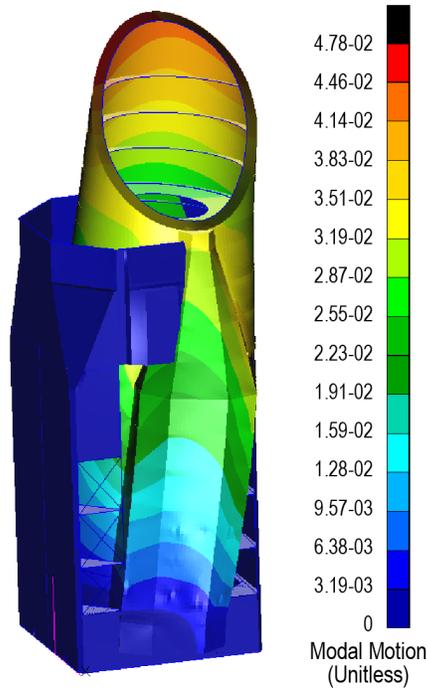

**Table 6.9-4.** Integrated model details.

| Element | Telescope FEM | Spacecraft FEM |
|---|---|---|
| Degrees of Freedom | 221,658 | 282,390 |
| Number of Elements | 42,953 | 57,813 |
| Element Types | CQUAD4, CTRIA3, CBAR, CBUSH, CONM2 | CQUAD4, CTRIA3, CBAR, CBUSH, CONM2 |
| Multipoint Constraints | 426 | 64 |
| Number of Grids | 36,928 | 47,065 |

**Figure 6.9-4.** Independent motion of the telescope barrel, 12.9 Hz first bending mode.

The other major structural model element, the spacecraft, is comprised almost entirely of a large box-like structure that functions as the telescope's sunshade and also as the load path for all spacecraft bus subsystems. This sunshade structure is constructed of aluminum/composite honeycomb panels with stiffening ribs along its longest dimension. All spacecraft subsystems are modeled as point masses within this structure. The microthruster locations are chosen to maximize thruster leverage against the solar pressure-induced torque on the spacecraft.

Because there are no deployables on the spacecraft, the launch configuration is essentially the same as the on-orbit operational configuration with one exception: the sun shade is temporarily tied to the telescope barrel during launch with a supporting lock to increase the overall flight system stiffness and to prevent the sun shade and telescope from contacting during launch. The lock is released after launch, which mechanically and thermally separates the telescope from the sun shade. This approach was used successfully on the Spitzer Space Telescope.

The integrated observatory finite element model (FEM) was created using the MSC Patran pre-processor and geometry created in Pro-Engineer CAD. The primary and secondary mirror FEMs were created independently using the NASA MSFC-developed Arnold Mirror Modeler. Using the integrated NASTRAN model, analyses were performed to ensure strength/stability and stiffness requirements were satisfied in accordance with NASA-STD5001B and the launch vehicle payload users guide (United Launch Alliance – Delta IV Heavy). Additionally, the integrated FEM was used to perform dynamic response and thermal analyses. **Table 6.9-4** summarizes the models.

Structural elements utilize composite construction where possible to provide a rigid and lightweight design. Where possible, M55J carbon composite material is used due to its excellent strength/stiffness and low mass density (1.58 g/cm³) specifications. Telescope structure skins, circumferential ribs, axial webs, and the forward contamination door utilize Honeycomb Sandwich Construction with M55J face sheets with Hexcel honeycomb core. Mirror support truss members assume M55J circular tube construction with titanium end fittings. Full advantage was taken to tailor the M55J unidirectional composite layup orientations for maximum performance and minimum mass. Structural damping is specified to be 0.0005 (0.05%).

### 6.9.5   Thermal Model

The telescope flight system's thermal model shares the same configuration design as the structural model, but uses a different finite element meshing scheme better suited to address areas critical to the thermal design. Temperature stability of the telescope optics and the telescope structure affects its optical alignment, optics deformation, and ultimately the telescope wavefront. The thermal model focuses on capturing the telescope thermal system's behavior in maintaining the temperature of the primary and secondary mirrors and minimizing changes in temperature following simulated observational maneuvers.

The thermal model includes the spacecraft, sunshade, and payload structures. The sunshade and spacecraft provide the first layer of isolation from the Sun and deep space. These components are passively controlled with multi-layer insulation (MLI) and low thermal conduction materials at structural interfaces (such as between the isolation ring and the payload truss). The telescope barrel provides the second layer of isolation. The barrel is cold-biased to about 240 K, with heaters to bring the temperature up to 260 K. The third layer is provided by thermal "cans" around each of the optics.

For the primary mirror, the interior of the can is divided into 24 zones, 8 on the sides and 16 on the base, with each zone containing platinum resistance thermometers (PRTs) for sensing and heaters for temperature control. Each zone in the PM thermal can is controlled to 270 K with proportional-integral-derivative (PID) controllers providing control authority. The secondary mirror temperature is similarly controlled. The tertiary mirror is smaller and more isolated than the primary mirror and is co-located on the anti-sun side of the telescope with the FGS and pick-off mirrors in a single, self-contained assembly. Thermal control of this assembly operates much like that of any other instrument, with its own active thermal control system.

The integrated observatory thermal model was created in Thermal Desktop using the geometry created in Pro-Engineer CAD. The Thermal Desktop model has 20,000 elements and calculates telescope's structure and mirror temperature distribution at 10,000 nodes. The temperature distribution for each node is mapped onto the NASTRAN FEM and the deflections created by each node's CTE is calculated using NASTRAN Solution 101. Rigid body motions (RBM) and mirror surface deformations are calculated from the NASTRAN deflections using SigFit. The primary and secondary mirror's mesh grids were sized to enable SigFit to fit thermally induced surface figure error (SFE) to higher order Zernike polynomials.

The model assumes MLI to control heat loss and to provide thermal isolation. Radiators remove heat from the science instruments and spacecraft electronics and the payload is passively cold-biased. Active heating is then used to maintain the operating temperature of the primary and secondary mirrors (see *Section 6.8.1.5*). These enclosures keep the PM/SM thermal environment at ~270 K. The model assumes TRL 9 capabilities for the enclosure specifications: thermal sensors with 50 mK measurement uncertainty; and PID operating with 30 s periods. The model has a total of 133 control zones. Of these, 36 are bang-bang survival heaters set at 212 K and 97 are PID control zones (**Table 6.9-5**). They are set to keep the primary and secondary mirror front face temperatures at ~270 K.

The current design does not actively control the structure temperature and, in the absence of the laser metrology system, would be highly sensitive to thermal changes. The thermal model predicts that the tube will have a gradient of over 100 K and the primary mirror truss will have a ~20 K gradient (**Figure 6.9-5**). Heaters could be added to the structure design in the future, but this seems unnecessary at this time.

Future improvements to the model would be to add detail such as adhesive joints, specifics about the mounts and athermalization components. Such detail would need a larger design effort than has been possible here.

**Table 6.9-5.** Thermal model details.

| Proportional Control Zones | |
|---|---:|
| Primary Mirror Thermal Enclosure | 82 |
| Primary Mirror Truss Hexapod Legs | 6 |
| Secondary Mirror Thermal Enclosure | 9 |
| **Bang-Bang Survival Heater Zones** | |
| Telescope Baffle Tube | 18 |
| Telescope Secondary Tower | 7 |
| Spacecraft Bus Structure | 3 |
| Spacecraft Fuel Tanks | 8 |

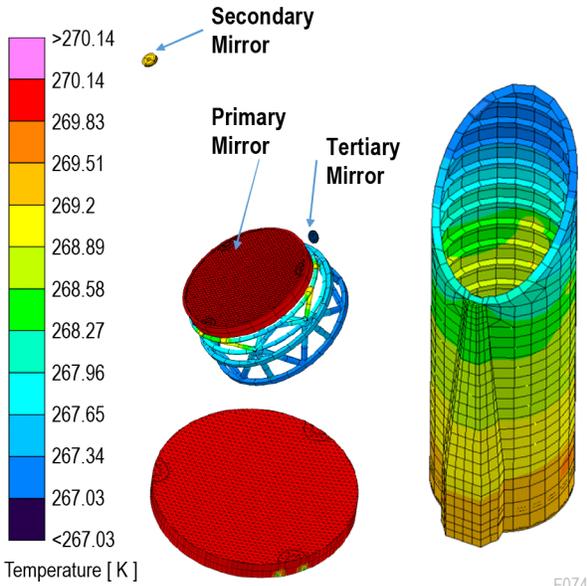

**Figure 6.9-5.** Temperature profiles of telescope barrel showing longitudinal gradient at a worst-case 180° sun angle, primary mirror showing mount effects, and primary mirror truss gradient.

### 6.9.6    MET Error

Tight relative position requirements between the telescope's three mirrors are needed to reach the demanding LOS and WFE stability requirements. The MET system easily meets this need. Each metrology gauge arm of the laser truss has a noise equivalent displacement of 150 pm RMS. The positional uncertainty of the mirrors can be calculated by modeling the metrology system (**Table 6.9-6**).

The WFE residual is calculated by placing these positions and angles into the optical model. These errors result in a negligible 0.025 mas per axis pointing variation on sky and negligible WFE of ~10 fm outside the "null space." These errors are insignificant in comparison to thermal distortions of the primary mirror.

### 6.9.7    Mechanical Jitter

Line-of-sight WFE instability occurs when jitter causes beamwalk on the secondary and tertiary mirrors. Since the mirrors are conics, beamwalk manifests itself as low-order astigmatism and coma.

The driving source of system-generated jitter is the microthrusters. The two thruster types attached to the spacecraft can act as sources of mechanical jitter. Both thruster types are attached to the spacecraft with long structural paths to the telescope mirrors, which introduces some isolation for the optics. Conventional thrusters located at the base of the telescope are used infrequently for large maneuvers; their effects are discussed in *Section 6.9.11*. Conversely, during science exposures, sets of microthrusters run continuously to maintain pointing and the noise (high frequency variability) on their thrust is a source of mechanical disturbance to the telescope. As shown in **Figure 6.9-6**, there are 4 microthruster modules located at the base of the observatory (aft) and another 4 modules located above the center of solar pressure (forward). Each forward module consists of 4 heads, each with 9 emitters, covering a 90° cone. Each aft module consists of 4 heads, with 18 emitters per head.

**Table 6.9-6.** Mirror position uncertainties generated by noise in the MET system, defining the mirror alignment capability of the HabEx telescope. The PM is considered fixed as the SM and TM are measured relative to it. The total PM-SM and PM-TM displacement misalignment are 0.94 nm and 10.51 nm, respectively. The respective rotational misalignments are 1.29 and 7.03 nrad.

| Mirror | Displacement [nm] | | | Rotation [mas] | | |
|--------|------|------|------|------|------|------|
| | ΔX | ΔY | ΔZ | Δθx | Δθy | Δθz |
| **PM** | 0.000 | 0.000 | 0.000 | 0.000 | 0.000 | 0.000 |
| **SM** | 0.655 | 0.655 | 0.131 | 0.055 | 0.063 | 0.253 |
| **TM** | 8.756 | 6.798 | 0.417 | 0.084 | 0.123 | 1.442 |

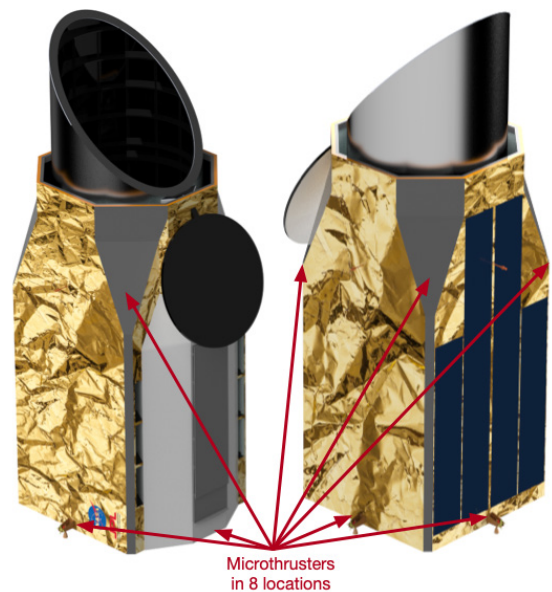

**Figure 6.9-6.** Locations of the microthrusters on HabEx.

Microthrusters provide variable thrust proportional to applied current. **Figure 6.9-7** shows a measured noise power spectrum density (PSD) for a colloidal microthruster used on LISA Pathfinder. The data indicates that these microthrusters have a maximum noise of about 0.05 µN/√Hz and may roll off starting at about 0.02 Hz (Ziemer et al. 2017). Because the data is noisy and has not been measured beyond 5 Hz, HabEx is assuming for its dynamic STOP analysis that each microthruster head has a flat, white noise spectrum specification of 0.1 µN/√Hz RMS, identified as the bold horizontal line in the figure, thereby providing analysis margin. Because the aft modules have twice as many emitters per head, the forward modules (with four heads) are specified to have a total flat noise spectrum of 0.4 µN/√Hz and the aft modules (with four eighteen-emitter heads) are specified to have a total noise of 0.8 µN/√Hz.

Finally, while the FEM's predicted performance is linear as a function of input, the physical system being modeled may not be linear. To mitigate this risk, a model uncertainty factor (MUF) is used. The microthruster specification provides at least a factor of two margin at low frequencies and more margin at higher frequencies—because the flat specification ignores mass damping—relative to the anticipated microthruster performance. Additionally, a MUF of 4 was applied to all amplitudes below 20 Hz, and a MUF of 2 was applied to all amplitudes above 20 Hz.

To predict mechanical LOS stability performance, the RBM of each mirror was calculated as a result of the structural response from 0–350 Hz to the microthruster noise applied to the structure from 0–10 Hz. **Figure 6.9-8** shows the predicted displacements and rotations for the PM, SM, and PM/SM when the baseline structure is exposed to the specified microthruster noise with all thruster heads firing. The graphs show the cumulative root sum square of the data, integrating from the high frequency end. The principal result (the product of the entire disturbance spectrum) is at the leftmost end of each curve. The steps in the curves show where the RMS value increases as a function (primarily) of the mirror assembly structural responses. The error budget (**Table 6.9-2**) allows 10 nm displacement (X and Y), 60 nm displacement (Z) and 0.5 mas (X and Y tilt) and 2 mas (Z rotation).

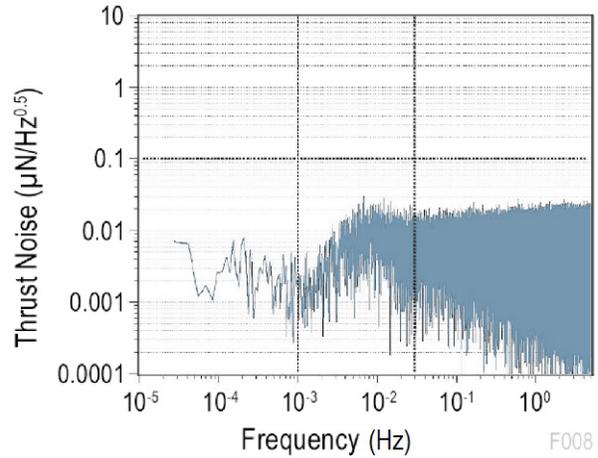

**Figure 6.9-7.** PSD noise plot for colloidal microthrusters on LISA Pathfinder (Ziemer et al. 2017).

These jitter-generated displacements of the optics are applied to the telescope optical model (**Table 6.9-7**) and yield negligible predicted resultant LOS and WFE. Both the jitter-driven LOS error of 0.012 mas and the MET mirror position measurement LOS error of 0.025 mas are small compared to the FGS pointing measurement capability of 0.7 mas, and all three combined are easily within the LOS stability requirement of 2 mas.

Like the MET measurement contribution to WFE, the jitter contribution to WFE is also around 0.01 pm, so their combined effect on WFE is insignificant.

From this and the preceding section, it is concluded that the combination of laser metrology and microthrusters provide for extremely high performance of the optical system. The following sections look at other possible sources of performance degradation.

**Table 6.9-7.** Jitter-generated displacements applied to the error budget. These errors result in a negligible 0.012 mas per axis pointing variation on sky and on the order of 10 fm WFE.

| Mirror | Displacement [nm] | | | Rotation [mas] | | |
|---|---|---|---|---|---|---|
| | ΔX | ΔY | ΔZ | Δθx | Δθy | Δθz |
| PM | 0.20 | 0.20 | 0.10 | 0.004 | 0.004 | 0.000 |
| SM | 0.50 | 0.50 | 0.01 | 0.004 | 0.004 | 0.004 |
| TM | 0.10 | 0.10 | 0.10 | 0.002 | 0.002 | 0.002 |

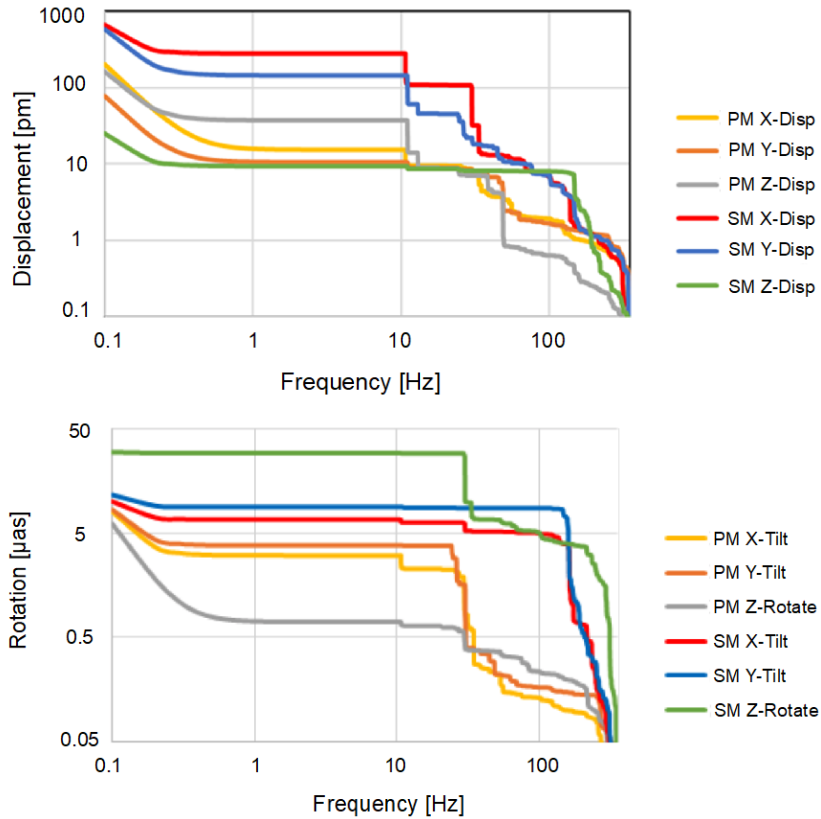

**Figure 6.9-8.** Displacement and rotation of the primary and secondary mirrors under the microthruster noise spectrum, shown as cumulative RSS, integrating from the high frequency end. The steps in the curves show where the RMS value increases as a function (primarily) of the mirror assembly structural responses. At higher frequencies, displacement is uncontrolled, whereas at lower frequencies, the MET and ZWFS systems will control the response. The flat regions below about 10 Hz indicate regions where there is zero influence on the mirrors.

### 6.9.8    Inertial Mirror Deformation

The preceding section covered positional motion of the mirrors. Deformation of the primary mirror surface is now considered as another source of wavefront error. Inertial WFE instability occurs when the primary mirror is accelerated by mechanical disturbances, causing it to react against its mounts and in so doing to elastically deform. **Figures 6.8-6** and **6.8-7** illustrated two such mechanical disturbance reaction modes for the primary mirror, 43.5 Hz rocking and 50 Hz bouncing modes. **Figure 6.9-9** shows how the mirror bends as it reacts against the hexapod mount for these two modes. A simple analysis approach is to scale the predicted (or measured, given a test article) gravity sag of the primary mirror when supported in a face up orientation and two other orthogonal orientations, as shown in **Figure 6.8-5**, by the micro-thruster noise. With the FEM, the mirror acceleration in three axes was calculated using the microthruster disturbance spectrum. The estimate of inertial RMS WFE caused by microthruster acceleration of the primary mirror is 0.42 pm, 0.53 pm, and 0.78 pm for the three orientations. The total WFE is the RSS of the X, Y, and Z components: 1.0 pm. Thus, the microthruster contribution to the WFE is small.

Further analysis using NASTRAN (**Figure 6.9-10**) shows the microthruster disturbances decomposed into Zernike modes, again using the cumulative rss curves. The WFE error budget allocations are shown in **Table 6.9-3** and exceed the calculated. The contrast performance is most sensitive to trefoil due to the VVC-6 suppression of lower order aberrations and the primary mirror's three-point mounting design. Trefoil has an allocation of 4.8 pm RMS while the accumulated expected value is ~1.1 pm, which is therefore comfortably lower.

### 6.9.9    CTE Inhomogeneity

CTE inhomogeneity (ΔCTE) due to thermal drift during observations will cause mirror surface distortions and hence, WFE. The AMTD project's 1.2 m ELZM was built using Zerodur® (like the HabEx primary) and tested specifically to determine its ΔCTE characteristics. That mirror was measured to have ~26 nm RMS surface deformation over a 62 K thermal range from 292 to 230 K. **Figure 6.9-11** shows the measured deformation and part of its decomposition into Zernike polynomials (Brooks et al. 2017). The deformation is converted into an inhomogeneity distribution which may then be used with the predicted temperature distribution (over the much smaller flight temperature range) to predict the surface figure. A prediction of the allowable HabEx mirror $\Delta T$ can be made using the data in the table.

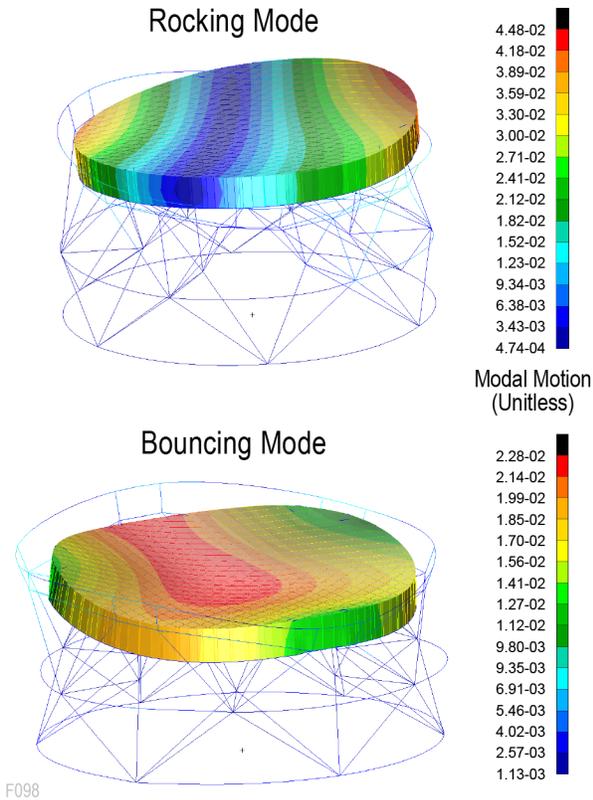

**Figure 6.9-9.** Mirror deformations from rocking mode (upper) and bouncing mode (lower).

A prediction of the allowable HabEx mirror $\Delta T$ can be made using the data in the table.

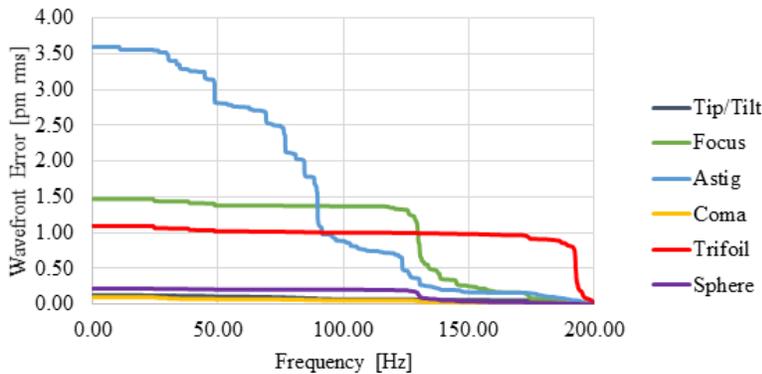

**Figure 6.9-10.** Inertial mirror deformation broken into Zernike modes. Of the terms shown, trefoil has an error budget allowance of 6.3 pm, so this term has significant margin. All other terms are negligible compared to the error budget.

Referring to **Table 6.9-3** the RMS wavefront allowances are used to set the maximum allowable temperature range based on the ELZM measured performance. For the given Zernike WFE budget allocations, a temperature stability of approximately 1.1 mK is needed to meet the WFE allocations based on the most limiting term, assuming that all WFE is generated from thermal instability alone which is obviously not the case. The WFE budget can be rebalanced by taking some WFE allocation from the power term and redistributing to Zernike terms with tight requirements. A WFE budget requiring a thermal stability of 10 mK in all terms can be constructed with little reduction in the margin on the power term, but this reallocation of WFE margin is not necessary for the HabEx baseline design as will be shown in the integrated STOP modeling (*Section 6.9.12*).

### 6.9.10 Temperature Gradients

The model predicts that the primary mirror front surface will have ~200 mK trefoil temperature gradient, shown in **Figure 6.9-5**. The source of this gradient is thermal conduction into the hexapod struts. The mirror will also have a ~3 K gradient between the front and the back caused by direct radiation from the front surface to space. The analysis of these effects is made as part of the integrated STOP modeling.

### 6.9.11 Response to Propulsion Impulse during Slew Events

Following a slew maneuver there is a ring-down time or impulse response time that measures how long it takes for the LOS and WFE to stabilize. Transient dynamic analysis was performed to predict ring down time via Finite Element Analysis (FEA) using MSC Patran as the pre/post processor and MSC NASTRAN as the solver. To simulate a nominal pitch maneuver, an 8.8 N thrust was applied as a 20.5 step function at the Y-axis ACS thrusters and after 368 seconds another was applied in the opposite direction to stop the slew (**Figure 6.9-12**). The relative motion

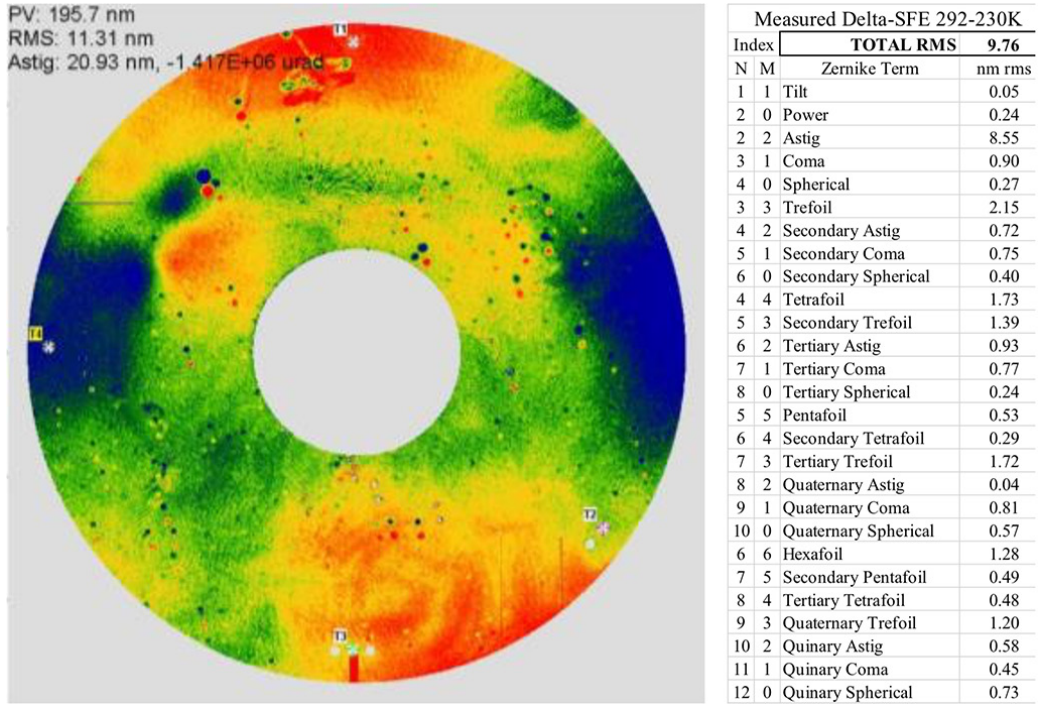

PV: 195.7 nm
RMS: 11.31 nm
Astig: 20.93 nm, -1.417E+06 urad

| Measured Delta-SFE 292-230K | | | |
|---|---|---|---|
| Index | **TOTAL RMS** | | **9.76** |
| N | M | Zernike Term | nm rms |
| 1 | 1 | Tilt | 0.05 |
| 2 | 0 | Power | 0.24 |
| 2 | 2 | Astig | 8.55 |
| 3 | 1 | Coma | 0.90 |
| 4 | 0 | Spherical | 0.27 |
| 3 | 3 | Trefoil | 2.15 |
| 4 | 2 | Secondary Astig | 0.72 |
| 5 | 1 | Secondary Coma | 0.75 |
| 6 | 0 | Secondary Spherical | 0.40 |
| 4 | 4 | Tetrafoil | 1.73 |
| 5 | 3 | Secondary Trefoil | 1.39 |
| 6 | 2 | Tertiary Astig | 0.93 |
| 7 | 1 | Tertiary Coma | 0.77 |
| 8 | 0 | Tertiary Spherical | 0.24 |
| 5 | 5 | Pentafoil | 0.53 |
| 6 | 4 | Secondary Tetrafoil | 0.29 |
| 7 | 3 | Tertiary Trefoil | 1.72 |
| 8 | 2 | Quaternary Astig | 0.04 |
| 9 | 1 | Quaternary Coma | 0.81 |
| 10 | 0 | Quaternary Spherical | 0.57 |
| 6 | 6 | Hexafoil | 1.28 |
| 7 | 5 | Secondary Pentafoil | 0.49 |
| 8 | 4 | Tertiary Tetrafoil | 0.48 |
| 9 | 3 | Quaternary Trefoil | 1.20 |
| 10 | 2 | Quinary Astig | 0.58 |
| 11 | 1 | Quinary Coma | 0.45 |
| 12 | 0 | Quinary Spherical | 0.73 |

**Figure 6.9-11.** 1.2 m SCHOTT ELZM: a 62 K thermal change: resulting surface change decomposed into Zernikes.

between the primary and secondary mirror was calculated for 300 s beyond the termination of the second thrust. No MUF was applied to this analysis.

Because the baseline telescope structure is very stiff, after 300 s of ring-down, the largest predicted relative motion between the primary and secondary mirrors is less than 1 pm (**Figure 6.9-13**). Thus, telescope slews will have negligible impact on the settling time before beginning science observations.

### 6.9.12 Integrated STOP Modeling

The coronagraph's sensitivity to distortions in the overall optical train necessitates examining the effects of thermal and mechanical distortions within the telescope, on the instrument's contrast performance. Results of this analysis have been iterated with the coronagraph, telescope and spacecraft designs, to ensure a final configuration capable of meeting the science-driven contrast levels specified in the STM and coronagraph error budget. To understand how expected thermal and mechanical loads on the flight system impact coronagraph performance, thermal and mechanical finite element models were used to estimate distortions of the telescope optics, which were then used to estimate changes in contrast performance using optical models. The process is commonly referred to as STOP modeling.

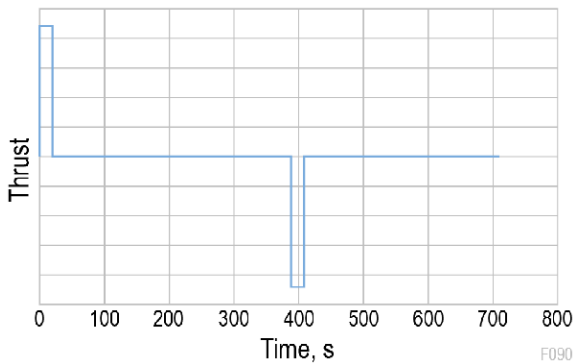

**Figure 6.9-12.** Slew impulses applied to Y-axis by ACS thruster.

The goal of STOP modeling is to simulate the thermal and mechanical effects experienced by the telescope flight system during observing conditions, and assess how those effects impact the coronagraph's contrast performance. This analysis differs from the work done in the preceding sub-section in a couple of ways. First, the simulations are based around the temporal evolution of observational scenarios rather than quasi-static situations. The results enable moment by moment assessment of the observatory's performance and demonstrate that the planned observations are feasible. Second, the data from the simulations can be directly transformed to coronagraph contrast through the coronagraph PROPER model, so the total system performance can be readily evaluated.

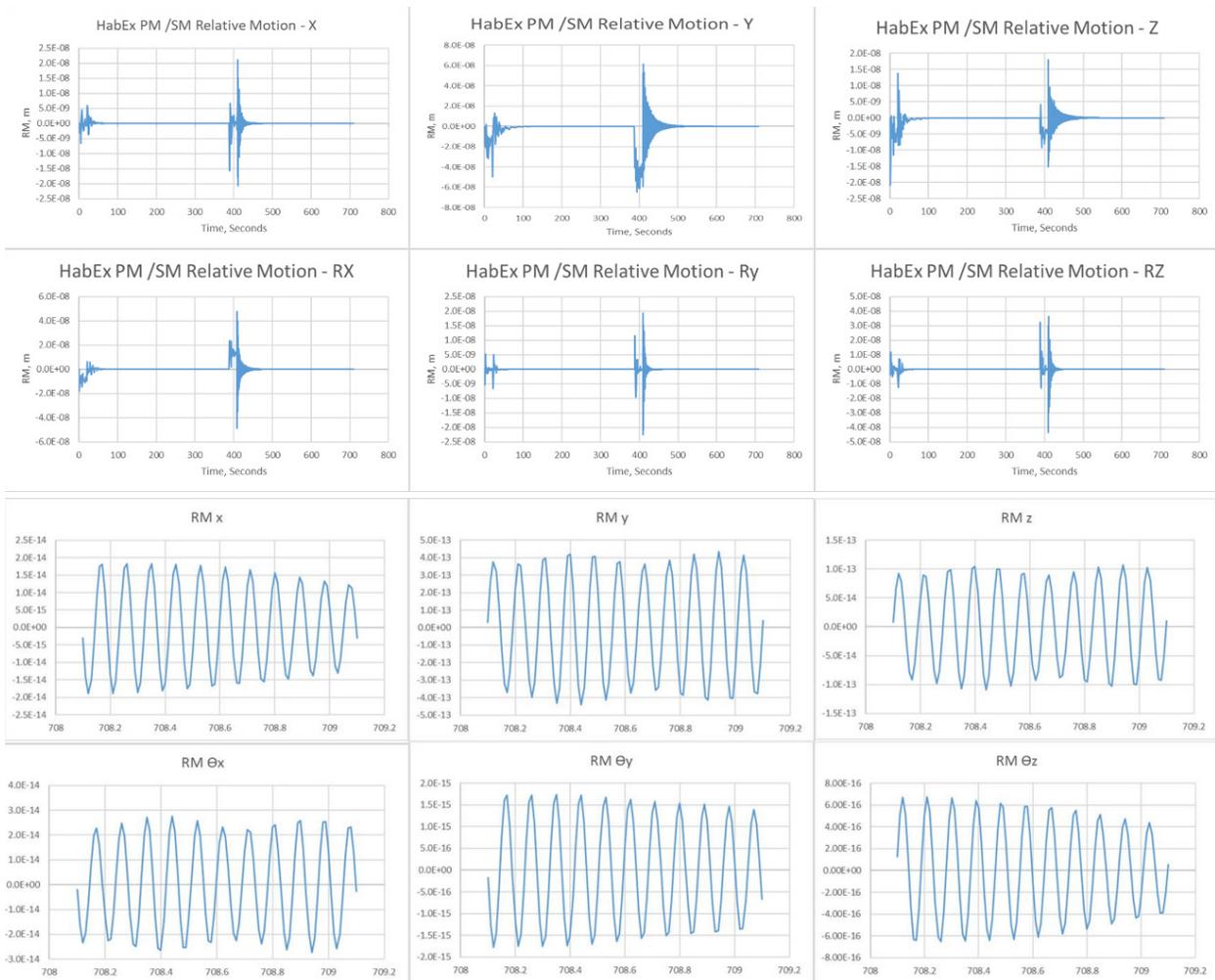

**Figure 6.9-13.** *Upper*: Relative motion between PM and SM caused by thruster impulses (infrequent large slew maneuvers: e.g., reference star to target star). *Lower*: After 300 s, relative motion between PM and SM caused by thruster impulses is less than 1 pm.

For the STOP modeling effort, HabEx leveraged a model pipeline developed for the WFIRST Coronagraph Instrument. The pipeline is built around an open-source Python-based workflow management software. This framework allows for a complex pipeline of batch jobs. The integrated modeling pipeline automates each model in sequence to produce time-dependent outputs reflecting the effect of temperature changes during an observation scenario. The models in the pipeline include Thermal Desktop®,

MX NASTRAN® for thermal and structural finite element modeling of the telescope flight system, and Sigmadyne SigFit® for translating rigid body motions and surface deformations into optical model input. Additional optical model inputs come from a model of the MET system and models of the bulk and ΔCTE of the primary and secondary mirrors. Synopsys' CodeV® optical design software was used for telescope optical modeling. Finally, the PROPER coronagraph model takes the calculated telescope output wavefront and determines the contrast of the coronagraph instrument. A block diagram of the pipeline flow is shown in **Figure 6.9-14**.

The modeling took into consideration mirror rigid-body thermal displacements, mirror thermal deformations, mirror thermal inhomogeneity, and laser metrology measurement error. However, unlike the prior section, the effects of these error contributions were assessed as an interacting system rather than individually. Zerodur® mirrors were assumed to have a non-zero CTE of 20 ppm to allow for uncertainty in the post-launch temperature offset. The assumptions for the thermal modeling, laser metrology measurement error and material inhomogeneity and are the same as in the previous section's modeling analyses and can be found in *Sections 6.9.5, 6.9.6,* and *6.9.9,* respectively.

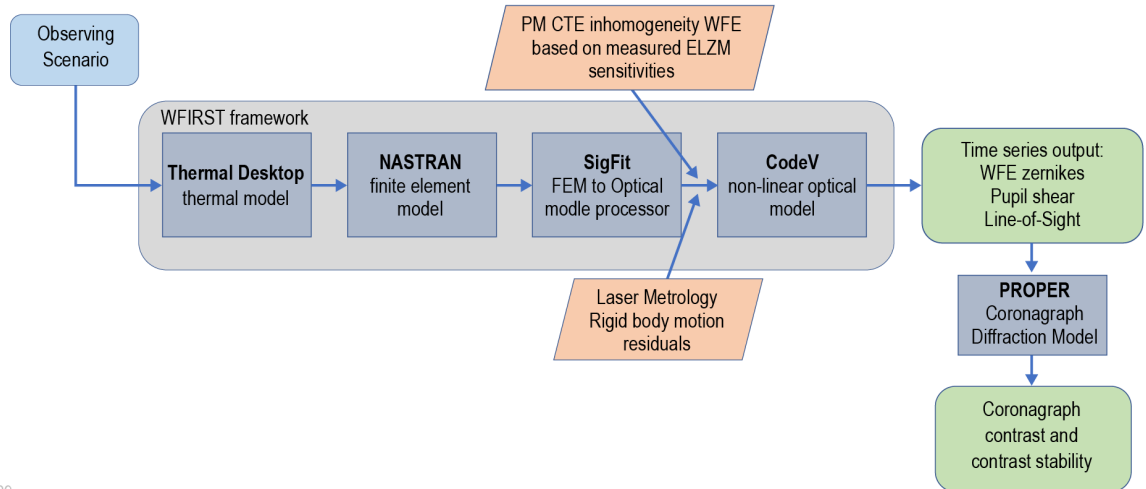

**Figure 6.9-14.** Showing the STOP modeling process from the observing scenario through the telescope and optical models, to a time series output that is fed to PROPER to calculate the coronagraph contrast as it evolves in time.

### 6.9.12.1  Observation Scenario

A single observational scenario was modeled to represent a coronagraph science observation using both reference differential imaging (RDI) and angular differential imaging (ADI). In RDI, the coronagraph first observes a bright reference star and calibrates the observatory with the deformable mirrors to create a deep contrast region of interest—referred to as "digging a dark hole." After the DMs are set and the "dark hole" generated, the observatory slews to a target star. In ADI, the coronagraph observes a target star then rotates 30° around the boresight and observes the target star a second time. Since speckle is an artifact of imperfections in the telescope, it should not change with the rotation and can be subtracted out.

In the RDI part of the observation scenario (shown in **Figure 6.9-15**), the telescope starts at L2 with a 100° sun angle and thermally equilibrates for 90 hrs. At 90 hrs, the observatory pitches +10° to a sun angle of 110° and holds for 10 hrs. This represents digging a dark hole on a reference star. The observatory then pitches another +10° to a target star and holds for 50 hrs.

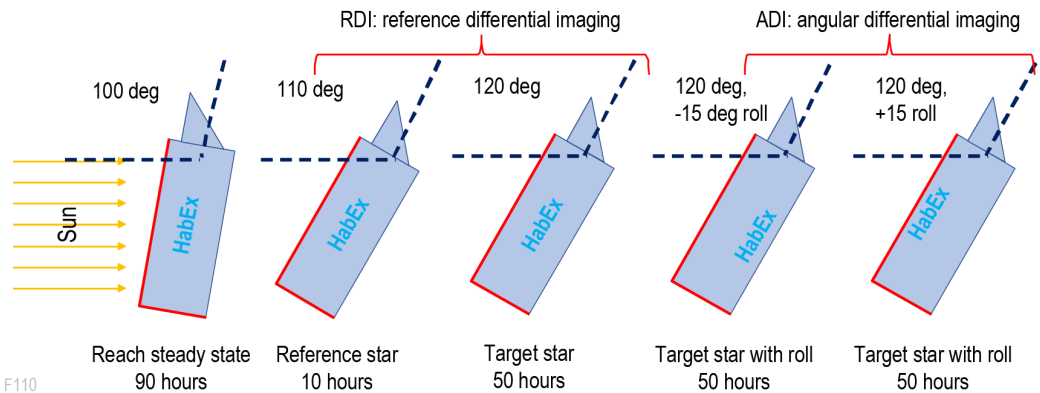

**Figure 6.9-15.** Observing scenario. Left: equilibration followed by RDI. Right: RDI followed by ADI. Note that roll is represented by "HabEx" moved on the bus.

The ADI observational scenario then follows, as shown in **Figure 6.9-15**. Following the RDI target star observation, the telescope rolls 15° around its boresight while remaining at the RDI 120° pitch. The flight system remains in this position for 50 hrs, then rolls -30° to come to an orientation of -15°. The ADI scenario continues with the telescope remaining in this position for another 50 hrs.

This particular observing scenario operates entirely with sun angles greater than 90°, exposing the bottom of the spacecraft to the Sun as well as the side. For sun angles less than 90°, only the sun-side of the sunshade is exposed and the changing thermal loads are more benign because of the smaller exposed area.

### 6.9.12.2 Results: Equilibration and Reference Differential Imaging

During the initial 90 hrs the observatory is acclimating to the L2 environment with active thermal control maintaining mirror temperatures near 270 K. The temperature change of the PM is shown in **Figure 6.9-16**. The scenario starts with an increase in temperature as the observatory slews to the reference star, then again at 100 hrs as it slews to the target star. The minuscule thermal response of the mirror is evident as it slowly settles back towards the temperature set point. Once equilibrated, over the next 60 hrs of reference and target star observations the maximum temperature variation of the primary mirror during the simulated RDI observation is about 0.3 mK from the reference star acquisition through the target star temperature maximum.

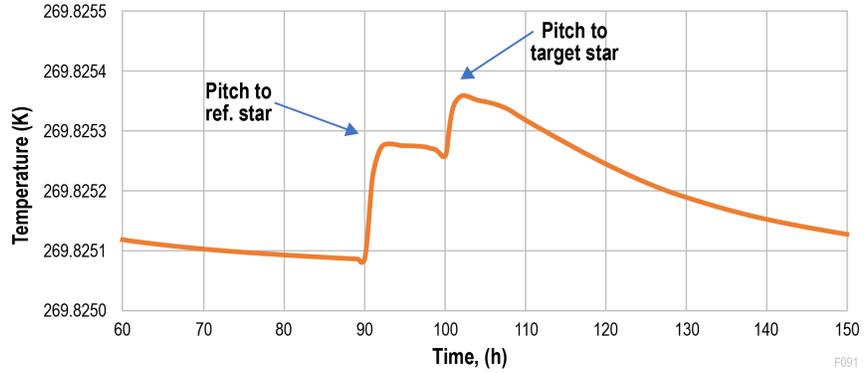

**Figure 6.9-16.** Average temperature change of the primary mirror under PID control. The significant events shown are the start of the pitch to the target star followed by the two ADI rolls.

A closer look at the stability of the PM and SM during target star imaging is shown in **Figure 6.9-17**. The PM, with its massive thermal inertia, is stable to better than 0.15 mK from the starting temperature during the target star acquisition through the 50-hour observation. The SM, which is smaller and has considerably less mass, is stable to about 5 mK.

Small surface deformations arise as the mirrors, with their small, but non-zero CTE, respond to the change in average temperature and thermal gradients. **Figures 6.9-18** and **6.9-19** show the resultant surface deformation separated into the most prominent Zernike coefficients. Of these Zernike terms, trefoil is of most interest as it is outside the null space (i.e., not one of the Zernike terms of the wavefront that are highly attenuated by the charge 6 vortex mask). The PM exhibits the most change in trefoil with about 0.75 pm RMS, while the SM exhibits about 0.4 pm RMS stability. The other low order terms—power, astigmatism, and coma—are within the VVC6 null space where the coronagraph can tolerate ~100s pm each without degrading instrument contrast.

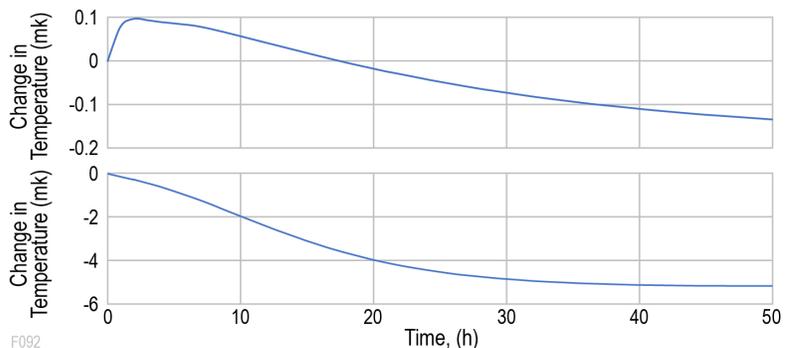

**Figure 6.9-17.** Change in average temperature of the primary and secondary mirrors in the first 50 hours on the target star in *top* and *bottom panels*, respectively.

The change in wavefront error at the entrance pupil of the coronagraph instrument is shown in **Figure 6.9-20**. The figure shows the Zernike coefficients that contribute most to the error. The majority of the changing wavefront is in the coronagraph null space, that includes astigmatism, coma, and spherical aberrations, and the total (RSS) peaks at just under 3 pm RMS. The RMS total wavefront error outside the null-space peaks at about 1.25 pm RMS. Trefoil, the dominant Zernike term outside of the null space, exhibits 1.2 pm of WFE. **Figure 6.9-21** shows that the thermally-driven LOS error is negligibly small.

### 6.9.12.3  Results: Angular Differential Imaging

The angular differential imaging response of the telescope is shown from hour 150 onwards. To recapitulate, after completion of the 50 hr RDI target star observation, the ADI observation begins with the telescope rolling 15° while remaining pointed at the target star. The telescope remains in this attitude for 50 hrs then rolls in the opposite direction 30°. This new position is held for another 50 hrs. The change between the first and the second roll results in the same insolation of the underside, and the side of the telescope facing the sun, while the "left" or "right" sides of the telescope become illuminated at an oblique angle. The response, decomposed into Zernike terms is shown in **Figure 6.9-22**. The roll has minimal effect (sub-picometer) on the most important trefoil term after 50 hrs, and continues to improve, promising very high contrast stability in this mode of operation. The 1.8 pm wavefront change in trefoil is expected to result in ~$5 \times 10^{-12}$ contrast change. **Figure 6.9-23** shows that trefoil is responsible for the lion's share of the contrast change and that 2.0 pm of trefoil yields a contrast change of $<10^{-11}$.

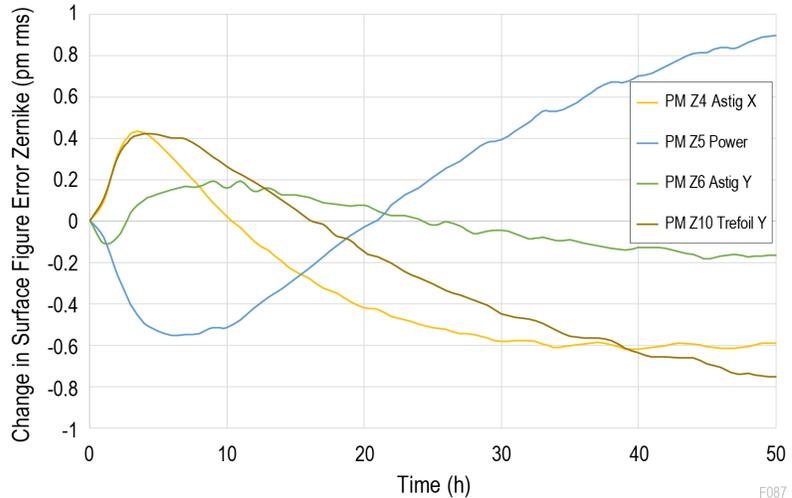

**Figure 6.9-18.** Primary mirror surface error stability after digging dark hole, +10° pitch.

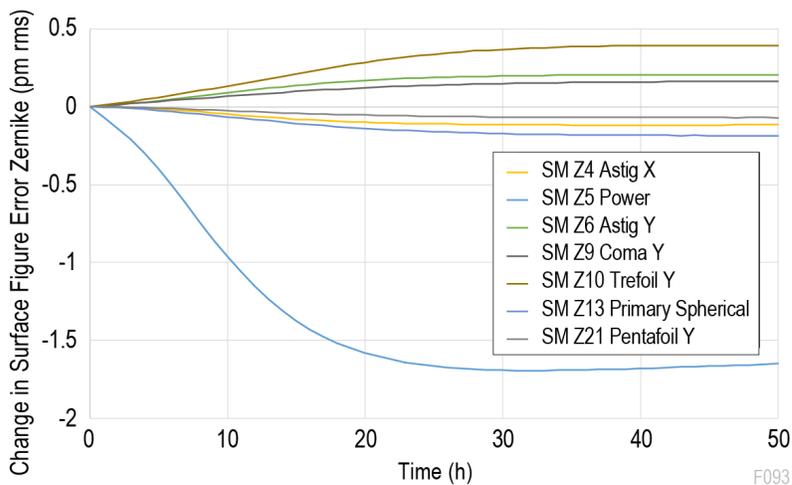

**Figure 6.9-19.** Secondary mirror surface error stability after digging dark hole, +10° pitch.

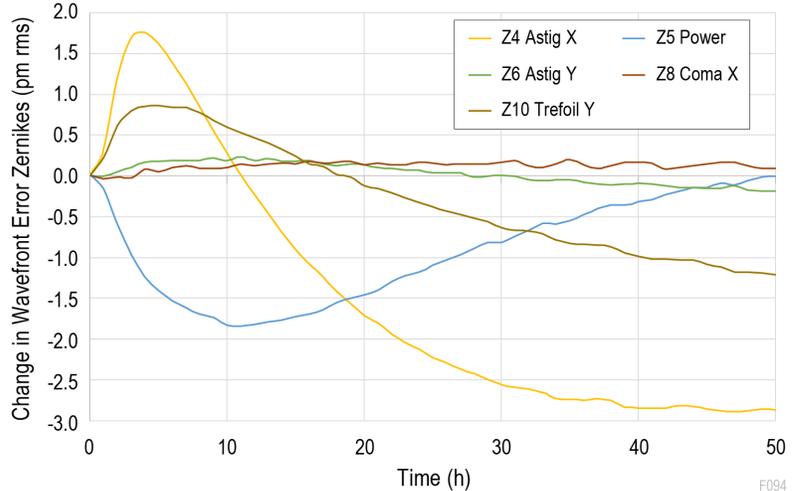

**Figure 6.9-20.** Telescope wavefront error stability after digging dark hole, 10° pitch

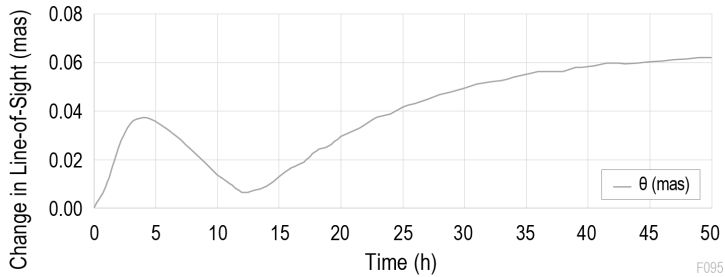

**Figure 6.9-21.** Telescope pointing stability after digging dark hole, +10° pitch showing the evolving pointing errors in the two orthogonal axes. The grey curve shows the net pointing error. Both rigid body and mirror thermal distortion effects are contributing to the result.

Finally, a contrast analysis is made based on a worst case using the PROPER model. From the STOP results after the pitch to the target star, for each aberration and pupil shift, the largest change was taken and added to the model. No additional wavefront sensing and control was applied. This represents a case notably worse than at any one particular time in the simulation. The resulting RMS contrast change in the dark hole as a function of angle is as shown in **Figure 6.9-24**; below $10^{-11}$ at the IWA$_{0.5}$ of $2.4\lambda/D$ and decreasing further with angle. Referring to the more general case, rather than the worst possible, the analysis shows a maximum contrast change of $2 \times 10^{-12}$ during the initial 50 hr observation of the target star.

This excellent contrast stability, combined with the excellent static raw contrast also predicted by end-to-end simulations (**Figure 6.3-9**), enables coronagraph direct imaging of Earth-sized planets in the HZ of nearby stars. It is primarily the result of four specific design decisions. First, the VVC-6 coronagraph reduces contrast sensitivity to the largest aberrations in the telescope system.

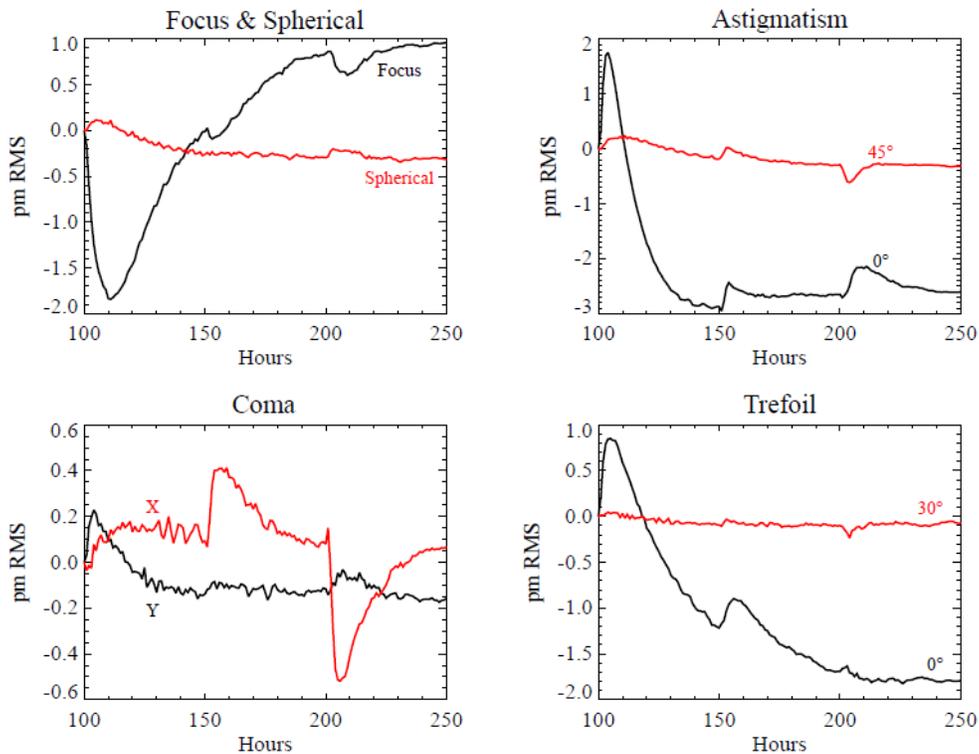

**Figure 6.9-22.** Low order thermal drift WFE stability: settling on the target star at hour 100, and then two roll maneuvers at 150 and 200 hrs. Bottom right: the differential roll is a sub-picometer disturbance to the trefoil (which has >6 pm allocation) after sufficient settling time.

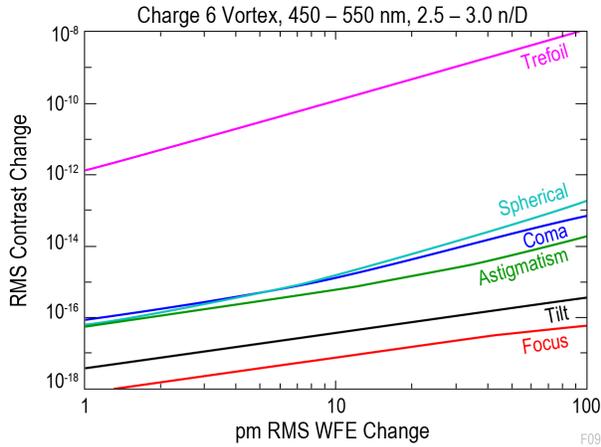

**Figure 6.9-23.** Contributions to contrasts change of low order Zernike modes.

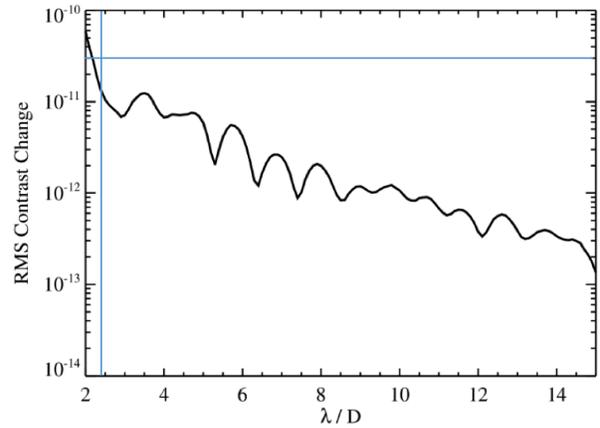

**Figure 6.9-24.** RMS contrast change as a function of angle for the worst case. *Blue lines* are drawn at 2.4 $\lambda/d$ and 30 $10^{-12}$ contrast change (stability).

Second, the heavy 4 m monolithic mirror eliminates mirror-segment edge scatter and provides high CTE stability and a great deal of thermal inertia to help stabilize contrast variations. This design choice is enabled by the use of the SLS, or possibly the SpaceX BFR, which deliver the launch mass and volume required for the mirror. Third, SIM-developed laser metrology essentially eliminates rigid body motion within the telescope mirrors which greatly benefits both contrast and LOS performance. Lastly, the adoption of ESA-heritage microthrusters and the removal of reaction wheels reduces the telescope's self-generated vibrational environment to nearly nothing. In concert, these design choices will create the largest, most stable, UV-to-near-IR telescope to have ever flown in space.

## 6.10 Telescope Flight System

This section describes the key design features of the telescope flight system (shown in **Figure 6.10-1**) including a description of the spacecraft bus, and discussions of some driving parameters for the flight system design. The OTA design is detailed in *Section 6.8*, while the instruments are discussed in *Sections 6.2–6.6*.

The HabEx telescope is a Class A system with redundant subsystems. The telescope bus and components benefit from high heritage for most of its functions, including telecom, CDH, power generation and distribution, thermal design, and monopropellant propulsion. This high heritage engineering, combined with the benign environment at L2, comparatively low data rates for the mission, and large launch vehicle mass margin lead to a bus design that is not expected to pose any significant challenges. Slewing is handled by monopropellant thrusters. To meet demanding observational requirements, HabEx has followed the Gaia example and removed the reaction wheels and employs microthrusters to counteract torques, primarily imparted by solar pressure. Microthruster technology has flown on Gaia and LISA Pathfinder, and is undergoing further development and maturation for use on other missions. Its future development program is summarized in *Chapter 11*. The propulsion systems and propellant loadings for both station keeping and slewing are sized for an initial 5-year mission plus a 5-year extended mission before needing to be refueled.

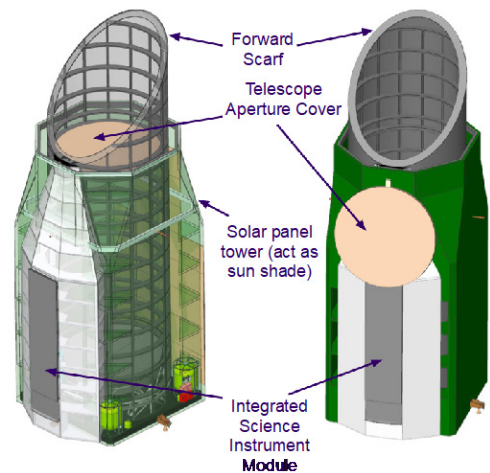

**Figure 6.10-1.** The HabEx telescope spacecraft concept.

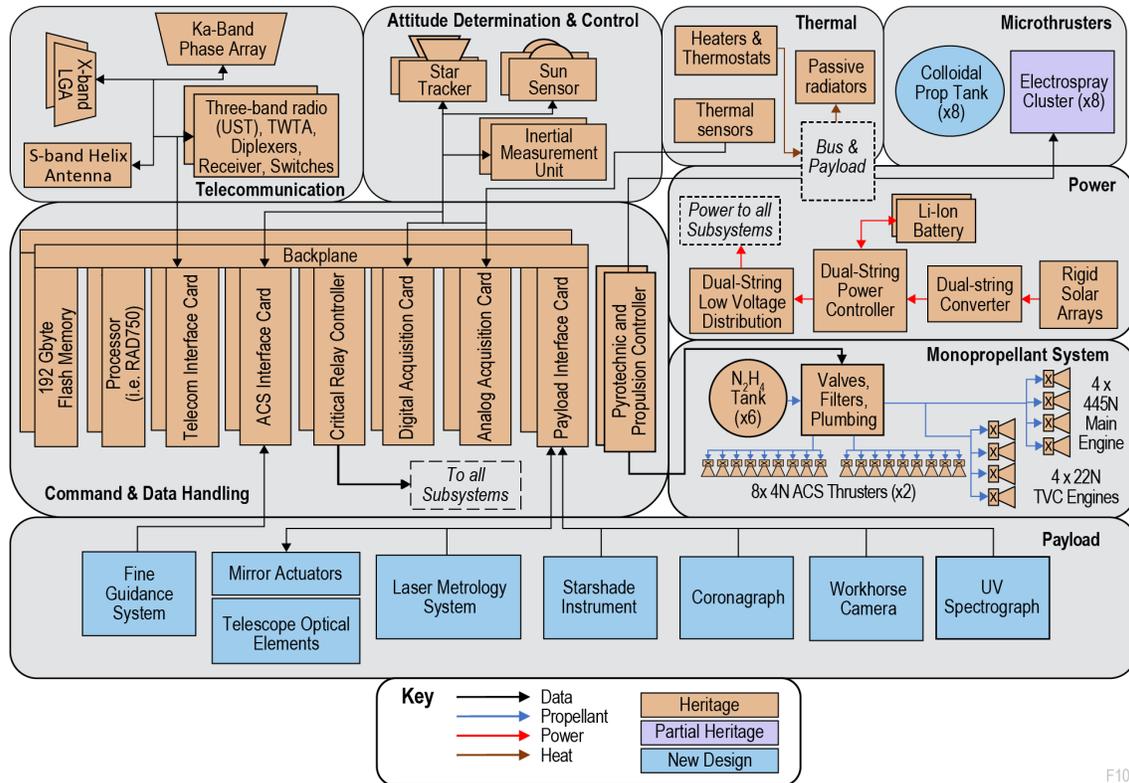

**Figure 6.10-2.** Block diagram of the telescope flight system identifying heritage and other components.

**Figure 6.10-2** is a block diagram describing the subsystems and **Table 6.10-1** offers a mass breakdown of the telescope flight system; the total mass (CBE) of the telescope spacecraft flight system is estimated to be 10,160 kg, with 28% average contingency and an additional 15% system margin, leading to a total margined dry mass and wet mass of 14,530 kg and 17,045 kg, respectively.

### 6.10.1 Structures & Mechanisms

The spacecraft bus structure supports the sunshade and bus subsystems including power, propulsion, communications, avionics, and guidance and control. In order to minimize risk and sources of vibration, there are no deployables or mechanisms on the telescope flight system with the exception of the telescope door, which can be opened and closed to protect the optics during servicing.

The bulkhead baseplate of the bus supports most spacecraft components. It is constructed of a thick aluminum honeycomb core with zero CTE carbon composite face sheets. This bulkhead encompasses a strong interface ring that both connects the payload to the spacecraft bus and serves as the interface of the entire telescope flight system to the launch vehicle. By using this common interface ring, payload launch loads are transferred directly into the ring without passing through the bus structure, greatly reducing the required mass of the bus structure.

**Table 6.10-1.** HabEx telescope flight system mass breakdown per subsystem. CBE: current best estimate. MEV: maximum expected value.

| | CBE (kg) | Cont. % | MEV (kg) |
|---|---|---|---|
| **Payload** | | | |
| Telescope and Instruments | 6080 | 30% | 7900 |
| Payload Thermal | 265 | 30% | 345 |
| **Spacecraft Bus** | | | |
| ACS | 20 | 1% | 20 |
| CDH | 20 | 10% | 25 |
| Power | 240 | 27% | 300 |
| Propulsion: Monoprop | 300 | 30% | 325 |
| Propulsion: Electrospray | 160 | 44% | 235 |
| Structures & Mechanisms | 2690 | 30% | 3490 |
| Spacecraft side adaptor | 45 | 30% | 60 |
| Telecom | 35 | 28% | 45 |
| Thermal | 350 | 30% | 460 |
| **Bus Total** | **3820** | **28%** | **4900** |
| **Spacecraft Total (dry)** | **10160** | **43%** | **14530** |
| Subsystem heritage contingency | 2980 | | |
| System margin | 1390 | | |
| Monoprop and pressurant | 2280 | | |
| Colloidal Propellant | 240 | | |
| **Total Spacecraft Wet Mass** | | | **17045** |
| Launch Vehicle Side Adaptor | | | 1500 |
| **Total Launch Mass** | | | **18550** |

The vertical sunshade structure thermally isolates the payload from both the heat from the sun and the cold of deep space, and also supports the primary solar array and a set of microthrusters. The 12 m tall sunshade consists of thin panels of aluminum honeycomb core with M55J composite face sheets. The anti-sun side of the sunshade has open panels to allow some exposure of the telescope to space, providing a thermal bias for thermal control authority. As was done for Spitzer, the sunshade structure is connected to the telescope barrel during launch to help stabilize the large, lightweight barrel until launch vehicle separation. Once on orbit the two are separated which increases the thermal and vibrational isolation between the sunshade and the telescope.

As shown in **Figure 6.10-3**, the payload sits directly on the interface ring so that the load path leads directly to the launch vehicle adaptor. The bus attaches to the interface ring from the sides with its load path also leading directly to the launch vehicle adaptor. With this ring as the only interface, both the payload and the bus are mostly independent and each can go through vibrational testing separately prior to integration.

Most spacecraft subsystems are mounted to the bulkhead and accessible from the sides of the sunshade. Propellant tanks for the monoprop thrusters, along with pressurant tanks, reside at each corner of the bulkhead. Piping from each tank converges at one side of the spacecraft where fueling valves are co-located, simplifying the fueling process during both I&T and servicing. Other subsystems are contained in serviceable modules that can be removed from the sides of the spacecraft.

### 6.10.2  Power

The baseline power system uses a dual-string "cold spare" approach, and small-cell technology

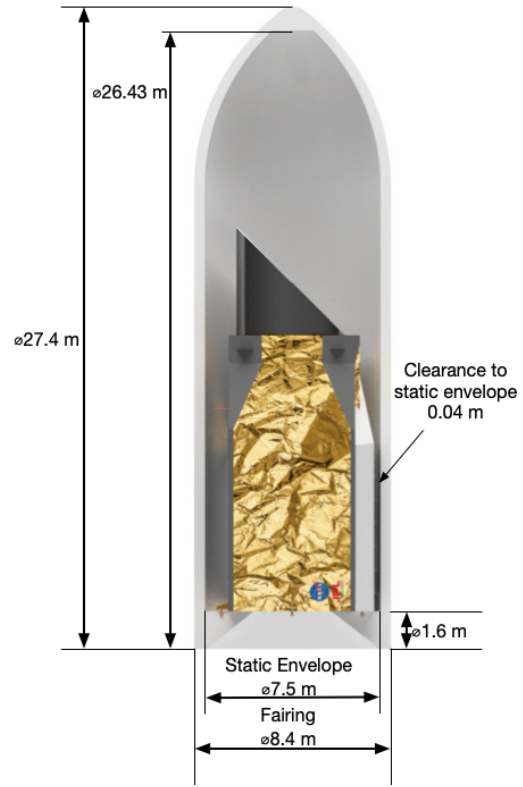

**Figure 6.10-3.** The HabEx telescope shown in the SLS Block 1B Cargo fairing.

Li-Ion batteries. The solar array is assumed not to be replaceable and is sized for a minimum lifetime of 20 years. Should the telescope need to operate beyond 20 years, a roll-out solar array (ROSA) can be attached on top of the original array during a servicing mission.

**Table 6.10-2** shows the power usage of spacecraft subsystems for different operational modes. The single largest flight system power user is the telescope's thermal system which must maintain the large primary mirror at room temperature while it is viewing deep space. This need requires 3,560 W of power during normal operations. The instruments themselves are the next largest power draw, around 450–550 W depending on observations being taken.

The solar arrays are sized for the simultaneous exo-planet and general astrophysics observations, which requires 7.0 kW of power including 2.1 kW for contingency and margin (see **Table 6.10-2**). This leads to a minimum array size of 39 m². This is at the end of a 20-year initial lifespan, and assumes GaAs triple junction (TJ) rigid solar cells, with a 29.5% efficiency and an off-sun angle of 40°. In additional to the primary array, there is a secondary array located on the base of the bus. The combination of these two body-mounted, non-articulating arrays allows for pointing the telescope up to 180° from the sun while providing sufficient power for most observational modes.

**Table 6.10-2.** Power equipment list for HabEx telescope flight system shows significant margin for all power modes.

| Subsystem | Unit | Launch | L2 Insertion | SSI-Only Science | UVS + HWC Science | Science Max | Down-link | Safe | Cruise |
|---|---|---|---|---|---|---|---|---|---|
| ACS | W | 0 | 15 | 10 | 15 | 15 | 20 | 2 | 2 |
| CDH | W | 40 | 40 | 40 | 40 | 40 | 40 | 40 | 40 |
| Instruments | W | 0 | 0 | 460 | 540 | 860 | 0 | 0 | 0 |
| Monoprop System | W | 30 | 360 | 1 | 1 | 1 | 1 | 30 | 1 |
| Electrospray Prop System | W | 0 | 25 | 25 | 25 | 25 | 25 | 25 | 25 |
| Telecom | W | 75 | 75 | 140 | 75 | 140 | 170 | 75 | 75 |
| Thermal | W | 410 | 810 | 3560 | 3560 | 3560 | 3560 | 810 | 410 |
| Power Subsystems | W | 60 | 80 | 140 | 150 | 150 | 130 | 70 | 60 |
| **SUBTOTAL** | **W** | **620** | **1410** | **4380** | **4410** | **4790** | **3950** | **1050** | **610** |
| Contingency and Margin | % | 43% | 43% | 43% | 43% | 43% | 43% | 43% | 43% |
| Contingency Power | W | 260 | 610 | 1880 | 1900 | 2060 | 1700 | 450 | 260 |
| Distribution Losses | W | 20 | 40 | 120 | 130 | 130 | 110 | 30 | 20 |
| **TOTAL** | **W** | **900** | **2060** | **6380** | **6440** | **6980** | **5760** | **1530** | **890** |

The batteries are sized to survive a 3-hour launch scenario while maintaining the depth-of-discharge above 70%. Two 66 Ah lithium ion batteries are needed.

### 6.10.3  Propulsion

Telescope propulsion is handled by two separate systems: a monopropellant propulsion system and a microthruster propulsion system The monopropellant system consists of one 445 N main engine, four 22 N thrust vector control engines, and sixteen 4.45 N attitude control engines; the monopropellant system is capable of 3-axis control and has been sized to perform trajectory maneuvers, station-keeping, slewing, attitude control, and disposal. A monopropellant system was selected over a bipropellant system since monopropellant is easier to refuel and is less complicated than biprop. The mass penalty for the lower specific impulse that comes with monopropellant was acceptable due to the large launch mass margin.

For a typical slew, the 4.45 N ACS thrusters fire with a 5% duty cycle capable of a slew rate around 0.15° per second. Total slew times assuming 5% duty cycle are plotted in **Figure 6.10-4**. While faster slews are possible by increasing the duty cycle, most science observations do not require such a high speed. The maximum tracking rate for non-sidereal objects within our planetary system is estimated to be 1 arcsecond per minute.

Contamination of telescope optics due to thruster firings had been previously studied by the Kepler mission (Sholes et al. 2004). It was found that the residual hydrazine pressure at the barrel opening due to chemical thruster firings was low enough as to be negligible. The HabEx thrusters are located even further away from the barrel opening than on Kepler.

HabEx baselines 2,280 kg of hydrazine, which budgets for TCMs, orbit maintenance, and slews. For TCMs and orbit maintenance over 10 years, 1,390 kg is budgeted. For slews, 890 kg is allocated. The slew propellant requirement of 360 kg covers 100% of Monte Carlo mission simulations of the first five years using the methods described in *Section 8.2*. The remaining 530 kg of hydrazine can be used for an extended mission while also leaving additional reserve propellant to conduct fast slews for target-of-opportunity follow-up observations during the entire HabEx mission.

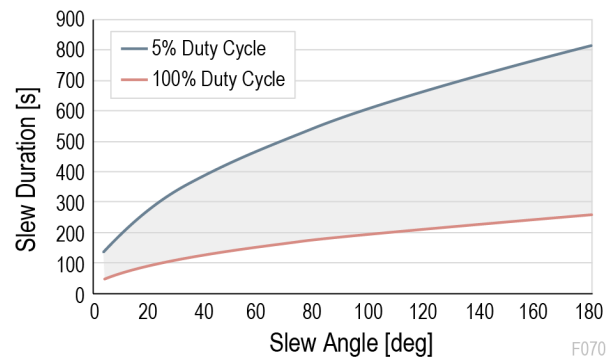

**Figure 6.10-4.** HabEx is capable of slewing 180° in less than five minutes and is a capable platform for multi-messenger observations. This figure shows the envelope of slew capability, shaded grey in between 5% and 100% duty cycle.

The second propulsion system—the microthruster system—consists of 8 colloidal electrospray microNewton thruster modules currently being developed for the ESA LISA mission and is solely responsible for maintaining fine-pointing, primarily by counteracting the effects of solar pressure.

Colloidal microthrusters use an organic colloidal suspension as a propellant. Electrostatic force expels the ionic colloidal molecules out the emitter nozzle at an effective velocity over 9,000 m/s. These thrusters have flown on LISA Pathfinder/ST7. New 18-emitter thruster heads currently in development for LISA have a maximum thrust of 100 µN, with a variable $I_{SP}$ of 1,000–1,800 s, and thrust resolution of 0.1 µN. For scale, 30 µN is about the weight of a mosquito on Earth, while 0.1 µN is the weight of a single mosquito antenna.

Since the primary purpose of the colloidal microthrusters is to compensate for the solar pressure induced torque on the spacecraft, and since the solar pressure is largely constant, or at least slowly varying on timescales much greater than the microthruster response times, the microthrusters will operate continuously with a near constant force. Enough propellant is included for 10 years of continuous operation.

Maximum thruster noise is <0.3 µN/√Hz up to about 1Hz and occurs when the thrust force level is in a state of change. When held constant, the thrust noise drops to <0.03 µN/√Hz. As found on LISA Pathfinder, the vibrational noise from microthrusters is up to 4 orders of magnitude less than reaction wheels (Ziemer et al. 2017).

Colloidal microthrusters are typically packaged in completely self-contained units including thruster heads, control electronics, and propellant tanks. This modularity makes these thrusters good candidates for servicing, requiring only an electrical connection for control, telemetry, and power, and a mechanical connection to secure the unit to the spacecraft. When serviced, a new, completely fueled unit can be installed to replace the original unit. Propellant sizing was designed assuming a 5-year nominal mission plus a 5-year extended mission. The mono-propellant system was also made to be serviceable by the use of a Vacco Type II interface resupply valve. 2,280 kg of hydrazine and 240 kg colloidal liquid are needed for the monopropellant and Busek thrusters, respectively, assuming the MEV spacecraft dry mass.

Although the colloidal microthrusters are preferred for HabEx, cold gas microthrusters are a viable alternative. Cold gas thrusters have flown on ESA's LISA Pathfinder mission and are currently flying on ESA's Gaia mission with over four years of nearly flawless operation. While these thrusters have a max thrust of 100 µN, they have an $I_{SP}$ of over 60 s and a thrust noise of less than 0.3 µN/√Hz, and would require carrying more propellant over the mission lifetime. Both cold gas thrusters and colloidal electrospray thrusters are under consideration for ESA's LISA mission.

### 6.10.4 Communications

The telescope telecommunication system was designed to support a crosslink between the telescope and the starshade, NASA's Deep Space Network (DSN) tracking for navigation, and downlinking of science data without disrupting on-going observations.

An S-band patch antenna is used for crosslinking between the telescope and starshade, while Ka- and X-band are used for telescope to DSN communications. Two Ka-band phase array antennas mounted on opposite sides of the telescope bus allow for simultaneous science observation and downlink. Unlike traditional steerable HGA, the phase array antennas do not require any mechanisms to steer the beam and thus do not introduce vibration. The two X-band low gain antennas (LGAs) offer nearly 4π steradian coverage. Command, engineering data, and navigation will be handled over the X-band link.

The S-band crosslink would allow for 100 bps communication with the starshade with 6.0 dB margin. The Ka-band would permit a downlink rate of 6.5 Mbps with 3.0 dB margin, while the X-band would permit downlink at 100 kbps with 9.0 dB margin.

### 6.10.5 Command and Data Handling

The telescope command & data handling (CDH) subsystem would be mostly built-to-print based on the JPL reference bus CDH design. The JPL reference bus provides standard CDH capabilities including spacecraft operations, communication, and data storage. Of particular note, the CDH subsystem was designed to provide 1 Tbit of storage, allowing the ability to minimize data downlinks to the DSN to 1 hour twice per week while maintaining ample memory margin. Furthermore, the CDH enables telescope-to-starshade S-band communications by adding a low-voltage differential signaling interface from the built-to-print design to the telescope transponders.

The flight software would be designed based on JPL Core flight software, which was designed to work with the JPL reference bus CDH subsystem. Some mission-specific changes would be made to accommodate science requirements, telescope-to-starshade communication, and attitude control integration.

### 6.10.6 Telescope Pointing Control

Direct imaging of exoplanets in the habitable zone of nearby sunlike stars with a coronagraph levies some of the most challenging pointing requirements ever met by a space telescope. With the LOS error requirement on the telescope set at 2 mas, HabEx would need to meet HST's best pointing performance on a routine basis. Fortunately, HabEx has three advantages. First, HabEx's diffraction-limited angular resolution is two-thirds that of HST, allowing for tighter angular sensing. Second, the environment at

Earth-Sun L2 has significantly less thermal and gravitational gradient disturbances than those experienced by HST. Third, without reaction wheels, HabEx's self-induced jitter is essentially nonexistent.

This section describes how HabEx would achieve the necessary LOS pointing for its telescope and instruments. The discussion covers the pointing requirements, pointing control architecture, operational modes and the expected pointing performance of the telescope and of the most demanding instrument: the coronagraph.

#### 6.10.6.1 Requirements

The telescope pointing requirement, levied by the instruments, is 2 mas RMS per axis at the FGS—about 1/10th of the 21 mas FWHM of its PSF at 0.4 μm wavelength. This amount of error reduces the Strehl ratio from the nominal 80% (diffraction limited) to 77.5%. Thus, the peak of the PSF for a chosen target will be reduced by only 3%: a small effect on observing efficiency. For the starshade instrument, the workhorse camera, and the UV spectrograph, this level of pointing is sufficient. For the coronagraph instrument, additional internal pointing refinement is required.

High-precision pointing is key to attaining the required levels of contrast in the HabEx coronagraph, and this drives the optical, mechanical, and ACS designs.

The pointing requirements arise from the coronagraph error budget (**Figure 5.2-1**) and are summarized in **Table 6.9-1**. In turn, these requirements derive fundamentally from the contrast degradation caused by small wavefront fluctuations as the telescope LOS drifts away from the target star. Internally, while the coronagraph FSM corrects the telescope pointing error, there is a wavefront error caused by the input beam "walking" across the optics. This error appears as a variation in the speckle pattern in the coronagraph dark field and drives the requirements on the telescope LOS error.

There are three key disturbance sources on the ACS pointing system ahead of the backend compensation by the coronagraph instrument.

- Quasi-static observatory drift (drift between telescope instrument boresight and FGS field star sensors), corrected by the FSM,

- Low-frequency observatory jitter, corrected by the FSM, and

- High-frequency observatory jitter (i.e., ACS residual control error near the ACS bandwidth frequency), not corrected by the FSM.

Both the corrected and uncorrected residual jitter must be very small, and the telescope must be designed to mitigate it. After the FSM, there remains WFE arising from residual control error from the ACS and from high frequency jitter. Ultimately, these requirements are driven by the need to maintain a stable speckle pattern and set the maximum coronagraph internal pointing error of 0.3 mas RMS per axis at 1σ.

### 6.10.6.2 Control Architecture

Pointing control is handled by a multistage control loop architecture (see **Figure 6.10-5**). A conventional stage uses star trackers and gyros as sensors, and monopropellant thrusters as the actuators. This loop is responsible for slewing and other typical ACS functions. Once the target is acquired and within the FOV of the telescope, a second loop takes control. This loop uses the telescope's fine-guidance system to sense position against field stars in the FGS, and the microthrusters to hold telescope position. The loop is responsible for counteracting environmental disturbances such as solar pressure and the L2 gravity gradient. The third stage is internal to the coronagraph. Pointing alignment is monitored using the HCG's ZWFS, which directly detects the tilt of the incoming wavefront. The tip/tilt is then corrected by the FSM.

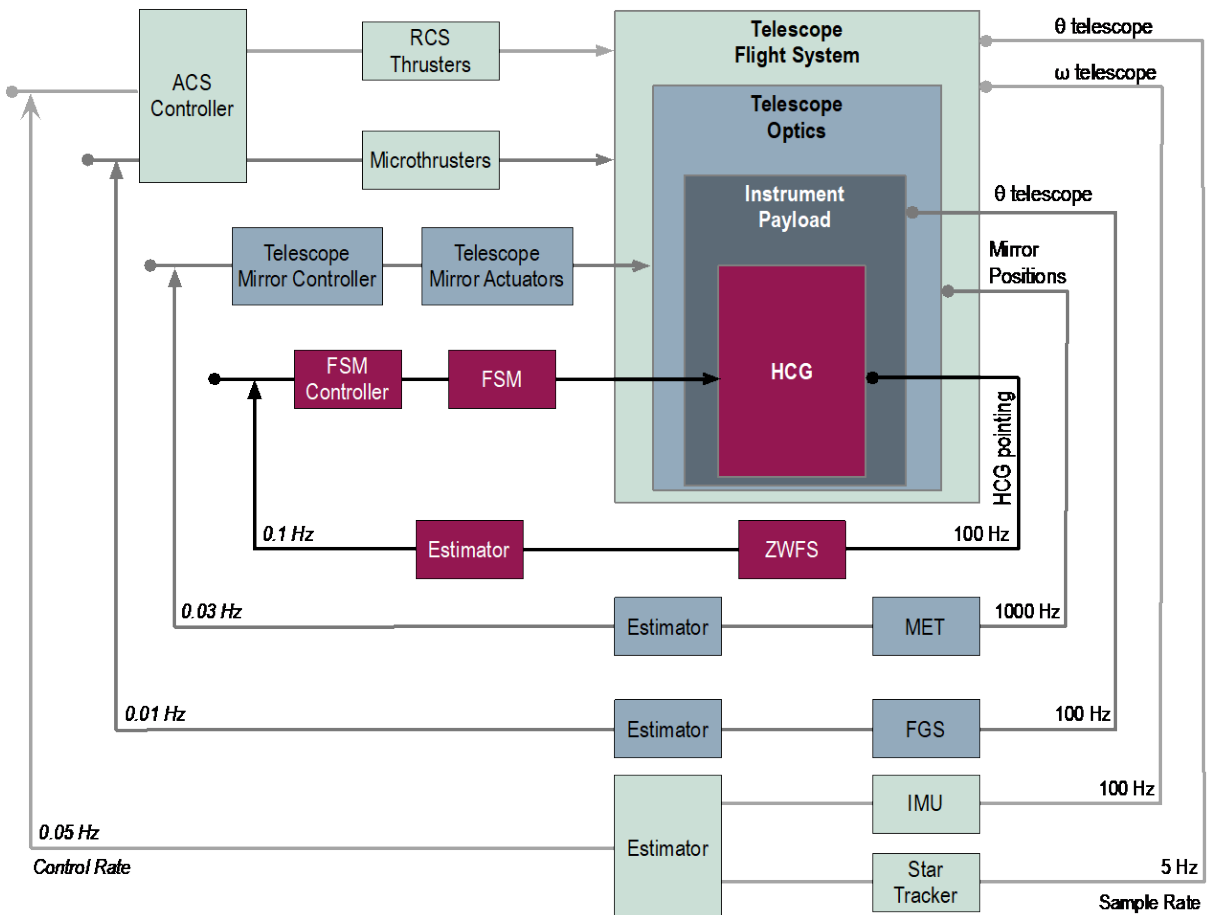

**Figure 6.10-5.** The HabEx pointing control architecture is defined by different control loops to provide pointing stability. Each control loop is defined its operational modes (slew, target acquisition, and science). In slew mode, IMUs and star trackers sense attitude, which is controlled the ACS controller and its commanding of RCS thruster firings. In target acquisition mode, attitude is sensed by FGS and controlled by microthrusters. In science mode, the FGS-microthruster control loop is still used. Additionally, the MET control loop maintains mirror alignment and the HWC ZWFS control loop maintains HWC alignment. Note that the MET control loop is independent of pointing modes discussed in this section.

In addition to the loops for sensing and correcting target position, HabEx also includes telescope thermal control and a laser truss to fix the relative positions of the first three mirrors. Details of the thermal control system are discussed in *Section 6.8.1.7* and the laser metrology system in *Section 6.8.5*. Both systems stabilize the telescope against slow thermal drift.

### 6.10.6.3 Pointing Modes

The primary pointing modes for HabEx are slew mode, target acquisition mode, and the science modes for each of the four instruments. Since the four instruments have separate fields of view, it would be possible to operate all of the instruments at the same time. Typically, the HWC and UVS instruments could both run opportunistic deep field observations while the starshade instrument or coronagraph is collecting spectra on a planetary system.

**Slew mode** begins by firing the monopropellant RCS thrusters to initiate rotation of the telescope toward the next target for observation. Thrusters are fired again to stop rotation once the target is reached and within the FOV of the telescope. The star trackers and the IMU are used to determine telescope orientation during this phase.

**Target acquisition mode** begins with the position sensing handoff from the star trackers to the FGS for refinement of the telescope's LOS error. Initial actuation is handled by the RCS thrusters until they reach near their minimum impulse bit limit, at which point the microthrusters take over actuation to bring the LOS error below the required 2 mas. The microthrusters continue to operate during instrument observations to counteract environmental disturbances, primarily solar pressure-induced torque, on the telescope.

Once the telescope has acquired the target and achieved 2 mas or better LOS error, **science modes** begin. Instruments remain in science mode until the science data has been collected, at which time, the telescope is ready to slew to the next target and the cycle repeats.

### 6.10.6.4 Slew Mode

Navigational needs for the HabEx spacecraft, slewing and repositioning, are handled by a combination of a multi-head star-tracker sensor system, an inertial measurement unit (IMU) with four fiber optic gyros for spacecraft position and attitude sensing, and a hydrazine monopropellant system for rotational and translational movement of the spacecraft. The star tracker and IMU are internally redundant, and the thrusters and gyros are numerically redundant. Fiber optic gyros, having no moving parts, thus producing no vibration, are also inherently more reliable than mechanical gyros. All components are flight-proven and commercially available.

Slew mode begins by firing the monopropellant thrusters to initiate rotation of the telescope toward the next target for observation. The star trackers are used to determine telescope orientation during this phase. The coarse guidance sensing system provides a telescope pointing accuracy of 40 mas, which would bring observational targets within the telescope's FGS FOV (>2 arcmin). The thrusters are fired again to stop rotation once the target is reached and within the FOV of the telescope. Sensing is then handed off to the telescope's fine-guidance system.

Design reference mission simulations of notional mission slew requirements found that 360 kg of hydrazine would meet 100% of Monte Carlo simulated missions over a 5-year period (see *Section 8.2* for details). HabEx has included 890 kg of propellant for observational slewing which covers the 5-year baseline mission, the 5-year extended mission and reserves. Nominal slewing is estimated at 0.15° per second but much faster slews are possible if needed, at a cost of greater fuel usage.

### 6.10.6.5 Acquisition Mode

Once in acquisition mode, sensing for telescope pointing is handed over to the telescope's FGS. The FGS is part of the telescope's payload and is described in *Section 6.8.6*. The system looks through the aperture of the telescope at bright, known stars in the FOV. Using actuated mirrors to position chosen field stars on the system's CCD detectors, the system can measure deviations of the telescope's LOS. That information is supplied to the fine guidance control loop, actuated with the microthrusters, to acquire the desired observational target and maintain the telescope's LOS on the target during science observations. As noted earlier, the FGS is composed of four CCD detector arrays looking through the aperture at four widely separated fields of view. This configuration allows the FGS to sense LOS errors of < 1 mas over most of the sky.

#### 6.10.6.6 Science Mode

In science mode, telescope pointing is maintained using the FGS and microthrusters in a control loop. Typically, with the starshade instrument operating, the HWC and UVS would also be operating. Over time, the relative boresights of the FGS and the SSI would evolve within the constraints imposed by the telescope thermal control system. Assuming a CTE of the structure of $10^{-6}$, a linear separation of 1 m between focal planes of the FGS and SSI, and 50 mK thermal control, the net result would be a maximum relative shift between the guide star and the science target of 0.004 pixels (12 µm pixel assumed) corresponding to 0.04 mas. This shift is extremely small relative to the angular pixel resolutions of the HWC, UVS, and starshade instruments (~12 to 30 mas) and is negligible. Even for the coronagraph this shift is small but would be corrected by the ZWFS control loop. Therefore, no special capability needs to be included to monitor relative boresights. Between the separate FGS optical paths, the relative boresights will also evolve, providing a verification of overall pointing stability over the time of an observation.

In the case of the coronagraph, internal LOS error must be further reduced to achieve the contrast levels needed for science observations. As shown in **Figure 5.2-1**, the internal LOS error must be reduced below 0.3 mas. To achieve this, a fine-pointing control loop internal to the coronagraph is engaged (**Figure 6.10-5**). Tip/tilt sensing is done with the ZWFS and corrected by the FSM. A small, low-noise, high-resolution focal-plane camera forms the sensor and supports high readout speeds (≥100 Hz). The FSM is a precision piezo-electric actuated steering mirror. The loop brings LOS error within the requirement and reduces any disturbance components (currently not expected) up to a ~5 Hz controller bandwidth, with a measurement error of ≤0.1 mas per tip/tilt axis. Higher ZWFS sampling rates are possible, which would allow for faster control rates as well as feedforward approaches. Current simulations show the settling time to be short—much less than a minute—with a steady state inertial pointing performance of ≤0.2 mas per tip/tilt axis, meeting the 0.3 mas requirement with significant margin. Additionally, dynamic laboratory testbed demonstration of LOS stabilization using a ZWFS (LOWFS) for the WFIRST coronagraph instrument (CGI) have already reached 0.36 mas rms stability per axis at photon fluxes equivalent to a V=5 star (Shi et al. 2018) and 0.2 mas RMS or better on brighter sources. This performance is already very close to the HabEx requirement, and importantly, it was demonstrated in the presence of WFIRST-like LOS and low-order wavefront disturbances which are an order of magnitude higher than those expected for HabEx ultra-stable telescope architecture.

### 6.11 Telescope I&T Plan

This section primarily discusses the Phase C–D integration and test flow for the baseline telescope spacecraft, including ground support equipment, testing, and facility considerations. A simplified overview of the sequential flow is shown in **Figure 6.11-1**. **Table 6.11-1** describes facilities required for telescope I&T and the necessary ground support equipment. Starshade integration and test is described in *Section 7.4*. Separately from this discussion, advancement of enabling technologies to TRL 5 is included in *Chapter 11*, detailed technology roadmaps are included in *Appendix E* the baseline schedule is discussed in *Chapter 9*. Several factors in the design of the HabEx baseline telescope, including its serviceability requirements and the interface design between the OTA and bus, have the added benefit of simplifying the integration and test of the spacecraft, which are explored in more depth in the following sections.

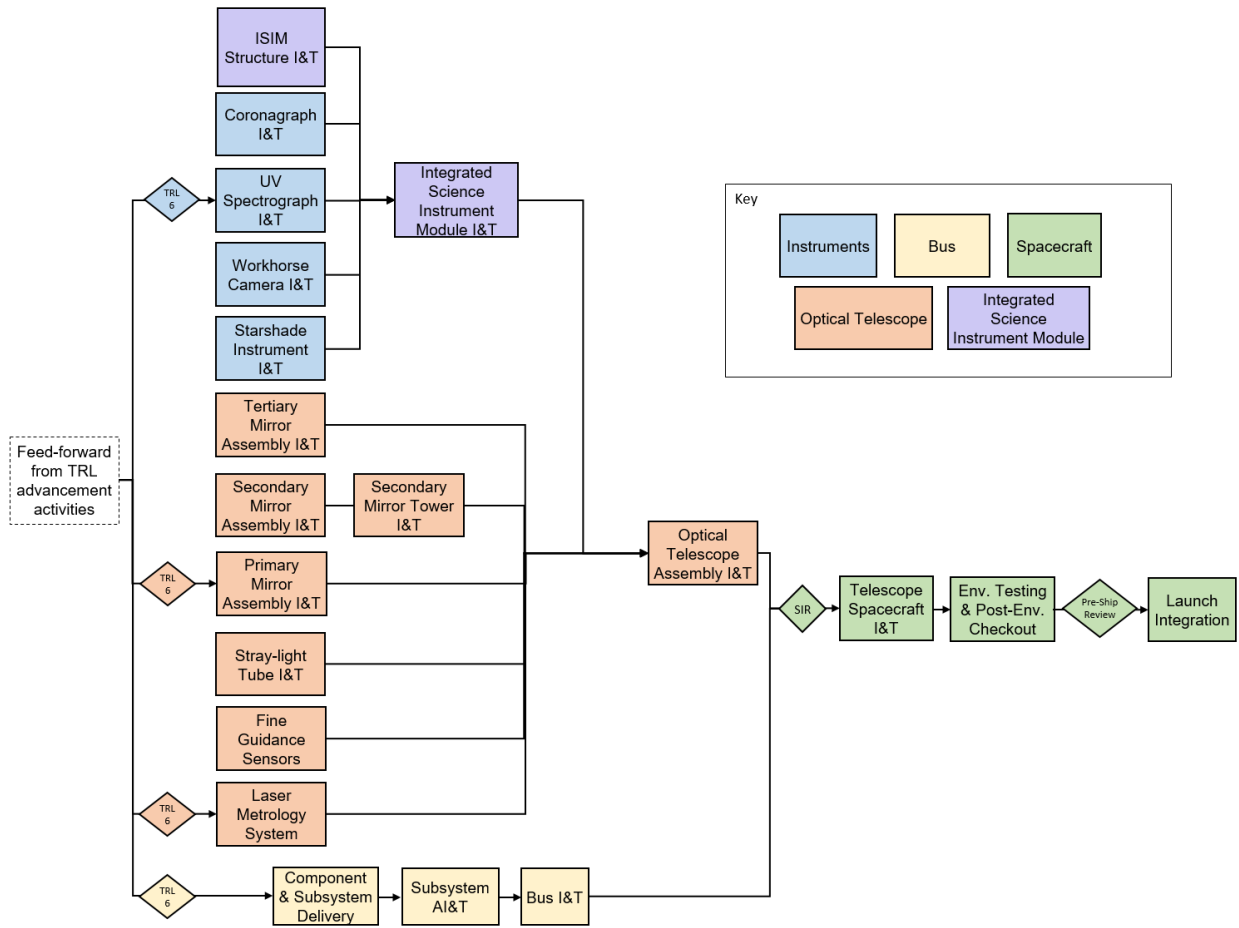

**Figure 6.11-1.** Telescope spacecraft integration and test flow, note: SIR is System Integration Review.

### 6.11.1 Facilities, Ground Support Equipment, and Testbeds

#### 6.11.1.1 Facilities

All integration and test activities for the telescope spacecraft can take place in conventional, existing facilities. Due to the large volume and mass of the spacecraft, the high bay used for final integration must be capable of accommodating the height, and supporting and moving the mass of the telescope system. As a general rule, the optical elements in the telescope and instruments are very sensitive to handling, volatiles, and particulates. There are high-heritage practices for handling these components that have been developed for other observatory spacecraft including Hubble, Spitzer, and JWST, and a review of materials and procedures for telescope integration will incorporate these methods and lessons learned. There are several existing thermal vacuum chambers that could accommodate the fully integrate spacecraft such as the Marshall Spaceflight Center X-Ray & Cryogenic Facility (XRCF) Chamber, and many more that could be used to test the bus and OTA separately.

#### 6.11.1.2 Mechanical Ground Support Equipment

The spacecraft and bus will require handling equipment and gantries in order to access all areas during integration. Handling fixtures must be appropriately designed for supporting the large mass of the system. A flat mirror external to the spacecraft, with the same diameter as the primary

**Table 6.11-1.** The required facilities and equipment for HabEx telescope flight system and payload I&T.

| Activity | Test Facility | Special Test Equipment |
|---|---|---|
| Instrument integration and test | Conventional I&T facility, cleaned and maintained to be safe for optical components | Spacecraft interface simulator |
| Instrument optical alignment and performance | Conventional I&T facility | Beam simulators tuned for each instrument's observation range |
| Mirror fabrication and test | Existing vendor facility | Described in *Section 11.3* |
| Optical telescope assembly integration, alignment, and test | Conventional I&T facility, cleaned and maintained to be safe for optical components | Beam simulator and flat mirror for alignment and end-to-end testing |
| Bus and spacecraft integration and test | Conventional I&T facility (with sufficient height and lifting capabilities) | Handling fixtures, spacecraft and ground station simulators, power supply equipment |
| Spacecraft environmental testing | Existing large thermal vacuum (TVAC) facility | Monitoring equipment for spacecraft health; measurement equipment and for optical element accuracy |

mirror, is required for end-to-end optical testing as well.

### 6.11.1.3 Electrical Support Equipment

Large power supplies will be required to simulate bus power output and test the fully assembled Integrated Science Instrument Module (ISIM). Similarly, in order to test the bus power distribution system, load simulators will be required as analogs for the instrument loads and telescope thermal loads.

### 6.11.1.4 Optical Test Equipment

At both the instrument and full telescope level, simulated beams will be required in order to both align the optical elements, and verify and validate the detector capabilities. This will involve several different pieces of equipment given the different observing bands of each, and different power levels in the beam depending on the level of assembly, i.e., simulated inputs directly into each instrument directly into the OTA, after assembly.

### 6.11.1.5 System Testbeds

For instrument I&T, a spacecraft simulator will be used to simulate the spacecraft bus and validate instrument interfaces, including command and data handling (CDH). For testing the bus and fully assembled spacecraft, two high fidelity testbeds will be built to test the bus avionics. The first is a testbed for the purpose of testing flight software using CDH hardware identical to the flight units. This testbed will be used during integration and test in order to validate procedures before execution on flight hardware, and to aid in trouble-shooting and regression testing. These capabilities extend beyond launch, giving ground operators a key ability to validate flight software updates and commands prior to uplink to the spacecraft, and providing a safe environment to test responses to in-flight anomalies. The second testbed is a mission system simulator. This includes environment models, simulated input data for spacecraft sensors, and simulation of telecom: both uplink/downlink with ground stations, and cross-link with the telescope. This simulator will be especially important for verification and validation of formation flying requirements, including response to simulated sensor input, cross-link communication, and GNC algorithms.

### 6.11.2 Payload I&T

The payload for the telescope system is the Optical Telescope Assembly, which includes the primary mirror assembly, secondary mirror assembly, tertiary mirror assembly, stray-light baffle tube with forward scarf, secondary mirror tower, and integrated science instrument module with four integrated instruments. Many items will be produced and tested in parallel before integration to form a complete payload.

### 6.11.3 Instrument I&T

Each of the four instruments, the HCG, SSI, UVS, and HWC will be assembled, integrated, and tested separately as standalone instruments. The specifics of testing for each will be different for each and depending on the final design trades completed. Despite their widely different functions and designs, each instrument will follow a similar general flow. Optical elements within each instrument will be individually subject to testing, including environmental and optical characterization and metrology prior to integration. Similarly, detectors will be tested in operational and survival environments, and their performance will be validated using simulated inputs. In parallel, instrument testbeds simulating a representative spacecraft interface will be used to validate the overall data flow and structure, and validate the general instrument support electronics. Once integrated with the instrument structure, more alignment of the optics will be verified, and characterization testing will be completed using the assembly optics and detectors. Finally, the fully integrated instrument will be tested using a simulated input beam and simulated spacecraft interface to validate the standalone instrument's end-to-end performance.

### 6.11.4 Integrated Science Instrument Module I&T

While the I&T diagram in **Figure 6.11-1** shows a linear flow of integration of instruments into the ISIM, in reality this process will be more complicated. Each instrument will likely be installed and removed several times for different checks, and for access to other parts. This process is simplified due to the fact that the ISIM is designed for serviceability; not only does this design enable servicing operations, it also eases integration on the ground. The ISIM, and the instruments within it, are installed on precision HST-style optical rails, greatly simplifying re-installation and alignment when a piece is removed for access or trouble-shooting. Because of this, the order of instrument integration and test has some degree of flexibility.

### 6.11.5 Mirror Assembly and Test

Detailed discussions of the mirror design and fabrication techniques, particularly for the large, monolithic primary mirror, are discussed further in *Sections 6.8, 11.3, and Appendix E*. In general, the telescope will be integrated, aligned, and tested using TRL 9 community best practices developed over many years using lessons learned from HST, Spitzer, and JWST (Stahl et al. 2010), which have been codified into a set of guiding principles (Stahl 2011). This begins with testing each optical component before integration into their individual assemblies using at least two independent methods. The optical prescription of every optical component will be controlled throughout its fabrication process using high-precision metrology tools, including: coordinate measuring machines; IR interferometers or Shack-Hartmann sensors; full and sub-aperture visible wavelength phase-measuring interferometers. Potential optical test geometries include center of curvature with computer generated holograms and infinite conjugate tests against a flat for the primary or tertiary mirror or against a Hindle shell or sphere for the secondary mirror. Verified methods will be used to 'back-out' gravity sag from all optical components to estimate the component's on-orbit optical figure. All optical component prescriptions will be verified at the operating temperature in a thermal-vacuum chamber. Once this testing is complete, the optical components can be formed with their respective mounts into full assemblies. Once integrated within their respective assemblies, each will be checked for functionality and specification compliance. Each assembly will be tested fully in a relevant thermal-vacuum environment in both survival and operational conditions to ensure its assembled surface figure has not changed by more than allowed by the error budget. All optical assemblies will also be tested over the full range of their hexapod motions, i.e., "pose" testing to ensure functionality and that there are no cable interfaces or unexpected mechanical deformations of the components.

### 6.11.6 Optical Telescope Assembly

Once all major telescope optical components are assembled and tested individually, they will be integrated and aligned with the ISIM and stray-light tube to form the OTA. As described in *Section 6.7*, the OTA is a complete structure, and interfaces with the bus mechanically only at the interface ring.

The complete OTA will be tested in a relevant thermal-vacuum environment, which serves three main purposes. First, this test validates that the integrated optical components comply with their specified prescriptions and can be aligned to create a telescope with the specified optical properties. Second, this test will also validate that all optical assemblies have an unencumbered full range of motion with no cable interferences or unexpected mechanical deformations of the components. Finally, this test will verify and validate the mirror thermal control systems and laser metrology systems. Due to the large spacecraft mass, vibration and acoustic testing may need to be performed separately for the OTA prior to integration with the bus.

### 6.11.7  Bus I&T

Bus integration and test will run in parallel to payload qualification and flight unit production and test. Due to the large spacecraft mass, vibration and acoustic testing may need to be performed separately for the bus prior to integration with the OTA.

#### 6.11.7.1  Subsystem Assembly, Integration, and Test

As shown in **Figure 6.10-1**, the system block diagram, most of the bus subsystem hardware is high heritage and based on existing, flight-proven designs. The only exception to this is the microthrusters used for stationkeeping; however, they are the subject of current TRL advancement activities described in *Section 11.3* from which HabEx will inherit. Most subsystems will be able to follow a build-to-print philosophy of reusing these designs, allowing more resources to be allocated to planning and completing difficult payload integration and test. Once components and box-level deliveries are completed and accepted, subsystems will be completed as necessary. Some of this work will be completed in parallel to the beginning of bus integration.

#### 6.11.7.2  Bus Integration and Test

As subsystem deliveries are completed, they will be integrated in the bus in a flow designed to minimize the number of disruptions and regression tests needed as more hardware is installed. Prior to full integration, both the subsystem and spacecraft will undergo safe-to-mate and safe-to-power procedures. After mechanical installation, the bus will undergo successively more and more complete functional testing until all components have been installed. Once successfully integrated, the spacecraft will undergo full system-level functional testing, to ensure all modes operate correctly, and sequence testing, to validate time critical sequences and autonomous functioning. The equipment described in *Section 6.11.1.5* is used to simulate the environment and stimulate the spacecraft during these tests. Especially critical in this testing will be the verification and validation of the flight software associated with formation flying, including sensing, crosslink, and control. At the bus level, this will include simulated input from the fine guidance system in order to exercise all parts of the control logic described in *Section 6.10.6*. This software will be tested at multiple levels and on testbeds prior to this point, and the bus-level sequence tests will confirm proper functioning on the flight hardware.

### 6.11.8  System I&T

Once the bus and payload are completed, a System Integration Review will assess hardware readiness to integrate. A series of fit checks and safe-to-mate procedures will be completed, followed by integration of the complete spacecraft. The OTA and bus are each complete structures, and mechanically connect only at their interface ring. Harness connections will be required for power and data flow between the bus and payload as well. It is also at this stage where the large rigid solar array is integrated. Once fully integrated, functional and sequence testing will be performed using the testbed electrical ground support equipment (EGSE) to verify that all spacecraft subsystems and modes operate correctly. This includes end-to-end testing of the fine guidance sensors, located on the payload, in conjunction with the Bus ACS system. This testing will also validate end-to-end data flow from instruments through the CDH and telecom systems.

#### 6.11.8.1  System-Level Environmental Testing

Because of the large size of the payload, it is likely that 3-axis vibration and acoustic testing will only be performed at the bus and payload levels. Additionally, if TVAC testing is performed separately on the bus and OTA, it may not be necessary to repeat at the fully integrated system level. Fully integrated electromagnetic interference (EMI) and radio frequency (RF) testing will be performed to assess compatibility of the bus and instruments.

### 6.11.8.2  Launch Integration

After successful post-environment testing, the spacecraft will be prepared for shipment to the launch site. A pre-ship review will be held to review comprehensive readiness and project status. Pending this gate, delivery will be completed, followed by post-ship inspection, functional testing, and checkout. The spacecraft will then be ready for integration with the launch vehicle, and coordinated operations with the launch service provider. After launch and initial deployment and checkout, the mission transfers to the operations phase; details of the DRM are included in *Chapter 8*.

# 7 BASELINE STARSHADE OCCULTER AND BUS

The HabEx starshade compliments the coronagraph in providing an additional starlight suppression technique that enables deep spectroscopic measurements through greater instantaneous bandwidth. In order to perform starlight suppression, the starshade flies in formation with the telescope. The starshade occulter creates a deep shadow, suppressing the light from the parent star and thereby revealing the reflected light from the exoplanets in the system. The optical design and position of the starshade occulter, along with the resolution and performance of the telescope and Starshade Instrument (SSI), determines the depth of the contrast in the dark field.

The starshade flight system is defined by its two parts. The starshade occulter is the payload, which is the part of the starshade responsible for blocking the starlight. The starshade bus is responsible for formation flight, propulsion, and typical spacecraft bus functions.

The optical performance of the starshade occulter is almost entirely an optomechanical attribute of the occulter design. Its size necessitates a deployable architecture that is passively shape controlled, both mechanically and thermally. The function of the starshade occulter's mechanical system is to reliably deploy on orbit, and meet the specified shape accuracy, shape stability, and solar edge scatter requirements. A 0.33 revolutions per minute (rpm) rotation reduces temperature gradients and improves shape stability.

The starshade must also be highly mobile, since it must move from target to target across the sky at a nominal separation of 76,600 km from the telescope. The starshade primarily operates in two modes. For about a quarter of the mission, the starshade flies in formation with the telescope to accomplish deep and broad survey programs. During this time, the starshade will participate in about 100 observations, with each target star system being observed at least twice. For the remainder of the mission, the starshade repositions from target to target using its electric propulsion system. Details of the HabEx baseline design reference mission are described in *Section 8.2*.

Accordingly, efficient propulsion and formation coordination are key capabilities required of the starshade bus. Starshade's electric propulsion system used to move between target stars is discussed in this chapter, while the formation flying control system, which positions starshade relative to the line of sight (LOS) of the telescope is discussed in *Section 8.1.7*. The starshade is also designed to be serviceable. Its serviceability is described in *Section 8.3*.

## 7.1 Starshade Occulter

The starshade's purpose is to create a deep shadow at the aperture of a space telescope by blocking starlight and limiting starlight diffracting into the shadow region. For this reason, the payload of the starshade is the occulter. The occulter is stowed at launch and mechanically deployed after launch.

### 7.1.1 Starshade Optical Design

The direct blockage of starlight with a simple, circular, opaque disk flying in formation with a telescope (such as that used in the upcoming ESA PROBA-3 [PRoject for OnBoard Autonomy] mission) is insufficient for exoplanet direct imaging due to starlight diffraction around the disk edge. A transition (or 'apodization') region, starting at the edge of the disk and extending radially outward, is required to mitigate diffraction. Ideally, the apodization region is a continuous gray-scale, but for the sake of a practical implementation, it is approximated as a binary function (all or none of the light passes at any point—an opaque mask). This yields the complex, yet distinctive starshade shape of a central disk with flower-like petals extending radially from the disk perimeter.

There is an infinite family of flower-like starshade shapes that produce a dark shadow suitable for planet hunting given a large enough starshade. To find these shapes, designers began by writing down analytic functions with a few parameters (e.g., Copi and Starkman 2000; Cash 2006). Later, Vanderbei et al. (2007) introduced more complex shapes with hundreds of parameters defining the edge shapes, and used linear optimization to choose the parameter values. Further design requirements beyond starlight suppression were set by other scientific and engineering considerations (e.g., disk diameter and petal length limitations, minimum feature sizes, bandpasses) constraining the many degrees of freedom in this optimization.

A three-step optical design process is employed in iterative fashion to find an optimal solution. First, parametric studies are conducted based on a large number of approximate solutions and curve fitting to illustrate trends. Second, tens of potential designs are run through the optimization scheme to identify candidates with high suppression and consistency with all imposed constraints. Finally, select designs are rigorously verified to provide the requisite starlight suppression at all points in the focal plane. Parameters are adjusted until the design is fully compliant with requirements imposed by scientific and engineering constraints.

Solutions for starshade designs are generated using the linear optimization tool described above, which finds the apodization petal shape that minimizes the modeled diffracted light over the full shadow region and wavelength range, subject to a predetermined maximum allowable light intensity within the shadow.

There is a small amount of freedom in selecting the number of petals used. The total number of petals is only bounded weakly by optical considerations—too few petals and terms ignored in the approximation slowly begin to become important. Conversely, an increased number of petals makes for smaller petal tips and smaller gaps between petals, as well as simply more hardware to manufacture, test, and deploy. Additional constraints include a minimum petal tip width and inter-petal gap of 1 mm, maximum petal lengths and widths that can be packaged for launch, and upper and lower bounds on the bandpass of operation. Inner working angle (IWA) was allowed to vary when generating families of designs for HabEx, with the smallest-IWA design with sufficient contrast being selected for the baseline.

Specific point designs are further evaluated for science performance based on the combination of parameters. Planet yield is evaluated for a target list constrained by a candidate starshade occulter's estimated IWA and contrast.

The occulter's optical shape is shown in **Figure 7.1-1**. The occulter consists of 24 petals, each 16 m long and approximately 2.6 m wide at the base. The occulter disk is about 20 m in diameter making for a starshade tip-to-tip width of 52 m. This 52 m design is exactly twice the scale of the 24-petal, 26 m Starshade to Technology Readiness Level 5 (S5) technology development starshade.

**Figure 7.1-2** shows the optical throughput of the starshade as a function of radial distance from the center. Interestingly enough the starshade exhibits a small amplification near the tips. However, for the purpose of contrast specification, this gain is ignored and the throughput unity in the wider field is used as the maximum. Some planet light can be seen between the petals of the starshade and this results in a gradual roll-off of throughput towards the center, closely approximating the geometric obscuration of the starshade as a function of radius. To specify the $IWA_{0.5}$ for the starshade, the $IWA_{0.5}$ for the longest wavelength of the science band is used; as can be seen in the inset to **Figure 7.1-2**, this is at 58 mas at 0.975 µm wavelength.

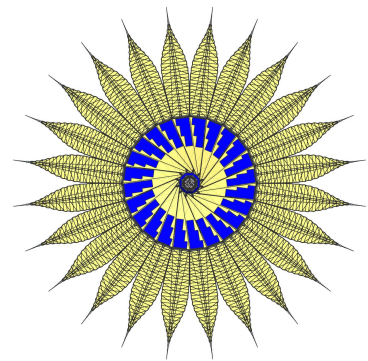

**Figure 7.1-1.** HabEx 52 m starshade occulter shape includes 24 petals. *Blue area* represents solar panels.

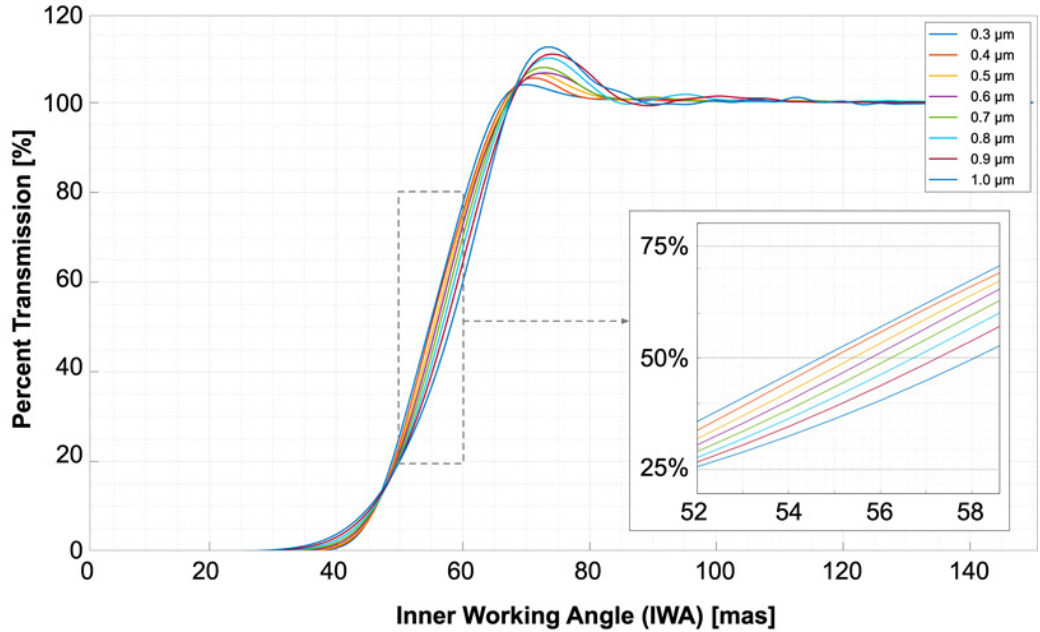

**Figure 7.1-2.** Same as Figure 6.4-3, this figure shows the starshade occulter throughput function with radial angle at a separation distance of 76,600 km. The inset identifies the small variation of IWA$_{0.5}$ as a function of wavelength.

### 7.1.2    Starshade Mechanical Design

The furled petal starshade design has been in development at JPL since the 2010 Astrophysics Decadal Survey identified the need to advance starshade technology. Great progress has been made in the design since then. This section explains the architecture and heritage of the HabEx design, and describes its mechanical features.

The approach of the furled petal architecture was to leverage existing heritage deployable structure technology to formulate a concept that would minimize uncertainty in technology development. The approach allows the starshade mechanical system to be functionally separated into two distinct subsystems that have separable requirements, can be developed in parallel, and validated with separate technology demonstrations.

**Figure 7.1-3** illustrates the two-stage deployment sequence of the starshade: unfurling of the petals followed by the inner disk deployment. Also highlighted are the two major subsystems: the inner disk and the petal.

The furled petal architecture draws on heritage from two flight-proven deployable

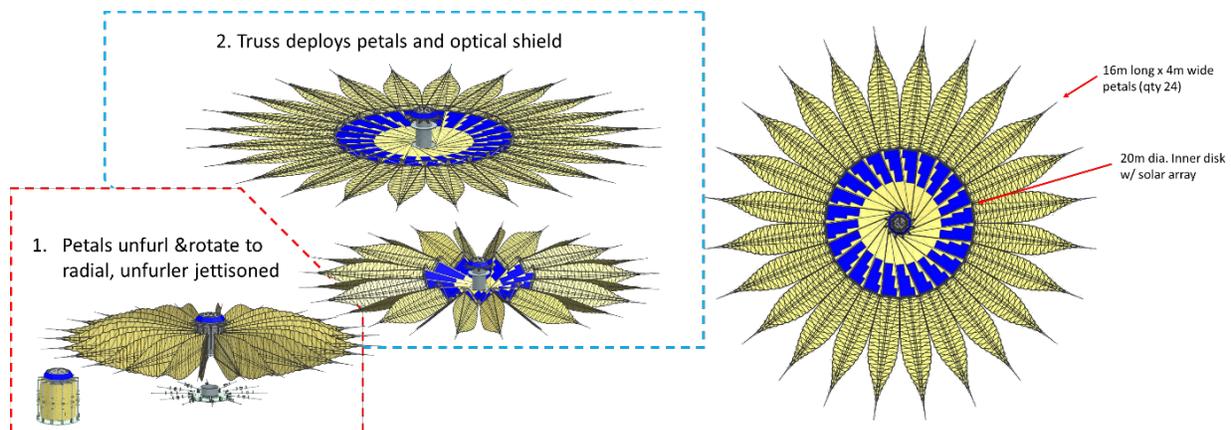

**Figure 7.1-3.** The deployment sequence for the HabEx 52 m starshade is identical to that being developed by the ongoing Starshade Technology Program (S5).

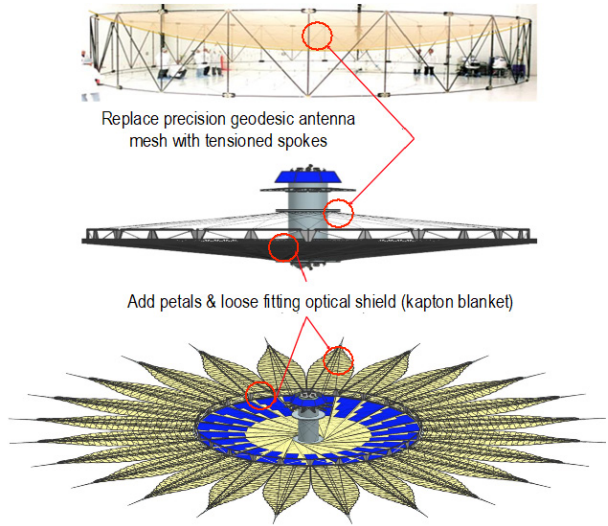

**Figure 7.1-4.** Traceability between Astromesh antenna technology and starshade design. The Astromesh antenna technology serves as the core to which the starshade petals are attached.

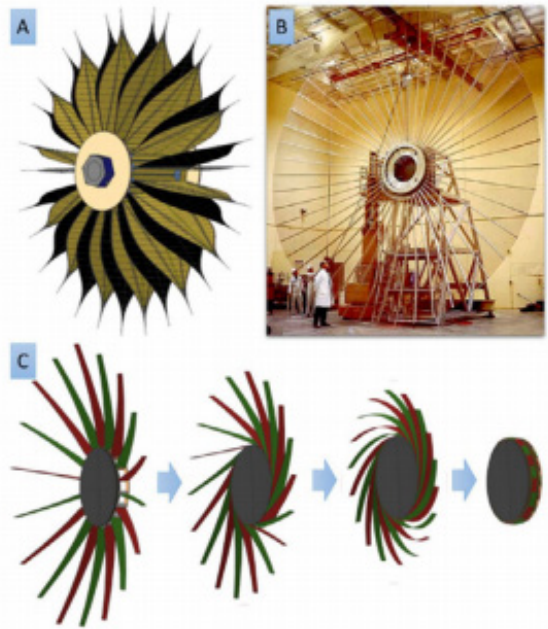

**Figure 7.1-5.** Comparison of Lockheed Martin (B) wrap-rib antenna to starshade petal wrapping architecture (*A, C*).

technologies—the Astromesh antenna (**Figure 7.1-4**) and the Lockheed Martin (LM) Wrap-rib antenna (**Figure 7.1-5**) (NRC 2010). The starshade inner disk is an adaptation of the Astromesh antenna, and is the core of the structure to which the petals attach. The Astromesh antenna is lightweight, precise, and has a high deployed-diameter-to-stowed-diameter ratio enabling very large deployed diameters to fit within a small launch vehicle fairing volume. Importantly, the Astromesh antenna has successfully deployed at least nine times on orbit, providing credibility to this deployment technology. The largest Astromesh antenna successfully deployed is a 16 m × 12 m ellipse, in comparison to the 20 m HabEx starshade inner disk diameter. The application of this technology to the starshade is illustrated in **Figure 7.1-4**, highlighting the replacement of the precision, gold-coated geodesic mesh that forms the antenna surface with the tensioned, linear spokes, resulting in a tensegrity ring formed by the perimeter truss that is rigid in plane. This ensures in-plane shape accuracy. The adaptation results in a ring that is less deep and better suited for attaching petals but retains the same deployment kinematics and mechanism upon which the antenna's perimeter truss architecture is based. The starshade perimeter truss is centered on a rigid stiff hub by the tensioned spokes. The hub houses the starshade spacecraft and propellant, as well as providing a stiff interface to the deployed starshade. In order to fit in the launch fairing, the petals and disk furl, or wrap, around the central hub.

The addition of the petals to the inner disk perimeter truss can be seen in **Figure 7.1-5**. To wrap the petals, the Lockheed Martin wrap-rib antenna approach was applied to the petals, resulting in the petals spirally wrapping around the stowed perimeter truss and central spacecraft for launch. The wrap-rib approach has been successfully deployed hundreds of times on orbit, again adding credibility to the use of this approach. **Figure 7.1-5A** illustrates the similarity between thin, radially oriented petals before wrapping for launch, and the thin, radial ribs of the wrap-rib rib antenna in **Figure 7.1-5B**. **Figure 7.1-5C** illustrates the wrapping of the ribs of the wrap-rib antenna around a large hub, which is the approach the petals follow. It is important to note that wrapping of the petals is in the out-of-plane direction, so as not to disturb the in-plane shape of the petal, the critical dimension for petal performance. Unfurling the petals is accomplished quasi-statically with a separate "unfurler." The unfurler is part of the Petal Launch Restraint & Unfurler Subsystem (PLUS), which is not considered a technology gap, but rather an engineering development.

The 52 m HabEx starshade is purposefully designed to be exactly twice the size of the 26 m S5 technology program design, greatly simplifying scaling and requirements traceability. The principle scaling challenge of the HabEx configuration compared to the S5 design is fitting an optical shape that is twice the diameter within the same 5 m launch fairing. The 5 m fairing is a design constraint because it allows the starshade to be launched separate of the telescope in any of a number of lower-cost commercially available vehicles. A series of configuration studies for the optical shield technology path has determined that for a given optical shield diameter and required thickness, the rigid thickness of the shield can be scaled such that it fits within the allowed radial diameter when stowed. The shield is designed such that any additionally thickness required by the shield for micrometeoroid mitigation is provided by Z-stringers (or one of various other options) that provide any additional shield layer separation.

As described above, the deployed starshade comprises two mechanical subsystems, the petals and the inner disk. The inner disk serves as the core to which the petals attach, and as stowed, forms the barrel-like structure around which the petals are wrapped for launch. Encaging the petals for launch restraint is the PLUS shown in **Figure 7.1-6** (*top*) HabEx design and (*bottom*) S5 engineering prototype. After launch, the PLUS quasi-statically unfurls the petals in a controlled fashion, ensuring the petals edges are not damaged.

### 7.1.3 Starshade Deployment

Deployment of the starshade involves multiple steps to transition from the compact, stowed system that fits within the launch vehicle fairing to the fully deployed operational system. Each step is described below.

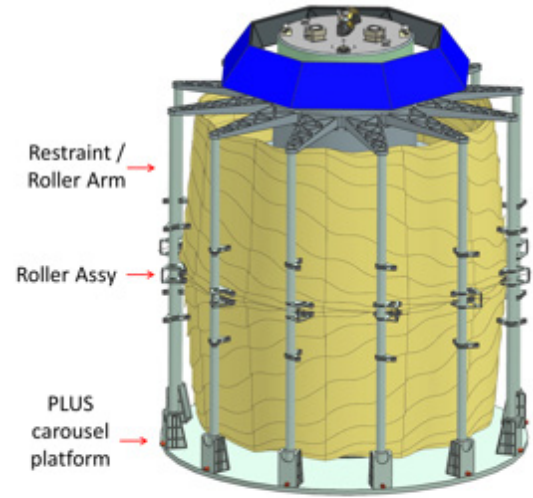

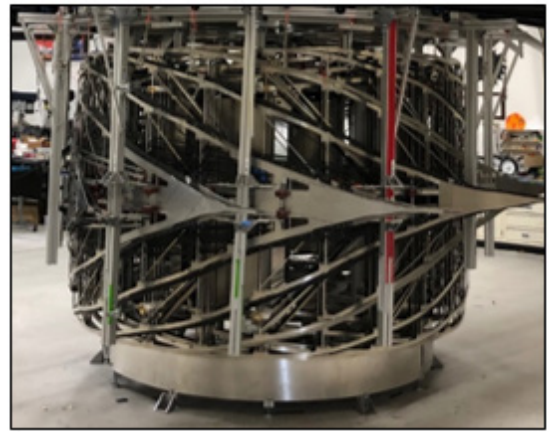

**Figure 7.1-6.** *Top:* HabEx PLUS design, which inherits from (*bottom*) the S5 PLUS engineering prototype in the furled configuration.

### Step One: Unfurling the Petals

The PLUS is a large carousel assembly that rotates about the starshade spacecraft hub's long axis. For launch, the PLUS is locked in rotation, and the vertical cage posts around its perimeter serve as an external boundary condition that preloads radially aligned launch-restraint interfaces on the spirally wrapped stack of petals.

Once on orbit, the petal preload mechanism on the cage posts is released and the petals then lightly preload a roller assembly on at the vertical center of the cage posts, which align with the petal centerline, **Figure 7.1-7A**. Petal unfurling is then controlled via two roller assemblies that extend tangentially from the vertical cage posts, allowing the furled strain energy in the petals

The roller assembly is centered vertically on the petal, aligning with the petal central spine.

The carousel rotational constraint is then released, and a single, redundant motor system slowly and deterministically rotates the carousel with respect to the wrapped petals, allowing for controlled release of the petal furled strain energy and ensuring no damage to the petal edges, as shown in **Figure 7.1-7B**.

### Step Two: Rotating the Petals

Once the petals have fully unfurled, they are passively rotated to a radial orientation via torsion springs in the hinges that attach the petals to the perimeter truss, shown in **Figure 7.1-7**. Once the petals are radial, and out of the way of the vertical cage posts, the cage posts are then rotated down and out of the way of the petals/truss system, allowing for the entire PLUS subsystem to be jettisoned before truss deployment (**Figure 7.1-7D**).

### Step Three: Truss and Petal Deployment

Once unfurled, the petals deploy passively from vertical to horizontal along with the active deployment of the perimeter truss. The perimeter truss design and deployment are fundamentally the same as those used on the Astromesh antenna, with deployment controlled via a braided steel cable that serpentines the diagonals of the truss, and is reeled in with a motor onto a spool, expanding the perimeter truss. This deployment technology has been used successfully more than nine times on orbit.

The truss is composed of thermally stable carbon fiber composite tubes, called longerons, which form a perimeter ring. This ring is placed in compression upon final deployment by the radial, thermally stable carbon fiber composite spokes that connect the ring to the central spacecraft hub. The tension and compression in the stiff and dimensionally accurate carbon fiber components creates a tensegrity structure that is precise and thermally stable, to which the petals are attached. By design, the perimeter truss deployment passively rotates the petals 90° into the plane of the starshade.

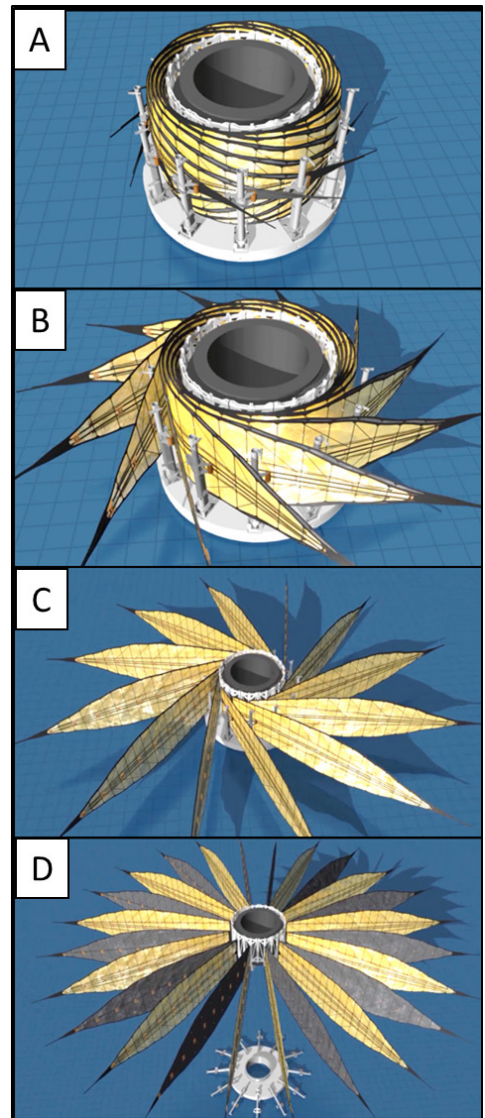

**Figure 7.1-7** Unfurling sequence as illustrated for S5 26 m design, first quasi-statically unfurling the petals (A–C), then rotating the petals 90 deg from tangentially oriented to radial (C–D), after which the PLUS is jettisoned (as shown in D).

The entire disk and all petals are covered with multiple layers of carbon impregnated black kapton—a material that intrinsically meets the HabEx opacity requirements—that unfolds as the truss deploys. Separation between the kapton layers mitigates the effect of micrometeoroid impacts by reducing the percentage of micometeoroid puncture holes that will provide a direct path for starlight to pass through the starshade and enter the telescope. The deployment of the truss pulls out the spirally wrapped opaque optical shield.

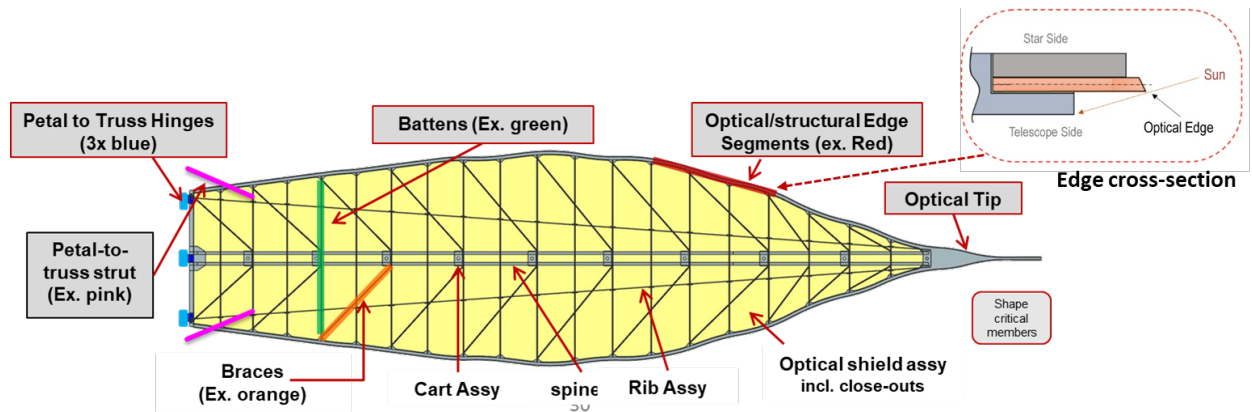

**Figure 7.1-8.** Details of a starshade petal including a cross-section of the optical edge. Petal shape is largely dominated by the width controlling elements, the battens. Solar edge scatter is minimized by reducing the edge radius of the optical edge as well as its reflectivity.

## Petal Structure

The starshade petal, unlike the inner disk, does not require the articulation of any joints or tensioned members to create its structure. Pictured in **Figure 7.1-8**, the petals are a lightweight planar carbon fiber composite structure that, as manufactured, meets the in-plane shape requirements of HabEx. The width of the petal, which is the critical dimensions, is provided by thermally stable carbon fiber composite rods, called "battens," that hold the petal structural edge at the periphery of the petal. The optical shape profile is able to meet shape requirements because it is produced in discrete 1 m segments that are precisely bonded to the structural edge in the correct location. The edge profile is formed by a thin, amorphous metal alloy, which is chemically etched to produce a sharp beveled edge that limits solar edge scatter from the edge into the telescope. The entire petal is then loosely covered with the same opaque optical shield as the inner disk. Because the petal structure is thin to allow for the petals to wrap for launch, out-of-plane stiffness of petals is provided via two piano-hinged ribs that passively deployed via a reliable and redundant over-center sprung hinge strut. These ribs are attached near the base of the petal to the perimeter truss, which provides a stiff out of plane connection from the petal to the perimeter truss ring.

## 7.2 Starshade Performance and Error Analysis

The starshade error budget, **Figure 5.2-2**, allocates contrast to the different effects that reduce raw contrast and stability required to achieve the driving science objective for the starshade observing case. The largest single allocation is mechanical shape error, and its contributions have been carefully evaluated by laboratory experience and simulation. Other contributing terms, such as solar edge scatter, sunlight leakage, and micrometeorite holes have also been evaluated. **Table 7.2-1** identifies key starshade flight system requirements, expected performance, and margin—identifying significant margin to error budget parameters. The expected raw contrast, $6.0 \times 10^{-11}$, and raw contrast stability, $1.0 \times 10^{-11}$, achieving the driving case's required signal-to-noise ratio with 91% of time margin.

### 7.2.1 Manufacturing and Deployment Tolerances

There are two driving requirements for manufacturing and deployment, petal shape and petal position.

The petal width profile must be manufactured to within a tolerance of ±140 µm. Compliance was demonstrated by test through a Technology Development for Exoplanet Missions (TDEM) activity (TDEM-09) led by Jeremy Kasdin of Princeton University.

**Table 7.2-1.** Starshade flight system requirements, expected performance, and margin.

| Parameter | Requirement | Expected Performance | Margin | Source |
|---|---|---|---|---|
| Observational band | 0.30–1.7 µm | 0.20–1.80 µm | Met by design | STM |
| IWA | ≤64 mas (0.87 µm)<br>≤80 mas (1.0 µm) | 57 mas (0.87 µm)<br>58 mas (1.0 µm) | 12% (0.87 µm)<br>38% (1.0 µm) | STM |
| Raw contrast | ≤1.0 × $10^{-10}$ | 6.0 × $10^{-11}$ | 67% | Error Budget |
| Raw contrast stability | ≤2.0 × $10^{-11}$ | 1.0 × $10^{-11}$ | 100% | Error Budget |
| Pointing control | ≤1° | <<1° | Met by design | Error Budget |
| Solar edge scatter | V > 25 mag/arcsec$^2$ | V > 25 mag/arcsec$^2$ | Met by design | Error Budget |
| Sunlight leakage | >32 $V_{mag}$ | >32 $V_{mag}$ | Met by design | Error Budget |
| Micrometeoroid holes | ≤500 ppm | 5 ppm | 9900% | Error Budget |
| Petal position (manufacture) | ≤±600 µm | ±340 µm | 76% | Error Budget |
| Petal shape (manufacture) | ≤±140 µm | ±80 µm | 75% | Error Budget |
| Petal position (thermal) | ≤±400 µm | ±62 µm | 545% | Error Budget |
| Petal shape (thermal) | ≤±160 µm | ±50 µm | 220% | Error Budget |

TDEM-09 petal prototype is 6 m long by 2.4 m wide and of flight-like carbon fiber composite construction (**Figure 7.2-1**). By comparison, the HabEx starshade petal is 16 m long by 4 m wide. Optical edge segments of matching carbon fiber construction were precisely positioned and bonded in place to define the petal width profile. The petal structure was assembled in a multistep process. It was populated with metrology targets and precisely measured using a large off-site coordinate measuring machine (CMM) with ±5 µm accuracy over the full petal length. This knowledge was used to precisely position optical edge segments relative to local metrology targets on the structure, using a small on-site CMM with ±10 µm accuracy over a few centimeters. After bonding all 10 optical edge segments in place, the petal was measured a final time with the large CMM. **Figure 7.2-2** shows resultant edge position errors relative to a best-fit nominal shape. The edge profile is within tolerance over 99% of edge length. TDEM-09 results fully demonstrate the ability to meet the allocated manufacturing tolerances for petal width profile.

The flight build will benefit from investment in an in-situ metrology tool. This tool will be mated to the assembly table (i.e., optical bench) and used for petal assembly, edge installation, and final shape measurement without moving the petal. One simplification for the TDEM was the use of square-cut optical edge segments. The flight unit requires a sharp bevel cut edge to limit

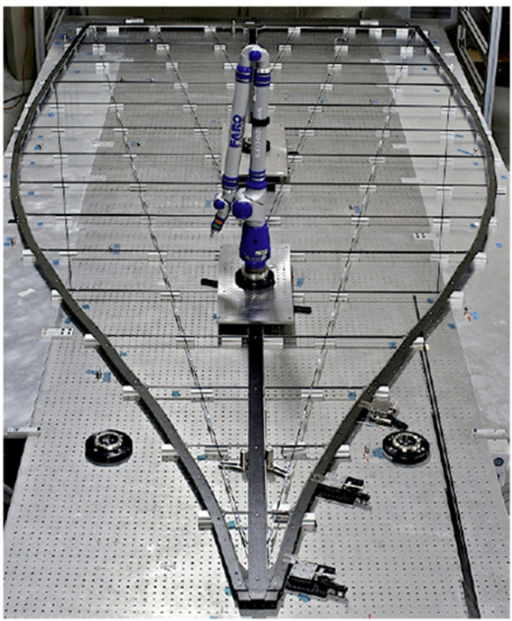

**Figure 7.2-1.** TDEM-09 petal prototype used to demonstrate manufacturing tolerance on petal width profile. Micrometer stages for positioning edge segments shown at bottom right.

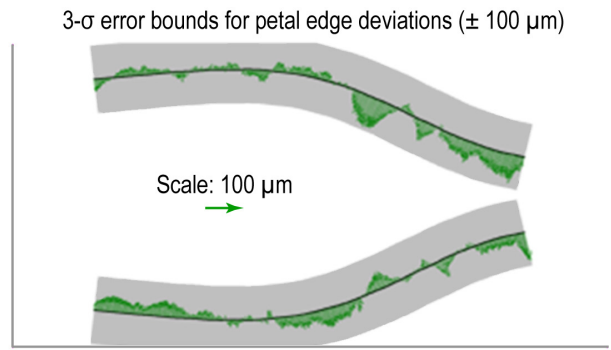

3-σ error bounds for petal edge deviations (± 100 µm)

Scale: 100 µm

**Figure 7.2-2.** Measured petal shape error (*green arrows*) vs. 100 µm tolerance for TDEM-09 (*gray band*) showed full compliance with the allocated tolerance.

scattered sunlight (**Figure 7.2-3**). This has led to the use of an optical metrology sensor (the Micro-Vu) that has a camera on a gantry CMM for the S5 prototype petal as shown in **Figure 7.2-3**.

Because the edges need to be sharp in order to limit solar scatter, the S5 prototype edges are chemically etched from a commercially available metallic glass nickel-iron alloy. **Figure 7.2-4** shows two of the 6 prototype edges produced to demonstrate the ±20 µm in-plane shape requirement, as well as the edge shape measurement vs the requirement. These edges were subjected to relevant environment testing and then verified for shape accuracy and edge scatter performance. These are the same construction of edge that are used on the S5 prototype petal.

The petal position (manufacture) requirement is defined by the ability of the perimeter truss and spokes to accurately deploy the petals to the correct location within ±600 µm. Each time the truss deploys, the petals are in a slightly different location with respect to each other. For this reason, the truss is deployed approximately 20 times to understand both accuracy and repeatability. This was first accomplished under TDEM-12, again, led Jeremy Kasdin to demonstrate the inherent capability of the perimeter truss. These results are discussed at length in the Exo-S report (Seager et al. 2015). More recently, in 2019, S5 built a medium-fidelity 10 m inner disk subsystem with low-fidelity optical shield that is currently undergoing deployment testing. The inner disk at Tendeg's facility in Louisville, CO along with preliminary results with a subset of the deployment data are shown in **Figure 7.2-5**. The plot in the lower image shows petal position error calculated as the difference between the measured mean and the design location. Data includes tolerance intervals that contain 99.73% of the underlying statistical population with 90% confidence, which fall well within a 300 µm radius. HabEx requirements is 600 µm, twice that of the 10 m disk, as the 20 m HabEx disk will have no more than 2× larger deployment error for a twice size inner disk.

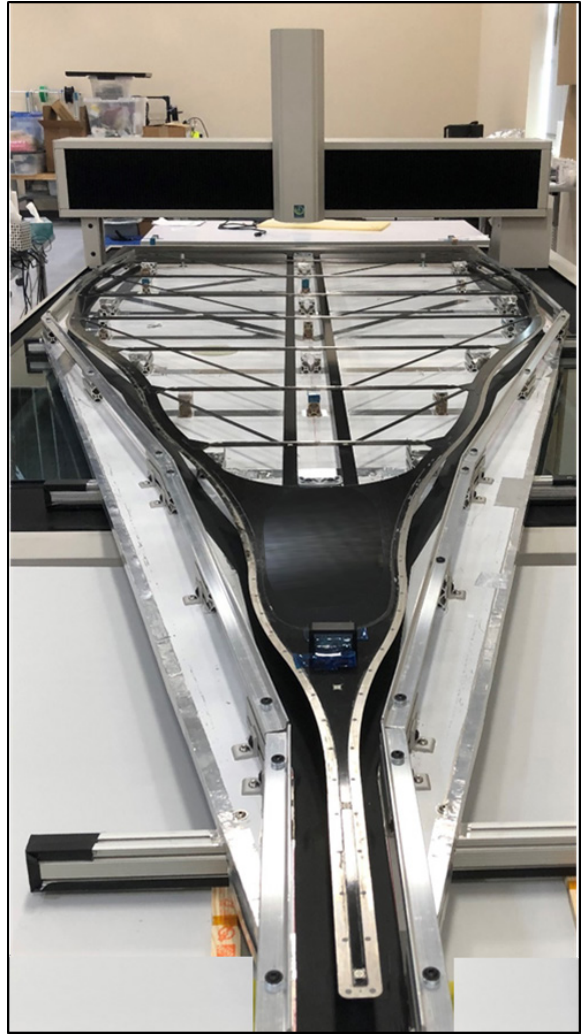

**Figure 7.2-3.** S5 Prototype 1 petal used to demonstrate shape accuracy after thermal and deploy cycles and to validate model of shape vs temperature (shape stability). Petal is sitting on Micro-Vu measurement machine at Tendeg's facility in Lousiville, CO. The overhead optical head on gantry CMM is used to measure edge shape profile.

### 7.2.2  Structural Analysis

In order to meet deployed shape performance requirements, the starshade structure must be sufficiently stiff and damped to ensure the structure maintains on-orbit shape during observations. To this end, the structure is first analyzed to determine the fundamental frequency, which is desired to be above 0.5 Hz, for position control of the spacecraft. Additionally, the lower-order mode shapes are assessed against critical performance error budget terms. Finally, the structure is assessed for the duration of its damping response to thruster firings during observations. In addition to meeting on-orbit shape, the structure must be suitable for ground handling during integration and test activities, as well as survive launch loads.

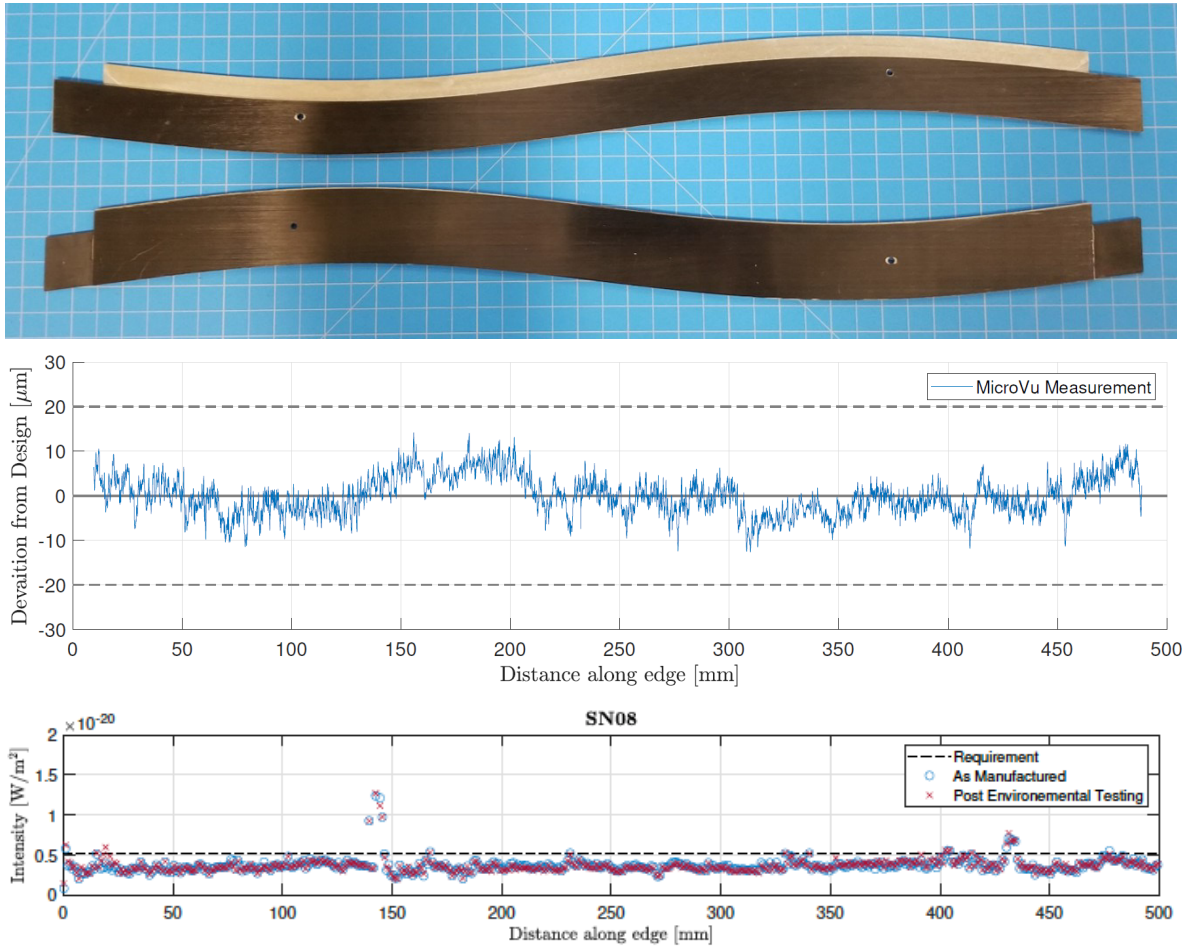

**Figure 7.2-4.** Half meter long S5 edge segments (*top*) with telescope side face up clearly showing metallic terminal edge material on upper edge, and target star facing side up on lower edge shown. These are two of six edges manufactured to demonstrate edge segment scatter performance and shape accuracy before and after relevant environmental testing. This new edge prototype has a sharp beveled metallic edge sandwiched by carbon fiber composite. *Middle:* In-plane shape profile for the edge meeting the ±20 μm requirement. *Bottom:* Laboratory solar scatter measurements for a prototype optical edge before and after thermal and deploy cycling. The edge meets the scatter requirements despite the isolated peaks in the scattered power.

### 7.2.3    Launch Structural Analysis

The launch configuration of the starshade is principally concerned with sizing of the central spacecraft cylinder to be stiff enough to meet the launch vehicle requirements and strong enough to support the mass of the stowed starshade that is attached to it. Because the stowed structure maintains a minimum inner diameter of 1.88 m to match that of the launch vehicle (as well as to provide sufficient volume for the propulsion tanks), the spacecraft cylinder is geometrically very stiff to begin with. The central cylinder is a honeycomb with carbon fiber composite facesheets that is sized to provide sufficient stiffness and strength for the given cylinder diameter. For the HabEx case, 1 inch honeycomb with 0.1 inch thick facesheets meets requirements with large margins. The Falcon 9 User's Guide states that a minimum payload primary mode of 10 Hz lateral and 25 Hz axial is required (SpaceX 2019). **Figure 7.2-6** shows that the first lateral (bending) mode for the starshade is 25 Hz and the first axial mode is 149 Hz, meeting the Falcon 9 requirements of 10 Hz lateral and 25 Hz axial. In addition to stiffness, the design was analyzed against strength and facesheet buckling which showed large margins as well.

The second consideration for launch configuration is the design of the PLUS launch restraint arms. Because the restraint arms are connected at the top and bottom of the starshade to the spacecraft, forming a very stiff and strong cage around the petals, sizing of the restraint arm diameter and tube wall thickness is all that is needed to ensure the required stiffness and strength for petal restraint. This will be determined when the detailed analysis of the petals in their launch configuration is performed. A sample of the detailed analysis performed by S5 for the 26 m starshade design is shown in **Figure 7.2-7**: detailed modal analysis of an entire starshade assembly (*top*), only one petal of the assembly shown (*middle*) and all petals hidden to show structure modes (*bottom*).

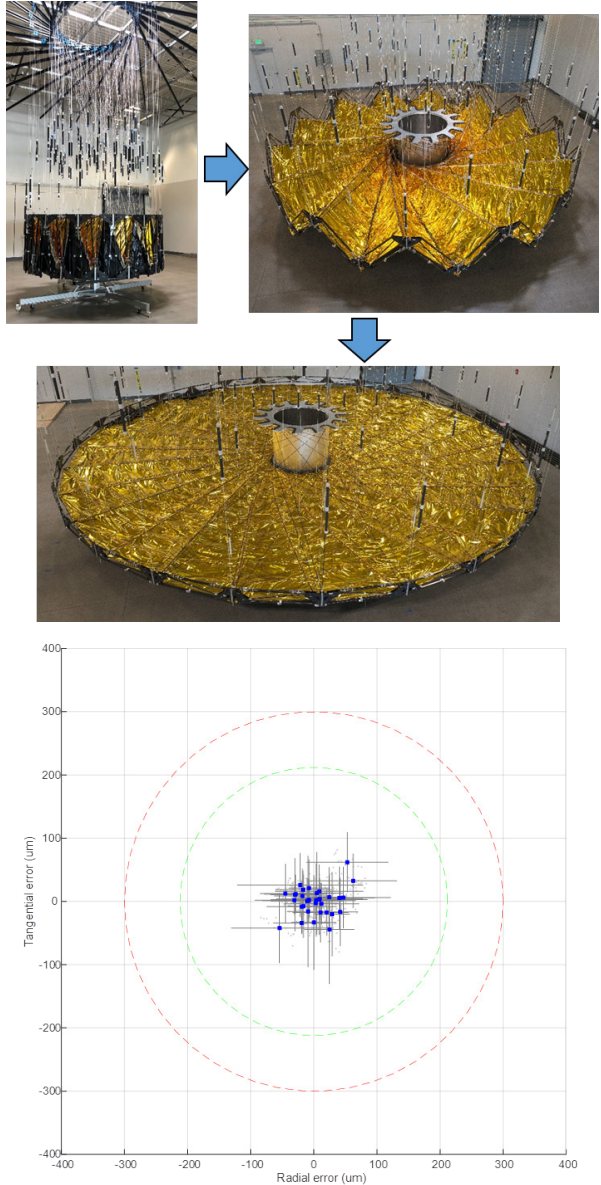

**Figure 7.2-5.** *Top:* S5 10 m inner disk with optical shield with ongoing deployments at Tendeg facility in Louisville, Colorado. *Bottom:* Preliminary data shows petal position error calculated as the difference between the measured mean and the design location. Data includes tolerance intervals that contain 99.73% of the underlying statistical population with 90% confidence, which fall well within a 300 μm radius. HabEx requirements is 600 μm, twice that of the 10 m disk, as the 20 m HabEx disk will have no more than a 2× larger deployment error.

### 7.2.4    On-Orbit Structural Analysis

The starshade structure has been analyzed for on-orbit structural performance. Because the starshade is designed to be very stiff in plane, the first mode is an anticlastic (potato chip) out-of-plane mode at 0.8 Hz shown in **Figure 7.2-8**, well above the 0.5 Hz goal for deployable structures that would not excite key error budget terms.

The first in-plane mode for the structure is approximately 15 Hz, well separated from the fundamental frequency of the structure and therefore not of concern for the key error budget deformations.

The starshade bus will periodically fire thrusters to hold lateral formation with the telescope. Conventional bipropellant thrusters are fired in pairs to apply a pure lateral translation. Analysis of the 26 m starshade for WFIRST with the same applied acceleration indicates a maximum in-plane shape deformation of only 11 μm, with a corresponding instrument contrast contribution of less than $1 \times 10^{-16}$, which is well within the allocated contrast. The disturbance is fully damped after about 10 seconds.

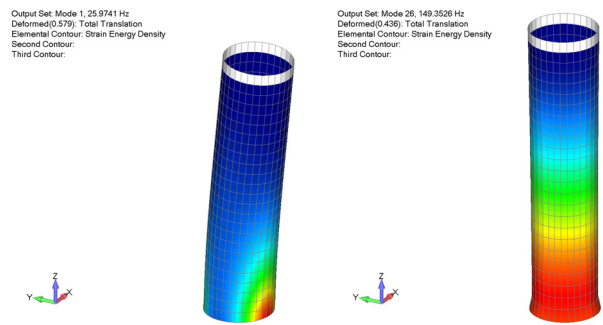

**Figure 7.2-6.** Finite element model modal results for the HabEx starshade spacecraft central cylinder to which the starshade is attached. The 1.88 m cylinder attached directly to the fairing interface at the bottom. With a 25 Hz first lateral (bending) mode and 149 HZ first axial mode meets the Falcon Heavy requirements of 10 and 25 Hz respectively.

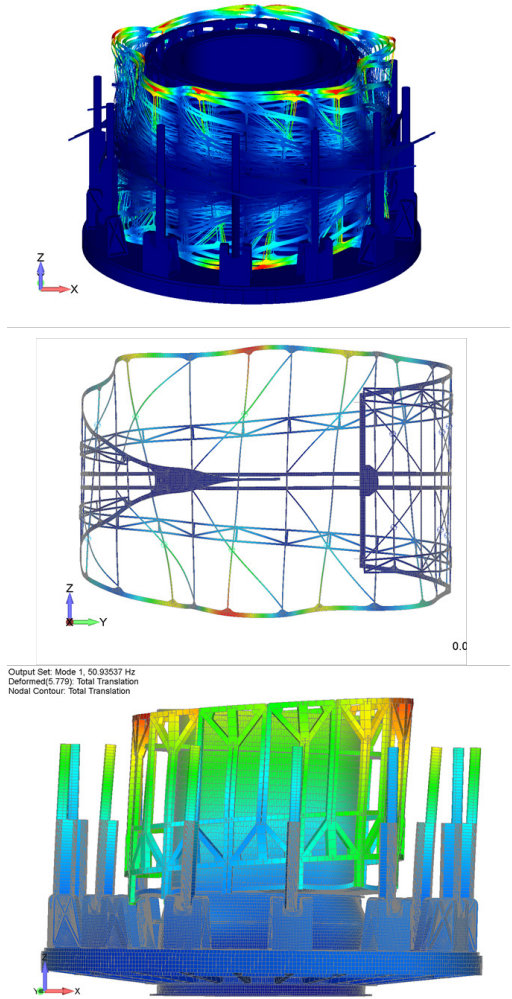

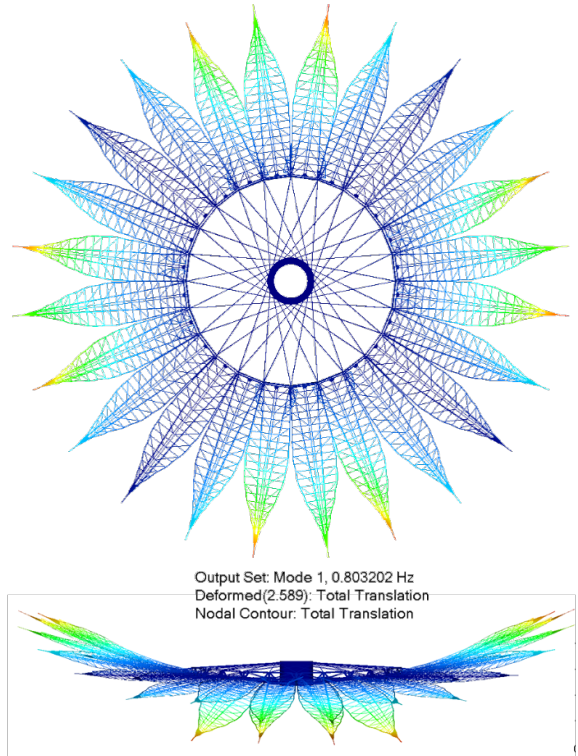

**Figure 7.2-8.** Structural modal analysis of the starshade showing representative mode shapes, all of which are out-of-plane shape deformations, which only indirectly contribute to degradation of starshade performance. First mode is at 0.8 Hz, well above the goal of 0.5 Hz.

**Figure 7.2-7.** Detailed finite element model with modal results for S5 technology project 26 m starshade. *Top:* shows entire assembly, *middle* shows just the petal (remainder hidden) and *bottom* shows just the structure (petals hidden).

While similar performance is expected for HabEx, the mission is not critically dependent on it. Science observation is separately interrupted by solar scatter from the thruster plume that can potentially saturate the science and interrupt science observations, but not permanently damage the detector. Pending a more detailed analysis, the preliminary plan is to avoid detector saturation by reading out the detector at a fast rate for a conservative duration of 10 s after receiving an imminent thruster firing alert from the starshade.

The dominant disturbance is the Earth gravity gradient between the two spacecraft, with an expected average value of about 2 μg. Given the oversized shadow diameter and a 1-sided dead-band strategy, the expected average period between thruster firings is about 10 minutes. This preliminary analysis conservatively bounds the observational overhead for holding lateral formation at less than 2%. The axial separation distance is allowed to drift for weeks at a time and the associated observational overhead time is negligible. The axial separation error for most observations is corrected as part of the repositioning maneuver, with essentially no observational overhead time.

### 7.2.5   On-Orbit Thermal Stability Analysis

The starshade instrument is sensitive to perturbations of in-plane occulter shape during observations. As a passively shape controlled instrument, the temperature of the structure varies based on the angle of the Sun with respect to the starshade, which is an unavoidable phenomenon. The starshade occulter structure is therefore designed to limit distortion of the shape due to the on-orbit thermal environment to within the error budget allocation for predicted temperatures across all sun angles for which observations are made. This is accomplished by utilizing very low thermal expansion materials, principally carbon fiber composites for all structural members, and invar (a low coefficient of temperature expansion [CTE] metal) for critical fittings and joints. The error budget terms under thermal stability are petal shape and petal position. Petal shape must be maintained to less than 160 μm from nominal and the petal position must be maintained to within 400 μm of nominal.

Petal shape thermal stability errors are principally caused by CTE differences between the varying structural members of the petal, which cause the petal assembly to distort non-uniformly as the petal changes temperature. The petal shape error is driven by the battens, which are the widthwise member of the petal. For this reason, the petal battens are constructed of a commercially available carbon fiber pultruded composite that, by virtue of the large-scale manufacturing process, has an extremely consistent CTE that has been tuned to almost zero across the temperature range of interest. The remaining petal structural elements are designed and manufactured to closely match the batten CTE to limit shape distortion due to interaction of the battens with the remaining structure.

The petals attach at their base to the perimeter truss, so petal position thermal stability error is principally caused by a deviation in the truss assembly CTE from the design nominal, which is defined as the CTE of the petal battens. The thermal stability of the starshade is therefore, by design, dependent on the ability to manufacture the carbon fiber composite structure to meet the on-orbit stability requirements. For this reason, the S5 technology program milestones focus on demonstrating the capability to manufacture the starshade components and demonstrate the stability requirement. The analysis presented here utilizes the validated models from the S5 program that are based on measurements of starshade prototype hardware, and is therefore representative of capability for future flight hardware. Some of this hardware is shown in **Figure 7.2-9**.

The process for validation starts the material level, where all materials used in the prototypes

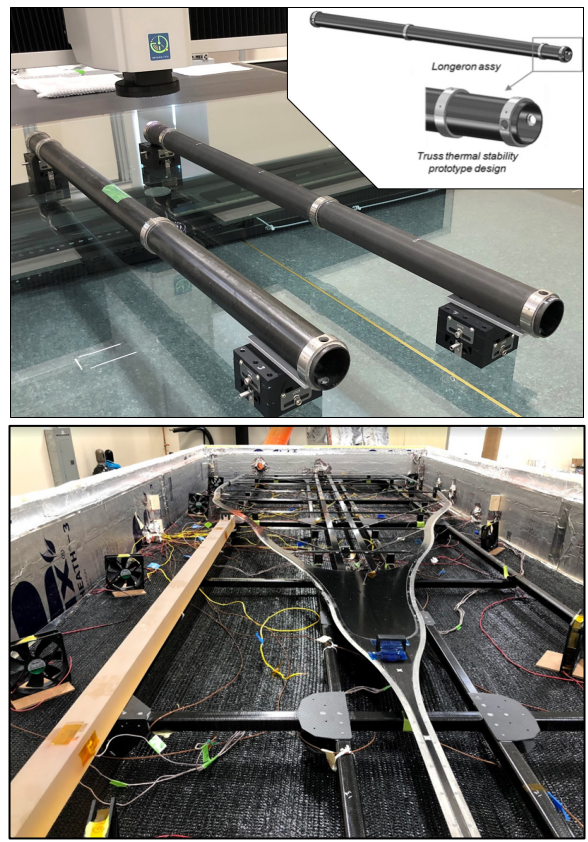

**Figure 7.2-9.** S5 thermal stability prototypes have raised confidence for HabEx. *Top:* Multiple truss component undergoing post thermal cycle dimensional stability inspection. The hardware was also measured for length as a function of temperature for the predicted on-orbit temperature range as an input to the thermal distortion model. *Bottom:* S5 prototype petal 1 in the thermal chamber at Tendeg's facility in Louisville, CO ready to undergo thermal stability testing led by NGIS-ATK in conjunction with Southern Research. The testing has produced data that is being utilized to fully validate S5 on-orbit performance models.

(e.g., carbon fiber composite) are measured at the coupon scale in a CTE facility at Northrop Grumman Innovation Systems (NGIS). These coupons are cut from the same sheets as the prototype hardware is constructed. The assembly hardware is then measured to validate the overall assembly, and understand and account for variations in the assembly finite element model and the actual hardware.

Thermal elastic deformation analysis has been performed on the 52 m HabEx design based on predicted temperatures for the starshade on-orbit. A thermal analysis model was created based on the starshade geometry and was used to determine on orbit starshade temperatures across the structure. These temperatures are then mapped to the structural model, which uses validated (as-measured) starshade material properties from the S5 technology milestone hardware to calculate the deformed shape of the starshade. The deformed starshade shape can then be compared to the error budget requirements to assess conformance.

The top panel of **Figure 7.2-8** shows the thermal analysis model and thermal distortions (*middle and bottom panel*) for the sun angle 83° case, where the sun is at a grazing angle with respect to the planar surface of the starshade. For this reason, the petals are relatively cold at an average temperature of approximately 223 K, whereas the truss is at room temperature, 293 K. The 83° sun angle case is driving for on-orbit performance with the largest errors compared to the requirements, namely petal shape error.

Raw distortions of the petals in inches are shown in **Figure 7.2-10**. Petal position is only changed by approximately 62 μm (*top panel*), well within the 400 μm requirement, and petal shape (width change, *bottom panel*) is only changes by a max amplitude of approximately 50 μm, well within the 160 μm requirement.

### 7.2.6    Other Analyses

#### 7.2.6.1    Solar Edge Scatter

Starshade optical edges are designed to limit solar scatter but a small amount of sunlight will scatter forward to the telescope. The diffuse component of scatter is very small and below the noise floor, but the diffracted and specular reflected components are more significant. The solar scatter that gets to the telescope originates from limited regions of the optical edge that present broadside to the Sun. As the starshade spins the integrated scatter appear to the telescope as two bright lobes that remain approximately fixed in space near the starshade tips. The optical edges are being designed to limit the brightness of these lobes to be dimmer than 25 visual magnitudes/arcsec[2]. Verification of this performance is a S5 technology milestone. The lobes can be calibrated as a function of Sun angle and the small amount of residual flux will only slightly limit the detection space at the IWA.

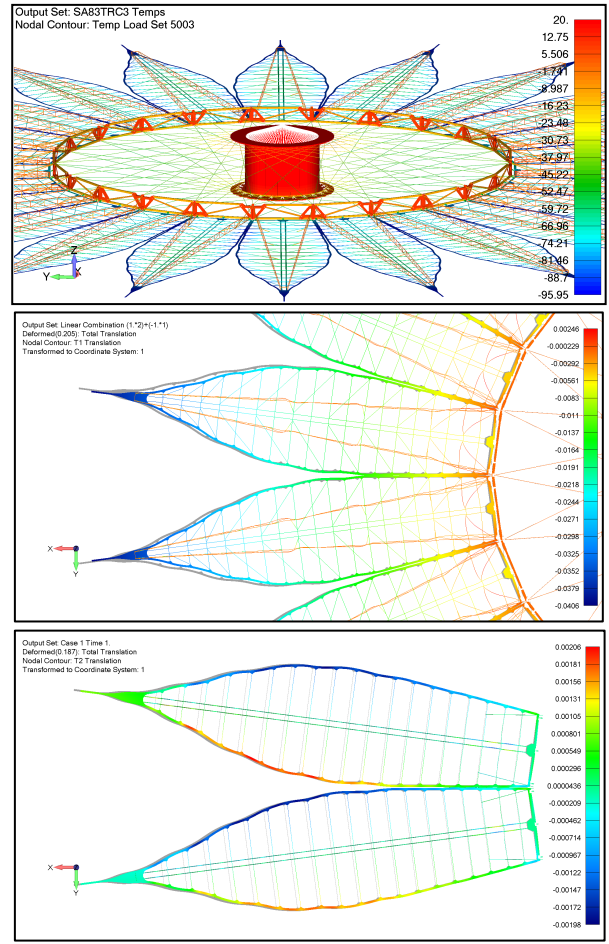

**Figure 7.2-10.** *Top:* The scenario driving thermal stability performance is an 83° sun angle with a warm, 20°C truss with a cold, -50°C average petal temperature. *Middle:* Petal position only changes by 62 μm, well within its 400 μm requirement. *Bottom:* Petal shape (width change) only changes by 50 μm, well within its 160 μm requirement. Raw deformation shown in inches.

### 7.2.6.2    Reflected Astronomical Objects

The starshade presents mostly Black Kapton to the telescope, with a known bidirectional reflectance distribution (BRDF). Reflected light from Jupiter, Mars, or the center of the Milky Way can appear as bright as 30 visual magnitudes for the worst-case geometry. Venus can also appear as bright as 31.3 visual magnitudes. By excluding the 1% of mission time when the worst-case geometry occurs, the brightness of these reflected bodies decreases by about 2 visual magnitudes. Most of this reflected energy comes from the inner disk which is well inboard of the IWA.

Reflected Earth-light can be much brighter but is only possible at the orbit extremes combined with pointing at stars just as they are entering the field of regard. Minor observational constraints can eliminate reflected Earth-light as a concern.

### 7.2.6.3    Micrometeroids

The starshade will accumulate micrometeoroid holes in the optical shield that covers most of the starshade with a small amount of starlight and sunlight transmitted forward to the telescope. The optical shield consists of three separated layers of black Kapton. The layer separation restricts the solid angle for starlight to transmit directly to the telescope and an instrument contrast allocation of $1.1 \times 10^{-12}$ (**Figure 5.2-2**) is achieved with large margins. The viewing geometry precludes direct sunlight transmission. However, sunlight entering an optical shield hole can reflect multiple times and exit another hole on a path to the telescope. Sunlight transmission through micrometeoroid holes is allocated a brightness dimmer than 31 visual magnitudes, which is estimated to translate to a hole area of at least 500 ppm in each of the three layers. To estimate the micrometeoroid flux, the Grun model is employed as specified in the James Webb Space Telescope (JWST) Environmental Requirements Document. Next, cumulative hole area is estimated after 5 years following guidelines in

NASA Standard PD-EC-1107 for particle speeds and shield stopping power. The result is an estimated hole area of <10 ppm in the two outer layers and less for the center layer. This analysis does not account for the steep rise in flux associated with seasonal meteor showers. Peaks in the micrometeoroid flux will require turning the starshade edge on to the peak flux. The loss in observation time can be mitigated to some extent by scheduling retarget maneuver coast periods to correspond to peak flux periods.

Starshade performance is relatively insensitive to direct micrometeoroid hits on the optical edge. Direct micrometeoroid hits on the optical edge are not expected to significantly degrade starlight diffraction, but could possibly manifest as a localized increase to scattered sunlight that can be confused as a planet. This occurs by exposing a larger surface area to the Sun. However, the particle size corresponding to an Earth-like planet is about 20 mm in diameter, while the largest particle with some reasonable probability of hitting the edge is much smaller at about 100 µm in diameter. Also, the edge damage will smear as the starshade spins. It is also worth noting that the nominal design already includes small gaps in between the approximately 1-m long edge segments and these gaps are longer in length than a 100 µm particle.

## 7.3    Starshade Bus

Despite the starshade occulter being a new system currently under development, the starshade bus, which assures successful delivery and operation of the occulter, relies on heritage hardware and proven designs. This section describes the key design features of each of the starshade bus subsystems, which are summarized in the system block diagram in **Figure 7.3-1**. **Table 7.3-1** presents the starshade mass breakdown; the total dry mass is estimated to be 5,080 kg, including 23% average contingency and an additional 20% system margin. Wet mass with contingency is 10,930 kg.

As described earlier, the HabEx starshade consists of a 20 m diameter disk surrounded by 24 petals with a tip-to-tip diameter of 52 m. The starshade has a 4.6 m diameter while stowed. This allows it to be launched separately from the telescope using a smaller faring. The starshade spacecraft bus is in the center of the starshade, fully within the hub. The total CBE mass of the petal and disk system (excluding the hub) is 1,620 kg, with an additional 800 kg allocated to the PLUS deployment system, which will be jettisoned after deployment. See *Section 7.1.2* for details on the starshade mechanical design.

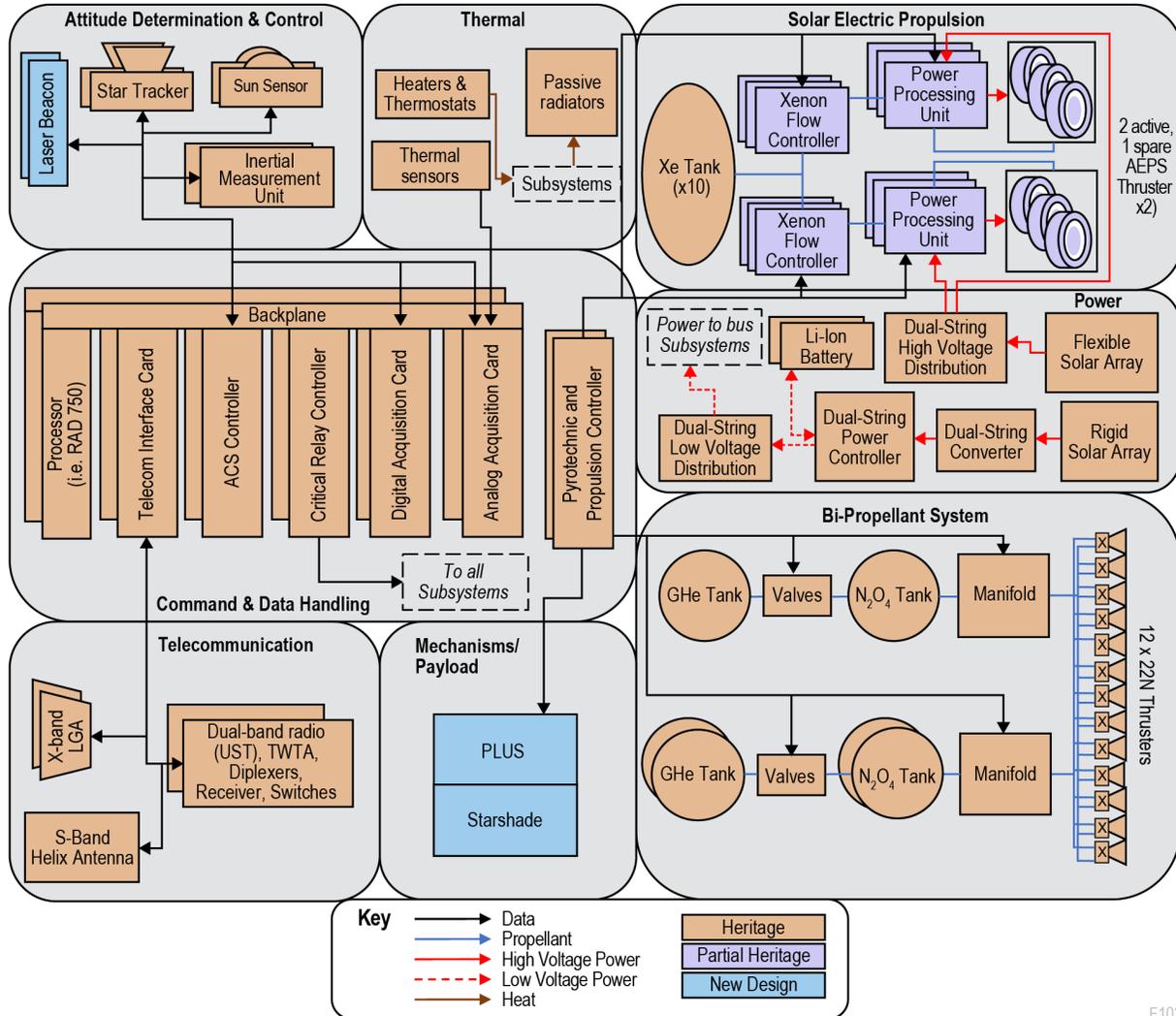

**Figure 7.3-1.** Starshade simplified block diagram illustrates the extensive use of heritage flight system components. Only starshade occulter and laser beacon represent new designs.

### 7.3.1 Structures and Mechanisms

The starshade hub structure was designed to fully contain the spacecraft bus subsystems as well as the hydrazine and Xenon propellants. The hub is a 4.6 m diameter honeycomb cylindrical structure with reinforced aluminum rings to prevent buckling and joints for reinforcements at attachment points. The inner core of the hub is 1.88 m in diameter. The starshade's hub structure is designed to attach directly to the launch vehicle's adapter ring, providing the best possible load path during launch. On either end of the center cylinder are two honeycomb flanges that reach out to the starshade truss structure to retain it for launch. The optical shield resides between the central cylinder and the truss.

Aside from the starshade payload deployment mechanisms, no additional mechanisms are needed on the flight system.

**Table 7.3-1.** Starshade flight system mass budget closed with ample margin to many launch vehicle lift capabilities. Current Best Estimate (CBE) and Maximum Expected Value (MEV).

| | CBE (kg) | Cont. % | MEV (kg) |
|---|---|---|---|
| **Payload** | | | |
| Petal and disk system | 1620 | 30% | 2100 |
| **Spacecraft Bus** | | | |
| ACS | 10 | 8% | 10 |
| C&DH | 15 | 20% | 20 |
| Power | 250 | 28% | 320 |
| Propulsion: Bipropellant | 210 | 3% | 215 |
| Propulsion: SEP | 890 | 21% | 1080 |
| Structures & Mechanisms | 460 | 30% | 600 |
| Telecom | 30 | 18% | 35 |
| Thermal | 180 | 30% | 230 |
| **Bus Total** | 2040 | 23% | 2500 |
| **Spacecraft Total (dry) CBE** | 3650 | 43% | 5230 |
| Subsystem Heritage Contingency | 960 | | |
| System Margin | 620 | | |
| Bipropellant | 2200 | | |
| Xenon SEP Propellant | 3500 | | |
| **Total Spacecraft Wet Mass** | | | 10930 |
| PLUS and Launch Adapter | | | 1220 |
| **Total Launch Mass** | | | 12150 |

### 7.3.2 Thermal

The purpose of the thermal subsystem is to maintain the starshade's bus subsystems' temperatures within their allowable flight temperatures.

The driving design issue for the bus thermal system is maintaining the temperature of the propellant tanks within operational limits. Approximately 150 W of heater power would be required to maintain the propellant temperatures during normal operations. However, only 30 W of make-up heater power would be required during the launch, downlink, and safe modes. The thermal design is an actively controlled system using thermistors to sense tank and subsystem temperatures, and strip heaters to add heat when needed. The subsystem also includes multilayer insulation (MLI) blankets, a set of variable conductance heat pipes to redistributed waste heat, and a 3.8 m² radiator to dissipate excess heat.

The starshade occulter's petals were designed to be passively thermally stable and do not need any active heating. See *Section 7.3.4* for details on the starshade payload thermal design and analysis.

### 7.3.3 Propulsion

The starshade bus would possess a hybrid propulsion system using a bipropellant chemical and solar electric propulsion (SEP). The chemical system is responsible for trajectory correction maneuvers (TCMs), station-keeping within the formation flying box, and slewing. The SEP system is used for retargeting, or transiting between target stars.

The chemical propulsion system uses twelve 22 N thrusters with an $I_{SP} \geq 280$ s, providing sufficient redundant control to meet TCM and formation flight requirements.

The SEP thrusters provide 0.52 N of thrust each and have an ISP of 3,000 s. Because the flexible array must be illuminated for the SEP engines to operate and the starshade must be able to translate in any direction in order to meet the observation needs, SEP engines are all gimbaled and placed on opposites sides of the starshade bus. This ensures sufficient pointing freedom is available to the spacecraft to power SEP and SEP to retarget the spacecraft. Two engines are required to achieve nominal thrust for retargeting. To mitigate risk of EP system failures, an additional engine be flown on each side in case of failure for a total of six SEP engines needed to support the HabEx mission.

Assuming a 5,230 kg dry mass, the 2,200 kg of bipropellant and 3,500 kg of xenon will yield at least 100 targeted starshade observations. The propulsion system was sized to meet the mission's 5-year requirement and was also designed to be refuellable. These tanks are refillable using a NASA Cooperative Servicing Valve (CSV) designed specifically for robotic refueling. This feature contributes to the vehicle's serviceability and provides the opportunity to extend the mission.

### 7.3.4 Power

**Table 7.3-2** shows the Power Equipment List (PEL) for the starshade flight system. The power requirements can be split into requirements on a 28 V bus to power bus functions and the bipropellant system, and an 800 V bus to power the electric propulsion

**Table 7.3-2.** Power estimates for starshade bus operational modes carry ample margin. Significant available area exists on the starshade to add solar panels, representing additional margin. *Note: 10% margin on SEP power, 43% margin on bus power.

| Subsystem | Unit | Launch | TCM 1 | Starshade Science | Transit/ Reposition | Down-link | Safe | Cruise |
|---|---|---|---|---|---|---|---|---|
| ACS | W | 35 | 50 | 80 | 50 | 50 | 40 | 10 |
| C&DH | W | 45 | 45 | 45 | 45 | 45 | 45 | 45 |
| Propulsion: Bipropellant | W | 30 | 260 | 30 | 30 | 3 | 30 | 3 |
| *Propulsion: SEP | W | 0 | 0 | 0 | 26,600 | 0 | 0 | 0 |
| Telecom | W | 75 | 75 | 65 | 75 | 75 | 75 | 75 |
| Thermal | W | 180 | 210 | 210 | 180 | 210 | 210 | 210 |
| Power Subsystems | W | 100 | 110 | 110 | 1400 | 110 | 110 | 100 |
| SUBTOTAL | W | 470 | 750 | 540 | 28,380 | 490 | 510 | 440 |
| Contingency and Margin | % | 43% | 43% | 43% | 13%* | 43% | 43% | 43% |
| Contingency Power | W | 200 | 320 | 230 | 3430 | 210 | 210 | 190 |
| Distribution Losses | W | 13 | 20 | 20 | 410 | 14 | 14 | 13 |
| TOTAL | W | 680 | 1090 | 790 | 32,220 | 710 | 730 | 640 |

system during repositioning. The most driving mode for the low voltage bus is a trajectory correction maneuver, which uses just over 1 kW of power. Other modes during operations, including with large off-points from the Sun, are bounded by this number. The electric propulsion system requires approximately 13.3 kW of power per engine during repositioning operations.

These requirements lead to a design using two different solar power arrays. A 28 V string is used to power the bus and an 800 V string is used to power the SEP system. It also contains two different types of arrays.

First, in order to power the bus before the starshade is deployed and during formation flight, a 28 V rigid array on top of the hub. The 1 kW array is sized to meet bus power requirements before starshade deployment, and during formation flight operations for Sun angles between 40–83°.

Second, a flexible array mounted directly onto the starshade inner disk will be used to power the SEP system. The flexible cells are strung together to form a 33.3 kW high-voltage array, which offers power margin to degradation and sun angle while delivering the 26.6 kW required to power two SEP thrusters simultaneously.

Battery sizing is set by the launch-phase power requirements, where it is assumed that the bus will be powered by batteries for up to 3 hours. The starshade requires two 66 Ah lithium ion batteries to avoid the battery depth-of-discharge dipping below 70% during that period of time.

### 7.3.5 Attitude Control System

The attitude control subsystem (ACS) requirements for the starshade are presented in **Table 7.2-1**. In addition to these requirements, the starshade must also carry a laser beacon to support formation flying (see *Section 8.1.7* for details).

The baseline starshade bus is currently designed to provide spin-stabilization using its chemical propulsion system. The starshade does not utilize reaction wheels and therefore no propellant is required to desaturate reaction wheels.

Attitude determination is achieved with star trackers and gyros, including additional gyros and Sun sensors as backup. Once the starshade is within sensor range of the telescope, formation flying control takes over to maintain position relative to the telescope.

### 7.3.6 Communications

The starshade does not directly generate any science products. Its telecommunication requirements are driven by its needs to communicate to the ground for commanding and ranging in X-band (1 kbps downlink requirement) and to communicate with the telescope in S-band for data transfer (100 bps) and ranging. The starshade telecommunication system would therefore be an exact replica of the telescope system, but without the Ka-band capability. It would be fully redundant and carry two universal space transponders (UST), two X-band low-gain antennas, and an S-band patch antenna. This system would easily meet the HabEx starshade downlink and cross-link requirements, with at least 6 dB of margin in all operational cases.

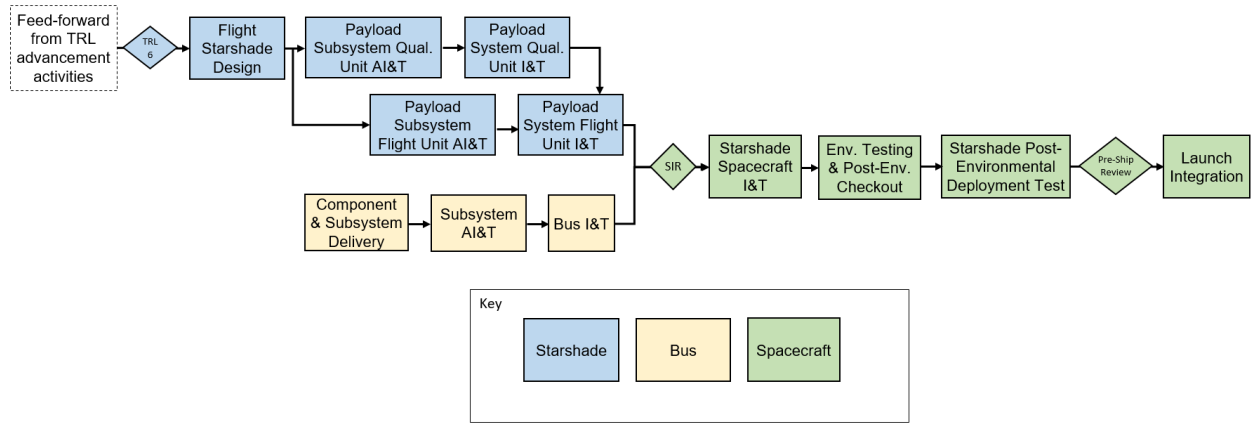

**Figure 7.4-1.** Starshade spacecraft integration and test flow. Note: SIR is System Integration Review.

### 7.3.7    Command & Data Handling

To help reduce cost, the starshade command and data handling (CDH) subsystem would be an exact replica of the telescope CDH, which is described in *Section 6.10.5*, without the expanded memory card. The starshade flight software would be somewhat simpler than that of the telescope since it only has a single deployment and lacks science instruments. However, the formation flying requirements, the spacecraft cross-link communication, and the ACS approach will all require some customization from JPL core software products. Nonetheless, the development risk associated with this type of software is expected to be low.

### 7.4    Starshade Integration and Test Plan

This section primarily discusses the Phase C–D integration and test flow for the baseline starshade spacecraft, including ground support equipment, testing, and facility considerations. An overview of the sequential flow is shown in **Figure 7.4-1**. Telescope integration and test is described in *Section 6.11*. **Table 7.4-1** summarizes facilities required for starshade integration and test (I&T) and the necessary ground support equipment. Separately from this discussion, advancement of enabling technologies to TRL 5 is discussed in *Section 11.2*, detailed technology roadmaps to TRL 6 are included in *Appendix E*, and the baseline schedule is shown in *Chapter 9*.

### 7.4.1    Facilities, Ground Support Equipment, and Testbeds

#### 7.4.1.1    Facilities

Almost all integration and test activities for the starshade can take place in conventional, existing facilities. However, in order to fully deploy the starshade including petal unfurling and disk deployment, a large, clean facility must be used, and in 2019 while there are buildings large enough, none are currently clean enough. To address this, a large facility, such as a hangar, may need to be retrofitted to support this test. This includes installing required handling equipment, gravity compensation equipment, power and control equipment, and taking steps to clean the facility to Class 10,000. This facility will be used a total of three times: to fully deploy a qualification starshade unit, to fully deploy the flight starshade payload prior to integration with the bus, and to fully deploy the fully integrated spacecraft after environmental testing.

#### 7.4.1.2    Mechanical Ground Support Equipment

The spacecraft and bus will require handling equipment and gantries in order to access all areas during integration. Prior to integration of the starshade to the bus, a dynamic simulator representative of the bus's mass and capable of reacting loads in a similar manner will be required to accurately assess deployment dynamics. During deployment testing at all stages of integration, custom gravity compensation equipment will be required to offset the weight of the starshade. This equipment will be reconfigured during deployment testing to test both the petal unfurling and disk deployment. The technology for this task is derived from solar panel and antenna deployment testing, and will be matured for this application during TRL advancement of the starshade itself.

Table 7.4-1. Starshade required facilities and equipment. *Only facility requiring special preparation efforts.

| Activity | Test Facility | Special Test Equipment |
|----------|---------------|------------------------|
| Petal assembly and shape verification before/after environmental tests | Conventional I&T facility | Optical bench with in-situ shape metrology and discrete angle scatterometry |
| Petal temperature and shape versus temperature model validation | Existing TVAC facility | Laser based shape metrology through chamber window |
| Inner disk assembly and deploy tests | Existing facility at deployable antenna vendor | Custom gravity compensation fixture, conventional laser trackers, spot solar simulator and SEP power supply simulator |
| Starshade payload system integration & partial deployment testing | Conventional I&T facility & environmental test facilities | Custom gravity compensation fixture, conventional photogrammetry, laser trackers |
| Starshade payload full deploy tests (payload level and post-environmental spacecraft level) | Retrofitted large facility (i.e., hangar), cleaned to Class 10,000* | Custom gravity compensation fixtures – 2 stage deployment (petals unfurl & disk deploys) |
| Bus integration and test | Conventional I&T facility | Handling fixtures, spacecraft and ground station simulators, power supply equipment |

### 7.4.1.3   Electrical Support Equipment

Because the flexible solar array is integrated into the starshade, special equipment will be required to test both pieces independently prior to integration. For the array itself, lighting and a SEP power electronics simulator will be required. For the bus, power supplies capable of delivering the large amounts of high voltage power will be required to assess functionality of the SEP power electronics. This high voltage, high power supply will require significant safety considerations in its design and operation in order to properly protect test personnel.

### 7.4.1.4   Optical Test Equipment

Petals will be assembled on an optical bench with in-situ shape metrology and discrete angle scatterometry. Shape metrology will be actively used to install optical edges, and to verify shape stability after a suite of environmental tests.

### 7.4.1.5   System Testbeds

Two high fidelity testbeds will be built to test the avionics. The first is a flight software testbed, using C&DH hardware identical to the flight units. This testbed will be used during integration and test in order to validate procedures before execution on flight hardware, and to aid in trouble-shooting and regression testing. These capabilities extend beyond launch, giving ground operators a key ability to validate flight software updates and commands prior to uplink to the spacecraft, and providing a safe environment to test responses to in-flight anomalies. The second testbed is a mission system simulator. This includes environment models, simulate input data for spacecraft sensors, and simulation of telecom: both uplink/downlink with ground stations, and cross-link with the telescope. This simulator will be especially important for verification and validation of formation flying requirements, including response to simulated sensor input, cross-link communication, and guidance, navigation, and control (GNC) algorithms.

In addition to these mechanical testbeds, once complete with testing, the qualification starshade unit will be available as a mechanical testbed for development and trouble-shooting activities.

### 7.4.2    Payload I&T

After the TRL advancing activities described in *Chapter 11* and *Appendix E* are complete, the starshade technology will be at TRL 6. At this point, the specific flight design for HabEx can be completed. A full qualification unit will be built of the starshade, and production of the flight unit will begin in parallel.

#### 7.4.2.1    Payload Subsystem Qualification Unit Assembly, Integration, and Test

A set of dedicated qualification petals and an inner disk will be assembled to flight specifications in order to validate the design for flight. Qualification testing is comprehensive, and may degrade hardware such that it is no longer suitable for flight. The petals are assembled on the optical bench described in *Section 7.4.1.4*. The units will be subjected individually to a series of tests, including deploy cycles, thermal cycles, vibration and long-term stowage. Petals will then be returned to the petal assembly facility for shape verification. In order to validate petal shape versus temperature models, a series of tests involving a pair of petals and representative truss attachments will be conducted in an existing thermal vacuum (TVAC) facility with a custom shape metrology system. In parallel to qualification of the flight petals, the inner disk qualification unit will be integrated, tested, and deployed in an existing deployable antenna vendor's facility with a custom gravity compensation fixture and conventional laser trackers.

#### 7.4.2.2    Payload System Qualification Unit Integration, and Test

Once qualification testing is complete on the disk and petals separately, they will be integrated together. The starshade payload system can be integrated in a conventional I&T facility, and may be partially deployed in this configuration. Once initial checkouts are completed, the fully integrated qualification unit will then be transported to the large deployment facility. A full deployment will be completed in two stages with a change to the gravity compensation configuration in between petal unfurling and disk deployment. The petal position is verified with conventional laser trackers and inter-petal clearances are verified with photogrammetry. Because the petals are rigid bodies once deployed, only their position, and not their shape, must be verified in the full system configuration. This activity will serve to test not only the flight design, but the gravity compensation fixtures and facility described in *Section 7.4.1.1*. These full deployment tests will be completed again after a suite of environmental tests on the assembled qualification starshade.

#### 7.4.2.3    Payload Subsystem Flight Unit Assembly, Integration, and Test

As qualification testing is ongoing, production will begin on the flight articles. These are subject to acceptance testing, which verifies that they are made to specification, but does not subject the units to excessive stress prior to flight. These will be assembled, integrated, and tested in a similar manner to the qualification units, while taking advantage of lessons learned in the process.

#### 7.4.2.4    Payload System Flight Unit Integration and Test

The flight petals and disk will be integrated in a conventional I&T facility, including both mechanical integration of the starshade, and electrical integration of the flexible solar array for SEP power. This array will be verified with spot solar simulation and a SEP power supply simulator during this phase. Once integrated, it will be deployed at the same large facility used for the qualification testing, before delivery to the spacecraft I&T flow and mating with the bus.

### 7.4.3    Bus I&T

Bus integration and test will run in parallel to payload qualification and flight unit production and test.

#### 7.4.3.1    Subsystem Assembly, Integration, and Test

As shown in **Figure 7.3-1**, the system block diagram, most of the bus subsystem hardware is high heritage and based on existing, flight-proven designs. Most subsystems will be able to follow a build-to-print philosophy of reusing these designs, allowing more resources to be allocated to planning and completing difficult system-level testing. Once components and box-level deliveries are completed and accepted, subsystems will be completed as necessary. Some of this work will be completed in parallel to the beginning of bus integration.

### 7.4.3.2 Bus Integration and Test

As subsystem deliveries are completed, they will be integrated in the bus in a flow designed to minimize the number of disruptions and regression testing needed as more hardware is installed. Prior to full integration, both the subsystem and spacecraft will undergo safe-to-mate and safe-to-power procedures. After mechanical installation, the bus will undergo successively more and more complete functional testing until all components have been installed. Once successfully integrated, the spacecraft will undergo full system-level functional testing, to ensure all modes operate correctly, and sequence testing, to validate time critical sequences and autonomous functioning. The equipment described in *Section 7.4.1.5* is used to simulate the environment and stimulate the spacecraft during these tests. Especially critical in this testing will be the verification and validation of the flight software associated with formation flying, including sensing, crosslink, and control. This software will be tested at multiple levels and on testbeds prior to this point, and the bus-level sequence tests will confirm proper functioning on the flight hardware.

### 7.4.4 System I&T

Once the bus and payload are completed, a System Integration Review will assess hardware readiness to integrate. A series of fit checks and safe-to-mate procedures will be completed, followed by integration of the complete spacecraft. The interface between the starshade and bus is primarily mechanical; electrical connections include the signal line to deploy the starshade and the power lines from the flexible solar array. Integrated functional and sequence testing will be performed using the testbed electronic ground support system (EGSE) to verify that all spacecraft subsystems and modes operate correctly. This includes partially deploying the starshade in a conventional I&T facility in order to validate commands from the bus to release the PLUS, and tests of the flexible arrays on the starshade while connected to the high voltage power system on the bus.

### 7.4.4.1 System-Level Environmental Testing

The spacecraft will then be prepared for environmental testing. This includes 3-axis random vibration, acoustic, electromagnetic interference (EMI), and TVAC testing. The full spacecraft will only undergo TVAC testing in a stowed configuration. Spacecraft health will be monitored during each testing, and following each the test engineers will perform detailed inspection and functional testing.

### 7.4.4.2 Post-Environmental Deployment Test

After environmental testing, the spacecraft will be sent to the hangar facility for final deployment testing. This process will be the final verification that the PLUS, starshade, and interface to the bus will not be damaged by flight environments.

### 7.4.4.3 Launch Integration

After successful post-environment deployment testing, the spacecraft will be cleaned and prepared for shipment to the launch site. A pre-ship review will be held to review comprehensive readiness and project status. Pending this gate, delivery will be completed, followed by post-ship inspection, functional testing, and checkout. The spacecraft will then be ready for integration with the launch vehicle, and coordinated operations with the launch service provider. After launch and initial deployment and checkout, the mission transfers to its operations phase; details of the design reference mission (DRM) are included in *Chapter 8*.

# 8 BASELINE MISSION CONCEPT

The baseline HabEx Observatory architecture is for an observatory that is comprised of telescope and starshade flight systems. While the telescope and starshade are unique, with different requirements and capabilities, they are co-managed, designed, and operated together to meet the HabEx science objectives specified in **Table 5.1-2**.

This chapter overviews aspects of the mission that are shared by the telescope and starshade: their concept of operations, design reference mission, ground systems, servicing concept, and management plan.

## 8.1 Baseline Observatory and Concept of Operations

The HabEx mission concept has three distinct phases—launch, cruise, and science operations—which are discussed in this section. HabEx's high-level architecture is summarized in **Figure 8.1-1**. At this high-level, the telescope and starshade are launched and transfer to Sun-Earth L2 independently. Telescope and starshade flight systems fly in formation for starshade observations only. Telescope and starshade flight systems operate independently during all other operations.

### 8.1.1 Launch

The telescope and starshade are launched independently, with the telescope designed to be launched on an SLS Block 1B with an 8.4 m diameter fairing and the starshade designed to be launched on a Falcon Heavy. Alternatives for both launch vehicles are in development and may exist in time for the HabEx mission so these selections are notional and for the purposes of this study only.

The HabEx baseline concept has selected an operations orbit at L2. Coronagraphy requires a very low disturbance environment and L2 offers a thermally stable orbit in proximity to Earth. Heliocentric drift away orbits are also possible but the observatory could not be serviced in the future and the starshade would need to launch at close to the same time as the telescope. L2's stable environment offers programmatic flexibility should a delayed starshade launch be seen as advantageous for funding or development reasons.

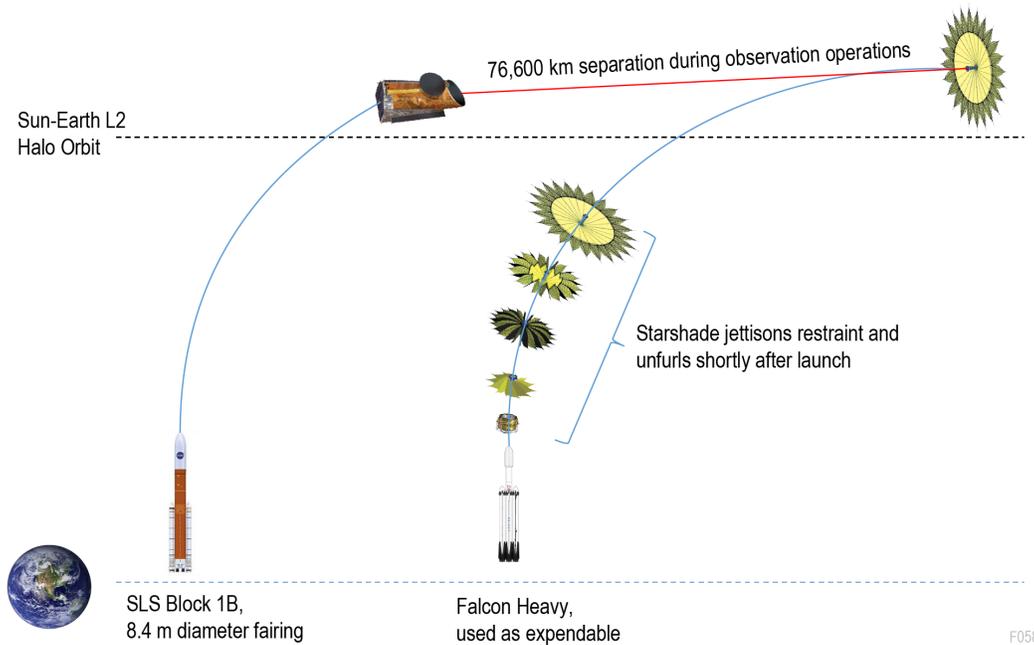

**Figure 8.1-1.** The HabEx baseline mission requires two launches of the independent telescope and starshade flight systems, permitting phasing in their development as discussed in *Chapter 9*. Following their cruise to L2, they operate independently except during formation flight used in starshade observations.

To reach HabEx's halo orbit at L2, a characteristic energy (C3) of -0.6 km$^2$/s$^2$ is required for HabEx's direct transfer. The maximum launch mass of the telescope flight system, including margin and contingency, is 18,550 kg including contingency and margin. Similarly, the maximum launch mass for the starshade flight system with margin and contingency is 12,150 kg. The telescope and starshade launch vehicles performance results in 20,000 kg and 3,000 kg of additional launch margin respectively (KSC LV Performance, assuming expendable Falcon Heavy launch). The high-level mass budgets of the HabEx two flight systems are described in **Tables 6.10-1** and **7.3-1**.

### 8.1.2   Cruise

Although the telescope and starshade launch separately, their cruise phases are similar. Three days after separation, both the telescope and the starshade perform their first trajectory correction maneuver (TCM 1), which begins their transfer to Earth-Sun L2. TCM-1's expected ΔV is approximately 50 m/s, depending on launch error and when it occurs following launch.

This cruise phase lasts 6 to 8 months during which starshade deployment and both spacecraft commissionings occur. Two-way Doppler and ranging with the Deep Space Network (DSN) using X-band for ephemeris reconstruction and trajectory updates will confirm both trajectories. Delta Differential One-way Ranging (DDOR) may also be used closer to the halo orbit insertion (HOI) to refine knowledge of the trajectory. **Table 8.1-1** provides the key orbit parameters for the mission.

Table 8.1-1. HabEx key L2 halo orbit parameters.

| Parameter | Value |
|---|---|
| Target/Destination | Earth-Sun L2 |
| Trajectory type | Direct transfer |
| Cruise duration | 3–5 months |
| Orbit diameter | ~780,000 km |
| Z amplitude | ~40,000 km |
| L2 Orbit period | 175 days |
| Eclipse time | 0 minutes |
| L2 Orbits/year | 2 |
| Max S/C-Sun distance | 1.012 AU |
| Max S/C-Earth range | 1,800,000 km |

### 8.1.3   Checkout and Commissioning

During cruise, both spacecraft undergo a range of commissioning and checkout activities. The telescope will open its aperture door 30 days after TCM-1. The telescope spacecraft commissions its subsystems and performs thruster calibrations for both the chemical propulsion system and the microthruster system, which are described in *Section 6.10.3*. In parallel, the starshade flight system commissions its own subsystems and unfurls its petals as soon as safely possible after TCM-1. Following unfurling, the starshade flight system will then jettison the Petal Launch Restraint & Unfurling System (PLUS) and the solar electric propulsion (SEP) system will be brought online for calibration. At this point, the precise mass properties of the deployed starshade system would be established.

### 8.1.4   Orbit Insertion

At the Launch+90 day, both the telescope and starshade perform a second TCM (TCM-2) of about 5 m/s ΔV, which prepares them for the HOI. Owing to the low-energy halo orbit about L2, there is a window starting at launch +180 days lasting to about launch +240 days where HOI can occur. The HOI burn is about 5 m/s ΔV for both telescope and starshade.

### 8.1.5   The L2 Orbit

HabEx's Sun-Earth L2 orbit is approximately 780,000 km in diameter. The telescope orbits the L2 point in 175 days so slightly more than 2 orbits are completed in a year. Since the telescope spacecraft is orbiting a point in space with no gravitational pull to define the orbit, the orbit must be maintained propulsively. Five trajectory correction maneuvers—requiring a total of about 5 m/s of ΔV—are executed every orbit to keep the telescope moving around the L2 point.

The L2 environment is very benign. The orbit is never in eclipse so the thermal variations are related to changes solar radiative emissions; as such these changes are small and slowly varying in comparison to an eclipsing Earth-orbiting environment. Solar pressure is the primary disturbance acting on the telescope, imparting a torque on the order of several micronewton-meters depending on the spacecraft's orientation toward the sun. Between the starshade and the telescope there is a small (on the order of ~ 1 μg) gravity gradient disturbance that the starshade must periodically correct. Overall, the environment is close to ideal for the operation of a large space telescope especially in conjunction with starshade formation flying.

### 8.1.6   Science Observational Modes

As discussed in *Chapter 6*, the telescope flight system is capable of fast slews when needed, but at the cost of greater fuel consumption. The HabEx telescope's high rigidity and thermal stability is expected to greatly reduce settling times, enabling ~90% telescope utilization. Together, these design characteristics make for highly capable and efficient observational platform.

During the 5-year primary mission, there are four types of science observation modes: coronagraph, starshade, parallel, and Guest Observer (GO). The coronagraph mode utilizes the coronagraph for exoplanet detection and orbital characterization. Starshade observations are used for imaging and spectral characterizations of planets, planetary systems and disks. Since each of the instruments has its own field of view (FOV) on the sky, parallel observations, utilizing the HabEx Workhorse Camera (HWC), UV Spectrograph (UVS), or both, can be conducted during the starshade or coronagraph observations. Given the long integration times in the starshade and coronagraph modes, deep field observations with the UVS and WHC are some of the most complementary uses in the parallel mode. Finally, fully half of the HabEx observational time is given to guest observer defined science and any of the four instruments can be used in the guest observer mode.

**Coronagraph Observations** There are two factors that contribute to overhead time associated with the Coronagraph (HCG) observations: system thermal and mechanical settling time following a slew, and "digging the dark hole"—the time required to set the coronagraph's deformable mirrors prior to an observation. The telescope structure damps out thruster firing vibrations to levels suitable for high contrast coronagraphy in less than 5 min (*Section 6.9*) and the thermal disturbance introduced by slewing never causes enough wavefront variation during the observation to prevent the coronagraph from reaching the required contrast level (with the exception of slews that move the base of the spacecraft in or out of shadow; see *Section 6.9* for a detailed analysis). This system-level thermal insensitivity is largely due to the use of a mirror laser truss system (MET), the large thermal inertia of the primary mirror, and a telescope thermal control system.

For the purpose of this study, HabEx digs the dark hole using reference differential imaging (RDI; *Section 6.9*). In RDI the telescope first align with a bright reference star near the target star and the HCG sets its deformable mirrors, maximizing the light suppression between the inner and outer working angle (IWA and OWA). The telescope then slews a small angle to the target star for the observation. The time on the reference star and the intermediate slew represent observational overhead; for this study that overhead was assumed to be 8 hrs per observation.

**Starshade Instrument Observations** During starshade observations, both the telescope and starshade rely on propulsion for station keeping. The telescope must maintain orbit around the L2 point resulting in an annual station keeping budget of 10 m/s ΔV per year and about five maintenance maneuvers per orbit.

During Starshade Instrument (SSI) observations, the starshade requires an average of 1.5 m/s ΔV per day to maintain line-of-sight formation flight. This ΔV requirement is met by bipropellant thrusting about every 600 s while observing. Before the ~1 s thruster firings, the starshade will signal to the telescope to pause observation. The time for the thruster plume to depart the maximum SSI OWA is less than 1 s. During these starshade thrustings, the total impulse imparted is ~1 N s and requires 10 s or less for settling time. During this time, the sensors in the SSI are continuously read out at a high rate to limit charge build up, but the data is dumped.

Between SSI observations, the starshade will spend most of its time repositioning for new observations, where ~130 m/s ΔV is required on average for each repositioning. Additionally, the starshade must offset solar pressure and maintain the halo orbit using about 0.01–0.2 m/s ΔV each day depending on the starshade's angle toward the Sun and the amount of propellant remaining on board. Repositionings are handled with high-$I_{SP}$ Hall-effect electric propulsion thrusters, utilizing Xenon for the propellant.

**Parallel Observations** Since none of the HabEx instruments share any portion of their FOVs, all have the potential to operate in parallel. While it is unlikely that either the HCG or SSI will be operated while the general astrophysics science is controlling the pointing of the telescope, there are general astrophysics science opportunities when the coronagraph or SSI is locked onto a target star. In particular, the long observation times required for exoplanet science can be used to generate deep field images on both the HWC and the UVS. Other general astrophysics science opportunities may exist if desirable targets are opportunistically within the HWC or UVS fields-of-view. During exoplanet science, the telescope has some ability to rotate about the boresight, particularly during starshade observations. This results in a partial annulus of potentially observable sky for the HWC. The HabEx exoplanet target stars are known (see *Appendix D* for the list of stars) so potential parallel observations can be identified and proposed as parallel observation science far in advance of the actual observation opportunity.

**GO Observations** During the primary mission, GO observations will be scheduled between starshade exoplanet direct imaging observations and in negotiation with HCG observations.

**GO Targets of Opportunity** Owing to HabEx's stable performance, needed to meet HCG requirements, settling time prior to GO observations is expected to be negligible. To conserve fuel, the telescope normally slews at about 0.15° per second. For targets of opportunity, such as gravitational wave event-related or gamma-ray burst observations, it can slew 180° in under 5 min if an observation requires it. These rapid slews burn significantly more fuel than the standard slew so the target must require the fast reposition and the science must merit the greater allocation of the limited resource. Additional fuel has been included in the design in anticipation of the periodic need to rapidly repoint to targets of opportunity.

**Field of Regard** The telescope allows a 40° minimum angle between the observation target, telescope and the Sun. This limit is set by the telescope scarf angle; observations closer to the sun-telescope line will not be possible since sunlight will be able to enter the telescope barrel. In the case of the starshade the upper limit to the field of regard is 83°. Beyond this limit, the telescope-facing surface of the starshade will be illuminated by the sun and light will reflect off the starshade and into the telescope. All other observations have a field of regard upper limit of 165°, set by solar array configuration and sizing, though the observatory can operate at non-driving power modes up to 180°. Dynamic keepouts associated with solar system bodies and how HabEx accounts for them are described in *Section 8.2*.

### 8.1.7 Formation Flying

Formation flying is defined as two or more spacecraft autonomously controlling relative position or attitude based on inter-spacecraft measurements. HabEx requires formation flight to align the starshade and telescope for science observations and to repoint this synthesized observatory at a new target star. Repointing is done by translating the starshade relative to the telescope, which is referred to as "retargeting."

The overall concept of operations (conops) for formation flying and the accompanying translational control requirements are shown in **Figure 8.1-1**. This conops leverages the extensive studies and engineering analyses that were performed for Exo-S (Seager et al. 2015; Scharf et al. 2004) and Probe-class Rendezvous concept. Each of the operational modes—initialization, retargeting, acquisition and science— are discussed subsequently.

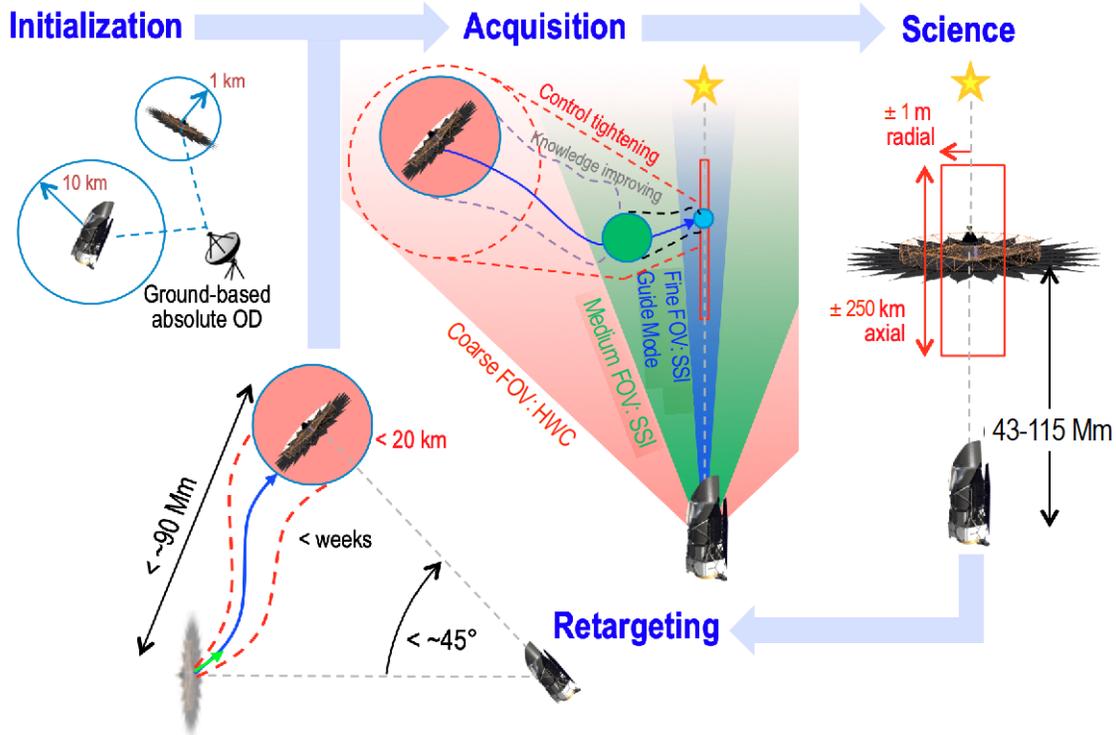

**Figure 8.1-1.** Starshade concept of operations for formation flying, detailed in *Section 8.1.7*. During (*re*)*targeting*, the starshade flight system will transit to the approximate LOS between the telescope and target star. During *acquisition*, multiple detection and control modes are employed to reduce lateral displacement to within ±1 m from LOS to enable *science mode* observations.

The formation flying architecture for HabEx is designed to conform to the requirements outlined in **Table 8.1-2**. The architecture is shown in **Figure 8.1-2** has the starshade maneuvering relative to the telescope. This arrangement allows the telescope to perform independent science during the days to weeks required for retargeting. Additionally, this architectural choice results in the so-called Leader/Follower formation control architecture that is commonly used in rendezvous and docking in low Earth orbit (LEO) and which makes control design, and stability and performance analyses straightforward (Scharf et al. 2004).

The driving requirement for formation flying is to align the starshade to within 1 m radially, that is, laterally, of the telescope-star line at spacecraft separations of 42,600–114,900 km for science mode. While aligning two spacecraft to 1 m at a separation of up to 114,900 km appears daunting, this formation flying problem is more tractable than might be expected for the following two reasons. First, even though inter-spacecraft distances are large, the relative dynamics remain benign: the gravity gradient at the Sun-Earth/Moon L2 point at the maximum planned separation is less than approximately $10^{-4}$ m/s² (Sirbu et al. 2010) and differential solar radiation pressure is orders of magnitude smaller. The HabEx gravity gradient is equivalent to that experienced at tens-of-meter separations in low Earth orbit, which is similar to the berthing distance used at the International Space Station (ISS). Although the telescope and starshade are far apart, they are not "flying apart."

**Table 8.1-2.** HabEx formation flying requirements and expected performance. See Table 5.4-10 for original requirements and context.

| Parameter | Requirement | Expected Performance | Margin | Source |
|---|---|---|---|---|
| LOS Separation Distance Accuracy | ≤ ±250 km | ≤ ±250 km | Met by design | MTM |
| Distance Sensing | ≤ ±25 km | ≤ ±25 km | Met by design | MTM |
| Lateral Displacement from LOS | ≤ ±1 m | < ±1 m | Met by design | Error Budget |
| Lateral Displacement Sensing Accuracy | ≤ ±0.3 m | <0.15 m | 50% | Error Budget |

This observation regarding thruster-firing intervals leads to the second reason that formation flying is tractable for HabEx, namely, controlling formation flying spacecraft to the submeter level is commonly done for rendezvous and docking with the ISS. A typical radial control requirement for the terminal docking phase is 10 cm (Kelly and Cryan 2010). For example, the European Space Agency's Automated Transfer Vehicle, which has a mass around 20,000 kg, controls to 10 cm. Formation control to 1 m is not only tractable but also commonly demonstrated in flight.

The principal challenge for HabEx formation flying is *sensing* the lateral offset of the starshade from the telescope-star line to a fraction of 1 m at tens of megameters, while the starshade is obstructing the star. A former HabEx technology gap (see **Table 11.1-1**) formalized this challenge as sensing the lateral offset of the starshade to 0.3 m. This gap has recently advanced to TRL 5 and is now ready for a mission start.

Note that inter-spacecraft range measurements with ~1 km precision will be made by a radio frequency (RF) inter-spacecraft link that also provides low-bandwidth communication for coordination. This RF link is not considered a technology challenge.

### 8.1.7.1 The Principal Formation Challenge: Fine Lateral Sensing

There has been extensive work on solving this lateral sensing challenge, which is referred to as fine lateral sensing. A short survey and further references can be found in (Scharf et al. 2016). The general approach is to utilize the telescope and the light of the target star that "leaks" around the starshade outside the wavelength of the science band, where the attenuation of the target star is only on the order of $10^{-3}$.

The fine lateral sensor uses the SSI in what is referred to as "guide" mode, producing pupil-plane images. The sensor's pupil-plane images are matched using least squares to a library of pre-generated images of the "shadow structure" of the starshade. Image matching is done on the telescope and the resulting lateral offset is sent to the starshade over the inter-spacecraft link (ISL). NASA's Starshade Technology Project (S5), managed by the Exoplanet Exploration Program Office, has demonstrated this sensing approach to TRL 5 earlier this year.

Current analyses show that performance of 15 cm $3\sigma$ is possible with ~5 s exposures of a mag 8 star when sensing in UV and operating within 1 m of alignment (Bottom et al. 2017). Sensing in the ultraviolet (UV) is done when science is being done in the infrared (IR). Conversely, when doing science in the visible or UV, sensing is done in the IR. Expected target stars have lower flux at UV wavelengths and instrument losses are greater. Hence, a 5 s exposure is considered worst case; tenths of a second is more typical. Even so, since thruster firings are needed only on the order of every hundreds of seconds, many formation measurements can be made, thereby improving relative velocity knowledge for efficient formation control.

Even in areas of low shadow structure (e.g., only smooth gradients), image matching can still be performed to ~25 cm $3\sigma$. As a result, the pupil-plane image-matching fine lateral sensor can be used out to lateral offsets equal to the radius of the starshade, 26 m.

Since the fine lateral sensor functions only when the starshade is within ~26 m of alignment, the acquisition mode is needed to move the starshade from the end of retargeting and initialization modes to within 1 m of alignment for science mode.

### 8.1.7.2 Initialization Mode

When the starshade first rendezvouses with the telescope after launch, it is being operated from the ground. Similarly, if during regular operations, either the telescope or the starshade enters safe mode and relative position knowledge is lost, the ground recovers the spacecraft. In both cases, the ground tracks both spacecraft and determines a trajectory for the starshade to align between the telescope and the target star at the desired range. The trajectory is uploaded to, and executed by, the starshade.

The position of the telescope is determined via standard tracking methods to better than 10 km 3σ (Truong et al. 2003), which can require ~30 min of tracking a day for 1–2 weeks; if DDOR is used, telescope position can be established in several days.

The position of the starshade can also be determined from the ground with much greater accuracy by utilizing the laser beacon it carries for use in acquisition mode. A visible, 1 W laser beacon with a 2.5° angular spread ensures the starshade appears as at least a 15th magnitude star at Earth while the 2.5° FOV subtends 5 Earth radii from Sun-Earth L2. Using the DSN radar to range to the starshade and astrometry with even just meter-class ground-based telescopes, the position of the starshade can be determined to 1 km 3σ in under an hour (Altmann et al. 2014). The relative position uncertainty is then the root sum-square of 10 km and 1 km, which is effectively 10 km. Back in space on the HabEx telescope, this level of uncertainty is sufficient to guarantee that the HWC will see the laser beacon of the starshade when it is pointed at the target star. The HWC FOV is 3 arcmin square (±18 km) at the minimum range of 42,000 km.

### 8.1.7.3   Retargeting Mode

At the end of the science mode, the lateral position and velocity of the starshade are known to better than 5 cm and 1 mm/s, and the axial position and velocity to better than 15 m and 4 cm/s, respectively (Scharf et al. 2016). This relative state knowledge is the initial condition for the retargeting trajectory.

The retargeting trajectory is planned on the ground and uploaded to the starshade. Using SEP, the starshade executes the planned trajectory.

Retargeting mode concludes with the starshade decelerating into its next observing position. After completing its deceleration, the starshade and its laser beacon should be within

~10 kilometers of lateral alignment, ready for acquisition by the FOV of the HWC.

### 8.1.7.4   Acquisition Mode

At the end of initialization or retargeting, the starshade laser beacon is turned down to tens of milliwatts to match the expected flux of the target star, and the starshade points the laser beacon at the telescope. The HWC images the laser beacon and the unobstructed target star on the same detector, producing bearing measurements with a resolution of ~$10^{-7}$ radians (a 4.4 m offset at 42,600 km).

The lateral position and velocity of the starshade will be known to better than 5 m and 1 cm/s, with less than an hour of measurements. This level of knowledge is enough to adjust the starshade velocity with chemical thrusters to achieve target final alignment. Since the bi-propellant thrusters are placed at various angles and can generate a thrust vector in any directions, the starshade does not need to reorient for this adjustment. A change in velocity of 0.5 m/s is enough to move the starshade to alignment in 5 hrs or less.

Any errors that accumulate (e.g., inexact knowledge of the gravity gradient or solar pressure) will be corrected using model predictive control throughout the acquisition mode. This control approach is illustrated in **Figure 8.1-2** (Garcia et al. 1989). After the thruster firing, HWC measurements continue, and relative position and velocity knowledge improve. If the starshade drifts sufficiently far off the planned trajectory, additional, much smaller adjustments are applied.

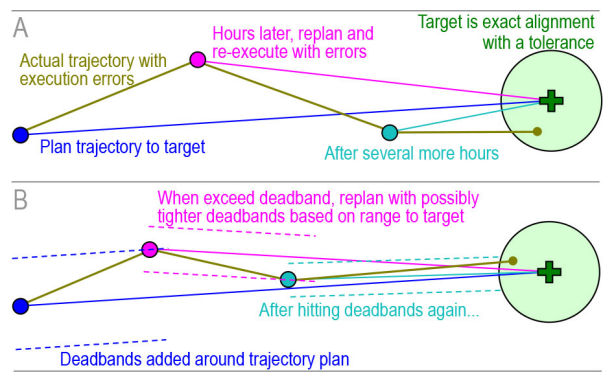

**Figure 8.1-2.** Example model predictive control approaches for controlling the starshade during acquisition.

The subsequent velocity changes are more efficient because both the relative state knowledge and the estimated differential acceleration between the starshade and telescope have improved.

When the starshade passes to within 2 km of alignment, the telescope slews to put SSI on the target star. The SSI half-FOV is 1.2 km at 42,600 km. The HWC's worst resolution is sufficient to steer the starshade to within 1 km of alignment—easily within the SSI's FOV. The SSI is in visible imaging mode, and like the HWC, measures the bearing between the starshade laser beacon and the target star on the same detector. The resolution of the SSI is 3.3 m at 42,600 km and 8.6 m at 114,900 km.

SSI measurements and model predictive control continue as the starshade approaches to within 26 m of alignment. At this point, the laser beacon is deactivated and the SSI switches to guide mode, becoming the fine lateral sensor. While the SSI resolution prior to switching to guide mode is ~14 m worst-case, previous analyses for Exo-S indicate that the position knowledge is generally 5 times better with an estimator. Estimator knowledge of ~3 m is sufficient to steer to within 26 m of alignment.

Once the fine lateral sensor acquires the starshade (indicated by a pupil-plane image match with low residual), model predictive control continues to steer the starshade until, finally, it is within 1 m of alignment, at which point science mode begins.

Autonomous logic is needed on the starshade to transition between sensors, switching estimators hand-off and coordinating with the telescope. Previous missions have demonstrated complex, sensor-based mode logic, such as Mars Science Laboratory and Orbital Express.

### 8.1.7.5   Science Mode

With the starshade within one meter of alignment, science observations can be performed. During science mode, the SSI is providing measurements to 30 cm (3σ). These measurements are used to estimate the relative position and velocity of the starshade and the differential acceleration between the starshade and telescope. When the starshade approaches the 1 m alignment limit, thrusters fire to correct the alignment. The starshade moves back into alignment until the gravity gradient and other environmental factors eventually move the starshade back to the one-meter limit, causing the cycle to repeat. This control method creates a "one-side" dead band where the starshade moves in a ballistic trajectory within the 1-meter alignment limit. Once within the 1 m alignment circle, the starshade fires thrusters at one edge of the circle, imparting just enough velocity to reach the other edge while opposing gravitational and other environmental forces. These environmental forces return the starshade to its approximate starting position where the cycle repeats. Each time the starshade fires its chemical thrusters to formation-keep the observation must be suspended to avoid corruption by the brightness of the thruster plumes. Therefore, thruster firing are coordinated between the telescope and starshade so that the observation data can be protected.

**Figure 8.1-3** shows an example of optimal deadbanding within the formation control requirement circle for the science mode. As an example, consider the starshade coasting into the final 1 m of alignment at point 1. As the starshade coasts across the circle, the estimate of the differential acceleration is continually updated. At point 2, the formation control algorithm uses its estimate at that time, $\mathbf{a}_1$, to fire the thrusters, targeting a drift to the "bottom" of the circle as indicated by the vector $\mathbf{a}_1$. The true differential

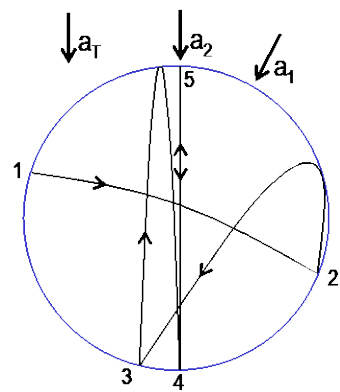

**Figure 8.1-3.** Example of optimal deadbanding for Science Mode.

acceleration in this example is the vector $\mathbf{a}_T$. During the ensuing coast, the differential acceleration estimate improves to $\mathbf{a}_2$.

When the boundary of the control requirement is encountered again at point 3, the thrusters are fired to coast to the bottom of the circle as indicated by $\mathbf{a}_2$. For the purposes of this example, assume $\mathbf{a}_2$ is close to $\mathbf{a}_T$. Thereafter, thrusters are fired to traverse the diameter of the circle aligned with $\mathbf{a}_T$ from point 4 to point 5 and back to point 4. The departing velocity at the bottom of the circle is sized to bring the starshade to zero relative velocity at the "top" of the circle and then "fall" back down again.

To execute the thruster firings, a thrust allocator uses the configuration of thrusters and the current estimate of the spacecraft attitude to compute thruster firing durations that give the desired force impulse. Thrust allocators function with a spinning starshade as well. An example thruster configuration is shown in **Figure 8.1-4**. Thrust allocators also handle motion of the starshade center of mass as propellant is expended.

### 8.1.7.6   Summary

While formation flying to 1 m at separations of up to ~14 Earth diameters initially appears daunting, the relative dynamics are similar to tens-of-meters separation in LEO, and the control performance has been previously demonstrated by much larger spacecraft in Earth orbit docking with the ISS. The principal challenge is sensing the lateral offset of a starshade from a target star to tens-of-centimeters while the target star itself is obscured by the starshade.

Several lateral sensing approaches exist, and NASA's Starshade Technology Project has recently matured one approach to TRL 5. The pupil-plane image-matching approach being matured does not require a laser beacon, has performance better than required with just seconds of exposure, functions even with secondary obscuration, and provides measurements from the edge of the starshade to the center.

### 8.2   Design Reference Mission

To show that the HabEx baseline mission design can successfully meet its science objectives within the primary mission, and to understand the relationship between the science mission, engineering design, and environment, a design reference mission (DRM) was developed. For the baseline GO science objectives, the DRM uses the on-line HabEx exposure time calculators to estimate required observation times. **Table 8.2-1** presents the estimated time required to fulfill the science objectives described in *Chapter 4*. Though rough, this table demonstrates that all the baseline GO science objectives could be accomplished in <0.5 yr of observatory time, i.e., significantly less than the 2.5 yr allocated for GO programs.

Verification of exoplanet science objectives and the time required to meet those objectives, were handled using the Exoplanet Open-Source Imaging Mission Simulator (EXOSIMS; Savransky and Garrett 2015; Savransky et al. 2017). EXOSIMS simulates the mission, observational targets, and environment using a Monte Carlo code with detail down to discrete replications of every individual exoplanet direct imaging observation. Post-processing of EXOSIMS results are also used to derive aspects of engineering design, such as telecom and propellant budgets based on modeled mission and systems performance.

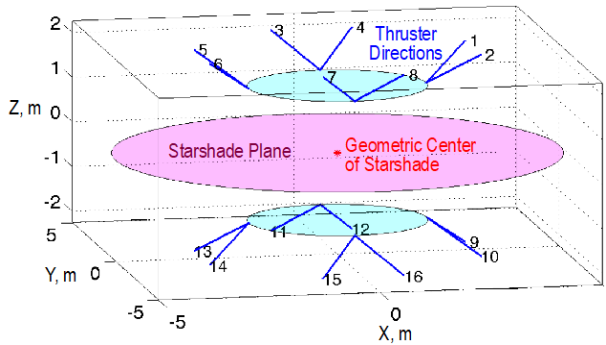

**Figure 8.1-4.** Example thruster configuration.

**Table 8.2-1.** HabEx will complete observations to meet GO science objectives in about 0.5 yr of the 2.5 yrs dedicated GO mission.

| Observing Program | Estimated Time Required |
|---|---|
| O9. Baryon lifecycle | ~6 weeks |
| O10. Metagalactic ionizing radiation | ~4 weeks |
| O11. Massive stars | ~3 weeks |
| O12. Hubble constant | ~3 weeks |
| O13. Dark matter in dwarf galaxies | ~2 weeks |
| O14. Globular clusters | ~2 weeks |
| O15. Exoplanet transit spectroscopy | ~1 week |
| O16. Transition disks | ~1 week |
| O17. Solar system auroral activity | ~1 week |

This full mission simulation approach allows for the encoding of the optical systems, the observatory and its dynamic constraints such as target-specific observing windows, telescope keep-out requirements, and variation in local zodiacal brightness as a function of observing direction. Dynamic constraints and the requirement of intelligently scheduling observations of specific targets becomes especially important in the case of starshades, where retargeting can have significantly different costs (in terms of propellant use and time).

By generating an ensemble of hundreds of full mission simulations for a particular population of planets, the full distributions of key science metrics are captured in addition to the number of planets observed and characterized, as the mission simulations can be used to generate statistics on essentially any value of interest. Examples of additional metrics of interest include the expected time to first planet detection, the expected posterior distributions of detected planet physical and orbital parameters, and starshade propellant use as functions of mission time.

A three-tier scheduling algorithm is implemented in EXOSIMS to optimize planning and refine yield estimates. The starshade is most constrained and scarcest resource, so it is scheduled as the first priority (Tier 1). A traveling salesman problem (TSP) optimizer plans the path that the starshade will take among the available targets, i.e. those not excluded by field of regard constraints, using a 2-step look ahead. When the starshade arrives on target, the telescope performs formation flying acquisition, which is modeled in EXOSIMS as a 6-hr overhead, and then begins the starshade spectral characterization observation. As soon as the integration time is complete, the starshade begins its transit to the next target and the telescope returns to observing with the coronagraph or general astrophysics instruments. The coronagraph observations are scheduled based on solar keepout constraints and prioritization of the targets, which are ranked by completeness divided by integration time. Revisits for the purpose of orbit determination are performed with a minimum cadence of 2 months. While coronagraph observations are scheduled and simulated against specific targets in Tier 2 of the schedule, the portion of time given to General Observatory (GO) science is monitored. When the time owed to GO exceeds 1 day, then GO time is allocated on Tier 3 of the schedule. The GO simulated as a time allocation and not specific targets and integration times. The tiers of the scheduler are shown in **Figure 8.2-1**.

The parameters used in the DRM simulation are the same astrophysical and instrument parameters used in the yield modeling. Some of the parameters more relevant to DRM scheduling are listed in **Table 8.2-2**. Placement of radiators was not considered a constraint in the anti-sun direction for DRM modeling and would need to be considered for a final mission design.

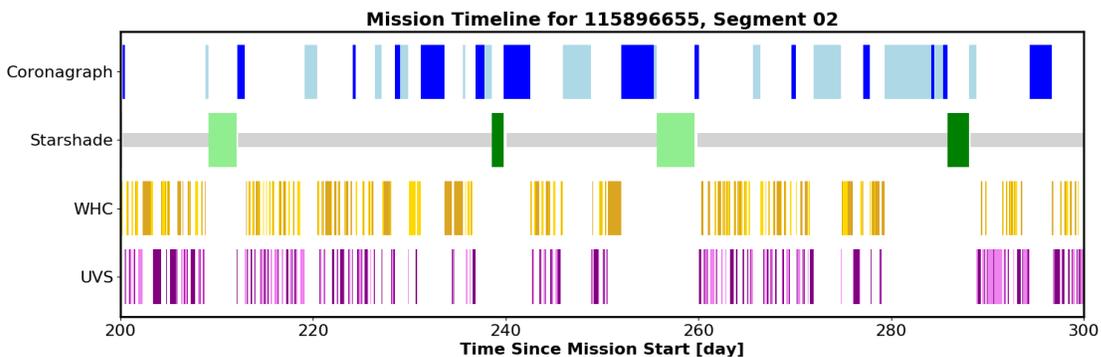

**Figure 8.2-1.** Characterization observations (green) and starshade retargeting transits (grey) are scheduled first in Tier 1. During starshade transits, coronagraph observations (blue) are scheduled in Tier 2 while the HWC (gold) and UVS (purple) observations are scheduled in Tier 3. Light and dark show a change of target.

**Table 8.2-2.** Parameters used in the Design Reference Mission simulation verifying HabEx's science operations conops.

| Parameter | Value |
|---|---|
| Minimum Sun angle | 40° (set by telescope) |
| Maximum Sun angle | 83° (set by starshade) |
| General Observatory portion of mission time | 50% |
| Characterization SNR | 10 |
| Characterization spectral resolution | 140 |
| Starshade central passband wavelength | 650 nm |
| Detection SNR | 7 |
| Detection bandwidth | 20% |
| Starshade diameter | 52 m |
| Starshade distance | 76,600 km |
| Dry mass of starshade | 5,230 kg |
| Fuel mass for starshade | 5,700 kg |
| Starshade transit $I_{SP}$ | 3,000 s |
| Starshade thrust | 1.04 N |
| Starshade station keeping specific impulse | 308 s |
| Formation flying acquisition overhead | 6 hrs |

The DRM simulation is run on hundreds of Monte Carlo cases. For each Monte Carlo case, a new set of planets is synthesized around the target stars. The planets are drawn from the probability distribution functions prescribed by the Science Analysis Group 13 (SAG-13) occurrence rates, as amended by Dulz et al. (in prep.), and physical property probability distributions the same as in the yield modeling, such as an albedo of 0.2 for rocky planets and 0.5 for gas giants. As exo-Earth candidate planets are discovered by the coronagraph blind search, they are promoted to targets for the starshade for spectral characterization. There are two criteria for promotion to a starshade target. First, that there must be at least three successful coronagraph detections of the planet spanning over half a period. Second, at the start of the mission, when the coronagraph is yet to discover earth candidates, exoplanet demographic science is performed by acquiring spectra on eight pre-selected deep dive targets. The deep dive targets have high completeness for each of the eight planet types and as such have a high probability of acquiring spectra of any of the planet types if such a planet exists around these eight. The starshade TSP scheduling begins with the deep dive targets and continues as newly discovered targets are added to the target pool. Each Monte Carlo case is different because different synthetic planets are 'discovered' and then characterized.

The result of one of many DRM Monte Carlo cases is shown in **Figure 8.2-2**. The figure shows the transits chosen by the Traveling Salesman Problem optimizer. The white region is the region observable to the starshade. The grey circle with yellow boundary is the keepout due to minimum sun angle that constrains both coronagraph and starshade targets. The gray and black dots are potential targets. Observations are indicated in green (habitable zone successful observation), purple (non-habitable zone successful observation), and red (insufficient SNR for detection); the size of the colored marker indicates the number of observations made.

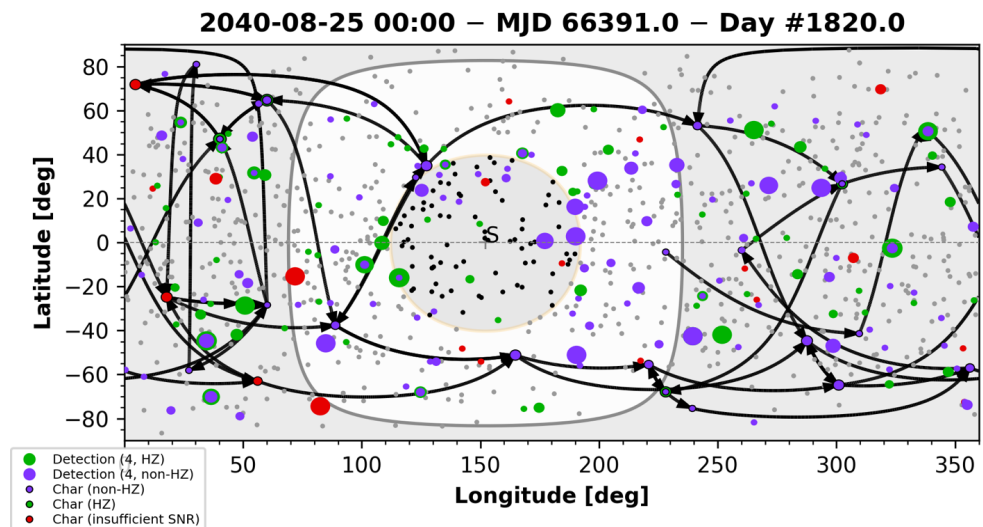

**Figure 8.2-2.** A 5-year mission simulation from 2035–2040 showing starshade transits (*black arrows*), starshade characterizations, and coronagraph observations.

The keep out constraints on the starshade create an interesting sky coverage and availability (**Figures 8.2-3** and **8.2-4**). Along the equatorial latitude, the observable window is 43 days, keep out for 80 days, and then observable for 43 days. The percent availability in sky coverage is 24% and the longest duration observable window, and hence the longest integration time for characterization, is 43 days. At 60° ecliptic latitude, the observable window avoids the solar keep-out constraint and results in a long ~150-day access window with ~40% sky coverage. It is important to note that 7° within the pole is not observable with the starshade due to the maximum sun angle constraint.

The location of the Sun, Earth, Moon, and other solar system planets are tracked throughout the mission lifetime using ephemerides from JPL Horizons. Interestingly, a retrograde motion of Jupiter in 2037 slightly diminishes sky coverage near 75° ecliptic longitude seen **Figure 8.2-3**. The keepout and observing windows of the top starshade targets is shown in **Figure 8.2-5**. Earth and moon keepouts of 45° are shown in

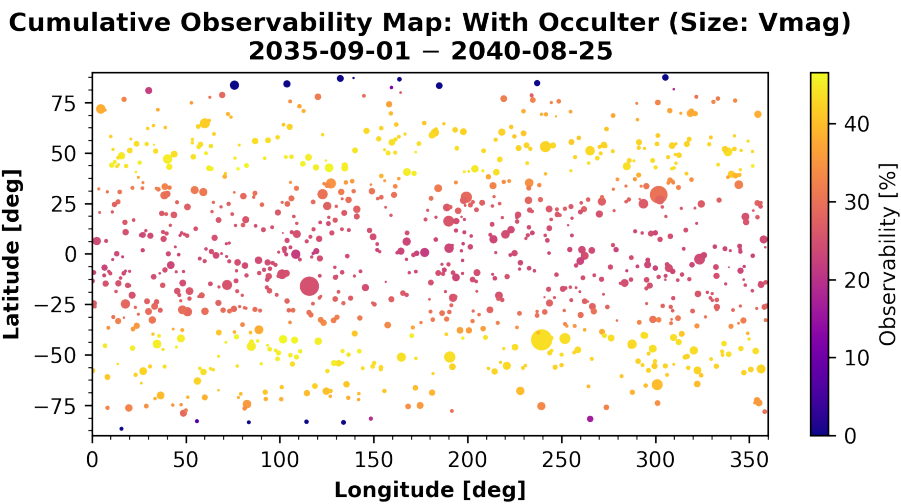

**Figure 8.2-3.** The sky coverage of the starshade shows the percentage of a year that various target stars are observable (*color of targets*). The size of the markers is the apparent magnitude.

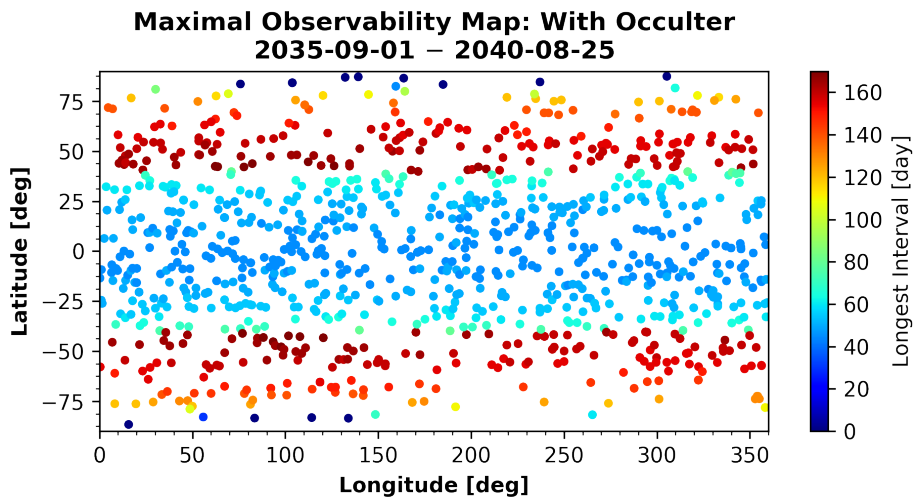

**Figure 8.2-4.** The maximum duration of an observation in days for the starshade is color coded. Within 7 degrees of the poles is not observable. The longest windows are between 44° and 60° ecliptic latitude.

**Figure 8.2-2** for illustration and are highly conservative. Earth glint off the starshade for some rare cases can occur and must be calculated on an observation by observation basis and has not yet been incorporated in the TSP optimizer for computational efficiency. Keepout for minimum angle to the earth and minimum angle to the moon was set to 1° for the simulations.

The ensemble of Monte Carlo DRMs provide many metrics for the mission design, as mean, standard deviation, or posterior distribution, including fuel use for retargeting transits and station keeping. The fuel used for station keeping depends on the lateral disturbance forces on the starshade which depend on the position in the L2 halo orbit and the pointing direction to the target. **Figure 8.2-6** illustrates of the variation of ΔV required to overcome disturbance forces over a sphere of pointing directions at various positions on the L2 halo orbit. The fuel used for retargeting transits and station keeping is shown cumulatively over mission elapsed time in **Figure 8.2-7**. Note that the variability due to station keeping is greater than the variability due to different transit paths. In this instance, the total fuel used by the starshade over the 5-year mission is about 5,000 kg, which leaves about 700 kg for a possible extended mission.

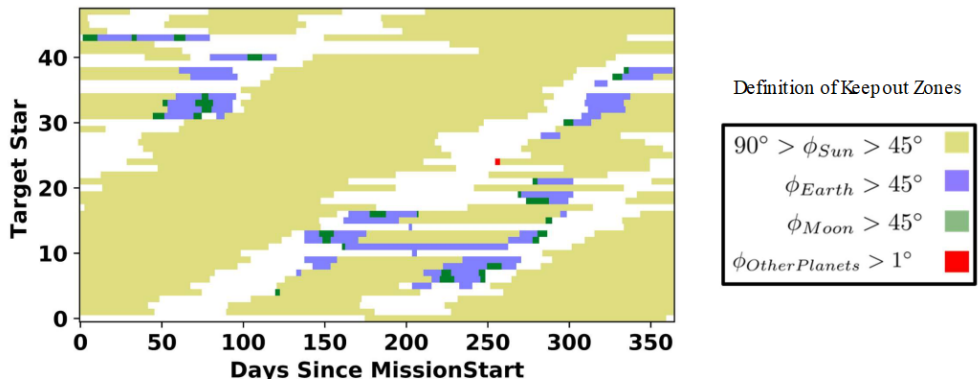

**Figure 8.2-5.** The binary keepout map shows days in which the stars on the y-axis are observable in white; the days when the star is in keepout are shaded with colors corresponding to different bright objects shown in the legend.

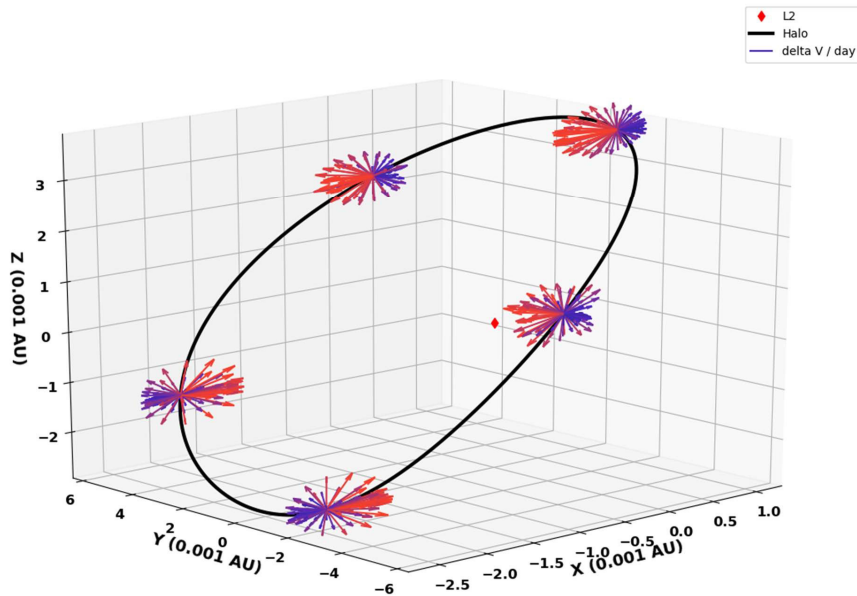

**Figure 8.2-6.** Five possible telescope locations on the halo orbit are plotted relative to the Sun-Earth L2. The arrows represent the varying magnitude of the daily ΔV depending on which target star the starshade needs to station keep with at that point. They also pertain to the highlighted columns on the mock table shown above. No real data is presented here, this is purely to demonstrate variability in daily ΔV.

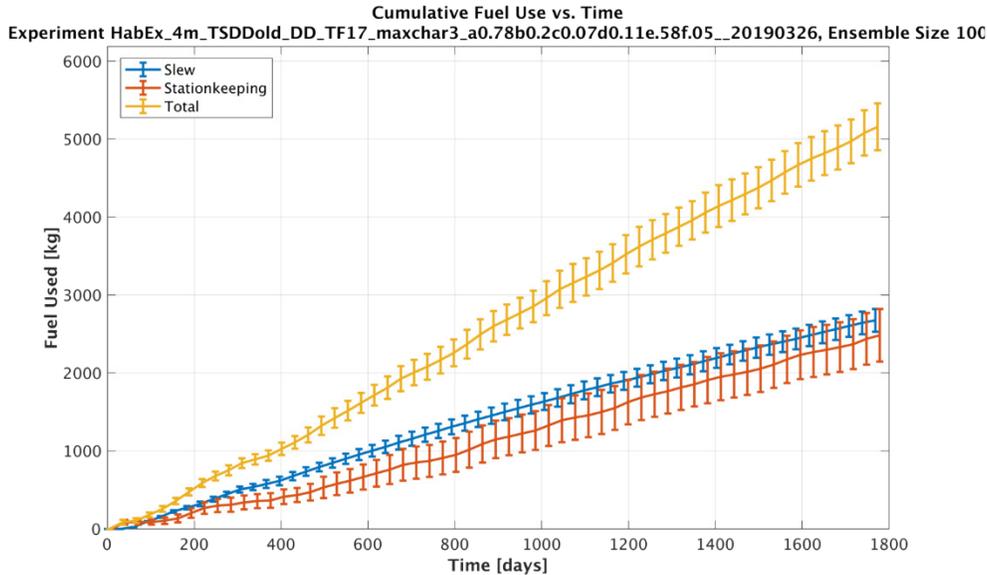

**Figure 8.2-7.** The DRM simulation tracks the propellant used for retargeting the starshade and for stationkeeping during formation flight. Post-processing over an ensemble of DRMs provide a mean and standard deviation for fuel use as a function of mission elapsed time. Note that in this figure "slew" references starshade retargeting transits.

## 8.3 Servicing

Servicing HabEx could significantly enhance its productivity, cost effectiveness, and extent its mission life. Accordingly, HabEx has investigated some options for future servicing to extend the lifetime of the observatory. While a future mission concept study will conduct trades on servicing operations, methods, and hardware, an initial high-level concept for servicing both the telescope and the starshade has been included in this report and is described in this section.

### 8.3.1 HabEx Serviced Elements

The telescope spacecraft carries sufficient propellant to carry out operations for 10 years but with careful management or reduced science operations the telescope can remain in L2 orbit much longer. Servicing events should target a cadence of under 10 years to maximize science utility. Servicing events will likely include the replacement and upgrade of the instruments and avionics, replacement the microthruster modules with new, fully-fueled units, replacement of the monopropellant thrusters and refueling the monopropellant tanks. The telescope solar array is sized for 20 years of operations so it will not need to be serviced at the same cadence.

During instrument replacement, the entire instrument module would be replaced en bloc. The module includes the four instruments, their thermal radiators, the tertiary mirror assembly, and the fine guiding system. Since HabEx is using a lateral mounting scheme for the instruments, the module is easily accessed from the anti-sun side of the telescope. The avionics—command and data handling (C&DH) subsystem, power control, inertial measurement unit (IMU), and telecom subsystem—are all attached to removable panels on the sun shield. Only the star trackers, sun sensors, and telecom antennas are located external to the sun shield. These panels, sensors and antenna can all be easily accessed on the spacecraft.

Microthrusters are installed in self-contained modules which are composed of three thrusters, a common support structure, and a shared fuel tank. The microthruster module will be replaced with a new, completely fueled unit.

Due to thruster throughput limits, new biprop thrusters will be required with a spacecraft refueling. Since the thrusters are all attached to the outside of the sunshield, access will be straightforward.

The telescope's solar array was designed to not require replacement before 20 years of on-orbit operation. If the telescope life is to be extended past 20 years, a new roll-out solar array (ROSA) could be attached over the old array. Care must be taken to design the original array with the proper mechanical, electrical, and thermal interfaces for this possibility.

The starshade spacecraft is also designed for limited servicing. The solar array is not replaceable, but the avionics can be accessed through the end of the hub's central tube. Refueling fittings are also located at one of these locations. SEP thrusters would need to be replaced with the refueling due to throughput limits.

One important trade will need to be conducted is to determine if it is cost effective to service the starshade rather than replace it with a new, fueled starshade. Replacing the entire spacecraft has the advantage of eliminating the micro-meteoroid damage that has accumulated on the original starshade. In addition, a complete replacement would allow improvements based on knowledge gained from operating the original starshade.

### 8.3.2  Servicing Method

**Figure 8.3-1** shows one way a servicing mission might be carried out for HabEx. Pieces are launched separately and aggregated at Earth-Moon L2 or some other such convenient intermediate orbit. Using an intermediate orbit has a number of benefits: first, the servicing vehicle requirements are relieved as it does not have to support a very large launch mass. Second, the list of available launch vehicles grows due to the more reasonable the trade between payload mass and volume of each launch package, and the number of launches. This makes the mission more robust and creates the potential for significant savings by use of reusable commercial launch options. Earth-Moon L2 has the further benefit of being within reach of the proposed Lunar Gateway space station. The servicer could take advantage of available ports, robotics, and extravehicular activity (EVA) crew to assist with its own aggregation activities, or could take advantage of its logistics chain as a lower cost method of hardware or fuel delivery. That being

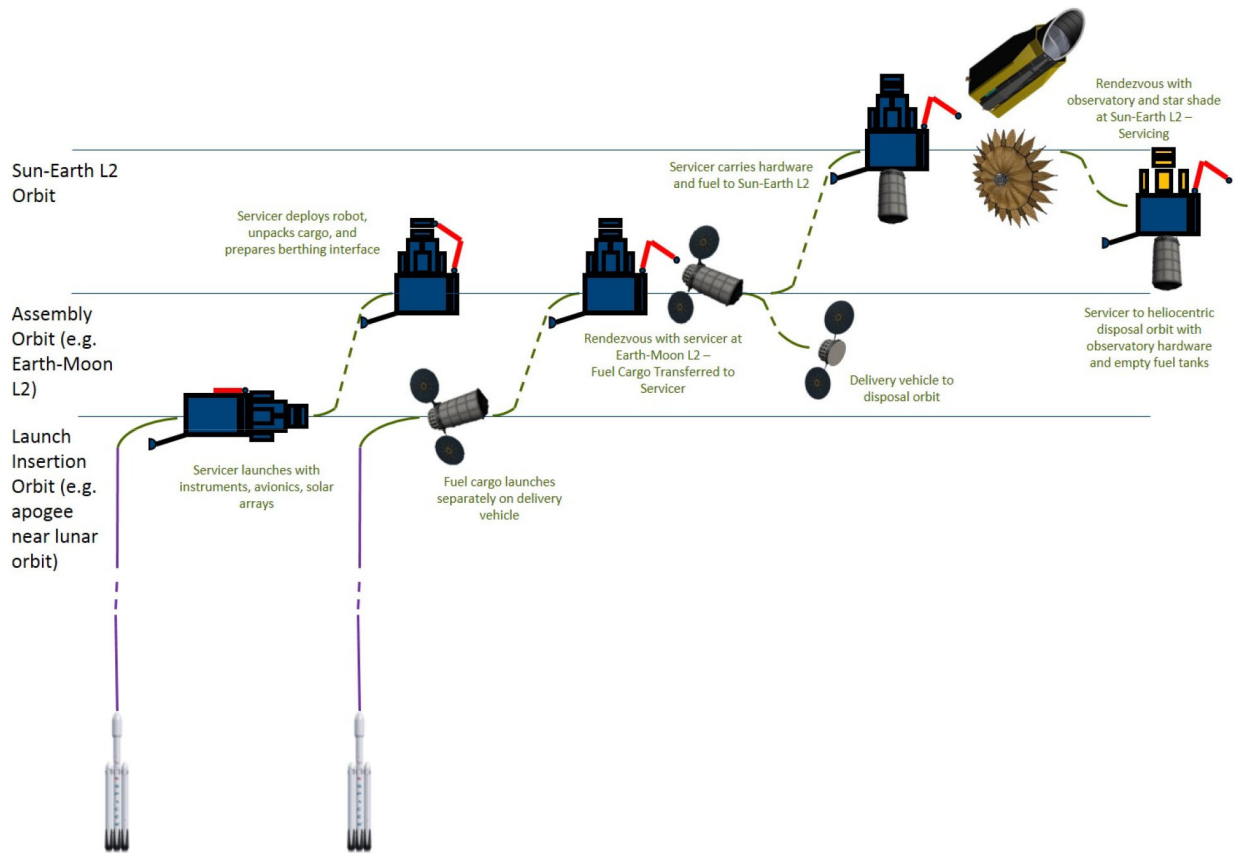

**Figure 8.3-1.** Notional mission profiles for HabEx telescope and starshade servicing.

said, many other mission profiles are possible; determining the best approach is left as a trade to future concept studies.

Once aggregation is complete at the intermediate orbit, servicer and supplies continue to L2 and perform rendezvous with the observatory.

### 8.3.3    Robotic System

The servicing vehicle envisioned for this mission would likely require two separate servicing robots. One heavy version with large end effector for the purpose of performing the initial capture of the observatory and starshade, and another lightweight, walking robot for the purpose of accessing the observatory microthrusters. Both would use seven rotary joints to provide kinematic redundancy, and would be of similar length, approximately 5 m from base to tip. **Figure 8.3-2** shows an overview of such a robot.

The end effectors associated with large and small robots would be suited to purpose. For the small servicing robot, the end effector would have umbilical connectors for walking, and a tool drive to actuate on-orbit interconnects. **Figure 8.3-3** shows a notional view of one such end effector. The large (capture) robot would likely have a larger, stronger end effector, with quick-capture mechanism. This might be similar to the Canadarm end effector, which was used for the capture and servicing of many satellites, including the Hubble Space Telescope (HST), by Space Shuttle crew, as well as for construction of the ISS.

### 8.3.4    Concept of Operations and Servicing Accommodations

Both the starshade and the telescope have an S-band telecom system which is nominally used for ranging and cross-spacecraft communication during formation flying. These radio systems can be used as a beacon for guiding the servicer at distances greater than 20 km.

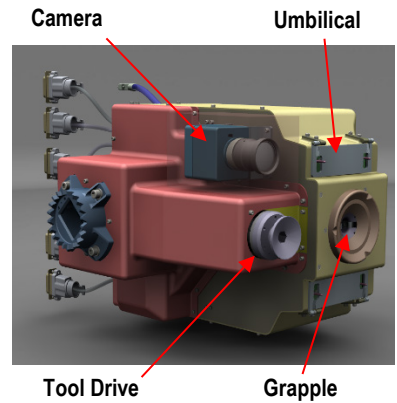

**Figure 8.3-3.** Notional end effector for walking/servicing robot with major components identified.

Between 20 km and 1 km, bearing and range are determined by camera (IR/visual) and laser rangefinder respectively. For this purpose, retro-reflectors could be added to the observatory to increase the effective range of the rangefinder. As the servicer approaches closer, a light detection and ranging (LIDAR) can be used initially for similar range and bearing measurements, and eventually to determine the spacecraft relative pose. Again, a pattern of retro-reflectors would improve the accuracy and stability of the pose solution. Similar items have been used by Orbital Express, European Automated Transfer Vehicle (ATV), and the NASA Sensor Test for the Orion Relav Risk Mitigation (STORRM) demonstration on ISS. Existing examples of laser retro-reflectors are shown in **Figures 8.3-4** and **8.3-5**.

During the final approach, the HabEx telescope aperture cover will be closed to completely avoid the possibility of contaminating the telescope mirrors by thruster plumes since the servicer will have to control its closing rate when approaching the telescope. Once the servicer had positioned itself relative to the observatory for capture, robot mounted cameras would provide guidance to maneuver the robot to capture the telescope spacecraft. In order to facilitate this track and capture, the telescope spacecraft would carry a grapple fixture and associated visual target, similar to those used on ISS visiting vehicles. For the terminal phase of capture, where the robot makes contact with the grapple hardware (**Figure 8.3-6**), the telescope spacecraft will need to suspend its attitude control so that the connected spacecraft attitude control systems do not begin to counteract each other. It would ideally remain passive in this sense, until servicing is complete and the servicer has released.

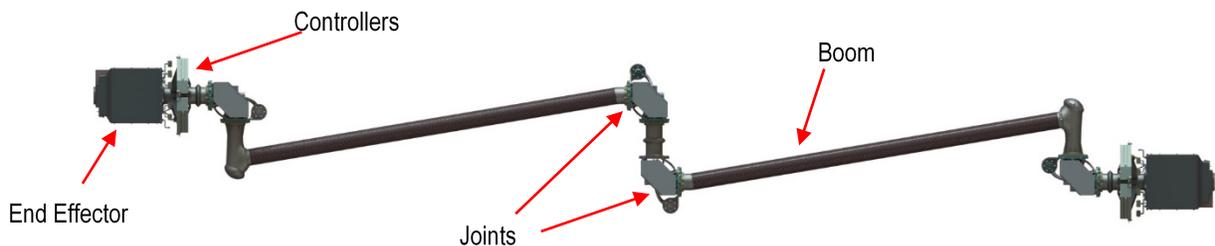

**Figure 8.3-2.** Notional servicing robot with major components identified. The servicing robot is about 5 m from tip-to-tip.

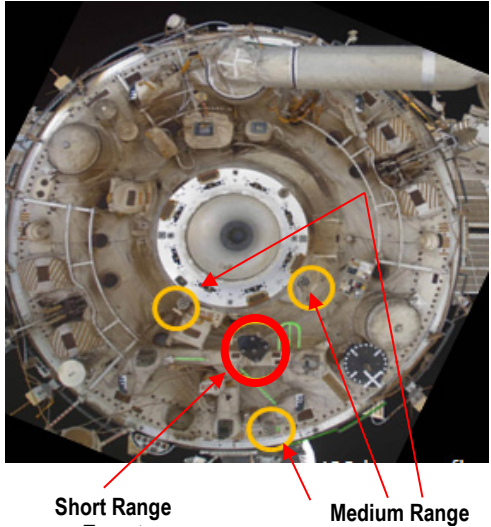

**Figure 8.3-4.** Arrangement of short and medium range retro-reflectors for rendezvous guidance: ISS Zvezda AFT.

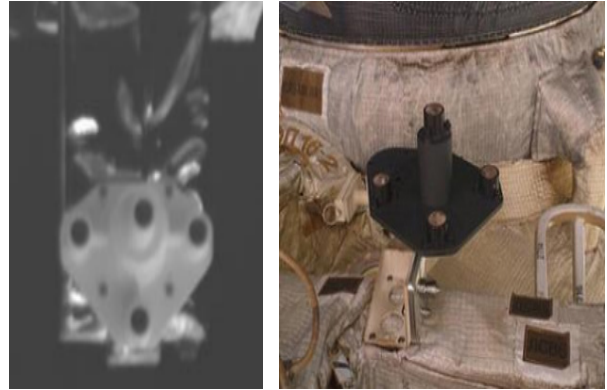

**Figure 8.3-5.** Retro-reflector arrangements used for short range rendezvous guidance: (*left*) Orbital Express (Heaton et al. 2008), (*right*) European ATV (Harding 2013).

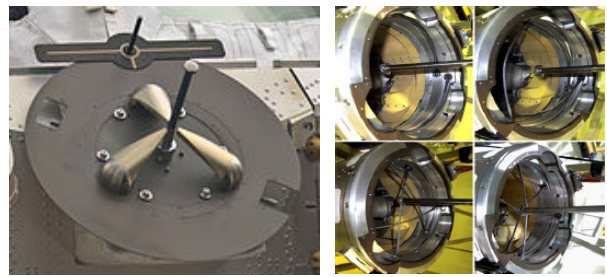

**Figure 8.3-6.** *Left:* ISS flight releasable grapple fixture; *Right:* Shuttle RMS end effector inner workings (Jorgensen and Bains 2011).

Once capture is complete, the robot would connect the telescope bus and servicer together, using a strong structural latching mechanism, such as the Common Attach System (CAS) used on ISS, or Soft Capture Mechanism (SCM) used on HST (**Figure 8.3-7**). This would require some passive structure on the telescope bus. Finally, during berthing, a blind mate umbilical connection would be mated to pass power and data between the two, with a separate connection for refueling.

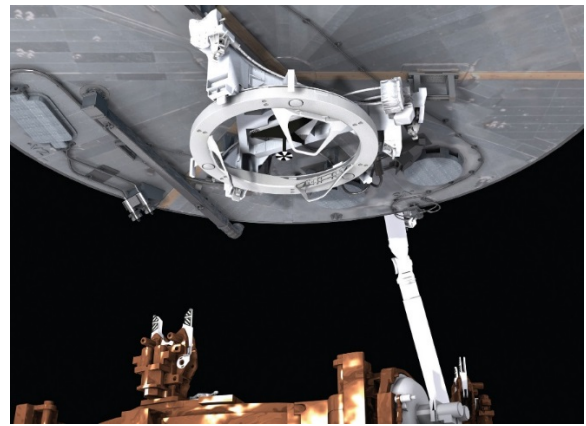

**Figure 8.3-7.** Passive side of Soft Capture Mechanism (SCM) installed on HST.

### 8.3.5    Refueling

Refueling could be simply accomplished by locating quick-disconnect ports on the observatory such that they mate with corresponding ports on the servicer during the berthing or docking process. This is how refueling of a hydrazine system was handled during the Orbital Express mission. NASA also has a

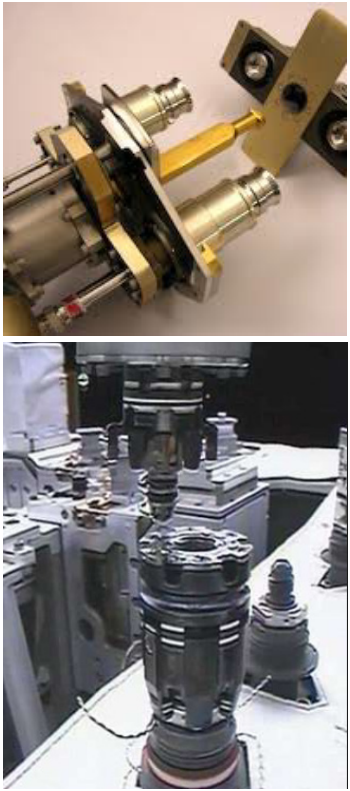

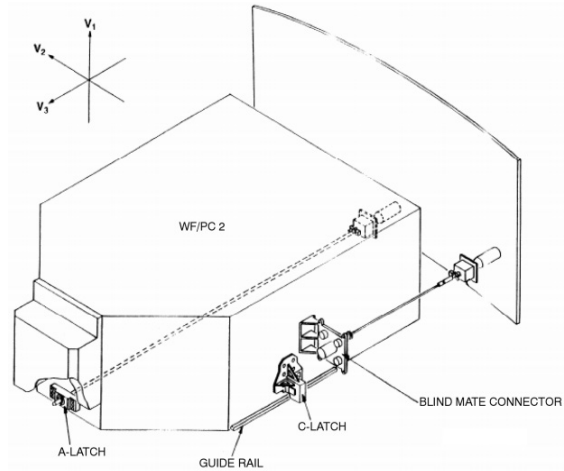

**Figure 8.3-9.** HST Wide Field Camera 3 (WF/PC2) overview, showing "A" and "C" latches, guide rail, and blind mate connector, along with EVA interfaces (NASA 2009).

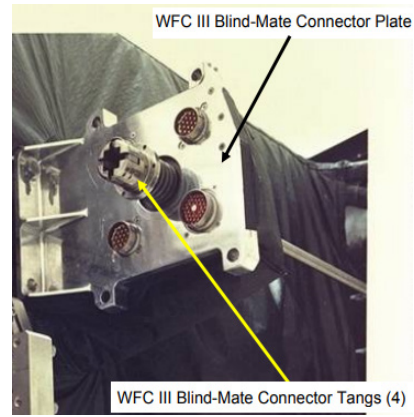

**Figure 8.3-10.** HST WFC3 blind mate connector (NASA 2009).

**Figure 8.3-8.** *Top:* Quick-disconnect port similar to that used on Orbital Express; *Bottom:* on orbit demonstration of fluid transfer using quick disconnect similar to NASA CSV (NASA 2013).

Cooperative Servicing Valve (CSV) (**Figure 8.3-8**), designed for robotic actuation, and qualified to transfer hydrazine, monomethylhydrazine (MMH), nitrogen tetroxide (NTO), pressurant, or Xenon.

### 8.3.6    Instrument Module Replacement

As noted earlier, the instrument module, including the detector radiator, would be replaced as a single unit. While its location is advantageous for servicing, its size may require careful consideration to ensure a successful change out. It is expected both robotic arms will be needed; the larger to handle the module and the smaller to reach attachment hardware. During replacement, the original module would be secured to a stowage site on the servicer, then the new module would be retrieved from the servicer and installed on the telescope. All this would be done using cooperative on-orbit interfaces, similar to those used by the HST instruments **Figure 8.3-9** and **Figure 8.3-10**). This includes some number of active and passive latches split across the module and the module mounting hardware. Active latches for HST were EVA crew actuated, by means of a torque driver. In the case of HabEx servicing, this torque driver might be integrated into the robot end effector. Additionally, guide features (e.g., rails) were added to HST instrument mounting assembly, will be included for HabEx instrument module installation as well.

### 8.3.7    Microthruster Replacement

Replacement of the microthruster modules will likely require the use of the small servicing robot since the modules are located far from the likely connection interface between telescope and servicer. One notional approach is to add grapple points and data bus repeaters to the telescope spacecraft so that the robot can move along the telescope's sun shield to the microthruster mounting locations. This would be done in a manner similar to ISS robots, with commands and telemetry being relayed to the servicer via an umbilical link at the berthing interface. The lightweight walking robot would then travel to the microthruster site, remove the old unit and hand it to the heavier robotic arm. The robotic arm might handle a carrier pallet holding several microthruster modules to reduce the travel distance for the walking robot (**Figure 8.3-11**). The large robot moves microthrusters from the servicer to the walking arm, and the walking arm would bring them to the installation sight and install. The modules do not require precision placement and only require a single electrical interface and simple mechanical attachments. An on-orbit electro-mechanical interface, similar to the interface for a Hall effect thruster (HET) module servicing could be used. An example of such an interface is shown in **Figure 8.3-12**. It should be noted that the starshade servicing will require the replacement HETs as part of its servicing. It should also be noted that unlike the HET replacement, the microthrusters do not require a propellant interface.

Alternately, the servicing robot might be carried by the larger robotic arm, which positions it back and forth between each desired servicing site, and the new hardware stowage site on the servicer. This also is similar to how the ISS

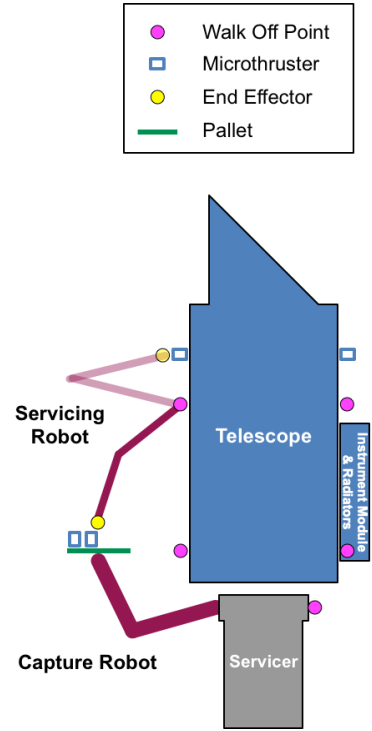

**Figure 8.3-11**. Concept of operations for microthruster change out. The servicing (walking) robot exchanges hardware with a pallet held in position by the robotic arm.

robotics are architected, as shown in **Figure 8.3-13**.

### 8.3.8 Avionics Replacement

Avionics replacement would be carried out in a manner similar to the microthruster replacement, except that without the reach issues involved, it can all be done by a single servicer-based robot arm. One difference however, is that most of the avionics are attached to the interior walls of the sun shield, so access panels will need to be opened as the first step of a servicing operation. This can be done in many ways, two of which that are commonly used on the ISS are shown in **Figure 8.3-14**.

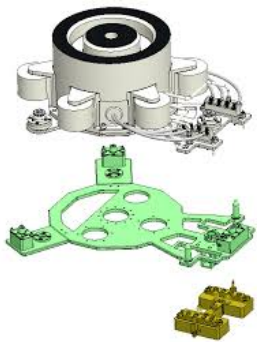

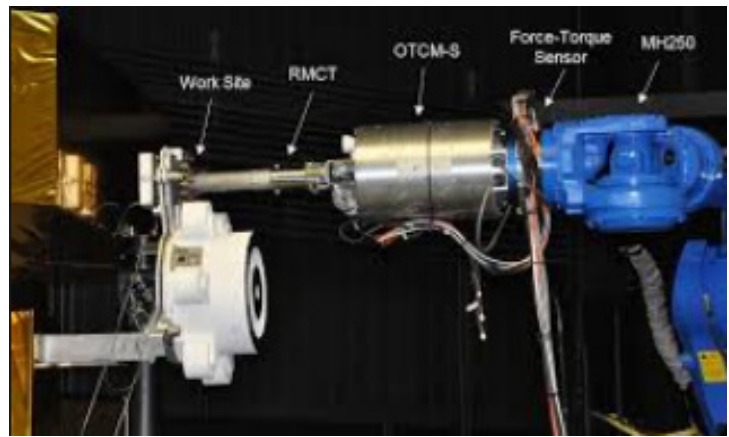

**Figure 8.3-12.** *Left:* A NASA concept for a Hall effect thruster (HET) module change-out. *Right:* NASA testing of robotic manipulation of that HET module (Martin et al. 2018).

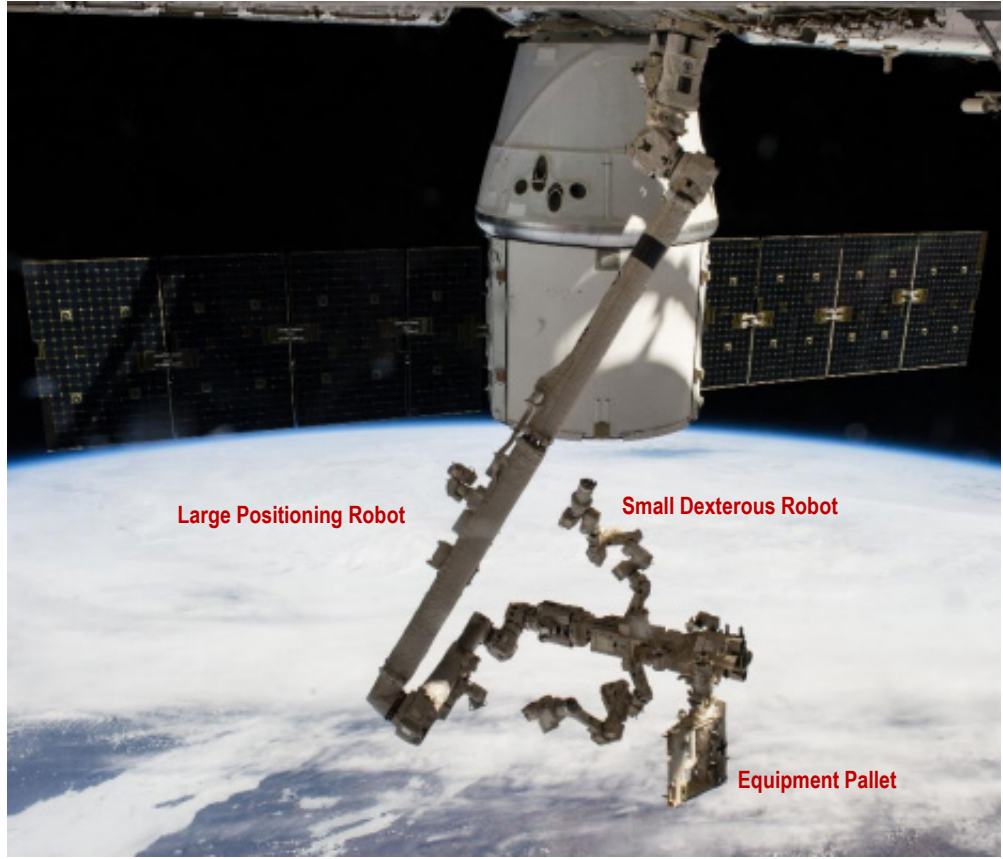

**Figure 8.3-13.** Architecture of ISS robotic systems that parallels the approach in Figure 8.3-11 (Visinsky 2017).

## 8.4 Operations and Ground Systems

The Ground Systems support for HabEx will be based on successful and typically extended organizational structures of other large NASA observatories such as the Spitzer Space Telescope, and interagency telescopes such as HST, Herschel, and Euclid. The principle elements of the ground segment consists of: (1) the Mission Operations Center (MOC), which carries out communications with the observatory, mission planning, orbit control, health and safety monitoring, and transfer of telemetry; (2) the Science Operations Center (SOC) responsible for observation schedule optimization, the data pipelines, archival storage and distribution of

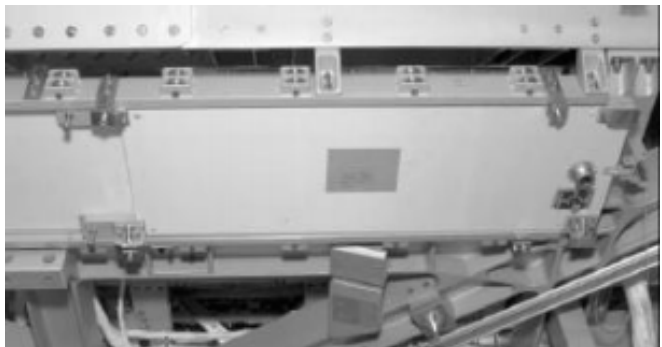 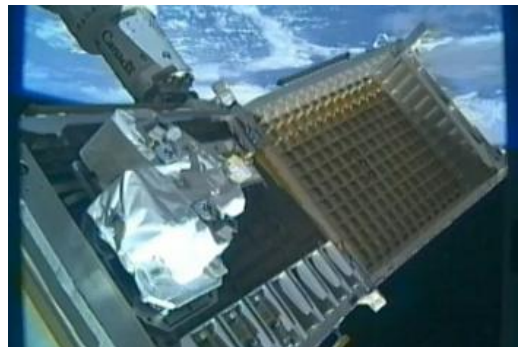

**Figure 8.3-14.** *Left:* ISS access hatches, sliding door. *Right:* Hinged lid (Harding 2011).

observations, community support and public outreach; and (3) Instrument Team centers at the relevant institutes, responsible for successful operation of the science payload by overseeing assembly and integration, commissioning, monitoring and calibration, and development of instrument-specific software and procedures for generating observing sequences and data processing. **Figure 8.4-1** shows the high-level elements of the ground system, showing two

DSN Ka/X and S-band stations. Operations support for the HabEx coronagraph and starshade are distinguished from the UVS and HWC, similar to the model for the WFIRST Coronagraph Instrument (CGI).

### 8.4.1 Mission Operations Systems

The development of the HabEx MOS will leverage existing multimission capabilities with systems and operational readiness achieved as early as possible by the MOC, reaching a fully operational state by launch. Given the broad MOS heritage at Caltech Infrared Processing & Analysis Center (IPAC), HabEx does not impose any unique requirements on MOS systems and procedure design. For example, even the most unusual aspect in commanding HabEx, the formation flight of two flight systems, is decomposed into commands for each individual flight system to independently enter a mode of operation for formation flight. As such, the HabEx requirements for MOS are no different than other missions and will leverage designs that maximize mission safety and minimizes risk of mission loss or compromise to mission objectives. Moreover, by leveraging existing MOS infrastructure HabEx preserves the flexibility in the ground system to support observatory-level integration and testing, pre-launch mission scenarios testing, and flight support through multiple mission phases (launch, on-orbit checkout, scientific performance verifications, and routine science operations), including potential extensions beyond the baseline lifetime of the observatory. Finally, the significant reuse and adaptation of software developed for similar missions, will help meet the growing demands and constraints placed on ground systems by technologically advanced instrumentation and high data collection rates.

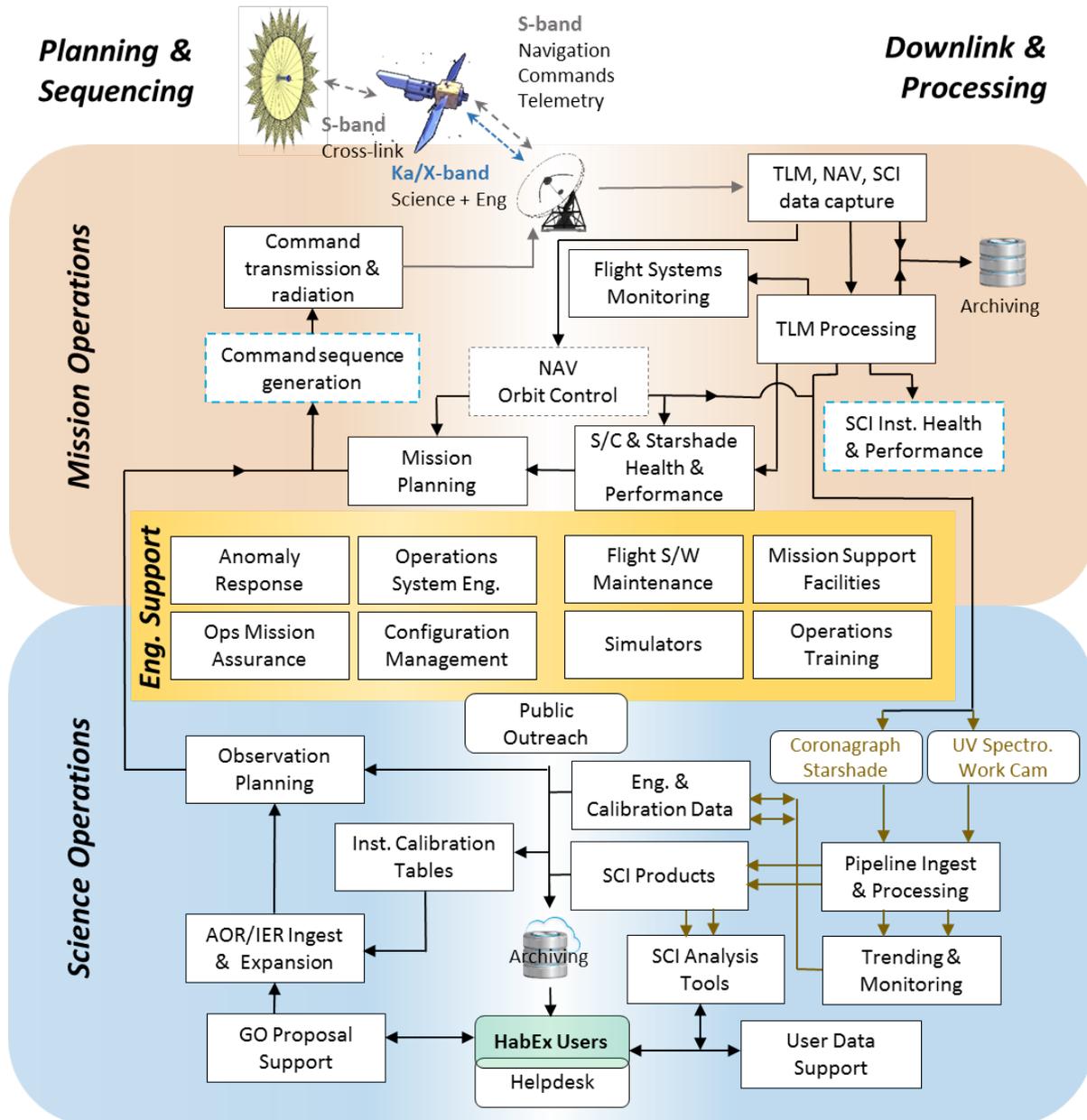

**Figure 8.4-1.** Schematic of the principle functions of the Mission and Science Operations Centers. Tasks bordered with blue dashed lines indicate where SOC participation is involved. Navigation (NAV, or Flight Dynamics) typically functions outside of the formal MOC.

### 8.4.2 Science Operations Support

The main functions of the HabEx SOC are to support the observing community through planning, processing, and storage, as well as building an efficient but flexible observing schedule. In the primary mission, there will be two tracks of science operations planning, for exoplanet direct imaging and general astrophysics observations that are merged into a universal observing and operating schedule for the HabEx telescope and starshade flight systems.

The exoplanet direct observation operations will impact telescope and starshade flight system operations. Coronagraph observations will be conducted with the telescope flight system alone and their results will feed-forward into the selection of follow-up observations with the starshade flight system and camera. Candidate follow-up observations will be scheduled alongside additional coronagraph observations by solving a "traveling salesman" problem that maximizes potential characterized exoplanet yield while minimizing use of fuel and as constrained by field of regard, candidate target location on the sky, and reserved GO time. Additionally, HabEx will develop science operations that maximizes the co-utilization of pointing to observe exoplanet direct and GO observation when the telescope can observe multiple requested fields simultaneously.

HabEx will select and plan GO observations based on long-standing practice. HabEx will use community support activities that are well-established for major observatories such as Spitzer, Herschel, HST, and the James Webb Space Telescope (JWST). HabEx SOC will include the development of graphical observation planning tools built around the instrument modes and Exposure Time Calculators that will be used by PIs in developing their observing plans. PI-requested observing schedules will be integrated with engineering, calibration, exoplanet direct observing, and GO programs in order to maximize observatory utilization. Final observing block command sequences will be transferred to the MOC for uplink and execution on-board the HabEx telescope flight system.

The SOC and MOC together can maximize observation efficiency for a pointed telescope like HabEx by adopting the concept of adaptive sequencing. This allows onboard software to allocate a general set of observations by their temporal validity range (set by purpose and telescope pointing avoidance zones) with minimum slewing, rather than follow a fixed timeline uplinked to the spacecraft. Such a system has been successfully applied to Spitzer, minimizing observatory down time and speeding recovery in the event of a contingency during a particular observation. This scheme can be particularly desirable given HabEx's reliance on finite propellant for telescope slewing.

The SOC will support the HabEx community's data needs by hosting the instrument data pipelines and development of analysis tools. The SOC will engage the community for observation planning and data processing with in-situ and web-based workshops, with an approach that reaches all levels of technical expertise on the complex suite of HabEx instrumentation. The methods to approach differing levels of expertise can be leveraged from SOC experience gained from similarly complicated observatories, such as Herschel. The HabEx user community will greatly benefit in their planning and data analysis by the expanding archival services of the NASA Exoplanet Archive at NASA Exoplanet Science Institute (NExScI) and the NASA InfraRed Science Archive hosted at Caltech/IPAC. These archives are key resources for upcoming exoplanet and astrophysics missions such as the Large Synoptic Survey Telescope (LSST) and Spectro-Photometer for the History of the Universe, Epoch of Reionization and Ices Explorer (SPHEREx).

### 8.4.3   Science Data Processing

The SOC will support the HabEx community's data needs by hosting the instrument data pipelines and development of science analysis tools. Observations obtained with the coronagraph require processing using techniques that can be leveraged from other space-based coronagraphs, such as the WFIRST CGI currently in development.

Storage of data ingested from the MOC and processed into final products is not expected to pose a serious challenge: HabEx is expected to generate a weekly maximum uncompressed volume of 184 GBytes from the UVS and HWC, and 26 GBytes from the exoplanet instruments, or ~11 TBytes of science instrument data per year. Accounting for engineering and pointing data, and expansion of the science data volume into final products including intermediate sandbox storage, the demands are still quite small compared to projections for JWST, WFIRST, and Origins Space Telescope (OST). Roughly 1.2 and 2.5 PBytes of space currently exists in the archives at Caltech/IPAC and Space Telescope Science Institute (STScI), and while these resources are expected to expand, new missions will benefit from the evolution of information technology, processing methods, and cloud storage potentials.

# 9 HabEx 4-Meter Baseline Management, Schedule, Cost, and Risk

Planning an ambitious NASA flagship mission has always been challenging, yet the lessons learned from past Decadal Surveys and projects such as the James Webb Space Telescope (JWST), Hubble Space Telescope (HST), Mars Science Laboratory (MSL), and others do allow the project to outline a management, design, and cost approach that bounds the development of HabEx in historical actuals (Bitten et al. 2019). HabEx is aware of the challenges of modeling costs for a great observatory in pre-Phase A development, and hence has applied conservative estimates, historical analogies, and used well-established models where possible to identify cost drivers. Possibly most importantly, the study team has taken the advice from past National Academies studies and focused primarily on an exploration of nine architecture options at a high level, showing the relative science gain and driving design requirements, rather than an overly detailed single design point (*Chapter 10*). In early formulation, the importance of fully understanding the design tradespace allows for decisions regarding technology focus, and early investments in engineering maturation without a premature commitment to a single point design. Additionally, HabEx recognizes the threat posed by concepts with overly optimistic schedule and cost estimates that are later exceeded during implementation, pulling resources from other astrophysics priorities and delaying future missions. From the start, and in keeping with guidance from the National Academies, HabEx has focused on creating an ambitious, large telescope mission while protecting balance across all astrophysics disciplines. This chapter presents sound management approaches, schedule and cost estimates that are in line with historic experience, and places them in context with current astrophysics commitments and other possible future missions.

## 9.1 Management Approach

HabEx will require the full partnership of multiple NASA centers, industry, and academia to implement the mission, much like the collaboration shown during this study phase. Although this report is not a detailed proposal with specific management plans, some observations and suggestions about the management of the future HabEx mission are provided. The HabEx study recommends that a single NASA center be assigned the overall mission management responsibility in order to optimize the decision-making authority and communication. During pre-Phase A and early Phase A, partnering arrangement will be developed through negotiated roles for NASA centers and foreign partners, where their excellence is best suited, and through competitive industry Requests for Proposals (RFPs) for some elements of the mission. Flagships, almost by definition for NASA, are one of a kind. However, industry has a broad range of experience, including non-NASA projects, so they can add unique, necessary value to developing a project like HabEx, in areas including telescope technology, large deployments, complete spacecraft, and instrument elements.

As envisioned, the HabEx baseline design is a single project with two flight elements: the telescope flight system and the starshade flight system. Each has a spacecraft and payload element. Managing such a development is certainly not without precedent. Though more tightly coupled, the Mars 2020, MSL, and Mars Exploration Rover (MER) missions are all composed of near stand-alone elements. Each has a cruise stage, entry-descent-landing stage and rover with different design teams controlling each element. Cassini offers maybe a better analogy, where the orbiter and probe were fairly decoupled designs, and were developed by completely different organizations (JPL and ESA, respectively) with the entire project managed under one organization. Dedicated project management for each flight system would report to the top-level project management.

A unique benefit of the starshade flight system is that it is almost uncoupled in development from the telescope flight system—more so than with any element in any of the analog examples. Naturally, the overall mission design and operations must account for the systems working together and being able to sense and communicate. However, the flexible approach with the starshade system allows it to launch separately, and later, than the telescope system. This gives significant management flexibility with regard to technology development, scheduling, budget profiling, descopes, launch vehicle availability, and launch windows.

One of the most significant management concerns in this Decadal Survey is likely to be how to prevent the kind of delays and cost growth experienced in past missions while still recommending an ambitious program. The HabEx study provides a proactive rather than reactive approach, explicitly using lessons learned from recent analysis (Bitten et al. 2019).

Optimism, or otherwise unrealistic initial cost estimates, has been a major contributor to cost run-up in previous large missions. This contributor has been addressed with the creation of the CATE/TRACE review. Now, with an independent assessment guiding the Decadal Survey's decision, studies like HabEx must be realistic about their expected costs when they are making design decisions. HabEx has been managing scope by using cost models and analogs to help in design decisions since the beginning of the study.

### 9.1.1 Technology Development

As noted in numerous analyses, a key focus of any flagship mission must be on the clear, early identification and maturation of enabling technologies (Bitten et al. 2019; Udomkesmalee and Hayati 2005; Laskin 2002). HabEx explicitly identified enabling and enhancing technologies in *Chapter 11*. A driving philosophy of the HabEx study has been to identify the highest technology readiness level (TRL) solutions available to meet the science requirements, hence, all HabEx technologies are currently at TRL 4 or higher. By the time the Decadal Survey report is released, HabEx will only have ten TRL 4 technologies remaining (due to existing funded activities), two of which are related to the 4 m mirror, which cannot advance without a recommendation from the Decadal Survey. The study team has also identified technologies to the lowest level reasonable in order to be able to address unique, specific development plans for each one, and avoid the obfuscation inherent in bundling many related technologies into a single technology "banner." Estimating technology development costs is always difficult, but accurate schedules and costs are more likely to be developed in pre-Phase A and Phase A if the technologies are broken into lower level elements. Naturally, system-level technology readiness remains and must also be addressed. The current HabEx plan allocates schedule and budget within the baseline, based on historically shown development of new technologies, even at the system level. Detailed technology plans and costing are useful and warranted early in pre-Phase A, but history has shown that they are often underestimated, and do not often produce the large "cost savings" down the road that are sometimes promised. HabEx has allocated a total percentage of project cost to development based on historical actuals of other flagships as another way to bound these costs.

Bitten et al. (2019) recommend providing consistent funding for the technologies to TRL 6 and including pass/fail gates. The HabEx study supports both of these ideas. The study also recommends dedicated technology funding under the project's direction (rather than through competed Strategic Astrophysics Technology [SAT] funding, for example). This allows for a system-wide assessment of technology maturation, impacting both science return and cost and schedule. It also allows for the creation of an independent technology review board (like JWST's Technology Non-Advocate Review) for a consistent, independent assessment of technology maturation across multiple areas. Another model for the gates already exists in the S5 technology task developing the starshade technologies to TRL 5. The NASA Exoplanet Exploration Program (ExEP) requires that a Technology Advisory Committee (ExoTAC) review the completion of each S5 milestone on the path to TRL 5. The ExoTAC is a non-NASA non-advocate team of subject matter experts so each milestone receives an independent review. HabEx recommends a similar process be put into place for all technologies to TRL 6. This type of cross-check will put attention on technologies needing maturation focus, and can also address system-level TRL maturity which is also critical.

### 9.1.2 Contributions

NASA and the worldwide astronomical community are best served by international partnerships that unite the broadest community, leverage expertise internationally, and share costs. HabEx, like the other three large mission studies, has a number of international observers on the study team representing ESA, JAXA, and other national agencies. There has been clear interest in exploring ways to participate in a future HabEx mission, which will need to be negotiated at an Agency level. As one example, ESA has already indicated a willingness to contribute at a level of ~€500M, and a number of potential areas for natural contributions have been identified. During the HabEx study the Max Planck Institute for Astronomy in Heidelberg, Germany, took the lead on designing the HabEx Workhorse Camera (HWC), and led a whitepaper submitted to the ESA Voyage 2050 call outlining potential contributions to HabEx. Further, the HabEx baseline monolithic primary mirror is from Schott, which could be a natural contribution from DLR given their national expertise. Another natural contribution is microthrusters, which are used for precision stability and have been demonstrated on ESA's Gaia mission. Both ESA member states and JAXA are planning to contribute to the Wide Field Infrared Survey Telescope (WFIRST) Coronagraph Instrument (CGI) and could naturally reprise or expand on those roles in a HabEx coronagraph. The interest in contributing to HabEx from the international community is clearly enthusiastic and there are more opportunities for technical contributions than funding nominally allocated. Consequently, early negotiations of contributions will fully identify how best to maximize international participation.

### 9.1.3 Systems Engineering and Design Management

Systems engineering for HabEx will follow the NASA NPR 7120.5 rules. As designed in this concept study, HabEx also complies with JPL's best practices and design principles (e.g., requiring an overall 30% mass margin). An important example of the HabEx systems engineering, and overall design approach, is in mass management. Given the driving requirements of exoplanet direct imaging, the telescope design is off-axis which significantly improves throughput over an on-axis aperture of the same size. With the goal to avoid costly telescope deployments, the off-axis design of the 4 m telescope then leads to driving the volume constraints of the launch vehicle fairing. The Space Launch System (SLS) Block 1B Cargo launch vehicle provides volume margin for the HabEx telescope, and also significant mass margin. Given NASA's commitment to the SLS, HabEx was designed to use the launch mass capacity of the SLS Block 1B Cargo to simplify the engineering design of the HabEx telescope system; simply put, more mass equals more stability and less engineering complexity. This does, however, sometimes cause mass-based cost model issues. The HabEx design assigns a current best estimate (CBE) for all mass elements based on design status, then assigns a growth contingency based on the American Institute of Aeronautics and Astronautics (AIAA) standards for maturity to achieve a maximum expected value (MEV). There is launch mass margin beyond that for the SLS Block 1B Cargo capability. Future Pre-Phase A efforts could reasonably spend time optimizing the design and reducing mass further, while still capitalizing on the benefits of the SLS for a smart use of system mass.

As a Class A project, HabEx has delineated the spares and EM counts in the Master Equipment List (MEL). Prototypes are developed for the most challenging elements (i.e., instruments, starshade petals), and selective redundancy is throughout the system for long-lead items, based on JPL best practices for other flagship-class missions.

The integration and test (I&T) approach for HabEx, particularly for the unique nature of the starshade element, is detailed in *Section 6.11* and *Section 7.4*. Model-based systems engineering, model validation approaches (like on JWST and MSL), and clear verification requirements are critical on facilities of this size, or complexity. There can be no physical test of the starshade flying in alignment at flight-like distances on the ground, of course. Yet, even currently funded technology maturation tasks under the S5 project are making progress showing a scaled starshade with the right flight-Fresnel number demonstrates and matches models of starlight suppression requirements (Harness et al. 2019). Formation flying milestones for TRL 5 lateral sensing and control have recently been demonstrated (Flinois et al. 2018). This model-validation approach with subscale testing will be critical throughout the project lifecycle. In addition, large observatories like HabEx will need to rely on national facilities for key thermal and vibration tests, and early assessments show that facilities at NASA's Johnson Space Center (JSC) and Glenn Research Center (GRC) are suitable for key thermal and vibration tests, respectively.

## 9.2 Risk List

As a part of the HabEx study, risks that affect development and operations were identified and assessed for their likelihood and consequence. The current highest impact risks and mitigation strategies were identified, along with their likelihood and consequence after mitigation is applied (**Table 9.2-1** and **Table 9.2-2**).

**Table 9.2-1.** Guidelines for defining HabEx risk consequence and likelihood.

| Rating | Consequence | Likelihood |
|--------|-------------|------------|
| 1 | Minimal | Remote |
| 2 | Small | Unlikely |
| 3 | Moderate | Possible |
| 4 | Significant | Likely |
| 5 | Complete Loss | Very Likely |

### 9.2.1 G-Release Error

The HabEx primary mirror is fabricated to its required on-orbit figure by characterizing and removing gravity sag from metrology data as discussed in *Section 6.8.1.3* and *Section 11.3.1.1*. HabEx will demonstrate a validated process for characterizing and compensating for gravity sag with the following activities:

- Predict the horizontal and vertical gravity-sag of a test-article mirror assembly using a high-fidelity finite element model of the test-article mirror created using 'as-built' dimensional measurements and calculated spatial stiffness data.
- Quantify the gravity-sag of the test-article mirror assembly using both the N-rotation and the face-up/face-down method.
- Correlate predicted and measured gravity sags of the test-article mirror assembly.
- Use measured gravity-sag data to estimate a 0-G surface.
- Demonstrate the ability to achieve a 0-G surface on a multipoint fabrication/metrology mount.

Moreover, the primary mirror also includes actuators that further mitigate the risk of excessive G-release error during operations.

### 9.2.2 Starshade Integration & Test

As discussed in *Section 7.1,* the starshade integration and test program is comprehensive with multiple activities occurring in parallel to distribute the risk of any single activity impacting the project schedule. Furthermore, the starshade's qualification for TRL 6 will be at full-scale and will include flight-like deployments and flight-like environmental testing, so the risk of an unexpected delay with the flight unit integration and test activities should be greatly reduced. If such a delay did occur, that delay would be first addressed with planned schedule reserve (as shown in **Figure 9.3-1**). In a more extreme delay, the starshade launch could be moved out without loss to the HabEx baseline science. In such a situation, the telescope could be launched months or even years ahead of the starshade and could begin Guest Observer (GO) science and coronagraph planet detections and orbit characterizations in advance of the starshade's spectral characterization activities.

**Table 9.2-2.** Highest impact HabEx risks and their mitigations.

| Risk ID | if | due to | then | Consequence | Likelihood | Mitigation | Consequence | Likelihood |
|---|---|---|---|---|---|---|---|---|
| | | | | Pre-Mitigation | | | Post-Mitigation | |
| 1. G-Release Error | The G-release error exceeds specifications | Inadequate characterization of gravity sag during fabrication | Static wavefront error degrades observations | 3 | 3 | Demonstrate ability to achieve 0 G surface during testing, corrective actuators during operations | 3 | 1 |
| 2. Starshade Integration & Test | Delivery of the starshade is delayed | Complications during integration and test | Late completion of baseline mission | 4 | 2 | Starshade can be launched after the telescope and still meet science requirements | 4 | 1 |
| 3. Starshade to TRL 5 | Late demonstration of Starshade to TRL 5 | Multiple development activities | The starshade spacecraft is delayed | 4 | 2 | Use slack in schedule, or delay the starshade development and launch | 4 | 1 |
| 4. EMCCD Development | Late demonstration of EMCCD to TRL 6 | Problems in development | direct imaging spectral band performance will be reduced | 3 | 2 | EMCCD not on the critical path – release technology development schedule slack | 3 | 1 |
| 5. Microthruster Lifetime | The telescope microthrusters are not qualified to expected lifetime | Problems in lifetime testing | The telescope cannot launch | 5 | 3 | Use schedule slack to resolve the problems, or manifest additional microthrusters | 5 | 1 |
| 6. SLS Launch Vehicle | The SLS Block 1B not available for HabEx | Unexpected development problems | The baseline telescope cannot be launched | 5 | 2 | Launch the telescope on an alternate launch vehicle | 5 | 1 |
| 7. Foreign Contribution | The ESA contribution does not materialize | Formal agreement not in place | The HabEx total cost increases | 3 | 2 | Release cost reserve equivalent to contribution | 3 | 1 |

### 9.2.3　Starshade to TRL 5

The Starshade to TRL 5 (S5) project is scheduled for completion in 2023 (*Section 11.2*). Development of prototypes that include all of the desired features has been stretched to 2023 due to S5 funding constraints. If the Astro2020 Decadal Survey recommends a starshade mission, then an accelerated funding profile could move S5 completion to 2022, and provide additional starshade technology development schedule margin

Without the S5 development acceleration, there is already a little more than three years of schedule slack before the start of Phase A for the starshade. Problems in completing S5 milestones would first use this slack. Delays greater than three years could still be addressed with a move out of the starshade launch date. Again, such a delay would not impact the overall baseline mission science return.

### 9.2.4　EMCCDs to TRL 6

The roadmap to develop the electron-multiplying charge-coupled devices (EMCCDs) to TRL 6 (**Figure E-3**) includes multiple activities that are performed sequentially, compounding the risk that delay in any one activity delays the entire development effort. However, the critical path runs through the telescope mirror development as discussed in *Section 9.1*, providing at least one year of slack for the EMCCD development to TRL 5 and three years of slack to TRL 6.

### 9.2.5　Microthruster Lifetime

Like conventional thrusters, there is a limit to the amount of propellant that can be exhausted by the microthrusters before the thrusters cease to function properly. Microthruster lifetime/throughput testing for the LISA mission has not yet reached the levels required for HabEx. If issues arise in the throughput testing, HabEx can follow two paths. First, there is sufficient slack to potentially resolve some testing issues, but if these throughput issues cannot be resolved for the HabEx operational case, then the telescope flight system carries sufficient mass margin for adding additional thrusters so that each individual thruster meets the lifetime specification. As the design for the microthruster manifold at each location has yet to be completed, multiple design options exist to meet this requirement.

### 9.2.6　SLS Launch Vehicle

The SLS Block 1B Cargo is baselined to launch the HabEx telescope, but has yet to launch. There are competing developments underway, most notably the SpaceX Big Falcon Rocket, so it is likely that there will be an alternative to launch the baseline telescope. This risk only applies to the telescope flight system since the starshade is baselined to launch on a Falcon Heavy, which is already operational.

### 9.2.7　Foreign Contribution

The HabEx total cost is offset by a contribution of $565M (FY2020) from ESA (see *Section 9.4*), but formal participation in the mission has yet to be established. This contribution is 43% of the $1309M HabEx reserves, representing a significant but unlikely cost risk.

## 9.3　Development Schedule

The HabEx schedule baseline is centered on minimizing programmatic risk and a realistic anchoring by historical analogies. The 138-month project schedule (Phases A–D) is in family with similar projects (**Table 9.3-1**). Schedule risk is minimized by allowing adequate time for technology maturation with sufficient schedule margin for a mission of this size.

HabEx is intended to follow WFIRST as the next large NASA astrophysics mission. The current WFIRST schedule shows a planned launch in the mid-2020s, which will open up the funding needed to begin the major portion of HabEx's development with formulation beginning shortly before WFIRST's launch.

### 9.3.1　Schedule Description

The HabEx baseline telescope schedule is shown in (**Figure 9.3-1**). This study anticipates a HabEx Phase A start at the beginning of FY2025, project Preliminary Design Review (PDR) in FY2029, and a launch date in FY2036. This will result in a Phase A–D duration of 138 months. Additionally, Pre-Phase A is planned at 36 months resulting in an overall duration for Pre-Phase A–D of 174 months.

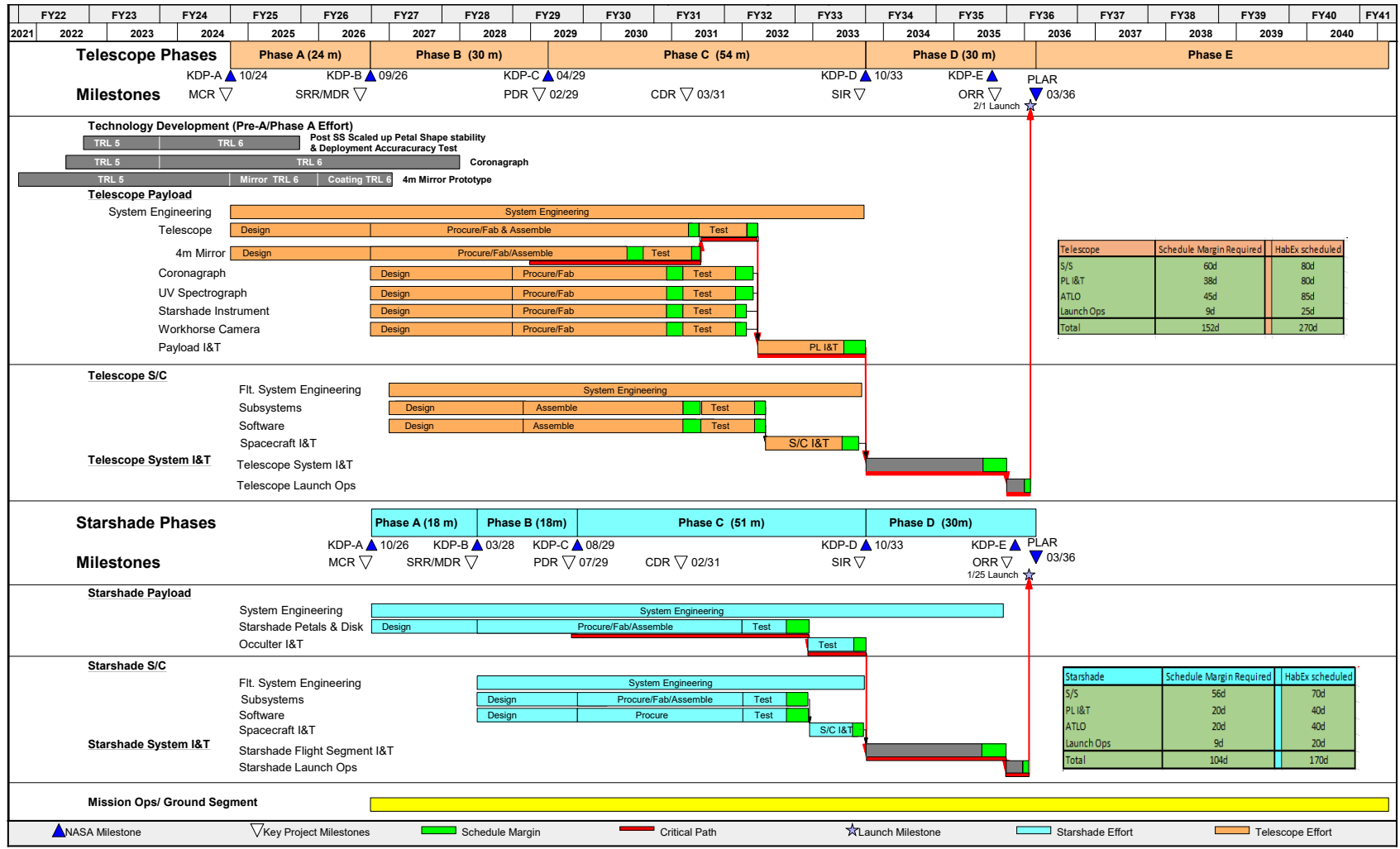

**Figure 9.3-1.** The HabEx baseline schedule has been well-bounded by historical analogs. The critical path flows through the 4 m monolith telescope development.

Project formulation (Phases A and B) runs for 54 months and includes requirements definition, system and subsystem design, and the start of procurements for long-lead items. The project formulation period encompasses the work needed to move all technologies to TRL 6 by PDR.

The flight system implementation (Phases C and D) takes 84 months and includes the fabrication, integration, and test of the two flight systems. The combined Phases C/D schedule also aligns well with historical analogies. Implementation ends with the launch and initial on-orbit checkout (launch date + 30 days). The schedule shows an overall margin of 270 days along the critical path, which exceeds JPL margin best practices for a schedule of this duration by 118 days.

The schedule's critical path runs through the telescope's mirror development. The 4 m mirror development will require building a prototype mirror to verify mirror design and manufacturing processes before fabricating the flight mirror. Following mirror development, the optical telescope assembly (OTA) must be built and tested.

The HabEx Telescope's planned launch vehicle is the SLS Block 1B. The starshade system will be launched separately with the SpaceX Falcon Heavy as the planned vehicle. There are at least 7 SLS Block 1B launches currently planned prior to HabEx. The Falcon Heavy completed its test flight on February 6, 2018, and presently has five planned launches prior to the HabEx launch.

The Starshade Phases A–D development is planned at 117 months with the start of Phase A at the beginning of FY2027. The starshade flight system schedule is largely driven by the starshade occulter's petal and disk development—the critical path for this flight system but not the critical path for the overall HabEx mission. The time required to develop the petals and disk is based on the S5 planned prototype development schedule, which is itself based on starshade expert judgement informed by first and second-generation starshade test units. Additional time has been added to the three-year prototype development to account for the increased size (52 m vs. 26 m) and the added complexity of a flight unit build. The starshade PDR would occur in FY2029. In addition, a 12-month Pre-Phase A is planned, resulting in an overall Starshade Pre-Phase A–D duration of 129 months.

This initial schedule makes no assumptions about available funding levels, and real funding profiles could dramatically affect the project duration and cost. This baseline HabEx development schedule and cost profile would cause some years to exceed projected funding levels if the astrophysics funding remains flat (see *Section 9.5, Affordability,* for details). However, longer development schedules, compliant with the available budget, are possible. Of particular interest, is the option of launching the telescope spacecraft first, then adding the starshade spacecraft to the observatory several years later. This alternative to the baseline mission is discussed in *Section 9.5.2.*

### 9.3.2 Comparison of Schedule to Analogs

**Table 9.3-1** provides a comparison of the HabEx project-level development schedule by phase to other large telescope missions. The telescope development schedule is equivalent to the project-level development schedule and was therefore the assessed duration. The Phase A–D duration for HabEx is 137 months (measured to launch; excludes 30 days of post launch activities).

**Table 9.3-1.** Comparison of the HabEx telescope schedule with historical mission schedule durations.

| Missions | Phase A | Phase B | Phase C | Phase D[(1)] | Total Start–LRD |
|----------|---------|---------|---------|--------------|-----------------|
| HabEx    | 24      | 30      | 54      | 29           | **137**         |
| Average  | 21[(6)] | 23      | 32      | 48[(5)]      | **123**[(5)]    |
| WFIRST   | 24      | 23[(4)] | 46[(4)] | 28[(4)]      | **121**         |
| JWST     | 21      | 20      | 25      | 132[(4)]     | **198**         |
| Spitzer  | 4       | 18      | 42      | 23           | **87**          |
| Chandra  | 19      | 36[(2)] | 13      | 42           | **110**         |
| HST      | 4       | 17      | 32      | 98[(3)]      | **151**         |

Normalization Notes:
1. Phase D durations assume the start of Phase D through launch (removed 30 days of post launch activities from HabEx Phase D duration)
2. Chandra Phase B was normalized to 36 months based on several replans
3. HST Phase D was not normalized due to insufficient understanding of the drivers for the 42-month Challenger accident delay
4. JWST and WFIRST are based on current estimated durations
5. JWST Phase D was removed from the average
6. HST And Spitzer Phase A durations appear to be outliers and were therefore removed from average

The HabEx combined Phases C and D durations (83 months) compares well to the historical mission average (80 months). The JWST Phase D duration was removed from the average calculation; the causes of the delays are not available at this time to enable the normalization of their schedule. The HST Phase D duration was not normalized for the 42-month Challenger accident slip (the original planned launch was October 1986 and the actual launch was April 1990). Separating the portion due to HST technical challenges from the amount that was entirely attributable to the stand down after the Challenger accident was not possible. The HabEx Payload and Spacecraft Integration and Test schedules provide adequate time to find and mitigate risks prior to delivery to System I&T (SI&T). The robust margins held by HabEx also protect against delays to SI&T.

## 9.4 Mission Cost

The HabEx baseline mission is currently estimated at $6.8B FY2020 and $9.1B in real year (RY) dollars (see **Table 9.4-1**). This estimate was developed from representative analogs and the best available cost models to ensure that HabEx scope decisions are seated in historic reality. The team recognizes that it can be difficult to extrapolate existing cost models to specific design challenges for a mission like HabEx, and have tried to adjust accordingly, erring on the side of conservatism, and bound with multiple models or historical actuals when possible. The primary source for the HabEx cost estimate was the JPL Institutional Cost Models (ICMs) from the Advanced Projects Design Team—Team X— which are based on an array of missions developed by JPL. These tools and their estimates are "owned" by the JPL internal organizations responsible for performing or overseeing the work represented in the estimates. HabEx also uses recognized external tools such as the NASA Instrument Cost Model (NICM), Marshall Space Flight Center (MSFC) generated space telescope cost models (Stahl et al. 2013; Stahl et al. 2019), and SEER-H for appropriate portions of the payload costs. Non-hardware portions of the Work Breakdown Structure (WBS) were estimated from average percentages from JPL-managed large mission analogs. Post-launch costs are based on HST and Spitzer actual annual operations budgets.

**Table 9.4-1.** HabEx cost summary.

| WBS Element | FY20$M | RY$M | Cost Basis |
|---|---|---|---|
| **Pre-Phase A** | **59** | **64** | Based on cost needed to advance technologies to TRL 5 |
| **Phase A** | **211** | **253** | Based on cost needed to advance technologies to TRL 6 |
| WBS 01–03 Proj Mgmt/Sys Eng (incl Mssn Design)/SMA | 444 | 589 | Percentage based on Flagship-class missions |
| WBS 04 Science | 113 | 150 | Percentage based on Flagship-class missions |
| WBS 05 Payload System | 1996 | 2643 | |
| P/L Mgmt/Sys Eng | 136 | 180 | Percentage based on Flagship-class missions |
| Coronagraph | 447 | 591 | NICM VIII System Model |
| Starshade Camera | 119 | 158 | NICM VIII System Model |
| UV Spectrograph | 257 | 340 | NICM VIII Subsystem Model |
| Telescope (OTA) | 659 | 872 | Average of Phil Stahl 2019 Multivariable and 2013 Single Variable equation |
| Fine Guider | 29 | 38 | NICM VIII System Model |
| Workhorse Camera | 180 | 238 | NICM VIII System Model |
| Starshade Petals and Disk | 170 | 227 | SEER-H Modeled Cost |
| WBS 06 Flight System + 10 ATLO | 1724 | 2291 | |
| Telescope Bus | 1045 | 1382 | Team X Study, includes Mgmt, SE and ATLO for Telescope Bus |
| Starshade Bus | 680 | 908 | Team X Study estimate for 72 m starshade bus, includes Mgmt, SE and ATLO for Telescope Bus |
| WBS 07/09 MOS/GDS | 85 | 113 | Team X Study |
| **Phase B-D Subtotal** | **4363** | **5785** | |
| Reserves (B–D) | 1309 | 1736 | 30% reserves |
| **Phase B-D w/ reserves** | **5672** | **7521** | |
| LV (Telescope) | 650 | 925 | Costs provided by NASA |
| LV (Starshade) | 300 | 429 | Costs provided by NASA |
| **Phase B-D w/ LV** | **6622** | **8875** | |
| ESA Contribution | -565 | -747 | |
| **Total Phase B-D w/ contribution** | **6057** | **8128** | |
| Operations (Phase E–F) | 400 | 609 | Based on average operations cost for HST and WFIRST |
| Phase E–F Reserves | 60 | 91 | 15% reserves |
| **Total Phase E-F** | **460** | **701** | |
| **Total Pre-Phase A-F** | **6786** | **9145** | |

Additionally, The HabEx total cost includes a contribution of $565M FY2020 ($751M RY) from ESA, which is shown as an offset to the Phases A–D cost. ESA participation in the mission has not yet been formalized, but contributions could include: the primary mirror, which already assumes a German manufactured material in the baseline design; one or two instruments in the payload; the microthruster control system which is based on those used on several ESA-led missions; participation in the U.S.-led coronagraph development; and science team membership.

An initial, high-level, analysis of the HabEx telescope estimate versus the telescope flight system (payload and spacecraft bus dry mass) was conducted to assess the reasonableness of the study development estimate (**Figure 9.4-1**). The starshade cost and mass were not included in this comparison due to the analogy missions being single flight element missions. For this comparison, the starshade occulter and bus were removed from the cost and the remaining wrap factor and level of effort derived costs were reduced proportionally using the same wrap factor percentages. The HabEx development cost (less the launch vehicle) with expended reserves compares well to analogous missions especially when considering that HST was a human-rated spacecraft and had a 4-year launch delay. Further detail in the following sections describe the derivation of, and analysis for, the cost estimate of each WBS element.

### Independent Cost Estimate (ICE)

The JPL Institutional Cost and Schedule Evaluation Office conducted an independent cost and schedule assessment of the HabEx study baseline design. Results of the assessment can be seen in *Appendix G*.

### Escalation Indices

The NASA Inflation Indices FY19 distribution was used to deescalate the costs from real year dollars to FY20. Escalation factors were available for fiscal years up to FY2028. For fiscal years beyond this, an average increase of 2.7% from the previous year was assumed. **Table 9.4-2** presents the escalation factors utilized for HabEx.

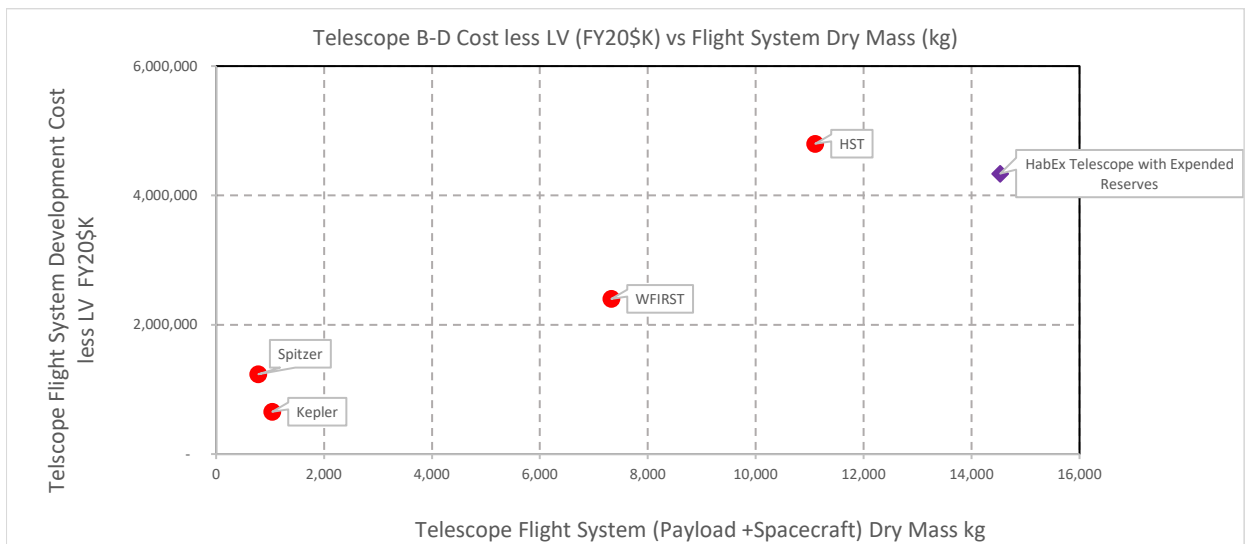

**Figure 9.4-1.** HabEx telescope Phase B–D cost estimate less launch vehicle costs plotted against telescope flight system (bus and payload dry mass) compares well to analog missions (starshade cost and mass were excluded as analogous missions were single flight element missions). HST data point is from Bitten et al. (2019). Note that the WFIRST cost is an approximation and was derived using publicly available data: $3.2B for the NASA directed total lifecycle cost less $500M for operations (approximately $100M per year for 5 years) and less $300M for the launch vehicle (assumed similar cost to Falcon Heavy).

### 9.4.1 Basis of Estimate

JPL's Team X was used to estimate most of the HabEx baseline design. Team X is a JPL concurrent engineering design environment created in 1995. Team members represent all JPL technical disciplines. The Team X Institutional Cost Model suite, developed using a combination of historical mission actuals and engineering expertise, has been approved by JPL implementing organizations and is consistent with JPL institutional guidelines. Model parameters, such as mission complexity, schedule, number of instruments, acquisition approach, parts class, and inheritance, are all considered in the estimate and are evaluated by engineers with expertise in systems being estimated. NICM was used to estimate the cost of the science instruments and two cost estimating relationships (CERs) developed by Philip Stahl were used to estimate the telescope cost. The starshade petals and disk were parametrically modeled in SEER-H. SEER-H inputs are consistent with other starshade estimates such as the Rendezvous Probe-class starshade and the Exo-S starshade. Pre-launch project management, systems engineering, mission assurance, and science support are based on JPL historical cost ratios from several large spacecraft built or managed by JPL.

The launch vehicle costs of $650M FY20 ($925M RY) for the SLS Block 1B and $300M FY20 ($429M RY) for the Falcon Heavy were assumed. The Pre-Phase A and Phase A estimates were derived from the cost of advancing technologies to TRL 5 and 6, respectively (see *Appendix E*, Technology Roadmap). Some of the technologies are/will be funded by current projects (e.g., S5, Technology Development for Exoplanet Missions [TDEM], WFIRST). However, for conservatism the total costs to advance these technologies were included in the HabEx cost estimate.

**Table 9.4-2.** NASA New Start inflation.

| FY2022 | FY2023 | FY2024 | FY2025 | FY2026 |
|--------|--------|--------|--------|--------|
| 1.058 | 1.087 | 1.116 | 1.146 | 1.176 |
| **FY2027** | **FY2028** | **FY2029** | **FY2030** | **FY2031** |
| 1.208 | 1.240 | 1.267 | 1.294 | 1.321 |
| **FY2032** | **FY2033** | **FY2034** | **FY2035** | **FY2036** |
| 1.348 | 1.375 | 1.402 | 1.429 | 1.456 |
| **FY2037** | **FY2038** | **FY2039** | **FY2040** | **FY2041** |
| 1.483 | 1.510 | 1.537 | 1.564 | 1.591 |

### Project Management, Systems Engineering, Mission Assurance, and Science

Phases A–D Project Management, Project Systems Engineering (including Mission Design), Mission Assurance, Science, Payload Management, and Payload Systems Engineering costs are estimated as percentages of all other development costs in the overall WBS, excluding the launch vehicle costs. These percentages were determined by averaging actual percentages from several completed large JPL missions. **Table 9.4-3** reflects the derivation of the percentages used for HabEx. As seen in **Table 9.4-3**, the project-level management of flagship missions (9–11%) tends to be lower than that of smaller missions (12–15%) due to the larger development cost base.

### Science Instruments

All science instruments were modeled using NICM, which is based on actual costs of over 150 completed flight instruments, including their expended reserves. The NICM System tool was used for conservatism to estimate the cost of the coronagraph, starshade instrument, workhorse camera, and fine guidance system. Due to the slow optical telescope feeding the instruments and the need to minimize the number of reflections in the UV spectrograph (UVS) to improve instrument throughput, the few UVS optical elements need to be physically far apart, requiring a great deal of structural mass within the

**Table 9.4-3.** HabEx non-hardware WBS percentages for Phases B–D are in family with other JPL large projects.

| WBS | WBS Element | HabEx | MSL | SMAP | Cassini | Juno |
|-----|-------------|-------|-----|------|---------|------|
| 01, 02, 03, 12 | PM, SE, MA, MD | 10.2% | 10.8% | 9.4% | 9.2% | 11.1% |
| 04 | Science | 2.6% | 1.0% | 3.3% | 2.6% | 3.3% |
| 05.01, 05.02 | PL Mgmt, SE* | 6.8% | 7.1% | 5.8% | 5.8% | 7.7% |

*PL Mgmt and SE are calculated as a cost ratio to the payload total cost.

instrument to support this configuration. As such, the NICM Subsystem tool was used to estimate the cost of the UVS to avoid overestimating the cost due to the large contribution of low-cost structure mass to the overall instrument mass.

Additionally, the reasonableness of the science instrument estimates was further supported by the analysis of the estimates against other historical data points. As seen in **Figure 9.4-2** and **Figure 9.4-3**, the payload instruments were compared against the trend line derived from the cost versus mass and cost versus power plots of various instruments.

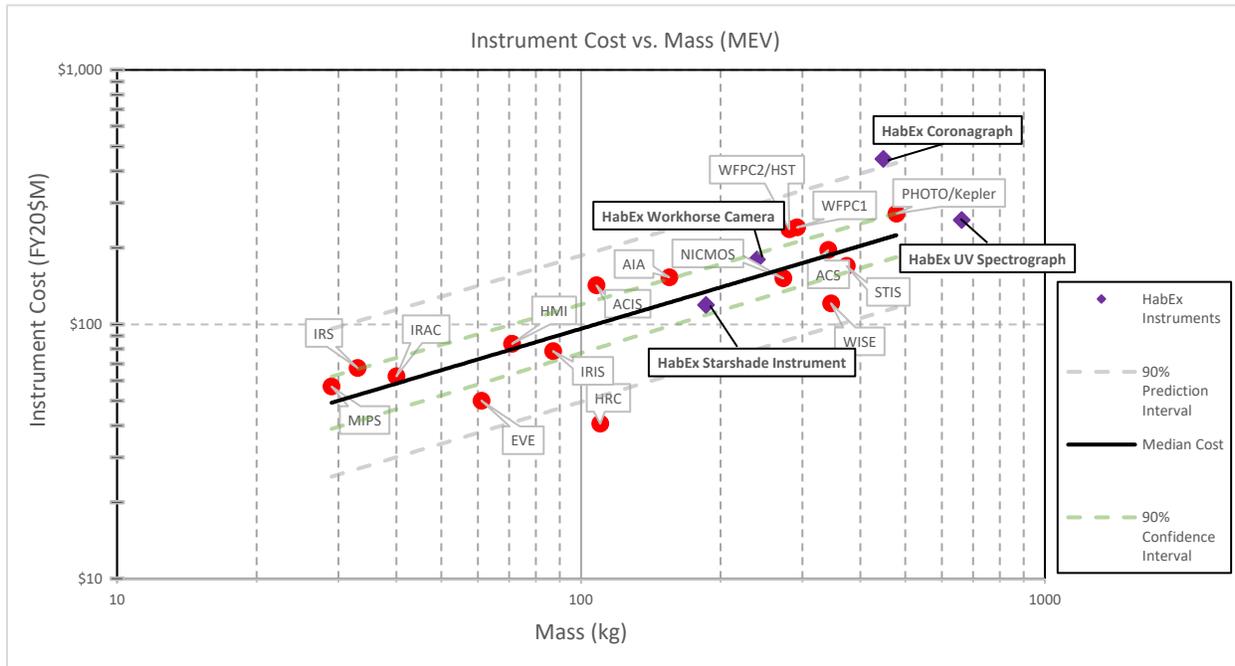

**Figure 9.4-2.** Instrument cost vs. mass (optical space observing instruments). The HabEx instruments clearly show they are in family with similar instruments, or even conservative (above the median line). Additionally, note that the HabEx instruments do not include potential expended reserves.

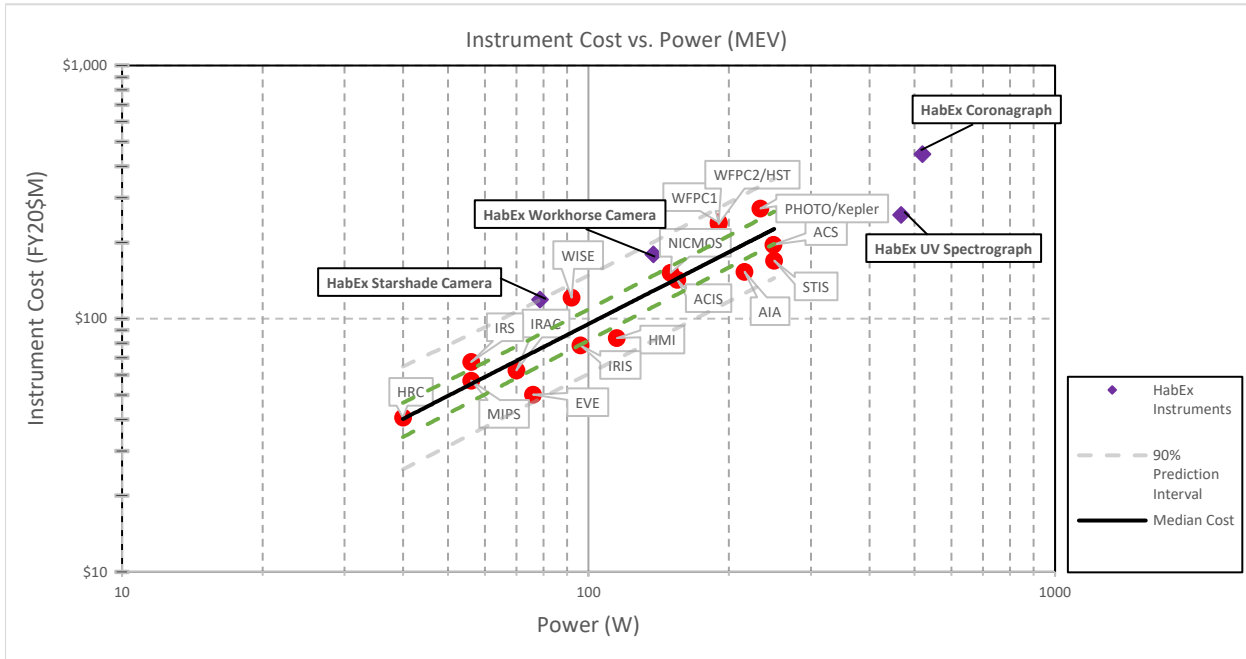

**Figure 9.4-3.** Instrument cost vs. power (optical space observing instruments). The power required of the UVS and coronograph are extended beyond past data, but appear to be along the trend line. Additional analysis of the instruments power (and mass) will continue to refine these values to help anchor them with respect to historical cost and mass/power relationships.

### Telescope OTA

The HabEx telescope OTA estimate is derived by averaging two CERs, which are calculated from statistical fits to historic actual costs. Both CERs were escalated to FY2020 dollars using the NASA Inflation Indices. The first is the single variable (aperture diameter) model given in Stahl et al. (2013). The second is the multi-variable cost estimating relationship given in Stahl et al. (2019). The latest telescope model includes more parameters so it is sensitive to several design characteristics that the older model could not capture, however, the 2013 model has been one of the main telescope cost estimating tools available to the space mission community since its first appearance. Since the two model cost estimates were significantly different, HabEx decided to use the average of the two for all concept estimates.

The telescope OTA cost was assessed against a trend line developed by plotting the OTA costs of various telescopes (Stahl et al. 2013) against their respective aperture diameters (as seen in **Figure 9.4-4**). In this comparison against a single variable, the cost appears to plot slightly low with respect to the trend line. However, when including the proportional expended reserves in the estimate, the HabEx OTA estimate plots on the trend line.

### Starshade Petals and Disk

The starshade payload was estimated using the parametric cost model, SEER-H 7.4. The starshade mass estimates—both for Team X and HabEx—were developed by JPL mechanical engineers with extensive experience in starshade and large deployable space antenna designs. All hardware was modeled as 100% new design. The Truss was modeled using the Spacecraft Structure knowledge base (K-base) while the Petals were modeled using the Secondary Structure K-base. The spokes were modeled using Precision Mechanism and the occulting disk was modeled using the Spacecraft Structure knowledge base for conservatism. The Certification Levels in SEER were increased to the highest settings to reflect Class A mission hardware.

### Spacecraft and ATLO (Telescope and Starshade Bus)

The Team X cost models were used to estimate WBS 06, Flight System, and WBS 10, ATLO (assembly, test, launch, operations), which includes the insight and oversight required to manage the telescope and starshade bus developments, the telescope and starshade bus hardware, and the testbed/integration costs. The telescope and starshade spacecraft buses were designed and costed by Team X with engineers representing each subsystem developing the estimates.

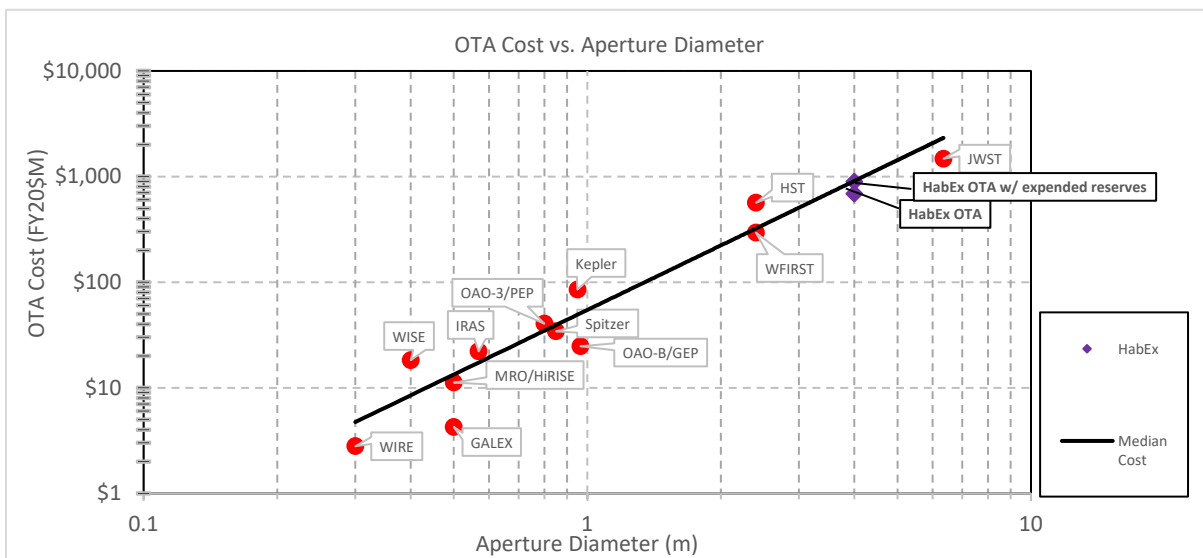

**Figure 9.4-4.** OTA cost vs. aperture diameter shows the detailed, modeled HabEx OTA is in family with historical actuals.

### Telescope Bus

The telescope spacecraft bus cost estimate was developed and vetted by Team X. The high mass of the spacecraft can be largely attributed to the mass of the structures subsystem (67% of total bus dry mass), which is required to support the OTA.

Historically, when comparing $/kg of various spacecraft subsystems, the structures subsystem tends to be the least costly. The SLS Block 1B Cargo launch capacity enables the HabEx telescope bus to increase the static structures mass which increases stability and decreases system complexity. Therefore, when comparing the $/kg of the HabEx telescope bus (including cost of ATLO) to mass constrained analogs, the HabEx telescope bus plots below the trend line, but is still in family with the comparison points (see **Figure 9.4-5**).

### Starshade Bus

The starshade spacecraft bus estimate was developed for a 72 m starshade design used in the HabEx interim report. Reducing the size of the starshade occulter to 52 m had very little effect on the overall bus design since the mass saved in the occulter was largely replaced with propellant making the overall mass about the same for both starshades (around the launch capacity for the Falcon Heavy). A $/kg comparison was conducted for this spacecraft bus (plus ATLO) as well (see **Figure 9.4-6**). It is important to note that the historical datapoints include expended reserves while the HabEx costs do not. Therefore, a second starshade bus cost was plotted to present the estimate with potential expended reserves (30%). The starshade bus cost estimate compares well even to spacecraft, which support science instrumentation, despite being a largely mechanical bus.

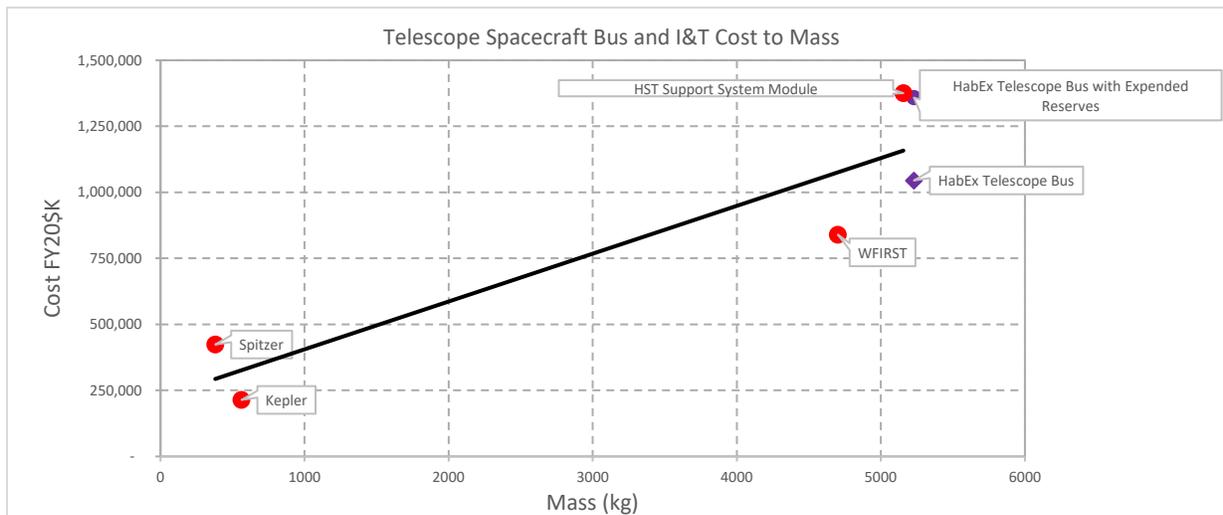

**Figure 9.4-5.** Cost to mass comparison of the HabEx telescope bus with historical telescope buses. Kepler and Spitzer data points were derived from CADRe data. HST cost and mass was gathered from the NASA REDSTAR database.

Note that the WFIRST data point was derived from publicly available data. The following sources were consulted for the approximate data point:

Cost:
- NASA directed lifecycle cost of $3.2B less $500M for 5 years of operations at $100/year and less $300M for LV (assumed cost similar to Falcon Heavy estimate). In order to derive the resulting spacecraft and ATLO cost from the development estimate of $2.4B, the cost percentage breakdown of typical telescope missions presented in Stahl et al. (2013) was employed to allocate 25% of the development cost to the spacecraft and 10% to ATLO.

Mass
- Total observatory dry mass of 7,324 kg was reported in the 2017 NASA WIETR report
- Removed 636 kg for the WFI instrument (reported in the 2015 WFIRST SDT report)
- Removed 1,763 kg for the OTA mass (reported in the 2015 WFIRST SDT report)
- Removed 224 kg for the CGI instrument (HabEx Coronagraph mass estimate divided in half)

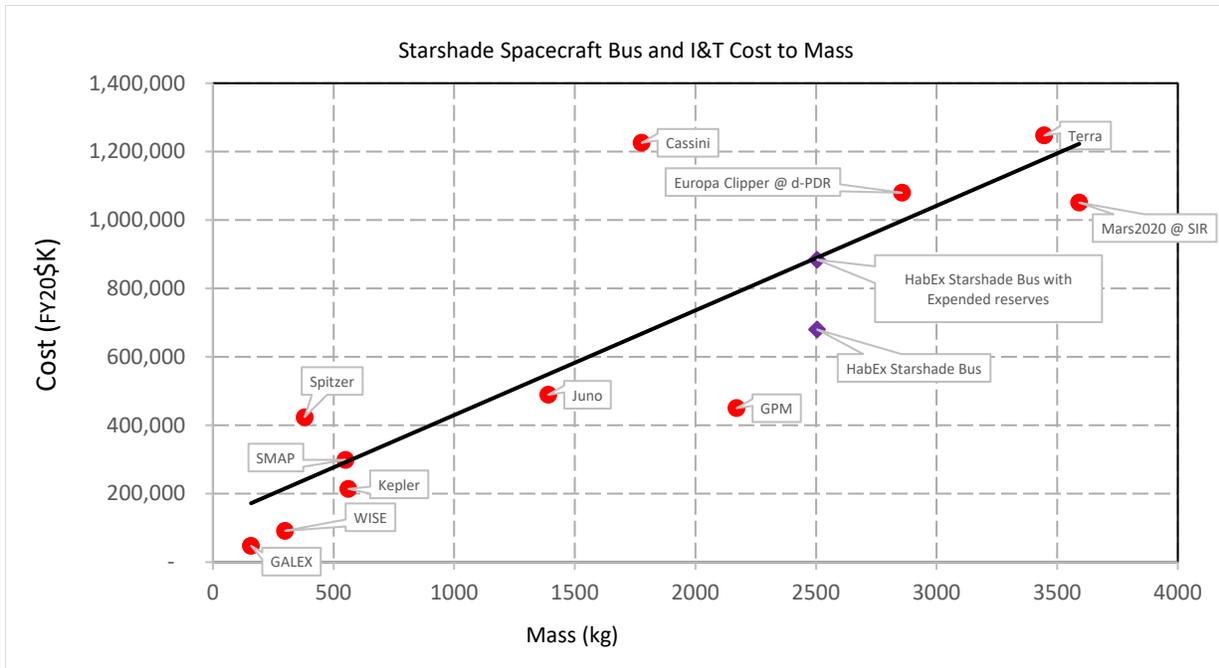

**Figure 9.4-6.** Starshade bus cost-to-mass comparison. All costs derived from CADRe data. Terra and GPM were NASA GSFC missions, but included here as *red points.*

***Mission Operations System and Ground Data System***

The Mission Operations System (MOS) and Ground Data System (GDS) cost estimate was developed by Team X. An in-depth analysis was conducted to evaluate the scope of work needed to meet the HabEx MOS/GDS requirements. The results of this analysis were used to derive the cost estimate. Cost-to-cost ratios for MOS and GDS were not used for this estimate due to the total development cost of this mission being much larger than other historical data points and would therefore result in an overly conservative estimate and would not capture a realistic workforce for this effort.

## 9.5    Affordability

Ambition and affordability can coexist with a balanced program that includes HabEx and a suite of Probe missions. Most major astrophysics missions are tied to a large telescope operating in some portion of the electromagnetic spectrum. The telescopes and their associated missions are usually ambitious, but even the most capable can only serve a subset of the nation's overall astrophysics science community. Preserving balance across all the astrophysics science disciplines remains a major concern for the Decadal Survey and factors into their decisions on which missions to recommend for development in the coming decade. HabEx has adopted this concern for balance in the overall NASA astrophysics program, and has built the concept – its schedule and budget – with an eye toward ensuring that space mission prospects remain available for astrophysics science beyond the fields addressed by HabEx. By adding two new $1B Probe missions per decade and with the existing four Explorer-class missions per decade, NASA's Astrophysics Division (APD) will be able to ensure that all astrophysics disciplines have opportunities to advance their science in a significant way, while a large mission is in development. This is one way to be both ambitious scientifically, and balanced for the community.

With both a large telescope and a separate starshade spacecraft, HabEx is challenged to stay within the assumed flat-projection of the current budget. The baseline concept requires slightly more than the present level of annual funding for development over a schedule only limited by technical issues. However, if the starshade launch follows the telescope by several years, the required funding more closely fits within current constraints without descoping any HabEx science. Both approaches are discussed in this chapter.

### 9.5.1 Baseline Launch Option

The HabEx baseline concept looks to reach its objectives for groundbreaking science, with a manageable overall mission cost, on a timeline consistent with other large space observatories. A profile based on historical data for the percentage of cost expended per phase was used to spread the HabEx costs over the 11.5 years of formulation and development, and the 5 years of operations. A Pre-Phase A start date of FY2022 was assumed based on the expected WFIRST launch date in 2025.

The need to develop the telescope spacecraft and the starshade spacecraft on roughly the same schedule to meet the baseline launch requirement pushes the additional funding need slightly above the current APD estimated funding levels for two years (see **Figure 9.5-1**). The total budget line and the currently planned commitments in **Figure 9.5-1** are based on the material presented by Paul Hertz, the Director of the NASA Astrophysics Division, to the Decadal Survey on Astronomy and Astrophysics on July 15, 2019. Additionally, the sandchart assumes retirement of HST after 2030. Note that in the early 2020s it may be possible to accelerate investments in HabEx, or Probes, to take advantage of the funding opportunities even with a flat projected budget.

### 9.5.2 Delayed Launch Variation

If current levels of APD funding cannot be raised to accommodate the HabEx baseline mission, the mission development could be handled as two non-concurrent launches: the telescope spacecraft developed and launched by 2036, followed by a separately launched starshade that would rendezvous with the telescope 4 years later. The HabEx delayed launch schedule is presented in **Figure 9.5-3**.

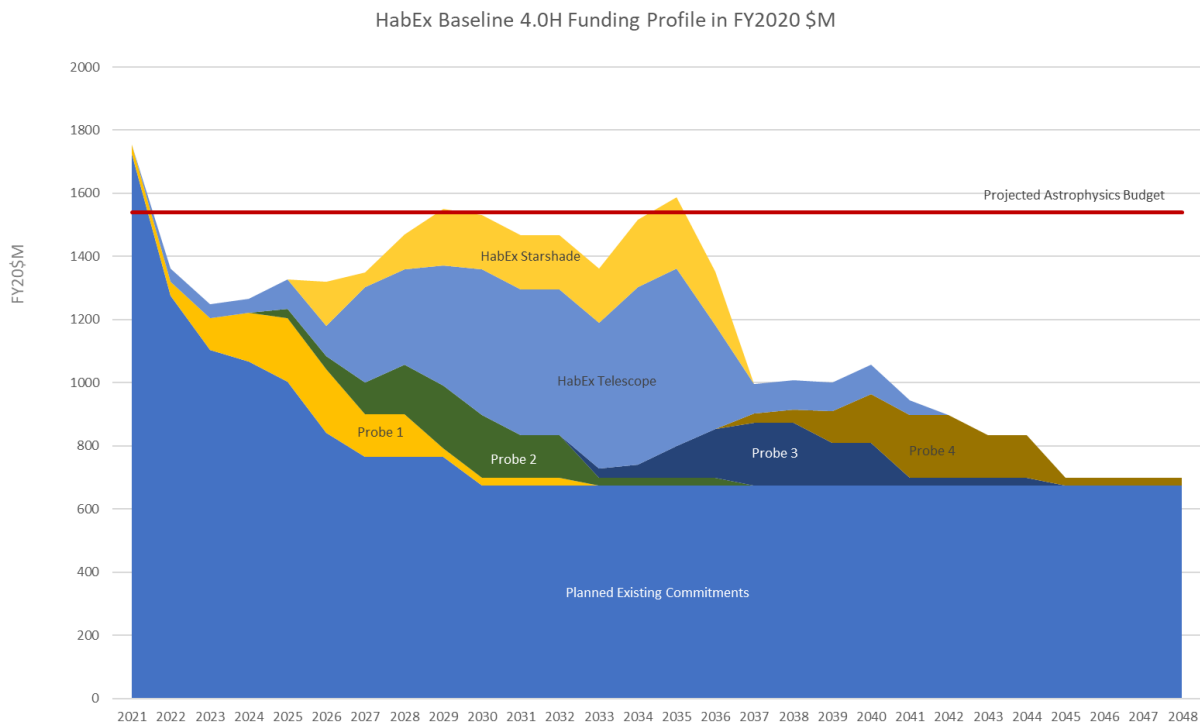

**Figure 9.5-1.** Baseline concept funding profile showing that, for a flat astrophysics budget projection, HabEx annual budgets would exceed current-level yearly allocations. However, such a profile would still assumes a diversified portfolio for NASA, including existing commitments (like Explorers, R&A), and the inclusion of a new line of Probes ($1B).

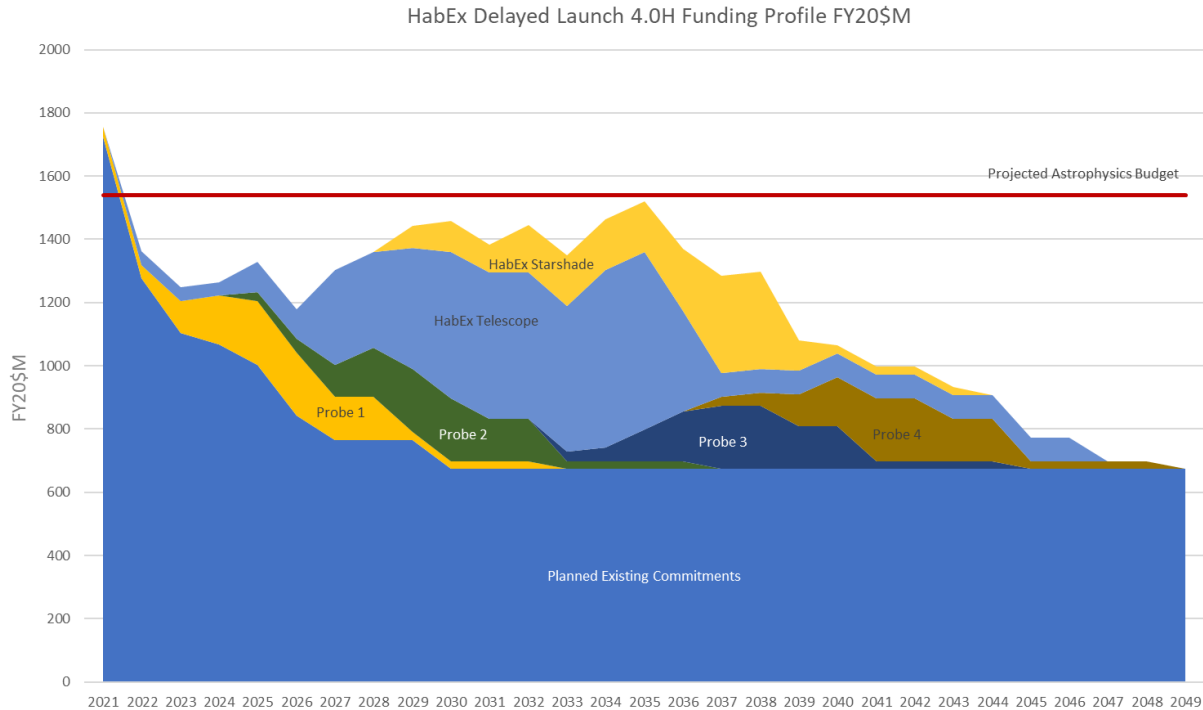

**Figure 9.5-2.** Delayed launch concept funding profile.

This approach has a number of advantages and disadvantages. First, the delayed launch approach will raise the overall mission cost by adding several more years of operations, but it will allow the total cost to be spread over more years, permitting the HabEx development cost profile to more closely fit into currently expected available funding levels (see **Figure 9.5-2**). Second, no baseline science is descoped with this method, although some science will be delayed until the starshade is available. This approach would focus on having the coronagraph detect planets and establish orbits prior to the starshade's arrival. Once in operation, the starshade would conduct characterizations of mostly known target planets. While not the preferred approach for this study, the delayed launch variation serves to illustrate the HabEx architecture's flexibility in addressing funding limitations.

### 9.5.3 Architecture Trades

In order to understand how major elements of the architecture drive cost, in addition to accommodating a less optimistic funding landscape, eight lower-cost variations on the 4H baseline design have been established. Two of these lower cost options, a 4-meter telescope excluding the starshade (option 4C) and a 3.2-meter segmented telescope, which includes the starshade but not the coronagraph (option 3.2S), have been developed in some detail including cost and schedule estimates. These two options are not the preferred back up architectures for HabEx; the HabEx STDT only recognizes a preference for the baseline 4H architecture discussed in this chapter. Instead, these two options are handled in more detail to allow the Decadal Survey's TRACE team to calibrate the HabEx estimates of all nine architectures to their own standards so that the Decadal Survey can consider all options within the HabEx architecture tradespace when determining the right balance of scientific impact and cost. Detailed design descriptions and associated costs and schedules of HabEx 4C and 3.2S are presented in *Appendices A* and *B*, respectively. All nine architecture options are discussed at a higher level in the architecture comparison in *Chapter 10*.

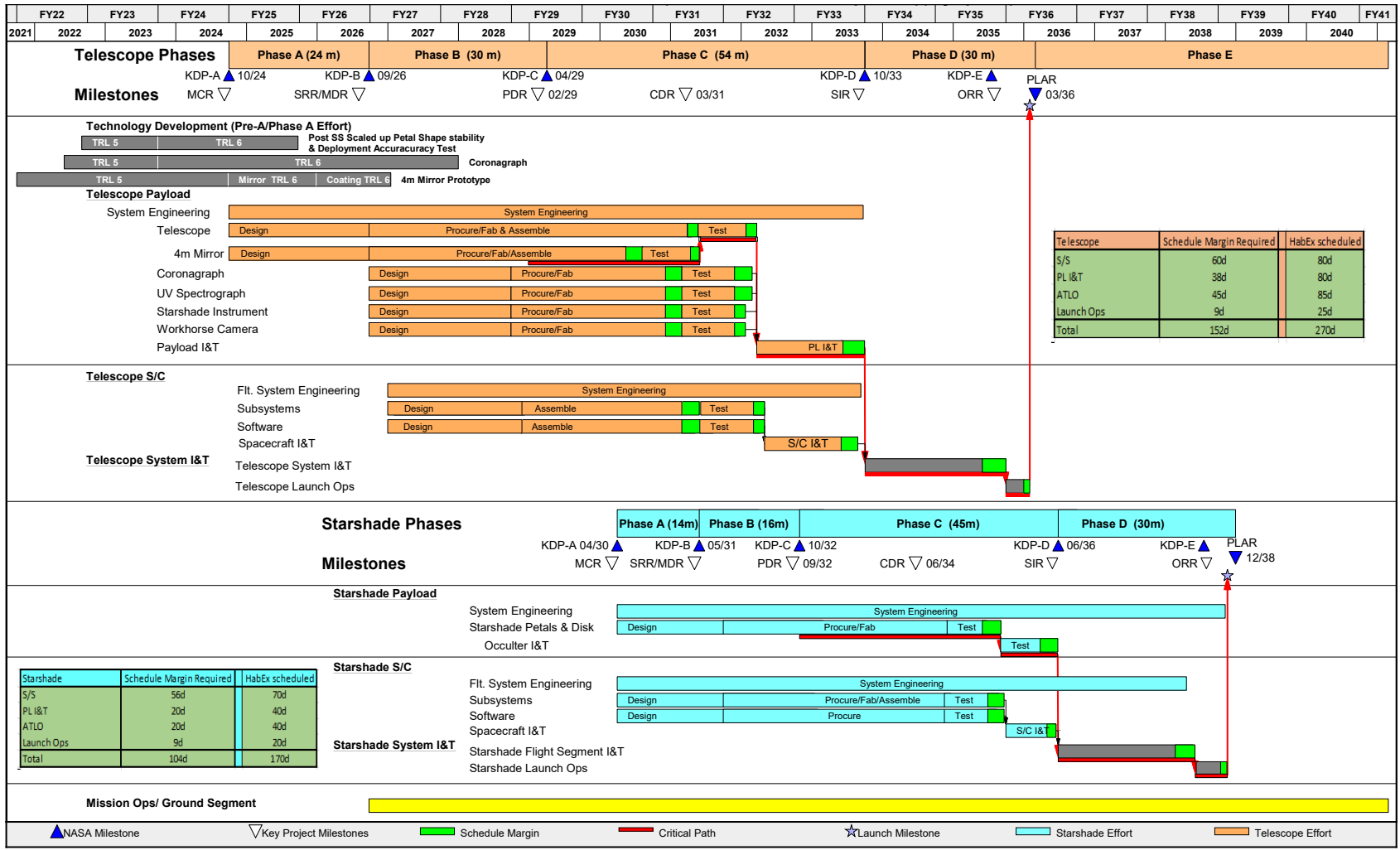

**Figure 9.5-3.** HabEx delayed launch concept schedule.

# 10 ARCHITECTURE TRADES

Science goals often can be reached through a number of different approaches. With each approach underlying a different mission architecture, evaluating the space of potential architectures is often the more useful information in initial mission studies since the programmatic landscape is not well defined and technological roadblocks still remain to be overcome.

Following the release of the large mission interim reports, NASA gave guidance to the four Science and Technology Definition Teams (STDTs) to look into additional lower-cost options. With mission cost closely coupled to aperture size, the HabEx STDT decided to examine the tradespace between the interim report's 4 m design and existing 2.4 m telescope capabilities. The starlight suppression method also had a significant impact on total project cost so it too helped define the dimensions of the architecture tradespace.

The HabEx Interim Report identified the STDT's preferred architecture for a 4 m telescope, one that included both a coronagraph and a starshade (**Figure 10-1**). While this architecture remains the favorite and is the baseline architecture for this report, this chapter presents other options that are responsive to NASA's request for lower cost designs. Two architecture design points, one for a 4 m coronagraph-only case (HabEx 4C) and a 3.2 m starshade-only case (HabEx 3.2S) developed in sufficient detail in *Appendix A* and *Appendix B*, respectively, to support independent cost and risk assessments. Both the general tradespace and two detailed assessments are described in order to help the Decadal Survey and their design assessment team evaluate the science, cost, and risk sensitivities over the full span of architecture options.

The HabEx 4H architecture is the preferred architecture and is baselined in this report. HabEx 4C and 3.2S are not second and third HabEx preferences. They were selected for further study to give a varied and detailed sampling across the studied options. The STDT sees all the options as being able to produce some level of valuable science but has identified the 4 m hybrid as the preferred architecture. With this in mind, apertures for the trade options were defined at the 4 m and 2.4 m range endpoints, as well as a 3.2 m mid-point. The 3.2 m represented some advancement over current space telescopes while carrying lower cost and less fabrication difficulty than a 4 m aperture. In addition to the variations in aperture, and as in the interim report's architecture study, this report's architecture trade also includes various starlight suppression techniques. Starshade-only, coronagraph-only, and a hybrid were defined for each of the three aperture sizes. The combination of the three distinct apertures with the three possible starlight suppression methods defined the nine different architecture options evaluated in this analysis. This chapter contains a description of the nine architectures, some comparative figures-of-merit (FoMs) representing science value, cost, and risk for each architecture, and a discussion of the sensitivities of these FoMs across the nine options.

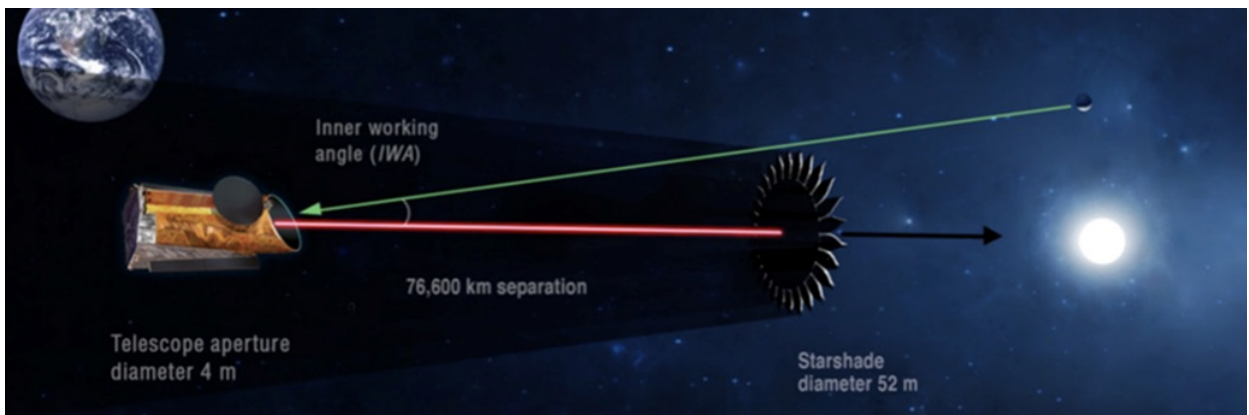

**Figure 10-1.** HabEx 4H baseline architecture.

## 10.1 Architecture Options

The nine HabEx architecture options are summarized in **Table 10.1-1**. Each is briefly described below:

**Option 4H.** The 4 m monolithic unobscured telescope with a coronagraph and a 52 m starshade.

This is the HabEx baseline option. The starshade flies at a 76,600 km separation from the telescope and has an $IWA_{0.5}$ of 58 mas. The telescope is launched on a Space Launch System (SLS) Block 1B due to both mass and volume requirements. The starshade is launched on a Falcon Heavy or equivalent launch vehicle.

**Table 10.1-1.** The HabEx study evaluated nine architectures, with this table summarizing high-level options of the HabEx architectures. Options shown in green are described in more detail outside of this chapter.

| | | Starlight Suppression Method | | |
|---|---|---|---|---|
| | | **H (Hybrid)** | **C (Coronagraph Only)** | **S (Starshade Only)** |
| **Telescope Aperture Diameter** | **4.0 m** | **A.** Off-axis 4.0 m monolithic telescope **B.** SLS launch for telescope, Falcon H for starshade **C.** Coronagraph operates in a number of narrow bands between 0.45–1.80 µm **D.** The Starshade Instrument operates in the 0.20–1.80 µm band **E.** Starshade is 52 m diameter and does 0.30–1.00 µm at 76,600 km with 58 mas IWA<br><br>Coronagraph is used primarily for narrow band detection and orbit determination. Starshade is used primarily for broadband spectral characterization.<br><br>**Baseline Option: HabEx 4H (details in _Chapters 6 & 7_)** | **A.** Off-axis 4.0 m monolithic telescope **B.** SLS launch for telescope **C.** Coronagraph operates in a number of narrow bands between 0.45–1.80 µm.<br><br>Coronagraph is used for both narrow band detection and orbit determination, as well as planet characterization through a series of different narrow band spectral observations.<br><br>**HabEx 4C (details in _Appendix A_)** | **A.** On-axis 4.0 m segmented telescope **B.** SLS or possibly New Glenn launch for telescope, Falcon H for starshade **C.** Starshade Instrument operates between 0.20–1.80 µm **D.** Starshade is 52 m diameter and does 0.30–1.00 µm at 76,600 km with 58 mas IWA<br><br>Starshade gets spectra of all targets first. Only revisits stars with EECs detected to get their orbits. |
| | **3.2 m** | **A.** Off-axis 3.2 m monolithic telescope **B.** Vulcan or New Glenn launch for the telescope, Falcon H for starshade **C.** Coronagraph operates in a number of narrow bands between 0.45–1.80 µm **D.** Starshade Instrument operates between 0.20–1.80 µm **E.** Starshade is 52 m diameter and does 0.30–1.00 µm at 76,600 km with 58 mas IWA<br><br>Coronagraph is used primarily for narrow band detection and orbit determination. Starshade is used primarily for broadband spectral characterization. | **A.** Off-axis 3.2 m monolithic telescope **B.** Vulcan or New Glenn launch for the telescope, Falcon H for starshade **C.** Coronagraph operates in a number of narrow bands between 0.45–1.80 µm<br><br>Coronagraph is used for both narrow band detection and orbit determination, as well as planet characterization through a series of different narrow band spectral observations. | **A.** On-axis 3.2 m segmented telescope **B.** Vulcan, New Glenn or possibly a Falcon H launch for telescope, Falcon H for starshade **C.** Starshade Instrument operates between 0.20–1.80 µm **D.** Starshade is 52 m diameter and does 0.30–1.00 µm at 76,600 km with 58 mas IWA<br><br>Starshade gets spectra of all targets first. Only revisits stars with EECs detected to get their orbits.<br><br>**HabEx 3.2S (details in _Appendix B_)** |
| | **2.4 m** | **A.** Off-axis 2.4 m monolithic telescope **B.** Vulcan, New Glenn or possibly Falcon H launch for telescope, Falcon H for starshade **C.** Coronagraph operates in a number of narrow bands between 0.45–1.80 µm **D.** Starshade Instrument operates between 0.20–1.80 µm **E.** Starshade is 30 m diameter and does 0.30–1.00 µm at 25,000 km with 103 mas IWA<br><br>Coronagraph is used primarily for narrow band detection and orbit determination. Starshade is used primarily for broadband spectral characterization. | **A.** Off-axis 2.4 m monolithic telescope **B.** Vulcan, New Glenn or possibly Falcon H launch for telescope **C.** Coronagraph operates in a number of narrow bands between 0.45–1.80 µm<br><br>Coronagraph is used for both narrow band detection and orbit determination, as well as planet characterization through a series of different narrow band spectral observations. | **A.** On-axis 2.4 m monolithic telescope **B.** Vulcan, New Glenn or Falcon H launch for telescope, Falcon H for starshade **C.** Starshade Instrument operates between 0.20–1.80 µm **D.** Starshade is 30 m diameter and does 0.30–1.00 µm at 25,000 km with 103 mas IWA<br><br>Starshade gets spectra of all targets first. Only revisits stars with EECs detected to get their orbits. |

**Option 4C.** The 4 m monolithic unobscured telescope with a coronagraph and without a starshade. Note that the coronagraph-only architectures do not return exoplanet science below 0.45 μm. Indeed, given the large number of reflecting surfaces required in the coronagraph optical path, the limited reflectivity of Al-based ultraviolet (UV) coatings, and the increased challenge of conducting high contrast coronagraphic observations in the UV (polarization effects and wavefront control in particular), only visible to near-infrared (IR) high contrast observations are available for the coronagraph-only architectures. There is no Starshade Instrument (SSI) in the telescope instrument payload, so the coronagraph must do planet detection, orbit determination, and spectral characterization by itself. The telescope is launched on an SLS Block 1B due to both mass and volume requirements. A more detailed discussion of this option is included in *Appendix A*.

**Option 4S.** The 4 m segmented on-axis telescope with a 52 m starshade and a starshade instrument (SSI). There is no coronagraph in the telescope instrument payload, requiring that the starshade does planet detection, orbit determination, and spectral characterization. The telescope is assumed to launch on a Vulcan launch vehicle. The starshade requires a Falcon Heavy or equivalent launch vehicle.

**Option 3.2H.** The 3.2 m monolithic unobscured telescope with a coronagraph and a 52 m starshade. The starshade flies at a 76,600 km separation from the telescope and creates an $IWA_{0.5}$ of 58 mas from 0.3 to 1 μm. The starshade design and capabilities are thus the same as in the 4H option. The telescope is assumed to launch on a Vulcan but may have mass or volume limitations that could later require a larger fairing or more lift capability. The starshade is launched on a Falcon Heavy or equivalent.

**Option 3.2C.** The 3.2 m monolithic unobscured telescope with a coronagraph. There is no starshade flight system and no SSI in the telescope instrument payload, so the coronagraph must do planet detection, orbit determination and spectral characterization by itself. Note that the coronagraph-only architectures cannot return exoplanet science below 0.45 μm as noted above for the 4C option. The telescope is assumed to launch on a Vulcan but may have mass or volume limitations that could later require a larger fairing or more lift capability.

**Option 3.2S.** The 3.2S option utilizes a segmented off-axis telescope with a 3.34 m diameter and is of equivalent area to a 3.2 m monolith. The architecture includes a 52 m starshade and SSI. There is no coronagraph in the telescope instrument payload, requiring that the starshade does planet detection, orbit determination, and spectral characterization. The starshade flies at a 76,600 km separation from the telescope with an $IWA_{0.5}$ of 58 mas. The starshade design and capabilities are hence the same as in the 4H option. The telescope requires a Vulcan or equivalent launch vehicle but cannot fit in a Falcon Heavy fairing due to the short height of the fairing. The starshade requires a Falcon Heavy or equivalent launch vehicle. A more detailed discussion of this option is included in *Appendix B.*

**Option 2.4H.** The 2.4 m monolithic unobscured telescope with a coronagraph and a 30 m starshade. The starshade flies at a 25,000 km separation from the telescope with an $IWA_{0.5}$ of 103 mas. The separation distance and the starshade size have been adjusted from the 4H design under two constraints. The first one is to maintain the same starshade Fresnel number as used in the 4H option. The second is to keep the coronagraph visible (0.5 μm) $IWA_{0.5}$ and the starshade (0.3–1 μm) $IWA_{0.5}$ roughly equal for effective collaboration in exoplanet identification and characterization using the two starlight suppression methods. The telescope can be launched on a Vulcan. The starshade fits on a Falcon Heavy.

**Option 2.4C.** The 2.4 m monolithic unobscured telescope with a coronagraph. There is no starshade flight system and no SSI in the telescope instrument payload, so the coronagraph must do planet detection, orbit determination, and spectral characterization by itself. Note that the coronagraph-only architectures cannot return exoplanet science below 0.45 μm, as noted above for the 4C option. The telescope can be launched on a Vulcan. The design is similar to the one explored in the Exo-Coronagraph (Exo-C) Extended Study.

**Option 2.4S.** The 2.4 m monolithic on-axis telescope with a 30 m starshade and SSI. Note that the primary mirror is monolithic since there is an extensive history of 2.4 m monolithic space telescopes developed for NASA and other U.S. government agencies. There is no coronagraph in the telescope instrument payload, which requires that the starshade conducts planet detection, orbit determination, and spectral characterization. The starshade flies at a 25,000 km separation from the telescope with an $IWA_{0.5}$ of 103 mas. The telescope is assumed to launch on a Vulcan. The starshade fits on a Falcon Heavy.

### 10.1.1 Fiducial Design Choices and Assumptions

Consistent fiducial design choices were needed across the nine competing architectures in the trade to truly assess the impact of aperture and star light suppression methods on science capability, cost, and risk. Except for the 4H (baseline), 4C (*Appendix A*) and 3.2S (*Appendix B*) options, these designs were not detailed; they only defined what was necessary to assess science yield, cost, and technical maturity (a surrogate for risk) at a coarse level. Engineering trades and operational scenario discussions provided enough insight to allow specification of several design requirements that would be needed no matter which option was selected.

**Orbit Location and Mission Duration.** The first consideration in the architecture trade concepts was where to locate the observatory (i.e., the combined telescope and starshade system). Various Earth orbits were unattractive due to the thermal variability of the orbits and its detrimental impact on coronagraph measurements. In addition, starshade operations were not possible due to the need for large-separation formation flying and long-period target tracking.

Heliocentric Earth trailing and leading orbits were not possible due to the need to make the observatory serviceable—a future servicing mission could not practically reach a telescope in such an orbit years after the initial launch. The ideal location for an exoplanet direct imaging mission would be at the Earth-Sun L2 point as chosen for the James Webb Space Telescope (JWST) and Wide Field Infrared Survey Telescope (WFIRST). This location provides a low disturbance environment, simplifies starshade formation flying and allows the possibility of future observatory servicing. Other Lagrange points are not as advantageous due to their distance from Earth (reduced data volume and more difficult servicing) or, in the case of L1, inferior observing field of regard.

For the architecture trade, a baseline mission of five years, with half the time going to exoplanet direct imaging and the other half going to observatory science, was assumed for the science yield calculations.

**Upper Limit on the Direct Imaging Observing Band.** Determining where to set the direct imaging spectral limit at the long wavelength end of the observing band is a trade between access to desired molecular spectral features and operating temperature for the telescope. 1.8 μm was adopted as the upper limit since it permits the telescope to operate near room temperature. Sensitive operation at wavelengths longer than 1.8 μm would require cooling the telescope well below room temperature, which adds cost and complexity to the integration and test of the payload. More concerning, a cold telescope primary mirror could condense contaminants on mirror surfaces, which would have detrimental consequences for the UV observatory science.

**Lower Limit on the Direct Imaging Observing Bands.** In the architecture trade, the blue-end limits for the coronagraph and SSI were set at different wavelengths. The SSI was set at 0.2 μm since such a limit would allow good characterization of exoplanet atmospheric Rayleigh scattering and disambiguate between different possible sources of absorption short of 0.33 µm (e.g., different ozone concentration levels, other UV absorbents). Starshade Instrument mirrors in one channel are coated in aluminum like the telescope primary and secondary mirrors. The coronagraph limit was complicated by high throughput losses due to its greater number of mirrors and the need to split orthogonal polarizations for high coronagraphic performance. The STDT saw throughput as a significant factor and decided on adopting silver coating for all mirrors within the coronagraph. Silver has a low-end reflectance drop-off starting around 0.45 µm so going down to 0.2 µm is not possible with the coronagraph. Additionally, increased polarization effects and tightened wavefront control requirements make high contrast coronagraphic observations more challenging in the UV than at visible or near infrared wavelengths.

**Direct Imaging Focal Plane Detectors.** The coronagraph and SSI need to have visible and IR detectors to cover the full spectral range. The chosen detectors are a mercury-cadmium-telluride (HgCdTe) avalanche photo diode (APD) device to cover in the near-IR to 1.8 µm, and an electron multiplying charge coupled device (EMCCD) to cover the visible spectrum. The EMCCD can be modified by a delta doping process to extend its sensitivity into the near-UV imaging. Both device types are in production with performance meeting HabEx's direct imaging requirements in the visible and near-IR; the EMCCD is baselined for the WFIRST coronagraph so it will have been flown in space before the HabEx mission. EMCCDs, HgCdTe APDs, and delta doping are all discussed in more detail in *Chapter 11*.

**Telescope Design.** The use of a coronagraph will levy significant requirements on the telescope. Throughput and starlight suppression are greatly improved with an unobscured aperture and monolithic mirror. In addition, contrast degradation due to polarization issues is mitigated to some degree by a design with a greater $f$/number. Accordingly, telescopes designed to support coronagraph imaging are longer, heavier, and carry tighter thermal and mechanical performance requirements than telescopes with the same sized aperture that just support starshade imaging or other astrophysical science. As such, the hybrid and coronagraph-only fiducial designs utilize $f$/2.5 unobscured telescopes with monolithic primary mirrors, while the starshade-only designs use on-axis, $f$/1.3 telescopes with segmented or monolithic primary mirrors. Though the starshade-only telescope could be implemented with either a monolithic or a segmented primary, the segmented was selected for the 4 m and the 3.2 m options as it is lighter weight and the smaller mirror segments would be easier to fabricate than a larger monolithic mirror. In addition, the inclusion of a segmented telescope in the tradespace study illustrates the variety of implementation possibilities that can be used to achieve the HabEx science goals. It should be noted that all nine options are non-deployed optical systems, regardless if they are monolithic or segmented, as deployment adds significant cost, complexity and technical risk to the design.

**Mirror Material.** Early design trades included a look at the best material for a large monolithic mirror to support coronagraphy. The options were Corning ULE® or Schott Zerodur®. The Zerodur® had an advantage in better thermal stability and homogeneity but the closed-back ULE® design made for a stiffer, lower mass mirror, which would be better for rejecting mechanical disturbances. The possibility of using microthrusters and eliminating the major mechanical disturbance shifted the decision toward Zerodur® for the monolithic mirror case. However, a starshade-only telescope does not need to meet the very tight thermal and mechanical tolerances required for coronagraphy. In fact, the mirror does not even need to be monolithic. In this case, the stiffer closed-back ULE® material simplifies fabrication and designing for the launch environment, so it is the preferred mirror material for the starshade-only architectures.

**Mirror Coatings.** Primary and secondary telescope mirrors are used by both the direct imaging instruments and the observatory science instruments. The STDT held extensive discussions on what coating material should be used on the two mirrors. Aluminum was traded against silver for the reflecting material. Silver did not permit observing in the UV and early simulations indicated that silver introduced polarization errors more readily than aluminum leading to inferior coronagraph contrast performance. Although silver offered slightly better reflectivity across the visible band, the STDT gave priority to contrast and UV performance and chose aluminum for the primary and secondary mirrors.

UV observatory science preferred an aluminum mirror coating with a protective overcoat that extended the UV spectral cut-off as far into the UV as possible. A number of overcoats were discussed including magnesium-fluoride, lithium-fluoride, and lithium-fluoride/magnesium-fluoride. Magnesium-fluoride is the overcoat used on the Hubble Space Telescope (HST) so it has a proven operational life approaching 30 years, but it also has a sharp observational cutoff at about 0.115 µm. The other coatings promise useable reflectance below 0.110 µm with some reaching down to 0.100 µm, but none have yet demonstrated the desired lifetime stability. For this study, the STDT elected not to add a new technological development with the mirror coating, and chose to use aluminum with a magnesium-fluoride overcoat like HST. More details on the mirror coatings can be found in *Chapter 11*.

**Observatory Science Instruments.** As noted earlier, the STDT identified a number of additional observatory science objectives (9 to 17) that could be realized with a large space telescope. The highest priority objectives were associated with two different instrument types: an ultraviolet spectrograph and a general purpose, or workhorse, camera with a spectrometer, operating in a spectral band from the visible to the near-IR. These instruments were compatible with each of the nine architecture options.

**Direct Imaging Instrumentation.** The two types of direct imaging instruments are coronagraphs using the internal occulting method of starlight suppression, and the camera supporting the external occulting starshade.

Within the coronagraphs there are a number of different internal occulting methods that can be used. Early simulations showed that the vector vortex design (VVC) would be less sensitive to telescope thermal and mechanical disturbances than the hybrid Lyot coronagraph (HLC). The VVC6 design was far less sensitive to telescope mirror rigid body motion but at the cost of a larger IWA (about 2.4 $\lambda$/D) than a lower charge VVC (VVC2 or VVC4). The VVC8 is even more immune to telescope disturbances than the VVC6 but again, the IWA will increase and there would be some decrease in the number of reachable habitable zones (HZs). The VVC6 was assumed for the final report architecture trades based on early results from simulations presented in the interim report. The SSI was not traded; it was designed to support a specific starshade size (52 m), a constant IWA over a 0.3–1.0 µm instantaneous band, with the option to move the starshade closer (near IR) or further away (UV) to cover the full 0.2 to 1.8 µm range.

**Starshade Size.** Starshade sizing is fundamentally a trade between diameter, IWA, and contrast level for fixed observational bands. HabEx set a primary objective of being able to characterize from 0.3–1 µm on a single target visit. Capturing such a broad spectrum reduces the number of starshade visits needed to complete detailed spectral characterization of a planetary system, and greatly increases the chance of finding evidence of atmospheric gases associated with life during the baseline 5-year mission. Early yield trade studies suggested an $IWA_{0.5}$ of about 80 mas would produce enough habitable zones to deliver a total exo-Earth candidate (EEC) completeness greater than 20 over the mission and a high enough (>98%) probability of spectrally characterizing at least one (*Section 3.1.1*), a number considered compelling by the STDT. Coupled with the need for $10^{-10}$ starlight suppression required to detect and characterize Earth-sized planets in the habitable zone, the architecture fiducial designs were driven toward starshades in the 50 m range. A 52 m design suitable for launch on a Falcon Heavy, or a more capable launch vehicle, was assumed for the 4.0 m and 3.2 m architecture options.

For the 2.4 m telescope diameter options, the approach was to keep the starshade (0.3–1 μm) $IWA_{0.5}$ equal to the coronagraph visible $IWA_{0.5}$ (at 0.5 μm), so that any planet detected by the coronagraph may be spectrally characterized by the starshade, while keeping the same Fresnel number at 1 μm, to preserve high contrast capabilities. As a result, a 30 m starshade flying at a separation of 25,000 km was assumed, producing an $IWA_{0.5}$ of 103 mas. This drove to a 30 m starshade flying 25,000 km away from the telescope.

The lower mass of the smaller starshade coupled with shorter retargeting distances allow for a greater number of retargets but at the cost of a larger IWA, which sacrifices the distant habitable zones that were already hard to observe due to the reduced telescope aperture.

**Starshade Propulsion System and Propellant Mass.** Launch vehicle delivered mass capability sets a constraint on overall starshade spacecraft mass. With a dry mass that is largely fixed and determined by the starshade size, launch capacity closely connects to available starshade propellant mass. Available propellant, starshade-telescope separation distance, propulsion system $I_{SP}$ and thrust capability are major factors in determining the maximum number of retargetings possible within the baseline 5-year mission, and accordingly, the science yield of a starshade-only options. For all options utilizing a starshade, the constraining launch vehicle was assumed to be a Falcon Heavy. The propulsion system is assumed to be a Hall-effect electric propulsion system with a specific impulse of 3,000 seconds and a maximum thrust of 0.52 N per thruster.

**Launch Vehicles.** In the timeframe of a future HabEx mission, the likely launch vehicles that will be available to the mission would be the SLS, SpaceX Big Falcon Rocket (BFR) and Falcon Heavy, the Blue Origin's New Glenn, and the United Launch Alliance's (ULA's) Vulcan. Of the new generation of launch vehicles, only the SpaceX Falcon Heavy is currently operational. The Delta IV Heavy is available today but that launch vehicle will be replaced by ULA's Vulcan launch vehicle long before any HabEx mission launch. For the architecture trade, the Vulcan, Falcon Heavy and the SLS Block 1B were adopted as the fiducial launch vehicles and set the mass and volume constraints for the trade.

## 10.2 Figures of Merit and Tradespace Evaluation

Once tasked with identifying lower cost options than the interim report baseline, the STDT decided against prioritizing the final report architecture options beyond the preferred 4 m hybrid baseline option. Each of the additional options could be viewed as viable under different budget conditions and future mission landscapes. HabEx decided the best way to support the Decadal review was to present the tradespace sensitivities and allow the Decadal Survey to identify which alternative architecture best fits their expectation of the future budget environment should the environment be seen as too restrictive to support the baseline option. As such, HabEx compiled several important FoMs for each architecture option so that the advantages and disadvantages of each could be readily compared. This chapter discusses these FoMs and the pros and cons for each architecture.

As in the interim report, the dimensions of the tradespace remain performance, risk, and cost. To gauge where each option stands within this space, a number of FoMs were defined and estimated. Exoplanet science performance was sized based on four parameters: the number of exo-Earth candidates (EECs)—Earth-sized planets in the habitable zone that were detected and spectrally measured—and the number of rocky, sub-Neptune, and giant planets detected over the course of the 5-year primary mission. It is important to note that for all apertures, the reported yields are based on optimizing the observation plan for the detection and characterization of EECs; a search optimized for gas giants could yield significantly more of these planets, especially for the smaller apertures considered. General observatory performance is more difficult to assess based on specific science goals since there are so many possible goals that can be addressed with a given space telescope. Instead of science goals, effective collecting area and spatial resolution were used to compare telescope performances as a surrogate to observatory science. Risk was evaluated using both the number of technologies at a low state of readiness and the number and type of launch vehicles. The NASA Technology Readiness Level (TRL) scale was used to assess TRLs of technologies used in all architectures, finding that no required technologies had TRLs lower than TRL 4. The number of TRL 4 required technologies was then included to help differentiate the relative risks between the different options. Architecture options requiring two launches rather than one carry some added risk. In addition, options requiring a unique (specific) launch vehicle have some programmatic risk since those options then become tied to that launch vehicle's development challenges. Finally, each architecture had a high-level cost estimate generated based on its fiducial design. A more detailed description of each of these FoMs and the methods used for their calculation follows.

**Habitable Zone Earth-sized Planet Detections and Characterizations.** The primary exoplanet science objectives for HabEx revolve around the detection and spectral characterization of EECs. While different starlight suppression systems and telescope apertures have different intrinsic strengths in terms of characterization potential (e.g., measuring orbits vs. spectra, inner vs. outer planets), a common FoM was adopted here for the purpose of architectures comparison. For each option, the first quoted FoM corresponds to the number of EECs detected, with orbits determined (through an average of 6 visits to guarantee detection at 4 epochs) and with $R = 70$ spectra obtained over the 0.45–1 µm region (with SNR = 10 per spectral bin). For all cases, the FoM was estimated using the Altruistic Yield Optimization (AYO) tool described in *Appendix C*, optimizing the observation plan *for the detection and characterization of EECs*. It is worth noting that optimizing the observing sequence and target list for different planet types would have resulted in different relative capabilities between habitable exoplanets characterization (Goal 1) and broad studies of exoplanetary systems (Goal 2). In the case of starshade options, EEC spectral characterization is obtained at once from 0.3–1 µm, hence including additional near UV characterization from 0.3–0.45 µm at $R = 7$. For starshade-only options, orbits are determined after spectral characterization. For target stars in the broad exoplanet survey, this means that multi-epoch visits and subsequent orbit determination will only occur for systems in which an EEC was detected, based on its single epoch visit location, flux and spectral features.

For coronagraph-only options, there is no UV capability, and the 0.45–1 µm visible spectra range are obtained through separate observations in four 20% bandwidth filters, one at a time. In all cases, yield estimates assumed 2.5 years of exoplanet detection and characterization science (time allocations from **Figure 3.3-4**), with no observations prior to HabEx. Eta-Earth used the 0.24 value produced by the Exoplanet Program Advisory Group (ExoPAG) SAG 13 report (Belikov 2017), as used in the HabEx interim report. Also, yield calculations assume a distribution of exozodi levels per star consistent with the Large Binocular Telescope Interferometer (LBTI) survey results.

**Overall Exoplanet Yields.** To enable the tradespace assessment to evaluate options on exoplanet science beyond just Earth-sized planets in the HZs of nearby stars, the total number of exoplanets detected was also used in the architecture sensitivity evaluation. AYO was also used to determine this FoM with the same Eta-Earth and exozodi assumptions as used in the HZ Earth-sized planet FoM. Detections assumed a signal-to-noise ratio (SNR) of 7 over a broadband in the visible. Each detection represents a unique planet. The total number of detections was further fragmented into three planet size bins: rocky planets (0.5–1.75 $R_\oplus$), sub-Neptunes (1.75–3.5 $R_\oplus$). and giant planets (3.5–14.3 $R_\oplus$). For architectures that include a coronagraph, a larger fraction of these detected planets will have orbits determined. For architectures that include a starshade, a larger fraction of these detected planets will have broadband spectra taken.

**Telescope Effective Collecting Area.** This parameter is derived from the telescope's collecting area multiplied by the ultraviolet spectrograph (UVS) instrument's throughput and detector quantum efficiency, with both assessed at 0.2 µm. This FoM is most useful in assessing the architecture's "goodness" with respect to ultraviolet astronomy.

**Telescope Spatial Resolution.** The angular resolution is set by the diffraction limit at 1.22 $\lambda/D$. For HabEx, the telescope is designed to be diffraction limited at 0.4 µm, so the angular resolution is simply a function of the primary mirror diameter.

**Number of Low TRL Required Technologies.** All new required technologies must be at TRL 5 by the start of Phase A for a new project. HabEx does not have any required technologies below TRL 4 for any of the options in the tradespace. Assessing technology risk was carried out by simply tabulating the number of technologies *currently at TRL 4* for each option.

**Number and Type of Launch Vehicles.** In the timeframe of the next large astrophysics mission following WFIRST, the current set of U.S. heavy lift launch vehicles will no longer be in service. A new generation of heavy lift and super heavy lift launch vehicles are currently in development and will be operational long before a launch date in the 2030s. While there is little risk that no launch vehicle will be ready, reliance on a specific single launch vehicle could pose some risk if that vehicle's development were to be delayed or halted.

Architectures with starshades require two launches, which adds some risk over the single-launch concepts since both launches must be successful to carry out the mission.

**Cost.** All costs are in $FY20 and are total lifecycle costs for the 5-year primary mission only. An initial estimate of all the architecture options in the tradespace was conducted using cost and technical evaluation-like (CATE-like) tools and data. Instruments were estimated using the latest, publicly released version of the NASA Instrument

Cost Model (NICM). NICM input parameters were taken from the report's assessment of instrument mass, power, and data rate. Telescope costs came from the averaging of cost models in "Update to single variable parametric cost models for space telescopes" model (Stahl et al. 2013) and "Optical Telescope Assembly Cost Estimating Model" (Stahl et al. 2019). Starshade petals and disk costs were developed from a SEER-H model run of the hardware. Spacecraft costs came from JPL's Team X's estimate of the HabEx baseline design at the time of the interim report. The telescope bus cost was scaled down for the 3.2 m and the 2.4 m apertures. Operations costs were assumed to be $80M per year (excluding reserves) based on HST and Spitzer actual costs. Other ancillary project costs using percentages taken from past CATE estimates. The SLS block 1B was assumed to cost $650M and the other launch vehicles were assumed to cost $300M based on NASA guidance. Technology development costs were not included in the estimates for this architecture tradespace sensitivity study. NASA has received communication from the European Space Agency (ESA) that ESA would like to participate in the next large NASA Astrophysics mission and is prepared to contribute up to €500M (~$565M) toward the mission. The cost estimates for all options assume this contribution and have been reduced by the $565M figure. The entire amount was assumed to be contributed toward the telescope spacecraft and not the starshade spacecraft.

## 10.3  Results

Overall results for the nine architecture options considered and all figures of merit adopted are summarized in **Table 10.3-1**.

### 10.3.1  The Hybrid Architectures

The hybrid options include the baseline option, HabEx 4H, and HabEx 3.2H and 2.4H options. They offer the most efficient exoplanet science strategies for each aperture size in the tradespace since these architectures utilize both exoplanet imaging systems. With both the coronagraph and the starshade, the hybrid architectures can handle detections and orbit determinations with the agile coronagraph, and spectral characterizations with the broadband, high-throughput starshade system, producing the highest planet yields at a given telescope size.

Each of the hybrid options consist of a telescope and a starshade spacecraft in a halo orbit at L2. The two flight systems are launched to L2 on separate launch vehicles. The telescope supports four instruments: the coronagraph, the SSI, the HabEx Workhorse Camera, and the Ultraviolet Spectrograph. Most of the telescope design requirements are driven by the HabEx

**Table 10.3-1.** Rough estimates of the exoplanet science yields, cost, and technological development attached to each of the HabEx evaluated architectures. *Note that for exo-Earth yield, the number count describes exo-Earths with orbits and spectra characterized. In all cases, a 5-year mission was assumed, with a 50/50 time split between exoplanet surveys and observatory science.

| | | Starlight Suppression Method | | |
|---|---|---|---|---|
| | | **H (Hybrid)** | **C (Coronagraph Only)** | **S (Starshade Only)** |
| **Telescope Aperture Diameter** | **4.0 m** | Exoplanet Yield:<br>　Exo-Earths: **8\***<br>　Rocky: **55**<br>　Mini-Neptunes: **60**<br>　Giants: **63**<br><br>Effective Collecting Area (0.2 μm): **$10^4$ cm² (14× HST)**<br><br>Spatial Resolution (0.4 μm): **25 mas**<br><br># of TRL 3 Technologies: **0**<br># of TRL 4 Technologies (2019): **13**<br># of Launch Vehicles: **2**<br>Unique New Launch Vehicle: **Yes**<br><br>Cost ($B FY20): **$6.8B** | Exoplanet Yield:<br>　Exo-Earths: **5\***<br>　Rocky: **34**<br>　Mini-Neptunes: **39**<br>　Giants: **41**<br><br>Effective Collecting Area (0.2 μm): **$10^4$ cm² (14× HST)**<br><br>Spatial Resolution (0.4 μm): **25 mas**<br><br># of TRL 3 Technologies: **0**<br># of TRL 4 Technologies (2019): **10**<br># of Launch Vehicles: **1**<br>Unique New Launch Vehicle: **Yes**<br><br>Cost ($B FY20): **$4.8B** | Exoplanet Yield:<br>　Exo-Earths: **5\***<br>　Rocky: **29**<br>　Mini-Neptunes: **48**<br>　Giants: **63**<br><br>Effective Collecting Area (0.2 μm): **$1.3 \times 10^4$ cm² (18× HST)**<br><br>Spatial Resolution (0.4 μm): **25 mas**<br><br># of TRL 3 Technologies: **0**<br># of TRL 4 Technologies (2019): **9**<br># of Launch Vehicles: **2**<br>Unique New Launch Vehicle: **No**<br><br>Cost ($B FY20): **$5.7B** |
| | **3.2 m** | Exoplanet Yield:<br>　Exo-Earths: **5\***<br>　Rocky: **33**<br>　Mini-Neptunes: **36**<br>　Giants: **36**<br><br>Effective Collecting Area (0.2 μm): **$6.4 \times 10^3$ cm² (9× HST)**<br><br>Spatial Resolution (0.4 μm): **31 mas**<br><br># of TRL 3 Technologies: **0**<br># of TRL 4 Technologies (2019): **12**<br># of Launch Vehicles: **2**<br>Unique New Launch Vehicle: **No**<br><br>Cost ($B FY20): **$5.7B** | Exoplanet Yield:<br>　Exo-Earths: **3\***<br>　Rocky: **24**<br>　Mini-Neptunes: **29**<br>　Giants: **30**<br><br>Effective Collecting Area (0.2 μm): **$6.4 \times 10^3$ cm² (9× HST)**<br><br>Spatial Resolution (0.4 μm): **31 mas**<br><br># of TRL 3 Technologies: **0**<br># of TRL 4 Technologies (2019): **9**<br># of Launch Vehicles: **1**<br>Unique New Launch Vehicle: **No**<br><br>Cost ($B FY20): **$3.7B** | Exoplanet Yield:<br>　Exo-Earths: **4\***<br>　Rocky: **23**<br>　Mini-Neptunes: **40**<br>　Giants: **56**<br><br>Effective Collecting Area (0.2 μm): **$8.2 \times 10^3$ cm² (11.5× HST)**<br><br>Spatial Resolution (0.4 μm): **30 mas**<br><br># of TRL 3 Technologies: **0**<br># of TRL 4 Technologies (2019): **9**<br># of Launch Vehicles: **2**<br>Unique New Launch Vehicle: **No**<br><br>Cost ($B FY20): **$5.0B** |
| | **2.4 m** | Exoplanet Yield:<br>　Exo-Earths: **3\***<br>　Rocky: **19**<br>　Mini-Neptunes: **27**<br>　Giants: **30**<br><br>Effective Collecting Area (0.2 μm): **$2.3 \times 10^4$ cm² (3.2× HST)**<br><br>Spatial Resolution (0.4 μm): **42 mas**<br><br># of TRL 3 Technologies: **0**<br># of TRL 4 Technologies (2019): **11**<br># of Launch Vehicles: **2**<br>Unique New Launch Vehicle: **No**<br><br>Cost ($B FY20): **$4.8B** | Exoplanet Yield:<br>　Exo-Earths: **1\***<br>　Rocky: **7**<br>　Mini-Neptunes: **10**<br>　Giants: **10**<br><br>Effective Collecting Area (0.2 μm): **$2.3 \times 10^4$ cm² (3.2× HST)**<br><br>Spatial Resolution (0.4 μm): **42 mas**<br><br># of TRL 3 Technologies: **0**<br># of TRL 4 Technologies (2019): **8**<br># of Launch Vehicles: **1**<br>Unique New Launch Vehicle: **No**<br><br>Cost ($B FY20): **$3.1B** | Exoplanet Yield:<br>　Exo-Earths: **2\***<br>　Rocky: **14**<br>　Mini-Neptunes: **25**<br>　Giants: **28**<br><br>Effective Collecting Area (0.2 μm): **$3.0 \times 10^4$ cm² (4.1× HST)**<br><br>Spatial Resolution (0.4 μm): **42 mas**<br><br># of TRL 3 Technologies: **0**<br># of TRL 4 Technologies (2019): **8**<br># of Launch Vehicles: **2**<br>Unique New Launch Vehicle: **No**<br><br>Cost ($B FY20): **$4.0B** |

Coronagraph and include: an unobscured, off-axis configuration; aluminum-coated monolithic primary and secondary mirrors; a $f/2.5$ optical design; and very tight mechanical and thermal stability. None of the telescope flight systems include reaction wheels. Slewing is handled propulsively with a hydrazine monopropellant system and fine pointing stability control is provided by microthrusters. The coronagraph can observe in the 0.45–1.8 µm band but at small fractions of the band for each observation due to contrast requirements. The starshade flies in formation with the telescope when observing. The SSI can observe from 0.3–1.8 µm with the 0.3–1 µm band as the primary observing spectral range; the 0.3–1 µm band can be characterized in a single observation.

### 10.3.1.1 HabEx 4H: The Baseline Option

**Design.** *Chapters 6, 7,* and *8* give a detailed description of the HabEx baseline design. For the purposes of this architecture tradespace study, a simplified description of the design was employed. The baseline includes a 52 m starshade flying at a separation of 76,600 km from a 4 m unobscured telescope with a monolithic primary mirror. The telescope flight system (**Figure 10.3-1**) must launch on an SLS Block 1B or the SpaceX BFR launch vehicle due to volume and mass considerations. The starshade can launch on the Falcon Heavy or any other launch vehicle that can meet or exceed the its deliverable mass capability; for the purposes of the trade a Falcon Heavy was assumed since it is the most mature of the new generation of launch vehicles.

**Science Yield.** Using the assumptions defined in *Appendix C* and instrument performance models defined in *Chapter 6*, it is estimated that HabEx will detect and characterize the orbits of 55 rocky planets (radii between 0.5–1.75 $R_\oplus$), with about 15 of them located in the HZ and around 8 small enough (<1.4 $R_\oplus$) to be possible HZ Earth analogs, 60 sub-Neptunes (1.75–3.5 $R_\oplus$), and 63 gas giants (3.5–14.3 $R_\oplus$). As can be seen in **Figure 3.3-6**, all of the planets discovered by HabEx occupy a region currently unexplored of the radius vs. separation parameter space. The yields are based on optimizing the observation plan for the detection and characterization of exo-Earth candidates; a search optimized for gas giants could yield significantly more of these planets.

**Mission Cost.** The estimated cost for the baseline concept is $6.8B FY20 including a $565M ESA contribution (see *Chapter 9* for programmatic and cost details). The baseline is the most expensive of the options in the tradespace mainly because it is the largest and most complex of the architectures evaluated.

**Required Technology Development.** Thirteen required technologies at TRL 4 were identified for the baseline design. The baseline has the most technologies requiring development since it includes both starlight suppression methods and the largest telescope being assessed as part of this study, however, since all technologies are already at TRL 4 and need only advance to TRL 5 to be ready for a new mission start, the work involved in advancing this option is manageable within the time available before WFIRST's launch opens up the funding for the next large mission. Details on the baseline technology gaps can be found in *Chapter 11*.

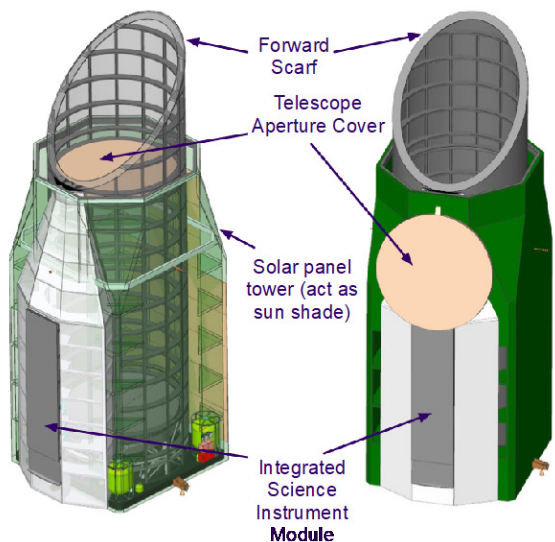

**Figure 10.3-1.** Telescope spacecraft used in the hybrid and coronagraph-only architectures.

### 10.3.1.2 Option 3.2H and 2.4H Differences from the Baseline Option

*Option 3.2H*

**Design Differences.** The primary difference between option 3.2H and the HabEx 4H baseline option is the reduction in the aperture size and, consequently, the overall telescope and telescope flight system sizes. This reduction permits the telescope flight system to possibly fit on other non-SLS launch vehicles (although the mass and volume may be tight).

**Science Difference (Table 10.4-1).** The number of EECs with orbits determined and 0.3–1 μm spectra measured is down by ~40% with respect to the 4H option. The reduction over the different planet types is 30–40%, but all exoplanet science baseline requirements are still met (with little margin). In terms of observatory science, **Table 10.4-1** shows that the 3.2H architecture achieves all Goal 3 science (i.e., Objectives O9 through O17) at the threshold mission level or better. Notably, a subset of objectives even meet the baseline mission requirements.

**Mission Cost.** The cost difference between HabEx 3.2H and the baseline design come from the reduction in telescope size, the subsequent reduction in the telescope bus size, and the utilization of a smaller launch vehicle for the telescope flight system. The total cost reduction is about $1.1B, with launch costs accounting for $0.4B of the total reduction, and telescope costs, spacecraft costs and their associated impact on management, systems engineering, mission assurance, and estimate reserves making up the remaining $0.7B.

**Required Technology Development.** HabEx 3.2H has 12 TRL 4 technology gaps. There is one fewer technology gap than the baseline, related to the large mirror fabrication. With the reduction in the size of the mirror, the current state of the art is close enough to the required size to constitute a medium fidelity model qualified in the relevant environment.

*Option 2.4H*

**Design Differences.** Like HabEx 3.2H, the aperture reduction is a major design difference between HabEx 2.4H and the baseline option, HabEx 4H. The reduction is a larger percentage of aperture diameter than that of 3.2H and puts the telescope flight system into the scale of HST of WFIRST. Most of the new heavy lift launch vehicles can handle the mass and volume requirements of the 2.4H telescope. In addition, to keep the starshade's IWA in line with the coronagraph's IWA, the starshade diameter has been reduced to 30 m and the separation distance reduced to 25,000 km.

**Science Difference (Table 10.4-1).** The number of EECs with orbits determined and 0.3–1 μm spectra measured is down by ~60% with respect to the 4H option. The reduction over the different planet types is 50–60%. Exoplanet science goals are now only met at threshold levels, due to reduced search and characterization completeness for all planet types. Only the debris disk exoplanet science goal (O8) is still met at baseline level. In terms of observatory science, no objectives meet the baseline level, most meet the threshold level, and a subset, such as O13, do not meet the threshold level.

**Mission Cost.** Reducing the telescope from 4 m to 2.4 m and the subsequent telescope bus size reductions, changing the SLS for a smaller launch vehicle, and reducing the size of the starshade contribute to a lower cost. The overall cost reduction from the baseline is about $2.0B. Launch costs comprise $0.4B of that amount. The reduction in the telescope size and its impact on the telescope bus contributed $0.7B to the total. The starshade resizing subtracted another $0.3B, and reserves and other non-hardware project costs composed the rest of the cost decrease.

**Required Technology Development.** HabEx 2.4H has only 11 TRL 4 technology gaps. There are two fewer technology gaps than the baseline; one related to the large mirror fabrication and one related to the large mirror coating. With the reduction in the size of the mirror, the mirror requirements are met by the current state of the art. Likewise, coating chambers suitable for 2.4 m mirrors for space applications exist and can likely be upgraded to meet uniformity requirements.

### 10.3.2 Coronagraph without a Starshade Architecture

The coronagraph-only architectures, options 4C, 3.2C, and 2.4C, avoid some of the programmatic complexity of the dual launch architectures. At the same time, for a fixed aperture size, these architectures also avoid some of the cost and technical complexity of the hybrid options but at the price of diminished science yield when compared to the hybrids, and limited spectral range and spectral characterization capability when compared to options including starshades.

For the architecture trade study, the coronagraph-only options are essentially the hybrid design without the starshade flight system. The orbit and launch vehicle remain the same. The payload difference is small, simply the SSI is removed. Operations are less complicated since there is only one spacecraft and no need for formation flying.

#### 10.3.2.1 Payload Differences from Baseline Option

The payload (i.e., the telescope and its associated instruments) is only slightly simplified from the payload in the hybrid options. The telescope remains the same $f/2.5$ primary mirror, unobscured design with a monolithic primary as used in the baseline design. The coronagraph, UVS, and workhorse camera instruments remain unchanged, but the SSI is eliminated. This means that the direct imaging spectral coverage is reduced from 0.3–1.8 μm to 0.45–1.8 μm. This change also requires the coronagraph to handle all spectral characterization science whereas the hybrid designs largely handled spectral measurements through the SSI's integral field spectrograph (IFS) in one 0.3–1 μm broadband observation. The coronagraph will cover the 0.45–1 μm band in 20% increments requiring longer duration observations than the hybrid options. Since the hybrid designs used the coronagraph for planet detection, there is no change in the number of close-in planets detected from the hybrid options. However, the coronagraph outer working angle (OWA) is significantly smaller than the starshade OWA, resulting in less distant giant planets detected than in the hybrid designs.

#### 10.3.2.2 Other Differences from Baseline Option

The removal of the starshade flight system eliminates the need for a second launch.

#### 10.3.2.3 Unique Architecture Differences from the Baseline Option

*Option 4C*

**Design Differences.** The primary design change from the baseline design is the elimination of the starshade flight system, SSI, and the second launch vehicle. This change also simplifies operations since formation flying is no longer required but it also adds complexity to the coronagraph science data acquisitions since the coronagraph must now do both detection and spectral characterization.

**Science Difference (Table 10.4-1).** The number of EECs with orbits determined and 0.45–1 μm spectra measured is down by ~40% with respect to the 4H option, and ocean glint studies (Objective 4) can only be carried out at threshold level due to the increased IWA at 0.87 μm. Another major difference is that access to the near UV part of EEC—and other planet type—spectra (Objective 3 baseline) is lost, making ozone detection more challenging at low concentration levels (ozone visible absorption feature is significantly weaker). The reduction in detection yield over the different planet types is 30–40% and access to the starshade-provided large (6") OWA is lost, meaning that Goal 2 objectives are only met at threshold level (smaller OWA). There is no impact in general observatory science, covered in *Chapter 4*, with respect to the 4H case.

**Mission Cost.** HabEx 4C cost savings over the baseline option largely stem from the elimination of the starshade, the SSI and the Falcon Heavy launch vehicle. The overall cost reduction is about $2.0B when compared to the baseline design. $0.3B comes from the removal of the Starshade flight system launch vehicle. The deletion of the SSI removes $0.2B from the total. The starshade occulter portion of the total is about $0.9B, and the reserves, management, systems engineering, and mission assurance costs related to the starshade comprise the remaining $0.6B.

**Required Technology Development.** HabEx 4C has 10 TRL 4 technology gaps. There are three fewer technology gaps than the baseline, all related to the starshade.

<u>Option 3.2C</u>

**Design Differences.** The design changes are the same as in HabEx 4C above plus a reduction in the size of the telescope. This reduced telescope will have rippling reductions through the telescope flight system. Less structure and propellant will be needed. The power required to heat the smaller telescope will decrease from the baseline, as will the telescope thermal losses. These mass savings will make the use of a non-SLS launch vehicle possible.

**Science Difference (Table 10.4-1).** The number of EECs with orbits determined and 0.45–1 μm spectra measured is down by ~60% with respect to the 4H option. Another major difference is that access to the near UV part of EEC—and other planet type spectra—is lost, making ozone detection more challenging at low concentration levels (ozone visible-band absorption feature is significantly weaker). The reduction over the different planet types is 50–60%. Due to the degraded IWA and reduced spectral capabilities compared to architecture 4H, ocean glint studies (Objective O4) and broad exoplanet atmospheric studies (Objective O6) cannot be conducted at threshold requirement levels. All other exoplanet science objectives are only met at threshold level, except for broad-band EEC searches and orbit determinations (Objective O1) which can still be conducted at baseline level. In terms of observatory science, the 3.2C option achieves all Goal 3 science at the threshold mission level or better. Specifically, since observatory science is not expected to use the limited resource of the starshade, the observatory science performance for the 3.2C option is identical to the 3.2H option.

**Mission Cost.** The cost difference between HabEx 3.2C and the baseline design come from the removal of the starshade flight system and the SSI, the elimination of the starshade launch vehicle, the reduction in telescope size and the utilization of a smaller telescope launch vehicle. The total estimated savings over the baseline costs are around $3.1B. The two launch vehicle changes reduce costs by $0.7B, elimination of the starshade flight system and SSI lowers costs by $1.1B, the telescope-related hardware reductions total $0.4B, and reserves, management, systems engineering, mission assurance, and other non-hardware related costs contribute another $0.9B toward the overall savings.

**Required Technology Development.** HabEx 3.2C has 9 TRL 4 technology gaps, with one attributable to removing large mirror fabrication and three to starshade removal. With the reduction in the size of the mirror, the current state of the art is close enough to the required size to constitute a medium fidelity model qualified in the relevant environment.

<u>Option 2.4C</u>

**Design Differences.** Like HabEx 3.2C, the design changes are the same as in HabEx 4C above plus a reduction in the size of the telescope. The reduction is larger than that of 3.2C and puts the telescope flight system into the scale as HST or WFIRST, although with an unobscured aperture.

**Science Difference (Table 10.4-1).** The number of EECs with orbits determined and 0.45–1 μm spectra measured is down by ~90% with respect to the HabEx 4H. Another major difference is that access to the near-UV part of planet type spectra is lost, making ozone detection more challenging at low concentration levels since the ozone visible absorption feature is significantly weaker than the ozone cut-off in the UV. The reduction over the baseline different planet types is ~80–90%. Due to degradation on IWA and collecting area, not on the exoplanet science objectives can be met at the threshold levels defined by the HabEx STDT except for debris disk broad-band imaging studies (Objective O8). In terms of observatory science, no objectives meet the baseline level, most meet the threshold level, and a subset do not meet the threshold level.

**Mission Cost.** Reducing the telescope from 4 m to 2.4 m, changing the SLS for a smaller launch vehicle, removing the starshade flight system and SSI, and eliminating the starshade launch vehicle reduces the overall costs by $3.7B from the baseline option. The reductions are the same as the $3.0B in option 3.2C plus an additional $0.7B for further telescope and telescope bus reductions, as well as associated non-hardware costs and reserves. This is the lowest cost option in the trade study.

**Required Technology Development.** HabEx 2.4C has only eight TRL 4 technology gaps. There are five fewer technology gaps than the baseline; three related to the starshade, one related to the large mirror fabrication, and one related to the large mirror coating. With the reduction in the size of the mirror, the mirror requirements are met by the current state of the art. Likewise, coating chambers suitable for 2.4 m mirrors for space applications exist and can likely be upgraded to meet uniformity requirements.

### 10.3.3 Starshade-Only Architectures

The starshade with no coronagraph architectures, options 4S, 3.2S, and 2.4S, utilize a starshade for both exoplanet detection and characterization. While easing requirements on telescope performance, this option is limited by the amount of propellant on the starshade, and the speed at which it can retarget. Additionally, the long-duration retargeting maneuvers also limit the number of targets that can be reached within the 5-year mission.

While the starshade and starshade spacecraft remain the same as in the baseline option, the concept takes advantage of the relaxed telescope wavefront stability requirements to reduce telescope size, and mass. Orbit and mission duration remain the same as in the baseline option.

#### 10.3.3.1 Payload Differences from Baseline Option

Like the baseline option, the starshade-only telescope instrumentation includes an SSI, UVS, and near-UV/visible/near-IR camera. No coronagraph was included in the architecture trade. With the more common on-axis design, the instruments were located behind the primary mirror.

Without the coronagraph, operations will be very different for the starshade-only architecture in comparison to the baseline. In the baseline, the coronagraph handled most of the planet detections and orbit determinations. These require blind searches of target systems to detect new exoplanets and repeat visits to establish planet orbits. While this work is possible with the starshade-only architecture, the time and fuel required to reposition the starshade for each visit and revisit will limit their numbers. Orbit determination in particular is a challenge for this architecture option, and only the systems with EECs detected will be visited multiple times for orbit determination. Modeled number of orbits (all planet types) determined by a starshade in the architecture trade, were less than a third of the number of orbits captured with options that include a coronagraph.

#### 10.3.3.2 Other Differences from Baseline Option

With the more compact and lighter telescope, use of a non-SLS launch vehicle is possible, while the baseline design and the 4C option must use the SLS or BFR for both mass and volume reasons.

#### 10.3.3.3 Unique Architecture Differences from the Baseline Option

*Option 4S*

**Design Differences.** Unlike the baseline option, the 4S starshade-only architecture uses a non-deployed, on-axis telescope design with a segmented primary mirror (**Figure 10.3-2**). Without needing to meet coronagraph requirements, the telescope can include a central obscuration and faster $f$/number ($f$/1.3 for option 4S verses $f$/2.5 for the HabEx 4H) creating a more compact telescope and flight system. The reduction in size and the use of a lighter-weight segmented primary result in a lower overall flight system mass.

**Science Difference (Table 10.4-1).** The number of EECs with orbits determined and 0.45–1 µm spectra measured is down by ~40% with respect to the HabEx 4H. The reduction over the different planet types detected is a function of planet size. It ranges from no change for giant planets) to ~20% reduction for sub-Neptunes and ~40% for rocky planets. Another major difference with respect to the baseline is that only the deep survey targets and the broad survey targets with EECs detected, have multi-epoch visits to constrain planetary orbits and radii. As a result, the number of non-EEC planets with radii estimated does not meet the threshold requirement for nearby systems architectural studies (Objective O5) and only meets threshold requirements for the studies of exoplanet atmosphere variations vs planetary distance and size (Objective O6). General observatory science is largely unaffected by the removal of the coronagraph, with the one exception of reduced protoplanetary disks science (Objective O16) due to the limited number of starshade slews expected to be available for observatory science.

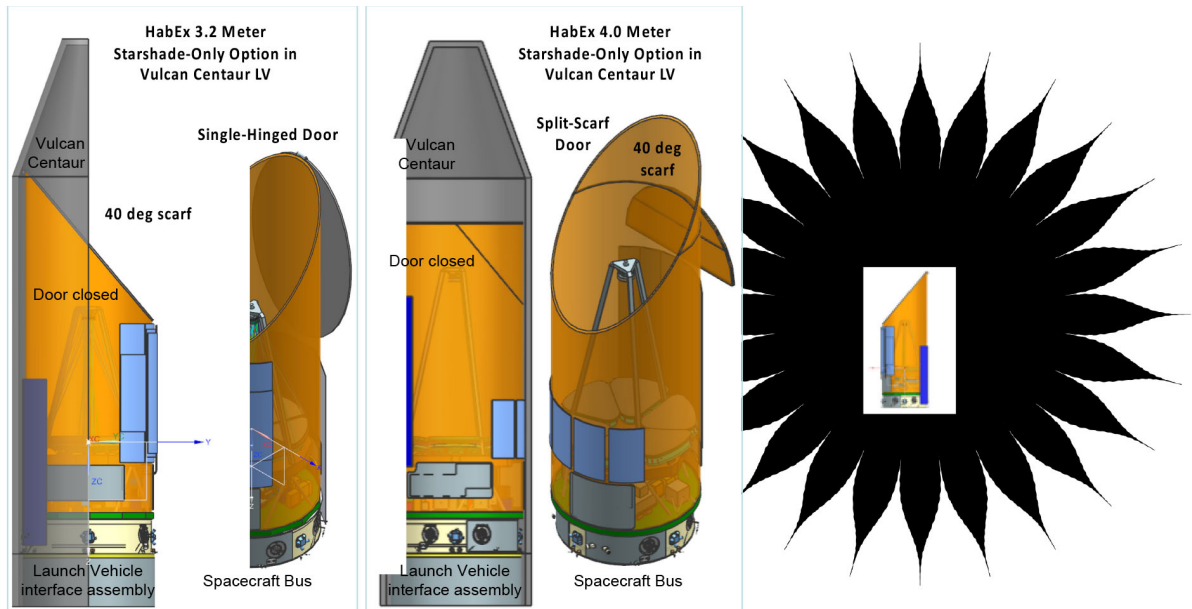

**Figure 10.3-2.** The 4.0 m and 3.2 m starshade-only options use a segmented telescope and a 52 m starshade.

**Mission Cost.** Option 4S cost savings over the baseline option come from the removal of the coronagraph, use of a smaller launch vehicle for the telescope flight system, and associated non-hardware costs and reserves. The total savings were estimated at $1.1B with $0.4B attached to the coronagraph hardware and $0.4B for use of a Vulcan launch vehicle for the telescope rather than the SLS. The remaining $0.3B is composed of reserves and non-hardware costs.

**Required Technology Development.** HabEx 4S has nine TRL 4 technology gaps. There are four fewer gaps than the HabEx 4H baseline, all relating to the elimination of the coronagraph and large monolithic mirror. An additional TRL 5 technology for active mirror segments, described in *Section B.3*.

### Option 3.2S

**Design Differences.** The design changes for the 3.2S option are the same as in option 4S above plus a reduction in the size of the telescope. This reduced-sized telescope will have rippling reductions through the telescope flight system. Less structure and propellant will be needed. The power required to heat the smaller telescope will decrease from the baseline, as will the telescope thermal losses.

**Science Difference (Table 10.4-1).** The number of EECs with orbits determined and 0.45–1 µm spectra measured is down by ~50% with respect to the 4H option. The reduction over the different planet types is a function of planet size. It ranges from ~10% (giant planets) to ~30% (sub-Neptunes) to ~50% (rocky planets). The reduction in EEC completeness results in this architecture only meeting Goal 1 science objectives at threshold level. Another major difference with respect to the baseline is that only the deep survey targets and the broad survey targets with EECs detected, have multi-epoch visits to constrain planetary orbits. As a result, the number of non-EEC planets with radii estimated does not meet the threshold requirement for nearby systems architectural studies (Objective O5) nor for the studies of exoplanet atmosphere variations vs planetary distance and size (objective O6). General observatory science is affected by the loss in aperture size with only a few objectives still meeting the baseline requirement level, and all the others met at threshold levels. We note that Objective O16, involving high-contrast imaging of transition disks, is down-graded for all starshade-only architectures, even 4S. This is because the starshade is not assumed to be available as a GO instrument given its limited fuel supply and associated number of allowed slews. So O16 science would be achievable by the hardware design, it is not clear that the mission design would have the capability for additional GO high-contrast imaging targets.

**Mission Cost.** The cost difference between option 3.2S and the baseline design come from the removal of the coronagraph, reduction in the telescope size, use of a smaller launch vehicle for the telescope flight system, and a reduction in the size of the telescope flight system. The total cost savings over the baseline option is estimated at about $1.8B. The telescope and telescope flight system reductions account for about $0.4B of this total. The coronagraph removal adds another $0.4B to the savings, Reductions in reserves and non-hardware costs related to the telescope and coronagraph changes contribute another $0.6B, and the reduction in the telescope launch vehicle makes up the remaining $0.4B.

**Required Technology Development.** HabEx 3.2S has nine TRL 4 technology gaps. There are four few gaps than the HabEx 4H baseline, all relating to the elimination of the coronagraph and large monolithic mirror. An additional TRL 5 technology for active mirror segments, described in *Section B.3.*

## Option 2.4S

**Design Differences.** The design changes from the baseline in the 2.4S option are the same as the design changes in the 3.2S option except: 1) the telescope is smaller, 2) the primary mirror is monolithic, and 3) the starshade size is reduced from 52 m to 30 m. The shift to monolithic was made to leverage the extensive flight experience with 2.4 m space telescopes. The smaller starshade was assumed to help off-set decreased exoplanet science performance stemming from the smaller aperture, with a lighter and more maneuverable starshade.

**Science Difference.** The number of EECs with orbits determined and 0.45–1 µm spectra measured is down by ~75% with respect to the 4H option. The reduction over the different planet types is a function of planet size. It ranges from ~60–70% (giant planets and sub-Neptunes) to 75% (rocky planets). Another major difference with respect to the baseline is that to the exception of the deep survey targets, only the target stars with EECs detected have multi-epoch visits to constrain planetary orbits. Due to degradation of IWA and collecting area, none the exoplanet science objectives can be met at the threshold levels defined by the HabEx STDT, except for debris disk broad-band imaging studies (Objective O8).

General observatory science is severely affected by the loss in aperture size with most objectives met at threshold requirement level only, and some objectives not even meeting this threshold level.

**Mission Cost.** Reducing the telescope from 4 m to 2.4 m, eliminating the coronagraph, changing the SLS for a smaller launch vehicle, and reducing the starshade all contribute to a lower cost. For this option, the cost savings when compared to the baseline is expected to be about $2.8B. Telescope and telescope bus resizing accounts for about $0.7B. The telescope launch vehicle change contributes about $0.4B. Another $0.4B comes from the removal of the coronagraph. The starshade resizing lowers the cost by about $0.5B and the remaining $0.8B is composed of reserves and non-hardware costs.

**Required Technology Development.** HabEx 2.4S has eight TRL 4 technology gaps. There are five fewer technology gaps than in the baseline, all relating to the elimination of the coronagraph, and the reduction in sizes of the primary mirror. With the reduction in the size of the mirror and shift to monolithic, the mirror is in line with many past space telescopes. Likewise, coating chambers suitable for 2.4 m mirrors for space applications exist and have coated many similar primary mirrors.

## 10.4 Architecture Tradespace Sensitivities

The purpose of this sensitivity exercise was not to find a "best" option; that decision has already been made by the STDT in the interim report and it is the baseline 4H option. This study was intended to uncover how science, risk, and cost vary across the tradespace to assist the Decadal Survey in assessing alternative designs if budgetary and programmatic limitations make the preferred baseline option untenable. Key parameters indicative of the science, risk, and cost of each of the nine architectures are summarized in **Table 10.3-1**.

A performance summary is shown in **Table 10.4-1**, indicating, for each of the nine architectures, whether a science objective is expected to be met at the baseline level (green cell), threshold level (yellow) or neither (orange) during HabEx prime 5-year mission, assuming a 50/50 time split between exoplanet surveys and observatory science. Some science capabilities are still retained in the orange-colored cells, but they fell below the threshold science requirements established by the HabEx STDT, e.g., in terms of the minimum number of objects characterized, depth of characterization or probability of success.

### Science

With respect to exoplanet science return (Goals 1 and 2), the baseline option is clearly the best with nearly double the number of EECs with orbits precisely constrained and broad spectra taken (at least from 0.3–1 μm) as for any of the other options. Overall exoplanet detections for the different planet types are also significantly higher than for the other options. This superior exoplanet science performance is due to the complementary interaction of the two starlight suppression techniques and the large telescope aperture. In particular, the 3.2H option performs better than either the 4C or 4S options.

While the exoplanet science is best with the combined hybrid configuration, observatory science capabilities are, to a large degree,

independent of the starlight suppression system implemented, and clearly favor the larger 4 m aperture options.

Both the exoplanet yields and the telescope performance characteristics driving observatory science are strongly coupled to the telescope aperture size. From a science-only perspective, a larger aperture is better.

At the low end of the science performance are the 2.4 m options. All represent some improvement in telescope performance, and hence General Observatory science return, over the venerable HST. All also enable direct imaging of mature exoplanet systems for the first time. However, the exo-Earth candidate yield estimates are low for all three options, with some real risk of zero candidates being characterized in a 5-year mission.

### Risk

Of the nine configurations examined, none require a technology at TRL 3 or lower. TRL 5 is required at the start of a new mission so all technologies needing further development are close to reaching this critical milestone. The hybrids require the most new technologies, with the 4 m baseline option requiring 13 TRL 4 technologies. The hybrids present the most required new technologies because they utilize both starlight suppression methods. All of these technologies have demonstrated performance in a laboratory environment but now require a performance verification of a medium fidelity test unit in a flight-like environment. Plans to advance these technologies are presented in *Chapter 11* and *Appendix E* and most of this work is currently funded and underway, with projected TRL 5 completion well before a likely mission Phase A start date. While the number of required technologies is high, the rigor in technology-gap identification likely exceeds that experienced in past Decadal Surveys, with most of the gaps being tracked, funded, and advanced since the last Decadal Survey.

**Table 10.4-1.** All HabEx architectures are capable of meeting a subset of the science measurement requirements flowing down from the HabEx science goals and objectives defined in Table 5.1-1, indicated in the left column. Each architecture's ability to meet the HabEx science objectives is shown in color-code. Color evaluations are made against the baseline and threshold requirements defined by the HabEx STDT for a 5-year prime mission evenly split between Exoplanet Surveys (Goals 1 and 2) and Observatory Science (Goal 3). *Green* identifies that baseline requirement is met, yellow identifies that threshold requirement is met, and *orange* identifies that neither is met. For Exoplanet Surveys, the observing strategy of each architecture is optimized to meet Goal 1 (Habitable Exoplanets science objectives). Optimizing it for Goal 2 instead (broader Exoplanetary Systems science objectives) would improve characterization of larger and/or more distant planets at the expense of exo-Earth candidates.

| HabEx Science Goals & Objectives | | 4H | 4C | 4S | 3.2H | 3.2C | 3.2S | 2.4H | 2.4C | 2.4S |
|---|---|---|---|---|---|---|---|---|---|---|
| **Habitable Exoplanets** | **O1** Exo-Earth candidates around nearby sunlike stars? | green | green | green | green | green | yellow | orange | orange | orange |
| | **O2** Water vapor in rocky exoplanet atmospheres? | green | green | green | yellow | yellow | yellow | orange | orange | orange |
| | **O3** Biosignatures in rocky exoplanet atmosphere? | green | green | yellow | yellow | yellow | yellow | orange | orange | orange |
| | **O4** Surface liquid water on rocky exoplanets? | green | yellow | green | orange | orange | green | orange | orange | orange |
| **Exoplanetary Systems** | **O5** Architectures of nearby planetary systems? | green | yellow | orange | green | green | green | orange | orange | orange |
| | **O6** Exoplanet atmospheric variations in nearby systems? | green | green | green | orange | orange | orange | orange | orange | orange |
| | **O7** Water transport mechanisms in nearby planetary systems? | green | green | green | green | green | green | green | green | green |
| | **O8** Debris disk architectures in nearby planetary systems? | green | green | green | green | green | green | yellow | yellow | green |
| **Observatory Science** | **O9** Lifecycle of baryons? | green | green | green | yellow | yellow | yellow | yellow | yellow | yellow |
| | **O10** Sources of reionization? | green | green | green | yellow | yellow | yellow | yellow | yellow | yellow |
| | **O11** Origins of the elements? | green | green | green | yellow | yellow | yellow | yellow | yellow | yellow |
| | **O12** Discrepancies in measurements of the cosmic expansion rate? | green | green | green | yellow | yellow | yellow | yellow | yellow | yellow |
| | **O13** The nature of dark matter? | green | green | green | yellow | yellow | yellow | orange | orange | orange |
| | **O14** Formation and evolution of globular clusters? | green | green | green | green | green | green | yellow | yellow | yellow |
| | **O15** Habitable conditions on rocky planets around M-dwarfs? | green | green | green | yellow | yellow | yellow | orange | orange | orange |
| | **O16** Mechanisms responsible for transition disk architectures? | green | green | yellow | green | green | green | green | green | green |
| | **O17** Physics driving star-planet interactions, *e.g.* auroral activity? | green | green | green | green | green | green | green | green | green |

At the very low end of the risk scale is the 2.4C option. With a 2.4 m monolithic mirror, which has been built and flown many times, and no starshade, the seven TRL 4 technology gaps are connected to the coronagraph and detectors.

Beyond the technical risks connected with technologies still in development, the launch vehicles also add some risk. The hybrid options and starshade-only options all require two successful launches to establish a fully operational observatory. While most of the new launch vehicles are in development, all have at least one alternative that could be considered if the primary choice does not deliver on time. As such, only the multiple launch risk differentiates between the options.

## Cost

With the inclusion of both direct imaging systems and the largest of the telescope apertures within the tradespace, the baseline architecture carries the largest cost of the nine options. At $6.8B FY20, while high in comparison to claimed costs for past Decadal Survey concepts, this is significantly less than the JWST realized costs and estimate to complete (~$10.5B when converted into $FY20 for comparison), and may be achievable at current NASA Astrophysics funding levels even after the addition of a new Probe line of missions, with the delay of the HabEx baseline starshade launch (see *Chapter 9* for details on the baseline cost estimate and how to implement it within the current funding level).

Aperture and multiple spacecraft are largely driving the cost of the options. Accordingly, the 2.4 m coronagraph-only option does best with cost at $3.1B. A number of options (4C, 3.2C, 3.2S, 2.4H, 2.4C, and 2.4S), are near or below $5B FY20 which should be manageable at current funding levels within a reasonable pre-launch development schedule of 10 to 12 years. These options even include the 4 m coronagraph-only architecture so all telescope sizes can be fitted into the available money at current funding levels.

# 11 TECHNOLOGY MATURATION

Since the 2010 Decadal Survey, the technologies needed for direct imaging of Earth-like exoplanets orbiting nearby sunlike stars using a HabEx-like architecture have advanced significantly. The investment in NASA's Wide Field Infrared Survey Telescope (WFIRST) Coronagraph Instrument (CGI) has matured many coronagraph-related technologies, including deformable mirrors (DMs), and electron multiplying charge coupled device (CCD) detectors (EMCCDs). The Starshade to Technology Readiness Level 5 (S5) became a NASA-funded program to develop starshade technologies to Technology Readiness Level (TRL) 5 by 2023. The Laser Interferometer Space Antenna (LISA)-Pathfinder mission demonstrated microthrusters in space, while the Gravity Recovery And Climate Experiment (GRACE) Follow-On mission utilized laser heterodyne metrology. Monolithic mirror fabrication using extremely low thermal expansion materials has reached 4 m with the Daniel K. Inoue Solar Telescope (DKIST) and European Extremely Large Telescope (E-ELT) secondary mirror. These technologies collectively enable a HabEx project start as early as 2026 with very modest additional investment. For the first time since the discovery of exoplanets, an exo-Earth direct imaging mission can be conceived and designed such that it is feasible to start in less than 10 years (<2030); possibly as soon as 6 years (~2026).

All of the enabling technologies required for HabEx have been carefully identified and their maturity critically assessed. From a technology standpoint, HabEx, as a whole, has a moderate risk level. Currently, there are no HabEx enabling technologies that are TRL 3 or less; HabEx enabling technologies are at TRL 4, 5, or higher. All but two of the HabEx enabling technologies are predicted to be at TRL 5 by the end of 2023—a full year before the nominal start of Phase A for the HabEx mission—and most via currently funded tasks. The two technologies that will not be at TRL 5 until 2024 are the microchannel plate (MCP) detector and the primary mirror fabrication. Since it is not enabling for any other proposed mission, the cost of a 4-meter monolithic primary mirror prototype is precluded from work ahead of the Decadal Survey recommendation. Thus, the time required to cast, grind, and polish the mirror from the time that it may be prioritized by the Decadal Survey limits the TRL completion date. Similarly, the MCP also assumes a technology development start in 2022, leading to TRL 5 completion in 2024. Earlier funding could advance this date by a year. However, even by the release of the 2020 Decadal Survey report, the current number of TRL 4 technologies should be reduced to 10, based on *currently funded* development efforts. Three of the technologies currently in development will be at TRL 6 in 2023.

There are currently 3 HabEx technologies at TRL 5 and 13 at TRL 4. While this number of technologies may appear large, it is not a reflection of the immaturity or complexity of the general HabEx system. Rather, it is a product of the maturity of the design and of the in-depth understanding of the technology maturation paths. HabEx has chosen to define an individual technology as one that requires its own separate technology maturation path, rather than one that is superficially related (in a systems engineering sense) to other technologies. HabEx could have chosen to roll up the number of technologies by such superficial similarity, which would have ultimately reduced the reported number of nascent technologies. However, the HabEx team considered this approach disingenuous, given that technologies that appear to be superficially similar often require quite different maturation paths. For example, the number of HabEx technologies could be artificially reduced by, for example, rolling up the three starshade mechanical technologies into one category. However, inspecting these technologies at a finer fidelity allows more precise tracking of the method by which these technologies will be matured, and makes it clear that they are, in fact, distinct technologies. For example, one of the starshade mechanical technologies, the Starshade Scattered Sunlight for Petal Edges, completed TRL 5 in the lab during summer of 2019 and the official milestone report to the Exoplanet Technical Advisory Committee (ExoTAC) is in progress.

Thus, while HabEx recognizes a large number of new technologies, these technologies are advancing rapidly, will be qualified at both the component and system levels, and will be qualified at full scale where applicable, with all the technology work completed before most of the NASA investment in the mission is made. The necessary technologies are close to ready and the risks are manageable. In comparison to NASA's JWST just before submittal to the 2000 Decadal Survey (Coulter 1998), the HabEx technologies are more mature (**Figure 11-1**).

## 11.1 Table of HabEx Technologies

In NASA mission lifecycles, TRL 5 is required for the start of Phase A and TRL 6 is required by Preliminary Design Review (PDR) and the start of Phase C. The official NASA TRL definitions are found in NPR 7123.1B, *Appendix E* and are abridged in **Table 11-1** for reference.

**Table 11.1-1** summarizes the HabEx technology challenges and shows the TRL expected by the end of 2019 and the end of 2023.

The technology discussion follows the optical path: the discussion begins with starshade technologies, moves to telescope technologies, next to instrument technologies, then spacecraft technologies and ends with alternative/enhancing technologies. Within each HabEx system, technologies are presented from lowest to highest current TRL. Each technology is discussed in its own section which includes the path to TRL 5. Roadmaps and plans to TRL 6 are in *Appendix E*, including a timeline of technology development (**Figure E-1**). The timeline shows that all technologies will be TRL 5 before the start of Phase A and all technologies will be TRL 6 a couple of years before the start of PDR.

**Table 11-1.** Definitions of Technology Readiness Level (TRL).

| TRL | Abridged Definition |
|-----|---------------------|
| 1 | Basic principles observed and reported |
| 2 | Technology concept and/or application formulated |
| 3 | Analytical and experimental critical function and/or characteristic proof-of-concept |
| 4 | Low-fidelity component or breadboard in lab demonstrates functionality and critical test environments |
| 5 | Medium fidelity component/system brassboard demonstrates overall performance in relevant environment |
| 6 | High fidelity system/subsystem model or prototype demonstrates critical performance in relevant environment; critical scaling issues understood |
| 7 | High fidelity engineering unit demonstrates performance in operational environment and platform |
| 8 | System is flight qualified |
| 9 | System is successfully operated in an actual mission. |

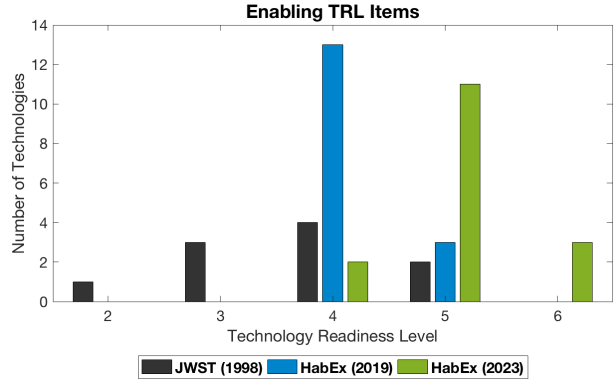

**Figure 11-1.** HabEx technology maturity is significantly higher than JWST maturity at the 2000 Decadal Survey.

## 11.2 Starshade

HabEx will qualify the starshade based on performance model validation and key subsystem ground testing. In November 2016, the Starshade Readiness Working Group (SSWG) recommended to the NASA Astrophysics Director a plan to validate starshade technologies to TRL 6 "that is both necessary and sufficient prior to building and flying" a starshade science mission. With the full concurrence of an independent Technical Advisory Committee, it was determined that "a ground-only development strategy exists to enable a starshade science flight mission" and "a prior flight technology demonstration is not required" (Blackwood 2016).

NASA organized its various starshade technology development activities into a central activity to drive the starshade technology gaps to TRL 5. This activity is "Starshade to TRL 5", or S5. The S5 Technology Development Plan (Willems 2018) will close the starshade gaps to TRL 5 by June 2023. The plan is phased to retire the highest risk elements, perform initial model validations, and demonstrate critical mechanical environments to TRL 4 for the Decadal Survey. The gaps are:

1. Petal position accuracy and stability (*Section 11.2.1.1*)

**Table 11.1-1.** HabEx 4 m baseline architecture technology gap list. Note that the cell coloring reflects TRL: TRL 3, *red*; TRL 4, *yellow*; TRL 5, *green*, and TRL 6, *blue*.

| Title | Description | Section | State of the Art | Capability Needed | TRL 2019 | Expected 2023 TRL |
|---|---|---|---|---|---|---|
| | | | **Enabling Technologies** | | | |
| Starshade Petal Position Accuracy and Stability | Deploy and maintain petal position accuracy in L2 environment | *Section 11.2.1.1* | • Petal position deployment tolerance (≤150 µm) verified with multiple deployments of 12 m flight-like perimeter truss and no optical shield<br>• No environmental testing | • Petal position deployment accuracy on 20 m perimeter truss: ±600 µm (3σ) bias<br>• Position stability in operational environment: ±400 µm (3σ) random | 4 | 5 |
| Starshade Petal Shape Accuracy and Stability | Starshade petal shape maintained after deployment, thermal at L2 | *Section 11.2.1.2* | • Manufacturing tolerance (≤100 µm) verified with low-fidelity 6 m long by 2.3 m prototype; no environmental tests<br>• Petal deployment tests conducted on prototype petals to demonstrate rib actuation; no post-deploy cycle and petal shape stability measurements | • Petal 16 m long by 4 m wide<br>• Petal shape manufacture: ±140 µm (3σ)<br>• Post-deploy cycle and petal shape thermal stability ≤ ±160 µm (3σ) | 4 | 5 |
| Starshade Scattered Sunlight for Petal Edges | Limit edge-scattered sunlight and diffracted starlight with petal optical edges | *Section 11.2.1.3* | • Chemically etched amorphous metal edges limit solar glint flux to 25 visual magnitudes in two main lobes, verified at coupon level<br>• In-plane shape tolerance of ±20 µm met at half meter length after integration onto prototype petal<br>• In plane shape stability demonstrated post-deploy and thermal cycle<br>• Scatter performance on half meter edge verified post environment | • 1 m length edges assembled precisely onto petal<br>• Petal edge in-plane shape tolerance: ±66 µm (3σ)<br>• Petal edge in-lane placement tolerance: ±55 µm (3σ)<br>• Solar edge scatter: 25 visual magnitudes in two main lobes | 5 | 5 |
| Starshade Contrast Performance Modeling and Validation | Validate at flight-like Fresnel numbers the equations that predict the contrasts | *Section 11.2.1.4* | • 1.5 × 10⁻¹⁰ contrast demonstrated at Fresnel Number$_{R=1}$ ~13 (monochromatic)<br>• Expect 1 × 10⁻¹⁰ contrast demonstrated at Fresnel Number$_{R=1}$ ~13 (10% bandwidth) in March 2019 | • Experimentally validated models with scaled flight-like geometry and Fresnel Number$_{R=1}$ ≥12 across a broadband optical bandpass. Validated models are traceable to 1 × 10⁻¹⁰ contrast system performance in space. | 4 | 5 |
| Starshade Lateral Formation Sensing | Lateral formation flying sensing to keep telescope in starshade's dark shadow | *Section 11.2.2.1* | • Simulations have shown centroid to ≤1/10th aperture with ample flux to support control loop<br>• Control algorithms demonstrated control ≤1 m radius within line of sight of the star for durations representative of typical starshade observation times | • Demonstrate sensing lateral errors ≤0.40 m accuracy (≤1/10th aperture) at scaled flight separations<br>• Control algorithms demonstrated with scaled lateral control corresponding to ≤1 m of the line of sight | 5 | 6 |
| Large Mirror Fabrication | Large monolith mirror that meets tight surface figure error and thermal control requirements at visible wavelengths | *Section 11.3.1.1* | • 4.2 m diameter, 420 mm thick blanks standard<br>• Schott demonstrated computer-controlled-machine lightweighting to pocket depth of 340 mm, 4 mm rib thickness on E-ELT M5 and 240 mm deep/2 mm thick rib on Schott 700 mm diameter test unit | • 4.04 m diameter substrate<br>• 3–4 mm ribs, 14 mm facesheet, and pocket depth of 290 mm for 400 mm thick blank<br>• Aerial density 110 kg/m²<br>• <5 ppb/K CTE homogeneity<br>• First mode ≥60 Hz | 4 | 4 |

| Title | Description | Section | State of the Art | Capability Needed | TRL 2019 | Expected 2023 TRL |
|-------|-------------|---------|------------------|-------------------|----------|-------------------|
| | | | • State-of-the-practice (SOP) lightweighting has yielded large mirrors of aerial density 70 kg/m$^2$<br>• Zerodur® can achieve 2.83 parts per billion/K CTE homogeneity (DKIST mirror)<br>• Wavefront stability: 25 nm RMS for HST in LEO<br>• Wavefront error of WFIRST-like primary mirror (spatial frequency cycles/beam diameter: nm RMS):<br>  ▪ 0–7 cy/$D$: 6.9 nm RMS<br>  ▪ 7–100 cy/$D$: 6.0 nm RMS<br>  ▪ >100 cy/$D$: 0.8 nm RMS | • Wavefront stability of 100s to a few picometers RMS (depending on spatial frequency) over 100s of seconds<br>• Wavefront error (spatial frequency cycles/beam diameter: nm RMS):<br>  ▪ 0–7 cy/$D$: 6.9 nm RMS<br>  ▪ 7–100 cy/$D$: 6.0 nm RMS<br>  ▪ >100 cy/$D$: 0.8 nm RMS | | |
| Large Mirror Coating Uniformity | Mirror coating with high spatial uniformity over the visible spectrum | *Section 11.3.1.2* | • Reflectance uniformity <0.5% of protected Ag on 2.5 m TPF Technology Demonstration Mirror<br>• IUE, HST, and GALEX used MgF$_2$ on Al to obtain >70% reflectivity from 0.115 µm to 2.5 µm<br>• Operational life: >28 years on HST | • Reflectance uniformity <1% over 0.45–1.0 µm<br>• Reflectivity comparable to HST:<br>  ▪ 0.115–0.3 µm: ≥70%<br>  ▪ 0.3–0.45 µm: ≥88%<br>  ▪ 0.45–1.0 µm: ≥85%<br>  ▪ 1.0–1.8 µm: ≥90%<br>• Operational life >10 years | 4 | 5 |
| Laser Metrology | Sensing for control of rigid body alignment of telescope front-end optics | *Section 11.3.2.1* | • Nd:YAG ring laser and modulator flown on LISA-Pathfinder<br>• Phase meters flown on LISA-Pathfinder and Grace Follow-On<br>• Sense at 1 kHz bandwidth<br>• Thermally stabilized Planar Lightwave Circuit at TRL 6. Thermal stability measured, which could provide uncorrelated per gauge error of 0.1 nm | • Sense at 100 Hz bandwidth<br>• Uncorrelated per gauge error of 0.1 nm | 5 | 6 |
| Coronagraph Architecture | Suppress starlight by a factor of ≤1E-10 at visible and near-IR wavelengths | *Section 11.4.1.1* | • Hybrid Lyot: 6 × 10$^{-10}$ raw contrast at 10% bandwidth across angles of 3–16 λ/$D$ demonstrated with a linear mask and an unobscured pupil in a static vacuum lab environment<br>• Vector vortex charge 4: 5 × 10$^{-10}$ raw contrast monochromatic across angles of 2–7 λ/$D$<br>• Lyot: 3.6 × 10$^{-10}$ raw contrast at 10% bandwidth over 3–7 λ/$D$ in a static lab environment (DST)<br>• Vector vortex charge 6: 8.5 × 10$^{-9}$ coherent contrast at 10% bandwidth across angles of 3–8 λ/$D$ demonstrated with an unobscured pupil in a static lab environment | • Vortex Charge 6<br>• Raw contrast of ≤2 × 10$^{-10}$<br>• Raw contrast stability of ≤2 × 10$^{-11}$<br>• Inner working angle (IWA) ≤ 2.4 λ/$D$<br>• Coronagraph throughput ≥10%<br>• Bandwidth ≥20% | 4 | 5 |

| Title | Description | Section | State of the Art | Capability Needed | TRL 2019 | Expected 2023 TRL |
|-------|-------------|---------|------------------|-------------------|----------|-------------------|
| Zernike Wavefront Sensing and Control (ZWFS) | Sensing and control of low-order wavefront drift; monitoring of higher order Zernike modes | *Section 11.4.2* | • <0.36 mas rms per axis LOS residual error demonstrated in lab with a fast-steering mirror attenuating a 14 mas LOS jitter and reaction wheel inputs on Mv = 5 equivalent source; ~26 pm RMS sensitivity of focus (WFIRST Coronagraph Instrument Testbed)<br>• WFE stability of 25 nm/orbit in low Earth orbit (HST). Higher low-order modes sensed to 10–100 nm WFE RMS on ground-based telescopes | • LOS error <0.2 mas RMS per axis<br>• Wavefront stability:≤~100 pm RMS over 1 second for vortex<br>• WFE <0.76 nm rms | 4 | 5 |
| Deformable Mirrors | Flight-qualified large-format deformable mirror | *Section 11.4.3* | • Micro-electromechanical DMs available up to 64 × 64 actuators, 400 μm pitch with 6 nm RMS flattened surface; 3 nm RMS demonstrated on 32 × 32 DM<br>• $8.5 \times 10^{-9}$ coherent contrast at 10% bandwidth in a static test achieved with smaller 32 x 32 MEMS DMs<br>• Contrast drift of ~1 × $10^{-12}$/hour over 4 hr, ~1 × $10^{-8}$ drift over 42 hr<br>• Drive electronics in DST provide 16-bit resolution, which contributes ~1 × $10^{-10}$ to contrast floor | • 64 × 64 actuators<br>• Enable coronagraph raw contrasts of ≤3 × $10^{-10}$ at ~20% bandwidth and raw contrast stability ≤3 × $10^{-11}$<br>• <3.3 nm RMS flattened surface figure error (SFE)<br>• Drive electronics of at least 18 bits | 4 | 5 |
| Delta Doped UV and Visible Electron Multiplying CCDs | Low-noise UV and visible detectors for exoplanet characterization | *Section 11.4.4.1* | • 1k × 1k EMCCD detectors (WFIRST)<br>  ▪ Dark current of 7 × $10^{-4}$ e-/px/s<br>  ▪ CIC of 2.3 × $10^{-3}$ e-/px/frame<br>  ▪ Read noise ~0 e- rms (in EM mode)<br>  ▪ Irradiated to equivalent of 6-year flux at L2<br>  ▪ Updated design for cosmic ray tolerance under test<br>• 4k × 4k EMCCD dark current 1.5 × $10^{-4}$ e-/px/s | • 0.45–1.0 μm response;<br>• Dark current <$10^{-4}$ e-/px/s<br>• CIC <3 × $10^{-3}$ e-/px/frame<br>• Effective read noise <0.1e- RMS<br>• Tolerant to a space radiation environment over mission lifetime at L2<br>• 4k × 4k format for Starshade IFS | 4 | 5 |
| Deep Depletion Visible Electron Multiplying CCDs | Low-noise detectors with improved QE at 940 nm for exoplanet characterization | *Section 11.4.4.2* | • Under investigation. e2V claims dark current is on boundary surface and not throughout volume<br>• CCD-201 is not currently made in deep depletion<br>• CCD-220 (regular CCD) dark current < 0.02 e-/px/s | • QE >80% at 0.940 μm<br>• thicker silicon (up to 200 μm thick layer), deep depletion devices<br>• 4k × 4k format for starshade IFS | 4 | 5 |
| Linear Mode Avalanche Photodiode Sensors | Near-infrared wavelength (0.9 μm to 2.5 μm), extremely low noise detectors for exo-Earth IFS | *Section 11.5.1.1* | • HgCdTe photodiode arrays have read noise <~2 e rms with multiple non-destructive reads; dark current <0.001 e-/pix; very radiation tolerant (JWST)<br>• HgCdTe APDs have dark current ~10–20 e-/s/pix, read noise <<1 e- rms, and < 1k × 1k format<br>• LMAPD have 0.0015 e-/pix/s dark current, <1 to 0.1 e- RMS readout noise (SAPHIRA) for 320×256, 24 μm pixels. This format is TRL 5.<br>• LMAPD 1k × 1k formats of 15 μm pixels have 0.04 e- RMS dark current at gain of 25 | • Read noise <<1 e- RMS<br>• Dark current <0.002 e-/pix/s<br>• In a space radiation environment over mission lifetime<br>• 320 × 256 format array, 24 μm pixels<br><br><br>• 1k × 1k pixel array, 15 μm pixels | 4 | 5 |

| Title | Description | Section | State of the Art | Capability Needed | TRL 2019 | Expected 2023 TRL |
|-------|-------------|---------|------------------|-------------------|----------|-------------------|
| UV Microchannel Plate (MCP) Detectors | Low-noise detectors for general astrophysics as low as 115 nm | *Section 11.4.4.3* | • MCPs: QE 44% 0.115–0.18 μm with alkalai photocathode, 20% with GaN; dark current ≤0.1–1 counts/cm²/s with ALD activation and borosilicate plates | • Dark current <0.001 e·/pix/s (173.6 counts/cm²/s), in a space radiation environment over mission lifetime, • QE>50% for 115–300 nm wavelengths | 4 | 4 |
| Microthrusters | Jitter is mitigated by using microthrusters instead of reaction wheels during exoplanet observations | *Section 11.6.1.1* | • Colloidal microthrusters 5–30 μN thrust with a resolution of ≤0.1 μN, 0.05 μN/√Hz, 100 days on-orbit on LISA-Pathfinder • Colloidal microthrusters with 100 μN thrust and 10-year lifetime under development • Cold-gas micronewton thrusters flown on Gaia (TRL 9), 0.1 μN resolution, 1 mN max thrust, 0.1 μN/sqrt (Hz), 4 years of on-orbit operation | • Thrust capability: 350 μN with 16 thruster cluster • Thrust resolution 4.35 μN • Thrust noise: 0.1 μN/√Hz • Operating life: 5 years | 4 | 5 |
| **Enhancing Technologies** | | | | | | |
| Far-UV Mirror Coating | Observatory Science imaging as low as 0.1 μm | *Section 11.7.1.1* | • For a ~0.1 μm cutoff, Al + LiF + AlF₃ has been demonstrated at the lab proof-of-concept level with test coupons achieving reflectivities <br> ▪ for >200 nm: 80% <br> ▪ for 103–200 nm: 70% <br> ▪ Lifetime: no loss of reflectivity after 3-year lab storage | • Reflectivity from 0.3–1.8 μm: >90% • Reflectivity from 115–300 nm: >80% • Reflectivity from 103–115 nm: >50% • Operational life: >10 years | 3 | 3 |
| Delta-Doped UV Electron Multiplying CCDs | Low-noise detectors for general astrophysics as low as 0.1 μm | *Section 11.7.3.1* | • Delta-doped EMCCDs: Same noise performance as visible with addition of high UV QE ~60–80% in 0.1–0.3 μm, dark current of 3 × 10⁻⁵ e·/pix/s beginning of life. 4k × 4k EMCCD fabricated. Dark current <0.001 e·/pix/s, in a space radiation environment over mission lifetime, ≥4k × 4k format fabricated | • Dark current <0.001 e·/pix/s, in a space radiation environment over mission lifetime, • ≥4k × 8k format for spectrograph run in full frame mode, • High QE for 100–300 nm wavelengths | 4 | 4 |
| Microshutter Arrays | An array of apertures for the UV spectrometer | *Section 11.7.2* | • 171 × 365 shutters with electrostatic and magnetic actuation (JWST NIRSpec, TRL 7) • 128 × 64 electrostatic actuated array at TRL 4; will fly in FORTIS sounding rocket summer 2019 • 840 × 420 electrostatic, buttable array developmental model with partial actuation | • 300 × 300 shutters needed | 3 | 5 |

2. Petal shape accuracy and stability (*Section 11.2.1.2*)

3. Scattered sunlight for petal edges (*Section 11.2.1.3*)

4. Contrast Performance Modeling and Validation (*Section 11.2.1.4*)

5. Lateral Formation sensing and control (*Section 11.2.2.1*)

At the core of the S5 activity, "starshade shape accuracy and stability requirements are derived from a comprehensive error budget that will be verified by mechanical and optical performance models anchored to subscale ground tests" (Ziemer 2018b).

### 11.2.1  Starshade: Currently TRL 4

#### 11.2.1.1  Starshade Petal Deployment Position Accuracy and Stability

The starshade must have the ability to stow, launch, and deploy the petals to within the deployment tolerances budgeted to meet the shape, and ultimately, the contrast requirements. The optical shield within the inner disk must deploy without damage and meet the opacity requirements. The inner disk perimeter truss and optical shield are separable with defined interfaces which enables parallel assembly and straight forward integration including pre- and post-optical shield integration verification of perimeter truss assembly deployment tolerances.

The starshade inner disk perimeter truss is an adaptation of the Astromesh antenna perimeter truss and is the structure to which the petals attach. The Astromesh antennas have successfully deployed at least nine times on orbit. In the launch stowed configuration, the petals are spirally wrapped around the stowed perimeter truss and central spacecraft in a similar fashion to the Lockheed Martin Wrap-rib Antenna concept. Wrap-rib antennae have successfully deployed hundreds of times on orbit.

The accuracy of deployment of the perimeter truss with linear spokes was measured on both a modified 12 m Astromesh antenna and a purpose-built truss of 10 m over multiple deployments. Both truss assemblies included four petals and deployed more than 15 times. The average position accuracy was less than 150 μm, well within the 500 μm accuracy required for HabEx for a 20 m truss.

The initial 10 m perimeter truss test took place in the Advanced Large Precision Structures (ALPS) Laboratory at JPL (**Figure 11.2-1**, top).

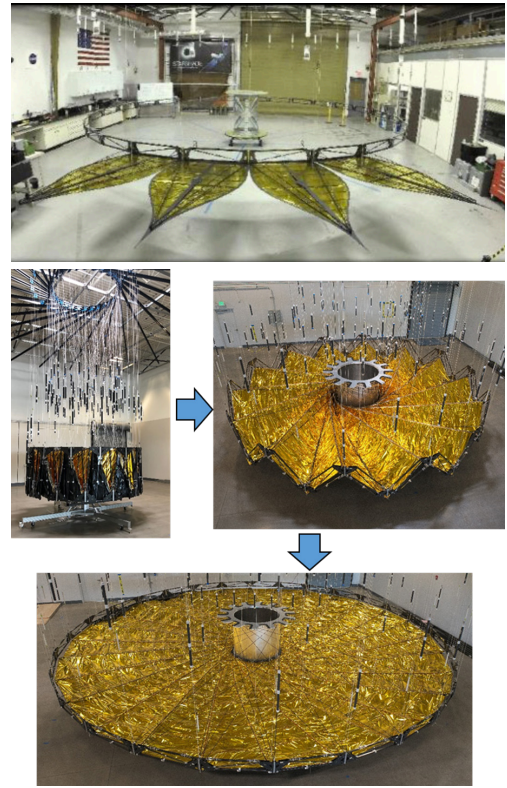

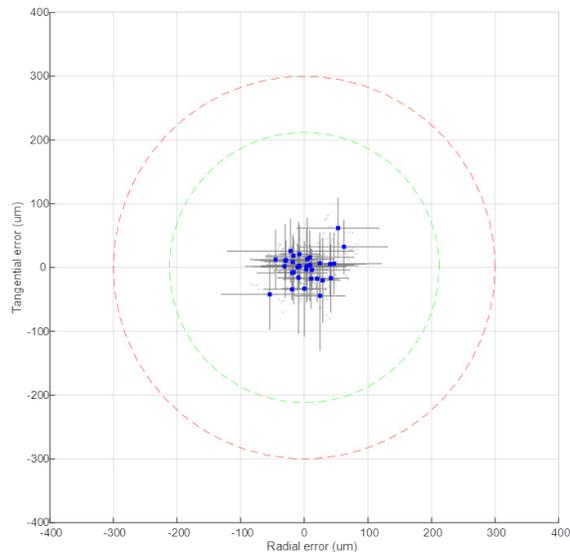

**Figure 11.2-1.** *Top:* Starshade 10 m perimeter truss with 3.5 m petals, four representative petals attached. *Middle:* Starshade 10 m perimeter truss with optical shield undergoing deployment. *Bottom:* Preliminary data for a subset of deployments shows petal position error well within a 300 μm radius.

The ALPS testbeds are used to develop the deployment architecture and gravity offload strategy for the starshade system.

More recently, in 2019, S5 built a medium fidelity 10 m inner disk subsystem with low fidelity optical shield that is currently undergoing deployment testing. Preliminary results show that deployment accuracy is well within 300 μm. The error scaled to the HabEx 20 m inner disk would double; the scaled error is firmly within the HabEx 600 μm requirement. The inner disk and shield testing is led by JPL and being performed in partnership with Tendeg at their Louisville, Colorado facility.

With the technology demonstrations to date, the Petal Deployment Position Accuracy technology item is at TRL 4.

### Path to TRL 5

S5 will achieve TRL 5 using an inner disk subsystem and truss bay assemblies. These allow for parallel paths of testing and model validation.

The truss bay assembly is the top image in **Figure 11.2-2** consisting of the longerons (purple horizontal bars) and node sub-assemblies (blue) and is the repeating element that comprises the perimeter truss; its critical dimensions determine the petal position. As an early risk reduction effort and an early milestone, the shape-critical components of the truss bay, the longeron and node assemblies, were built and tested for their critical dimensions due to thermal strain to validate the models at a component level; this demonstrates the largest contributor to petal position on orbit. Measurements of a longeron assembly length as a function of temperature have been made and are well within the requirements for S5. The longeron assembly and node assembly have been measured for dimensional stability post thermal cycling and have been shown to be well within requirements. Node assembly model validation is in process; preliminary model validation on the data to date show large margins on the S5 requirements. This effort has been mostly completed with some ongoing testing (Webb et al. 2019).

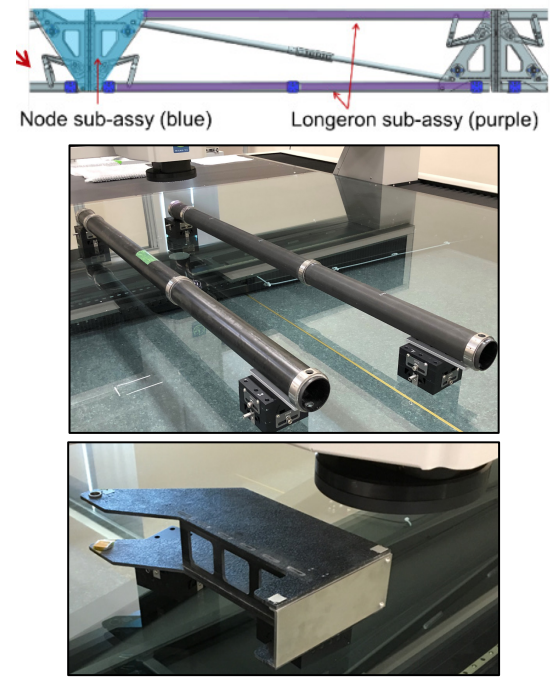

**Figure 11.2-2.** The shape-critical components of the inner disk truss bay assembly are the node assembly (*blue*) and the longeron assembly (*purple*). The longeron thermal stability test articles (*middle panel*) met requirements. The node assembly (*bottom* panel) met post-thermal cycle requirements; shape vs. temperature testing is in progress.

After component-level testing and model validation, the truss bay assembly will be tested. The truss bay is the critical level at which to test relevant environments for the perimeter truss. A medium-fidelity truss bay assembly that includes all features required to interface to the petal will undergo thermal cycling and be subjected to stowed stresses to validate models of load and thermal strain. Lastly a model of critical truss bay dimension change as a function of temperature will be validated to within 200 μm in a hot box over the full operating temperature range, similar to the thermal deformation testing of the Surface Water and Ocean Topography (SWOT) antenna engineering model truss at Northrop Grumman Aerospace Systems San Diego facility. It is scheduled to be completed by June 2023.

S5 will demonstrate deployment tolerances and validate load models and deployment kinematic models with the 10 m diameter medium fidelity disk subsystem pictured in **Figure 11.2-1** (middle). The article is half-scale for HabEx, which is sufficient for TRL 5 according to the consensus of the SSWG and the Chief Technologists of the Exoplanet Exploration Program (ExEP), Physics of the Cosmos (PCOS), and Cosmic Origins (COR) (Lisman 2019). As an early milestone, petal position accuracy will be demonstrated with the 10 m perimeter truss with shield (**Figure 11.2-1**, middle) with the full set of data for petal position accuracy by December 2019. The optical shield will then be upgraded to medium fidelity and four full petals, including petal optical shields, as well as 24 partial petals to show interface loads at all locations, will be added to the perimeter truss. Laboratory deployment of this article will validate the modeled deployment kinematics by measuring the shape as a function of deployment, and as a function of deploying and retarding forces. This will mature the Petal Deployment Position Accuracy and Stability gap to TRL 5 by June 2023.

### 11.2.1.2 Starshade Petal Shape and Stability

The starshade petal is designed to be a thermally stable structure without break points or hinges along the length of the petal. It unfurls passively to its deployed shape. It must maintain its shape, particularly the petal width, to enable deep starlight suppression. The carbon-fiber battens maintain a stable structure while the optical edge provides a precise shape. The carbon-fiber battens control the width of the petal and are manufactured by die pultrusion which results in very consistent, near zero, stable coefficient of thermal expansion (CTE); the CTE stability has been measured and characterized by S5. The optical edge is then attached to this thermally stable structure to provide the precise shape at the terminal edge. The optical edge is made of discrete ~ 1 m long sections, precisely bonded to the petal structure, made of amorphous metal alloy that has a precise, chemically etched bevel. The bevel limit solar scatter from the petal edge (see *Section 11.2.2.1*).

A large-scale test article has been fabricated which demonstrated the required in-plane shape accuracy as well as edge bevel. A petal 2.3 m wide by 6 m long was constructed and the shape measured in a lab environment using a coordinate measuring machine (Kasdin et al. 2012). The low-fidelity petal demonstrated petal manufacturing accuracy and therefore places the Petal Shape Accuracy technology at TRL 4.

#### *Path to TRL 5*

S5 will demonstrate requirements with two petal test articles. The first article is 1.5 m wide × 4 m long, will include amorphous metal edges, and will validate models of critical dimensions versus temperature as well as perform deployment cycles and thermal cycles (**Figure 11.2-3**, right). All thermal testing is complete; preliminary model validation shows performance with large margins. Post-thermal cycle shape stability is well within requirements. Deploy cycle and shape verification will complete the testing schedule, and all data will be reported on by December 2019. Optical shields will then be added to the petal. As a development activity and early risk reduction effort, the widest, central subsection of the petal has already been constructed and thermally cycled. The pre/post shape measurements show shape stability to within a few micrometers, adding confidence and reducing risk to the effort.

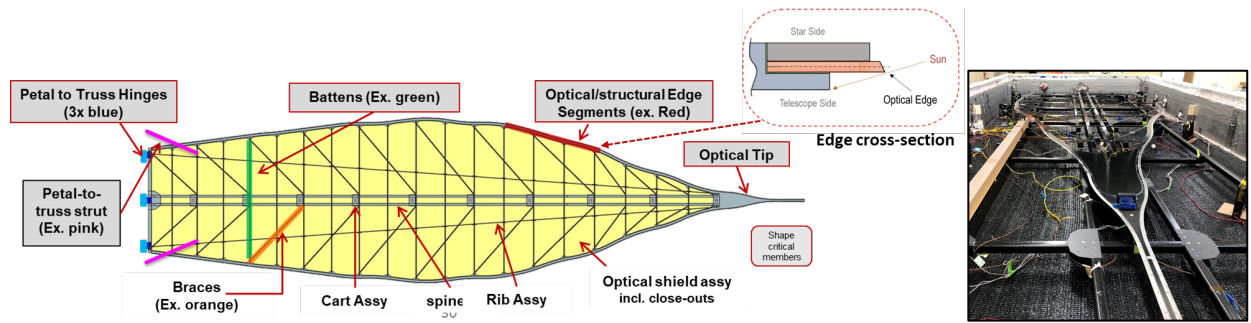

**Figure 11.2-3.** *Left*: Detail of starshade petal with inset of cross section of optical edge. *Right:* Petal test article #1 in the thermal chamber at Tendeg ready to undergo thermal stability testing.

The second article is 1.5 m wide and 6 m long and will include all features and interfaces at medium fidelity, including the precision bevel on the optical edges. Article #2 will repeat the testing. Both test articles are 1/2 scale in width, which is the critical dimension and sufficient scaling to be considered TRL 5 for HabEx.

### 11.2.1.3 Starshade Scattered Sunlight for Petal Edges

The primary function of the starshade optical edges is to provide the correct apodization function needed to suppress starlight to levels suitable for exoplanet direct imaging. To do this, light emanating from other sources—principally edge-scattered sunlight—must be mitigated. The intensity of this scattered light must be limited to below the exozodiacal background.

Light scatter is driven by the product of the area and reflectivity of the scattering surface. As such, to mitigate edge-scatter, the starshade optical edges must limit the product of these two parameters. An S5 trade study (Steeves et al. 2018) demonstrated that sharp, highly reflective edges resulted in much better solar scatter performance than low reflectivity edges due to the inherently much larger reflecting area of the low reflectivity coatings. The low reflectivity options tested did not meeting solar scatter performance, while the sharp and reflective edges did.

The edges must demonstrate shape accuracy, shape stability and low edge scatter to meet performance requirements. To resolve this technology gap, JPL led an effort to produce prototype optical edges. These edges were constructed using thin strips of amorphous metal alloy. Amorphous metal was used because the absence of material grain structure allows extremely sharp edges at the sub-micron level. Chemical etching techniques produced the necessary beveled edge and can be implemented on meter-scale edge segments with 10s of microns-level in-plane tolerances and sub-micron terminal edge radius (reducing the effective area for solar edge scatter). Multiple coupon samples were constructed and their geometries were characterized using scanning electron microscope (SEM) images (**Figure 11.2-4**). A terminal radius of <0.5 µm was achieved with low levels of variability across each coupon (Steeves et al. 2018). The solar glint performance of these coupons was also established using a custom scattered-light testbed; measurements indicate that the scattered flux is dimmer than the predicted intensity of the background zodiacal light over a broad range of Sun angles (Steeves et al. 2018).

*Path to TRL 5*

Suitable solar glint performance was demonstrated at the coupon level, and recently at half-meter-scale. S5 demonstrated TRL 5 performance in June 2019 with a half-meter-scale test sample that meets scattering and shape requirements after deployment cycles (**Figure 11.2-4**, bottom). The in-plane shape profile for the edge met the ±20 µm shape requirement. The edge meets the scatter requirements post-thermal and post-deploy cycling (Hilgemann et al. 2019). The half-meter-long optical edge is half-scale in length and exactly in-scale for bevel requirements for HabEx, so the half-meter S5 optical edge is considered TRL 5 for HabEx. A TRL 5 milestone report to the ExoTAC is in progress.

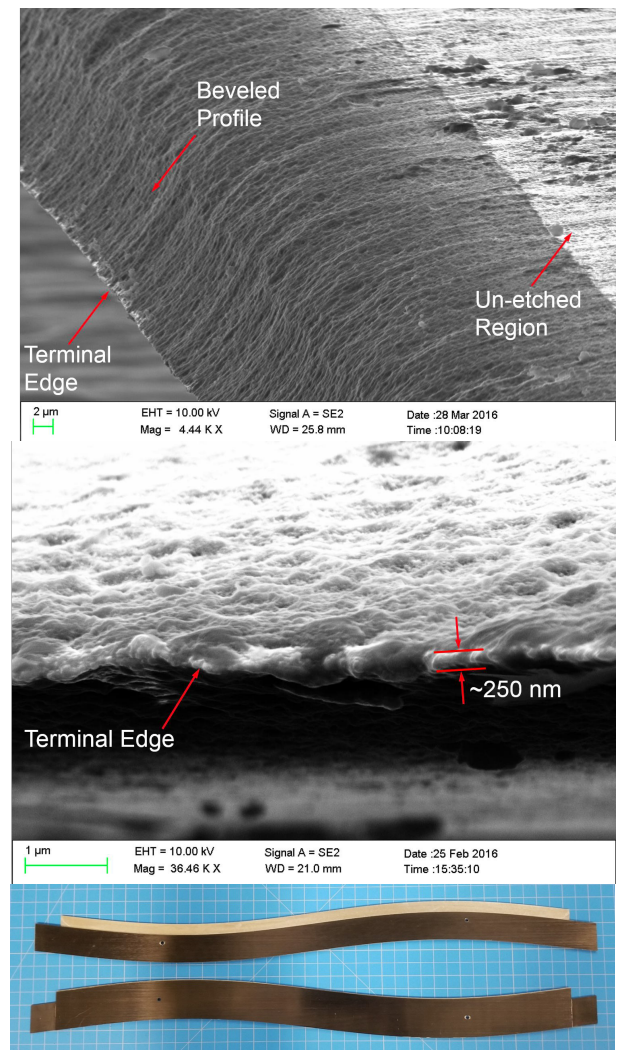

**Figure 11.2-4.** *Top and middle:* SEM images of the starshade beveled edge and terminal edge of coupons. *Bottom:* Half-meter-long S5 edge segments.

### 11.2.1.4 Starshade Contrast Performance Modeling and Validation

The optical performance of the starshade can only be tested on the ground at a small scale due to practical considerations: near full scale testing is not possible on the ground due to the large starshade/sensor separation distances required. Scalar diffraction theory shows that the starshade diameter and distance to aperture is invariant to scale, as long as the scale is significantly larger than the wavelength of light.

To discuss the scale invariance in scalar diffraction theory, the Fresnel number is a convenient concept.

The Fresnel number is defined for an electromagnetic wave passing an aperture as:

$$F = \frac{r^2}{\lambda L}$$

where $r$ is the aperture (or starshade) radius, $L$ is the distance from the aperture (or starshade) to observation plane (or telescope), and $\lambda$ is the wavelength of the light of interest. The scalar diffraction integral preserves the Fresnel number and the propagating optical field is invariant to this number.

Thus, the light suppression achieved in a 1 cm telescope aperture using a 10 cm diameter starshade at 100 m distance will also be achieved in a 1 m telescope aperture using a 10 m diameter starshade at 10,000 m distance, if the starshades have identical shapes. This property makes it possible to reproduce and verify, in a relatively small ground-based testbed, the optical performance expected for a 52 m starshade operating in formation with HabEx. The masks studied in the Princeton testbed are close to the same Fresnel number of the HabEx starshade.

Sub-scale starshade optical performance testing and model verification are now underway at a Princeton University testbed (**Figure 11.2-5**) (Kim et al. 2017). The starshade is a 50 mm diameter mask etched into a silica substrate by the JPL Microdevices Lab. A diverging 0.638 µm laser beam illuminates the starshade mask from

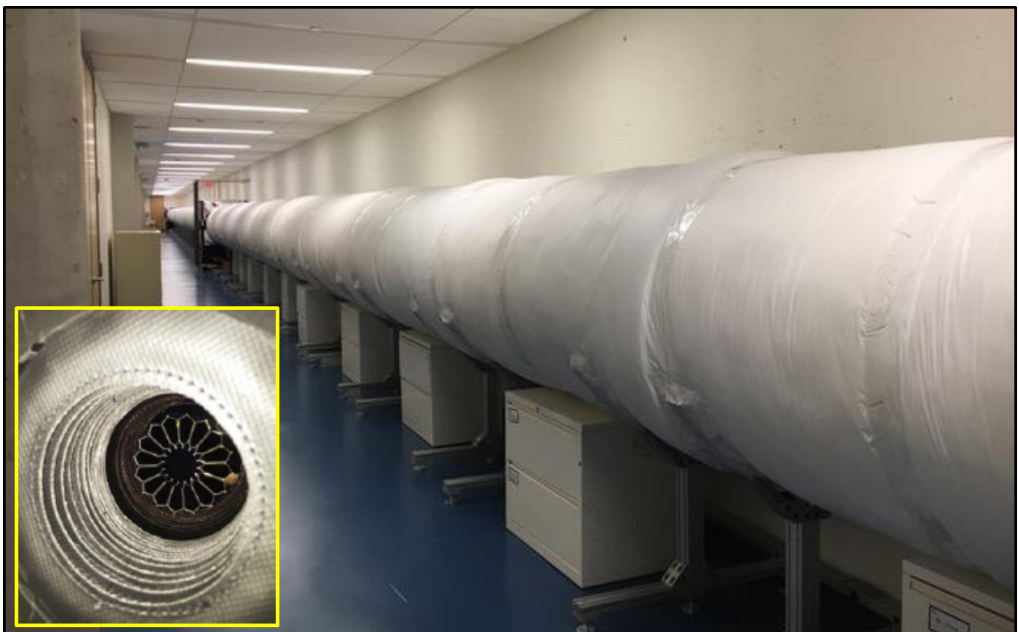

**Figure 11.2-5.** S5's starshade model validation testbed at Princeton. Model starshade shown in inset.

27 m away. The starshade and a 5 mm wide aperture are separated by 50 m. Contrast levels are detected with an EMCCD. Contrast performance has consistently improved as mask fabrication precision improved.

The testbed achieved an average contrast of $1.5 \times 10^{-10}$ at the inner working angle (IWA) for $F$=14. At the IWA, ~44% of the search space was below $1 \times 10^{-10}$ while the rest was contaminated by non-scalar diffraction related to the microscopic openings at the inter-petal valleys of the laboratory-scale mask, an effect caused by the scale of the test article and not relevant to full scale starshades (Harness et al. 2019a). Despite the contamination, the annular contrast was below $1 \times 10^{-10}$ immediately outside the IWA and beyond, with a contrast floor of ~$2 \times 10^{-11}$, as shown in **Figure 11.2-6**. The experiment was repeated at four wavelengths spanning a 10% bandwidth with similar results (Harness et al. 2019b). These results advance the starshade modeling technology gap to TRL 4.

*Path to TRL 5*

S5 will achieve TRL5 for HabEx via one additional milestone: optical model validation through measurement of contrast degradation induced by shape errors deliberately built into starshade masks (S5 Milestone 2). Milestone 2 is expected to complete by January 2020 (Willems 2018).

## 11.2.2 Starshade: Currently TRL 5

### 11.2.2.1 Starshade Lateral Formation Sensing

S5 demonstrated lateral sensing of the starshade to TRL 5 in Flinois et al. 2018 with formal approval by the Exoplanet Technical Analysis Committee (Boss et al. 2019).

Lateral formation sensing is needed during science observations with the starshade to keep the starshade centered on the target star. The sensing approach uses pupil-plane images at a wavelength outside the science band, where the starshade occulter's attenuation of the starlight is only ~$10^{-4}$ or less (see **Figure 11.2-7**). The shadow has sufficient structure that by matching de-trended pupil-plane images to a library of pre-generated images the lateral position of the starshade can be determined to 15 cm ($3\sigma$) with short exposures. See *Section 8.1.7* for more details on the design and operation of the HabEx formation flying system.

S5 matured lateral sensing to TRL 5 with three tasks: the first verified the optical model out-of-band suppression patterns in a testbed, the second used an algorithm to infer lateral offset with testbed images at flight signal-to-noise ratio (SNR), and the third demonstrated lateral position control with a Matlab-simulated lateral control servo.

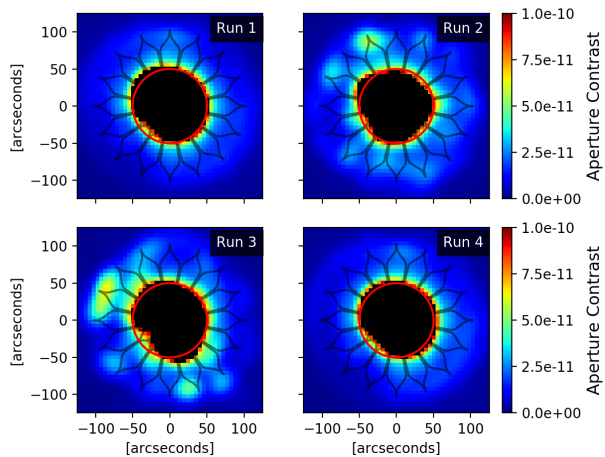

**Figure 11.2-6.** S5 Milestone 1A monochromatic contrast better than $10^{-10}$, averaged in $\lambda/D$ wide photometric aperture centered on each pixel. Black pixels show contrasts $3\sigma$ worse than $10^{-10}$.

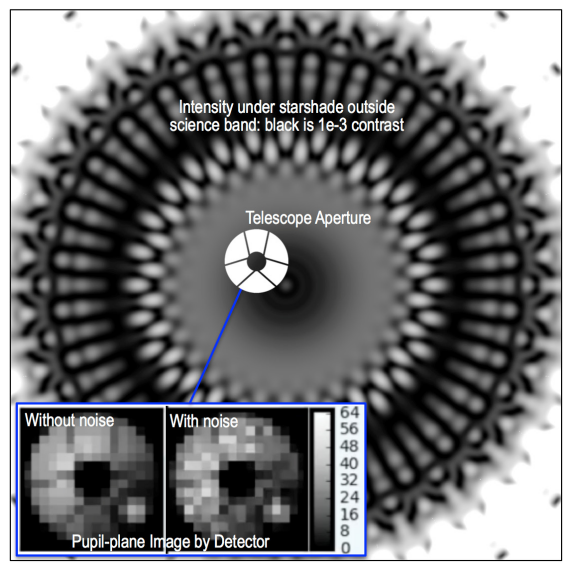

**Figure 11.2-7.** The starshade precision lateral sensing testbed used a WFIRST-like pupil for pupil-plane image matching.

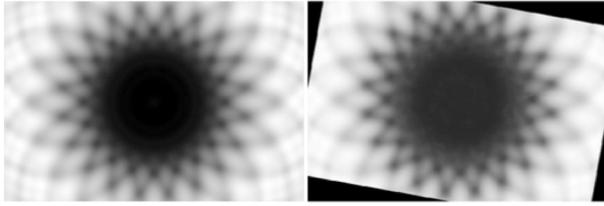

**Figure 11.2-8.** Out-of-band starshade shadow images from the Low-Contrast Testbed. *Left:* Simulated image. *Right:* Testbed image.

The lateral sensing testbed was similar to the Princeton testbed but scaled down with lower fidelity and lower contrast. The testbed is 2 m long with a Fresnel number close to the HabEx number, and includes a 6 mm diameter starshade, and a detector mounted on a motion stage. The detector was placed at various positions, including position extremes, and the out-of-band images recorded. An example comparison of a predicted image and an initial image from the low-contrast testbed is shown in **Figure 11.2-8**.

Formation flying control was demonstrated in simulation using a noise model for the lateral sensor (**Figure 11.2-9**). The control simulation included the validated sensor model, realistic thruster dynamics that require thrust allocation, thruster minimum impulse, and errors in attitude knowledge of the starshade. The dynamics used a representative maximum gravity gradient and an optimal circular deadbanding algorithm, including representative drift times between thruster firings. In addition, an estimator combining the lateral sensor and radio frequency (RF) ranging measurements was developed. Realistic actuator misalignments and mass property uncertainties were also included.

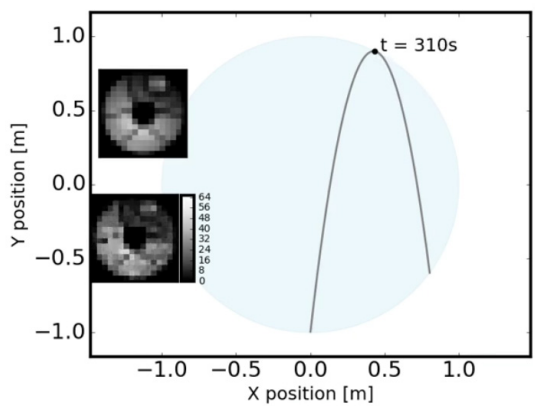

**Figure 11.2-9.** Formation flying simulation showing path of starshade within 1-meter disk (*blue*) and corresponding sensing images: *top inset* is a simulated image; *bottom inset* is testbed measured image.

The high-fidelity formation flying simulation, using images from the testbed, demonstrated lateral control to within a 1 m radius disk. The algorithms were found to provide an approximately periodic drift time of about 10 minutes on average for worst-case acceleration conditions for HabEx (Ziemer 2018a). The control provides high observation efficiency for exoplanet science.

## 11.3   Telescope

### 11.3.1   *Telescope: Currently TRL 4*

Ground-based 4-meter-class monolithic primary mirrors first came into use in 1948 with the 200-inch Hale telescope. Today, there are more than two dozen monolithic telescopes with apertures between 3.5 m and 8.2 m operating worldwide. Magnesium-fluoride over coated aluminum mirrors has been used successfully in space applications for decades; the Hubble Space Telescope (HST) being the most notable example. Building a 4 m mirror for space applications is not a technology development, but building a 4 m mirror capable of supporting coronagraphy at the $10^{10}$ contrast level is. It is the demanding requirements levied on the mirror by the coronagraph that turns an engineering development into a technology development. This section describes the technology development work required to ensure that the HabEx primary mirror will be able to deliver the performance needed to support high contrast coronagraphy.

#### 11.3.1.1   Large Aperture Monolithic Primary Mirror Fabrication

Building a 4 m primary mirror for a space telescope presents a number of challenges: size and mass, stiffness, thermal stability, and thermal homogeneity. Size and mass can constrain the entire telescope flight system architecture, but this has largely been mitigated by the next generation of super heavy lift launch vehicles currently in development. Both the Space Launch System (SLS) and the Big Falcon Rocket (BFR) can easily accommodate a HabEx-like 4 m space telescope with ample mass and volume margin. Stiffness affects both manufacturability (gravity sag removal and surface figure) and mechanical stability through the coupling of undamped vibrations. HabEx has mitigated the latter through the removal of reaction wheels and the introduction of microthrusters, but the former issue—manufacturability—along with the thermal issues, remain areas of attention in the mirror's technology development.

The 4 m monolith mirror design is currently considered at TRL 4 by analysis of the design and assessment of fabrication capabilities and prior sub-scale mirrors. TRL 5 would be achieved with a demonstration of thermal uniformity and surface figure error of a full-scale mirror first in a ground support equipment (GSE) mount and then in a proto-flight mount. TRL 6 performance would be demonstrated on a proto-flight mirror assembly, which includes the mirror, mount actuators, and launch locks, that includes environmental testing for launch shock and vibration, as well as thermal balance. This section discusses the TRL 4 assessment and the path to TRL 5. The path from TRL 4 to TRL 6 is presented in *Appendix E*.

As described in *Chapter 6*, the primary mirror is an open-back Zerodur® design with some lightweighting, but not extreme lightweighting. With the launch mass capability of the SLS, there is no need to push the state of the art in lightweighting, and the extra mass adds thermal inertia which actually benefits the telescope thermal design. Zerodur® has flown in space over 30 times (Döhring et al. 2009) so the material is at TRL 9.

Both SCHOTT's Zerodur® and Corning's ULE may be acceptable materials for the primary mirror, but Zerodur® is manufactured out of a single boule and is expected to produce better CTE homogeneity. Homogeneity reduces thermal-induced focus error (Jedamzik and Westerhoff 2017) so it is an important quality in space telescope mirrors. Zerodur® Extreme achieves CTE homogeneity of better than 7 ppb/K. CTE homogeneity of 1–5 ppb/K has been measured through the thickness of a sample boule. Zerodur® CTE homogeneity was verified at 5 ppb/K using SCHOTT's extremely lightweight Zerodur® mirror (ELZM) (**Figure 11.3-1**) via thermal testing at Marshall Space Flight Center (Brooks et al. 2017). SCHOTT's 4.2 m meniscus secondary mirror for the European Extremely Large Telescope (E-ELT) has a measured CTE homogeneity of 2.83 ppb/K. A CTE homogeneity of 5 ppb/K with 5 mK thermal control provides wavefront stability better than 2 pm rms for the primary mirror design, which is sufficient for the HabEx wavefront stability requirement. In addition to the excellent CTE homogeneity, the Zerodur® coefficient of thermal expansion (CTE) can be 'tuned' to provide zero-CTE over a range of operational temperatures.

Manufacturing techniques have now advanced to a sufficient capability to enable the development of a 4 m space-qualified mirror. SCHOTT has the capacity to make 4.2 m diameter × 42 cm thick mirror blanks (SCHOTT white paper in *Appendix F*). SCHOTT manufactured the recent 4.26 m Daniel K. Inouye Advanced Solar Telescope (DKIST) primary mirror (Jedamzik et al. 2014; **Figure 11.3-3**) and the E-ELT secondary mirror (**Figure 11.3-3**; ESO 2019). In addition, SCHOTT regularly manufactures 2 m × 40 cm lightweight ultra-stiff structures from Zerodur® with ultra-CTE homogeneity for its lithography bench product line (Westerhoff and Werner 2017). SCHOTT uses computer-controlled-machining to produce ribs as thin as 2 mm (**Figure 11.3-4**) (see the SCHOTT white paper on manufacturability of a 4 m monolith for HabEx in *Appendix F* for further details).

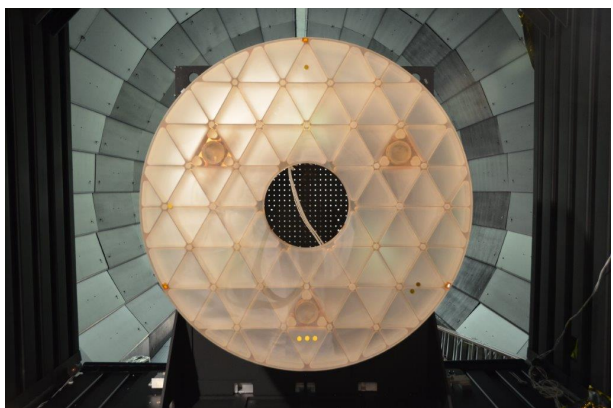

**Figure 11.3-1.** SCHOTT 1.2 m diameter and 125 mm thick Zerodur ELZM mirror in MSFC XRCF thermal/vacuum test chamber (Brooks et al. 2017).

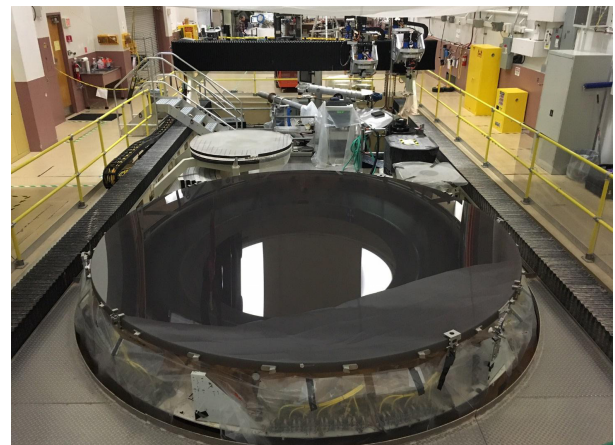

**Figure 11.3-2.** 4.2-meter Daniel K. Inouye Solar Telescope primary mirror (Oh et al. 2016).

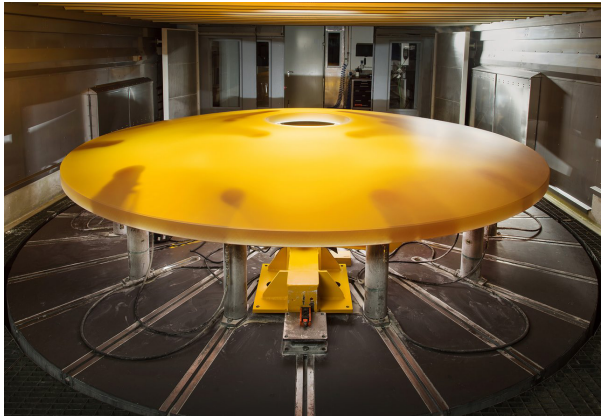

**Figure 11.3-3.** 4.2 meter European Extremely Large Telescope secondary mirror blank. Credit: SCHOTT.

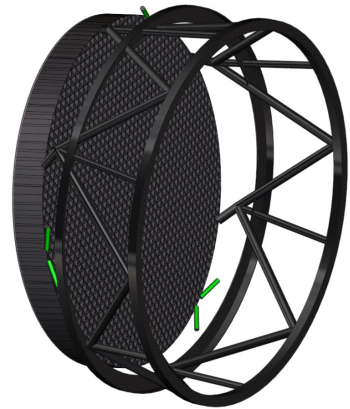

**Figure 11.3-5.** Baseline 4 m × 40 cm thick flat-back open-back isogrid core Zerodur® mirror.

Industry has already demonstrated other important mirror manufacturing capabilities (**Figure 11.3-5** and **Figure 11.3-6**). Collins has fabricated mirrors with total surface figure errors below 6 nm rms, mid-spatial-frequency error under 2 nm rms, and surface roughness of less than 1 nm rms. On Chandra, Collins produced Zerodur® mirror surfaces with a roughness of 0.2 nm rms. Other mirror fabricators capable of meeting this level precision include L3 Brashear, University of Arizona Optical Sciences Center, Harris, and REOSC.

With respect to gravity sag, Collins has demonstrated TRL 9 ability to back-out these errors in mirrors as large as 2.5 m to an accuracy of less than 3 nm rms (Yoder and Vukobratovich 2015) using the gravity flip metrology method. This method allows empirical determination of gravity deformation by comparing results from different mirror gravity flip orientations. Traditionally stiff space mirrors show a few interference fringes of gravity sag; the less stiff HabEx mirror will show up to 300 interference fringes. A computer-generated hologram (CGH) could compensate for the predicted gravity sag with each orientation. CGHs have been used to test aspheric mirrors with 2 nm rms uncertainties (Stahl and Morgan 2019). Demonstration of gravity sag back-out with the HabEx lower stiffness mirror is on the path planned for TRL 5 maturation of the primary mirror fabrication.

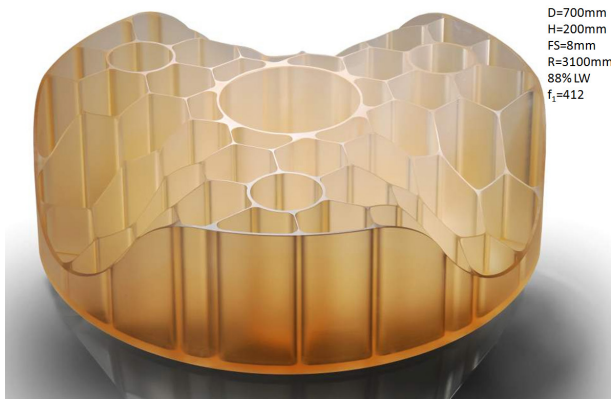

**Figure 11.3-4.** SCHOTT 700 mm diameter and 200 mm high Zerodur® demonstration piece showing advanced light-weighting, cells with 2 mm machined walls, and contouring of the back. The back of the facesheet within each pocket is conformal to the facesheet. Credit: SCHOTT.

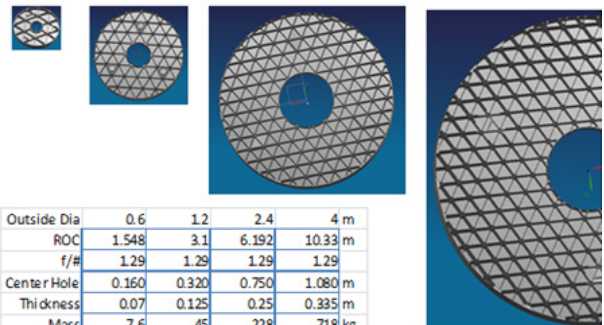

| Outside Dia | 0.6 | 1.2 | 2.4 | 4 | m |
|---|---|---|---|---|---|
| ROC | 1.548 | 3.1 | 6.192 | 10.33 | m |
| f/# | 1.29 | 1.29 | 1.29 | 1.29 | |
| Center Hole | 0.160 | 0.320 | 0.750 | 1.080 | m |
| Thickness | 0.07 | 0.125 | 0.25 | 0.335 | m |
| Mass | 7.6 | 45 | 228 | 718 | kg |
| fo | 419 | 213 | 115 | ~80 | Hz |

**Figure 11.3-6.** Results of analysis of 0.6, 1.2, 2.4, and 4 m lightweight Zerodur® mirror substrates by the SCHOTT process. Masses represented are consistent with most present and anticipated OTAs for spaceborne missions. Each case was constrained to satisfy launch load with strength margin, although launch locks are assumed for the 4 m case (Hull et al. 2013). Credit: SCHOTT.

Other ways to guide the optical surface finishing and remove gravity surface figure error are possible. Optical metrology accuracy, dynamic range, and spatial resolution are critical. White papers in *Appendix F* by AOS and Harris Corporation assess methods to meet surface figuring to better than 18 nm rms and surface roughness of about 2 nm rms.

Actuators could be used on the back of the primary mirror to reduce risk with gravity sag back out. The actuators could be optimally placed to give control authority to the low-order WFEs that could arise from gravity back-out errors. Modeling shows that optimal placement of actuators can reduce mirror errors to 2.5 nm rms. These actuators could serve dual duty as launch locks.

*Path to TRL 5*

TRL 5 would be achieved with a demonstration of thermal stability and surface figure error of a full-scale mirror in a ground support equipment (GSE) mount. A sequence of tests can minimize cost and risk. A thermal test of the mirror blank, polished to a sphere, would show that the CTE homogeneity requirement over the surface of the mirror is met. Next the blank would be CNC machined for lightweighting, still with a spherical surface polished so that the gravity sag can be measured; the spherical surface makes optical alignment of the test faster and does not require an additional, custom null corrector element. Then the mirror would be ground to the aspheric prescription and final polishing performed. Note that additional facesheet thickness will be required in the spherical surface of the mirror that will be ground away in the figuring of the aspheric surface. Finally, the surface figure error of the final polished surface is tested and gravity sag backed-out. To verify the mechanical stiffness of the mirror, a mechanical ping test will be performed. Shock and vibration are not considered part of the TRL 5 maturity because they are dependent on the mirror assembly, particularly the mirror mount design, and are more appropriately tested at the assembly level for TRL 6.

### 11.3.1.2 Large Mirror Coating and Uniformity

All HabEx instruments are affected by the telescope mirror coating performance, so all instruments must be considered when defining the mirror reflective properties. Telescope mirror coatings for the HabEx mission require the following fundamental properties:

- Spectral coverage with high throughput from 0.115 to 1.8 μm
- Uniformity of reflectivity—both amplitude and phase—of ≥99% over the full aperture
- Consistent reflective properties for at least 10 years. Since HabEx is serviceable but cannot replace its mirrors, a coating that can last 20 or 30 years is highly desirable.

**Table 11.3-1.** State-of-the-art coatings for large aperture space telescope primary mirrors.

| | HST | Kepler | JWST | FUSE |
|---|---|---|---|---|
| **PM Size** | 2.4 m monolith | 1.4 m monolith; 950 mm entrance aperture | 18 hexagonal Be mirror segments (~1.52-m wide) with total collecting area of 25 sq m | Four mirrors of 38.7 × 35.2 cm each |
| **Spectral Range** | 0.115–2.5 μm | 0.3–1.2 μm | 0.7–20 μm | 0.0905–0.1187 μm |
| **Operational Lifetime** | >27 years | > 9 years | 10 years max (Design Life) | 8 years (had significant Al coating degradation) |
| **Coating** | Protected Al (MgF$_2$ on Al) | Protected Ag (multilayer on Ag) | Protected Au | LiF on Al on 2 mirrors and SiC on other 2 |
| **Uniformity** | No ground measurement | <30 nm PV; Reflectivity variation <2% | <1% thickness variation among the 18 segments. <10 nm pv; Reflectance variation <0.5% in the IR | Not measured |

This section describes both the high-heritage baseline coating, and better-UV-performing alternatives that could be considered if they are technologically mature at the time of the future mission.

### 11.3.1.2.1 Baseline Al+MgF₂ Coating

The current state-of-the-art for space telescope mirror coatings is summarized in **Table 11.3-1**. HabEx selected an HST–like coating: an aluminum reflecting surface with a magnesium-fluoride protective overcoat. The materials and processes have been flight-proven by HST over the last 29 years and are at TRL 9. Silver and gold coatings do not meet the spectral range needed by HabEx, and though the lithium-fluoride overcoat used on the Far Ultraviolet Spectroscopic Explorer (FUSE) went below 0.1 μm in spectral coverage, the coating had degradation issues during the FUSE mission (Fleming et al. 2017). Work to develop a stable LiF protected aluminum coating for spectral coverage below 0.115 μm continues. Should one be developed in time for a future HabEx mission, the improved coating would offer a significant enhancement to the current ultraviolet science case.

Aluminum mirrors overcoated with MgF₂ have been used on space telescopes since the 1970s. Most notable is the mirror coating for the long-operating HST observatory. **Figure 11.3-7** shows a model of reflectance performance of a HST-like mirror coating in comparison with ideal Al with no overcoat. The coating on HST provides high reflectivity at wavelengths greater than ~0.12 μm. Below 0.115 μm, the reflectivity drops sharply to less than 20% due to the absorption edge of MgF₂. This level of performance is sufficient to meet HabEx baseline requirements.

Uniformity of the 4 m mirror coating is the primary coating issue for HabEx. Coating uniformity—specifically, reflectivity phase and amplitude—is mainly a result of the coating process controls relevant to the specific chamber geometry. As such, engineering development is needed to build a sufficiently large chamber for the 4 m primary, and to optimize manufacturing processes to ensure a coating with less than 1% variability, as desired for HabEx.

The Kepler 1.4 m primary mirror has a protected silver coating generated using ion assisted deposition with a moving source, resulting in a thickness uniformity of about 30 nm peak-to-valley with about 2.5% reflectivity variation (Sheikh et al. 2008). Better uniformity has been achieved on JWST. The JWST gold mirrors showed <10 nm peak-to-valley thickness non-uniformity with <0.5% reflectance non-uniformity in the infrared among its 18 hexagonal segments (Lightsey et al. 2012).

In 2004, Kodak (now Harris Corp, Rochester) demonstrated reflectivity variability of less than 0.5% for a high reflectivity protected silver coating over a 2.5 m diameter optic as part of the Terrestrial Planet Finder Technology Demonstration Mirror project (Cohen and Hull 2004).

These historical examples of large space mirrors with highly uniform protected metal coatings are subscale manufacturing demonstrations for a future 4 m HabEx mirror with an HST-like Al+MgF$_2$ coating.

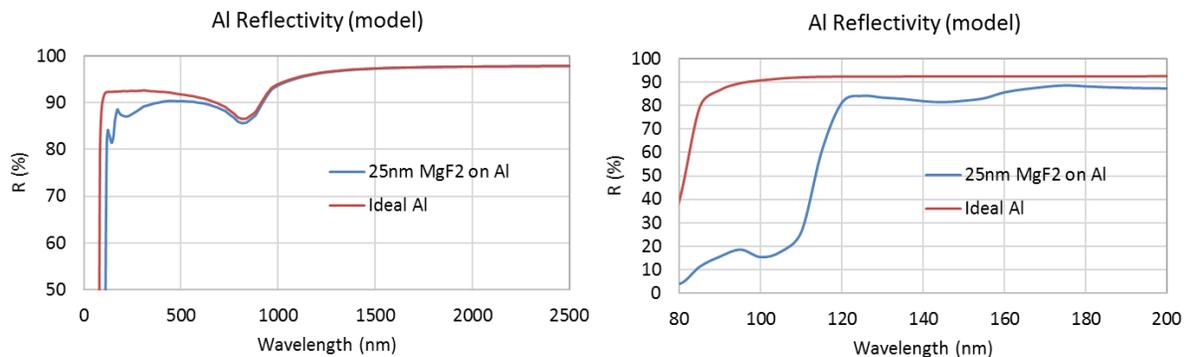

**Figure 11.3-7.** Aluminum reflectivity with and without a protective layer of MgF₂ (**HST-like model prediction**); the spikes and dips between 0.09 and 0.2 μm are a consequence of interference effects and absorption due to the protective layer and depends critically on the optical constants of the material, which depend on the coating process. The dip at ~0.83 μm is due to the native absorption property of Al.

### Path to TRL 5

The main element needed to advance the TRL of mirror coating uniformity for a 4 m mirror is a coating chamber of sufficient size; such a chamber does not currently exist. ZeCoat plans to build a 6 m coating chamber at their new facility in St. Louis which could create a uniform AL+MgF$_2$ coating on a mirror up to 5 m in diameter (see ZeCoat white paper in *Appendix F*). ZeCoat is currently developing their coating approach in a 2.4 m chamber via NASA Astrophysics Research and Analysis (APRA). The Al process uses a network of many sources evaporating quickly and simultaneously for uniformity. The MgF$_2$ coating is produced similar to the Kepler mirror with motion-controlled sources. The goal of the APRA is to advance their manufacturing process for both the Al and protective MgF$_2$ coatings to TRL 5 by 2021 for a 2.3 m diameter mirror.

For TRL 5, the coating uniformity would need to be demonstrated on coupons representing a 4 m diameter mirror. The development process would begin with coupons in a single line across the diameter of the chamber. These provide inexpensive iterations in the engineering of the coating process. Then coupons would be placed in an orthogonal line for additional tuning of the process. Finally, many coupons would be placed at many locations over the entire surface of the mirror. Such a task would not likely be funded unless HabEx is prioritized in the Decadal Survey.

Demonstrating the coating on a full-scale 4 m mirror to the required uniformity would achieve TRL 6.

### 11.3.2  Telescope: Currently TRL 5

Laser metrology for the sensing and control of the rigid body positions of the secondary mirror and tertiary mirror relative to the primary mirror is currently at TRL 5.

#### 11.3.2.1  Laser Metrology

As noted in *Section 6.8.5*, a laser metrology truss provides the sensing end of a Laser Metrology Subsystem (MET) rigid body control loop for the telescope optics. Using rigid body actuators on the secondary and tertiary mirrors, MET actively maintains alignment of the front-end optics, removing the primary source of telescope wavefront drift. This breakthrough technology operates at high bandwidth and can maintain control throughout all phases of the mission, effectively creating a near perfect, infinitely stiff, telescope truss.

The backbone of MET is the laser metrology gauge which monitors any changes in the distance to a retroreflector. Each planar lightwave circuit (PLC) beam launcher, or gauge, requires a stable laser source and a phase meter to operate. Each of these components are at TRL 6 or higher.

The laser source for MET at JPL has historically been a Nd:YAG non-planar ring oscillator (NPRO). A similar, TRL 9, Nd:YAG ring laser has flown on LISA Pathfinder for the laser metrology system monitoring the test masses (Voland et al. 2017). A RIO PLANEX laser module with ~1.5 µm wavelength, which is better matched to the PLC beam launchers, was developed as a Grace Follow-On candidate via Small Business Innovation Research (SBIR) (Stolpner 2010) and independently assessed at TRL 6 for GSFC (Piccirilli 2011). Since laser metrology operates as a heterodyne system, a thermally stabilized PLC beam launcher (**Figure 11.3-8**) is sufficient for the purposes of HabEx.

The phase meter monitors the heterodyne measurement signal and compares it to the reference signal. Changes in the phase between the two signals are directly related to the change in the distance between the PLC beam launcher and the retroreflector. The LISA-Pathfinder phase meter is an example of a suitable phase meter that is at TRL 9. In addition, the same type of phasemeter has recently flown on the Grace Follow-On mission (Heinzel et al. 2017).

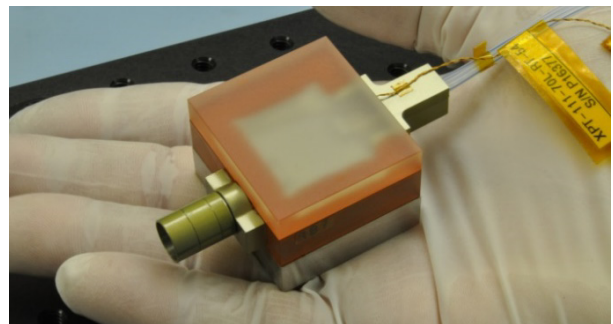

**Figure 11.3-8.** PLC beam launcher.

The final component of the MET system is the beam launcher. For HabEx, the beam launcher is the beam splitting/combining system that must be mounted to the telescope optics and therefore must be small and of similar construction. The PLC beam launcher evolved from the large, external metrology beam launchers developed for the Space Interferometry Mission and has been refined into a very compact, stable component. The PLC beam launcher has been fully qualified and 10 s of units have been built in order to refine manufacturing process and improve performance. Recent testing at JPL (dynamics test, thermal cycling, 100 krad radiation) has brought the PLC beam launchers to TRL 6.

*Path to TRL 5+*

The individual components of the metrology system are at TRL 6 or greater. The PLC BL thermal sensitivity was tested and the results indicate, via analysis, that the PLC BL would enable 0.1 nm uncorrelated noise for a gauge. However, the specific components of the HabEx metrology gauge design (PLC BL, Grace Follow-On phase meter, and laser) have not yet been assembled to a metrology system and tested.

The path to TRL 6 will be to assemble a single metrology gauge and demonstrate the uncorrelated noise at 0.1 nm, measure the PSD of the noise, and measure the PSD of the noise correlated to the thermal sensitivity of the gauge. This test can be easily performed at JPL using the GFO engineering model phasemeter, existing PLC BL, and an engineering model of the laser.

## 11.4 Instruments

### 11.4.1 Instruments: Currently TRL 4

Several technologies relevant to the HabEx instruments are currently TRL 4: coronagraph architecture, Zernike wavefront sensing and control (ZWFS), deformable mirrors, and detectors including UV Microchannel Plate detectors, delta doped UV-EMCCDs and deep depleted EMCCDs for enhanced NIR response.

#### 11.4.1.1 Coronagraph Architecture

The 2010 Decadal report recommended medium investment in direct imaging technology including coronagraphs (NRC 2010). Tremendous progress in coronagraph performance has been made in the last decade. Through the efforts on the WFIRST technology demonstration coronagraph and several strategic technology investments by the NASA Exoplanet Exploration Program, exoplanet direct imaging contrast performance is nearing the levels required to detect Earth-sized planets in the habitable zone of nearby stars.

This section covers the state of the art for the two coronagraph architectures—the vortex coronagraph (VC) which is the baseline for HabEx, and the hybrid Lyot coronagraph (HLC) which is the alternate—and the work needed to advance these technologies to TRL 5 with respect to HabEx's requirements.

##### 11.4.1.1.1 VVC and HLC Architectures

As noted earlier, the current HabEx design uses the VVC as the baseline design and the HLC as a backup option. Details of the coronagraph design and decision rationale are discussed in *Section 6.3.1*.

The block diagram in **Figure 11.4-1** identifies the major coronagraph elements common to both the VVC and the HLC: a fine-steering mirror (FSM) to control pointing and mitigate jitter; two $64 \times 64$ DMs to correct wavefront error (WFE); and a low order wavefront sensor (LOWFS) to detect WFE. These architectures have nearly the same optical layout so they are of similar size and footprint, and could be exchanged even in a fairly advanced design with minimal impact, adding flexibility to any future mission development.

The vortex coronagraph uses a focal-plane phase mask (**Figure 11.4-2**) to redirect the on-axis starlight to the outside of a subsequent pupil image, where it is blocked. The vortex phase pattern consists of an azimuthal phase ramp that reaches an even multiple of $2\pi$ radians in one circuit about the center of the mask. The very center of the vortex mask is usually covered by a small opaque spot, in order to mask defects near the phase pattern's central singularity, where the desired spatial orientation gradient is too large. In the HabEx design, a dichroic coating is placed on the VC mask and the mask is slightly tilted so that the reflected, out-of-band light can be used by the ZWFS style LOWFS.

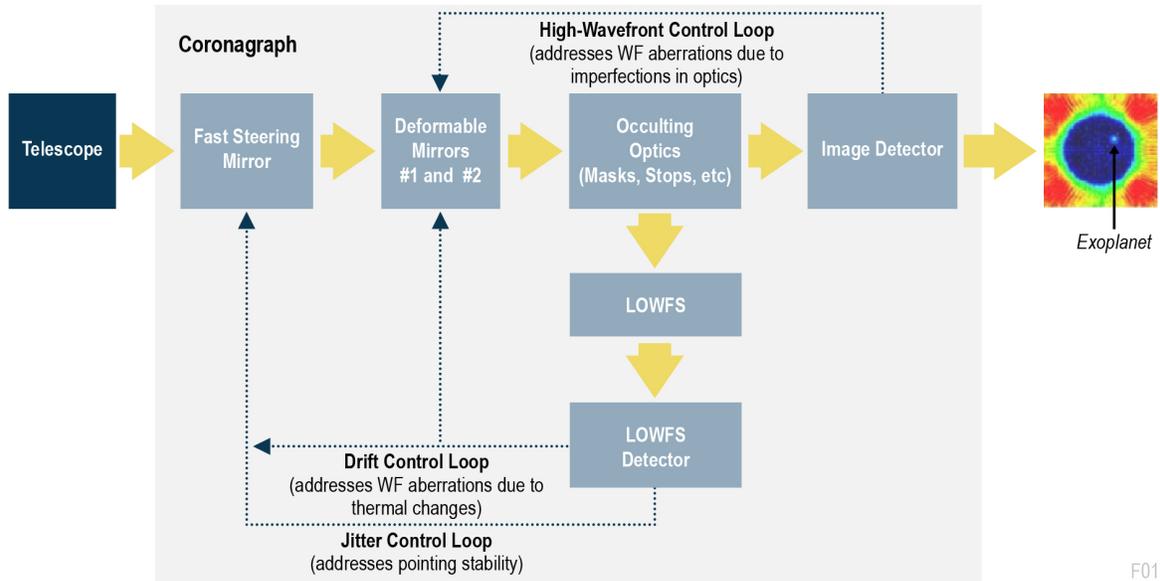

**Figure 11.4-1.** Coronagraph control loop block diagram for vector vortex and hybrid Lyot coronagraphs.

The HLC mask uses a partially opaque spot to block the majority of the target starlight and an overlaid phase modulation pattern provided by an optimized dielectric layer. The HLC design includes optimized DM shapes that help make the coronagraph achromatic and mitigate sensitivity to low-order aberrations. The HLC mask is slightly tilted and the central obscuration is reflective to send on-axis starlight to the LOWFS.

### 11.4.1.1.2 State of the Art

In the course of development of coronagraph masks and architectures, a series of deep nulls have been accomplished over a variety of bandwidths and working angles. Here, we survey the deepest contrasts and most relevant achievements.

The HLC has demonstrated the deepest starlight suppression to date—$6 \times 10^{-10}$ raw contrast over 10% bandwidth from 3 to $16\,\lambda/D$—and is one of the two baselined coronagraphs on the WFIRST coronagraph instrument (Trauger et al. 2015). While this is close to the HabEx requirement ($1 \times 10^{-10}$ contrast over a 20% bandwidth with an inner working angle at $2.4\,\lambda/D$), work is still needed.

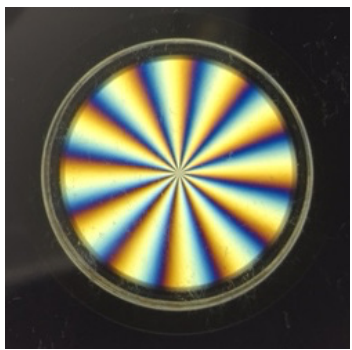

**Figure 11.4-2.** A charge 6 liquid crystal polymer vector vortex mask as seen through crossed polarizers. Credit: E. Serabyn

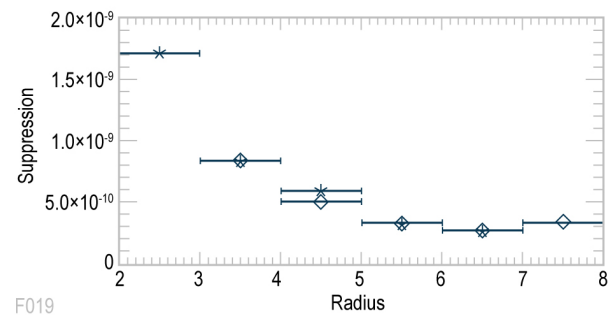

**Figure 11.4-3.** Cross-cuts through vortex dark holes of 2 to $7\,\lambda/D$ (*asterisks*) and 3 to $8\,\lambda/D$ (*diamonds*) show $5 \times 10^{-10}$ contrast (Serabyn et al. 2013).

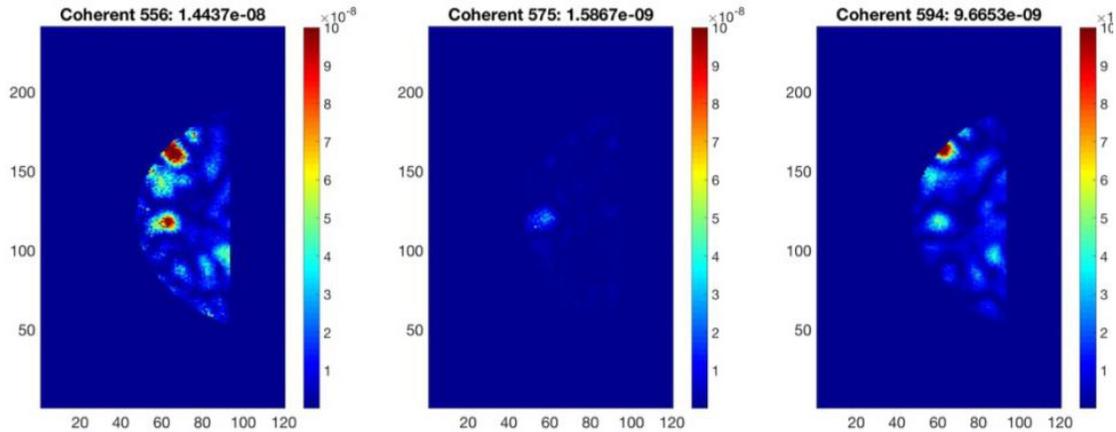

**Figure 11.4-4.** A vortex coronagraph 10% bandwidth dark hole at the 8.5 × 10⁻⁹ level covering 3 to 8 $\lambda/D$ shown in a series of 3.3% bandwidths (Serabyn et al. 2019).

In the first vortex-related Technology Demonstration for Exoplanet Mission (TDEM), carried out in the original high-contrast imaging testbed (HCIT) chamber, monochromatic raw contrasts of $5 \times 10^{-10}$ were demonstrated (Serabyn et al. 2013) for dark holes extending both from 3 to 8 $\lambda/D$, and from 2 to 7 $\lambda/D$ (**Figure 11.4-3**), demonstrating very good performance all the way into 2 $\lambda/D$.

Since then, the VC goal has shifted to broadband performance. For broadband testing under a second TDEM, a new HCIT at JPL was used with an unobscured aperture and BMC DMs. The best broadband result achieved to date is $8.5 \times 10^{-9}$ contrast over 10% bandwidth for the coherent light (light that responds to DM changes). A charge 4 vortex mask was used and the dark hole was from 3 to 8 $\lambda/D$. Over the central 3.33% bandwidth the contrast was $2 \times 10^{-9}$ (Serabyn et al. 2019). Incoherent light leakage of a uniform background was at the level of $4 \times 10^{-9}$ and limited the overall contrast to $1.3 \times 10^{-8}$. The dark hole shown in **Figure 11.4-4** is dominated by the red spots which are likely due to contamination on the vortex mask. The incoherent light leakage is likely due to systematics of the new HCIT testbed and are being addressed with continued experience on the testbed.

The Decadal Studies Testbed (DST) at JPL strives to get coronagraph performance required for direct imaging of exo-Earths. The best performance to date in the DST was achieved with a plain Lyot coronagraph: $3.6 \times 10^{-10}$ raw contrast at 10% bandwidth over 3–7 $\lambda/D$ in a static lab environment.

### Path to TRL 5

To achieve TRL 5 for HabEx, the vortex coronagraph mask at charge 6 would need to be demonstrated at $1 \times 10^{-10}$ coherent contrast at 10% bandwidth over a dark hole 3 to 13 $\lambda/D$ in a testbed with an unobscured aperture and static vacuum environment. The vortex mask would need to be shown to maintain performance under radiation exposure

### 11.4.2 ZWFS and Control

The coronagraph ZWFS uses the out-of-band starlight reflected from the coronagraph mask to sense the low order WFE, which includes line-of-sight (LoS) pointing error and thermal-induced low-order wavefront drift.

The ZWFS is based on the Zernike phase contrasting principle where a small (~1–2 $\lambda/D$) phase dimple with phase difference of ~$\lambda/2$ is placed at center of the rejected starlight PSF. The modulated PSF light is then collimated and forms a pupil image at the ZWFS camera. The interferences between the light passing inside and outside the phase dimple convert the wavefront phase error into the measurable intensity variations in the pupil image on the ZWFS camera. The spatial sampling of the pupil image on the ZWFS camera depends on the spatial frequency of WFE to be sensed.

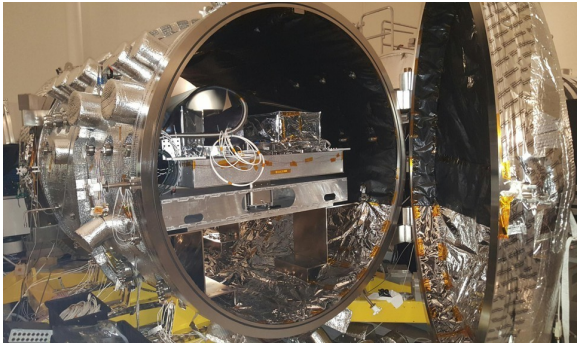

**Figure 11.4-5.** WFIRST coronagraph instrument testbed.

The ZWFS-sensed tip-tilt errors are used to control the FSM for pointing control and, if needed, jitter suppression. Similar to the WFIRST coronagraph instrument, the control contains a feedback loop to correct the telescope's LOS drift and a feedforward loop to suppress LOS jitter. The ZWFS-sensed low-order wavefront errors beyond tip-tilt could be corrected using the DMs (*Section 6.3*).

A ZWFS-based LOWFS has been developed, designed, and demonstrated for the WFIRST coronagraph instrument (**Figure 11.4-5**) at JPL's LOWFS testbed and occulting mask coronagraph (OMC) dynamic testbed. Testbed results have shown that ZWFS is very sensitive, capable of sensing LOS tilt less than 0.2 mas and low-order WFE as small as 12 picometers (RMS). OMC dynamic test results demonstrated that with the LOWFS FSM and DM control loops closed, the HLC maintains contrasts to better that $10^{-8}$ with the presence of WFIRST-like LoS variations (~14 mas drift and ~2 mas jitter) and slow-varying low-order WFE disturbances (~1 nm rms at ~1 mHz) (Shi et al. 2017).

The testbed results remain constant as the source brightness is varied by nearly 4 orders of magnitude, with the faintest level equivalent to stellar magnitude MV ~6. With WFIRST-like line-of-sight jitter disturbances injected by the testbed OTA Simulator's Jitter Mirror the LOWFS LOS FSM loops have been demonstrated to be able to maintain the contrast stability well below $10^{-8}$ for source as faint as MV = 5. The post correction residual LOS error, measured by broadband coronagraph contrast, for a source equivalent to $M_V$ = 5, was 0.36 mas and low order

WFE (focus) to 26 pm averaged over 5 s (Shi et al. 2018). This demonstration exceeds the HabEx required low order WFE of ~100 pm over 1 second and is close to the LOS residual requirement of 0.2 mas. The WFIRST LOWFS result is close to what HabEx requires and will need some slight additional work to meet the HabEx requirement.

<u>*Path to TRL 5*</u>

The development of LOWFS for WFIRST has matured the technology to TRL 4 level for HabEx. The next phase of development is a planned upgrade to the Decadal Studies Testbed (DST) that will include a ZWFS with a Lyot mask having a dichroic phase structure similar to the HabEx ZWFS design. The DST will use the ZWFS to achieve $10^{-9}$ contrast over a 10% bandwidth. While a valuable development step, this does not reach TRL 5 for HabEx.

TRL 5 development of ZWFS will be done in a dedicated ZWFS testbed before being integrated into a coronagraph architecture testbed. The testbed will use a dichroic phase structure on a VC mask. The will demonstrate low order wavefront sensitivity to the required level for a low brightness source and achieve $10^{-10}$ contrast over a 10% bandwidth.

ZWFS will be needed in the VC coronagraph testbed to control lab environment jitter for a contrast stability of $1 \times 10^{-11}$. The demonstration of the VC mask in a HabEx-like coronagraph architecture testbed that achieves the $1 \times 10^{-10}$ raw contrast and $1 \times 10^{-11}$ contrast stability during injection of the expected dynamic wavefront disturbances, including LOS and thermal WFE drift, that uses LOWFS, would qualify ZWFS at TRL 6 for HabEx.

### 11.4.3  Deformable Mirrors

The baseline DMs are 64-actuator by 64-actuator DMs developed by Boston Micromachines Corporation (BMC) and shown in **Figure 11.4-6**. The DMs are microelectromechanical systems (MEMS) made using semiconductor device fabrication technologies. The DM has a continuous facesheet for the surface of the mirror; the actuators pull on the back of the mirror using capacitance with an electrode in the back plane. The 4,096-actuator DM has been used in ground-based coronagraphy on the Gemini Planet Imager (Macintosh et al. 2014). The 4,096 DM has a 3.5 μm stroke and 400 μm pitch.

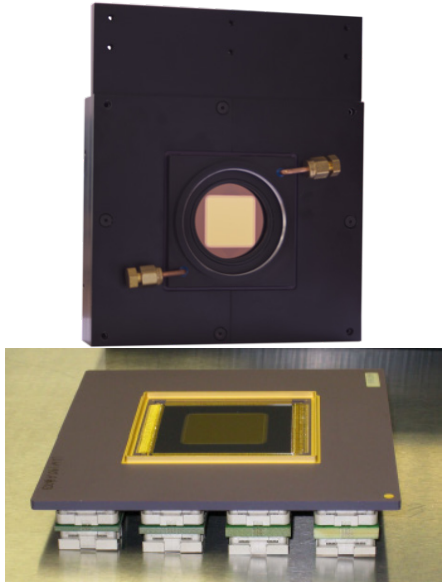

**Figure 11.4-6.** BMC 64 × 64 deformable mirror in vacuum package (*top*, credit: BMC) and embedded in chip carrier (*bottom*, credit: GPI; Hill et al. 2008).

BMC DMs have proven useful for ground-based coronagraph instruments and have been demonstrated in a suborbital sounding rocket (Douglas et al. 2018). Additional use of the BMC DMs is underway in more ground-based instruments, a high precision testbed, and in space on a CubeSat (**Table 11.4-1**).

An initial deep contrast was achieved by a coronagraph using the BMC DM to $2 \times 10^{-7}$ raw contrast over $2 \cdot 10^{-11} \lambda/D$ at 0.65 μm central wavelength and 10% bandwidth in the Exoplanetary Circumstellar Environments and Disk Explorer

(EXCEDE) proposal testbed (Sirbu et al. 2016). Recent coronagraph testing with a single 32 × 32 actuator BMC DM was performed in the Exoplanet Exploration Program Office General Purpose Coronagraph Testbed: a coherent contrast of $8.5 \times 10^{-9}$ was achieved with a charge 4 vortex coronagraph mask at 10% bandwidth over $3\text{–}8 \lambda/D$ in a static environment. The BMC DM was measured to have a contrast drift rate of $1 \times 10^{-12}$/hour RMS over 280 minutes and contrast drift of an order of magnitude over 42 hours due to a single DM pattern (Prada et al. 2019). Further investigation is required to determine the source of the drift, such as the drive electronics or the MEMs device. Potentially, monitoring with the ZWFS or an internal source and the ZWFS (Moore and Redding 2018) could stabilize the slow drifts over long integration times.

BMC DMs of the 34 × 34 actuator size were environmentally tested for shock and vibe in 2018 via a TDEM (Bierden 2013). A comparison of pre-test to post-test performance characterization by BMC is promising. JPL post-performance characterization is awaiting testbed availability and will provide a cross validation of successful environmental survival.

SBIR and WFIRST investments have improved the surface figure error (SFE) of BMC DMs. An SBIR in 2016 reduced rms figure error due to quilting and scalloping from 6 nm rms to 3.3 nm rms, which is sufficient for HabEx. Continued development would be needed to achieve 3.3 nm rms SFE with high production yield. Under WFIRST, the unpowered mirror deformation was improved by using a thicker

**Table 11.4-1.** Boston Micromachines Corporation DMs in current and planned astronomical use.

|  | Actuators | Instrument | Location |
|---|---|---|---|
| Ground | 140 | ROBO-AO | Palomar 2012, Kitt Peak 2015 |
| Ground | 1,024 | Shane-AO | Lick Observatory 2013 |
| Ground | 2,040 | SCExAO | Subaru 2013 |
| Ground | 4,092 | GPI | Gemini South 2013 |
| Space | 1,024 | PICTURE-B | Sounding Rocket 2015 |
| Ground | 2,040 | MagAO-X | U of Az, *In work* |
| Ground | 492 | Rapid Transit Surveyor | U of H, *In work* |
| Ground | 952 | Keck Planet Imager and Characterizer | Keck, *In work* |
| Testbed | 1,000 segments | Caltech HCST | Testbed, *In work* |
| Testbed | 952 | Princeton HCIL | Testbed, In work |
| Space | 140 | DEMI | CubeSat, *scheduled for launch in 2019* |

mirror substrate; this reduces the amount of stroke required for self-correction.

_Path to TRL 5_

NASA investment through TDEM, SBIR, and WFIRST has brought BMC DMs close to TRL 5. Facets of the HabEx required capabilities have been met in various devices: lower SFE periodic pattern, thicker substrate, correct thickness of actuator membrane stiffness for full stroke and stroke resolution, and format size.

Environmental test for shock and vibe was performed under a TDEM on $34 \times 34$ actuator DMs and will be performed by WFIRST on $50 \times 50$ actuator engineering model BMC DMs. The WFIRST risk reduction DMs are undergoing development for thickened substrate, packaging, connectorization, wire bonding static pull test, and environmental testing including thermal survival. The environmental testing by the TDEM and WFIRST will be sufficient for TRL 5 for HabEx.

TRL 5 optical performance for HabEx DMs requires demonstration of a coherent contrast of $1 \times 10^{-10}$ with $2 \times 10^{-11}$ contrast stability at 10% bandwidth in a laboratory environment This contrast will be achieved in the HabEx coronagraph testbed. The contrast is anticipated to require for the DMs the 18-bit electronics drivers planned for the DST upgrade in FY20.

### 11.4.4 Detectors

HabEx can achieve its primary exoplanet scientific objectives with detectors that operate within the 0.3–1.0 μm spectral range. High performance in this range can be achieved using existing silicon-based detectors (e.g., CCD arrays) with high TRL. Extending the spectral range at both ends enables a greater return for the exoplanet science and is required to meet the observatory science requirements. This extended spectral coverage necessitates a closer look at existing detector capabilities in the UV down to 0.115 μm and in the near-IR out to 1.8 μm for both the exoplanet and general astrophysics science.

This section introduces detector candidates that have been selected by careful examination of the performance and latest status of the available technologies. The principles of operation for these detectors are briefly described and information is provided on the performance of their major relevant parameters, TRL status, and the path of further development.

#### 11.4.4.1 Delta Doped UV-EMCCDs

EMCCDs are otherwise conventional CCDs that possess high SNR by the virtue of having an additional serial register. This so-called "gain" register produces gain via avalanche multiplication in a stochastic process. Gains of greater than 1,000 can be achieved and photon counting can be performed. The Teledyne e2v's CCD 201, which has been baselined for the WFIRST coronagraph, has also been optimized for high efficiency and high stability in the 0.4–1.0 μm range using a delta doping process.

The delta-doping utilizes JPL's low temperature (<450°C) molecular beam epitaxy (MBE) growth process to inject dopant atoms in a highly localized layer. "Delta-doping creates very high electric fields near the surface that drive photo-generated charge away from the back surface and suppresses the generation of excess dark current from the exposed silicon surface" (Hoenk et al. 2009).

Extensive radiation testing for WFIRST has been carried out as part of the WFIRST coronagraph instrument technology development program (Harding et al. 2015; 2018). The CCD201 is currently at TRL 5 for WFIRST, which has a nearly identical environment as HabEx. The end-of-life dark current measured by WFIRST indicates that the HabEx EMCCD will need to be cooled to 165 K. A larger format of the EMCCD ($4k \times 8k$, CCD 282) has been demonstrated (Daigle et al. 2018) and is baselined for HabEx.

_Path to TRL 5_

To achieve TRL5, delta doping will be performed with an EMCCD on a subscale, $1k \times 1k$ format detector. The same radiation mitigation approaches used for the WFIRST EMCCDs will be used for the HabEx EMCCDs and may be further improved on the road to TRL5. The flow to TRL 5 and TRL 6 is shown in _Appendix E_.

#### 11.4.4.2 EMCCDs with Enhanced Near-IR QE

For the EMCCD used in the starshade visible IFS, enhanced QE performance for the water line at 940 nm is desired. The response of conventional back-illuminated silicon sensors is reduced for increasing wavelength as the silicon band edge is approached at ~1,100 nm. Lower energy photons are increasingly likely to pass through the detector substrate undetected as silicon becomes more transparent. Some high TRL, photon-counting detectors can have a QE of less than 20% at wavelengths greater than 940 nm due to a typically reduced substrate thickness (~15 µm).

A significantly improved QE can be achieved by thickening the silicon for a region of deep depletion. Conventional thick CCDs have been demonstrated with QE >90% at 940 nm (multiple ground and space demonstrations including e2v CCD 261 and CCD220 (Downing et al. 2013; Downing et al. 2018). Additional engineering work is required in order to match the noise performance of the thinner devices. The influence of fringing, dark-current, and QE vs. thickness will be considered with respect to the desired long wavelength cutoff.

*Path to TRL 5*

The QE, deep depletion, and delta doping have been demonstrated in various detectors, making EMCCDs TRL 4 for HabEx. WFIRST has performed significant development on EMCCDs which achieve radiation hardness and dark current levels that meet the HabEx requirements. The additional performance desired by delta doping for the UV and deep depletion for the NIR require additional development to achieve TRL 5. The required QE and dark current will need to be demonstrated on engineering models for the UV delta doped EMCCD and the deep depletion EMCCDs. Development of these devices is not currently funded. A flow for a development to TRL 5 is included in *Appendix E* with the detector TRL 6 roadmap.

#### 11.4.4.3 UV Microchannel Plate Detectors

Microchannel plates (MCPs) have been the workhorse of ultraviolet instruments for several decades. The detector systems are comprised of three main technological components: a photocathode material that absorbs photons in the desired spectral range and emits an electron; a microcapillary array, which accelerates this electron under an applied voltage in order to create large signal gain; and a readout to convert this electron cloud into a resulting image. Photocathode materials in the UV range considered by HabEx are typically alkali materials like KBr (FUSE, New Horizons-Alice), CsI (HST Cosmic Origins Spectrograph [COS], HST Space Telescope Imaging Spectrograph [STIS], Juno Ultraviolet Spectrograph [UVS]) or $Cs_2Te$ (Galaxy Evolution Explorer [GALEX], HST-STIS). Recent work has also explored the development of gallium nitride photocathodes, which may be able to cover the entire HabEx band and operate with higher quantum efficiency (Siegmund et al. 2013).

Recent MCP developments include atomic layer deposition (ALD) on borosilicate microcapillary arrays. An ALD MCP detector has flown on the Limb-imaging Ionospheric and Thermospheric Extreme-UV Spectrograph (LITES) International Space Station (ISS) instrument (Siegmund et al. 2017). Larger format 200 × 200 mm detectors with ALD borosilicate channels have been developed for the Dual-channel Extreme Ultraviolet Continuum Spectrograph (DEUCE) and Integral Field Ultraviolet Spectroscopic Experiment (INFUSE) sounding rocket missions at the University of Colorado, Boulder.

Continued development of UV MCP detectors and associated readouts will improve performance and packaging in the coming years. In 2012, a NASA Strategic Astrophysics Technology (SAT) grant was awarded to raise the TRL of a 50 mm square cross-strip MCP detector from 4 to 6. The team was also funded in 2016 with a follow-on SAT to scale this detector to a flight qualified 100 × 100 mm format (Vallerga et al. 2016). Even larger formats (200 × 200 mm) are also being developed, though the 100 × 100 mm format meets the HabEx design requirement.

*Path to TRL 5*

MCP detectors currently baselined for HabEx use elements that are individually TRL 5 but have not yet been combined into a single detector; this makes the HabEx baseline microchannel plate detector TRL 4. To achieve TRL 5, the baseline MCP would need to be fabricated and pass performance testing for the desired QE and dark current. A roadmap to TRL 5 is presented with the TRL 6 roadmap in *Appendix E*.

## 11.5 Instrument: Currently TRL 5 or Higher

The linear mode avalanche photodiode near-infrared detectors are at TRL 5 for the 320×256 device and TRL 4 for the 1k × 1k pixel device.

### 11.5.1.1 Linear Mode Avalanche Photodiode Near-IR Detectors

Leonardo-ES Ltd in Southampton, UK, has been developing HgCdTe avalanche photo diode (APD) sensors for astronomy in collaboration with the European Southern Observatory and the University of Hawaii since 2008. The devices use metalorganic vapor phase epitaxy (MOVPE) grown on gallium arsenide (GaAs) substrates. This, in combination with a mesa device structure, produces a detector that achieves a noiseless avalanche gain, very low dark current (due to band gap engineering) and a near-ideal spatial frequency response. A device identified as "Saphira"—a 320 × 256, 24 µm pixel detector—has been developed for wavefront sensors, interferometry, and transient event imaging and is currently in use in a number of ground-based telescopes including Subaru and NASA's Infrared Telescope Facility (Atkinson et al. 2018). Larger 1k × 1k arrays with 15 µm pixels have been fabricated and have begun preliminary testing. The full detector characterization testing is expected in summer 2019.

Saphira has demonstrated read noise as low as 0.26 electrons rms and single photon imaging with avalanche gains of up to 500. An avalanche gain of 25 can be achieved with dark current of less than 0.04 electrons per second per pixel. This dark current translates into nearly a factor of five improvement in SNR for signals of the order of 100 photons, or a factor of 25 improvement in observation time.

*Path to TRL 5*

The Saphira detector has been assessed at TRL 5 for the standard size 320 × 256 of 24 µm pixels, which means they have been tested in a relevant environment. The newer, large format of 1k × 1k of 15 µm pixels has undergone preliminary testing with promising results. This large size is baselined for the HabEx starshade infrared IFS and is assessed at TRL 4 for that format size. Completion of the planned testing program for this format size may show the HabEx required performance, at which time the TRL assessment would increase to 5.

There is a current European Space Agency (ESA) program to assess radiation (gamma and proton) resilience and, to date, there has been no change in the detector performance after exposures of 50 krads of gamma radiation and $1 \times 10^{11}$ cm$^{-2}$ fluence of energetic protons.

Currently, there is a NASA program funding development of a custom MOVPE design for low-background/high-gain imaging aimed at extending the gain and reducing dark current of the Saphira detectors even further. The continued development of the Saphira detectors is also funded by commercial interest in photon-counting IR detectors. The current level of funding is expected to mature the 1k × 1k format device to TRL 5 by the end of 2022.

## 11.6 Spacecraft

The microthrusters are the sole technology maturation item for the HabEx spacecraft. The microthrusters flew on ESA's LISA-Pathfinder. HabEx requires a larger microthruster than the LISA-Pathfinder model, so the microthruster is considered TRL 5 for the HabEx spacecraft. A larger microthruster is currently being developed to TRL 6 as part of the NASA possible contributions to ESA/LISA.

### 11.6.1 Spacecraft: Currently TRL 4

#### 11.6.1.1 Microthrusters

Colloidal microthrusters provide low-noise, precision throttleable thrust and drag-free operation of spacecraft against disturbances—mainly solar pressure—for ultra-fine pointing and telescope stability control. Microthrusters are a breakthrough in astrophysics observatory technology allowing the removal of reaction wheels on the space telescope platform, and in so doing, providing an extremely low-disturbance environment for the telescope. An additional benefit for the HabEx concept: without significant self-generated vibrational disturbances, a less stiff, monolithic primary mirror—ideal for coronagraphy—is not only feasible but even preferable to a segmented mirror for thermal and design simplicity reasons.

Busek Co., Inc. has worked with JPL to provide two clusters of 4 colloidal microthrusters (**Figure 11.6-1**) for the NASA Space Technology 7 Disturbance Reduction System (ST7-DRS) mission in 2008. ST7-DRS was launched on board the ESA's LISA-Pathfinder Spacecraft in December 2016, and accumulated over 100 days of operation on orbit. The LISA-Pathfinder colloidal microthrusters were single string designs, intended for only 90 days of operation. Each thruster emits a finely controlled electrospray (electrostatically accelerated charged droplets) using an ionic liquid propellant, producing between 5–30 µN of thrust with 100 nN resolution All eight thrusters demonstrated full thrust range and controllability after 8 years of ground storage. As a system, thrust noise has been measured using ESA's inertial sensor on the LISA Technology Package at levels ≤0.1 µN/√Hz (average per thruster).

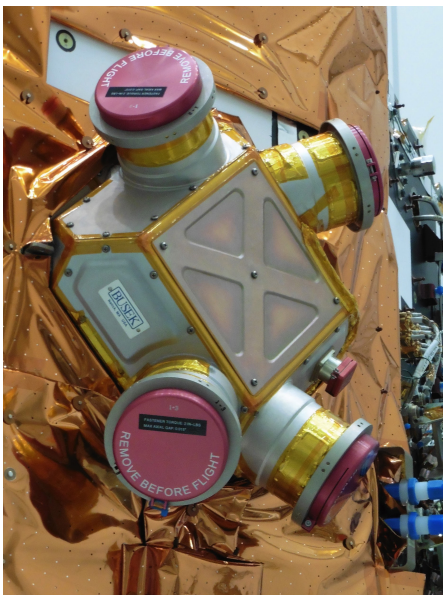

**Figure 11.6-1.** A single cluster of four Busek Co. colloidal microthrusters integrated on the LISA-Pathfinder Spacecraft just prior to launch. Image courtesy ESA / Airbus.

NASA's Physics of the Cosmos (PCOS) Program is currently developing the colloidal microthruster technology as a potential contribution to the ESA-led LISA mission. During the next three years, the colloidal microthrusters will be redesigned to be fully redundant with sufficient capacity to support a 12-year mission (Ziemer 2017). The effort completed a successful peer review of all the design changes from ST7/LISA Pathfinder to LISA to improve redundancy and lifetime (Ziemer 2018b). The colloidal microthruster technology development program has entered Phase 2, which consists of building and testing breadboard and brassboard-level hardware with the expectation of reaching TRL 5 on all components by the end of 2020. The effort is on schedule to reach TRL 6 by the end of 2022.

*Path to TRL 5*

For HabEx, the reliability and lifetime technology development activities for LISA would provide a strong basis for colloidal microthruster use. The LISA design change towards a heavy spacecraft that increased the maximum thrust required to 150 µN benefits HabEx. The PCOS technology development program will fully define the microthruster requirements for LISA.

The maturation to TRL 5 and TRL 6 by the PCOS technology development program will make the colloidal microthrusters TRL 5 and could make them TRL 6 for HabEx if the LISA requirements suffice for HabEx, which is likely. Otherwise, HabEx will reperform the TRL 6 test program with a microthruster designed to meet HabEx requirements.

## 11.7  Enhancing/Alternative Technologies

Technologies that would improve the science performance of the mission if adopted, but are not necessary to meet the science goals, are enhancing technologies. These technologies may be at too low a maturity to warrant adopting currently, but if matured to TRL 5 in time to be adopted by the mission would be worthwhile inclusions. Far UV coatings and next-generation microshutter arrays are discussed in such a context.

Alternative technologies offer a risk reduction option or performance benefits to a baseline technology. Alternatives could be selected during Phase A if the alternative matured into a more desirable solution or if unfeasible challenges were met in the baseline technology. Delta-doped EMCCDs for the UVS instead of MCPs is one such technology. Electrostrictive deformable mirrors is another. Both are discussed in detail in their own sections.

### 11.7.1  Currently TRL 3

#### 11.7.1.1  Far-UV Coatings

Extending the coating performance down to ~0.1 μm in the far-UV (FUV) requires technology development. Significant research and development work is underway at JPL and Goddard Space Flight Center (GSFC) to accomplish FUV spectral coverage combined with long-term stability. At present, one of the coatings capable of this spectral range is at TRL 3.

One of the more promising candidates is a new lithium-fluoride evaporation technique with an added thin layer of aluminum-fluoride to protect the lithium-fluoride layer. Work at GSFC has explored evaporation of LiF at elevated substrate temperatures, which has been shown to improve performance and environmental stability over legacy LiF coatings such as those used on the FUSE mission. Although improved, these coatings still exhibit degradation of reflectance in moderate humidity storage conditions. This has motivated JPL research into a stacked approach where the GSFC LiF coating is itself protected by a second layer of $AlF_3$.

Early lifetime stability tests of the $Al+LiF+AlF_3$ are encouraging. Samples have been tested for reflectivity changes over a 3-year period. The samples were stored in normal laboratory conditions with relative humidity ranging from 20 to 50% at nominally 68°F. Measured performance is shown in **Figure 11.7-1**

(Pham and Neff 2016), and (Balasubramanian et al. 2015). Flight demonstrations of these new LiF coatings including this stacked approach are being pursued at the University of Colorado, Boulder. The sounding rocket mission Suborbital Imaging Spectrograph for Transition region Irradiance from Nearby Exoplanet host stars (SISTINE) will utilize these coatings to enable imaging spectroscopy down to 100 nm (Fleming et al. 2016). Expected to launch in summer 2019, the instrument includes a 0.5 m diameter primary mirror that will demonstrate coating uniformity on a larger scale than the current coupon samples. The CubeSat mission SPRITE will implement similar coatings and evaluate their stability in a space environment, with an anticipated mission lifetime of one year and launch in the 2021 timeframe.

*Path to TRL 5*

Extending the spectral range of the HabEx telescope optics down to ~0.1 μm is not in the current baseline design due to the technological maturation needed. The technology could be matured through optimization of existing processes in both the GSFC evaporation process for LiF and the JPL ALD process for the thin top layer, repeated demonstrations of the optimized coating meeting reflectance requirements, demonstration of reflectance uniformity on coupons representing a 4 m diameter mirror, and accelerated lifetime testing for stability. Should additional investment in this technology result in a demonstrably stable coating able to meet uniformity requirements, then a future HabEx mission could elect to use such a coating in place of the current baseline HST-like coating.

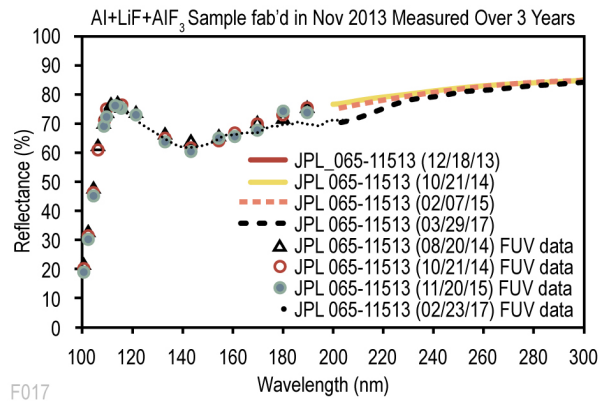

**Figure 11.7-1.** $AlF_3$ overcoat prevents LiF moisture degradation in lab environment (Balasubramanian et al. 2017).

### 11.7.2 Next Generation Microshutter Array

The workhorse camera (WHC) provides both high-resolution imaging and multi-object spectroscopy. Multi-object spectroscopy (MOS) is enabled by an array of slits or apertures that allow the spectra from multiple sources or objects in a field to be separated. The JWST Near Infrared Spectrometer NIRSpec instrument uses an actuated microshutter array (MSA) for MOS. The array can be programmed to provide various arrangements of apertures to suite the field being imaged. The array is actuated by electrostatic and magnetic means. The array size is 171 × 356 shutters. NIRSpec used a 2 × 2 grid of the MSAs, as seen in **Figure 11.7-2**; the large volume surrounding the MSA is due to the magnetic actuation.

HabEx is baselining the JWST MSA for the workhorse camera. A 1 × 2 grid of arrays will be used for HabEx instead of the 2 × 2 grid used in NIRSpec. The JWST MSA meets HabEx requirements and is TRL 7.

A next generation microshutter array (NGMSA) is under development at GSFC via SAT funding. The NGMSA is a larger format (840 × 420) array, shown in **Figure 11.7-3**, and actuated electrostatically, shedding the bulky magnetic actuation component of the previous generation. The intent of the SAT is to mature the large format MSA from TRL 3 to TRL 5. The SAT is scheduled to complete environmental testing and achieve TRL 5 by end of FY21. If the SAT finishes as expected, then HabEx could utilize a single array of the NGMSA in the workhorse camera. The packaging of the NGMSA has a significantly streamlined volume and a lower mass, making it the preferred option once it reaches TRL 5.

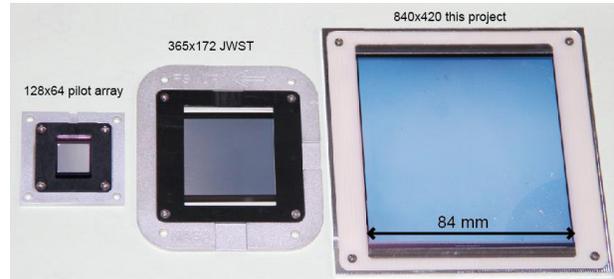

**Figure 11.7-3.** The Next Generation MSA (*right*) is over 4× the size of the JWST MSA.

### 11.7.3 Currently TRL 4

#### 11.7.3.1 Delta-Doped UV EMCCDs

Delta-doped UV EMCCDs offer an alternative to current MCPs in the 0.1–0.3 μm wavelength range. High efficiency (>60% quantum efficiency) in the 0.1–0.4 μm range has also been demonstrated on EMCCDs. JPL has been working closely with e2v to develop the end-to-end processing for CCD 201 and has focused on the high efficiency and photon-counting performance of the detector. A delta-doped EMCCD with coatings to optimize the performance at 0.205 μm (**Figure 11.7-4**) was delivered to FIREBall, a balloon-based UV experiment Nikzad et al. 2017. The detectors performed nominally during the 2018 flight. Final analysis of the science data is expected in 2019 Another EMCCD that is optimized for 0.120–0.150 μm range is baselined on the sounding rocket SHIELDS, which is expected to fly in early 2019 and would advance to TRL 6.

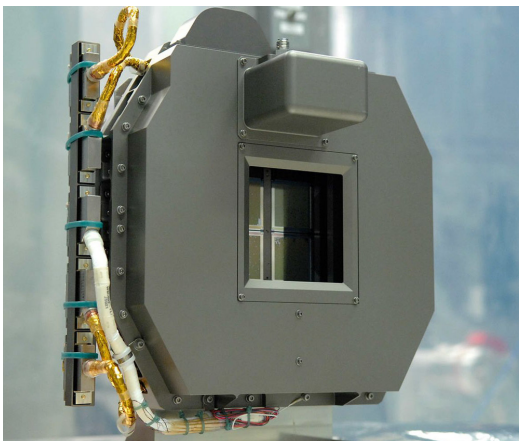

**Figure 11.7-2.** The JWST NIRSpec microshutter array consists of a 22 grid of 172×365 shutters. Credit: JWST.

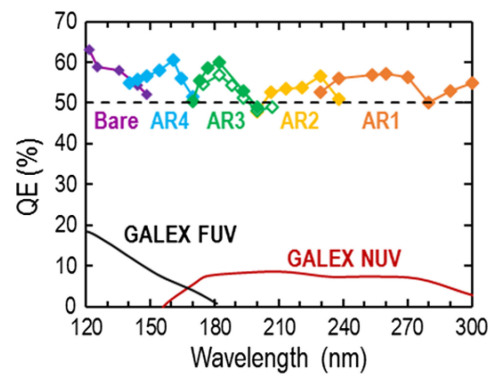

**Figure 11.7-4.** Demonstrations of >50% QE with customized AR-coated, delta-doped CCDs (*closed diamonds*) and delta-doped EMCCDs (*open diamonds*). The GALEX MCP detector QEs are shown for comparison.

The delta-doped EMCCDs being developed for the starshade IFS (*Section 11.4.4.1*) will be used in the 0.2–0.45 μm range. The EMCCDs discussed in this section would be used in the 0.12–0.35 μm range for the UV Spectrometer. Development of delta-doped EMCCDs for use in the UVS includes optimization of an AR coating to achieve overall >50% QE over the 0.12–0.3 μm range.

### 11.7.3.2 Electrostrictive Deformable Mirrors

NASA has been investing in electrostrictive deformable mirrors for over a decade through work in the High Contrast Imaging Testbed with deformable mirrors made by AOA Xinetics (Trauger et al. 2003). WFIRST is further developing the Xinetics DMs for flight implementation in a 48 × 48 actuator, 1 mm pitch format (an engineering model is shown in **Figure 11.7-5**). The Xinetics DMs are composed of a block of electrostrictive Lead-Magnesium-Niobate (PMN) actuators behind a fused silica face sheet. The actuators are surface normal and the continuous facesheet ensures a smoothly varying wavefront.

Xinetics DMs have been produced in 11 × 11, 24 × 24, 36 × 36, and 48 × 48 actuator formats. A Xinetics DM for HabEx with format 64 × 64 actuators would most likely be made by mosaicking together a 2 × 2 array of 32 × 32 actuator modules. Two 64 × 64 actuator DMs were fabricated using a mosaic of 32 × 32 DMs behind a continuous face sheet. This 64 × 64 actuator DM helped achieve an important contrast milestone for the vortex coronagraph mask: $3.5 \times 10^{-9}$ monochromatic and $2.6 \times 10^{-8}$ broadband at 10% bandwidth (Mawet et al. 2011). The flattened surface figure error was 2.2 nm rms (focus removed), shown in **Figure 11.7-6**. The boundaries between the 32 × 32 modules are visible but slight. The boundaries occur due to material property mismatch between epoxy and PMN material and could be improved with further development, if needed. The rms surface figure of this mirror is sufficient for HabEx.

The WFIRST 48 × 48 actuator DM assemblies will undergo thermal and vibration tests before the end of 2019. The WFIRST DMs have a redesigned electrical interconnect (an earlier design passed a general vibration test) and the DM assemblies have a revised design to push the lowest DM vibration mode above 500 Hz to better comply with the WFIRST launch environment. The electrical interconnect would need to be scaled up for the HabEx 64 × 64 actuator DM.

**Figure 11.7-5.** The surface figure error of a Xinetics 64 × 64 actuator DM is 2.2 nm rms (focus removed) and shows slight print through at the boundaries of the mosaicked 32 × 32 actuator modules.

**Figure 11.7-6.** The WFIRST coronagraph instrument is developing 48 × 48 actuator DMs by Xinetics for flight.

One potential drawback of the electrostrictive DMs is intrinsic drift of the PMN material. The PMN shows a $dZ\log(t)$ drift behavior, where $dZ$ is the size of the last change, and $t$ is the time since that step. The timescale of drifts after initial setup and flattening is on the scale of days. Afterwards, the commanded DM changes to maintain contrast may be small and the size of drift may be inconsequential. This would need to be analyzed carefully if electrostrictive DMs were used, because it could drive requirements on thermal stability or decrease observing efficiency due to long settle times. Possibly a feed-forward control approach could be explored to mitigate impact.

The pitch of the Xinetics DMs is 1 mm. This is larger than the 400 μm pitch of the baseline BMC. Xinetics DMs and would increase the size of the coronagraph optics proportionately. The coronagraph instruments would increase in volume and mass, an important aspect of the trade.

# 12 HABEX EXOPLANET SCIENCE ENHANCEMENTS

One of the strengths of the HabEx exoplanet survey is that it is self-contained. Specifically, it does not rely on any prior knowledge or contemporaneous independent observations provided by other ground- or space-based facilities. The broad survey has been designed such that HabEx will use the coronagraph to detect and measure the orbits of ~15 habitable zone (HZ) rocky planets around sunlike stars, including ~8 exo-Earth candidates (EECs), given the planet occurrence rates assumed (*Appendix C*). In addition, for nearly all of these it will use the starshade to obtain the spectra of these planets from the ultraviolet (UV) (0.2 µm) to the near-infrared (NIR) (1.8 µm), with spectral resolution of $R = 140$ in the 0.45–1 µm range (*Section 3.1*). This will allow HabEx to capture the absorption bands of key molecular species, which can be used to distinguish between different types of exoplanets. These features include, but are not limited to, water vapor bands, oxygen and ozone features, carbon dioxide, and methane bands. All of these features are critical to assessing the habitability of and searching for life on these worlds. Thus, HabEx will empirically define the habitable zone, without requiring any exterior information about these planets.

Similarly, for the ~8 deep survey targets, HabEx will detect planets as small as Mars, create complete family portraits of these neighboring systems, including obtaining spectra in the wavelength range of 0.2–1.8 µm, and reasonably precise orbits for planets with periods of <15 years (*Section 3.2*), and well as detect analogs to our zodiacal and Kuiper dust belts.

While HabEx exoplanet science yield is already impressive, this chapter reviews different opportunities for further improving it using ancillary observations from separate instruments or improved high-contrast imaging techniques currently in development that HabEx might be able to leverage if proven successful. These opportunities to enhance HabEx science return range from precursor or contemporaneous observations to determine planet mass, mainly using high precision radial velocity measurements, to improved precursor exozodi observations and multi-star wavefront control algorithms. The potential gain of each development, together with its current state-of-the art is also briefly examined. Finally, this chapter concludes with recommendations for observations or technology development efforts that would help make these potential enhancements a reality over the next decade.

## 12.1 Ancillary Physical Information on Nearby Exoplanetary Systems

There are several additional physical parameters of the planets that HabEx will detect, which would be extremely useful to measure or further constrain through ancillary observations, e.g., to aid in interpreting HabEx spectra. These are specifically the planetary radius, mass (and thus surface gravity and density), and surface temperature. As planetary radius is concerned, broad-band direct imaging alone at multiple epochs can only estimate it within a factor of ~2 due to the albedo size degeneracy (*Section 3.1*). Better accuracy can potentially be achieved through spectral observations over a broad wavelength range and subsequent spectral retrieval of planet parameters (e.g., Feng et al. 2018). But for visible spectra, accuracies will remain limited to >30–60% depending on exact planet type and spectral information available (*Section 3.1*). Given the rarity of transiting events for Earth size planets in the HZ of sunlike stars (~1% probability) and for planets further out, accurate radii measurements of HabEx detected exoplanets would have to wait for follow-up mid-infrared detections, as measuring both the visible and thermal fluxes would break the degeneracy between albedo and radius. Estimating the planet temperature will also ultimately require detecting and taking spectra of the planet in the thermal infrared, which in turn will likely require a mid-infrared space interferometer (e.g., Léger et al. 1996; Beichman et al. 1999; Quanz et al. 2018), or may be possible for a handful of planets from the ground, e.g., using the European Extremely Large Telescope (E-ELT) (Quanz 2014). These concepts are out of the scope of this report, and so will not be discussed further, except to note that there are mid-infrared interferometry space mission concepts currently being considered, such as the Large Interferometry For Exoplanets (LIFE) mission concept recently submitted in response to the European Space Agency (ESA) Voyage 2050 call for science whitepapers. As ancillary measurements of nearby planetary systems are concerned, this chapter continues by concentrating on mass determination and the need for additional observations to constrain the amount of dust in the terrestrial planet regions of other stars, e.g., the exozodi background.

## Mass Determination

The mass of an exoplanet is a fundamental quantity, particularly for terrestrial planets, as it determines the amount of latent heat from formation, the amount of radiogenic heating, and, together with the radius, the cooling rate, surface gravity, and density.

These properties may then be important for determining the habitability of the planet, although sometimes in fairly complicated ways. The density informs the planet's bulk properties and allows one to distinguish terrestrial planets from water-rich planets and mini-Neptunes (e.g., Grasset et al. 2009). An estimate of the planet's surface gravity improves the retrieval of atmospheric abundances from atmospheric spectra. The surface area to mass ratio largely determines cooling rate as a function of time, and thus the internal thermodynamic state of the planet. This, in turn, determines whether it is internally active and can support plate tectonics, and/or can replenish the atmosphere via volcanoes. The internal thermodynamic state together with the planetary rotation rate also influences whether the planet can support a dynamo and thus a magnetic field, which itself (along with the surface gravity), effects the ability of the planet to retain its atmosphere (e.g., Zahnle and Catling 2013).

Direct imaging and spectrophotometric measurements of exoplanets at optical wavelengths will generally not be able to inform planetary masses.[1] Furthermore, the astrometric precision needed to detect the reflex motion of the star on the planet is generally well out of the reach of a mission like HabEx, as it has not been designed to achieve this goal. Therefore, planet masses must be estimated by other observational methods.

There are essentially only two practical and generic methods of measuring the mass of directly imaged planets: radial velocities (RVs) and astrometry of the host star. Both methods have their advantages and drawbacks, and both have significant technological challenges that must be met before they can be applied to true Earth analogs, i.e., Earth-mass planets orbiting within the habitable zones of solar-type stars. In addition, the two methods are most sensitive to different regions of parameter space, have different selection biases and completeness, and can be applied more or less successfully to different types of stars.

The next sections (*Sections 12.2 and 12.3*) summarize the general application of astrometry and RVs to measuring the orbits and masses of planets, focusing on the properties of each method and the challenges that must be overcome before they can be applied to Earth analogs. These sections focus on the requirements placed on these methods to measure the masses of Earth analogs,[2] without focusing on the *timing* of when (and if) these measurements will take place. This is an important distinction, because a precursor program (*Section 12.4*) that identifies target stars that potentially host Earth analogs is likely to be more expensive and time consuming than a program that attempts to measure the masses of planets discovered by HabEx whose orbits are at least partially constrained.

---

[1] It may be possible to measure the mass of a planet via the dynamical perturbations it has on nearby planetary companions and faint debris disk structures. However, we expect such situations to be relatively rare.

[2] Note we focus on Earth analogs in this chapter because these are likely to have the smallest radial velocity and astrometric signals, and thus are the most difficult to detect using these methods.

Similarly, cotemporaneous observations with RVs (*Section 12.5*) or astrometry, may relax the requirements on the coronagraph to determine complete orbits during the mission.

## 12.2 High Precision Astrometry: Methodology and Requirements

A planetary companion to a star can be indirectly detected by measuring the astrometric (i.e., angular) shift of the star on the sky as it orbits the center-of-mass of the planet/system (i.e., the orthogonal components to those that are measured via RVs). The astrometric method requires narrow-angle astrometry, e.g., measuring the (generally very small) centroid shifts of the target star relative to angularly proximate but presumably much more distant stars on the sky. Astrometry-based planet mass measurements offer, in principle, a number of advantages relative to RV measurements. The astrometric signal amplitude is much less dependent on system inclination, and is much less affected by stellar activity than other techniques (**Table 12.2-1**). This opens the possibility of precise planetary mass measurements around more host stars, including A and early F types, which are favorable for RV studies, because they typically have fewer lines, and these lines are typically rotationally broadened. Since there is no inclination degeneracy, astrometric measurements alone can provide an accurate estimate of the planet mass (as long as the mass of the host well-known). Furthermore, astrometry is able to measure the coplanarity of multiple systems.

In the case of planets with a mass similar to the Earth, however, the corresponding astrometric signal is so exceptionally small that it can *only be detected from space*. Indeed, the astrometric semi-amplitude α of the motion of a star due to a planet orbiting it is:

$$\alpha = 3 \left( \frac{m_p}{M_{Earth}} \right) \left( \frac{a}{1\ AU} \right) \left( \frac{M_*}{M_{Sun}} \right)^{-1} \left( \frac{D}{1\ pc} \right)^{-1} \mu as$$

with same notations as above and with *a* representing the semi-major axis of the planet orbit, and *D* the distance to the star. For an Earth-like planet orbiting a sunlike star at 10 pc (roughly the distance out to which HabEx can detect EECs), the astrometric semi-amplitude is then 0.3 μas. In comparison, the ESA astrometric space mission Gaia will provide global astrometry at an accuracy of "only" 10 μas for stars with V magnitude 6 to 12 (Perryman 2014). This performance may extend down to V = 3 or brighter, depending on calibration techniques still being developed to cope with very bright stars, such as those considered by HabEx (V = 0 to 7, with a median of V = 4; see *Section 3.3*). Gaia's capability enables detection and mass determination of giant exoplanets and possibly Neptune-mass exoplanets in the HabEx sample, and this information shall be readily available by 2030. But Gaia's precision and current time line is insufficient to measure the mass of Earth analogs or even sub-Neptunes around our nearest neighbors.

Simply put, measuring the mass of rocky planets with astrometry would require a mission with ~1 cm/s astrometric accuracy and be able to detect or follow-up ~40–50 Earthlike candidates orbiting nearby sunlike stars. At the moment, no such mission is currently being planned.

**Table 12.2-1.** The measurement drifts that would affect different exoplanet detection techniques when observing a solar twin as seen edge-on from 10 pc. Variable spots and bright solar structures cause position shifts of the Sun's photocenter (astrometric position root-mean-squared variability) over time, and drifts in the measured radial velocity (RV in m/s) and transit stellar intensity (TSI) signals. Indicated drift values are based on solar data obtained between 1996–2007. The entire cycle (referred as "all") as well as low- and high-activity periods are considered. As shown in the bottom row, only the astrometric drifts stay at or below the expected signals from an Earth seen around a solar twin at 10 pc. Table adapted from Lagrange et al. (2011).

| Period | Position rms (mas) | RV rms without Convection (m/s) | RV rms with Convection (m/s) | Transit Signal rms Intensity |
|---|---|---|---|---|
| Full cycle | 0.08 | 0.33 | 2.40 | $3.6 \times 10^{-4}$ |
| High activity 1 | 0.11 | 0.42 | 1.42 | $4.5 \times 10^{-4}$ |
| High activity 2 | 0.09 | 0.37 | 1.62 | $3.9 \times 10^{-4}$ |
| Low activity | 0.02 | 0.08 | 0.44 | $1.2 \times 10^{-4}$ |
| Earth signal | ±0.3 | ±0.09 | ±0.09 | $8 \times 10^{-5}$ |

## 12.3 High Precision Radial Velocities: Methodology and Requirements

A planetary companion to a star can be indirectly detected by measuring the reflex RVs (or Doppler shifts) of the star as it orbits the center-of-mass of the planet/system. The method requires measuring (generally very small) centroid shifts of lines in the spectrum of the host star. As such, the radial velocity method requires high resolution (>60,000) spectra to resolve the stellar spectral lines for typical thin-disk, solar-like stars (Beatty and Gaudi 2015), which typically rotate at a few to tens of kilometers per second and have photospheric temperatures of ~4,000 K to 6,000 K. These spectra must also span a relatively large wavelength range, as the precision with which one can measure the centroid of a single line with width of a few kilometers per second is far too poor to detect planetary companions. Thus, one must measure the centroids of many spectral lines, and average these to achieve the final per-measurement precision. These requirements combine to essentially require echelle spectrographs on relatively large aperture telescopes of greater than roughly 3 m, although of course this depends on the mass of the planet and the brightness of the host star, amongst other properties. Furthermore, in general the spectrographs must be very stable in order to disentangle instrumental drifts (which create non-common path offsets) from true RV signals[3].

The difficulty of achieving a given RV precision depends on the mass, inclination, and period of the planet, as well as the mass of the host star. The radial velocity semi-amplitude of a star of mass $M_*$ due to an orbiting planet of significantly smaller mass $m_p$ and period $P$ (in years) is:

$$K \sim 9 \text{ cm/s} \left(\frac{P}{yr}\right)^{-1/3} \frac{(m_p/m_{Earth})\sin i}{(M_*/M_{Sun})^{2/3}}$$

where $i$ is the inclination of the orbit, such that $i = 90°$ is edge-on. For a Jupiter analog orbiting a solar-type star, $K \sim 12$ m/s, well within the reach of current instrumentation. It is expected that for all but the lowest mass or most distant planets, it should be possible to measure the masses with currently-achievable RV precision. However, for an Earth analog orbiting a solar-type star, the above equation indicates $K \sim 10$ cm/s, which about an order of magnitude smaller than the precision and accuracy of the current state of the art of precision RV instruments. Furthermore, a robust measurement of such a signal would require control of systematics (e.g., accuracy) at closer to 1 cm/s over a timescale of at least a year.

Additionally, the requirement of a large number of spectral lines and a large photon count generally favors bright stars with many spectral lines. This effectively eliminates the detection of low-mass planets to main-sequence stars considerably more massive than the Sun, which generally have few spectral lines, and whose spectral lines tend to be highly broadened by rotation. Young and active stars are also generally poor targets for RV surveys, as they have 'intrinsic radial velocity jitter' due to stellar activity (starspots, plages, etc.) that complicates the interpretation of the RV signal.

Although the detection of a planet via RVs technically only measures $m_p \sin i$ (assuming knowledge of the mass of the star), in principle, a single direct imaging measurement of the planet with respect to the host star can resolve the inclination ambiguity (unless the planet is at quadrature). In practice, a few direct imaging positional measurements will likely be necessary.

The largest advantage of RVs over astrometry is that it may be possible to achieve both the ~10 cm/s precision *and* 1 cm/s accuracy needed to detect Earth analogs around sunlike stars from the ground, whereas astrometry at the level needed to detect such planets will almost certainly require a space-based mission. The ability to measure the masses of Earth analogs from the ground, should it be possible, can provide enormous cost savings over a space-based astrometry mission. The largest remaining challenge of RVs however, is likely due to the correlated and non-Gaussian RV variations due to the intrinsic stellar activity (often called "stellar jitter"), which causes variations of the shape of the stellar spectral lines over many timescales. This "stellar jitter" may present an irreducible systematics noise floor and thus limit the ability to measure the RV signal induced by the planet, regardless of the measurement precision reached on the total RV signal.

---

[3] Note that it is possible to pass the light of the star through a gas absorption cell, which imprints spectral lines of known wavelengths on the stellar spectra, thereby eliminating the need of extreme stability. Unfortunately, this method results in a loss in the total throughput, as well as requiring more sophisticated reduction algorithms.

## 12.4  Precursor Exoplanet Observations

Precursor observations of the target sample of a direct imaging mission may yield a number of advantages, depending the nature of the direct imaging mission. In general, the impact of precursor knowledge on which stars harbor EECs on the overall survey strategy and total EEC yield depends on whether the mission is in the *target-limited* regime, where additional resources would not yield a significantly larger yield of planets due to the limitations of the mission itself, or whether the mission is in the *resource-limited* regime, where additional starts could be included in the target sample, thereby improving the yield, if the limiting resources were greater. In general, the two most important limiting resources are total mission time and fuel for the starshade. However, it is essential to note that which category a certain mission falls into also depends on the goals of the mission. For example, if one is only interested in EECs, then one might run out of suitable targets sooner than if one is interested in all planets.

Henceforth this section will focus on RV precursor observations, for several reasons. First, the difference between the effect of astrometry and RV precursor observations is not likely to be large. Second, the US exoplanet community is currently largely focused on attempting to achieve the 10 cm/s precision and 1 cm/s accuracy need to detect EECs, and considerably less emphasis is currently being placed worldwide on the technology needed to achieve sub-microarcsecond astrometry from space.

Precise precursor radial velocities (PRVs) will provide multiple advantageous contributions to the scientific yield and optimization of HabEx. First, PRVs may aid in the optimization of the target sample. For example, if a system is known to already host a massive Jovian planet in the HZ, or if PRVs are able to demonstrate that no Earth-mass planet exists in the HZ, that particular system may be given a lower priority or excluded from the target list. Second, PRVs may aid in scheduling efficiency by only taking observations when a planet of interest is at maximum elongation and thus is known to be outside the direct imaging instruments inner working angle (IWA). Third, PRVs may reduce the number of revisits required by using PRVs to determine the orbits of the planets instead of direct imaging,[4] as well as to help disambiguate between different planets in a system.

### 12.4.1  A Fiducial Simulation of Precursor RV Observations Required to Inform HabEx Target List

In order to estimate the impact of precursor observations, a PRV survey of 53 HabEx direct imaging targets (subset of HabEx master target list, *Appendix D*) was simulated, with spectral types later than F2 and viewable from the Northern Hemisphere (Newman et al. in prep.; Newman et al. 2018). This simulated survey assumed a PRV instrument capable of achieving 3 cm/s instrumental stability. It further was assumed to be placed on the Large Binocular Telescope (LBT), an 8-m-class telescope. The simulation incorporated the known spectral types, brightnesses, rotational velocities, surface gravities, metallicities and coordinates of the HabEx targets. In order to estimate exposure times, the simulations used this information, along with the prescription for the intrinsic PRV spectral information content of each star from Beatty and Gaudi (2015). The properties of the spectrograph were considered, including spectral resolution, spectral grasp and throughput efficiency, detector noise and read out times, as well as the slew-rate and pointing limits of the telescope. Realistic weather losses were also accounted for, as well as sunrise and sunset times for Arizona, in order to simulate the optimized scheduling of a PRV survey using 25% of the telescope time available for five years.

---

[4] Note that, even with a complete orbit determined by RVs, it is still necessarily to take in principle at least one, and practically several, direct imaging observations to measure the inclination angle of the orbit, which is not constrained by RVs.

Hypothetical planet signals were injected into the simulated RV survey data using the ExoPAG Study Analysis Group 13 (SAG-13) demographics (Kopparapu et al. 2018). A random draw prescription for the mass and period of the exoplanets was followed that is identical to the exoplanet demographics used for the HabEx yield calculations, which in turn are based upon a slightly modified version of the SAG-13 occurrence rates (see *Appendix C*). The Forecaster mass-radius relation of Chen and Kipping (2017) was used to assign a radius to each of the simulated exoplanets according to this relation and its mass. System inclinations and longitudes of periastron were randomly drawn, and the eccentricities were drawn from a Beta distribution following Kipping (2013).

The simulation optimistically assumes that stellar activity is perfectly corrected in the simulated survey, in order to establish the best-case scenario. Finally, the optimistic assumption is made that the orbital elements of each individual planet are initially known. Due to computational and time limitations, it is also assumed that the longitude of periastron for eccentric orbits is known perfectly. Some of the simulated planets have eccentricities of up to ~0.65, but most are nearly circular; therefore, this assumption does not significantly impact our overall results. Nevertheless, it is an optimistic assumption.

Using a custom-modified version of the RadVel analysis software that makes use of the recent emcee v3.0rc2 Markov Chain Monte Carlo

(MCMC) sampler Python library (Fulton et al. 2018), the Bayesian posterior probability distribution evidence supporting the existence of each planet in the PRV time-series data is evaluated for a given system, performing a model comparison and evaluating the log-likelihoods with all combinations of planets removed. Favored models are considered recovered if the periods and velocity semi-amplitudes match the injected parameters within a factor of up to 50%, although often times the match is much better. If these criteria are not met, the signal is assumed to be a false positive. The other planets in the system are either noted as marginally recovered or excluded detections (**Figure 12.4-1** and **Figure 12.4-2**). There is not a significant dependence on the recovered status of injected planets on the number of planets per star.

Using the StarSIM 2.0 code (Herrero et al. 2016) that models stellar activity including starspots and plages, observations of one typical HabEx target, HIP 6171, are simulated for a 5-year PRV survey. For each observation, RV variability (or "jitter") arising from a Sun-like level of stellar activity of a few m/s is assumed, and 3 cm/s of photon noise and 3 cm/s of instrumental noise is added to each observation in quadrature. A total of 446 observations over the 5-year survey was assumed.

Five planets were injected into the simulated RV dataset, a Mercury analog (P=88d, K=2 cm/s), a Venus analog (P=204d, K=0.21 m/s), a HZ super-Earth (P=307d, K=0.81 m/s), a massive Jovian planet (P=3300d,K=130 m/s), and a Neptune-like analog (51000d, K=0.24 m/s), each with moderate e<0.25 eccentricities. This simulation does not assume perfect knowledge of the longitude of periastron or any orbital element, and these are modelled as a free parameter in our analysis.

No correction is made for stellar activity, rather this is modeling using a Gaussian process. The Bayesian log-likelihoods from RadVel support the accurate detection of all but the Mercury-like analog. This implies that the detection and masses of HZ Earth-mass analogs can be recovered and inform HabEx science with

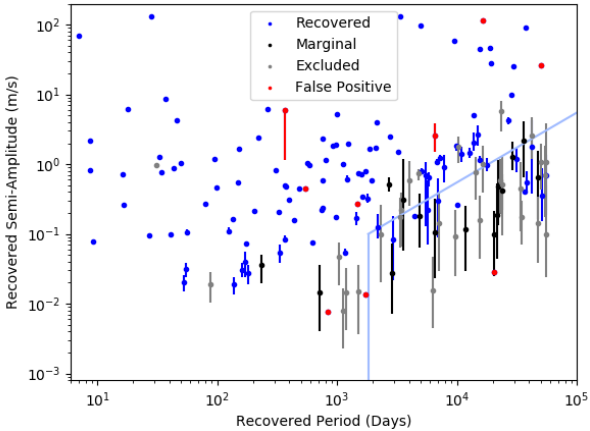

**Figure 12.4-1.** A dedicated precursor PRV survey will find Earth-mass HZ targets amenable to direct imaging with HabEx in multi-planet systems, provided that stellar activity is mitigated. Exoplanets from our simulated precursor PRV survey are plotted as a function of orbital period and velocity semi-amplitude. 9 cm/s corresponds to 1 Earth-mass for a one-year orbital period. Blue circles indicate recovered planets; black circles indicate marginal detections; grey circles indicated excluded detections, and red circles indicate false-positives. The light blue line indicates the survey duration at 5 years and approximate sensitivity limit.

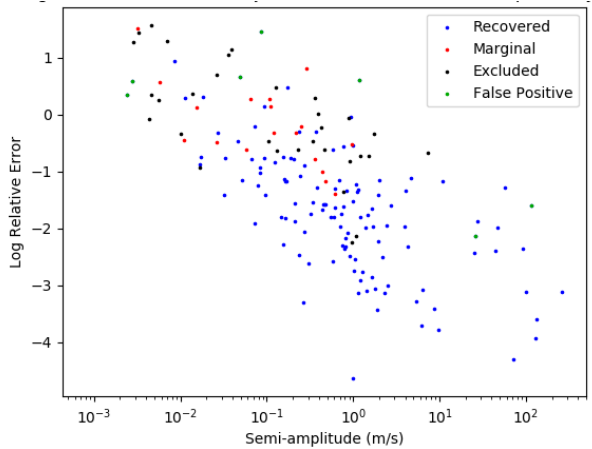

**Figure 12.4-2.** The PRV error in the recovered planet mass for an Earth-mass HZ planet (K = 9 cm/s) will range from 10% up to 100% from a 5-year, 25% time survey of HabEx direct imaging targets on a 8 m class telescope with a 3 cm/s PRV spectrograph "super-NEID", under the optimistic scenario outlined here. The log of the relative error is plotted for our simulated exoplanet systems for HabEx targets as a function of the exoplanet velocity semi-amplitude. Blue circles indicate recovered planets; black circles indicate marginal detections; grey circles indicated excluded detections, and red circles indicate false-positives. Many of the excluded and marginal detections with K < 10 cm/s are from more massive exoplanets at long orbital periods >5 years.

Gaussian process modeling of stellar activity. However, only one of the eccentricities is correctly recovered (the massive Jovian analog), whereas the rest "blow up" to e=0.5–0.65 (the maximum cutoff in our recovery tests). With an incorrect eccentricity for the Super-Earth analog, the orbit information will not be as useful for informing HabEx direct imaging observations.

Next, a correction to the stellar activity for our simulation of HIP 6171 is included, reducing the activity RV rms variability by a factor of 60%, with the residuals again modeled using a Gaussian process model. With the reduced stellar activity, the correct eccentricity and phase of the HZ super-Earth analog is also now recovered, but the eccentricities of the Neptune, Venus, and Mercury analogs are not.

The overall conclusion is that, in order for precursor PRVs to inform the HabEx target list and/or survey strategy by accurately identifying HZ Earth-mass analogs and measuring their orbits to sufficient precision, more than ~60% of the stellar activity must be removed.

Practically this implies the PRVs must be corrected for activity from spots and plages, but not necessary activity from convection, which contributes much less to the PRV rms due to stellar activity. Alternatively, fewer targets, or a PRV survey with more than 25% of the time for longer than five years, would yield better constraints on the masses and orbits. Although this scenario was not simulated here, it is explored in Hall et al. (2018). The results of these simulations and those of Hall et al. (2018) are similar, as they conclude that a 75% reduction in stellar activity is required to detect and accurately characterize the orbits of EECs.

### 12.4.2 Conclusions from the Precursor RV Simulations

Thus, in the optimistic scenario, precursor radial velocities provide several advantages for HabEx that are outlined in *Section 3.4*. The first advantage is knowing *a priori* which stars have HZ Earth analogs (the "where"). Depending on the value of $\eta_{\text{Earth}}$, and the number of stars for which the HZ is accessible via direct imaging, there may be a small reduction in the number of stars to be monitored for a fixed yield of imaged Exo-Earths. This may, in turn, impact the balance of the direct imaging mission time spent surveying versus characterization.

The second advantage of knowing the orbits of HZ Earth analogs *a priori* is that it may enable optimization of the timing of the observations such that observations are only made when the target is outside of the IWA and sufficiently bright to be detected (the "when"). This may also allow for a reduction in the required number of visits of a given system, since the orbit can be determined from PRVs and a handful of direct imaging detections, rather than completely by direct imaging. Additionally, exoplanets orbiting more distant host stars may be imaged whose maximum angular separation is just beyond the IWA, potentially increasing the yield. However, this advantage is diminished by any requirement to have observations of the exoplanet at multiple phases during the orbit.

Third, the *a priori* knowledge of the angular separation (modulo the unknown angle on the sky) can potentially aid in the rejection of false positives, both background objects or additional planets in the system in non-HZ orbits (the "which").

Combined, these three advantages of knowing the "where," "when," and "which" of the target sample may have a modest impact on terrestrial exoplanet yield, and a significant impact on survey efficiency for HabEx. This is because HabEx is largely in the target-limited regime. Assessing the impact of precursor PRVs is dependent upon the number of accessible HZs. In the target-limited regime (e.g., smaller telescope aperture and ~50 HZs accessible), the optimistic scenario will have only a modest impact on exoplanet yield since HabEx may likely survey all ~50 systems regardless of the precursor knowledge. In the resource-limited regime (e.g., larger telescope aperture that has more than ~200 HZs accessible, or a starshade-only mission that has a limited number of slews), the optimistic scenario may have a significant impact on terrestrial exoplanet yield. However, introducing any prior information from PRVs on target selection will introduce a sample bias, particularly in system inclination, and complicate any completeness or statistical analysis derived from the cumulative observations of the HabEx mission.

Given that the mitigation of stellar activity is currently an area of active research, the conservative assumption under the pessimistic scenario requires that a direct imaging mission design be robust against a lack of precursor information from PRVs—e.g., that the mission be able survey enough stars to both detect and characterize candidate HZ exoplanet systems. Nonetheless, the pessimistic scenario will still allow for the identification of Jovian and Saturn analogs beyond the ice-line of direct imaging targets. This may in turn relax constraints on the outer working angle (OWA) requirement for systems with which one desires a full characterization of exoplanet architecture.

Finally, precursor PRV/astrometry observations can go "stale." Uncertainties in the time of periastron and on the orbital elements will degrade the predictive ability of precursor RVs, and thus the direct imaging yield and orbit characterization will be benefit most by RVs that continue up until, during, and after the mission launch.

## 12.5 Contemporaneous Exoplanet Observations

RVs or astrometric observations contemporaneous with the HabEx mission may provide the following information for the planets directly detected by HabEx:

1. Measure or better constrain the exoplanet orbit, including confirmation of the HZ location for EECs. This would impact the number of direct imaging visits required to measure orbits and the mission survey strategy. Consequently, this impacts the trade between a coronagraph and starshade, where the starshade will be limited in the number of possible revisits to a system due to its limited number of slews.

2. Measure the exoplanet mass to higher precision than in the case of precursor observations. For RVs, the difference between contemporaneous and precursor observations can be summed up as follows. One or two direct imaging epochs may provide a constrained prior on the phase of the RV signal, allowing for more sensitive mass measurements with contemporaneous RVs, much like the transit ephemerides of a transiting exoplanet enables the determination of the RV quadrature times (characterized by maximum velocity deviation). Similarly, a large improvement in planet mass and orbit determination precision is expected when using simultaneous detection via astrometric and direct imaging observations (Guyon et al. 2013). This higher precision mass estimation would impact the characterization of the observed exoplanet atmosphere, and help distinguish between different atmospheric models.

## 12.6 Exozodi Precursor Observations: Current State of the Art, Limitations and Impact

The Large Binocular Telescope Interferometer (LBTI) Hunt for Observable Signatures of Terrestrial Systems (HOSTS) mid-infrared (~11 μm) exozodi survey of 38 nearby main sequence stars completed in 2018 (Ertel et al. 2018; Ertel et al. 2019). The HOSTS survey reached the best sensitivity to date for such observations, constraining the median exozodi level of sunlike stars to be $4.5^{+7.3}_{-1.5}$ times higher than in the solar system at 11 μm, and pointing to a bi-modal overall distribution of exozodi levels, with relatively few stars hosting extreme amounts of dust. However, the typical (1σ) HOSTS measurement uncertainty per individual star is ~20 zodis for early-type stars (A–F5) and ~50 zodis for sunlike stars, i.e., far above solar density levels. Out of 23 sunlike stars (all in HabEx master's target list, *Appendix D*), the survey only detected dust around four potential HabEx targets (Eps Eri, Tet Boo, 72 Her and 110 Her), all at exozodi brightness levels 100 to 500 times higher than observed around the Sun at the same wavelength.

For HabEx yield calculations (*Appendix C, Section C.2.1.2*), these four stars were assigned their LBTI-measured exozodi levels. All other potential HabEx target stars (*Appendix D*) were assigned exozodi levels randomly drawn from the distribution that best fits the LBTI data.

A first uncertainty in HabEx exoplanet science yield calculation, comes then from the actual exozodi levels of nearby *individual* HabEx target stars. This finite sampling uncertainty is correctly captured in HabEx yield analysis, and its impact remains fairly negligible for HabEx observations of ~50 targets, a result consistent with previous analysis by Stark et al. (2015) (Figure 15) which showed that *for a given exozodi level distribution*, knowing the exozodi brightness of individual stars provided little improvement in yield (10–20%). This is especially true as stars with high (>100) exozodi levels will be quickly identified during HabEx's first direct imaging visit to a given system.

Moreover, the LBTI survey was conducted in the mid-infrared and a solar dust density profile was assumed to derive exozodi level estimates. But what is of interest to HabEx surveys is the exozodi level at UV, optical, and NIR wavelengths. Because the basic dust properties (e.g., density profile, size distribution and albedo) cannot be uniquely derived from measurements over a narrow wavelength range. exozodi levels of individual targets at relevant HabEx wavelengths may be significantly different from those measured by LBTI at 11 μm. The difficulty to extrapolate exozodi measurements from one wavelength to another is well illustrated by the so-called "hot excess phenomenon" detected in the near infrared around a significant fraction of main sequence stars (Absil et al. 2013; Ertel et al. 2014), in most cases with no counterpart in the mid-IR (Mennesson et al. 2014; Ertel et al. 2018). It is generally believed, based on high contrast interferometric and polarimetric observations (Mennesson et al. 2011; Marshall et al. 2016) as well as modeling efforts (Rieke et al. 2016; Kral et al. 2017; Kirchschlager et al. 2018), that these hot excesses come primarily from sub-micron sized dust particles concentrating close to the sublimation of the host star and from thermal emission rather than scattering. A positive scenario for future missions, but definitely worth further detailed observations.

## 12.7 Technical State of the Art: PRVs Progress toward 1 cm/s Accuracy

The recent announcement of a roughly Earth-mass companion in the HZ of Proxima Centauri b (Anglada-Escudé et al. 2016) and two possible super-Earths at the HZ edges of Tau Ceti (Feng et al. 2017), all with RV semiamplitude $K < 1$ m/s, along with the controversial claim of an Earth-mass planet on a ~3-day orbit around Alpha Centauri B (Dumusque et al. 2012; Rajpaul et al. 2016) highlights both the success and the challenges of obtaining accurate and precise sub m/s RVs (see also Fischer et al. 2016).

The expected capabilities and impact of near-future radial velocity studies is assessed in Plavchan et al. (2015) and Fischer et al. (2016).

Current generation instruments and data analysis techniques are limited in sensitivity to a reflex motion of $K \sim 1$–$2$ m/s by stellar activity and instrument systematics. Over the next 5 years, the next generation of PRV systems will come online (e.g., EXPRESS, ESPRESSO, NEID), which are expected to have reduced instrumental systematics. Combined with improved data analysis techniques for mitigating stellar activity (Barnes et al. 2017), these instruments may or may not allow for the detection of reflex motion of significantly less than K ~ 1 m/s. If this increased radial velocity sensitivity is optimistically realized, the nearest several hundred stars later than ~F2 in spectral type may be surveyed for Earth and super-Earths in the HZ of sunlike stars.

Pushing even further in RV's capabilities, a planet mass determination with a precision of 10% would distinguish among different terrestrial planet composition models. For an Earth analog in a HZ orbit, determining the mass to ~10% requires ~1 cm/s radial velocity accuracy on timescales of a few year. Such a capability is not currently possible from the ground. Both High Resolution Spectrograph (HIRES; formerly CODEX) and Giant Magellan Telescope (GMT) Consortium Large Earth Finder (G-CLEF) for the E-ELT and GMT respectively are being designed for an instrument systematic uncertainty of ~2 cm/s (Plavchan et al. 2015).

Current generation radial velocity surveys can observe a single target for 5 minutes to reach a photon noise precision of ~1 m/s (e.g., HIRES on Keck), and thus allowing a single facility to observe on the order of 100 stars in a single night. However, a photon noise of 1 cm/s will require significantly longer integration times. For example, a ~1 cm/s photon noise is reached for with a total integration time of one hour on a 10 m telescope at V ~ 4 mag (Plavchan et al. 2015). Thus, mass determination of exoplanets at this precision will require significant telescope time.

The telescope time required will limit the number of targets that can be observed with competed facilities such as the E-ELT, Thirty Meter Telescope (TMT), and GMT. Developing the data analysis tools, cadence and wavelength coverage for stellar activity in radial velocity spectroscopic time-series is also an active area of research. For example, Calar Alto high-Resolution search for M dwarfs with Exoearths with Near-infrared and optical Echelle Spectrographs (CARMENES) has recently reported the detection of the "color" of radial velocities from stellar activity (Reiners 2016), consistent with simulations from the StarSIM 2.0 code (Plavchan 2016), showing that the RVs exhibit a color dependence that is also a function of a star's rotational phase. In the next several years, these preliminary results will be subject to peer review and published, along with other upcoming visible and NIR precision RV spectrographs (e.g., NEID, Habitable Planet Finder).

Finally, if there are no further improvements in radial velocity sensitivity from future generation instruments due to stellar activity, then pessimistically only Jovian analog companions (orbital periods up to 1–2 decades) orbiting the nearest stars will be known a priori from radial velocities and amenable to direct imaging.

## 12.8 Technical State of the Art: the Binary Star Opportunity

HabEx projected exoplanet science yield assumes no active cancellation if starlight scattered by nearby companions on multi-star systems. It has been extremely conservative in eliminating stars in binary systems where the companion angular separation and relative flux results in prohibitive scattered light contamination. Nevertheless, given that half of all stellar systems are binaries, this cut has removed some potentially excellent targets, including in particular alpha Cen A and B. Should it be possible to mitigate the effects of scattered light to still achieve HabEx primary exoplanet science goals and objectives in multiple star systems, they would be added to the target list. Hereafter, the yield and design impact of being able to control light scattered by stellar companions are considered more carefully.

Approximately half of all nearby Sun-like stars are located in multi-star systems, primarily in binaries (Raghavan et al. 2010). Many of these multi-star systems have dynamically stable circumstellar regions that could host planetary systems (Quarles and Lissauer 2016). Many promising multi-star systems are left out of the original HabEx target lists (*Appendix D*) and excluded in all planet yield calculations presented in *Chapter 3*. That is because the technology to suppress the light of both stars, while very promising, remains fairly new. It still under laboratory development, and presented here as a potential enhancement to the HabEx baseline design presented in the previous chapters.

Multi-Star Wavefront Control (MSWC) is a technology that has been maturing over the past 4 years (Belikov et al. 2015). It uses a deformable mirror in the wavefront control system of a coronagraph to simultaneously and independently remove both the on-axis star and off-axis star leakage to enable imaging multi-star systems. MSWC is primarily an algorithmic method that is integrated as part of the nominal wavefront control loop. It requires no hardware modifications for binaries with small separations (MSWC-0), and a mild diffraction grating for larger separations

(MSWC-s, where "s" stands for "super-nyquist"). MSWC-s allows removing speckles at high-spatial frequencies that would normally be outside the deformable mirror's controllable region.

Multi-star imaging can be enabled on any of the proposed HabEx options by using either a MSWC-0 or MSWC-s, as long as a deformable mirror is available in the optical path. Thus, any of the HabEx configurations that have a coronagraph available could be used with MSWC. A starshade can also use MSWC-s as long as a deformable mirror is available in the optical path. (Since the starshade effectively removes the on-axis star, MSWC-s only needs to correct one off-axis star, in which case it reduces to a simpler algorithm called Super-Nyquist Wavefront Control, or SNWC).

### 12.8.1 Multi-Star Science Potential with HabEx

#### 12.8.1.1 Multi-Star Systems within 20 pc

There are 517 FGK stars within 20 pc, out of which 259 are in multi-star systems, and nearly 150 of them are affected by significant companion star leakage in the HZ of the central star of interest (Sirbu et al. 2017b). Therefore, the addition of multi-star systems could significantly increase the accessible targets for HabEx (Stark et al. 2019). In addition to increasing the quantity of targets, binary stars also provide more opportunities for high quality targets in terms of SNR as well as spatial and spectral resolution. In particular, Alpha Centauri is the nearest star system to us and contains two sunlike stars, making it a very favorable target. However, it is excluded in this study due to its binary nature.

In particular, the Alpha Centauri system is 2.4× closer than the next nearest non-M dwarf star, and thus represents a unique opportunity to study a planetary system with at least 2.4× greater spatial resolution, with host stars that are 2 magnitudes brighter than any planetary system around all other known nearby non-M dwarf stars. Put another way, imaging the planetary system of Alpha Centauri is equivalent to imaging the planetary system of any other non-M dwarf star with an aperture at least 2.4× larger than HabEx (in terms of photon flux and resolution).

Shown in **Figure 12.8-1** is the leakage resulting from the off-axis companion of each binary star out to 20pc for the HabEx coronagraph and wavefront errors based on the primary mirror's power spectral density specification (with an integrated total of 20 nm RMS of surface figure error following a spectral envelope of $f^{-2.5}$). The leakage is due partly to diffraction (which dominates for closely separated binaries) and partly due to optical aberrations on the HabEx primary mirror (which dominates for wider separated binaries). The dotted vertical line indicates the nominal control limit at 32 $\lambda/D$ (spatial Nyquist limit) of the 64×64 actuator deformable mirrors (DMs) baselined for HabEx. There are four quadrants indicated in **Figure 12.8-1**, defined by the $10^{-10}$ contrast target of HabEx and the DM Nyquist limit. Stars in Quadrants 3 & 4 have companion leakage below the $10^{-10}$ level and their companion leakage contribution can be safely ignored (i.e., they can be treated as single-stars). However, most binary stars appear in Quadrants 1 & 2, where the leakage is significant. For these stars, the companion leakage must be removed in order to meet HabEx contrast requirements.

### 12.8.1.2 List of Very Nearby Binary Systems

In addition to binary stars being abundant in general, they happen to be overrepresented among the nearest few FGK stars. For example, within 4 pc, Alpha Centauri, Procyon, 61 Cygni and Epsilon Indi, are all binaries, while only Epsilon Eridani and Tau Ceti are single stars (Belikov et al. 2016). Therefore, there is a high chance that the best (or at least closest) potentially habitable planet target for characterization by HabEx will reside in a binary star system. **Table 12.8-1** summarizes some of the nearest Sun-like star targets that could be imaged with MSWC. Alpha Centauri appears to be the best target for almost any direct imaging as already described above. 61 Cygni has a wide enough separation that MSWC may not be required, depending on the exact HabEx primary mirror (PM) wavefront errors power spectral density (PSD) and stability, with 61 Cyg B being more affected by the leak of 61 Cyg A. For all other binary targets listed, MSWC is needed.

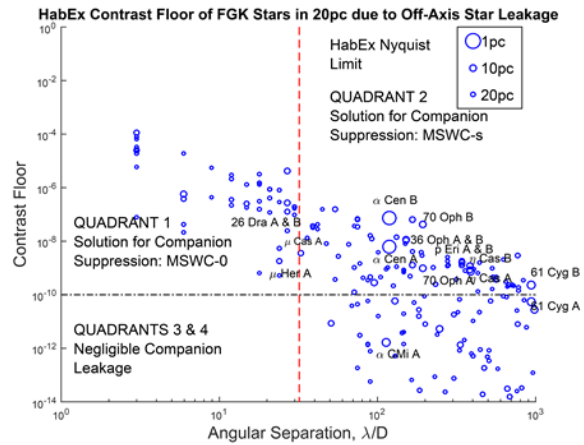

**Figure 12.8-1.** Visible contrast floor at primary star location due to off-axis star leakage from companion star (from Sirbu et al. 2017b) at a given angular separation. Each blue circle corresponds to a star in a multiple stellar system. For instance, the *61 Cyg A point* at the extreme right indicates the contrast floor set by 61 Cyg B starlight spilled in the immediate vicinity of 61 Cyg A.

### 12.8.2 Multi-Star Imaging Technology Options

The main challenge when imaging multi-star systems is the off-axis leakage introduced by the companion star. This additional leakage is incoherently added to the typical speckle field due to diffraction and aberrations from the on-axis star. To solve this problem, a set of wavefront control solutions have been developed to address different imaging scenarios which are described and summarized here. These also represent different operation regimes for multi-star imaging.

**Table 12.8-1.** List of the nearest sunlike binary star targets and their binary companion properties.

| Star | Type | Dist. (pc) | V-mag | Comp. V-mag | Comp. Sep(") |
|------|------|-----------|-------|-------------|--------------|
| α Cen A | G2V | 1.3 | 0.0 | 1.3 | 4 |
| α Cen B | K1V | 1.3 | 1.3 | 0.0 | 4 |
| 61 Cyg A | K5V | 3.5 | 5.2 | 6.1 | 31 |
| 61 Cyg B | K7V | 3.5 | 6.1 | 5.2 | 31 |
| 70 Oph A | K0V | 4.1 | 5.1 | 6.2 | 7 |
| 70 Oph B | K4V | 4.1 | 6.2 | 5.1 | 7 |
| 36 Oph A | K2V | 5.5 | 5.1 | 5.1 | 5 |
| 36 Oph B | K1V | 5.5 | 5.1 | 5.1 | 5 |
| η Cas A | G0V | 6.0 | 3.5 | 7.5 | 13 |
| η Cas B | K7V | 6.0 | 7.5 | 3.5 | 13 |

### 12.8.2.1 Multi-Star Wavefront Control in Sub-Nyquist Regime (MSWC-0), for Closely Separated Stars

The main principle of MSWC is to use two non-redundant sets of modes on the DM to independently suppress the leak from two stars. MSWC is coronagraph agnostic, and therefore is compatible with almost any existing direct imaging mission concept that has a DM. The MSWC-0 version requires no changes to HabEx hardware, but works only when the binary separation is within the Nyquist controllable field of the DM—the Sub-Nyquist regime (quadrants I and III in **Figure 12.8-1**).

### 12.8.2.2 Super-Nyquist Wavefront Control (SNWC), for the Case Where the On-axis Star is Completely Suppressed

Wavefront Control (SNWC) to extend the DM's controllable region (Thomas et al. 2015). The main principle behind this method is that the diffraction grating creates a set of point spread function (PSF) replicas, and conventional (sub-Nyquist) wavefront control can create a dark zone within the Nyquist range of each of the replicas, thus extending the range to create a dark hole arbitrarily far from the star. A single diffraction grating whose periodicity matches the DM Nyquist limit could in theory enable the creation of a dark zone anywhere (although the total area of the zone will still be Nyquist-limited). SNWC by itself is a single-star algorithm, but could be used to image multi-star systems with a starshade blocking the on-axis star (which does not require active DM control), and the off-axis speckles removed with SNWC.

### 12.8.2.3 Multi-Star Wavefront Control in Super-Nyquist Regime (MSWC-s), for Widely Separated Stars

SNWC and MSWC-0 can be combined into "MSWC-s", enabling high-contrast regions for multi-star systems at any angular separations (Thomas et al. 2015 and Sirbu et al. 2017a). In this case, the MSWC principle of simultaneously and independently removing speckles from both stars is used. Additionally, the SNWC principle of replicating the PSF and using diffraction orders to extend the controllable region of the DM to eliminate speckles at high-frequencies can also be used.

## 12.8.3 HabEx Multi-Star Imaging Modes and Instrument Compatibility

The set of multi-star imaging scenarios and the applicability of one of the wavefront control solutions described earlier is summarized in **Table 12.8-2**. This is a comprehensive list that lists all the possible techniques to block the on-axis and off-axis stars.

For HabEx specifically, the high-level options that are most relevant are Options #1 and #4. Option #1 uses the on-axis coronagraph in conjunction with MSWC to reduce off-axis leakage without requiring any additional hardware beyond the diffraction grating for wide separation stars. Option #4 uses the starshade to block the on-axis star and a DM available through the starshade optical channel with a diffraction grating to remove the off-axis leakage due to the companion. Option #4 can also be useful to potentially achieve deeper contrast than option #3.

For the case of MSWC-s it is necessary to have a mild grating, which creates dim replicas of the companion within the DM sub-Nyquist control region (Bendek et al. 2017; Bendek et al. 2013). The grating could be the result of the quilting of the Boston Micromachines DM. In the case where a featureless DM is used, such as Xinetics, then a mild grating with dots or lines can be added in any plane conjugated to the DM upstream of the coronagraph's focal plane. For example, **Figure 12.8-2** shows a possible implementation where the mild grating is placed on the surface of the FSM. Alternatively, the grating could be placed on the surface of the dichroic or the polarizers.

## 12.8.4 Computer Demonstrations of Multi-Star Imaging with HabEx using MSWC

### 12.8.4.1 Multi-Star Wavefront Control with the HabEx Vortex Coronagraph

HabEx has baselined the vector vortex charge 6 as the default coronagraph option (for those configurations which use a coronagraph). The vector vortex coronagraph consists of an

**Table 12.8-2.** Summary of multi-star imaging scenarios and applicability of wavefront control (WC) solutions.

| # | Scenario On-axis blocker | Off-axis blocker | WC Solutions Star Separation at < N/2 λ/D* | Star Separation at > N/2 λ/D* | Notes |
|---|---|---|---|---|---|
| 1 | Coronagraph 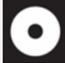 | None (WC only) | **MSWC-0** | **MSWC-s** | Existing coronagraphic mission concepts are already capable of MSWC-0 with no hardware modifications. MSWC-s requires quilting on the DM or a mild grating in the pupil plane. |
| 2 | Coronagraph 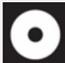 | 2nd Coronagraph | **MSWC-0** | **MSWC-s** | The second (off-axis) coronagraph would require an additional mask. It can be helpful if diffraction rings from the off-axis star are significant. MSWC is still required if wavefront error is significant. |
| 3 | Coronagraph 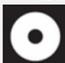 | Starshade 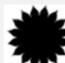 | SSWC (i.e., standard WC) | SSWC (i.e., standard WC) | Adding a starshade effectively reduces binaries to single-star suppression problem, at a cost of adding a starshade. |
| 4 | Starshade 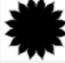 | None (WC only) | SSWC (i.e., standard WC) | **SNWC** | Adding a deformable mirror (without a coronagraph) to a starshade mission theoretically enables double-star suppression. |
| 5 | Starshade 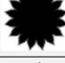 | Coronagraph 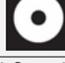 | SSWC (i.e., standard WC) | **SNWC** | The off-axis coronagraph is not necessary for a well-baffled telescope, but may relax the stroke requirement on the DM for close stars. |
| 6 | Starshade 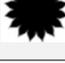 | 2nd Starshade 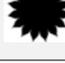 | No WC required | No WC required | Adding a starshade for the off-axis star effectively reduces binaries to a single-star suppression problem, but at a cost of adding a second starshade. |

SSWC = Single Star Wavefront Control (WC), SNWC = Super-Nyquist WC, MSWC-0 = Multi-Star WC (0th order, or sub-Nyquist),
MSWC-s = Multi-Star WC (super-Nyquist)
*Assuming DM = NxN actuators

achromatic focal plane mask that in theory is well suited for usage with MSWC.

This is demonstrated by simulations (**Figure 12.8-3**) showing dark holes achieved with MSWC using the HabEx Vector Vortex Charge 6 coronagraph at separations, optical PSD, and brightness ratios representative of the Alpha Centauri system. The left-panel shows imaging of Alpha Centauri A on-axis, and the right-panel shows Alpha Centauri B on-axis. In both cases target contrasts within a factor of 2 or better of the $10^{-10}$ contrast levels are demonstrated in a 10% band. The diffraction orders used in this simulation are representative of the natural quilting on the current generation of Boston Micromachine DMs. Alternatively, the diffraction order strength and location can be designed with pupil-plane masks featuring non-reflective dots whose spacing and size can be specified.

## 12.8.4.2 Super-Nyquist Wavefront Control with HabEx Starshade

SNWC can also be used with the HabEx starshade configuration. In the baseline HabEx scenario, both a coronagraph and a starshade will be available. In this case, if the optical path of the starshade channel has access to the coronagraph DM (or a separate DM) this can be utilized to enable imaging of binaries with a starshade. To demonstrate this, an imaging scenario was simulated in **Figure 12.8-4**, using SNWC with an on-axis starshade, assuming HabEx-provided optical PSDs and brightness ratios representative of Alpha Centauri system. The imaging strategy with a starshade is to first create larger area dark hole in a narrower band to enable detection, followed by a smaller but deeper and broader-band dark zone for characterization.

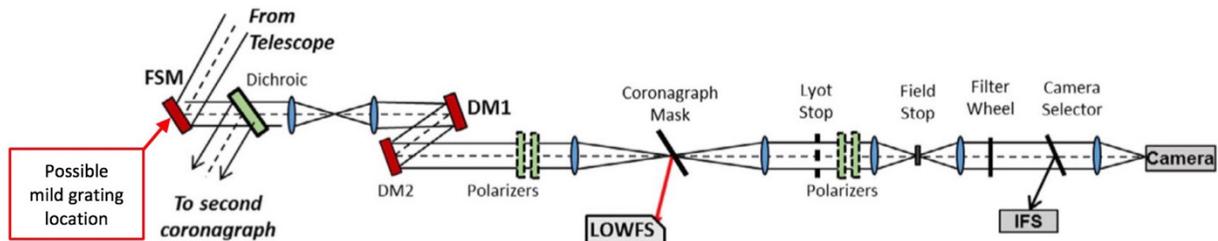

**Figure 12.8-2.** A grating can be printed on the surface of the fine steering mirror (FSM) to enable multi-star wavefront control (MSWC) with a featureless, deformable DM.

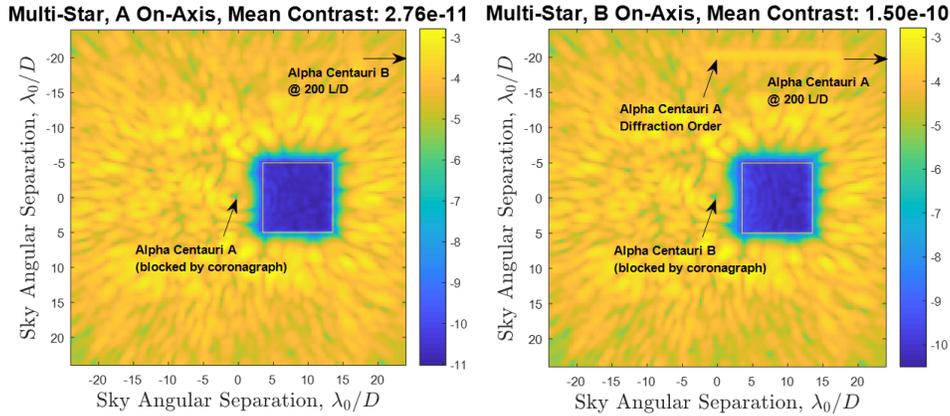

**Figure 12.8-3.** Simulation of speckle suppression in a binary star system with HabEx using MSWC-s and the Vector Vortex Coronagraph, in a 10% band: (*Left*) Imaging Alpha Centauri A on-axis, and (*Right*) Imaging Alpha Centauri B on-axis.

### 12.8.5 Summary of TRL

The technology readiness level (TRL) of MSWC is currently TRL 3. A funded technology effort is currently under way at the NASA Ames Coronagraph Experiment (ACE) laboratory to advance TRL to 4 by the end of 2019.

The left panel of **Figure 12.8-5** shows a NASA-funded feasibility laboratory demonstration (at TRL 3) for both MSWC-0 and SNWC (Belikov et al. 2017b). These demonstrations used a simple imaging system with a DM and no coronagraph, in order to study MSWC by itself as a first step. In **Figure 12.8-5**, a computer model is also shown (albeit without wavefront errors) with behavior at least qualitatively very similar to the lab demonstrations. An MSWC-generalized version of classical speckle nulling was used for this demonstration.

The four right plots of **Figure 12.8-5** shows a laboratory demonstration where a speckle region at 100 $\lambda/D$ away from the star was suppressed by a factor of 10 (from roughly $10^{-7}$ to $10^{-8}$) in monochromatic light. Similar to **Figure 12.8-3**, no coronagraph was used for this demonstration to study pure wavefront control. Validated models show that SNWC works with coronagraphs in broadband light (**Figures 12.8-4 and 12.8-5**), although the size of the dark region shrinks as the band goes up. To date, MSWC-0 and −s has been successfully tested with multiple coronagraphs in simulation and found that the performance does not strongly depend on coronagraph architecture.

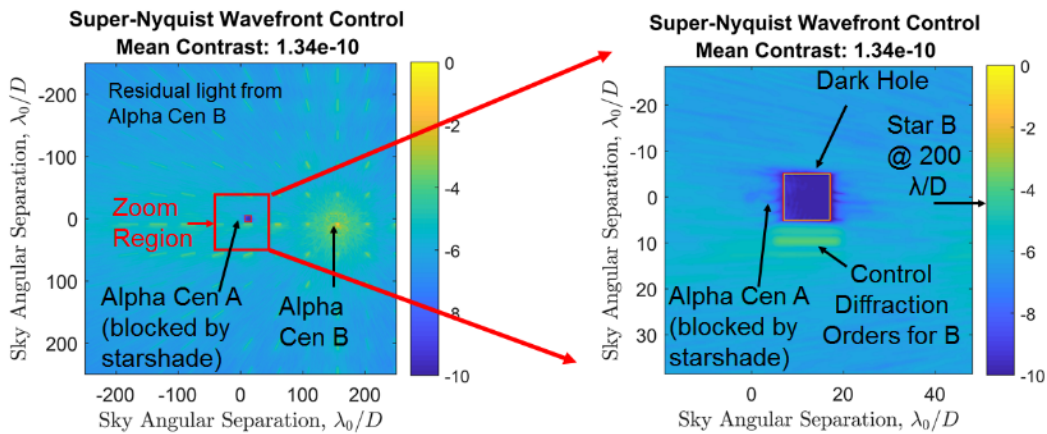

**Figure 12.8-4.** Simulation of SNWC with HabEx Starshade: (*Left*) Wide-field image showing the field with the companion star and its diffraction orders, (*Center*) Resulting dark hole with the on-axis star suppressed by the starshade and the off-axis leakage suppressed by SNWC in 10% band, (*Right*) Follow-up characterization dark hole in 25% band (spatial scale is zoomed).

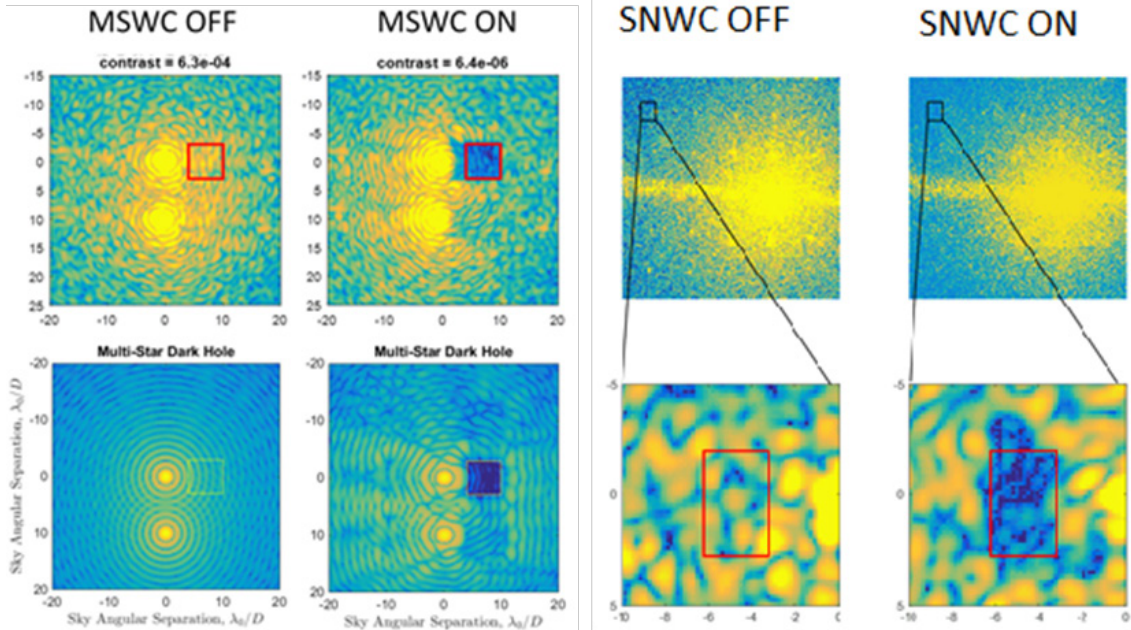

**Figure 12.8-5.** TRL 3 Laboratory demonstration (TRL 3) of MSWC-0 (*4 panels on the left*) and SNWC (*4 panels on the right*). MSWC demonstration: light from each star is independently suppressed by at least a factor of 10 in the red square area, both in the laboratory (*top*) and in simulations (*bottom*). Contrast and size of the dark zone are likely limited by absence of coronagraph and model errors. SNWC demonstration: a dark zone is successfully created at ~100 $\lambda/D$ with suppression factor of 10 achieved in the zoomed-in region (*bottom*).

The upcoming TRL 4 demonstration (Belikov et al. 2017a) includes a test with the Subaru Coronagraphic Extreme Adaptive Optics (SCExAO) Instrument, using its calibration source; as well as a $10^{-7}$ contrast demonstration with a coronagraph in monochromatic light at the Ames Coronagraph Experiment testbed, using an aperture for either WFIRST, LUVOIR, or HabEx.

## 12.9   Recommendations

### 12.9.1   Further Development of Extreme Precision Radial Velocities

*Section 12.1* argued that it is vital that the masses of planets, particularly EECs, that are detected by HabEx be measured. Although precursor observations that aim to detect Earth analogs around nearby stars can either improve the yield of the HabEx survey, or (more likely) improve the efficiency, the masses will need to be measured regardless. There are essentially two methods of measuring the masses of EECs: RVs and astrometry. However, both would need considerable development to enable the detection of Earth analogs. Attempts to develop the RV technique to the point where it can achieve accuracies of ~1 cm/s (also known as Extreme Precision Radial Velocities, or EPRVs) are currently favored over astronomy, primarily for the reason that it may be possible to reach this precision from the ground, whereas reaching an astrometric precision of ~0.1 μas needed to detect Earth analogs almost certainly required going to space. Furthermore, RVs are needed for many other applications, including, e.g., transit surveys such as TESS and PLATO. For this reason, the National Academies of Sciences, Engineering, and Medicine (NAS) Exoplanet Science Strategy (ESS) report found that "The radial velocity method will continue to provide essential mass, orbit, and census information to support both transiting and directly imaged exoplanet science for the foreseeable future."

This report also found that "Radial velocity measurements are currently limited by variations in the stellar photosphere, instrumental stability and calibration, and spectral contamination from telluric lines. Progress will require new instruments installed on large telescopes, substantial allocations of observing time, advanced statistical methods for data analysis informed by theoretical modeling, and collaboration between observers, instrument builders, stellar astrophysicists, heliophysicists, and statisticians."

To maximally enhance the exoplanet science achieved with HabEx, we strongly endorse the top-level recommendation of the NAS ESS report that "NASA and NSF should establish a strategic initiative in extremely precise radial velocities (EPRVs) to develop methods and facilities for measuring the masses of temperate terrestrial planets orbiting sunlike stars."

We note that, because the efforts needed to determine whether or not it is possible to reach a systematic precision of ~1 cm/s from the ground will almost certainly take many years, and that the stars that will be targeted by HabEx will need to be monitored for many years if and when this precision is demonstrated, it is critical that NASA and NSF begin planning for this initiative as soon as possible.

### 12.9.2 Technology Development for High Contrast Imaging of Binary Stars

As demonstrated in *Section 12.5*, there are many promising technologies that can suppress the scattered light from nearby binary companions to target stars, thereby allowing high-contrast imaging and potentially the detection and characterization of planets orbiting these stars. Given that roughly half of all solar type stars are in binary systems, these technologies have the potential to significantly increase the viable target sample and thus the number of planets that can detected and characterized within a given total observing time or IWA limit. Of particular importance is the fact that some of these technologies have been shown to be applicable to the Alpha Centauri system, which, if not for the fact that it is a binary, would easily be the best target for direct imaging searches for planets.

However, many of these technologies are at relatively low TRL 3 levels. There are plans to bring several of these technologies to TRL 4 and we encourage continued investment in these technologies.

### 12.9.3 Improved Characterization of Exozodi Levels around Nearby Sunlike Stars

In order to accurately estimate the exozodi background to be faced by HabEx around *individual* targets, i.e., to go beyond statistical knowledge and current model-dependent wavelength extrapolations (*Section 12.6*), high contrast resolved exozodi observations are required in the actual HabEx visible to near infrared wavelength range, with sensitivity down to $\leq 10$ times the solar zodi level. More generally, the next breakthrough in measuring the radial and azimuthal structures of exozodiacal clouds, understanding their origins and connection to planet properties, requires spatially resolved visible and infrared observations at higher contrast, angular and temporal resolutions than currently available from space and from the ground (e.g., Mennesson et al. 2019).

As precursor exozodi observations are concerned, the main recommendation is therefore to foster instrumentation developments for (i) high-contrast space-based imaging systems in the visible, and (ii) ground-based high-contrast near-IR interferometric systems, using separate telescopes and aperture masking on ELTs. For instance, visible observations with $\gtrsim 1$ m space-based telescopes at contrast levels below $\sim 10^{-7\text{-}8}$ per spatial resolution element—as specified for the WFIRST coronagraph instrument (Mennesson et al. 2018)—would be of high interest to optimize the target selection and observing efficiency of HabEx (or LUVOIR) direct imaging surveys. Such measurements would also cross an important threshold in debris disks physics, detecting and mapping exozodi structures at low enough optical depths ($\lesssim 10 \times$ solar) that they will be dominated by transport phenomena rather than collisions (Kuchner and Stark 2010).

# 13 SUMMARY AND OUTLOOK

In this report, we have described the science motivation and preferred architecture of HabEx, the Habitable Exoplanet Observatory. As envisioned, HabEx will be a large strategic mission that would launch mid-2030s. With its large aperture and suite of four highly capable instruments, HabEx has three primary science goals: (1) Seek out nearby worlds and explore their habitability, (2) Map out nearby planetary systems and understand the diversity of the worlds they contain, and (3) Enable new explorations of astrophysical systems from our own solar system to galaxies and the universe by extending our reach in the ultraviolet (UV) through near-infrared (IR). We have motivated these three primary science goals, which were developed to be responsive to the extraordinary revolutions in the fields of Galactic and extragalactic astrophysics, exoplanets, and planetary science over the past few decades. These revolutions have raised a large number of scientific questions, from which we have defined the scientific objectives of the mission. The telescope, instrument, and mission functional requirements ultimately flow directly from these science objectives.

HabEx was designed to be a Great Observatory that can be realized in the 2030s. To achieve this, the guiding philosophy of this study was the recognition that any recommendation by the Astro2020 Decadal Survey must balance scientific ambition with programmatic and fiscal realities, while simultaneously considering the impact of its development schedule on the greater astronomy community and the need for a broad portfolio of science investigations. The HabEx study therefore aimed to develop a mission capable of the most compelling science possible, while still adhering to likely cost, technology, risk, and schedule constraints. The preferred HabEx architecture was thus chosen to be technically achievable within this time frame, leveraging the maturation of several enabling technologies over the last decade or more. HabEx was further conservatively designed with substantial margins, and as a result suffers from relatively low levels of risk and high levels of technological maturity.

The preferred HabEx architecture is a 4 m, f/2.5, monolithic, off-axis telescope that employs two starlight suppression technologies, a coronagraph and a starshade. As we describe in detail, the hybrid design greatly enhances the hybrid, dual technology direct imaging science capabilities of HabEx. The preferred architecture also provides the community with imaging and spectroscopic capabilities orders of magnitude better than the Hubble Space Telescope, which uniquely complement current and planned space and ground-based observatories.

The time for HabEx is now. A confluence of several relatively recent events has made the HabEx mission not only scientifically and technologically viable within the next few decades, but also timely. Revolutions in our understanding of the contents of our own solar system, in our understanding of the properties and architectures of other planetary systems, and in our understanding of the contents, geometry, and indeed entire history of our universe, have made the time ripe for a mission with the capabilities of HabEx.

Perhaps more important is our newfound knowledge that potentially habitable planets—rocky worlds that may have thin atmospheres like the Earth and with orbits about their parent stars that permit liquid water on their surface—are likely relatively common, with roughly 25% of sunlike stars hosting such planets. Since the minimum aperture required to detect and characterize a given number of Earth-like planets scales as the frequency of habitable planets, this implies that relatively small apertures, such as the 4 m aperture architecture detailed in this report, are capable of achieving the ancient and lofty goals of searching for habitable conditions and even providing evidence for life outside our solar system. Equally important is the relatively recent discovery that the majority of sunlike stars likely do not harbor bright habitable zone (HZ) dust disks that could impede our ability to detect and characterize exoplanets as faint as the Earth.

Simultaneously, rapid advances of the starlight suppression technologies have made the contrast and resolution requirements to detect and characterize Earth-like planets around the nearest stars achievable within our immediate horizon. Other technological developments, in particular in the areas of very sensitive detectors, large mirror fabrication, and spacecraft pointing and thermal and vibration control, have also enabled many of the science applications of HabEx described in this report.

The preferred HabEx 4H architecture design and implementation are detailed extensively in *Chapters 6*, *7*, and *8*. Here "4" stands for the primary diameter of 4 m and "H" stands for hybrid, denoting its use of both the coronagraph and starshade starlight suppression technologies. HabEx 4H is a monolithic, off-axis, 4 m aperture telescope equipped with a suite of four instruments. Two of these instruments are dedicated to high-contrast direct imaging of exoplanets, while the other two instruments are designed to maximize the unique strengths of HabEx for studies of general astrophysics and solar system studies: a large aperture, and thus high resolution and a large photon collection area, combined with exceptional UV sensitivity incorporating cutting-edge technologies.

The HabEx 4H architecture relies on the two most mature starlight suppression technologies. Coronagraphs allow the starlight to enter the telescope aperture but suppress the starlight to the requisite contrast levels using a sophisticated optical system, while starshades use a large, opaque structure located roughly a hundred thousand kilometers away from the telescope to block the starlight before it enters the telescope aperture. Both technologies have intrinsic strengths and weaknesses.

Coronagraphs are much nimbler than starshades since the starlight suppression happens within the telescope. However, they have narrow bandwidths and so require multiple channels or serial observations to obtain broadband spectra. Thus, obtaining spectra covering a wide wavelength range is expensive or time consuming, or both. Also, coronagraphs generally have lower throughput, require more complicated instruments, and require highly stable telescopes.

Starshades have traditionally been considered less mature than coronagraphs, but have recently 'caught up' to the technical readiness levels of coronagraph technologies due to focused investments by NASA. Starshades have the advantage that the starlight never enters the telescope aperture, as it is occulted by the starshade that is located roughly a hundred thousand kilometers away. As a result, the instrument and telescope design are dramatically simpler than that for a coronagraph. For a starshade, there are more relaxed requirements on the telescope stability and the fidelity of the instrument, besides the obvious requirements for a low dark current and read noise for the detector (which also exist for a coronagraph). The primary difficulty with a starshade is that, because of the distance of the starshade from the telescope, the starshade requires time and propellant in order to transit between targets. These resources are limited and constrain the number of exoplanetary systems that can be observed with the starshade. For HabEx, roughly 100 slews are currently being baselined.

Fortunately, the two methods of starlight suppression are very complementary, which is one of our primary motivations for preferring this architecture. The nimbleness of the coronagraph complements the starshade's ability to achieve deep and wide-wavelength spectra of planets with relatively short integration times. The strength of the coronagraph is to detect planets that are potentially inside the habitable zone, characterize their colors, and measure their orbits. The strength of the starshade is the ability to acquire precise spectra over a wide wavelength range in a relatively short integration time. Together these two technologies allow simultaneously for the detection, orbit characterization, and spectral characterization of a large variety of planets and a large number of systems. The HabEx 4H architecture takes advantage of the strengths and complementarity of both methods to maximize its science capabilities.

Technology gaps have been identified and evaluated in detail, and the path, cost, and timeline to close those gaps to Technology Readiness Level (TRL) 6 has been detailed in *Chapter 11*. The result is a technology plan that is fully integrated into the overall HabEx cost and schedule estimates.

Acknowledging that the constraints that must be considered by the Astro2020 Decadal Survey are difficult to anticipate, or may even change over time, this study considers eight other architectures, as described in detail in *Chapter 10*. These are denoted in a shorthand way using the aperture diameter and starlight suppression technology, and include, along with our preferred 4 m hybrid architectures (4H), two other hybrid designs at 3.2 m and 2.4 m (3.2H and 2.4H, respectively); three architectures that only include a coronagraph: 4C, 3.2C, and 2.4C; and three architectures that only include a starshade: 4S, 3.2S, and 2.4S. While all of the architectures were considered in some detail, the 4C (*Appendix A*) and the 3.2S (*Appendix B*), along with our preferred architecture (the 4H), anchor this nine-architecture trade as they were studied at a much more detailed level. All nine of the architectures are compelling and can all achieve at least some subset of the objectives outlined in the Science Traceability Matrix (*Chapter 5*). Thus, all the HabEx architectures considered in this report enable groundbreaking science, while at the same time offering greater flexibility in budgeting and phasing. In this way, HabEx can still be compatible with a balanced portfolio, even for the most pessimistic fiscal projections. HabEx aims to provide the Astro2020 Decadal Survey with additional flexibility in its decision making.

HabEx represents a comparatively affordable, low-risk, large strategic mission that can be developed and launched in the mid-2030s. As such, HabEx will be the next large strategic mission in the spirit of the current great observatories, such as Hubble, Chandra, and Spitzer, and following the next great observatories of the James Webb Space Telescope (JWST) and Wide Field Infrared Survey Telescope (WFIRST). HabEx will provide unique and unprecedented capabilities, including the highest resolution images at wavelengths from roughly 0.3 to 1 μm, the largest UV collecting area of any previous satellite, and multi-object spectroscopy from the UV to the near-IR over a relatively large field of view. These capabilities will enable a broad range of exciting science that we can envision today, as well as science we cannot imagine as of yet. In addition, HabEx will revolutionize our understanding of planetary systems, offering nearly complete "family portraits" and associated spectra of the menagerie of gas giants, Neptunes, super-Earths, and terrestrial worlds orbiting our nearest neighbors. Finally, HabEx will enable, for the first time, direct spectra in reflected light of terrestrial worlds in the habitable zones of the nearest stars, perhaps finding not only habitable worlds, but potentially even signatures of biological activity. The technology required to accomplish all these goals is either in hand, or there is a well-developed path toward full readiness (*Chapter 11*).

The time is now. For the first time in human history, if we so choose, we have the scientific knowledge and technological ability to seek the answer to one of humankind's most fundamental, profound, and enduring questions: Are we alone in the universe?

# A CORONAGRAPH-ONLY ARCHITECTURES

## A.1 Introduction

While the HabEx 4H architecture is the preferred architecture and is, by design, capable of achieving all science objectives—eight other architectures are capable of achieving a subset of the science objectives at lower costs as summarized in *Chapter 10*. One key fiducial design choice that distinguishes the nine architectures is the selection of direct imaging instruments. In the case of hybrid architectures and architectures without a starshade, the telescope's payload and flight system design are in large part driven by the requirements from the coronagraph. However, because a coronagraph-only architecture does not require the development and launch of a separate starshade flight element, the overall architecture is simplified into a single launch and a single telescope flight element.

This appendix defines the telescope payload, flight system, and mission for a coronagraph-only architecture. In contrast, the HabEx 3.2S, and 4S architectures, which are starshade-only and does not use a coronagraph, are described in *Appendix B*.

The removal of a starshade changes HabEx's observing program and total exoplanet science yield, including both the number of targets characterized, and the extent of characterization possible for each target. The starshade transit is no longer a factor in observatory scheduling. Yet without the starshade, the deeper characterization of each identified target becomes more difficult. The coronagraph requires multiple images in order to characterize across different spectral bands, and for each band, a new "dark hole" must be dug. This process is described in *Section 6.3*. In addition, the lower throughput of the instrument requires longer integration times. As a result of these two factors, more time is required per target using the coronagraph for spectroscopy compared to using a starshade-supported instrument. In addition to the differences in scheduling and observing times, the extent of characterization possible is changed. The lowest observable wavelength using a coronagraph is 0.45 μm, compared to 0.20 μm with the starshade. This precludes the observation of ozone, an important marker of potential habitability.

As detailed in *Section 10.3*, the coronagraph-only architectures require less technology development than the hybrid architectures. In the case of HabEx 4C, there are four fewer required Technology Readiness Level (TRL) 4 technologies.

In no HabEx architectures are there any required TRL 3 technologies.

## A.2 HabEx 4C Architecture

This primary focus of this appendix is the HabEx 4C architecture. Because the coronagraph is a major driver of flight system requirements, the designs of the payload and flight system are almost identical to the baseline described in *Chapter 6*, and as much of this appendix is framed as a delta to telescope design for HabEx 4H. In contrast, an new design point was evaluated for the starshade-only architectures described in *Appendix B*.

The high-level HabEx 4C architecture is shown in **Figure A.2-1**. The most impactful changes for the HabEx coronagraph-only architectures are the removal of: the starshade flight element, the need for a second launch, and, the starshade instrument. Consquently, a different the exoplanet observing program is adapted to perform direct spectroscopy with the coronagraph. With lower end-to-end throughput than the starshade and time to "dig the dark hole" for different bands, coronagraph spectropscopy will take longer to achieve the same signal-to-noise ratio.

However, HabEx 4C retains the coronagraph's driving optical requirements, met by the ultra stable monolithic primary mirror and stiff structure. Their mass requires HabEx 4C to launch on a NASA Space Launch System (SLS).

### A.2.1 Payload Differences from 4H

The coronagraph is the primary driver of the unobscured, off-axis, monolithic mirror and the

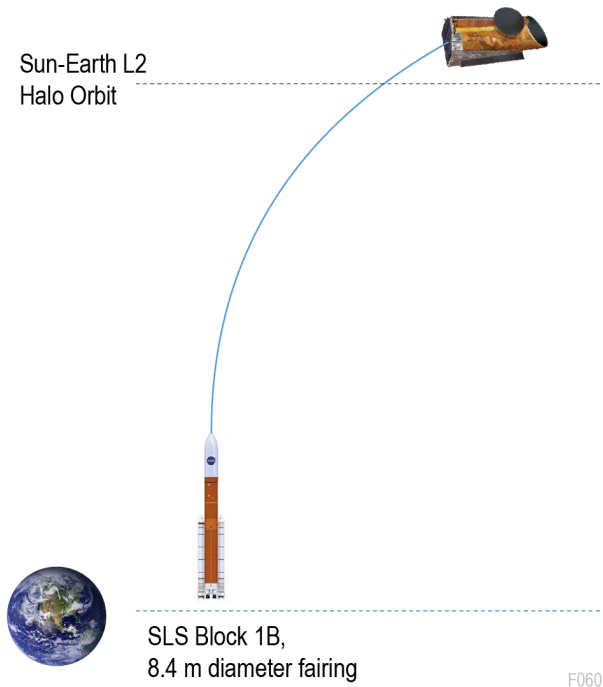

Sun-Earth L2
Halo Orbit

SLS Block 1B,
8.4 m diameter fairing

F060

**Figure A.2-1.** The HabEx 4C architecture only requires one launch for the telescope flight system.

optical telescope assembly (OTA) described in *Chapter 6*. With the same driving requirements, the OTA for the HabEx 4C architecture is effectively identical to the HabEx 4H baseline design.

The HabEx 4C architecture includes the HabEx Coronagraph (HCG) for exoplanet science and the UV Spectrograph (UVS) and the HabEx Workhorse Camera (HWC) for general astrophysics. All three of these instruments are described in *Chapter 6*. The HabEx 4C architecture requires no major changes in their design from their description in *Chapter 6*.

Without the need for the Starshade Instrument (SSI) at launch, the HabEx 4C telescope flight system can manifest an additional payload, whose options are summarized in **Table A.2-1**. The first option is to expand the scope of the instruments in order to take advantage of more room in the instrument bay. The second option adds an additional instrument for either general astrophysics or exoplanet science. A third option installs a starshade instrument while at Sun-Earth L2 post-launch, which would "upgrade" the coronagraph-only telescope in a hybrid mission. This option would, in effect, spread out the cost of a hybrid architecture, allowing for the launch of the additional instrument and necessary starshade flight element at an unspecified later date. This would coincide with a servicing mission to refuel and make other repairs to the telescope flight system.

### A.2.2 Flight System Differences from Baseline

Many of the driving requirements for the 4H flight system are also applicable for the 4C architecture, resulting in a similar flight system design. A MEL and PEL for the 4C are shown in **Tables A.2-2** and **A.2-3**, respectively. These tables both reflect the design option that includes three instruments: the coronagraph, UVS, and HWC. Additionally, for each subsystem, a summary of design differences compared to the baseline is included in **Table A.2-4**. The biggest change is the potential to eliminate the S-band system used for formation flying; however, it may be beneficial to retain this part of the design. S-band would be required if a starshade is added later as described in *Section A.2.1*, and an S-band crosslink could be used to aid in rendezvous for servicing, though it is not required. As a note, the masses here reflect structural and thermal point

**Table A.2-1.** Options for a fourth instrument in the HabEx 4C architecture.

| Option | Impacts |
|---|---|
| No additional instrument | • Lowest cost 4.0 m telescope option<br>• Could change configuration or scope of existing instruments |
| Additional instrument | • Exoplanet or general astrophysics instrument<br>• New requirements levied on system; dependent on which instrument chosen |
| Starshade Instrument installed after launch | • Upgrades 4C to 4H during the course of the mission<br>• Allows for 4H science with a spread-out cost, and flexibility to de-scope during development<br>• Additional complexity of installing an instrument in deep space<br>• Requires S-band for later cross-link |

Table A.2-2. Mass Equipment List (MEL) for 4C architecture with three instruments.

| | CBE (kg) | Cont. % | MEV (kg) |
|---|---|---|---|
| **Payload** | | | |
| Telescope and Instruments | 5730 | 30% | 7450 |
| Payload Thermal | 265 | 30% | 345 |
| **Spacecraft Bus** | | | |
| ACS | 20 | 3% | 20 |
| C&DH | 20 | 11% | 25 |
| Power | 240 | 28% | 300 |
| Propulsion: Monoprop | 300 | 5% | 325 |
| Propulsion: Microthruster | 160 | 44% | 235 |
| Structures & Mechanisms | 2690 | 30% | 3490 |
| Spacecraft side adaptor | 45 | 30% | 60 |
| Telecom | 35 | 28% | 45 |
| Thermal | 350 | 30% | 460 |
| **Bus Total** | 3820 | 29% | 4900 |
| **Spacecraft Total (dry)** | 9815 | 43% | 14035 |
| Subsystem heritage contingency | 2880 | | |
| System margin | 1340 | | |
| Monoprop and pressurant | 2280 | | |
| Colloidal propellant | 240 | | |
| **Total Spacecraft Wet Mass** | | | 16550 |
| Launch vehicle side adaptor | | | 1500 |
| **Total Launch Mass** | | | 18050 |

design that has not been re-optimized for the removal of the starshade instrument (SSI).

### A.2.3 Differences to Mission from Baseline

While the HabEx 4C flight system is very similar to the HabEx 4H telescope flight system, there are significant differences in operations. Because the exoplanet observations are no longer jointly scheduled with the starshade and the coronagraph maintaining a greater field of regard, there is more flexibility in how all observations are scheduled. The coronagraph and fast slew speed of the telescope, detailed in *Chapter 6*, enable a quick survey of potential targets, including detection and orbit determination. However, compared to the starshade instrument, characterization of the exoplanets—in particular spectroscopy—becomes more difficult with a coronagraph. The overall observable band is cut-off at the lower wavelengths, with coronagraph observations down to 450 nm compared to starshade observations as low as 200 nm in the hybrid and starshade architectures. This cutoff is above the wavelength of ozone and removes the ability to detect an important marker of potential habitability. Importantly, the coronagraph can only observe one narrow band at a time, meaning multiple images must be taken in order to fully characterize each target. For each image in a new band, the process of digging the dark hole, described in *Chapter 6*, must be performed again. Additionally, because of the lower overall throughput of light, the integration times for the coronagraph will be longer than they would be for a starshade instrument. Both of these provide increase the dwell time required at each characterization target, which could instead be used for surveying or general astrophysics in other architectures.

Table A.2-3. Power Equipment List (PEL) for 4C architecture.

| Subsystem | Unit | Launch | L2 Insertion | Corona-graph Science | UVS + HWC | Science Max | Down-link | Safe | Cruise |
|---|---|---|---|---|---|---|---|---|---|
| ACS | W | 0 | 15 | 10 | 15 | 15 | 20 | 2 | 2 |
| C&DH | W | 40 | 40 | 40 | 40 | 40 | 40 | 40 | 40 |
| Instruments | W | 0 | 0 | 400 | 540 | 860 | 0 | 0 | 0 |
| Monoprop System | W | 30 | 360 | 1 | 1 | 1 | 1 | 30 | 1 |
| Electrospray Prop System | W | 0 | 25 | 25 | 25 | 25 | 25 | 25 | 25 |
| Telecom | W | 75 | 75 | 140 | 75 | 140 | 170 | 75 | 75 |
| Thermal | W | 410 | 810 | 3560 | 3560 | 3560 | 3560 | 810 | 410 |
| Power Subsystems | W | 60 | 80 | 140 | 150 | 150 | 130 | 70 | 60 |
| **SUBTOTAL** | W | 620 | 1410 | 4320 | 4410 | 4790 | 3950 | 1050 | 610 |
| Contingency and Margin | % | 43% | 43% | 43% | 43% | 43% | 43% | 43% | 43% |
| Contingency Power | W | 260 | 610 | 1860 | 1900 | 2060 | 1700 | 450 | 260 |
| Distribution Losses | W | 20 | 40 | 120 | 130 | 130 | 110 | 30 | 20 |
| **TOTAL** | W | 900 | 2060 | 6300 | 6440 | 6980 | 5760 | 1530 | 890 |

**Table A.2-4.** Description of changes between 4H and 4C architectures by subsystem.

| Subsystem | 4C Differences compared to 4H |
|---|---|
| Structure and Mechanisms | • Different instrument bay design depending on payload option chosen<br>• No changes to barrel, mirror support, or bus structures |
| Thermal | • No change to OTA thermal design<br>• Different thermal design for instrument bay depending on instrument chosen; results in similar or decreased power usage |
| Power | • No changes: general astrophysics is the most driving power mode, resulting in similar arrays and power system to 4H. |
| Propulsion | • No changes: difference in dry mass is less than 2% without starshade instrument, so no change to propulsion design or propellant budgets. |
| Communications | • No changes to X-band or Ka-Band systems<br>• Potential to eliminate S-band system. Should be retained if a starshade is to be added later. Could be used to assist in servicing, but not required. |
| C&DH | • No changes: total data volume is reduced without SSI, but general astrophysics instruments and coronagraphs are more driving for data volume requirements. |
| Pointing Control | • No change to stability and pointing requirements; driven primarily by coronagraph<br>• No need for formation flying control if no starshade is used |

While the HabEx 4C architecture does offer cost and complexity advantages compared to the baseline, the difficulty of performing spectroscopy with a coronagraph, and its impact on the overall scientific return, should not be underestimated.

## A.3   Technologies for HabEx Coronagraph-Only Architectures

The HabEx 4C architecture eliminates five technology gaps compared to the baseline, these gaps are associated with the starshade. Most of the remaining TRL 4 and 5 technologies are related to the large monolithic mirror and the coronagraph. All gaps for HabEx 4C are shown in **Table A.3-1**. This architecture adds no new technologies as compared to the baseline. A more detailed discussion of all technology gaps is included in *Chapter 11*.

**Table A.3-1.** HabEx 4C Enabling Technologies at TRL 4 or 5.

| Title | Section |
|---|---|
| **Enabling Technologies** | |
| Large Mirror Fabrication | 11.3.1.1 |
| Large Mirror Coating Uniformity | 11.3.1.2 |
| Laser Metrology | 11.3.2.1 |
| Coronagraph Architecture | 11.4.1.1 |
| LOWFS | 11.4.2 |
| Deformable Mirrors | 11.4.3 |
| Delta Doped UV and Visible Electron Multiplying CCDs | 11.5.1.1 |
| Deep Depletion Visible Electron Multiplying CCDs | 11.5.1.1 |
| Linear Mode Avalanche Photodiode Sensors | 11.5.1.2 |
| UV Microchannel Plate (MCP) Detectors | 11.4.4 |
| Microthrusters | 11.6.1.1 |
| **Enhancing Technologies** | |
| Far-UV Mirror Coating | 11.7.1.1 |
| Delta-Doped UV Electron Multiplying CCDs | 11.7.3.1 |
| Microshutter Arrays | 11.7.2 |

# B  STARSHADE-ONLY ARCHITECTURES

While the HabEx 4H architecture is the preferred architecture and is, by design, capable of achieving all science objectives—eight other architectures are capable of achieving a subset of the science objectives at lower costs as summarized in *Chapter 10*. One key fiducial design choice that distinguishes the nine architectures is the selection of direct imaging instruments. In the case of hybrid architectures, the payload and flight system design are in large part driven by the requirements from the HabEx Coronagraph (HCG). A coronagraph capable of HabEx science requires an unobscured, off-axis telescope architecture, leading to a large and heavy spacecraft. Moreover, coronagraphs need ultra-stable optics to preserve high contrast. Starshade-only missions do not carry these requirements, resulting in a new starshade-only mission architecture optimized without the need for an unobscured, off-axis optical design and the ultra-stability necessary for a coronagraph mission.

This appendix defines the telescope payload, flight system, and mission for the HabEx 3.2S and 4S architectures. Specifically, the concept for HabEx 3.2S was the subject of a system study and multiple trades studies, and is reported here in detail. The HabEx 4S architecture is evaluated and presented as an upscaling of the HabEx 3.2S architecture.

The starshade flight system defined in *Chapter 7* for the baseline, HabEx 4H mission is essentially identical to the starshade for HabEx 4S and 3.2S. The only modification being in the case of the starshade for HabEx 3.2S, where the petal shape design is reoptimized for the smaller aperture. The starshade occulter deployment, flight system, and concept of operations are not modified and are thus not described in this chapter. In contrast, the HabEx 4C architecture, that is coronagraph-only and does not utilize a starshade, is described in *Appendix A*.

Removal of HCG changes HabEx's observing program and exoplanet yield. The starshade is fuel limited to about 100 total starshade observations, with each transit taking on the order of days. While this constraint is shared by the HabEx 4H architecture, the HCG carries the burden during the broad survey phase, performing about 280 visits to detect and determine the orbits of exoplanets. Over five years, this results in lower total EEC yield for the starshade-only mission, while the number of exo-Earth candidates (EECs) spectrally characterized is about the same as compared to the hybrid mission. Conversely, HabEx's Observatory Science program benefits from the removal of the coronagraph. The additional time spent waiting for the starshade to transit and retarget is now reserved for additional observations with the observatory science instruments.

The EEC yield for a starshade-only mission could go up by as much as a factor of 2 if the starshade is refueled, or if a second starshade is provided. Starshade-only missions would also benefit from precursor RV observations sensitive enough to identify high-value targets and/or measure exoplanet orbits.

Notably, as already discussed in *Section 10.3*, these starshade-only architectures do not require the SLS for launch owing to the reduced launch mass resulting from the design choice of a structurally efficient segmented aperture and on-axis telescope. As also discussed in *Section 10.3*, the number of required TRL 4 technologies is less for starshade-only architectures. This is primarily due to the removal of the coronagraph. However, these starshade-only architectures introduce one new TRL 5 technology, the active ultra-low expansion (ULE) mirror segment.

In no HabEx architectures are there any required TRL 3 technologies.

## B.1  HabEx 3.2S Architecture

The HabEx 3.2S mission concept meets all but one of the HabEx science baseline objectives at the threshold level, achieving some at the goal level. An assessment of science objectives that can be met with the HabEx 3.2S is shown in **Table 10.4-1**. The cost reduction is the result of the removal of the coronagraph and reduction of aperture size. By removing the coronagraph, the observatory's wavefront stability requirements are relaxed from <100 pm to 10 nm, permitting the use of a lower-mass segmented primary mirror. The primary limitation incurred by removing the coronagraph is dependence on the starshade occulter for all exoplanet direct detection and characterization. In terms of the concept of operations, observations are limited by the starshade transits between targets. For exoplanet science goals 1 and 2, the HabEx 3.2S telescope payload architecture requires a different observational strategy from the HabEx 4H to account for the lack of a coronagraph. Namely, the coronagraphic broad survey from HabEx 4H is replaced by an augmented starshade broad survey. However, the starshade remains propellant constrained, reducing the breadth of the broad survey. This is reflected in the HabEx 3.2S ability to meet HabEx science objectives as summarized in **Table 10.4-1**. Nonetheless, all HabEx 3.2S science instruments can observe simultaneously and are capable of performing the Parallel Observation Program described in *Section 4.10.9*.

Some aspects of the HabEx 3.2S architecture duplicate design choices made for the HabEx 4H mission concept, choices that enable the low disturbance environment needed to accomplish sensitive measurements. First, as summarized in **Figure B.1-1**, both telescope and starshade flight systems will launch separately and operate at Earth-Sun L2, preserving a stable thermal environment and greater field of regard than available in Earth orbit. The scarfed telescope barrel acts as a sun shade, permitting Guest Observer (GO) science observations in directions ranging from anti-Sun to within 40° of the Sun, while keeping sunlight out of the barrel interior. Starshade observations are limited to 40–83° sun angle to keep the Sun away from the obscuring disc face. Both architectures also use low-disturbance colloidal microthrusters instead of reaction wheels or control-moment gyros, for ultra-low image jitter. For payload thermal control, both telescopes use similar passive architectures, avoiding any risk of disturbance and thermal design complexity from active thermal control with cryocoolers. Both telescope power system designs use fixed solar arrays, for rigidity and simplicity.

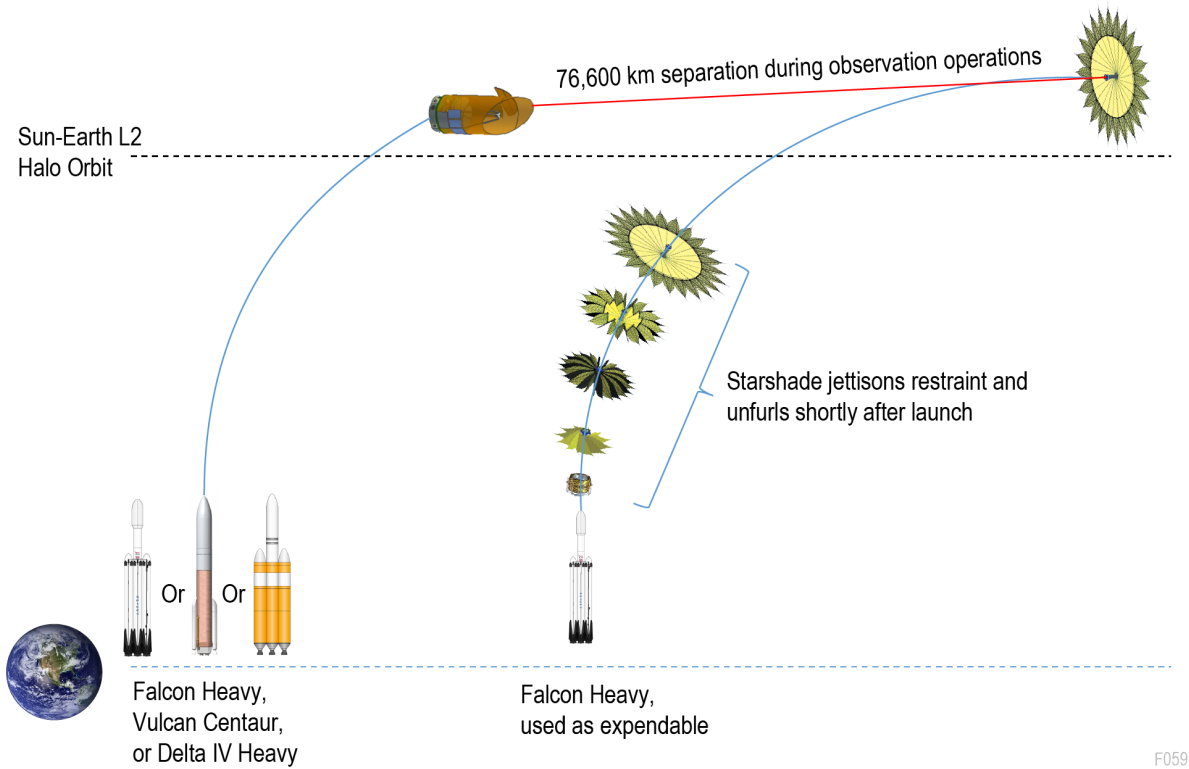

**Figure B.1-1.** The HabEx starshade-only architectures maintain are similar to the hybrid architectures insofar as they baseline two separate launches. However, lower telescope flight system mass for the point design of HabEx 3.2S permits use of a Falcon 9 or similar class launch vehicle in place of an SLS, reducing mission cost.

Serviceability is also a consideration for HabEx 3.2S and is described in *Section 8.3*. The instruments are designed to be accessed by a manipulator approaching from the side of the spacecraft, so that key components can be replaced. The telescope bus propulsion system can be refueled. Servicing of the starshade is identical across both architectures.

### B.1.1 HabEx 3.2S Telescope Payload

This section describes the driving requirements and design of the HabEx 3.2S payload (**Figure B.1-2**). The HabEx 3.2S payload consists of the optical telescope assembly (OTA) and three scientific optical instruments: the Starshade Instrument (SSI), the Ultraviolet Spectrograph

(UVS; Mooney et al. 2018b), and the HabEx Workhorse Camera (HWC), as shown in **Figure B.1-3**. The payload also includes two fine guidance system cameras (FGS A and B), which are used to measure the telescope line of sight for attitude control purposes. These instruments are similar in concept to the HabEx 4H instruments, including all features and modes, while their layout has been adapted for the architecture's different optical design.

#### B.1.1.1 Optical Telescope Assembly

As illustrated in **Figure B.1-3**, the HabEx 3.2S telescope uses an actively controlled on-axis telescope configuration. The secondary mirror (SM) and SM supports are within the beam footprint so that the width of the optics is defined by the primary mirror (PM) aperture. HabEx 3.2S's on-axis design is capable of meeting any of the non-coronagraph HabEx

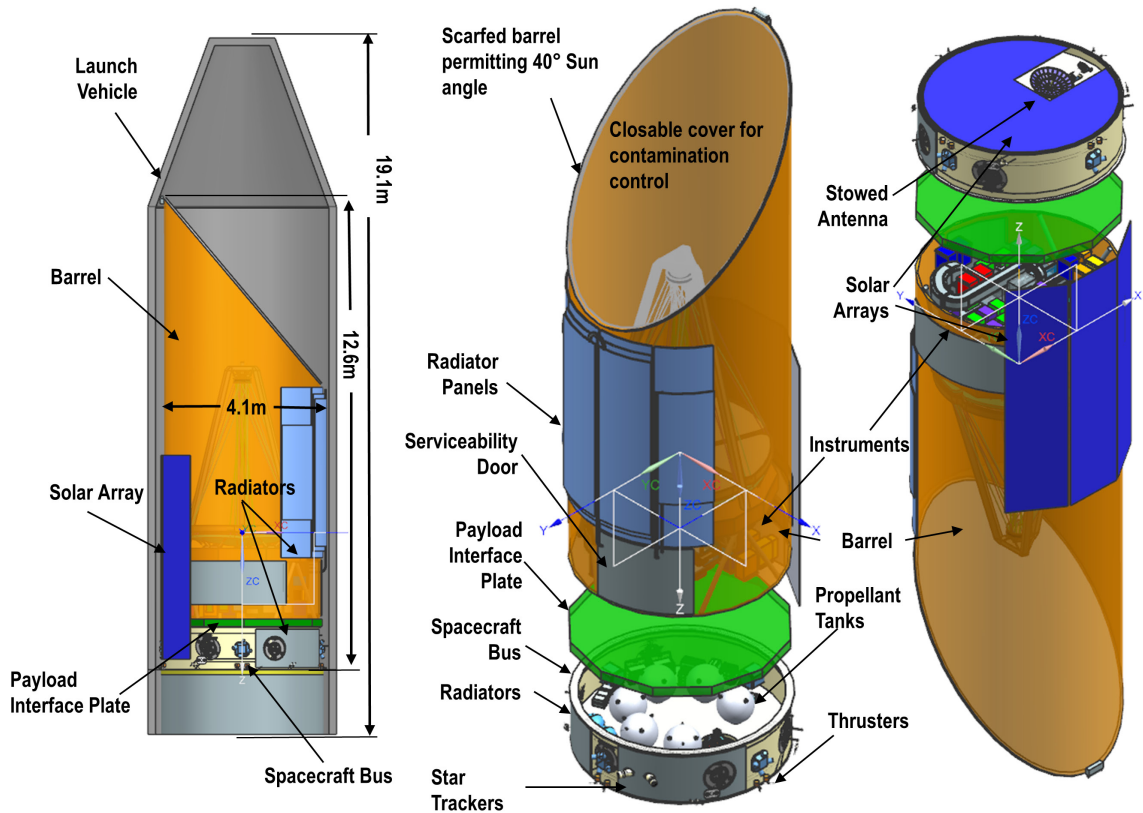

**Figure B.1-2.** *Left:* Overall view of the HabEx 3.2S Observatory in the Vulcan Centaur shroud. *Right:* Exploded diagrams show the payload and bus assemblies. The payload includes the telescope, optical instruments, and supporting structures, thermal and electronics subsystems. The bus provides the flight systems, including avionics, power, communications and propulsion. The Payload Interface Plate (PIP) provides a modular interface, allowing the payload and bus to be assembled separately, then brought together before launch.

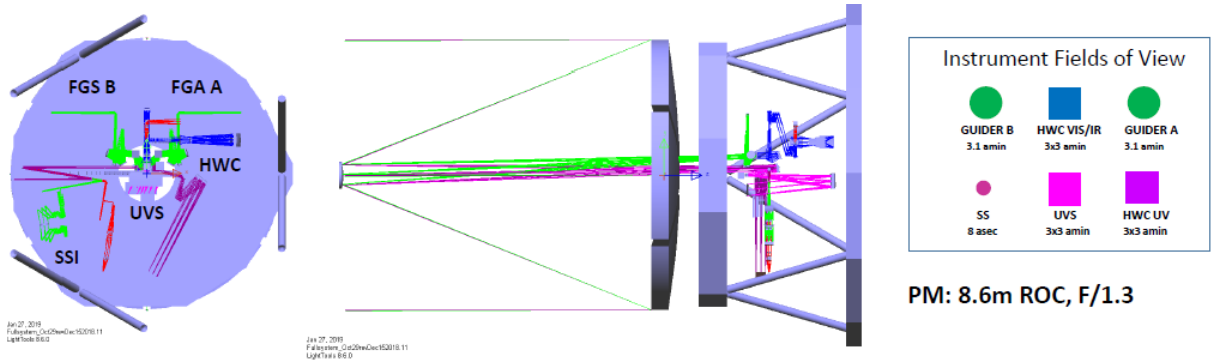

**Figure B.1-3.** Optical layout, showing the optical Telescope Assembly (OTA), the three science instruments, the two guiders, and the instrument fields of view. The total telescope field of view is about 0.26×0.14 degrees (16×8 arcmin).

requirements despite obscuration from the SM and SM supports.

The basic configuration of the HabEx 3.2S OTA and instruments is sketched in **Figure B.1-3**. Light entering the telescope is reflected from its 3.34 m aperture diameter, *f*/1.3 PM to the SM, placed 4 m above the PM. From the SM it is reflected through a gap in the middle of the PM into the instrument bay. There it encounters six separate tertiary mirrors (TMs), defining six separate instrument fields of view: four science fields and two FGS fields. The TMs collimate the light entering each instrument, so that the telescope optical configuration is afocal, with 1/150×-class beam magnification. The TMs are contained in a central core optics structure, keeping them rigidly aligned to the aft metering structure to provide an alignment reference for the rest of the optics. Fold mirrors direct the light from each field into its respective instrument, except for the UVS. The afocal beam interfaces make the instruments more robust to misalignment to the core optics, which is important since most are designed to be removable and replaceable for on-orbit servicing.

### Active Optics and the Segmented PM

HabEx 3.2S uses active optics, with mirrors that can be controlled, to ensure that its optical performance goals are met, even if there are unanticipated problems on orbit—such as occurred on the Hubble Space Telescope (Allen et al. 1990; Redding et al. 1995). Key elements in the active control design are: a controlled, segmented PM, controlled PM, wavefront sensing and control (WFSC), and laser truss metrology (MET). The PM segments are equipped with rigid body actuators to align and phase the six separate mirrors into one coherent whole, and figure control actuators, to ensure that each segment has the correct optical figure. The SM will have rigid body actuators to allow the telescope to be focused and collimated after its rough ride to orbit. Star images and spectra from the science instruments will be computer processed on the ground to measure the WFE across the fields of view (FOVs) of each camera. Then PM and SM actuators will be set to optimize optical quality across all the instruments. WFSC is primarily used to initialize the telescope after launch, correcting the expected large initial errors. Then occasional recalibration observations will be done to monitor and correct long-term drifts. To ensure that the telescope remains precisely in the optimal configuration as determined by WFSC, MET continuously measures the position of each major optical assembly, closing the loop with the rigid body actuators to keep the telescope aligned.

The HabEx 3.2S segmented PM can be built within the current space-qualified mirror state-of-practice. Its segments, at 1.4 m size, are within the 2.4 m capability proven by previous missions. The PM segments have a first fundamental vibration mode of 300 Hz. Segment control increases in operational complexity as WFSC is needed to phase the mirrors after launch. Nonetheless, WFSC and MET give HabEx resilience to the sorts of problems that have limited other missions in the past: optical testing errors; unexpected environmental conditions; gravity release prediction errors; changes induced by launch loads; misalignments between instruments; and mirror fabrication errors. All these can be corrected on orbit by the HabEx 3.2S WFSC system, reducing mission performance risk, and relaxing fabrication and alignment tolerances.

*Wavefront Error Budget*

The top-level optical performance requirement for HabEx 3.2S is that it be diffraction limited at a wavelength of 0.40 μm in the UV and visible bands. This corresponds to a requirement that the total system Wavefront Error (WFE) not exceed 30 nm in the UV and visible instruments. WFE is the deviation of the optical phase from the ideal: a perfect spherical wavefront converging to a point at the final focus of each instrument. WFE is caused by factors such as: instrument and telescope design compromises; as-built mirror figure and alignment errors; drifting mirror figure and alignment induced by thermal changes or material desorption; and for systems like HabEx, with actively controlled mirrors, wavefront sensing and control errors. The HabEx 3.2S WFE budget of **Figure B.1-4** flows down the top-level 30 nm requirement to the key observatory subsystems, setting measurable performance requirements for each subsystem, and a margin (RSS) of 14 nm.

The 30 nm total WFE requirement is the same as specified for the coronagraph-equipped HabEx H and C architectures, but the allocation of the errors to the various subsystems differs. In particular, the starshade-only HabEx architectures do not need picometer-level wavefront stability. The HabEx 3.2S WFE budget therefore allocates a more relaxed 10 nm WFE to drift terms, such as thermally driven deformations and misalignments. This is to be compared to the HabEx H and C requirements: ≤1 nm RMS for wavefront drift effects slower than 1 mHz; and ≤5 pm RMS for wavefront effects that occur faster than 1 mHz.

**Figure B.1-4** is a post-WFSC budget, where the allocated levels of WFE include the compensating effects of the WFSC system, which measures the system WFE and corrects it using PM and SM actuators. Residual errors, due to imperfections in the WFSC control process, are bookkept in the WFSC errors column, which totals 11 nm. Similarly, the drift WFE includes the compensating effects of the Laser Truss Metrology system, which makes continuous nm-precision measurements of the optics, and feeds back low bandwidth corrections to the WF Control actuators, to preserve the alignments established by the WFSC system.

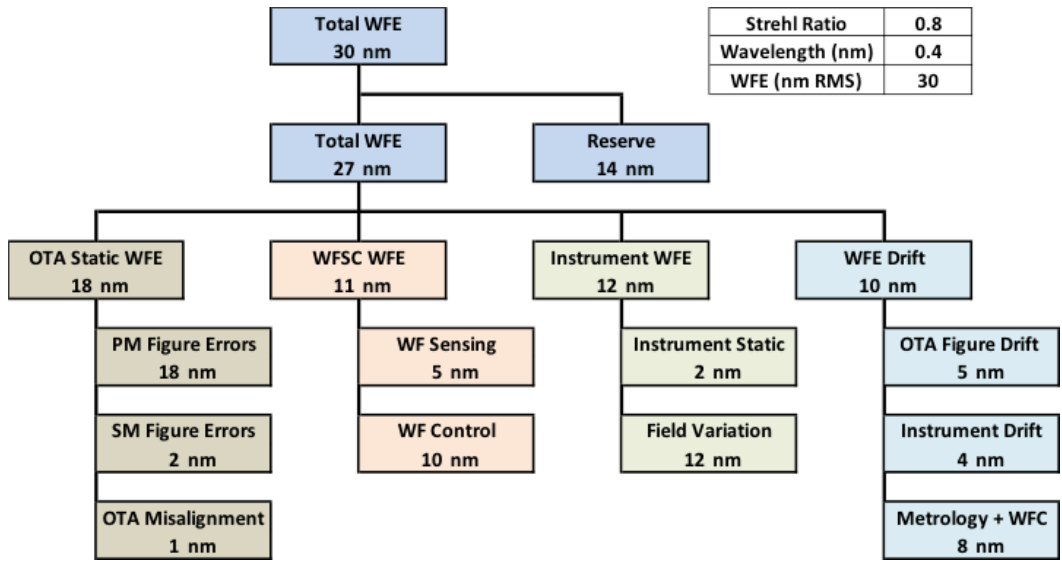

**Figure B.1-4.** Wavefront error (WFE) budget, showing allocations to the major subsystems of the observatory, after Wavefront Sensing and Control (WFSC). The 30 nm total wavefront error is the same as that specified for HabEx 4H, but the sub-allocations differ. HabEx 3.2S has a much-relaxed WFE drift requirement, since it does not have a coronagraph.

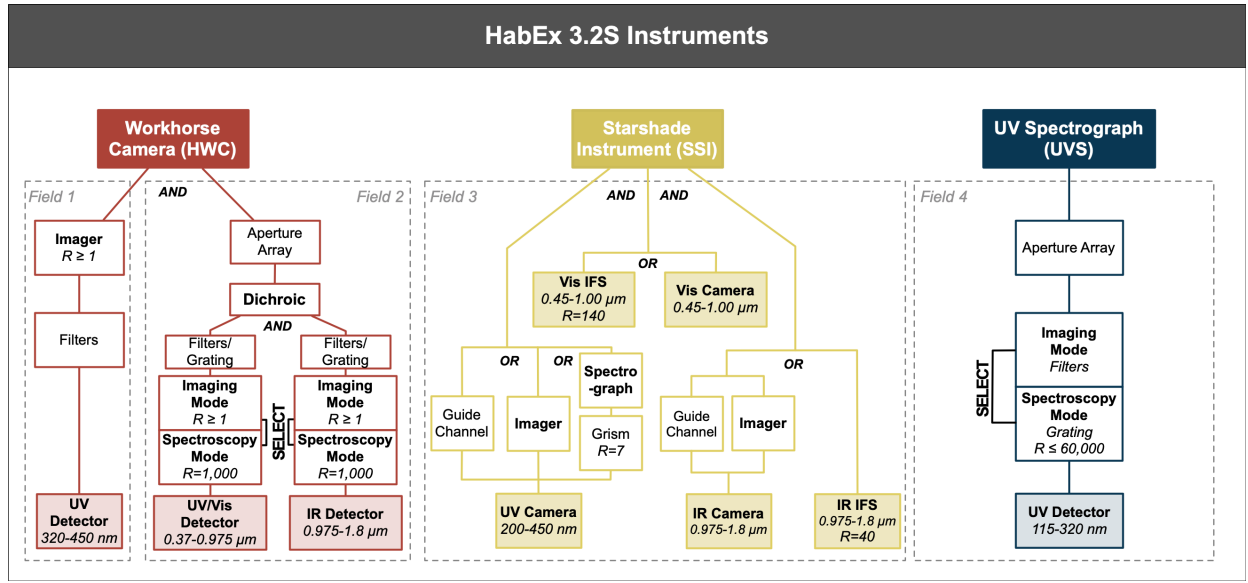

**Figure B.1-5.** HabEx 3.2S payload block diagram, showing the scientific optical instruments, including the various channels and modes. To compare with the HabEx 4H payload, see Figure 6.2-5.

## Optical Telescope Assembly

For the HabEx 4H architecture, the optical design is primarily driven by the WFE requirement associated with coronagraph observations. Without the coronagraph, the HabEx 3.2S optical design is motivated by engineering consideration. Specifically, telescope optics were designed together with the UVS, whose sensitivity depends on having a minimum number of optical surfaces and requiring spot sizes under 30 μm at the telescope Cassegrain focus, where a Micro Shutter Array (MSA) is placed, and at the focal plane itself, for the full 3 arcmin square field.

Requirements can be met with a 2-mirror telescope fore-optics configuration, achieving the 30 μm spot size requirement at the MSA in the UVS channel. Then, in the UVS, a tertiary mirror collimates the beam and sends it to the UVS gratings. Since the beam is collimated, the gratings can all be straight-ruled on flat substrates, which are easy to fabricate. A final focusing mirror provides the needed 30 μm spot size image quality across the full 3 arcmin field at the UVS focal plane. The light path through the OTA and UVS has only 5 reflections total, the minimum number capable of meeting all UVS requirements.

The same foreoptics provide equal optical quality to the other instruments, in their respective fields of view, shown in **Figure B.1-5**. As is described below, each instrument has its own TM, all of which are rigidly aligned to the OTA core optics assembly. This provides an afocal interface to each instrument, significantly easing the alignment of the instrument optics to the OTA. Other key requirements for the OTA are listed in **Table B.1-1**.

Requirements for the telescope structures are to survive launch, to provide high stiffness during imaging operations, to be thermally very stable, and not to introduce any microdynamics: small pops or lurches that can occur at joints or latches during changes in strain. The structure is engineered together with the thermal system to meet these requirements through five design decisions. First, to use high stiffness, low-expansion composite materials. Second, the scarfed outer barrel structure shades optical structures from direct sunlight while providing support for solar cells and radiators. Third, modular payload-to-spacecraft bus interfaces using a shared payload interface plate hold the barrel sunshade and various payload electronics assemblies away from the critical optical structures. Fourth, the aft metering structure supports the OTA optics and the instrument assemblies. The aft metering structure is isolated using flexured struts, to prevent barrel or bus strains from distorting the optical train. Fifth, there are no deployed structures in the optical load path. Only the outer barrel cover is deployable, so that it can be closed to avoid contamination during servicing.

**Table B.1-1.** OTA requirements. Note that coronagraph-based science measurement requirements requiring a ≥3.7 m, unobscured monolithic, f/2.50 PM are not applicable to a starshade-only mission. The 3.34 m PM size (effective area of a 3.2 m PM) is driven by architectural choice rather than design to meet objective requirements. For comparison to HabEx 4H, please see Table 6.2-1.

| Parameter | Requirement | Expected Performance | Margin |
|---|---|---|---|
| Angular Resolution | 50 mas | 30 mas | 67% |
| Bandpass | ≤0.115 μm to ≥1.70 μm | 0.115–2.50 μm | Met by design |

**Figure B.1-6** outlines the OTA structure and its components. The structural core of the telescope is the aft metering structure, a thick, stiff composite plate that is thermally controlled to remain extremely stable. The aft metering structure top side directly supports the Secondary support structure and the PM segment assemblies. Its underside supports the core optics

assembly, which hold the fold mirrors and TMs for the instruments, and the struts and rails that support the instruments themselves.

The aft metering structure is itself connected to the payload interface plate by long struts with a flexured interface, so that the aft metering structure does not change shape if the payload interface plate bends or deforms. The payload interface plate provides the mechanical interface to the bus, as it forms the top plate of the bus structure when attached in the final spacecraft integration operations. The payload interface plate serves as the support plate for non-instrument payload electronics. Most instrument electronics boxes are attached to their respective instrument enclosures, so that they can be removed and replaced along with the instruments during on-orbit servicing. The payload interface plate also supports the outer barrel assembly, the scarfed 4.3 m wide cylindrical tube that keeps sunlight from illuminating any optically sensitive structures or components.

The outer barrel assembly will be a large, stiff, honeycomb composite structure. It serves as a sunshield and light baffle, and as support for the

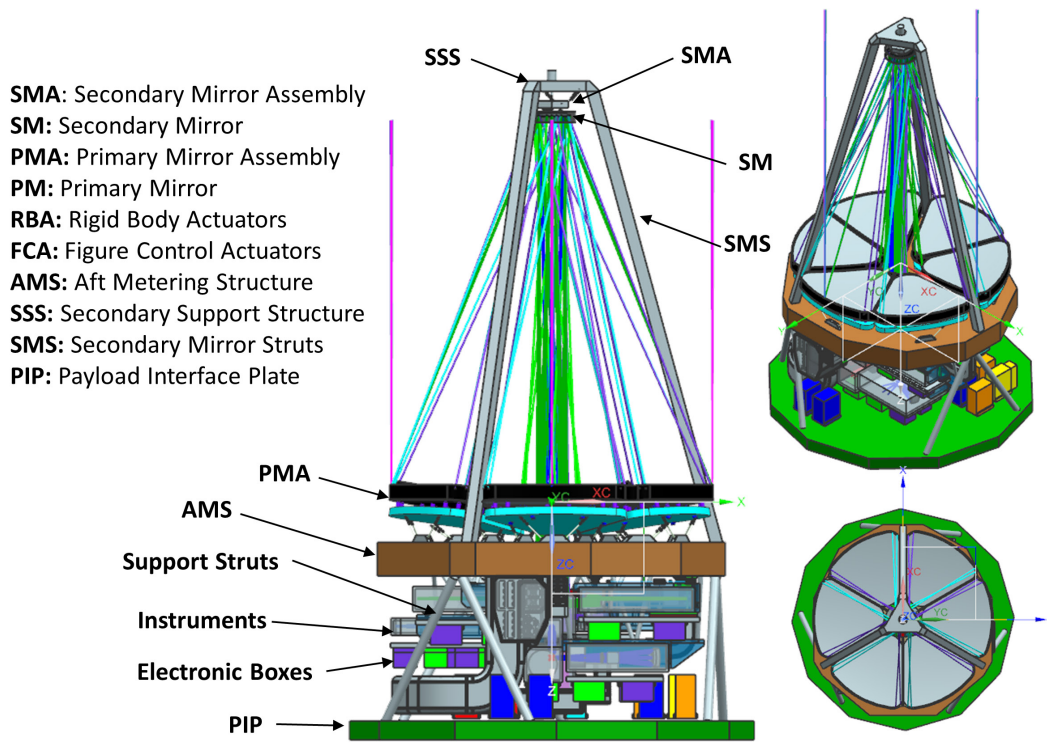

**SMA**: Secondary Mirror Assembly
**SM**: Secondary Mirror
**PMA**: Primary Mirror Assembly
**PM**: Primary Mirror
**RBA**: Rigid Body Actuators
**FCA**: Figure Control Actuators
**AMS**: Aft Metering Structure
**SSS**: Secondary Support Structure
**SMS**: Secondary Mirror Struts
**PIP**: Payload Interface Plate

**Figure B.1-6.** OTA structures, shown with the instruments, electronics, and thermal hardware in place.

radiator and solar array assemblies, as described below. It is also a contamination protection shroud, equipped with a door which is closed during integration and launch. The door is opened early in on-orbit checkout operations, to allow volatiles to escape the OTA. It remains open during operations, but can be reclosed for servicing, to prevent propulsion byproducts from the servicing vehicle from contaminating the optics.

*Primary Mirror*

The HabEx 3.2S PM segment subsystems include a number of components. As sketched in **Figure B.1-7**, each PM segment assembly consists of an ultra low expansion (ULE) glass substrate mounted to an intermediate reaction structure by flexured struts. Also connecting the reaction body and mirror substrate are 23 figure control actuators, which are used to adjust the mirror optical figure in the WFSC process. The figure control actuators use force feedback methods to ensure forces on the substrate are constant, and that any thermal deformations of the reaction body are not transmitted to the substrate. The reaction structure-mirror substrate assembly is mounted to the payload baseplate so that each segment can be commanded with a six degree-of-freedom motion envelope. When installed, adjacent pairs of segments will be surrounded by a thermal enclosure, with heaters keeping the assembly temperatures stable at the required operating temperatures. Laser MET beam launchers, used by the WFSC system to continuously measure the optical alignments, are attached to the substrate. The glass substrate consists of a lightweight ULE core, sandwiched between a front plate and a back plate. Mass areal density of the substrate is designed to be 20 kg/m², with the first bending mode occurring at a frequency of 300 Hz.

PM segment figure error is the largest contributor to the WFE budget of **Figure B.1-4**, at 18 nm WFE rms, even after the effect of WFSC. Note that 18 nm WFE corresponds to 9 nm surface figure error, since these errors are doubled upon reflection. The segment WFE budget is presented in **Figure B.1-7**. The expected WFE performance of the HabEx 3.2S segment design, assuming current fabrication processes and figure control actuator correction, is 16 nm rms, meeting the 18 nm rms allocation with 8 nm margin. The residual errors reflect current fabrication processes and experience, and are dominated by gravity removal errors, radius of curvature matching errors, coating stresses, temperature change effects, and mounting effects.

*PM Segment Thermal Control*

Each pair of PM segment assemblies is surrounded by a thermal enclosure, or can, as shown in **Figure B.1-8**. The thermal can is a thin, very lightweight composite structure, surrounding the sides and back of the segment substrates. It is equipped with heaters and temperature sensors, and covered by multi-layer insulation (MLI), to form a tightly-controlled thermal environment. Closed-loop thermal control maintains the nominal 270 K mirror temperature to ±1 mK. Each thermal can includes a rim strip, used to mask the edges of the segments, so that the uncoated, edges of the glass substrate are not directly exposed to cold space. The rim strip also serves as an optical mask for the segment edges, extending 5 mm over the reflective surface. This small masked part of the segments can therefore have slightly relaxed figure specifications. The heater power required to maintain segments and SM at 270 K is 520 W.

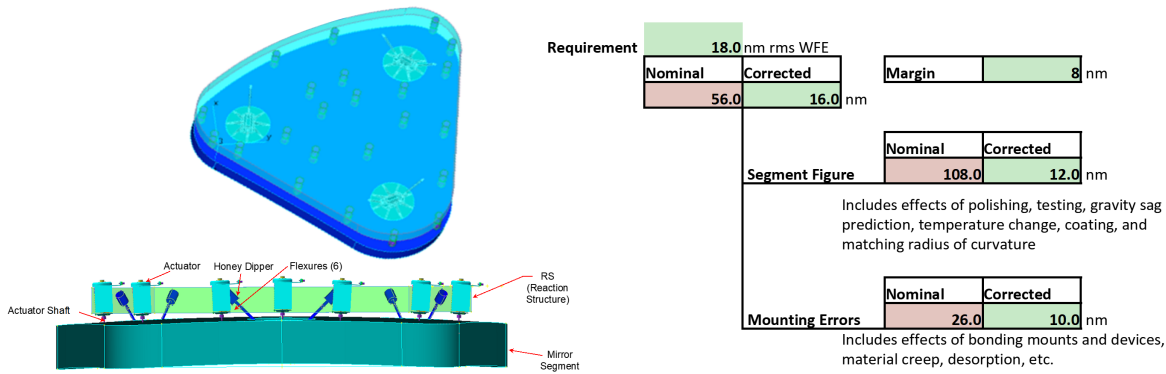

**Figure B.1-7**. *Left:* HabEx 3.2S primary mirror segment design, showing the glass mirror substrate attached to its reaction body support structure via struts and figure control actuators. The reaction body is in turn attached to Aft Metering Structure (AMS) by rigid body actuators. *Right:* Mirror surface figure error budget, showing both nominal (uncorrected by WFSC), and WFSC-corrected performance, which sums to 16 nm rms WFE, meeting the 18 nm allocation of Figure B.1-4 with 8 nm margin.

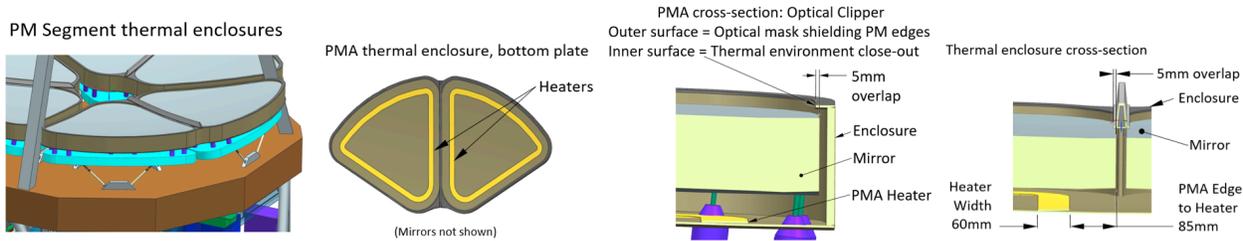

**Figure B.1-8.** Primary mirror segment thermal "can" enclosures provide the stable thermal environment needed to preserve segment figure during observations. Maximum heater power required is 520 W.

### *Laser Truss Metrology (MET)*

The MET system is a network of 42 laser distance gauges. As illustrated in **Figure B.1-9**, each distance gauge consists of a beam launcher at one end of the measured distance, and a corner cube at the other (Mooney et al. 2015; Zhao 2004). A collimated probe beam from each laser gauge beam launcher propagates through free space and is reflected back from the corner cube, which may be many meters away. The beam is reflected back to the beam launcher where it couples into fiber optics. There it is mixed with a reference beam that is reflected within the beam launcher, to generate interference fringes. Phasemeter electronics using heterodyne detection methods measure the phase of the probe beam relative to the reference beam—thus measuring the distance between the beam launcher and the corner cube. The light source for all of the beams comes from a single redundant laser, feeding a fiber distribution network. The laser wavelength is about 1.5 μm, and the transmitted beam power is 1 μW per gauge. The laser is stabilized to a molecular line. The phasemeter electronics are derived from the flight-proven, picometer accuracy laser gauge systems demonstrated on orbit by LISA Pathfinder and GRACE Follow-On.

The laser gauges are configured in an optical truss, with six beam launchers on each segment illuminating three corner cubes on the SM, and six more beams between the instrument bench and the SM, as shown on **Figure B.1-9**. This configuration allows measurement of all motions between the PM, SM and back end assembly, with a precision of 1 nm or better. The measurements are used to stabilize the alignments in a closed loop, with any alignment drift removed using the PM and SM rigid body actuators. PM and SM

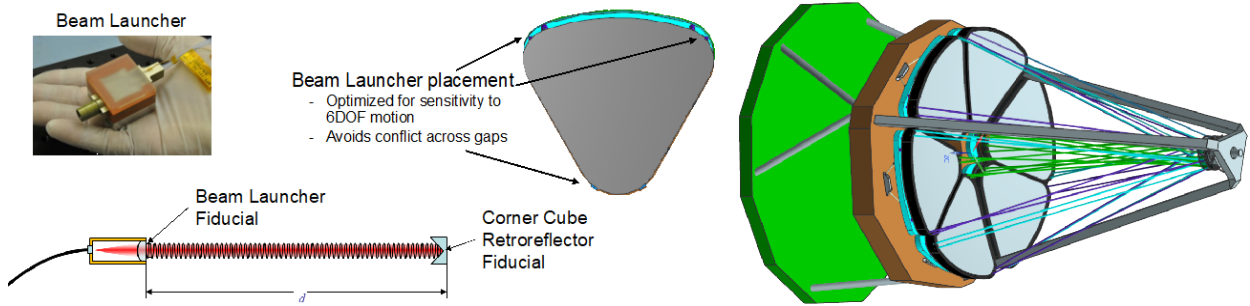

**Figure B.1-9.** Laser truss metrology (MET) components and configuration. Fiber-fed beam launchers attached to the PM segments (and to the back-end instrument core optics assembly) illuminate corner cubes mounted to the SM, measuring all optical alignments continuously to a precision of <1 nm. Alignment drift is measured and controlled using the RBAs, to stabilize the observatory beam train, and to precisely implement commanded PM or SM motions.

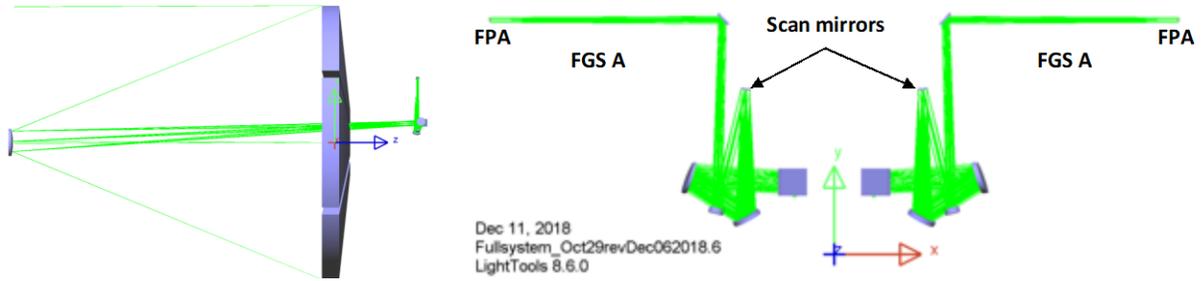

**Figure B.1-10.** Fine Guidance Sensors A and B, 2 separate instruments equipped with scan mirrors, each covering a circular field of regard of 3.1 arcmin diameter, with a detector field of view of 17.1 arcsec. The fields are separated by 13 arcmin.

motion commands are also executed in closed loop, to ensure precise and stable control.

### Telescope Fine Guidance Sensors

Without the coronagraph driving the stability requirement, HabEx 3.2S has two identical fine guidance sensors (FGS) as opposed to HabEx 4H's four FGSs. With the two FGSs, shown in **Figure B.1-10**, there is ≥95% probability that two or more guide stars will be in the field of view.

The HabEx 3.2S FGSs are otherwise similar to those of HabEx 4H. They are $f/48$ imagers observing in the visible band, 0.5–0.9 μm. Its requirements and performance are identified in **Table B.1-2**. Each FGS views a total field of 7.5 arcmin² with a 3.1 arcmin diameter. The FGS FPA subtends only 0.285 arcmin in far field, so the FGS optics includes a scan mirror in collimated space that sweeps the focal plane FOV across the full field of regard.

The optical design of each guider channel, other than the common telescope primary and secondary mirrors, includes a conic tertiary that creates a collimated beam, demagnified by 115× from the primary mirror. At the exit pupil following the tertiary is located the scan mirror, which tilts in a range of ±1.3° to direct the beam into the final focuser. In the beam space after the scan mirror, the field angle range is only 0.79° total, so a simple off-axis paraboloid is sufficient to form the $f/48$ focus, with RMS WFE of 42 nm or less across the field. FGS requirements are summarized in **Table B.1-2**.

### Telescope Operations

The WFSC system assures that the telescope optics meet WFE requirements. It uses star images and spectra to measure the large initial errors in the system, and then commands actuators to correct those errors. During operations, it uses laser distance gauges to continuously measure the position of each optic relative to its commanded position, and corrects any drifts using the rigid body actuators. WFSC compensates errors everywhere in the beam train, establishing and preserving the WFE performance specified in the WFE budget of **Figure B.1-4**.

The WFSC process begins after launch, when the PM segments and the SM are released from their launch supports, with initial misalignments that can be a mm or more. The first step is to turn on the MET system, moving the SM and PM segments as needed to peak up the laser power in each beam launcher. This establishes closed-loop servo control for all subsequent operations.

Telescope initialization then commences, with the telescope pointed to a bright, isolated star, and imaged in one of the science cameras. The star image on the focal plane will be broken up into blobs, or subimages—one for each segment—as illustrated in **Figure B.1-11**. Coarse alignment algorithms command motions of each segment to identify which blob corresponds to which segment. Then the segments are aligned and brought to a common focus—but not yet a common phase. WFE after coarse alignment operations will be ~10 μm, almost entirely in segment piston mode.

**Table B.1-2.** HabEx 3.2S FGS camera requirements and compliance.

| Parameter | Requirement | Performance | Margin |
|---|---|---|---|
| Waveband, Imaging | 0.50–0.90 μm | 0.50–0.90 μm | Met by design |
| Field of Regard | ≥2.5 × 2.5 arcmin | 3.1 arcmin circular | 0.6 arcmin |
| Field of View | ≥15 arcsec | 17.1 × 17.1 arcsec | 2.1 arcsec |
| Centroiding Accuracy | <2.5 mas | 1.7 mas (1/10 pixel) | 0.8 mas |

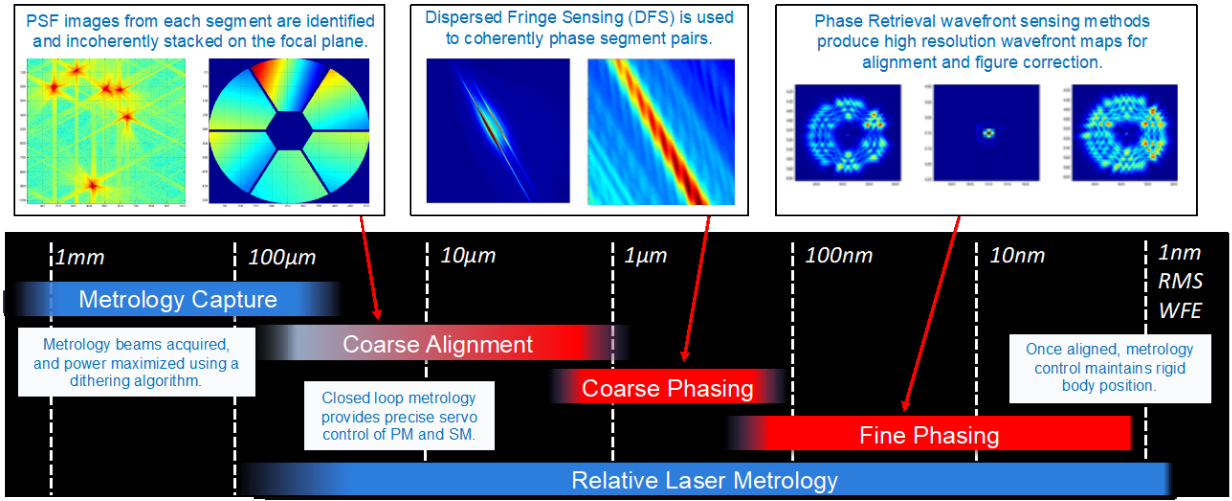

**Figure B.1-11.** WFSC Initialization process, run at the beginning of the mission to establish the needed high optical quality, and provide alignment set points for the MET maintenance system.

Coarse phasing operations use spectra from a science camera to modulate the fixed phase differences between pairs of segments, for Dispersed Fringe Sensing. Piston differences between segments generate distinctive fringe patterns allowing easy absolute piston measurement. These piston errors are then removed using the PM segment rigid body actuators.

The final step in telescope initialization—fine phasing—uses defocused star images processed on the ground to generate high-resolution, high-accuracy WFE maps. These in turn are used to set both the rigid body and fine control actuators to match the desired final wavefront target. Fine Phasing uses images from all of the cameras, taken in 3–5 field points, to determine the global best setting for all actuators, to collimate the telescope and meet requirements in all cameras. At the conclusion of Fine Phasing the telescope will be aligned, with WFE consistent with the WFE budget. This condition will be maintained by continuous operation of the MET system. Periodic wavefront calibration fine phasing in a single field of a single camera will be performed to ensure that no drifts have occurred, with a frequency of once per week, to once per month. **Figure B.1-12** shows an overall block diagram of the WFSC system.

### B.1.1.2 Starshade Occulter and Starshade Instrument (SSI)

The HabEx 3.2S starshade occulter and the Starshade Instrument (SSI) are used together to provide high dynamic range observations, of exoplanets around other stars and potentially other extended scenes surrounding bright central objects, such as disks around QSOs. The HabEx 3.2S starshade occulter is almost identical in design to the Starshade described in *Chapter 7*. It is 52 m in diameter, with 16 m long petals and a 20 m hub, but the edges of the petals are shaped to be optimized for the smaller 3.34 m aperture while providing the same <10⁻¹⁰ contrast ratio and nearly the same inner working angle (IWA) for the design bandwidth of 0.30–1.00 μm. The left panel of **Figure B.1-13** is a simulated SSI image and shows the expected contrast in the SSI focal plane, for the starshade placed at its nominal distance, at the two extremes of its bandpass. Note that the contrast performance does not suffer from the segmented HabEx 3.2S aperture.

To observe an exoplanetary system, the 52 m starshade will be maneuvered into position directly on the line of sight between the telescope and the target star, at a distance of about 76,600 km from the telescope. This will create a conical shadow, where the shadow is about 1 m wider than the telescope on all sides of the telescope. The shadow keeps starlight out of the telescope aperture, providing the ≤10$^{-10}$ contrast over an IWA$_{0.5}$ of 58 mas needed for exoplanet imaging and spectroscopy in the 0.30–1.00 μm waveband. Yet dim features of the shadow can be seen at wavelengths outside the science band using one of the SSI UV or IR channels in starshade guider mode, with a pupil imaging lens inserted. This provides measurements of the position of the shadow relative to the stellar line of sight, which the telescope communicates to the starshade, to help it keep the shadow on the telescope, while the other channels are observing the scene. The right panel of **Figure B.1-13** also provides an illustration of HabEx 3.2S imaging performance, with a simulated 24-hour observation of a representative solar system as seen from a distance of 7.2 pc.

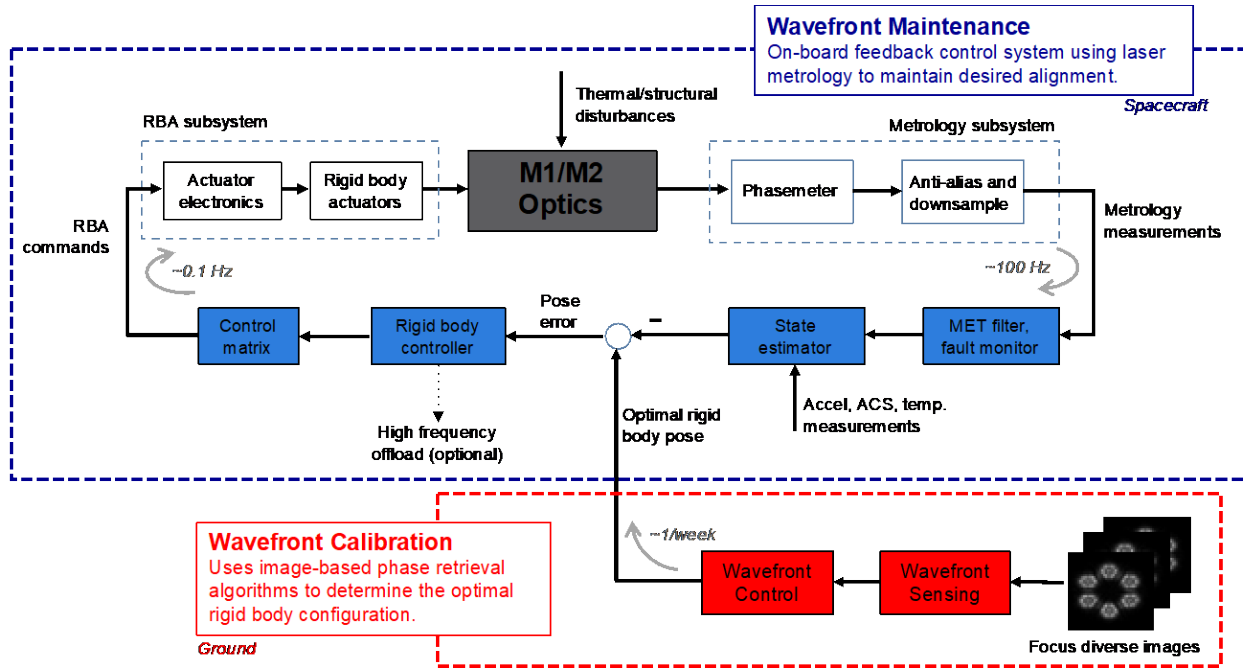

**Figure B.1-12.** WFSC wavefront maintenance on-board closed-loop control, using MET to preserve the alignments ("pose") that maximize optical quality, as determined by WF sensing in the wavefront calibration/fine phasing observations.

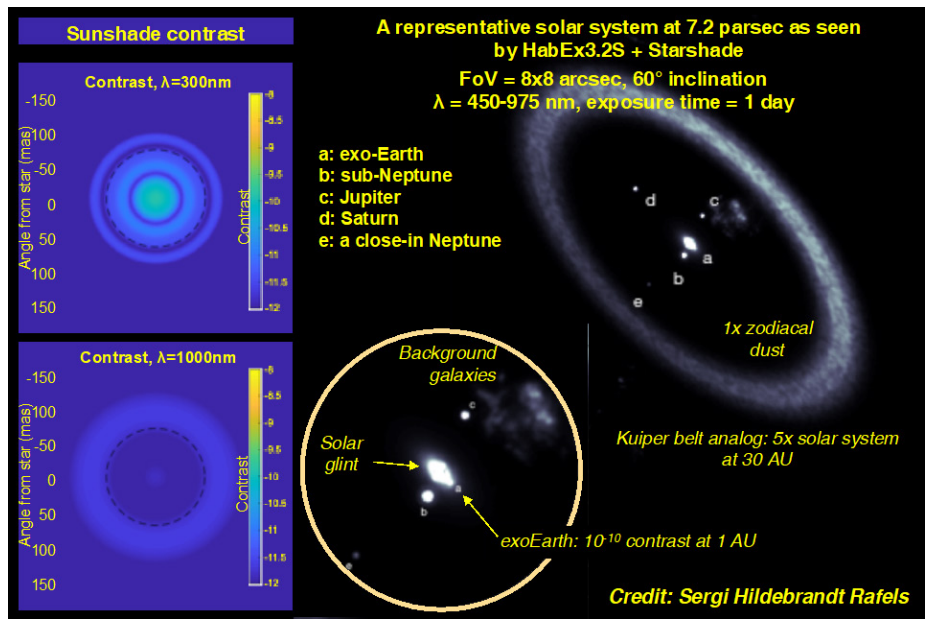

**Figure B.1-13.** *Left:* SSI contrast performance at 0.3 μm and 1.0 μm, when placed at a distance of 76,600 km from the telescope. *Right:* Simulated image of a representative system at a distance of 7.2 pc, as seen in a 24 hour exposure by HabEx 3.2S. This broadband image clearly shows an inner exo-Earth planet, plus others extending to 40 AU. It also shows the impact of sunlight glinting off of the edges of the sunshade petals. Other features include the exozodi, a Kuiper belt analog and background galaxies.

**Table B.1-3.** HabEx 3.2S starshade parameters. For comparison with HabEx 4H, please see Table 6.4-2.

| Diameter | 52 m | Baseline Mode | | UV Mode, small IWA | | IR Mode, large IWA | |
|---|---|---|---|---|---|---|---|
| Petal Length | 16 m | Bandpass | 0.30–1.00 μm | Bandpass | 0.20–0.667 μm | Bandpass | 0.54–1.80 μm |
| Minimum Fresnel # | 8.8 | $IWA_{0.5}$ | 58 mas | $IWA_{0.5}$ | 39 mas | $IWA_{0.5}$ | 104 mas |
| | | Separation Distance | 76,600 km | Separation Distance | 114,900 km | Separation Distance | 42,600 km |

As in the HabEx 4H architecture, different bands will be observable at different Telescope-Occulter separation distances, resulting in different IWAs, as summarized in **Table B.1-3**.

### B.1.1.3 Design

The HabEx 3.2S SSI field of view was chosen to be 8 arcsecond, meeting Science Objective 5's threshold requirement for OWA. The SSI optical layout is shown in **Figure B.1-14**, with its requirements and performance defined in Key in **Table B.1-4** while its specifications are defined in **Table B.1-5**. The separate UV, visible, and near-IR channels share the same 8 arcsec diameter field, so that they all can observe the same objects at the same time. Each channel has imaging and spectroscopic (within a narrower, 1.3–2.6 arcsec field) modes. Additionally, the UV and IR channels have starshade guiding modes. A typical operational configuration might utilize the UV channel in guider mode, with the pupil imaging lens inserted in the beam to create an image of the shadow on the UV detector for starshade guiding, while the IR and visible channels perform imaging or spectroscopy.

The optical design of the SSI channel includes the telescope PM and SM, adding a conic tertiary to create a collimated beam, demagnified by 150× from the PM. The 8 arcsec field on the sky translates to a field angle range in the small-beam space of only 0.52°, which allows single-mirror reimaging mirrors rather than the more complex 3-mirror relays used in wider-field instruments, keeping UV sensitivity high while still forming a high quality $f/80$ converging beam, with RMS WFE of 9 nm across the field. This $f/80$ beam propagates directly to the FPA in the UV path, and it is the input beam to the Vis and IR paths.

The three SSI channels are separated by dichroic beamsplitters. The first beamsplitter reflects the UV to its $f/80$ focus, transmitting the Vis and IR. The second beamsplitter reflects the Vis and transmits the IR into their respective channels. Selection of modes within each channel is done using insertable devices: selector mirrors for the visible and IR channels to send the light to the imaging FPA or through the respective IFS optics; pupil imaging lenses in the IR and UV channels, and an insertable prism in the UV channel. Single mirrors within each channel are sufficient to form the final images on the focal plane arrays (FPAs), at $f/70$ for the visible imager and $f/51$ for the IR imager. Wavefront errors meet allocations derived from the error budget for each channel, preserving diffraction-limited performance, at 0.40 μm for the UV and visible channels, and at 1.00 μm for the IR channel.

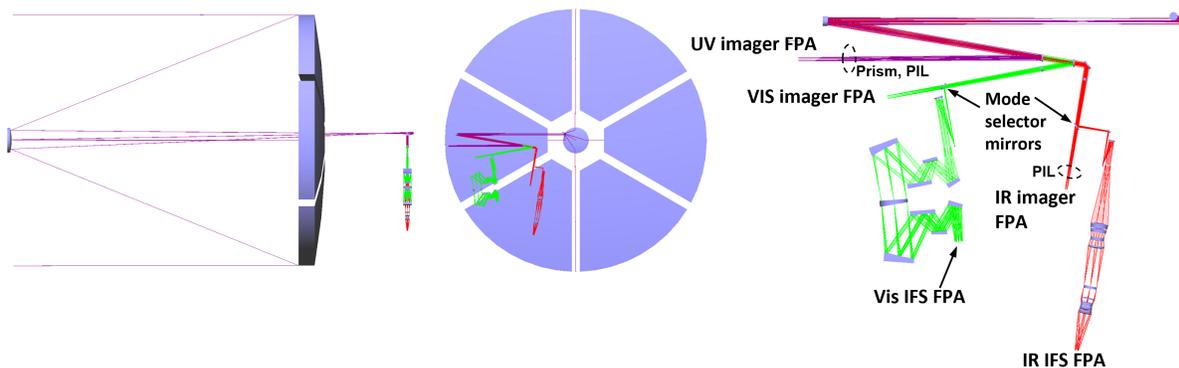

**Figure B.1-14.** HabEx 3.2S telescope and SSI, showing near-IR, visible, and UV channels, with imaging and spectroscopic modes. The IR and UV channels also are equipped with flip-in pupil imaging lenses.

**Table B.1-4.** HabEx 3.2S SSI requirements and compliance. For comparison with HabEx 4H, please see Table 6.4-1.

| Parameter | Requirement | Expected Performance | Margin | Source |
|---|---|---|---|---|
| Spectral Range | ≤0.3 μm to ≥1.70 μm | 0.20–1.80 μm | Met by design | STM |
| Spectral Resolution | ≥5 (0.3–0.35 μm)<br>≥40 (0.63 μm)<br>≥70 (0.75–0.78 μm)<br>≥8 (0.80 μm)<br>≥35 (0.82 μm)<br>≥100 (0.87 μm)<br>≥32 (0.89 μm)<br>≥17 (0.94 μm)<br>≥20 (1.06 μm)<br>≥19 (1.13 μm)<br>≥12 (1.15 μm)<br>≥10 (1.40 μm)<br>≥11 (1.59–1.60 μm)<br>≥10 (1.69–1.70 μm) | 7 (0.20–0.45 μm)<br>140 (0.45–0.975 μm)<br>37 (0.975–1.80 μm) | Met by design | STM |
| OWA (0.5 μm) | Baseline: ≥6 arcsec<br>Threshold: ≥0.5 arcsec | 4 arcsec | Met by design | STM |
| End-to-End Throughput (0.450 μm) | 22% | 30% | 36% | Error Budget |

**Table B.1-5.** HabEx 3.2S SSI specifications. For comparison with HabEx 4H, please see Table 6.4-3.

| Cameras | UV Channel | Visible Channel | IR Guide Channel |
|---|---|---|---|
| FOV | 10" | 12" | - |
| Wavelength Bands (nm) | 200–450 | 450–975 | 975–1800 |
| Pixel Resolution | 14.2 mas | 14.2 mas | 12 cm |
| Telescope Resolution | 21 mas | 21 mas | - |
| Detector | 1×1 CCD201 | 1×1 CCD201 | 1×1 LMAPD |
| Array Width (pixels) | 1024 | 1024 | 256 |

| Spectrometers | UV Channel | Visible Channel | IR Channel |
|---|---|---|---|
| FOV | 10" | 2" | 4" |
| Wavelength Bands (nm) | 200–450 | 450–975 | 975–1800 |
| Spectrometer Resolution | 7 | 140 | 40 |
| Spectrometer Type | Slit/grism | IFS | IFS |
| Detector | 1×1 CCD201 | 1×1 CCD282 | 2×2 LMAPD |
| Array Width (pixels) | 1024 | 4,096 | 2,048 |

### B.1.1.4 Ultraviolet Spectrograph (UVS)

The HabEx 3.2S UVS has sensitivity nearly matching the HabEx 4H UVS though increased throughput, compensating for a smaller aperture. The UVS accepts light from a field of view centered 0.060° off of the primary mirror axis, encompassing a FOV of 3×3 arcmin, illustrated in **Figure B.1-3**. Selection of objects within that field is made using the MSA, an array of commandable shutters of 200×100 μm in size, that open to pass light from the selected areas of the FOV. To match the curved Cassegrain focus, each quadrant of the MSA will be a separate flat array, slightly tilted to better match the ideal focal surface. The maximum diffracted PSF size (50% encircled energy diameter) at the MSA is 31 μm at 300 nm wavelength, and the likelihood that any random object in the field is not vignetted ranges from 60 to 74% (Scowen et al. 2018).

### Design

Following the light path past the MSA, shown in **Figure B.1-15**, is the UVS tertiary mirror, which collimates the beam onto the gratings. After the gratings is the quaternary mirror, which creates a f/50-class beam converging on the FPA. Both the tertiary and quaternary mirrors are freeform aspheric shapes (Standard Zernike) in order to achieve the stringent 12 nm RMS instrument wavefront error budget allocation.

Dispersion is generated using gratings in a linear sliding mechanism located in the collimated beam after the tertiary mirror. On the mechanism are 20 dispersing elements and a weakly curved spherical mirror, to provide selectable bandpass and spectral resolution, enabling coverage over the full 115–320 nm waveband at resolutions from R = 500 to 60,000 (**Figure B.1-16**). (The definition of R is $R = \lambda/\Delta\lambda$, were $\lambda$ is the central wavelength in the band, and $\Delta\lambda$ is the wavelength interval that deflects the image point by 2 pixels).

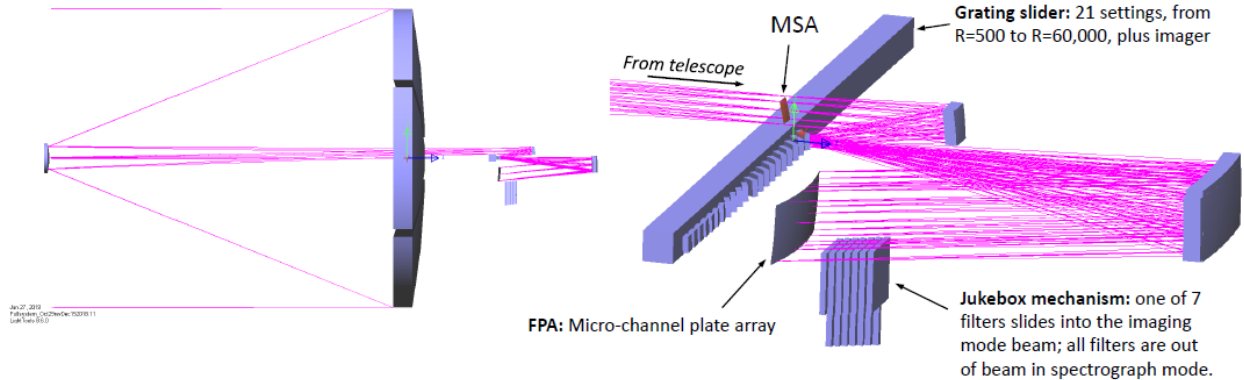

**Figure B.1-15.** UV Spectrograph (UVS) optical layout. A 5-bounce design intended to meet stringent optical quality requirements while maximizing UV throughput over a 8 arcmin field. The grating slider enables spectral resolutions from R = 500 to 60,000 covering the full 115–300 nm waveband, and includes a weak mirror for imaging. The jukebox filter mechanism enables photometry in the imaging mode.

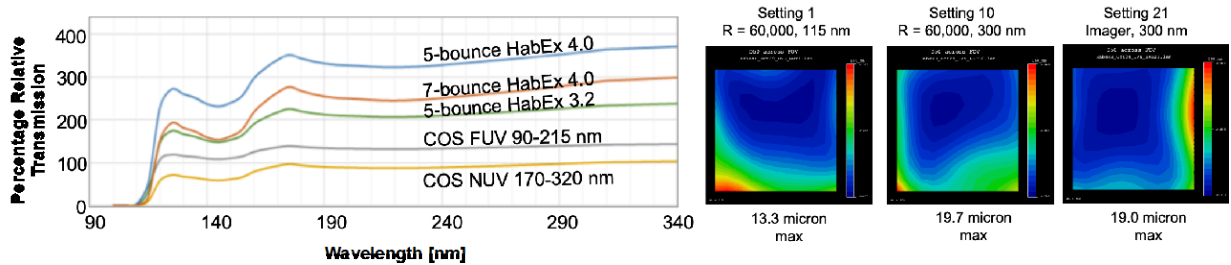

**Figure B.1-16.** UVS design performance. *Left:* Sensitivity (given by the *5-bounce HabEx 3.2* curve) is approximately double that of the Hubble Cosmic Origins Spectrograph. *Right:* Design optical quality is less than 20 µm encircled energy diameter, consistent with overall requirement of 30 µm.

A conventional wheel mechanism to carry the 20 gratings and 1 mirror was not practical, due to clearance constraints between other channels. Because each grating is on a flat substrate with uniformly spaced straight lines, in a collimated beam, the setting precision can be achieved with a linear slider. The grating spacing on the 46×36 mm clear aperture grating elements range from 0.093 µm at R = 60,000 to 14 µm at R = 500. See the requirements and specification tables, **Table B.1-6** and **Table B.1-7**, respectively, for full details.

The final optical element in the UVS imaging path (only) is a filter, located between the quaternary mirror and the FPA. At any one time, one of seven filters can be inserted into the beam in imaging mode; the filters are entirely out of the beam in the 20 spectrograph modes, as illustrated in **Figure B.1-3**. Due to the limited available space for this filter mechanism, the only sufficiently compact filter mechanism concept was a set of seven filters, each moving independently in one direction in or out of the beam, similar to how a jukebox flips one disk from a stack of disks into play mode. The thickness of the filter will introduce a different optical path into the imaging mode than the spectrograph mode, mostly defocus. To correct for this defocus, the imaging mirror on the slider mechanism will have a weak spherical curvature. (Without the filter, the imaging mirror could have been flat). Given the filter clear aperture of about 150×150 mm, the filter thickness might be 15 mm. Assuming a Fused Silica substrate, the defocused RMS WFE is up to 85 nm without focus correction. A maximum RMS WFE of 15 nm can be restored by placing a concave radius of 427 m on the imaging mirror. This corresponds to a surface sag of 0.42 μm at the edge of the clear aperture of the imaging mirror.

**Table B.1-6.** UVS requirements and compliance. For comparison with HabEx 4H, please see Table 6.5-1.

| Parameter | Requirement | Expected Performance | Margin | Source |
|---|---|---|---|---|
| Spectral Range | ≤115 nm to ≥320 nm | 115–320 nm | Met by design | STM |
| Spectral Resolution, *R* | Up to ≥ 60,000 depending on the measurement | 1 (imaging), 500, 1,000, 3,000, 6,000, 12,000, 24,000, 60,000 | Met by design | STM |
| Angular Resolution | 50 mas | 25 mas | 100% | STM |
| FOV | >2.5 × 2.5 arcmin² | 3 × 3 arcmin² | 20% | STM |
| Multi-object Spectroscopy | Yes | Yes | Met by design | MTM |

**Table B.1-7.** UVS design parameters. For comparison with HabEx 4H, please see Table 6.5-2.

| Parameter | Specification |
|---|---|
| FOV | 3 × 3 arcmin² |
| Wavelength Bands | 20 bands covering 115 to 320 nm |
| Spectral Resolutions | 60,000; 25,000; 12,000; 6,000; 3,000; 1,000; 500; 1 (imaging) |
| Telescope Resolution | Diffraction limited at 400 nm |
| Detector | 3×5 MCP array, 100 mm² each |
| Array Width | 17,000 × 30,000 pixels (pores) |
| Microshutter Aperture Array | 2×2 array of 171×365 200×100 μm apertures |

The illuminated extent of the UVS image surface in the current design is about 390×160 mm, encompassing all 21 settings. The FPA will be populated with MCP detectors or with Delta-Doped CCDs, and will require about 26,000 x 11,000 15 μm pixels to cover the full FOV. The ideal shape of the FPA would be toroidal, so different FPA sections will be arranged at different tilts to approximate the toroidal surface of best focus.

### B.1.1.5 HabEx Workhorse Camera (HWC)

The HabEx 3.2S Workhorse Camera (HWC) design differs from the HabEx 4H HWC concept described in *Section 6.6*. The HabEx 3.2S HWC observes in three bands instead of HabEx 4H HWC's two bands, with each being: UV (0.32–0.45 μm); visible (0.45–0.95 μm); and, near IR (0.95–1.80 μm) (Martin 2018; Acton et al. 2012). Its design is overviewed in **Figure B.1-17** with its requirements and specification overviewed in **Table B.1-8** and **Table B.1-9**, respectively. Its transmission is compared to the HabEx 4H HWC in **Figure B.1-18**. The Vis and IR channels share a common MSA. The UV channel is separate from the visible/IR and does not have an MSA, allowing the channel to have fewer mirrors and higher UV transmission than if it shared the optics with the visible/IR channel through the MSA. The visible and IR channels observe the same 3×3 arcmin² field simultaneously, as the field is shared using a dichroic beamsplitter. The UV channel views a separate field of view. The switch between imaging and spectrometry modes is implemented using flip-in grisms located in the filter wheel assemblies, and an insertable MSA. The spectral resolution when in spectroscopic mode is R = 1,000 in Vis and IR channels. When in imaging mode, the MSA is commanded out of the beam to avoid vignetting. All three HWC channels are equipped with filter wheels for use when imaging.

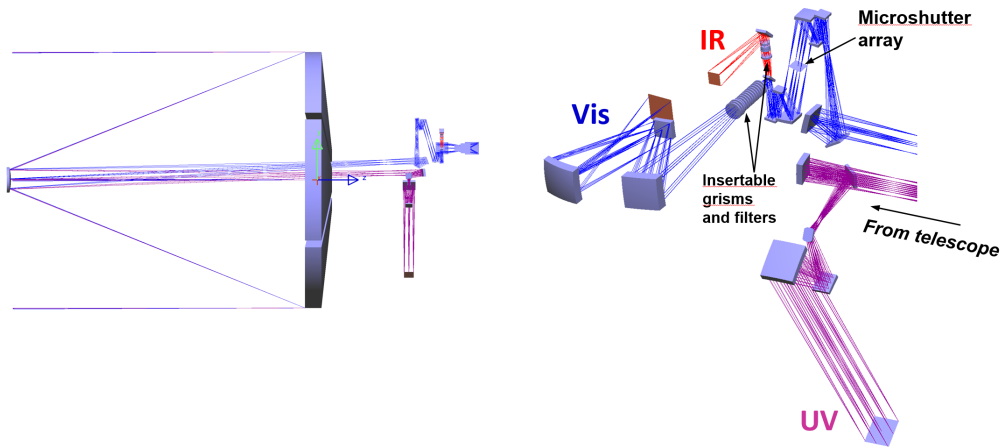

**Figure B.1-17.** The HabEx telescope and Workhorse Camera (HWC) light path diagram, showing UV, visible, and IR channels, with imaging and spectroscopic modes.

**Table B.1-8.** HWC requirements and compliance. For comparison with HabEx 4H, please see Table 6.6-1.

| Parameter | Requirement | Expected Performance | Margin | Source |
|---|---|---|---|---|
| Spectral Range | ≤0.37 µm to ≥1.70 µm | 0.32–1.80 µm | Met by design | STM |
| Spectral Resolution, R | Up to ≥1,000 depending on the measurement | ≤2,000 | Met by design | STM |
| Angular Resolution | 50 mas | 25 mas | 100% | STM |
| FOV | ≥ 2 × 2 arcmin² | 3 × 3 arcmin² | 50% | STM |
| Multi-object Spectroscopy | Yes | 342 × 730 apertures | Met by design | STM |
| Noise Floor | ≤10 ppm | 10 ppm | Met by design | STM |

**Table B.1-9.** HWC design specifications. For comparison with HabEx 4H; please see Table 6.6-2.

| | VIS Channel | IR Channel |
|---|---|---|
| FOV | 3'×3' | 3'×3' |
| Wavelength Bands | 0.37–0.975 µm | 0.95–1.80 µm |
| Pixel Resolution | 15.5 mas | 24.5 mas |
| Telescope Resolution | 30.9 mas | 49 mas |
| Design Wavelength | 0.6 µm | 0.95 µm |
| Detector | 3×3 CCD203 | 2×2 H4RG10 |
| Detector Array Width | 12,288 pixels | 8,192 pixels |
| Spectrometer | R = 1,000 | R = 1,000 |
| Microshutter Array | 2×2 arrays; 180×80 µm aperture size; 171×365 apertures | |

The HWC UV and Vis channels utilize delta-doped CCDs (Stern 2018), with 12 µm pixel size, mosaiced to provide a total of 12k × 12k pixels. The FPA is passively cooled to 153 K. The NIR channel would utilize a low-noise hybrid HgCdTe/CMOS detector such as the Teledyne H4RG, with 12 µm pixels in a 4k × 4k format (Nikzad et al. 2012).

### Design

The optical design of the HWC UV channel, other than the common telescope primary and secondary mirrors, includes a conic tertiary that creates a collimated beam, demagnified by 160× from the primary mirror. Following the tertiary is a fold mirror, a filter, another fold mirror, and a 2-mirror imager (rotationally symmetric aspheric shapes) forming a $f/41$ final focus. The RMS WFE across the $3 \times 3$ arcmin² FOV is less than 15 nm, consistent with the instrument WFE allocation (**Figure B.1-4**).

In the visible and IR channels of the HWC, as in the UV channel, a conic tertiary mirror creates a collimated beam at about 150× demagnification from the primary mirror, slightly different from the magnification in the UV channel. After the visible/IR tertiary is a three-mirror focuser that creates a well-corrected telecentric $f/20$ focus where the MSA is located. The MSA is followed by the inverse of the focuser, creating another collimated beam. A dichroic beamsplitter is placed in the collimated beam to reflect visible and transmit IR. Following the beamsplitter in each channel is a filter wheel assembly which holds several wheels carrying filters or dispersing elements. After the filter assembly is the final focusing section. In the IR channel, this is a relatively simple $f/20$ refractive group consisting of five rotationally symmetric elements of common near-IR materials. A refractive solution is compact and is adequate to correct chromatic aberration in this band. In the Vis channel, the *f*/41 focusing group consists of three mirrors (one of which is a rotationally symmetric aspheric, and two of which are Zernike aspherics). The maximum nominal RMS WFE of the full channel is about 16 nm in the Vis, and 35 nm in the near IR across the 3×3 arcmin$^2$ FOV.

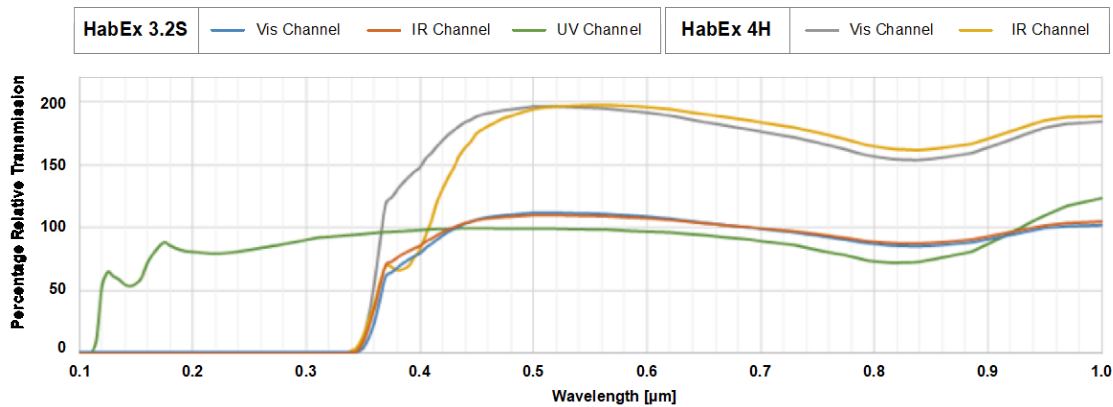

**Figure B.1-18.** Comparison of HabEx Workhorse Camera transmission for HabEx 4H and HabEx 3.2S.

### B.1.1.6   Payload Thermal System

Similar to HabEx 4H, the payload thermal system uses a combination of active heating and passive cooling methods to provide mK-level stability for optical structures, cooling for electronics, and cryocooling for relevant detectors. The payload thermal architecture is identical to that of HabEx 4H, described in *Section 6.7*, with the exception a reduced heat lift requirement due to the removal of the coronagraph. This results in smaller radiators, where 5.5 m$^2$ of surface area is required to radiate to meet margined 240 K electronics requirements; 4.3 m$^2$ to meet margined 130 K focal plane requirements; and, 9 m$^2$ to meet margined 55 K focal plane requirements. Similar to HabEx 4H, the payload thermal subsystem is capable of lifting and rejecting heat across the field of regard and over 180° of telescope rotation along the optical axis.

### B.1.2   HabEx 3.2S Telescope Flight System

By removing the coronagraph from HabEx the requirements on the Telescope flight system are relaxed, permitting a simpler HabEx 3.2S design point that can be achieved at lower cost. Specifically, the design point can be achieved through the provision of a standard bus from a spacecraft vendor, e.g., Ball BCP5000. Thus, HabEx 3.2S design emphasizes high-heritage processes, components and subsystems, drawing on experience with previous space observatories and spacecraft. This heritage is identified in the discussion of the respective bus subsystems. The bus is designed following commercial practice, using standardized subsystems packaged into a mission-specific but generic structure. There are only two subsystems that deviate from high-heritage hardware and practice: the colloidal microthrusters used for precision attitude control and the refuellable propulsion system.

**Figure B.1-19** shows an exploded view of the major bus subsystems. The structure is a simple cylinder 1.1 m long by 4.2 m wide, supporting all bus systems and the payload in a format compatible with the launch vehicle. The propulsion system includes hydrazine monopropellant thrusters used for station keeping and large-angle attitude maneuvers, and micro-g colloidal thrusters used to maintain precision pointing during science observations. The attitude control system uses star trackers, gyros, and a signal provided by the FGS cameras in the payload to measure spacecraft attitude, and command the propulsion system to maneuver and point the telescope. The power system utilizes fixed solar arrays to provide power in all observing conditions without deployed elements. The communications system receives commands from the ground station and transmits data and telemetry. It also provides formation flying signals to the starshade during joint operations.

The HabEx 3.2S Telescope flight system mass equipment list (MEL) is shown on **Table B.1-10**. The bus subsystems are also illustrated in the block diagram of **Figure B.1-20**, which indicates the chief interfaces.

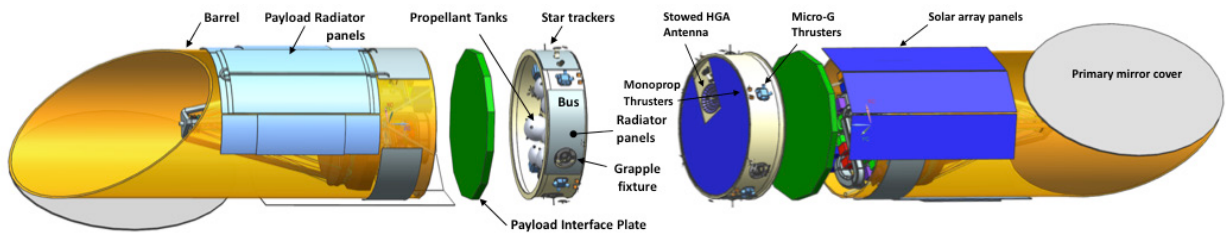

**Figure B.1-19.** Spacecraft bus, showing bus structure, propellant tanks, thrusters, star trackers, solar panels and bus radiators. The Payload Interface Plate (PIP) provides the mechanical interface to the payload. Grapple fixtures would be used by servicing spacecraft to dock prior to on-orbit servicing.

**Table B.1-10.** MEL for 3.2S architecture with three instruments.

| | CBE (kg) | Cont. % | MEV (kg) |
|---|---|---|---|
| **Payload** | | | |
| Telescope and Instruments | 1360 | 30% | 1770 |
| Payload Thermal | 230 | 30% | 300 |
| **Spacecraft Bus** | | | |
| ACS | 15 | 15% | 20 |
| C&DH | 75 | 15% | 85 |
| Power | 210 | 15% | 240 |
| Propulsion: Monoprop | 200 | 30% | 270 |
| Structures & Mechanisms | 1390 | 32% | 1845 |
| Spacecraft side adaptor | 75 | 30% | 100 |
| Telecom | 65 | 15% | 75 |
| Thermal | 70 | 30% | 90 |
| **Bus Total** | 2030 | 28% | 2610 |
| **Spacecraft Total (dry)** | 3625 | 43% | 5185 |
| Subsystem heritage contingency | 1060 | | |
| System margin | 500 | | |
| Monoprop and pressurant | 756 | | |
| **Total Spacecraft Wet Mass** | | | 5940 |
| Launch vehicle side adaptor | | | 1350 |
| **Total Launch Mass** | | | 7290 |

## B.1.2.1 Structures & Mechanisms

The spacecraft primary structure consists of an equipment mounting panel, payload interface panel, cylindrical/barrel structure, and an interface to connect with the launch vehicle adapter. A rendering of the spacecraft primary structure is shown in **Figure B.1-21**.

The 4.2 m diameter cylindrical structure design provides high stiffness-to-weight for supporting the payload and optimizing load paths. The cylindrical structure is a lightweight sandwich-structured composite that will be constructed from aluminum face sheets and aluminum honeycomb core. The thickness of the face sheets and honeycomb core will be tailored to meet static and dynamic load requirements.

The cylindrical bus structure enables direct transfer of structural loads from the payload cylindrical structure, through the payload interface plate connected to the bus structure, to the similarly sized launch vehicle adapter. This matching of diameters minimizes the stresses on the structure and decreases the complexity and risk of the bus structure design. This structural design approach is common for large commercial busses.

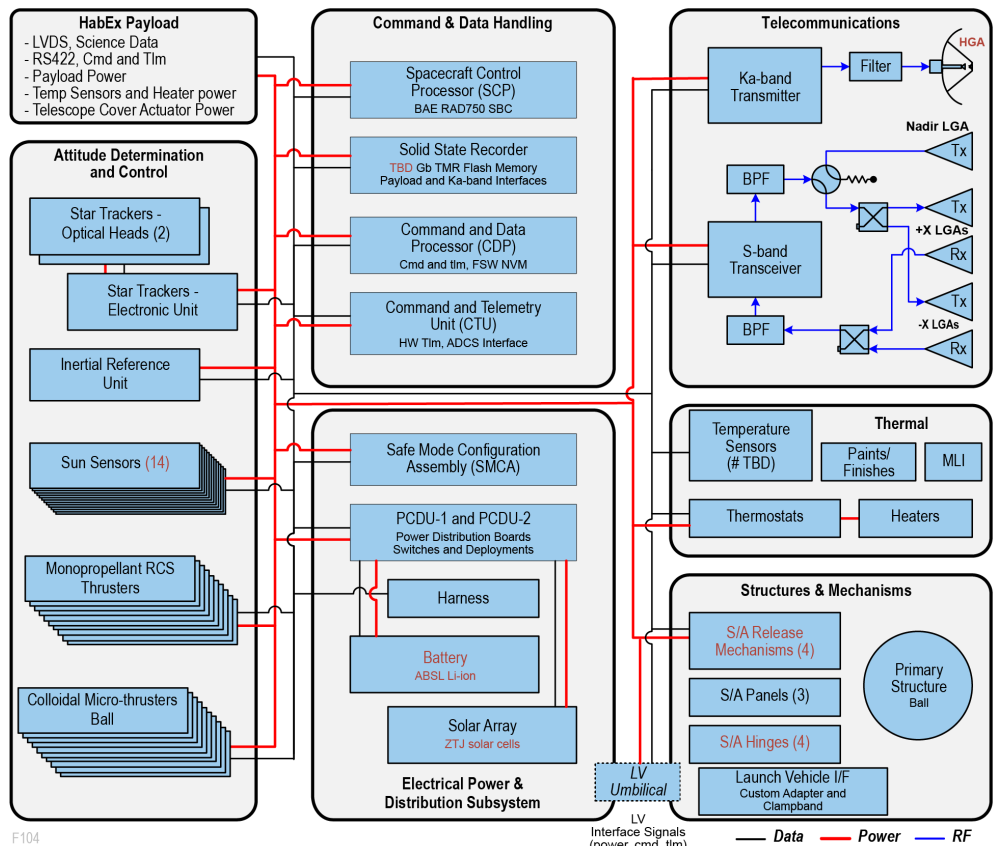

**Figure B.1-20.** HabEx 3.2S bus functional block diagram, showing bus subsystems, with signal, power and communications interfaces indicated. For comparison with HabEx 4H, please see Figure 6.10-2.

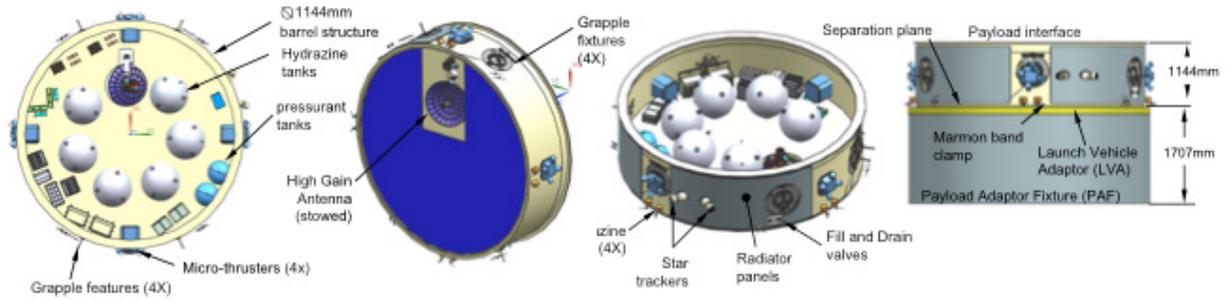

**Figure B.1-21.** HabEx 3.2S spacecraft primary structure, showing major components: cylindrical barrel structure, equipment mounting panel, PIP, LVA and PAF. The PAF height depends on launch vehicle, and may be a custom build. For the Delta IV Heavy, the 4393-5 PAF is assumed, with a height of 2.16 m. For the Vulcan Centaur, a similar design is assumed, but with a reduced height of 1.56 m, consistent with the Delta 4394 PAF.

The bus equipment is mounted on a 4 m diameter circular equipment mounting panel that closes out the bottom of the cylinder. There is sufficient area on the panel for mounting all the required bus equipment. The equipment mounting panel is a sandwich-structured composite of aluminum face sheets and aluminum honeycomb core. The thickness of the honeycomb core is sized to provide the needed out-of-plane stiffness. Secondary structures include struts and bracketry to structurally support the solar panels, propellant tanks, and antenna gimbal. The antenna gimbal mechanism will be actuated during spacecraft maneuvers, to keep the antenna pointed toward the Earth. It will be held rigid during observations to avoid disturbing the observatory.

The bus will be attached to the launch vehicle using a standard payload attachment fixture (PAF) specific to the available enhanced expendable launch vehicle-class (EELV-class). For instance, a standard 4293-5 PAF can be used with the Delta IV Heavy. The actual interface will be made through a detaching Launch Vehicle Adaptor (LVA) device: a cylinder matching the bus outer shell to the PAF. This will provide a direct load path from the spacecraft to the launch vehicle. Since the diameters are closely matched, there will be efficient load transfer from the LVA to the PAF without significant tapering to minimize stress concentrations at the interfaces.

### B.1.2.2 Power

The power system consists of solar arrays, batteries, and hardware to support the power distribution network. The peak power consumption mode is simultaneous science observation and downlink, totaling 4,000 W with 3,300 W consumed by the payload and 700 W by the bus. Aft- and barrel-mounted solar panels, 10 m$^2$ and 21 m$^2$, respectively, are fixed and do not require mechanisms. They are sized and configured to meet HabEx power requirements through all sun angles, 40–180°, as demonstrated in **Figure B.1-22**.

A direct-energy power architecture, to include flat panel solar arrays, batteries, and hardware to support the power distribution network, will be used to maintain a positive power balance in the form of 28 V unregulated bus voltage during all spacecraft modes. Solar array strings will be managed by a charge control algorithm to provide the necessary power to the bus and payload during nominal operations and to maintain adequate charge in the battery system for use during HabEx any contingency scenarios.

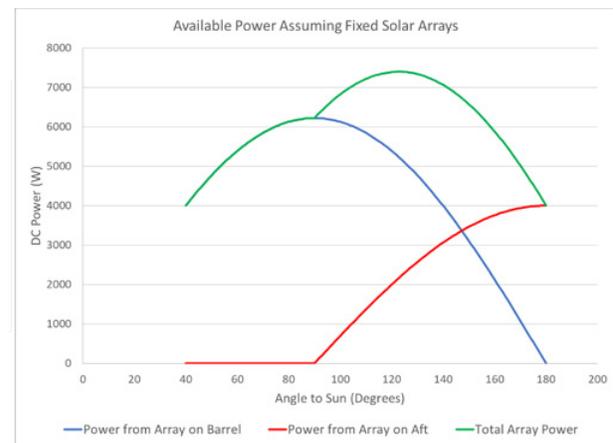

**Figure B.1-22.** HabEx 3.2S solar panels are configured to provide sufficient power across all operational sun angles.

### B.1.2.3 Propulsion

Like HabEx 4H, HabEx 3.2S's propulsion is handled by two separate systems: a mono-propellant propulsion system and a microthruster propulsion system. The low-complexity, monopropellant hydrazine blow-down design based on previous commercial flight systems, including Deep Impact and Kepler, is used to perform all Telescope slews and propulsive maneuvers. The microthruster design will be used to mitigate solar pressure torques and to point the observatory down to the milli-arcsecond level.

The hydrazine monopropellant RCS thruster architecture consists of eight 1 N thrusters. The thrusters are used for large maneuvers between science data collections, implementing angular rate profiles for tracking solar system objects, and for attitude control in safe mode. The thrusters are located at the aft end of the bus and are oriented to provide a high amount of authority in the cross-boresight axes. Their location at the aft end minimizes the likelihood of contamination of optical surfaces and provides a large offset from the spacecraft center-of-gravity (CG) to maximize torque, reducing the propellant requirement for telescope slews. Thruster nozzles are oriented perpendicular to the moment arm to maximize torque as well.

HabEx 3.2S uses identical microthrusters to HabEx 4H, which are described in *Section 6.10.3*. Microthrusters are located in four locations, 90° apart, on the bus similar to the monopropellant RCS system. As L2, solar radiation pressure torques are expected to be 400–500 µN-m, based on scaling up observed solar radiation pressure torques on Kepler. At that level, the four microthrusters will be able to balance the disturbance torque with less than 100% of their maximum thrust value. The attitude control algorithm would formulate a torque command that is limited to the maximum capability of the electrospray heads. This torque command is then sent to the thruster control algorithm, which uses knowledge of the torque available to formulate thrust commands for each thruster.

### B.1.2.4 Communications

Removal of the coronagraph has two implications for the design of the HabEx 3.2S telecommunications subsystem. First, without the coronagraph's stability requirement a gimbaled high gain antenna (HGA) can be used in place of a phased array antenna, reducing cost. Second, with the starshade now performing the broad survey, there is more observational time available for general astrophysics instruments per week, resulting in a larger data volume requirement.

The HabEx 3.2S telecommunications design thus differs as the phased array antennas are replaced with a gimbaled 65 cm diameter Ka-band HGA driven by a 70 W Traveling Wave Tube Amplifier (TWTA). The mission uses the near-Earth 26 GHz Ka-band downlink spectrum for high-rate mission data downlink at a nominal rate of 150 Mbps, yielding a 130.4 Mbps. This design accommodates for any increase in weekly data volume.

### B.1.2.5 Command and Data Handling

HabEx 3.2S's command and data handling subsystem does not drive the design. Its commanding and data requirements can be met by high heritage design and components.

### B.1.2.6 Telescope Pointing Control

The pointing control goal of HabEx 3.2S is identical to that of HabEx: slew to a target and maintain pointing stability during observation. However, the HabEx architecture utilized the coronagraph in the pointing control loop in order to meet the coronagraph's pointing stability requirements. The pointing control architecture for HabEx 3.2S is identical to the HabEx 4H pointing control architecture described in *Section 6.10.6* and **Figure 6.10-5**, with the exception of the inner control loop for the coronagraph.

### B.1.2.7 Thermal

HabEx 3.2S's spacecraft thermal architecture includes the components shown in

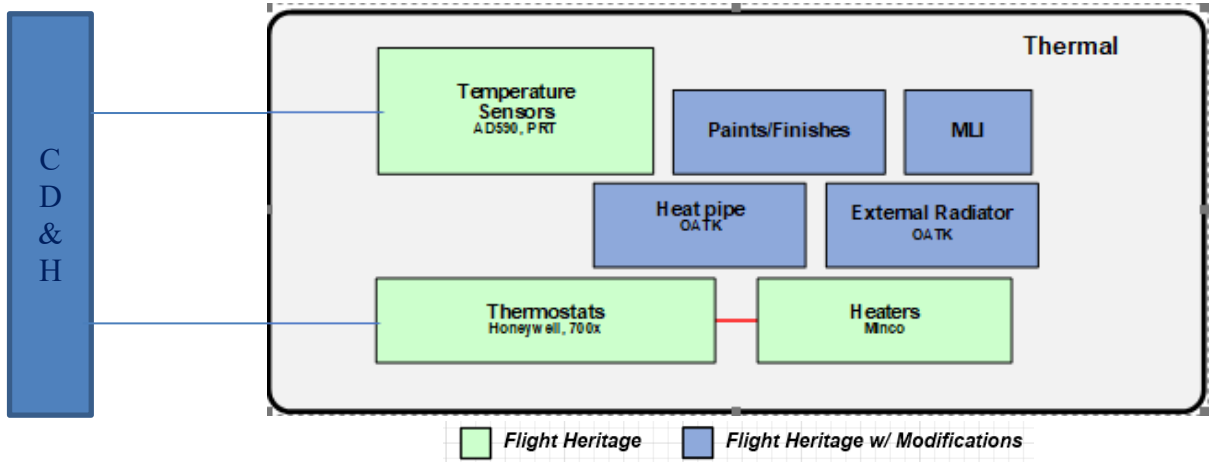

**Figure B.1-23.** HabEx 3.2S thermal block diagram.

**Figure B.1-23**. Heaters will be used to control baseplate temperature. Cooling for heat-generating units such as electronics boxes mounted on the bus lower deck will be provided by a 3.4 m² radiator. A set of embedded heat pipes will be directly coupled to units mounted to the bus lower deck. From the embedded heat pipes, a set of externally mounted "jumper" heat pipes will provide the thermal connection to the radiator. For this design, ammonia filled Constant Conductance Heat Pipes are baselined.

To isolate the internal spacecraft bus from the external environment and to minimize thermal gradients, the majority of the spacecraft bus will be blanketed. MLI blankets and thermal isolation washers will be utilized in between the sun facing solar panel, the aft mounted solar panel, and the external radiator to thermally isolate these components from the spacecraft. Blankets will be used with conductive isolation to ensure an adiabatic interface to the instrument deck, shown in **Figure B.1-24**.

A circular heat pipe embedded into the base mounting panel rings the inside of the bus, providing easy access to cooling for the various electronics boxes on the panel. External jumper heat pipes bent in a 90° configuration would be used to connect the internal thermal system to the external radiator, which is mounted to the anti-sun side of the spacecraft bus. The thermal analysis assumes two jumper heat pipe connections working at 20 W/°C.

### B.1.2.8  Formation Flight

HabEx 3.2S's concept of operations for formation flight is identical to HabEx 4H, and is described in *Section 8.1.7*.

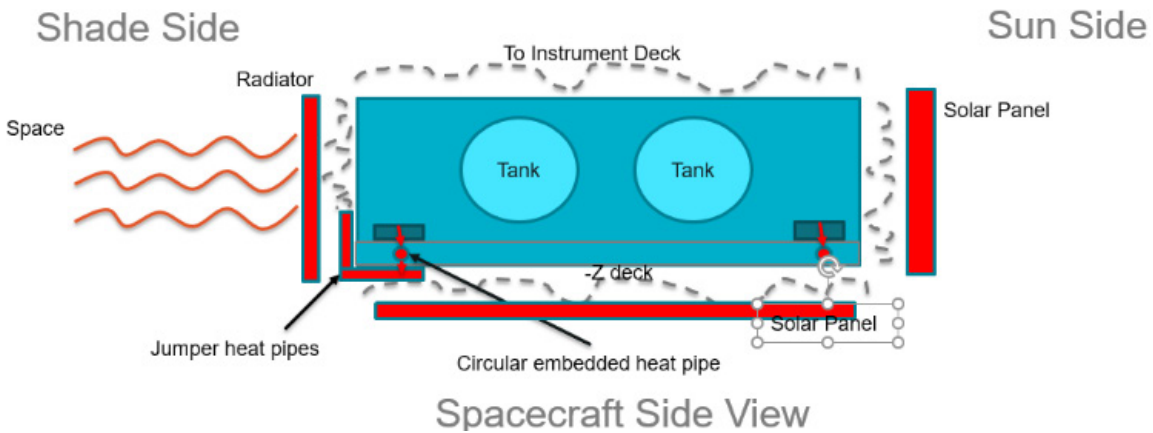

**Figure B.1-24.** Heat flow Illustration describing thermal flows for the HabEx 3.2S bus.

## B.2   HabEx 4S Architecture

While a HabEx 4S architecture can be defined for monolithic or segmented apertures, the approach taken in this section is to scale up the HabEx 3.2S architecture reported in *Section B.1*. The design laid out in *Section B.1.1* is scalable up to 4 m aperture, for launch with the Vulcan Centaur or Delta IV Heavy, with their 5-meter shrouds. While the concept of operations and observing program remain the same, total throughput and angular resolution are improved with the larger aperture. As summarized in **Table 10.4-1**, HabEx 4S meet 14/17 of HabEx 4H's science objective baseline requirements and 2/17 at threshold requirements.

To evaluate a HabEx 4S configuration, the HabEx 3.2S optical design was scaled up by a factor of 1.25. This grew the aperture to 4 m, and the SM-PM distance to 5.4 m. The instruments, their enclosures and devices were grown in proportion. The outer barrel assembly grew only slightly wider, since it is oversized for the HabEx 3.2S aperture—it grew to an outer dimension of 4.4 m, leaving 8.6 cm clearance to the segments. To fit the longer SM-PM distance, the optical bench assembly grew longer as well, to

9.4 m. The spacecraft bus was assumed to be the same for HabEx 4S as for HabEx 3.2S, for this first estimate. These changes are indicated in **Figure B.2-1**, which shows that the overall size of HabEx 4S is compatible with both launch vehicles, provided the split-scarf payload door is used.

Some architectural changes were required to keep the payload within the 4.6 m width of the dynamic envelope of the launch vehicles' 5 m shrouds. The instruments and telescope were rotated within the barrel, to provide clearance between instrument structures and the aft metering structure support struts. The payload thermal system's three stage radiator does not fit within the wider outer barrel assembly, so a single-stage radiator augmented by cryocoolers was used instead. This reduces the size, mass, and especially the thickness of the payload radiator, and adds the mass needed for cryocoolers and their vibration isolators. Similarly, a more conformal and larger solar array is required, so the solar array is reconfigured to use narrower panels.

This upscaled design resulted in a margined payload mass of 4,970 kg, for a HabEx 4S observatory total margined mass of 9,320 kg. This is much less than the 13,400 kg Vulcan Centaur lift capability to L2, and it is also less than that of the Delta IV Heavy (10,000 kg). Further design work would be required to finalize the design for HabEx 4S, to refine the instrument packaging, to improve clearances, and to reoptimize the bus design for the larger payload. Nonetheless, this first look indicates that a 4 m starshade-only HabEx is feasible within current and projected commercial launch vehicle capabilities, and would not require an SLS.

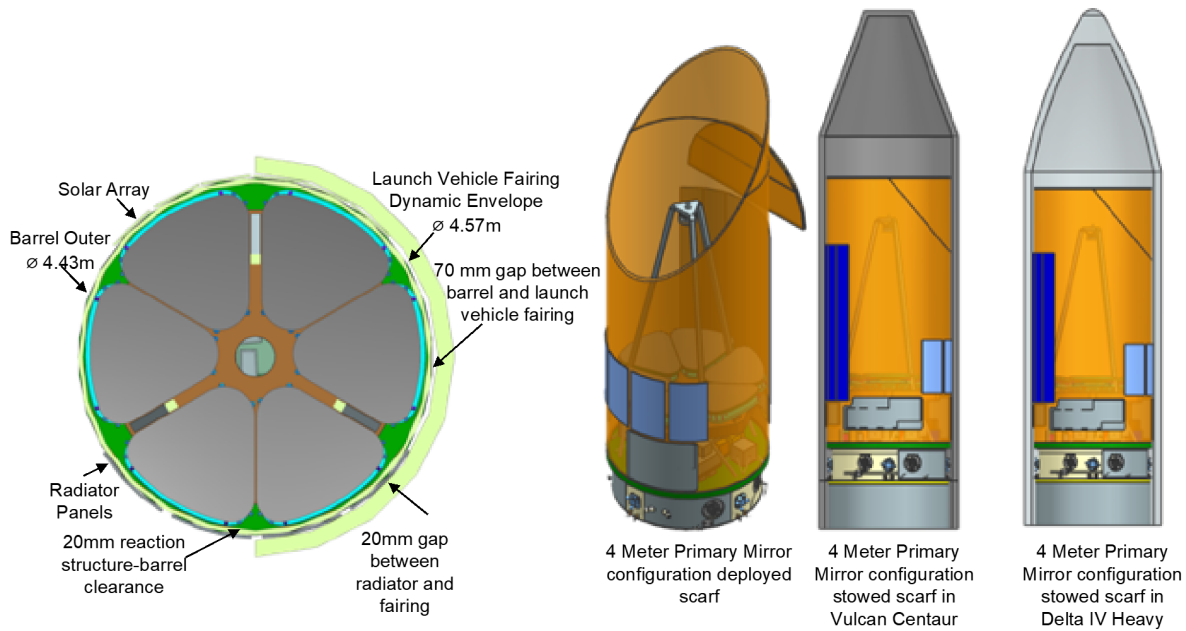

**Figure B.2-1.** The 4-meter aperture, starshade only HabEx4S design, scaled from HabEx 3.2S. The smaller radiator was adopted to provide clearance between the OBA and the launch vehicle fairing, and is enabled by also adopting cryocoolers for the IR focal planes. Mass and volume, with the split scarf barrel door, are within Vulcan and Delta IV launch vehicle capabilities.

## B.3 Technology Readiness of HabEx Starshade-Only Architectures

HabEx 3.2S has 12 technologies, currently at TRL 4 or TRL 5, that need further development. These technologies are summarized in **Table 11.1-1**. Referring to this table, the needed technologies are:

- Starshade Petal Position Accuracy and Stability, currently TRL 4, expected to be TRL 5 by 2023
- Starshade Petal Shape Accuracy and Stability, currently TRL 4, expected to be TRL 5 by 2023
- Starshade Scattered Sunlight for Petal Edges, currently TRL 5
- Starshade Contrast Performance Modeling and Validation, currently TRL 4, expected to be TRL 5 by 2023
- Starshade Lateral Formation Sensing, currently TRL 5
- Laser Metrology, currently TRL 5, expected to be TRL 6 by 2023

- Delta Doped UV and Visible Electron Multiplying CCDs, currently TRL 4, expected to be TRL 5 by 2023
- Deep Depletion Visible Electron Multiplying CCDs, currently TRL 4, expected to be TRL 5 by 2023.
- Linear Mode Avalanche Photodiode Sensors, currently TRL 4, expected to be TRL 5 by 2023
- UV Microchannel Plate (MCP) Detectors, currently TRL 4
- Microthrusters, currently TRL 4, expected to be TRL 5 by 2023

HabEx 3.2S would also benefit from the three enhancing technologies noted in **Table 11.1-1**:

- Far-UV Mirror Coating, currently TRL 4, expected to be TRL 4 by 2023
- Delta-Doped UV Electron Multiplying CCDs, currently TRL 4, expected to be TRL 4 by 2023
- Microshutter Arrays with higher shutter count, currently TRL 3, expected to be TRL 5 by 2023

The specific developments that are required, and the technology maturation path forward, for these items, is described in *Chapter 11*.

In addition to these technologies, the HabEx 3.2S mirror segments should be noted as an enabling technology. The current state of the art is represented by the Harris AMSD/MMSD mirrors referenced in *Section F.3*, **Figure B.3-1**.

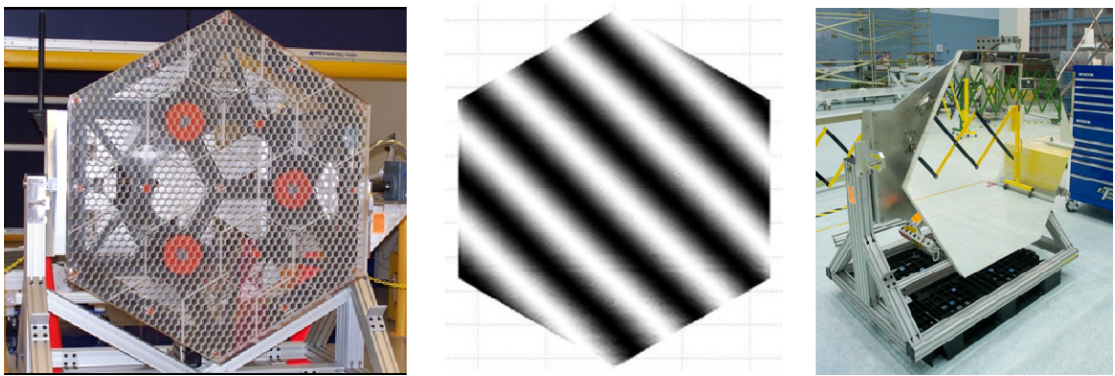

**Figure B.3-1.** ULE glass mirrors fabricated by Harris, Inc., following the methods described in Mooney et al. (2018a); Matthews (2017); Mooney et al. (2018b). These mirrors have demonstrated WFE performance without actuation (single mirror, with gravity effects backed out) of 15 nm RMS, and have passed shock and vibration tests sufficient to demonstrate launch survivability for the glass substrate.

Development from TRL 5 to TRL 6 would involve fabrication and test of a complete HabEx 3.2S-traceable segment subsystem, including figure control actuators, demonstrating <18 nm RMS wavefront error. This is noted in **Table B.3-1**.

**Table B.3-1.** HabEx starshade-only enabling technology gap list addition.

| Title | Description | State of the Art | Capability Needed | TRL 2019 | Expected 2023 TRL |
|---|---|---|---|---|---|
| Active ULE Mirror Segment | PM mirror segments, 1.4–1.7 m size, equipped with figure control actuators to meet UV wavefront requirements | • 2.4 m diameter closed-back ULE mirrors standard (HST)<br>• 1.4 m closed-back ULE hexagonal segments with areal density <20 kg/m² (AMSD, MMSD)<br>• Force-feedback figure control actuation (AMSD, MMSD) | • 1.4 m wedge-shaped closed-back ULE mirror<br>• Surface figure error to meet 18 nm RMS WFE allocation in 0g<br>• Aerial density <40 kg/m²<br>• 5 ppb/K CTE homogeneity<br>• Substrate first mode ≥ 200 Hz | 5 | 5 |

# C SCIENCE YIELD ASSUMPTIONS AND COMPUTATIONS

This appendix details the methodology, instrumental, and astrophysical assumptions used to derive the planet yield estimates summarized in *Section 3.3*. It also provides further information about the exoplanet surveys' operation concept, and presents the full yield results obtained for different planet types under a broad range of planet occurrence rate and exozodi level assumptions, ranging from pessimistic to optimistic.

## C.1 Methodology for Estimating the Yield of the HabEx Direct Imaging Surveys for Exoplanets

The estimate of the yield of directly imaged planets assumed that HabEx must conduct a blind survey to search for and characterize potentially Earth-like exoplanets. While the efficiency of the HabEx exoplanet survey and the quality of its data products would benefit from a precursor survey identifying potentially Earth-like planets, this study conservatively assumed such a survey does not exist at the time of launch. Thus, the yield of such a blind survey is a probabilistic quantity, which depends on HabEx's capabilities using the coronagraph and starshade, the occurrence rate of planets of various types, their detectability, and the unknown distribution of planets around individual nearby stars.

To calculate expected exoplanet yields, the Altruistic Yield Optimization (AYO) yield code of Stark et al. (2014) was used, which employs the completeness techniques introduced by (Brown 2005). For each star in the HabEx master target list (*Appendix D*), a random distribution of a large number of synthetic planets of a given type was made, forming a "cloud" of synthetic planets around each star, as shown in **Figure C.1-1**. Planet types are defined by a range of radii, albedo, and orbital elements. The reflected light flux was calculated for each synthetic planet, given its properties, orbit, and phase, and then determining the exposure time required to detect it at signal-to-noise ratio (SNR) = 7. Based on these detection times and the exposure time of a given observation, the fraction of the synthetic planets that are detectable (i.e., the completeness, as a function of exposure time) was calculated. The completeness simply expresses the probability of detecting that planet type, if such a planet exists. The average yield of an observation is the product of the completeness and the occurrence rate of a given planet type. This process is repeated for every observation until the total mission lifetime is exceeded, arriving at an average total mission yield. In reality, yields may vary from this average due to the random distribution of planets and exozodi around individual stars; this source of uncertainty was incorporated in the study's yield calculations by accounting for the Poisson probability distribution of planets and exozodi levels for each star.

The techniques of Stark et al. (2015) and Stark et al. (2016), which optimize the observation plan to maximize the yield of potentially Earth-like planets, were employed. For a coronagraph-based search, this involves optimizing the targets selected for observation, the exposure time of each observation, the delay time between each observation of a given star, and the number of observations of each star (Stark et al. 2015). For a starshade-based search, a similar optimization was made, but the time between observations was not allowed to be optimized due to expected scheduling constraints; instead the balance between fuel use and exposure time was optimized (Stark et al. 2016).

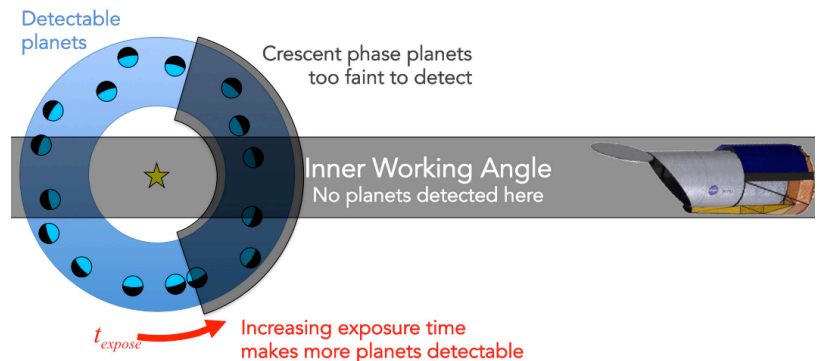

**Figure C.1-1.** The completeness of an observation is the fraction of detectable planets to total planets and is a function of exposure time. The yield of an observation is the product of completeness and the probability that such a planet actually exists (the occurrence rate).

Observations are not directly scheduled. As discussed in *Section C.3.5*, the baseline 4 m HabEx architecture would detect planets and measure orbits with a coronagraph, then measure spectra with a starshade. The ability to schedule the observations is expected to have a negligible impact on the coronagraph-based search given HabEx's large field of regard. However, the ability to schedule the observations is more of an issue for the starshade, which has a smaller field of regard and requires direct scheduling with realistic mission dynamic elements, such as solar angle constraints, to optimally schedule starshade observations and more realistically determine the quality and/or quantity of spectra obtained with the starshade, as shown in *Section C.3.8*.

## C.2    Inputs and Assumptions

Yield estimates require simulating the execution of a mission at a high level. They are therefore dependent on a large number of assumptions about the target stars, planetary systems they host, and the capabilities of the mission. Given the inherent uncertainties in many of these assumptions, consistency between yield analyses is of primary importance. This study adopts inputs and assumptions that are consistent with the choices made by the Exoplanet Standard Definition and Evaluation Team and those made by the Large Ultraviolet/Optical/Infrared Surveyor (LUVOIR) Science and Technology Definition Team (STDT). Fiducial assumptions about the parameters that affect the yield of both the coronagraph and the starshade are now reviewed and justified.

### C.2.1    Astrophysical Assumptions

Astrophysical assumptions include planet types and associated occurrence rates, the brightness and extent of exozodiacal and zodiacal dust that will affect observational performance, and the quality of the data in the target catalog.

### C.2.1.1    Planet Types and Occurrence Rates

HabEx followed the planet categorization scheme of Kopparapu et al. (2018), which consists of a 3 by 5 grid of planets (**Figure C.2-1**) binned by temperature (hot, warm, and cold) and planet radius: small rocky planets (0.5–1 $R_\oplus$), rocky super-Earths (1–1.75 $R_\oplus$), sub-Neptunes (1.75–3.5 $R_\oplus$), Neptune-size planets (3.5–6 $R_\oplus$), and giant planets (6–14.3 $R_\oplus$). Each planet was assigned an albedo (listed in **Figure C.2-1**), a Lambertian scattering phase function, and all planets were assumed to be on circular orbits. The semi-major axis boundaries that define the temperature bins of each planet type are assumed to scale with the bolometric stellar insolation, such that they scale with the square root of the bolometric stellar luminosity.

For exo-Earth candidates, this study adopted the region within the green outline in **Figure C.2-1**. By this definition, exo-Earth candidates are on circular orbits and reside within the conservative habitable zone (HZ), spanning 0.95–1.67 AU for a solar twin (Kopparapu et al. 2013). Only planets with planetary radii smaller than 1.4 $R_\oplus$ and orbital radii larger than or equal to $0.8a^{-0.5}$, where $a$, expressed in AU, is the semi-major axis for a solar twin, were included. The lower limit on this definition of the radius of exo-Earth candidate is derived from an empirical atmospheric loss relationship derived from solar system bodies (Zahnle and Catling 2017). The upper limit on planet radius is a *conservative* interpretation of an empirically measured transition between rocky and gaseous planets at smaller semi-major axes (Rogers 2015). All exo-Earth candidates were assigned Earth's geometric albedo of 0.2, assumed to be valid at all wavelengths of interest.

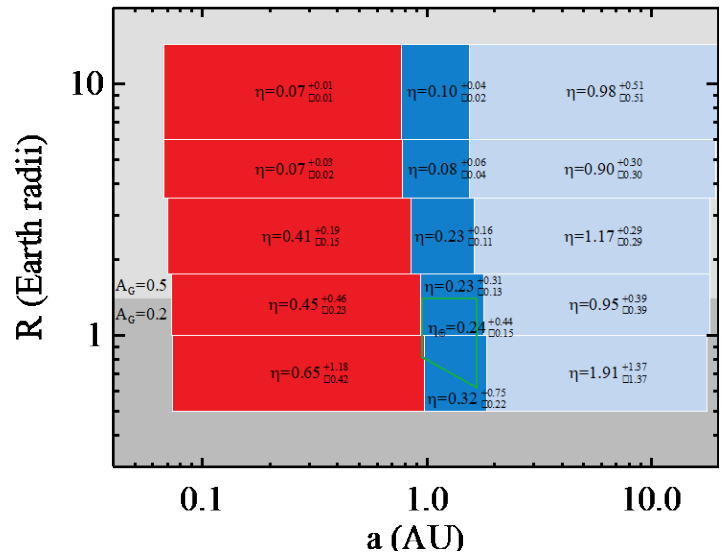

**Figure C.2-1.** Planet classifications for a solar twin used for yield modeling, including bin-integrated occurrence rates (η) and geometric albedos ($A_G$). Planets are binned into hot (*red*), warm (*blue*), and cold (*ice blue*) temperature bins and 5 size bins ranging from small rocky planets to giant planets. The green outline indicates the boundaries of exo-Earth candidates. The semi-major axis boundaries shown are for a solar twin; semi-major axis boundaries are scaled to maintain a constant bolometric insolation.

The occurrence rate values were adopted from the analysis by Dulz et al. (2019). Dulz et al. (2019) is based on the SAG-13 meta-analysis of Kepler data Kopparapu et al. 2018, given by

$$\frac{d^2N(R,P)}{d\ln R \, d\ln P} = \Gamma R^\alpha P^\beta,$$

where $N(R,P)$ is the number of planets per star in a bin centered on radius $R$ and period $P$, $R$ is in $R_E$ and $P$ is in years, and $[\Gamma, a, \beta] = [0.38, -0.19, 0.26]$ for $R < 3.4 \, R_\oplus$ and $[\Gamma, a, \beta] = [0.73, -1.18, 0.59]$ for $R \geq 3.4 \, R_\oplus$. Dulz et al. (2019) update the SAG-13 occurrence rates to address two notable limitations. First, the SAG-13 occurrence rates of planets larger than 10 $R_\oplus$ are uncertain and are roughly a factor of 2 less than measured radial velocity (RV) occurrence rates; Dulz et al. (2019) adopt the occurrence rates of Fernandes et al. (2019) for these planets. Second, extrapolating the Study Analysis Group 13 (SAG-13) fit to our cold planets results in dynamically unstable systems; Dulz et al. (2019) impose simple stability criteria to constrain the occurrence rates of cold planets assuming maximally packed systems. **Figure C.2-1** lists the occurrence rates when integrating over the boundaries of each planet type. Within each planet type, the Dulz et al. (2019) radius and period distribution were used. With this distribution, within a given planet radius and temperature bin, small planets usually outnumber large planets.

The adopted occurrence rates of Dulz et al. (2019) are based on the SAG-13 meta-analysis, which is a crowd-sourced average of published and unpublished occurrence rates, averaged over FGK spectral types. Uncertainties on the SAG-13 occurrence rates are not well understood, and are simply set to the standard deviation of the crowd-sourced values. Because of the large uncertainties in the SAG-13 occurrence rates, this analysis has weak constraints on how occurrence rates change with spectral type. Thus, the analysis simply assumes that the occurrence rates for each planet type bin are independent of spectral type.

In particular, for exo-Earths in the HZ of sunlike stars, the resulting occurrence rate estimate is $\eta_{\text{Earth}} = 0.24^{+0.46}_{-0.16}$. This value is consistent with what is arguably the most careful estimate of $\eta_{\text{Earth}}$ (and its statistical and systematic uncertainties) by the Kepler team itself (Burke et al. 2015). Burke et al. (2015) notes, however, that different but equally plausible methods of treating various systematic errors can change this value by factors of several in either direction. Partly this is due to the fact that any estimate of $\eta_{\text{Earth}}$ from the Kepler survey is necessarily an extrapolation. Nevertheless, pending a more robust estimate of $\eta_{\text{Earth}}$ accounting for all Kepler data, this study adopts the SAG-13 value and uncertainty.

**Table C.2-1** summarizes the key astrophysical assumptions underlying the exo-Earth candidate yield calculations.

### C.2.1.2 Exozodiacal and Zodiacal Dust

Exozodiacal dust adds background noise, thereby reducing the SNR of a planet detection relative to the case of no exozodiacal dust. Recent results from the LBTI HOSTS survey for exozodiacal dust provide constraints on the exozodi distribution. Yield calculations herein adopt the freeform distribution that best fits Large Binocular Telescope Interferometer (LBTI) Hunt for Observable Signatures of Terrestrial Systems (HOSTS) data (Ertel et al. 2019), which has a median of 4.5 zodis of dust and appears bi-modal, with relatively few stars hosting extreme amounts of dust Ertel et al. 2018. Stars were assigned an exozodi level randomly drawn from this distribution.

The LBTI HOSTS survey detected dust around four potential HabEx targets: 297 zodis around Eps Eri, 148 around Tet Boo, 588 zodis around 72 Her, and 235 zodis around 110 Her. For yield calculations, these stars were assigned their LBTI-measured exozodi levels.

The definition of 1 zodi is a uniform (optically-thin) optical depth producing a V band surface brightness of 22 mag arcsec$^{-2}$ at a projected separation of 1 AU around a solar twin. Thus, the exozodi surface brightness drops off as the inverse square of the projected separation (Stark et al. 2014). Because the HZ boundaries scale by the *bolometric* stellar insolation, the V band scattered light surface brightness of 1 zodi of exozodi varies with spectral type (Stark et al. 2014).

**Table C.2-1.** Summary of astrophysical assumptions.

| Parameter | Value | Description |
|---|---|---|
| $\eta_\oplus$ | 0.24 | Fraction of sunlike stars with an exo-Earth candidate |
| $R_p$ | [0.6, 1.4] $R_\oplus$ | Planet radius[a] |
| $a$ | [0.95, 1.67] AU | Semi-major axis[b] |
| $e$ | 0 | Eccentricity (circular orbits) |
| $Cos\ i$ | [−1, 1] | Cosine of inclination (uniform distribution) |
| $\omega$ | [0, 2π] | Argument of pericenter (uniform distribution) |
| $M$ | [0, 2π] | Mean anomaly (uniform distribution) |
| $\Phi$ | Lambertian | Phase function |
| $A_G$ | 0.2 | Geometric albedo of planet from 0.55–1 µm |
| $z_c$ | 23 mag arcsec$^{-2}$ | Average V band surface brightness of zodiacal light for coronagraph observations[c] |
| $z_s$ | 22 mag arcsec$^{-2}$ | Average V band surface brightness of zodiacal light for starshade observations[c] |
| $x$ | 22 mag arcsec$^{-2}$ | V band surface brightness of 1 zodi of exozodiacal dust[d] |
| $n$ | Drawn from LBTI best fit distribution | Number of zodis for all stars |

[a] Lower boundary is a function of $a$ according to the SAG-13 occurrence.
[b] $a$ given for a solar twin. The habitable zone is scaled by $\sqrt{L_*/L_\odot}$.
[c] Local zodi calculated based on ecliptic pointing of telescope. On average, starshade observes into brighter zodiacal light.
[d] For solar twin. Varies with spectral type, as zodi definition fixes optical depth.

The solar system's zodiacal brightness varies with ecliptic latitude and longitude; the closer one observes toward the Sun, the brighter the zodiacal cloud will appear. The zodiacal brightness for each target star is calculated by making simple assumptions about typical telescope pointing (Leinert et al. 1998). For the coronagraph, HabEx assumed that observations could be made near where the local zodi is minimized and adopted a solar longitude of 135 degrees for all targets. For the starshade, the field of regard is limited to solar elongations between 40 and 83 degrees; a constant solar elongation of ~60 degrees was adopted. As a result, the starshade's line of sight through the zodiacal cloud is ~2.5 times brighter than that of the coronagraph.

### C.2.1.3   Target Catalog

The input star catalog was formed using the methods of Stark et al. (2019). Briefly, the target list (*Appendix D*) is equivalent to the union of the Hipparcos New Reduction catalog and the Gaia TGAS catalog. For each star, HabEx adopted the most measured parallax value from the Hipparcos, Gaia TGAS, and GAIA DR2 catalogs, then down-selected to stars within 50 pc. BVI photometry and spectral types were obtained from the Hipparcos catalog. Additional bands and missing spectral types were supplemented using SIMBAD. All stars identified as luminosity class I-III were filtered out, leaving only main sequence stars, sub-giants, and few unclassified luminosity classes. Binary parameters were retrieved from the Washington Double Star catalog, which was cross-referenced with our catalog via SIMBAD.

While the accuracy of any individual star's parameters may be important when planning actual observations, yield estimates can be very robust to these inaccuracies, as their effects average out when considering a large target sample. Accordingly, the blind search portion of HabEx's broad exoplanet survey is expected to be fairly robust to these uncertainties. The exo-Earth yield for the deep survey portion of HabEx's exoplanet search, on the other hand, would be much more sensitive to the uncertainties of the eight individual stars observed.

### C.2.1.4   Propagation of Astrophysical Uncertainties

All major known sources of astrophysical uncertainty are propagated through yield calculations and shown in **Figure 3.3-5**: uncertainty in occurrence rates, uncertainty in exozodi level, and the Poisson noise associated with the planet population and exozodi level of individual stars.

Yield is calculated using the nominal best-fit exozodi distribution from LBTI data, as well as the ±1σ distributions. For each exozodi distribution, 20 yield calculations are performed to sample the Poisson noise associated with individual exozodi levels.

For each exozodi distribution, yield calculations are performed for the pessimistic, nominal, and optimistic occurrence rates, for a total of 180 calculations. Each of the 180 yield calculations results in a list of optimized observations with associated completeness. Monte Carlo draws are then performed for each individual simulation to determine whether a planet is detected in a given observation, sampling the Poisson noise associated with planet populations of individual stars. The number of draws is weighted by the probability distribution of occurrence rate values and exozodi distribution. The detailed shape of these probability distributions are unknown. Normal distributions are assumed and the nominal and ±1σ values are weighted accordingly; the tails of the distributions beyond ±1.5σ are ignored, which are expected to minimally impact the estimated uncertainties.

### C.2.2  Binary Stars

Detecting exoplanets in binary star systems presents additional challenges. Light from companion stars outside of the coronagraph's field of view, but within the field of view of the telescope, will reflect off the primary and secondary mirrors. Due to high-frequency surface figure errors and contamination, some of this light is scattered into the coronagraph's field of view. For some binary systems, this stray light can become brighter than an exo-Earth.

The stray light from binary stars in the final image plane was directly calculated. The numerical stray light models of Sirbu et al. (in prep) were utilized. These models predict the power in the wings of the point spread function (PSF) at large separations assuming a λ/20 root-mean-square

(RMS) surface roughness and an $f^3$ envelope, where $f$ is the spatial frequency of optical aberrations. Stray light was assumed to be measureable, or able to be modeled; it was included simply as an additional source of background noise. This study made no artificial cuts to the target list based on binarity, and allowed the benefit-to-cost optimization in the AYO yield code to determine whether or not stray light noise makes a target unobservable. In practice, the AYO prioritization does reject a number of binary systems with contrast ratios close to unity and/or close separations. It should be noted that including the full amount of light scattered by the companion is actually conservative, as the companion scattered starlight could be actively reduced with specialized observation methods (*Section 12.8*). For example, HabEx could use the starshade to block the companion starlight while observing with the coronagraph (Sirbu et al. 2017a), or use multi-star or super-Nyquist wavefront control coronagraphic techniques (Thomas et al. 2015; Sirbu et al. 2017b).

### C.2.3  Mission Parameters

For all mission concepts investigated, a total lifetime of 5 years was assumed. For the baseline hybrid mission architecture, 2.5 years of total exoplanet science time (including overheads) was allocated, leaving 2.5 years for dedicated observatory science (not counting parallel observations). Of these 2.5 years, 2.25 years are devoted to a broad exoplanet survey optimized for potentially Earth-like planets, and 3 months are devoted to deep survey observations of 8 nearby stars using the starshade (depending on individual exozodi levels). Each coronagraph observation was assigned a 3 min overhead for slewing, a 5 min overhead for dynamic/thermal settling, a 5 h overhead for initial dark hole generation on a reference star, and a 10% tax on science time to account for a single touch-up iteration of WFSC on the science target. Starshade observations were assigned a 3 min overhead for telescope slewing and a 6-hour overhead for SS-telescope alignment. Total exposure time and overheads were required to fit within the exoplanet science time budget. Retargeting transit and slew time of the starshade did not count against the exoplanet science time; it was assumed that during the retargeting either the coronagraph or general astrophysics instruments would be observing targets.

For planet detections, this analysis required an SNR = 7 evaluated over the full bandpass of the detection instrument, where both signal and noise are evaluated in a simple photometric aperture of $0.7\,\lambda/D$ in radius. The SNR was evaluated according to Eq. 7 in Stark et al. (2014), which includes a conservative factor of 2 on the background Poisson noise to account for a simple background subtraction. A background term for detector noise was also included and is discussed in *Section C.2.4.3*. For spectral characterizations, HabEx required a spectrum with $R = 140$ and SNR = 10 per spectral channel, which was evaluated at a wavelength of 0.975 micron.

### C.2.4  Instrument Performance Assumptions

#### C.2.4.1  Coronagraph Assumptions

Coronagraph performance was estimated via end-to-end simulations of the full optical system, using a wave propagation model, incorporating realistic wavefront errors, polarization effects and wavefront control (Krist 2019). The baseline architecture for this report used the vector vortex coronagraph 6 (VVC6) described in detail in *Section 6.3*. Leaked starlight spatial distribution was simulated as a function of stellar diameter and given the estimated pointing jitter before and after FSM correction. Off-axis PSFs were computed as a function of angular separation, providing inputs to the yield calculations according to the standards of Stark (2017).

The detailed end-to-end wave propagation and STOP analyses (*Chapter 6*) include all known effects. However, unknown – and unmodeled - effects may still impact the raw contrast achievable in practice with the vortex coronagraph. As a result, this analysis included a raw contrast floor of $10^{-10}$, which sets the level of shot noise coming from leaked starlight in the SNR calculations. A constant $10^{-10}$ floor is assumed all over the coronagraphic dark hole up to the outer working angle (OWA), meaning that the local level of raw contrast is always set to the worse of $10^{-10}$ and any value predicted by detailed simulations.

**Table C.2-2** summarizes the coronagraph performance that HabEx adopted. Note that although these metrics may provide a useful high-level understanding of coronagraph performance, some metrics should be interpreted with caution. For example, the inner working angle (IWA) estimates where the planet's throughput reaches 50% of the maximum value, but this does not mean that there is no planet signal interior to the IWA. On the contrary, the VVC does provide useful (albeit lower) throughput down to $\sim 2\,\lambda/D$, such that bright, short-period planets may be detectable interior to the quoted $2.4\,\lambda/D$ IWA (62 mas at 0.5 μm). However, only the planets detected by the coronagraph outside of 58 mas can be spectrally characterized by the starshade between 0.3 and 1.0 μm, and only those planets count towards the yields computed here and summarized in *Section 3.3*.

**Table C.2-2.** Summary of adopted vortex coronagraph performance. Listed contrast is for a theoretical point source; contrasts used in simulations included the effects of finite stellar diameter. While only the spatially averaged raw contrast and coronagraph throughput are indicated, AYO simulations used their actual values at the planet angular separation.

| Parameter | Value | Description |
|---|---|---|
| $\zeta$ | $2.5 \times 10^{-10}$ | Raw contrast[a] |
| $\Delta\text{mag}_{floor}$ | 26.5 | Systematic noise floor (faintest detectable point source) |
| $T_{core}$ | 0.50 | Coronagraphic core throughput[b] |
| $T$ | 0.18 | End-to-end facility throughput, including QE but excluding core throughput |
| $IWA_{0.5}$ | $2.4\,\lambda/D$ | Inner working angle[c] |
| $OWA$ | $32\,\lambda/D$ | Outer working angle |
| $\Delta\lambda$ | 20% | Bandwidth |

[a] Value at IWA, assuming a 1 mas diameter star and 0.2 mas rms jitter per axis. Raw contrast curves are calculated based on target star diameter.
[b] Fraction of palent light capatured in photometric region (0.7 $\lambda/D$ in radius)
[c] Separation at which off-axis throughput reaches half the maximum value.

The OWA is the maximum angular separation where planets can be detected, which is set by the size of the dark hole generated by the deformable mirrors (DMs) in the coronagraph. The angular radius of the dark hole is limited to $(N_{act}/2) \times (\lambda/D)$, where $N_{act}$ is the number of DM actuators across the beam diameter; the assumed 32 $\lambda/D$ OWA is consistent with the baseline HabEx design with 64×64 actuators DMs.

The bandpass of the VC is theoretically unlimited, but in practice is limited by the wavefront control system architecture. High-Contrast Imaging Testbed results indicate that surpassing a bandwidth of $\Delta\lambda/\lambda = 0.2$ is challenging with a conventional dual DM coronagraph layout, thereby justifying the adopted bandwidth of 20%.

The total throughput of the system in **Table C.2-2** is evaluated at visible wavelengths and includes the reflectivity of all optical surfaces, the detector quantum efficiency (QE), IFS throughput, and a 5% contamination budget. This throughput metric does not include the core throughput of the coronagraph, which was taken into account separately via the off-axis PSF simulations discussed above. Detector parameters are discussed below.

### C.2.4.2 Starshade Assumptions

Starshade optical performance was also estimated using a wave propagation model. Using the standardized yield metrics of Stark et al. (2019), these performance metrics were then translated into the leaked starlight and off-axis

PSFs for the starshade. **Table C.2-3** summarizes the performance metrics of the starshade adopted in this study.

Sunlight glints, both through reflection and diffraction, from the shape-defining edges of the starshade petals. The edges are nearly razor-sharp to minimize the glint, and are highly specular, resulting in a two-lobe pattern on the sky. The magnitude of the glint was calculated from laboratory measurements of scatter from edge coupons (Martin et al. 2013; Steeves et al. 2018). The surface brightness distribution of solar glint is expected to be repeatable and measurable; it is included as a source of photon noise in exposure time calculations. The surface brightness of glint is faint enough compared to the exozodi surface brightness that it does not impact the yield estimates significantly.

**Table C.2-3** also lists the assumed starshade propulsion parameters. For yield calculations, HabEx assumed that 100 starshade targetings were available for the overall exoplanet surveys (*Section C.3.8*). The surveys consist of a (2.25-year-long) broad survey of 42 stars (about 280 coronagraph observations, and about 60 (up to 75) starshade observations on the most interesting systems) and an additional 3 months of deep survey observations of a select group of 8 stars (about 25 starshade observations). The starshade fuel mass was then computed consistently, to allow a total of observations 100 over the 5-year prime mission, minus the starshade total observing time.

**Table C.2-3.** Summary of adopted starshade performance.

| Parameter | Value | Description |
|---|---|---|
| $\zeta$ | $10^{-10}$ | Raw contrast[a] |
| $\Delta mag_{floor}$ | 26.5 | Systematic noise floor (faintest detectable point source) |
| $T_{core}$ | 0.69 | Starshade core throughput |
| $T$ | 0.20 | End-to-end facility throughput, including QE but excluding core throughput |
| $IWA$ | 58 mas | Inner working angle (constant from 0.3–1 μm) |
| $OWA$ | $\infty$ | Outer working angle |
| $\Delta\lambda$ | 0.7 μm | Instantaneous spectral bandwidth |
| $D_{ss}$ | 52 m | Diameter of starshade |
| $z_{ss}$ | 76,600 km | Telescope-starshade separation |
| $m_{dry}$ | 5,230 kg | Dry mass of starshade spacecraft including contingency |
| $m_{fuel}$ | 5,700 kg | Total mass of starshade propellant |
| $I_{sk}$ | 308 s | Specific impulse of station keeping propellant (chemical) |
| $I_{slew}$ | 3,000 s | Specific impulse of propulsion (see *Section 7.2.3*) |
| $\epsilon_{sk}$ | 0.8 | Efficiency of station keeping fuel noise |
| $T$ | 1.04 N | Thrust |

### C.2.4.3  Detector and Other Performance Assumptions

**Table C.2-4** lists the detector noise parameters that were assumed for yield calculations. The total detector noise count rate in the photometric aperture was calculated as

$$\mathrm{CR_{b,detector}} \approx n_{pix}\left(\xi + \mathrm{RN}^2/\tau_{expose} + 6.73 f\,\mathrm{CIC}\right),$$

where $f$ is the photon counting rate and $n_{pix}$ is the number of pixels contributing to the signal and noise. The parameter $f$ was tuned to each individual target, such that the photon-counting detector time-resolves photons from sources 10 times as bright as an Earth-twin at quadrature.

The analysis assumed that the IFS splits the core of the PSF into 4 lenslets at the shortest wavelength, each of which are dispersed into 6 pixels per spectral channel for a total of 24 pixels per spectral channel at the shortest wavelength. For spectral characterization with the starshade, a larger average—$n_{pix} = 72$ per spectral channel—was adopted, assuming that the starshade (extremely IFS must Nyquist sample at its shortest wavelength). For broadband coronagraphic detections using the imager, the analysis assumed 4 pixels for the core of the planet. Note that the assumed detector noise is sufficiently low that small changes to the number of pixels have a negligible impact on yield.

**Table C.2-4.** Photon-counting CCD noise parameters adopted for yield modeling.

| Parameter | Value | Description |
|-----------|-------|-------------|
| $\xi$ | 3×10⁻⁵ counts pix⁻¹ sec⁻¹ | Dark current |
| *RN* | 0 counts pix⁻¹ read⁻¹ | Read noise (N/A) |
| $\tau_{read}$ | N/A | Read time |
| *CIC* | 1.3×10⁻³ counts pix⁻¹ clock⁻¹ | Clock induced charge |

### C.3  Operations Concepts

Yield is commonly thought of as the number of planets detected and/or characterized. As shown by Stark et al. (2016), the yield of a mission is very sensitive to precisely what measurements are required for "characterization," and how the mission goes about making those measurements. Thus, the yield depends on the science products desired and how the mission conducts the observations.

### C.3.1  Desired Science Products

HabEx is designed to obtain three primary data products on planets identified as exo-Earth candidates:

1. Photometry: to detect planets and measure brightness and color
2. Spectra: to assess chemical composition of atmospheres
3. Orbit measurement: to determine if planet resides in HZ and measure spectro-photometric phase variations

In the following sections, this appendix describes how HabEx would obtain these data products in an efficient manner to maximize the yield of the mission. While the mission observing strategy and scheduling are optimized for the exo-Earth candidate characterizations, many other planets would be observed in these systems and their yields are also calculated in the direct imaging planet yield analysis.

### C.3.2  Dealing with Confusion

Upon initial detection of a possible companion, the nature of the source may be unclear. The mission would have only photometry, possibly one color, and a stellocentric separation to determine the nature of the object. Color, brightness, and the fact the source is unresolved may allow us to discriminate between background galaxies and exoplanets. However, recent work has shown that other planets can mimic the color of exo-Earth candidates (e.g., Krissansen-Totton et al. 2016). Furthermore, planets that most easily mimic Earth are small, hot terrestrial planets that have even higher occurrence rates than exo-Earth candidates (van Gorkom & Stark, in prep), so planet-planet confusion may be common. However, performing costly characterizations on all planets mimicking an Earth could decrease the efficiency of the exoplanet survey and reduce the yield of exo-Earth candidates; there may be a need to disambiguate point sources to identify high priority planets.

HabEx, with its dual coronagraph and starshade design, is capable of dealing with these expected sources of confusion without significantly impacting the yield. As shown by Stark et al. (2016), coronagraphs excel at orbit determination, but take longer to provide a spectrum with broad wavelength coverage. Starshades on the other hand, excel at quickly providing spectra, but can only constrain the orbits for a handful of targets due to the cost of slewing the starshade. Combined, these two instruments provide HabEx with multiple and complementary ways to characterize a system.

### C.3.3    Order of Operations

The order in which observations are conducted and the instrument used to perform those observations would impact the final yield of the mission. For example, performing all initial detection and proper motion follow-up with the starshade would be far from ideal, as this requires many costly slews of the starshade. A more efficient order of operations would play to the strengths of each instrument, e.g., by using the coronagraph for initial detections and then to establish orbits, followed by using the starshade to obtain spectra of interesting systems when planets are known to be at advantageous phases.

Ultimately these decisions would depend on uncertain quantities, like $\eta_{Earth}$ for nearby FGK stars and the rate of confusion with background objects. In a low $\eta_{Earth}$ scenario (~0.1), and because of finite search completeness per visit, HabEx would have to search tens of stars to detect 1 exo-Earth candidate. Because of the fuel cost associated with slewing the starshade, initial detection and orbit determination with the coronagraph would likely be better in this scenario, especially if the confusion rate is high. If $\eta_{Earth}$ is high (~1) and the rate of confusion is low, or if precursor observations with other facilities have already revealed which stars host exo-Earth candidates, it may be better to just search with the starshade and immediately take spectra.

A precise operations concept will require further detailed study and will surely be adapted "on the fly" during mission operations. HabEx's dual instrument design would allow maximum flexibility to adapt to these unavoidable astrophysical uncertainties.

### C.3.4    Simulating Operations Concepts

To simulate a given operations concept, this study would need to generate a fictitious universe and model the execution of the mission one observation at a time, adapting to the detections, non-detections, and false positives as the simulated mission progresses, with decision-making logic. While current yield codes are capable of dynamically scheduling with realistic mission constraints to desired decision making logic Morgan et al. 2017; Savransky and Garrett 2015(Morgan et al. 2017; Savransky and Garrett 2015, *Section C.3.8*), a static time-budgeting approach is more agile for exploration of a variety of operations concepts and is used here.

HabEx approximated the impact of different operations concepts with the AYO yield code by adopting general rules that define the observation plan. For example, to include orbit determination, a system was required to be observed at least six times to a depth consistent with detecting an exo-Earth. To include the effects of confusion, the problem was bounded by assuming that either all systems have a source of confusion in the HZ with the expected flux of an exo-Earth, or none of the systems had a source of confusion in the HZ.

Ten different operations concepts for a variety of HabEx mission designs, for both high and low $\eta_{Earth}$ scenarios, were studied. Each concept produced varying amounts and types of data, ranging from broadband detections only to orbits and spectra. The rows in **Figure C.3-1** show some of the operations concepts considered, with the order of operations for each concept proceeding from left to right. After each step in the operation plan, it is assumed that the information obtained up to that point allows us to perfectly disambiguate exo-Earth candidates (no confusion), or does not provide any disambiguation (100% confusion). Under the assumption of perfect disambiguation, subsequent measurements are only made on the expected yield of exo-Earth, i.e., a fraction of the target sample. Under the assumption of no disambiguation, subsequent measurements must be made on all systems.

### C.3.5 Adopted Broad Survey Operations Concept

By studying all of the operations concepts listed above, this study determined an operations concept that achieved all desired science products listed in **Figure C.3-1** and maximized the exo-Earth yield of the broad survey exoplanet search. Importantly, this operations concept is realistic, flexible, and does not require advanced autonomous decision making on board the spacecraft.

The following operations scenario for the broad survey 2.25-year exoplanet search was adopted:

1. Detect planets using the coronagraph in broad-band filter 1 (450–550 nm) and filter 4 (700–860 nm), providing color information for planets detected in both but only requiring detection in filter 1

2. Revisit *all* systems as necessary with the coronagraph until the orbits of high-priority planets are sufficiently constrained (likely more than 6 times each over the course of months to years)

3. Based on the color (in favorable cases), orbit, and brightness, identify high-priority targets for spectral characterization

4. Schedule and conduct starshade spectral characterization observations (from 0.3– 1 μm) at an advantageous exo-Earth orbital phase, if possible, for all planetary systems with EECs detected. Repeat spectral measurements an average of 3 times, extending wavelength coverage to full 0.2– 1.8 μm region in most favorable cases.

5. Schedule and conduct starshade spectral characterization observations (from 0.3– 1 μm) of all planetary systems with no EECs detected

This operations scenario is both realistic and robust to error. By requiring orbit measurement regardless of what is detected, the operations concept is straightforward, does not rely on any confusion mitigation immediately after a detection, and proper motion would be established for free for all detected planets. Because the HabEx coronagraph's field of regard is nearly the full sky at any given time, the revisit schedule for each star can easily be optimized to maximize detections and constrain orbits without detailed consideration of whether or not the targets are inaccessible. Finally, with an expected yield of ~8 exo-Earth candidates and up to 75 starshade targetings available over 5 years for the broad survey, HabEx expects to be able to measure the spectrum of every exo-Earth candidate multiple times, in addition to every other planetary system at least once. In other words, the yield of characterized exo-Earth candidates is not expected to be limited by the starshade's fuel constraints.

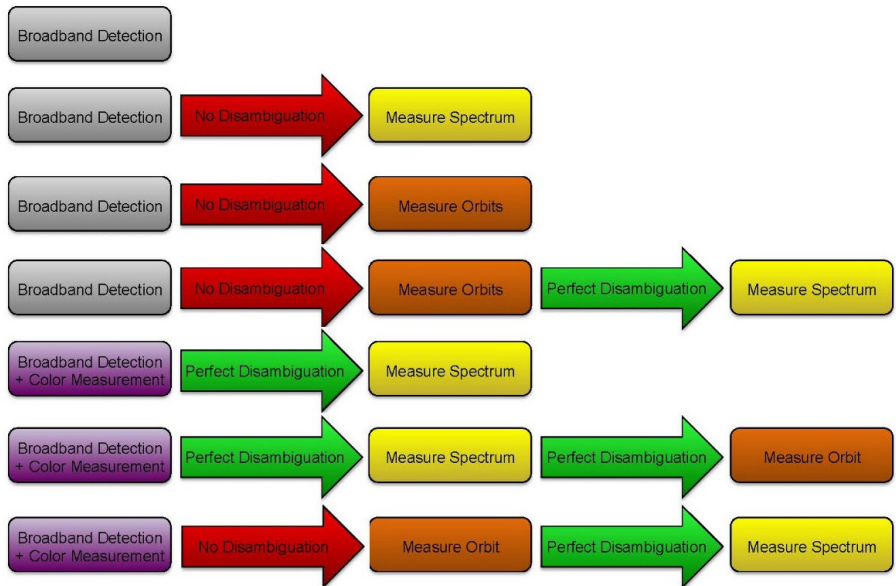

**Figure C.3-1.** A sample of the operations concepts studied for HabEx. Each row presents a simplified operations concept in which a measurement is made, which then provides either perfect disambiguation of exo-Earth candidates, or no disambiguation. The adopted HabEx operations concept is the bottom one, where planetary orbits are measured with the coronagraph before spectra are taken with the starshade, providing the highest science yield (see text for details).

### C.3.6 Adopted Deep Survey Operations Concept

Three months of mission time and 25 starshade targetings are assumed to be devoted to deep survey observations of 8 high-priority nearby stars, separate from the broad survey target list. HabEx would perform spectroscopic observations of each of these systems an average of ~3 times over the course of the mission using the starshade and IFS instruments. Of these 8 targets, more favorable targets may be observed up to 5 times while less favorable targets may be observed only twice.

For each deep survey observation, HabEx would use the starshade to obtain: a) a deep broadband image limited by the systematic noise floor (~26.5 mags fainter than the host star, consistent with detection of a Mars-sized planet in the HZ of a sunlike star); and b) a visible wavelength $R = 140$ IFS spectrum sufficient to obtain SNR = 10 per spectral channel on an Earth-twin at quadrature. These deep exposures and spectra would allow the first detailed understanding of Earth's nearest neighbors.

Deep survey targets are selected based on the high completeness of all planet types that can be achieved with relatively short exposure times and will depend on the real-world exozodi levels of individual nearby stars. **Table C.3-1** lists a representative deep dive target list, which range from late G to late K type stars. **Table C.3-1** also lists the total expected exposure time for each target and the expected detection yields for exo-Earth candidates.

### C.3.7 Combined Exoplanet Survey and Overall Planet Yields

The overall 2.5-year exoplanet survey consists of:

- 2.25 years of broad survey operations, including coronagraph multi-epoch

**Table C.3-1.** Example "deep survey" target list for HabEx; stars may vary depending on real-world exozodi levels. Columns are, from left to right, star name, distance, and spectral type, followed by the total exposure time devoted to this target (sum of 3 visits), and the total expected exo-Earth candidate yield. For each of these stars the exo-Earth HZ search completeness is close to 100%, and the expected yield is then close to the assumed occurrence rate of exo-Earths around sunlike stars (0.24).

| Name | Dist (pc) | Type | $\Sigma_T$ (days) | $\Sigma Y_{EEC}$ |
|---|---|---|---|---|
| τ Ceti | 3.69 | G8V | 0.87 | 0.24 |
| 82 Eri | 6.04 | G8V | 1.02 | 0.24 |
| ε Indi | 3.62 | K5V | 1.69 | 0.24 |
| 40 Eri | 4.98 | K1V | 1.93 | 0.24 |
| σ Dra | 5.76 | K0V | 2.13 | 0.24 |
| GJ570 | 5.80 | K4V | 5.05 | 0.24 |
| 61 Cyg A | 3.49 | K5V | 12.48 | 0.24 |
| 61 Cyg B | 3.49 | K7V | 16.44 | 0.21 |

detections (1.1 yrs) followed by multi-epoch starshade spectra of all planetary systems with exo-Earths candidates detected (0.65 yr) and single epoch spectra of all other planetary systems (0.5 yr);

- 3 months of deep survey, using the starshade only for broad-band large OWA imaging and spectral characterizations at 3 different epochs.

The characteristics of these broad and deep surveys are summarized in **Figure C.3-2.** The overall planet yield expected from the 2 surveys is summarized in **Table C.3-2**, using the nominal instrumental parameters but a variety of planet occurrence rate assumptions, ranging from pessimistic, to nominal to optimistic. For each planet type, the nominal case refers to the mean occurrence rate derived by Dulz et al. (2019). The pessimistic and optimistic yield estimates assume the $\pm 1\sigma$ limits on planet occurrence rates (**Figure C.2-1**).

### C.3.8 Scheduling of Exoplanet Surveys Observations

The adopted broad and deep surveys operation concepts were also simulated with EXOSIMS (Savransky et al. 2017), detailed in *Section 8.2*, a design reference mission (DRM) code that uses a different approach to planet yield estimation than the AYO algorithm statistical completeness approach discussed in the previous sections. With EXOSIMS, many realizations of the universe are drawn, each with a different planet distribution around individual target stars, resulting in a different scheduling scheme and planet yield for each draw. As illustrated in **Figure C.3-2**, EXOSIMS was also used to check that the target observations prioritized by the AYO algorithm were indeed schedulable using realistic mission factors such as solar keep-out (grey circle with yellow edge centered on "S" for Sun), starshade glint constraints (field of regard is the white region where the sun angle <83°), slew times and fuel use. The starshade slew path (black arrows) is scheduled with a three-step look-ahead Traveling Salesman Problem optimizer, and spectral characterization occurs at the end of each arrow, as prescribed in the broad and deep dive operations concepts.

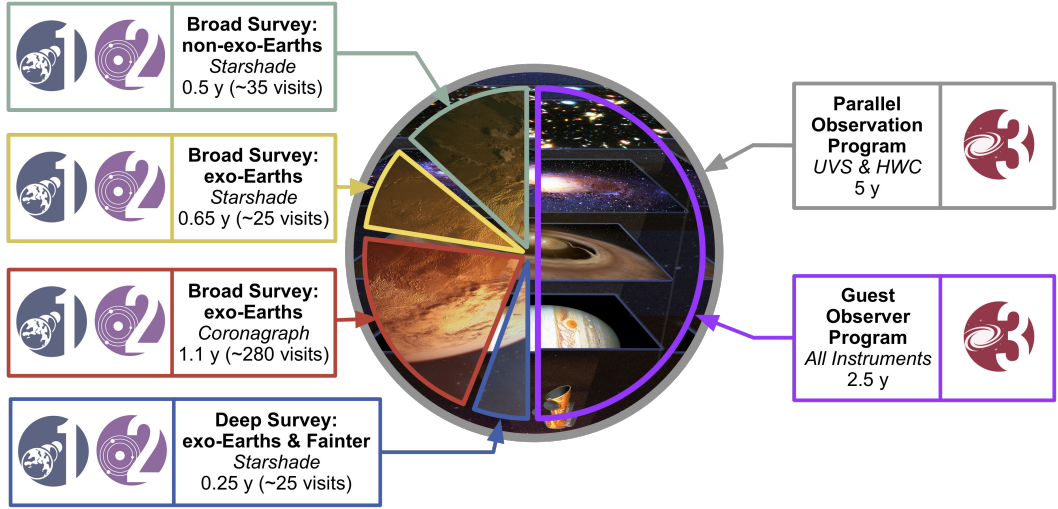

**Figure C.3-2.** HabEx time allocation for a nominal 5-year mission. The broad-survey uses both the coronagraph (for multi-epoch imaging) and the starshade (for spectroscopy). The deep survey only uses the starshade.

**Table C.3-2**. HabEx yield estimates for different planet types. As indicated in Figure C.2-1, planets are categorized by a range of surface temperatures (hot, warm, and cold) and planetary radii: small rocky planets (0.5–1 $R_\oplus$), large rocky planets (super-Earths 1–1.75 $R_\oplus$), sub-Neptune size (1.75–3.5 $R_\oplus$), Neptune-size (3.5–6 $R_\oplus$) and giant planets (6–14.3 $R_\oplus$). HZ exo-Earth candidates occupy a subset of the rocky planets bins, and their yield is given in the 2nd column. Planet yields are indicated for the broad survey, the deep survey and the combination of both. Assumed occurrence rates from Dulz et al. (2019) are consistent with estimates from the SAG-13 meta-analysis of Kepler data for hot and warm planets; occurrence rate upper limits from Dulz et al. (2019) are lower than SAG-13 upper limits for cold planets due to dynamic stability constraints. "Nominal", "pessimistic" and "optimistic" planet yield estimates are given from top to bottom. They correspond to the nominal, +1σ and -1σ planet occurrence rates (e.g., for Earth-like planets, $\eta_{Earth}$ = 0.24, 0.08, and 0.70, respectively) and also account for Poisson noise uncertainty.

| Planet Types | exo- Earths | Hot Small Rocky | Warm Small Rocky | Cold Small Rocky | Hot Super Earths | Warm Super Earths | Cold Super Earths | Hot Sub-Neptunes | Warm Sub-Neptunes | Cold Sub-Neptunes | Hot Neptune | Warm Neptune | Cold Neptune | Hot Jupiter | Warm Jupiter | Cold Jupiter |
|---|---|---|---|---|---|---|---|---|---|---|---|---|---|---|---|---|
| **Nominal Planet Occurrence Rates** | | | | | | | | | | | | | | | | |
| Planet Yields from Broad Survey | 6.4 | 6.9 | 3.1 | 0.6 | 11.3 | 8.9 | 11.1 | 10.9 | 9.9 | 31.6 | 2.1 | 3.4 | 22.0 | 2.2 | 4.5 | 20.5 |
| Planet Yields from Deep Survey | 1.4 | 1.6 | 1.8 | 2.6 | 1.1 | 1.3 | 3.9 | 0.9 | 1.3 | 6.3 | 0.2 | 0.5 | 4.4 | 0.2 | 0.6 | 3.7 |
| *Planet Yields from Both Surveys* | *7.8* | *8.6* | *4.9* | *3.7* | *12.4* | *10.2* | *15.1* | *11.8* | *11.2* | *37.3* | *2.3* | *3.8* | *26.5* | *2.4* | *5.1* | *24.2* |
| **Pessimistic Planet Occurrence Rates** | | | | | | | | | | | | | | | | |
| Planet Yields from Broad Survey | 2.6 | 2.5 | 1.0 | 0.1 | 5.8 | 4.2 | 4.8 | 7.6 | 6.3 | 18.7 | 1.5 | 2.1 | 11.9 | 2.1 | 4.1 | 15.5 |
| Planet Yields from Deep Survey | 0.5 | 0.5 | 0.6 | 0.7 | 0.5 | 0.5 | 1.5 | 0.6 | 0.7 | 3.4 | 0.1 | 0.2 | 2.2 | 0.2 | 0.5 | 2.2 |
| Planet Yields from Both Surveys | 3.2 | 3.0 | 1.6 | 0.8 | 6.3 | 4.8 | 6.3 | 8.2 | 7.1 | 22.1 | 1.6 | 2.4 | 14.1 | 2.3 | 4.6 | 17.7 |
| **Optimistic Planet Occurrence Rates** | | | | | | | | | | | | | | | | |
| Planet Yields from Broad Survey | 14.3 | 16.2 | 8.1 | 1.1 | 18.8 | 16.1 | 11.7 | 13.9 | 12.6 | 23.5 | 2.7 | 4.7 | 20.1 | 2.1 | 4.7 | 23.6 |
| Planet Yields from Deep Survey | 4.0 | 4.6 | 6.0 | 6.6 | 2.4 | 3.1 | 4.8 | 1.7 | 2.3 | 5.7 | 0.3 | 0.8 | 5.3 | 0.3 | 0.8 | 6.2 |
| Planet Yields from Both Surveys | 18.3 | 20.8 | 14.1 | 7.7 | 21.2 | 19.2 | 16.5 | 15.6 | 14.8 | 29.2 | 3.0 | 5.5 | 25.3 | 2.3 | 5.6 | 29.8 |

During the starshade repositionings, coronagraph observations are scheduled and time allocated to other observatory science (i.e., using the HWC and UVS instruments). The synthetic planets are 'observed' and considered detected or characterized if the goal SNR is reached: green for rocky planets in the HZ, purple for all other planets including rocky planets not in the HZ, red for insufficient SNR to detect any planets, grey for an unobserved star, all from a broad list of ~760 potential target stars. The size of the circle indicates the number of repeat detections or characterizations, with the case of 4 detections shown in the legend for scale. Spectral characterizations with the starshade are distinguished by a black edge to the circle and are at the tip of a black slew arrow. The simulated 5-year DRM shown here, one of a Monte Carlo ensemble of DRMs, performed the deep survey and the follow-up characterization of coronagraph-discovered planets (broad survey) with 100 starshade slews (**Figure C.3-3**). These current EXOSIMS results make us highly confident that the observations above are indeed schedulable when taking dynamics mission constraints into account.

Preliminary cross-checks of th yields predicted by the AYO and EXOSIMS algorithms is showing reasonable agreement. A full cross-check of yields and cross-validation of physics modeling will be presented in the final report of Standards Definition and Evaluation Team (Morgan in prep.).

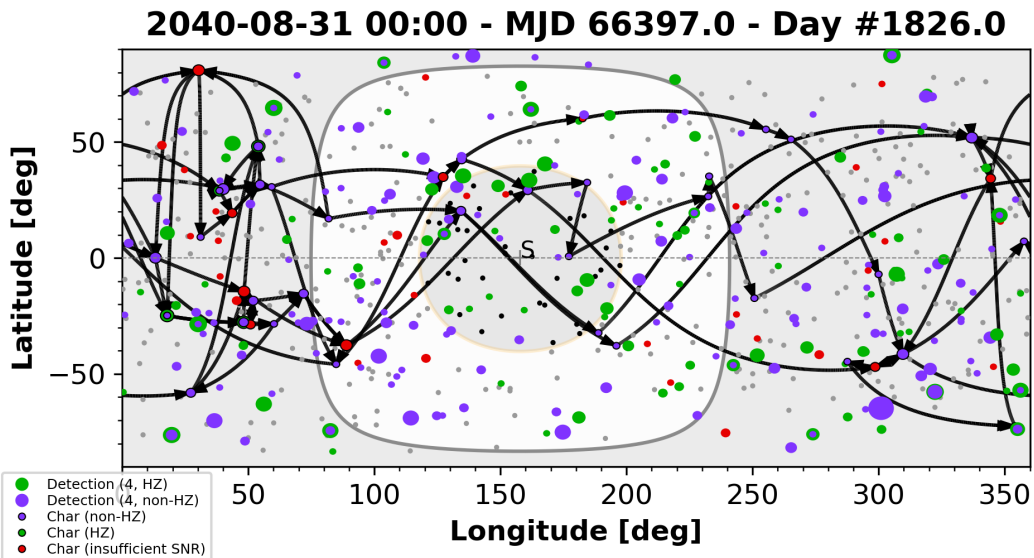

**Figure C.3-3.** EXOSIMS design reference mission simulation scheduling observations planned for the broad and deep survey operations concept. Starshade retargeting transits over a nominal 5-year mission (2035–2040) are indicated by black arrows. Using realistic mission factors such as solar field-of-regard, transit times and fuel usage, 100 starshade transits can be accommodated with fuel margin.

# D TARGET LISTS FOR EXOPLANET SURVEYS

We show hereafter two different lists of prime targets for HabEx exoplanet direct imaging surveys.

The first, larger "master list" of 150 stars (**Table D-1**) is obtained assuming that the full 5-year duration of the HabEx prime mission is devoted to exo-Earths, and is split between exo-Earth candidate (EEC) searches with the coronagraph (around 0.5 μm) and spectral characterization of EECs with the starshade (0.30–1.00 μm). Target selection and overall Design Reference Mission (DRM) optimization (visiting epochs and duration) are obtained by applying the altruistic yield optimization (AYO) algorithm (*Appendix C*) to maximize the EEC yield of the HabEx baseline architecture (4H), assuming that there is no exozodi emission around any of the targets. This list provides an absolute upper limit to the EEC yield achievable over 5 years under ideal conditions. It extends to ~22 pc and represents the main overall sample that HabEx exoplanet "broad" and "deep" EEC surveys can draw from.

The second shorter list of 50 stars (**Table D-2**, all stars within ~15 pc) represents a more realistic subset of targets to be observed in 2 years of total time under non-ideal conditions. This list is obtained by randomly assigning exozodi levels to each target in the previous list (consistent with the distribution inferred by the Large Binocular Telescope Interferometer (LBTI) exozodi survey, *Section 3.3*) and using the AYO algorithm to optimize the total number of EECs that can be detected by the coronagraph (at 0.5 μm) and spectrally characterized by the starshade (0.30–1.0 μm) over 2 years. In this case, the optimum sequence of observations found by AYO depends on the actual exozodi level assigned to each star. This assumes that the exozodi level is known around each star prior to the HabEx mission, or, more realistically that high exozodi levels can be precisely measured after a single observation of the system: both scenarios produce very similar EEC characterization yields (Stark et al. 2015).

In both tables, target stars are ordered by increasing distance and the 8 deep survey targets are highlighted in *purple*.

*Column Headings:*

A. Star number, ranked by increasing distance
B. Star HIP number
C. Visible apparent magnitude
D. Distance in parsec
E. Spectral type
F. Earth equivalent insolation distance in milliarcsec (mas)
G. Inner edge of the habitable zone (HZ) (in mas)
H. Outer edge of the habitable zone (in mas)
H2. Randomly drawn exozodi level (relative to the solar zodi level) (For **Table D-2** only)
I. Total exposure time used for broadband imaging summed over all visits (in days)
J. Total completeness for exo-Earths in the habitable zone, summed over all visits
K. Total number of exo-Earth candidates statistically detected, assuming an exo-Earth occurrence rate of 0.243
L. Cumulative completeness (CC) for exo-Earths in the habitable zone, summed overall all target stars up to the current one. For example (**Table D-2**), observing the nearest 26 stars in the list (targets closer than HIP 61317 = β CVn located at 8.6 pc) yields a cumulative completeness of ~20, meeting Objective O1 baseline requirement.
M. Average completeness (AC) for HZ exo-Earths searches, computed for all target stars up to the current one. For example (**Table D-2**), the nearest 26 in the list (targets within 8.6 pc) are observed with an average completeness per star of ~80%.

**Table D-1.** List of target stars obtained when optimizing the detection and spectral characterization of EECs with HabEx baseline architecture (4H), assuming 5 years of observations and no exozodi emission.

| A | B | C | D | E | F | G | H | I | J | K | L | M |
|---|---|---|---|---|---|---|---|---|---|---|---|---|
| Star Number | HIP | Vmag | Dist (pc) | Type | EEID (mas) | HZ Inner Edge (mas) | HZ Outer Edge (mas) | BB Imaging Time (days) | EEC Complete-ness | EEC Yield | CC | AC |
| 1 | 54035 | 7.49 | 2.55 | M2V | 61 | 58 | 102 | 3.95 | 0.987 | 0.245 | 0.987 | 0.987 |
| 2 | 16537 | 3.72 | 3.20 | K2V | 184 | 175 | 307 | 3.43 | 1.000 | 0.248 | 1.987 | 0.994 |
| 3 | 114046 | 7.35 | 3.29 | M2/M3V | 61 | 58 | 102 | 5.06 | 0.963 | 0.239 | 2.950 | 0.983 |
| 4 | 104217 | 6.05 | 3.49 | K7V | 85 | 81 | 143 | 3.51 | 0.997 | 0.247 | 3.947 | 0.987 |
| 5 | 104214 | 5.2 | 3.50 | K5V | 103 | 98 | 172 | 2.57 | 1.000 | 0.248 | 4.947 | 0.989 |
| 6 | 37279 | 0.4 | 3.51 | F5IV-V | 757 | 719 | 1265 | 3.24 | 0.345 | 0.086 | 5.291 | 0.882 |
| 7 | 1475 | 8.09 | 3.56 | M1V | 56 | 53 | 93 | 6.37 | 0.795 | 0.197 | 6.086 | 0.869 |
| 8 | 8102 | 3.49 | 3.60 | G8V | 194 | 184 | 323 | 2.65 | 0.999 | 0.248 | 7.085 | 0.886 |
| 9 | 108870 | 4.69 | 3.64 | K5V | 129 | 123 | 216 | 3.65 | 1.000 | 0.248 | 8.085 | 0.898 |
| 10 | 105090 | 6.69 | 3.97 | M1/M2V | 70 | 67 | 117 | 5.66 | 0.980 | 0.243 | 9.064 | 0.906 |
| 11 | 49908 | 6.6 | 4.87 | K8V | 67 | 64 | 112 | 6.47 | 0.964 | 0.239 | 10.029 | 0.912 |
| 12 | 19849 | 4.43 | 5.04 | K1V | 129 | 122 | 215 | 4.84 | 0.999 | 0.248 | 11.028 | 0.919 |
| 13 | 97649 | 0.76 | 5.13 | A7IV-V | 629 | 597 | 1050 | 1.63 | 0.219 | 0.054 | 11.247 | 0.865 |
| 14 | 25878 | 7.97 | 5.70 | M1V | 45 | 43 | 75 | 3.15 | 0.125 | 0.031 | 11.371 | 0.812 |
| 15 | 96100 | 4.67 | 5.77 | K0V | 115 | 109 | 192 | 5.34 | 0.998 | 0.248 | 12.370 | 0.825 |
| 16 | 3821 | 3.46 | 5.84 | G0V | 189 | 180 | 316 | 11.60 | 0.766 | 0.190 | 13.136 | 0.821 |
| 17 | 73184 | 5.72 | 5.88 | K4V | 79 | 75 | 132 | 4.08 | 0.543 | 0.135 | 13.678 | 0.805 |
| 18 | 84478 | 6.33 | 5.95 | K5V | 67 | 63 | 112 | 6.75 | 0.903 | 0.224 | 14.581 | 0.810 |
| 19 | 15510 | 4.26 | 6.00 | G8V | 135 | 128 | 226 | 6.13 | 0.998 | 0.248 | 15.579 | 0.820 |
| 20 | 99240 | 3.55 | 6.10 | G5IV-Vvar | 189 | 180 | 316 | 5.60 | 0.986 | 0.245 | 16.566 | 0.828 |
| 21 | 114622 | 5.57 | 6.53 | K3Vvar | 83 | 79 | 139 | 5.82 | 0.976 | 0.242 | 17.542 | 0.835 |
| 22 | 12114 | 5.79 | 7.24 | K3V | 72 | 68 | 120 | 3.84 | 0.334 | 0.083 | 17.876 | 0.813 |
| 23 | 3765 | 5.74 | 7.44 | K2V | 73 | 69 | 121 | 7.83 | 0.938 | 0.233 | 18.814 | 0.818 |
| 24 | 2021 | 2.82 | 7.46 | G2IV | 256 | 243 | 427 | 4.07 | 0.770 | 0.191 | 19.584 | 0.816 |
| 25 | 7981 | 5.24 | 7.61 | K1V | 89 | 85 | 149 | 7.69 | 0.958 | 0.238 | 20.542 | 0.822 |
| 26 | 113283 | 6.48 | 7.61 | K4Vp | 58 | 55 | 97 | 9.19 | 0.806 | 0.200 | 21.347 | 0.821 |
| 27 | 113368 | 1.17 | 7.70 | A3V | 530 | 504 | 886 | 1.79 | 0.061 | 0.015 | 21.408 | 0.793 |
| 28 | 22449 | 3.19 | 8.04 | F6V | 210 | 200 | 351 | 5.27 | 0.849 | 0.211 | 22.258 | 0.795 |
| 29 | 64924 | 4.74 | 8.51 | G5V | 108 | 103 | 181 | 8.13 | 0.950 | 0.236 | 23.207 | 0.800 |
| 30 | 1599 | 4.23 | 8.53 | F9V | 133 | 126 | 221 | 8.76 | 0.978 | 0.243 | 24.186 | 0.806 |
| 31 | 61317 | 4.24 | 8.61 | G0V | 132 | 126 | 221 | 8.21 | 0.971 | 0.241 | 25.157 | 0.812 |
| 32 | 32984 | 6.58 | 8.75 | K3V | 55 | 52 | 91 | 5.24 | 0.494 | 0.123 | 25.651 | 0.802 |
| 33 | 99825 | 5.73 | 8.80 | K3V | 73 | 69 | 121 | 9.33 | 0.877 | 0.217 | 26.528 | 0.804 |
| 34 | 23311 | 6.22 | 8.85 | K3V | 64 | 60 | 106 | 7.14 | 0.725 | 0.180 | 27.253 | 0.802 |
| 35 | 27072 | 3.59 | 8.88 | F7V | 175 | 166 | 292 | 6.91 | 0.896 | 0.222 | 28.149 | 0.804 |
| 36 | 17378 | 3.52 | 9.07 | K0IV | 204 | 194 | 340 | 7.96 | 0.780 | 0.193 | 28.929 | 0.804 |
| 37 | 15457 | 4.84 | 9.15 | G5Vvar | 102 | 97 | 171 | 7.11 | 0.912 | 0.226 | 29.841 | 0.807 |
| 38 | 57939 | 6.42 | 9.18 | G8Vp | 51 | 48 | 84 | 8.69 | 0.647 | 0.160 | 30.487 | 0.802 |
| 39 | 64394 | 4.23 | 9.18 | G0V | 133 | 126 | 221 | 9.58 | 0.964 | 0.239 | 31.452 | 0.806 |
| 40 | 105858 | 4.21 | 9.27 | F6V | 132 | 125 | 220 | 9.74 | 0.972 | 0.241 | 32.424 | 0.811 |
| 41 | 57443 | 4.89 | 9.29 | G3/G5V | 99 | 94 | 166 | 7.62 | 0.943 | 0.234 | 33.367 | 0.814 |
| 42 | 56452 | 5.96 | 9.54 | K0V | 64 | 60 | 106 | 10.63 | 0.881 | 0.219 | 34.248 | 0.815 |
| 43 | 56997 | 5.31 | 9.58 | G8Vvar | 84 | 79 | 140 | 8.57 | 0.902 | 0.224 | 35.151 | 0.817 |
| 44 | 81300 | 5.77 | 9.92 | K2V | 70 | 66 | 116 | 6.62 | 0.778 | 0.193 | 35.928 | 0.817 |
| 45 | 8362 | 5.63 | 10.04 | K0V | 74 | 70 | 124 | 10.58 | 0.894 | 0.222 | 36.823 | 0.818 |
| 46 | 68184 | 6.49 | 10.08 | K3V | 56 | 53 | 93 | 5.47 | 0.434 | 0.108 | 37.257 | 0.810 |

Table D-1. List of target stars obtained when optimizing the detection and spectral characterization of EECs with HabEx baseline architecture (4H), assuming 5 years of observations and no exozodi emission.

| A | B | C | D | E | F | G | H | I | J | K | L | M |
|---|---|---|---|---|---|---|---|---|---|---|---|---|
| Star Number | HIP | Vmag | Dist (pc) | Type | EEID (mas) | HZ Inner Edge (mas) | HZ Outer Edge (mas) | BB Imaging Time (days) | EEC Complete-ness | EEC Yield | CC | AC |
| 47 | 86400 | 6.53 | 10.09 | K3V | 52 | 50 | 87 | 4.34 | 0.319 | 0.079 | 37.575 | 0.799 |
| 48 | 29271 | 5.08 | 10.21 | G5V | 93 | 88 | 155 | 10.55 | 0.947 | 0.235 | 38.522 | 0.803 |
| 49 | 13402 | 6.05 | 10.36 | K1V | 62 | 59 | 104 | 10.96 | 0.754 | 0.187 | 39.276 | 0.802 |
| 50 | 14632 | 4.05 | 10.51 | G0V | 144 | 137 | 241 | 10.22 | 0.889 | 0.221 | 40.165 | 0.803 |
| 51 | 57632 | 2.14 | 11.00 | A3Vvar | 343 | 326 | 572 | 2.62 | 0.097 | 0.024 | 40.263 | 0.789 |
| 52 | 10644 | 4.84 | 11.01 | G0V | 101 | 96 | 168 | 11.70 | 0.891 | 0.221 | 41.154 | 0.791 |
| 53 | 88972 | 6.38 | 11.10 | K2V | 54 | 51 | 90 | 4.39 | 0.371 | 0.092 | 41.524 | 0.783 |
| 54 | 57757 | 3.59 | 11.12 | F8V | 176 | 167 | 294 | 8.04 | 0.726 | 0.180 | 42.251 | 0.782 |
| 55 | 3093 | 5.88 | 11.14 | K0V | 67 | 64 | 112 | 5.88 | 0.589 | 0.146 | 42.840 | 0.779 |
| 56 | 12777 | 4.1 | 11.15 | F7V | 139 | 132 | 232 | 16.26 | 0.767 | 0.190 | 43.606 | 0.779 |
| 57 | 78072 | 3.85 | 11.18 | F6V | 155 | 147 | 259 | 9.20 | 0.813 | 0.202 | 44.419 | 0.779 |
| 58 | 42808 | 6.58 | 11.19 | K2V | 50 | 47 | 83 | 4.40 | 0.219 | 0.054 | 44.639 | 0.770 |
| 59 | 47080 | 5.4 | 11.20 | G8IV-V | 82 | 77 | 136 | 3.66 | 0.168 | 0.042 | 44.806 | 0.759 |
| 60 | 72848 | 6 | 11.38 | K2V | 63 | 60 | 106 | 10.69 | 0.676 | 0.168 | 45.482 | 0.758 |
| 61 | 67927 | 2.68 | 11.40 | G0IV | 271 | 257 | 452 | 3.88 | 0.242 | 0.060 | 45.725 | 0.750 |
| 62 | 23693 | 4.71 | 11.62 | F7V | 105 | 100 | 176 | 12.24 | 0.915 | 0.227 | 46.639 | 0.752 |
| 63 | 109176 | 3.77 | 11.80 | F5V | 160 | 152 | 268 | 8.41 | 0.747 | 0.185 | 47.387 | 0.752 |
| 64 | 77257 | 4.42 | 11.82 | G0Vvar | 122 | 116 | 204 | 12.37 | 0.867 | 0.215 | 48.254 | 0.754 |
| 65 | 15330 | 5.53 | 12.04 | G2V | 74 | 70 | 123 | 9.93 | 0.812 | 0.201 | 49.066 | 0.755 |
| 66 | 15371 | 5.24 | 12.05 | G1V | 84 | 79 | 140 | 11.82 | 0.878 | 0.218 | 49.943 | 0.757 |
| 67 | 80686 | 4.9 | 12.18 | F9V | 97 | 92 | 162 | 12.22 | 0.886 | 0.220 | 50.829 | 0.759 |
| 68 | 41926 | 6.38 | 12.18 | K0V | 52 | 50 | 87 | 4.24 | 0.226 | 0.056 | 51.055 | 0.751 |
| 69 | 26779 | 6.21 | 12.28 | K1V | 57 | 55 | 96 | 5.04 | 0.332 | 0.082 | 51.387 | 0.745 |
| 70 | 24813 | 4.69 | 12.48 | G0V | 108 | 103 | 181 | 11.06 | 0.764 | 0.190 | 52.151 | 0.745 |
| 71 | 40693 | 5.95 | 12.56 | K0V | 63 | 60 | 105 | 9.40 | 0.573 | 0.142 | 52.724 | 0.743 |
| 72 | 43587 | 5.96 | 12.59 | G8V | 65 | 62 | 109 | 7.41 | 0.448 | 0.111 | 53.172 | 0.739 |
| 73 | 58576 | 5.54 | 12.70 | K0IV | 76 | 72 | 127 | 13.36 | 0.677 | 0.168 | 53.849 | 0.738 |
| 74 | 85235 | 6.44 | 12.79 | K0V | 50 | 48 | 84 | 4.37 | 0.171 | 0.042 | 54.020 | 0.730 |
| 75 | 10798 | 6.33 | 12.83 | G8V | 52 | 50 | 87 | 7.12 | 0.252 | 0.063 | 54.272 | 0.724 |
| 76 | 80337 | 5.37 | 12.91 | G3/G5V | 79 | 75 | 132 | 12.59 | 0.743 | 0.184 | 55.015 | 0.724 |
| 77 | 51459 | 4.82 | 12.91 | F8V | 100 | 95 | 168 | 12.02 | 0.828 | 0.205 | 55.843 | 0.725 |
| 78 | 22263 | 5.49 | 13.24 | G3V | 75 | 71 | 125 | 14.69 | 0.796 | 0.197 | 56.639 | 0.726 |
| 79 | 98036 | 3.71 | 13.38 | G8IVvar | 183 | 174 | 306 | 11.87 | 0.471 | 0.117 | 57.110 | 0.723 |
| 80 | 7513 | 4.1 | 13.41 | F8V | 140 | 133 | 233 | 12.18 | 0.740 | 0.184 | 57.850 | 0.723 |
| 81 | 116771 | 4.13 | 13.43 | F7V | 137 | 130 | 229 | 13.33 | 0.751 | 0.186 | 58.601 | 0.723 |
| 82 | 107556 | 2.85 | 13.63 | A5mF2(IV) | 242 | 230 | 404 | 4.03 | 0.186 | 0.046 | 58.787 | 0.717 |
| 83 | 544 | 6.07 | 13.78 | K0V | 59 | 56 | 99 | 4.95 | 0.250 | 0.062 | 59.037 | 0.711 |
| 84 | 53721 | 5.03 | 13.80 | G0V | 93 | 88 | 155 | 12.96 | 0.729 | 0.181 | 59.766 | 0.712 |
| 85 | 16852 | 4.29 | 13.96 | F9V | 129 | 123 | 215 | 13.62 | 0.724 | 0.180 | 60.490 | 0.712 |
| 87 | 79672 | 5.49 | 14.13 | G1V | 75 | 71 | 125 | 15.35 | 0.676 | 0.168 | 61.167 | 0.711 |
| 86 | 12843 | 4.47 | 14.14 | F5/F6V | 117 | 111 | 195 | 13.67 | 0.770 | 0.191 | 61.937 | 0.712 |
| 88 | 102485 | 4.13 | 14.24 | F5V | 136 | 129 | 227 | 13.32 | 0.704 | 0.175 | 62.642 | 0.712 |
| 89 | 102422 | 3.41 | 14.35 | K0IV | 214 | 203 | 357 | 7.42 | 0.228 | 0.056 | 62.869 | 0.706 |
| 90 | 70497 | 4.04 | 14.39 | F7V | 143 | 135 | 238 | 10.94 | 0.654 | 0.162 | 63.524 | 0.706 |
| 91 | 42438 | 5.63 | 14.45 | G1.5Vb | 70 | 67 | 117 | 11.98 | 0.605 | 0.150 | 64.129 | 0.705 |
| 92 | 28103 | 3.71 | 14.51 | F1V | 163 | 154 | 272 | 7.68 | 0.512 | 0.127 | 64.641 | 0.703 |

**Table D-1.** List of target stars obtained when optimizing the detection and spectral characterization of EECs with HabEx baseline architecture (4H), assuming 5 years of observations and no exozodi emission.

| A | B | C | D | E | F | G | H | I | J | K | L | M |
|---|---|---|---|---|---|---|---|---|---|---|---|---|
| Star Number | HIP | Vmag | Dist (pc) | Type | EEID (mas) | HZ Inner Edge (mas) | HZ Outer Edge (mas) | BB Imaging Time (days) | EEC Complete-ness | EEC Yield | CC | AC |
| 93 | 84862 | 5.38 | 14.54 | G0V | 79 | 75 | 132 | 13.62 | 0.699 | 0.173 | 65.340 | 0.703 |
| 94 | 25278 | 5 | 14.59 | F8V | 93 | 88 | 155 | 14.28 | 0.684 | 0.170 | 66.024 | 0.702 |
| 95 | 75181 | 5.65 | 14.69 | G2V | 70 | 66 | 116 | 14.76 | 0.622 | 0.154 | 66.645 | 0.702 |
| 96 | 47592 | 4.93 | 14.82 | G0V | 95 | 91 | 159 | 14.10 | 0.734 | 0.182 | 67.380 | 0.702 |
| 97 | 49081 | 5.37 | 14.93 | G1V | 80 | 76 | 134 | 16.18 | 0.646 | 0.160 | 68.026 | 0.701 |
| 98 | 59199 | 4.02 | 14.95 | F0IV/V | 141 | 134 | 235 | 10.90 | 0.624 | 0.155 | 68.649 | 0.701 |
| 99 | 95447 | 5.17 | 14.96 | G8IVvar | 90 | 86 | 151 | 17.10 | 0.681 | 0.169 | 69.330 | 0.700 |
| 100 | 3583 | 5.8 | 14.99 | G5IV | 65 | 62 | 109 | 16.47 | 0.606 | 0.150 | 69.937 | 0.699 |
| 101 | 82860 | 4.88 | 15.08 | F6Vvar | 97 | 92 | 161 | 11.96 | 0.708 | 0.176 | 70.645 | 0.699 |
| 102 | 5862 | 4.97 | 15.18 | F8V | 94 | 90 | 157 | 12.44 | 0.663 | 0.165 | 71.308 | 0.699 |
| 103 | 27435 | 5.97 | 15.25 | G4V | 60 | 57 | 100 | 8.16 | 0.299 | 0.074 | 71.607 | 0.695 |
| 104 | 95501 | 3.36 | 15.53 | F0IV | 191 | 182 | 319 | 6.14 | 0.269 | 0.067 | 71.876 | 0.691 |
| 105 | 107649 | 5.57 | 15.56 | G2V | 72 | 68 | 120 | 7.16 | 0.447 | 0.111 | 72.323 | 0.689 |
| 106 | 86796 | 5.12 | 15.61 | G5V | 90 | 86 | 150 | 14.40 | 0.618 | 0.153 | 72.941 | 0.688 |
| 107 | 71284 | 4.47 | 15.66 | F3Vwvar | 115 | 109 | 192 | 13.40 | 0.696 | 0.173 | 73.638 | 0.688 |
| 108 | 88745 | 5.05 | 15.74 | F7V | 90 | 86 | 150 | 14.28 | 0.693 | 0.172 | 74.331 | 0.688 |
| 109 | 77760 | 4.6 | 15.83 | F9V | 112 | 106 | 186 | 15.35 | 0.703 | 0.174 | 75.033 | 0.688 |
| 110 | 3909 | 5.17 | 15.88 | F7IV-V | 85 | 81 | 142 | 14.05 | 0.599 | 0.149 | 75.633 | 0.688 |
| 111 | 98767 | 5.73 | 16.01 | G6IV+... | 69 | 66 | 116 | 5.24 | 0.281 | 0.070 | 75.914 | 0.684 |
| 112 | 112447 | 4.2 | 16.03 | F7V | 133 | 126 | 221 | 17.25 | 0.529 | 0.131 | 76.444 | 0.683 |
| 113 | 50954 | 3.99 | 16.14 | F2IV | 143 | 136 | 240 | 9.80 | 0.516 | 0.128 | 76.960 | 0.681 |
| 114 | 38908 | 5.59 | 16.20 | G2V... | 71 | 67 | 118 | 15.10 | 0.595 | 0.148 | 77.554 | 0.680 |
| 115 | 32480 | 5.24 | 16.65 | G0V | 83 | 79 | 139 | 14.89 | 0.545 | 0.135 | 78.099 | 0.679 |
| 116 | 35136 | 5.54 | 16.87 | G0V | 73 | 69 | 121 | 13.72 | 0.452 | 0.112 | 78.551 | 0.677 |
| 117 | 86736 | 4.86 | 17.12 | F6/F7V | 97 | 92 | 162 | 14.05 | 0.513 | 0.127 | 79.065 | 0.676 |
| 118 | 12653 | 5.4 | 17.33 | G3IV | 77 | 73 | 129 | 14.28 | 0.542 | 0.135 | 79.607 | 0.675 |
| 119 | 7978 | 5.52 | 17.34 | F8V | 73 | 69 | 122 | 15.40 | 0.531 | 0.132 | 80.138 | 0.673 |
| 120 | 76829 | 4.64 | 17.39 | F5IV-V | 107 | 102 | 179 | 13.09 | 0.535 | 0.133 | 80.673 | 0.672 |
| 121 | 100017 | 5.91 | 17.47 | G3V | 61 | 58 | 103 | 12.88 | 0.315 | 0.078 | 80.988 | 0.669 |
| 122 | 78459 | 5.39 | 17.48 | G2V | 78 | 74 | 131 | 17.13 | 0.575 | 0.143 | 81.563 | 0.669 |
| 123 | 64792 | 5.19 | 17.54 | G0Vs | 85 | 81 | 142 | 16.00 | 0.458 | 0.114 | 82.021 | 0.667 |
| 124 | 17651 | 4.22 | 17.64 | F3/F5V | 130 | 124 | 218 | 13.17 | 0.505 | 0.125 | 82.526 | 0.666 |
| 125 | 89042 | 5.47 | 17.75 | G0V | 75 | 71 | 125 | 13.38 | 0.421 | 0.104 | 82.947 | 0.664 |
| 126 | 46509 | 4.59 | 17.76 | F6V | 110 | 104 | 184 | 13.17 | 0.495 | 0.123 | 83.442 | 0.662 |
| 127 | 65721 | 4.97 | 17.91 | G5V | 98 | 93 | 163 | 18.17 | 0.434 | 0.108 | 83.877 | 0.660 |
| 128 | 61174 | 4.3 | 17.96 | F2V | 125 | 119 | 209 | 13.05 | 0.463 | 0.115 | 84.340 | 0.659 |
| 129 | 910 | 4.89 | 17.99 | F5V | 96 | 91 | 161 | 14.83 | 0.494 | 0.123 | 84.834 | 0.658 |
| 130 | 32439 | 5.44 | 18.21 | F8V | 75 | 71 | 126 | 15.35 | 0.504 | 0.125 | 85.338 | 0.656 |
| 131 | 40843 | 5.13 | 18.23 | F6V | 86 | 82 | 144 | 15.80 | 0.453 | 0.112 | 85.790 | 0.655 |
| 132 | 26394 | 5.65 | 18.28 | G3IV | 69 | 66 | 116 | 16.67 | 0.450 | 0.112 | 86.241 | 0.653 |
| 133 | 109422 | 4.94 | 18.35 | F6V | 94 | 89 | 157 | 14.78 | 0.480 | 0.119 | 86.721 | 0.652 |
| 134 | 18859 | 5.38 | 18.77 | F5V | 77 | 73 | 129 | 14.65 | 0.385 | 0.096 | 87.106 | 0.650 |
| 135 | 67153 | 4.23 | 18.80 | F3V | 129 | 123 | 216 | 14.04 | 0.441 | 0.109 | 87.546 | 0.648 |
| 136 | 4151 | 4.8 | 18.80 | F8V | 101 | 96 | 169 | 15.26 | 0.501 | 0.124 | 88.048 | 0.647 |
| 137 | 48113 | 5.08 | 18.90 | G2V | 90 | 86 | 151 | 16.92 | 0.449 | 0.111 | 88.497 | 0.646 |
| 138 | 73996 | 4.93 | 19.35 | F5V | 94 | 89 | 157 | 15.19 | 0.486 | 0.121 | 88.983 | 0.645 |

**Table D-1.** List of target stars obtained when optimizing the detection and spectral characterization of EECs with HabEx baseline architecture (4H), assuming 5 years of observations and no exozodi emission.

| A | B | C | D | E | F | G | H | I | J | K | L | M |
|---|---|---|---|---|---|---|---|---|---|---|---|---|
| Star Number | HIP | Vmag | Dist (pc) | Type | EEID (mas) | HZ Inner Edge (mas) | HZ Outer Edge (mas) | BB Imaging Time (days) | EEC Complete-ness | EEC Yield | CC | AC |
| 139 | 29800 | 5.04 | 19.53 | F5IV-V | 89 | 85 | 149 | 14.97 | 0.380 | 0.094 | 89.363 | 0.643 |
| 140 | 97675 | 5.12 | 19.53 | F8V | 88 | 83 | 147 | 17.11 | 0.430 | 0.107 | 89.793 | 0.641 |
| 141 | 92043 | 4.19 | 19.56 | F6V | 133 | 126 | 222 | 13.47 | 0.374 | 0.093 | 90.167 | 0.639 |
| 142 | 45333 | 5.18 | 19.66 | F9V | 86 | 82 | 144 | 15.90 | 0.408 | 0.101 | 90.574 | 0.638 |
| 143 | 27321 | 3.85 | 19.75 | A3V | 153 | 145 | 256 | 8.44 | 0.228 | 0.057 | 90.802 | 0.635 |
| 144 | 39903 | 4.74 | 19.98 | F5V | 103 | 97 | 171 | 15.07 | 0.456 | 0.113 | 91.258 | 0.634 |
| 145 | 64408 | 4.85 | 20.29 | G3V | 102 | 97 | 170 | 12.54 | 0.333 | 0.083 | 91.591 | 0.632 |
| 146 | 34834 | 4.49 | 20.86 | F0IV | 114 | 108 | 190 | 13.87 | 0.394 | 0.098 | 91.984 | 0.630 |
| 147 | 97295 | 5 | 20.90 | F5 | 91 | 87 | 152 | 11.54 | 0.275 | 0.068 | 92.259 | 0.628 |
| 148 | 25110 | 5.08 | 20.97 | F6V | 88 | 84 | 148 | 13.09 | 0.307 | 0.076 | 92.566 | 0.625 |
| 149 | 86486 | 4.76 | 21.23 | F3IV | 102 | 97 | 170 | 15.08 | 0.323 | 0.080 | 92.889 | 0.623 |
| 150 | 16245 | 4.71 | 21.75 | F5IV-V | 104 | 99 | 174 | 14.21 | 0.331 | 0.082 | 93.220 | 0.621 |

**Table D-2.** Illustrative list of target stars obtained when optimizing the detection and spectral characterization of EECs with HabEx baseline architecture (4H), randomly assigning individual stars exozodi levels and assuming 2 years of observations.

| A | B | C | D | E | F | G | H | H2 | I | J | K | L | M |
|---|---|---|---|---|---|---|---|---|---|---|---|---|---|
| Star Number | HIP | Vmag | Dist (pc) | Type | EEID (mas) | HZ inner edge (mas) | HZ outer edge (mas) | Exozodi Level (zodis) | BB Imaging Time (days) | EEC Complete-ness | EEC Yield | CC | AC |
| 1 | 54035 | 7.49 | 2.55 | M2V | 61 | 58 | 102 | 1.63 | 5.42 | 0.893 | 0.222 | 0.893 | 0.893 |
| 2 | 16537 | 3.72 | 3.20 | K2V | 184 | 175 | 307 | 297.00 | 9.87 | 0.792 | 0.197 | 1.686 | 0.843 |
| **3** | **104217** | **6.05** | **3.49** | **K7V** | **85** | **81** | **143** | **0.89** | **3.93** | **0.997** | **0.247** | **2.683** | **0.894** |
| **4** | **104214** | **5.2** | **3.50** | **K5V** | **103** | **98** | **172** | **150.64** | **16.42** | **0.871** | **0.216** | **3.553** | **0.888** |
| 5 | 37279 | 0.4 | 3.51 | F5IV-V | 757 | 719 | 1265 | 4.31 | 3.45 | 0.346 | 0.086 | 3.899 | 0.780 |
| 6 | 1475 | 8.09 | 3.56 | M1V | 56 | 53 | 93 | 1.07 | 12.84 | 0.646 | 0.160 | 4.545 | 0.757 |
| **7** | **8102** | **3.49** | **3.60** | **G8V** | **194** | **184** | **323** | **2.92** | **1.00** | **0.994** | **0.247** | **5.539** | **0.791** |
| **8** | **108870** | **4.69** | **3.64** | **K5V** | **129** | **123** | **216** | **7.37** | **3.35** | **0.985** | **0.244** | **6.524** | **0.815** |
| 9 | 105090 | 6.69 | 3.97 | M1/M2V | 70 | 67 | 117 | 1.19 | 12.67 | 0.969 | 0.240 | 7.493 | 0.833 |
| 10 | 49908 | 6.6 | 4.87 | K8V | 67 | 64 | 112 | 2.12 | 10.41 | 0.966 | 0.240 | 8.458 | 0.846 |
| **11** | **19849** | **4.43** | **5.04** | **K1V** | **129** | **122** | **215** | **150.76** | **11.96** | **0.448** | **0.111** | **8.907** | **0.810** |
| 12 | 97649 | 0.76 | 5.13 | A7IV-V | 629 | 597 | 1050 | 7.14 | 2.22 | 0.211 | 0.052 | 9.118 | 0.760 |
| **13** | **96100** | **4.67** | **5.77** | **K0V** | **115** | **109** | **192** | **10.88** | **9.78** | **0.943** | **0.234** | **10.061** | **0.774** |
| 14 | 3821 | 3.46 | 5.84 | G0VSB | 189 | 180 | 316 | 6.41 | 4.16 | 0.270 | 0.067 | 10.331 | 0.738 |
| **15** | **73184** | **5.72** | **5.88** | **K4V** | **79** | **75** | **132** | **1.47** | **14.49** | **0.767** | **0.190** | **11.098** | **0.740** |
| 16 | 84478 | 6.33 | 5.95 | K5V | 67 | 63 | 112 | 3.46 | 15.37 | 0.912 | 0.226 | 12.010 | 0.751 |
| **17** | **15510** | **4.26** | **6.00** | **G8V** | **135** | **128** | **226** | **6.86** | **2.27** | **0.978** | **0.243** | **12.988** | **0.764** |
| 18 | 99240 | 3.55 | 6.10 | G5IV-Vvar | 189 | 180 | 316 | 0.66 | 0.90 | 0.940 | 0.233 | 13.928 | 0.774 |
| 19 | 114622 | 5.57 | 6.53 | K3Vvar | 83 | 79 | 139 | 2.26 | 5.29 | 0.962 | 0.239 | 14.890 | 0.784 |
| 20 | 3765 | 5.74 | 7.44 | K2V | 73 | 69 | 121 | 2.73 | 13.31 | 0.894 | 0.222 | 15.784 | 0.789 |
| 21 | 2021 | 2.82 | 7.46 | G2IV | 256 | 243 | 427 | 0.06 | 4.17 | 0.770 | 0.191 | 16.554 | 0.788 |
| 22 | 7981 | 5.24 | 7.61 | K1V | 89 | 85 | 149 | 2.01 | 11.74 | 0.944 | 0.234 | 17.498 | 0.795 |
| 23 | 22449 | 3.19 | 8.04 | F6V | 210 | 200 | 351 | 1.75 | 6.06 | 0.800 | 0.198 | 18.298 | 0.796 |
| 24 | 64924 | 4.74 | 8.51 | G5V | 108 | 103 | 181 | 2.18 | 10.33 | 0.868 | 0.215 | 19.167 | 0.799 |
| 25 | 1599 | 4.23 | 8.53 | F9V | 133 | 126 | 221 | 6.43 | 8.15 | 0.659 | 0.164 | 19.826 | 0.793 |

**Table D-2.** Illustrative list of target stars obtained when optimizing the detection and spectral characterization of EECs with HabEx baseline architecture (4H), randomly assigning individual stars exozodi levels and assuming 2 years of observations.

| A | B | C | D | E | F | G | H | H2 | I | J | K | L | M |
|---|---|---|---|---|---|---|---|---|---|---|---|---|---|
| Star Number | HIP | Vmag | Dist (pc) | Type | EEID (mas) | HZ inner edge (mas) | HZ outer edge (mas) | Exozodi Level (zodis) | BB Imaging Time (days) | EEC Complete-ness | EEC Yield | CC | AC |
| 26 | 61317 | 4.24 | 8.61 | G0V | 132 | 126 | 221 | 1.94 | 7.82 | 0.832 | 0.206 | 20.658 | 0.795 |
| 27 | 99825 | 5.73 | 8.80 | K3V | 73 | 69 | 121 | 4.82 | 17.48 | 0.680 | 0.169 | 21.338 | 0.790 |
| 28 | 23311 | 6.22 | 8.85 | K3V | 64 | 60 | 106 | 1.11 | 20.74 | 0.810 | 0.201 | 22.148 | 0.791 |
| 29 | 27072 | 3.59 | 8.88 | F7V | 175 | 166 | 292 | 5.59 | 6.57 | 0.593 | 0.147 | 22.741 | 0.784 |
| 30 | 105858 | 4.21 | 9.27 | F6V | 132 | 125 | 220 | 4.21 | 7.53 | 0.644 | 0.160 | 23.385 | 0.780 |
| 31 | 56452 | 5.96 | 9.54 | K0V | 64 | 60 | 106 | 2.29 | 15.31 | 0.607 | 0.151 | 23.992 | 0.774 |
| 32 | 56997 | 5.31 | 9.58 | G8Vvar | 84 | 79 | 140 | 7.95 | 13.61 | 0.453 | 0.112 | 24.445 | 0.764 |
| 33 | 29271 | 5.08 | 10.21 | G5V | 93 | 88 | 155 | 4.79 | 11.55 | 0.508 | 0.126 | 24.954 | 0.756 |
| 34 | 14632 | 4.05 | 10.51 | G0V | 144 | 137 | 241 | 1.94 | 7.91 | 0.607 | 0.151 | 25.561 | 0.752 |
| 35 | 10644 | 4.84 | 11.01 | G0V | 101 | 96 | 168 | 2.86 | 11.02 | 0.574 | 0.142 | 26.135 | 0.747 |
| 36 | 57757 | 3.59 | 11.12 | F8V | 176 | 167 | 294 | 4.98 | 7.01 | 0.315 | 0.078 | 26.450 | 0.735 |
| 37 | 12777 | 4.1 | 11.15 | F7V | 139 | 132 | 232 | 1.54 | 7.57 | 0.422 | 0.105 | 26.872 | 0.726 |
| 38 | 78072 | 3.85 | 11.18 | F6V | 155 | 147 | 259 | 0.84 | 6.90 | 0.625 | 0.155 | 27.497 | 0.724 |
| 39 | 72848 | 6 | 11.38 | K2V | 63 | 60 | 106 | 1.10 | 4.59 | 0.277 | 0.069 | 27.774 | 0.712 |
| 40 | 67927 | 2.68 | 11.40 | G0IV | 271 | 257 | 452 | 1.62 | 5.86 | 0.228 | 0.056 | 28.001 | 0.700 |
| 41 | 109176 | 3.77 | 11.80 | F5V | 160 | 152 | 268 | 2.72 | 6.63 | 0.362 | 0.090 | 28.363 | 0.692 |
| 42 | 15371 | 5.24 | 12.05 | G1V | 84 | 79 | 140 | 2.04 | 10.40 | 0.467 | 0.116 | 28.830 | 0.686 |
| 43 | 80686 | 4.9 | 12.18 | F9V | 97 | 92 | 162 | 0.23 | 7.76 | 0.684 | 0.170 | 29.515 | 0.686 |
| 44 | 24813 | 4.69 | 12.48 | G0V | 108 | 103 | 181 | 0.60 | 9.48 | 0.601 | 0.149 | 30.116 | 0.684 |
| 45 | 51459 | 4.82 | 12.91 | F8V | 100 | 95 | 168 | 2.05 | 8.70 | 0.345 | 0.086 | 30.461 | 0.677 |
| 46 | 98036 | 3.71 | 13.38 | G8IVvar | 183 | 174 | 306 | 0.37 | 11.19 | 0.394 | 0.098 | 30.855 | 0.671 |
| 47 | 116771 | 4.13 | 13.43 | F7V | 137 | 130 | 229 | 1.46 | 7.29 | 0.306 | 0.076 | 31.161 | 0.663 |
| 48 | 12843 | 4.47 | 14.14 | F5/F6V | 117 | 111 | 195 | 1.59 | 8.34 | 0.284 | 0.070 | 31.444 | 0.655 |
| 49 | 102485 | 4.13 | 14.24 | F5V | 136 | 129 | 227 | 0.31 | 6.65 | 0.355 | 0.088 | 31.799 | 0.649 |
| 50 | 82860 | 4.88 | 15.08 | F6Vvar | 97 | 92 | 161 | 0.09 | 7.77 | 0.490 | 0.122 | 32.289 | 0.646 |

HabEx exoplanet yield described in *Section 3.3* and *Appendix C* is based on observations of the representative draw of 50 stars listed in **Table D-2**. The distribution of target stars physical characteristics (spectral type, apparent V magnitude, diameter), target priority and HZ completeness as a function of star luminosity and distance are shown in **Figure D-1**. These results are typical of the many DRM draws conducted to evaluate HabEx mean exoplanet yield—and its uncertainties—assuming a 2-year EEC survey, including detection, orbital determination through multi-epoch observations, and spectral characterization (*Appendix C*).

The number of EECs detected, with orbits determined and spectrally characterized by baseline HabEx 4H architecture is shown in

**Figure D-2** as a function of stellar distance, total broad-band imaging time and number of stars observed. For all plots, the left y-axis indicates the cumulative completeness of the survey, i.e., the number of EECs that would be characterized if every star surveyed had exactly one. The right y-axis indicates the number of EECs that would be characterized under the nominal occurrence rate assumed for EECs ($\eta = 0.243$).

Plots on the left column are based on the larger 5-year EEC survey target list (**Table D-1**): the dotted line assumes no exozodi and hence provides an upper bound to the expected science yield. The solid line shows for comparison the yield achievable over 5 years, assuming that all stars in **Table D-1** have instead a common exozodi level of 4.5 zodis, i.e., set to the median level derived from LBTI observations. Spectral characterization assumes a resolution $R = 70$ (minimum HabEx requirement) across the 450–1,000 nm region, and $R = 7$ between 300 nm and 450 nm. In this more realistic case, exozodi creates a constant background source that becomes dominant at larger distances, and only 89 stars are now observable within the 5 years allocated. The diminishing science return of observing more stars at larger distances, especially when exozodi emission is present, can be readily seen in the middle and bottom plots. Over 5 years, 14 EECs can be detected, with measured orbits and spectra.

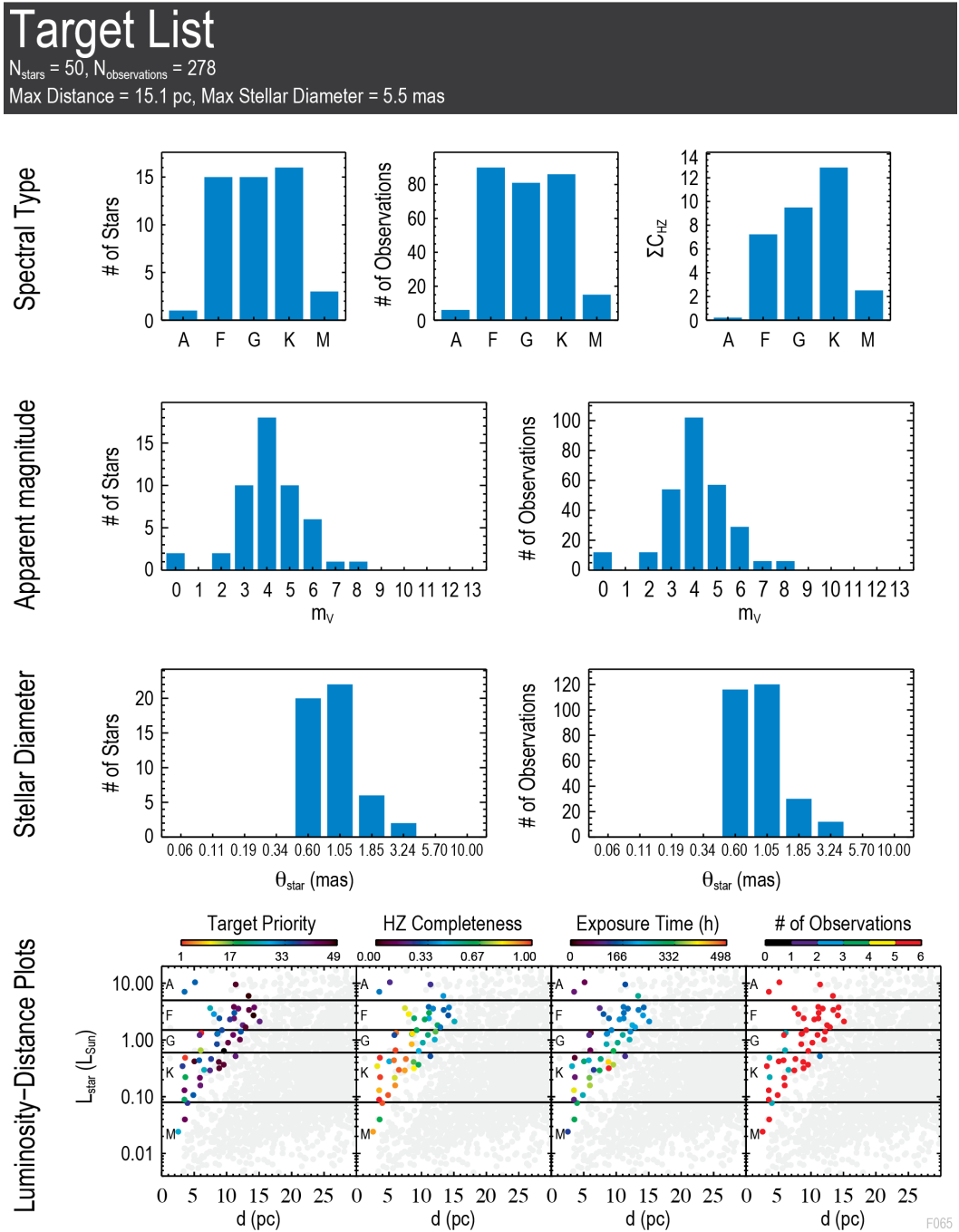

**Figure D-1.** Summary of HabEx target physical characteristics and observability for the smaller 2-year sample of 50 stars shown in Table D-2, with individual exozodi levels randomly assigned.

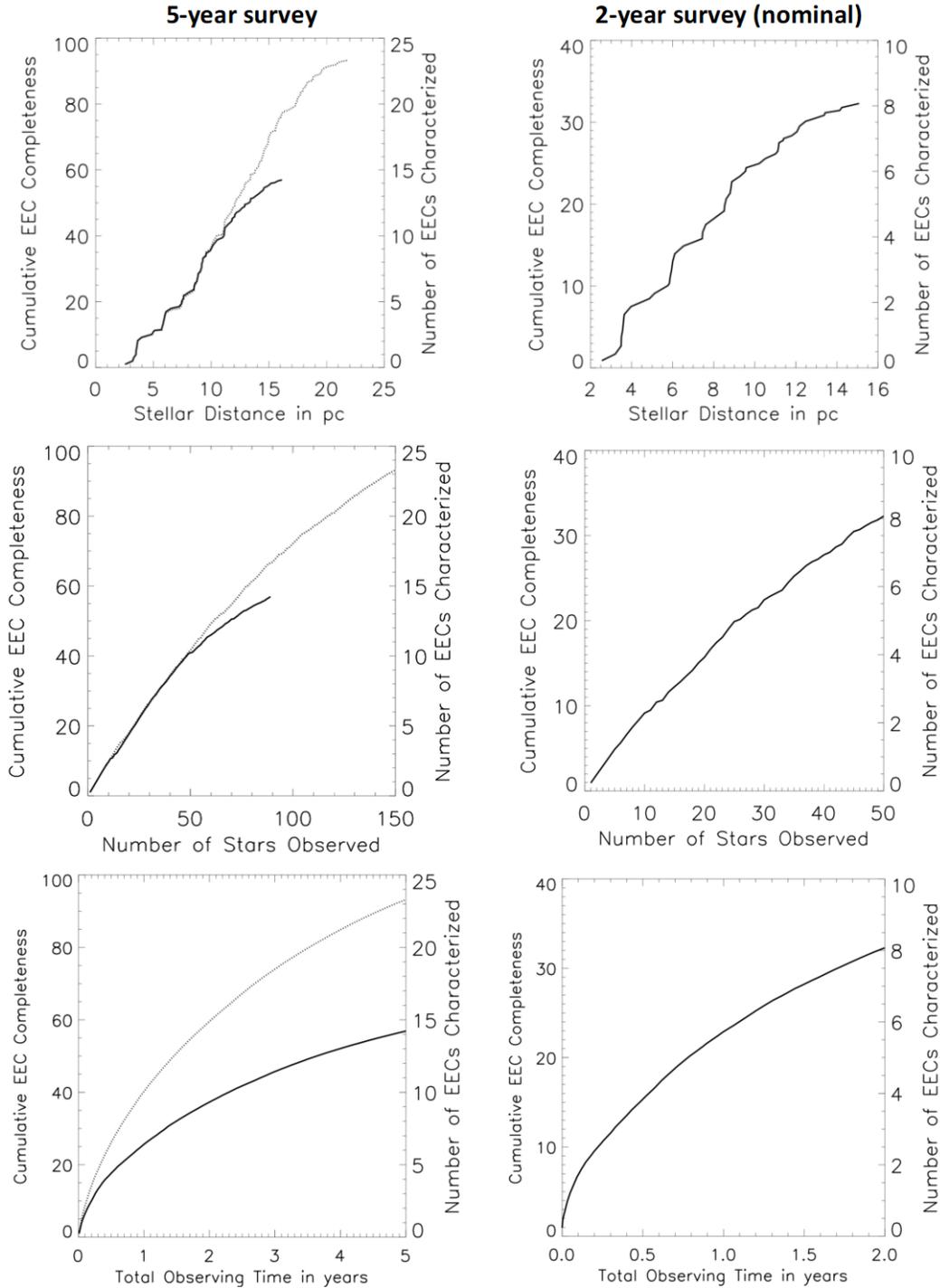

**Figure D-2.** HabEx 4H projected number of EECs detected, with orbit determined and spectrally characterized over at least the 300nm - 1000nm region. Left column plots are based on the larger 5-year EEC survey target list. The solid line uses a realistic constant exozodi level of 4.5 zodis per star and provides a more realistic estimate, while the dotted line is the upper limit achieved assuming no exozodi emission. Right column plots are based on the smaller 2-year EEC survey target list and assume a random draw of exozodi levels from a distribution for each star. For each plot, the left y-axis ("cumulative EEC completeness") indicates the number of EECs characterized by HabEx if all targets stars had exactly one, while the right y-axis (number of EECs characterized) assumes an EEC occurrence rate of 0.243 across the sample. See text for details.

Plots on the right column are based on the smaller 2-year EEC survey target list (**Table D-2** and *Section 3.3*), with a realistic draw of individual exozodi levels. In that case, 1.1 years are dedicated to the broad survey of EECs with the coronagraph (multi-epoch broadband imaging) and 0.9 year is used for spectral characterization with the starshade of all systems with EECs detected by the coronagraph plus the starshade deep dive targets. Spectral characterization assumes a resolution $R = 140$ (nominal HabEx 4H architecture) across the 450-1,000 nm region, and $R = 7$ between 300 nm and 450 nm. Spectral observations of systems with no EECs would take an extra 0.5 year, so that the total time devoted to exoplanet direct imaging and characterization is 2.5 years, as assumed in HabEx prime mission notional time allocation (**Figure 3.3-4**). Over 2 years, 8 EECs can be detected, with measured orbits and spectra.

It is worth noting that for a given total observing time, the EEC yield is slightly higher in the left column plots than in the right ones (e.g., 9.5 vs. 8.1 EECs characterized after 2 years). This is because in the left column case, a constant exozodi level is assumed per target rather than an actual distribution with high exozodi outliers, and spectral measurements are conducted at lower spectral resolution at visible wavelengths ($R = 70$ vs. $R = 140$).

Finally, plots showing EEC yield as a function of the number of stars observed or total observing time (coronagraph multi-epoch broadband imaging + starshade spectral characterization), stars are no longer ordered by increasing distance. They are instead ordered by decreasing EEC yield per hour of observation, resulting in smoother yield curves.

# E  TECHNOLOGY ROADMAP

The technology roadmaps show the steps of development of current technology readiness level (TRL) to TRL 6. Flow from one technology to another, such as into the coronagraph architecture testbed, are shown. The dates shown assume currently funded development to continue as planned and assume that the first year of funding for HabEx technology development is FY2022, should HabEx be prioritized by the Decadal Committee.

The roadmaps presented are for enabling technologies:

- Starshade Mechanical Technologies: one figure to TRL 5 (from the S5 Technology Plan) and one figure to TRL 6

- Starshade Contrast Performance Modeling and Validation

- Starshade Lateral Formation Sensing

- Large Aperture Monolith Mirror Fabrication and Mirror Coating Uniformity

- Laser metrology components are individually TRL 6 or higher. The laser metrology system, as a single metrology gauge, will be assembled and the noise performance and thermal sensitivity will be measured.

- Coronagraph architecture, Zernike wavefront sensor (ZWFS), and deformable mirrors (DMs) are assembled into a coronagraph testbed to achieve $1 \times 10^{-10}$ contrast at 20% bandwidth, the required performance for TRL 6.

- Detectors: linear mode avalanche photodiode (LMPAD), delta-doped electron-multiplying charge-coupled devices (EMCCD), Deep Depleted EMCCD, and microchannel plate detectors

- Microthrusters

One technology does not have a road map to TRL 6:

- Starshade Scattered Sunlight from Petal Edges will be matured with the Starshade Mechanical Technology gaps. Half-meter length optical edges will be attached to two proto-flight petals; the petals will undergo thermal cycling, deployment cycles, shape accuracy measurements and strain measurements. With the optical edges attached throughout these activities, the optical edges will mature to TRL 6.

The timeline to mature enabling technologies to TRL 6 (**Figure E-1**) shows that all technologies will reach TRL 5 before the end of FY24 in time for project start. Enabling technologies will reach TRL 6 by the end of FY27, well before the telescope and starshade Preliminary Design Reviews (PDRs), which occur in FY29.

## E.1  Starshade Mechanical Technologies

The starshade mechanical technologies are comprised of the Starshade Petal Position Accuracy and Stability, Starshade Petal Shape Accuracy and Stability, and Starshade Scattered Sunlight for Petal Edges. The development approach mirrors the S5 approach to achieve TRL 5 (**Figure E-1**). To achieve TRL 6 for HabEx, full-scale high-fidelity test articles will be used. While TRL 6 could be achieved with half-scale test articles, the full-scale test article becomes the engineering model for the mission. This creates an overall cost and schedule efficiency for the mission.

The starshade mechanical technologies will be in the process of maturing to TRL 5 when the Decadal Survey is released. Should HabEx be prioritized, increasing funding to the starshade mechanical technologies in FY22 would accelerate the schedule by a year for completion of TRL 5 by end of FY22.

### E.1.1  Starshade Petal Shape Accuracy and Stability

The Starshade Petal Shape Accuracy and Stability will be developed with a full scale, 2.4 m wide and 16 m long starshade petal. The two-dimensional edge profile of the petal will be measured repeatedly throughout development, such as after deployment cycles and after thermal cycling. The current S5 petal undergoing

TRL 5 maturation is measured with a Micro-Vu precision measuring machine. The full-scale petal exceeds the size of the largest Micro-Vu machine. A larger machine could be custom produced; there are no technological advances required for a larger machine.

The precision optical edges, 1-meter-long, will be integrated onto the high-fidelity petal test article. This will mature the Starshade Scattered Sunlight for Petal Edges technology to TRL 6 simultaneously with the Petal Shape Accuracy and Stability.

### E.1.2    Starshade Petal Position Accuracy and Stability

The starshade petal position accuracy is concerned with the deployment of the central truss and the unfurling of the petals. The full-scale test articles include a 20 m diameter truss, a full-scale Petal Launch & Unfurl Subsystem (PLUS), four high-fidelity petals and twenty and medium fidelity petals. The optical shields for the central truss and the petals are also included. The PLUS is not a technological advancement but a critical piece of ground support equipment (GSE) for the test activities. The optical shields will be tested for opacity and for appropriate, non-contacting overlap at the joints of the petals.

The unfurling tests will ensure that the PLUS is performing correctly to prevent the petal edges from contacting each other. The kinematics models will also be validated. Stowage creep can be measured at this level or with a single petal stowed in a GSE mount.

The high-fidelity petal test articles will be mounted on the full-scale starshade once the petal-level testing is complete. This is a cost and schedule saving measure in the roadmap. The high-fidelity petals on the deployment truss are important for model validation and verifying the position accuracy with respect to the truss bay.

The deployment of the 52 m starshade test article will require a large facility with clean room capability and gravity off-load harness. There are multiple facilities large enough to accommodate the starshade, and most of them will require upgrade to a cleanroom (**Table E-3**).

### E.2    Starshade Contrast Performance Modeling and Validation

The starshade optical contrast is being developed to TRL 5 in a subscale testbed at Princeton (*Section 11.2.1.4*). The subscale masks have some micron-sized features that induce polarization effects. Current modeling matches the optical performance including the induced polarization. The polarization effects are an artifact of the scale of the testbed and will not be present in the full-scale starshade.

To achieve TRL 6, the starshade optical contrast milestones in the S5 Technology Plan (Milestones 1A, 1B, and 2) will be repeated on a testbed with masks that are twice the diameter of the masks in the Princeton testbed. The TRL 6 masks will be 10 cm in diameter and will allow for sufficiently large features that polarization effects will not be induced.

The testbed will require a tunnel 308 m in length and about a meter in diameter. The tunnel should be lightly evacuated to eliminate Rayleigh scattering. The tunnel should be thermally stable at the location of the light source, the mask, and the detector, to ensure stability of alignment of the optical components. Access to the tunnel where the masks are inserted will need to be a cleanroom environment to prevent contamination on the starshade masks. The facility may need to be newly built and is approximated at a cost of $7 million.

### E.3    Starshade Lateral Formation Sensing

S5 matured lateral sensing to TRL 5 with three tasks: the first verified the optical model out-of-band suppression patterns in a testbed, the second used an algorithm to infer lateral offset with testbed images at flight signal-to-noise ratio (SNR), and the third demonstrated lateral position control with a Matlab-simulated lateral control servo. The lateral sensing testbed was similar to the Princeton testbed but scaled down with lower fidelity and lower contrast and with a Fresnel number close to the HabEx Fresnel number.

The testbed will be reincarnated with higher quality optics and higher fidelity starshade mask to mature the lateral formation sensing to TRL 6. The out-of-band suppression patterns will be measured in the ultraviolet (UV) and the near-infrared. The detector will be placed on an actuated translation stage and the control algorithms demonstrated with closed-loop servo control.

## E.4    Large Mirror Fabrication

Large mirror fabrication will be matured to TRL 6 with a full-scale 4 m mirror. The flow was sequenced to retire risks as early as possible in the process and to be efficient in schedule. The full-scale test article could become a proto-flight or flight mirror for HabEx.

The 4 m mirror will be cast in Zerodur® as a solid boule. A thermal test of the mirror blank, polished to a sphere, would show that the CTE homogeneity requirement over the surface of the mirror is met. This serves as the acceptance test of the boule. Next, the blank will be CNC machined for lightweighting, still with a spherical surface polished so that the gravity sag can be measured; the spherical surface makes optical alignment of the test faster and does not require an additional custom null corrector element. Then, the mirror will be ground to the aspheric prescription and final polishing performed. Note that additional facesheet thickness will be required in the spherical surface of the mirror that will be ground away in the figuring of the aspheric surface.

Finally, the surface figure error of the final polished surface will tested and gravity sag backed out. To verify the mechanical stiffness of the mirror, a mechanical ping test will be performed. Shock and vibration are not considered part of the TRL 5 maturity because they are dependent on the mirror assembly, particularly the mirror mount design, and are more appropriately tested at the assembly level for TRL 6.

The roadmap to TRL 6 considers the large mirror and mirror cell as an assembly. Testing occurs at the assembly level. TRL 6 includes vibe and shock. The measurement of the mirror wavefront to the specified requirement after shock and vibe is the final activity.

### E.4.1    Mirror Coating Uniformity

Demonstrating uniformity over a 4 m aperture is the goal of the mirror coating uniformity activity. A coating chamber capable of accommodating a 4 m mirror has been constructed at Harris and is being populated with emitter sources.

A uniform coating is difficult to achieve over a large area because many source emitters are placed over the area. Active control is often employed, such as geometric shadowing of the emitters or motion of the emitters, to uniformly distribute the coating material over the substrate. The complexity of the calibration and achieving 1% uniformity in a new large chamber brings this technology beyond the realm of engineering to technology development.

The coating uniformity can be developed subscale at 2.4 m. ZeCoat has an Astrophysics Research and Analysis Program (APRA) to develop coating advances in Al+MgF$_2$ on 2.4 m substrates. These advances could be applied to a 4 m coating. The 4 m coating can be developed with coupons spread over the 4 m shape. The final TRL 6 demonstration will be done on a full-scale 4 m mirror. Testing for longevity and environmental exposure will be conducted on witness samples.

## E.5    Coronagraph Testbed

The coronagraph architecture mask, the ZWFS, and the DMs achieve critical performance when they enable a coronagraph contrast of $1 \times 10^{-10}$ at 20% bandwidth. To mature each of these technologies to TRL 6, they will be assembled into a HabEx coronagraph testbed. The testbed will use the Decadal Studies Testbed. The first task will be do reduce the noise floor of the DST to less than $1 \times 10^{-10}$. Next, the contrast will be demonstrated monochromatically, then at 10% bandwidth, and finally at 20% bandwidth with a controlled dynamic disturbance.

Some of these technologies will undergo individual maturation. The ZWFS will be developed in an independent testbed to facilitate model validation before being integrated into the HabEx coronagraph testbed. The DMs will have engineering models that will go through environmental testing and include potentially destructive testing of flight-like connectorization.

## E.6    Detectors

All of the detectors are currently at TRL 4. The detectors follow similar maturation paths. After achieving TRL 5, the path to TRL 6 concerns the detector assembly, which is the detector in flight-packaging with flight-like connectorization. The thermal, vibe, and shock tests will be performed on the detector assembly. The radiation dosing will include total ionizing dose (TID) and displacement damage dose (DDD) for HabEx's L2 halo orbit.

## E.7    Microthrusters

Microthrusters will mature to TRL 6 via NASA funded technology development as a candidate for NASA contribution to the

European Space Agency's (ESA's) LISA mission. The development includes improvements such as larger nozzles and switching for redundant fuel supply and power supply channels, a design change for increasing maximum thrust to support a more massive spacecraft, and extensive lifetime testing (accelerated testing representing a 10-year lifetime) at the University of California, Los Angeles. The Microthruster Technology Development Plan is owned by John Ziemer (JPL), lead of microthruster Technology Development as part of Physics of the Cosmos (PCOS) / Cosmic Origins (COR) development of candidate NASA contributions to ESA's LISA mission. The effort plans to achieve TRL 6 by end of FY2022. Part of the development is the full definition of the LISA microthruster requirements. If the LISA microthruster requirements do not meet the HabEx microthruster requirements, then HabEx will repeat the TRL 6 activities with microthrusters designed for the HabEx requirements.

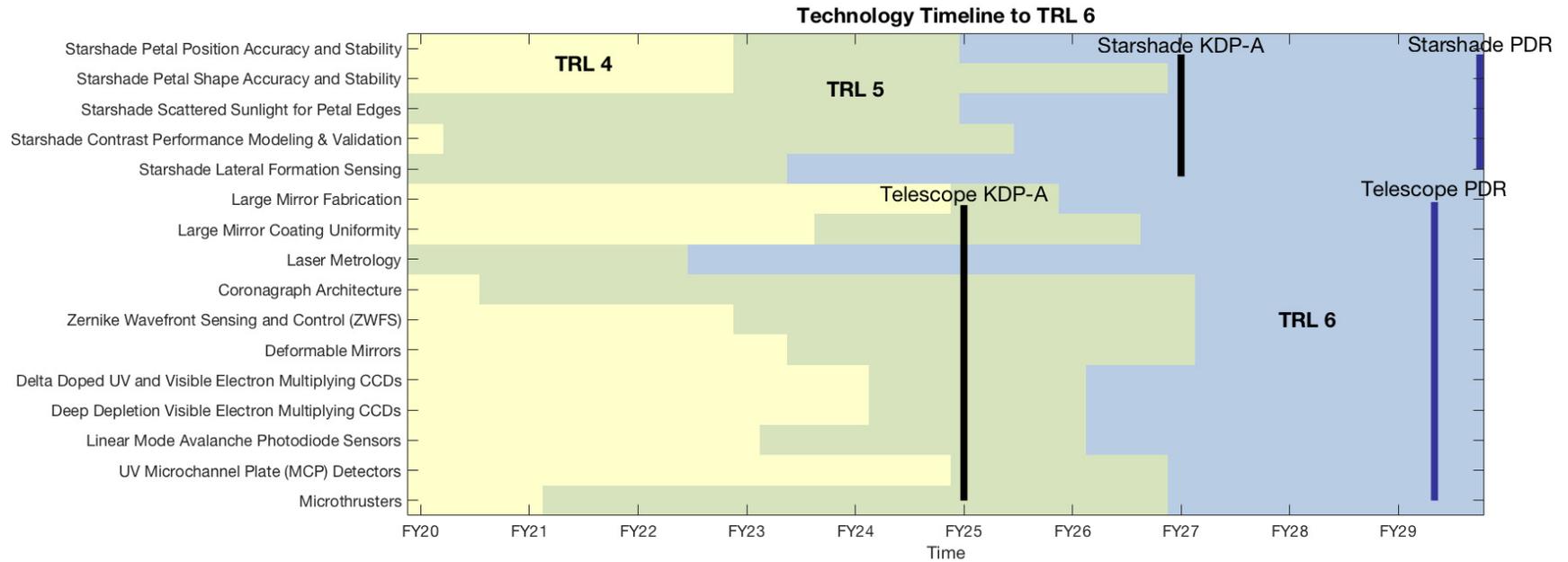

**Figure E-1.** The timeline to mature enabling technologies to TRL 6 shows that all technologies will reach TRL 5 before end of FY24 in time for project start. Enabling technologies will reach TRL 6 by end of FY27, well before the telescope and starshade Preliminary Design Reviews (PDRs), which occur in FY29. (KDP: Key Decision Point.)

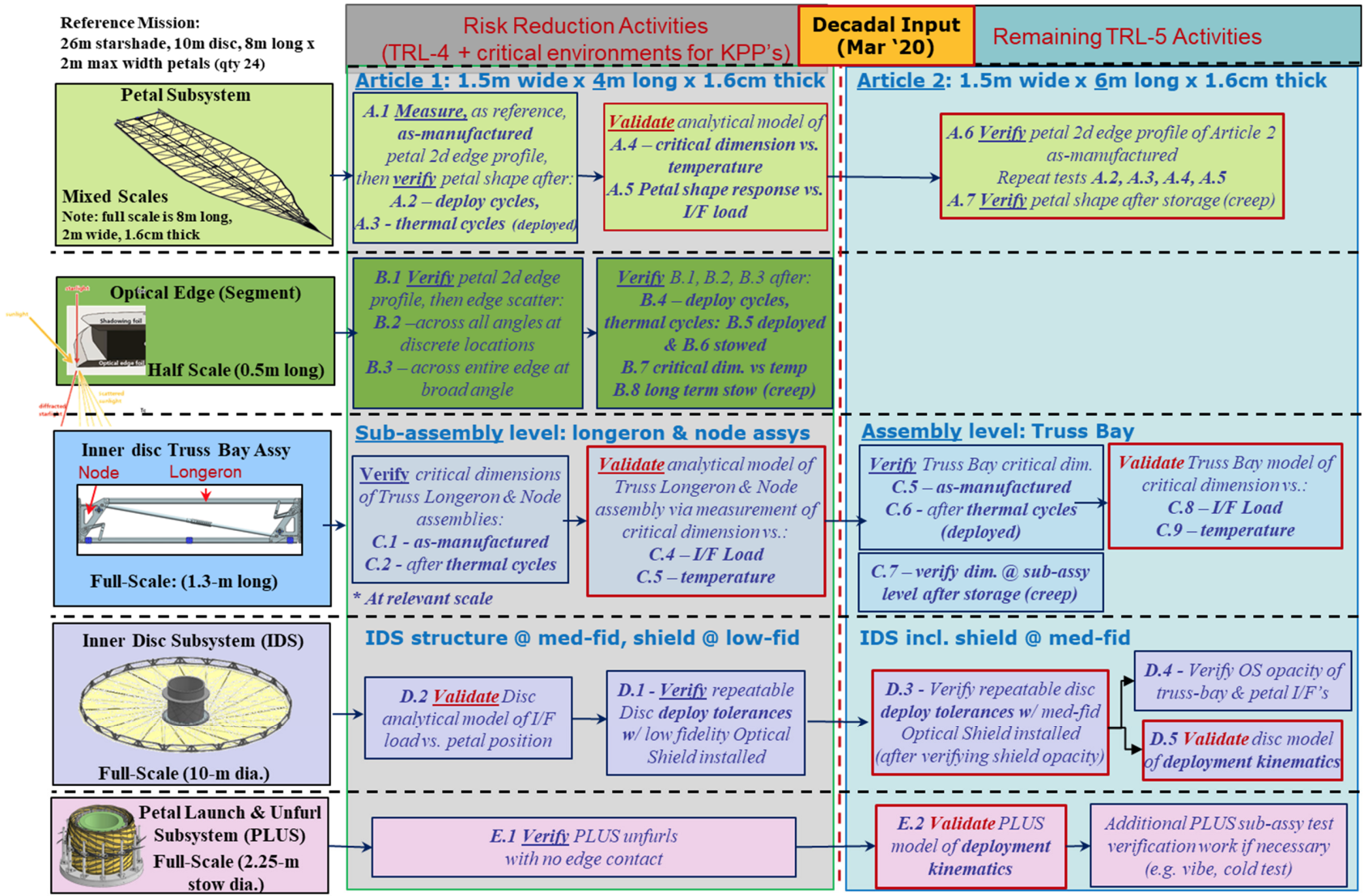

**Figure E-2.** Timeline of TRL 5 to TRL 6 maturation of key starshade technologies. Top-level summary of key activities that mature the starshade mechanical technologies to TRL 5. Credit: S5 Technology Development Plan.

## Petal Shape & Stability

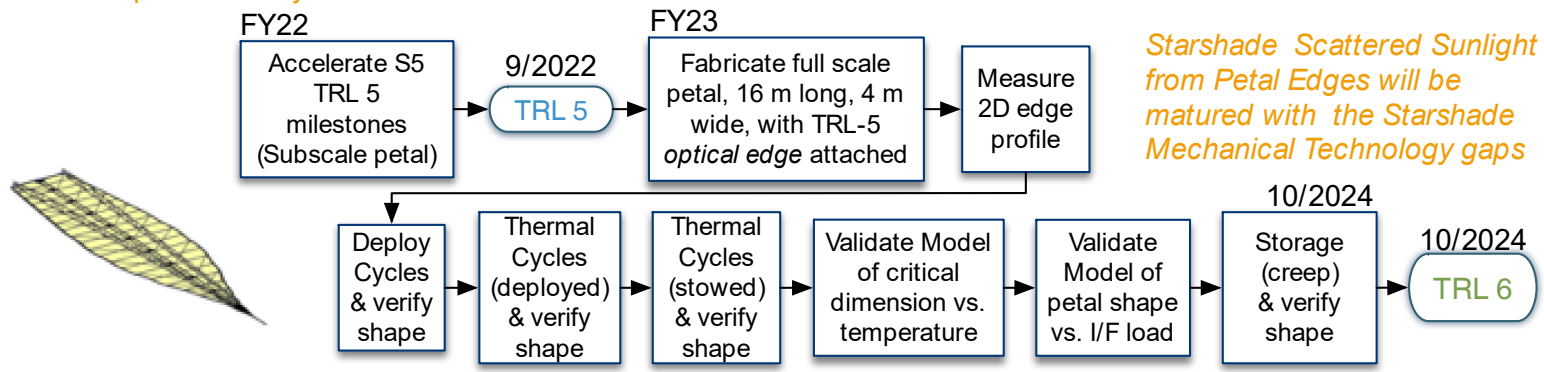

FY22 — Accelerate S5 TRL 5 milestones (Subscale petal) — 9/2022 TRL 5 — FY23 Fabricate full scale petal, 16 m long, 4 m wide, with TRL-5 *optical edge* attached → Measure 2D edge profile

*Starshade Scattered Sunlight from Petal Edges will be matured with the Starshade Mechanical Technology gaps*

Deploy Cycles & verify shape → Thermal Cycles (deployed) & verify shape → Thermal Cycles (stowed) & verify shape → Validate Model of critical dimension vs. temperature → Validate Model of petal shape vs. I/F load → 10/2024 Storage (creep) & verify shape → 10/2024 TRL 6

## Petal Unwrapping

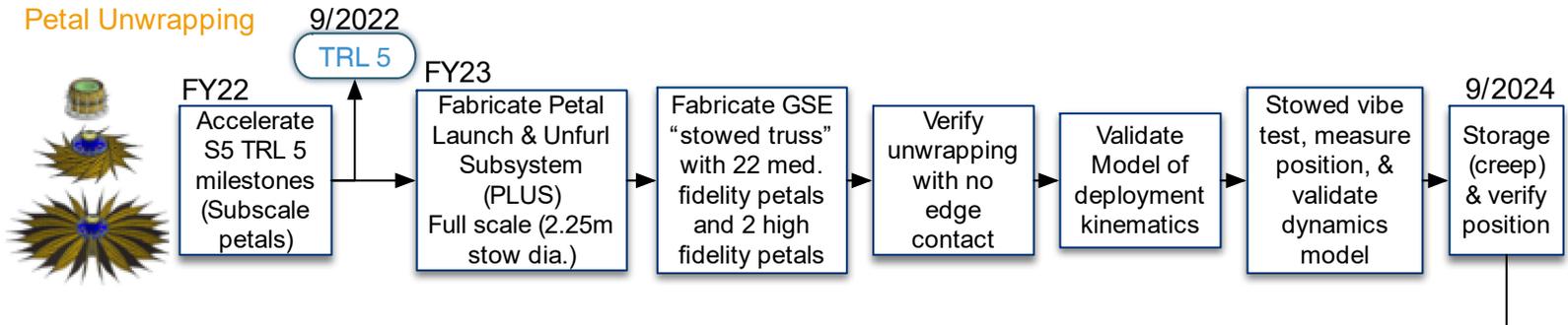

9/2022 TRL 5

FY22 — Accelerate S5 TRL 5 milestones (Subscale petals) → FY23 Fabricate Petal Launch & Unfurl Subsystem (PLUS) Full scale (2.25m stow dia.) → Fabricate GSE "stowed truss" with 22 med. fidelity petals and 2 high fidelity petals → Verify unwrapping with no edge contact → Validate Model of deployment kinematics → Stowed vibe test, measure position, & validate dynamics model → 9/2024 Storage (creep) & verify position

## Truss Deploy

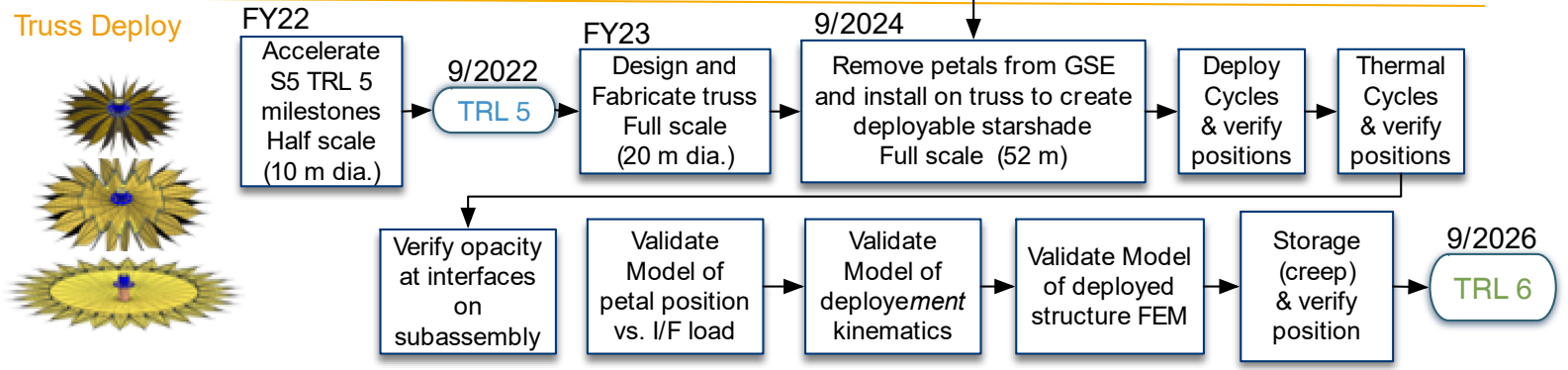

FY22 Accelerate S5 TRL 5 milestones Half scale (10 m dia.) → 9/2022 TRL 5 → FY23 Design and Fabricate truss Full scale (20 m dia.) → 9/2024 Remove petals from GSE and install on truss to create deployable starshade Full scale (52 m) → Deploy Cycles & verify positions → Thermal Cycles & verify positions

Verify opacity at interfaces on subassembly → Validate Model of petal position vs. I/F load → Validate Model of deploye*ment* kinematics → Validate Model of deployed structure FEM → Storage (creep) & verify position → 9/2026 TRL 6

**Figure E-3.** Items marked 'full-scale' are full-scale for a 26 m S5 reference mission and half-scale for the HabEx 52 m starshade. Half-scale, high-fidelity test articles are sufficient for TRL 6.

## Starshade Lateral Formation Sensing

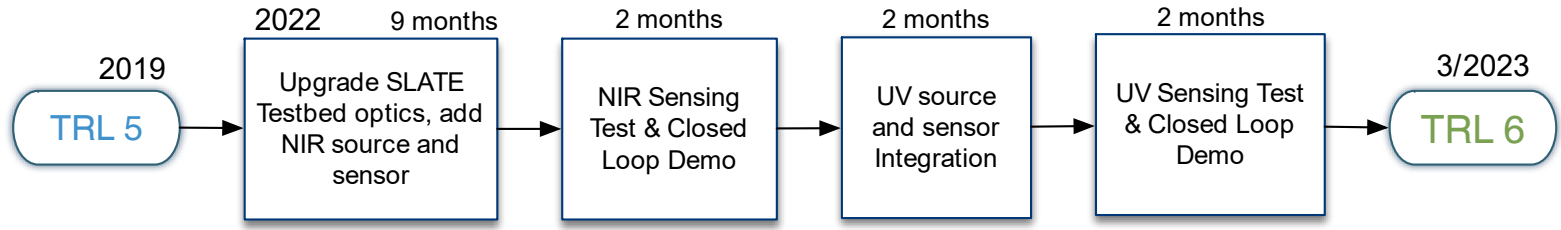

## Starshade Contrast Performance and Model Validation

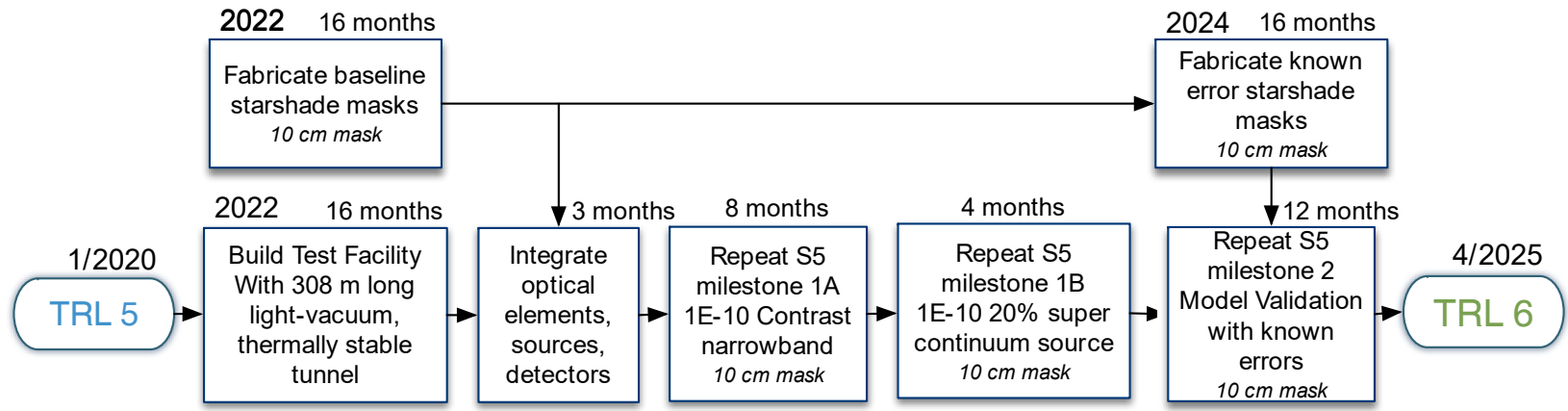

**Figure E-4.** The Roadmap to TRL 6 for starshade lateral formation sensing includes a high-fidelity upgrade to the previous low-fidelity Starshade Lateral Alignment Testbed (SLATE) testbed. The Starshade Contrast Performance and Model Validation testbed is twice the scale of the S5 testbed to ensure performance in the scalar regime of diffraction.

## Large Mirror Fabrication

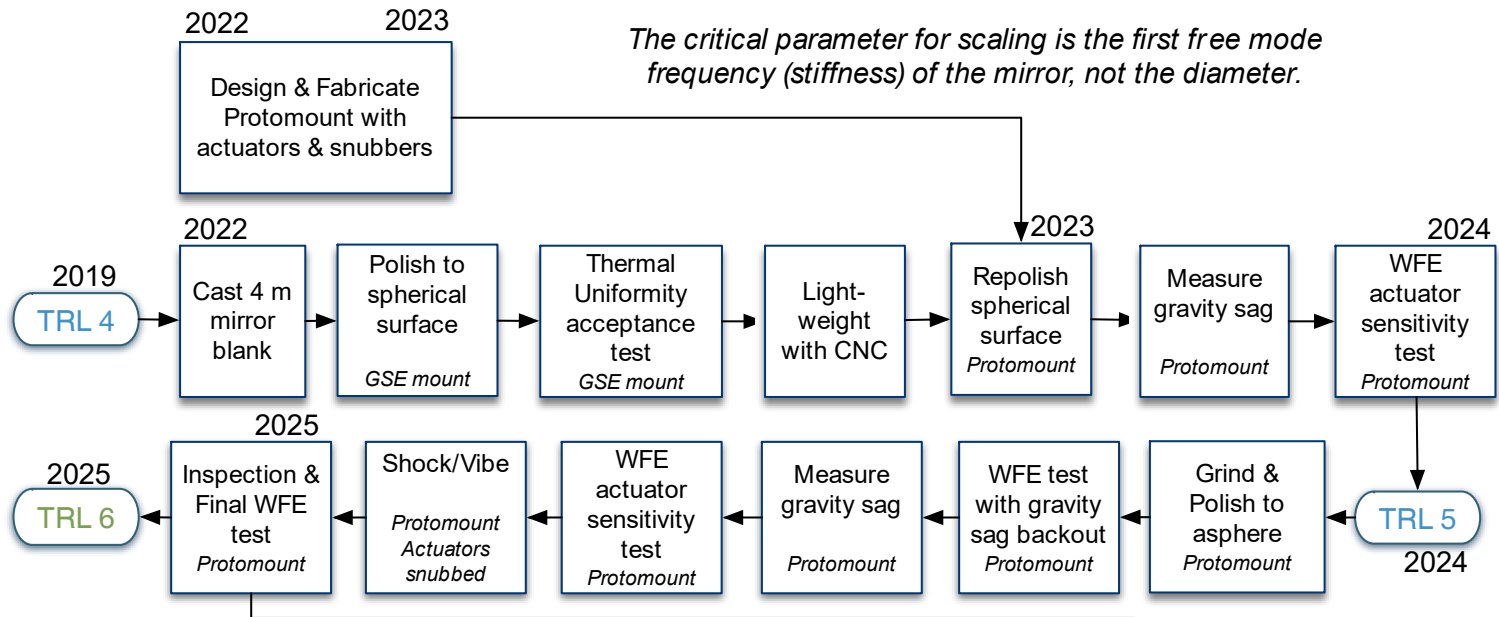

## Large Mirror Coating Uniformity

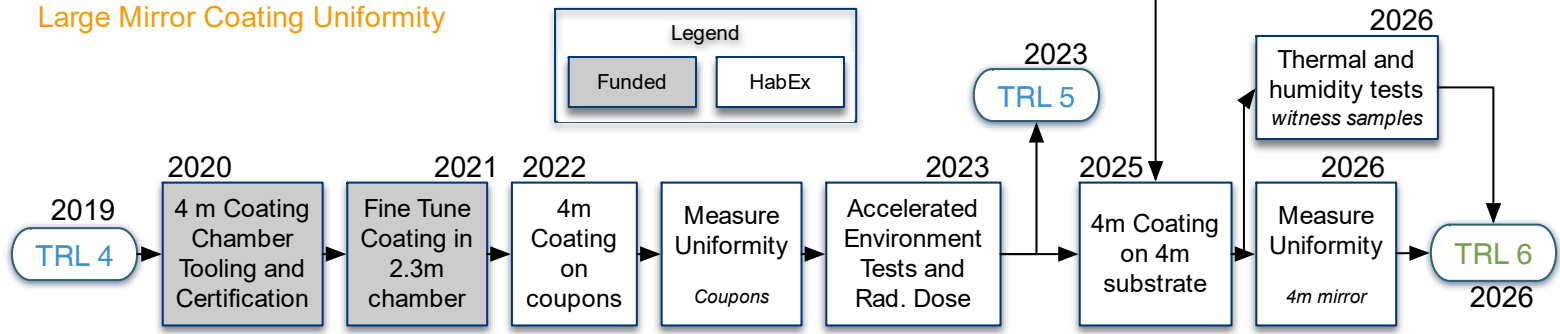

**Figure E-5.** Roadmap to TRL 6 for large mirror fabrication of a 4 m monolithic mirror and for large mirror coating uniformity.

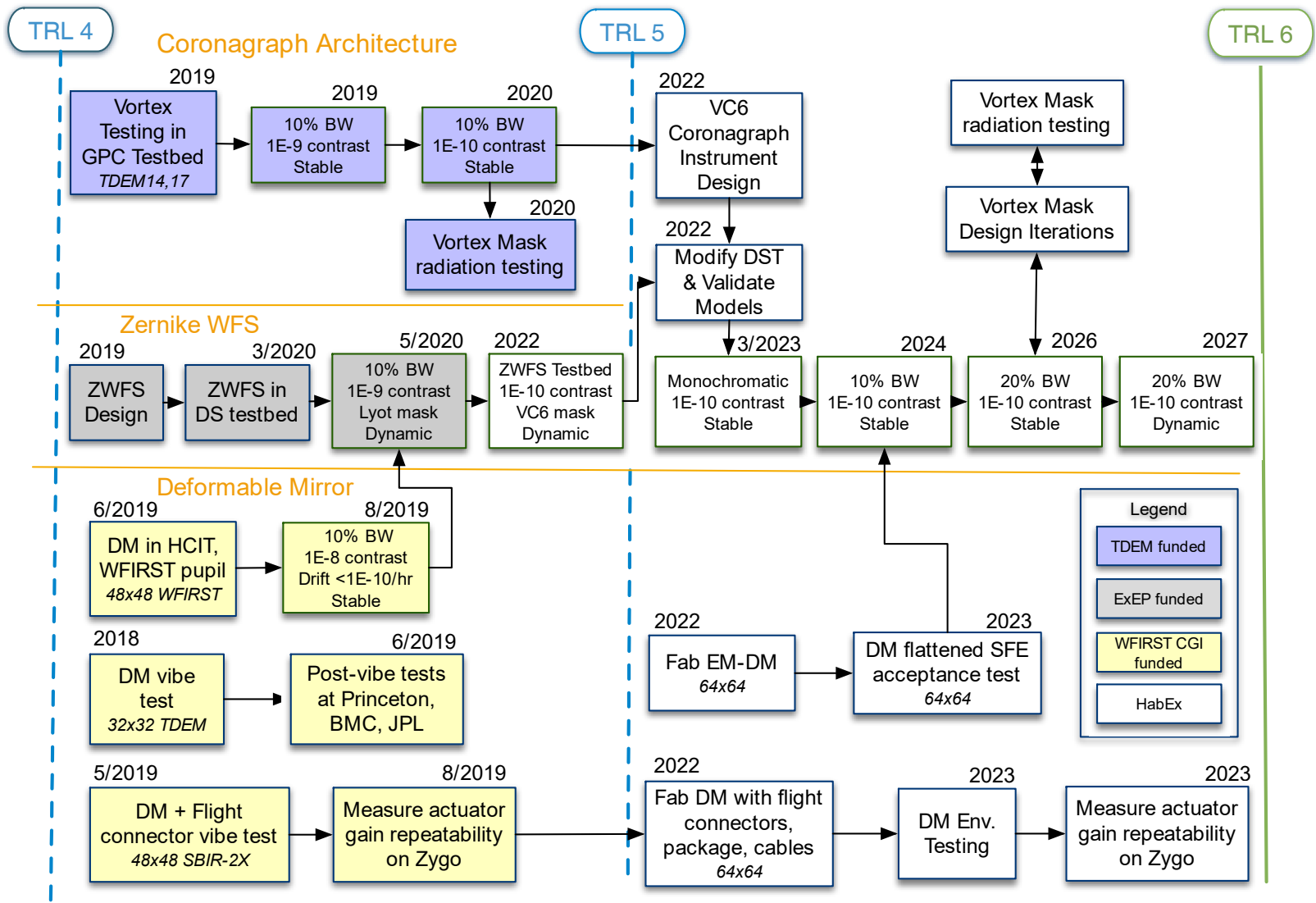

**Figure E-6.** The roadmaps to TRL 6 for the coronagraph architecture, Zernike wavefront sensor, and deformable mirrors (DMs) feed into a single coronagraph architecture testbed. DM flight like packaging and connectorization undergoes environmental testing and post-testing verification separate from the engineering unit DMs in the coronagraph testbed.

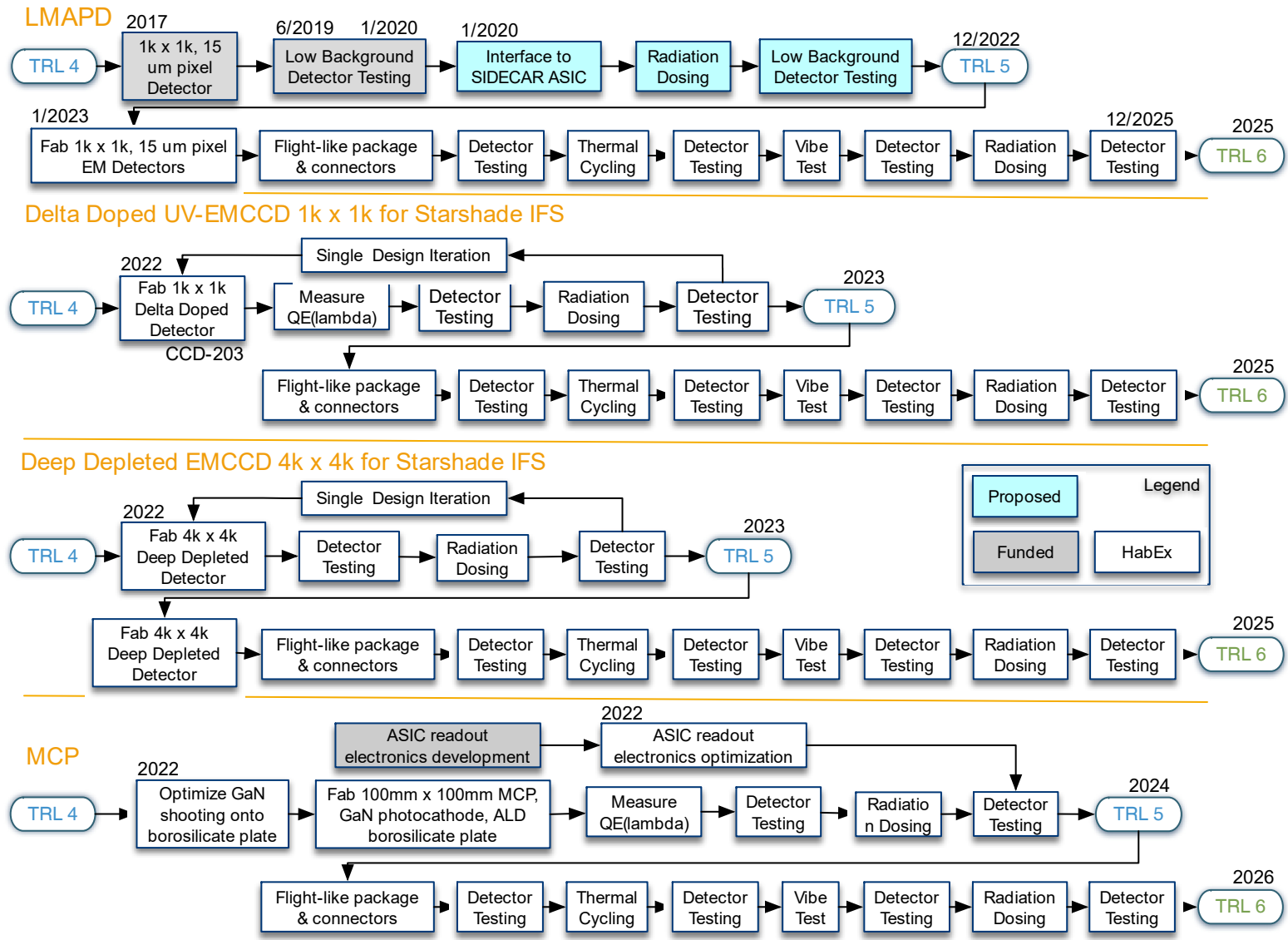

**Figure E-7.** Roadmap to TRL 6 for detectors including linear mode avalanche photodiode (LMPAD), delta doped electron multiplying charge coupled devices (EMCCDs), deep depleted EMCCDs, and microchannel plate detectors.

## Microthrusters

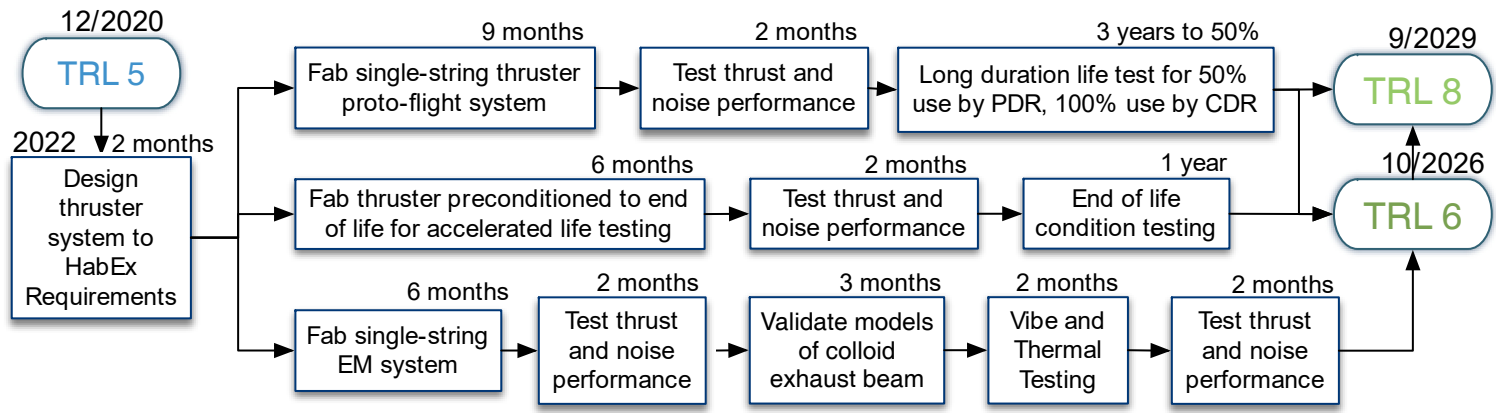

## Laser Metrology

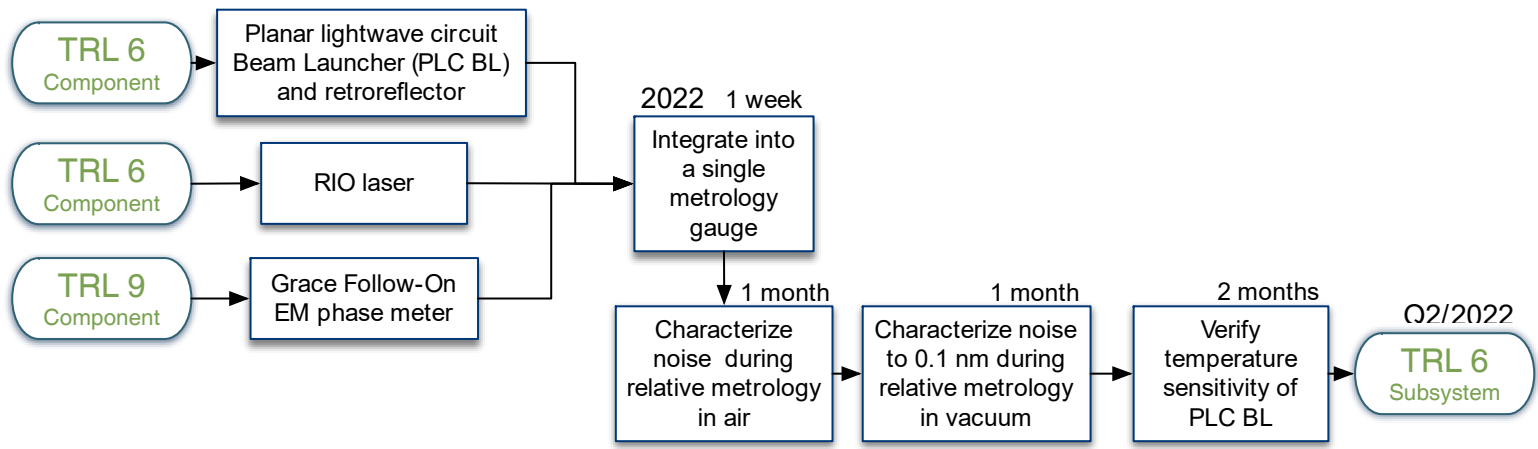

**Figure E-8.** Roadmap to TRL 6 for microthrusters utilizes three parallel paths to minimize schedule. The laser metrology roadmap integrates TRL 6 components into a single laser metrology gauge to demonstrate performance at TRL 6.

**Table E-1.** Technology cost estimates to TRLs 5 and 6.

| Technology Gap Title | Cost Estimate ($FY20) | | Basis of Estimate | |
|---|---|---|---|---|
| | To TRL 5 | To TRL 6 | To TRL5 | To TRL6 |
| Starshade Lateral Formation Sensing | N/A | $1.1M | At TRL 5 | Based on S5 SLATE testbed actual costs |
| Starshade Contrast Performance Modeling and Validation | N/A | $10M | Funded by S5 | Based on S5 Contrast Performance testbed actual costs and grassroots estimate |
| Starshade Scattered Sunlight | N/A | N/A | Funded by S5 | Included in starshade mechanical technologies |
| Starshade Mechanical Technologies | N/A | $147M | Funded by S5 | Exo-S report and S5 estimates for TRL5 |
| Large Mirror Fabrication | $40M | $40M | Quote from industry | |
| Large Mirror Coating Uniformity | $5M | $2.5M | Quote from industry | |
| Laser Metrology | N/A | $0.5M | At TRL 5 | Based on IRAD testbed actual costs |
| Coronagraph Architecture | N/A | $7M | Funded by TDEM | Based on WFIRST CGI actual costs |
| ZWFS | N/A | $1M | Funded by DST | Included in coronagraph architecture |
| Deformable Mirrors | N/A | $1.5M | Funded by WFIRST | Based on WFIRST DM actual costs |
| On Board Computing | N/A | $4 M | Funded by SAT | Based on SAT planned costs |
| Delta Doped UV EMCCDs | $4.6M | $2.3M | Vendor quote and grassroots estimate | |
| Deep Depletion EMCDDS | $3.8M | $2.3M | Vendor quote and grassroots estimate | |
| Linear Mode Avalanche Photodiode Sensors | $1.5M | $1M | Vendor quote and grassroots estimate | |
| Microchannel Plate Detectors | $3.7M | $2.6M | Vendor quote and grassroots estimate | |
| Microthrusters | N/A | ($10M) | At TRL 5 | Covered by NASA candidate for LISA contribution, pending LISA microthruster requirement definition. |
| TOTAL | $58.6M | $232.8M | | |

**Table E-2.** Technology cost risk estimate for currently funded technologies to TRL 5.

| Technology Gap Title | Cost Estimate ($FY20) | Basis of Estimate | |
|---|---|---|---|
| | To TRL 5 | To TRL5 | To TRL6 |
| Starshade Contrast Performance Modeling and Validation | TRL 5 by 1/2020 | TRL 5 by 1/2020 | S5 Technology Development Plan |
| Starshade Mechanical Technologies | $26M | Funded by S5 | S5 Technology Development Plan |
| Coronagraph Architecture | $1.1M | Funded by TDEM | WFIRST CGI and TDEM plan |
| ZWFS | $0.03M | Funded by DST | DST plan |
| Deformable Mirrors | TRL 5 by 10/2019 | Funded by WFIRST | WFIRST CGI plan and TDEM |
| TOTAL | $27.13M | | |

**Table E-3.** HabEx required facilities. Most facilities required by HabEx already exist.

| Technology Title | Facility | Location | Currently Exists |
|---|---|---|---|
| **Starshade Lateral Formation Sensing** | For SLATE testbed lab space and optical bench | JPL | Yes |
| **Starshade Contrast Performance Modeling and Validation** | 308 m long, lightly evacuated, thermally stable tube with clean room surrounding mask insertion area | TBD | No |
| **Starshade Mechanical Technologies** | Vacuum chamber to thermally cycle 2.3 m × 8 m petal | JPL | Yes |
| **Starshade Mechanical Technologies** | Room for 10 m perimeter truss deployment | Northrop Grumman, Aerojet | Yes |
| **Starshade Mechanical Technologies** | Deployment test of 52 m starshade | NASA Ames, Northrup, Vandenburg AFB, Edwards AFB, Point Mugu Naval Base | Requires upgrade to clean room |
| **Large Mirror Fabrication** | Mirror blank casting, machining, testing | SCHOTT | Yes |
| **Large Mirror Fabrication** | Final mirror figuring and testing | Collins, L3 Harris, AOS | Yes |
| **Mirror Coating Uniformity** | 4 m coating chamber | L3 Harris | Yes |
| **Coronagraph Architecture** | Coronagraph Testbed in HCIT-1 or HCIT-2 | JPL | Yes |
| **ZWFS** | Coronagraph Testbed in HCIT-1 or HCIT-2 | JPL | Yes |
| **Deformable Mirrors** | Coronagraph Testbed in HCIT-1 or HCIT-2, Vacuum Surface Gauge | JPL | Yes |
| **Delta Doped EMCCD** | Delta doping facility at Microdevices Lab | JPL | Yes |
| **Detectors** | Detector testing lab | JPL, Berkeley | Yes |

# F  INDUSTRY WHITE PAPERS ON TECHNOLOGY CAPABILITIES AND FACILITIES

*Appendix F* has been withheld in this version of the HabEx report due to U.S. Export Regulations.

# G BASELINE INDEPENDENT COST ESTIMATE

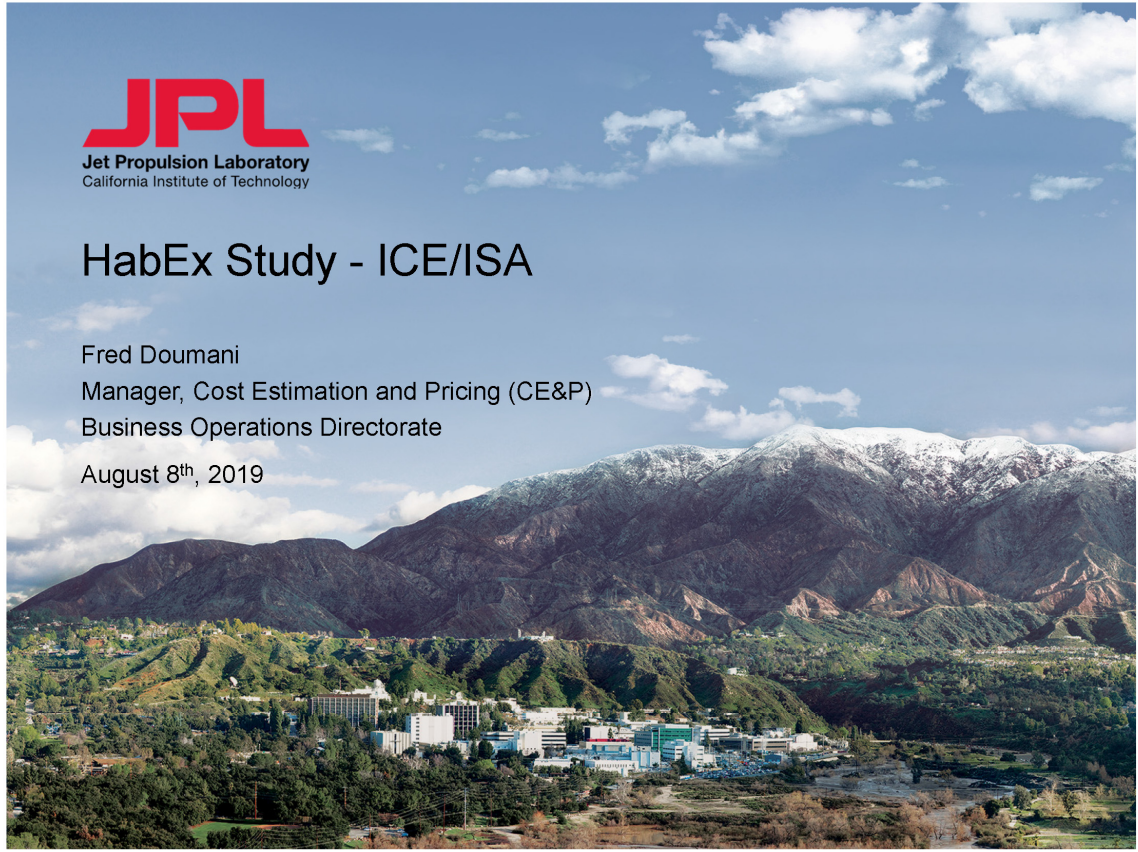

COST ESTIMATION AND PRICING SECTION

# Independent Cost Estimate and Independent Schedule Assessment

## Executive Summary
### *HabEx Study*

- Performed an Independent Cost Estimate and Schedule Assessment (ICE/ISA) based on the HabEx study
- The primary objectives of the ICE/ISA are to:
  - Assess the overall completeness and adequacy of the concept's Pre-Phase A-F cost and Phase A-D schedule
  - Evaluate potential cost and schedule risk through model benchmarking and historical analogies

# ICE/ISA Process Overview

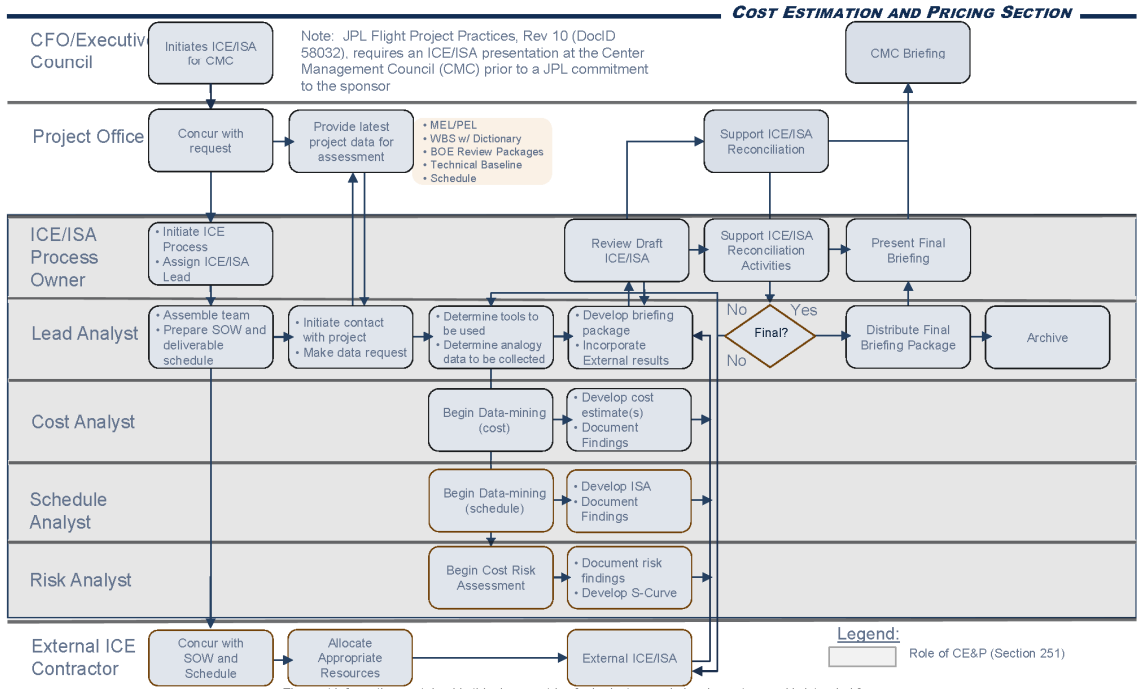

# Steps to develop an ICE/ISA

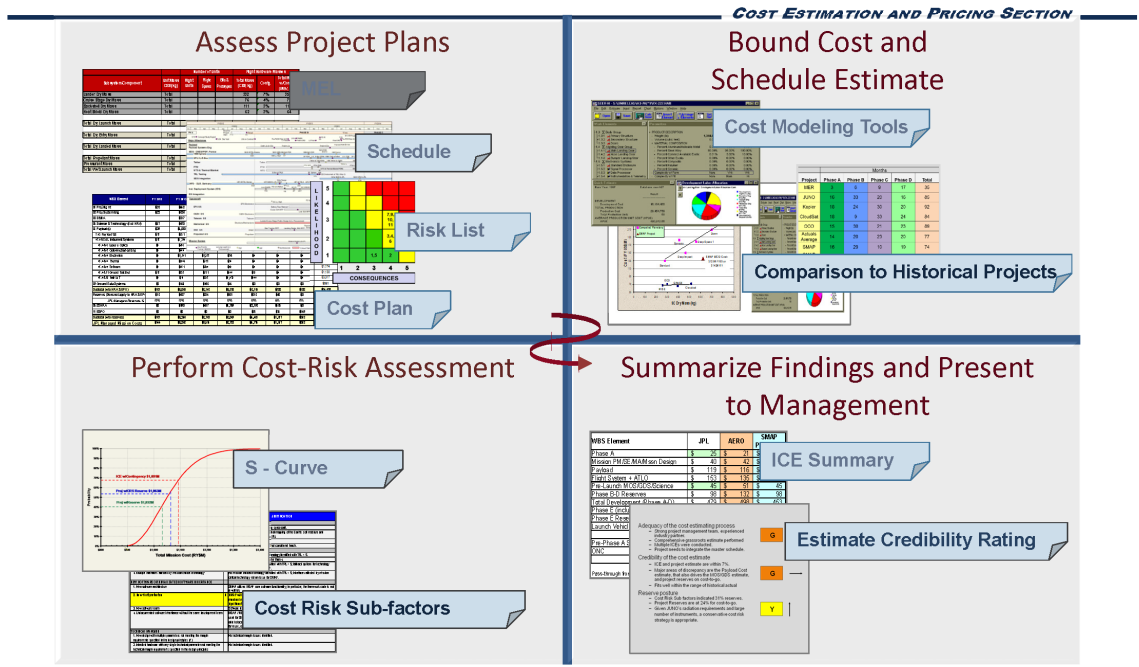

# HabEx Overview – Tier 1 Schedule
## HabEx Study

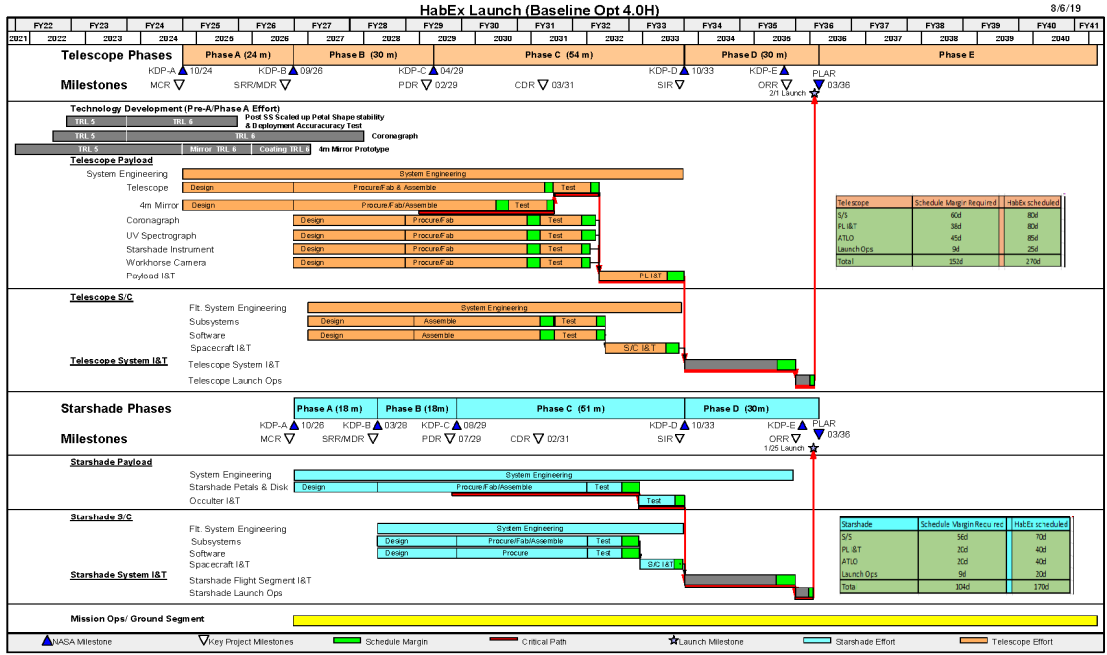

# HabEx Overview – Cost Summary FY20 $M
## *HabEx Study*

- HabEx baseline option
- Concept estimate based on TeamX, NICM, and Telescope CER
- Phases B-D UFE is set to 30%
- Phases E-F UFE is set to 15%
- Current total Pre-Phase A-F estimate is $6,786M FY20

| WBS | HabEx 08/06/2019 |
|---|---|
| Pre-Phase A | 59 |
| Phase A | 211 |
| 01 PM / 02 PSE (w/ MD) / 03 S&MA | 444 |
| 04 Science | 113 |
| 05 Payload System | 1,996 |
| 05.01 PL Mgmt / 05.02 PL SE | 136 |
| 05.04 Coronagraph | 447 |
| 05.05 Starshade Camera | 119 |
| 05.06 UV Spectrograph | 257 |
| 05.07 Optical Telescope System | 659 |
| 05.08 Workhorse Camera | 29 |
| 05.09 Fine Guidance System | 180 |
| 05.10 Starshade Petals and Disk | 170 |
| 06 Spacecraft System + 10 ATLO | 1,724 |
| 06.01 FS Mgmt / 06.02 FS SE | In 06.16 |
| 06.16A Telescope Bus | 1,045 |
| 06.16B Starshade Bus | 680 |
| 07 MOS / 09 GDS | 85 |
| Subtotal (Ph B-D) | 4,363 |
| Reserves | 1,309 |
| Ph B-D w/ Reserves | 5,672 |
| LV (Telescope) | 650 |
| LV (Starshade) | 300 |
| Ph B-D w/ Reserves and LV | 6,622 |
| ESA Contributions | (565) |
| Ph B-D w/ Reserves, LV, and Contribution | 6,057 |
|  |  |
| Operations | 400 |
| Subtotal (Ph E-F) | 400 |
| Reserves | 60 |
| Ph E-F w/ Reserves | 460 |
| Total Pre-Phase A-F | 6,786 |

# ICE Assumptions (1 of 2)
## *HabEx Study*

═══════════════════════════════ COST ESTIMATION AND PRICING SECTION ═══

- ICE estimate is for LCC (Pre-Phase A-F)
  - Pre-Phase A, Phase A, Launch Vehicle, and Contributions are pass-through from study
- ICE Phases B-F validation tools include: NASA PCEC, SEER-H, NICM VIII, Trend line, SOCM, and wrap factors
  - No fees and burdens applied to cost models (if applicable) because the study is not center specific
- HabEx MEL/PEL dated 07/31/19 from study
- ICE Average suggests 41% UFE on Phases B-D and 15% on Phase E to reach the 70th %-tile confidence level

# ICE Assumptions (2 of 2)
## *HabEx Study*

*COST ESTIMATION AND PRICING SECTION*

- **Model Settings:**
  - NASA PCEC (Method 1) and SEER-H (Method 2)
    - Class A Mission – Earth Orbiting
    - Sparing philosophy based on the MEL
  - Method 1 is based on the worst case results of the 2 PCEC variations by WBS
    - 1) Telescope and Starshade buses as 2 separate missions
    - 2) Telescope and Starshade buses as 1 mission with two flight elements
  - Optical Telescope System – Trend line (Cost vs. Aperture Diameter)
  - Other payloads are modeled with NICM VIII and SEER-H
    - There are some overlaps with study team also using NICM

The cost information contained in this document is of a budgetary and planning nature and is intended for informational purposes only. It does not constitute a commitment on the part of JPL and/or Caltech.

# ICE Methodologies by Element
## *HabEx Study*

| WBS | Method 1 | Method 2 |
|---|---|---|
| Pre-Phase A | Pass-thru | |
| Phase A | Pass-thru | |
| Phase B-D | | |
| 01 PM | PCEC | MSL Wrap |
| 02 PSE (w/ MD) | PCEC | MSL Wrap |
| 03 S&MA | PCEC | MSL Wrap |
| 04 Science | PCEC | MSL Wrap |
| 05 Payload | NICM VIII / SEER-H / Trend line | NICM VIII / SEER-H / Trend line |
| 06 Flight System | PCEC | SEER-H |
| 07 MOS | PCEC | MSL Wrap |
| 09 GDS | PCEC | MSL Wrap |
| 10 ATLO | PCEC | SEER-H |
| Phase E/F | SOCM | |

 The cost information contained in this document is of a budgetary and planning nature and is intended for informational purposes only. It does not constitute a commitment on the part of JPL and/or Caltech. 

# Key Findings

*HabEx Study*

- HabEx study estimate is 4% to 15% below the ICE range, deltas by WBS are:
  - LOE Tasks, study estimate is ~$50M to $150M below the ICE range:
    - As hardware cost base increases, LOE wrap factors tend to decrease. MSL wraps are used to conservatively model the LOE efforts.
  - Optical Telescope System, study estimate of $659M versus the ICE $802M. Estimation and validation is a challenge:
    - 4m UVOIR Telescope has limited analogies
    - NICM is not applicable, the HabEx aperture at 4m is out of the NICM range: IR aperture range is 0.3m to 0.85m
  - Starshade Petals and Disk, study estimate of $170M versus ICE $258M:
    - ICE assumes a combination of precision mechanism and primary structure. Need further breakout of MLI from mechanism and structure to fine tune the ICE estimate. (Modeled using SEER-H)
  - UFE, study estimate is ~$400M to $700M below the ICE range

# Key Findings
### *HabEx Study*

COST ESTIMATION AND PRICING SECTION

- HabEx UFE of 30% during development is at the lower range of UFE assessments (30% to 47%)
  - JPL statistical cost-growth analysis resulted in a 70th %-tile confidence level at $7,528M.
    - For HabEx to reach 70th %-tile confidence, study will need to increase UFE to 47% during development.
  - JPL Cost Risk Subfactors analysis yields a UFE of 33% for Phases B-D
  - Institutional requirement for UFE is 30% for Phases B-D at this stage in lifecycle

# ICE Summary – Cost in FY20 $M
## *HabEx Study*

HabEx Cost and Validation Summary (FY20 $M)

| WBS | HabEx 08/06/2019 | ICE Method 1 | ICE Method 2 | ICE Average |
|---|---|---|---|---|
| Pre-Phase A | 59 | 59 | 59 | 59 |
| Phase A | 211 | 211 | 211 | 211 |
| 01 PM / 02 PSE (w/ MD) / 03 S&MA | 444 | 462 | 538 | 500 |
| 04 Science | 113 | 118 | 50 | 84 |
| 05 Payload System | 1,996 | 2,140 | 2,201 | 2,170 |
| 05.01 PL Mgmt / 05.02 PL SE | 136 | 81 | 142 | 111 |
| 05.04 Coronagraph | 447 | 460 | 460 | 460 |
| 05.05 Starshade Camera | 119 | 114 | 114 | 114 |
| 05.06 UV Spectrograph | 257 | 286 | 286 | 286 |
| 05.07 Optical Telescope System | 659 | 802 | 802 | 802 |
| 05.08 Workhorse Camera | 180 | 111 | 111 | 111 |
| 05.09 Fine Guidance System | 29 | 28 | 28 | 28 |
| 05.10 Starshade Petals and Disk | 170 | 258 | 258 | 258 |
| 06 Spacecraft System + 10 ATLO | 1,724 | 1,483 | 1,957 | 1,720 |
| 06.01 FS Mgmt / 06.02 FS SE | In 06.16 | In 06.16 | In 06.16 | In 06.16 |
| 06.16A Telescope Bus | 1,045 | 886 | 1,088 | 987 |
| 06.16B Starshade Bus | 680 | 597 | 868 | 733 |
| 07 MOS / 09 GDS | 85 | 109 | 220 | 165 |
| Subtotal (Ph B-D) | 4,363 | 4,312 | 4,965 | 4,639 |
| Reserves | 1,309 | 1,765 | 2,032 | 1,900 |
| Ph B-D w/ Reserves | 5,672 | 6,078 | 6,997 | 6,538 |
| LV (Telescope) | 650 | 650 | 650 | 650 |
| LV (Starshade) | 300 | 300 | 300 | 300 |
| Ph B-D w/ Reserves and LV | 6,622 | 7,028 | 7,947 | 7,488 |
| ESA Contributions | -565 | -565 | -565 | -565 |
| Ph B-D w/ Reserves, LV, and Contribution | 6,057 | 6,463 | 7,382 | 6,923 |
| | | | | |
| Operations | 400 | 291 | 291 | 291 |
| Subtotal (Ph E-F) | 400 | 291 | 291 | 291 |
| Reserves | 60 | 44 | 44 | 44 |
| Ph E-F w/ Reserves | 460 | 335 | 335 | 335 |
| Total Pre-Phase A-F | 6,786 | 7,067 | 7,987 | 7,528 |

# ICE Trend Line Analysis – OTS

*HabEx Study – OTS ICE Analysis*

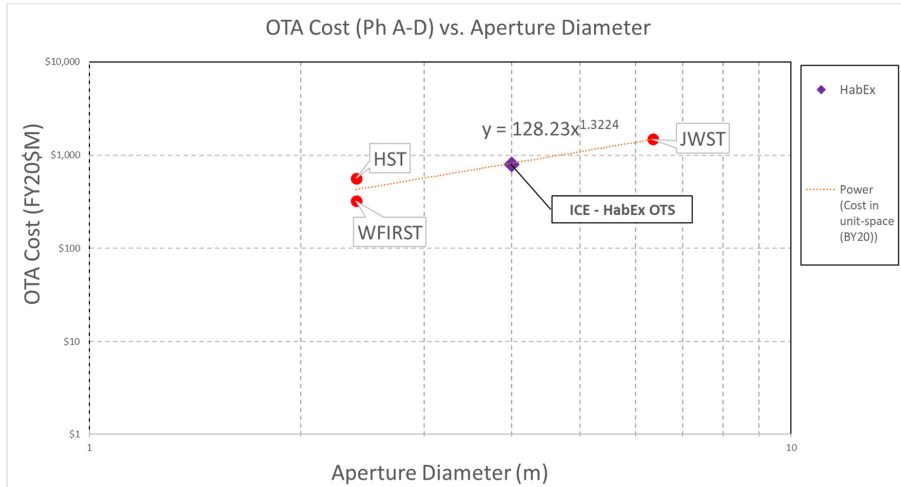

- Cost versus Aperture Diameter
  - Trend line based on HST, WFIRST, and JWST
    - ICE - HabEx OTS @ 4m: $802M FY20

 The cost information contained in this document is of a budgetary and planning nature and is intended for informational purposes only. It does not constitute a commitment on the part of JPL and/or Caltech. 14

# ICE S-Curve and Summary – FY20 $M
### HabEx Study

COST ESTIMATION AND PRICING SECTION

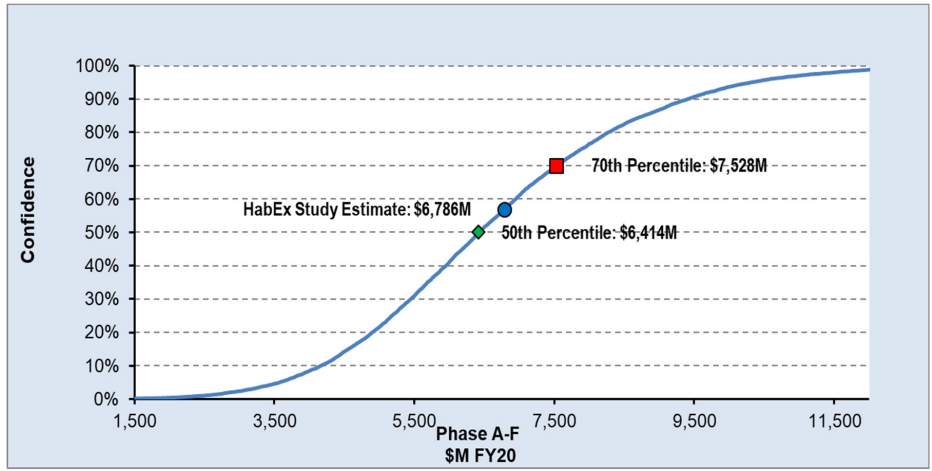

HabEx Phases A-F cost of $6,786M FY20 (with 30% UFE during development) plots on the 57th %-tile of the ICE S-Curve. To get to 70th %-tile ($7,528M), HabEx will need an additional $742M.

The cost information contained in this document is of a budgetary and planning nature and is intended for informational purposes only. It does not constitute a commitment on the part of JPL and/or Caltech

# HabEx Schedule Duration Comparison

## *HabEx Study - Phases A-D*

COST ESTIMATION AND PRICING SECTION

| Missions | Phase A | Phase B | Phase C | Phase D | Total Start-LRD |
|----------|---------|---------|---------|---------|-----------------|
| *HabEx* | *23* | *31* | *54* | *28* | *136* |
| Average | 15 | 23 | 32 | 42 | 111 |
| JWST | 21 | 20 | 25 | 132* | 198 |
| Chandra | 19 | 36 | 13 | 42 | 110 |
| WFIRST | 24 | 23 | 46 | 28 | 121 |
| Hubble | 4 | 17 | 32 | 98 | 151 |
| Kepler | 18 | 24 | 31 | 18 | 92 |
| Spitzer | 4 | 18 | 42 | 23 | 87 |

- Normalization Notes:
  - Phase D durations for all Missions was normalized using the start of Phase D through Launch (the duration between Launch and end of Phase D varied across this mission set)
  - Chandra Phase B duration has been normalized to 36 months based on several replans and scope changes
  - Hubble Space Telescope Phase D duration was not normalized due to insufficient understanding of the drivers for the 42 month slip, how much was driven by technical challenges versus the shuttle program stand down after the Challenger accident
  - JWST and WFIRST are based on current estimated durations
  - JWST Phase D duration is not included in the average calculation, the cause of the delays are not available to enable normalization

HabEx Phase A-D schedule is in alignment with historical analogies.  Phases C and D should be combined in order to compare to historical missions. The duration from start of Phase C to Launch of 82 months is robust compared to the historical mission average of 74 months.  The Payload and Spacecraft Integration and Test schedules provide adequate time to find and mitigate risks prior to delivery to System Integration and Test (SI&T).  The robust margins held by HabEx also protect against delays to SI&T.

The cost information contained in this document is of a budgetary and planning nature and is intended for informational purposes only. It does not constitute a commitment on the part of JPL and/or Caltech.

# ICE/ISA Assessment Summary

## *HabEx Study - Conclusions*

COST ESTIMATION AND PRICING SECTION

- HabEx study cost estimate is 4% to 15% below the ICE range
  - Study plots at 57th %-tile on the ICE S-curve and is in line with missions at this stage in life cycle
  - Cost deltas in the Optical Telescope System and Starshade Petals and Disks
    - Size of telescope is significantly outside of the model range
    - Limited number of analogous missions
- Schedule has adequate duration compared to history
  - Critical path funded schedule margin exceeds institutional requirements by ~6 months on the Telescope System and ~3 months on the Starshade System

**Overall Assessment:** HabEx cost and schedule estimates are reasonable given a Pre-Phase A level of maturity

# H GLOSSARY

| Detecting and Characterizing Exoplanets | |
|---|---|
| **Term** | **Definition** |
| 2D EEC Zone | See *Exo-Earth candidates (EEC)* |
| Arago spot | Also known as a Poisson spot or Fresnel bright spot, is the bright spot appearing in a circular object's shadow due to Fresnel diffraction. The HabEx Starshade Instrument senses the shape of the Arago spot to guide and lock formation with the external starshade occulter. |
| Coronagraphy | Science observations enabled by blocking light from a bright source, e.g., star. |
| Coronagraph | In general, a device used to block light from a bright source. This report describes an *internal* coronagraph, the HabEx coronagraph. |
| "Digging the Dark Hole" | The process by which the HabEx coronagraph achieves the required raw contrast by modulating its deformable mirrors and observing results using the low order wavefront sensor. |
| Exo-Earth Candidates (EEC) | Exoplanets located in the 2-Dimensional EEC zone defined in the planet radius vs star-planet separation space as:<br>• $0.8\ R_\oplus\ a^{-0.5} \leq r \leq 1.4\ R_\oplus$, where r is the radius of the planet, $a$ is the semi-major axis of the planet, $R_\oplus$ is the radius of the Earth; and,<br>• $0.95\ \text{AU} \leq a \leq 1.67\ \text{AU}$ for a sunlike star, corresponding to the conservative HZ. |
| Inner Working Angle at 50% Transmission ($IWA_{0.5}$) | For the coronagraph, $IWA_{0.5}$ is the angular separation from the star at which system transmission reaches 50% of the maximum off-axis transmission.<br><br>For the starshade, $IWA_{0.5}$ is the angular separation from the star at which system transmission reaches 50% of the transmission without the starshade. The $IWA_{0.5}$ is slightly smaller than the angular separation between the starshade petal tip and the starshade center ($IWA_{tip}$), as seen from the telescope. For the HabEx starshade design, the $IWA_{0.5} = 0.83 \times IWA_{tip.}$<br><br>While planets can still be detected interior to the $IWA_{0.5}$, albeit with reduced transmission, the $IWA_{0.5}$ can be used as a rough proxy for the closest separation from the star at which a planet may be detected. |

| Detecting and Characterizing Exoplanets | |
| --- | --- |
| Term | Definition |
| Ocean Glint | In exoplanet direct observation, sunlight reflected from exo-oceans creates observable ocean glint. Ocean glint results in a substantial ($\geq$ 50%) increase in apparent albedo compared to the case where only the effect of clouds is taken into account. Ocean glint also causes an apparent reddening of the planet. |
| Outer Working Angle (OWA) | The outer working angle specifies the maximum angular separation between a star and exoplanet, where the exoplanet will be observable. For a coronagraph, the size of the high contrast (dark hole) region is limited by the number of actuators per deformable mirror. For a starshade, it is limited by the size of the detector. |
| Planet-to-Star Flux Ratio | Planet-to-star flux ratio is an intrinsic property of the astronomical source. It can be computed at a single wavelength or over some spectral range. |
| Raw (Instrument) Contrast | Raw contrast is an optical instrument property. It measures the ability of a starlight suppression instrument to reject on-axis starlight while letting off-axis light go through with high efficiency. At any location in the final focal plane, the raw contrast (a unitless number $\ll$ 1) is defined as the ratio of the residual starlight signal detected within a spatial region centered at that location with the star on the optical axis, to the starlight signal that would be detected within the same spatial region if the star was centered at that location. |
| Starshade | The starshade occulter is an external coronagraph that is used in six of the nine architectures summarized in *Chapter 10*. In this report, the term "starshade" is used to describe both the starshade external occulter and starshade flight system, where the former is the payload of the latter. |
| Suppression (starshade) | Suppression is a metric defining the quality of *starlight* suppression. It is the ratio of the integrated light in the telescope pupil with the starshade in place relative to the integrated light in the telescope pupil without the starshade. When this residual light is focused inside the telescope it will usually result in regions of higher and lower raw contrast. |

| HabEx Systems and Their Design | |
|---|---|
| **Term** | **Definition** |
| Aperture Array | *See Microshutter Array* |
| Colloidal Microthrusters | Microthrusters that operate by electrostatically accelerating ionized colloidal liquid propellant, permitting low disturbance and high specific impulse operation. |
| Figure Control | The practice of changing telescope figure to meet optical requirements. |
| Formation Flight | Coordinated spaceflight of multiple spacecraft. In this report, it refers to the maintaining starshade attitude and position with respect to telescope line-of-sight. |
| Halo Orbit | Orbits at Lagrange Points-1 and -2 that are inherently unstable, but require small adjustments (~10 m/s per year) to maintain. |
| Margin | For this report, margin is defined as \|Requirement − CBE\|/min(Requirement, CBE). |
| Metrology | In general, the science of measurement. In this report, metrology refers to measurements that are used to establish optical quality by the optical telescope assembly. |
| Microshutter Array | Also described as an "Aperture Array", a plane positioned ahead of the focal plane detector in the light path where shutters can be actuated to open or close over portions of the focal plane. |
| Phased Array Antenna | A planar antenna comprised of multiple emitting elements, individually controlled to be differently phased. By phasing elements individually, different gain patterns can be achieved to "steer the beam" without requiring antenna articulation. |
| Segmented Telescope | In contrast to telescopes with a monolithic primary mirror, a segmented telescope uses a primary mirror that is comprised of multiple segments. |
| Servicing | In this report, servicing refers to operations to refuel and/or exchange modules between a servicing spacecraft and the HabEx flight systems. |
| Structural, Thermal, and OPtical (STOP) Analysis | STOP analysis verifies that HabEx performance meets requirements by modeling structural, thermal, and optical parameters over the range of HabEx operational conditions. |

# I ACRONYMS

| | | | |
|---|---|---|---|
| AC | average completeness | BAO | baryon acoustic oscillation |
| ACE | Ames Coronagraph Experiment | BFR | Big Falcon Rocket |
| ACS | Advanced Camera for Surveys | BMC | Boston Micromachines Corporation |
| ACS | attitude control system | | |
| ADC | Analog-to-digital converter | BOSS | Big Occulting Steerable Satellite |
| ADCS | attitude determination and control subsystem | CARMENES | Calar Alto high-Resolution search for M dwarfs with Exoearths with Near-infrared and optical Echelle Spectrographs |
| ADI | angular differential imaging | | |
| AE | Archean Earth | | |
| AFRL | Air Force Research Laboratory | | |
| AFT | allowable flight temperature | CAS | Common Attach System |
| AFTA | Astrophysics Focused Telescope Asset | CAST | Control Analysis Simulation Testbed |
| AG | geometric albedo | CATE | Cost Appraisal and Technical Evaluation |
| AGN | active galactic nuclei | | |
| ALD | atomic layer deposition | CBE | current best estimate |
| ALMA | Atacama Large Millimeter/ submillimeter Array | CC | cumulative completeness |
| | | CCD | charge coupled device |
| ALPS | Advanced Large Precision Structure | CDH | command and data handling |
| | | CDM | cold dark matter |
| AMS | Aft Metering Structure | CEL | collisionally excited emission line |
| AMTD | Advanced Mirror Technology Development | | |
| | | CER | cost estimating relationship |
| AMTD-2 | Advanced Mirror Technology Development Phase 2 | CG | center-of-gravity |
| | | CG | coronagraph |
| AO | adaptive optics | CGH | computer generated hologram |
| AOM | acousto-optic modulator | CGI | (WFIRST) coronagraph instrument |
| APD | (NASA) Astrophysics Division | | |
| APD | avalanche photodiode | CGM | circumgalactic matter |
| APLC | apodized pupil Lyot coronagraph | CMB | cosmic microwave background |
| | | CMOS | complementary metal-oxide semiconductor |
| ARIEL | Atmospheric Remote-sensing Infrared Exoplanet Large-survey | | |
| | | CMT | colloidal microthruster |
| | | CNES | Centre National d'Etudes Spatiales |
| ASIC | application specific integrated circuit | | |
| | | CNO | carbon-nitrogen-oxygen |
| ASMCS | Astrophysics Strategic Mission Concept Study | CONOPS | concept of operations |
| | | COR | cosmic origins |
| ATLAST | Advanced Technology Large Aperture Space Telescope | COS | Cosmic Origins Spectrograph |
| | | CPCM | center of pressure/center of mass |
| ATV | Automated Transfer Vehicle | | |
| AU | astronomical unit | CRIRES | CRyogenic high-resolution InfraRed Echelle Spectrograph |
| AUI | Associated Universities, Inc. | | |
| AYO | altruistic yield optimization | | |
| CSA | Canadian Space Agency | ExoTAC | Exoplanet Technical Analysis Committee |
| CSV | Cooperative Servicing Valve | | |
| CTE | coefficient of thermal expansion | FEA | Finite element analysis |
| DDOR | Delta Differential One-way Ranging | FEM | Finite element model |
| | | FGC | Formation guidance channel |
| DI | directly imaged | FGS | Fine guidance system |
| DLR | Deutschen Zentrums für Luft- und Raumfahrt | FINESSE | Fast INfrared Exoplanet Spectroscopic Survey Explorer |
| DM | deformable mirror | FIREBALL | Faint Intergalactic-medium Redshifted Emission Balloon |
| DOF | degree-of-freedom | | |
| DQE | Differential Quantum Efficiency | FoM | figure of merit |
| DRM | design reference mission | FOV | field of view |
| DSN | Deep Space Network | FPA | focal plane array |
| ee | encircled energy | FSM | Fine steering mirror |
| EEC | exo-Earth candidate. See *Appendix H* for definition. | FSW | flight software |
| | | FUSE | Far Ultraviolet Spectroscopic Explorer |
| E-ELT | European Extremely Large Telescope | FUV | far ultraviolet |
| EGSE | electrical ground support equipment | FWHM | full width at half maximum |
| | | FY | fiscal year |
| ELT | Extremely Large Telescope | GaAs | gallium arsenide |
| ELZM | Extreme Lightweight Zerodur Mirrors | GALEX | Galaxy Evolution Explorer |
| | | GCA | glass capillary array |
| EMCCD | electron multiplying charge coupled device | G-CLEF | GMT-Consortium Large Earth Finder |
| EMI | electromagnetic interference | GDS | ground data system |
| EOR | Epoch of Reionization | GMT | Giant Magellan Telescope |
| EPRV | Extremely Precise Radial Velocity | GN&C | guidance, navigation, and control |
| eROSITA | extended Roentgen Survey with an Imaging Telescope Array | GO | Guest Observer |
| | | GP-B | Gravity Probe B |
| E-S | Earth-Sun | GPI | Gemini Planet Imager |
| ESA | European Space Agency | GSE | ground support equipment |
| ESO | European Southern Observatory | GSFC | Goddard Space Flight Center |
| ESPRESSO | Echelle SPectrograph for Rocky Exoplanet and Stable Spectroscopic Observations | Gyr/Gy | gigayear |
| | | HabEx | Habitable Exoplanet Imaging Mission |
| EXCEDE | Exoplanetary Circumstellar Environments and Disk Explorer | HCG | HabEx Coronagraph |
| | | HCIT | High Contrast Imaging Testbed |
| ExEP | Exoplanet Exploration Program | HCST | High Contrast High-Resolution Spectroscopy for Segmented Telescopes Testbed |
| Exo-C | Exo-Coronagraph | | |
| EXOPAG | Exoplanet Program Advisory Group | HD | Henry Draper (Catalogue) |
| | | HEC | HabEx Earth Candidate |
| Exo-S | Exo-Starshade | HET | Hall Effect Thruster |
| EXOSIMS | Exoplanet Open Source Imaging Mission Simulator | HGA | high gain antenna |
| | | HgCdTe | mercury cadmium telluride |
| | | HIP | Hipparcos Catalog |
| HIRES | (E-ELT) High Resolution Spectrograph | LBT | Large Binocular Telescope |
| HITRAN | high-resolution transmission molecular absorption (database) | LBTI | Large Binocular Telescope Interferometer |
| | | LBV | Luminous Blue Variable |
| | | LCP | liquid crystal polymer |
| HLC | hybrid Lyot coronagraph | LEO | low Earth orbit |
| HOI | halo orbit insertion | LGA | low-gain antenna |
| HOSTS | Hunt for Observable Signatures of Terrestrial Systems | LIFE | Large Interferometry For Exoplanets |
| HPSC | High Performance Space Computing | LISA | Laser Interferometer Space Antenna |
| HST | Hubble Space Telescope | LM | Lockheed Martin |
| HWC | HabEx Workhorse Camera | LMAPD | linear mode avalanche photodiode |
| HZ | habitable zone | | |
| I&T | integration and test | LMC | Large Magellanic Cloud |
| ICE | independent cost estimate | LOS | line of sight |
| ICM | institutional cost model | LOWFS(C) | low order wavefront sensing (and control) |
| ICM | intracluster medium | | |
| ICSO | International Conference on Space Optics | LSST | Large Synoptic Survey Telescope |
| IFS | integral field spectrograph | LTF | low temperature fusion |
| IGM | intergalactic medium | LUVOIR | Large UV Optical Infrared Surveyor |
| IMBH | intermediate mass black hole | | |
| IMF | initial mass function | LV | launch vehicle |
| IMU | inertial measurement unit | LVA | Launch Vehicle Adaptor |
| IPAC | Infrared Processing Analysis Center | LVDS | low-voltage differential signaling |
| | | LyC | Lyman continuum |
| IPAG | Institut de Planétologie et d'Astrophysique de Grenoble | mas | milliarcsecond |
| | | MBE | molecular beam epitaxy |
| IR | infrared | MCMC | Markov Chain Monte Carlo |
| IRAC | Infrared Array Camera | MCP | microchannel plate |
| IRAS | Infrared Astronomical Satellite | MEL | Master Equipment List |
| ISIM | integrated science instrument module | MEMS | microelectromechanical systems |
| | | MET | laser METrology and control |
| ISL | interspacecraft link | MEV | maximum expected value |
| ISM | interstellar medium | MGSE | mechanical ground support equipment |
| ISS | International Space Station | | |
| IUE | International Ultraviolet Explorer | MIB | metagalactic ionizing background |
| IWA | inner working angle. See *Appendix H* for definitions of $IWA_{0.5}$ of and $IWA_{tip}$. | MIDEX | Mid-sized Explorer |
| | | MIT | Massachusetts Institute of Technology |
| JAXA | Japan Aerospace Exploration Agency | MLA | microlens array |
| | | MLI | multilayer insulation |
| JPL | Jet Propulsion Laboratory | MOC | Mission Operations Center |
| JWST | James Webb Space Telescope | MOS | mission operations system |
| KT | Kepner-Tregoe | MOS | multi-object spectroscopy |
| L2 | Earth-Sun Lagrange Point-2 | | |
| MOVPE | metalorganic vapor phase epitaxy | PIAACMC | phase-induced amplitude apodization complex mask coronagraph |
| MPC | multi-point constraint | | |
| MPV | maximum possible value | PICTURE-B | Planet Imaging Coronagraphic Technology Using a Reconfigurable Experimental Base |
| MSA | microshutter array | | |
| MSFC | Marshall Space Flight Center | | |
| MSWC | Multi-Star Wavefront Control | | |
| MTM | Mission Traceability Matrix | PID | proportional-integral-derivative |
| MUF | model uncertainty factor | PIP | Payload Interface Plate |
| Myr/My | million years | PLATO | PLAnetary Transits and Oscillations of stars |
| NAOJ | National Astronomical Observatory of Japan | | |
| | | PLC | planar lightwave circuit |
| NASA | National Aeronautics and Space Administration | PLF | payload fairing |
| | | PLUS | Petal Launch Restraint & Unfurler Subsystem |
| NAV | Navigation | | |
| near-IR | near-infrared | PM | primary mirror |
| NGAS | Northrop Grumman Aerospace Systems | PMA | primary mirror assembly |
| | | PMN | lead-magnesium-niobate |
| NGMSA | next generation microshutter array | PROBA | PRoject for OnBoard Autonomy |
| NICM | NASA Instrument Cost Model | PRT | platinum resistance thermometer |
| NICMOS | Near Infrared Camera and Multi-Object Spectrometer | PRV | precursor radial velocity |
| | | PSD | power spectral density |
| NIR | near-infrared | PSF | point spread function |
| NIRCam | Near Infrared Camera | PTCS | Predictive Thermal Control Study |
| NPRO | Nd:YAG non-planar ring oscillator | | |
| | | QE | quantum efficiency |
| NRAO | National Radio Astronomy Observatory | QSO | quasi-stellar object |
| | | R&D | research and development |
| NRC | National Research Council | RBM | rigid body motion |
| NSF | National Science Foundation | RCS | reaction control system |
| NUV | near ultraviolet | RDI | reference differential imaging |
| NWNH | New Worlds New Horizons | RECONS | REsearch Consortium On Nearby Stars |
| OAP | off-axis parabola | | |
| OFTI | Orbits for the Impatient | RF | radio frequency |
| OMC | occulting mask coronagraph | RFP | Request for Proposal |
| OTA | optical telescope assembly | RMS | radio-millimeter-submillimeter |
| OWA | outer working angle | RMS | root sum squared |
| PAF | payload adapter fixture | ROSA | roll-out solar array |
| pc | parsec | ROSES | Research Opportunities in Space and Earth Sciences |
| PC | photonics crystal | | |
| PCOS | Physics of the Cosmos | RSG | Red Super Giant |
| PDR | Preliminary Design Review | RV | radial velocity |
| PE | Proterozoic Earth | RWA | reaction wheel assembly |
| PEL | Power Equipment List | RY | real year |
| PIAA | phase-induced amplitude apodization | S/C | spacecraft |
| | | S5 | Starshade to Technology 5 project |
| SAG | Study Analysis Group | SSWG | Starshade Readiness Working Group |
| SAT | Strategic Astrophysics Technology | ST7-DRS | Space Technology 7 Disturbance Reduction System |
| SBIR | Small Business Innovation Research | STDT | Science and Technology Definition Team |
| SCExAO | Subaru Coronagraphic Extreme Adaptive Optics | STEM | Storable Tubular Extendible Member |
| SCM | Soft Capture Mechanism | STIS | Space Telescope Imaging Spectrograph |
| SDSS | Sloan Digital Sky Survey | | |
| SED | spectral energy distribution | STM | Science Traceability Matrix |
| SEM | scanning electron microscope | STOP | structural thermal and optical performance |
| SEP | solar electric propulsion | | |
| SFE | surface figure error | STORRM | Sensor Test for the Orion Relav Risk Mitigation |
| SHHLLV | super heavy-lift launch vehicle | | |
| SI&T | System Integration and Test | SWOT | Surface Water and Ocean Topography |
| SIDM | self-interacting dark matter | | |
| SIMBAD | Set of Identifications, Measurements, and Bibliography for Astronomical Data | TCM | trajectory correction maneuver |
| | | TDEM | Technology Development for Exoplanet Missions |
| SIR | System Integration Review | TDP | Technology Development Plan |
| SLATE | Starshade Lateral Alignment Testbed | TESS | Transiting Exoplanet Survey Satellite |
| SLS | Space Launch System | TGAS | Tycho-Gaia astrometric solution |
| SM | secondary mirror | THEIA | Telescope for Habitable Exoplanets and Interstellar/ Intergalactic Astronomy |
| SMA | secondary mirror assembly | | |
| SMC | Small Magellanic Cloud | | |
| SN | supernova | TJ | Triple Junction |
| SNe | supernovae | TM | tertiary mirror |
| SNR | signal-to-noise ratio | TMA | tertiary mirror assembly |
| SNWC | Super-Nyquist Wavefront Control | TMA | three-mirror anastigmat |
| | | TMT | Thirty Meter Telescope |
| SOA | state of the art | TPF | Terrestrial Planet Finder |
| SOC | Science Operations Center | TPF-I | Terrestrial Planet Finder Interferometer |
| SOFIA | Stratospheric Observatory for Infrared Astronomy | | |
| | | TRACE | Technical, Risk, and Cost Evaluation |
| SOP | state-of-the-practice | | |
| SP | shaped pupil | TRL | Technology Readiness Level |
| SPECULOOS | Search for habitable Planets EClipsing ULtra-cOOl Stars | TSI | transit stellar intensity |
| | | TSP | traveling salesman problem |
| SPHERE | (VLT) Spectro-Polarimetric High-contrast Exoplanet Research (instrument) | TWTA | Traveling Wave Tube Amplifier |
| | | ULA | United Launch Alliance |
| | | ULE | ultra low expansion |
| SRON | Netherlands Institute for Space Research | UMBRAS | Umbral Missions Blocking Radiating Astronomical Sources |
| SS | starshade | | |
| SSI | starshade imager | UST | universal space transponder |
| SSI | (HabEx) Starshade Instrument | UTAS | UTC Aersopace Systems |
| UTC | United Technologies Corp. | WFPC2 | Wide-Field Planetary Camera 2 |
| UV | ultraviolet | WFPC3 | Wide Field Camera 3 |
| UVOIR | near-UV to far-infrared | WFSC | wavefront sensing and control |
| UVS | ultraviolet spectrograph | WG | Working Group |
| VC | vortex coronagraph | WHC | Workhorse Camera |
| Vis | visible | WISE | Wide-field Infrared Survey Explorer |
| VLT | Very Large Telescope | | |
| VV | vector vortex | WMAP | Wilkinson Microwave Anisotropy Probe |
| VVC | vector vortex coronagraph | | |
| WBS | Work Breakdown Structure | XRCF | X-ray and Cryogenic Facility |
| WC | wavefront control | $z$ | redshift (value) |
| WFE | wavefront error | ZWFS | Zernike wavefront sensor |
| WFIRST | Wide-Field Infrared Survey Telescope | | |

# J  REFERENCES

* There are no endnotes/citations in the *Executive Summary*, *Chapter 13*, and *Appendices E–I*.

## Chapter 1

## Chapter 2

## Chapter 3

## Chapter 4

## Chapter 5

## Chapter 7

## Chapter 8

## Chapter 10

# Chapter 11

## Chapter 12

## Appendix B

## Appendix D